\documentclass[graybox,envcountchap]{svmono}

\usepackage{mathptmx}
\usepackage{helvet}
\usepackage{courier}
\usepackage{type1cm}        

\usepackage{makeidx}         
\usepackage{graphicx}        
\usepackage{multicol}        
\usepackage[bottom]{footmisc}

\usepackage{fancyhdr}
\usepackage{eucal}
\usepackage{amsmath}
\usepackage{mathrsfs}
\usepackage{amsfonts}
\usepackage{amssymb}
\usepackage{amscd}
\usepackage{bbm}
\usepackage{graphicx}
\usepackage{graphics}
\usepackage{latexsym}
\usepackage{color}
\usepackage{float}
\usepackage{enumerate}
\usepackage{calc}
\usepackage[T1]{fontenc} 

\usepackage{bm}

\usepackage{pifont}
\usepackage[caption=false]{subfig}

\newcommand{\ud}{\mathrm{d}}

\newcommand{\xx}{\mathbf{x}}
\newcommand{\yy}{\mathbf{y}}
\newcommand{\zz}{\mathbf{z}}
\newcommand{\pp}{\mathbf{p}}
\newcommand{\qq}{\mathbf{q}}

\newcommand{\ii}{\mathrm{i}}
\newcommand{\cH}{\mathcal{H}}
\newcommand{\ran}{\mathrm{ran}}

\newcommand{\op}{
  \mathop{
    \vphantom{\bigoplus} 
    \mathchoice
      {\vcenter{\hbox{\resizebox{\widthof{$\displaystyle\bigoplus$}}{!}{$\boxplus$}}}}
      {\vcenter{\hbox{\resizebox{\widthof{$\bigoplus$}}{!}{$\boxplus$}}}}
      {\vcenter{\hbox{\resizebox{\widthof{$\scriptstyle\oplus$}}{!}{$\boxplus$}}}}
      {\vcenter{\hbox{\resizebox{\widthof{$\scriptscriptstyle\oplus$}}{!}{$\boxplus$}}}}
  }\displaylimits 
}

\makeindex             

\begin{document}

\author{Matteo Gallone, Alessandro Michelangeli}
\title{Self-adjoint extension schemes and modern applications to quantum Hamiltonians}

\date{23 February 2023}
\maketitle

\frontmatter

%
%

\foreword

The ways mathematics and its applications evolve are manifold and difficult to describe. Often progress has been achieved by finding models that are both simple enough to permit a thorough, detailed mathematical analysis and are in a sense emblematic for what can happen in more general and realistic situations.

In classical mechanics this is the case, for instance, of the pendulum, the harmonic oscillator, or two point particles moving under Newton gravitational force. In all such examples the equations of motion can be solved ``explicitly'', getting insight into what might happen in more general situations. Already in the case of three point particles with Newton's forces, as one knows, more complicated solutions may arise, even the ``chaotic'' ones described originally by H.~Poincar\'e. Classical mechanics managed to obtain nevertheless considerable insights into this and even more complicated problems by different methods, qualitative and quantitative. One major technique is the use of perturbation schemes by splitting the total system to be investigated into a ``main part'', manageable by a relatively simple study, and a ``perturbation'', considered to be small in a certain sense.

A similar circumstance arises in quantum mechanics. As well known, the dynamics is here given by the one-parameter unitary groups generated by Hamiltonians realised as (linear) self-adjoint operators acting in a (complex) Hilbert space. In Schr\"{o}dinger's framework the Hilbert space for the $N$-particle problem in three dimensions is the Lebesgue space of all square integrable, complex-valued functions over $\mathbb{R}^{3N}$, and the Hamiltonian $H$ has essentially the form of a kinetic energy operator $H_0$ (the prototype is the negative of the Laplace operator $\Delta$ on $\mathbb{R}^{3N}$) plus an ``interaction term'' $V$, the quantum version of the classical $N$-particle problem (namely, the sum of two-body Newton potentials). Since both the ``kinetic'' and the ``interaction'' part are unbounded operators, a first issue is to identify a non-trivial domain in the Hilbert space on which the sum $H_0+V$ is defined. A second task, often interrelated with the first, is to associate to the sum a \emph{self-adjoint} operator $H$ extending $H_0+V$, and generating the unitary group giving the dynamics. Such problems, in particular the latter, are not present in the classical mechanical case, and solving them is not at all simple, as we shall discuss below. Once the self-adjoint realisation of the quantum Hamiltonian is identified, the question of computing relevant quantities arises. Here again one is confronted with a rather complex task and perturbation theoretical methods had to be devised.

The treatment of these basic problems was a main motivation for the development of a large part of the theory of linear operators in Hilbert spaces. Historically, this subject was founded by D.~Hilbert (and E.~Schmidt and F.~Riesz in the second half of the 1910's), in relation to the theory of integral equations, and, more abstractly, by J.~von Neumann (1929), in connection with the development of a rigorous approach to quantum mechanics. Perturbation theory in Hilbert spaces had its origin in the works by Lord Rayleigh (1927) and E.~Schr\"{o}dinger (1926-1928) and much further developed mathematically especially by F.~Rellich (1937-1944) and K.~O.~Friedrichs (1934, 1938).

There are two main sources of this field. One is the theory of second order differential operators, going back, in the case of singular ordinary differential operators and equations (Sturm-Liouville problems),s to the work of H.~Weyl (1910). Here the special dependence on the boundary conditions and singularities of the coefficients was already stressed. Another source is to be found in consideration with problems of quantum field theory, as they appeared in works aimed at building ``models'' of quantum fields (works by V.~Fock, F.~A.~Berezin, L.~Faddeev, R.~Minlos in the Soviet Union, and K.~O.~Friedrichs, I.~Segal, and others in the West). The main point here is that the natural splitting of a Hamiltonian in a ``simple part'' and an ``interaction part'' is not possible, staying on a given domain in a given Hilbert space, but rather the space itself and the interaction part should emerge as a suitable limit of regularised problems. Such ``truly singular perturbations'' are kind of hidden in a heuristic splitting of the Hamiltonian into a free and an interaction part (no bound of the perturbation part in terms of the unperturbed one is possible).

The mathematical framework for these ``truly singular perturbation'' phenomena are part of the theory of self-adjoint extensions, which is the main topic of the present book by Gallone and Michelangeli. Its purpose is to present in a unified way approaches that have been developed mainly by J.~von Neumann on the one hand, and on the other by the Soviet Union school of M.~\v{S}.~Birman, M.~G.~Kre{\u\i}n, and M.~I.~Vi\v{s}ik. For the case of self-adjoint operators in Hilbert space it represents a long waited synthesis along the way of those provided before for more regular situations by classical books such as those of T.~Kato in the early 1950's, and others including the series by M.~Reed and B.~Simon (1972-1979) and, in a more abstract setting, the series by N.~Dunford and J.~T.~Schwartz (1957-1971). Concerning the case of Schr\"{o}dinger operators, the present book goes also beyond the method of quadratic forms developed particularly by B.~Simon (1971).

 In fact, the book by Gallone and Michelangeli manages to provide a combined discussion of both the general ``abstract'' von Neumann approach of self-adjoint extensions of linear (not necessarily semi-bounded) symmetric operators in Hilbert space, and the more ``concrete'' Kre{\u\i}n-Vi\v{s}ik-Birman approach for lower semi-bounded symmetric operators, comparing them, and stressing their differences and the advantages in using them in a complementary way. Besides, the book offers striking examples of important applications of these theories to long standing open problems developed essentially by the confluence of methods developed especially for Schr\"{o}dinger-type operators on one hand by the Soviet Union school (F.~A.~Berezin, L.~Faddeev, R.~Minlos, and their followers) and by the Italian school (G.~Dell'Antonio, R.~Figari, A.~Teta, and their followers, notably including newer developments by A.~Michelangeli and his co-workers). Among other results, this has led, through recent works that are discussed in this book, to the solution of long standing conjectures concerning the quantum mechanical three-body problem with two-body translation-invariant point interactions (such three-body problems play an important role in contemporary nuclear and cold atom physics).

Let me comment on the single parts of this beautiful monograph.

The preface already describes well in precise and concise terms the subject matter and the problems to be presented and handled, putting them in a proper context.

The first part starts with a first chapter presenting a very useful and clear reminder of basic facts about Hilbert spaces, and operator theory on them, in particular symmetric and self-adjoint operators. As opposite to all other chapters, where very detailed proofs are presented and commented upon, the focus of Chapter \ref{chaper-preliminaries} is in results and illustrations, rather than accurate demonstrations, placing the former in the appropriate context and with motivations. The presentation goes from basic general facts and discussions about linear operators and their spectrum, to symmetric and self-adjoint operators, spectral theory, perturbations, quadratic forms. Of special interest is the concise survey of Weyl's limit-point limit-circle analysis (for Schr\"{o}dinger and Dirac operators).

In Chapter \ref{chaper-extension-schemes} the theoretical constructions of the von Neumann and the Kre{\u\i}n-Vi\v{s}ik-Birman extension schemes are beautifully described and compared. Important topics such as the Friedrichs extension are justly given much weight, due to their importance in many interesting applications. von Neumann's approach to the extension of symmetric operators via Cayley's transform is presented, and its relation with the Dunford-Schwartz framework in terms of abstract boundary problems is discussed. Then Kre{\u\i}n's extension theory of \emph{positive} symmetric operators is very clearly delineated, together with its variant due to T.~Ando and K.~Nishio. Then the Vi\v{s}ik-Birman parametrisation of self-adjoint extensions is analysed and its consequences, including spectral results, are discussed in depth. This presentation is based on original contributions, particularly by the authors and co-workers, which constitute a major achievement in this area in recent years, including the detailed study of relations between the von Neumann and the  Kre{\u\i}n-Vi\v{s}ik-Birman approach. This by itself should make the book to a standard reference for any further work in the area of singular perturbation theory.

My strong statement is reinforced by the content on the second part, on which I shall briefly comment. This second part is dedicated to applications and illustrations of the general extension schemes brilliantly presented in the first part. It consists of four chapters, all dealing with the treatment of singular partial differential operators that arise in quantum mechanics. They also serve as illustrations and motivations for the use of the same methods in other areas (including, among others, acoustics, electromagnetism, stochastic analysis).

Chapter \ref{chapter-Hydrogenoid} treats spectral problems arising in connection with ``Hydrogenoid Hamiltonians'', thus describing an atom with atomic number $Z>0$, perturbed by a potential concentrated at the origin. The physical literature is first recalled, including Sommerfeld's analysis of the influence of relativistic effects on the spectrum. A rigorous formulation of the problem on the basis of the extension theory presented in Chapter \ref{chaper-extension-schemes} is provided. The detailed results in previous works which made use of von Neumann's extension theory are summarised in Theorem \ref{thm:3Dclass}. These results are re-proved through an alternative path by means of the Kre{\u\i}n-Vi\v{s}ik-Birman scheme, following recent works by the authors. These beautiful results are stated in Theorems \ref{thm:1Fried}-\ref{thm:EV_corrections} and proved in all details in Sections \ref{sec:Section_of_Classification} and \ref{sec:III-perturbations}. A complete discussion of the location of the eigenvalues is provided. 

Chapter \ref{chapter-Dirac-Coulomb} treats Dirac-Coulomb Hamiltonians for heavy nuclei. The study of such Hamiltonians started with the groundbreaking work by A.~Sommerfeld on the ``fine structure formula''. It has been known since the 1970's that for any sufficiently large number $Z$ of nucleons there exist a large number of possible Hamiltonians. This led to the discussion, and eventually the solution, of the problem of finding the distinguished self-adjoint realisation for which both the kinetic and the potential energy have finite expectations, but 
the determination of the location of the spectrum for each admissible Hamiltonian (not only the distinguished one) has only been possible through the work of the present authors, on the basis of their presentation of the Kre{\u\i}n-Vi\v{s}ik-Birman extension scheme. The chapter first summarises the corpus of classical results for the various regimes of magnitude of $Z$ (Theorem \ref{thm:recap}), and then focusses on the critical regime of large $Z$'s. In Sections \ref{sec:mainresults}-\ref{sec:IV-proofs} a whole class of self-adjoint extensions arising for large $Z$ is identified, and their spectra are characterised. The interest in these extension is related to the experimental possibility of testing the influence of point-like central perturbations (the self-adjoint extensions being characterised by local boundary conditions at the origin where the Coulomb potential has its source). This leads to quite explicit results, as well as estimates on the spectral gap. The chapter ends with an interesting discussion of the classical methods to derive Sommerfeld's ``fine structure formula'', which is valid for the distinguished extension, and its relation with the eigenvalues of the generic Dirac-Coulomb Hamiltonian.

In Chapter \ref{chapter-Grushin} a problem of geometric global analysis is treated. It concerns a quantum Schr\"{o}dinger particle moving on a degenerate Riemannian manifold, and its possible confinement in regions determined by the singularity of the metric (``Grushin structures'', in the nomenclature of sub-Riemannian geometry). This chapter presents extended knowledge on the family of self-adjoint realisations of the Laplace-Beltrami operator on such manifolds, obtained by the authors using the Kre{\u\i}n-Vi\v{s}ik-Birman theory. The cases of the Grushin(-type) planes, respectively, cylinders are analysed in detail, and the notions of geometric quantum confinement and transmission protocol are very clearly discussed, in relation to those self-adjoint realisations of the Laplace-Beltrami operator which induce either one or the other phenomenon. Systematic original results of Michelangeli and his co-workers are explained, and connections with partial outcomes obtained by previous methods are discussed. As a by-product, the whole theory on one-sided and two-sided fibres is included, which has interest per se and testifies an amazing richness of such a detailed and appealing analysis. Besides, relevant features on positive extensions are given (e.g., Theorem \ref{thm:positivity}), ground states and spectra are studied (Theorems \ref{thm:MainSpectral}-\ref{thm:GroundStateCH}), as well the corresponding scattering theory (Section \ref{sec:V-scattering}).

In Chapter \ref{chapter-bosonic_trimer} both the Kre{\u\i}n-Vi\v{s}ik-Birman and the von Neumann extension theory are applied to the long standing problem of finding a suitable mathematical language for models of multi-particle systems with very short range interactions like those treated in nuclear physics, or, nowadays, in ``unitary gases'' of cold atom physics. The pairing among particles (interpreted as nucleons or atoms at ultra-low temperature) is described by a sum of two-body translation-invariant interactions, ideally represented by a Dirac measure depending on the difference of the particle coordinates. These models have a long history, both in the physical and the mathematical literature. They are also prototypes of problems of singularities arising in relativistic quantum field theory, and in fact they conceptually emerged at the origin of many ideas of renormalisation theory (as illustrated in the works by L.~Faddeev and F.~A.~Berezin -- further references and discussion may be found in \cite{Albeverio-Figari-2018}, written by myself in collaboration with R.~Figari). More recently this kind of models have also formed a renaissance coming from the area of cold atom physics, related to, e.g., Einstein-Bose condensation physics and Feshbach resonance techniques in heteronuclear gaseous mixtures, that have developed new powerful experimental techniques. By these, certain theoretical results have found concrete applications and new precise methods have to be developed to better connect theoretical predictions with computational and experimental data. 
This final chapter beautifully describes the background for such investigations from a mathematical point of view, maintaining also close attention to the underlying physical modelling.

As it turns out, the ``Efimov effect'' (the formation, in systems of three or more particles, even with regular short-range two-body interaction, of an infinite number of bound states accumulating from below at the bottom of the continuous spectrum, when the two-body forces induce an effective long-range interaction between the three particles) is related to an effect discussed in nuclear physics as early as 1935 by L.~H.~Thomas, and this in turn is related to the study of point interactions as idealised models of short range interactions between particles, introduced both in connection with molecular physics by H.~Bethe and R.~Peierls in the same year, and further studied by E.~Fermi in 1936. L.~H.~Thomas argued that already for a three-particle system in three dimensions with translation-invariant two-body interactions infinitely many negative energy bound state can arise accumulating at minus infinity (the ``Thomas effect''). Further quantum-mechanical models with point interactions were studied in the 1950's, as part of the newly developing scattering theory in the Soviet Union by the work by physicists such as K.~A.~Ter-Martirosyan, G.~S.~Danilov, and others from the 1950's (and later, in the 1960's, by fundamental works by L.~Faddeev), and in the West under the influence of W.~Heisenberg's $S$-matrix approach by R.~Jost (1955), E.~Lieb and H.~Koppe (1959), and H.~M.~Nussenzveig (1961). A first more mathematical study of models with point interactions in three dimensions dates from the works by F.~A.~Berezin and L.~Faddeev (1961), and R.~Minlos and L.~Faddeev (1962). Let me also stress that the connections with renormalisation theory and quantum field theory were already present at this stage (e.g., by works of Berezin on the Lee model). These mathematical references used self-adjoint extension theory of symmetric operators (mainly the von Neumann scheme).

In atomic physics the computational discovery in 1970 of the Efimov effect has generated a lot of investigations that are still vigorously ongoing (see the recent reviews cited in Chapter \ref{chapter-bosonic_trimer}). By a heuristic argument one can in principle understand this effect by scaling the two-body potentials as to take the form of delta-like profiles, and hence relate the Efimov effect with the above-mentioned Thomas effect. 
Already for the problem of a single particle with point potential at the origin, with a Hamiltonian of the heuristic form $-\Delta+\lambda\delta(x)$ in $L^2(\mathbb{R}^d)$, one can see that for dimension $d\geqslant 2$ difficulties arise (for $d=1$ at least the Hamiltonian can be defined in the sense of quadratic forms). In fact, following the above-mentioned scaling reasoning, for $d=2$ or $d=3$ an approximation of $\delta$ by smooth functions does yield a limit as $\varepsilon\downarrow 0$ in the sense of norm resolvent convergence, provided that $\lambda$ above is taken to be $\varepsilon$-dependent and vanishes with a certain rate. The limit belongs to a one-parameter family of Hamiltonians $H_\alpha$ depending on a real parameter $\alpha$, which is essentially the inverse ``scattering length'' for the scattering associated with $H_\alpha$. Now, the family of the $H_\alpha$'s can also be obtained by the von Neumann or the Kre{\u\i}n-Vi\v{s}ik-Birman extension theory, starting from the restriction of $-\Delta$ to the dense domain $C^\infty_0(\mathbb{R}^d\setminus\{0\})$. Actually, the latter procedure also yielded the solution to the problem of defining Hamiltonians for the corresponding three-particle systems with pairwise two-body translation-invariant point interaction, with $L^2(\mathbb{R}^d)$ replaced by $L^2(\mathbb{R}^{3d})$ and the singular set $\{0\}$ replaced by the union $\Gamma$ of the hyperplanes of coincidence among pairs of particles, a problem where the approximation techniques are still not quite well developed. The exposition of this solution on the basis of a smart combination of both the von Neumann and the Kre{\u\i}n-Vi\v{s}ik-Birman extension schemes exposed in Chapter \ref{chaper-extension-schemes} makes the reading in Chapter \ref{chapter-bosonic_trimer} most interesting and passionate.

 Sections \ref{sec:generalextscheme} and \ref{sec:two-body-short-scale-sing} discuss the form of mathematically admissible Hamiltonians for the mentioned problem, hence self-adjoint realisations of the minimally defined free Hamiltonians restricted on functions that vanish away from the configurations of coincidence among particles. This is done by implementing the Kre{\u\i}n-Vi\v{s}ik-Birman extension scheme, upon characterising the closure, adjoint, deficiency space, and Friedrichs extension of the minimal operator. Among these, it becomes relevant to select those that have a more direct physical interpretation, i.e., discarding boundary conditions of self-adjointness that correspond to unwanted behaviour as particles come on top of each other. In Section \ref{sec:TMSextension-section} the physically meaningful short-scale Ter-Martirosyan Skornyakov condition is specified, based on which one has to select actual self-adjoint realisations of physical interest. It is shown in Theorem \ref{thm:globalTMSext} and its corollary that self-adjoint extensions of the minimal operator that do satisfy the Ter-Martirosyan Skornyakov condition in their domain are characterised in terms of an inverse scattering length real parameter $\alpha$ and their existence is boiled down to the technically simpler self-adjointness of certain Birman operators in certain negative Sobolev spaces over $\mathbb{R}^3$. This self-adjointness is then reduced, exploiting rotation invariance, to a corresponding self-adjointness problem in a definite angular momentum sector. 
 In Section \ref{sec:higherell} sectors of positive angular momenta are investigated in detail, whereas in Section \ref{sec:lzero} the most important and subtle sector or zero angular momentum is discussed. The analysis provided is particularly detailed, since it implies the lower unboundedness of the relevant Ter-Martirosyan-Skornyakov symmetric realisations of the formal Hamiltonian in the zero angular momentum sector. Special attention is dedicated (using the von Neumann scheme, and preparation from previous subsections) to characterise the family of self-adjoint extensions of the latter Ter-Martirosyan-Skornyakov symmetric operator (Proposition \ref{prop:Aellzero-selfadj}). In Section \ref{sec:canonicalmodel} the canonical model for a Hamiltonian in the case of three bosonic particles and for $\alpha=0$ (the ``unitary case'') is constructed (Theorem \ref{thm:H0beta}) and proved to satisfy the Ter-Martirosyan-Skornyakov condition, and not to be lower semi-bounded. Its negative spectrum consists of eigenvalues as made precise in Theorem \ref{thm:spectralanalysis} and Corollary \ref{cor:spectralanalysis}. These correspond to those expected from the Efimov computation analysis (Efimov effect) as well as from the analysis by L.~H.~Thomas (Thomas effect). The associated eigenfunctions are also exhibited. Section \ref{sec:canonicalmodel} closes with several remarks that explain how from the obtained results a deeper understanding is possible of the Thomas effect (Remark \ref{rem:eigenfunctions}), the location of the essential spectrum on the positive half-axis for all elements of the constructed one-parameter family of self-adjoint realisations (Remark \ref{rem:whereas}, following previous similar reasonings by S.~Becker, A.~Michelangeli, and A.~Ottolini), as well as the mechanism by which the self-adjoint extensions are sorted out from the family of Ter-Martirosyan-Skornyakov symmetric extensions (Remark \ref{rem:choicedomains}). Moreover, variants for the canonical models are analysed (Section \ref{sec:variants}), in the sense of changing the realisation in the sectors of positive angular momentum. Sections \ref{sec:illposed} and \ref{sec:regularisedmodels} present an instructive discussion of the relations between the chosen realisation of the Hamiltonian and others that have appeared in the mathematical and, implicitly, in the physical literature. Additionally, the connection to previous results on three-particle models with particles with Fermi or mixed exchange symmetry is scrutinised. Sections \ref{sec:MFregularisation} and \ref{sec:MinFadzero} are devoted to the Minlos-Faddeev regularisation (also discussed in works by G.~Flamand and by R.~H\o{}egh-Krohn, T.~T.~Wu, and myself): they show, in particular, that such a regularisation leads to a positive Hamiltonian, without negative eigenvalues. In Section \ref{sec:highenergycutoff} a Hamiltonian with high-energy regularisation is studied, related to the one analysed by means of quadratic forms by G.~Basti, R.~Figari, and A.~Teta. It is also shown to have essential spectrum $[0,+\infty)$, with no negative eigenvalues.

The book ends with two appendices. Appendix \ref{appendix_SA_in_physics} is concerned with the requirement of self-adjointness in the formalisation of quantum mechanics, based on a recent important discussion in a work by A.~Cintio and A.~Michelangeli. The accent is on mathematical requirements following directly from physics, letting self-adjointness, rather than symmetry alone, emerge as a natural property. The appendix is written in lucid style and provides a useful discussion with many examples clarifying what happens if some requirements are abandoned -- for instance, the non-uniqueness of the Schr\"{o}dinger dynamics when self-adjointness is not made explicit. In Appendix \ref{appendix_pedagogical_examples} basic references to pedagogical examples are provided; it is to be noted that many of these examples had a strong impact on research and on the development of more realistic case studies.

In conclusion let me state that the authors have managed to present at the same time a beautiful synthesis of the basic theories of self-adjoint extensions of symmetric, respectively, positive symmetric operators in Hilbert space, as well as striking applications to various mathematical and physical problems with singular perturbations of operators of Schr\"{o}dinger and Dirac type. For this, abstract Hilbert space methods are combined with concrete partial differential and advanced analytic methods. Among the applications are forefront problems in areas like global and spectral analysis (with applications in control theory, stochastic analysis, and mathematical physics), as well as present days few-body nuclear and cold atom physics.

\vspace{\baselineskip}
\begin{flushright}\noindent
Bonn, December 2021 \hfill {\it Sergio Albeverio}\\
\end{flushright}

%
%

\preface

 This monograph presents revised and enlarged materials from previous lecture notes of undergraduate and graduate courses and seminars delivered by both authors over the last years at the Ludwig Maximilian University (LMU) of Munich, the International School for Advanced Studies (SISSA) in Trieste, the University of Milan, and the University of Bonn, as well as in several international meetings and workshops. It also includes, into a unified perspective, revised parts of recent articles written by the two authors, separately or jointly, also together with other collaborators.

 The main subject is central both in abstract operator theory and in applications to quantum mechanics: to decide whether a given densely defined and symmetric operator on Hilbert space admits a \emph{unique} self-adjoint realisation, namely its operator closure, or whether instead it admits an \emph{infinite} multiplicity of distinct self-adjoint extensions, and in the latter case to classify them and characterise their main features (operator and quadratic form domains, spectrum, etc.)

 The intimate connection to physics is due to the fundamental circumstance that quantum observables, in particular the Hamiltonians of quantum systems, must be self-adjoint operators on Hilbert space, owing to general principles and interpretation requirements.

 This is at the same time a very classical, well established subject, and a territory of novel, modern applications. We indeed tried at our best to reflect such two-fold nature into this monograph as well.
 
 \begin{center}
\rule{0.5\textwidth}{.8pt}
\end{center}
 
 The self-adjoint extension problem, in its explicit setting, is almost as old as the very mathematical formulation of quantum mechanics as it emerged at the end of the 1920's: it is posed by J.~von Neumann in his seminal work announced in 1928 and published in 1930, which already included the general criteria of self-adjointness and classification scheme as we still know them nowadays, as well as the then-open question of the existence of the `highest' extension (the Friedrichs extension).\index{Friedrichs extension} 
 It has of course also ante litteram precursors, significantly in H.~Weyl's limit-point limit-circle theory\index{Weyl limit-point/limit-circle} for Sturm-Liouville operators,\index{Sturm-Liouville operators} which dates back to the early years of the 20th century.

 von Neumann's picture\index{von Neumann's extension theory} can be summarised as follows. A linear operator $S$ on Hilbert space, which is symmetric, namely whose expectations (the mean value with respect to states belonging to the domain of $S$) are all real, and whose domain is dense, is necessarily closable and admits the adjoint: both $\overline{S}$, the operator closure, and $S^*$, the adjoint, are closed extensions of $S$; in particular, $\overline{S}$ is the minimal one, namely the closed extension with smallest domain, and $S^*$ is the maximal one. For short,
 \[
  S\;\subset\;\overline{S}\;\subset\; S^*\,.
 \]
Generically, the domain of $S^*$ may be strictly larger than that of $\overline{S}$; when, instead,
\[
 \overline{S}\;=\; S^*\,,
\]
the operator $S$ is said to be essentially self-adjoint, and when even more
\[
 S\;=\;S^*\,,
\]
$S$ is self-adjoint. Beside, in the lack of (essential) self-adjointness, only two possibilities may occur: either there is no self-adjoint operator $\widetilde{S}$ that extends $S$, or a whole family of $\infty^{d^2}$ distinct self-adjoint extensions $\widetilde{S}$ of $S$ exist, for some cardinal number $d$ called the deficiency index of $S$, and they all satisfy $\overline{S}\subset\widetilde{S}\subset S^*$. In fact, $d$ is the dimension of either subspaces $\ker(S^*\pm\ii\mathbbm{1})$, and the self-adjoint extensions of $S$ are in one-to-one correspondence with the possible unitary isomorphisms $U$ between such two subspaces, through a constructive formula that links each extension $S_U$ to the unitary $U$. The whole picture is non-trivial for unbounded symmetric operators: a bounded and everywhere defined symmetric operator on Hilbert space is always self-adjoint.

 For about two decades after von Neumann's work on self-adjoint operators, one major focus in the newborn self-adjoint extension theory was the extension problem of densely defined semi-bounded operators. For them, the existence and the distinguished features of the extremal (Friedrichs) extension were established, and a final picture was eventually completed, with the seminal work of M.~G.~Kre{\u\i}n\index{Kre{\u\i}n's extension theory} in the mid of the 1940's, which showed that the self-adjoint extensions of, say, a lower semi-bounded, densely defined operator are naturally ordered in the sense of the usual operator ordering (smaller form domains and larger expectations), below the highest (Friedrichs) one.

 Whereas von Neumann's and Kre{\u\i}n's analyses were completely abstract and general (the one by Kre{\u\i}n limited to semi-bounded operators), such notions soon found applications, among others, in two main fields: quantum mechanics on the one hand, with the self-adjointness problem of quantum Hamiltonians, and partial differential equations on the other, specifically the solution theory of elliptic boundary value problems on domains. Along the latter direction, in particular, M.~I.~Vi\v{s}ik studied the existence and uniqueness of solutions in relation to boundary conditions (of self-adjointness, or more general of closedness) for the differential operator, and established a convenient parametrisation of all such operator realisations in terms of convenient, and simpler, boundary operators. Right after, spanning the first half of the 1950's, M.~\v{S}.~Birman produced a complete operator-theoretic classification of the self-adjoint extensions of a lower semi-bounded and densely defined (symmetric) operator, moving from Vi\v{s}ik's parametrisation, and characterising the properties of the quadratic form, the spectrum, and the ordering of those extensions that were lower semi-bounded themselves, in terms of analogous operator-theoretic, spectral, and ordering features of the `boundary' operator that labels each extension. This coherent and comprehensive corpus of knowledges based on the works of Kre{\u\i}n, Vi\v{s}ik, and Birman,\index{Kre{\u\i}n-Vi\v{s}ik-Birman theory} underwent various new analyses and partial reformulations that are by now classical themselves, in particular, at the end of the 1960's and beginning of 1970's, by T.~Ando and K.~Nishio (on the `smallest positive' Kre{\u\i}n-von Neumann extensions) and by G.~Grubb (on the generalisation to closed extensions). It is also the subject line that has led in more modern times to the theory of boundary triplets,\index{boundary triplets} as we shall mention in a moment.

 On the side of applications to quantum mechanics, von Neumann himself had realised by 1928 that the critical question was to build Hamiltonians as self-adjoint operators, and it was not before 1951 that T.~Kato established his celebrated proof of the essential self-adjointness, on a natural domain, of atomic Hamiltonians
 \[
  -\sum_{j=1}^N\frac{\hslash^2}{\,2m_j}\Delta_{x_j}-\sum_{j=1}^N\frac{Ze}{\,|x|\,}+\sum_{1\leqslant j<k\leqslant N}\frac{e^2}{\,|x_j-x_k|\,}\,.
 \]
 Since then, a series of classical self-adjointness results had appeared, and an industry developed in understanding the presence of self-adjointness, for differential operators of quantum-mechanical relevance, including Schr\"{o}dinger operators with external magnetic fields and Dirac operators. Next to such developments, a typical circumstance one faces in modelling a quantum-mechanical system occurs when heuristics and first-quantisation-like reasoning provide the formal (differential, usually) action of the candidate Hamiltonian, which can be also minimally defined as a symmetric operator on a natural dense domain, and yet turns out to have equal and non-zero deficiency indices. This is the signature of the existence of an infinity of distinct self-adjoint extensions, each of which carries a different physical model and is generally characterised by additional boundary-type conditions, namely additional information that was missing in the minimal domain. The goal is then to identify all such self-adjoint realisations of the initial candidate Hamiltonian and possibly select among them, even if they are all mathematically well-posed, only those that are physically meaningful in that they implement realistic conditions `at the boundary' (for example extensions with local boundary conditions as opposed to non-local ones).

 The above procedure of initial minimal definition, subsequent general self-adjoint-extension, and physical selection, is omnipresent in quantum mechanical modelling, and relies precisely on abstract self-adjoint extension schemes such as the one by von Neumann\index{von Neumann's extension theory} and the one by Kre{\u\i}n, Vi\v{s}ik, and Birman.\index{Kre{\u\i}n-Vi\v{s}ik-Birman theory} Historically, the former attained a more central status, not just for its applicability to non-semi-bounded minimal operators (the vast majority of candidate Hamiltonians are minimal symmetric operators that are lower semi-bounded or have real points in their regularity domains), rather for a more limited circulation of the Russian works across the Western and the Eastern blocks of the twentieth century. The customary textbook playground one teaches in class may be, for instance, the Hamiltonian of a free quantum particle on interval, initially defined on smooth functions compactly supported away from the edges, and self-adjointly extended by means of linear conditions at the boundary linking the value of the function and its derivative; however, this should not obfuscate the fact that non-trivial, real-world applications had attracted interest and energies over the last six decades! A paradigmatic example is the modelling of three-dimensional multi-particle quantum systems where particles move freely but for when they come on top of each other (`contact interaction'),\index{contact interaction} thus with a minimal Hamiltonian that is the free kinetic energy operator on smooth functions supported away from the coincidence configurations: the problem of identifying and classifying physically meaningful self-adjoint realisations was posed first in two celebrated works by F.~Berezin and L.~Faddeev, and by R.~Minlos and L.~Faddeev, at the beginning of the 1960's, and this is still an active research field nowadays.

 It is in the above sense of \emph{classical} subject with \emph{modern} applications that we shaped the present monograph and its very title.

 \begin{center}
\rule{0.5\textwidth}{.8pt}
\end{center}

 The overall material is cast into two main parts, the first one with the general theory and the second with a spectrum of applications, together with side contents in two final appendices.

 The general theory focusses on the above-mentioned classical extension schemes. While we do not claim any special originality -- von Neumann's theory is widely accessible throughout beautiful classical and modern monographs and reviews, and so is, albeit to a lesser extent, Birman's classification scheme -- we reckon that the value of our presentation relies in providing a full-fledged discussion, including all proofs, of all the founding aspects and tools of such classical extension schemes.

 We start, making a small derogation from the natural historical flow, with the discussion of the Friedrichs extension,\index{Friedrichs extension} which is in fact an independent quadratic form construction, and after that we present the theory of the Cayley transform\index{Cayley transform} for symmetric operators, on which von Neumann's extension scheme\index{von Neumann's extension theory} was originally based and is indeed presented here. We also supplement this with the reinterpretation made by N.~Dunford and J.~T.~Schwartz, where the emphasis is rather put on the abstract boundary value problem. Next, we discuss the Kre{\u\i}n transform of positive operators and its role in Kre{\u\i}n's theory of self-adjoint extensions, that parallels the role of the Cayley transform in von Neumann's extension theory. We then present in its entirety Kre{\u\i}n's extension theory\index{Kre{\u\i}n's extension theory} of symmetric semi-bounded operators, following the original work by Kre{\u\i}n, and also discussing the Ando-Nishio variant for the characterisation of the Kre{\u\i}n-von Neumann extension. After this, the Vi\v{s}ik-Birman parametrisation of self-adjoint extensions is analysed in its original form and in its more typical modern re-parametrisation, and all its fundamental ancillary results are discussed on invertibility of the extensions, their semi-boundedness, their spectrum, and Kre{\u\i}n formula type resolvent identities. A detour on that sub-class of extensions that retain the same Friedrichs lower bound concludes such a review.

 All this is preceded by a chapter of preliminaries, that is meant to collect all the necessary theoretical pre-requisites from Hilbert space operator theory, and self-adjoint operator theory in particular, and to which systematic reference is made throughout the monograph. At the price of not presenting proofs therein, as they can be easily tracked down across standard textbooks, that chapter has the merit of outlining all such pre-requisites in logical order (the omitted proofs would indeed unfold each one relying on the previous ones) and in a straightforwardly searchable form, much as a hitchhiker's kind of guide across all the basic generalities on symmetric and self-adjoint operators on Hilbert space (including the spectral theorem). In fact, that chapter reflects a separate course delivered by one of us for several years both in Munich and in Trieste for both master and doctoral students.

 In the second part of the monograph, the applications concern various classes of quantum models that are receiving today new or renewed interest in mathematical physics, in particular from the point of view of self-adjointly realising certain operators of interests, classifying their self-adjoint extensions as actual quantum Hamiltonians, studying their spectral and scattering properties, and the like, but also from the point of view of intermediate technical questions that have theoretical interest per se, such as characterising the corresponding operator closures and adjoints.

 Our choice of applications is obviously subjective to some extent, but on the other hand it is driven by a pedagogical criterion. Indeed, all considered models provide instructive playgrounds for the concrete usage of general self-adjoint extension theories. Moreover, the first two that we discuss, one-body Dirac-Coulomb systems with heavy nuclei and Hydrogenoid atoms\index{Hydrogenoid Hamiltonian} with singular point-like perturbation at the centre, are somewhat classical in that for them the self-adjoint extension and spectral problem was already studied and solved in the past, however by means of von Neumann's extension theory, whereas here the analysis is re-done with the instructive and often much cleaner alternative of the Kre{\u\i}n-Vi\v{s}ik-Birman extension scheme. This also allows for an illuminating comparison between the two approaches, with strengths on both sides.

 Instead, the next two classes of models discussed in the applications, the geometric confinement and transmission of a quantum particle constrained on a Grushin manifold, and the bosonic trimer with zero-range interaction, beside representing two further excellent playgrounds for self-adjoint extension schemes, are in addition part of particularly active research lines (in the first case, also a particularly novel subject) that in these very days are producing beautiful new results and difficult open problems. The choice of the bosonic trimer is further pedagogically motivated by the circumstance that the full comprehension of the model requires a subtle and somewhat inescapable synergy between the von Neumann and the Kre{\u\i}n-Vi\v{s}ik-Birman extension schemes.

 A final appendix is inserted to revisit the emergence of self-adjointness in the foundations of quantum mechanics. Again, the material here is classic, but instead of laying down the usual quantum-mechanical principles, which include the self-adjointness of observables, and deduce from them physically plausible and relevant consequences, we proceed the opposite way, showing how certain standard requirements that quantum physics demand imply necessarily the observables to be self-adjoint, and not merely symmetric, operators.

 One further appendix collects references to standard pedagogical examples. In fact, whereas in the present monograph the applications of the classical self-adjoint extension schemes are all part of active ongoing research, it maybe beneficial for the non-specialist reader to also be re-directed to those segments of the literature where basic, yet instructive pedagogical examples are discussed in depth.

 Throughout the presentation we kept a style that makes the material easily accessible at the graduate level: the subject is surely for experts, but we deliberately maintained a large amount of details and extensive and systematic reference to the literature, including the historical perspective.

  \begin{center}
\rule{0.5\textwidth}{.8pt}
\end{center}

  Beside what is part of the content of this monograph, it is worth briefly commenting on one important subject that is not part of it, instead, and is deliberately left out: the modern theory of boundary triplets.\index{boundary triplets}

  As previously alluded to, in more recent times core aspects of self-adjoint theories were revisited and rephrased, and largely generalised, among others by Y.~M.~Arlinski\u{\i}, J.~Behrndt, V.~A.~Derkach, S.~Hassi, P.~Kurasov, M.~M.~Malamud, Z.~Sebesty\'{e}n, H.~S.~V.~de Snoo, E.~Tsekanovski{\u\i}, and others, in the form of a new theoretical framework initially introduced by A.~N.~Ko\v cube\u\i~and V.~M.~Bruk in the mid 1970's, with ideas that can be traced back to J.~W.~Calkin. This has led to the theory of boundary triplets\index{boundary triplets} and their Weyl functions,\index{Weyl function} in application to the analysis of boundary value problems for differential equations and general operators in Hilbert spaces.

  As far as self-adjoint extensions are concerned, the theory of boundary triplets\index{boundary triplets} is conceptually equivalent to the classical ones and indirectly modelled on the old Vi\v{s}ik-Birman classification. It puts emphasis on the extension mechanism induced by abstract boundary conditions expressed by certain boundary maps that implement the abstract Green identity of the considered symmetric operator, much in analogy to the role of the Birman extension parameter.

  Luckily, very recent and beautiful reviews are available, including last year's comprehensive book by Behrndt, Hassi, and de Snoo on boundary value problems, Weyl functions,\index{Weyl function} and differential operators, which surely do not make the subject missing here. Moreover, and most importantly, whereas natural playgrounds for boundary triplets are Sturm-Liouville operators,\index{Sturm-Liouville operators} Schr\"{o}dinger operators on bounded domains, and canonical systems of differential equations, as a matter of fact the classical extension schemes a la von Neumann\index{von Neumann's extension theory} and a la Kre{\u\i}n-Vi\v{s}ik-Birman\index{Kre{\u\i}n-Vi\v{s}ik-Birman theory} turn out to be the most natural, informative, and efficient, for  many modern quantum-mechanical applications.

 \begin{center}
\rule{0.5\textwidth}{.8pt}
\end{center}

 A considerable portion of the work and research projects that eventually led to this monographs were developed in the course of both authors' activity at the SISSA Trieste: there, we benefited from the steady external support of L.~D\k{a}browski, whom we gratefully thank here.

 We should also like to tribute our personal and scientific acknowledgements to several colleagues with whom we had many valuable exchanges on multiple topics of this monographs: our co-authors on this subject A.~Cintio, A.~Ottolini, and E.~Pozzoli, as well as our colleagues S.~Albeverio, N.~Arrizabalaga, J.~Behrndt, I.~Beshastnyi, L.~D\k{a}browski, D.~Dimonte, M.~Erceg, V.~Georgiev, V.~Lotoreichik, G.~Morchio, D.~Noja, A.~Posilicano, G.~Ruzza, N.~Santamaria, R.~Scandone, T.~Turgut, K.~Yajima.

 In particular, we are indebted to S.~Albeverio, 
 who kindly accepted to write the Foreword to this monograph,  
 for his overall support and scientific advice, being indeed one of the world's leading figures, among many other fields, in the area of self-adjoint solvable models in quantum mechanics.

 Last, we gratefully acknowledge the support of the Italian National Institute for Higher Mathematics (INdAM), the Hausdorff Center for Mathematics, and the Alexander von Humboldt Foundation. 

%
%
%
%

\vspace{\baselineskip}
\begin{flushright}\noindent
Milan and Bonn,\hfill {\it Matteo Gallone}\\
December 2021\hfill {\it Alessandro Michelangeli}\\
\end{flushright}

\tableofcontents

\mainmatter
%
%
%

\begin{partbacktext}
\part{Theory}

\end{partbacktext}

%
%
%
\chapter{Generalities on symmetric and self-adjoint operators on Hilbert space}
\label{chaper-preliminaries} 

This chapter collects an amount of basic definitions and properties from operator theory on Hilbert space, which we use in the next chapters. The reader is assumed to have already some familiarity with all such notions and hence the presentation is kept in the form of a self-consistent overview for later reference from subsequent chapters. Throughout this survey, proofs are omitted: they are all classical and can be easily tracked down across the literature. The materials presented here are discussed in depth, among others, in Amrein \cite{Amrein-HilberSpMethods-2009}, Berezans$'$ki\u{\i} \cite{Berezanskii-1968}, Berezans$'$ki\u{\i} and Kondrat$'$ev \cite{Berezansky-Kondratiev7-vol1-orig1988}, Birman and Solomjak \cite{Birman-Solomjak-1980book}, De Oliveira \cite{DeOliveira-SpectralTheory-Book2009}, Davies \cite{Davies_LinearOpSpectra-book2007}, Dunford and Schwartz \cite{Dunford-Schwartz-I,Dunford-Schwartz-II}, Ji\v{r}\'{\i}, Exner, and Havl\'{\i}\v{c}ek \cite{Jiri-Exner-Havlicek-2008}, Kato \cite{Kato-perturbation}, Reed and Simon \cite{rs1}, Riesz and N\`agy \cite{Riesz-Nagy_FA-1955_Eng}, Schm\"{u}dgen \cite{schmu_unbdd_sa}, Smirnov \cite{Smirnov-HigherMaths-5-orig1959}, Weidmann \cite{Weidmann-LinearOperatosHilbert,Weidmann-book1987}.

\section{Preliminary notions}\label{sec:I-preliminaries}

A fundamental underlying structure throughout this book is a (complex) \emph{Hilbert space}, \index{Hilbert space} namely a linear vector space on the complex field equipped with a positive definite inner product and complete with respect to the norm topology induced by the inner product. A notion of real Hilbert spaces is possible as well, but in application to quantum mechanics as in this book the field will always be $\mathbb{C}$.

Customarily, a Hilbert space shall be denoted with $\mathcal{H}$, its zero vector with $0$, the inner product with $\langle\cdot,\cdot\rangle$, and the norm with $\|\cdot\|$, with the convention that the inner product is anti-linear in the first entry and linear in the second. Thus, $\|\psi\|=\sqrt{\langle\psi,\psi\rangle}$ and $\overline{\langle\psi,\varphi\rangle}=\langle\varphi,\psi\rangle$. The scalar product is continuous in the norm topology separately in each entry, and also jointly in both entries. Notation such as $\langle\cdot,\cdot\rangle_{\cH}$ and $\|\cdot\|_{\cH}$ are used when it is necessary to stress the reference to the specific Hilbert space under consideration. Finite-dimensional Hilbert spaces are essentially of the form $\mathbb{C}^n$ for some $n\in\mathbb{N}$, with the ordinary inner product. Familiarity is also assumed with standard concrete examples of infinite-dimensional Hilbert spaces: in particular, two prototypical infinite-dimensional Hilbert spaces are $\cH=L^2(\Omega,\ud\mu)$ for some measure space $(\Omega,\mu)$ and with scalar product
\[
 \langle f,g\rangle_{L^2}\;:=\;\int_\Omega \overline{f(\omega)}\,g(\omega)\,\ud\mu(\omega)\,,
\]
and the Hilbert space
\[
 \ell^2(\mathbb{N})\;:=\;\Big\{ x\equiv(x_n)_{n\in\mathbb{N}}\,\Big|\,x_n\in\mathbb{C}\;\forall n\in\mathbb{N}\textrm{ and }\sum_{n\in\mathbb{N}}|x_n|^2<+\infty\Big\}
\]
with scalar product
\[
 \langle x,y\rangle_{\ell^2}\;:=\;\sum_{n\in\mathbb{N}}\overline{x_n}\,y_n
\]
(and analogously $\ell^2(\mathbb{Z})$ and the like).

Hilbert norm and scalar product satisfy the \emph{Cauchy-Schwarz inequality} \index{Cauchy-Schwarz inequality}
\[
 |\langle\psi,\varphi\rangle|\;\leqslant\;\|\psi\|\|\varphi\|
\]
(with equality if and only if $\psi$ and $\varphi$ are linearly dependent), the \emph{polarisation identity} \index{polarisation identity}
\[
 \langle\psi,\varphi\rangle\;=\;\frac{1}{4}\big(\|\psi+\varphi\|^2-\|\psi-\varphi\|^2\big)-\frac{\ii}{4}\big(\|\psi+\ii\,\varphi\|^2-\|\psi-\ii\,\varphi\|^2\big)\,,
\]
the \emph{parallelogram identity} \index{parallelogram identity}
\[
 \|\psi+\varphi\|^2+\|\psi-\varphi\|^2\;=\;2\big(\|\psi\|^2+\|\varphi\|^2\big)\,,
\]
and the \emph{Ptolemaic inequality} \index{Ptolemaic inequality}
\[
 \|\psi-\varphi\|\|\xi-\eta\|\;\leqslant\;\|\psi-\xi\|\|\varphi-\eta\|+\|\xi-\varphi\|\|\psi-\eta\|
\]
for every $\psi,\varphi,\xi,\eta\in\cH$.

A linear subspace of $\cH$ will be ordinarily referred to simply as a subspace. A closed subspace of $\cH$ is naturally a Hilbert space itself, with the scalar product and norm topology induced by $\cH$, and for this reason is called \emph{Hilbert subspace} \index{Hilbert subspace} of $\cH$. The \emph{closure} of a subspace $V\subset\cH$ in the Hilbert norm topology of $\cH$ will be denoted by $\overline{V}$. 
For two \emph{orthogonal} vectors, namely $\psi,\varphi\in\cH$ with $\langle\psi,\varphi\rangle=0$, one also writes $\psi\perp\varphi$. The \emph{orthogonal complement} \index{orthogonal complement} of a subset $V\subset\cH$ is the subspace
\[
 V^\perp\;:=\;\{\varphi\in\cH\,|\,\langle\varphi,\psi\rangle=0\;\forall\psi\in V\}\,,
\]
which is closed in $\cH$. For any subspace $V\subset\cH$ one has $V^\perp=\overline{V}^\perp$ and $(V^\perp)^\perp=\overline{V}$.

Given two Hilbert spaces $\cH_1$ and $\cH_2$ (in particular, two closed subspaces of a common Hilbert space), $\cH:=\cH_1\oplus\cH_2$ defines a new Hilbert space, called the \emph{orthogonal direct sum} \index{orthogonal direct sum of Hilbert spaces} of $\cH_1$ and $\cH_2$, consisting of pairs $(\psi_1,\psi_2)$ with $\psi_j\in\cH_j$, $j\in\{1,2\}$, with the natural component-wise sum and multiplication by scalars, and with inner product
\[
 \langle (\psi_1,\psi_2),(\varphi_1,\varphi_2)\rangle_{\cH}\;:=\;\langle\psi_1,\varphi_1\rangle_{\cH_1}+\langle\psi_2,\varphi_2\rangle_{\cH_2}\,.
\]
When $\cH=\cH_1\oplus\cH_2$ as above, one also writes $\cH_1=\cH\ominus\cH_2$ to mean the orthogonal complement in $\cH$ of the Hilbert subspace $\cH_2$, and analogously $\cH_2=\cH\ominus\cH_1$.
This extends trivially to finite orthogonal direct sums, whereas the countable orthogonal direct sum of the collection $(\cH_j)_{j\in\mathbb{N}}$ of Hilbert spaces is the Hilbert space 
\[
 \cH\;\equiv\;\bigoplus_{j\in\mathbb{N}}\cH_j\;:=\;
 \left\{
 \begin{array}{c}
  \psi\equiv(\psi_j)_{j\in\mathbb{N}}\;\textrm{ with }\;\psi_j\in\cH_j \\
  \textrm{and such that }\sum_{j\in\mathbb{N}}\|\psi_j\|^2_{\cH_j}<+\infty
 \end{array}
 \right\}
\]
with the natural linear space structure induced component-wise and with inner product
\[
 \langle\psi,\varphi\rangle_{\cH}\;:=\;\sum_{j\in\mathbb{N}}\langle\psi_j,\varphi_j\rangle_{\cH_j}\,.
\]

In particular, the definition above is applicable to a (finite or countably infinite) collection $(V_j)_j$ of closed subspaces, respectively, of the Hilbert spaces of the collection $(\cH_j)_j$ (including the case when all the $\cH_j$'s are the same), in which case $\bigoplus_j V_j$ is a closed subspace of $\bigoplus \cH_j$. If instead not all the subspaces $V_j$ are (known to be) closed, it is convenient in the applications that will be discussed in this book to use the modified symbol $\boxplus_j V_j$ for the (not necessarily closed) subspace
\[
 \op_j V_j\;:=\;
 \left\{
 \begin{array}{c}
  \psi\equiv(\psi_j)_{j\in\mathbb{N}}\;\textrm{ with }\;\psi_j\in V_j \\
  \textrm{and such that }\sum_{j\in\mathbb{N}}\|\psi_j\|^2_{\cH_j}<+\infty
 \end{array}
 \right\}
\]
of $\bigoplus \cH_j$. Thus, for a subspace $V$ of a Hilbert space $\cH$, $V\boxplus V^\perp$ is a subspace of $\cH$ and $\overline{V}\oplus V^\perp=\cH$. The latter identity is the \emph{orthogonal direct sum decomposition} \index{orthogonal direct sum decomposition} of $\cH$ induced by $V$.

 The above construction of countable orthogonal direct sum of Hilbert (sub-)spa\-ces has a further generalisation to the continuum. For a given Hilbert space $\mathfrak{h}$ and the corresponding Hilbert space $L^2(\mathcal{M},\mathfrak{h})$ of $\mathfrak{h}$-valued functions on a $\sigma$-finite measure space $(\mathcal{M},\mu)$, one also writes
  \[
  L^2(\mathcal{M},\mathfrak{h})\;\equiv\;\int_{\mathcal{M}}^{\oplus}\mathfrak{h}\,\ud \mu
 \]
  and refers to the r.h.s.~above as the \emph{constant-fibre direct integral}\index{constant-fibre!direct integral of Hilbert spaces} Hilbert space over $(\mathcal{M},\mu)$ modelled on the \emph{fibre}\index{fibre!of constant-fibre direct integral} $\mathfrak{h}$. From this perspective, $\cH:=\int_{\mathcal{M}}^{\oplus}\mathfrak{h}\,\ud \mu$ is the Hilbert space
  \[
   \cH\;=\;\int_{\mathcal{M}}^{\oplus}\mathfrak{h}\,\ud \mu\;=\;
   \left\{ 
   \begin{array}{c}
    \psi\equiv(\psi_m)_{m\in\mathcal{M}}\;\textrm{such that }\psi_m\in\mathfrak{h}\textrm{ for a.e.}\, m \\
    \textrm{and }\int_{\mathcal{M}}\|\psi_m\|_{\mathfrak{h}}^2\,\ud\mu(m)\,<\,+\infty
   \end{array}
   \right\}
  \]
  with the natural fibre-wise vector space structure and scalar product
  \[
   \langle \psi,\varphi\rangle_{\cH}\;=\;\int_{\mathcal{M}}\langle \psi_m,\varphi_m\rangle_{\mathfrak{h}}\,\ud\mu(m)\,.
  \]
  (Recall that `a.e.' stands for `almost everywhere' with respect to the measure $\mu$.)  
  In particular, if $\mu(m)=\sum_k\delta(m-m_k)$, namely a (finite or infinite) sum of point measures at given points $m_k$ of $\mathcal{M}$, then $\int_{\mathcal{M}}^{\oplus}\mathfrak{h}\,\ud \mu$ is nothing but the orthogonal direct sum of copies of the same $\mathfrak{h}$.

Given a subset $V\subset\cH$, $\mathrm{span}V$ denotes as usual the subspace of finite linear combinations of elements from $V$. A subset $V\subset\cH$ is dense in $\cH$, in the norm topology, if and only if $\overline{\mathrm{span}V}=\cH$, equivalently, $V^\perp=\{0\}$.

A collection $(\psi_\alpha)_{\alpha\in I}$ ($I$ being a generic index set) in a Hilbert space $\cH$ is called an \emph{orthonormal system} \index{orthonormal system} if $\|\psi_\alpha\|=1$ and $\langle\psi_\alpha,\psi_\beta\rangle=0$ for any distinct $\alpha,\beta\in I$, and an \emph{orthonormal basis} \index{orthonormal basis} of $\cH$ if it is an orthonormal system such that its finite linear combinations are dense in $\cH$. A Hilbert space $\cH$ is \emph{separable} \index{separable Hilbert space} (as a topological space with respect to its norm topology) if and only if $\cH$ has a countable orthonormal basis. The `only if' direction is established by thinning out a countable dense set in $\cH$ until it becomes linearly independent and then orthonormalising the resulting collection via the customary Gram-Schmidt algorithm. Thus, the existence of an orthonormal basis in a separable Hilbert space is constructive; in fact, an orthonormal basis always exists, which in the non-separable case follows by the axiom of choice.

Given an orthonormal basis $(\psi_\alpha)_{\alpha\in I}$ in a Hilbert space $\cH$, the following properties hold, that are collectively referred to as the \emph{orthonormal basis representation}. \index{orthonormal basis representation} The index set $I_\psi:=\{\alpha\in I|\langle\psi_\alpha,\psi\rangle\neq 0\}$ associated to a vector $\psi\in\cH$ is at most countable, and independently of the order of the summands
\[
 \begin{split}
  \psi\;&=\;\sum_{\alpha\in I}\langle\psi_\alpha,\psi\rangle\psi_\alpha\,, \\
  \|\psi\|^2\;&=\;\sum_{\alpha\in I}|\langle\psi_\alpha,\psi\rangle|^2\qquad (\textrm{\emph{Parseval identity}})\,, \index{Parseval identity}
 \end{split}
\]
the first series converging in norm, where $\alpha$ above actually only runs over $I_\psi$ and hence the series consists of no more than countably many terms. Moreover, for any other $\varphi\in\cH$,
\[
 \langle\psi,\varphi\rangle\;=\;\sum_{\alpha\in I}\overline{\langle\psi_\alpha,\psi\rangle}\langle\psi_\alpha,\varphi\rangle\,,
\]
where the series converges absolutely, and in fact all summands labelled by $\alpha\notin I_\psi\cap I_\varphi$ are zero. Conversely, if scalars $c_\alpha$'s are given with $\sum_{\alpha\in I}|c_\alpha|^2<+\infty$ (in particular, at most countably many of them are non-zero), then $\xi:=\sum_{\alpha\in I} c_\alpha\psi_\alpha$ converges in norm independently of the order and $\|\xi\|^2=\sum_{\alpha\in I}|c_\alpha|^2$. The above basis representation is unique, in the sense that the coefficients are uniquely determined by the vector expanded along the basis. If the initial $(\psi_\alpha)_{\alpha\in I}$ is only an orthonormal system in $\cH$, all facts above remain valid except the Parseval identity that is modified as
\[
  \|\psi\|^2\;\geqslant \;\sum_{\alpha\in I}|\langle\psi_\alpha,\psi\rangle|^2\qquad (\textrm{\emph{Bessel inequality}})\,. \index{Bessel inequality}
\]

A Hilbert space $\cH$ is self-dual ($\cH\cong\cH^*$). In fact, the map $\psi\mapsto\langle\psi,\cdot\rangle$ is an anti-linear Hilbert space isomorphism between $\cH$ and the dual space\index{dual space} $\cH^*$ of the bounded linear $\cH\to\mathbb{C}$ functionals (the \emph{Riesz lemma}\index{Riesz lemma}\index{theorem!Riesz (lemma)}). In particular, the Hilbert norm is also characterised variationally by
\[
 \|\psi\|\;=\;\sup_{ \substack{ \varphi\in\cH \\ \varphi\neq 0  }}\frac{|\langle\varphi,\psi\rangle|}{\|\varphi\|}\;=\;\sup_{ \substack{ \varphi\in\mathcal{D} \\ \varphi\neq 0  }}\frac{|\langle\varphi,\psi\rangle|}{\|\varphi\|}\,,
\]
where $\mathcal{D}$ is a dense subspace of $\cH$. As such, $\cH$ is reflexive, i.e., isomorphic to $\cH^{**}$ via the isomorphism $\psi\mapsto e_\psi$ with $e_\psi(\ell):=\ell(\psi)$ $\forall\ell\in\cH^*$. Moreover, beside the metric topology induced by the norm, $\cH$ is also naturally equipped with the weaker, non-metric topology called \emph{(Hilbert space) weak topology}\index{weak topology} (in fact, the weak-$*$ topology, in the sense of Banach space topologies) characterised by being the weakest topology making the functionals $\cH\ni\psi\mapsto\langle\psi,\cdot\rangle$ continuous. Norm (also said: strong) topology and weak topology on $\cH$ are equivalent only if $\dim\cH<\infty$. As $\cH$ is reflexive, the norm-closed unit ball $B_\cH:=\{\psi\in\cH\,|\,\|\psi\|\leqslant 1\}$ is compact in the weak topology (the \emph{Banach-Alaoglu theorem}\index{theorem!Banach-Alaoglu}). If $\cH$ is separable, then there exists a norm $\|\cdot\|_w$ on $\cH$ (and hence a metric $\varrho_w(\psi,\varphi):=\|\psi-\varphi\|_w$) such that $\|\psi\|_w\leqslant\|\psi\|$ and whose metric topology restricted to the closed unit ball $B_\cH:=\{\psi\in\cH\,|\,\|\psi\|\leqslant 1\}$ of $\cH$ is precisely the Hilbert space weak topology. For concreteness one may define
 \[
  \|\psi\|_w\;:=\;\sum_{n=1}^\infty \frac{1}{\:2^n}|\langle \xi_n,\psi\rangle|
 \]
 for a dense countable collection $(\xi_n)_{n\in\mathbb{N}}$ in $B_\cH$ which identifies the norm $\|\cdot\|_w$. Thus, as a metric space $(B_\cH,\varrho_w)$ is also complete in the weak topology, beside being weakly compact. Instead, $B_\cH$ is norm-compact only if $\dim\cH<\infty$.
 Given $(\psi_n)_{n\in \mathbb{N}}$ and $\psi$ in $\cH$, one writes $\psi_n\rightharpoonup\psi$ to indicate the weak convergence of the sequence (whereas $\psi_n\to\psi$ stands for $\|\psi_n-\psi\|\to 0$). One then has:
 \[
  \begin{split}
     \psi_n\rightharpoonup\psi\quad&\Leftrightarrow\quad \langle\varphi,\psi_n\rangle\xrightarrow{n\to\infty}\langle\varphi,\psi_n\rangle\;\;\forall\varphi\in\cH\,, \\
     \psi_n\rightharpoonup\psi\quad&\Rightarrow\quad\exists\,C>0\textrm{ such that }\|\psi_n\|\,\leqslant\,C\,, \\
     \psi_n\rightharpoonup\psi\quad&\Rightarrow\quad\|\psi\|\,\leqslant\,\liminf_{n\to\infty}\|\psi_n\|\,, \\
     \left. 
     \begin{array}{c}
      \psi_n\rightharpoonup\psi \\
      \|\psi_n\|\to\|\psi\|
     \end{array}
     \right\}\quad&\Leftrightarrow\quad\psi_n\to\psi\,.
  \end{split}
 \]

A \emph{linear operator} \index{linear operator} acting on a Hilbert space $\mathcal{H}$ is a linear mapping $A$ of a linear subspace $\mathcal{D}(A)$ of $\mathcal{H}$, called the \emph{domain}\index{domain}\index{operator!domain} of $A$, into $\mathcal{H}$ itself. Ordinarily in the present context one only writes `operator' for `linear operator'. In the lack of a universal standard, the expression `operator $A$ \emph{in} a space $\cH$' shall be conventionally used instead of `operator $A$ \emph{on} $\cH$' to emphasise that $A$ has values in $\cH$, in order to avoid potential ambiguities occurring for example when $\cH$ is a (Hilbert) subspace of a larger (Hilbert) space $\mathcal{K}$. The \emph{kernel}\index{kernel}\index{operator!kernel} $\ker A$ and the \emph{range}\index{range}\index{operator!range} $\mathrm{ran}\,A$ of $A$ are, respectively, the subspaces
\[
 \begin{split}
  \ker A\;&:=\;\{\psi\in\mathcal{D}(A)\,|\,A\psi=0\}\,, \\
  \mathrm{ran}\,A\;&:=\;\{A\psi\,|\,\psi\in\mathcal{D}(A)\}\,.
 \end{split}
\]
The cardinal number
\[
 \mathrm{rank}\,A\;:=\;\dim\mathrm{ran}\,A
\]
is called the \emph{rank}\index{rank}\index{operator!rank} of the operator $A$.

All this (as well as the material in the following) generalises straightforwardly for linear operators between two possibly distinct Hilbert spaces $\mathcal{H}_1$ and $\mathcal{H}_2$. 
In particular, two Hilbert spaces $\cH_1$ and $\cH_2$ are said to be \emph{isomorphic} when there exists an operator $U$ from $\cH_1$ onto $\cH_2$, namely with $\mathcal{D}(U)=\cH_1$ and $\mathrm{ran}\,U=\cH_2$, such that $\langle U\psi,U\varphi\rangle_{\cH_2}=\langle \psi,\varphi\rangle_{\cH_1}$ $\forall\psi,\varphi\in\cH_1$: such $U$ is called \emph{unitary operator}, \index{operator!unitary} and one writes $\cH_1\cong\cH_2$, or also $\cH_1\xrightarrow[U]{\cong}\cH_2$, a linear map that is also referred to as \emph{Hilbert space isomorphism} \index{Hilbert space isomorphism} between $\cH_1$ and $\cH_2$. 
Finite-dimensional Hilbert spaces are isomorphic to $\mathbb{C}^n$ for some $n\in\mathbb{N}$ and separable infinite-dimensional Hilbert spaces are isomorphic to $\ell^2(\mathbb{N})$, $\ell^2(\mathbb{Z})$,  $L^2(\mathbb{R},\ud x)$, and the like, in all case via a natural isomorphism that connects two orthonormal bases.

Given two operators $A$ and $B$ on a Hilbert space $\cH$, by $A\subset B$ one means that $\mathcal{D}(A)\subset\mathcal{D}(B)$ and $A\psi=B\psi$ $\forall\psi\in\mathcal{D}(A)$: in this case $A$ is said to be a \emph{restriction} \index{operator restriction} of $B$ and $B$ an \emph{extension} \index{operator extension} of $A$. The \emph{restriction} of $A$ to a subspace $\mathcal{D}\subset\mathcal{D}(A)$ is the operator, denoted by $A|_{\mathcal{D}}$ or also $A\upharpoonright\mathcal{D}$, with domain $\mathcal{D}$ and action $A|_{\mathcal{D}}\psi=A\psi$ $\forall\psi\in\mathcal{D}$.

Linear operators acting on the same Hilbert space undergo a notion of sum, multiplication, and multiplication by scalar, defined in the natural way, i.e., point-wise in the vectors of the Hilbert space and compatibly with the respective domains. Thus,
\[
 \begin{split}
  (A+B)\psi\;&=\;A\psi+B\psi\,,\qquad \mathcal{D}(A+B)=\mathcal{D}(A)\cap\mathcal{D}(B)\,, \\
  (AB)\psi\;&=\;A(B\psi)\,,\qquad \qquad\,\mathcal{D}(AB)=\{\psi\in\mathcal{D}(B)|B\psi\in\mathcal{D}(A)\}\,,
 \end{split}
\]
etc. This defines in particular powers $A^n$, $n\in\mathbb{N}_0$, with the convention $A^0=\mathbbm{1}$, the identity on $\cH$. Such operations satisfy the distributive laws
\[
 (A+B)C\;=\;AC+BC\,,\qquad A(B+C)\;\supset\; AB+AC\,.
\]
If the operator $A$ on $\cH$ is injective ($\ker A=\{0\}$), then $A$ is \emph{invertible on its range} and its \emph{inverse} \index{operator inverse} $A^{-1}$ is the operator defined by $\mathcal{D}(A^{-1}):=\mathrm{ran}\,A$ and $A^{-1}(A\psi):=\psi$ $\forall\psi\in\mathcal{D}(A)$, and one has $\mathrm{ran}(A^{-1})=\mathcal{D}(A)$. For two injective operators, $(AB)^{-1}=B^{-1}A^{-1}$.

For a generic operator $A$ on $\cH$ and vectors $\psi,\varphi\in\mathcal{D}(A)$, one has
\[
\begin{split}
 4\langle\psi,A\varphi\rangle\;&=\;\langle(\psi+\varphi),A(\psi+\varphi)\rangle-\langle(\psi-\varphi),A(\psi-\varphi)\rangle \\
 &\qquad - \ii\langle(\psi+\ii\varphi),A(\psi+\ii\varphi)\rangle+\ii\langle(\psi-\ii\varphi),A(\psi-\ii\varphi)\rangle\,,
\end{split}
\]
which is also called \emph{polarisation identity} (for operators). \index{polarisation identity} The quantity $\langle\psi,A\psi\rangle$ is customarily referred to as the \emph{expectation} \index{expectation} of the operator $A$ in the state $\psi$. Thus, as a consequence, a densely defined operator with all zero expectations is necessarily the zero operator (on its domain).

The \emph{graph} \index{graph} $\Gamma(A)$ of an operator $A$ on $\cH$ is the subspace
\[
 \Gamma(A)\;:=\;\{(\psi,A\psi)\,|\,\psi\in\mathcal{D}(A)\}
\]
of $\cH\oplus\cH$. On the other hand, a subspace $\Gamma$ of $\cH\oplus\cH$ is the graph of an operator $A$ on $\cH$ if and only if the condition $(0,\varphi)\in\Gamma$ for some $\varphi\in\cH$ implies $\varphi=0$: in this case $\mathcal{D}(A)=\{\psi\in\cH\,|\,\exists\varphi\in\cH\textrm{ with }(\psi,\varphi)\in\Gamma\}$ and $A\psi=\varphi$ for $(\psi,\varphi)\in\Gamma$. The operator inclusion $A\subset B$ is therefore equivalent to $\Gamma(A)\subset\Gamma(B)$. The assignment
\[
 \langle\psi,\varphi\rangle_{\Gamma(A)}\;:=\;\langle\psi,\varphi\rangle+\langle A\psi,A\varphi\rangle\,,\qquad\psi,\varphi\in\mathcal{D}(A)\,,
\]
defines a scalar product on $\mathcal{D}(A)$ whose associated norm
\[
 \|\psi\|_{\Gamma(A)}\;:=\;\big(\|\psi\|_{\cH}^2+\|A\psi\|_{\cH}^2\big)^{\frac{1}{2}}\;\approx\;\|\psi\|_{\cH}+\|A\psi\|_{\cH}\,,\qquad\psi\in\mathcal{D}(A)
\]
(where `$\approx$' above denotes the equivalence of norms) is called the \emph{graph norm} \index{graph norm} of $A$.

 Given two Hilbert spaces $\cH$ and $\cH'$, a \emph{Hilbert space tensor product}\index{tensor product (of Hilbert spaces)!finite} of them is a Hilbert space $\cH\otimes\cH'$, together with a \emph{bi-linear} map $\otimes:\cH\times\cH'\to\cH\otimes\cH'$, $(\psi,\psi')\mapsto\psi\otimes\psi'$, such that the finite linear combinations of elements of the form $\psi\otimes\psi'$ are dense in $\cH\otimes\cH'$ and
 \[
  \langle \psi\otimes\psi' , \varphi\otimes\varphi'\rangle_{\cH\otimes\cH'}\;=\;\langle\psi,\varphi\rangle_{\cH}\langle\psi',\varphi'\rangle_{\cH'}
 \]
 for any possible choice of the vectors. In particular, $\|\psi\otimes\psi'\|_{\cH\otimes\cH'}=\|\psi\|_{\cH}\|\psi'\|_{\cH'}$. A tensor product between two Hilbert spaces $\cH$ and $\cH'$ always exists, and if $\cH\otimes\cH'$ and $\cH\widetilde{\otimes}\cH'$ are two such tensor products, then there exists a unique unitary operator $U:\cH\otimes\cH'\xrightarrow{\cong}\cH\widetilde{\otimes}\cH'$ such that $U(\psi\otimes\psi')=\psi\widetilde{\otimes}\psi'$ $\forall\psi\in\cH$ and $\forall\psi'\in\cH'$, whence the legitimate expression `\emph{the}' tensor product between $\cH$ and $\cH'$. Moreover, if $(\psi_\alpha)_{\alpha\in I}$ and $(\psi'_{\alpha'})_{\alpha'\in I'}$ are orthonormal bases, respectively, of $\cH$ and $\cH'$, where $I$ and $I'$ are generic (possibly uncountable) index sets, then the collection $\{\psi\otimes\psi'\,|\,\alpha\in I,\alpha'\in I'\}$ is an orthonormal basis of $\cH\otimes\cH'$. In particular, $\cH\otimes\cH'$ is separable if and only if both factors $\cH$ and $\cH'$ are.

 An obvious generalisation of all definitions and properties above holds for the tensor product of a finite number $N\in\mathbb{N}$ of Hilbert spaces $\cH_1,\dots,\cH_N$, now equipped with a \emph{multi-linear} map $(\psi_1,\dots,\psi_N)\mapsto\psi_1\otimes\cdots\otimes\psi_N$.

\section{Bounded, closed, closable operators}\label{sec:I-bdd-closable-closed}

An operator $A$ on a Hilbert space $\cH$ is said to be \emph{bounded} (on its domain) \index{operator!bounded} if there is a constant $c>0$ such that $\|A\psi\|\leqslant c\|\psi\|$ $\forall\psi\in\mathcal{D}(A)$, in which case the quantity
\[
 \|A\|_{\mathrm{op}}\;:=\;\sup_{\substack{\psi\in\mathcal{D}(A) \\ \psi\neq 0}}\frac{\|A\psi\|}{\|\psi\|}\,,
\]
also denoted simply as $\|A\|$, is called the \emph{norm} \index{operator norm} of $A$. Otherwise, $A$ is said to be \emph{unbounded}. \index{operator!unbounded} The operator norm satisfies
\[
 \begin{split}
  \|A+B\|_{\mathrm{op}}\;&\leqslant\;\|A\|_{\mathrm{op}}+\|B\|_{\mathrm{op}} \qquad \textrm{(\emph{triangular inequality})} \index{operator!triangular inequality}\,, \index{operator triangular inequality}\\
  \|AB\|_{\mathrm{op}}\;&\leqslant\;\|A\|_{\mathrm{op}}\|B\|_{\mathrm{op}} \qquad\quad \textrm{(\emph{product inequality})} \index{operator!product inequality}\,. \index{operator product inequality}
 \end{split}
\]

An operator $A$ on $\cH$ is bounded if and only if it is continuous\index{operator!continuous}, and if and only if it is sequentially continuous: thus, the boundedness of $A$ is equivalent to the property that $\psi_n\to\psi$ in $\mathcal{D}(A)$ implies $A\psi_n\to A\psi$ in $\cH$, both limits in the norm of $\cH$ as $n\to\infty$. For this equivalence, linearity of $A$ is crucial. If $A$ is bounded and $\mathcal{D}(A)$ is dense in $\cH$, then $A$ has a unique bounded extension $\widetilde{A}$ on the whole $\cH$, i.e., $\mathcal{D}(\widetilde{A})=\cH$ and $\widetilde{A}\psi=\lim_{n\to\infty}A\psi_n$ for any $\psi\in\cH$ and any sequence $(\psi_n)_{n\in\mathbb{N}}\subset\mathcal{D}(A)$ with $\psi_n\to\psi$, the limit being independent of the approximating sequence. It is customary to denote $\widetilde{A}$ simply with $A$, namely to replace $A$ with its canonical bounded extension to the whole space.

An operator $A$ on a Hilbert space $\cH$ is said to be \emph{closed} \index{operator!closed} if any of the following equivalent conditions hold:
\begin{enumerate}[(i)]
 \item if a sequence $(\psi_n)_{n\in\mathbb{N}}$ in $\mathcal{D}(A)$ satisfies $\psi_n\to\psi$ and $A\psi_n\to\varphi$ as $n\to\infty$ in the norm of $\cH$ for some $\varphi\in\cH$, then $\psi\in\mathcal{D}(A)$ and $A\psi=\varphi$;
 \item $(\mathcal{D}(A),\langle\cdot,\cdot\rangle_{\Gamma(A)})$ is a Hilbert space, equivalently, $(\mathcal{D}(A),\|\cdot\|_{\Gamma(A)})$ is complete;
 \item the graph $\Gamma(A)$ is closed in $\cH\oplus\cH$.
\end{enumerate}
For any closed operator $A$ and $\lambda\in\mathbb{C}$, the subspace $\ker(A-\lambda\mathbbm{1})$ is closed in $\cH$.

An operator $A$ on a Hilbert space $\cH$ is said to be \emph{closable} \index{operator!closable} if any of the following equivalent conditions hold:
\begin{enumerate}[(i)]
 \item[(i)] $A\subset B$ for some closed operator $B$ on $\cH$;
 \item[(ii)] if a sequence $(\psi_n)_{n\in\mathbb{N}}$ in $\mathcal{D}(A)$ satisfies $\psi_n\to 0$ and $A\psi_n\to\varphi$ as $n\to\infty$ in the norm of $\cH$ for some $\varphi\in\cH$, then $\varphi=0$;
 \item[(iii)] the closure of $\Gamma(A)$ in $\cH\oplus\cH$ is the graph of an operator.
\end{enumerate}
When this is the case, the operator $\overline{A}$ such that $\overline{\Gamma(A)}=\Gamma(\overline{A})$ is uniquely determined by $A$, satisfies
\[
 \begin{split}
  \mathcal{D}(\overline{A})\;&=\;\left\{\psi\in\cH\left|
  \begin{array}{c}
   \exists\,(\psi_n)_{n\in\mathbb{N}}\subset\mathcal{D}(A)\textrm{ such that }\psi_n\xrightarrow[]{n\to\infty}\psi \\
   \textrm{and }(A\psi_n)_{n\in\mathbb{N}}\textrm{ converges in }\cH
  \end{array}
  \right.\right\}, \\
  \overline{A}\psi\;&=\;\lim_{n\to\infty}A\psi_n\,,
 \end{split}
\]
where the limit defining $\overline{A}\psi$ is actually independent of the approximating sequence, and is the smallest closed extension of $A$, in the sense of operator inclusion. $\overline{A}$ is called the (operator) \emph{closure} \index{operator closure} of $A$. In particular,
\[
 \mathcal{D}(\overline{A})\;=\;\overline{\,\mathcal{D}(A)\,}^{\|\cdot\|_{\Gamma(A)}}.
\]

If $A$ is closable on $\cH$, $\ker (\overline{A}-\lambda\mathbbm{1})$ is a closed subspace of $\cH$ (in the norm of $\cH$) for any $\lambda\in\mathbb{C}$. If $A$ is closable on $\cH$, and in addition for some $\lambda\in\mathbb{C}$ there is $c_\lambda>0$ such that  $\|(A-\lambda\mathbbm{1})\psi\|\geqslant c_\lambda\|\psi\|$ $\forall\psi\in\mathcal{D}(A)$, then
 \[
  \mathrm{ran}(\overline{A}-\lambda\mathbbm{1})\;=\;\overline{\mathrm{ran}(A-\lambda\mathbbm{1})}
 \]
 is a closed subspace of $\cH$. In particular, if $A$ is closed and $\|(A-\lambda\mathbbm{1})\psi\|\geqslant c_\lambda\|\psi\|$ $\forall\psi\in\mathcal{D}(A)$ for some $\lambda\in\mathbb{C}$ and $c_\lambda>0$, then $\mathrm{ran}(A-\lambda\mathbbm{1})$ is a closed subspace.

A bounded operator $A$ is necessarily closable (but, in general, not the other way around), and when $A$ is bounded the norms $\|\cdot\|_{\cH}$ and $\|\cdot\|_{\Gamma(A)}$ are equivalent on $\mathcal{D}(A)$. A bounded operator $A$ is closed if and only if $\mathcal{D}(A)$ is closed in $\cH$. On the other hand, if $A$ is closed and $\mathcal{D}(A)$ is closed in $\cH$, then (by the closed graph theorem, actually) $A$ is bounded. As a consequence, an unbounded and closed operator cannot be defined on the whole Hilbert space it acts on.

If $A$ is a closed operator on $\cH$ and $B$ is everywhere defined and bounded on $\cH$, then $A+B$ and $AB$ are closed, whereas in general $BA$ is not.

Given an operator $A$ on a Hilbert space $\cH$, a subspace $\mathcal{D}\subset\mathcal{D}(A)$ is called a \emph{core}\index{core} for $A$ if $\mathcal{D}$ is dense in $(\mathcal{D}(A),\langle\cdot,\cdot\rangle_{\Gamma(A)})$, i.e., if for any $\psi\in\mathcal{D}(A)$ there is a sequence $(\psi_n)_{n\in\mathbb{N}}$ in $\mathcal{D}$ with $\psi_n\to\psi$ and $A\psi_n\to A\psi$ in $\cH$. If $A$ is closed, a subspace $\mathcal{D}\subset\mathcal{D}(A)$ is a core for $A$ if and only if $\overline{A\upharpoonright\mathcal{D}}=A$.

\section{Adjoint operators}\label{sec:I-adjoint}

To any \emph{densely defined} operator $A$ on a Hilbert space $\cH$ one associates its \emph{adjoint} $A^*$, which is the operator acting on $\cH$ defined by
\[
 \begin{split}
  \mathcal{D}(A^*)\;&:=\;\{\varphi\in\cH\,|\,\exists\xi_{\varphi}\in\cH\textrm{ such that }\langle\xi_\varphi,\psi\rangle=\langle\varphi,A\psi\rangle\;\forall\psi\in\mathcal{D}(A)\} \,,\\
  A^*\varphi\;&:=\;\xi_{\varphi}
 \end{split}
\]
(owing to the density of $\mathcal{D}(A)$, each such $\xi_{\varphi}$ is indeed uniquely defined, given $\varphi$), in which case one has
\[
 \langle \varphi,A\psi\rangle\;=\;\langle A^*\varphi,\psi\rangle\qquad\forall\psi\in\mathcal{D}(A)\,,\;\forall\varphi\in\mathcal{D}(A^*)\,.
\]
In general $\mathcal{D}(A^*)$ may well not be dense, and it may even be just $\{0\}$.

Apart from the usual pedagogical examples, that may be found in the references cited at the beginning of this chapter, the determination of the adjoint, in particular of the precise domain of the adjoint, is a hard task.

 Here below an amount of standard properties of the adjoint are listed. Assume that $A$ and $B$ are linear operators on a Hilbert space $\cH$, with $\mathcal{D}(A)$ dense in $\cH$.
\begin{enumerate}[(i)]
 \item $A^*$ is closed.
 \item $\ker A^*=(\mathrm{ran}\,A)^\perp$, equivalently, $\cH=\overline{\mathrm{ran}\,A}\oplus \ker A^*$.
 \item If $A\subset B$, then $A^*\supset B^*$.
 \item $(\alpha A)^*=\overline{\alpha} A^*$ $\forall\alpha\in\mathbb{C}$.
 \item If $\mathcal{D}(A+B)$ is dense, then $(A+B)^*\supset A^*+B^*$. If $B$ is bounded and everywhere defined on $\cH$, then $(A+B)^*= A^*+B^*$.
 \item If $\mathcal{D}(AB)$ is dense, then $(AB)^*\supset B^*A^*$. If $\mathcal{D}(AB)$ is dense and $A$ is bounded and everywhere defined, then $(AB)^*= B^*A^*$.
 \item $A$ is closable if and only if $\mathcal{D}(A^*)$ is dense too, in which case $\overline{A}=A^{**}(:=(A^*)^*)$ and $(\overline{A})^*=A^*$.
 \item $A$ is closed if and only if $A=A^{**}$.
 \item If $\ker A=\{0\}$ and $\mathrm{ran}\,A$ is dense, then $(A^*)^{-1}=(A^{-1})^*$.
 \item If $\ker A=\{0\}$ and $A$ is closable, then $A^{-1}$ is closable if and only if $\ker\overline{A}=\{0\}$, in which case $(\overline{A})^{-1}=\overline{A^{-1}}$.
 \item[(xi)] If $A$ is invertible, then $A$ is closed if and only if so is $A^{-1}$.
\end{enumerate}

 For a densely defined and closed operator $A$ on $\cH$, the subspaces $\mathrm{ran}\,A$ and $\mathrm{ran}\,A^*$ of $\cH$ are not closed in general, but $\mathrm{ran}\,A$ is closed if and only if so too is $\mathrm{ran}\,A^*$. Instead, $\ker A^*$ is always closed (consistently with the general property stated in Section \ref{sec:I-bdd-closable-closed}, since $A^*$ is a closed operator, by (i) above).

 From (ii) above, any densely defined and closable operator $A$ on $\cH$ yields the orthogonal decompositions
 \[
  \begin{split}
   \cH\;&=\;\overline{\mathrm{ran}\,A}\oplus \ker A^*\;=\;\overline{\mathrm{ran}\,A^*}\oplus \ker \overline{A} \\
   &=\;\overline{\mathrm{ran}(A-\lambda\mathbbm{1})}\oplus \ker (A^*-\overline{\lambda}\mathbbm{1}) \\
   &=\;\overline{\mathrm{ran}(A^*-\overline{\lambda}\mathbbm{1})}\oplus \ker (\overline{A}-\lambda\mathbbm{1})
  \end{split}
 \]
 for any $\lambda\in\mathbb{C}$.

 If $A$ is closable and densely defined, and in addition for some $\lambda\in\mathbb{C}$ there is $c_\lambda>0$ such that  $\|(A-\lambda\mathbbm{1})\psi\|\geqslant c_\lambda\|\psi\|$ $\forall\psi\in\mathcal{D}(A)$, then $\mathrm{ran}(\overline{A}-\lambda\mathbbm{1})=\overline{\mathrm{ran}(A-\lambda\mathbbm{1})}$ (Sect.~\ref{sec:I-bdd-closable-closed}), and therefore the above decomposition takes the simplified form
  \[
   \cH\;=\;\mathrm{ran}(\overline{A}-\lambda\mathbbm{1})\oplus \ker (A^*-\overline{\lambda}\mathbbm{1})\,.
 \]

 \section{Minimal and maximal realisations of linear differential operators}\label{sec:MinimalAndMaximalRealisations}

 For linear differential operators with smooth coefficients on domains of $\mathbb{R}^d$, the above notions of closure and adjoint with respect to the underlying $L^2$-Hilbert space (with the Lebesgue measure induced from $\mathbb{R}^d$) are formulated through the specific (and classical) terminology of minimal and maximal realisation of the differential operator.

 A \emph{formal linear differential operator}\index{formal differential operator}\index{operator!formal differential} (or also \emph{formal linear differential expression}) of order $r\in\mathbb{N}_0$ on an open $\Omega\subset\mathbb{R}^d$, $d\in\mathbb{N}$ is the expression
 \[
  L\;=\;\sum_{|\alpha|\leqslant r}a_\alpha(x)\, D^{\alpha}\,,\qquad x\equiv(x_1,\dots,x_d)\in\Omega\,,
 \]
 written with the customary multi-index notation, namely with \emph{multi-index}\index{multi-index} $\alpha\equiv(\alpha_1,\dots,\alpha_d)\in\mathbb{N}_0^d$ and 
 \[
  \begin{split}
     & D^\alpha\,:=\;(-\ii)^{|\alpha|}\partial^\alpha\,,\qquad \partial^\alpha\,:=\,\frac{\partial^{\alpha_1}}{\partial {x_1^{\alpha_1}}}\cdots\frac{\partial^{\alpha_d}}{\partial {x_d^{\alpha_d}}}\,,\qquad |\alpha|\,:=\,\alpha_1+\cdots+\alpha_d\,, \\
     & \qquad\qquad \alpha\leqslant\beta\qquad\Leftrightarrow\qquad \alpha_1\,\leqslant\beta_1\,,\dots,\,\alpha_d\,\leqslant\beta_d\,,
  \end{split}
 \]
 
 As a linear map, $L$ above is not associated with any functional domain, whence the terminology `formal'.
 When $L$ has smooth coefficients, one associates to it its \emph{formal adjoint}\index{formal adjoint}\index{operator!formal adjoint}, namely the formal differential operators
 \[
  \begin{split}
    L^\dagger\;&:=\;\sum_{|\alpha|\leqslant r} D^{\alpha}\,\overline{a_\alpha(x)}\;\equiv\;\sum_{|\alpha|\leqslant r} a_\alpha^\dagger(x)\, D^\alpha\,, \\
    a_\alpha^\dagger(x)\;&:=\;\sum_{ \substack{ \alpha\leqslant\beta \\ |\alpha|,|\beta|\leqslant r}  }\begin{pmatrix} \beta \\ \alpha \end{pmatrix} D^{\beta-\alpha}\,\overline{a_\beta(x)}
  \end{split}
 \]
 with the convention
 \[
  \begin{pmatrix} \beta \\ \alpha \end{pmatrix}\;:=\;\begin{pmatrix} \beta_1 \\ \alpha_1 \end{pmatrix}\cdots\begin{pmatrix} \beta_d \\ \alpha_d \end{pmatrix}
 \]
 for the binomial coefficients.
 The coefficients $a_\alpha^\dagger$ are chosen so as, integrating by parts and using Green's formula,\index{Green's formula} 
 \[
  \langle \varphi,L\psi\rangle_{L^2(\Omega)}\;=\;\langle L^\dagger\varphi,\psi\rangle_{L^2(\Omega)}\qquad\forall \psi\in C^r(\Omega)\,,\;\forall\varphi\in C^r_\mathrm{c}(\Omega)\,,
 \]
 and also so as, by the Leibniz rule,
 \[
  L^\dagger \psi\;=\;\sum_{|\alpha|\leqslant r} D^{\alpha}\,\big(\overline{a_\alpha}\psi)\;=\;\sum_{|\alpha|\leqslant r} a_\alpha^\dagger\, D^\alpha\psi\qquad\forall \psi\in C_c^r(\Omega)\,.
 \]
 In fact, the formal adjoint $L^\dagger$ is uniquely determined by imposing $\langle \varphi,L\psi\rangle_{L^2(\Omega)}=\langle L^\dagger\varphi,\psi\rangle_{L^2(\Omega)}$ for any $\psi,\varphi\in C_c^\infty(\Omega)$, whence also $(L^\dagger)^\dagger=L$. 
 By construction, in particular,
 \[
  a_\alpha^\dagger\;=\;\overline{a_\alpha}\qquad \textrm{for }\;|\alpha|\,=\,r\,.
 \]
 When $a_\alpha^\dagger=a_\alpha$ for all $|\alpha|\leqslant r$, i.e., $L=L^\dagger$, one says that $L$ is \emph{formally self-adjoint}.\index{operator!formally self-adjoint}

 A generic formal differential operator $L$ as above, with smooth coefficients, is naturally realised as an operator in the Hilbert space $L^2(\Omega)$ (with Lebesgue measure induced from $\mathbb{R}^d$) in various realisations, namely with certain canonical domains.

 The \emph{maximal realisation}\index{maximal realisation of differential operator}\index{operator!maximal realisation} (or also \emph{weak realisation}) of $L$ in $L^2(\Omega)$ is the operator $A_\mathrm{max}$ defined as
 \[
  \begin{split}
   \mathcal{D}(A_\mathrm{max})\;&:=\;\left\{ 
   g\in L^2(\Omega)\,\left|
   \begin{array}{c}
    \exists \,h_g\in L^2(\Omega)\;\textrm{ such that} \\
    \langle L^\dagger\varphi,g\rangle_{L^2(\Omega)}\,=\,\langle \varphi,h_g\rangle_{L^2(\Omega)}\quad\forall \varphi\in C^\infty_c(\Omega) \\
    \textrm{i.e.,\;\; $Lg=h_g$\;\; weakly} 
   \end{array}
   \right.\right\} \\
   &\,\,=\;\big\{ g\in L^2(\Omega)\,\big|\, Lg\in\mathcal{D}'(\Omega)\textrm{ satisfies } Lg\in  L^2(\Omega)\big\}\,, \\
   A_\mathrm{max} g\;&:=\;Lg\;=\;h_g\,.
  \end{split}
 \]
 (In the definition above $h_g$ is clearly uniquely determined by $g$, owing to the density of $C^\infty_c(\Omega)$ in $L^2(\Omega)$.) In other words, $A_\mathrm{max}$ is the largest possible operator in $L^2(\Omega)$ associated to the differential action $L$ by distribution theory. On a generic function in its domain, $A_\mathrm{max}$ acts with weak derivatives.

 The maximal realisation\index{maximal realisation of differential operator}\index{operator!maximal realisation} of $L$ satisfies
 \[
  A_\mathrm{max}\;=\;\big( L^\dagger|_{C^\infty_c(\Omega)}\big)^*\,,
 \]
 meaning $L^\dagger|_{C^\infty_c(\Omega)}$ as the operator in $L^2(\Omega)$ with domain $C^\infty_c(\Omega)$ and action $\psi\mapsto L^\dagger\psi$. As such, $A_\mathrm{max}$ is a (densely defined and) closed operator. In particular, $\mathcal{D}(A_\mathrm{max})$ is closed with respect to the graph norm $\psi\mapsto\|\psi\|_{\Gamma(L)}\equiv(\|\psi\|^2_{L^2}+\|L\psi\|^2_{L^2})^\frac{1}{2}$.

 The above identity $\langle \varphi,L\psi\rangle_{L^2(\Omega)}=\langle L^\dagger\varphi,\psi\rangle_{L^2(\Omega)}$, valid for any $\psi,\varphi\in C_c^\infty(\Omega)$, shows, in view of the definition of the adjoint, that
 \[
  L|_{C^\infty_c(\Omega)}\;\subset\;A_\mathrm{max}\,,
 \]
 thus implying that $L|_{C^\infty_c(\Omega)}$ is closable as an operator in $L^2(\Omega)$. The \emph{minimal realisation}\index{minimal realisation of differential operator}\index{operator!minimal realisation} (or also \emph{strong realisation} of $L$ in $L^2(\Omega)$) is the operator
 \[
  A_{\mathrm{min}}\;:=\;\overline{\,L|_{C^\infty_c(\Omega)}\,}\,,
 \]
 that is,
 \[
  \begin{split}
   \mathcal{D}(A_{\mathrm{min}})\;&:=\;\overline{\,C^\infty_c(\Omega)\,}^{\|\cdot\|_{\Gamma(L)}}\,, \\
   A_{\mathrm{min}}\phi\;&:=\; L\phi\qquad \forall\phi\in\mathcal{D}(A_{\mathrm{min}})\,.
  \end{split}
 \]
 Thus,
 \[
  A_{\mathrm{min}}\;\subset\;A_{\mathrm{max}}
 \]
 and $A_{\mathrm{min}}$ is the smallest closed restriction of $A_{\mathrm{max}}$ whose domain contains $C^\infty_c(\Omega)$.  On a generic function in its domain, $A_\mathrm{min}$ acts with classical derivatives.

 One defines in a completely analogous manner the maximal and minimal realisation of the formal adjoint $L^\dagger$ of $L$. In order to compare all such operators it is convenient to switch to the self-explanatory notation
 \[
  L_{\mathrm{max}}(\equiv A_{\mathrm{max}})\,,\qquad L_{\mathrm{min}}(\equiv A_{\mathrm{min}})\,,\qquad L^\dagger_{\mathrm{max}}\,,\qquad L^\dagger_{\mathrm{min}}
 \]
 that treats all such realisations on the same footing. Thus,
 \[
 \begin{array}{rlcrl}
  L_{\mathrm{max}}\;=&\big( L^\dagger|_{C^\infty_c(\Omega)}\big)^*\,, & \qquad &  L^\dagger_{\mathrm{max}}\;=& \big( L|_{C^\infty_c(\Omega)}\big)^*\,,  \\
   L|_{C^\infty_c(\Omega)}\;\subset& L_{\mathrm{max}}\,, & \qquad &  L^\dagger|_{C^\infty_c(\Omega)}\;\subset& L^\dagger_{\mathrm{max}}\,, \\
   L_{\mathrm{min}}\;=&\overline{\,L|_{C^\infty_c(\Omega)}\,}\,, & \qquad & L^\dagger_{\mathrm{min}}\;=& \overline{\,L^\dagger|_{C^\infty_c(\Omega)}\,}\,.
 \end{array}
 \]
%
%
 In fact, it turns out that
 \[
  \begin{split}
   L_{\mathrm{max}}\;&=\;\big(L^\dagger_{\mathrm{min}}\big)^*\,,\qquad L^\dagger_{\mathrm{min}}\;=\;(L_{\mathrm{max}})^*\,, \\
   L^\dagger_{\mathrm{max}}\;&=\;\big(L_{\mathrm{min}}\big)^*\,,\qquad L_{\mathrm{min}}\;=\;\big(L^\dagger_{\mathrm{max}}\big)^*\,.
  \end{split}
 \]

 If the formal differential operator $L$ considered at the beginning is formally self-adjoint, then
 \[
  L|_{C^\infty_c(\Omega)}\;\subset\;A_{\mathrm{min}}\;\subset\;(A_{\mathrm{min}})^*\;=\;A_{\mathrm{max}} \qquad\qquad \textrm{($L$ formally self-adjoint)}\,.
 \]
 In the standard language introduced in Section \ref{sec:I-symmetric-selfadj} below, in this case one says that $L|_{C^\infty_c(\Omega)}$ is a densely defined and symmetric operator, and $A_{\mathrm{min}}$ is a densely defined, closed, and symmetric operator. A typical example of this case occurs when $L$ has real-valued smooth coefficients.


%
%
%
%
%
%
%
%
%
%
%
%
%
%

\section{Bounded operators. Compacts. Unitaries. Orthogonal projections.}\label{sec:bdd-compacts-unitaries-orthproj}

By $\mathcal{B}(\cH)$ one denotes the collection of all bounded operators on $\cH$, with the natural structure of normed, complex, involutive, associative algebra over $\mathbb{C}$ given by operator sums, products, and multiplications by scalars defined point-wise on the vectors of $\cH$, and where the norm is the operator norm with the above-mentioned triangular and product inequality, and the involution \index{involution} is the (anti-linear) adjoint map $A\mapsto A^*$ with the properties
\[
 A^{**}=A\,,\qquad (AB)^*=B^*A^*\,,\qquad (\alpha A+\beta B)^*=\overline{\alpha}A^*+\overline{\beta}B^*
\]
for arbitrary $A,B\in\mathcal{B}(\cH)$ and $\alpha,\beta\in\mathbb{C}$. Moreover, $\mathcal{B}(\cH)$ is complete in the operator norm topology and hence is a Banach space, and
\[
 \|A^*\|_{\mathrm{op}}\;=\;\|A\|_{\mathrm{op}}\qquad \forall A\in\mathcal{B}(\cH)\,,
\]
that is, the involution  $A\mapsto A^*$ is an anti-linear automorphism of $\mathcal{B}(\cH)$. By definition, such properties qualify $\mathcal{B}(H)$ as a Banach $*$-algebra (with identity), which is non-commutative as long as $\dim\cH>1$. In addition,
\[
 \|A^*A\|_{\mathrm{op}}\;=\;\|A\|_{\mathrm{op}}^2\qquad \forall A\in\mathcal{B}(\cH)\qquad (C^*\textrm{-\emph{condition}})\,,
\]
which, by definition, makes $\mathcal{B}(\cH)$ a $C^*$-algebra \index{operator!$C^*$-algebra} (with identity).


An element $U\in\mathcal{B}(\cH)$ is called \emph{unitary operator}\index{operator!unitary} if it is surjective and norm-preserving, i.e., $\mathrm{ran}\,U=\cH$ and $\|U\psi\|=\|\psi\|$ $\forall\psi\in\cH$. In this case, $\ker U=\{0\}$ and $U^{-1}=U^*$. Thus, equivalently, $U$ is unitary if $UU^*=U^*U=\mathbbm{1}$. Given an operator $A$ and a unitary $U$ on $\cH$, the two operators $A$ and $UAU^*$ are said to be \emph{unitarily equivalent}.\index{operator!unitarily equivalent} If $A$ is closed, closable, or bounded, so too is any unitarily equivalent version of it.

Given a closed subspace $V$ of a Hilbert space $\cH$, the map $P_V$ such that $P_V\psi:=\psi_V$ $\forall\psi\in\cH$, defined in view of the orthogonal decomposition $\cH=V\oplus V^\perp$, $\psi=\psi_V+\psi_{V^\perp}$, is an everywhere defined and bounded operator on $\cH$ called the \emph{orthogonal projection onto $V$}. In particular, $\|P_V\|_{\mathrm{op}}=1$, $\ker P_V=V^\perp$, $\mathrm{ran}\,P_V=V$, $P_V^2=P_V=P_V^*$.

In general, an \emph{orthogonal projection}\index{orthogonal projection}\index{operator!orthogonal projection} is an operator $P\in\mathcal{B}(\cH)$ such that $P=P^*=P^2$. An orthogonal projection $P$ induces the orthogonal decomposition $\cH=P\cH\oplus(\mathbbm{1}-P)\cH$. It is customary to denote with $|\psi\rangle\langle\psi|$ the orthogonal projection onto $\mathrm{span}\{\psi\}$ for any $\psi\in\cH$.

For an operator $A\in\mathcal{B}(\cH)$ it is equivalent that
\begin{itemize}
 \item for every bounded sequence $(\psi_n)_{n\in\mathbb{N}}$ in $\cH$, the sequence $(A\psi_n)_{n\in\mathbb{N}}$ has a convergent subsequence;
 \item every sequence $(\psi_n)_{n\in\mathbb{N}}$ in $\cH$ that is weakly convergent to some $\psi\in\cH$ is mapped by $A$ into the norm-convergent $(A\psi_n)_{n\in\mathbb{N}}$ to $A\psi$, i.e., $\psi_n\rightharpoonup\psi\Rightarrow A\psi_n\to A\psi$;
\end{itemize}
in which case $A$ is said to be a \emph{compact operator}.\index{operator!compact} If in addition $\cH$ is separable, the above two conditions, hence the compactness of $A$, are equivalent to
\begin{itemize}
 \item there is a sequence $(A_n)_{n\in\mathbb{N}}$ in $\mathcal{B}(\cH)$ of finite-rank operators such that $\|A-A_n\|_{\mathrm{op}}\to 0$.
\end{itemize}
In particular, any finite-rank operator is compact. The simplest rank-one operator is the operator denoted by $|\psi\rangle\langle\varphi|$, for given $\psi,\varphi\in\cH$, where the notation stands for the map $\cH\ni\xi\stackrel{|\psi\rangle\langle\varphi|}{\longmapsto}\langle\varphi,\xi\rangle\,\psi$. Actually, $\||\psi\rangle\langle\varphi|\|_{\mathrm{op}}=\|\psi\|\|\varphi\|$, and kernel and range of $|\psi\rangle\langle\varphi|$ are, respectively, $\{\varphi\}^\perp$ and $\mathrm{span}\{\psi\}$. The adjoint of $|\psi\rangle\langle\varphi|$ is $|\varphi\rangle\langle\psi|$. In particular, $|\psi\rangle\langle\psi|$ is a rank-one orthogonal projection.

Clearly, if $\dim\cH=\infty$, then the identity operator $\mathbbm{1}\in\mathcal{B}(\cH)$ is not compact.

For any two operators $A,C\in\mathcal{B}(\cH)$, if $C$ is compact then so too are $AC$ and $CA$. Compact operators on a Hilbert space $\cH$ constitute a sub-$C^*$-algebra of $\mathcal{B}(\cH)$. 
When $\dim\cH<\infty$, it coincides with $\mathcal{B}(\cH)$. When instead $\dim\cH=\infty$, compact operators on $\cH$ form the only non-trivial ideal\index{ideal} of $\mathcal{B}(\cH)$ that is closed in the operator norm topology. The deep and rich structure and properties of this ideal and its (non-closed) sub-ideals has no special impact on the matter of this monograph, and hence no further preliminary material is reviewed here in this respect.

The sole exception that is relevant to mention is that class of compact operators on $L^2(\mathbb{R}^d,\ud x)$, $d\in\mathbb{N}$, known as \emph{Hilbert-Schmidt operators}\index{operator!Hilbert-Schmidt} (on $L^2(\mathbb{R}^d)$): they are all the $A$'s in $\mathcal{B}(L^2(\mathbb{R}^d))$ for which there exists a function $K_A\in L^2(\mathbb{R}^d\times\mathbb{R}^d,\ud x\,\ud y)$, uniquely identified by $A$ and called the \emph{integral kernel}\index{integral kernel}\index{operator!integral kernel} of $A$, such that
\[
 (A\psi)(x)\;=\;\int_{\mathbb{R}^3}K_A(x,y)\psi(y)\,\ud y\qquad\textrm{for a.e.~$x\in\mathbb{R}^d$, }\quad\forall\psi\in L^2(\mathbb{R}^d,\ud x)\,.
\]
Hilbert-Schmidt operators (on $L^2(\mathbb{R}^d)$) are a non-closed ideal of $\mathcal{B}(L^2(\mathbb{R}^d))$, and any Hilbert-Schmidt operator\index{operator!Hilbert-Schmidt} $A$ on $L^2(\mathbb{R}^d)$ satisfies
\[
 \|A\|_{\mathrm{op}}\:\leqslant\;\|K_A\|_{L^2(\mathbb{R}^d\times\mathbb{R}^d)}\,.
\]
The r.h.s.~above is called the \emph{Hilbert-Schmidt norm}\index{Hilbert-Schmidt norm} of $A$.

\section{Invariant and reducing subspaces}\label{sec:I_invariant-reducing-ssp}

Given two operators $A_1$ and $A_2$, acting respectively on the Hilbert spaces $\cH_1$ and $\cH_2$, their \emph{operator orthogonal sum} \index{operator orthogonal sum} is the operator $A_1\oplus A_2$ acting on $\cH_1\oplus\cH_2$ defined by
\[
 \begin{split}
  \mathcal{D}(A_1\oplus A_2)\;&:=\;\mathcal{D}(A_1)\boxplus\mathcal{D}(A_2)\,, \\
  (A_1\oplus A_2)(\psi_1,\psi_2)\;&:=\;(A_1\psi_1,A_2\psi_2)\quad \forall \psi_1\in\mathcal{D}(A_1)\,,\forall\psi_2\in\mathcal{D}(A_2)\,.
 \end{split}
\]
It is equivalent that both $A_1$ and $A_2$ are closed on the respective Hilbert space, and that $A_1\oplus A_2$ is closed on $\cH_1\oplus\cH_2$.

Analogously, from the collection $(A_j)_{j\in\mathbb{N}}$ of operators acting on the respective Hilbert space of the collection $(\cH_j)_{j\in\mathbb{N}}$ one defines
\[
 \begin{split}
  \mathcal{D}\bigg(\bigoplus_{j\in\mathbb{N}}A_j\bigg)\;&:=\;\displaystyle\op_{j\in\mathbb{N}}\mathcal{D}(A_j)\;=\;
\left\{
 \begin{array}{c}
  \psi\equiv(\psi_j)_{j\in\mathbb{N}}\;\textrm{ with }\;\psi_j\in\mathcal{D}(A_j) \\
  \textrm{and such that }\sum_{j\in\mathbb{N}}\|\psi_j\|^2_{\cH_j}<+\infty \\
  \textrm{and }\sum_{j \in \mathbb{N}} \|A_j \psi_j\|_{\cH_j}^2 < + \infty 
 \end{array} 
 \right\}, \\
 \bigoplus_{j\in\mathbb{N}}A_j\;\psi\;&:=\;(A_j\psi_j)_{j\in\mathbb{N}}\,.
 \end{split}
\]
By construction, each component domain $\mathcal{D}(A_j)$ is invariant under the direct sum operator, as it is mapped into the corresponding $\cH_j$.

More generally, if $A$ is an operator on a Hilbert space $\cH$, a closed subspace $\cH_0\subset\cH$ is called \emph{invariant subspace} \index{invariant subspace} for $A$ when $A$ maps $\mathcal{D}(A)\cap\cH_0$ into $\cH_0$. Associated to an invariant subspace $\cH_0$ for $A$, the operator $A_0:=A\upharpoonright\mathcal{D}(A)\cap\cH_0$ is called the \emph{part of the operator} \index{operator!part on invariant subspace} $A$ on $\cH_0$. An invariant subspace $\cH_0$ for $A$ is called in addition a \emph{reducing subspace}\index{operator!reducing subspace}\index{reducing subspace} if there exist operators $A_0$ on $\cH_0$ and $A_1$ on $\cH_0^\perp$ such that $A=A_0\oplus A_1$ with respect to the decomposition $\cH=H_0\oplus \cH_0^\perp$. When $A$ admits only the trivial reducing subspaces $\{0\}$ and $\cH$ it is said to be \emph{irreducible}. \index{operator!irreducible}

For an operator $A$ on a Hilbert space $\cH$ and a closed subspace $\cH_0\subset\cH$, with associated orthogonal projection $P_0$, the following properties are equivalent:
\begin{enumerate}[(i)]
 \item $\cH_0$ is a reducing subspace for $A$;
 \item $\cH_0^\perp$ is a reducing subspace for $A$;
 \item both $\cH_0$ and $\cH_0^\perp$ are invariant subspaces for $A$ and $P_0\mathcal{D}(A)\subset\mathcal{D}(A)$;
\end{enumerate}
when any of the above is matched, $A=A_0\oplus A_1$, where $A_0:=A\upharpoonright P_0\mathcal{D}(A)$ and $A_1:=A\upharpoonright(\mathbbm{1}-P_0)\mathcal{D}(A)$.

In fact, upon defining the \emph{commutant} \index{commutant} \index{operator commutant} $\{A\}'$ of $A$ as the set
\[
 \{A\}'\;:=\;\{B\in\mathcal{B}(\cH)\,|\,BA\subset AB\}\,,
\]
namely the set of bounded operators on $\cH$ that map $\mathcal{D}(A)$ into itself and such that $AB\psi=BA\psi$ $\forall\psi\in\mathcal{D}(A)$, one has that $\cH_0$ is a reducing subspace for $A$ if and only if $P_0\in\{A\}'$, and $A$ is irreducible if and only the only orthogonal projections belonging to $\{A\}'$ are $\mathbb{O}$ and $\mathbbm{1}$.

 Given a countable collection of Hilbert spaces $(\cH_j)_{j\in\mathbb{N}}$ and a collection $(A_j)_{j\in\mathbb{N}}$ of operators acting on the respective Hilbert space, for the orthogonal sum operator
 \[
  A\;=\;\bigoplus_{j\in\mathbb{N}}A_j\qquad\textrm{on}\qquad \cH\;=\;\bigoplus_{j\in\mathbb{N}}\cH_j
 \]
  one has
  \[
   \begin{split}
      \overline{A}\;&=\;\bigoplus_{j\in\mathbb{N}}\,\overline{A_j}\,, \\
      A^*\;&=\;\bigoplus_{j\in\mathbb{N}}\,A_j^*\,, \\
      \ker A^*\;&=\;\bigoplus_{j\in\mathbb{N}}\,\ker A_j^*
   \end{split}
 \]
  tacitly understanding each operator closure, adjoint, or kernel as taken in the respective Hilbert space. Hence, $A$ is closed (respectively, injective) if and only if all the $A_j$'s are.

\section{Spectrum}\label{sec:I-spectrum}

To a \emph{closed} operator $A$ acting on a Hilbert space $\cH$ one associates two subsets of $\mathbb{C}$, the \emph{resolvent set}\index{resolvent set} $\rho(A)$ and the \emph{spectrum}\index{spectrum} $\sigma(A)$, defined respectively by
\[
 \begin{split}
  \rho(A)\;&:=\;\left\{\lambda\in\mathbb{C}\,\left|\,
  \begin{array}{c}
    A-\lambda\mathbbm{1}\textrm{ is invertible with bounded inverse}  \\
    \textrm{defined everywhere on }\cH
  \end{array}
  \right.\right\}, \\
  \sigma(A)\;&:=\;\mathbb{C}\setminus\rho(A)\,.
 \end{split}
\]
When the dependence on the underlying Hilbert space need be emphasised, one writes $\rho_\cH(A)$, $\sigma_\cH(A)$.
The same definition applied to non-closed operators would yield the trivial notion $\rho(A)=\emptyset$, $\sigma(A)=\mathbb{C}$. When $\lambda\in\rho(A)$,  $A-\lambda\mathbbm{1}$ is a $\mathcal{D}(A)\to\cH$ bijection and the operator $(A-\lambda\mathbbm{1})^{-1}$ is called the \emph{resolvent}\index{resolvent operator} of $A$, which therefore satisfies
\[
 (A-\lambda\mathbbm{1})(A-\lambda\mathbbm{1})^{-1}\;=\;\mathbbm{1}\,,\qquad (A-\lambda\mathbbm{1})^{-1}(A-\lambda\mathbbm{1})\;\subset\;\mathbbm{1}\,.
\]

In the definition of the resolvent set $\rho(A)$, the specification that the $\mathcal{D}(A)\to\cH$  bijection $A-\lambda\mathbbm{1}$ has bounded inverse is customary, but actually redundant, and one can equivalently define $\rho(A)$ and $\sigma(A)$ through the properties
\[
 \begin{split}
  \rho(A)\;&=\;\{\lambda\in\mathbb{C}\,|\,A-\lambda\mathbbm{1}\textrm{ is a }\mathcal{D}(A)\to\cH\textrm{ bijection}\}\,, \\
  \sigma(A)\;&=\;\{\lambda\in\mathbb{C}\,|\,A-\lambda\mathbbm{1}:\mathcal{D}(A)\to\cH\textrm{ is not bijective}\}\,.
 \end{split}
\]
Indeed, as $A$ is closed and invertible, $(A-\lambda\mathbbm{1})^{-1}$ is closed too, and being also everywhere defined it is necessarily bounded owing to the closed graph theorem.

The set $\rho(A)$ is open in $\mathbb{C}$; $\sigma(A)$ is closed. In fact, any closed subset $\Sigma\subset\mathbb{C}$, including the whole $\mathbb{C}$ itself, is the spectrum of a closed operator on some Hilbert space. For a non-empty $\Sigma$ this is standard to see by taking a countable dense $(\lambda_n)_{n\in\mathbb{N}}$ in $\Sigma$ and constructing the diagonal operator $A$ on $\ell^2(\mathbb{N})$ such that the $n$-th basis vector (with respect to the canonical basis) is an eigenvector with eigenvalue $\lambda_n$. Examples of closed operators with empty spectrum are also fairly standard: for instance, the operator $A$ on $\cH=L^2(0,1)$ with $\mathcal{D}(A)=\{f\in H^1(0,1)\,|\,f(0)=0\}$ and $Af=f'$. A more concrete example of closed operator $A$ with $\sigma(A)=\mathbb{C}$ is $A=-\ii\frac{\ud}{\ud x}$ on $\cH=L^2(0,1)$ with $\mathcal{D}(A)=H^1(0,1)$: every $\lambda\in\mathbb{C}$ is eigenvalue of $A$ with eigenvector $e^{\ii\lambda x}$.

If $A$ 
is a closed operator on $\cH$ and $\lambda\in\rho(A)$, then $\mathrm{ran}(A-\lambda\mathbbm{1})$ is closed in $\cH$. 
If in addition $A$ is densely defined and closed, then
%
\[
 \begin{split}
  \sigma(A)^*\;&\equiv\;\{\overline{\lambda}\,|\,\lambda\in\sigma(A)\}\;=\;\sigma(A^*)\,, \\
  \big((A-\lambda\mathbbm{1})^{-1}\big)^*\;&=\;(A^*-\overline{\lambda}\mathbbm{1})^{-1}\qquad  \forall\lambda\in\rho(A)\,, \\
  \cH\;&=\;\mathrm{ran}(A-\lambda\mathbbm{1})\oplus\ker(A^*-\overline{\lambda}\mathbbm{1})\qquad\forall\lambda\in\rho(A)\,.
 \end{split}
\]
(The latter identity above improves the general decomposition $\cH=\overline{\mathrm{ran}(A-\lambda\mathbbm{1})}\oplus\ker(A^*-\overline{\lambda}\mathbbm{1})$ valid, as discussed previously, for any densely defined and closable $A$ and any $\lambda\in\mathbb{C}$.)


For a generic closed operator $A$ on Hilbert space $\cH$, the condition $\lambda\in\sigma(A)$ amounts to the lack of bijectivity of $A-\lambda\mathbbm{1}$ as a $\mathcal{D}(A)\to\cH$ map. The regimes of lack of injectivity or of surjectivity are characterised by introducing the \emph{point spectrum}\index{spectrum!point spectrum}\index{point spectrum} $\sigma_{\mathrm{p}}(A)$, the \emph{continuous spectrum}\index{continuous spectrum}\index{spectrum!continuous spectrum} $\sigma_{\mathrm{c}}(A)$ and the \emph{residual spectrum}\index{residual spectrum}\index{spectrum!residual spectrum} $\sigma_{\mathrm{res}}(A)$ of $A$ as
\[
 \begin{split}
 \sigma_{\mathrm{p}}(A)\;:=&\;\big\{\lambda\in\mathbb{C}\,|\,\ker(A-\lambda\mathbbm{1})\neq\{0\}\big\}\;=\;\{\textrm{eigenvalus of $A$}\}\,, \\
 \sigma_{\mathrm{c}}(A)\;:=&\;\left\{\lambda\in\mathbb{C}\left| 
  \begin{array}{c}
   A-\lambda\mathbbm{1}\textrm{ is injective on $\mathcal{D}(A)$ and not surjective on $\cH$}, \\
   \textrm{and }\,\overline{\mathrm{ran}(A-\lambda\mathbbm{1})}=\cH
  \end{array}
 \right.\!\!\right\} \\
 =&\;\left\{
 \lambda\in\sigma(A)\setminus\sigma_{\mathrm{p}}(A)\left|\,
 \begin{array}{c}
  \mathrm{ran}(A-\lambda\mathbbm{1})\varsubsetneq\overline{\mathrm{ran}(A-\lambda\mathbbm{1})}=\cH
 \end{array}
 \right.\!\!\right\}, \\
   \sigma_{\mathrm{res}}(A)\;:=&\;\left\{
 \lambda\in\mathbb{C}\left|
 \begin{array}{c}
  A-\lambda\mathbbm{1}\textrm{ is injective and }\,\overline{\mathrm{ran}(A-\lambda\mathbbm{1})}\varsubsetneq\cH
 \end{array}
 \right.\!\!\right\} \\
 =&\;\left\{
 \lambda\in\sigma(A)\setminus\sigma_{\mathrm{p}}(A)\left|\,
 \begin{array}{c}
  \overline{\mathrm{ran}(A-\lambda\mathbbm{1})}\varsubsetneq\cH
 \end{array}
 \right.\!\!\right\}.
 \end{split}
\]
%
One has the disjoint union
\[
 \sigma(A)\;=\;\sigma_{\mathrm{p}}(A)\,\,\dot{\cup}\,\,\sigma_{\mathrm{c}}(A)\,\,\dot{\cup}\,\,\sigma_{\mathrm{res}}(A)\,.
\]
Thus, the lack of injectivity of $A-\lambda\mathbbm{1}$ amounts to $\lambda\in\sigma_{\mathrm{p}}(A)$, and the injectivity with lack of surjectivity amounts to $\lambda\in\sigma_{\mathrm{c}}(A)\cup \sigma_{\mathrm{res}}(A)$. The set $\sigma_{\mathrm{p}}(A)$ is the part of the spectrum consisting of eigenvalues of $A$, and for any such $\lambda$ $\ker(A-\lambda\mathbbm{1})$ is the \emph{eigenspace} relative to $\lambda$, its elements are the \emph{eigenvectors} of $A$ relative to $\lambda$, and its dimension is the multiplicity of $\lambda$. Unlike the whole $\sigma(A)$, $\sigma_{\mathrm{p}}(A)$ is not necessarily closed in $\mathbb{C}$: eigenvalues may accumulate to a point that is not an eigenvalue.

One should be cautioned that alternative definitions of point, continuous, and residual spectrum are sometimes adopted in the literature, replacing the ones above and denoted here as
\[
 \begin{split}
 \widehat{\sigma}_{\mathrm{c}}(A)\;:=&\;\left\{\lambda\in\mathbb{C}\left| 
  \begin{array}{c}
   \,\mathrm{ran}(A-\lambda\mathbbm{1})\textrm{ is not closed in $\cH$}
  \end{array}
 \right.\!\!\right\}, \\
   \widehat{\sigma}_{\mathrm{res}}(A)\;:=&\;\left\{
 \lambda\in\mathbb{C}\left|
 \begin{array}{c}
  A-\lambda\mathbbm{1}\textrm{ is injective and }\,\mathrm{ran}(A-\lambda\mathbbm{1})=\overline{\mathrm{ran}(A-\lambda\mathbbm{1})}\varsubsetneq\cH
 \end{array}
 \right.\!\!\right\} \\
 =&\;\left\{
 \lambda\in\sigma(A)\setminus\sigma_{\mathrm{p}}(A)\left|\,
 \begin{array}{c}
  (A-\lambda\mathbbm{1})\textrm{ has bounded inverse} \\
  \textrm{not defined on the whole $\cH$}
 \end{array}
 \right.\!\!\right\}.
 \end{split}
\]
 By definition, $\sigma_{\mathrm{c}}(A)\subset\widehat{\sigma}_{\mathrm{c}}(A)$ and $\sigma_{\mathrm{res}}(A)\supset\widehat{\sigma}_{\mathrm{res}}(A)$, with inclusions that may be strict, and in general $\sigma_{\mathrm{p}}(A)\cap\widehat{\sigma}_{\mathrm{c}}(A)\neq\emptyset$.


For any two closed operators $A_1$ and $A_2$ acting, respectively, on the Hilbert spaces $\cH_1$ and $\cH_2$,
\[
 \begin{split}
  \sigma_{\cH_1\oplus\cH_2}(A_1\oplus A_2)\;&=\;\sigma_{\cH_1}(A_1)\cup\sigma_{\cH_2}(A_2)\,, \\
  \sigma_{\cH_1\oplus\cH_2,\mathrm{p}}(A_1\oplus A_2)\;&=\;\sigma_{\cH_1,\mathrm{p}}(A_1)\cup\sigma_{\cH_2,\mathrm{p}}(A_2)\,.
 \end{split}
\]

Given $\lambda_0\in\rho(A)$ and $\lambda\in\mathbb{C}$ with $\lambda\neq\lambda_0$,
\begin{itemize}
 \item $\lambda\in\rho(A)$ if and only if $(\lambda-\lambda_0)^{-1}\in\rho((A-\lambda_0\mathbbm{1})^{-1})$,
 \item $\lambda\in\sigma_{\mathrm{p}}(A)$ if and only if $(\lambda-\lambda_0)^{-1}\in\sigma_{\mathrm{p}}((A-\lambda_0\mathbbm{1})^{-1})$, in which case both eigenvalues have the same multiplicity.
\end{itemize}


For closed operators $A$ and $B$ with $\mathcal{D}(A)\supset\mathcal{D}(B)$, one has the \emph{resolvent identities}\index{resolvent identities}
\[
 \begin{split}
  R_\lambda(A)-R_\lambda(B)\;=\;R_\lambda(A)(B-A)R_\lambda(B)&\qquad\forall \,\lambda\in\rho(A)\cap\rho(B)\,, \\
  R_\lambda(A)-R_\mu(A)\;=\;(\lambda-\mu)R_\lambda(A)R_\mu(A)&\qquad\forall \,\lambda,\mu\in\rho(A)\,,
 \end{split}
\]
where the customary shorthand $R_\lambda(A):=(A-\lambda\mathbbm{1})^{-1}$ is adopted. In particular, $R_\lambda(A)$ and $R_\mu(A)$ above commute.

When in addition $A$ is everywhere defined and bounded on $\cH$, i.e., $A\in\mathcal{B}(\cH)$, one has the further \emph{resolvent series}\index{resolvent series}
\[
 \begin{split}
  R_\lambda(A)\;&=\;\sum_{n=0}^\infty(\lambda-\lambda_0)^n R_{\lambda_0}(A)^{n+1}\qquad 
  \begin{array}{l}
   \forall\lambda_0\in\rho(A)\,,\;\forall\lambda\in\mathbb{C} \\
   \textrm{with }|\lambda-\lambda_0|<\|R_{\lambda_0}(A)\|_{\mathrm{op}}^{-1}\,,
  \end{array} \\
  R_\lambda(A)\;&=\;-\frac{1}{\lambda}\sum_{n=0}^\infty\Big(\frac{A}{\lambda}\Big)^n\qquad\qquad\qquad 
  \forall\lambda\in\mathbb{C}\textrm{ with }|\lambda|>\|A\|_{\mathrm{op}}\,,
 \end{split}
\]
both series converging in operator norm.
The former shows that the map $\lambda\mapsto R_\lambda(A)$ is analytic on $\rho(A)$ with values in the Banach space $(\mathcal{B}(\cH),\|\cdot\|_{\mathrm{op}})$, and it implies in particular
\[
 \begin{split}
  &\textrm{$R_\lambda(A)\to R_{\lambda_0}(A)$ in operator norm as $\lambda\to\lambda_0\in\rho(A)$,} \\
  &\lim_{\delta\to 0}\,\frac{R_{\lambda+\delta}-R_{\lambda}}{\delta}\;=:\;\frac{\ud}{\ud\lambda}R_{\lambda}(A)\;=\;R_{\lambda}(A)^2\quad\forall\lambda\in\rho(A)
 \end{split}
\]
 (the $\delta$-limit being in the operator norm).

 The latter series above, also known as \emph{Neumann series},\index{Neumann series} implies that $\|R_\lambda(A)\|_{\mathrm{op}}\to 0$ as $|\lambda|\to +\infty$, whence the conclusion that $\sigma(A)$ is never empty when $A\in\mathcal{B}(\cH)\setminus\{\mathbb{O}\}$ (for, otherwise, $\lambda\mapsto R_\lambda(A)$ would be an entire bounded analytic function on the whole $\mathbb{C}$, hence the zero operator, a contradiction).

The Neumann series also implies that $\sigma(A)$, when $A\in\mathcal{B}(\cH)$, is contained in the closed disc of radius $\|A\|_{\mathrm{op}}$ centred at the origin of $\mathbb{C}$. In fact, the smallest disc centred at the origin and containing $\sigma(A)$ has radius
\[
 r(A)\;:=\;\sup_{\lambda\in\sigma(A)}|\lambda|\;=\;\lim_{n\to\infty}\|A^n\|^{\frac{1}{n}}\qquad (A\in\mathcal{B}(\cH))\,,
\]
where the quantity $r(A)$ ($\leqslant \|A\|_{\mathrm{op}}$) is called the \emph{spectral radius}\index{spectral radius} of $A$.

For everywhere defined and bounded operators $A$ and $B$,
\[
 \sigma(AB)\cup\{0\}\;=\;\sigma(BA)\cup\{0\}\qquad (A,B\in\mathcal{B}(\cH))\,.
\]

Spectrum and point spectrum are both invariant under unitary transformations of the Hilbert space $\cH$, that is,
\[
 \sigma(A)\;=\;\sigma(UAU^*)\,,\qquad \sigma_{\mathrm{p}}(A)\;=\;\sigma_{\mathrm{p}}(UAU^*)\qquad (\textrm{$U$ unitary on $\cH$})
\]
for any closed operator $A$ on $\cH$.

 Compact operators exhibit special properties of their resolvent and spectrum. If $A$ is a compact operator on the Hilbert space $\cH$, then either $(\mathbbm{1}-A)^{-1}$ exists in $\mathcal{B}(\cH)$ or there exists $\psi\in\cH\setminus\{0\}$ such that $A\psi=\psi$ (the \emph{Fredholm alternative});\index{Fredholm alternative}\index{theorem!Fredholm alternative} moreover, $\sigma(A)$ is a discrete set having no accumulation point except possibly the value zero, and every non-zero $\lambda\in\sigma(A)$ is an eigenvalue of finite multiplicity (the \emph{Riesz-Schauder theorem}).\index{theorem!Riesz-Schauder} If $\mathrm{dim}\cH=\infty$ and $A$ is compact, then necessarily $0\in\sigma(A)$ (although not necessarily $0\in\sigma_{\mathrm{p}}(A)$). If instead $\mathrm{dim}\cH<\infty$ and $A$ is compact, then $\sigma(A)$ has no accumulation point.

 A closed operator $A$ on Hilbert space $\cH$ is said to have \emph{purely discrete spectrum}\index{spectrum!purely discrete spectrum} (see also Sect.~\ref{sec:I-parts-of-spectrum} below) if $\sigma(A)$ only consists of isolated eigenvalues with finite multiplicity (equivalently, eigenvalues with finite multiplicity and no finite accumulation points). For a closed operator $A$,
 \[
  \begin{array}{c}
   (A-\lambda_0\mathbbm{1})^{-1}\textrm{ is compact} \\
   \textrm{for some }\lambda_0\in\rho(A)
  \end{array}\qquad\Rightarrow\qquad
  \begin{array}{c}
   (A-\lambda\mathbbm{1})^{-1}\textrm{ is compact }\forall \lambda\in\rho(A) \\
   \textrm{and $A$ has purely discrete spectrum\,.}
  \end{array}
 \]

\section{Symmetric and self-adjoint operators}\label{sec:I-symmetric-selfadj}

An operator $A$ on a complex Hilbert space $\cH$ is said to be \emph{symmetric}\index{operator!symmetric} or \emph{Hermitian}\index{operator!Hermitian} (with respect to $\cH$) when
\begin{equation*}
 \langle \psi,A\phi\rangle\;=\;\langle A\psi,\phi\rangle\qquad\forall\psi,\phi\in\mathcal{D}(A)\,,
\end{equation*}
which, by polarisation, is equivalent to
\begin{equation*}
 \langle \psi,A\psi\rangle\;=\;\langle A\psi,\psi\rangle\qquad\forall\psi\in\mathcal{D}(A)
\end{equation*}
and also equivalent to
\begin{equation*}
 \langle \psi,A\psi\rangle\,\in\,\mathbb{R}\qquad\forall\psi\in\mathcal{D}(A)\,.
\end{equation*}
If in addition $\mathcal{D}(A)$ is dense, symmetry of $A$ is tantamount as
\begin{equation*}
 \mathcal{D}(A)\,\subset\,\mathcal{D}(A^*)\qquad\textrm{and}\qquad A\psi=A^*\psi\quad\forall\psi\in\mathcal{D}(A)\,,
\end{equation*}
i.e., $A\subset A^*$. In particular, a densely defined symmetric operator admits a densely defined adjoint.
The symmetric operators on the same Hilbert space clearly form a real vector space (with respect to the ordinary operator sum and multiplication by a scalar).

A symmetric operator is not necessarily closable. For example, given a non-closable operator $B$ on a Hilbert space $\cH$, the operator $A$ on $\cH\oplus\cH$ with $\mathcal{D}(A):=\mathcal{D}(B)\boxplus\{0\}$ and $A(\psi,0):=(0,B\psi)$ is symmetric (because its expectations are all zero) and not closable. Instead, a densely defined symmetric operator $A$ is always closable (because $A\subset A^*$ and $A^*$ is closed), and its closure $\overline{A}$ is symmetric too (because $A\subset A^*$ implies $(\overline{A}^*=)A^*\supset A^{**}=\overline{A}$). If furthermore $A$ is symmetric and everywhere defined on the whole $\cH$, then $A$ is necessarily closed and hence bounded, a result called the \emph{Hellinger-Toeplitz theorem}.\index{theorem!Hellinger-Toeplitz}

The eigenvalues of a symmetric operator $A$ on $\cH$ are necessarily real, and eigenvectors corresponding to distinct eigenvalues are necessarily orthogonal.

A densely defined operator $A$ on $\cH$ is said to be \emph{self-adjoint}\index{operator!self-adjoint} (with respect to $\cH$) if $A=A^*$, and is \emph{essentially self-adjoint}\index{operator!essentially self-adjoint} if $\overline{A}$ is self-adjoint. Thus, for a densely defined $A$,
\[
 \begin{split}
  \textrm{symmetry:} &\qquad A\;\subset\;\overline{A}\;\subset\;A^* \\
  \textrm{essential self-adjointness:} &\qquad A\;\subset\;\overline{A}\;=\;A^* \\
  \textrm{self-adjointness:} &\qquad A\;=\;\overline{A}\;=\;A^*\,,
 \end{split}
\]
where $\overline{A}=A^{**}$, owing to the closability of $A$ and the density of $\mathcal{D}(A^*)$.

By definition, self-adjointness is stronger than essential self-adjointness, which is stronger than symmetry. For a densely defined operator $A$ on $\cH$, since $A\subset A^*$, the self-adjointness of $A$ is equivalent to $\mathcal{D}(A)=\mathcal{D}(A^*)$, and the essential self-adjointness of $A$ is equivalent to $\mathcal{D}(\overline{A})=\mathcal{D}(A^*)$.
If $A$ is everywhere defined and bounded on $\cH$, symmetry and self-adjointness of $A$ are equivalent: indeed, if $A\in\mathcal{D}(\cH)$ and $A\subset A^*$, then necessarily $\mathcal{D}(A^*)=\mathcal{D}(A)=\cH$, so the extension $A^*$ of $A$ must be necessarily $A$ itself. Self-adjoint operators, in particular, are closed.

Any self-adjoint operator is \emph{maximally symmetric}\index{operator!maximally symmetric}, meaning that if $A$ is self-adjoint and $A\subset B$ for some symmetric $B$, then $A=B$. (Indeed, $B\subset B^*$, $B^*\subset A^*$ (which follows from $A\subset B$), and $A^*=A$, whence $B\subset A$.) In particular, if $A$ is essentially self-adjoint, then $\overline{A}$ is the only self-adjoint extension of $A$.

\textbf{Self-adjointness and normality.} A densely defined operator $A$ on Hilbert space $\cH$ is said to be \emph{normal}\index{operator!normal} when $\mathcal{D}(A)=\mathcal{D}(A^*)$ and $\|A\psi\|=\|A^*\psi\|$ $\forall\psi\in\mathcal{D}(A)$. Equivalently, a densely defined operator $A$ is normal if and only if it is closed and satisfies $AA^*=A^*A$ (the latter identity includes the condition $\mathcal{D}(AA^*)=\mathcal{D}(A^*A)$, which in practice is often difficult to check; this is why the previous characterisation is practically more manageable). Any self-adjoint operator is normal, and a densely defined symmetric operator is normal if and only if it is self-adjoint. In particular, for $A\in\mathcal{B}(\cH)$ the normality of $A$ is equivalent to the property $AA^*=A^*A$, and also equivalent to the property $\|A\psi\|=\|A^*\psi\|$ $\forall\psi\in\cH$. Normal operators $A,B$ in $\mathcal{B}(\cH)$ satisfy Fuglede's theorem \index{theorem!Fuglede} $AB=BA$ $\Rightarrow$ $A^*B=BA^*$ and Putnam's \index{theorem!Putnam-Fuglede} corollary to Fuglede's theorem $AT=TB$ $\Rightarrow$ $A^*T=TB^*$ $\forall T\in\mathcal{B}(\cH)$.

\textbf{Self-adjointness and invertibility.} If $A$ is self-adjoint and injective, then its inverse $A^{-1}$ is self-adjoint too. If $A$ is symmetric and surjective ($\mathrm{ran}\,A=\cH$), then $A$ is invertible and self-adjoint and $A^{-1}\in\mathcal{B}(\cH)$.


\textbf{Symmetry / self-adjointness and operator orthogonal sum.} For a finite or countably infinite collection $(A_n)_n$ of operators, each one in the corresponding Hilbert space of the collection $(\cH_n)_n$, their operator orthogonal sum $A:=\bigoplus_n A_n$ on the Hilbert space orthogonal sum $\cH:=\bigoplus_n\cH_n$ is, respectively, symmetric, essentially self-adjoint, or self-adjoint, if and only if so too are all the $A_n$'s.

\textbf{Self-adjointness and reducibility.} For any self-adjoint operator $A$ in $\mathcal{B}(\cH)$, an invariant subspace $\cH_0$ for $A$ is also reducing. This is not true, in general, when $A$ is self-adjoint and unbounded: for example, with respect to the Hilbert space $L^2(\mathbb{R})$ the operator $A=-\ii\frac{\ud}{\ud x}$ with domain $H^1(\mathbb{R})$ is self-adjoint and unbounded, and the Hilbert subspace $\cH_0$ consisting of $L^2$-functions supported in $[0,1]$ is invariant for $A$, and so too is $\cH_0^\perp$; however, from any $\psi\in H^1(\mathbb{R})$ with $\psi(0)\neq 0$ one obtains $P_{\cH_0}\psi\notin H^1(\mathbb{R})$, so $A$ fails to be reduced over $\cH=\cH_0\oplus\cH_0^\perp$ (Sect.~\ref{sec:I_invariant-reducing-ssp}). 
A closed subspace $\cH_0$ of a Hilbert space $\cH$, which is invariant for a (possibly unbounded) self-adjoint $A$ on $\cH$, is also reducing if $A\upharpoonright\cH_0\cap\mathcal{D}(A)$ is self-adjoint in the Hilbert subspace $\cH_0$, in which case
\[
 \overline{\,\cH_0\cap\mathcal{D}(A)\,}^{\|\,\|_{\Gamma(A)}}=\;P_{\cH_0}\mathcal{D}(A)\,.
\]
More generally, if $A$ is a closed and symmetric operator on $\cH$ such that $A\mathcal{D}_0\subset\cH_0$ for a closed subspace $\cH_0\subset\cH$ and a dense subspace $\mathcal{D}_0\subset\cH_0$ with $\mathcal{D}_0\subset\mathcal{D}(A)$, and if in addition $A_0:=A\upharpoonright\mathcal{D}_0$ is essentially self-adjoint with respect to $\cH_0$, then $\cH_0$ is indeed a reducing subspace for $A$, and
\[
\begin{split}
  \overline{A_0}\;&=\;\overline{A\upharpoonright\mathcal{D}_0}\;=\;A\upharpoonright P_{\cH_0}\mathcal{D}(A)\,, \\
  \mathcal{D}(\overline{A_0})\;&=\;\overline{\mathcal{D}_0}^{\|\,\|_{\Gamma(A)}}=\;P_{\cH_0}\mathcal{D}(A)\,.
\end{split}
\]
%

\textbf{Symmetry and spectrum.} Recall once again that eigenvalues of a symmetric operator are real and eigenvectors belonging to distinct eigenvalues are orthogonal. For a densely defined, closed, and symmetric operator $A$, $\sigma(A)$ is \emph{one} of the following:
\begin{itemize}
 \item the closed upper or lower complex plane ($\overline{\mathbb{C}^+}$ or $\overline{\mathbb{C}^-}$), 
 \item the entire complex plane,
 \item a subset of the real axis.
\end{itemize}

\textbf{Self-adjointness and spectrum -- I.} 
For a closed and symmetric operator $A$, $A$ is self-adjoint if and only if $\sigma(A)\subset\mathbb{R}$, in which case 
\[
 \|A\|_{\mathrm{op}}\;=\;\sup_{\lambda\in\sigma(A)}|\lambda|\,.
\]
Thus, in particular, a self-adjoint $A$ is bounded if and only if $\sigma(A)$ is bounded (hence also compact) in $\mathbb{R}$, in which case the spectral radius amounts to $r(A)= \|A\|_{\mathrm{op}}$. 
For a self-adjoint operator $A$ on a Hilbert space $\cH$ the resolvent set $\rho(A)$ is characterised by a less restrictive condition (than the general definition prescribing the points $\lambda$ of $\rho(A)$ to be those for which $A-\lambda\mathbbm{1}$ has bounded inverse defined everywhere on $\cH$), and actually 
\[
 \begin{split}
  \rho(A)\;&=\;\{\,\lambda\in\mathbb{C}\,|\,A-\lambda\mathbbm{1}\textrm{ has bounded inverse on its range}\,\} \\
  &=\;\left\{\lambda\in\mathbb{C}\,\left|
  \begin{array}{c}
   \|(A-\lambda\mathbbm{1})\psi\|\geqslant c_\lambda\|\psi\| \\
   \textrm{for some }c_\lambda>0 \textrm{ and }\forall\psi\in\mathcal{D}(A)
  \end{array}
  \right.\!\right\} \\
  &=\;\{\,\lambda\in\mathbb{C}\,|\,\mathrm{ran}(A-\lambda\mathbbm{1})=\cH\,\}\qquad\qquad\qquad\qquad\qquad (A=A^*)\,.
 \end{split}
\]
Moreover, when $A=A^*$, $\rho(A)\supset\mathbb{C}\setminus\mathbb{R}$ and
\[
 \|(A-\lambda\mathbbm{1})^{-1}\|_{\mathrm{op}}\;\leqslant\;|\mathfrak{Im}\lambda|^{-1}\qquad (A=A^*,\;\lambda\in\mathbb{C}\setminus\mathbb{R})\,.
\]
Thus, for a self-adjoint operator $A$ on a Hilbert space $\cH$ and for $\lambda\in\mathbb{R}$,
\[
 \lambda\in\sigma(A)\qquad \Leftrightarrow\qquad 
 \left\{
  \begin{array}{c}
   \exists\,(\psi_n)_{n\in\mathbb{N}}\subset\mathcal{D}(A)\textrm{ with }\|\psi_n\|=1\,\forall n\in\mathbb{N} \\
   \textrm{such that }\|A\psi_n-\lambda\psi_n\|\xrightarrow{n\to\infty}0
  \end{array}
 \right.
\]
(\emph{Weyl criterion} for the spectrum).\index{spectrum!Weyl criterion}\index{Weyl criterion!for spectrum}\index{theorem!Weyl (criterion for spectrum)} A sequence $(\psi_n)_{n\in\mathbb{N}}$ of the type above is called \emph{Weyl sequence}\index{Weyl sequence} for $A$ at the spectral point $\lambda$.

\textbf{Self-adjointness and spectrum -- II.} If $A$ is self-adjoint on $\cH$, then $\sigma_{\mathrm{res}}(A)=\emptyset$, because if $A-\lambda\mathbbm{1}$ is injective, then by self-adjointness $\overline{\mathrm{ran}(A-\lambda\mathbbm{1})}=(\ker(A-\lambda\mathbbm{1}))^\perp=\cH$.
%
%
%
%
%

\textbf{Self-adjointness and spectrum -- III: spectrum of a self-adjoint orthogonal sum.} For a given collection $(\cH_j)_{j\in\mathbb{N}}$ of Hilbert spaces, and a given collection $(A_j)_{j\in\mathbb{N}}$ of self-adjoint operators, the $j$-th of which acts in the Hilbert space $\cH_j$, 
 \[
     \sigma\Big(\bigoplus_{j\in\mathbb{N}}A_k\Big) \;=\; \overline{\,\bigcup_{j \in \mathbb{N}} \sigma(A_k)}\,, 
 \]
 where the spectrum on the l.h.s.~is referred to the Hilbert space $\bigoplus_{j\in\mathbb{N}}\cH_j$ and each $j$-th spectrum on the l.h.s.~is referred to $\cH_j$. In particular, for a finite self-adjoint orthogonal sum,
 \[
  \sigma_{\cH_1\oplus\cH_2}(A_1\oplus A_2)\;=\;\sigma_{\cH_1}(A_1)\cup\sigma_{\cH_2}(A_2)\,.
 \]

 \textbf{Self-adjointness and non-emptiness of spectrum.}
 As mentioned above, self-adjoint operators have only real spectrum. Actually, the spectrum of a self-adjoint operator is never empty. In the bounded case, this is consistent with the previously discussed property that $\sigma(A)\neq\emptyset$ for generic (even non-symmetric) $A\in\mathcal{B}(\cH)$. In the unbounded case (where unbounded close operators may have empty spectrum), the non-emptiness of $\sigma(A)$ when $A=A^*$ is a consequence of the spectral theorem presented in the following.

\textbf{Semi-boundedness.} A symmetric operator $A$ on a Hilbert space $\cH$ is said to be \emph{lower semi-bounded}\index{operator!lower semi-bounded} (respectively, \emph{upper semi-bounded}\index{operator!upper semi-bounded}) if
\[
 \langle\psi,A\psi\rangle\;\geqslant\;\mathfrak{m}\|\psi\|^2\qquad(\textrm{respectively, }\langle\psi,A\psi\rangle\;\leqslant\;\mathfrak{m}\|\psi\|^2\,)\qquad\forall\psi\in\mathcal{D}(A)
\]
for some $\mathfrak{m}>0$ which is called a \emph{lower bound}\index{operator!lower bound} (respectively, an \emph{upper bound}\index{operator!upper bound}) for $A$. A lower or upper semi-bounded operator is called \emph{semi-bounded}\index{operator!semi-bounded}. For a lower semi-bounded symmetric $A$, the quantity
\[
 \mathfrak{m}(A)\;:=\;\inf_{\substack{\psi\in\mathcal{D}(A) \\ \psi\neq 0}}\frac{\langle\psi,A\psi\rangle}{\;\|\psi\|^2}\;>\;-\infty
\]
is called the \emph{greatest lower bound}\index{operator!greatest lower bound} of $A$. A lower semi-bounded symmetric operator $A$ is called \emph{positive}\index{operator!positive} (or \emph{non-negative}\index{operator!non-negative}), and one writes $A\geqslant\mathbb{O}$, when $\mathfrak{m}(A)\geqslant 0$, i.e., $\langle\psi,A\psi\rangle\geqslant 0$ $\forall\psi\in\mathcal{D}(A)$, and in particular \emph{strictly positive}\index{operator!strictly positive} when $\langle\psi,A\psi\rangle> 0$ $\forall\psi\in\mathcal{D}(A)\setminus\{0\}$. One also writes $A\geqslant\mathfrak{m}\mathbbm{1}$, respectively $A\leqslant\mathfrak{m}\mathbbm{1}$, for some $\mathfrak{m}\in\mathbb{R}$, to mean that $A-\mathfrak{m}\mathbbm{1}\geqslant\mathbb{O}$, respectively $A-\mathfrak{m}\mathbbm{1}\leqslant\mathbb{O}$. A strictly positive $A$ may still have $ \mathfrak{m}(A)=0$ (for example, $A:=-\Delta$ on $\cH=L^2(\mathbb{R}^d)$, $d\in\mathbb{N}$, with $\mathcal{D}(A):=H^2(\mathbb{R}^d)$).

 For a positive symmetric operator $A$ the \emph{Cauchy-Schwarz inequality}\index{Cauchy-Schwarz inequality} takes the form
 \[
  |\langle\phi,A\psi\rangle|^2\;\leqslant\;\langle\phi,A\phi\rangle\,\langle\psi,A\psi\rangle\qquad\forall\phi,\psi\in\mathcal{D}(A)\,.
 \]
 In particular, if $A$ is densely defined and positive, the condition $\langle\psi,A\psi\rangle=0$ for some $\psi\in\mathcal{D}(A)$ implies $A\psi=0$.

 If an operator $A$ is symmetric (respectively, self-adjoint), injective, and positive, then $A^{-1}$ is symmetric (respectively, self-adjoint) and positive. The same holds expunging `positive' from both the hypothesis and the thesis.
 
 For a densely defined and closed operator $A$,
 \begin{itemize}
  \item $A^*A$ is self-adjoint and positive, and $\mathcal{D}(A^*A)$ is a core for $A$;
  \item $\mathbbm{1}+A^*A$ is a self-adjoint $\mathcal{D}(A^*A)\to\cH$ bijection, whose inverse $(\mathbbm{1}+A^*A)^{-1}$ is self-adjoint in $\mathcal{B}(\cH)$ and satisfies $\mathbb{O}\leqslant(\mathbbm{1}+A^*A)^{-1}\leqslant\mathbbm{1}$.
 \end{itemize}

%

\textbf{Symmetry and range.} For a symmetric operator $A$ on a Hilbert space $\cH$,
\[
 \|(A-\lambda\mathbbm{1})\psi\|\;\geqslant\;|\mathfrak{Im}\lambda|\,\|\psi\|\qquad \forall\psi\in\mathcal{D}(A)\,,\;\forall\lambda\in\mathbb{C}\,,
\]
and for a symmetric and lower semi-bounded $A$ on $\cH$,
\[
 \|(A-\lambda\mathbbm{1})\psi\|\;\geqslant\;(\mathfrak{m}(A)-\lambda)\,\|\psi\|\qquad \forall\psi\in\mathcal{D}(A)\,,\;\forall\lambda\in\mathbb{R}\,.
\]
In particular, 
\begin{itemize}
 \item[--] if $A$ is symmetric and $\lambda\in\mathbb{C}\setminus\mathbb{R}$,
 \item[--] or if $A$ is symmetric and lower semi-bounded, and $\lambda<\mathfrak{m}(A)$,
\end{itemize}
then $A-\lambda\mathbbm{1}$ has bounded inverse on $\mathrm{ran}(A-\lambda\mathbbm{1})$. 
If $A$ in addition is symmetric and closable, and $\lambda\in\mathbb{C}\setminus\mathbb{R}$ or $\lambda<\mathfrak{m}(A)$, then the above bounds for $\|(A-\lambda\mathbbm{1})\psi\|$ imply 
\begin{itemize}
 \item $A-\lambda\mathbbm{1}$ has bounded inverse on $\mathrm{ran}(A-\lambda\mathbbm{1})$,
 \item $\overline{A}-\lambda\mathbbm{1}$ has bounded inverse on $\mathrm{ran}(\overline{A}-\lambda\mathbbm{1})$,
 \item $\mathrm{ran}(\overline{A}-\lambda\mathbbm{1})=\overline{\mathrm{ran}(A-\lambda\mathbbm{1})}$.
\end{itemize}
(The latter property is consistent with the analogous general property stated at the end of Section \ref{sec:I-adjoint}.) In particular, if $A$ is symmetric and densely defined, and $\lambda\in\mathbb{C}\setminus\mathbb{R}$ or $\lambda<\mathfrak{m}(A)$, then all three conclusions right above are valid and 
\[
 \cH\;=\;\mathrm{ran}(\overline{A}-\lambda\mathbbm{1})\oplus\ker(A^*-\overline{\lambda}\mathbbm{1})
\]
(compare also this with analogous general property stated at the end of Section \ref{sec:I-adjoint}.)

\textbf{Deficiency indices.} The \emph{Krasnosel'ski\u{\i}-Kre{\u\i}n theorem}\index{theorem!Krasnosel'ski\u{\i}-Kre{\u\i}n} states that for a closable operator $A$ (thus including a densely defined and symmetric $A$) on a Hilbert space $\cH$, the cardinal number $\dim(\mathrm{ran}(A-\lambda\mathbbm{1}))^\perp$ is constant in $\lambda$ on each connected component of the open set
\[
 \left\{\lambda\in\mathbb{C}\left|
 \begin{array}{c}
  \exists\, c_\lambda>0\textrm{ such that} \\
  \| (A-\lambda\mathbbm{1})\psi\|\geqslant c_\lambda\|\psi\|\quad\forall\psi\in\mathcal{D}(A) \\
  \textrm{(i.e., $A-\lambda\mathbbm{1}$ has bounded inverse)}
 \end{array}
\right.\!\!\right\}\supset\;\rho(A)\,,
\]
and in particular for a closable and symmetric operator $A$ on $\cH$ $\dim(\mathrm{ran}(A-\lambda\mathbbm{1}))^\perp$ is constant in $\lambda$ on each upper and lower complex half-plane $\mathbb{C}^\pm$. This justifies the definition of the cardinal numbers
\[
 \begin{split}
  d_+(A)\;&:=\;\dim(\mathrm{ran}(A-\overline{\lambda}\mathbbm{1}))^\perp\,,\qquad \mathfrak{Im}\lambda>0\,, \\
  d_-(A)\;&:=\;\dim(\mathrm{ran}(A-\overline{\lambda}\mathbbm{1}))^\perp\,,\qquad \mathfrak{Im}\lambda<0\,,
 \end{split}
\]
called the \emph{deficiency indices}\index{deficiency indices} of the symmetric and closable $A$.
In the more restrictive case when $A$ is densely defined and symmetric, one also has
\[
 \begin{split}
  d_+(A)\;&=\;\dim(\mathrm{ran}(A-\overline{\lambda}\mathbbm{1}))^\perp\;=\;\dim\ker(A^*-\lambda\mathbbm{1})\,,\qquad \mathfrak{Im}\lambda>0\,, \\
  d_-(A)\;&=\;\dim(\mathrm{ran}(A-\overline{\lambda}\mathbbm{1}))^\perp\;=\;\dim\ker(A^*-\lambda\mathbbm{1})\,,\qquad \mathfrak{Im}\lambda<0\,.
 \end{split}
\]
A conventional choice in this setting is $d_\pm(A)=\dim\ker(A^*\mp\ii\mathbbm{1})$. As a further part of the Krasnosel'ski\u{\i}-Kre{\u\i}n theorem\index{theorem!Krasnosel'ski\u{\i}-Kre{\u\i}n}, if $A$ is densely defined and symmetric on $\cH$ with the additional property that $A$ is lower semi-bounded, then
\[
\begin{split}
 & d_+(A)\;=\;d_-(A)\;=\;\dim(\mathrm{ran}(A-\lambda\mathbbm{1}))^\perp\;=\;\dim\ker(A^*-\lambda\mathbbm{1}) \\
 & \qquad \qquad\forall \lambda\in\mathbb{C}\setminus[\mathfrak{m}(A),+\infty)\,,
\end{split}
\]
 and similarly, if $A$ is densely defined and symmetric on $\cH$ with the additional property that $A-\lambda_0\mathbbm{1}$ has bounded inverse on $\mathrm{ran}(A-\lambda_0\mathbbm{1})$ for some $\lambda_0\in\mathbb{R}$, then
  \[
 d_+(A)\;=\;d_-(A)\;=\;\dim(\mathrm{ran}(A-\lambda_0\mathbbm{1}))^\perp\;=\;\dim\ker(A^*-\lambda_0\mathbbm{1})\,.
\]
 In the latter two cases considered above, the cardinal number number $d_+(A)\;=\;d_-(A)$ is called \emph{the} deficiency index of $A$. For a given finite or countably infinite collection $(\cH_j)_{j}$ of Hilbert spaces, and a given collection $(A_j)_{j}$ of closable symmetric operators, the $j$-th of which acts in the Hilbert space $\cH_j$, 
 \[
  d_\pm\Big(\bigoplus_{j}A_j\Big)\;=\;\sum_j d_\pm(A_j)\,,
 \]
 each deficiency index above being computed in the respective Hilbert space.

\textbf{Basic criterion of self-adjointness -- I.} The domain of the adjoint of a densely defined and symmetric operator $A$ on a Hilbert space $\cH$ has the following structure:
\[
 \begin{split}
  & \mathcal{D}(A^*)\;=\;\mathcal{D}(\overline{A})\dotplus\ker(A^*-\lambda\mathbbm{1})\dotplus\ker(A^*-\overline{\lambda}\mathbbm{1})\qquad \forall\lambda\in\mathbb{C}\setminus\mathbb{R}\,, \\
  & \dim\big(\mathcal{D}(A^*)/\mathcal{D}(\overline{A})\big)\;=\;d_+(A)+d_-(A)
 \end{split}
\]
(\emph{von Neumann's formula}\index{von Neumann's formula}\index{theorem!von Neumann (formula)}). This determines the following \emph{basic criterion for (essential) self-adjointness}\index{basic criterion for self-adjointness}: if $A$ is a densely defined and symmetric operator on $\cH$, and if $\lambda_\pm\in\mathbb{C}$ with $\mathrm{Im}\lambda_\pm\gtrless 0$, then
\[
 \begin{split}
  \begin{array}{c}
   \textrm{$A$ is essentially} \\
   \textrm{self-adjoint}
  \end{array}
   \;&\quad\Leftrightarrow\quad\;
  \left\{ 
   \begin{array}{c}
    \ker(A^*-\lambda_\pm\mathbbm{1})=\{0\} \\
    \textrm{i.e., }d_+(A)=d_-(A)=0
   \end{array}
  \right.\;\quad\Leftrightarrow\quad\;
  \overline{\mathrm{ran}(A-\overline{\lambda_\pm}\mathbbm{1})}=\cH\,, \\
  \begin{array}{c}
   \textrm{$A$ is self-adjoint}
  \end{array}
   \;&\quad\Leftrightarrow\quad\;
  \left\{ 
   \begin{array}{c}
    \textrm{$A$ is closed, and} \\
    \ker(A^*-\lambda_\pm\mathbbm{1})=\{0\} \\
    \textrm{i.e., }d_+(A)=d_-(A)=0
   \end{array}
  \right.\;\quad\Leftrightarrow\quad\;
  \mathrm{ran}(A-\overline{\lambda_\pm}\mathbbm{1})=\cH\,.
 \end{split}
\]
Somehow more strongly:
\begin{itemize}
 \item for a densely defined, symmetric, and lower semi-bounded $A$ on $\cH$, $A$ is essentially self-adjoint if and only if $\ker(A^*-\lambda\mathbbm{1})=\{0\}$, or equivalently $\overline{\mathrm{ran}(A-\lambda\mathbbm{1})}=\cH$, for one and hence for all $\lambda\in\mathbb{C}\setminus[\mathfrak{m}(A),+\infty)$;
 \item for a densely defined and symmetric $A$ on $\cH$ such that $A-\lambda_0\mathbbm{1}$ has bounded inverse on $\mathrm{ran}(A-\lambda_0\mathbbm{1})$ for some $\lambda_0\in\mathbb{R}$, $A$ is essentially self-adjoint if and only if $\ker(A^*-\lambda_0\mathbbm{1})=\{0\}$, or equivalently $\overline{\mathrm{ran}(A-\lambda_0\mathbbm{1})}=\cH$.
\end{itemize}

\textbf{Basic criterion of self-adjointness -- II.} For symmetric operators whose domain's density is not known a priori, but whose closedness is known, self-adjointness is characterised as follows. First, for an operator $A$ on Hilbert space $\cH$,
\[
 \left\{
  \begin{array}{c}
   \textrm{$A$ is symmetric and, for some }\lambda\in\mathbb{C}, \\
   \mathrm{ran}(A-\lambda\mathbbm{1})=\cH\,,\; \overline{\mathrm{ran}(A-\overline{\lambda}\mathbbm{1})}=\cH
  \end{array}
 \right.
 \quad \Rightarrow \quad
  \left\{
   \begin{array}{c}
     \textrm{$A$ is self-adjoint} \\
     \textrm{and }\lambda,\overline{\lambda}\in\rho(A)\,,
   \end{array}
  \right.
\]
whence, in particular,
\[
\textrm{$A$ is symmetric and }\mathrm{ran}\,A\,=\,\cH
 \quad \Rightarrow \quad
 \textrm{$A$ is self-adjoint and }A^{-1}\in\mathcal{B}(\cH)\,. 
\]
As a consequence, if $A$ is a closed and symmetric operator on the Hilbert space $\cH$ and if $\lambda_\pm\in\mathbb{C}$ with $\mathrm{Im}\lambda_\pm\gtrless 0$, then
\[
   \begin{array}{c}
   \textrm{$A$ is self-adjoint}
  \end{array}
   \;\quad\Leftrightarrow\quad\;
  \left\{ 
   \begin{array}{c}
    \mathrm{ran}(A-\lambda_\pm\mathbbm{1})\,=\,\cH \\
    \textrm{i.e., }d_+(A)=d_-(A)=0
   \end{array}
  \right.\quad \Leftrightarrow \quad \sigma(A)\,\subset\,\mathbb{R}\,.
\]

 \section{Weyl limit-point limit-circle analysis}\label{sec:WeylsCriterion}

 For special classes, relevant in applications, of differential operators on finite or infinite sub-intervals of the real line, the control of the essential self-adjointness, or, in the lack-of, the computation of the deficiency indices, can be made within a framework that predates by at least two decades the very notion of self-adjointness and self-adjoint extension theory, and was developed by Weyl before 1910.

 The first type of operators covered in the Weyl scheme is the minimally defined scalar, second-order, differential (Schr\"{o}dinger) operator
 \[
  A\;=\;-\frac{\ud^2}{\ud x^2}+V(x)\,,\qquad\mathcal{D}(A)\;=\;C^\infty_c(a,b)
 \]
 on the Hilbert space $L^2(a,b)$, for given $a,b$ such that $-\infty\leqslant a<b\leqslant+\infty$ and for a given real-valued, continuous function $V$ on $(a,b)$. $A$ is densely defined and symmetric, hence closable. Its closure $\overline{A}$ is is the minimal realisation (in a completely analogous sense as that of Section \ref{sec:MinimalAndMaximalRealisations})\index{minimal realisation of differential operator}\index{operator!minimal realisation} of the formal differential operator\index{formal differential operator}\index{operator!formal differential} $-\frac{\ud^2}{\ud x^2}+V$, and its adjoint is the maximal realisation,\index{maximal realisation of differential operator}\index{operator!maximal realisation} explicitly,
 \[
  \begin{split}
     \mathcal{D}(A^*)\;&=\;\left\{ 
  g\in L^2(a,b)\,\left|
  \begin{array}{c}
   g,g'\in AC[\alpha,\beta] \\
   \textrm{for any compact sub-interval }\;[\alpha,\beta]\subset(a,b)\,, \\
   \textrm{and }\;-g''+Vg\in L^2(a,b)
  \end{array}
  \!\!\!\right.\right\}, \\
  Ag\;&=\;-g''+Vg\,.
  \end{split}
 \]
 In particular, $A^*=\overline{A}^*$. In the above expression, $AC[\alpha,\beta]$ denotes the space of \emph{absolutely continuous functions}\index{absolutely continuous functions} on $[\alpha,\beta]$, namely those functions $\psi$ on $[\alpha,\beta]$ such that
 \[
  \psi(x)\;=\;\psi(\alpha)+\int_\alpha^x\eta(t)\,\ud t\qquad x\in[\alpha,\beta]
 \]
 for some $\eta\in L^1(\alpha,\beta)$: for any such $\psi$, $\eta$ is uniquely determined and $\psi'=\eta$ almost everywhere on $[\alpha,\beta]$.

 For arbitrary $\lambda\in\mathbb{C}$, the differential problem $-g''+Vg=\lambda g$ on $(a,b)$ has a two-dimensional space of solutions. The \emph{Weyl alternative}\index{Weyl alternative}\index{theorem!Weyl (alternative)} states that precisely one of the following two possibilities is valid at any of the two end points $a,b$, here generally denoted with $\mathsf{p}$:
 \begin{itemize}
  \item For each $\lambda\in\mathbb{C}$, all solutions to $-g''+Vg=\lambda g$ on $(a,b)$ are square-integrable near $\mathsf{p}$ (that is, respectively, on $(a,b)\cap(-\infty,c)$ or  $(a,b)\cap(c,+\infty)$ for some $c$ with $a<c<b$, depending on whether $\mathsf{p}=a$ or $\mathsf{p}=b$). This case is called the \emph{limit-circle case at $\mathsf{p}$}, and one also says that the operator $A$ \emph{is in the limit-circle case at $\mathsf{p}$}.\index{Weyl limit-point/limit-circle}\index{limit-point/limit-circle}
  \item For each $\lambda\in\mathbb{C}$, there exists one solution to $-g''+Vg=\lambda g$ on $(a,b)$ that is not square-integrable near $\mathsf{p}$, in which case, for any $\lambda\in\mathbb{C}\setminus\mathbb{R}$, there is a unique (up to a constant pre-factor) non-zero solution to the differential problem which is square-integrable near $\mathsf{p}$. This case is called the \emph{limit-point case at $\mathsf{p}$}, and one also says that  $A$ \emph{is in the limit-point case at $\mathsf{p}$}.
 \end{itemize}

 When $b=+\infty$, a typical and useful sufficient condition for $A$ to be in the limit-point case at infinity is that for some positive differentiable function $M$ and some $c>a$ one simultaneously has
 \begin{itemize}
  \item $V(x)\,\geqslant\, -M(x)$ $\;\forall x\geqslant c$,
  \item $\displaystyle\int_c^{+\infty}M(x)^{-\frac{1}{2}}\,\ud x\,=\,+\infty$,
  \item the function $M'M^{-\frac{3}{2}}$ is bounded near infinity.
 \end{itemize}
 In particular, if $V$ is bounded from below near infinity, or if $V(x)\geqslant-C x^2$ near infinity for some $C>0$, then $A$ is in the limit-point case at infinity.

 When $a=0$, a typical and useful criterion to decide whether $A$ is in the limit-circle or limit-point case at zero is the following:
 \begin{itemize}
  \item If $V(x)\geqslant\frac{3}{4}x^{-2}$ near zero, then $A$ is in the limit-point case at zero.
  \item If $|V(x)|\leqslant(\frac{3}{4}-\varepsilon)x^{-2}$ near zero, then $A$ is in the limit-circle case at zero.
 \end{itemize}

 The second type of operators covered in the Weyl scheme is the minimally defined spinor, first-order, differential (Dirac) operator
 \[
  A\;=\;-\ii\sigma_2\frac{\ud}{\ud x}+W(x)\,,\qquad\mathcal{D}(A)\;=\;C^\infty_c(a,b)\otimes\mathbb{C}^2
 \]
 on the Hilbert space $L^2(a,b)\otimes\mathbb{C}^2\cong L^2((a,b),\mathbb{C}^2)$, for given $a,b$ such that $-\infty\leqslant a<b\leqslant+\infty$ and for a given $2\times 2$ vector-valued, continuous function $W$ on $(a,b)$ such that for every $x\in(a,b)$ the matrix $W(x)$ is Hermitian and $\sigma_2 W(x)$ is trace-less, where 
 \[
   \sigma_2\;=\;\begin{pmatrix}
               0 & -\ii \\ \ii & 0
              \end{pmatrix}
 \]
 is the second Pauli matrix.\index{Pauli matrices}

 For arbitrary $\lambda\in\mathbb{C}$ the differential problem $-\ii\sigma_2g'+Wg=\lambda g$ on $(a,b)$ has a two-dimensional space of solutions (the order of the derivative is 1, but the problem is now vector-valued).
  The \emph{Weyl alternative}\index{Weyl alternative}\index{theorem!Weyl (alternative)} in this case states that precisely one of the following two possibilities is valid at any of the two end points $a,b$, generally denoted with $\mathsf{p}$:
  \begin{itemize}
   \item For each $\lambda\in\mathbb{C}$, all solutions to $-\ii\sigma_2g'+Wg=\lambda g$ on $(a,b)$ are square-integrable near $\mathsf{p}$. This case is called the \emph{limit-circle case at $\mathsf{p}$}, and one also says that the operator $A$ \emph{is in the limit-circle case at $\mathsf{p}$}.\index{Weyl limit-point/limit-circle}\index{limit-point/limit-circle}
  \item For each $\lambda\in\mathbb{C}$, there exists one solution to $-\ii\sigma_2g'+Wg=\lambda g$ on $(a,b)$ that is not square-integrable near $\mathsf{p}$. This case is called the \emph{limit-point case at $\mathsf{p}$}, and one also says that  $A$ \emph{is in the limit-point case at $\mathsf{p}$}.
  \end{itemize}

  In either settings above for the operator $A$ (Schr\"{o}dinger and Dirac), the following \emph{Weyl limit-point limit-circle criterion}\index{Weyl criterion!limit-point limit-circle}\index{Weyl limit-point/limit-circle}\index{limit-point/limit-circle}\index{theorem!Weyl (limit-point limit-circle criterion)} holds: $A$ has deficiency indices
  \begin{itemize}
   \item $(2,2)$ if $A$ is in the limit-circle case at both end points,
   \item $(1,1)$ if $A$ is in the limit-circle case at one end point and in the limit-point case at the other,
   \item $(0,0)$ if $A$ is in the limit-point case at both end points,
  \end{itemize}
 and no other possibility can occur.
%
%
%
%
%
%
%
%
%
%
%
%
%
%
%
%
%
%
%

\section{Spectral theorem}\label{sec:I_spectral_theorem}

The \emph{spectral theorem}\index{theorem!spectral theorem}\index{spectral theorem} is a complex of results, with various equivalent or related formulations and consequences, whose main statement asserts that given a self-adjoint operator $A$ on a Hilbert space $\cH$ there exists a unique spectral measure $E^{(A)}$ in $\cH$ on the Borel $\sigma$-algebra\index{Borel $\sigma$-algebra} $\mathfrak{B}(\mathbb{R})$ such that
\[\tag{$\bigstar$}\label{eq:spectdecomp}
 A\;=\;\int_\mathbb{R}\lambda \,\ud E^{(A)}(\lambda)\,.
\]
The r.h.s.~of the above identity is referred to as the \emph{spectral decomposition}\index{spectral decomposition} of $A$, and the meaning of its various symbols is going to be explained here below.

\textbf{Spectral measures.} A \emph{spectral measure}\index{spectral measure}, or \emph{projection-valued measure}\index{projection-valued measure}\index{p.v.m.} (`p.v.m.') in the Hilbert space $\cH$ on the Borel $\sigma$-algebra $\mathfrak{B}(\mathbb{R})$ (for short: on $\mathbb{R}$) is a map
\[
 E\,:\,\mathfrak{B}(\mathbb{R})\,\longrightarrow\,\{\textrm{orthogonal projections on $\cH$}\}\,,
\]
the orthogonal projections being those operators $P\in\mathcal{B}(\cH)$ with $P=P^*=P^2$, and $\mathfrak{B}(\mathbb{R})$ denoting the collection of all subsets of $\mathbb{R}$ obtained from the open sets (in the ordinary Euclidean topology) through the operations of countable union, countable intersection, and relative complement, which satisfies the two properties
\begin{enumerate}[(i)]
 \item $E(\mathbb{R})=\mathbbm{1}$,
 \item $E$ is \emph{countably additive}\index{countable additivity}, meaning that for any sequence $(\Lambda_n)_{n\in\mathbb{N}}$ of pairwise disjoint Borel subsets of $\mathbb{R}$, $E(\bigcup_{n\in\mathbb{N}}\Lambda_n)=\sum_{n\in\mathbb{N}}E(\Lambda_n)$, such series converging in the strong operator sense.
\end{enumerate}
Thus, in particular,
\[
 E(\emptyset)=\mathbb{O}\,,\qquad E(\Lambda_1\cap\Lambda_2)=E(\Lambda_1)E(\Lambda_2)\,.
\]
Equivalently, a map $E$ from $\mathfrak{B}(\mathbb{R})$ to the orthogonal projections on $\cH$ is a spectral measure if and only if $E(\mathbb{R})=\mathbbm{1}$ and for each $\psi\in\cH$ the set function $\mathfrak{B}(\mathbb{R})\ni\Lambda\mapsto E_\psi(\Lambda):=\langle\psi,E(\Lambda)\psi\rangle$ is a (real, positive, finite) measure. Such a measure is called the \emph{scalar spectral measure associated with $E$ and $\psi$}\index{scalar spectral measure}, is customarily denoted also by $\mu_\psi^{(E)}$, and satisfies
\[
 \mu_\psi^{(E)}(\Lambda)\;=\;\|E(\Lambda)\psi\|^2\,,\qquad \mu_\psi^{(E)}(\mathbb{R})\;=\;\|\psi\|^2\:<\:+\infty\,.
\]
In turn, fixed $\psi,\varphi\in\cH$, $\mu_{\psi,\varphi}^{(E)}(\Lambda):=\langle\psi,E(\Lambda)\varphi\rangle$ defines a complex measure on $\mathfrak{B}(\mathbb{R})$ and by polarisation\index{polarisation identity}
\[
 \mu_{\psi,\varphi}^{(E)}\;=\;\frac{1}{4}\Big(\mu_{\psi+\varphi}^{(E)}-\mu_{\psi-\varphi}^{(E)}-\ii\mu_{\psi+\ii\varphi}^{(E)}+\ii\mu_{\psi-\ii\varphi}^{(E)}\Big)\,.
\]
Equivalent symbols that are used interchangeably are
\[
 \ud\mu_{\psi,\varphi}^{(E)}(\lambda)\;\equiv\;\ud\langle\psi,E(\lambda)\varphi\rangle\;\equiv\;\langle\psi,\ud E(\lambda)\varphi\rangle\,,\qquad \ud\mu_{\psi}^{(E)}(\lambda)\;\equiv\;\ud\mu_{\psi,\psi}^{(E)}(\lambda)\,,
\]
with the meaning that
\[
 \int_{\Lambda}\ud\mu_{\psi,\varphi}^{(E)}(\lambda)\;=\;\langle\psi,E(\Lambda)\varphi\rangle\,.
\]
To each projection-valued measure $E$ one associates the set 
\[
 \mathrm{supp}\,E\;:=\;\mathbb{R}\;\setminus\!\!\bigcup_{\substack{\Lambda $\textrm{\small open\normalsize}$ \\ E(\Lambda)=\mathbb{O}}}\Lambda\;\;=\;\;
 \left\{
 \begin{array}{c}
  \textrm{the smallest closed subset $C\subset \mathbb{R}$} \\
  \textrm{such that }E(C)=\mathbbm{1}\,,
 \end{array}\right.
\]
called the \emph{support} of $E$. Observe that, by construction,
\[
 \left. 
 \begin{array}{c}
  \Lambda_0\in\mathfrak{B}(\mathbb{R}) \\
  \psi_0\in\mathrm{ran}\,E(\Lambda_0)
 \end{array}
 \right\}\quad\Rightarrow\quad
 \langle\psi_0,E(\Omega)\psi_0\rangle\;=\;0\quad \forall\Omega\in\mathfrak{B}(\mathbb{R})\textrm{ with }\Omega\cap\Lambda_0=\emptyset\,,
\]
thus meaning that $\mu_{\psi}^{(E)}$ has only support inside $\Lambda_0$.
Furthermore, the notion of projection-valued measure is equivalent to the following notion of resolution of identity. If $E$ is a p.v.m.~on $\mathbb{R}$ with respect to the Hilbert space $\cH$, then the collection $\{E(\lambda)\,|\,\lambda\in\mathbb{R}\}$ defined by
\[
 E(\lambda)\;:=\;E((-\infty,\lambda])
\]
is a \emph{resolution of the identity on $\cH$}\index{resolution of the identity}, which by definition means that the $E(\lambda)$'s are orthogonal projections on $\cH$ satisfying the properties of
\begin{enumerate}[(i)]
 \item \emph{monotonicity}, namely the subspace inclusion $E(\lambda_1)\cH\subset E(\lambda_2)\cH$ (equivalently, the operator ordering $E(\lambda_1)\leqslant E(\lambda_2)$) for $\lambda_1\leqslant\lambda_2$,
 \item \emph{strong right continuity:} $E(\lambda)\psi\xrightarrow[]{\lambda\downarrow\lambda_0}E(\lambda_0)\psi$ (in $\cH$-norm) $\forall \psi\in\cH$, $\forall\lambda_0\in\mathbb{R}$,
 \item \emph{normalisation:} $E(\lambda)\psi\xrightarrow[]{\lambda\to -\infty}0$, $E(\lambda)\psi\xrightarrow[]{\lambda\to +\infty}\psi$ $\forall\psi\in\cH$.
\end{enumerate}
Concerning (ii) above, since
\[
 \| E(\lambda)\psi-E(\mu)\psi\|^2 \;=\;\big| \| E(\lambda)\psi\|^2-\| E(\mu)\psi\|^2\big|\qquad \forall\lambda,\mu\in\mathbb{R}\,,
\]
then $\lambda\mapsto E(\lambda)$ is strongly (right) continuous if and only if $\lambda\mapsto\| E(\lambda)\psi\|^2$ is (right) continuous. Conversely, given a resolution of the identity $\{E(\lambda)\,|\,\lambda\in\mathbb{R}\}$ on $\cH$ there exists a unique p.v.m.~$E$ on $\mathbb{R}$ with respect to $\cH$ such that $E((-\infty,\lambda])=E(\lambda)$ $\forall\lambda\in\mathbb{R}$.

A prototypical example is $\cH=L^2(\mathbb{R},\ud x)$, $(E(\Lambda)\psi)(x):=\mathbf{1}_\Lambda(x)\psi(x)$ for almost every $x\in\mathbb{R}$, and for any $\psi\in L^2(\mathbb{R},\ud x)$ and any Borel subset $\Lambda\subset\mathbb{R}$, where $\mathbf{1}_\Lambda$ denotes as usual the characteristic function of $\Lambda$ -- that is, each $E(\Lambda)$ is the multiplication by $\mathbf{1}_\Lambda$. In this case the corresponding resolution of the identity $\{E(\lambda)\,|\,\lambda\in\mathbb{R}\}$ is given by the operators $E(\lambda)$ of multiplication by $\mathbf{1}_{(-\infty,\lambda]}$.

\textbf{Spectral integrals.}
Given a spectral measure $E$ on $\mathbb{R}$ with respect to a Hilbert space $\cH$, one constructs \emph{spectral integrals}\index{spectral integral}, namely operators on $\cH$ of the form $\int_{\mathbb{R}}f(\lambda)\ud E(\lambda)$ as the expression in \eqref{eq:spectdecomp} above, defined as follows. The class of functions for which such a construction is possible is the space $\mathcal{S}(\mathbb{R},\mathfrak{B}(\mathbb{R}),E)$ of the $E$-a.e.~finite, $\mathfrak{B}(\mathbb{R})$-measurable functions $f:\mathbb{R}\to\mathbb{C}\cup\{\infty\}$, meaning that measurability of $f$ is with respect to the Borel sets of $\mathbb{R}$ and that $E(\{\lambda\in\mathbb{R}\,|\,f(\lambda)=\infty\})=\mathbb{O}$. The spectral integral of any such $f$ with respect to $E$ is defined along these steps of increasing generality.
\begin{enumerate}
 \item If $f$ is a \emph{really simple function},\index{really simple functions} i.e., $f=\sum_{j=1}^n c_j \mathbf{1}_{\Lambda_j}$ for some $n\in\mathbb{N}$, $c_1,\dots,c_n\in\mathbb{C}$, and $\Lambda_1,\dots\Lambda_n$ pairwise disjoint Borel subsets of $\mathbb{R}$, then 
 \[
  \int_{\mathbb{R}}f(\lambda)\,\ud E(\lambda)\;=\;\int_{\mathbb{R}}\sum_{j=1}^n c_j \mathbf{1}_{\Lambda_j}(\lambda)\,\ud E(\lambda)\;:=\;\sum_{j=1}^n c_j E(\Lambda_j)\;\in\;\mathcal{B}(\cH)\,.
 \]
 The finite additivity of $\ud E(\lambda)$ is enough to guarantee that the above definition is independent of the particular representation of the function $f$.
 \item The really simple functions on $\mathbb{R}$ are dense in the space  $\mathcal{B}(\mathbb{R},\mathfrak{B}(\mathbb{R}))$ of complex-valued, bounded, $\mathfrak{B}(\mathbb{R})$-measurable functions on $\mathbb{R}$ with the  $\|\cdot\|_{\mathrm{sup}}$-norm,
 and $\|\int_{\mathbb{R}}f(\lambda)\,\ud E(\lambda)\|_{\mathrm{op}}\leqslant\|f\|_{\mathrm{sup}}$ for any bounded and really simple $f$. This allows to define, for generic $f\in\mathcal{B}(\mathbb{R},\mathfrak{B}(\mathbb{R}))$,
 \[
  \int_{\mathbb{R}}f(\lambda)\,\ud E(\lambda)\;:=\;\|\cdot\|_{\mathrm{op}}\!-\!\lim_{n\to\infty}\int_{\mathbb{R}}f_n(\lambda)\,\ud E(\lambda)\;\in\;\mathcal{B}(\cH)
 \]
 irrespective of the sequence $(f_n)_{n\in\mathbb{N}}$ of $\|\cdot\|_{\mathrm{sup}}$-approximants of $f$ from the class of really simple functions.
 \item For generic $f\in\mathcal{S}(\mathbb{R},\mathfrak{B}(\mathbb{R}),E)$ one defines
 \[
  \mathcal{D}\Big({\textstyle  \int_{\mathbb{R}}f(\lambda)\,\ud E(\lambda)} \Big)\;:=\;\Big\{\psi\in\cH\,\Big|\,\int_{\mathbb{R}}|f(\lambda)|^2\,\ud\mu_\psi^{(E)}(\lambda)\,<\,+\infty\Big\}\,,
 \]
 which is in fact a dense subspace of $\cH$ equivalently characterised as
 \[
  \begin{split}
  \mathcal{D}\Big({\textstyle  \int_{\mathbb{R}}f(\lambda)\,\ud E(\lambda)} \Big)\;=&\;
  \left\{ 
   \begin{array}{c}
    \psi\in\cH\textrm{ such that} \\
     \displaystyle\int_{\mathbb{R}}f(\lambda)\mathbf{1}_{M_n}(\lambda)\,\ud E(\lambda)\,\psi\textrm{ converges in $\cH$} \\
     \textrm{as }n\to\infty
   \end{array}
  \right\} \\
  M_n\;:=&\;\:\{\lambda\in\mathbb{R}\,|\,|f(\lambda)|\leqslant n\}\,,\quad n\in \mathbb{N}\,,
  \end{split}
 \]
 the above spectral integral of $f\mathbf{1}_{M_n}\in \mathcal{B}(\mathbb{R},\mathfrak{B}(\mathbb{R}))$ 
 being defined in step 2. On such domain one defines
 \[
  \int_{\mathbb{R}}f(\lambda)\,\ud E(\lambda)\;:=\;\textrm{strong--}\!\lim_{\!\!\!\!\!\!\!\!\!\!\!\!\!\!\!\!\!\!\!\!\!\!\!\!\!\!\!\!n\to\infty}\int_{\mathbb{R}}f(\lambda)\mathbf{1}_{M_n}(\lambda)\,\ud E(\lambda)\,.
 \]
\end{enumerate}

By construction,
\[
 \int_{\mathbb{R}}f(\lambda)\,\ud E(\lambda)\;=\;\int_{\mathrm{supp}\,E}f(\lambda)\,\ud E(\lambda)\qquad\forall f\in\mathcal{S}(\mathbb{R},\mathfrak{B}(\mathbb{R}),E)\,. 
\]

Concerning the above spectral decomposition \eqref{eq:spectdecomp}, one concrete expression to compute the action of the operator $\int_{\mathbb{R}}\lambda\,\ud E(\lambda)$ is therefore
\[
 \int_{\mathbb{R}}\lambda\,\ud E(\lambda)\psi\;=\;\lim_{n\to\infty}\;\sum_{j=-n^2}^{n^2-1}\frac{j}{n}\,E\Big(\Big(\frac{j}{n},\frac{j+1}{n}\Big]\Big)\psi
\]
for all $\psi\in\cH$ for which such limit exists in $\cH$.

\textbf{Spectral theorem.}\index{spectral theorem} The spectral theorem then states that for a given self-adjoint operator $A$ on $\cH$, $A$ undergoes the spectral decomposition \eqref{eq:spectdecomp} for a unique projection-valued measure $E^{(A)}$, uniqueness meaning that if $\int_\mathbb{R}\lambda\ud E_1(\lambda)=A=\int_\mathbb{R}\lambda\ud E_2(\lambda)$ for two p.v.m.~$E_1$ and $E_2$, then $E_1=E_2$. Moreover,
\[
 \mathcal{D}(A)\;=\;\Big\{\psi\in\cH\,\Big|\,\int_\mathbb{R}|\lambda|^2\ud\langle\psi,E^{(A)}(\lambda)\psi\rangle<+\infty\Big\}.
\]
For the scalar spectral measure associated with the (p.v.m.~relative to the) operator $A$ and the vector $\psi\in\cH$ one writes $\ud\mu_\psi^{(A)}\equiv\ud\langle\psi,E^{(A)}(\lambda)\psi\rangle$ etc.

\textbf{Spectral theorem for compacts.}\index{spectral theorem} In the special case when the self-adjoint operator $A$ is compact (in particular, a Hermitian matrix), the spectral theorem takes a simplified form that is in fact conceptually independent from the preceding spectral measure construction, meaning that it can be established by more basic functional-analytic methods (specifically: the Fredholm alternative\index{Fredholm alternative}\index{theorem!Fredholm alternative} and the Riesz-Schauder theorem\index{theorem!Riesz-Schauder} for compact operators, or elementary arguments for finite matrices), while of course it fits the scheme of the general spectral theorem. 
Recall (Sect.~\ref{sec:I-spectrum}) that if $A=A^*\in\mathcal{B}(\cH)$ is compact, then $\sigma(A)$ consists of a discrete collection of real eigenvalues, the non-zero ones having finite multiplicity, with possible accumulation if and only if $\dim\cH=\infty$, in which case their limit point is necessarily zero, although zero would not necessarily be an eigenvalue itself. The statement of the spectral theorem in this case takes the form of the \emph{Hilbert-Schmidt theorem}:\index{theorem!Hilbert-Schmidt}\index{spectral theorem}\index{theorem!spectral theorem} if $A=A^*\in\mathcal{B}(\cH)$ is compact, then there exist an at most countable orthonormal basis $(\psi_n)_{n}$ of $\overline{\mathrm{ran}\,A}=(\ker A)^\perp$ and correspondingly a sequence $(\lambda_n)_n$ in $\mathbb{R}\setminus\{0\}$ such that
\[
 A\;=\;\sum_n\lambda_n |\psi_n\rangle\langle\psi_n|\qquad (\textrm{thus, }A\psi_n=\lambda_n\psi_n)\,,
\]
and in addition $\lambda_n\xrightarrow{n\to\infty}0$ if $\dim\overline{\mathrm{ran}\,A}=\infty$. The above sum, if consisting of an infinite number of summands, converges in the operator norm.
In terms of the general spectral theorem, 
\[
 \begin{split}
  E^{(A)}(\Lambda)\;&=\;\sum_{n\,|\,\lambda_n\in\Lambda}|\psi_n\rangle\langle\psi_n|\qquad\forall\Lambda\in\mathfrak{B}(\mathbb{R})\,, \\
  \ud\mu_\psi^{(A)}\;&=\;\sum_{n}|\langle\psi_n,\psi\rangle|^2\delta(\lambda-\lambda_n)\ud\lambda+\|P_{\ker A}\psi\|^2\delta(\lambda)\ud\lambda \qquad\forall\psi\in\cH\,.
 \end{split}
\]
In particular,
\[
 \dim\mathrm{ran}\,E^{(A)}(\mathbb{R}\setminus(-\varepsilon,\varepsilon))\,<\,\infty\quad\forall\varepsilon>0\,,
\]
thus, $E^{(A)}(\Lambda)$ has finite rank for every Borel $\Lambda\in\mathbb{R}\setminus(-\varepsilon,\varepsilon)$.

\section{Functional calculus}\label{sec:I_funct_calc}\index{functional calculus}

Given a self-adjoint operator $A$ on Hilbert space $\cH$, with spectral measure $E^{(A)}$, the assignment $f\mapsto f(A)$, where $f\in\mathcal{S}(\mathbb{R},\mathfrak{B}(\mathbb{R}),E^{(A)})$ and
\[
 f(A)\;:=\;\int_{\mathbb{R}}f(\lambda)\,\ud E^{(A)}(\lambda)
\]
is called the \emph{functional calculus}\index{functional calculus} of $A$. By definition of spectral integral, each such $f(A)$ is a (normal, in fact) operator on $\cH$ with domain
\[
 \mathcal{D}(f(A))\;=\;\Big\{\psi\in\cH\,\Big|\,\int_{\mathbb{R}}|f(\lambda)|^2\,\ud\mu_\psi^{(A)}(\lambda)<+\infty\Big\}\,,
\]
i.e.,
\[
 \psi\,\in\, \mathcal{D}(f(A))\qquad\Leftrightarrow\qquad f\,\in\, L^2(\mathbb{R},\ud\mu_\psi^{(A)})\,.
\]
In particular (as follows from the previous properties of spectral integrals and from the properties of functional calculus\index{functional calculus} listed in the following), 
\begin{itemize}
 \item if $f$ is the function $f(\lambda)=\lambda$, then $f(A)=A$;
 \item for any Borel subset $\Lambda\subset\mathbb{R}$, $\mathbf{1}_\Lambda(A)=E^{(A)}(\Lambda)$;
 \item if $f$ is a polynomial $f(\lambda)=\sum_{n=1}^N a_n\lambda^n$, then $f(A)=\sum_{n=1}^N a_n A^n$ in the ordinary sense;
 \item if $A$ is bounded and self-adjoint, then for any $z\in\mathbb{C}$ the operators $\sum_{n=0}^{\infty}\frac{z^n}{n!}A^n$ (as convergent series in operator norm) and $f_z(A)$ given by the functional calculus with $f_z(\lambda):=e^{z\lambda}$ are the same element of $\mathcal{B}(\cH)$, and for it one writes $e^{zA}$; moreover, the map $\mathbb{C}\ni z\mapsto e^{z\lambda}\in\mathcal{B}(\cH)$ is continuous.
\end{itemize}

Prototypical examples of functional calculus:\index{functional calculus}
\begin{enumerate}[1.]
 \item For given sequences $(\lambda_n)_{n\in\mathbb{N}}$ in $\mathbb{R}$ and $(P_n)_{n\in\mathbb{N}}$ of orthogonal projections onto pairwise orthogonal subspaces of a Hilbert space $\cH$ (i.e., $P_nP_m=\mathbb{O}$ when $n\neq m$) and such that $\sum_{n\in\mathbb{N}} P_n=\mathbbm{1}$, the self-adjoint operator
 \[
  \mathcal{D}(A)\;:=\;\Big\{\psi\in\cH\,\Big|\,\sum_{n\in\mathbb{N}}|\lambda_n|^2\|P_n\psi\|^2<+\infty\Big\}\,,\qquad A\psi\;:=\;\sum_{n\in\mathbb{N}}\lambda_n P_n\psi 
 \]
 acting on $\cH$ has spectral measure
 \[
  E^{(A)}(\Lambda)\;=\;\sum_{\lambda_n\in\Lambda}P_n\,,\qquad \Lambda\subset\mathfrak{B}(\mathbb{R})
 \]
 and functional calculus
  \[
  \mathcal{D}(f(A))\;=\;\Big\{\psi\in\cH\,\Big|\,\sum_{n\in\mathbb{N}}|f(\lambda_n)|^2\|P_n\psi\|^2<+\infty\Big\}\,,\qquad f(A)\psi\;=\;\sum_{n\in\mathbb{N}}f(\lambda_n) P_n\psi\,.
 \]
 \item For a given positive regular Borel measure $\mu$ on an interval $J\subset\mathbb{R}$, the self-adjoint operator $A$
 \[
  \mathcal{D}(A)\;:=\;\Big\{\psi\in L^2(J,\ud\mu)\,\Big|\,\int_J x^2\,|\psi(x)|^2\,\ud\mu(x)<+\infty\Big\}\,,\qquad A \psi\;:=\;x\psi
 \]
 acting on the Hilbert space $L^2(J,\ud\mu)$ 
 has spectral measure
 \[
  E^{(A)}(\Lambda)\psi\;=\;\mathbf{1}_{\Lambda}\psi\,,\qquad \Lambda\subset\mathfrak{B}(\mathbb{R})
 \]
 and functional calculus
 \[
 \begin{split}
    \mathcal{D}(f(A))\;&=\;\Big\{\psi\in L^2(J,\ud\mu)\,\Big|\,\int_J |f(x)|^2|\psi(x)|^2\,\ud\mu(x)<+\infty\Big\}\,,\\
    f(A) \psi\;&=\;f\psi\,.
 \end{split}
 \]
 \item For a given positive regular Borel measure $\mu$ on an interval $J\subset\mathbb{R}$ and a (Borel-) measurable function $\varphi:J\to\mathbb{R}$, the self-adjoint operator $A$
 \[
  \mathcal{D}(A)\;=\;\Big\{\psi\in L^2(J,\ud\mu)\,\Big|\,\int_J|\varphi(x)|^2|\psi(x)|^2\,\ud\mu(x)<+\infty\Big\}\,,\quad A \psi\;=\;\varphi\psi\,.
 \]
 acting on the Hilbert space $L^2(J,\ud\mu)$ 
 has spectral measure
 \[
  E^{(A)}(\Lambda)\psi\;=\;\mathbf{1}_{\varphi^{-1}(\Lambda)}\psi\,,\qquad \Lambda\subset\mathfrak{B}(\mathbb{R})
 \]
 and functional calculus
 \[
  \begin{split}
     \mathcal{D}(f(A))\;&=\;\Big\{\psi\in L^2(J,\ud\mu)\,\Big|\,\int_J |f(\varphi(x))|^2|\psi(x)|^2\,\ud\mu(x)<+\infty\Big\}\,, \\
     f(A) \psi\;&=\;f(\varphi(x))\psi\,.
  \end{split}
 \]
\end{enumerate}

The functional calculus\index{functional calculus} has a number of useful properties that eventually follow from the properties of the basic spectral integral defined for really simple functions. The most relevant are listed here below. In the following $A$ is a self-adjoint operator on a Hilbert space $\cH$.

\textbf{Functional calculus: matrix elements and norms}.\index{functional calculus} For $A=A^*$ on $\cH$, $f,g\in\mathcal{S}(\mathbb{R},\mathfrak{B}(\mathbb{R}),E^{(A)})$, $\psi\in\mathcal{D}(f(A))$, $\varphi\in\mathcal{D}(g(A))$, one has
\[
 \begin{split}
  \langle f(A)\psi,g(A)\varphi\rangle\;&=\;\int_{\mathbb{R}}\overline{f(\lambda)}\,g(\lambda)\,\ud\mu_{\psi,\varphi}^{(A)}(\lambda)\,, \\
  \langle \psi,f(A)\psi\rangle\;&=\;\int_{\mathbb{R}}f(\lambda)\,\ud\mu_{\psi}^{(A)}(\lambda)\,, \\
  \|f(A)\psi\|^2\;&=\;\;\int_{\mathbb{R}}|f(\lambda)|^2\,\ud\mu_{\psi}^{(A)}(\lambda)\,, \\
  f(\lambda)\,=\,g(\lambda)\;E^{(A)}\textrm{-a.e.}\;\;&\Leftrightarrow\;\;f(A)=g(A)\,.
 \end{split}
\]
  The Cauchy-Schwartz inequality\index{Cauchy-Schwarz inequality} in the first identity above yields
 \[
  \Big| \int_{\mathbb{R}}f(\lambda)\,g(\lambda)\,\ud\mu_{\psi,\varphi}^{(A)}(\lambda)\Big|\;\leqslant\Big( \int_{\mathbb{R}}|f(\lambda)|^2\,\ud\mu_{\psi}^{(A)}(\lambda)\Big)^{\!\frac{1}{2}}\Big( \int_{\mathbb{R}}|g(\lambda)|^2\,\ud\mu_{\varphi}^{(A)}(\lambda)\Big)^{\!\frac{1}{2}}.
 \]

\textbf{Functional calculus: algebraic properties}.\index{functional calculus} For $A=A^*$ on $\cH$, $f,g\in\mathcal{S}(\mathbb{R},\mathfrak{B}(\mathbb{R}),E^{(A)})$, $\alpha,\beta\in\mathbb{C}$, one has
\[
 \begin{split}
  f(A)^*\;&=\;\overline{f}(A)\,, \\
  \textrm{in particular, }f(A)=f(A)^*\;&\Leftarrow\;f(\lambda)\in\mathbb{R}\;\;E^{(A)}\textrm{-a.e.~on $\mathbb{R}$}\,, \\
  (\alpha f+\beta g)(A)\;&=\;\overline{\alpha f(A)+\beta g(A)}\,, \\ 
  (fg)(A)\;&=\;\overline{f(A)g(A)}\,,\quad\textrm{in particular, }f(A)=\overline{f(A)}\textrm{ is closed}\,, \\
  \mathcal{D}(f(A)g(A))\;&=\;\mathcal{D}(g(A))\cap\mathcal{D}((fg)(A))\,, \\
  f(A)f(A)^*\;=\;|f|^2(A)\;&=\;f(A)^*f(A)\,,\quad\textrm{in particular, $f(A)$ is normal}\,,\index{operator!normal} \\
  f(\lambda)\neq 0\;\;E^{(A)}\textrm{-a.e.~on $\mathbb{R}$}\;&\Rightarrow\;f(A)\textrm{ is invertible, and }f(A)^{-1}=\frac{1}{f}(A)\,, \\
  f(\lambda)\geqslant 0\;\;E^{(A)}\textrm{-a.e.~on $\mathbb{R}$}\;&\Rightarrow\;f(A)\,\geqslant\,\mathbb{O}\,.
 \end{split}
\]
 The above information for $(\alpha f+\beta g)(A)$ and $(fg)(A)$ can be strengthened as
 \[
  \left. 
  \begin{array}{c}
   \textrm{in addition, for some $c>0$}, \\
   |f(\lambda)+g(\lambda)|\geqslant c |f(\lambda) | \;\;E^{(A)}\textrm{-a.e.}
  \end{array}
  \right\}\quad \Rightarrow \quad  (\alpha f+\beta g)(A)\;=\;\alpha f(A)+\beta g(A)
 \]
 (which includes the special case $f\geqslant 0$, $g\geqslant 0$ $E^{(A)}$-a.e.), and 
  \[
  \left. 
  \begin{array}{c}
   \textrm{in addition} \\
   f(\lambda)\geqslant c > 0 \;\;E^{(A)}\textrm{-a.e.}
  \end{array}
  \right\}\quad \Rightarrow \quad  (fg)(A)=f(A)g(A)\,.
 \]
 The same strengthened formulas are valid in the bounded functional calculus summarised here below.

\textbf{Functional calculus: bounded case}.\index{functional calculus} For $A=A^*$ on $\cH$ one defines
\[
 \begin{split}
  L^\infty(\mathbb{R},E^{(A)})\;&:=\;\big\{f\in\mathcal{S}(\mathbb{R},\mathfrak{B}(\mathbb{R}),E^{(A)})\,|\,\|f\|_{E^{(A)},\infty}<+\infty\big\}\,, \\
  \|f\|_{E^{(A)},\infty}\;&:=\;\sup_{\substack{N\in \mathfrak{B}(\mathbb{R}) \\ E^{(A)}(N)=\mathbb{O} }}\{|f(x)|\,|\,x\in\mathbb{R}\setminus N\}\,.
 \end{split}
\]
Clearly, $\mathcal{B}(\mathbb{R},\mathfrak{B}(\mathbb{R}))\subset L^\infty(\mathbb{R},E^{(A)})$.
Then, for $f\in\mathcal{S}(\mathbb{R},\mathfrak{B}(\mathbb{R}),E^{(A)})$,
\[
 f(A)\;\in\;\mathcal{B}(\cH)\quad\Leftrightarrow\quad f\in L^\infty(\mathbb{R},E^{(A)})\,,\quad\textrm{in which case } \|f(A)\|_{\mathrm{op}}\,=\,\|f\|_{E^{(A)},\infty}\,.
\]
This establishes that the \emph{bounded functional calculus}\index{functional calculus!bounded}
\[
 L^\infty(\mathbb{R}, E^{(A)})\to\mathcal{B}(\cH)\,,\qquad f\mapsto f(A)
\]
is a norm-continuous (actually, norm-preserving) $*$-homomorphism between unital Banach $*$-algebras -- in fact, unique, owing to the uniqueness of the spectral measure $E^{(A)}$. 
The special case of bounded functional calculus where $A=A^*\in\mathcal{B}(\cH)$ and $f\in C(\sigma(A))$ is referred to as \emph{continuous functional calculus}\index{functional calculus!continuous} (for bounded self-adjoint operators): clearly, in this case $\|f(A)\|_{\mathrm{op}}=\|f\|_{E^{(A)},\infty}=\|f\|_{\mathrm{sup}}=\sup_{\lambda\in\sigma(A)}|f(\lambda)|$.
Further notable properties of the bounded functional calculus in the special sub-case of bounded Borel-measurable functions are:
\[
\begin{split}
  \left. 
 \begin{array}{c}
  \textrm{$(f_n)_{n\in\mathbb{N}}$ and $f$ in $\mathcal{B}(\mathbb{R},\mathfrak{B}(\mathbb{R}))$} \\
  \textrm{with }|f_n(x)|\leqslant\mathrm{const.}\;\forall x\in\mathbb{R}\,,\forall n\in\mathbb{N} \\
  \textrm{and }f_n(x)\xrightarrow{n\to\infty}f(x)\;\textrm{$E^{(A)}$-a.e.~on $\mathbb{R}$}
 \end{array}
 \right\}\quad &\Rightarrow\quad \textrm{strong--}\!\lim_{\!\!\!\!\!\!\!\!\!\!\!\!\!\!\!\!\!\!\!\!\!\!\!\!\!\!\!\!n\to\infty}f_n(A)=f(A)\,, \\
    f,g\in\mathcal{B}(\mathbb{R},\mathfrak{B}(\mathbb{R}))\,\textrm{ and }\,\alpha,\beta\in\mathbb{C}
 \qquad &\Rightarrow\quad 
 \begin{cases}
  (\alpha f+\beta g)(A)\,=\,\alpha f(A)+\beta g(A)\,, \\
  (fg)(A)=f(A)g(A)\,.
 \end{cases}
\end{split}
\]

\textbf{Functional calculus: spectrum}.\index{functional calculus} For $A=A^*$ on $\cH$, $f\in\mathcal{S}(\mathbb{R},\mathfrak{B}(\mathbb{R}),E^{(A)})$, $\psi_0\in\mathcal{D}(A)$, $\lambda_0\in\mathbb{C}$, $\mathfrak{m}\in\mathbb{R}$, one has	
\[
 \begin{split}
  \sigma(f(A))\;&=\;\mathrm{ess}\,\mathrm{ran}f \;:=\;
  \left\{
  \begin{array}{c}
   \lambda\in\mathbb{C}\textrm{ such that, }\forall\varepsilon>0, \\
   E^{(A)}\big(\{t\in\mathbb{R}\,|\,|f(t)-\lambda|<\varepsilon\}\big)\neq\mathbb{O}
  \end{array}
  \right\}, \\
  \sigma(A)\;&=\;\{\lambda\in\mathbb{R}\,|\,E^{(A)}((\lambda-\varepsilon,\lambda+\varepsilon))\neq\mathbb{O}\;\,\forall\varepsilon>0\} \\
  &=\;\{\lambda\in\mathbb{R}\,|\,E^{(A)}(\lambda-\varepsilon)\neq E^{(A)}(\lambda+\varepsilon)\;\,\forall\varepsilon>0\} \\
  &=\;\mathrm{supp}\,E^{(A)}\,, \\
  \sigma(f(A))\;&\subset\;\overline{f(\sigma(A))}\,,\textrm{ where }f(\sigma(A)):=\{f(\lambda)\,|\,\lambda\in\sigma(A)\}\,, \\
  \sigma(f(A))\;&=\;\overline{f(\sigma(A))}\,,\textrm{ if in addition $f|_{\sigma(A)}$ is continuous,} \\
  \sigma(f(A))\;&=\;f(\sigma(A)),\textrm{ if in addition }
   \begin{cases}
    \textrm{$f|_{\sigma(A)}$ is continuous, and} \\
    \textrm{$\sigma(A)$ or $\mathrm{supp}f$ is bounded}, \\
   \end{cases} \\
     E^{(A)}(f^{-1}(\{\lambda_0\}))\neq\mathbb{O}\;&\Leftrightarrow\; \lambda_0\textrm{ is an eigenvalue of }f(A)\,, \\
     &\;\textrm{in which case } E^{(A)}(f^{-1}(\{\lambda_0\}))\cH\,=\textrm{ eigenspace of $f(A)$ at $\lambda_0$} \\
     &\;\textrm{($E^{(A)}(f^{-1}(\{\lambda_0\}))$ being the eigenspace's orthogonal projection)}, \\
    E^{(A)}(\{\lambda_0\})\neq\mathbb{O}\;&\Leftrightarrow\; \lambda_0\textrm{ is an eigenvalue of }A \\
    &\;\textrm{($E^{(A)}(\{\lambda_0\})$ being the eigenspace's orthogonal projection)}, \\
    &\;\textrm{therefore, each isolated point of $\sigma(A)$ is eigenvalue for $A$}, \\
    A\psi_0=\lambda_0\psi_0 \;&\Rightarrow\;f(A)\psi_0=f(\lambda_0)\psi_0\,,\quad\textrm{if in addition $f|_{\sigma(A)}$ is continuous}, \\
    A\in\mathcal{B}(\cH)\;&\Rightarrow\;\sigma(A)\subset[-\|A\|_{\mathrm{op}},\|A\|_{\mathrm{op}}]\,, \\
    A\geqslant\mathfrak{m}\mathbbm{1}\;&\Rightarrow\;\sigma(A)\subset[\mathfrak{m},+\infty)\,.
 \end{split}
\]
The above inclusion $\sigma(f(A))\subset\overline{f(\sigma(A))}$ and its special cases are referred to as the \emph{spectral mapping theorem}.\index{theorem!spectral mapping}\index{spectral mapping theorem}

\textbf{Non-emptiness of the spectrum of a self-adjoint operator}. Since $\sigma(A)=\mathrm{supp}\,E^{(A)}$ and hence
\[
 A\;=\;\int_\mathbb{R}\lambda \,\ud E^{(A)}(\lambda)\;=\;\int_{\mathrm{supp}\,E^{(A)}}\lambda \,\ud E^{(A)}(\lambda)\;=\;\int_{\sigma(A)}\lambda \,\ud E^{(A)}(\lambda)\,,
\]
then $\sigma(A)\neq\emptyset$.

\textbf{Functional calculus: resolvent}.\index{functional calculus} For $A=A^*$ on $\cH$, $f\in\mathcal{S}(\mathbb{R},\mathfrak{B}(\mathbb{R}),E^{(A)})$, $z\in\mathbb{C}$,
\[
 \begin{split}
  z\in\rho(f(A))\;&\Rightarrow\;(f(A)-z\mathbbm{1})^{-1}=\int_{\mathbb{R}}\frac{1}{\,f(\lambda)-z\,}\,\ud E^{(A)}(\lambda)\,, \\
  z\in\rho(A)\;&\Rightarrow\;(A-z\mathbbm{1})^{-1}=\int_{\mathbb{R}}\frac{1}{\,\lambda-z\,}\,\ud E^{(A)}(\lambda)\,, \\
  z\in\rho(A)\;&\Rightarrow\;\|(A-z\mathbbm{1})^{-1}\|_{\mathrm{op}}\,=\,\frac{1}{\mathrm{dist}(z,\sigma(A))}\,, \\
     z\in\mathbb{C}\setminus\mathbb{R},\;\mathfrak{Im}z\gtrless 0\;&\Rightarrow\;(A-z\mathbbm{1})^{-1}\psi=\pm\ii\int_0^{+\infty}e^{\pm\,\ii\, z\, t} e^{\mp\,\ii\, t\, A}\psi\,\ud t\quad\forall\psi\in\cH 
 \end{split}
\]
 (the latter integral being meant in the Riemann sense). The above expression for $\|(A-z\mathbbm{1})^{-1}\|_{\mathrm{op}}$ strengthens the previously discussed inequalities  $\|(A-z\mathbbm{1})^{-1}\|_{\mathrm{op}}\leqslant|\mathfrak{Im}\lambda|^{-1}$, valid when $A$ is symmetric and $z\in\mathbb{C}\setminus\mathbb{R}$, as well as $\|(A-z\mathbbm{1})^{-1}\|_{\mathrm{op}}\leqslant(\mathfrak{m}(A)-z)^{-1}$, valid when $A$ is symmetric and lower semi-bounded, and $z<\mathfrak{m}(A)$. 
 Spectral projections are conveniently expressed in terms of suitable integrals of resolvents through identities collectively referred to as \emph{Stone's formula}\index{Stone's formula}: for $A=A^*$ on $\cH$,  $a,b\in\mathbb{R}\cup\{\pm\infty\}$ with $a<b$, $c,d\in\mathbb{R}$ with $c<d$, $\lambda_0\in\mathbb{R}$, one has
  \[
 \begin{split}
  E^{(A)}([a,b])+E^{(A)}((a,b))\;&=\;\textrm{strong--}\!\lim_{\!\!\!\!\!\!\!\!\!\!\!\!\!\!\!\!\!\!\!\!\!\!\!\!\!\!\!\!\varepsilon\downarrow 0}\frac{1}{\,\pi\ii\,}\int_{a}^b\big((A-(t+\ii\varepsilon)\mathbbm{1})^{-1}-(A-(t-\ii\varepsilon)\mathbbm{1})^{-1}\big)\,\ud t\,, \\
  E^{(A)}((c,d])\;&=\;\textrm{strong--}\!\lim_{\!\!\!\!\!\!\!\!\!\!\!\!\!\!\!\!\!\!\!\!\!\!\!\!\!\!\!\!\delta\downarrow 0}\;\;\textrm{strong--}\!\lim_{\!\!\!\!\!\!\!\!\!\!\!\!\!\!\!\!\!\!\!\!\!\!\!\!\!\!\!\!\varepsilon\downarrow 0}\,\frac{1}{\,2\pi\ii\,}\;\times \\
  &\qquad\times\int_{c+\delta}^{d+\delta}\big((A-(t+\ii\varepsilon)\mathbbm{1})^{-1}-(A-(t-\ii\varepsilon)\mathbbm{1})^{-1}\big)\,\ud t\,, \\
    E^{(A)}((-\infty,d])\;&=\;\textrm{strong--}\!\lim_{\!\!\!\!\!\!\!\!\!\!\!\!\!\!\!\!\!\!\!\!\!\!\!\!\!\!\!\!\delta\downarrow 0}\;\;\textrm{strong--}\!\lim_{\!\!\!\!\!\!\!\!\!\!\!\!\!\!\!\!\!\!\!\!\!\!\!\!\!\!\!\!\varepsilon\downarrow 0}\,\frac{1}{\,2\pi\ii\,}\;\times \\
  &\qquad\times\int_{-\infty}^{d+\delta}\big((A-(t+\ii\varepsilon)\mathbbm{1})^{-1}-(A-(t-\ii\varepsilon)\mathbbm{1})^{-1}\big)\,\ud t\,, \\
  E^{(A)}(\{\lambda_0\})\;&=\;\textrm{strong--}\!\lim_{\!\!\!\!\!\!\!\!\!\!\!\!\!\!\!\!\!\!\!\!\!\!\!\!\!\!\!\!\varepsilon\downarrow 0}\;(-\ii)\varepsilon(A-(\lambda_0+\ii\varepsilon)\mathbbm{1})^{-1}\,.
 \end{split}
\]
 The ordering of the double strong-lim is crucial. The integrals in Stone's formula are meant as Riemann integrals; when the integration interval therein is infinite, the corresponding improper integrals are understood as strong limits of the analogous integrals on a finite interval. Moreover (\emph{mean ergodic theorem}),\index{theorem!mean ergodic}
 \[
  \begin{split}
   E^{(A)}(\{\lambda_0\})\;&=\;\textrm{strong--}\!\lim_{\!\!\!\!\!\!\!\!\!\!\!\!\!\!\!\!\!\!\!\!\!\!\!\!\!\!\!\!\!\!\!\!M\to+\infty}\;\frac{1}{M}\int_0^M e^{\,\ii\,t\,\lambda_0}e^{-\ii\, t\, A}\,\ud t\;=\;\textrm{strong--}\!\lim_{\!\!\!\!\!\!\!\!\!\!\!\!\!\!\!\!\!\!\!\!\!\!\!\!\!\!\!\!\!\!\!\!M\to+\infty}\;\frac{1}{2M}\int_{-M}^M e^{\,\ii\,t\,\lambda_0}e^{-\ii\, t\, A}\,\ud t
  \end{split}
 \]
 (again in the Riemann sense). In particular, if $\ker A\neq\{0\}$, the projection $P_{\ker A}$ onto $\ker A$ is
 \[
  P_{\ker A}\;=\;\textrm{strong--}\!\lim_{\!\!\!\!\!\!\!\!\!\!\!\!\!\!\!\!\!\!\!\!\!\!\!\!\!\!\!\!\!\!\!\!M\to+\infty}\;\frac{1}{M}\int_0^M e^{-\ii\, t\, A}\,\ud t
 \]
 and for any $\psi\perp\ker A$ one has $\big\|M^{-1} \int_0^M e^{-\ii\, t\, A}\psi\,\ud t\big\|_{\cH}\xrightarrow{\;M\to+\infty\;}0$\,.

  \textbf{Functional calculus: Riesz projection}.\index{functional calculus} For a self-adjoint operator $A$ on Hilbert space $\cH$ admitting a non-empty compact subset $\Lambda\in\sigma(A)$ separated by the rest of $\sigma(A)$, let $\Gamma_\Lambda$ be a closed, piece-wise smooth, positively oriented curve in $\mathbb{C}$ contained $\rho(A)$ and encircling $\Lambda$. Then
  \[
   \mathbf{1}_\Lambda(A)\;=\;E^{(A)}(\Lambda)\;=\;\frac{\ii}{2\pi}\oint_{\Gamma_\Lambda}(A-z\mathbbm{1})^{-1}\,\ud z\,.
  \]
  The expression on the r.h.s.~is called the \emph{Riesz projection}\index{Riesz projection} for $A$ relative to $\Lambda$. This construction covers also the case when $\Lambda$ consists of an isolated point $\lambda_0\in\sigma(A)$, in which case $\lambda_0$ is an eigenvalue for $A$ and $P_{\{\lambda_0\}}\equiv P_\Lambda$ is the orthogonal projection onto the corresponding eigenspace.

 \textbf{Functional calculus: commutativity}. For given $T\in\mathcal{B}(\cH)$ and $A=A^*\in\mathcal{B}(\cH)$ for some Hilbert space $\cH$, 
 \[
 \begin{split}
    [T,A]\,=\,\mathbb{O}\qquad &\Leftrightarrow\qquad \big[T,E^{(A)}(\Lambda)\big]\,=\,\mathbb{O}\quad\forall\,\Lambda\subset \mathfrak{B}(\mathbb{R}) \\
    &\Leftrightarrow\qquad 
     \begin{cases}
      \;[T,(A-z\mathbbm{1})^{-1}]\,=\,\mathbb{O} \\
      \;\textrm{for one (hence for all) $z\in\rho(A)$}
     \end{cases} \\
     &\Leftrightarrow\qquad \big[ T,(\mathbbm{1}+A^*A)^{-\frac{1}{2}} \big]\,=\,\mathbb{O}
 \end{split}
 \]
 (where $[T,A]:=TA-AT$ etc.). For  $T\in\mathcal{B}(\cH)$ and for (possibly unbounded) $A=A^*$ on $\cH$, the above chain of equivalent implications remains valid with $[T,A]=\mathbb{O}$ replaced by $TA\subset AT$.
 The everywhere defined and bounded self-adjoint operator $(\mathbbm{1}+A^*A)^{-\frac{1}{2}}$ (Sect.~\ref{sec:I-symmetric-selfadj}) is called the \emph{bounded transform}\index{bounded transform}\index{operator!bounded transform} of $A$.
%
%
%
%
%
 For any two self-adjoint operators $A_1$ and $A_2$ on Hilbert space $\cH$,
 \[
  \begin{split}
   \left.\begin{array}{c}
    \big[E^{(A_1)}(\Lambda),E^{(A_2)}(\Omega)\big]\,=\,\mathbb{O} \\
    \forall\,\Lambda,\Omega\subset \mathfrak{B}(\mathbb{R})
   \end{array}\right\}\quad &\Leftrightarrow \quad 
   \left\{\begin{array}{c}
    \big[(A_1-z_1\mathbbm{1})^{-1},(A_2-z_2\mathbbm{1})^{-1}\big]\,=\,\mathbb{O} \\
    \textrm{for one choice (hence for all)} \\
    z_1\in\rho(A_1),\,z_2\in\rho(A_2)
   \end{array}\right. \\
   &\Leftrightarrow \quad
   \begin{array}{c}
    \big[ e^{\,\ii\, t\, A},e^{\,\ii\, s\, B}\big]\,=\,\mathbb{O}\quad\forall t,s\in\mathbb{R}
   \end{array} \\
   &\Leftrightarrow \quad \big[A_1(\mathbbm{1}+A_1^*A_1)^{-\frac{1}{2}},A_2(\mathbbm{1}+A_2^*A_2)^{-\frac{1}{2}}\big]\,=\,\mathbb{O} \\
   &\Leftrightarrow \quad 
   \left\{\begin{array}{c}
    (A_1-z_1\mathbbm{1})^{-1}A_2\,\subset\, A_2(A_1-z_1\mathbbm{1})^{-1} \\
    \textrm{for one (hence for all) } z_1\in\rho(A_1)\,,
   \end{array}\right.
  \end{split}
 \] 
 in which case $A_1$ and $A_2$ are said to \emph{strongly commute}.\index{strong commutativity}\index{operator!strong commutativity} When the above $A_1$ and $A_2$ are strongly commuting, then there exists a dense subspace $\mathcal{D}\subset\cH$ such that
 \[
  \begin{split}
   & \mathcal{D}\textrm{ is a common core for $A_1$ and $A_2$}\,, \\
   & A_1\mathcal{D}\,\subset\,\mathcal{D}\,,\;\;A_2\mathcal{D}\,\subset\,\mathcal{D}\,, \\
   & A_1A_2\psi\,=\,A_2A_1\psi\;\;\forall\psi_1,\psi_2\in\mathcal{D}\,.
  \end{split}
 \]
 When in particular the self-adjoint operators $A_1$ and $A_2$ are bounded, strong commutativity is equivalent to ordinary commutativity $[A_1,A_2]=\mathbb{O}$. For any (possibly unbounded) strongly commuting\index{strong commutativity}\index{operator!strong commutativity} self-adjoint operators $A_1$ and $A_2$ on Hilbert space $\cH$, the operator $N=A_1+\ii\,A_2$ is normal\index{operator!normal} and $N^*=A_1-\ii\,A_2$; each normal operator has this structure, that is, if $N$ is normal on $\cH$, then $A_1:=\frac{1}{2}(\overline{N+N^*})$ and $A_2:=\frac{1}{2\ii}(\overline{N-N^*})$ are strongly commuting self-adjoint operators such that $A_1+\ii\,A_2=N$.

 \textbf{Functional calculus: square root}.\index{functional calculus} For $A=A^*\geqslant\mathbb{O}$ acting on $\cH$, there is a unique $B=B^*\geqslant\mathbb{O}$ on $\cH$ such that $B^2=A$, and one writes $B\equiv A^{\frac{1}{2}}$. Such $B$ is actually $B:=f(A)$ with $f(\lambda):=\sqrt{\lambda}$, and indeed, since $\mathrm{supp}\,E^{(A)}=\sigma(A)\subset[0,+\infty)$, one has $f\in \mathcal{S}(\mathbb{R},\mathfrak{B}(\mathbb{R}),E^{(A)})$. If in addition $\ker A=\{0\}$, then
 \[
  \big(A^{\frac{1}{2}}\big)^{-1}\;=\;(A^{-1})^{\frac{1}{2}}\;=\;A^{-\frac{1}{2}}\,.
 \]
 In the special case when $A=A^*\geqslant\mathbb{O}$ and $A$ is bounded, one has $\|A^{\frac{1}{2}}-p_n(A)\|_{\mathrm{op}}\xrightarrow{n\to\infty} 0$ for any sequence $(p_n)_{n\in\mathbb{N}}$ of polynomials such that $p_n\to\sqrt{x}$ uniformly in $x\in[0,a]$ with $a\geqslant\|A\|_{\mathrm{op}}$, i.e., with $[0,a]\supset\sigma(A)$.

 \textbf{Functional calculus: fractional powers}.\index{functional calculus} For $A=A^*\geqslant\mathbb{O}$ acting on $\cH$ and for $\alpha\in(0,1)$, the self-adjoint operator $A^{\alpha}$ satisfies
 \[
  \begin{split}
   \mathcal{D}(A)\;&\subset\;\mathcal{D}(A^{\alpha})\,, \\
   \mathcal{D}(A^{\frac{\alpha}{2}})\;&=\;\Big\{\psi\in\cH\,\Big|\,\int_0^{+\infty} t^{\alpha-1}\big\langle\psi, A(A+t\mathbbm{1})^{-1}\psi\big\rangle\,\ud t<+\infty\Big\}, \\
    \|A^{\frac{\alpha}{2}}\psi\|^2\;&=\;\frac{\sin\pi\alpha}{\pi}\,\int_0^{+\infty} t^{\alpha-1}\big\langle\psi, A(A+t\mathbbm{1})^{-1}\psi\big\rangle\,\ud t\qquad\forall\psi\in \mathcal{D}(A^{\frac{\alpha}{2}})\,, \\
   A^{\alpha}\psi\;&=\;\frac{\sin\pi\alpha}{\pi}\,\displaystyle\lim_{\substack{ \varepsilon\downarrow 0 \\ R\to +\infty }}\int_\varepsilon^R t^{\alpha-1}A(A+t\mathbbm{1})^{-1}\psi\,\ud t\qquad\forall\psi\in \mathcal{D}(A)\,,
  \end{split}
 \]
 as well as the \emph{McCarty inequalities}\index{McCarty inequality}
  \[
  \begin{split}
   \langle\psi,A^\alpha\psi\rangle\;&\leqslant\;\langle\psi,A\psi\rangle^\alpha\,\|\psi\|^{2(1-\alpha)}\qquad\forall\psi\in \mathcal{D}(A)\,,\;\forall\alpha\in(0,1)\,, \\
   \langle\psi,A^\beta\psi\rangle\;&\geqslant\;\langle\psi,A\psi\rangle^\beta\,\|\psi\|^{2(1-\beta)}\qquad\,\forall\psi\in \mathcal{D}(A)\,,\;\forall\beta>1\,, \\
   \langle\psi,A^{-\gamma}\psi\rangle\;&\geqslant\;\langle\psi,A\psi\rangle^{-\gamma}\,\|\psi\|^{2(1+\gamma)}\!\qquad\forall\psi\in \mathcal{D}(A)\,,\;\forall\gamma>0 \;\textrm{ (when $\ker A=\{0\}$)}\,.
  \end{split}
 \]

%

\section{Spectral theorem in multiplication form}

The spectral theorem for self-adjoint operators has a convenient, equivalent re-formulation essentially stating that every self-adjoint operator has the form, up to unitary equivalence, of a multiplication operator by a real function. Such a construction proceeds as follows.

For a self-adjoint operator $A$ on a Hilbert space $\cH$, and a vector $\psi_0\in\cH$, it is equivalent that
\begin{itemize}
 \item $\mathrm{span}\big\{E^{(A)}(\Lambda)\psi_0\,|\,\Lambda\subset \mathfrak{B}(\mathbb{R})\big\}$ is dense in $\cH$,
 \item $\psi_0\in\displaystyle\bigcap_{n=0}^\infty\mathcal{D}(A^n)$ and $\mathrm{span}\{A^n \psi_0\,|\,n\in\mathbb{N}_0\}$  is dense in $\cH$,
 \item $\big\{f(A)\psi_0\,\big|\,f\in L^2(\mathbb{R},\ud\mu_{\psi_0}^{(A)}) \big\}=\cH$\,,
 \item $\cH$ is the only closed subspace $V\subset\cH$ such that $\psi_0\in V$ and $V$ is reducing for $A$ \\
 (`reducing' meaning: $A(\mathcal{D}(A)\cap V)\subset V$, $A(\mathcal{D}(A)\cap V^\perp)\subset V^\perp$, and $P_V\mathcal{D}(A)\subset\mathcal{D}(A)$, where $P_V$ is the orthogonal projection from $\cH$ onto $V$ -- see Section \ref{sec:I_invariant-reducing-ssp}).
\end{itemize}
The elements of the subspace $C^\infty(A):=\displaystyle\bigcap_{n=0}^\infty\mathcal{D}(A^n)$ are called \emph{smooth vectors} for $A$ (or \emph{$A$-smooth vectors}\index{smooth vectors}). The subspace $\mathrm{span}\{A^n \psi_0\,|\,n\in\mathbb{N}_0\}$ is also called the \emph{Krylov subspace}\index{Krylov subspace} relative to $A$ and $\psi_0$, denoted also as $\mathcal{K}(A,\psi_0)$. A (smooth) vector $\psi_0$ satisfying any of the above conditions is called \emph{cyclic vector}\index{cyclic vector} for $A$. A self-adjoint operator admitting a cyclic vector is said to be a \emph{cyclic operator}\index{cyclic operator}\index{operator!cyclic} or also to have \emph{simple spectrum}.\index{simple spectrum}\index{operator!with simple spectrum} If a self-adjoint operator has simple spectrum, the Hilbert space it acts on is necessarily separable (for, $\cH$ admits the countable dense consisting of vectors $A^n\psi_0$, in the notation above). For a self-adjoint operator $A$ on $\cH$, the occurrence of admitting a cyclic vector is equivalent to any of the two conditions here below:
\begin{itemize}
 \item the commutant\index{commutant} $\{A\}'=\{B\in\mathcal{B}(\cH)\,|\,BA\subset AB\}$ of $A$ is commutative,
 \item each $B\in\{A\}'$ is a bounded function $f(A)$ of $A$.
\end{itemize}

A prototypical operator with simple spectrum is the self-adjoint multiplication operator $M_x$ on $L^2(\mathbb{R},\ud\mu)$, with $\mu$ a positive regular Borel measure on $\mathbb{R}$, defined as usual by
\[
 \mathcal{D}(M_x)\,:=\,\{\psi\in L^2(\mathbb{R},\ud\mu)\,|\,x\psi\in L^2(\mathbb{R},\ud\mu)\}\,,\qquad M_x\psi\,:=\,x\psi\,.
\]
If $\mu$ is finite, then a cyclic vector is the function $\mathbf{1}$, as for every Borel subset $\Lambda\in\mathbb{R}$ the spectral projection $E^{(M_x)}(\Lambda)$ is the multiplication by $\mathbf{1}_\Lambda$ and hence $\mathrm{span}\{E^{(M_x)}(\Lambda)\mathbf{1}\,|\,\Lambda\subset \mathfrak{B}(\mathbb{R})\}$ is the subspace of simple functions, dense in $L^2(\mathbb{R},\ud\mu)$. For generic $\mu$, a cyclic vector is the function 
\[
 \psi_0\,:=\,\sum_{n\in\mathbb{Z}}\frac{1}{\:2^{|n|}}\,\frac{1}{\,\mu([n,n+1))^{\frac{1}{2}}}\,\mathbf{1}_{[n,n+1)}\,;
\]
indeed, $\mathrm{span}\{E^{(M_x)}(\Lambda)\psi_0\,|\,\Lambda\subset \mathfrak{B}(\mathbb{R})\}$ contains the dense subspace of characteristic functions of each bounded Borel subset.

If a self-adjoint operator $A$ on Hilbert space $\cH$ admits a cyclic vector $\psi$, then the map $f(A)\psi\mapsto f$ defined for the $f$'s belonging to the dense subspace $L^2(\mathbb{R},\ud\mu_{\psi}^{(A)})\cap\mathcal{S}(\mathbb{R},\mathfrak{B}(\mathbb{R}),E^{(A)})$ of $L^2(\mathbb{R},\ud\mu_{\psi}^{(A)})$ lifts to a unitary operator
\[\tag{$\bigstar'$}
 U:\cH\xrightarrow{\cong}L^2\big(\mathbb{R},\ud\mu_{\psi}^{(A)}\big)\quad\textrm{ such that } UAU^{-1}\,=\,M_x\,,
\]
where $M_x$ is the multiplication by $x$ considered above. One has $U\psi=\mathbf{1}$. Each eigenvalue of $A$ has multiplicity one: indeed, $\lambda\in\mathbb{R}$ is an eigenvalue of $M_x$ if and only if $\mu_{\psi}^{(A)}(\{\lambda\})\neq 0$ in which case the associate eigenfunctions are complex multiples of $\mathbf{1}_{\{\lambda\}}$. Thus, self-adjoint operators with simple spectrum are unitarily equivalent to self-adjoint multiplications by the real variable on an explicit $L^2(\mathbb{R},\ud\mu)$-space, and their eigenvalues are non-degenerate. 

In the general case when a self-adjoint operator $A$ on Hilbert space $\cH$ does not necessarily have cyclic vectors, the above unitary equivalence is generalised as follows. For given $\psi\in\cH$, the subspace
\[
 \cH_\psi\;:=\;\overline{\,\mathrm{span}\big\{E^{(A)}(\Lambda)\psi\,|\,\Lambda\subset \mathfrak{B}(\mathbb{R})\big\}\,}\;=\;\big\{f(A)\psi\,\big|\,f\in L^2(\mathbb{R},\ud\mu_{\psi}^{(A)}) \big\} 
\]
is called the \emph{cyclic subspace}\index{cyclic subspace} for $A$ relative to $\psi$. A family $(\psi_\alpha)_{\alpha\in\mathcal{I}}$ (for some index set $\mathcal{I}$) of normalised vectors of $\cH$ with $\cH_{\psi_\alpha}\perp\cH_{\psi_\beta}$ for $\alpha\neq\beta$, and maximal with respect to such properties, is called a \emph{spectral basis}\index{spectral basis} for $A$. A standard application of Zorn's lemma shows that, given the self-adjoint $A$, a spectral basis for $A$ always exists and, in terms of the spectral basis $(\psi_\alpha)_{\alpha\in\mathcal{I}}$,
\[
 \cH\;=\;\bigoplus_{\alpha\in\mathcal{I}}\cH_{\psi_\alpha}\,.
\]
Moreover, with respect to the above orthogonal decomposition, each subspace $\cH_{\psi_\alpha}$ is reducing for $A$ and therefore
\[
 A\;=\;\bigoplus_{\alpha\in\mathcal{I}} \;A\big|_{\cH_{\psi_\alpha}}\,.
\]
In the above r.h.s.~by definition $\mathcal{D}(A\big|_{\cH_{\psi_\alpha}})=\mathcal{D}(A)\cap\cH_{\psi_\alpha}$ and $\mathcal{D}(A)=\op_{\alpha\in\mathcal{I}}\mathcal{D}(A\big|_{\cH_{\psi_\alpha}})$. By construction each $\psi_\alpha$ from the spectral basis is cyclic for $A\big|_{\cH_{\psi_\alpha}}$ with respect to the Hilbert space $\cH_{\psi_\alpha}$, hence $A\cong M_x$ on $L^2\big(\mathbb{R},\ud\mu_{\psi_\alpha}^{(A)}\big)$. This establishes the unitary (Hilbert space isomorphism)
\[
 \begin{split}
   U:\cH\;\to\:&\:\bigoplus_{\alpha\in\mathcal{I}}\;L^2\big(\mathbb{R},\ud\mu_{\psi_\alpha}^{(A)}\big)\,, \\
   \displaystyle\bigoplus_{\alpha\in\mathcal{I}}f_\alpha(A)\psi_\alpha \;\mapsto\:&\:\displaystyle\bigoplus_{\alpha\in\mathcal{I}}\,f_\alpha\:\equiv\:(f_\alpha)_{\alpha\in\mathcal{I}}\,,
 \end{split}
\]
and 
\[
 (UAU^{-1}f)_\alpha(\lambda)\;=\;\lambda f_\alpha(\lambda)\quad \forall f\,\equiv\,(f_\alpha)_{\alpha\in\mathcal{I}}\,\in\,\mathcal{D}(UAU^{-1})\,.
\]
The above expression of the unitary equivalent version $UAU^{-1}$ of $A$ is called the \emph{spectral representation}\index{spectral representation} of $A$. One also has
\[
 \sigma(A)\;=\;\overline{\,\bigcup_{\alpha\in\mathcal{I}}\mathrm{supp}\,\mu_{\psi_\alpha}^{(A)}\,}\,.
\]

If, in addition, the considered Hilbert space $\cH$ the self-adjoint $A$ acts on is separable, and hence any spectral basis for $A$ is necessarily at most countable, say, $(\psi_\alpha)_{\alpha\in\mathcal{I}}\equiv(\psi_n)_{n\in\mathcal{N}}$ with $\mathcal{N}:=\{1,2,\dots,N\}$, $N\in\mathbb{N}\cup\{\infty\}$, the above unitary equivalence for $\cH$ and $A$ can be conveniently pushed further as follows. One considers the additional Hilbert space $L^2(\Upsilon,\ud\mu)$, where $\Upsilon:=\mathbb{R}\times\mathcal{N}$ and $\mu$ is defined by the requirement that its restriction to the $n$-th copy of $\mathbb{R}$ is $2^{-n}\mu_{\psi_n}^{(A)}$. Thus,
%
%
%
\[
 \mu(\Lambda_1,\Lambda_2,\Lambda_3,\dots)\;:=\;\sum_{n\in\mathcal{N}}2^{-n}\mu_{\psi_n}^{(A)}(\Lambda_n) \qquad\forall(\Lambda_n)_{n\in\mathcal{N}}\:\textrm{ with }\Lambda_n\in\mathfrak{B}(\mathbb{R})\,.
\]
The measure $\mu$ is finite, as $\mu(\Upsilon)=\sum_{n\in\mathcal{N}} 2^{-n}\mu_{\psi_n}^{(A)}(\mathbb{R})=\sum_{n\in\mathcal{N}} 2^{-n}\leqslant 1$.
%
%
%
%
%
 One has the Hilbert space isomorphism
\[
 \begin{split}
  U'\;:\;\bigoplus_{n\in\mathcal{N}}\;L^2\big(\mathbb{R},\ud\mu_{\psi_n}^{(A)}\big)&\xrightarrow{\;\;\cong\;\;	}\;L^2(\Upsilon,\ud\mu),,  \\
  f \equiv(f_n)_{n\in\mathcal{N}}\;&\longmapsto\;F\,, \\
  \textrm{with } \;\;F(\lambda)\,:=\,\sum_{n\in\mathcal{N}}2^{\frac{n}{2}}&f_n(\lambda_n)\,,\quad \lambda\equiv(\lambda_n)_{n\in\mathcal{N}}\in\Upsilon\,.
 \end{split}
\]
Indeed,
\[
 \begin{split}
   \|F\|^2_{L^2(\Upsilon,\ud\mu)}\;&=\;\int_\Upsilon  |F(\lambda)|^2 \ud\mu(\lambda)\;=\;\sum_{n\in\mathcal{N}}\int_\mathbb{R}|2^{\frac{n}{2}}\varphi_n(\lambda_n)|^2 \,2^{-n}\mu_{\psi_n}^{(A)}(\lambda_n) \\
   &=\;\sum_{n\in\mathcal{N}}\int_\mathbb{R}|\varphi_n(\lambda_n)|^2\mu_{\psi_n}^{(A)}(\lambda_n)\,.
 \end{split}
\]
The composition $\mathcal{U}:=U\circ U'$ of the two previously considered unitaries is therefore a Hilbert space isomorphism $\cH\xrightarrow{\cong}L^2(\Upsilon,\ud\mu)$. By construction, for $F$ in a suitable dense of $L^2(\Upsilon,\ud\mu)$, $\mathcal{U}^{-1} F=\bigoplus_{n\in\mathcal{N}}f_n(A)\psi_n\in \bigoplus_{n\in\mathcal{N}}\cH_{\psi_n}=\cH$ for some $f_n\in L^2(\mathbb{R},\ud\mu_{\psi_n}^{(A)})\cap\mathcal{S}(\mathbb{R},\mathfrak{B}(\mathbb{R}),E^{(A)})$; then, $A\mathcal{U}^{-1}F=\bigoplus_{n\in\mathcal{N}}g_n(A)\psi_n$ with $g_n(t):=tf_n(t)$, and also $\mathcal{U}A\mathcal{U}^{-1}F=G$ with $G(\lambda)=\sum_{n\in\mathcal{N}}2^{\frac{n}{2}}g_n(\lambda_n)=\sum_{n\in\mathcal{N}}2^{\frac{n}{2}}\lambda_n f_n(\lambda_n)$ $\forall\lambda\equiv(\lambda_n)_{n\in\mathcal{N}}\in\Upsilon$. Since $F(\lambda)=\sum_{n\in\mathcal{N}}2^{\frac{n}{2}}f_n(\lambda_n)$, then $G(\lambda)=\Phi(\lambda)F(\lambda)$ $\forall\lambda\in\Upsilon$.
Summarising, for the given self-adjoint operator $A$ on $\cH$, there exist
\begin{itemize}
 \item a finite measure space $(\Upsilon,\ud\mu)$,
 \item a real-valued measurable function $\Phi$ on $\Upsilon$,
 \item and a unitary $\mathcal{U}:\cH\xrightarrow{\cong}L^2(\Upsilon,\ud\mu)$
\end{itemize}
such that
\[\tag{$\bigstar''$}
 \mathcal{U} A \mathcal{U}^{-1}\;=\;\textrm{operator of multiplication by $\Phi$}\,.
\]
The unitary equivalence $(\bigstar'')$, as well as its particular case $(\bigstar')$ when $A$ has simple spectrum, is an equivalent formulation of the spectral theorem \eqref{eq:spectdecomp} of Section \ref{sec:I_spectral_theorem} which goes under the name of \emph{spectral theorem in multiplication form}.\index{spectral theorem}\index{theorem!spectral theorem} In the above notation, if $F$ and $\widetilde{F}$ are the representative in $L^2(\Upsilon,\ud\mu)$ of two vectors in $\cH$, one also has the (rigorously justified) Dirac-type formulas
\[
 \begin{split}
   \big\langle F,\widetilde{F}\big\rangle_{L^2(\Upsilon,\ud\mu)}\;&=\;\sum_{n\in\mathcal{N}}\int_{\mathbb{R}}\overline{f_n(t)}\,\widetilde{f}_n(t)\,\ud\mu_{\psi_n}(t)\,, \\
   \big\langle F,\mathcal{U} A \mathcal{U}^{-1}\widetilde{F}\big\rangle_{L^2(\Upsilon,\ud\mu)}\;&=\;\sum_{n\in\mathcal{N}}\int_{\mathbb{R}}\overline{f_n(t)}\,t\widetilde{f}_n(t)\,\ud\mu_{\psi_n}(t)\,.
 \end{split}
\]

\section{Parts of the spectrum}\label{sec:I-parts-of-spectrum}

For a given self-adjoint operator $A$ on a Hilbert space $\cH$ one characterises the following \emph{spectral subspaces}\index{spectral subspaces} of $A$.
\begin{enumerate}[(i)]
 \item \emph{Pure point spectral subspace}\index{spectral subspaces!pure point subspace}:
 \[
  \begin{split}
   \cH_{\mathrm{p}}(A)\;&:=\;
   \begin{cases}
    \overline{\,\mathrm{span}\{\psi\in\cH\,|\,\psi\textrm{ is an eigenvector for }A\}\,,} & \textrm{if $A$ has eigenvalues,} \\
    \{0\}\,, & \textrm{if $A$ has no eigenvalue}
   \end{cases} \\
   &\;=\;\big\{\psi\in\cH\,\big|\,\exists\,N\subset\mathbb{R},\textrm{ at most countable, such that }E^{(A)}(N)\psi=\psi\big\} \\
   &\;=\;\big\{\psi\in\cH\,\big|\,\exists\,N\subset\mathbb{R},\textrm{ at most countable, such that }\mu_{\psi}^{(A)}(\mathbb{R}\setminus N)=0\big\}\,.
  \end{split}
 \]
 The set $N$ above can be taken as the collection of the eigenvalues of $A$. 
 $\cH_{\mathrm{p}}(A)$ is a closed subspace of $\cH$ and has orthogonal projection
 \[
  P_{\cH_{\mathrm{p}}(A)}\;=\;E^{(A)}(N)\;=\;\sum_{\lambda\textrm{ eigenv.}}E^{(A)}(\{\lambda\})\,.
 \]
 \item \emph{Continuous spectral subspace}\index{spectral subspaces!continuous subspace}:
 \[
  \begin{split}
   \cH_{\mathrm{c}}(A)\;&:=\;\big\{\psi\in\cH\,\big|\,\lambda\mapsto\langle\psi,E^{(A)}(\lambda)\psi\rangle\textrm{ is continuous on $\mathbb{R}$}\big\} \\
    &\;=\;\big\{\psi\in\cH\,\big|\,\lambda\mapsto E^{(A)}(\lambda)\psi\textrm{ is a continuous $\mathbb{R}\to\cH$ map}\big\} \\
    &\;=\;\big\{\psi\in\cH\,\big|\,\lambda\mapsto \|E^{(A)}(\lambda)\psi\|\textrm{ is continuous}\big\} \\
    &\;=\;\big\{\psi\in\cH\,\big|\,E^{(A)}(\{\lambda\})\psi=0\;\forall\lambda\in\mathbb{R}\big\} \\
    &\;=\;\big\{\psi\in\cH\,\big|\,E^{(A)}(N)\psi=0\textrm{ for every at most countable subset }N\subset\mathbb{R}\big\} \\
    &\;=\;\big\{\psi\in\cH\,\big|\,\mu_{\psi}^{(A)}\textrm{ is a continuous measure}\big\}\,.
  \end{split}
 \]
 Recall: $E^{(A)}(\lambda)=E^{(A)}((-\infty,\lambda])$; a measure is continuous\index{measure!continuous} when it has no atoms (no pure points). $\cH_{\mathrm{c}}(A)$ is a closed subspace of $\cH$.
 \item \emph{Absolutely continuous spectral subspace}\index{spectral subspaces!absolutely continuous subspace}:
 \[
  \begin{split}
   \cH_{\mathrm{ac}}(A)\;&:=\;\left\{ \psi\in\cH\left|
   \begin{array}{c}
    \textrm{the measure $\mu_\psi^{(A)}$ is absolutely continuous} \\
    \textrm{with respect to the Lebesgue measure}
   \end{array}
   \right.\right\} \\
   &\;=\;\big\{\psi\in\cH\,\big|\,\lambda\mapsto\langle\psi,E^{(A)}(\lambda)\psi\rangle\textrm{ is a function in }AC(\mathbb{R})\big\} \\
   &\;=\;\big\{\psi\in\cH\,\big|\,\lambda\mapsto\|E^{(A)}(\lambda)\psi\|^2\textrm{ is a function in }AC(\mathbb{R})\big\}\,.
  \end{split}
 \]
 Recall: absolute continuity\index{measure!absolutely continuous} of $\mu_\psi^{(A)}$ with respect to the Lebesgue measure $\mu_\mathcal{L}$ means that $\mu_\psi^{(A)}(N)=0$ for every Borel subset
 $N\subset\mathbb{R}$ with $\mu_\mathcal{L}(N)=0$; $AC(\mathbb{R})$ is the space of absolutely continuous functions\index{absolutely continuous functions} on $\mathbb{R}$ (Sect.~\ref{sec:WeylsCriterion}). $\cH_{\mathrm{ac}}(A)$ is a closed subspace of $\cH$.
 \item \emph{Singular spectral subspace}\index{spectral subspaces!singular subspace}:
  \[
  \begin{split}
   \cH_{\mathrm{s}}(A)\;&:=\;\left\{ \psi\in\cH\left|
   \begin{array}{c}
    \textrm{the measure $\mu_\psi^{(A)}$ and the Lebesgue measure $\mu_{\mathcal{L}}$} \\
    \textrm{on $\mathbb{R}$ are mutually singular}
   \end{array}
   \right.\right\} \\
   &\;=\;\big\{\psi\in\cH\,\big|\,\exists N\subset\mathbb{R}\textrm{ with }\mu_\mathcal{L}(N)=0\textrm{ and }E^{(A)}(N)\psi=\psi\big\}\,.
  \end{split}
 \]
 Recall: $\mu_\psi^{(A)}$ and $\mu_{\mathcal{L}}$ being mutually singular\index{measure!mutually singular} (for which one writes also $\mu_\psi^{(A)}\perp\mu_{\mathcal{L}}$) means that there exists a Borel subset
 $N\subset\mathbb{R}$ with $\mu_\mathcal{L}(N)=0$ and $\mu_\psi^{(A)}(\mathbb{R}\setminus N)=0$. $\cH_{\mathrm{s}}(A)$ is a closed subspace of $\cH$.
 \item \emph{Singular continuous spectral subspace}\index{spectral subspaces!singular continuous subspace}:
   \[
  \begin{split}
   \cH_{\mathrm{sc}}(A)\;&:=\;\left\{ \psi\in\cH\left|
   \begin{array}{c}
    \textrm{the measure $\mu_\psi^{(A)}$ is singular continuous} \\
    \textrm{with respect to the Lebesgue measure $\mu_{\mathcal{L}}$}
   \end{array}
   \right.\right\} \\
   &\;=\;\big\{ \psi\in\cH_{\mathrm{c}}(A)\,\big|\,\textrm{$\mu_\psi^{(A)}$ and $\mu_{\mathcal{L}}$ are mutually singular}
   \big\} \\
    &\;=\;\cH_{\mathrm{c}}(A)\cap\cH_{\mathrm{s}}(A)\,.
  \end{split}
 \]
 Recall: $\mu_\psi^{(A)}$ being singular continuous\index{measure!singular continuous} with respect to $\mu_{\mathcal{L}}$ means that $\mu_\psi^{(A)}$ has no atoms and $\mu_\psi^{(A)}\perp\mu_{\mathcal{L}}$. $\cH_{\mathrm{sc}}(A)$ is a closed subspace of $\cH$.
 \item Orthogonal decompositions:
 \[
  \begin{split}
      \cH_{\mathrm{c}}(A)\;&=\;\cH_{\mathrm{ac}}(A)\oplus\cH_{\mathrm{sc}}(A)\,, \\
   \cH_{\mathrm{s}}(A)\;&=\;\cH_{\mathrm{p}}(A)\oplus\cH_{\mathrm{sc}}(A)\,, \\
   \cH_{\mathrm{sc}}(A)\;&=\;\cH_{\mathrm{c}}(A)\cap\cH_{\mathrm{s}}(A)\,, \\
   \cH\;&=\;\cH_{\mathrm{p}}(A)\oplus\cH_{\mathrm{c}}(A) \\
   &=\;\cH_{\mathrm{ac}}(A)\oplus\cH_{\mathrm{s}}(A) \\
   &=\;\cH_{\mathrm{p}}(A)\oplus\cH_{\mathrm{ac}}(A)\oplus\cH_{\mathrm{sc}}(A)\,.
  \end{split}
 \]
 \item Invariance: for any $\lambda\in\mathbb{R}$
 \[ 
  \begin{split}
    E^{(A)}(\lambda)\cH_{\mathrm{p}}(A)\;&\subset\;\cH_{\mathrm{p}}(A)\,, \\
    E^{(A)}(\lambda)\cH_{\mathrm{ac}}(A)\;&\subset\;\cH_{\mathrm{ac}}(A)\,, \\
    E^{(A)}(\lambda)\cH_{\mathrm{sc}}(A)\;&\subset\;\cH_{\mathrm{sc}}(A)\,,
  \end{split}
 \]
 where as usual $E^{(A)}(\lambda)=E^{(A)}((-\infty,\lambda])$.
\end{enumerate}

 The (absolutely) continuous spectral subspace of $A$ is also characterised in terms of the long-time behaviour of the unitary group generated by $A$. More precisely (the \emph{Ruelle-Amrein-Georgescu-En\ss{} (RAGE) theorem}),\index{theorem!RAGE}
 \[
 \begin{split}
  \psi\in\cH_{\mathrm{c}}(A)\qquad&\Leftrightarrow\qquad  \lim_{T\to\infty}\,\frac{1}{T}\int_0^T\big|\langle\psi,e^{-\,\ii\, t\, A}\psi\rangle\big|^2\;=\;0 \\
  &\Leftrightarrow\qquad \lim_{T\to\infty}\,\frac{1}{T}\int_0^T\big\|\,C e^{-\,\ii\,t\,A}\psi\big\|^2\,\ud t\;=\;0  \\
    &\qquad \qquad \textrm{for any compact $C$ on $\cH$}\,, \\
  \psi\in\cH_{\mathrm{ac}}(A)\qquad&\Rightarrow\qquad  \lim_{t\to\infty} \:\big\|\,C e^{-\,\ii\,t\,A}\psi\big\|\;=\;0 \\
  &\qquad \qquad \textrm{for any compact $C$ on $\cH$}\,. \\
 \end{split}
 \]
 The same holds with $\frac{1}{2T}\int_{-T}^T$ instead of $\frac{1}{T}\int_0^T$.

 Each of the subspaces $\cH_{\mathrm{p}}(A)$, $\cH_{\mathrm{c}}(A)$, $\cH_{\mathrm{ac}}(A)$, $\cH_{\mathrm{s}}(A)$,$\cH_{\mathrm{sc}}(A)$ reduces $A$. Correspondingly, the operators $A\big|_{\cH_{\mathrm{p}}(A)}$, $A\big|_{\cH_{\mathrm{c}}(A)}$, $A\big|_{\cH_{\mathrm{ac}}(A)}$, $A\big|_{\cH_{\mathrm{s}}(A)}$, $A\big|_{\cH_{\mathrm{sc}}(A)}$ are self-adjoint on the respective Hilbert spaces. This leads to defining the following parts of the spectrum $\sigma(A)$:
 \[
  \begin{array}{lcrl}
   \textrm{\emph{point spectrum}\index{point spectrum}\index{spectrum!point spectrum} of $A$}& \qquad & \sigma_{\mathrm{p}}(A)&:=\;\{\textrm{eigenvalues of $A$}\}\,, \\
   \textrm{\emph{continuous spectrum}\index{continuous spectrum}\index{spectrum!continuous spectrum} of $A$}& \qquad & \sigma_{\mathrm{c}}(A)&:=\;\sigma_{\cH_{\mathrm{c}}(A)}\big( A\big|_{\cH_{\mathrm{c}}(A)}\big)\,, \\
   \textrm{\emph{absolutely continuous spectrum}\index{absolutely continuous spectrum}\index{spectrum!absolutely continuous spectrum} of $A$}& \qquad & \sigma_{\mathrm{ac}}(A)&:=\;\sigma_{\cH_{\mathrm{ac}}(A)}\big( A\big|_{\cH_{\mathrm{ac}}(A)}\big)\,, \\
   \textrm{\emph{singular spectrum}\index{singular spectrum}\index{spectrum!singular spectrum} of $A$}& \qquad & \sigma_{\mathrm{s}}(A)&:=\;\sigma_{\cH_{\mathrm{s}}(A)}\big( A\big|_{\cH_{\mathrm{s}}(A)}\big)\,, \\ 
   \textrm{\emph{singular continuous spectrum}\index{singular continuous spectrum}\index{spectrum!singular continuous spectrum} of $A$}& \qquad & \sigma_{\mathrm{sc}}(A)&:=\;\sigma_{\cH_{\mathrm{sc}}(A)}\big( A\big|_{\cH_{\mathrm{sc}}(A)}\big)\,,
  \end{array}
 \]
 with the additional tacit definition that if $\dim\cH_{\mathrm{c}}(A)=0$, then $\sigma_{\mathrm{c}}(A)=\emptyset$, etc.

 Since $\lambda_0\in\sigma_{\mathrm{p}}(A)$ $\Leftrightarrow$ $E^{(A)}(\{\lambda_0\})\neq \mathbb{O}$ (Sect.~\ref{sec:I_funct_calc}), and since for any $\lambda\in\mathbb{R}$ $E^{(A)}(\{\lambda\})=E^{(A)}((-\infty,\lambda])-E^{(A)}((-\infty,\lambda))$ (Sect.~\ref{sec:I_spectral_theorem}), then $\sigma_{\mathrm{p}}(A)$ consists of all points of discontinuity for the resolution-of-identity map $\lambda\mapsto E^{(A)}(\lambda)$, equivalently, all points of failure of strong left continuity (strong right continuity is always true by the general assumption made in the definition of resolution of the identity).

 One says that
 \[
  \textrm{$A$ has \emph{purely} }\;
  \begin{cases}
   \textrm{point}\index{spectrum!purely point spectrum} \\
   \textrm{continuous}\index{spectrum!purely continuous spectrum} \\
   \textrm{absolutely continuous}\index{spectrum!purely absolutely continuous spectrum} \\
   \textrm{singular}\index{spectrum!purely singular spectrum} \\
   \textrm{singular continuous}\index{spectrum!purely singular continuous spectrum}
  \end{cases}
  \;\;\textrm{spectrum if }\quad
  \left. 
  \begin{array}{l}
 \cH_{\mathrm{p}}(A) \\ 
  \cH_{\mathrm{c}}(A) \\
   \cH_{\mathrm{ac}}(A) \\
    \cH_{\mathrm{s}}(A) \\
     \cH_{\mathrm{sc}}(A) \\
  \end{array}
  \right\}
  =\;\cH\,.
 \]
 All the subsets $\sigma_{\mathrm{c}}(A)$, $\sigma_{\mathrm{ac}}(A)$, $\sigma_{\mathrm{s}}(A)$, $\sigma_{\mathrm{sc}}(A)$ are closed in $\mathbb{R}$, whereas $\overline{\sigma_{\mathrm{p}}(A)}= \sigma_{\cH_{\mathrm{p}}(A)}\big( A\big|_{\cH_{\mathrm{p}}(A)}\big)$, and 
 \[
  \begin{split}
     \sigma_{\mathrm{c}}(A)\;&=\;\sigma_{\mathrm{ac}}(A)\cup\sigma_{\mathrm{sc}}(A)\,, \\
     \sigma_{\mathrm{s}}(A)\;&=\;\overline{\sigma_{\mathrm{p}}(A)}\cup\sigma_{\mathrm{sc}}(A)\,, \\
     \sigma(A)\;&=\;\overline{\sigma_{\mathrm{p}}(A)}\cup\sigma_{\mathrm{c}}(A) \\
     &=\;\sigma_{\mathrm{ac}}(A)\cup\sigma_{\mathrm{s}}(A) \\
     &=\;\overline{\sigma_{\mathrm{p}}(A)}\cup\sigma_{\mathrm{ac}}(A)\cup\sigma_{\mathrm{sc}}(A)\,.
  \end{split}
 \]
 Moreover, the spectrum of $A$ and its spectral parts above are all unitary invariants, that is, 
 \[
  \sigma(UAU^{-1})\,=\,\sigma(A)\,,\qquad \sigma_\alpha(UAU^{-1})\,=\,\sigma_\alpha(A)\;\;\forall\alpha\in\{\textrm{p,c,ac,s,sc}\}
 \]
 for any Hilbert space unitary isomorphism $U:\cH\to\mathcal{K}$.

 An additional and useful decomposition of the spectrum is the following. Let $A$ be a self-adjoint operator on a Hilbert space $\cH$. One defines the \emph{discrete spectrum $\sigma_{\mathrm{disc}}(A)$ of $A$}\index{spectrum!discrete spectrum}\index{discrete spectrum} and the \emph{essential spectrum $\sigma_{\mathrm{ess}}(A)$ of $A$}\index{spectrum!essential spectrum}\index{essential spectrum} as
 \[
  \begin{split}
   \sigma_{\mathrm{disc}}(A)\;&:=\;\{ \textrm{isolated eigenvalues of $A$ with finite multiplicity } \} \\
   \sigma_{\mathrm{ess}}(A)\;&:=\;\sigma(A)\setminus\sigma_{\mathrm{disc}}(A) \\
   &\;\,=\;\{  \textrm{accumulation points of $\sigma(A)$} \} \;\cup\;
   \left\{ 
   \begin{array}{c}
    \textrm{eigenvalues of $A$} \\
    \textrm{with infinite multiplicity}
   \end{array}
   \right\}.
  \end{split}
 \]
  The essential spectrum $ \sigma_{\mathrm{ess}}(A)$ is always closed.
  One has
  \[
   \begin{split}
    \sigma_{\mathrm{disc}}(A)\:&\subset\:\overline{\sigma_{\mathrm{p}}(A)}\,, \\
     \sigma_{\mathrm{ess}}(A)\:&=\:\sigma_{\mathrm{c}}(A)\,\cup\,\big( \overline{\sigma_{\mathrm{p}}(A)}\setminus\sigma_{\mathrm{disc}}(A)\big) \,, \\
    \sigma_{\mathrm{ac}}(A) \,\cup\,\sigma_{\mathrm{sc}}(A)\:=\:\sigma_{\mathrm{c}}(A)\:&\subset\: \sigma_{\mathrm{ess}}(A)\,.
   \end{split}
  \]
  Discrete and essential spectra too are unitary invariants, that is, 
 \[
  \sigma_{\mathrm{disc}}(UAU^{-1})\,=\,\sigma_{\mathrm{disc}}(A)\,,\qquad \sigma_{\mathrm{ess}}(UAU^{-1})\,=\,\sigma_{\mathrm{ess}}(A)
 \]
 for any Hilbert space unitary isomorphism $U:\cH\to\mathcal{K}$.
  One says that $A$ has \emph{purely discrete},\index{spectrum!purely discrete spectrum} respectively \emph{purely essential}\index{spectrum!purely essential spectrum} spectrum if $\sigma(A)=\sigma_{\mathrm{disc}}(A)$, respectively $\sigma(A)=\sigma_{\mathrm{ess}}(A)$. Operators of both types exist, but:
  \[
   \begin{split}
    A=A^*\in\mathcal{B}(\cH)\textrm{ and }\dim \cH<\infty & \quad \Rightarrow\quad\sigma(A)=\sigma_{\mathrm{disc}}(A)\,,\;\; \sigma_{\mathrm{ess}}(A)=\emptyset\,, \\
    A=A^*\in\mathcal{B}(\cH)\textrm{ and }\dim \cH=\infty & \quad \Rightarrow\quad\sigma_{\mathrm{ess}}(A)\neq\emptyset\,.
   \end{split}
  \]
 Moreover, in the infinite-dimensional case,
  \[
   \begin{split}
    \dim \cH=\infty \quad &\Rightarrow\quad \sigma(\mathbbm{1})\,=\,\{1\}\,=\,\sigma_{\mathrm{ess}}(\mathbbm{1})\,, \\
    \left.
    \begin{array}{c}
     \dim \cH=\infty\,, \\
     \textrm{$A=A^*$ is compact on $\cH$}
    \end{array}
\right\}\quad &\Rightarrow\quad 0\,\in\,\sigma_{\mathrm{ess}}(A)\,.
   \end{split}
  \]
 The occurrence of purely discrete self-adjoint operators (emptiness of the essential spectrum) is characterised as
 \[
  A=A^*\textrm{ and }\sigma_{\mathrm{ess}}(A)=\emptyset\qquad\Leftrightarrow\qquad 
  \begin{cases}
   \quad\; (A-z\mathbbm{1})^{-1}\textrm{ is compact} \\
   \;\textrm{for one (hence for all) }z\in\rho(A)\,.
  \end{cases}
 \]

 As an equivalent characterisation, for $A=A^*$ on $\cH$ and $\lambda\in\mathbb{R}$,
 \[ 
  \begin{split}
 \lambda\in \sigma_{\mathrm{disc}}(A)\quad&\Leftrightarrow\quad0<\dim\mathrm{ran}\,E^{(A)}((\lambda-\varepsilon,\lambda+\varepsilon))<\infty\textrm{ for some }\varepsilon>0\,, \\
   \lambda\in\sigma_{\mathrm{ess}}(A)\quad&\Leftrightarrow\quad\dim\mathrm{ran}\,E^{(A)}((\lambda-\varepsilon,\lambda+\varepsilon))=\infty\quad\forall\varepsilon>0 \\
   &\Leftrightarrow\quad
   \left\{ 
   \begin{array}{l}
    \exists\,(\psi_n)_{n\in\mathbb{N}}\subset\mathcal{D}(A)\textrm{ with} \\
    \bullet\;\;\;\displaystyle\liminf_{n\to\infty}\|\psi_n\|>0 \\
    \bullet\;\;\;\psi_n\rightharpoonup 0\textrm{ as }n\to\infty \\
    \bullet\;\;\;\|A \psi_n - \lambda \psi_n\|\xrightarrow{n\to\infty} 0
   \end{array}
   \right. \;\Leftrightarrow\quad
   \left\{ 
   \begin{array}{l}
    \exists\,(\psi_n)_{n\in\mathbb{N}}\subset\mathcal{D}(A)\textrm{ with} \\
    \bullet\;\;\;\|\psi_n\|=1\;\forall n\in\mathbb{N} \\
    \bullet\;\;\;\psi_n\perp \psi_m\textrm{ for }n\neq m \\
    \bullet\;\;\;\|A \psi_n - \lambda \psi_n\|\xrightarrow{n\to\infty} 0
   \end{array}
   \right.
   \end{split}
 \]
(\emph{Weyl criterion} for the essential spectrum).\index{spectrum!Weyl criterion}\index{Weyl criterion!for essential spectrum}\index{theorem!Weyl (criterion for essential spectrum)}  A sequence $(\psi_n)_{n\in\mathbb{N}}$ of the first type above is called \emph{singular Weyl sequence}\index{Weyl sequence} for $A$ at the spectral point $\lambda$. A sequence of the second type above is therefore an example of orthogonal singular Weyl sequence. For $a,b\in\mathbb{R}\cup\{\pm\infty\}$ with $a<b$,
 \[
  \begin{split}
   0<\dim\mathrm{ran}\,E^{(A)}((a,b))<\infty\quad&\Rightarrow\quad \sigma(A)\cap(a,b)\subset\sigma_{\mathrm{disc}}(A)\,, \\
   \dim\mathrm{ran}\,E^{(A)}((a,b))=\infty\quad&\Rightarrow\quad\sigma_{\mathrm{ess}}(A)\cap(a,b)\neq\emptyset\,.
   \end{split}
 \]

 With respect to the orthogonal sum of self-adjoint operators, spectrum, essential spectrum, and discrete spectrum have the following general behaviour. For a given collection $(A_k)_{k\in\mathbb{N}}$ of self-adjoint operators, the $k$-th of which acts in the Hilbert space $\cH_k$, 
 \[
  \begin{split}
     \sigma\Big(\bigoplus_{k\in\mathbb{N}}A_k\Big) \;&=\; \overline{\,\bigcup_{k \in \mathbb{N}} \sigma(A_k)}\,, \\
  \sigma_{\mathrm{ess}}\Big(\bigoplus_{k\in\mathbb{N}}A_k\Big) \;&\supset\; \bigcup_{k \in \mathbb{N}}\sigma_{\mathrm{ess}}(A_k)\,, \\
   \sigma_{\mathrm{disc}}\Big(\bigoplus_{k\in\mathbb{N}}A_k\Big) \;&\subset\; \bigcup_{k \in \mathbb{N}}\sigma_{\mathrm{disc}}(A_k)\,,
  \end{split}
 \]
 where the spectra on the l.h.s.~are referred to the Hilbert space $\bigoplus_{k\in\mathbb{N}}\cH_k$ and each $k$-th spectrum on the l.h.s.~is referred to $\cH_k$.

\section{Perturbations of self-adjoint operators}\label{sec:perturbation-spectra}

 \textbf{Relatively bounded perturbation.}\index{perturbation!relatively bounded perturbation} For two operators $A$ and $B$ on Hilbert space $\cH$ one says that $B$ is \emph{relatively bounded with respect to $A$}, or also \emph{relatively $A$-bounded},\index{operator!relatively bounded} if, for some $a,b\geqslant 0$,
 \[
  \mathcal{D}(A)\subset\mathcal{D}(B)\qquad\textrm{ and }\qquad \|B\psi\|_{\cH}\,\leqslant\,a\|A\psi\|_{\cH}+b\|\psi\|_{\cH}\qquad\forall\psi\in\mathcal{D}(A)\,,
 \]
 or, equivalently, if, for some $a,b\geqslant 0$,
 \[
  \mathcal{D}(A)\subset\mathcal{D}(B)\qquad\textrm{ and }\qquad \|B\psi\|^2_{\cH}\,\leqslant\,a\|A\psi\|^2_{\cH}+b\|\psi\|^2_{\cH}\qquad\forall\psi\in\mathcal{D}(A)\,.
 \]
 In other words, $B$ is relatively $A$-bounded if and only if $\mathcal{D}(A)\subset\mathcal{D}(B)$ and $B$ maps $(\mathcal{D}(A),\|\cdot\|_{\Gamma(A)})$ continuously into $\cH$. Recall that $\|\cdot\|_{\Gamma(A)}$ is the graph norm for $A$ (Sect.~\ref{sec:I-preliminaries}.) The infimum of the non-negative $a$'s for which there is $b\geqslant 0$ such that the above inequality holds is called the \emph{relative bound}\index{operator!relative bound} of $B$ with respect to $A$ (or \emph{$A$-bound}). When $B$ is $A$-bounded with $A$-bound equal to zero, then $B$ is said \emph{infinitesimally small with respect to $A$}, or also \emph{infinitesimally $A$-small}.\index{operator!infinitesimally relatively small}\index{perturbation!infinitesimally small perturbation}
 
 Standard cases of relative (or infinitesimally small) boundedness:
 \begin{itemize}
  \item every everywhere defined and bounded operator $B$ on $\cH$ is infinitesimally small with respect to any other operator on $\cH$;
  \item every symmetric operator $B$ is infinitesimally small with respect to $B^2$, since $\|B\psi\|_{\cH}\leqslant\varepsilon\|B^2\psi\|_{\cH}+\varepsilon^{-1}\|\psi\|_{\cH}$ $\forall\psi\in\mathcal{D}(B^2)\subset\mathcal{D}(B)$ and $\forall\varepsilon>0$;
  \item every densely defined and closable operator $B$ on $\cH$ is relatively bounded by any closed operator $A$ on $\cH$ such that $\mathcal{D}(A)\subset\mathcal{D}(B)$;
  \item every closed operator $B$ on $\cH$ is relatively bounded by any closed operator $A$ on $\cH$ such that $\mathcal{D}(A)\subset\mathcal{D}(B)$ and $\rho(A)\neq\emptyset$, in which case the relative bound of $B$ with respect to $A$ is majorised by $\|B(A-z\mathbbm{1})^{-1}\|_{\mathrm{op}}$ for any $z\in\rho(A)$.
 \end{itemize}

 If $B$ is relatively $A$-bounded on the Hilbert space $\cH$ with $A$-bound strictly smaller than 1, then $A+B$ is closed if and only if $A$ is closed, and if in addition $A$ is closable, so too is $A+B$. Thus, relatively bounded (additive) perturbations with relative bound strictly smaller then $1$ preserve closedness and closability.

 If $A$ is a densely defined symmetric operator and the operator $B$ is infinitesimally small with respect to $A$, then $d_\pm(A+B)=d_\pm(A)$, i.e., infinitesimally small perturbations preserve the deficiency indices.

 If $A$ is self-adjoint on $\cH$ and the operator $B$ on $\cH$ satisfies $\mathcal{D}(B)\supset\mathcal{D}(A)$, then $B$ is relatively $A$-bounded if and only if $B(A-z\mathbbm{1})^{-1}\in\mathcal{B}(\cH)$ for one, hence for all $z\in\rho(A)$, in which case the relative $A$-bound of $B$ amounts to
 \[
  \inf_{z\in\rho(A)}\big\|B(A-z\mathbbm{1})^{-1}\big\|_{\mathrm{op}}\;=\;\lim_{\mathbb{R}\ni\lambda\to\infty}\big\|B(A-\ii\lambda\mathbbm{1})^{-1} \big\|_{\mathrm{op}}\,.
 \]

\noindent 
\textbf{Kato-Rellich perturbation.}\index{perturbation!Kato-Rellich perturbation} The \emph{Kato-Rellich theorem}\index{theorem!Kato-Rellich} covers the case of a self-adjoint operator $A$ on Hilbert space $\cH$ and a symmetric and relatively $A$-bounded operator $B$ on $\cH$ with $A$-bound strictly lower than 1: then, $A+B$ is self-adjoint on $\mathcal{D}(A+B)=\mathcal{D}(A)$, and moreover, if a subspace $\mathcal{D}\subset\mathcal{D}(A)$ is a domain of essential self-adjointness for $A|_{\mathcal{D}}$, then $\mathcal{D}$ is also a domain of essential self-adjointness of $(A+B)|_{\mathcal{D}}$. Thus, any symmetric and relatively bounded perturbation of a self-adjoint operator, with relative bound strictly smaller than one (for short: any \emph{Kato-Rellich perturbation}\index{Kato-Rellich perturbation}), preserves self-adjointness.

 In the special case where in addition $A$ is lower semi-bounded, the Kato-Rellich perturbation $B$ preserves lower semi-boundedness too, in the quantitative sense
 \[
  \mathfrak{m}(A+B)\;=\;\mathfrak{m}(A) - \max\Big\{ \frac{b}{1-a}\,,\,a|\mathfrak{m}(A)|+b\Big\}
 \]
 (where $\|B\psi\|\leqslant a\|A\psi\|+b\|\psi\|$ $\forall\psi\in\mathcal{D}(A)\subset\mathcal{D}(B)$ and $0<a<1$).

 \textbf{W\"{u}st perturbation.}\index{perturbation!W\"{u}st perturbation} Analogously, the \emph{W\"{u}st theorem}\index{theorem!W\"{u}st} states that given an essentially self-adjoint operator $A$ on $\cH$ and a symmetric operator $B$ on $\cH$ with $\mathcal{D}(B)\supset\mathcal{D}(A)$ and such that, for some $b>0$, $\|B\psi\|_{\cH}\leqslant\|A\psi\|_{\cH}+b\|\psi\|_{\cH}$ $\forall\psi\in\mathcal{D}(A)$, then $(A+B)|_{\mathcal{D}(A)}$ is essentially self-adjoint, and so too is $(A+B)|_{\mathcal{D}}$ for any core $\mathcal{D}\subset\mathcal{D}(A)$ for $A$.

 \textbf{Compact perturbation in the resolvent sense.}\index{perturbation!compact perturbation} Any two self-adjoint operators $A_1$ and $A_2$ on Hilbert space $\cH$ admitting a common point $z\in\rho(A_1)\cap\rho(A_2)$ at which the resolvent difference 
 \[
  (A_2-z\mathbbm{1})^{-1}-(A_1-z\mathbbm{1})^{-1}
 \]
 is a compact operator on $\cH$, necessarily satisfy the property $\sigma_{\mathrm{ess}}(A_1)=\sigma_{\mathrm{ess}}(A_2)$. In particular, if $A_1$ and $A_2$ are two self-adjoint extensions of the a densely defined and symmetric operator $S$ having finite deficiency indices, then $\sigma_{\mathrm{ess}}(A_1)=\sigma_{\mathrm{ess}}(A_2)$, because the two resolvents $(A_1-\ii\mathbbm{1})^{-1}$ and $(A_2-\ii\mathbbm{1})^{-1}$ coincide by assumption on $\mathrm{ran}(S-\ii\mathbbm{1})$ and therefore the range of $(A_2-\ii\mathbbm{1})^{-1}-(A_1-\ii\mathbbm{1})^{-1}$ is contained in the finite-dimensional subspace $(\mathrm{ran}(S-\ii\mathbbm{1}))^\perp$, meaning that such resolvent difference is indeed compact.

  \textbf{Relatively compact perturbation.}\index{perturbation!relatively compact perturbation}  For two operators $A$ and $C$ on Hilbert space $\cH$ such that $A$ is closed, $\mathcal{D}(A)\subset\mathcal{D}(C)$, and $C$ maps $(\mathcal{D}(A),\|\cdot\|_A)$ compactly into $\cH$, one says that $C$ is \emph{relatively compact with respect to $A$}, or also \emph{relatively $A$-compact}.\index{operator!relatively compact} In other words, the operator $C$ is relatively compact with respect to a closed operator $A$ whenever $\mathcal{D}(A)\subset\mathcal{D}(C)$ and any sequence $(\psi_n)_{n\in\mathbb{N}}$ in $\mathcal{D}(A)$ satisfying $\sup_{n\in\mathbb{N}}(\|\psi_n\|_{\cH}+\|A\psi_n\|_{\cH})<+\infty$ is mapped into a sequence $(C\psi_n)_{n\in\mathbb{N}}$ that admits a convergent sub-sequence in $\cH$. 
  Equivalently, the operator $C$ is relatively compact with respect to a closed operator $A$ whenever for any sequence $(\psi_n)_{n\in\mathbb{N}}$ in $\mathcal{D}(A)$ satisfying $\psi_n\rightharpoonup 0$ and $A\psi_n\rightharpoonup 0$ weakly in $\cH$ as $n\to\infty$, one has $\|Cf_n\|_{\cH}\to 0$. 

 A compact operator $C$ on $\cH$ is relatively compact with respect to any other closed operator on $\cH$.
  If, for three operators $A,B,C$ on $\cH$, $B$ is relatively $A$-bounded and $C$ is relatively $B$-compact, then $C$ is also relatively $A$-compact. 

  When $A$ is self-adjoint, $C$ is relatively $A$-compact if and only if the operator $C E^{(A)}(I)$ is compact for any bounded interval $I\subset\mathbb{R}$, in which case $B$ is infinitesimally $A$-small. 
 
When a closed operator $A$ has non-empty resolvent set $\rho(A)$ (in particular, when $A$ is self-adjoint), an operator $C$ on the same Hilbert space $\cH$ is relatively $A$-compact if and only if $\mathcal{D}(A)\subset\mathcal{D}(C)$ and $C(A-z\mathbbm{1})^{-1}$ is a compact operator on $\cH$ for one and hence for all $z\in\rho(A)$. 

If $A$ is densely defined and closed on $\cH$ and $C$ is relatively $A$-compact, then $C$ is infinitesimally $A$-small. 

 

 \textbf{Relatively compact and symmetric perturbation.} Any relatively compact and symmetric perturbation of a self-adjoint operator preserves self-adjointness and the essential spectrum. More precisely: if $A$ is a self-adjoint operator on Hilbert space $\cH$ and $C$ is a symmetric and relatively $A$-compact operator on $\cH$, then $A+C$ is self-adjoint on $\mathcal{D}(A+C)=\mathcal{D}(A)$ and $\sigma_{\mathrm{ess}}(A+C)=\sigma_{\mathrm{ess}}(A)$. This holds in particular when $A$ is self-adjoint and $C$ is compact and self-adjoint, in which case the result is the classical theorem of Weyl on compact self-adjoint perturbations.\index{theorem!Weyl (compact self-adjoint perturbations)}

 Given three operators $A,B,C$ on the Hilbert space $\cH$ such that $A$ is self-adjoint, $B$ is symmetric and relatively $A$-bounded with $A$-bound strictly smaller than 1, and $C$ is relatively $A$-compact, then $C$ is also relatively $(A+B)$-compact.

 The RAGE theorem\index{theorem!RAGE} (Sect.~\ref{sec:I-parts-of-spectrum}) has a version for relatively compact operators: if $A$ is a self-adjoint operator on Hilbert space $\cH$, then 
 \[
 \begin{split}
  \psi\in\cH_{\mathrm{c}}(A)\qquad&\Rightarrow\qquad  \lim_{T\to\infty}\,\frac{1}{T}\int_0^T\big\|\,C e^{-\,\ii\,t\,A}\psi\big\|_{\cH}^2\,\ud t\;=\;0 \\
  &\qquad \qquad \textrm{for any relatively $A$-compact $C\in\mathcal{B}(\cH)$\,,} \\
  \psi\in\cH_{\mathrm{ac}}(A)\qquad&\Rightarrow\qquad  \lim_{t\to\infty}\,\big\|\,C e^{-\,\ii\,t\,A}\psi\big\|_{\cH}\;=\;0 \\
  &\qquad \qquad \textrm{for any relatively $A$-compact $C\in\mathcal{B}(\cH)$\,.}
 \end{split}
 \]

%
%
%
%
%
%
%

%
%
%
%

%
%

\section{Quadratic forms and self-adjoint operators}\label{sec:I_forms}

A \emph{sesquilinear form}\index{sesquilinear form}\index{form!sesquilinear} on a Hilbert space $\cH$ is a mapping $q[\cdot,\cdot]:\mathcal{D}[q]\times\mathcal{D}[q]\to\mathbb{C}$, where $\mathcal{D}[q]$ is a subspace of $\cH$ called the \emph{domain of the form}\index{form!domain}, which is (conventionally) linear in the second entry and anti-linear in the first. A \emph{quadratic form}\index{quadratic form}\index{form!quadratic} on $\cH$ with domain $\mathcal{D}[q]$ is a mapping $q[\,\cdot\,]:\mathcal{D}[q]\to\mathbb{C}$ such that, for all $\psi\in\mathcal{D}[q]$ $q[\psi]=q[\psi,\psi]$ for some sesquilinear form $q[\cdot,\cdot]$ on $\cH$ with domain $\mathcal{D}[q]$. The sesquilinear form can be recovered from its quadratic form by means of the \emph{polarisation identity}\index{polarisation identity}\index{form!polarisation}
\[
 q[\psi,\varphi]\;=\;\frac{1}{4}\Big(q[\psi+\varphi]-q[\psi-\varphi]-\ii q[\psi+\ii\varphi]+\ii q[\psi-\ii\varphi] \Big)\,,\quad\psi,\varphi\in\mathcal{D}[q]\,,
\]
so that one can speak unambiguously of the \emph{form}\index{form} $q$.
Forms undergo a natural point-wise sum of forms and multiplication by scalars:
\[
 \begin{split}
  (q_1+q_2)[\psi,\varphi]\;&:=\;q_1[\psi,\varphi]+q_2[\psi,\varphi]\,,\qquad\mathcal{D}[q_1+q_2]:=\mathcal{D}[q_1]\cap\mathcal{D}[q_2]\,, \\
  (z\,q)[\psi,\varphi]\;&:=\;z\,q[\psi,\varphi]\,,\;\;z\in\mathbb{C}\,,\qquad\qquad\quad \mathcal{D}[z\, q]:=\mathcal{D}[q]\,, \\
  (q+z)[\psi,\varphi]\;&:=\;q[\psi,\varphi]+z\langle\psi,\varphi\rangle\,,\;\;z\in\mathbb{C}\,,\;\;\;\mathcal{D}[q+z]:=\mathcal{D}[q]\,.
 \end{split}
\]

A form $q$ is said to be \emph{bounded}\index{bounded form}\index{form!bounded} when
\[
 \sup_{\psi,\varphi\in\mathcal{D}[q]\setminus\{0\}}\,\frac{q[\psi,\varphi]}{\;\|\psi\|_{\cH}\|\varphi\|_{\cH}}\;<\;+\infty\,.
\]
Everywhere defined sesquilinear forms are just expectations of uniquely associated everywhere defined operators, that is, if $q:\cH\times\cH\to\mathbb{C}$ is an everywhere defined and bounded sesquilinear form on the Hilbert space $\cH$, then there exists a unique $A_q\in\mathcal{B}(\cH)$ such that $q[\psi,\varphi]=\langle\psi,A_q\varphi\rangle_{\cH}$ $\forall \psi,\varphi\in\cH$.

A form $q$ is called \emph{symmetric}\index{symmetric form}\index{form!symmetric} (or \emph{Hermitian}\index{Hermitian form}\index{form!Hermitian}) if
\[
 q[\varphi,\psi]\;=\;\overline{q[\psi,\varphi]}\qquad\forall \psi, \varphi\in\mathcal{D}[q]\,,
\]
in which case, equivalently, the associated quadratic form is real-valued. For an everywhere defined and bounded form which is also symmetric, the corresponding $A_q\in\mathcal{B}(\cH)$ is symmetric. A symmetric form is said to be \emph{lower semi-bounded}\index{lower semi-bounded form}\index{form!lower semi-bounded} if in addition there is $\mathfrak{m}\in\mathbb{R}$, called \emph{lower bound}\index{form!lower bound} for the form, such that 
\[
 q[\psi]\;\geqslant\;\mathfrak{m}\,\|\psi\|^2\qquad\forall \psi\in\mathcal{D}[q]\,,
\]
in which case one writes $q\geqslant\mathfrak{m}$. For a lower semi-bounded form $q$ the quantity 
\[
 \mathfrak{m}(q)\;:=\;\inf_{ \substack{ \psi\in\mathcal{D}[q] \\ \psi\neq 0}  }\frac{\,q[\psi]\,}{\;\|\psi\|_\cH^2}\;>\;-\infty
\]
is called the \emph{greatest lower bound}\index{form!greatest lower bound} (or \emph{bottom}\index{form!bottom}) of the form.
In particular, $q$ is a \emph{positive}\index{positive form}\index{form!positive} form when $\mathfrak{m}(q)\geqslant 0$, i.e., $q\geqslant 0$, and a positive form $q$ satisfies the \emph{Cauchy-Schwarz inequality}\index{Cauchy-Schwarz inequality}
\[
 |q[\psi,\varphi]|\;\leqslant\;\sqrt{q[\psi]\,}\,\sqrt{q[\varphi]\,}\qquad \forall\psi,\varphi\in\mathcal{D}[q]\qquad (q\geqslant 0)\,.
\]
This implies, in particular,
\[
 \begin{split}
  \sqrt{q[\psi+\varphi]\,}\;&\leqslant\;\sqrt{q[\psi]\,}+\sqrt{q[\varphi]\,} \,, \\
  \big| \sqrt{q[\psi]\,}-\sqrt{q[\varphi]\,}\,\big|\;&\leqslant\;\sqrt{q[\psi-\varphi]\,}
 \end{split}\qquad \forall\psi,\varphi\in\mathcal{D}[q]\qquad (q\geqslant 0)\,.
\]

A lower semi-bounded symmetric form $q\geqslant\mathfrak{m}$ induces a natural scalar product and norm on its domain,
\[
 \begin{split}
  \langle \psi,\varphi\rangle_q\;&:=\;q[\psi,\varphi]+(1-\mathfrak{m})\langle\psi,\varphi\rangle\,, \\
  \|\psi\|_q\;&:=\;\sqrt{\langle \psi,\psi\rangle_q}\;=\;\sqrt{q[\psi]+(1-\mathfrak{m})\|\psi\|_\cH^2}\;\geqslant\;\|\psi\|_\cH\,,
 \end{split}
\]
for all $\psi,\varphi\in\mathcal{D}[q]$, and norms $\|\cdot\|_q$ defined with different lower bounds of $q$ are all equivalent, indeed
\[
 \begin{split}
   a\,\big(q[\psi]&+(1-\mathfrak{m}')\|\psi\|_\cH^2\big) \;\leqslant\;q[\psi]+(1-\mathfrak{m})\|\psi\|_\cH^2 \\
   &\leqslant\;q[\psi]+(1-\mathfrak{m}')\|\psi\|_\cH^2\qquad\forall\mathfrak{m}'\leqslant\mathfrak{m}\leqslant\mathfrak{m}(q) \\
   \textrm{for }\,a\;&:=\; \frac{1}{2} \,\inf_{\psi\in\mathcal{D}[q]\setminus\{0\}}\,\frac{\,q[\psi]+(1-\mathfrak{m})\|\psi\|_{\cH}^2}{\,q[\psi]+(1-\mathfrak{m}')\|\psi\|_{\cH}^2}\:\leqslant\:\frac{1}{2}\,.
 \end{split}
\]
For a given lower semi-bounded symmetric form $q$, the Cauchy-Schwarz inequality\index{Cauchy-Schwarz inequality} for the positive form $q-\mathfrak{m}$ and the bound $\|\cdot\|\leqslant\|\cdot\|_q$ imply
\[
 |q[\psi,\varphi]|\;\leqslant\;(1+|\mathfrak{m}|)\|\psi\|_q\|\varphi\|_q\,;
\]
therefore, $q$ is bounded on the (a priori non-complete) inner product space $(\mathcal{D}[q],\langle\cdot,\cdot\rangle_q)$. A linear subspace $\mathcal{D}\subset\mathcal{D}[q]$ is called a \emph{(form) core}\index{core}\index{form!core} of $q$ if $\mathcal{D}$ is dense in $(\mathcal{D}[q],\langle\cdot,\cdot\rangle_q)$.

A lower semi-bounded form $q$ on Hilbert space $\cH$ with domain $\mathcal{D}[q]$ is said to be \emph{closed}\index{closed form}\index{form!closed} if any of the following equivalent conditions holds:
\begin{enumerate}[(i)]
 \item $(\mathcal{D}[q],\langle\cdot,\cdot\rangle_q)$ is a Hilbert space;
 \item if a sequence $(\psi_n)_{n\in\mathbb{N}}$ in $\mathcal{D}[q]$ satisfies $\psi_n\to\psi$ in the $\cH$-norm for some $\psi\in\cH$, and $q[\psi_n-\psi_m]\to 0$, then $\psi\in\mathcal{D}[q]$ and $q[\psi_n-\psi]\to 0$\,;
 \item the map $q':\cH\to\mathbb{R}\cup\{+\infty\}$ defined as 
 \[
  q'[\psi]\;:=\;
  \begin{cases}
   q[\psi]\,, & \textrm{if }\psi\in\mathcal{D}[q]\,, \\
   +\infty\,,  & \textrm{if }\psi\in\cH\setminus\mathcal{D}[q]\,,
  \end{cases}
 \]
 is lower semi-continuous, i.e., $\displaystyle q'\big[\lim_{n\to\infty}\psi_n\big]\leqslant\liminf_{n\to\infty}q'[\psi_n]$\,;
 \item if a sequence $(\psi_n)_{n\in\mathbb{N}}$ in $\mathcal{D}[q]$ satisfies $\psi_n\to\psi$ in the $\cH$-norm for some $\psi\in\cH$, and for some $\kappa>0$ one has $q[\psi_n]\leqslant\kappa$ $\forall n\in\mathbb{N}$, then $\psi\in\mathcal{D}[q]$ and $q[\psi]\leqslant\displaystyle\liminf_{n\to\infty}q[\psi_n]$\,.
\end{enumerate}
The finite sum of lower semi-bounded closed forms is also closed.

A lower semi-bounded form $q$ on Hilbert space $\cH$ with domain $\mathcal{D}[q]$ is said to be \emph{closable}\index{closable form}\index{form!closable} if any of the following equivalent conditions holds:
\begin{enumerate}[(i)]
 \item there exists a lower semi-bounded, closed form on $\cH$ that extends $q$;
 \item if a sequence $(\psi_n)_{n\in\mathbb{N}}$ in $\mathcal{D}[q]$ satisfies $\psi_n\to 0$ in $\cH$ and $q[\psi_n-\psi_m]\to 0$, then $q[\psi_n]\to 0$\,;
 \item the unique bounded linear $\cH_q\to\cH$ map obtained by continuity extension of the continuous embedding map $(\mathcal{D}[q],\|\cdot\|_q)\to(\cH,\|\cdot\|)$ to the Hilbert space $\cH_q$ given by the completion of $(\mathcal{D}[q],\|\cdot\|_q)$, is actually injective.
\end{enumerate}
When this is the case, setting 
\[
 \begin{split}
  \mathcal{D}[\overline{q}]\;&:=\;
  \left\{ 
  \psi\in\cH\left|
  \begin{array}{c}
   \textrm{there is }\big(\psi^{(\psi)}_n\big)_{n\in\mathbb{N}}\textrm{ in }\mathcal{D}[q]\textrm{ such that} \\
   \psi^{(\psi)}_n\to\psi\,,\quad q\big[\psi^{(\psi)}_n-\psi^{(\psi)}_m\big]\to 0
  \end{array}
  \right.\right\}\,, \\
  \overline{q}[\psi,\varphi]\;&:=\;\lim_{n\to\infty} q\big[\psi^{(\psi)}_n,\varphi^{(\varphi)}_n\big]\,,
 \end{split}
\]
defines, independently of the above sequences of approximants, a lower semi-bounded and closed quadratic form $\overline{q}$, called the \emph{closure}\index{form!closure}\index{form closure} of $q$, and characterised by the property of being the smallest closed extension of $q$. Moreover, $\mathfrak{m}(\overline{q})=\mathfrak{m}(q)$, i.e., $\overline{q}$ has the same greatest lower bound as $q$.

To every self-adjoint operator $A$ acting on a Hilbert space $\cH$ one associates the quadratic form $A[\cdot]$ defined, in terms of the spectral representation of $A$, as
\[
 \begin{split}
  \mathcal{D}[A]\;&:=\;\mathcal{D}\big(|A|^{\frac{1}{2}}\big)\;=\;\Big\{\psi\in\cH\,\Big|\,\int_{\mathbb{R}}|\lambda|\,\ud \mu_\psi^{(A)}(\lambda)<+\infty\Big\}\,, \\
  A[\psi]\;&:=\;\int_{\mathbb{R}}\lambda\,\ud \mu_\psi^{(A)}(\lambda)\qquad \Big(\textrm{i.e.,}\;\;\;A[\psi,\varphi]\;=\;\int_{\mathbb{R}}\lambda\,\ud \mu_{\psi,\varphi}^{(A)}(\lambda)\Big)\,,
 \end{split}
\]
and called the \emph{(quadratic) form of (or: associated to)} $A$.\index{form!of self-adjoint operator} 
In contrast to the \emph{operator domain}\index{operator!domain}\index{domain} $\mathcal{D}(A)$ of $A$, the subspace $\mathcal{D}[A]$ is called the \emph{form domain}\index{operator!form domain} of $A$. One has
\[
 \mathcal{D}(A)\;=\;\mathcal{D}(|A|)\;\subset\;\mathcal{D}\big(|A|^{\frac{1}{2}}\big)\;=\;\mathcal{D}[A]\,.
\]
In terms of its quadratic form, $A$ satisfies
\[
 \begin{split}
  \mathcal{D}(A)\;&=\;\big\{\psi\in\mathcal{D}[A]\,\big|\,\exists\,\xi_\psi\in\cH\textrm{ such that }A[\varphi,\psi]=\langle\varphi,\xi_\psi\rangle\;\forall\varphi\in\mathcal{D}[A]\big\}\,, \\
  A\psi\;&=\;\xi_\psi
 \end{split}
\]
(with $\xi_\psi$ above uniquely identified, owing to the density of $\mathcal{D}[A]$), and
\[
 A[\varphi,\psi]\;=\;\langle\varphi,A\psi\rangle\qquad\forall\varphi\in\mathcal{D}[A]\,,\quad\forall\psi\in\mathcal{D}(A)\,.
\]

The quadratic form of a lower semi-bounded self-adjoint operator $A$ on $\cH$ satisfies the following properties.
\begin{enumerate}[(i)]
 \item The greatest lower bounds of $A$ and of $A[\cdot]$ coincide:
 \[
  \mathfrak{m}(A)\;=\;\mathfrak{m}(A[\cdot])\;=\;\inf\sigma(A)\,.
 \]
 \item $\mathcal{D}[A]\;=\;\mathcal{D}\big((A-\mathfrak{m}\mathbbm{1})^{\frac{1}{2}}\big)$  $\forall\,\mathfrak{m}\leqslant\mathfrak{m}(A)$.
 \item For any $\mathfrak{m}\leqslant\mathfrak{m}(A)$ and any $\psi,\varphi\in\mathcal{D}[A]=\mathcal{D}\big((A-\mathfrak{m}\mathbbm{1})^{\frac{1}{2}}\big)$,
 \[
  A[\psi,\varphi]\;=\;\big\langle (A-\mathfrak{m}\mathbbm{1})^{\frac{1}{2}}\psi,(A-\mathfrak{m}\mathbbm{1})^{\frac{1}{2}}\varphi\big\rangle+\mathfrak{m}\langle\psi,\varphi\rangle\,.
 \]
  In particular, 
  \[
   A=A^*\geqslant\mathbb{O}\quad\Rightarrow\quad A[\psi,\varphi]\;=\;\big\langle A^{\frac{1}{2}}\psi,A^{\frac{1}{2}}\varphi\big\rangle\quad \forall\psi,\varphi\in \mathcal{D}[A]=\mathcal{D}\big((A^{\frac{1}{2}}\big)\,.
  \]
  \item $
 \left. 
 \begin{array}{c}
  A=A^*\geqslant\mathbb{O}\,, \\
  \ker A\,=\,\{0\}
 \end{array}
 \right\}
 $\quad $\Rightarrow$ \quad  $A^{-1}[\psi]\;=\!\displaystyle\sup_{\varphi\in\mathcal{D}[A]\setminus\{0\}}\frac{\:|\langle\varphi,\psi\rangle|^2}{A[\varphi]}$ \quad $\forall\psi\in\mathcal{D}[A^{-1}]$.
 \item $A[\cdot]$ is a lower semi-bounded, closed form.
 \item For any $\mathfrak{m}<\mathfrak{m}(A)$, $\mathcal{D}[A]$ is the completion of $\mathcal{D}(A)$ in the norm associated with the scalar product $(\psi,\varphi)\mapsto\langle\psi,A\varphi\rangle-\mathfrak{m}\langle\psi,\varphi\rangle$. $\mathcal{D}(A)$ is a core for the form $A[\cdot]$.
 \item For any $\mathfrak{m}\leqslant\mathfrak{m}(A)$, the form norm $\|\cdot\|_{A[\cdot]}$ and the graph norm $\|\cdot\|_{\Gamma((A-\mathfrak{m}\mathbbm{1})^{1/2})}$ are equivalent norms on $\mathcal{D}[A]=\mathcal{D}\big((A-\mathfrak{m}\mathbbm{1})^{\frac{1}{2}}\big)$, and therefore a linear subspace of $\mathcal{D}[A]$ is a core for the form $A[\cdot]$ if and only if it is a core for the operator $(A-\mathfrak{m}\mathbbm{1})^{\frac{1}{2}}$.
 \item For any $\mathfrak{m}<\mathfrak{m}(A)$,
 \[
  \begin{split}
   \left. 
   \begin{array}{c}
    \textrm{the (linear, bounded, injective) embedding} \\
   (\mathcal{D}[A],\|\cdot\|_{A[\cdot]})\to(\cH,\|\cdot\|_\cH)\textrm{ is compact}
   \end{array}
   \right\}\quad &\Leftrightarrow\quad (A-\mathfrak{m}\mathbbm{1})^{-\frac{1}{2}}\textrm{ is compact} \\
    \Leftrightarrow\quad \left.
    \begin{array}{c} 
    (A-z\mathbbm{1})^{-1}\textrm{ is compact} \\
    \textrm{for one, hence for all }z\in\rho(A)
    \end{array}\right\}
    \quad &\Leftrightarrow\quad
    \sigma(A)\textrm{ is purely discrete}\,.
  \end{split}
 \]
 \item For $A=A^*\geqslant\mathbb{O}$ and $B=B^*\geqslant\mathbb{O}$, $\mathcal{D}(A)\subset\mathcal{D}(B)$ $\Rightarrow$ $\mathcal{D}[A]\subset\mathcal{D}[B]$, and  $\mathcal{D}(A)=\mathcal{D}(B)$ $\Rightarrow$ $\mathcal{D}[A]=\mathcal{D}[B]$.
\end{enumerate}

Along the converse direction, to every densely defined form $q$ on $\cH$ one associates the operator $A_q$ \emph{associated with the form} $q$\index{operator!associated with a form} defined as
\[
 \begin{split}
  \mathcal{D}(A_q)\;&:=\;\big\{\psi\in\mathcal{D}[q]\,\big|\,\exists\,\xi_\psi\in\cH\textrm{ such that }q[\varphi,\psi]=\langle\varphi,\xi_\psi\rangle\;\forall\varphi\in\mathcal{D}[q]\big\}\,, \\
  A_q\psi\;&:=\;\xi_\psi
 \end{split}
\]
(with $\xi_\psi$ above uniquely identified, owing to the density of $\mathcal{D}[A]$).
In particular,
\[
 q[\varphi,\psi]\;=\;\langle\varphi,A_q\psi\rangle\qquad\forall\varphi\in\mathcal{D}[q]\,,\;\;\forall\psi\in\mathcal{D}(A_q)\,.
\]
Moreover, $A_{q+z}=A_q+z\mathbbm{1}$ $\forall z\in\mathbb{C}$. $A_q$ is well defined and linear, but not necessarily densely defined. If $A$ is a self-adjoint operator, the operator associated to its quadratic form $A[\cdot]$ is precisely $A$ itself.

If $q$ is a densely defined, lower semi-bounded, closed form on $\cH$, then the associated operator $A_q$ is self-adjoint, and $q\equiv A_q[\cdot]$. Moreover, the mapping
\[
 \begin{split}
  \left\{ 
  \begin{array}{c}
  \textrm{lower semi-bounded,} \\
  \textrm{self-adjoint operators on $\cH$}
  \end{array}
  \right\} \; &\longrightarrow \; 
  \left\{ 
  \begin{array}{c}
  \textrm{densely defined, lower semi-bounded,} \\
  \textrm{closed forms on $\cH$} \\
  \end{array}
  \right\} \\
  A \; & \longmapsto \; q_A
 \end{split}
\]
is a bijection with inverse $q\mapsto A_q$.

Self-adjoint operators on a common Hilbert space are naturally equipped with an order relation defined in terms of the associated quadratic forms, by setting, for any two $A=A^*$, $B=B^*$ on $\cH$,
\[
 A\,\geqslant\,B\quad \stackrel{\textrm{(def)}}{\Longleftrightarrow}\quad 
 \begin{cases}
  \mathcal{D}[A]\,\subset\,\mathcal{D}[B]\,, \\
  A[\psi]\,\geqslant\,B[\psi]\;\;\forall \psi\in\mathcal{D}[A]\,.
 \end{cases}
\]
When $A\geqslant B$ one equivalently writes $B\leqslant A$. Thus, if simultaneously $A\geqslant B$ and $A\leqslant B$, then $A=B$. The condition $A=A^*\geqslant\mathfrak{m}\mathbbm{1}$, coincides with the semi-boundedness relation defined in general for symmetric operators. For lower semi-bounded self-adjoint operators $A$ and $B$, the above order relation takes the equivalent, simplified form 
\[ 
 \begin{split}
   A\,\geqslant\,B
  \quad &\Leftrightarrow\quad 
  \begin{cases}
  \mathcal{D}(A)\,\subset\,\mathcal{D}[B]\,, \\
  \langle\psi, A\psi\rangle \,\geqslant\,B[\psi]\;\;\forall \psi\in\mathcal{D}(A)
 \end{cases} \\
 &\!\!\!\!\!\!\!\!\!\!\!\!\!\!\!\!(A=A^*,B=B^*\textrm{ lower semi-bounded})\,.
 \end{split}
\]
Further properties of the order relation for any two lower semi-bounded, self-adjoint operators $A$ and $B$ on $\cH$ are:
\begin{enumerate}[(i)]
 \item if $A,B\geqslant\mathbb{O}$, then \quad $
  A\,\geqslant\,B$\quad $\Leftrightarrow$ \quad 
  $\begin{cases}
  \mathcal{D}(A^{\frac{1}{2}})\,\subset\,\mathcal{D}(B^{\frac{1}{2}})\,, \\
  \big\|A^{\frac{1}{2}}\psi\big\| \,\geqslant\,\big\|B^{\frac{1}{2}}\psi\big\|\;\;\forall \psi\in\mathcal{D}(A^{\frac{1}{2}})\,;
 \end{cases}$
 \item $
 \left. 
 \begin{array}{c}
  A\,\geqslant\,B\,\geqslant\,\mathbb{O}\,, \\
  \ker B\,=\,\{0\}
 \end{array}
 \right\}
 $\quad $\Rightarrow$ \quad $\ker A\,=\,\{0\}\;\textrm{ and }\; B^{-1}\,\geqslant\,A^{-1} $\,;
 \item $A\,\geqslant\,B\,\geqslant\,\mathbb{O}$ \quad $\Rightarrow$ \quad $A^\alpha\,\geqslant\,B^\alpha$ \;\;$\forall\alpha\in[0,1]$ \quad (\emph{Heinz inequality}),\index{Heinz inequality}
 
 whereas in general it is false that $A^2\geqslant B^2$ (or with any other exponent $\alpha>1$), for example this fails with $A=\begin{pmatrix} 2 & 1 \\ 1 & 1 \end{pmatrix}$, $B=\begin{pmatrix} 1 & 0 \\ 0 & 0 \end{pmatrix}$;
 \item if $\lambda<\min\{\mathfrak{m}(A),\mathfrak{m}(B)\}$, then $\lambda\in \rho(A)\cap\rho(B)$ and 
 \[
  A\,\geqslant\,B\quad\Leftrightarrow\quad (B-\lambda\mathbbm{1})^{-1}\,\geqslant\,(A-\lambda\mathbbm{1})^{-1}\,.
 \]
\end{enumerate}

 \section{Min-Max principle}\label{sec:minmax}

 To a lower semi-bounded self-adjoint operator $A$ on an infinite-dimensional Hilbert space $\cH$ one associates the sequences $(\mu_n(A))_{n\in\mathbb{N}}$, $(\widetilde{\mu}_n(A))_{n\in\mathbb{N}}$, and $(\lambda_n(A))_{n\in\mathbb{N}}$ in $\mathbb{R}$ defined as
 \[
  \begin{split}
   \mu_n(A)\;&:=\;\; \sup_{ \substack{ \textrm{linear subspace }\mathcal{D} \\ \textrm{with }\dim\mathcal{D}=n-1   } } \;\;\inf_{ \substack{ \psi\in\mathcal{D}(A)\setminus\{0\} \\ \psi\perp\mathcal{D}}}\;\;\frac{\langle\psi,A\psi\rangle}{\langle\psi,\psi\rangle}\,, \\
   \widetilde{\mu}_n(A)\;&:=\;\; \sup_{ \substack{\textrm{linear subspace }\mathcal{D} \\ \textrm{with }\dim\mathcal{D}=n-1   } } \;\;\inf_{ \substack{ \psi\in\mathcal{D}[A]\setminus\{0\} \\ \psi\perp\mathcal{D}}}\;\frac{A[\psi]}{\langle\psi,\psi\rangle}\,, \\
   \lambda_n(A)\;&:=\;
   \begin{cases}
    \begin{array}{l}
     \textrm{the $n$-th eigenvalue of $A$, in increasing order} \\
     \textrm{and counting multiplicity, if $\lambda_n<\inf\sigma_{\mathrm{ess}}(A)$}\,, 
    \end{array} \\ \\
    \begin{array}{l}
     \textrm{$\inf\sigma_{\mathrm{ess}}(A)$, if $A$ has a finite number $k\in\mathbb{N}_0$} \\
     \textrm{of eigenvalues strictly below $\sigma_{\mathrm{ess}}(A)$, and $n>k$}\,. 
    \end{array}
    \end{cases}
  \end{split}
 \]
 (It is understood that $\inf\sigma_{\mathrm{ess}}(A)=+\infty$ when $\sigma_{\mathrm{ess}}(A)=\emptyset$.)
 In particular, $\mu_1(A)=\widetilde{\mu}_1(A)=\inf\sigma(A)=\mathfrak{m}(A)$, and moreover $\sigma(A)=\sigma_{\mathrm{disc}}(A)$ (equivalently, $\sigma_{\mathrm{ess}}(A)=\emptyset$) if and only if $\lim_{n\to\infty}\lambda_n(A)=+\infty$. All three sequences $(\mu_n(A))_{n\in\mathbb{N}}$, $(\widetilde{\mu}_n(A))_{n\in\mathbb{N}}$, and $(\lambda_n(A))_{n\in\mathbb{N}}$ are non-decreasing, and
 \[
  \begin{split}
   \dim E^{(A)}((-\infty,\lambda))\cH\,<\,n \quad & \quad \textrm{if } \lambda\,<\,\mu_n(A)\,, \\
   \dim E^{(A)}((-\infty,\lambda))\cH\,\geqslant\,n \quad & \quad \textrm{if } \lambda\,>\,\mu_n(A)\,.
  \end{split}
 \]

 The above sequences (and their properties discussed in the following) are particularly meaningful when $A$ does admit eigenvalues below the bottom of the essential spectrum.

 The \emph{min-max principle}\index{min-max principle}\index{theorem!min-max} for the considered lower semi-bounded self-adjoint operator $A$ on an infinite-dimensional Hilbert space $\cH$ expresses the identities
 \[
  \lambda_n(A)\;=\;\mu_n(A)\;=\;\widetilde{\mu}_n(A)\;=\;\inf\{\lambda\in\mathbb{R}\,|\,\dim E^{(A)}((-\infty,\lambda))\cH\geqslant n\}\,,\quad n\in\mathbb{N}\,.
 \]
 This theorem characterises variationally the eigenvalues of $A$ below the bottom of the essential spectrum only in terms of the expectations $\langle\psi,A\psi\rangle$ or $A[\psi]$, with no references to the eigenvectors.
 Thus, 
 \begin{itemize}
  \item if, for two integers $k,n\in\mathbb{N}$ with $k>n$ one has $\mu_k(A)>\mu_n(A)$, then $A$ has at least $n$ eigenvalues (counting multiplicity) strictly below the bottom of $\sigma_{\mathrm{ess}}(A)$, and $\mu_n(A)$ itself is an eigenvalue;
  \item if, instead, for some $n\in\mathbb{N}$ one has $\mu_k(A)=\mu_n(A)$ $\forall k\in\mathbb{N}$ with $k>n$, then $A$ has $n-1$ eigenvalues (counting multiplicity) strictly below the bottom of $\sigma_{\mathrm{ess}}(A)$, and $\mu_n(A)=\inf\sigma_{\mathrm{ess}}(A)$.
 \end{itemize}

 The `min-max' terminology is due to the fact that in special circumstances the above definition of $\mu_n(A)$ and $\widetilde{\mu}_n(A)$ `inf' and `sup' can be replaced with `min' and `max'. For concreteness, if $A$ is a lower semi-bounded self-adjoint operator with purely discrete spectrum, and hence the $\lambda_n(A)$'s defined above form the non-increasing sequence of the eigenvalues of $A$, then  `inf' and `sup' in the definition of $\mu_n(A)$ and $\widetilde{\mu}_n(A)$ are indeed replaced, respectively, with `min' and `max'.

%

 As a direct consequence of the min-max principle, if $A$ and $B$ are two lower semi-bounded self-adjoint operators acting, respectively, on the infinite-dimensional Hilbert spaces $\cH$ and $\mathcal{K}$, with $\mathcal{H}\subset\mathcal{\mathcal{K}}$ and $A\geqslant B$, then
 \begin{itemize}
  \item $\lambda_n(A)\geqslant\lambda_n(B)$ $\forall n\in\mathbb{N}$,
  \item $N(A;\lambda)\leqslant N(B;\lambda)$  $\forall\lambda\in\mathbb{R}$, \\ where $N(A,\lambda)$ (respectively, $N(B,\lambda)$) denotes the number of those $n$'s in $\mathbb{N}$ for which $\lambda_n(A)<\lambda$ (respectively, $\lambda_n(B)<\lambda$).
  \item if $B$ has purely discrete spectrum, so has $A$.
 \end{itemize}

 The min-max principle allows to estimate the eigenvalues of $A$ below the essential spectrum by finite-dimensional approximation. Explicitly (the \emph{Rayleigh-Ritz method}),\index{Rayleigh-Ritz method} for the compression $A_V:=P_VA\upharpoonright V$ of $A$ onto a $d$-dimensional subspace $V\subset\mathcal{D}(A)$, where $d\in\mathbb{N}$ and $P_V$ is the orthogonal projection from $\cH$ onto $V$, one has $\lambda_n(A)\leqslant\lambda_n(A_V)$ for each $n\in\{1,\dots, d\}$ (the $\lambda_n(A_V)$'s are computed with respect to the Hilbert space $V$). In particular, if $\sigma_{\mathrm{disc}}(A)$ consists of $M$ eigenvalues $\lambda_1,\dots,\lambda_M$ (counting multiplicity), then $\lambda_n\leqslant\lambda_n(A_V)$ for $n=1,\dots,\min\{d,M\}$.
 
 For any self-adjoint $C\in\mathcal{B}(\cH)$ one has $|\lambda_n(A+C)-\lambda_n(A)|\leqslant\|C\|_{\mathrm{op}}$ $\forall n\in\mathbb{N}$.

 As a further consequence of the min-max principle, if $A$ and $B$ are two self-adjoint operators acting on the same Hilbert spaces $\cH$, such that $A\geqslant\mathbb{O}$, $B$ is infinitesimally $A$-small (Sect.~\ref{sec:perturbation-spectra}), and $\sigma_{\mathrm{ess}}(A+\beta B)=[0,+\infty)$ for every $\beta\in[0,+\infty)$, then the function
 \[
  [0,+\infty)\;\ni\;\beta\,\mapsto\,\lambda_n(A+\beta B)
 \]
 is non-increasing and continuous for each $n\in\mathbb{N}$, and strictly monotone at all points $\beta$ where $\lambda_n(A+\beta B)<0$. In particular, the number of negative eigenvalues of $A+\beta B$ is increasing.

 Should the Hilbert space $\cH$ have finite dimension $\dim\cH=d\in\mathbb{N}$, then the numbers $\mu_n(A)$ and $\widetilde{\mu}_n(A)$ are clearly not defined for $n>d$, yet the above min-max principle and its consequences are still valid for each $n\in\{1,\dots,d\}$.

%
%
%
\chapter{Classical self-adjoint extension schemes}
\label{chaper-extension-schemes} 

The main theme throughout this book is the identification of self-adjoint extensions of a given symmetric and densely defined operator on Hilbert space. A symmetric and densely defined operator that is not essentially self-adjoint admits either no self-adjoint extensions, or infinitely many, and the basic questions are when such extensions occur and how to construct and classify them. This Chapter is devoted to present the by now well-established classical answers to such general problem, and to prepare the tools needed in the applications discussed in the second part of the book. 


Notably, whereas the self-adjoint extension problem was posed and understood in all its essential aspects quite long ago, it has been revisited at the theoretical level in several phases and with different mathematical languages that span the last 90 years, a fact that is explained mainly by the variety and the historical evolution of applications (to quantum mechanics, quantum fields, partial differential equations, probability, operator-theoretic frameworks, spectral theory, etc.) and also by the interesting circumstance that many advances had for a long period a limited circulation across the Western and the Eastern blocks of the twentieth century.

In this respect, it is fair to speak of self-adjoint extension \emph{schemes}, as the plural used in the title of the present Chapter, with specific reference to those two main schemes that are by now \emph{classical}, in that they were the first to be elaborated and were given a self-consistent, comprehensive form soon after the self-adjoint extension problem was formulated: the ``western'' one by von Neumann, which came first and in principle can be applied to the extension of any symmetric operator, and the one by the ``Russian school'' of Kre{\u\i}n, Vi\v{s}ik, and above all Birman, which is tailored on a relevant subclass of symmetric operators, for concreteness the semi-bounded ones. 

The presentation here is a revised and considerably expanded version of \cite{GMO-KVB2017} and \cite{M-2020-Friedrichs}, and further relevant materials and references on this subject are in \cite{Krasnoselskii-1949,Smirnov-HigherMaths-5-orig1959,Glazman-1963,Grubb-1968,faris-1975,Alonso-Simon-1980,Alonso-Simon-1980-2,Gorbachuk-Gorbachuk-1984,Albeverio-HK-Fenstad-Lindstrom_nonstandard1986,albeverio-solvable,Koshmanenko-1993,albeverio-karwowski-koshmanenko-1995,Kuzhel-Kuzhel-1998,Albeverio-Koshmanenko-1999,albverio-kurasov-2000-sing_pert_diff_ops,Posilicano2000_Krein-like_formula,Posilicano-2003-additivPerturb,Kurasov-Kuroda-2004,Albeverio-Kuzhel-Nizhnik-2008,MathChallZR2021}.


\section{Friedrichs extension}\label{sec:II-Friedrichs}

The self-adjoint extension problem was posed first by von Neumann at the end of 1920's: in particular, in \cite[Satz 43]{vonNeumann1930} (announced in 1928, published in 1930) he established that a lower semi-bounded symmetric operator always admits a self-adjoint extension with largest lower bound strictly smaller than, but arbitrarily close to the largest lower bound of the original operator. von Neumann also conjectured the existence of an extension with precisely the same largest lower bound, a fact that was proved by Stone \cite[Theorem 9.21]{Stone1932} in 1932. Then Friedrichs \cite{Friedrichs1934} in 1934 (and, in a considerably simplified form, Freudenthal \cite{Freudenthal-1936} in 1936) presented the construction of one such extension, with distinguished properties and named since after as the \emph{Friedrichs extension}.\index{Friedrichs extension}

Even though, chronologically, the Friedrichs-Freudenthal construction is not the first result in self-adjoint extension theory, and moreover it is inherently a quadratic form construction that is independent from the general operator-theoretic schemes for identifying and classifying the extensions, it is often convenient to discuss it first for at least three reasons: it reflects the relevance of the problem of the existence of lower semi-bounded self-adjoint extensions, which was indeed central at the birth of self-adjoint extension theory; it provides a distinguished extension that is often used as a reference with respect to which to express other self-adjoint extensions; it also provides a fundamental tool in quantum mechanics, where ordinarily the actual Hamiltonian for the system under consideration is precisely the Friedrichs extension of a `formal' Hamiltonian dictated by physical heuristics such as first quantisation and initially realised as a densely defined, symmetric, lower semi-bounded operator on Hilbert space \cite{CM-selfadj-2021}.

\begin{theorem}[Friedrichs extension]\label{thm:Friedrichs-ext}\index{theorem!Friedrichs} Let $S$ be a lower semi-bounded and densely defined symmetric operator on a Hilbert space $\cH$.
\begin{enumerate}[(i)]
 \item The form $(\psi,\varphi)\mapsto\langle \psi,S\varphi\rangle$ with domain $\mathcal{D}(S)$ is lower semi-bounded and closable. Its closure is the form whose domain, denoted by $\mathcal{D}[S]$, is given by the completion of $\mathcal{D}(S)$ with respect to the norm
 \[
  \psi\,\mapsto\,\langle \psi,S\psi\rangle+(1-\mathfrak{m}(S))\|\psi\|_\cH^2\,,
 \]
where $\mathfrak{m}(S)$ is the bottom\index{operator!greatest lower bound} (i.e., the greatest lower bound) of $S$, and whose value $S[\psi,\varphi]$ on any two $\psi,\varphi\in\mathcal{D}[S]$ is given by $S[\psi,\varphi]=\lim_{n\to\infty}\langle \psi_n,S\varphi_n\rangle$, where $(\psi_n)_n$ and $(\varphi_n)_n$ are any two sequences in $\mathcal{D}(S)$ that converge, respectively, to $\psi$ and $\varphi$ in the above norm.
 \item The form $(S[\cdot],\mathcal{D}[S])$ is lower semi-bounded and closed. Therefore, the operator associated with $(S[\cdot],\mathcal{D}[S])$ is self-adjoint. It is called the \emph{Friedrichs extension} of $S$ and denoted by $S_\mathrm{F}$. By definition,
 \begin{equation}\label{eqII:-Fr-dom}
 \begin{split}
  \mathcal{D}(S_\mathrm{F})\;&=\;\left\{\psi\in\mathcal{D}[S]\,\left|
  \begin{array}{c}
  \exists\,\xi_\psi\in\cH\textrm{ such that } \\
  S[\varphi,\psi]=\langle \varphi,\xi_\psi\rangle\;\:\forall \varphi\in\mathcal{D}[S]
  \end{array}\!\right.\right\}, \\
  S_\mathrm{F} \psi\;&=\;\xi_\psi\,.
 \end{split}
 \end{equation}
  \item $S_\mathrm{F}$ is a lower semi-bounded self-adjoint extension of $S$ with the same greatest lower bound as $S$, i.e.,
 \begin{equation}\label{eq:II-mSFequalmS}
 \mathfrak{m}(S_\mathrm{F})\;=\;\mathfrak{m}(S)\,,
 \end{equation}
and whose associated quadratic form coincides with the closure if the form $(\psi,\varphi)\mapsto\langle \psi,Sg\rangle$ considered in (i)-(ii), i.e., 
 \begin{equation}\label{DfSF}
 \mathcal{D}[S_\mathrm{F}]\;=\;\mathcal{D}[S]\,,\qquad S_\mathrm{F}[\psi,\varphi]\;=\;S[\psi,\varphi]\,.
 \end{equation}
 \item $\mathcal{D}(S_\mathrm{F})=\mathcal{D}(S^*)\cap\mathcal{D}[S]$ and $S_\mathrm{F}=S^*\upharpoonright\mathcal{D}[S]$.
 \item $S_\mathrm{F}$ is the \emph{only} self-adjoint extension of $S$ whose operator domain is contained in $\mathcal{D}[S]$. 
 \item If $\widetilde{S}$ is another lower semi-bounded self-adjoint extension of $S$, then $S_\mathrm{F}\geqslant \widetilde{S}$.
 \item $(S+\lambda\mathbbm{1})_{\mathrm{F}}=S_\mathrm{F}+\lambda\mathbbm{1}$ for $\lambda\in\mathbb{R}$.
\end{enumerate}
\end{theorem}

\begin{proof}
 (i) and (ii). Obviously the form $q[\psi,\varphi]:=\langle\psi,S\varphi\rangle$, $\mathcal{D}[q]:=\mathcal{D}(S)$ is lower semi-bounded. To check its closability one has to show that for an arbitrary sequence $(\psi_n)_{n\in\mathbb{N}}\subset\mathcal{D}(S)$ with $\|\psi_n\|_{\cH} \to 0$ and $\langle\psi_n-\psi_m,S(\psi_n-\psi_m)\rangle\to 0$ necessarily $\langle\psi_n,S\psi_n\rangle\to 0$. In terms of form norm $\|\psi\|_q:=(\langle\psi,S\psi\rangle+(1-\mathfrak{m})\|\psi\|_\cH^2)^{\frac{1}{2}}$, where $\mathfrak{m}\leqslant\mathfrak{m}(S)$, $\|\psi_n-\psi_m\|_q^2=\langle\psi_n-\psi_m,S(\psi_n-\psi_m)\rangle+(1-\mathfrak{m})\|\psi_n-\psi_m\|_\cH^2\to 0$, meaning that $(\psi_n)_{n\in\mathbb{N}}$ is $\|\,\|_q$-Cauchy and hence $\|\,\|_q$-converges to some point $\psi$ in the $q$-norm completion $\cH_q$ of $\mathcal{D}(S)$. As
 \[
  \langle\psi,\varphi\rangle_q\,=\,\lim_{n\to\infty}\langle\psi_n,\varphi\rangle_q\,=\,\lim_{n\to\infty}\langle\psi_n,(S+(1-\mathfrak{m})\mathbbm{1})\varphi\rangle\,=\,0\qquad\forall\varphi\in\mathcal{D}(S)\,,
 \]
 and as $\mathcal{D}(S)$ is dense in $\cH_q$, then $\psi=0$, whence $\|\psi_n\|_q\to 0$. Therefore,
 \[
  \langle\psi_n,S\psi_n\rangle\,=\,\|\psi_n\|_q^2-(1-\mathfrak{m})\|\psi_n\|_\cH^2\,\to\,0\,.
 \]
 The remaining part of the statement of part (i), and part (ii), are immediate applications of the general mechanism of closure of a closable form (Sect.~\ref{sec:I_forms}).

 (iii) By standard properties (Sect.~\ref{sec:I_forms}), the bottom of $S_\mathrm{F}$ equals the bottom of its form $S_\mathrm{F}[\cdot]\equiv S[\cdot]$, which is the closure of the form $q$ and hence has the same bottom of $q$, and in turn the latter coincides by construction with the bottom of $S$. By construction $\mathcal{D}(S)\subset\mathcal{D}[S_\mathrm{F}]$ and $S_\mathrm{F}[\varphi,\psi]=\langle\varphi,S\psi\rangle$ for any $\psi,\varphi\in\mathcal{D}(S)$. Moreover (Sect.~\ref{sec:I_forms}), $\mathcal{D}(S)$ is a core for $S_\mathrm{F}[\cdot]$. Thus, at fixed $\varphi$, since both $S_\mathrm{F}[\cdot]$ and $\|\cdot\|_\cH$ are $\|\,\|_q$-continuous, the latter identity lifts to $S_\mathrm{F}[\varphi,\psi]=\langle\varphi,S\psi\rangle$ for any $\varphi\in\mathcal{D}[S_\mathrm{F}]$. Hence, owing to \eqref{eqII:-Fr-dom}, $\psi\in\mathcal{D}(S_\mathrm{F})$ and $S_F\psi=S\psi$, i.e., $S\subset S_\mathrm{F}$.

 (iv) The fact that $\mathcal{D}(S_\mathrm{F})\subset \mathcal{D}(S^*)\cap\mathcal{D}[S]$ follows from the obvious inclusion $\mathcal{D}(S_\mathrm{F})\subset \mathcal{D}[S_\mathrm{F}]=\mathcal{D}[S]$ (see \eqref{eqII:-Fr-dom} and \eqref{DfSF} above) and from the implication $S\subset S_\mathrm{F}\Rightarrow S_\mathrm{F}=S_\mathrm{F}^*\subset S^*$. For the converse, let $\psi\in \mathcal{D}(S^*)\cap\mathcal{D}[S]=\mathcal{D}(S^*)\cap\mathcal{D}[S_\mathrm{F}]$: then, for any $\varphi\in\mathcal{D}(S)\subset\mathcal{D}(S_\mathrm{F})\subset\mathcal{D}[S_\mathrm{F}]$,
 \[
     S_\mathrm{F}[\varphi,\psi]\;=\;\overline{S_\mathrm{F}[\psi,\varphi]}\;=\;\overline{\langle\psi,S_\mathrm{F}\varphi\rangle}\;=\;\langle S_\mathrm{F}\varphi,\psi\rangle\;=\;\langle S\varphi,\psi\rangle\;=\;\langle\varphi,S^*\psi\rangle\,,
 \]
  having used \eqref{eqII:-Fr-dom} in the second identity. As $\mathcal{D}(S)$ is a core for $S_\mathrm{F}[\cdot]$ and both $S_\mathrm{F}[\cdot]$ and $\|\cdot\|_\cH$ are $\|\,\|_q$-continuous, the latter identity lifts to $S_\mathrm{F}[\varphi,\psi]=\langle\varphi,S^*\psi\rangle$ for any $\varphi\in\mathcal{D}[S_\mathrm{F}]$. Hence, owing to \eqref{eqII:-Fr-dom}, $\psi\in\mathcal{D}(S_\mathrm{F})$. Thus, $\mathcal{D}(S_\mathrm{F})=\mathcal{D}(S^*)\cap\mathcal{D}[S]$. Since $S_\mathrm{F}\subset S^*$, and hence $S_\mathrm{F}=S^*\upharpoonright\mathcal{D}(S_\mathrm{F})$, and since $\mathcal{D}(S_\mathrm{F})=\mathcal{D}(S^*)\cap\mathcal{D}[S]$, then obviously $S_\mathrm{F}=S^*\upharpoonright\mathcal{D}[S]$.

  (v) If for some $\widetilde{S}=(\widetilde{S})^*\supset S$ one has $\mathcal{D}(\widetilde{S})\subset\mathcal{D}[S]=\mathcal{D}[S_\mathrm{F}]$, then $\widetilde{S}\subset S^*$ and $\mathcal{D}(\widetilde{S})\subset \mathcal{D}(S^*)\cap\mathcal{D}[S]=\mathcal{D}(S_\mathrm{F})$. As the self-adjoint $\widetilde{S}$ is maximally symmetric (Sect.~\ref{sec:I-symmetric-selfadj}), then $\widetilde{S}=S_\mathrm{F}$.

  (vi) Let $\widetilde{S}=(\widetilde{S})^*\supset S$ with $\mathfrak{m}(\widetilde{S})>-\infty$. The quadratic form $\langle\cdot,\widetilde{S}\,\cdot\,\rangle$ with domain $\mathcal{D}(\widetilde{S})$, extends the quadratic form $\langle\cdot,S\,\cdot\,\rangle$ with domain $\mathcal{D}(S)$. The same inclusion therefore holds for the respective closure forms, namely $\widetilde{S}[\cdot]$ extends $S_\mathrm{F}[\cdot]$. That is, $\widetilde{S}[\psi]=S_\mathrm{F}[\psi]$ $\forall\psi\in\mathcal{D}[S_\mathrm{F}]\subset\mathcal{D}[\widetilde{S}]$. Which means precisely $S_\mathrm{F}\geqslant\widetilde{S}$.

  (vii) One has the quadratic form identity $\langle\cdot,(S+\lambda\mathbbm{1})\,\cdot\,\rangle=\langle\cdot,S\,\cdot\,\rangle+\lambda\|\cdot\|_\cH^2$, which lifts to the corresponding form closures, hence $(S+\lambda\mathbbm{1})_{\mathrm{F}}[\cdot]=S_\mathrm{F}[\cdot]+\lambda \| \cdot \|^2=(S_\mathrm{F}+\lambda\mathbbm{1})[\cdot]$. The one-to-one form-operator correspondence discussed in Sect.~\ref{sec:I_forms} then implies $(S+\lambda\mathbbm{1})_{\mathrm{F}}=S_\mathrm{F}+\lambda\mathbbm{1}$.  
\end{proof}

  As straightforward as the following consequence of Theorem \ref{thm:Friedrichs-ext} is, it is convenient to highlight it for its standard usage in applications.

  \begin{lemma}\label{lem:sumofFriedrichs}
  Let $S=\bigoplus_{k\in\mathbb{Z}}S(k)$ be a direct sum operator acting on the orthogonal direct sum Hilbert space $\cH=\bigoplus_{k\in\mathbb{Z}}\cH_k$, where each $S(k)$ is densely defined, symmetric, and semi-bounded from below on $\cH_k$, with uniform lower bound
  \[
   \mathfrak{m}\;:=\;\inf_{k\in\mathbb{Z}}\;\inf_{\substack{u\in\mathcal{D}(S(k)) \\ u\neq 0}}  \frac{\langle u,S(k)u\rangle_{\cH_k}}{\|u\|^2_{\cH_k}}\;>\;-\infty\,.
  \]
  Denote by $S_\mathrm{F}$, resp.~$S_\mathrm{F}(k)$, the Friedrichs extension of $S$, resp.~$S(k)$. Then
  \[
   S_\mathrm{F}\;=\;\bigoplus_{k\in\mathbb{Z}}\,S_\mathrm{F}(k)\,.
  \]
 \end{lemma}

 \begin{proof}
  It is clear that $\bigoplus_{k\in\mathbb{Z}}S_\mathrm{F}(k)$ is a self-adjoint extension of $S$ (Sect.~\ref{sec:I_invariant-reducing-ssp} and \ref{sec:I-symmetric-selfadj}). To recognise it as the Friedrichs extension, it suffices to check (Theorem \ref{thm:Friedrichs-ext}(v)) that the \emph{operator} domain $\mathcal{D}(\bigoplus_{k\in\mathbb{Z}}S_\mathrm{F}(k))$ is an actual subspace of the \emph{form} domain $\mathcal{D}[S]$. To this aim, one observes that
  \[
   \begin{split}
    \mathcal{D}\bigg(\bigoplus_{k\in\mathbb{Z}}S_\mathrm{F}(k)\bigg)\;&=\;\op_{k\in\mathbb{Z}}\mathcal{D}(S_\mathrm{F}(k))\;\subset\;\op_{k\in\mathbb{Z}}\mathcal{D}[S(k)]\,,
   \end{split}
  \]
  the inclusion following from $\mathcal{D}(S_\mathrm{F}(k))\subset\mathcal{D}[S(k)]$, the Friedrichs-extension characterising property valid for each $k$. On the other hand (Sect.~\ref{sec:I_forms}), denoting with $\mathbbm{1}_k$ the identity on $\cH_k$,
  \[
   \mathcal{D}[S]\;=\;\mathcal{D}((S-\mathfrak{m}\mathbbm{1})^{\frac{1}{2}})\,,\qquad \mathcal{D}[S(k)]\;=\;\mathcal{D}((S(k)-\mathfrak{m}\mathbbm{1}_k)^{\frac{1}{2}})
  \]
whence
  \[
   \begin{split}
    \mathcal{D}[S]\;&=\;\mathcal{D}\Big[\bigoplus_{k\in\mathbb{Z}}S(k)\Big]\;=\;
    \mathcal{D}\Big(\Big(\bigoplus_{k\in\mathbb{Z}}\,(S(k)-\mathfrak{m}\mathbbm{1}_k)\Big)^{\!\frac{1}{2}}\,\Big) \\
    &=\;\mathcal{D}\Big(\bigoplus_{k\in\mathbb{Z}}\,(S(k)-\mathfrak{m}\mathbbm{1}_k)^{\frac{1}{2}}\Big)\;=\;\op_{k\in\mathbb{Z}}\mathcal{D}\big((S(k)-\mathfrak{m}\mathbbm{1}_k)^{\frac{1}{2}}\big)\;=\;\op_{k\in\mathbb{Z}}\mathcal{D}[S(k)]\,.
   \end{split}
  \]
 This proves the desired inclusion.
 \end{proof}

\section{Cayley transform of symmetric operators}

Given a Hilbert space $\cH$, a number $z\in\mathbb{C}\setminus\mathbb{R}$, and a densely defined and symmetric operator $S$, the \emph{Cayley transform}\index{Cayley transform} of $S$ (at the point $z$) is the operator 
\begin{equation}
V_S\;:=\;(S-z\mathbbm{1})(S-\overline{z}\mathbbm{1})^{-1}\,,\qquad\mathcal{D}(V_S)\,:=\,\ran(S-\overline{z}\mathbbm{1})\,.
\end{equation}
As $\overline{z}\notin\mathbb{R}$ and by symmetry $S$ can only have real eigenvalues (Sect.~\ref{sec:I-symmetric-selfadj}), $S-\overline{z}\mathbbm{1}$ is indeed invertible on its range. (In fact, by symmetry, $\|(S-\overline{z}\mathbbm{1})\psi\|\geqslant |\mathfrak{Im}\,z|\,\|\psi\|$ $\forall\psi\in\mathcal{D}(S)$ (see Sect.~\ref{sec:I-symmetric-selfadj}), and $\mathfrak{Im}\,z\neq 0$, therefore $S-\overline{z}\mathbbm{1}$ has bounded inverse on its range.) By construction,
\begin{equation}\label{eq:II-Cayley-equiv}
 V_S(S-\overline{z}\mathbbm{1})\psi\;=\;(S-z\mathbbm{1})\psi\qquad\forall\psi\in\mathcal{D}(S)
\end{equation}
and
\begin{equation}\label{sec:I-domranCayley}
 \begin{split}
  \mathcal{D}(V_S)\;&=\;\ran(S-\overline{z}\mathbbm{1})\,, \\
   \mathrm{ran}\,V_S\;&=\;\mathrm{ran}\,(S-z\mathbbm{1})\,.
 \end{split}
\end{equation}

\begin{theorem}[Cayley transform]\label{thm:II-Cayley_transf}\index{Cayley transform}~
 \begin{enumerate}[(i)]
  \item The Cayley transform $S\mapsto V_S$ is a bijective map of the set of densely defined symmetric operators $S$ on $\cH$ onto the set of all isometric (i.e., norm-preserving) operators $V$ on $\cH$ for which $\ran(\mathbbm{1}-V)$ is dense in $\cH$. Moreover, 
\begin{equation}\label{eq:Cayley-DomS}
\ran(\mathbbm{1}-V_S)\;=\;\mathcal{D}(S)\,.
\end{equation}
  \item For given $z\in\mathbb{C}\setminus\mathbb{R}$, and given isometric operator $V$ on $\cH$ with $\mathrm{ran}(\mathbbm{1}-V)$ dense in $\cH$, the operator
  \begin{equation}\label{eq:inverse-Cayley}
S_V\;:=\;(z\mathbbm{1}-\overline{z}V)(\mathbbm{1}-V)^{-1}\,,\qquad \mathcal{D}(S_V)\;:=\;\mathrm{ran}\,(\mathbbm{1}-V)
\end{equation}
is well defined, and the map $V \mapsto S_V$ inverts the Cayley transform $S\mapsto V_S$ at the point $z$.
 \item $S$ is closed if and only if so is $V_S$, in which case $V_S$ is an isometry between the \emph{closed} subspace $\ran(S-\overline{z}\mathbbm{1})$ to the \emph{closed} subspace $\mathrm{ran}\,(S-z\mathbbm{1})$.
 \item If $S'$ is another symmetric operator on $\cH$, then $S\subset S'$ if and only if $V_S\subset V_{S'}$.
 \end{enumerate}
\end{theorem}

The map $V\mapsto S_V$ defined in \eqref{eq:inverse-Cayley} is called the \emph{inverse Cayley transform}.\index{inverse Cayley transform}

\begin{proof}[Proof of Theorem \ref{thm:II-Cayley_transf}]~

(i) and (ii). Let $\psi\in\mathcal{D}(S)$. By symmetry,
\[
 \left. 
 \begin{array}{c}
  \|(S-\overline{z}\mathbbm{1})\psi\|^2 \\
  \|(S-z\mathbbm{1})\psi\|^2
 \end{array}
 \right\}\,=\,\|(S-\mathfrak{Re}\,z\mathbbm{1})\psi\pm\mathfrak{Im}\,z\,\psi\|^2\,=\,\|(S-\mathfrak{Re}z\,\mathbbm{1})\psi\|^2+|\mathfrak{Im}\,z|^2\,\|\psi\|^2\,.
\]
This and \eqref{eq:II-Cayley-equiv} imply
\[
 \begin{split}
  \|V_S(S-\overline{z}\mathbbm{1})\psi\|\;=\;\|(S-z\mathbbm{1})\psi\|\;=\;\|(S-\overline{z}\mathbbm{1})\psi\|\,.
 \end{split}
\]
This proves the isometric property of $V_S$. Beside, the inequality $\|(S-\overline{z}\mathbbm{1})\psi\|\geqslant |\mathfrak{Im}\,z|\,\|\psi\|$ is established.

Furthermore, since \eqref{eq:II-Cayley-equiv} implies $(\mathbbm{1}-V_S)(S-\overline{z}\mathbbm{1})\psi=(z-\overline{z})\psi$, then \eqref{eq:Cayley-DomS} is true. Since $z-\overline{z}\neq 0$, one also deduces
\[
 (\mathbbm{1}-V_S)(S-\overline{z}\mathbbm{1})\psi\,=\,0\quad\Rightarrow\quad \psi\,=0\,\quad\Rightarrow\quad(S-\overline{z}\mathbbm{1})\psi\,=\,0\,,
\]
that is, $\ker(\mathbbm{1}-V_S)=\{0\}$ and $(S-\overline{z}\mathbbm{1})\psi=(z-\overline{z})(\mathbbm{1}-V_S)^{-1}\psi$. On the other hand, using \eqref{eq:II-Cayley-equiv} again, one finds $(z\mathbbm{1}-\overline{z}V_S)(S-\overline{z}\mathbbm{1})\psi=(z-\overline{z})S\psi$, whence 
\[
 \begin{split}
    S\psi\;&=\;(z-\overline{z})^{-1}(z\mathbbm{1}-\overline{z}V_S)(S-\overline{z}\mathbbm{1})\psi   \\
    &=\;(z-\overline{z})^{-1}(z\mathbbm{1}-\overline{z}V_S)(z-\overline{z})(\mathbbm{1}-V_S)^{-1}\psi 
 \end{split}
\]
So far, this shows that $S\subset(z\mathbbm{1}-\overline{z}V_S)(\mathbbm{1}-V_S)^{-1}$. But if conversely $\psi$ is in the domain of $(z\mathbbm{1}-\overline{z}V_S)(\mathbbm{1}-V_S)^{-1}$, then $\psi\in\mathcal{D}((\mathbbm{1}-V_S)^{-1})=\mathrm{ran}\,(\mathbbm{1}-V_S)=\mathcal{D}(S)$, owing to \eqref{eq:Cayley-DomS}. Therefore $S=(z\mathbbm{1}-\overline{z}V_S)(\mathbbm{1}-V_S)^{-1}$.

It remains to prove that the map $V\mapsto S_V$ defined in \eqref{eq:inverse-Cayley} is well defined and actually inverts $S\mapsto V_S$ among the mentioned spaces. So now $V$ is a norm-preserving operator such that $\mathrm{ran}(\mathbbm{1}-V)$ is dense in $\cH$. In fact, the density of $\mathrm{ran}(\mathbbm{1}-V)$ implies \emph{injectivity} of $\mathbbm{1}-V$: for, if $\varphi\in\ker(\mathbbm{1}-V)$ (hence $V\varphi=\varphi$), then 
\[
 \langle (\mathbbm{1}-V)\xi,\varphi\rangle\;=\;\langle \xi,\varphi\rangle-\langle V\xi,\varphi\rangle\;=\;\langle \xi,\varphi\rangle-\langle V\xi,V\varphi\rangle\;=\;\langle \xi,\varphi\rangle-\langle \xi,\varphi\rangle\;=\;0
\]
for any $\xi\in\mathcal{D}(V)$, whence $\varphi=0$ because $\mathrm{ran}(\mathbbm{1}-V)$ is dense. Therefore, $\mathbbm{1}-V$ is invertible on its range, and the operator
\[
 S_V\;:=\;(z\mathbbm{1}-\overline{z}V)(\mathbbm{1}-V)^{-1}\,,\qquad \mathcal{D}(S_V)\;:=\;\mathrm{ran}\,(\mathbbm{1}-V)
\]
is well-defined and with dense domain. By definition, $S_V (\mathbbm{1}-V)\xi=(z\mathbbm{1}-\overline{z}V)\xi$ for any $\xi\in\mathcal{D}(V)$.

$S_V$ is also symmetric, because for every $\psi=(\mathbbm{1}-V)\varphi\in\mathcal{D}(S_V)$ one has
\[
 \begin{split}
  \langle\psi,S_V\psi\rangle\;&=\;\langle (\mathbbm{1}-V)\varphi,S_V (\mathbbm{1}-V)\varphi\rangle\;=\; \langle (\mathbbm{1}-V)\varphi,  (z\mathbbm{1}-\overline{z}V)\varphi \rangle \\
  &=\;z\|\varphi\|^2+\overline{z}\|V \varphi\|^2-z\langle V\varphi, \varphi\rangle-\overline{z}\langle \varphi,V\varphi\rangle \\
  &=\;2\,\|\varphi\|^2\mathfrak{Re}\,z-2\,\langle V\varphi,\varphi\rangle\,\;\in\;\mathbb{R}\,.
 \end{split}
\]

Last, one checks that the Cayley transform of $S_V$ is precisely $V$. The above identity $S_V (\mathbbm{1}-V)\xi=(z\mathbbm{1}-\overline{z}V)\xi$, valid for $\xi\in\mathcal{D}(V)$, allows to compute
\[
 (S_V-\overline{z}\mathbbm{1})(\mathbbm{1}-V)\xi\;=\;(z-\overline{z})\xi\,,\quad\textrm{ whence }\quad (\mathbbm{1}-V)\xi\;=\;(z-\overline{z})(S_V-\overline{z}\mathbbm{1})^{-1}\xi\,,
\]
and also $(S_V-z\mathbbm{1})(\mathbbm{1}-V)\xi=(z-\overline{z})V\xi$, whence 
\[
 V\xi\;=\;(z-\overline{z})^{-1}(S_V-z\mathbbm{1})(\mathbbm{1}-V)\xi\;=\;(z-\overline{z})^{-1}(z-\overline{z})(S_V-z\mathbbm{1})(S_V-\overline{z}\mathbbm{1})^{-1}\xi\,.
\]
This shows that $V\subset (S_V-z\mathbbm{1})(S_V-\overline{z}\mathbbm{1})^{-1}$. From $\mathcal{D}(S_V-\overline{z}\mathbbm{1})=\mathcal{D}(S_V)=\mathrm{ran}(\mathbbm{1}-V)$ and $(S_V-\overline{z}\mathbbm{1})(\mathbbm{1}-V)\xi=(z-\overline{z})\xi$ $\forall\xi\in\mathcal{D}(V)$, one also deduces $\mathrm{ran}\,(S_V-\overline{z}\mathbbm{1})=\mathcal{D}(V)$.
On the other hand, 
\[
 \mathcal{D}\big( (S_V-z\mathbbm{1})(S_V-\overline{z}\mathbbm{1})^{-1}\big)\;\subset\;\mathcal{D}\big( (S_V-\overline{z}\mathbbm{1})^{-1}\big)\;=\;\mathrm{ran}\,(S_V-\overline{z}\mathbbm{1})\;=\;\mathcal{D}(V)\,,
\]
whence finally $V = (S_V-z\mathbbm{1})(S_V-\overline{z}\mathbbm{1})^{-1}$. This proves that $V$ is the Cayley transform of $S_V$.

(iii) The closedness of the bounded operator $V_S$ is equivalent to the closedness in $\cH$ of the subspace $\mathcal{D}(V_S)=\mathrm{ran}(S-\overline{z}\mathbbm{1})$ (Sect.~\ref{sec:I-bdd-closable-closed}). Concerning the subspace $\mathrm{ran}\,(S-\overline{z}\mathbbm{1})$, assume that $S$ is closed and let $\varphi\in\overline{\mathrm{ran}\,(S-\overline{z}\mathbbm{1})}$, i.e., $(S-\overline{z}\mathbbm{1})\psi_n\to\varphi$ as a limit in $\cH$ for some sequence $(\psi_n)_{n\in\mathbb{N}}$ in $\mathcal{D}(S)$. As $\|(S-\overline{z}\mathbbm{1})(\psi_n-\psi_m)\|\geqslant |\mathfrak{Im}\,z|\,\|\psi_n-\psi_m\|$ and $\mathfrak{Im}\,z\neq 0$, the sequence $(\psi_n)_{n\in\mathbb{N}}$ is a Cauchy sequence in $\cH$, hence $\psi_n\to\psi$ for some $\psi\in\cH$. Owing to the closedness of $S$ (Sect.~\ref{sec:I-bdd-closable-closed}), then $\psi\in\mathcal{D}(S)$ and $\varphi= (S-\overline{z}\mathbbm{1})\psi\in\mathrm{ran}\,(S-\overline{z}\mathbbm{1})$, which proves that $\mathrm{ran}\,(S-\overline{z}\mathbbm{1})$ is closed in $\cH$. Conversely, assume that $\mathrm{ran}\,(S-\overline{z}\mathbbm{1})$ is closed in $\cH$ and assume also that $(\psi_n)_{n\in\mathbb{N}}$ is a sequence in $\mathcal{D}(S)$ such that $\psi_n\to\psi$ and $(S-\overline{z}\mathbbm{1})\psi_n\to\varphi$ for some $\psi,\varphi\in\cH$. In particular, $\varphi\in\overline{\mathrm{ran}\,(S-\overline{z}\mathbbm{1})}=\mathrm{ran}\,(S-\overline{z}\mathbbm{1})$, thus, $\varphi=(S-\overline{z}\mathbbm{1})\widetilde{\psi}$ for some $\widetilde{\psi}\in\mathcal{D}(S)$. Then
\[
 \begin{split}
   \|\psi-\widetilde{\psi}\|\;&\leqslant\;\|\psi-\psi_n\|+\|\psi_n-\widetilde{\psi}\| \\
   &\leqslant\;\|\psi-\psi_n\|+|\mathfrak{Im}\,z|^{-1}\,\|(S-\overline{z}\mathbbm{1})(\psi_n-\widetilde{\psi})\|\;\longrightarrow\;0\,,
 \end{split}
\]
i.e., $\psi=\widetilde{\psi}\in\mathcal{D}(S)$ and $\varphi=(S-\overline{z}\mathbbm{1})\psi$. This means (Sect.~\ref{sec:I-bdd-closable-closed}) that $S$ is closed. The above reasoning can be repeated verbatim with $z$ in place of $\overline{z}$, thereby proving that $S$ is closed if and only if the subspace $\mathrm{ran}(S-z\mathbbm{1})$ is closed.

(iv) From \eqref{eq:II-Cayley-equiv}, namely
\[
 \begin{split}
   V_S(S-\overline{z}\mathbbm{1})\psi\;&=\;(S-z\mathbbm{1})\psi\qquad\forall\psi\in\mathcal{D}(S)\,, \\
   V_{S'}(S'-\overline{z}\mathbbm{1})\psi'\;&=\;(S'-z\mathbbm{1})\psi'\quad\;\;\,\forall\psi'\in\mathcal{D}(S')\,,
 \end{split}
\]
the claimed property follows directly.   
\end{proof}

\begin{corollary}\label{cor:II-caley1}~
\begin{enumerate}[(i)]
 \item A densely defined symmetric operator is self-adjoint if and only if its Cayley transform\index{Cayley transform} is unitary on the whole Hilbert space.
 \item A unitary operator $V$ is the Cayley transform\index{Cayley transform} of a densely defined symmetric operator if and only if $\ker(\mathbbm{1}-V)=\{0\}$.
\end{enumerate}
\end{corollary}

\begin{proof}
 (i) For a densely defined symmetric operator $S$ (i.e., $S\subset S^*$), $S=S^*$ if and only if $\cH=\ran(S-z\mathbbm{1})=\ran(S-\overline{z}\mathbbm{1})$ (Sect.~\ref{sec:I-symmetric-selfadj}). The latter condition, on account of \eqref{sec:I-domranCayley},
 is the same as  $\cH=\ran(V_S)=\mathcal{D}(V_S)$. That is, the norm-preserving $V_S$ is a unitary map on the whole $\cH$.
 
 (ii) On account of part (i) and Theorem \ref{thm:II-Cayley_transf}(i), $V$ is the Cayley transform of a densely defined symmetric operator on a Hilbert space $\cH$ if and only if $\mathrm{ran}\,(\mathbbm{1}-V)$ is dense in $\cH$. The latter condition and the identity, valid for any $\xi\in\cH$ and $\varphi\in\ker(\mathbbm{1}-V)$,
 \[
 \langle (\mathbbm{1}-V)\xi,\varphi\rangle\;=\;\langle \xi,\varphi\rangle-\langle V\xi,\varphi\rangle\;=\;\langle \xi,\varphi\rangle-\langle V\xi,V\varphi\rangle\rangle\;=\;\langle \xi,\varphi\rangle-\langle \xi,\varphi\rangle\;=\;0\,,
\]
 is equivalent to $\ker(\mathbbm{1}-V)=\{0\}$.  
\end{proof}

\begin{corollary}\label{cor:II-caley2}
 Let $S$ be a densely defined symmetric operator on a Hilbert space $\cH$, and let $V_S$ be its Cayley transform.\index{Cayley transform} 
 There is a one-to-one correspondence between the symmetric (respectively, closed symmetric; respectively, self-adjoint) extensions $\widetilde{S}$ of $S$ in $\cH$ and the isometric (respectively, closed isometric; respectively, unitary) extensions $\widetilde{V}$ of $V_S$ in $\cH$.
 Such a correspondence is explicitly given by $\widetilde{S}\stackrel{1:1}{\longmapsto}\widetilde{V}=V_{\widetilde{S}}$\,.
\end{corollary}

\begin{proof}
 If $\widetilde{S}$ is a symmetric extension of $S$, then (Theorem \ref{thm:II-Cayley_transf}(i) and (iv)) $V_{\widetilde{S}}$ is an isometric extension of $V_S$.
 Conversely, if $\widetilde{V}$ is an isometric extension of $V_S$, then $\mathrm{ran}(\mathbbm{1}-\widetilde{V})\supset \mathrm{ran}(\mathbbm{1}-V_{S})$, hence $\mathrm{ran}(\mathbbm{1}-\widetilde{V})$ is dense in $\cH$ because (Theorem \ref{thm:II-Cayley_transf}(i)) so is $\mathrm{ran}(\mathbbm{1}-V_{S})$. As such, $\widetilde{V}=V_{\widetilde{S}}$ for some symmetric extension $\widetilde{S}$ of $S$ (Theorem \ref{thm:II-Cayley_transf}(i) and (iv)). This establishes the claimed correspondence $\widetilde{S}\stackrel{1:1}{\longmapsto}\widetilde{V}=V_{\widetilde{S}}$ in the general case of symmetric extensions of $S$. The same correspondence holds between closed such $\widetilde{S}$'s and closed such $\widetilde{V}$'s, on account of Theorem \ref{thm:II-Cayley_transf}(iii). And the same correspondence holds between self-adjoint such  $\widetilde{S}$'s and unitary such $\widetilde{V}$'s on account of Corollary \ref{cor:II-caley1}.  
\end{proof}

\section{von Neumann's extension theory of symmetric operators}\label{sec:II-vN-theory}\index{von Neumann's extension theory}

 A convenient starting point is \emph{von Neumann's formula} for the domain of the adjoint of a densely defined and symmetric operator on Hilbert space. It was stated in Section \ref{sec:I-symmetric-selfadj} as part of standard preliminary materials, but it is convenient to present also the proof here, as the proof proceeds through reasonings that are quite typical of self-adjoint extension theories.

 \begin{proposition}[von Neumann's formula]\label{prop:II-vNformula}\index{von Neumann's formula}\index{theorem!von Neumann (formula)}\index{von Neumann's formula} \noindent Let $S$ be a densely defined symmetric operator on a Hilbert space $\cH$, and let $z\in\mathbb{C}\setminus\mathbb{R}$. Then
  \begin{equation}\label{eq:II-vNformula}
    \begin{split}
  & \mathcal{D}(S^*)\;=\;\mathcal{D}(\overline{S})\dotplus\ker(S^*-z\mathbbm{1})\dotplus\ker(S^*-\overline{z}\mathbbm{1})\,,\\
  & \dim\big(\mathcal{D}(S^*)/\mathcal{D}(\overline{S})\big)\;=\;d_+(S)+d_-(S)\,.
 \end{split}
  \end{equation}
 \end{proposition}

 \begin{proof}
  Obviously, $\mathcal{D}(S^*)\supset\mathcal{D}(\overline{S})+\ker(S^*-z\mathbbm{1})+\ker(S^*-\overline{z}\mathbbm{1})$. For the converse inclusion, let $g\in\mathcal{D}(S^*)$. With respect to the decomposition $\cH=\mathrm{ran}(\overline{S}-z\mathbbm{1})\oplus\ker(S^*-\overline{z}\mathbbm{1})$ (Sect.~\ref{sec:I-symmetric-selfadj}), one has $(S^*-z\mathbbm{1})g=(\overline{S}-z\mathbbm{1})f\oplus\widetilde{u}_-$ for some $f\in\mathcal{D}(\overline{S})$ and $\widetilde{u}_-\in \ker(S^*-\overline{z}\mathbbm{1})$. Set $u_-:=(\overline{z}-z)^{-1}\widetilde{u}_-$, so that $(S^*-z\mathbbm{1})u_-=\widetilde{u}_-$. Therefore, $(S^*-z\mathbbm{1})g=(\overline{S}-z\mathbbm{1})f+\widetilde{u}_-=(\overline{S}-z\mathbbm{1})f+(S^*-z\mathbbm{1})u_-$, from which one obtains $(S^*-z\mathbbm{1})(g-f-u_-)=0$, i.e., $u_+:=g-f-u_-\in \ker(S^*-z\mathbbm{1})$. Thus, $g=f+u_+ +u_-$, which proves the inclusion $\mathcal{D}(S^*)\subset\mathcal{D}(\overline{S})+\ker(S^*-z\mathbbm{1})+\ker(S^*-\overline{z}\mathbbm{1})$. The first identity in \eqref{eq:II-vNformula} is established up to checking that the sums are direct, which is going to be shown now. In fact, if $f+u_+ +u_-=0$ for $f\in\mathcal{D}(\overline{S})$, $u_+\in\ker(S^*-z\mathbbm{1})$, $u_-\in \ker(S^*-\overline{z}\mathbbm{1})$, then $0=(S^*-z\mathbbm{1})(f+u_+ +u_-)=(\overline{S}-z\mathbbm{1})f+(\overline{z}-z)u_-$, whence
  \[
  \begin{split}
    u_-\;&=\;-(\overline{z}-z)^{-1}(\overline{S}-z\mathbbm{1})f \\
    &\in\; \mathrm{ran}(\overline{S}-z\mathbbm{1})\cap\ker(S^*-\overline{z}\mathbbm{1}) \;=\;\mathrm{ran}(\overline{S}-z\mathbbm{1})\cap\big(\mathrm{ran}(\overline{S}-z\mathbbm{1})\big)^\perp=\{0\} \, .
  \end{split}
  \]
One thus deduces that $u_-=0$ and $\overline{S}f=zf$. Owing to the symmetry of $\overline{S}$, with $z\notin\mathbb{R}$, necessarily $f=0$. Then also $u_+=0$. The sum in \eqref{eq:II-vNformula} indeed direct. The second line in \eqref{eq:II-vNformula} follows at once from the first and from the definition of deficiency indices (Sect.~\ref{sec:I-symmetric-selfadj}).    
 \end{proof}

  Along the same route, one proves other von Neumann-like decomposition formulas.

  \begin{proposition}\label{prop:II-vNlike-decomp}\index{theorem!von Neumann (formula)}
   Let $S$ be a densely defined symmetric operator on a Hilbert space $\cH$, let $\widetilde{S}$ be a self-adjoint extension of $S$, and let $z\in\rho(\widetilde{S})$ (in particular, $z\in\mathbb{C}\setminus\mathbb{R}$). Then
   \begin{eqnarray}
    \label{eq:II-vNlike1} \mathcal{D}(S^*)&=&\mathcal{D}(\overline{S})\dotplus(\widetilde{S}-z\mathbbm{1})^{-1}\ker(S^*-\overline{z}\mathbbm{1})\dotplus\ker(S^*-z\mathbbm{1})\,,  \\
    \label{eq:II-vNlike2} \mathcal{D}(S^*)&=&\mathcal{D}(\widetilde{S})\dotplus\ker(S^*-z\mathbbm{1})\,, \\
    \label{eq:II-vNlike3} \mathcal{D}(\widetilde{S})&=&\mathcal{D}(\overline{S})\dotplus(\widetilde{S}-z\mathbbm{1})^{-1}\ker(S^*-\overline{z}\mathbbm{1})\,.
   \end{eqnarray}
  \end{proposition}

  \begin{proof}
   Since $S$ is closable and densely defined (Sect.~\ref{sec:I-symmetric-selfadj}), and since $z\in\rho(\widetilde{S})$ and therefore $(\widetilde{S}-z\mathbbm{1})^{-1}$ exists everywhere defined and bounded on $\cH$, then (Sect.~\ref{sec:I-adjoint}) $\cH=\ran\,(\overline{S}-z\mathbbm{1})\oplus\ker(S^*-\overline{z}\mathbbm{1})$. Owing to such a decomposition, for any $g\in\mathcal{D}(S^*)$ there are $f\in\mathcal{D}(\overline{S})$ and $u_-\in\ker(S^*-\overline{z}\mathbbm{1})$ such that $(S^*-z\mathbbm{1})g=(\overline{S}-z\mathbbm{1})f\oplus u_-$. From this identity and from $S\subset\widetilde{S}=(\widetilde{S})^*\subset S^*$ one deduces that the vector $u_+:=g-f-(\widetilde{S}-z\mathbbm{1})^{-1}u_-$ satisfies $(S^*-z\mathbbm{1})u_+\;=\;(S^*-z\mathbbm{1})g-(\overline{S}-z\mathbbm{1})f-u_-=0$, that is, $u_+\in\ker(S^*-z\mathbbm{1})$. Thus, $g=f+(\widetilde{S}-z\mathbbm{1})^{-1}u_-+u_+$, which shows that $\mathcal{D}(S^*)\subset\mathcal{D}(\overline{S})\dotplus(\widetilde{S}-z\mathbbm{1})^{-1}\ker(S^*-\overline{z}\mathbbm{1})\dotplus\ker(S^*-z\mathbbm{1})$. The opposite inclusion is trivial, hence \eqref{eq:II-vNlike1} is established, but for the proof that the sum is direct, which is checked as follows. If for some  $f\in\mathcal{D}(\overline{S})$, $u_+\in\ker(S^*-z\mathbbm{1})$, and $u_-\in\ker(S^*-\overline{z}\mathbbm{1})$ one has $g:=f+(\widetilde{S}-z\mathbbm{1})^{-1}u_-+u_+=0$, then $0=(S^*-z\mathbbm{1})g=(\overline{S}-z\mathbbm{1})f+u_-$, whence $(\overline{S}-z\mathbbm{1})f=u_-=0$ because $(\overline{S}-z\mathbbm{1})f\perp u_-$. This means that $u_-=0$ and $f=(\widetilde{S}-z\mathbbm{1})^{-1}(\overline{S}-z\mathbbm{1})f=0$, whence also $u_+=0$: the sum is direct. The identity \eqref{eq:II-vNlike2} follows at once by comparing \eqref{eq:II-vNlike1} and \eqref{eq:II-vNlike3}, so it remains to prove \eqref{eq:II-vNlike3}. The sum in \eqref{eq:II-vNlike1} is direct, and the inclusion $\mathcal{D}(\widetilde{S})\supset\mathcal{D}(\overline{S})+(\widetilde{S}-z\mathbbm{1})^{-1}\ker(S^*-\overline{z}\mathbbm{1})$ is obvious, so it suffices to show that $\mathcal{D}(\widetilde{S})\subset\mathcal{D}(\overline{S})+(\widetilde{S}-z\mathbbm{1})^{-1}\ker(S^*-\overline{z}\mathbbm{1})$. Let $g\in\mathcal{D}(\widetilde{S})$. Owing to \eqref{eq:II-vNlike1}, there are $f\in\mathcal{D}(\overline{S})$, $u_+\in\ker(S^*-z\mathbbm{1})$, and $u_-\in\ker(S^*-\overline{z}\mathbbm{1})$ such that $g=f+(\widetilde{S}-z\mathbbm{1})^{-1}u_-+u_+$. Since $f+(\widetilde{S}-z\mathbbm{1})^{-1}u_-\in\mathcal{D}(\widetilde{S})$, then also $u_+\in\mathcal{D}(\widetilde{S})$, and $(\widetilde{S}-z\mathbbm{1})u_+=(S^*-z\mathbbm{1})u_+=0$, meaning that $u_+=0$, because $z\in\rho(\widetilde{S})$. Thus, $g=f+(\widetilde{S}-z\mathbbm{1})^{-1}u_-\in\mathcal{D}(\overline{S})+(\widetilde{S}-z\mathbbm{1})^{-1}\ker(S^*-\overline{z}\mathbbm{1})$.   
  \end{proof}

  Since for a closed symmetric (in particular self-adjoint) extension $\widetilde{S}$ of the densely defined symmetric operator $S$ one has $S\subset\widetilde{S}=(\widetilde{S})^*$ and hence $\widetilde{S}=\overline{\widetilde{S}}=\widetilde{S}^{**}\subset(\widetilde{S})^*$,
  any such $\widetilde{S}$ must be a closed restriction of $S^*$. In view von Neumann's formula \eqref{eq:II-vNformula} it is natural to expect that such a restriction is to be formulated in terms of some additional condition among the summand subspaces $\mathcal{D}(\overline{S})$, $\ker(S^*-z\mathbbm{1})$, $\ker(S^*-\overline{z}\mathbbm{1})$. Instead, it is not obvious, a priori, that the actual restriction condition only involves the deficiency subspaces $\ker(S^*-z\mathbbm{1})$, $\ker(S^*-\overline{z}\mathbbm{1})$.

\begin{theorem}\label{thm:vonN_thm_symm_exts}\index{theorem!von Neumann (symmetric extensions)}
  Let $S$ be a densely defined symmetric operator on a Hilbert space $\cH$. Any closed symmetric extension $\widetilde{S}$ of $S$ is of the form $S_U$ for some isometry $U:\mathcal{V}_+\to\mathcal{V}_-$ between closed subspaces $ \mathcal{V}_+\subset\ker(S^*-z\mathbbm{1})$ and $ \mathcal{V}_-\subset\ker(S^*-\overline{z}\mathbbm{1})$ for one, hence for all $z\in\mathbb{C}$ with $\mathfrak{Im}\,z>0$, such that $\dim \mathcal{V}_+=\dim \mathcal{V}_-$. Any such $S_U$ is given by
  \begin{equation}\label{eq:II-symm-ext}
   \begin{split}
    \mathcal{D}(S_U)\;&:=\;\mathcal{D}(\overline{S})\dotplus(\mathbbm{1}-U)\mathcal{V}_+\,, \\
    S_U(f+(\mathbbm{1}-U)u)\;&:=\;\overline{S}f+z u-\overline{z} Uu\,.
   \end{split}
  \end{equation}
  Moreover, $S$ and $\widetilde{S}$ have deficiency indices, respectively $d_\pm(\widetilde{S})$ and $d_\pm(S)$, satisfying
  \begin{equation}\label{eq:II-symm-ext-def}
   d_\pm(\widetilde{S})\;=\;d_\pm(S)-\dim\mathcal{V}_\pm\,.
  \end{equation}
\end{theorem}

 A frequent choice in Theorem \ref{thm:vonN_thm_symm_exts} is $z=\ii$, in which case
  \begin{equation}
   \begin{split}
    \mathcal{V}_\pm\;&\subset\;\ker(S^*\mp\ii\mathbbm{1})\,, \\
    \mathcal{D}(S_U)\;&=\;\mathcal{D}(\overline{S})\dotplus(\mathbbm{1}-U)\mathcal{V}_+\,, \\
    S_U(f+(\mathbbm{1}-U)u)\;&=\;\overline{S}f+\ii u+\ii \,Uu\,.
   \end{split}
  \end{equation}

 \begin{proof}[Proof of Theorem \ref{thm:vonN_thm_symm_exts}]
  Since any closed symmetric extension of $S$ is a closed symmetric extension of $\overline{S}$, one can assume that $S$ is closed in the first place.

  As seen in Corollary \ref{cor:II-caley2}, the Cayley transform $\widetilde{S}\mapsto \widetilde{V}=V_{\widetilde{S}}$ gives a one-to-one correspondence between the closed symmetric extensions $\widetilde{S}$ of $S$ and the closed isometric extensions $\widetilde{V}$ of $V_S$, equivalently (Sect.~\ref{sec:I-bdd-closable-closed}), isometric extensions $\widetilde{V}$ of $V_S$ for which $\mathcal{D}(\widetilde{V})$ is a closed subspace of $\cH$. Recall also (as seen in the proof of Theorem \ref{thm:II-Cayley_transf}(ii)) that $\mathbbm{1}-\widetilde{V}$ is injective, owing to the density of $\mathrm{ran}(\mathbbm{1}-\widetilde{V})$.

  Now, $\mathcal{D}(V_S)=\mathrm{ran}(S-\overline{z}\mathbbm{1})$ (see \eqref{sec:I-domranCayley} above), the subspace $\mathcal{D}(\widetilde{V})$ must extend the \emph{closed} subspace $\mathcal{D}(V_S)=\mathrm{ran}(S-\overline{z}\mathbbm{1})$, and (see Sect.~\ref{sec:I-spectrum}) $\cH=\mathrm{ran}(S-\overline{z}\mathbbm{1})\oplus\ker(S^*-z\mathbbm{1})$: therefore, $\mathcal{D}(\widetilde{V})=\mathcal{D}(V_S)\oplus \mathcal{V}_+$ for some \emph{closed} subspace $\mathcal{V}_+\subset \ker(S^*-z\mathbbm{1})$. Moreover, $\widetilde{V}$ must map $\mathcal{V}_+$ isometrically to $\widetilde{V}\mathcal{V}_+\subset\mathrm{ran}\,\widetilde{V}=\mathrm{ran}\,V_{\widetilde{S}}=\mathrm{ran}(\widetilde{S}-z\mathbbm{1})$ (the latter identity following from \eqref{sec:I-domranCayley}), and $\mathrm{ran}(\widetilde{S}-z\mathbbm{1})\supset \mathrm{ran}(S-z\mathbbm{1})=\mathrm{ran}\,V_S$ (again owing to \eqref{sec:I-domranCayley}): due to the injectivity of $\widetilde{V}$,  $\mathcal{V}_-:=\widetilde{V}\mathcal{V}_+$ must be a \emph{closed} subspace in $\mathrm{ran}(S-z\mathbbm{1})^\perp=\ker(S^*-\overline{z}\mathbbm{1})$. Summarising, each such $\widetilde{V}$ is necessarily of the form
  \[
   \begin{split}
       \mathcal{D}(\widetilde{V})\;&=\;\mathcal{D}(V_S)\oplus\mathcal{V}_+\,, \\
       \widetilde{V} (v\oplus u)\;&=\; V_S \,v\oplus U\,u\,,
   \end{split}
  \]
  for some isometric operator $U:\mathcal{V}_+\to\mathcal{V}_-$.

  As $\widetilde{V}=V_{\widetilde{S}}$ \,, then $\widetilde{S}=(z\mathbbm{1}-\overline{z}\widetilde{V})(1-\widetilde{V})^{-1}$ (Theorem \ref{thm:II-Cayley_transf}(ii)); using this identity, the analogous identity $S=(z\mathbbm{1}-\overline{z}V_S)(1-V_S)^{-1}$, and $\widetilde{V}\supset V_S$, one deduces
  \[
   \begin{split}
       \mathcal{D}(\widetilde{S})\;&=\;\mathrm{ran}(\mathbbm{1}-\widetilde{V})\;=\;(\mathbbm{1}-\widetilde{V})\mathcal{D}(\widetilde{V})\;=\;(\mathbbm{1}-\widetilde{V})\mathcal{D}(V_S)\dotplus(\mathbbm{1}-\widetilde{V})\mathcal{V}_+ \\
       &=\; \mathcal{D}(S)\dotplus(\mathbbm{1}-\widetilde{V})\mathcal{V}_+\,,
   \end{split}
  \]
  the last two sums being \emph{direct} owing to the injectivity of $\mathbbm{1}-\widetilde{V}$. This reproduces the first line in \eqref{eq:II-symm-ext}. In turn, the second line follows from the first using the fact that $\widetilde{S}=S^*|_{\mathcal{D}(\widetilde{S})}$:
  \[
   \begin{split}
       \widetilde{S}(f+(\mathbbm{1}-U)u)\;&=\;S^*(f+(\mathbbm{1}-U)u)\;=\;Sf+S^*(\mathbbm{1}-U)u \\
       &=\;Sf+zu-\overline{z}\, Uu\,,
   \end{split}
  \]
  having used $S^*u=zu$ and $S^*Uu=\overline{z}\,Uu$ in the last identity.

  Last, from $\mathcal{D}(\widetilde{V})=\mathcal{D}(V_S)\oplus\mathcal{V}_+$, one easily deduces $\mathcal{D}(V_S)^\perp=\mathcal{D}(\widetilde{V})^\perp\oplus\mathcal{V}_+$; moreover, since $\mathfrak{Im}\,z>0$, then (Sect.~\ref{sec:I-symmetric-selfadj})
  \[
   \begin{split}
   \dim\mathcal{D}(V_{S})^\perp\;&=\;\dim\big(\mathrm{ran}(S-\overline{z}\mathbbm{1})\big)^\perp\;=\; d_+(S)\,, \\
    \dim\mathcal{D}(\widetilde{V})^\perp\;&=\;\dim\mathcal{D}(V_{\widetilde{S}})^\perp\;=\;\dim\big(\mathrm{ran}(\widetilde{S}-\overline{z}\mathbbm{1})\big)^\perp\;=\;d_+(\widetilde{S})\,.
   \end{split}
  \]
 Therefore, $d_+(S)=d_+(\widetilde{S})+\dim \mathcal{V}_+$. Analogously, $d_-(S)=d_-(\widetilde{S})+\dim \mathcal{V}_-$. Thus, \eqref{eq:II-symm-ext-def} is proved.    
 \end{proof}

 \begin{theorem}[von Neumann's theorem on self-adjoint extensions]\label{thm:vonN_thm_selfadj_exts}\index{theorem!von Neumann (self-adjoint extensions)}
A densely defined symmetric operator $S$ on a Hilbert space $\cH$ admits self-adjoint extensions if and only if $S$ has equal deficiency indices. In this case there is a one-to-one correspondence between the self-adjoint extensions of $S$ and the unitary isomorphisms between $\ker(S^*-z\mathbbm{1})$ and $\ker(S^*-\overline{z}\mathbbm{1})$, where $z\in\mathbb{C}\!\setminus\!\mathbb{R}$ is fixed and arbitrary. Each self-adjoint extension is of the form $S_U$ for some unitary $U:\ker(S^*-z\mathbbm{1})\xrightarrow[]{\;\cong\;}\ker(S^*-\overline{z}\mathbbm{1})$, where
\begin{equation}
\begin{split}
\mathcal{D}(S_U)\;&=\;\mathcal{D}(\overline{S})\,\dotplus\,(\mathbbm{1}-U)\ker(S^*-z\mathbbm{1})\,, \\
S_U(f+u-Uu)\;&=\;\overline{S}u+zu-\overline{z}\,Uu\;=\;S^*(f+u-Uu)\,.
\end{split}
\end{equation}
\end{theorem}

\begin{proof}
 The family of self-adjoint extensions of $S$ in $\cH$ is a sub-family of all closed symmetric extensions, hence a subfamily of the operators $S_U$ characterised in Theorem \ref{thm:vonN_thm_symm_exts}.

 On account of Corollary \ref{cor:II-caley1}(i), each such $S_U$ is self-adjoint if and only if its Cayley transform $V_{S_U}$ is a unitary operator on the whole $\cH$. As seen in the proof of Theorem \ref{thm:vonN_thm_symm_exts},
 \[
  \begin{split}
   \mathcal{D}(V_{\overline{S}})\;&=\;\mathrm{ran}(\overline{S}-\overline{z}\mathbbm{1})\,, \\
   \mathcal{D}(V_{S_U})\;&=\;\mathcal{D}(V_{\overline{S}})\oplus\mathcal{V}_+\;=\;\mathrm{ran}(\overline{S}-\overline{z}\mathbbm{1})\oplus\mathcal{V}_+
  \end{split}
 \]
 for some closed subspace $\mathcal{V}_+\subset\ker(S^*-z\mathbbm{1})$, the subspace $\mathrm{ran}(\overline{S}-\overline{z}\mathbbm{1})$ (as well as $\mathrm{ran}(\overline{S}-z\mathbbm{1})$) being closed too because $\overline{S}$ is closed (Theorem \ref{thm:II-Cayley_transf}(iii)). So now the unitarity of $V_{S_U}$, in view of
 \[
  \mathcal{D}(V_{S_U})\;=\;\cH\;=\;\mathrm{ran}(\overline{S}-\overline{z}\mathbbm{1})\oplus\ker(S^*-z\mathbbm{1})\;=\;\mathrm{ran}(\overline{S}-z\mathbbm{1})\oplus\ker(S^*-\overline{z}\mathbbm{1})\,,
 \]
  is equivalent to the fact that $\mathcal{V}_+=\ker(S^*-z\mathbbm{1})$, in which case the component of $V_{S_U}$ on $\ker(S^*-z\mathbbm{1})$ maps this space unitarily onto $\ker(S^*-\overline{z}\mathbbm{1})$. The latter operator is precisely $U:\ker(S^*-z\mathbbm{1})\stackrel{\cong}{\longrightarrow}\ker(S^*-\overline{z}\mathbbm{1})$. Therefore, the self-adjointness of the extension $S_U$ is tantamount as $\ker(S^*-z\mathbbm{1})\stackrel{\cong}{\longrightarrow}\ker(S^*-\overline{z}\mathbbm{1})$, i.e, $d_+(S)=d_-(S)$. When this is the case, the labelling operator $U$ is a unitary mapping between the two isomorphic deficiency subspaces, and conversely any such unitary labels a self-adjoint $U$.  
\end{proof}

 \begin{remark}
   For short, von Neumann's construction of (all) self-adjoint extensions $\widetilde{S}$ of $S$ provided by Theorem \ref{thm:vonN_thm_selfadj_exts}, which exist under the necessary and sufficient condition $d_+(S)=d_-(S)$, consists of extending the isometry
 \[
  V_S:\ran(S-\overline{z}\mathbbm{1})\to \ran(S-z\mathbbm{1})\,,
 \]
 the Cayley transform\index{Cayley transform} of $S$, to a unitary operator on the whole $\cH$, the Cayley transform $V_{\widetilde{S}}$ of the actual self-adjoint extension $\widetilde{S}$, so as 
 \begin{equation}
  V_{\widetilde{S}}\;=\;\overline{V_S}\oplus U
 \end{equation}
 as a unitary on
 \[
  \cH\,=\,\overline{\ran(S-z\mathbbm{1})}\oplus\ker(S^*-z\mathbbm{1})\;\stackrel{\cong}{\longrightarrow}\;\cH\,=\,\overline{\ran(S-z\mathbbm{1})}\oplus\ker(S^*-z\mathbbm{1})\,,
 \]
 where the unitary $U:\ker(S^*-z\mathbbm{1})\stackrel{\cong}{\longrightarrow}\ker(S^*-z\mathbbm{1})$ labels the extension $\widetilde{S}\equiv S_U$.
  For each $S_U$, the unitary $U$ is the restriction to $\ker(S^*-z\mathbbm{1})$ of the Cayley transform of $S_U$. 
 \end{remark}

 For a given densely defined self-adjoint operator $S$ on the Hilbert space $\cH$, and $z\in\mathbb{C}\setminus\mathbb{R}$, Theorems \ref{thm:vonN_thm_symm_exts}-\ref{thm:vonN_thm_selfadj_exts} show that there are precisely three alternatives:
 \begin{itemize}
  \item[1.] $d_+(S)=d_-(S)=0$, in which case $\overline{S}$ is the unique self-adjoint extension of $S$, i.e.,   
  $S$ is essentially self-adjoint ($\overline{S}=S^*$);
  \item[2.] $d_+(S)=d_-(S)>0$, in which case $S$ admits an infinite multiplicity of distinct self-adjoint extensions $S_U$, parametrised by the unitaries $U:\ker(S^*-z\mathbbm{1})\stackrel{\cong}{\longrightarrow}\ker(S^*-\overline{z}\mathbbm{1})$;
  \item[3.] $d_+(S)\neq d_-(S)$, in which case $\ker(S^*-z\mathbbm{1})$ and $\ker(S^*-\overline{z}\mathbbm{1})$ are not isomorphic and hence $S$ does not admit self-adjoint extensions.
 \end{itemize}
 In case 2.~and case 3.~with non-zero deficiency indices, $S$ admits an infinite multiplicity of distinct closed symmetric extensions $S_U$ parametrised by the isometries $U:\mathcal{V}_+\to\mathcal{V}_-$ between closed subspaces $:\mathcal{V}_+\subset\ker(S^*-z\mathbbm{1})$ and $:\mathcal{V}_-\subset\ker(S^*-\overline{z}\mathbbm{1})$, which shows that in case 2.~the self-adjoint extensions are maximal closed symmetric extensions, in the sense of operator inclusion.

 The above formulation of von Neumann's extension theory of symmetric operators (Theorems \ref{thm:vonN_thm_symm_exts}-\ref{thm:vonN_thm_selfadj_exts}) is the original one from $\sim$1930  \cite{vonNeumann1930}, built upon the use of the Cayley transform. Here the presentation is chosen so as to prepare the parallel with the analogous use of the Kre{\u\i}n transform for extending lower semi-bounded symmetric operators (Sect.~\ref{sec:II-Kreintransform}-\ref{sec:II-KreinExtTheory}).

 An alternative, equivalent, and widespread derivation of Theorems \ref{thm:vonN_thm_symm_exts} and \ref{thm:vonN_thm_selfadj_exts}, appeared some 30 years later, due to Dunford and Schwartz \cite[Sect.~XII.4]{Dunford-Schwartz-II}, and designed so as to highlight the underlying abstract boundary value problem. 
 The Dunford-Schwartz approach makes use of these two sesquilinear forms, both with domain $\mathcal{D}(S^*)$: the scalar product 
 \begin{equation}
   (g,h)_{\Gamma(S^*)}\;:=\;\langle g,h\rangle+\langle S^* g,S^*g\rangle\,,
 \end{equation}
  which induces the graph norm $\|\cdot\|_{\Gamma(S^*)}$ associated with $S^*$, and the form
  \begin{equation}
   [g,h]_{S^*}\;:=\;\langle S^*g,h\rangle-\langle g,S^*h\rangle\,,
  \end{equation}
  also referred to as the \emph{boundary form}.\index{boundary form}

  \begin{lemma}\label{lem:II-AclosedAsymmetric}
   Let $S$ be a densely defined and symmetric operator on a Hilbert space $\cH$.
   \begin{enumerate}[(i)]
    \item One has
    \begin{equation}\label{eq:II-DSstaroplus}
     \mathcal{D}(S^*)\;=\;\mathcal{D}(\overline{S})\oplus_{S^*}\ker(S^*-\ii\mathbbm{1})\oplus_{S^*}\ker(S^*+\ii\mathbbm{1})\,,
    \end{equation}
    where the orthogonality $\oplus_{S^*}$ is referred to the scalar product $(\cdot,\cdot)_{\Gamma(S^*)}$, meaning in particular that each summand in the above r.h.s.~is a $\|\cdot\|_{\Gamma(S^*)}$-closed subspace of $\mathcal{D}(S^*)$.
     \item $\widetilde{S}$ is a closed symmetric extension of $S$ if and only if $\widetilde{S}=S^*\!\upharpoonright{\mathcal{D}}$ for some subspace $\mathcal{D}\subset\mathcal{D}(S^*)$ that contains $\mathcal{D}(\overline{S})$, is closed in the graph topology (i.e, the $\|\cdot\|_{\Gamma(S^*)}$-topology), and on which the $[\cdot,\cdot]_{S^*}$-form vanishes.
   \item There is a one-to-one correspondence $\mathcal{U}\stackrel{1:1}{\,\longmapsto\,}\mathcal{U}_0$
    \[
     \left\{ 
     \begin{array}{c}
      \textrm{subspaces $\mathcal{U}\subset\mathcal{D}(S^*)$} \\
      \textrm{that are $\|\cdot\|_{\Gamma(S^*)}$-closed,} \\
      \textrm{annihilate the $[\cdot,\cdot]_{S^*}$-form,} \\
      \textrm{and contain $\mathcal{D}(\overline{S})$}
     \end{array}
     \right\}\stackrel{1:1}{\,\longrightarrow\,}
      \left\{ 
     \begin{array}{c}
      \textrm{subspaces} \\
      \textrm{$\mathcal{U}_0\subset\ker(S^*-\ii\mathbbm{1})\oplus_{S^*}\ker(S^*+\ii\mathbbm{1})$} \\
      \textrm{that are $\|\cdot\|_{\Gamma(S^*)}$-closed} \\
      \textrm{and annihilate the $[\cdot,\cdot]_{S^*}$-form}
     \end{array}
     \right\}
    \]
    such that $\mathcal{U}=\mathcal{D}(\overline{S})\oplus_{S^*}\mathcal{U}_0$.
   \end{enumerate}
  \end{lemma}

  \begin{proof} Recall that the closedness of an operator is equivalent to the closure of its domain in its graph norm (Sect.~\ref{sec:I-bdd-closable-closed}).

  (i) $\mathcal{D}(\overline{S})$ is $\|\cdot\|_{\Gamma(\overline{S})}$-closed, hence $\|\cdot\|_{\Gamma(S^*)}$-closed. As for the subspaces $\ker(S^*\pm\ii\mathbbm{1})$, they are $\|\cdot\|_{\cH}$-closed owing to the closedness of $S^*$ (Sect.~\ref{sec:I-bdd-closable-closed} and \ref{sec:I-adjoint}), hence closed also in the stronger $\|\cdot\|_{\Gamma(S^*)}$-norm. The mutual orthogonality of such subspaces with respect to the $(g,h)_{\Gamma(S^*)}$-inner product then follows straightforwardly. For instance, if $f\in\mathcal{D}(\overline{S})$ and $u\in\ker(S^*-\ii\mathbbm{1})$, then
  \[
   \begin{split}
       (f,u)_{\Gamma(S^*)}\;&=\;\langle f,u\rangle+\langle S^*f,S^*u\rangle\;=\;\langle f,u\rangle+\ii\langle\overline{S}f,u\rangle\;=\;\langle f,u\rangle+\ii\langle f,S^*u\rangle \\
       &=\;\langle f,u\rangle+\ii^2\langle f,u\rangle\;=\;0\,.
   \end{split}
  \]
  The other checks are done analogously. Thus, 
  \[
   \mathcal{D}(S^*)\;\supset\;\mathcal{D}(\overline{S})\oplus_{S^*}\ker(S^*-\ii\mathbbm{1})\oplus_{S^*}\ker(S^*+\ii\mathbbm{1})\,.
  \]
   In view of von Neumann's formula (Proposition \ref{prop:II-vNformula}), with $z=\ii$ in \eqref{eq:II-vNformula}, the latter inclusion is an actual identity.
   
  (ii) From  $\widetilde{S}=\overline{\widetilde{S}}=\widetilde{S}^{**}\subset(\widetilde{S})^*$ one deduces $S\subset\widetilde{S}=\overline{\,\widetilde{S}\,}\subset(\widetilde{S})^*$, so $\widetilde{S}$ is an actual restriction of $S^*$ onto some subspace $\mathcal{D}=\mathcal{D}(\widetilde{S})$. The closedness of $\widetilde{S}$ amounts to the closedness of $\mathcal{D}$ in the $\Gamma(\widetilde{S})$-graph norm, hence in the $\|\cdot\|_{\Gamma(S^*)}$-norm. The symmetry of $\widetilde{S}$ amounts to the vanishing of the $[\cdot,\cdot]_{S^*}$-form on $\mathcal{D}$.

  (iii) For a given $\|\cdot\|_{\Gamma(S^*)}$-closed subspace $\mathcal{U}_0\subset\ker(S^*-\ii\mathbbm{1})\oplus_{S^*}\ker(S^*+\ii\mathbbm{1})$ annihilating the $[\cdot,\cdot]_{S^*}$-form, set
  \[
   \mathcal{U}\;:=\;\mathcal{D}(\overline{S})\oplus_{S^*}\mathcal{U}_0\;\subset\; \mathcal{D}(\overline{S})\oplus_{S^*}\ker(S^*-\ii\mathbbm{1})\oplus_{S^*}\ker(S^*+\ii\mathbbm{1})=\mathcal{D}(S^*)\,.
  \]
$\mathcal{U}$ is $\|\cdot\|_{\Gamma(S^*)}$-closed, as $\oplus_{S^*}$-orthogonal sum of $\|\cdot\|_{\Gamma(S^*)}$-closed subspaces.
  The $[\cdot,\cdot]_{S^*}$-form vanishes on $\mathcal{D}(\overline{S})$ by construction and on $\mathcal{U}_0$ by assumption; therefore, for generic $g_j=f_j+u_j$, $j\in\{1,2\}$, with $f_j\in\mathcal{D}(\overline{S})$ and $u_j\in\mathcal{U}_0$,
  \[
   \begin{split}
   [g_1,g_2]_{S^*}\;&=\;[f_1,f_2]_{S^*}+[f_1,u_2]_{S^*}+[f_2,u_1]_{S^*}+[u_1,u_2]_{S^*}  \\
   &=\;[f_1,u_2]_{S^*}+[f_2,u_1]_{S^*}\,.
   \end{split}
  \]
  On the other hand, $[f_1,u_2]_{S^*}=\langle S^* f_1,u_2\rangle-\langle f_1, S^* u_2\rangle=\langle\overline{S} f_1,u_2\rangle-\langle\overline{S} f_1,u_2\rangle=0$, and analogously $[f_2,u_1]_{S^*}=0$. Thus, $[g_1,g_2]_{S^*}=0$, which proves that the $[\cdot,\cdot]_{S^*}$-form vanishes on the whole $\mathcal{U}$.

  Conversely, let now $\mathcal{U}\subset \mathcal{D}(S^*)$ be a $\|\cdot\|_{\Gamma(S^*)}$-closed subspace annihilating the $[\cdot,\cdot]_{S^*}$-form and containing $\mathcal{D}(\overline{S})$, and set $\mathcal{U}_0:=\mathcal{U}\cap\big( \ker(S^*-\ii\mathbbm{1})\oplus_{S^*}\ker(S^*+\ii\mathbbm{1})\big)$.
   Owing to \eqref{eq:II-DSstaroplus}, the subspace $\mathcal{U}$ must have the structure $\mathcal{U}=\mathcal{D}(\overline{S})\oplus_{S^*}\mathcal{U}_0$ for some $\|\cdot\|_{\Gamma(S^*)}$-closed subspace $\mathcal{U}_0\subset\ker(S^*-\ii\mathbbm{1})\oplus_{S^*}\ker(S^*+\ii\mathbbm{1})$. Besides, $\mathcal{U}_0$ necessarily annihilates the $[\cdot,\cdot]_{S^*}$-form.  
  \end{proof}

 \begin{proof}[Second proof of Theorem \ref{thm:vonN_thm_symm_exts} for $z=\ii$]
  On account of Lemma \ref{lem:II-AclosedAsymmetric}(ii)-(iii), \emph{any} closed symmetric extension $\widetilde{S}$ of the densely defined symmetric operator $S$ is of the form $\widetilde{S}=S^*\upharpoonright\mathcal{D}(\widetilde{S})$ with $\mathcal{D}(\widetilde{S})=\mathcal{D}(\overline{S})\oplus_{S^*}\mathcal{U}_0$ for some subspace $\mathcal{U}_0\subset\ker(S^*-\ii\mathbbm{1})\oplus_{S^*}\ker(S^*+\ii\mathbbm{1})$ that is $\|\cdot\|_{\Gamma(S^*)}$-closed and annihilates the $[\cdot,\cdot]_{S^*}$-form. 
  Since an arbitrary $u_0\in \mathcal{U}_0$ has the form $u_0=u_+ \oplus_{S^*} u_-$ for some $u_\pm\in\ker(S^*\mp\ii\mathbbm{1})$, the subspaces 
   \[
   \mathcal{V}_\pm:=\{ u_\pm\in \ker(S^*\mp\ii\mathbbm{1})\,|\,\exists u_\mp\in \ker(S^*\pm\ii\mathbbm{1})\textrm{ with } u_+ \oplus_{S^*} u_-\in \mathcal{U}_0\}
  \]
  are $\|\cdot\|_{\Gamma(S^*)}$-closed, and $\mathcal{U}_0=\mathcal{V}_+ \oplus_{S^*}\mathcal{V}_-$, whence also $\mathcal{D}(\widetilde{S})=\mathcal{D}(\overline{S})\oplus_{S^*}\mathcal{V}_+ \oplus_{S^*}\mathcal{V}_-$.
  Moreover, for arbitrary $u_0=u_+ \oplus_{S^*} u_-\in \mathcal{U}_0$, $u_\pm\in\mathcal{V}_\pm$,
  \[
   0\;=\;[u_0,u_0]_{S^*}\;=\;\langle S^* u_0,u_0\rangle-\langle  u_0,S^*u_0\rangle\;=\;2\ii\|u_-\|^2-2\ii\|u_+\|^2\,,
  \]
  i.e., $\|u_+\|=\|u_-\|$. This means that the map $U:\mathcal{V}_+\to\mathcal{V}_-$, $u_+\mapsto u_-$, defined by $u_+ \oplus_{S^*} u_-\in \mathcal{U}_0$ is an isometry. In particular, the domain $\mathcal{V}_+$ of $U$ is closed in $\cH$. One re-writes
  \[
       \mathcal{D}(\widetilde{S})\;=\;\mathcal{D}(\overline{S})\oplus_{S^*}\mathcal{V}_+ \oplus_{S^*}\mathcal{V}_-\;=\;\mathcal{D}(\overline{S})\oplus_{S^*}(\mathbbm{1}+U)\mathcal{V}_+
  \]
  and, applying $\widetilde{S}=S^*\upharpoonright\mathcal{D}(\widetilde{S})$,
  \[
   \widetilde{S}(f+u+U u)\;=\;\overline{S}f+\ii u-\ii\, U u\,.
  \]
 Conversely, if $U:\mathcal{V}_+\to\mathcal{V}_-$ is an isometry between $\cH$-closed, hence also $\|\cdot\|_{\Gamma(S^*)}$-closed subspaces 
  $\mathcal{V}_\pm\subset\ker(S^*\mp\ii\mathbbm{1})$ and $\widetilde{S}$ is defined as above, then $\mathcal{D}(\widetilde{S})$ is a $\|\cdot\|_{\Gamma(S^*)}$-closed subspace of $\mathcal{D}(S^*)$ and has the form $\mathcal{D}(\widetilde{S})=\mathcal{D}(\overline{S})\oplus_{S^*}\mathcal{U}_0$ with $\mathcal{U}_0:=\mathcal{V}_+ \oplus_{S^*}\mathcal{V}_-$. The isometric condition implies that
  \[
   [u+Uu,u+Uu]_{S^*}\;=\;2\ii\|Uu\|^2-2\ii\|u\|^2\;=\;0\qquad\forall u\in \mathcal{V}_+\,,
  \]
 that is, $\mathcal{U}_0$ annihilates the $[\cdot,\cdot]_{S^*}$-form. Applying Lemma \ref{lem:II-AclosedAsymmetric}(ii)-(iii) one deduces that $\widetilde{S}$ is a closed symmetric extension of $S$. One can then identify and rename 
 $\widetilde{S}\equiv S_U$. The final reasoning for the claim concerning the deficiency indices is the same as in the first proof. Theorem \ref{thm:vonN_thm_symm_exts} is thus proved up to changing $U$ in $-U$.   
 \end{proof}

 One last result that is central in von Neumann's analysis of self-adjoint extensions concerns the existence of self-adjoint extensions of a densely defined symmetric operator commuting with a conjugation. In this context, by \emph{conjugation}\index{operator!conjugation}\index{conjugation operator} on a given Hilbert space $\cH$ one means an \emph{anti-linear} map $C:\cH\to\cH$ that is norm-preserving and satisfies $C^2=\mathbbm{1}$. In particular, a conjugation $C$ satisfies $\langle C\varphi,C\psi\rangle=\overline{\langle\varphi,\psi\rangle}$ $\forall\varphi,\psi\in\cH$.

 \begin{theorem}[von Neumann's conjugation criterion]\label{thm:II-vNconj}\index{theorem!von Neumann (conjugation criterion)}
  Let $S$ be a densely defined and symmetric operator on a Hilbert space $\cH$. If there is a conjugation $C$ on $\cH$ such that $C\mathcal{D}(S)\subset\mathcal{D}(S)$ and $SC\psi=CS\psi$ for all $\psi\in\mathcal{D}(S)$, then $d_+(S)=d_-(S)$, i.e., $S$ has equal deficiency indices, and therefore $S$ has self-adjoint extensions.  
 \end{theorem}

 \begin{proof}
  From $C\mathcal{D}(S)\subset\mathcal{D}(S)$ and $C^2=\mathbbm{1}$ one deduces $C\mathcal{D}(S)=\mathcal{D}(S)$. For fixed $z\in\mathbb{C}\setminus\mathbb{R}$, let $\varphi\in\ker(S^*-z\mathbbm{1})$. Then, for every $\psi\in\mathcal{D}(A)$,
  \[
   0\;=\;\overline{\langle(S^*-z\mathbbm{1})\varphi,\psi\rangle}\;=\;\overline{\langle\varphi,(S-\overline{z}\mathbbm{1})\psi\rangle}\;=\;\langle C\varphi,C(S-\overline{z}\mathbbm{1})\psi\rangle\;=\;\langle C\varphi,(S-z\mathbbm{1})C\psi\rangle\,.
  \]
  As $\psi$ runs over the whole $\mathcal{D}(S)$, so too does $C\psi$, and the above identity then implies $(S^*-\overline{z}\mathbbm{1})C\varphi=0$. This proves that $C\ker(S^*-z\mathbbm{1})\subset\ker(S^*-\overline{z}\mathbbm{1})$. An analogous reasoning proves that $C\ker(S^*-\overline{z}\mathbbm{1})\subset\ker(S^*-z\mathbbm{1})$. Thus, using $C^2=\mathbbm{1}$, $\ker(S^*-\overline{z}\mathbbm{1})=C\ker(S^*-z\mathbbm{1})$. Since $C$ preserves norms, necessarily $\dim\ker(S^*-z\mathbbm{1})=\dim\ker(S^*-\overline{z}\mathbbm{1})$.  
 \end{proof}

 \section{Kre{\u\i}n transform of positive operators}\label{sec:II-Kreintransform}

 The Kre{\u\i}n transform was introduced by Kre{\u\i}n \cite{Krein-1947} in 1946. Cayley and Kre{\u\i}n transforms can be regarded, respectively, as the ``complex'' and the ``real'' version of the same idea, that consists
 \begin{itemize}
  \item in the complex (Cayley) case, of exploiting the property $|\frac{w-\ii}{w+\ii}|=1$ $\Leftrightarrow$ $w\in\mathbb{R}$ and the bijection $\mathbb{R}\to(\mathbb{S}^1\setminus\{1\})$, $w\mapsto\frac{w-\ii}{w+\ii}$,
  \item and in the real (Kre{\u\i}n) case, of exploiting the property $\frac{w-1}{w+1}\in[-1,1)$ $\Leftrightarrow$ $w\geqslant 0$ and the bijection $[0,+\infty)\to[-1,1)$, $w\mapsto\frac{w-1}{w+1}$.
 \end{itemize}
 In either case, the check that number $w$ may run over the \emph{unbounded} interval of $\mathbb{R}$ is reduced to the check that through a bijective function $f$, the number $f(w)$ runs on a \emph{bounded} set, thus making the second check ``easier'', at least conceptually; moreover, the bijectivity of $f$ reconstructs uniquely $w$ from $f(w)$. This idea is then lifted from scalars to the self-adjoint extensions of symmetric operators.

 Given a Hilbert space $\cH$ and a positive symmetric operator $S$ (not necessarily closed or densely defined), the \emph{Kre{\u\i}n transform}\index{Kre{\u\i}n transform} of $S$ is the operator
 \begin{equation}\label{eq:II-Kreintransf}
  K_S\;:=\;(S-\mathbbm{1})(S+\mathbbm{1})^{-1}\,,\qquad\mathcal{D}(K_S)\;:=\;\mathrm{ran}(S+\mathbbm{1})\,.
 \end{equation}
 By positivity, $S+\mathbbm{1}$ is indeed invertible on its range. By construction,
 \begin{equation}\label{eq:II-Kreintransf2}
   K_S(S+\mathbbm{1})\psi\;=\;(S-\mathbbm{1})\psi\qquad\forall\psi\in\mathcal{D}(S)
 \end{equation}
and 
 \begin{equation}\label{eq:II-Kreintransf3}
  \begin{split}
   \mathcal{D}(K_S)\;&=\;\mathrm{ran}(S+\mathbbm{1})\,, \\
   \mathrm{ran}\,K_S\;&=\;\mathrm{ran}(S-\mathbbm{1})\,.
  \end{split}
 \end{equation}
 Another convenient expression, following directly from the definition, is
 \begin{equation}\label{eq:II-Kreintransf4}
  K_S\;=\;\mathbbm{1}-2(S+\mathbbm{1})^{-1}\,.
 \end{equation}

 \begin{theorem}[Kre{\u\i}n transform]\label{thm:II-KreinTransf}\index{Kre{\u\i}n transform}\index{theorem!Kre{\u\i}n (Kre{\u\i}n transform)}~
 \begin{enumerate}[(i)]
  \item The Kre{\u\i}n transform $S\mapsto K_S$ is a bijective map of the set of positive symmetric operators on $\cH$ onto the set of symmetric operators $K$ on $\cH$ with $\|K\|_{\mathrm{op}}\leqslant 1$ and $\ker(\mathbbm{1}-K)=\{0\}$. Its inverse is the map $K\mapsto S_K$, with
  \begin{equation}\label{eq:II-Kreintransf-inv}
   S_K\;:=\;(\mathbbm{1}+K)(\mathbbm{1}-K)^{-1}\,,\qquad \mathcal{D}(S_K)\;:=\;\mathrm{ran}(\mathbbm{1}-K)\,.
  \end{equation}
  \item If a positive symmetric operator $S$ is unbounded, then $\|K_S\|_{\mathrm{op}}=1$.
  \item For any two positive symmetric operators $S_1,S_2$ on $\cH$, $S_1\subset S_2$ $\Leftrightarrow$ $K_{S_1}\subset K_{S_2}$. 
  \item For any two positive self-adjoint operators $S_1,S_2$ on $\cH$, $S_1\geqslant S_2$ $\Leftrightarrow$ $K_{S_1}\geqslant K_{S_2}$. 
  \item A positive symmetric operator $S$ on $\cH$ is self-adjoint if and only if $\mathcal{D}(K_S)=\cH$, i.e., if and only if $K_S$ is self-adjoint.
 \end{enumerate}
 \end{theorem}

 Theorem \ref{thm:II-KreinTransf}(i) establishes in particular that the Kre{\u\i}n transform\index{Kre{\u\i}n transform} $K_S$ is a symmetric \emph{contraction} (i.e., symmetric and with $\|K_S\|_{\mathrm{op}}\leqslant 1$) such that $\mathbbm{1}-K_S$ is injective. The map $K\mapsto S_K$ defined in \eqref{eq:II-Kreintransf-inv} is called the \emph{inverse Kre{\u\i}n transform}.\index{inverse Kre{\u\i}n transform} By construction,
 \begin{equation}\label{eq:II-Kreintransf-inv2}
  S_K(\mathbbm{1}-K)\varphi\;=\;(\mathbbm{1}+K)\varphi\qquad\forall\varphi\in\mathcal{K}
 \end{equation}
 and 
  \begin{equation}\label{eq:II-Kreintransf-inv3}
  \begin{split}
   \mathcal{D}(S_K)\;&=\;\mathrm{ran}(\mathbbm{1}-K)\,, \\
   \mathrm{ran}\,S_K\;&=\;\mathrm{ran}(\mathbbm{1}+K)\,.
  \end{split}
 \end{equation}

 \begin{proof}[Proof of Theorem \ref{thm:II-KreinTransf}]
  (i) Let $S$ be positive symmetric. For generic $\varphi\in\mathcal{D}(K_S)$, hence $\varphi=(S+\mathbbm{1})\psi$ with $\psi\in\mathcal{D}(S)$, one has
  \[
   \begin{split}
       \langle \varphi, K_S\,\varphi\rangle\;&=\;\langle(S+\mathbbm{1})\psi,(S-\mathbbm{1})\psi\rangle\;=\;\|S\psi\|^2-\|\psi\|^2\;\in\;\mathbb{R}
   \end{split}
  \]
 (having used \eqref{eq:II-Kreintransf2} in the first identity and the symmetry of $S$ in the second), so $K_S$ is symmetric. Moreover, owing to the positivity and symmetry of $S$, and to \eqref{eq:II-Kreintransf2} again,
  \[
   \|K_S\varphi\|^2\;=\;\|S\psi\|^2+\|\psi\|^2-2\langle\psi, S\psi\rangle\;\leqslant\;\|S\psi\|^2+\|\psi\|^2+2\langle\psi, S\psi\rangle\;=\;\|\varphi\|^2\,,
  \]
  which shows that $\|K_S\|_{\mathrm{op}}\leqslant 1$. And since 	
  \[
   0\;=\;(\mathbbm{1}-K_S)\varphi\;=\;2\psi\quad\Rightarrow\quad \psi=0\quad\Rightarrow\quad\varphi=0\,,
  \]
  then $\mathbbm{1}-K_S$ is injective. Thus, $K_S$ belongs to the claimed target set.

  Conversely, let now $K$ be a symmetric operator with $\|K\|_{\mathrm{op}}\leqslant 1$ and $\ker(\mathbbm{1}-K)=\{0\}$, and let $\psi=(\mathbbm{1}-K)\varphi$ be a generic element in $\mathcal{D}(S_K)$, where $\varphi\in\mathcal{D}(K)$. One has
  \[
   \langle \psi,S_K\psi\rangle\;=\;\langle (\mathbbm{1}-K)\varphi,(\mathbbm{1}+K)\varphi\rangle\;=\;\|\varphi\|^2-\|K\varphi\|^2\;\geqslant\;0\,,
  \]
  having used \eqref{eq:II-Kreintransf-inv2} in the first identity, and the symmetry and the norm property of $K$ in the second. Thus, $S_K$ is positive symmetric.

  It remains to show that the two maps $S\mapsto K_S$ and $K\mapsto S_K$ invert each other. Let us start with determining, for given $K_S$ defined as above, its inverse Kre{\u\i}n transform $S_{K_S}$. If $\psi\in\mathcal{D}(S)$, then $\varphi:=(S+\mathbbm{1})\psi\in\mathcal{D}(K_S)$ (owing to \eqref{eq:II-Kreintransf3}) and  $(\mathbbm{1}-K_S)\varphi=2 \psi$ (owing to \eqref{eq:II-Kreintransf2}), whence $\psi\in\mathrm{ran}(\mathbbm{1}-K_S)$. On the other hand, if  $\psi\in\mathrm{ran}(\mathbbm{1}-K_S)$, i.e., $\psi=(\mathbbm{1}-K_S)\varphi$ for some $\varphi\in\mathcal{D}(K_S)=\mathrm{ran}(S+\mathbbm{1})$ (owing to \eqref{eq:II-Kreintransf3}), then $\varphi=(S+\mathbbm{1})\widetilde{\psi}$ for some $\widetilde{\psi}\in\mathcal{D}(S)$ and $\psi=(\mathbbm{1}-K_S)(S+\mathbbm{1})\widetilde{\psi}=2\widetilde{\psi}$ (owing to \eqref{eq:II-Kreintransf2}), whence $\psi\in\mathcal{D}(S)$. This proves that $\mathcal{D}(S)=\mathrm{ran}(\mathbbm{1}-K_S)$, whence also $\mathcal{D}(S)=\mathcal{D}(S_{K_S})$ (owing to \eqref{eq:II-Kreintransf-inv3}). In addition, since for $\psi\in\mathcal{D}(S)$ and $\varphi:=(S+\mathbbm{1})\psi$ one has (owing to \eqref{eq:II-Kreintransf2}) $(\mathbbm{1}+K_S)\varphi=2 S\psi$ and $(\mathbbm{1}-K_S)\varphi=2 \psi$, then $S\psi=\frac{1}{2}(\mathbbm{1}+K_S)\varphi=\frac{1}{2}(\mathbbm{1}+K_S)2(\mathbbm{1}-K_S)^{-1}\psi=S_{K_S}\psi$ (owing to \eqref{eq:II-Kreintransf-inv}). The conclusion is $S_{K_S}=S$.

  The check that conversely $K_{S_K}=K$ is done in a completely analogous way, this time using the identities $(S_K+\mathbbm{1})\psi=2\varphi$ and $(S_K-\mathbbm{1})\psi=2K\varphi$ for $\varphi\in\mathcal{D}(K)$ and $\psi=(\mathbbm{1}-K)\varphi\in\mathcal{D}(S_K)$.
  
  (ii) As $S$ is unbounded, $\|S\psi_n\|\xrightarrow{n\to\infty}0$ for some sequence $(\psi_n)_{n\in\mathbb{N}}$ of unit vectors ($\|\psi_n\|=1$) in $\mathcal{D}(S)$. Because of \eqref{eq:II-Kreintransf3}, $\varphi_n:=(S+\mathbbm{1})\psi_n\in\mathcal{D}(K_S)$, and because of \eqref{eq:II-Kreintransf2}, $K_S\varphi_n=S\psi_n-\psi_n$. Then
  \[
   1\;\geqslant\;\|K_S\|_{\mathrm{op}}\;\geqslant\;\frac{\|K_S\varphi_n\|}{\|\varphi_n\|}\;\geqslant\;\frac{\,\big|\|S\psi_n\|-1\big|\,}{\|S\psi_n\|+1}\;\xrightarrow{n\to\infty}\;1\,,
  \]
  whence $\|K_S\|_{\mathrm{op}}=1$.

  (iii) The claim is obvious from the definition \eqref{eq:II-Kreintransf} of $K_{S_1}$ and $K_{S_2}$.

  (iv) The ordering $S_1\geqslant S_2$ is equivalent to $(S_2+\mathbbm{1})^{-1}\geqslant(S_1+\mathbbm{1})^{-1}$ (Sect.~\ref{sec:I_forms}). The latter, owing to \eqref{eq:II-Kreintransf4}, is equivalent to $K_{S_1}\geqslant K_{S_2}$.

  (v) Let $S$ be a positive symmetric operator. If, by assumption and owing to \eqref{eq:II-Kreintransf3}, $\cH=\mathcal{D}(K_S)=\mathrm{ran}(S+\mathbbm{1})$, then $S$ is self-adjoint and $-1\in\rho(S)$ (Sect.~\ref{sec:I-symmetric-selfadj}). Conversely, if $S$ is self-adjoint, then $\ker(S+\mathbbm{1})=\{0\}$, whence (Sect.~\ref{sec:I-symmetric-selfadj}) $\cH=\mathrm{ran}(\overline{S}+\mathbbm{1})\oplus\ker(S^*+\mathbbm{1})=\mathrm{ran}(S+\mathbbm{1})$, i.e., (owing to \eqref{eq:II-Kreintransf3}) $\cH=\mathcal{D}(K_S)$. Thus, for a positive symmetric $S$, its self-adjointness is equivalent to $\mathcal{D}(K_S)=\cH$.    
 \end{proof}

 The following consequence of Theorem \ref{thm:II-KreinTransf} stays on the same conceptual footing for the Kre{\u\i}n transform as Corollary \ref{cor:II-caley2} for the Cayley transform.

 \begin{corollary}\label{cor:II-krein}\index{theorem!Kre{\u\i}n (Kre{\u\i}n transform)}\index{Kre{\u\i}n transform}
  Let $S$ be a densely defined, positive, symmetric operator on a Hilbert space $\cH$ and let $K_S$ be its Kre{\u\i}n transform.   There is a one-to-one correspondence between the positive self-adjoint extensions $\widetilde{S}$ of $S$ in $\cH$ and the contractive self-adjoint extensions $\widetilde{K}$ of $K_S$ in $\cH$, i.e., the 
  bounded self-adjoint extensions $\widetilde{K}$ of $K_S$ with $\|\widetilde{K}\|_{\mathrm{op}}\leqslant 1$. Such correspondence is explicitly given by $\widetilde{S}\mapsto\widetilde{K}=K_{\widetilde{S}}$. In particular, for any bounded self-adjoint extensions $\widetilde{K}$ of $K_S$ with $\|\widetilde{K}\|_{\mathrm{op}}\leqslant 1$ one necessarily has $\ker(\mathbbm{1}-\widetilde{K})=\{0\}$.
 \end{corollary}

 \begin{proof}
  Let $\widetilde{K}$ be a bounded self-adjoint extensions of $K_S$ with $\|\widetilde{K}\|_{\mathrm{op}}\leqslant 1$.
  Since $\mathrm{ran}(\mathbbm{1}-\widetilde{K})\supset \mathrm{ran}(\mathbbm{1}-K_S)=\mathcal{D}(S)$, the subspace $\mathrm{ran}(\mathbbm{1}-\widetilde{K})$ is dense in $\cH$. On account of the symmetry of $\widetilde{K}$, $\ker(\mathbbm{1}-\widetilde{K})\perp\mathrm{ran}(\mathbbm{1}-\widetilde{K})$, therefore $\ker(\mathbbm{1}-\widetilde{K})=\{0\}$. Then $\widetilde{K}=K_{\widetilde{S}}$ for some positive symmetric operator $\widetilde{S}$ on $\cH$ (Theorem \ref{thm:II-KreinTransf}(i)) which is also an extension of $S$ (Theorem \ref{thm:II-KreinTransf}(iii)), and in fact a self-adjoint extension (Theorem \ref{thm:II-KreinTransf}(v)). Conversely, if $\widetilde{S}$ is a positive self-adjoint extension of $S$, then $\widetilde{K}:=K_{\widetilde{S}}$ is bounded and symmetric with $\|\widetilde{K}\|_{\mathrm{op}}\leqslant 1$ (Theorem \ref{thm:II-KreinTransf}(i)), extends $K_S$ (Theorem \ref{thm:II-KreinTransf}(iii)), and is self-adjoint (Theorem \ref{thm:II-KreinTransf}(v)).  
 \end{proof}

 \section{Kre{\u\i}n's extension theory of symmetric semi-bounded operators}\label{sec:II-KreinExtTheory}\index{Kre{\u\i}n's extension theory}

 Kre{\u\i}n's theory of self-adjoint extension of symmetric semi-bounded operators was developed in Kre{\u\i}n's 1946 work \cite{Krein-1947} (published in 1947), following a previous short announcement of the main results without proofs published in 1945 \cite{Krein-1945}. It was then applied in the same author's 1947 work \cite{Krein-1947-II} to differential boundary value problems. In more recent times (2000-2009), core aspects of the theory were revisited, among others, by Arlinski\u{\i}, Hassi, Sebesty\'{e}n, and de Snoo \cite{Arlinski-YHassi-Seb-Shoo-2001}, Hassi, Malamud, and de Snoo \cite{Hassi-Malamud-Snoo-2004}, Arlinski{\u\i} and Tsekanovski{\u\i} \cite{arlinski-tsekanoviski-2005}, Kurasov \cite{Kurasov-2009} from a perspective oriented towards the new theory of boundary triplets.\index{boundary triplets}

 In this Section Kre{\u\i}n's theory is presented by combining the scheme of the original version \cite{Krein-1947} with the elegant reformulation given in 1969 by Ando and Nishio \cite{ando-nishio-1969}.
 
 
 Even though the special case of extension of semi-bounded symmetric operators is clearly covered by von Neumann's general theory\index{von Neumann's extension theory} (Sect.~\ref{sec:II-vN-theory}), several important aspects remain undetected when the latter theory is applied to the semi-bounded setting. For one thing, von Neumann's Theorem \ref{thm:vonN_thm_selfadj_exts} is not elucidative concerning the semi-bounded extensions of a semi-bounded symmetric operator, in particular the extension parametrisation $U\leftrightarrow S_U$ therein does not identify a priori which $U$ labels the distinguished Friedrichs extension. 

 In fact, in the almost \emph{twenty years} between the seminal works by von Neumann \cite{vonNeumann1930} (from 1928) and by Kre{\u\i}n \cite{Krein-1947} (from 1946), one major focus in the newborn self-adjoint extension theory was the extension problem of densely defined \emph{lower semi-bounded} operators (of course there is no conceptual difference between lower and upper semi-boundedness).

 As commented already, for any such $S$ the \emph{existence} of self-adjoint extensions of $S$ had been established by von Neumann \cite[Satz 43]{vonNeumann1930} (with also the proof that there are extensions whose bottom is arbitrarily close to $\mathfrak{m}(S)$), and in fact also in \cite[Satz 42]{vonNeumann1930}, as will be commented in a moment, then by Stone \cite[Theorem 9.21]{Stone1932} in 1932 (with also the non-constructive proof that there is one extension with bottom equal to $\mathfrak{m}(S)$), and finally by Friedrichs \cite{Friedrichs1934} in 1934, followed by Freudenthal \cite{Freudenthal-1936} in 1936. None of the above-mentioned proofs passed through the check that owing to its semi-boundedness $S$ has necessarily equal deficiency indices: such an implication was established by Calkin \cite[Theorem 3]{Calkin-1940} in 1940 under the additional assumption that $S$ is closed (today the typical reasoning for that result is, e.g., \cite[Theorem X.1 and Corollary]{rs2}). In fact Calkin proved the more general result \cite[Theorem 2]{Calkin-1940} that if $S$ is symmetric \emph{and closed}, and $(S-z\mathbbm{1})^{-1}$ exists and is bounded for some $z\in\mathbb{C}$, then $S$ admits self-adjoint extensions having $z$ in the their resolvent set: the semi-bounded case is covered with $z<\mathfrak{m}(S)$. The equality of the deficiency indices of a symmetric \emph{and closable} $S$ that is also semi-bounded follows by the Krasnosel'ski\u{\i}-Kre{\u\i}n theorem proved in \cite{Krein-Krasnoselskii-Milman-1948} in 1947.

 Kre{\u\i}n's theory \cite{Krein-1947} was developed as an extension theory for \emph{positive} and densely defined operators: it is indeed non-restrictive to shift the lower semi-bounded $S$ to $S+\lambda\mathbbm{1}$ with $\lambda\geqslant -\mathfrak{m}(S)$ and thus obtain a positive operator.

 A first explicit motivation of Kre{\u\i}n \cite{Krein-1947} was the then-open question of \emph{uniqueness} of the Friedrichs extension (which in modern terms is Theorem \ref{thm:Friedrichs-ext}(v) above). Already in Freudenthal's elegant refinement \cite{Freudenthal-1936} of Friedrichs' construction, a simple explicit example was given of multiple extensions enjoying the same property of having the highest possible bottom. Eventually Kre{\u\i}n's theory could establish (among further additional information) that the Friedrichs extension is the unique one with operator domain contained in the form domain of the initial symmetric, densely defined, lower semi-bounded operator (\cite[Theorem 10]{Krein-1947}, namely the present Theorem \ref{thm:Friedrichs-ext}(v)).

 The other major focus of Kre{\u\i}n \cite{Krein-1947} was the \emph{structure} of the family of positive self-adjoint extensions of the positive and densely defined $S$. It was known that this family contains the Friedrichs extension $S_\mathrm{F}$, with the property $\mathfrak{m}(S_\mathrm{F})=\mathfrak{m}(S)$ and the then-open question of uniqueness. One further positive self-adjoint extension was introduced in 1928 by von Neumann in \cite[Satz 42]{vonNeumann1930} under the additional assumption that $S$ has strictly positive bottom ($\mathfrak{m}(S)>0$), by now referred to as the \emph{Kre{\u\i}n-von Neumann extension},\index{Kre{\u\i}n-von Neumann extension} henceforth denoted as $S_\mathrm{N}$, for which one may have instead $\mathfrak{m}(S_\mathrm{N})=0$. ($S_\mathrm{N}$ is customarily named also after Kre{\u\i}n because in Kre{\u\i}n's work \cite{Krein-1947} it was re-obtained and characterised under the more general assumption $\mathfrak{m}(S)\geqslant 0$.) If $S$ is densely defined and with strictly positive bottom, $S_\mathrm{N}$ is defined as 
 \begin{equation}\label{eq:II-def-SN-first}
  \begin{array}{rcl}
    \mathcal{D}(S_\mathrm{N})\;&:=&\;\mathcal{D}(\overline{S})\dotplus\ker S^* \\
    S_\mathrm{N}(f+u)\;&:=&\;\overline{S}f
  \end{array}\qquad\quad (\mathfrak{m}(S)>0)\,.
 \end{equation}
 Later this definition will be extended to the general positive case, namely $\mathfrak{m}(S)\geqslant 0$.

 \begin{lemma}\label{lem:II-SN-first}
  Let $S$ be a densely defined and lower semi-bounded symmetric operator with $\mathfrak{m}(S)>0$ and let $S_\mathrm{N}$ be the operator \eqref{eq:II-def-SN-first}. Then $S_\mathrm{N}$ is a lower semi-bounded and positive self-adjoint extension of $S$. Moreover, $\mathfrak{m}(S_\mathrm{N})=0$ if and only if $\ker S^*\neq\{0\}$.
 \end{lemma}

 \begin{proof}
  By construction $S_\mathrm{N}$ has dense domain and extends $S$ (it restricts $S^*$). Moreover, $\mathcal{D}(\overline{S})\cap\ker S^*=\{0\}$, and therefore $\mathcal{D}(\overline{S})\dotplus\ker S^*$ is indeed a direct sum, because if $f\in\mathcal{D}(\overline{S})\cap\ker S^*$ then $0=S^*f=\overline{S}f$, whence $f=0$, as $\overline{S}$ is injective ($\mathfrak{m}(\overline{S})=\mathfrak{m}(S)>0$). For generic $f+u\in \mathcal{D}(\overline{S})\dotplus\ker S^*=\mathcal{D}(S_\mathrm{N})$, one has $\langle f+u,S_\mathrm{N}(f+u)\rangle=\langle f+u,\overline{S}f\rangle=\langle f,\overline{S} f\rangle\in\mathbb{R}$, owing to the symmetry of $\overline{S}$ and $S^*u=0$, therefore $S_\mathrm{N}$ is symmetric. This also shows that $S_\mathrm{N}$ is positive, because so is $\overline{S}$. Next, it will be now shown that $S_\mathrm{N}$ is reduced with respect to the decomposition $\cH=\mathrm{ran}\,\overline{S}\oplus\ker S^*$. (Observe that, since $S$ is symmetric and densely defined, hence closable, and since $\mathfrak{m}(S)>0$, then $\mathrm{ran}\,\overline{S}=\overline{\mathrm{ran}\,S}$ is a closed subspace, see Section \ref{sec:I-symmetric-selfadj}.) In fact, $\mathcal{D}(S_\mathrm{N})\cap \mathrm{ran}\,\overline{S}=\mathcal{D}(\overline{S})\cap \mathrm{ran}\,\overline{S}$, whence $S_\mathrm{N}(\mathcal{D}(S_\mathrm{N})\cap \mathrm{ran}\,\overline{S})=\overline{S}(\mathcal{D}(\overline{S})\cap \mathrm{ran}\,\overline{S})\subset \mathrm{ran}\,\overline{S}$, implying that the Hilbert subspace $\mathrm{ran}\,\overline{S}$ is invariant for $S_\mathrm{N}$. Moreover, $S_\mathrm{N}\ker S^*=S^*\ker S^*=\{0\}$, hence trivially the Hilbert subspace $\ker S^*$ is invariant for $S_\mathrm{N}$. And if $P_{\ker S^*}$ is the orthogonal projection onto $\ker S^*$, then $P_{\ker S^*}\mathcal{D}(S_\mathrm{N})\subset\ker S^*\subset \mathcal{D}(S_\mathrm{N})$. All conditions are therefore matched that allow to conclude that $\mathrm{ran}\overline{S}$ and $\ker S^*$ are reducing subspaces for $S_\mathrm{N}$ (Sect.~\ref{sec:I_invariant-reducing-ssp}). Thus, $S_\mathrm{N}=S_1\oplus S_2$ with $S_1:=S_\mathrm{N}\upharpoonright(\mathcal{D}(S_\mathrm{N})\cap \mathrm{ran}\,\overline{S})=\overline{S}\upharpoonright(\mathcal{D}(\overline{S})\cap \mathrm{ran}\,\overline{S})$ and $S_2:=S_\mathrm{N}\upharpoonright\ker S^*=\mathbb{O}\upharpoonright\ker S^*$. $S_2$ is clearly self-adjoint in $\ker S^*$. As for $S_1$, it has real expectations because so has $S_\mathrm{N}$, hence is symmetric, and moreover $\mathrm{ran}\,S_1=\mathrm{ran}\,\overline{S}$: then (Sect.~\ref{sec:I-symmetric-selfadj}) $S_1$ is self-adjoint in $\mathrm{ran}\,\overline{S}$. As a consequence, $S_\mathrm{N}=S_1\oplus S_2$ is self-adjoint in $\cH$. Concerning the last claim, if $\ker S^*\neq\{0\}$, then there exist normalised vectors $u\in\ker S^*$ and for them $\langle u,S_\mathrm{N}u\rangle=0$, whence $\mathfrak{m}(S_\mathrm{N})=0$. Conversely, if $\ker S^*=\{0\}$, then \eqref{eq:II-def-SN-first} implies $\overline{S}=S_\mathrm{N}$, that is, $S$ is essentially self-adjoint, and $\mathfrak{m}(S_\mathrm{N})=\mathfrak{m}(\overline{S})=\mathfrak{m}(S)>0$.  
 \end{proof}

 Kre{\u\i}n \cite{Krein-1947} thus moved from the then-available knowledge of existence of \emph{two} positive self-adjoint extensions of a given $S$ that is symmetric, densely defined, and with $\mathfrak{m}(S)>0$: the extensions $S_\mathrm{F}$ and $S_\mathrm{N}$. The possible occurrence $\mathfrak{m}(S_\mathrm{N})=0$ (the lowest possible bottom among the positive self-adjoint extensions of $S$) and the fact that $\mathfrak{m}(S_\mathrm{F})=\mathfrak{m}(S)$ (the largest possible bottom) posed the question on whether $S_\mathrm{N}$ and $S_\mathrm{F}$ are the two edges of a whole range of positive extensions: eventually Kre{\u\i}n's theory confirmed that structure of the family of positive extensions ``between'' $S_\mathrm{N}$ and $S_\mathrm{F}$.

 The actual starting point of \cite{Krein-1947} was, slightly more generally, a densely defined and positive symmetric operator $S$ (thus with $\mathfrak{m}(S)\geqslant 0$, not necessarily $\mathfrak{m}(S)> 0$) on a given Hilbert space $\cH$. The first crucial result of Kre{\u\i}n's theory is Corollary \ref{cor:II-krein}, that reduces the quest of self-adjoint extensions of such $S$ to the quest of contractive self-adjoint extensions of the Kre{\u\i}n transform\index{Kre{\u\i}n transform} $K_S$ of $S$. The latter problem, equivalent to the former, is conceptually easier, as it concerns bounded operators.

 The second crucial result of Kre{\u\i}n's theory \cite{Krein-1947} was to establish that the contractive self-adjoint extensions of the Kre{\u\i}n transform\index{Kre{\u\i}n transform} $K_S$ of $S$ are all included, in the sense of the self-adjoint operator ordering, between a highest and a lowest extension.

 \begin{theorem}\label{thm:II-KreinThm-Kext}
  Let $K$ be a bounded symmetric operator on a Hilbert space $\cH$ (with no a priori assumption on closedness or density of $\mathcal{D}(K)$). Then the family 
  \[
   \mathscr{B}(K)\;:=\;\left\{
   \begin{array}{c}
    \textrm{bounded self-adjoint extensions $\widetilde{K}$ of $K$ in $\cH$} \\
    \textrm{with $\| \widetilde{K}\|_{\mathrm{op}}=\|K\|_{\mathrm{op}}$}
   \end{array}
   \right\}
  \]
  is non-empty and contains elements $K_m$ and $K_M$ such that any self-adjoint operator $\widetilde{K}$ on $\cH$ belongs to $\mathscr{B}(K)$ if and only if $K_m\leqslant\widetilde{K}\leqslant K_M$.
 \end{theorem}

 Theorem \ref{thm:II-KreinThm-Kext} is essentially Kre{\u\i}n's \cite[Theorem 3]{Krein-1947}, but instead of presenting here Kre{\u\i}n's ingenious construction, an alternative route shall be followed in a moment, consisting of the very elegant refinement of the theory made two decades later (in 1969) by Ando and Nishio \cite{ando-nishio-1969}. For the time being Kre{\u\i}n's argument can be completed: indeed, combining the two pillars of the theory, namely Theorem \ref{thm:II-KreinTransf} - Corollary \ref{cor:II-krein} and Theorem \ref{thm:II-KreinThm-Kext}, one obtains the following.

 \begin{theorem}[Kre{\u\i}n's theorem on positive self-adjoint extensions]\label{thm:II-Krein-final-thm}\index{theorem!Kre{\u\i}n (self-adjoint extension)}
  Let $S$ be a densely defined, symmetric, positive ($\mathfrak{m}(S)\geqslant 0$) operator on a Hilbert space $\cH$. If $S$ is bounded, then its closure $\overline{S}$ is the only self-adjoint extension of $S$, and has the same norm. If additionally $S$ is unbounded, then $S$ admits two positive self-adjoint extension, $S_\mathrm{F}$ and $S_\mathrm{N}$, with $S_\mathrm{F}\geqslant S_\mathrm{N}$, and a self-adjoint operator $\widetilde{S}$ on $\cH$ is a positive self-adjoint extension of $S$ if and only if $S_\mathrm{N}\leqslant S\leqslant S_\mathrm{F}$, equivalently, $(S_\mathrm{F}+\lambda\mathbbm{1})^{-1}\leqslant(S+\lambda\mathbbm{1})^{-1}\leqslant(S_\mathrm{N}+\lambda\mathbbm{1})^{-1}$ for one, hence for all $\lambda>0$. $S_\mathrm{F}$ and $S_\mathrm{N}$ are therefore the unique largest and unique smallest among the positive self-adjoint extensions of $S$. $S_\mathrm{F}$ is the Friedrichs extension of $S$ (Theorem \ref{thm:Friedrichs-ext}). $S_\mathrm{N}$ is the \emph{Kre{\u\i}n-von Neumann extension},\index{Kre{\u\i}n-von Neumann extension} of $S$, and when $\mathfrak{m}(S)>0$ it is precisely the operator \eqref{eq:II-def-SN-first}. If $S_\mathrm{F}=S_\mathrm{N}$, then $S$ is essentially self-adjoint, with $\overline{S}=S_\mathrm{F}=S_\mathrm{N}$.
 \end{theorem}

 \begin{proof}
  The bounded case is obvious, so let us assume that $S$ is densely defined, symmetric, positive, and unbounded. Unboundedness implies that the Kre{\u\i}n transform of $S$ has norm $\|K_S\|_{\mathrm{op}}=1$ (Theorem \ref{thm:II-KreinTransf}(ii)). $K_S$ admits two bounded self-adjoint extensions $K_m$ and $K_M$ on $\cH$ with $\|K_m\|_{\mathrm{op}}=\|K_M\|_{\mathrm{op}}=1$ and the properties described in Theorem \ref{thm:II-KreinThm-Kext}. Then, owing to Corollary \ref{cor:II-krein}, their inverse Kre{\u\i}n transforms 
  \[
   S_{K_m}\;=\;(\mathbbm{1}+K_m)(\mathbbm{1}-K_m)^{-1}\,,\qquad S_{K_M}\;=\;(\mathbbm{1}+K_M)(\mathbbm{1}-K_M)^{-1}
  \]
  are positive self-adjoint extensions of $S$. Since $K_m\leqslant K_M$ (Theorem \ref{thm:II-KreinThm-Kext}), correspondingly $S_{K_m}\leqslant S_{K_M}$ (Theorem \ref{thm:II-KreinTransf}(iv)). Moreover, along the same reasoning, a self-adjoint operator $\widetilde{S}$ is a positive self-adjoint extension of $S$ if and only if its Kre{\u\i}n transform $K_{\widetilde{S}}$ is a bounded self-adjoint extension of $K_S$ with $\|K_{\widetilde{S}}\|_{\mathrm{op}}=1$ (Corollary \ref{cor:II-krein} and Theorem \ref{thm:II-KreinTransf}(ii)), i.e., if and only if $K_{\widetilde{S}}\in \mathscr{B}(K_S)$, in the notation of Theorem \ref{thm:II-KreinThm-Kext}. The latter condition is equivalent to $K_m\leqslant K_{\widetilde{S}}\leqslant K_M$ (Theorem \ref{thm:II-KreinThm-Kext}), which is in turn equivalent to $S_{K_m}\leqslant\widetilde{S}\leqslant S_{K_M}$ (Theorem \ref{thm:II-KreinTransf}(iv)). The equivalence of the latter chain of inclusions with the corresponding inclusions for resolvents was recalled in Section \ref{sec:I_forms}. The above findings and Theorem \ref{thm:Friedrichs-ext}(vi) imply $S_{K_M}=S_\mathrm{F}$, the Friedrichs extension, which is the largest positive self-adjoint extension of $S$ and satisfies $\mathfrak{m}(S_\mathrm{F})=\mathfrak{m}(S)$, and imply also that $S_\mathrm{N}:=S_{K_m}$ is the smallest positive self-adjoint extension of $S$. 
  The property that $S_\mathrm{N}$ is precisely the operator \eqref{eq:II-def-SN-first} when $\mathfrak{m}(S)>0$ will be proved in Theorem \ref{thm:II-KvNextension}(iv).  
 \end{proof}

 The terminology introduced by Kre{\u\i}n for the two distinguished extensions $S_\mathrm{F}$ and $S_\mathrm{N}$ of Theorem \ref{thm:II-Krein-final-thm}, nowadays universally referred to, respectively, as Friedrichs and Kre{\u\i}n-von Neumann extension, was the \emph{rigid} (\includegraphics[scale=0.16]{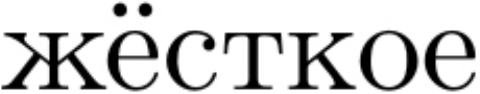}) extension ($S_\mathrm{F}$, originally denoted as $S_\mu$) and the \emph{soft} (\includegraphics[scale=0.16]{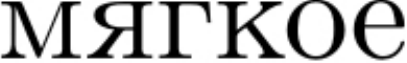}) extension ($S_\mathrm{N}$, originally denoted as $S_M$).\index{soft extension}\index{rigid extension}

  A detour is now taken through the approach by Ando and Nishio \cite{ando-nishio-1969}, upon which Theorem \ref{thm:II-KreinThm-Kext} will be proved. In fact, the Ando-Nishio approach reproduces more generally the conclusions of Theorem \ref{thm:II-Krein-final-thm}, hence the whole final results of Kre{\u\i}n's extension theory.

 For a given positive symmetric operator $S$ on the Hilbert space $\cH$, not necessarily densely defined, and for given $g\in\cH$, one defines
 \begin{equation}\label{eq:nSbound}
  \mathfrak{n}_S(g)\;:=\;\sup_{f\in\mathcal{D}(S)}\frac{\;|\langle g,Sf\rangle|^2}{\langle f,Sf\rangle}
 \end{equation}
 with the convention, here, that $\frac{0}{0}:=0$, and
 \begin{equation}\label{eq:II-ES}
  \begin{split}
   \mathcal{E}(S)\;&:=\;\{ g\in\cH\,|\, \mathfrak{n}_S(g)\,<\,+\infty\} \\
   &\;\,=\;\{g\in\cH\,|\,\exists \kappa_g\geqslant 0\textrm{ with }|\langle g,Sf\rangle|^2\leqslant\kappa_g\langle f,Sf\rangle\;\forall\,f\in\mathcal{D}(S)\}\,,
  \end{split}
 \end{equation}
 where obviously 
 \begin{equation}
  \mathfrak{n}_S(g)\;=\;\inf\{\kappa_g\,|\,g\in\mathcal{E}(S)\}\,.
 \end{equation}

 \begin{theorem}[Ando-Nishio theorem]\label{thm:Ando-Nishio}\index{theorem!Ando-Nishio}
  Let $S$ be a positive symmetric operator on the Hilbert space $\cH$.
  \begin{enumerate}[(i)]
   \item Any positive symmetric extension $\widetilde{S}$ of $S$ satisfies $\mathcal{D}(\widetilde{S})\subset \mathcal{E}(S)$.
   \item $S$ admits positive self-adjoint extensions on $\cH$ if and only if $\mathcal{E}(S)$ is dense in $\cH$.
   \item When any of the conditions (ii) is matched, $S$ has a unique smallest positive self-adjoint extension $S_\mathrm{N}$: its quadratic form is explicitly given by
  \begin{equation}\label{eq:II_SNgeneralform}
   \begin{split}
    \mathcal{D}[S_\mathrm{N}]\;&:=\;\mathcal{D}(S_\mathrm{N}^{\frac{1}{2}})\;=\;\mathcal{E}(S)\,, \\
    S_\mathrm{N}[g]\;&:=\;\|S_\mathrm{N}^{\frac{1}{2}}g\|^2\;=\;\mathfrak{n}_S(g)\,.
   \end{split}
  \end{equation}
  \end{enumerate}
  \end{theorem}

 \begin{proof} 
 (i) For $g\in\mathcal{D}(\widetilde{S})$ the Cauchy-Schwarz inequality yields
 \[
  |\langle g,Sf\rangle|^2\;=\;|\langle g,\widetilde{S}f\rangle|^2\;\leqslant\;\langle f,\widetilde{S}f\rangle\,\langle g,\widetilde{S}g\rangle\;=\;\langle f,Sf\rangle\,\langle g,\widetilde{S}g\rangle \qquad \forall\,f\in\mathcal{D}(S)\,,
 \]
 whence $\mathfrak{n}_S(g)\leqslant\langle g,\widetilde{S}g\rangle<+\infty$, i.e., $g\in \mathcal{E}(S)$.

  (ii) If $\widetilde{S}$ is a positive self-adjoint extension of $S$, then the density of $\mathcal{E}(S)$ follows from part (i).

  For the converse implication, assume that $\mathcal{E}(S)$ is dense in $\cH$. Assume further that for some $h\in\cH$ there is a sequence $(f_n)_{n\in\mathbb{N}}$ in $\mathcal{D}(S)$ such that $\|Sf_n- h\|\to 0$ and $\langle f_n,Sf_n\rangle\to 0$ as $n\to\infty$. Owing to the density of $\mathcal{E}(S)$, for arbitrary $\varepsilon>0$ there is $g\in\mathcal{E}(S)$ with $\|h-g\|\leqslant\varepsilon$, and for such $g$
  \[
   |\langle g,Sf_n\rangle|^2\;\leqslant\;\mathfrak{n}_S(g)\,\langle f_n,Sf_n\rangle\,.
  \]
  The limit $n\to\infty$ then yields $|\langle g,h\rangle|=0$. Thus, $\|h\|^2=\langle h,h\rangle=|\langle h,h-g\rangle|\leqslant\varepsilon\|h\|$, implying $\|h\|\leqslant\varepsilon$. Since $\varepsilon$ is arbitrary, $h=0$. It is therefore proved, thanks to the density of $\mathcal{E}(S)$, that \emph{for any} $h\in\cH$
  \[\tag{*}\label{eq:II-positclos}
   \left.  
   \begin{array}{c}
    Sf_n\to h \\
    \langle f_n,Sf_n\rangle\to 0
   \end{array}
   \right\}\qquad\Rightarrow\qquad\;h\,=\,0\,.
  \]

  Next, let $P:\cH\to\cH$ be the orthogonal projection onto $\overline{\mathrm{ran}\,S}$ and consider the operator $A$ on the Hilbert space $\overline{\mathrm{ran}\,S}$ (equipped with the inherited Hilbert structure from $\cH$)
  defined by
  \[
   \mathcal{D}(A)\;:=\;\mathrm{ran}\,S\,,\qquad A(Sf)\;:=\;Pf\quad\forall f\in\mathcal{D}(S)\,.
  \]
  Such a definition is well-posed because $Sf=0$ implies $Pf=0$: indeed, if $Sf=0$, then $0=\langle h,Sf\rangle=\langle Sh,f\rangle$ $\forall h\in\mathcal{D}(S)$, whence $f\perp\mathrm{ran}\,S$, and therefore $Pf=0$. Moreover, $A$ does act within the Hilbert space $\overline{\mathrm{ran}\,S}$, because $\mathrm{ran}\,A\subset\mathrm{ran}\,P=\overline{\mathrm{ran}\,S}$. It also follows from definition that
  \[\tag{**}\label{eq:II-SAS}
   \langle Sf,A\,Sf\rangle\;=\;\langle Sf,Pf\rangle\;=\;\langle PSf,f\rangle\;=\;\langle f,Sf\rangle\qquad\forall f\in\mathcal{D}(S)\,.
  \]

  The operator $A$ is densely defined in $\overline{\mathrm{ran}\,S}$ by construction, and is positive symmetric, because of \eqref{eq:II-SAS} and the positivity of $S$.
Therefore (Theorem \ref{thm:Friedrichs-ext}), the Friedrichs extension $A_{\mathrm{F}}$ of $A$ exists and is a positive self-adjoint operator in $\overline{\mathrm{ran}} S$. 

  Now, $A_\mathrm{F}$ is actually injective. To see this, let $g\in\mathcal{D}(A_\mathrm{F})$ with $A_\mathrm{F} g=0$. Then also $A_\mathrm{F}[g]=0$. On account of Theorem \ref{thm:Friedrichs-ext}(i), there is a sequence $(Sf_n)_{n\in\mathbb{N}}$ in $\mathcal{D}(A)$ (hence a sequence $(f_n)_{n\in\mathbb{N}}$ in $\mathcal{D}(S)$) such that
  \[
   \|Sf_n-g\|_{\mathcal{D}[A]}\,\xrightarrow{\,n\to\infty\,}\,0\qquad\textrm{and}\qquad 0\;=\;A_\mathrm{F}[g]\;=\;\lim_{n\to\infty}\langle Sf_n, A\,Sf_n\rangle\,,
  \]
  where the above norm is the usual form norm on $\mathcal{D}[A]$: in this case, since $A$ is positive, the above limits and \eqref{eq:II-SAS} imply $Sf_n\to g$ in $\cH$ and $\langle f_n,S f_n\rangle\to 0$. Then property \eqref{eq:II-positclos} implies $g=0$, which proves injectivity.

  $A_\mathrm{F}$ is therefore invertible on its range, and it makes sense to set
  \[
   S_\mathrm{N}\;:=\;P\,A_\mathrm{F}^{-1}P\,,
  \]
  as an operator on $\cH$ (thus, with domain $\mathcal{D}(S_\mathrm{N})=\{g\in\cH\,|\,Pg\in\mathrm{ran}\,A_\mathrm{F}\}$). 
  In other words, with respect to the Hilbert space decomposition (Sect.~\ref{sec:I-adjoint})
  \[
   \cH\;=\;\overline{\mathrm{ran}\,S}\oplus\ker S^*\,,
  \]
  one has that $S_\mathrm{N}$ satisfies the operator orthogonal sum (Sect.~\ref{sec:I_invariant-reducing-ssp}) $S_\mathrm{N}=A_\mathrm{F}^{-1}\oplus\mathbb{O}$ (recall, indeed, that $A_\mathrm{F}$ and $A_\mathrm{F}^{-1}$ are operators in $\overline{\mathrm{ran}\,S}$ and $P$ projects onto $\overline{\mathrm{ran}\,S}$), that is, $S_\mathrm{N}$ is the extension by zero of $A_\mathrm{F}^{-1}$, originally defined on $\overline{\mathrm{ran}\,S}$, also to the orthogonal complement $\ker S^*$. Since $\mathrm{ran}\,A_\mathrm{F}^{-1}=\mathcal{D}(A_\mathrm{F})\subset\overline{\mathrm{ran}\,S}$, we can also write $S_\mathrm{N}=PA_\mathrm{F}^{-1}P=A_\mathrm{F}^{-1}P$.

  The operator $A_\mathrm{F}^{-1}$ is self-adjoint (Sect.~\ref{sec:I-symmetric-selfadj}) and positive (Sect.~\ref{sec:I_funct_calc}) on $\overline{\mathrm{ran}\,S}$, therefore $S_\mathrm{N}=A_\mathrm{F}^{-1}\oplus\mathbb{O}$ is positive and self-adjoint on the whole $\cH$. Besides, since $Pf=ASf$ for any $f\in\mathcal{D}(S)$, then
  \[
   S_\mathrm{N} f\;=\;A_\mathrm{F}^{-1}Pf\;=\;A_\mathrm{F}^{-1}ASf\;=\;Sf\qquad \forall f\in\mathcal{D}(S)\,,
  \]
  meaning that $S_\mathrm{N}$ is a positive and self-adjoint \emph{extension} of $S$. This completes the proof of part (ii).

  (iii) Let us show that the self-adjoint extension $S_\mathrm{N}$ is precisely the one with the claimed properties.

  The functional calculus for $A_\mathrm{F}$ (Sect.~\ref{sec:I_funct_calc}) and the above operator orthogonal sum $S_\mathrm{N}=A_\mathrm{F}^{-1}\oplus\mathbb{O}$ imply that $S_\mathrm{N}^{\frac{1}{2}}=A_\mathrm{F}^{-\frac{1}{2}}\oplus\mathbb{O}=PA_\mathrm{F}^{-\frac{1}{2}}P=A_\mathrm{F}^{-\frac{1}{2}}P$. Let now $g\in\mathcal{D}[S_\mathrm{N}]$. By positivity and self-adjointness (Sect.~\ref{sec:I_forms}), $\mathcal{D}[S_\mathrm{N}]=\mathcal{D}\big(S_\mathrm{N}^{\frac{1}{2}}\big)$ and $S_\mathrm{N}[g]=\big\|S_\mathrm{N}^{\frac{1}{2}}g\big\|^2=\big\|A_\mathrm{F}^{-\frac{1}{2}}Pg\big\|^2$. Thus, $\mathcal{D}[S_\mathrm{N}]$ consists of all $g$'s in $\cH$ with $\big\|A_\mathrm{F}^{-\frac{1}{2}}Pg\big\|<+\infty$ and
  \[
   S_\mathrm{N}[g]\;=\;\sup_{\substack{ \varphi\in\mathcal{D}(A_\mathrm{F}^{-1/2}) \\ \varphi\neq 0  }}\frac{\:\big|\big\langle \varphi,A_\mathrm{F}^{-\frac{1}{2}} Pg  \big\rangle \big|^2}{\|\varphi\|^2}\;=\;\sup_{\substack{ h\in\mathcal{D}(A_\mathrm{F}^{1/2}) \\ h\neq 0  }}\frac{\:|\langle h,Pg\rangle|^2}{\big\| A_\mathrm{F}^{\frac{1}{2}}h\big\|^2}\,,
  \]
  the first identity following from the variational characterisation of the Hilbert norm (Sect.~\ref{sec:I-preliminaries}) using the fact that the non-zero elements of $\mathcal{D}\big(A_\mathrm{F}^{-\frac{1}{2}}\big)$ form a dense in $\overline{\mathrm{ran}\,S}$, and the second identity following from having set $h:=A_\mathrm{F}^{-\frac{1}{2}}\varphi$. Now, for any $h\in \mathcal{D}\big(A_\mathrm{F}^{\frac{1}{2}}\big)=\mathcal{D}[A_\mathrm{F}]$, on account of Theorem \ref{thm:Friedrichs-ext}(i), and owing to the positivity of $A_\mathrm{F}$, there is a sequence $(h_n)_{n\in\mathbb{N}}$ in $\mathcal{D}(A)=\mathrm{ran}\,S$, hence $h_n=S f_n$ for some sequence $(f_n)_{n\in\mathbb{N}}$ in $\mathcal{D}(S)$, such that $h_n\to h$ in $\cH$ and $\langle h_n,A h_n\rangle\to \big\| A_\mathrm{F}^{\frac{1}{2}}h\big\|$. The latter limit, owing to \eqref{eq:II-SAS}, is the same as $\langle f_n, Sf_n\rangle\to \big\| A_\mathrm{F}^{\frac{1}{2}}h\big\|$, and the former limit implies $\langle Sf_n,g\rangle=\langle Sf_n,Pg\rangle=\langle h_n,Pg\rangle\to\langle h,Pg\rangle$. Therefore,
  \[
   S_\mathrm{N}[g]\;=\;\sup_{ f\in\mathcal{D}(S) }\frac{\:|\langle g,Sf\rangle|^2}{\langle f,Sf\rangle}\;=\;\mathfrak{n}_S(g)\,.
  \]
  Since $h \neq 0$ and hence also $A_\mathrm{F}^{\frac{1}{2}}h\neq 0$, in the limit $\langle f_n, Sf_n\rangle\to \big\| A_\mathrm{F}^{\frac{1}{2}}h\big\|$ one has eventually $\langle f_n, Sf_n\rangle>0$: this means that the first identity above is initially derived under the condition that $\langle f,Sf\rangle>0$, and then is written in its final form setting conventionally $\frac{0}{0}:=0$. This establishes \eqref{eq:II_SNgeneralform}.

  It remains to prove that $S_\mathrm{N}$ is the smallest positive self-adjoint extension of $S$. To this aim, let $\widetilde{S}$ be any other positive self-adjoint extension. For $f\in\mathcal{D}(S)$ and $g\in\mathcal{D}[\widetilde{S}]=\mathcal{D}\big(\widetilde{S}^{\frac{1}{2}}\big)$ one has 
  \[
     |\langle g,Sf\rangle|^2\;=\;|\langle g,\widetilde{S}f\rangle|^2\;=\;\big|\big\langle \widetilde{S}^{\frac{1}{2}}g,\widetilde{S}^{\frac{1}{2}}f\big\rangle\big|^2 \;\leqslant\;\langle f,\widetilde{S}f\rangle\,\big\|\widetilde{S}^{\frac{1}{2}}g\big\|^2\;=\;\langle f,Sf\rangle\,\widetilde{S}[g]\,.
  \]
  This implies, in view of \eqref{eq:II-ES}, that $g\in\mathcal{E}(S)=\mathcal{D}[S_\mathrm{N}]$ and that $S_\mathrm{N}[g]\leqslant \widetilde{S}[g]$.    
 \end{proof}

 The Ando-Nishio theorem (Theorem \ref{thm:Ando-Nishio}) comes with a number of notable consequences.

 \begin{corollary}\label{cor:II_Ando-Nishio-cor}\index{theorem!Ando-Nishio}
   Let $S$ be a positive symmetric operator on the Hilbert space $\cH$.
   \begin{enumerate}[(i)]
    \item A further reference operator $R$ be given, which be positive and self-adjoint on $\cH$. $S$ has a positive self-adjoint extension $\widetilde{S}$ on $\cH$ satisfying $\widetilde{S}\leqslant R$ if and only if 
    \begin{equation}\label{eq:II_cond_RS}
     |\langle g,Sf\rangle|^2\;\leqslant\;\langle f,Sf\rangle\,\langle g,Rg\rangle\qquad\forall f\in\mathcal{D}(S)\,,\;\forall g\in\mathcal{D}(R)\,.
    \end{equation}
    \item For given $b>0$ the operator $S$ admits a positive, bounded, and self-adjoint extension $\widetilde{S}$ satisfying $\|\widetilde{S}\|_{\mathrm{op}}\leqslant b$ if and only if 
    \begin{equation}\label{eq:II_cond_RS-2}
     \|Sf\|^2\;\leqslant\;b\,\langle f,Sf\rangle\qquad\forall f\in\mathcal{D}(S)\,.
    \end{equation}
   \end{enumerate}
 \end{corollary}

 \begin{proof}
  (i) Assume that \eqref{eq:II_cond_RS} holds true. Then, in view of \eqref{eq:II-ES}, $\mathcal{D}(R)\subset\mathcal{E}(S)$: so $\mathcal{E}(S)$ is dense and $S$ admits the Kre{\u\i}n-von Neumann extension $S_\mathrm{N}$ (Theorem \ref{thm:Ando-Nishio}(ii)-(iii)). Formulas \eqref{eq:nSbound}, \eqref{eq:II_SNgeneralform}, and \eqref{eq:II_cond_RS} then imply
  \[
   S_\mathrm{N}[g]\;=\;\mathfrak{n}_S(g)\;\leqslant\;\langle g,Rg\rangle\;=\;R[g]\qquad \forall g\in\mathcal{D}(R)\subset\mathcal{E}(S)=\mathcal{D}[S_\mathrm{N}]\,,
  \]
  whence $S_\mathrm{N}\leqslant R$ (Sect.~\ref{sec:I_forms}). Thus, $S$ admits the self-adjoint extension $\widetilde{S}:=S_\mathrm{N}$ that satisfies $\widetilde{S}\leqslant R$. For the converse implication, assume now that $\widetilde{S}=(\widetilde{S})^*\supset S$ and $\widetilde{S}\leqslant R$. Owing to Theorem \ref{thm:Ando-Nishio}(iii), $S_\mathrm{N}\leqslant \widetilde{S}\leqslant R$, whence $\mathcal{D}[R]\subset\mathcal{D}[S_\mathrm{N}]$: therefore, if $g\in\mathcal{D}(R)$, then $g\in \mathcal{D}[S_\mathrm{N}]$ and $\mathfrak{n}_S(g)=S_\mathrm{N}[g]\leqslant R[g]=\langle g,Rg\rangle$ (having used \eqref{eq:II_SNgeneralform} again). The inequality $\mathfrak{n}_S(g)\leqslant\langle g,Rg\rangle$ $\forall g\in\mathcal{D}(R)$, on account of \eqref{eq:nSbound}, yields \eqref{eq:II_cond_RS}.
  
  (ii) With the reference operator $R=b\mathbbm{1}$, part (i) implies that $S$ admits a self-adjoint extension extension $\widetilde{S}$ with $\mathbb{O}\leqslant\widetilde{S}\leqslant b\mathbbm{1}$, equivalently, a positive self-adjoint extension $\widetilde{S}\in\mathcal{B}(\cH)$ with $\|\widetilde{S}\|_{\mathrm{op}}\leqslant b$, if and only if
  \[
   \frac{\:|\langle g,Sf\rangle|^2}{\:\|g\|^2}\;\leqslant\;b\,\langle f,Sf\rangle\qquad\forall f\in\mathcal{D}(S)\,,\;\forall g\in\mathcal{D}(R)\,.
  \]
  The latter formula is equivalent to \eqref{eq:II_cond_RS-2}.    
 \end{proof}

  As a further corollary, that deserves the status of a theorem on its own, one has the complete picture for the Kre{\u\i}n-von Neumann extension.

  \begin{theorem}[Kre{\u\i}n-von Neumann extension]\label{thm:II-KvNextension}\index{Kre{\u\i}n-von Neumann extension}
   Let $S$ be a symmetric and positive ($\mathfrak{m}(S)\geqslant 0$) operator on the Hilbert space $\cH$ such that the subspace $\mathcal{E}(S)$ defined in \eqref{eq:II-ES} is dense in $\cH$. (A sufficient condition for the density of $\mathcal{E}(S)$ is the density of $\mathcal{D}(S)$.)
   \begin{enumerate}[(i)]
    \item There exists a unique smallest positive self-adjoint extension $S_\mathrm{N}$, called the \emph{Kre{\u\i}n-von Neumann extension}\index{Kre{\u\i}n-von Neumann extension} of $S$. 
     \item The quadratic form of $S_\mathrm{N}$ is given by
      \begin{equation}\label{eq:II_SNgeneralform2}
   \begin{split}
    \mathcal{D}[S_\mathrm{N}]\;&=\;\mathcal{D}(S_\mathrm{N}^{\frac{1}{2}})\;=\;\mathcal{E}(S)\,, \\
    S_\mathrm{N}[g]\;&=\;\|S_\mathrm{N}^{\frac{1}{2}}g\|^2\;=\;\mathfrak{n}_S(g)\,,
   \end{split}
  \end{equation}
    where the functional $\mathfrak{n}_S$ is defined in \eqref{eq:nSbound}.
 
 \hspace{-0.8cm} For the remaining parts (iii)-(vi) assume furthermore that $S$ is densely defined.
   \item One has 
   \begin{equation}\label{eq:II_DSNcorollary}
    \begin{split}
     \mathcal{D}(S_\mathrm{N})\;&=\;
     \left\{ 
     g\in\mathcal{D}(S^*)\left|
     \begin{array}{c}
           \exists\,(f_n)_{n\in\mathbb{N}}\textrm{ in }\mathcal{D}(S)\textrm{ such that} \\
           Sf_n\xrightarrow{n\to\infty}S^*g \\
           \textrm{ and } \langle f_n-f_m,S(f_n-f_m)\rangle\xrightarrow{n,m\to\infty}0
     \end{array}
     \right.\right\}, \\
     S_\mathrm{N} g\;&=\;\lim_{n\to\infty}Sf_n\,.
    \end{split}
   \end{equation}
   \item If in addition $\mathfrak{m}(S)> 0$, then 
    \begin{equation}\label{eq:II-def-SN-first2}
  \begin{array}{rcl}
    \mathcal{D}(S_\mathrm{N})\;&=&\;\mathcal{D}(\overline{S})\dotplus\ker S^* \\
    S_\mathrm{N}(f+u)\;&=&\;\overline{S}f
  \end{array}\qquad\quad (\mathfrak{m}(S)>0)
 \end{equation}
  and  
      \begin{equation}\label{eq:II-def-SN-first3}
  \begin{array}{rcl}
    \mathcal{D}[S_\mathrm{N}]\;&=&\;\mathcal{D}[\overline{S}]\dotplus\ker S^*\;=\;\mathcal{D}[S_\mathrm{F}]\dotplus\ker S^* \\
    S_\mathrm{N}[f+u]\;&=&\;\overline{S}[f]\;=\;S_\mathrm{F}[f]
  \end{array}\quad (\mathfrak{m}(S)>0)\,.
 \end{equation}
   Moreover, $\mathfrak{m}(S_\mathrm{N})=0$ if and only if $\ker S^*$ is non-trivial.
  \item $\ker S_\mathrm{N}=\{0\}$ if and only if $\mathrm{ran}\,S$ is dense in $\cH$. When this is the case, $S^{-1}$ is a densely defined and positive symmetric operator, and
  \begin{equation}
   S_\mathrm{N}^{-1}\;=\;(S^{-1})_{\mathrm{F}}\,.
  \end{equation}
  \item Assume that $\ker S=\{0\}$. Then $S^{-1}$ has a positive self-adjoint extension on $\cH$ if and only if $\ker S_\mathrm{F}=\{0\}$. When this is the case, 
  \begin{equation}
  (S^{-1})_{\mathrm{N}}\;=\;S_\mathrm{F}^{-1}\,.
  \end{equation}
   \end{enumerate}
  \end{theorem}

 \begin{proof}
  One deduces directly from  Theorem \ref{thm:Ando-Nishio} both (i) and (ii), as well as the fact that the density of $\mathcal{D}(S)$ implies the density of $\mathcal{E}(S)$.

  (iii) In the notation of the proof of Theorem \ref{thm:Ando-Nishio}, $g\in\mathcal{D}(S_\mathrm{N})$ if and only if $Pg\in\mathrm{ran}\,A_\mathrm{F}$, i.e., $Pg=A_\mathrm{F}h_{Pg}$ for some $h_{Pg}\in\mathcal{D}(A_\mathrm{F})$ (which is actually unique for given $Pg$, as $A_\mathrm{F}$ is injective). On the other hand, on account of Theorem \ref{thm:Friedrichs-ext}(i), every $h\in\mathcal{D}[A_\mathrm{F}]$ is the limit, in the $\mathcal{D}[A]$-norm, of a sequence $(h_n)_{n\in\mathbb{N}}$ from $\mathcal{D}(A)$. This means that $h_n\to h$ in $\cH$ and $\langle h_n-h_m,A(h_n-h_m)\rangle\to 0$, owing to the positivity of $A$. Thus, for any $h\in \mathcal{D}[A_\mathrm{F}]$ there is a sequence $(f_n)_{n\in\mathbb{N}}$ from $\mathcal{D}(S)$ with $h_n=Sf_n$ and hence such that $Sf_n\to h$ and $\langle f_n-f_m,S(f_n-f_m)\rangle\to 0$ (Here we used property \eqref{eq:II-SAS} of Theorem \ref{thm:Ando-Nishio}'s proof, namely $\langle h_n-h_m,A(h_n-h_m)=\langle f_n-f_m,S(f_n-f_m)\rangle$). The above vector $h_{Pg}$ is therefore characterised by the condition of belonging to $\mathcal{D}[A_\mathrm{F}]$ \emph{and} the additional condition $Pg=A_\mathrm{F}h_{Pg}=A^*h_{Pg}$. Summarising, $g\in\mathcal{D}(S_\mathrm{N})$ if and only if
  \[
   Pg\,=\,A^*\big(\lim_{n\to\infty} Sf_n\big)\,,\qquad \langle f_n-f_m,S(f_n-f_m)\rangle\,\xrightarrow{n,m\to\infty}\,0
  \]
  for some sequence $(f_n)_{n\in\mathbb{N}}$ in $\mathcal{D}(S)$. 
  The latter identity is an identity in the Hilbert subspace $\overline{\mathrm{ran}\,S}$ and can be therefore equivalently re-written by testing it against the dense $\mathrm{ran}\,S$ of $\overline{\mathrm{ran}\,S}$, namely
  \[
   \langle S\varphi,Pg\rangle\;=\;\langle S\varphi,A^* h\rangle\qquad\forall\varphi\in\mathcal{D}(S)\,.
  \]
  On the l.h.s.~$\langle S\varphi,Pg\rangle=\langle\varphi,S^*g\rangle$, on the r.h.s.~$\langle S\varphi,A^* h\rangle=\langle AS\varphi,h\rangle=\langle P\varphi,h\rangle=\langle\varphi,h\rangle$. Thus, $S^*g=h$, whence $S_\mathrm{N}g=S^*g=h=\lim_{n} Sf_n$. Formula \eqref{eq:II_DSNcorollary} is thus established.
  

 (iv) Let $\widetilde{S}$ be the operator defined as in \eqref{eq:II-def-SN-first}, namely     
 \begin{equation*}
  \begin{array}{rcl}
    \mathcal{D}(\widetilde{S})\;&:=&\;\mathcal{D}(\overline{S})\dotplus\ker S^*  \, ,\\
    \widetilde{S}(f+u)\;&:=&\;\overline{S}f\,.
  \end{array}
 \end{equation*}
  $\widetilde{S}$ is a positive self-adjoint extension of $S$, and $\mathfrak{m}(\widetilde{S})=0$ if and only if $\ker S^*$ is non-trivial (Lemma \ref{lem:II-SN-first}). Now, obviously $\mathcal{D}(\overline{S})\subset\mathcal{D}(S_\mathrm{N})$. Moreover, $\ker S^*\subset\ker S_\mathrm{N}^{\frac{1}{2}}$, because if $g\in\ker S^*$, then $\langle g,Sf\rangle=\langle S^*g,f\rangle=0$, and then \eqref{eq:nSbound} and \eqref{eq:II_SNgeneralform2} imply $S_\mathrm{N}^{\frac{1}{2}}g=0$. In turn, this also implies $S_\mathrm{N}g=0$, thus concluding that $\ker S^*\subset\ker S_\mathrm{N}$. Therefore, $\mathcal{D}(\widetilde{S})\subset\mathcal{D}(S_\mathrm{N})$. As both $\widetilde{S}$ and $S_\mathrm{N}$ are self-adjoint extensions of $S$, necessarily 
  $\widetilde{S}=S_\mathrm{N}$. This establishes \eqref{eq:II-def-SN-first2}. Concerning \eqref{eq:II-def-SN-first3}, consider $g\in\mathcal{D}[S_\mathrm{N}]$. Since $S_\mathrm{N}$ is self-adjoint and therefore its form is closed (Sect.~\ref{sec:I_forms}), $\mathcal{D}(S_\mathrm{N})$ is dense in $\mathcal{D}[S_\mathrm{N}]$ in the form norm of $S_\mathrm{N}$ (Sect.~\ref{sec:I_forms}), hence there is a sequence $(g_n)_{n\in\mathbb{N}}$ in $\mathcal{D}(S_\mathrm{N})$ such that $\|g_n-g\|_{S_\mathrm{N}[\cdot]}\to 0$. In particular, $(g_n)_{n\in\mathbb{N}}$ is a Cauchy sequence in the form norm of $S_\mathrm{N}$, meaning  
  \[
   S_\mathrm{N}[g_n-g_m]+\|g_n-g_m\|_{\cH}^2\;=\;\|g_n-g_m\|_{S_\mathrm{N}[\cdot]}^2\;\xrightarrow{\,n,m\to\infty\,}\;0\,.
  \]
  Since $S_\mathrm{N}$ is positive, this implies $S_\mathrm{N}[g_n-g_m]\to 0$, and re-writing $g_n=f_n+u_n$ according to \eqref{eq:II-def-SN-first2}, with $f_n\in\mathcal{D}(\overline{S})$ and $u_n\in\ker S^*$, one also has (using $\mathfrak{m}(S)>0$)
  \[
   \begin{split}
       \|f_n-f_m\|_{\cH}^2\;&\leqslant\;\frac{1}{\mathfrak{m}(S)}\,\langle f_n-f_m,\overline{S}(f_n-f_m)\rangle\;=\;\frac{1}{\mathfrak{m}(S)}\,\langle g_n-g_m,S_\mathrm{N}(g_n-g_m)\rangle \\
       &=\;\frac{1}{\mathfrak{m}(S)}\,S_\mathrm{N}[g_n-g_m]\;\xrightarrow{\,n,m\to\infty\,}\;0\,.
   \end{split}
  \]
  Thus, $(f_n)_{n\in\mathbb{N}}$ is a Cauchy sequence in $\mathcal{D}(S)$ and hence $f_n\to f$ for some $f\in\cH$.
  In turn, from $f_n\to f$ and $\langle f_n-f_m,\overline{S}(f_n-f_m)\rangle\to 0$ and from the closedness of the form of $\overline{S}$ one concludes that $f\in\mathcal{D}[\overline{S}]$ (Sect.~\ref{sec:I_forms}), or equivalently $f\in\mathcal{D}[S_\mathrm{F}]$ (Theorem \ref{thm:Friedrichs-ext}(iii)). Moreover, $u_n=g_n-f_n\to g-f=:u$ and $u\in\ker S^*$ because this subspace is closed in $\cH$. Thus, $g=f+u\in\mathcal{D}[\overline{S}]+\ker S^*$ and (Sect.~\ref{sec:I_forms}) $S_\mathrm{N}[g]=\lim_{n}S_\mathrm{N}[g_n]=\lim_n\overline{S}[f_n]=\overline{S}[f]$. The sum $\mathcal{D}[\overline{S}]+\ker S^*$ is direct because if $f\in\mathcal{D}[\overline{S}]$, $u\in\ker S^*$, and $f+u=0$, then
  $0=S_\mathrm{N}[f+u]=\overline{S}[f]$, whence $f=0$ and also $u=0$. This way, one concludes $\mathcal{D}[S_\mathrm{N}]\subset\mathcal{D}[S_\mathrm{F}]\dotplus\ker S^*$. For the opposite inclusion, \eqref{eq:II-def-SN-first2} implies $\ker S^*\subset\mathcal{D}(S_\mathrm{N})\subset\mathcal{D}[S_\mathrm{N}]$, and the property $S_\mathrm{F}\geqslant S_\mathrm{N}$ (Theorem \ref{thm:Friedrichs-ext}(vi)) implies $\mathcal{D}[S_\mathrm{F}]\subset\mathcal{D}[S_\mathrm{N}]$ (Sect.~\ref{sec:I_forms}), whence indeed $\mathcal{D}(S_\mathrm{F})\dotplus\ker S^*\subset\mathcal{D}[S_\mathrm{N}]$.

  (v) Assume that $\ker S_\mathrm{N}=\{0\}$. Since $S_\mathrm{N}\geqslant\mathbb{O}$, then $\ker S_\mathrm{N}=\ker S_\mathrm{N}^{\frac{1}{2}}$. Thus, if for some $g$ one has $S_\mathrm{N}^{\frac{1}{2}}g=0$, then necessarily $g=0$. On account or \eqref{eq:nSbound} and \eqref{eq:II_SNgeneralform2}, and owing to the positivity of $S$, this is re-phrased by saying that if $\langle g,Sf\rangle=0$ $\forall f\in\mathcal{D}(S)$, then $g=0$. Therefore, $\mathrm{ran}\,S$ is dense in $\cH$.
  Conversely, if $\overline{\mathrm{ran}\,S}=\cH$, then also $\overline{\mathrm{ran}\,S_\mathrm{N}}=\cH$, whence $\ker S_\mathrm{N}=\{0\}$ (because $\ker S_\mathrm{N}=(\mathrm{ran}\,S_\mathrm{N})^\perp$, Sect.~\ref{sec:I-adjoint}).
  
  Let now the condition $\overline{\mathrm{ran}\,S}=\cH$ be satisfied and let $\widetilde{S}$ be an arbitrary positive self-adjoint extension of $S$. Reasoning as above, also $\overline{\mathrm{ran}\,\widetilde{S}}=\cH$, whence $\ker S=\ker \widetilde{S}=\{0\}$. Both $S$ and $\widetilde{S}$ are thus invertible on their ranges, and $\widetilde{S}^{-1}$ is self-adjoint and positive (Sect.~\ref{sec:I_funct_calc}). Beside, $\widetilde{S}^{-1}$ extends $S^{-1}$. Therefore, $(S^{-1})_{\mathrm{F}}\geqslant\widetilde{S}^{-1}\geqslant\mathbb{O}$ (Theorem \ref{thm:Friedrichs-ext}(vi)). On account of the inversion property presented in the end of Sect. \ref{sec:I_forms}, $\widetilde{S}\geqslant((S^{-1})_{\mathrm{F}})^{-1}$, as an inequality between two positive self-adjoint operators. It is not difficult to realise that $((S^{-1})_{\mathrm{F}})^{-1}$ is an actual extension of $S$. The latter inequality then says that $((S^{-1})_{\mathrm{F}})^{-1}$ is the \emph{smallest} positive self-adjoint extension of $S$: then necessarily $((S^{-1})_{\mathrm{F}})^{-1}=S_\mathrm{N}$ (owing to part (i)). Whence finally $(S^{-1})_{\mathrm{F}}=S_\mathrm{N}^{-1}$.

  (vi) Assume that $\ker S_\mathrm{F}=\{0\}$. Then both $S$ and $S_\mathrm{F}$ are invertible on their ranges. Moreover, $S_\mathrm{F}^{-1}$ is self-adjoint and positive (Sect.~\ref{sec:I_funct_calc}), and extends $S^{-1}$. Conversely, assume that $S^{-1}$ admits a positive self-adjoint extension $\widetilde{T}$. Since $\mathrm{ran}\,\widetilde{T}\supset \mathrm{ran}\,S^{-1}=\mathcal{D}(S)$, the range of $\widetilde{T}$ is dense, whence $\ker\widetilde{T}=(\mathrm{ran}\,\widetilde{T})^\perp=\{0\}$. The operator $\widetilde{T}^{-1}$ is therefore well-defined, self-adjoint and positive (Sect.~\ref{sec:I_funct_calc}), and it extends $S$. As such, $S_\mathrm{F}\geqslant \widetilde{T}^{-1}$ (Theorem \ref{thm:Friedrichs-ext}(vi)). On account of the inversion property presented in the end of Section \ref{sec:I_forms}, $\ker S_\mathrm{F}=\{0\}$ and $\widetilde{T}\geqslant S_\mathrm{F}^{-1}$. The latter inequality then says that $S_\mathrm{F}^{-1}$  is the \emph{smallest} positive self-adjoint extension of $S^{-1}$ (owing to part (i)). Whence finally $S_\mathrm{F}^{-1}=(S^{-1})_{\mathrm{N}}$.  
 \end{proof}

  One can finally get back to the proof of Theorem \ref{thm:II-KreinThm-Kext}.

 \begin{proof}[Proof of Theorem \ref{thm:II-KreinThm-Kext}]
  The operators
  \[
   K_\pm\;:=\;\|K\|_{\mathrm{op}}\mathbbm{1}\pm K\qquad (\mathcal{D}(K_\pm)\,=\,\mathcal{D}(K))
  \]
  are bounded and positive symmetric: boundedness and symmetry are obvious, and from $|\langle\varphi,K\varphi\rangle|\leqslant\|K\|_{\mathrm{op}}\|\varphi\|^2$ $\forall\varphi\in\mathcal{D}(K)$ one deduces also
  \[
   \langle\varphi,K_\pm\varphi\rangle\;=\;\|K\|_{\mathrm{op}}\|\varphi\|^2\pm\langle\varphi,K\varphi\rangle\;\geqslant\; 0\,.
  \]

  For $\varphi\in\mathcal{D}(K)$ one has 
  \[
   \begin{split}
   \|K_\pm\varphi\|^2\;&=\;\|K\|_{\mathrm{op}}\|\varphi\|^2+\|K\varphi\|^2\pm 2\|K\|_{\mathrm{op}}\langle\varphi, K\varphi\rangle \\
   &\leqslant\;2\|K\|_{\mathrm{op}}\langle\varphi,(\|K\|_{\mathrm{op}}\mathbbm{1}\pm K)\varphi\rangle\;=\;2\|K\|_{\mathrm{op}}\langle\varphi,K_\pm\varphi\rangle\,.
   \end{split}
  \]
  Corollary \ref{cor:II_Ando-Nishio-cor}(ii) then implies that each operator $K_\pm$ admits a bounded and positive self-adjoint extension $\widetilde{K}_\pm$ satisfying $\|\widetilde{K}_\pm\|_{\mathrm{op}}\leqslant 2\|K\|_{\mathrm{op}}$. As a consequence (Theorem \ref{thm:Ando-Nishio}(iii)), there is a smallest positive self-adjoint extension $K_{\pm,N}$ of $K_\pm$. Since $\mathbb{O}\leqslant K_{\pm,N}\leqslant\widetilde{K}_\pm$ and $\|\widetilde{K}_\pm\|_{\mathrm{op}}\leqslant 2\|K\|_{\mathrm{op}}$, then $K_{\pm,N}\leqslant 2\|K\|_{\mathrm{op}}\mathbbm{1}$ and each $K_{\pm,N}$ is bounded.

  Set 
  \[
   \begin{split}
    K_m\;&:=\;K_{+,N}-\|K\|_{\mathrm{op}}\mathbbm{1}\,, \\
    K_M\;&:=\;\|K\|_{\mathrm{op}}\mathbbm{1}-K_{-,N}\,.
   \end{split}
  \]
  By construction $K_m$ and $K_M$ are bounded and self-adjoint, and in fact they both extend $K$ because
  \[
   \begin{split}
    K_m\;\supset\;K_+-\|K\|_{\mathrm{op}}\mathbbm{1}\;=\;K\,, \\
    K_M\;\supset\;\|K\|_{\mathrm{op}}\mathbbm{1}-K_-\;=\;K\,.
   \end{split}
  \]

  Since $K_{\pm,N}\geqslant 0$, one has $K_m\geqslant-\|K\|_{\mathrm{op}}\mathbbm{1}$ and $K_M\leqslant\|K\|_{\mathrm{op}}\mathbbm{1}$; moreover, since $K_{\pm,N}\leqslant 2\|K\|_{\mathrm{op}}\mathbbm{1}$, one also has $K_m\leqslant \|K\|_{\mathrm{op}}\mathbbm{1}$ and $K_M\geqslant-\|K\|_{\mathrm{op}}\mathbbm{1}$. Summarising,
  \[
   \begin{split}
    -\|K\|_{\mathrm{op}}\mathbbm{1}\;&\leqslant\;K_m\;\leqslant\;\|K\|_{\mathrm{op}}\mathbbm{1}\,, \\
        -\|K\|_{\mathrm{op}}\mathbbm{1}\;&\leqslant\;K_M\;\leqslant\;\|K\|_{\mathrm{op}}\mathbbm{1}\,,
   \end{split}
  \]
  which proves that $\|K_m\|_{\mathrm{op}}\leqslant\|K\|_{\mathrm{op}}$ and $\|K_M\|_{\mathrm{op}}\leqslant\|K\|_{\mathrm{op}}$. Combining this with the trivial bounds $\|K\|_{\mathrm{op}}\leqslant\|K_m\|_{\mathrm{op}}$ and $\|K\|_{\mathrm{op}}\leqslant\|K_M\|_{\mathrm{op}}$ yields $\|K_m\|_{\mathrm{op}}=\|K_M\|_{\mathrm{op}}=\|K\|_{\mathrm{op}}$. This shows that $\mathscr{B}(K)$ is non-empty and contains both extensions $K_m$ and $K_M$.

  Let $\widetilde{K}\in \mathscr{B}(K)$, namely a bounded self-adjoint extension of $K$ with $\|\widetilde{K}\|_{\mathrm{op}}=\|K\|_{\mathrm{op}}$. Then the operators $\|K\|_{\mathrm{op}}\mathbbm{1}\pm \widetilde{K}$ are bounded self-adjoint extensions, respectively, of $K_\pm=\|K\|_{\mathrm{op}}\mathbbm{1}\pm K$ and in fact they are also positive, as one sees reasoning as for the positivity of $K_\pm$ above. As argued already, $K_{\pm,N}$ is the smallest positive self-adjoint extension of $K_\pm$, whence $K_{\pm,N}\leqslant\|K\|_{\mathrm{op}}\mathbbm{1}\pm \widetilde{K}$. This yields
  \[
   K_m\;=\;K_{+,N}-\|K\|_{\mathrm{op}}\mathbbm{1}\;\leqslant\;\widetilde{K}\;\leqslant\;\|K\|_{\mathrm{op}}\mathbbm{1}-K_{-,N}\;=\;K_M\,.
  \]

  It remains to prove the converse to the last property, that is: if $\widetilde{K}$ is a self-adjoint operator with $K_m\leqslant\widetilde{K}\leqslant K_M$, then $\widetilde{K}\in \mathscr{B}(K)$. The assumed inequality and the fact that both $K_M$ and $K_m$ extend $K$ imply that $\mathcal{D}(K)\subset\mathcal{D}(K_M)\subset\mathcal{D}(\widetilde{K})\subset\mathcal{D}(K_m)$ and that for $\varphi\in\mathcal{D}(K)$ one has $K\varphi=K_M\varphi\geqslant\widetilde{K}\varphi\geqslant K_M\varphi=K\varphi$, whence $K\varphi=\widetilde{K}\varphi$. This shows that $\widetilde{K}$ is a self-adjoint extension of $K$. The same assumed inequality implies $\|K_m\|_{\mathrm{op}}\leqslant\|\widetilde{K}\|_{\mathrm{op}}\leqslant\|K_M\|_{\mathrm{op}}$, and since $\|K_m\|_{\mathrm{op}}=\|K_M\|_{\mathrm{op}}=\|K\|_{\mathrm{op}}$ one concludes $\|\widetilde{K}\|_{\mathrm{op}}=\|K\|_{\mathrm{op}}$. Thus, $\widetilde{K}\in \mathscr{B}(K)$.  
 \end{proof}

  To complete the discussion on the core results of Kre{\u\i}n's extension theory \cite{Krein-1945,Krein-1947}, one further property is worth being included, apparently simple but deep, and brilliantly proved in \cite[Theorem 15 and Lemma 8]{Krein-1947}, which is fundamental in the subsequent development of the theory by Vi\v{s}ik and Birman (it will be used in Lemma \ref{lem:domB-1indomSBform}(ii) below).

  \begin{theorem}\label{thm:II-S[f,u]-Krein-2}\index{theorem!Kre{\u\i}n (self-adjoint extension)}
   Let $S$ be a densely defined and lower semi-bounded symmetric operator on a Hilbert space $\cH$ with greatest lower bound $\mathfrak{m}(S)>0$, let $S_\mathrm{F}$ be its Friedrichs extension, and let $\widetilde{S}$ be a lower semi-bounded self-adjoint extension of $S$. Then:
   \begin{eqnarray}
    \mathcal{D}[\widetilde{S}]\;&=&\;\mathcal{D}[S_\mathrm{F}]\dotplus \mathcal{D}[\widetilde{S}]\cap\ker S^*\,, \label{eq:II-S[f,u]-Krein} \\
    \widetilde{S}[f,u]\;&=&\;0\qquad\forall f\in\mathcal{D}[S_\mathrm{F}]\,,\;\forall u\in \mathcal{D}[\widetilde{S}]\cap\ker S^*\,,\label{eq:II-S[f,u]-Krein-2} \\
    \widetilde{S}[f+u,f'+u']\;&=& S_\mathrm{F}[f,f']+\widetilde{S}[u,u'] \label{eq:II-S[f,u]-Krein-3} \\
    & & \forall f,f'\in\mathcal{D}[S_\mathrm{F}]\,,\; \forall u,u'\in \mathcal{D}[\widetilde{S}]\cap\ker S^*\,. \nonumber
   \end{eqnarray}
  \end{theorem}

  \begin{proof}
   Concerning \eqref{eq:II-S[f,u]-Krein}, one distinguishes two cases, depending on the sign of $\mathfrak{m}(\widetilde{S})$. If $\mathfrak{m}(\widetilde{S})\geqslant 0$, then $S_\mathrm{N}\leqslant\widetilde{S}\leqslant S_\mathrm{F}$ (Theorem \ref{thm:II-Krein-final-thm}), whence also $\mathcal{D}[S_\mathrm{F}]\subset\mathcal{D}[\widetilde{S}]\subset\mathcal{D}[S_\mathrm{N}]$ (Sect.~\ref{sec:I_forms}). On the other hand, $\mathcal{D}[S_\mathrm{N}]=\mathcal{D}[S_\mathrm{F}]\dotplus\ker S^*$ (Theorem \ref{thm:II-KvNextension}(iv)), whence
   \[\tag{*}\label{eq:II-inclusionDS}
    \mathcal{D}[\widetilde{S}]\;\subset\;\mathcal{D}[S_\mathrm{F}]\dotplus\ker S^*\,.
   \]
  The inclusion \eqref{eq:II-inclusionDS} implies that any $g\in \mathcal{D}[\widetilde{S}]$ can be represented as $g=f+u$ for some $f\in\mathcal{D}[S_\mathrm{F}]\subset\mathcal{D}[\widetilde{S}]$ and $u\in\ker S^*$, therefore $u=g-f\in \mathcal{D}[\widetilde{S}]$, meaning that $u\in \mathcal{D}[\widetilde{S}]\cap\ker S^*$. Thus, \eqref{eq:II-inclusionDS} implies $\mathcal{D}[\widetilde{S}]\subset\mathcal{D}[S_\mathrm{F}]\dotplus \mathcal{D}[\widetilde{S}]\cap\ker S^*$, whence also \eqref{eq:II-S[f,u]-Krein}, since $\mathcal{D}[S_\mathrm{F}]\subset\mathcal{D}[\widetilde{S}]$ and obviously $ \mathcal{D}[\widetilde{S}]\cap\ker S^*\subset \mathcal{D}[\widetilde{S}]$. When instead $\mathfrak{m}(\widetilde{S})< 0$, let $\lambda>-\mathfrak{m}(\widetilde{S})$ and consider the densely defined symmetric operator $T:=S+\lambda\mathbbm{1}$, that has strictly positive bottom, and its positive self-adjoint extension $\widetilde{T}:=\widetilde{S}+\lambda\mathbbm{1}$ (indeed, $\mathfrak{m}(T)\geqslant\mathfrak{m}(\widetilde{T})=\mathfrak{m}(\widetilde{S})+\lambda>0$). Then, reasoning as above,
   \[
    \mathcal{D}[\widetilde{S}]\;=\;\mathcal{D}[\widetilde{T}]\;\subset\;\mathcal{D}[T]\dotplus\ker T^*\;=\;\mathcal{D}[S_\mathrm{F}]\dotplus\ker(S^*+\lambda\mathbbm{1})\,.
   \]
  On the other hand,
   \[
    \ker(S^*+\lambda\mathbbm{1})\;\subset\;\mathcal{D}(S^*)\;=\;\mathcal{D}(S_\mathrm{F})\dotplus\ker S^*\;\subset\;\mathcal{D}[S_\mathrm{F}]\dotplus\ker S^*\,;
   \]
  the above direct sum decomposition of $\mathcal{D}(S^*)$ follows from \eqref{eq:II-vNlike2} of Proposition \ref{prop:II-vNlike-decomp} (taking $\widetilde{S}=S_\mathrm{F}$ and $z=0$ therein), or also from \eqref{eq:II-KreinDecomp} below. The two latter chains of inclusions yield \eqref{eq:II-inclusionDS} again, whence \eqref{eq:II-S[f,u]-Krein} also in this second case.

  Concerning \eqref{eq:II-S[f,u]-Krein-2}, let $f\in\mathcal{D}[S_\mathrm{F}]$ and $u\in \mathcal{D}[\widetilde{S}]\cap\ker S^*$. Since $\mathcal{D}(S)$ is dense in $\mathcal{D}[S_\mathrm{F}]$ in the form norm of $S$ (Theorem \ref{thm:Friedrichs-ext}(i)), there is a sequence $(f_n)_{n\in\mathbb{N}}$ in $\mathcal{D}(S)$ such that $S_\mathrm{F}[f_n-f]+\|f_n-f\|^2\to 0$ (Theorem \ref{thm:Friedrichs-ext}(i)). This implies separately  $S_\mathrm{F}[f_n-f]\to 0$ and  $\|f_n-f\|\to 0$, owing to the positivity of $S_\mathrm{F}$. Fixed $\lambda\geqslant-\mathfrak{m}(\widetilde{S})$, consider the operators $T:=S+\lambda\mathbbm{1}$ and $\widetilde{T}:=\widetilde{S}+\lambda\mathbbm{1}$: since $\widetilde{T}$ is self-adjoint and positive,
  \[
   \widetilde{S}[g,g']+\lambda\langle g,g'\rangle\;=\;\widetilde{T}[g,g']\;=\;\big\langle \widetilde{T}^{\frac{1}{2}}g,\widetilde{T}^{\frac{1}{2}}g'\big\rangle\qquad\forall g,g'\in\mathcal{D}[\widetilde{S}]
  \]
  (Sect.~\ref{sec:I_forms}), which is true in particular for the elements of $\mathcal{D}[S_\mathrm{F}]$, because $S_\mathrm{F}\geqslant\widetilde{S}$ (Theorem \ref{thm:Friedrichs-ext}(vi)) and hence $\mathcal{D}[S_\mathrm{F}]\subset\mathcal{D}[\widetilde{S}]$ (Sect.~\ref{sec:I_forms}). Therefore, 
  \[
   S_\mathrm{F}[f_n-f]+\lambda\|f_n-f\|^2\;\geqslant\;\widetilde{S}[f_n-f]+\lambda\|f_n-f\|^2\;=\;\widetilde{T}[f_n-f]\;=\;\big\|\widetilde{T}^{\frac{1}{2}}(f_n-f)\big\|^2
  \]
  (having used again $S_\mathrm{F}\geqslant\widetilde{S}$), whereby
  \[
   0\;=\;\lim_{n\to\infty}\big(S_\mathrm{F}[f_n-f]+\lambda\|f_n-f\|^2\big)\;\geqslant\;\lim_{n\to\infty}\big\|\widetilde{T}^{\frac{1}{2}}(f_n-f)\big\|^2\,.
  \]
  Thus, $\widetilde{T}^{\frac{1}{2}}f_n\to\widetilde{T}^{\frac{1}{2}}f$. As a consequence,
  \[
    \widetilde{S}[f,u]+\lambda \langle f,u\rangle\;=\;\widetilde{T}[f,u]\;=\;\big\langle \widetilde{T}^{\frac{1}{2}}f,\widetilde{T}^{\frac{1}{2}}u\big\rangle\;=\;\lim_{n\to\infty}\big\langle \widetilde{T}^{\frac{1}{2}}f_n,\widetilde{T}^{\frac{1}{2}}u\big\rangle\,.
  \]
  By construction, $\widetilde{T}$ is an extension of $T$, therefore $f_n\in\mathcal{D}(S)=\mathcal{D}(T)\subset\mathcal{D}(\widetilde{T})$; thus, 
  \[
   \begin{split}
    \widetilde{S}[f,u]+\lambda \langle f,u\rangle\;&=\;\lim_{n\to\infty}\big\langle \widetilde{T}^{\frac{1}{2}}f_n,\widetilde{T}^{\frac{1}{2}}u\big\rangle\;=\;\lim_{n\to\infty}\langle \widetilde{T}f_n,u\rangle \\
    &=\;\lim_{n\to\infty}\big(\langle Sf_n,u\rangle +\lambda \langle f_n,u\rangle\big)\;=\;\lambda \langle f,u\rangle\,.
   \end{split}
  \]
  This finally implies \eqref{eq:II-S[f,u]-Krein-2}.

  Clearly, \eqref{eq:II-S[f,u]-Krein-3} follows from \eqref{eq:II-S[f,u]-Krein-2}.
  \end{proof}

 \section{Vi\v{s}ik-Birman parametrisation of self-adjoint extensions}\label{sec:II-VBparametrisation-orig}

 Kre{\u\i}n's extension theory provides the structure, in particular, the ordering between $S_\mathrm{F}$ and $S_\mathrm{N}$, of the positive self-adjoint extensions of a densely defined and positive symmetric operator $S$. Further results of Kre{\u\i}n's 1946 work \cite{Krein-1947}, concerning the non-positive extensions and their negative spectrum, were established within the assumption that $S$ has finite deficiency index: they are included in the more general discussion of Section \ref{sec:II-spectralKVB}.

 Soon after, that perspective was generalised and completed by Vi\v{s}ik, first in announcement-only form \cite{Vishik-1949} in 1949, then with full-fledged analysis \cite{Vishik-1952} in 1952, and by Birman, in announcement-only form \cite{Birman-1953} in 1953 and with thorough analysis \cite{Birman-1956} communicated in 1954 and published in 1956. (We refer to \cite{KM-2015-Birman} for the translation and adaptation of \cite{Birman-1953} in English; a previous translation was made available by Albeverio in the 1970's -- see, e.g., \cite{Skau-1979}). The seminal works \cite{Vishik-1952} and \cite{Birman-1956}, respectively by Vi\v{s}ik and Birman extended Kre{\u\i}n's theory to general semi-bounded (say, positive) and densely defined symmetric operators, irrespectively of the finiteness or infinity of their deficiency indices, and to all self-adjoint extensions (that is, not just the positive ones), providing also complete parametrisations and characterisations of the whole family of self-adjoint extensions, alternative to von Neumann's Theorem \ref{thm:vonN_thm_selfadj_exts}.

 Vi\v{s}ik, who was motivated by the study of elliptic boundary value problems on domain, investigated more generally closed extensions of a closed operator; Birman focussed for concreteness on the symmetric/self-adjoint case with methods that can be naturally adapted to the scenario of closed extensions -- a perspective later carried on by Grubb \cite{Grubb-1968} (see \cite[Chapter 13]{Grubb-DistributionsAndOperators-2009}). In retrospect, it is fair to consider the Vi\v{s}ik-Birman contribution as an \emph{autonomous} corpus of fundamental results that complement Kre{\u\i}n's ones to a more general Kre{\u\i}n-Vi\v{s}ik-Birman extension theory \cite{GMO-KVB2017}.\index{Kre{\u\i}n-Vi\v{s}ik-Birman theory}

 In this Section, following Birman's original work \cite{Birman-1956} and the recent analysis of it made in \cite{KM-2015-Birman,GMO-KVB2017}, the Vi\v{s}ik-Birman parametrisation will be discussed for the family of self-adjoint extensions of a densely defined symmetric operator $S$ on the Hilbert space $\cH$, which be lower semi-bounded with positive with greatest lower bound $\mathfrak{m}(S)>0$. In this case $S$ is closable and its deficiency index is $d(S)=\mathrm{dim}\ker S^*$ (Sect.~\ref{sec:I-symmetric-selfadj}). Moreover, $\mathrm{ran}\,\overline{S}=\overline{\mathrm{ran}\,S}$ (Sect.~\ref{sec:I-symmetric-selfadj}), and the Friedrichs extension $S_\mathrm{F}$ has everywhere defined and bounded inverse on $\cH$ (Theorem \ref{thm:Friedrichs-ext}(iii)).

 Any self-adjoint extension of $S$ is a self-adjoint restriction of $S^*$, the domain of which one has the following convenient representation, completely analogous to von Neumann's formula from Proposition \ref{prop:II-vNformula}.

 \begin{proposition}[Kre{\u\i}n and Vi\v{s}ik-Birman decomposition formulas]\label{prop:II-KVB-decomp-of-Sstar}
  Let $S$ be a densely defined and positive symmetric operator with bottom $\mathfrak{m}(S)>0$ on the Hilbert space $\cH$, and let $S_\mathrm{F}$ be the Friedrichs extension of $S$. Then \index{Kre{\u\i}n decomposition formula}\index{Vi\v{s}ik-Birman decomposition formula}
  \begin{eqnarray}
   \mathcal{D}(S^*)&=&\mathcal{D}(S_\mathrm{F})\dotplus\ker S^*\,, \label{eq:II-KreinDecomp} \\
   \mathcal{D}(S^*)&=&\mathcal{D}(\overline{S})\dotplus S_\mathrm{F}^{-1}\ker S^*\dotplus \ker S^* \,, \label{eq:II-VishBirDecomp} \\
   \mathcal{D}(S_\mathrm{F})&=&\mathcal{D}(\overline{S})\dotplus S_\mathrm{F}^{-1}\ker S^*\,. \label{eq:II-SF_VB_decomp}
  \end{eqnarray}
 \end{proposition}

 Formula \eqref{eq:II-KreinDecomp} is due to Kre{\u\i}n \cite[Lemma 7]{Krein-1947}; formula \eqref{eq:II-VishBirDecomp} is due to Vi\v{s}ik \cite[Eq.~(1.41)]{Vishik-1952} and was re-proved by Birman \cite[Theorem 1]{Birman-1956}.

 \begin{proof}[Proof of Proposition \ref{prop:II-KVB-decomp-of-Sstar}] The whole thesis follows at once from Proposition \ref{prop:II-vNlike-decomp} by taking $\widetilde{S}=S_\mathrm{F}$ and $z=0$ therein. Owing to the independent origin of formulas \eqref{eq:II-KreinDecomp}-\eqref{eq:II-SF_VB_decomp}, their proof will be presented also here.  
 Concerning \eqref{eq:II-KreinDecomp},   $\mathcal{D}(S^*)\supset\mathcal{D}(S_\mathrm{F})+\ker S^*$ because each summand is a subspace of $\mathcal{D}(S^*)$. For the opposite inclusion one can always decompose an arbitrary $g\in\mathcal{D}(S^*)$ as $g=S_\mathrm{F}^{-1}S^*g+(g-S_\mathrm{F}^{-1}S^*g)$, where $S_\mathrm{F}^{-1}S^*g\in\mathcal{D}(S_\mathrm{F})$ and $g-S_\mathrm{F}^{-1}S^*g\in\ker S^*$. Furthermore, the sum on the r.h.s.~of \eqref{eq:II-KreinDecomp} is direct because any $g\in\mathcal{D}(S_\mathrm{F})\cap \ker S^*$ is necessarily in $\ker S_\mathrm{F}$ ($S_\mathrm{F} g=S^*g=0$), and from the injectivity of $S_\mathrm{F}$ one has $g=0$. Concerning \eqref{eq:II-VishBirDecomp}, again it is obvious that $\mathcal{D}(S^*)\supset\mathcal{D}(\overline{S})\dotplus S_\mathrm{F}^{-1}\ker S^*\dotplus \ker S^*$. For the converse, any $g\in\mathcal{D}(S^*)$ can be written as  $g=S_\mathrm{F}^{-1}S^*g+u$ for some $u\in\ker S^*$. In turn, in view of $\cH=\overline{\ran\,S}\oplus\ker S^*=\ran\,\overline{S}\oplus\ker S^*$ one writes $S^*g=\overline{S}f+\widetilde{u}$ for some $f\in\mathcal{D}(\overline{S})$ and $\widetilde{u}\in\ker S^*$, whence $S_\mathrm{F}^{-1}S^*g=S_\mathrm{F}^{-1} \overline{S}f+S_\mathrm{F}^{-1}\widetilde{u}=f+S_\mathrm{F}^{-1}\widetilde{u}$, thus also $g=f+S_\mathrm{F}^{-1}\widetilde{u}+u$. To show that the sum in \eqref{eq:II-VishBirDecomp} is direct one argues as follows: if $g=f+S_\mathrm{F}^{-1}\widetilde{u}+u=0$, then $0=S^*g=\overline{S}f+\widetilde{u}$, which forces $\overline{S}f=\widetilde{u}=0$ because $\overline{S}f\perp\widetilde{u}$; then also $f=S_\mathrm{F}^{-1}\overline{S}f=0$ and, from $g=0$, also $u=0$. Last, \eqref{eq:II-SF_VB_decomp} is obtained through the very same reasoning as for \eqref{eq:II-VishBirDecomp}: it is clear that $\mathcal{D}(S_\mathrm{F})\supset\mathcal{D}(\overline{S})\dotplus S_\mathrm{F}^{-1}\ker S^*$, and conversely for $g\in\mathcal{D}(S_\mathrm{F})$ one writes $S_\mathrm{F}g=\overline{S}f+\widetilde{u}\in\ran\,S\oplus\ker S^*$, whence $g=S_\mathrm{F}^{-1}S_\mathrm{F}g=S_\mathrm{F}^{-1}\overline{S}f+S_\mathrm{F}^{-1}\widetilde{u}=f+S_\mathrm{F}^{-1}\widetilde{u}$.  
 \end{proof}

 \begin{remark}
  Analogously to (the proof of) \eqref{eq:II-SF_VB_decomp},
  \begin{equation}\label{eq:II-DomStilde}
\mathcal{D}(\widetilde{S})\;=\;\mathcal{D}(\overline{S})\dotplus \widetilde{S}^{-1} \ker S^*
\end{equation}
for any self-adjoint extension $\widetilde{S}$ of $S$ that is invertible everywhere on $\cH$. In fact, \eqref{eq:II-DomStilde} is nothing but \eqref{eq:II-vNlike3} from Proposition \ref{prop:II-vNlike-decomp} with $z=0$.
 \end{remark}

 \subsection{Vi\v{s}ik's $B$ operator}\index{Vi\v{s}ik extension parameter}\index{extension parameter!Vi\v{s}ik's}

 Let $\widetilde{S}$ be a self-adjoint extension of $S$ with respect to the underlying Hilbert space $\cH$ ($S$ is densely defined, positive symmetric, with $\mathfrak{m}(S)>0$). Correspondingly, let $U_0$ and $U_1$ be the two closed subspaces of $\ker S^*$ (and hence of $\cH$), and let $\cH_+$ be the closed subspace of $\cH$, uniquely associated to $\widetilde{S}$ through the definitions 
\begin{equation}\label{eq:U0-U1-H+}
U_0\;:=\;\ker\widetilde{S}\,,\qquad \ker S^*\;=\;U_0\oplus U_1\,,\qquad \cH_+\;:=\;\ran\,\overline{S}\oplus U_1\,.
\end{equation}
Thus,
\begin{equation}\label{eq:decompH}
\cH\;=\;\overline{\ran\,S}\oplus\ker S^*\;=\;\ran\,\overline{S}\oplus U_1\oplus U_0\;=\;\cH_+\oplus \ker\widetilde{S}\,.
\end{equation}
Let $P_+:\cH\to\cH$ be the orthogonal projection onto $\cH_+$. In terms of such objects, $\widetilde{S}$ satisfies the following properties.

\begin{lemma}\label{lemma:decompStilde}~
\begin{enumerate}[(i)]
 \item $\ran\,\overline{S}\oplus U_1=\cH_+=\overline{\ran\,\widetilde{S}}$, i.e., $\ran\,\widetilde{S}$ is dense in $\cH_+$.
\item $\ker\widetilde{S}=(\mathbbm{1}-P_+)\mathcal{D}(\widetilde{S})$.
 \item  $\mathcal{D}(\widetilde{S})=(\mathcal{D}(\widetilde{S})\cap \cH_+)\boxplus\ker\widetilde{S}=P_+\mathcal{D}(\widetilde{S})\boxplus\ker\widetilde{S}$ and also $\mathcal{D}(\widetilde{S})\cap\cH_+=P_+\mathcal{D}(\widetilde{S})$.
 \item $\mathcal{D}(\widetilde{S})\cap\cH_+$ is dense in $\cH_+$.
 \item $\widetilde{S}$ maps $\mathcal{D}(\widetilde{S})\cap\cH_+$ into $\cH_+$.
 \item $\ran\,\widetilde{S}=\ran\,\overline{S}\,\boxplus\,\widetilde{U}_1$ where $\widetilde{U}_1$ is a dense subspace of $U_1$ uniquely identified by $\widetilde{S}$.
\end{enumerate}
\end{lemma}

\begin{proof}
(i) follows by \eqref{eq:decompH}, because $\overline{\ran\,\widetilde{S}}$ is the orthogonal complement to $\ker\widetilde{S}$  in $\cH$ (owing to the self-adjointness of $\widetilde{S}$). In (ii) the ``$\supset$'' inclusion is obvious and conversely, if $u_0\in \ker\widetilde{S}\subset\mathcal{D}(\widetilde{S})$, then $u_0=(\mathbbm{1}-P_+)u_0\in (\mathbbm{1}-P_+)\mathcal{D}(\widetilde{S})$.
To establish (iii), decompose a generic $g\in\mathcal{D}(\widetilde{S})$ as $g=f_+ +u_0$ for some $f_+\in\cH_+$ and $u_0\in U_0=\ker\widetilde{S}$ (using $\cH=\cH_+\oplus\ker\widetilde{S}$): since $f_+=g-u_0\in\mathcal{D}(\widetilde{S})$, then $f_+\in\mathcal{D}(\widetilde{S})\cap \cH_+$ and therefore $\mathcal{D}(\widetilde{S})\subset (\mathcal{D}(\widetilde{S})\cap \cH_+)\boxplus\ker\widetilde{S}$. The opposite inclusion is obvious; thus, $\mathcal{D}(\widetilde{S})=(\mathcal{D}(\widetilde{S})\cap \cH_+)\boxplus\ker\widetilde{S}$. It remains to prove that $\mathcal{D}(\widetilde{S})\cap\cH_+=P_+\mathcal{D}(\widetilde{S})$: the inclusion $\mathcal{D}(\widetilde{S})\cap\cH_+\subset P_+\mathcal{D}(\widetilde{S})$ is obvious; as for the converse, if $h=P_+ g\in P_+\mathcal{D}(\widetilde{S})$ for some $g\in\mathcal{D}(\widetilde{S})$, decompose $g=f_+ + u_0$ in view of $\mathcal{D}(\widetilde{S})=(\mathcal{D}(\widetilde{S})\cap \cH_+)\boxplus\ker\widetilde{S}$, then $h= P_+ g = f_+\in\mathcal{D}(\widetilde{S})\cap \cH_+$, which completes the proof.
%
To establish (iv), for fixed arbitrary $h_+\in\cH_+$ let $(f_n)_{n\in\mathbb{N}}$ be a sequence in $\mathcal{D}(\widetilde{S})$ of approximants of $h_+$ (indeed $\mathcal{D}(\widetilde{S})$ is dense in $\cH$): then, as $n\to\infty$, $f_n\to h_+$ implies $P_+ f_n\to h_+$. (v) is an immediate consequence of (i), because $\widetilde{S}$ maps $\mathcal{D}(\widetilde{S})\cap\cH_+$ into $\ran \widetilde{S}$. Last, (vi) will be proved. Recall that $\overline{\ran S}=\ran\overline{S}$, because $S$ is closable and injective ($\mathfrak{m}(S)>0$). Set $\widetilde{U}_1:=\{u_g\in U_1\,|\,\widetilde{S}g=\overline{S}f_g+u_g\textrm{ for }g\in\mathcal{D}(\widetilde{S})\}$, where $f_g\in\mathcal{D}(\overline{S})$ and $u_g\in U_1$ are uniquely determined by the given $g\in\mathcal{D}(\widetilde{S})$ through the decomposition $\overline{\ran\,\widetilde{S}}=\cH_+=\ran\,\overline{S}\oplus U_1$. The inclusions $\ran\,\widetilde{S}\subset\ran\,\overline{S}\boxplus \widetilde{U}_1$ and $\ran\,\widetilde{S}\supset\ran\,\overline{S}$ are obvious; furthermore, $\ran\,\widetilde{S}\supset\widetilde{U}_1$ because each $\widetilde{u}_1\in\widetilde{U}_1$ is by definition the difference of two elements in $\ran\,\widetilde{S}$. Thus, $\ran\,\widetilde{S}=\ran\,\overline{S}\boxplus \widetilde{U}_1$. It remains to prove the density of $\widetilde{U}_1$ in $U_1$. Given an arbitrary $u_1\in U_1\subset\cH_+$ and a sequence $(\widetilde{S}g_n)_{n\in\mathbb{N}}\in\ran\,\widetilde{S}$ of approximants of $u_1$ (owing to the  density of $\ran\,\widetilde{S}$ in $\cH_+$), decompose $\widetilde{S}g_n=\overline{S}f_n+\widetilde{u}_n$, for some $f_n\in\mathcal{D}(\overline{S})$ and $\widetilde{u}_n\in \widetilde{U}_1$, in view of $\ran\,\widetilde{S}=\ran\,\overline{S}\boxplus \widetilde{U}_1$: denoting by $P_1:\cH_+\to\cH_+$ the orthogonal projection onto $U_1$, one has $u_1=P_1u_1=P_1\lim_n(\overline{S}f_n+\widetilde{u}_n)=\lim_n\widetilde{u}_n$, which shows that $(\widetilde{u}_n)_{n\in\mathbb{N}}$ is a sequence in $\widetilde{U}_1$ of approximants of $u_1$.
\end{proof}

 Since $\widetilde{S}$ maps $\mathcal{D}(\widetilde{S})\cap\cH_+$ into $\cH_+$ (Lemma \ref{lemma:decompStilde}(v)) 
and trivially $U_0$ into itself, and since $P_+$ maps $\mathcal{D}(\widetilde{S})$ into itself (Lemma \ref{lemma:decompStilde}(iii)), 
then $\cH_+$ and $U_0$ are reducing subspaces for $\widetilde{S}$ (Sect.~\ref{sec:I_invariant-reducing-ssp}). The non-trivial (i.e., non-zero) part of $\widetilde{S}$ in this reduction is the operator 
\begin{equation}
\widetilde{S}_+\;:=\;\widetilde{S}\upharpoonright\mathcal{D}(\widetilde{S}_+)\,,\qquad \mathcal{D}(\widetilde{S}_+)\;:=\;\mathcal{D}(\widetilde{S})\cap \overline{\ran\,\widetilde{S}}\;=\;P_+\mathcal{D}(\widetilde{S})\,,
\end{equation}
which is therefore a densely defined, injective, and self-adjoint operator in the Hilbert space $\cH_+$.
Explicitly,
\begin{equation}\label{eq:Stilde+}
 \begin{split}
\ran\,\widetilde{S}\;&=\;\{\widetilde{S}g\,|\,g\in\mathcal{D}(\widetilde{S})\}\;=\;\{\widetilde{S}P_+g\,|\,g\in\mathcal{D}(\widetilde{S})\}\,, \\
\widetilde{S}_+ P_+g\;&=\;\widetilde{S}P_+g\qquad\forall g\in\mathcal{D}(\widetilde{S})\,, \\
\ran \widetilde{S}_+\;&=\;\ran\,\widetilde{S}\,.
\end{split}
\end{equation}
The inverse of $\widetilde{S}_+$ (on $\cH_+$) is  the self-adjoint operator $\widetilde{S}_+^{-1}$ (on $\cH_+$)  with
\begin{equation}\label{eq:S+-1}
\begin{split}
\mathcal{D}(\widetilde{S}_+^{-1})\;& =\;\ran\,\widetilde{S}\,, \\
\widetilde{S}_+^{-1} (\widetilde{S} P_+g)\;&=\;P_+g\qquad\forall g\in\mathcal{D}(\widetilde{S})\,,
\end{split}
\end{equation}
and hence
\begin{eqnarray}
\ran(\widetilde{S}_+^{-1})\;=\;\widetilde{S}_+^{-1}\ran\,\widetilde{S}\; &=&\; P_+\mathcal{D}(\widetilde{S})\,, \label{eq:S+-1toranStilde} \\
\widetilde{S}_+^{-1}\ran\,\overline{S}\;&=&\;P_+\mathcal{D}(\overline{S})\,. \label{eq:S+-1toranS}
\end{eqnarray}
(\eqref{eq:S+-1toranS} follows from $\widetilde{S}_+^{-1} (\widetilde{S} P_+f)=P_+f$  in \eqref{eq:S+-1}, letting now $f$ run on $\mathcal{D}(\overline{S})$ only: the r.h.s.~gives $P_+\mathcal{D}(\overline{S})$, on the l.h.s.~one uses that $\widetilde{S}P_+f=\widetilde{S}f=\overline{S}f$ $\forall f\in\mathcal{D}(\overline{S})$ and hence $\{\widetilde{S}P_+f\,|\,f\in\mathcal{D}(\overline{S})\}=\ran\,\overline{S}$.)

Furthermore, by setting
\begin{equation}\label{eq:defStildetilde-1}
\begin{split}
\mathcal{D}(\mathscr{S}^{-1})\;&:=\;\mathcal{D}(\widetilde{S}_+^{-1})\,\boxplus\,U_0\;=\;\ran\,\widetilde{S}\,\boxplus\,\ker\widetilde{S}\,,  \\
\mathscr{S}^{-1}\widetilde S g\;&:=\;\widetilde{S}_+^{-1}\widetilde{S} g\;=\;g \qquad \forall g \in \mathcal{D}(\widetilde{S})\,,\\
\mathscr{S}^{-1}\upharpoonright\ker\widetilde{S}\;&:=\;\mathbb{O}\,,
\end{split}
\end{equation}
one defines a self-adjoint operator $\mathscr{S}^{-1}$ on the whole $\cH$, with reducing subspaces $\cH_+=\overline{\ran\,\widetilde{S}}$ and $U_0=\ker\widetilde{S}$.

Two further useful properties are the following.

\begin{lemma}\label{lemma:S-Stilde-Sbig}~
\begin{enumerate}[(i)]
 \item $\;\:$ $\mathcal{D}(\overline{S})+\ker\widetilde{S}\;=\;P_+\mathcal{D}(\overline{S})\,\boxplus\,\ker\widetilde{S}$\,.
 \item $\;\:$ $P_+\mathcal{D}(\widetilde{S})\;=\;P_+\mathcal{D}(\overline{S})+\mathscr{S}^{-1}\widetilde{U}_1$\,.
\end{enumerate}
\end{lemma}

\begin{proof}
The inclusion $\mathcal{D}(\overline{S})+U_0\subset P_+\mathcal{D}(\overline{S})\,\boxplus\,U_0$ in (i) follows from the fact that each summand on the l.h.s.~belongs to the sum on the r.h.s., in particular, $\mathcal{D}(\overline{S})\subset P_+\mathcal{D}(\overline{S})+(\mathbbm{1}-P_+) \mathcal{D}(\overline{S})$. Conversely, given a generic $h=P_+f\in P_+\mathcal{D}(\overline{S})$ for some $f\in\mathcal{D}(\overline{S})$ and $u_0\in U_0$, then $h+u_0=f+u_0'$ with $u_0':=u_0-(1-P_+)f\in U_0$, which proves the inclusion $\mathcal{D}(\overline{S})+U_0\supset P_+\mathcal{D}(\overline{S})\,\boxplus\,U_0$.
(ii) follows by applying $\mathscr{S}^{-1}$ to the decomposition $\ran\,\widetilde{S}=\ran\, \overline{S}\boxplus\widetilde{U}_1$ of Lemma \ref{lemma:decompStilde}(vi): indeed, on the l.h.s.~one gets $\mathscr{S}^{-1}\ran\,\widetilde{S}=P_+\mathcal{D}(\widetilde{S})$, owing to \eqref{eq:defStildetilde-1}, whereas on the r.h.s.~one gets $\mathscr{S}^{-1}\ran\,\overline{S}=P_+\mathcal{D}(\overline{S})$, owing to \eqref{eq:S+-1toranS}.
\end{proof}

 Summarising so far, the given operator $S$ and the given self-adjoint extension $\widetilde{S}$ determine  canonically (and, in fact, constructively) the closed subspace $U_1$ of $\ker S^*$, the dense subspace $\widetilde{U}_1$ in $U_1$, the closed subspace $\cH_+=\overline{\ran\,\widetilde{S}}=\ran\,\overline{S}\oplus U_1$ of $\cH$ (equivalently, the orthogonal projection $P_+$ onto $\cH_+$), and the self-adjoint operator $\mathscr{S}^{-1}$ on $\cH$, with the properties discussed above. In terms of these data, one defines
\begin{equation}\label{eq:defBbig}
\begin{split}
\mathscr{B}\;&:=\;\mathscr{S}^{-1}-P_+ S_\mathrm{F}^{-1} P_+ \,,\\
\mathcal{D}(\mathscr{B})\;&:=\;\mathcal{D}(\mathscr{S}^{-1})\;=\;\ran\,\widetilde{S}\,\boxplus\,\ker\widetilde{S}\,,
\end{split}
\end{equation}
a self-adjoint operator on $\cH$ with the following properties.

 \begin{lemma}\label{lemma:Bbig}~
\begin{enumerate}[(i)]
 \item $\mathscr{B}$ is self-adjoint on $\cH$ and it is bounded if and only if the inverse of the operator $\widetilde{S}\upharpoonright\big(\mathcal{D}(\widetilde{S})\cap \overline{\ran\,\widetilde{S}}\,\big)$ (i.e.,  $\widetilde{S}_+^{-1}$) is bounded as an operator on $\cH_+$.
 \item With respect to the decomposition $\cH=\ran\overline{S}\oplus U_1\oplus U_0$ from \eqref{eq:decompH}, one has $\mathcal{D}(\mathscr{B})=\ran\,\overline{S}\,\boxplus\,\widetilde{U}_1\,\boxplus\,U_0$,  $\mathscr{B}\,\ran\,\overline{S}=\{0\}$, $\mathscr{B}\,\widetilde{U}_1\subset U_1$, and $\mathscr{B}\,U_0=\{0\}$.
\end{enumerate}
\end{lemma}

\begin{proof}
(i) is obvious from the definition of $\mathscr{B}$ and of $\mathscr{S}^{-1}$: the former is bounded if and only if the latter is. As for (ii), $\mathcal{D}(\mathscr{B})=\ran\,\widetilde{S}\,\boxplus\,\ker\widetilde{S}=\ran\,\overline{S}\,\boxplus\,\widetilde{U}_1\,\boxplus\,U_0$ follows from \eqref{eq:defBbig} and  Lemma \ref{lemma:decompStilde}(vi). Moreover, $\mathscr{B}\,U_0=\{0\}$ is obvious from \eqref{eq:defStildetilde-1} and  \eqref{eq:defBbig}. To see that $\mathscr{B}\,\ran\,\overline{S}=\{0\}$ let $f\in\mathcal{D}(\overline{S})$, then
\[
 \mathscr{B} \overline{S}f\;=\;\mathscr{S}^{-1}\overline{S}f-P_+ S_\mathrm{F}^{-1}\overline{S}f\;=\;\widetilde{S}_+^{-1}\overline{S}f-P_+f\;=\;P_+f-P_+f\;=\;0\,,
\]
having used \eqref{eq:defBbig} and $\ran\,\overline{S}\subset\cH_+$ in the first equality,  \eqref{eq:defStildetilde-1} in the second equality, and \eqref{eq:S+-1toranS} in the third equality. 
%
%
In view of the decomposition $\cH=\ran\overline{S}\oplus U_1\oplus U_0$,  $\mathcal{D}(\mathscr{B})=\ran\,\overline{S}\,\boxplus\,\widetilde{U}_1\,\boxplus\,U_0$, the self-adjointness of $\mathscr{B}$ and the fact that  $\mathscr{B}\,\ran\,\overline{S}=\{0\}$ and $\mathscr{B}\,U_0=\{0\}$ imply necessarily  $\mathscr{B}\,\widetilde{U}_1\subset U_1$.
\end{proof}

 As a direct consequence of Lemma \ref{lemma:Bbig} above, the restriction of $\mathscr{B}$ to $\widetilde{U}_1$, i.e., the operator
\begin{equation}\label{eq:defB}\index{Vi\v{s}ik extension parameter}\index{extension parameter!Vi\v{s}ik's}
B\;:=\;\big(\mathscr{S}^{-1}-P_+ S_\mathrm{F}^{-1} P_+\big)\upharpoonright\mathcal{D}(B)\,,\qquad \mathcal{D}(B)\;:=\;\widetilde{U}_1\,.
\end{equation}
is a self-adjoint operator in the Hilbert space $U_1=\ker S^*\ominus\ker\widetilde{S}$ (with dense domain $\widetilde{U}_1$), which itself is canonically determined by $\widetilde{S}$. The interest on $B$ is motivated by the following fundamental property.

\begin{proposition}[$B$-decomposition formula]\label{prop:BirmanBformula}
\begin{equation}\label{eq:StildeB}
\mathcal{D}(\widetilde{S})\;=\;\mathcal{D}(\overline{S})\,\dotplus\,(S_\mathrm{F}^{-1}+B)\widetilde{U}_1\,\dotplus\,U_0\,.
\end{equation}
\end{proposition}

\begin{proof}
One has
\[
\begin{split}
\mathcal{D}(\widetilde{S})\;&=\;P_+\mathcal{D}(\widetilde{S})+U_0\qquad\qquad\qquad\qquad\quad\textrm{(Lemma \ref{lemma:decompStilde}(iii))} \\
& =\; P_+\mathcal{D}(\overline{S})+\mathscr{S}^{-1}\widetilde{U}_1+U_0 \qquad\qquad\;\;\;\,\textrm{(Lemma \ref{lemma:S-Stilde-Sbig}(ii))} \\
& =\;\mathcal{D}(\overline{S})+\mathscr{S}^{-1}\widetilde{U}_1+U_0 \qquad\qquad\qquad\textrm{(Lemma \ref{lemma:S-Stilde-Sbig}(i))} \\
& =\;\mathcal{D}(\overline{S})+(P_+S_\mathrm{F}^{-1}P_++B)\widetilde{U}_1+U_0 \qquad\textrm{(formula \eqref{eq:defB})}  \\
& =\;\mathcal{D}(\overline{S})+(P_+S_\mathrm{F}^{-1}+B)\widetilde{U}_1+U_0 \qquad\quad\textrm{($\widetilde{U}_1\subset\cH_+$)} \\
& =\;\mathcal{D}(\overline{S})+P_+(S_\mathrm{F}^{-1}+B)\widetilde{U}_1+U_0 \qquad\quad\textrm{($B\widetilde{U}_1\subset U_1\subset\cH_+$, Lemma \ref{lemma:Bbig}(ii)).}
\end{split}
\]
This identity, together with
\begin{equation}\tag{*} 
P_+(S_\mathrm{F}^{-1}+B)\widetilde{U}_1+U_0\;=\;(S_\mathrm{F}^{-1}+B)\widetilde{U}_1+U_0\,,
\end{equation}
yields $\mathcal{D}(\widetilde{S})=\mathcal{D}(\overline{S})+(S_\mathrm{F}^{-1}+B)\widetilde{U}_1+U_0$, and this sum is direct because if $\mathcal{D}(S^*)\supset\mathcal{D}(\widetilde S) \ni g := f + (S_\mathrm{F}^{-1}+B)\tilde u_1 +u_0=f+S_\mathrm{F}^{-1}\tilde u_1 + (B\tilde u_1 +u_0)$ then, according to \eqref{eq:II-VishBirDecomp}, $g=0$ implies $f=0$, $\tilde u_1=0$, and $u_0=0$. Thus, in order to prove \eqref{eq:StildeB} it only remains to prove (*). 
For the inclusion $P_+(S_\mathrm{F}^{-1}+B)\widetilde{U}_1+U_0\subset(S_\mathrm{F}^{-1}+B)\widetilde{U}_1+U_0$ pick $\psi:=P_+(S_\mathrm{F}^{-1}+B)\widetilde{u}_1+u_0$ for generic $\widetilde{u}_1\in\widetilde{U}_1$ and $u_0\in U_0$. From \eqref{eq:defB}, from the fact that $\widetilde{u}_1=P_+\widetilde{u}_1$, and from  $P_+\mathscr{S}^{-1}\widetilde{u}_1=\mathscr{S}^{-1}\widetilde{u}_1$ (which follows from \eqref{eq:defStildetilde-1}), one has 
\[
P_+(S_\mathrm{F}^{-1}+B)\widetilde{u}_1\;=\;P_+S_\mathrm{F}^{-1} P_+\widetilde{u}_1+P_+\mathscr{S}^{-1}\widetilde{u}_1-P_+S_\mathrm{F}^{-1} P_+\widetilde{u}_1\;=\;\mathscr{S}^{-1}\widetilde{u}_1
\]
as well as
\[
(S_\mathrm{F}^{-1}+B)\widetilde{u}_1\;=\;(S_\mathrm{F}^{-1}\widetilde{u}_1-P_+S_\mathrm{F}^{-1}\widetilde{u}_1)+\mathscr{S}^{-1}\widetilde{u}_1\;=\;u_0'+\mathscr{S}^{-1}\widetilde{u}_1\,,
\]
where $u_0':=S_\mathrm{F}^{-1}\widetilde{u}_1-P_+S_\mathrm{F}^{-1}\widetilde{u}_1\in\,\cH\ominus\cH_+=U_0$; therefore,
\[
\begin{split}
\psi\;&=\;P_+(S_\mathrm{F}^{-1}+B)\widetilde{u}_1+u_0\;=\;\mathscr{S}^{-1}\widetilde{u}_1+u_0 \\
&=\;u_0'+\mathscr{S}^{-1}\widetilde{u}_1+u_0-u_0'\;=\;(S_\mathrm{F}^{-1}+B)\widetilde{u}_1+(u_0-u_0')\,,
\end{split}
\]
which proves that $\psi\in(S_\mathrm{F}^{-1}+B)\widetilde{U}_1+U_0$. The opposite inclusion to establish (*),
that is, $P_+(S_\mathrm{F}^{-1}+B)\widetilde{U}_1+U_0\supset(S_\mathrm{F}^{-1}+B)\widetilde{U}_1+U_0$, is proved repeating the same argument in reverse order.
\end{proof}

 \subsection{Vi\v{s}ik-Birman extension representation}\label{sec:II-VBextparametrisation}

 \begin{theorem}[Vi\v{s}ik-Birman representation theorem]\label{thm:VB-representaton-theorem}\index{theorem!Vi\v{s}ik-Birman (self-adjoint extension)} Let $S$ be a densely defined and positive symmetric operator with greatest lower bound $\mathfrak{m}(S)>0$ on the Hilbert space $\cH$. 
  There is a one-to-one correspondence between the  family of the self-adjoint extensions of  $S$ in $\cH$ and the family of the self-adjoint operators in Hilbert subspaces of $\ker S^*$, that is, the collection of triples $(U_1,\widetilde{U}_1,B)$, where $U_1$ is a closed subspace of $\ker S^*$, $\widetilde{U}_1$ is a dense subspace of  $U_1$, and $B$ is a self-adjoint operator in the Hilbert space $U_1$ with domain $\mathcal{D}(B)=\widetilde{U}_1$.
  For each such triple, let $U_0$ be the closed subspace of $\ker S^*$ defined by $\ker S^*=U_0\oplus U_1$. Then, in this correspondence $S_B\leftrightarrow B$, each self-adjoint extension $S_B$ of $S$ is given by
\begin{equation}\label{eq:defSB}
\begin{split}
S_B\;&=\;S^*\upharpoonright\mathcal{D}(S_B)\,, \\
\mathcal{D}(S_B)\;&=\;\mathcal{D}(\overline{S})\,\dotplus\,(S_\mathrm{F}^{-1}+B)\widetilde{U}_1\,\dotplus\,U_0\,.
\end{split}
\end{equation}
  In particular,
   \begin{eqnarray}
    \ker S_B\:&=&\:U_0\;=\;\ker S^*\cap\mathcal{D}(B)^\perp\,, \label{eq:II-kerSB} \\
    \mathrm{ran}\:S_B\:&=&\,\mathrm{ran}\,\overline{S}\boxplus\widetilde{U}_1\;=\;\mathrm{ran}\,\overline{S}\boxplus\mathcal{D}(B)\,. \label{eq:II-ranSB}
   \end{eqnarray}
\end{theorem}

 \begin{proof}
  The fact that \emph{each} self-adjoint extension of $S$ is precisely of the form $S_B$ is the content of Proposition \ref{prop:BirmanBformula}, where $B$ is Vi\v{s}ik's $B$ operator associated with the considered self-adjoint extension. Conversely, one has to prove that each operator on $\cH$ of the form $S_B$ above is a self-adjoint extension of $S$, and that the correspondence $S_B\leftrightarrow B$ is one-to-one.
  
  Fixed $(U_1,\widetilde{U}_1,B)$ as in the statement, let $S_B$ be the corresponding operator. One sees from \eqref{eq:defSB} that $S_B$ is densely defined ($\mathcal{D}(S_B)\supset\mathcal{D}(S)$) and it is an operator extension of $S$ ($S_B f=\overline{S}f$ for all $f\in\mathcal{D}(\overline{S})$). In fact, $S_B$ is a \emph{symmetric} extension: indeed, for generic $g:=f+(S_\mathrm{F}^{-1}+B)\widetilde{u}_1+u_0\in\mathcal{D}(S_B)$, 
    \[
\begin{split}
\langle g,S_B\,g\rangle\;&=\;\langle f+(S_\mathrm{F}^{-1}+B)\widetilde{u}_1+u_0, \overline{S}f+\widetilde{u}_1\rangle\\
&=\;\langle f,\overline{S}f\rangle+\langle f,\widetilde{u}_1\rangle+\langle S_\mathrm{F}^{-1}\widetilde{u}_1,\overline{S}f\rangle+\langle S_\mathrm{F}^{-1}\widetilde{u}_1,\widetilde{u}_1\rangle+\langle B\widetilde{u}_1,\widetilde{u}_1\rangle
\end{split}
\]
  (having used $\langle B\widetilde{u}_1,\overline{S}f\rangle=\langle S^* B \widetilde{u}_1,f\rangle=0$, $\langle u_0,\overline{S}f\rangle=\langle S^*u_0,f\rangle=0$, and $\langle u_0,\widetilde{u}_1\rangle=0$), and moreover $f=S_\mathrm{F}^{-1}\overline{S}f$ (as a consequence of $\mathfrak{m}(S)>0$); thus,
  \[
   \langle g,S_B\,g\rangle\;=\;\langle f,\overline{S}f\rangle+\langle S_\mathrm{F}^{-1}\widetilde{u}_1,\widetilde{u}_1\rangle+\langle B\widetilde{u}_1,\widetilde{u}_1\rangle+2\,\mathfrak{Re}\langle f,\widetilde{u}_1\rangle\;\in\;\mathbb{R}
  \]
 (having used the symmetry, and hence the reality of expectations, of $\overline{S}$, $S_\mathrm{F}^{-1}$, and $B$).
 Therefore, $S\subset S_B\subset {S_B}^{\!*}\subset S^*$ and in order to show that $S_B={S_B}^{\!*}$ it suffices to prove that $\mathcal{D}(S_B)\supset\mathcal{D}({S_B}^{\!*})$.

 Let $h\in\mathcal{D}({S_B}^{\!*})$. Since  ${S_B}^{\!*}\subset S^*$, \eqref{eq:II-VishBirDecomp} implies $h=\varphi+S_\mathrm{F}^{-1}\overline{v}+v$ for some $\varphi\in\mathcal{D}(\overline{S})$ and $v,\overline{v}\in \ker S^*$, and ${S_B}^{\!*}h=\overline{S}\varphi+\overline{v}$. Actually $\overline{v}\in U_1$, because $\ker S^*=U_0\oplus U_1$ and $\langle\overline{v},u_0\rangle=\langle {S_B}^{\!*}h,u_0\rangle-\langle\overline{S}\varphi,u_0\rangle=\langle h,S_Bu_0\rangle-\langle\varphi,S^*u_0\rangle=0-0=0$ $\forall u_0\in U_0$. Thus, representing $v=v_0+v_1$, $v_0\in U_0$, $v_1\in U_1$, one writes
\[
h\;=\;\varphi+S_\mathrm{F}^{-1}\overline{v}+v\;=\;\varphi+(S_\mathrm{F}^{-1}\overline{v}+v_1)+v_0\,.
\]
 Now, when the identity $\langle h,S_B g\rangle=\langle S_B^*h,g\rangle$, clearly valid $\forall g\in\mathcal{D}(S_B)$, is specialised for the $g$'s of the form $g=f+(S_\mathrm{F}^{-1}+B)\widetilde{u}_1$, $f\in\mathcal{D}(\overline{S})$, $\widetilde{u}_1\in\mathcal{D}(B)$, on the one hand one finds
\[
\begin{split}
\langle h,S_B g\rangle\;&=\;\langle(\varphi+S_\mathrm{F}^{-1}\overline{v})+v_1+v_0,S^*(f+S_\mathrm{F}^{-1}\widetilde{u}_1+B\widetilde{u}_1)\rangle \\
&=\;\langle(\varphi+S_\mathrm{F}^{-1}\overline{v}),S^*(f+S_\mathrm{F}^{-1}\widetilde{u}_1)\rangle+\langle v_1+v_0,\overline{S}f+\widetilde{u}_1\rangle \\
&=\;\langle(\varphi+S_\mathrm{F}^{-1}\overline{v}),S_\mathrm{F}(f+S_\mathrm{F}^{-1}\widetilde{u}_1)\rangle+\langle v_1,\widetilde{u}_1\rangle
\end{split}
\]
(indeed, $S_B B\widetilde{u}_1=S^*B\widetilde{u}_1=0$,  $\varphi+S_\mathrm{F}^{-1}\overline{v}\in\mathcal{D}(S_\mathrm{F})$, $f+S_\mathrm{F}^{-1}\widetilde{u}_1\in\mathcal{D}(S_\mathrm{F})$, $\langle v_1+v_0,\overline{S}f\rangle=\langle S^*(v_1+v_0),f\rangle=0$, and $\langle v_0,\widetilde{u}_1\rangle=0$), and on the other hand
\[
\begin{split}
\langle S_B^*h,g\rangle\;&=\;\langle \overline{S}\varphi+\overline{v},(f+S_\mathrm{F}^{-1}\widetilde{u}_1)+B\widetilde{u}_1\rangle \\
&=\;\langle S_\mathrm{F}(\varphi+S_\mathrm{F}^{-1}\overline{v}),(f+S_\mathrm{F}^{-1}\widetilde{u}_1)\rangle+\langle\overline{v},B\widetilde{u}_1\rangle
\end{split}
\]
(indeed, $\langle\overline{S}\varphi,B\widetilde{u}_1\rangle=\varphi,S^*B\widetilde{u}_1\rangle=0$). Equating these two expressions and using the self-adjointness of $S_\mathrm{F}$ yields
\[
\begin{split}
\langle v_1,\widetilde{u}_1\rangle\;=\;\langle\overline{v},B\widetilde{u}_1\rangle\qquad\forall\,\widetilde{u}_1\in\mathcal{D}(B)
\end{split}
\]
which implies, owing to the self-adjointness of $B$, $\overline{v}\in\mathcal{D}(B)$ and $B\overline{v}=v_1$. Thus, the above decomposition for $h$ reads now
\[
h\;=\;\varphi+(S_\mathrm{F}^{-1}+B)\,\overline{v}+v_0
\]
for some $\overline{v}\in\mathcal{D}(B)=\widetilde{U}_1$ and $v_0\in U_0$. This shows that $h\in\mathcal{D}(S_B)$, and hence $S_B=S_B^*$.

 Last, one proves that the operator $B$ in the decomposition \eqref{eq:defSB} is unique and therefore  the correspondence between self-adjoint extensions of $S$ and operators of the form $S_B$ is one-to-one. Indeed, if 
\begin{equation*}
\mathcal{D}(S_B)\;=\;\mathcal{D}(\overline{S})\,\dotplus\,(S_\mathrm{F}^{-1}+B)\widetilde{U}_1\,\dotplus\,U_0\;=\;\mathcal{D}(\overline{S})\,\dotplus\,(S_\mathrm{F}^{-1}+B')\widetilde{U}_1'\,\dotplus\,U_0'
\end{equation*}
where $\ker S^*=U_0\oplus U_1=U_0'\oplus U_1'$ and where $B$ and $B'$ are self-adjoint operators, respectively on the Hilbert spaces $U_1$ and $U_1'$, with domain, respectively, $\widetilde{U}_1$ and $\widetilde{U}_1'$, then the action of $S_B$ on an arbitrary element $g\in\mathcal{D}(S_B)$ gives, in terms of the decomposition $g=f+(S_\mathrm{F}^{-1}+B)\widetilde{u}_1+u_0=f'+(S_\mathrm{F}^{-1}+B')\widetilde{u}'_1+u'_0$, $S_Bg=\overline{S}f+\widetilde{u}_1=\overline{S}f'+\widetilde{u}_1'$; each sum belongs to the orthogonal sum $\cH=\ran\,\overline{S}\oplus\ker S^*$, whence $\widetilde{u}_1=\widetilde{u}_1'$ and, by injectivity of $S$, $f=f'$ (whence also $u_0=u'_0$); thus, $\widetilde{U}_1=\widetilde{U}_1'$ and, after taking the closure $U_1=U_1'$ and $U_0=U_0'$; this also implies $B\widetilde{u}_1=B'\widetilde{u}_1$, whence $B=B'$. 
 \end{proof}

 It is customary to refer to the operator $B$ as the \emph{Vi\v{s}ik extension parameter}\index{Vi\v{s}ik extension parameter}\index{extension parameter!Vi\v{s}ik's} of the self-adjoint extension $S_B$.

 In retrospect, the whole construction that led to Theorem \ref{thm:VB-representaton-theorem} relies on the crucial role of $S_\mathrm{F}$, in terms of which one defines the Vi\v{s}ik parameter $B$ of the generic self-adjoint extension $S_B$ of $S$. (Proposition \ref{prop:parametrisation_SF_SN} below emphasises even more the role of `pivot' reference extension played by $S_\mathrm{F}$.) In fact, the characterising features of $S_\mathrm{F}$ of being the unique largest self-adjoint extension of $S$ and the unique self-adjoint extension whose operator domain is contained in $\mathcal{D}[S]$ are \emph{not} used in such construction: only the properties that $S_\mathrm{F}^{-1}$ is everywhere defined and bounded and that $S_\mathrm{F}$ extends $S$ are exploited.

 Along the same line, the assumption $\mathfrak{m}(S)>0$ technically speaking only enters in the identity $\ran\,\overline{S}=\overline{\ran\,S}$ and to apply Vi\v{s}ik's decomposition formula $\mathcal{D}(S^*)=\mathcal{D}(\overline{S})\dotplus S_\mathrm{F}^{-1}\ker S^*\dotplus \ker S^*$ (Proposition \ref{prop:II-KVB-decomp-of-Sstar}), beside of course in claiming the very existence of $S_\mathrm{F}$ and its bounded invertibility over the whole $\cH$.

 In this respect, Theorem \ref{thm:VB-representaton-theorem} can be reproduced in a \emph{generalised} form where $S_\mathrm{F}$ is replaced by a \emph{distinguished} self-adjoint extension $S_\mathrm{D}$ such that $S_\mathrm{D}^{-1}$ exists everywhere defined and bounded on $\cH$, and the assumption of semi-boundedness of $S$ is replaced by the assumption of existence of such distinguished extension $S_\mathrm{D}$.
 
%
%

 For this condition to occur, on account of the already mentioned result of Calkin \cite[Theorem 2]{Calkin-1940} (see the discussion at the beginning of Sect.~\ref{sec:II-KreinExtTheory}), it is sufficient that $S$ is a closed symmetric operator with zero in its resolvent set.

 \begin{theorem}[Generalised Vi\v{s}ik-Birman representation theorem]\label{thm:VB-representaton-theorem-GENER}\index{theorem!Vi\v{s}ik-Birman (self-adjoint extension)} Let $S$ be a densely defined symmetric operator on the Hilbert space $\cH$, which admits a self-adjoint extension $S_\mathrm{D}$ that has everywhere defined bounded inverse on $\cH$. There is a one-to-one correspondence between the  family of the self-adjoint extensions of  $S$ in $\cH$ and the family of the self-adjoint operators in Hilbert subspaces of $\ker S^*$, that is, the collection of triples $(U_1,\widetilde{U}_1,B)$, where $U_1$ is a closed subspace of $\ker S^*$, $\widetilde{U}_1$ is a dense subspace of  $U_1$, and $B$ is a self-adjoint operator in the Hilbert space $U_1$ with domain $\mathcal{D}(B)=\widetilde{U}_1$.
  For each such triple, let $U_0$ be the closed subspace of $\ker S^*$ defined by $\ker S^*=U_0\oplus U_1$. Then, in this correspondence $S_B\leftrightarrow B$, each self-adjoint extension $S_B$ of $S$ is given by
\begin{equation}\label{eq:defSB-new}
\begin{split}
S_B\;&=\;S^*\upharpoonright\mathcal{D}(S_B)\,, \\
\mathcal{D}(S_B)\;&=\;\mathcal{D}(\overline{S})\,\dotplus\,(S_\mathrm{D}^{-1}+B)\widetilde{U}_1\,\dotplus\,U_0\,.
\end{split}
\end{equation}
  In particular,
     \begin{eqnarray}
    \ker S_B\:&=&\:U_0\;=\;\ker S^*\cap\mathcal{D}(B)^\perp\,, \label{eq:II-kerSB-2} \\
    \mathrm{ran}\:S_B\:&=&\,\mathrm{ran}\,\overline{S}\boxplus\widetilde{U}_1\;=\;\mathrm{ran}\,\overline{S}\boxplus\mathcal{D}(B)\,. \label{eq:II-ranSB-2}
   \end{eqnarray}
\end{theorem}

 \begin{proof}
  It suffices to discuss those points, in the reasoning that unfolded through Lemmas \ref{lemma:decompStilde}-\ref{lemma:Bbig}, Proposition \ref{prop:BirmanBformula}, and Theorem \ref{thm:VB-representaton-theorem}, where the new assumptions require a modification of the proofs.

  First, one observes that the identity $\ran\,\overline{S}=\overline{\ran\,S}$ is still valid. Indeed, one the one hand $S$ is closable, being symmetric and densely defined (Sect.~\ref{sec:I-symmetric-selfadj}). Moreover, for any $f\in\mathcal{D}(S)$ one has $Sf=S_\mathrm{D}f$, whence $\|f\|=\|S_\mathrm{D}^{-1}S_\mathrm{D}f\|\leqslant\|S_\mathrm{D}^{-1}\|_{\mathrm{op}}\|Sf\|$. Such two properties, as stated in the end of Section \ref{sec:I-adjoint}, imply $\ran\,\overline{S}=\overline{\ran\,S}$.
  
  
  With this observation in mind, Lemmas \ref{lemma:decompStilde}-\ref{lemma:S-Stilde-Sbig} and their proofs clearly remain unaltered.

  The operator $\mathscr{B}$ is re-defined, instead of \eqref{eq:defBbig}, as  
  \begin{equation*}
\begin{split}
\mathscr{B}\;&:=\;\mathscr{S}^{-1}-P_+ S_\mathrm{D}^{-1} P_+ \,,\\
\mathcal{D}(\mathscr{B})\;&:=\;\mathcal{D}(\mathscr{S}^{-1})\;=\;\ran\,\widetilde{S}\,\boxplus\,\ker\widetilde{S}\,.
\end{split}
\end{equation*}
  Correspondingly, in the proof of Lemma \ref{lemma:Bbig} (whose statement remains unaltered), one only has to update the check that $\mathscr{B}\overline{S}f=0$ $\forall f\in\mathcal{D}(\overline{S})$, a fact that now is obtained from
  \[
 \mathscr{B} \overline{S}f\;=\;\mathscr{S}^{-1}\overline{S}f-P_+ S_\mathrm{D}^{-1}\overline{S}f\;=\;\widetilde{S}_+^{-1}\overline{S}f-P_+f\;=\;P_+f-P_+f\;=\;0\,.
\]
 Here one used that $S$ is injective (because $\|Sf\|\geqslant\|S_\mathrm{D}^{-1}\|_{\mathrm{op}}^{-1}\|f\|$ $\forall f\in\mathcal{D}(S)$), and so is $\overline{S}$, and $S_\mathrm{D}$ inverts $\overline{S}$ on its range: $S_\mathrm{D}^{-1}\overline{S}f=f$.

 Also the operator $B$ is re-defined, instead of \eqref{eq:defB}, as
 \begin{equation*}
B\;:=\;\big(\mathscr{S}^{-1}-P_+ S_\mathrm{D}^{-1} P_+\big)\upharpoonright\mathcal{D}(B)\,,\qquad \mathcal{D}(B)\;:=\;\widetilde{U}_1\,.
\end{equation*}

 Proposition \ref{prop:BirmanBformula} and its proof can be updated replacing $S_\mathrm{F}$ with $S_\mathrm{D}$ throughout. The only step to re-write is the check that the sum $\mathcal{D}(\overline{S})+(S_\mathrm{D}^{-1}+B)\widetilde{U}_1+U_0$ is direct: now, if $\mathcal{D}(S^*)\supset\mathcal{D}(\widetilde S) \ni g := f + (S_\mathrm{D}^{-1}+B)\tilde u_1 +u_0=f+S_\mathrm{D}^{-1}\tilde u_1 + (B\tilde u_1 +u_0)$ then, according to the decomposition formula $\mathcal{D}(S^*)=\mathcal{D}(\overline{S})\dotplus S_\mathrm{D}^{-1}\ker S^*\dotplus\ker S^*$ (Proposition \ref{prop:II-vNlike-decomp}), $g=0$ implies $f=0$, $\tilde u_1=0$, and $u_0=0$. (Of course, one could have used also in the proof of Proposition \ref{prop:BirmanBformula} this decomposition of $\mathcal{D}(S^*)$, instead of \eqref{eq:II-VishBirDecomp}.)

  Last, by inspection one sees that the whole proof of Theorem \ref{thm:VB-representaton-theorem} can be now repeated verbatim, replacing $S_\mathrm{F}$ with $S_\mathrm{D}$ everywhere: the distinguished extension is only used through the technical preliminaries that have been just updated and through the properties $\mathcal{D}(\overline{S})\subset\mathcal{D}(S_\mathrm{D})$ and $S_\mathrm{D}^{-1}=(S_\mathrm{D}^{-1})^*\in\mathcal{B}(\cH)$.  
 \end{proof}

 \subsection{Birman's operator}

 Out of Vi\v{s}ik's $B$ operator, one constructs the self-adjoint operator $B_\star^{-1}$ on the Hilbert space
\begin{equation}\label{eq:HB}
\cH_B\;:=\;\overline{\ran\,B}\,\oplus\,\ker S_B\,,
\end{equation}
which is a Hilbert subspace of $\ker S^*$ (recall the notation $\ker S_B\equiv U_0$, $\mathcal{D}(B)=\widetilde{U}_1$, $\overline{\mathcal{D}(B)}=U_1$ from \eqref{eq:U0-U1-H+} and \eqref{eq:defB}, and observe that $\overline{\ran\,B}\,\oplus\,\ker S_B\subset U_1\oplus U_0=\ker S^*$),
defined by
\begin{equation}\label{eq:Bto-1_original}
\begin{split}
\mathcal{D}(B_\star^{-1})\;&:=\;\ran\,B\,\boxplus\, \ker S_B\,, \\
B_\star^{-1}\upharpoonright\ran\,B\;&:=\;B^{-1}\,, \\
B_\star^{-1}\upharpoonright\ker S_B\;&:=\;\mathbb{O}\,.
\end{split}
\end{equation}
In other words, since  $U_1=\overline{\mathcal{D}(B)}=\overline{\ran\,B}\oplus\ker B$, $\widetilde{U}_1=\mathcal{D}(B)=(\mathcal{D}(B)\cap\overline{\ran\,B})\boxplus\ker B$, and $\ker S^*=\ker S_B\oplus\overline{\mathcal{D}(B)}$,
\begin{equation}\label{eq:Bto-1}
\begin{split}
\cH_B\;=&\;\;\overline{\ran\,B}\,\oplus\,(\ker S^*\cap\mathcal{D}(B)^\perp) \\
=&\;\;\overline{\ran\,B}\,\oplus\,(\ker S^*\ominus\overline{\mathcal{D}(B)}\,)\;=\;\overline{\mathcal{D}(B_\star^{-1})}\,, \\
\mathcal{D}(B_\star^{-1})\;=&\;\;\ran\,B\,\boxplus\, (\ker S^*\cap\mathcal{D}(B)^\perp)\,,  \\
B_\star^{-1}Bu_1\;=&\;u_1 \qquad\!\!\forall u_1\in\mathcal{D}(B)\cap\overline{\ran\,B}\,, \\
B_\star^{-1} u_0\;=&\;\;0\qquad\forall u_0\in\ker S^*\cap\mathcal{D}(B)^\perp\,.
\end{split}
\end{equation}
Thus, the action of $B_\star^{-1}$ on a generic element $B\widetilde{u}_1\in\ran\,B$ is given, in view of the decomposition $\widetilde{u}_1=u_1+u_0$ for some $u_1\in\mathcal{D}(B)\cap \overline{\ran\,B}$ and $u_0\in\ker B$, by $B_\star^{-1}B\widetilde{u}_1=u_1$.

It is fair to refer to $B_\star^{-1}$ as \emph{Birman's operator} or also the \emph{Birman extension parameter}\index{Birman extension parameter}\index{extension parameter!Birman's} (in analogy to the Vi\v{s}ik extension parameter $B$), for it is Birman who first determined and exploited its properties (Lemma \ref{lem:domB-1indomSBform} and Theorem \ref{thm:semibdd_exts_operator_formulation} below).

 \begin{lemma}\label{lem:domB-1indomSBform}~
\begin{enumerate}[(i)]
 \item If, with respect to the notation of \eqref{eq:defSB} and \eqref{eq:Bto-1}, $S_B$ is a self-adjoint extension of a given densely defined symmetric operator $S$ with positive bottom, then
 \begin{eqnarray}
  \label{eq:inclusion_DB_in_DF+DB-1} \mathcal{D}(S_B)\;&\subset&\;\mathcal{D}(S_\mathrm{F})\,\dotplus\,\mathcal{D}(B_\star^{-1})\,, \\
  \label{eq:DB-1_into_DSB} \mathcal{D}(B_\star^{-1})\;&\subset &\;\mathcal{D}[S_B]\cap\ker S^*\,.
 \end{eqnarray}
 \item If in addition $S_B$ is \emph{lower semi-bounded}, then
 \begin{equation}\label{eq:SB-and-B-1}
S_B[v_1,v_2]\;=\;\langle v_1,B_\star^{-1} v_2\rangle\qquad\quad\forall v_1,v_2\in\mathcal{D}(B_\star^{-1})\,.
\end{equation}
\end{enumerate}
\end{lemma}

 \begin{proof}
  Concerning \eqref{eq:inclusion_DB_in_DF+DB-1}, any $g\in\mathcal{D}(S_B)$ decomposes as $g=f_0+(S_\mathrm{F}^{-1}+B)\widetilde{u}_1+u_0=f+v$, according to \eqref{eq:defSB}, for some $f_0\in\mathcal{D}(\overline{S})$, $\widetilde{u}_1\in\widetilde{U}_1=\mathcal{D}(B)$, $u_0\in U_0=\ker S^*\cap\mathcal{D}(B)^\perp$, where $f:=f_0+S_\mathrm{F}^{-1}\widetilde{u}_1\in\mathcal{D}(S_\mathrm{F})$ and (by \eqref{eq:Bto-1}) $v:=B\widetilde{u}_1+u_0\in\mathcal{D}(B_\star^{-1})$. The sum on the r.h.s.~of \eqref{eq:inclusion_DB_in_DF+DB-1} is direct because if $v\in\mathcal{D}(S_\mathrm{F})\cap\mathcal{D}(B_\star^{-1})$, then
  $v\in \ker S^*$ by definition \eqref{eq:HB}-\eqref{eq:Bto-1_original} and hence
  $\|v\|^2=\langle S_\mathrm{F} S_\mathrm{F}^{-1}v,v\rangle=\langle S_\mathrm{F}^{-1}v,S_\mathrm{F}v\rangle=\langle S_\mathrm{F}^{-1}v,S^*v\rangle=0$. 
  Concerning \eqref{eq:DB-1_into_DSB}, the inclusion $\mathcal{D}(B_\star^{-1})\subset\ker S^*$ follows from the definition \eqref{eq:Bto-1_original}, so it remains to prove that $\mathcal{D}(B_\star^{-1})\subset\mathcal{D}[S_B]$.
  Decompose an arbitrary $v\in\mathcal{D}(B_\star^{-1})$, according to \eqref{eq:Bto-1}, as $v=B\widetilde{u}_1+u_0$ for some $\widetilde{u}_1\in\mathcal{D}(B)\cap\overline{\ran\,B}\subset\widetilde{U}_1$ and $u_0\in\ker S_B=U_0$. 
  As $v=\big((S_\mathrm{F}^{-1}+B)\widetilde{u}_1+u_0\big)-S_\mathrm{F}^{-1}\widetilde{u}_1$, $v$ is the difference between an element in $\mathcal{D}(S_B)$, according to  \eqref{eq:defSB}, and an element in $\mathcal{D}(S_\mathrm{F})$. 
  Since $\mathcal{D}(S_B)\subset\mathcal{D}[S_B]$ and $\mathcal{D}(S_\mathrm{F})\subset\mathcal{D}[S_\mathrm{F}]\subset\mathcal{D}[S_B]$ (the Friedrichs extension has the smallest form domain among all semi-bounded extensions, Theorem \ref{thm:Friedrichs-ext}(vii)), then $v\in\mathcal{D}[S_B]$, which proves $\mathcal{D}(B_\star^{-1})\subset\mathcal{D}[S_B]$ and completes the proof of \eqref{eq:DB-1_into_DSB}. 
  Concerning \eqref{eq:SB-and-B-1}, consider again an arbitrary 
$v=B\widetilde{u}_1+u_0$ in $\mathcal{D}(B_\star^{-1})$ as above: for $f:=S_\mathrm{F}^{-1}\widetilde{u}_1\in\mathcal{D}(S_\mathrm{F})$ and $g:=f+v=S_\mathrm{F}^{-1}\widetilde{u}_1+B\widetilde{u}_1+u_0\in\mathcal{D}(S_B)$, one has $S_B g=\widetilde{u}_1=S_\mathrm{F} f$, $B_\star^{-1}v=\widetilde{u}_1$, and
\[
\begin{split}
S_B[g,g]\;&=\;\langle g,S_B g\rangle\;=\;\langle f+v,\widetilde{u}_1\rangle\;=\;\langle f,S_\mathrm{F} f\rangle+\langle v,\widetilde{u}_1\rangle \\&=\;S_\mathrm{F}[f,f]+\langle v,B_\star^{-1}v\rangle\,.
\end{split}
\]
All this is valid \emph{irrespectively of the semi-boundedness} of $S_B$. On the other hand, if $S_B$ is bounded below, then $S_B[f,v]=0$, whereby
 \[
S_B[g,g]\;=\;S_\mathrm{F}[f,f]\,+\,S_B[v,v]
\]
 (Theorem \ref{thm:II-S[f,u]-Krein-2}, using that $f\in\mathcal{D}(S_\mathrm{F})$ and $v\in\mathcal{D}(B_\star^{-1})\subset\mathcal{D}[S_B]\cap\ker S^*$, owing to \eqref{eq:DB-1_into_DSB}). Thus, by comparison, $S_B[v,v]=\langle v,B_\star^{-1}v\rangle$ $\forall v\in\mathcal{D}(B_\star^{-1})$. \eqref{eq:SB-and-B-1} then follows by polarisation.
 \end{proof}

  \begin{remark}
    In general \eqref{eq:inclusion_DB_in_DF+DB-1} is only valid as a strict inclusion. Indeed, a generic $h\in\mathcal{D}(S_\mathrm{F})\dotplus\mathcal{D}(B_\star^{-1})$ can be re-written, according to \eqref{eq:II-SF_VB_decomp} and \eqref{eq:Bto-1}, as 
    \[
     h=\big(f+S_\mathrm{F}^{-1}u\big)+\big(B\widetilde{u}_1+u_0\big)\;=\;(f+S_\mathrm{F}^{-1}\widetilde{u}_1+B\widetilde{u}_1+u_0)+S_\mathrm{F}^{-1}(u-\widetilde{u}_1)
    \]
    for some  $f\in\mathcal{D}(\overline{S})$, $\widetilde{u}_1\in\mathcal{D}(B)$, $u_0\in \ker S^*\cap\mathcal{D}(B)^\perp$, $u\in\ker S^*$, and this is not enough to use \eqref{eq:defSB} and deduce that $h\in\mathcal{D}(S_B)$.
  \end{remark}

 \subsection{Semi-bounded extensions: operator parametrisation}
 
 The classification provided by Theorem \ref{thm:VB-representaton-theorem} is general and characterises \emph{all} self-adjoint extensions of $S$ (a densely defined, positive symmetric operator on $\cH$, with $\mathfrak{m}(S)>0$). In addition, it allows to monitor those extensions $S_B$ that are \emph{lower semi-bounded}, in terms of a special sub-class of extension parameters $B$. 

 $S$ may have also self-adjoint extensions that are \emph{unbounded} from below. A prototypical abstract example is the case of the infinite orthogonal Hilbert space sum $\cH=\bigoplus_{n\in\mathbb{N}}\mathfrak{h}_n$ (Sect.~\ref{sec:I-preliminaries}) that reduces an operator $S=\bigoplus_{n\in\mathbb{N}}S_n$ (Sect.~\ref{sec:I_invariant-reducing-ssp}), where each $S_n$ is densely defined and positive on $\mathfrak{h}_n$, and admits a lower semi-bounded self-adjoint extension $\widetilde{S}_n$ with $\mathfrak{m}(\widetilde{S}_n)=-n$: say (see, e.g., \cite[Section I.3.1]{albeverio-solvable}), $\mathfrak{h}_n\equiv L^2(\mathbb{R})$, $S_n\equiv-\frac{\ud^2}{\ud x^2}$ on the domain $C^\infty_0(\mathbb{R}\setminus\{0\})$, and with self-adjoint extension $\widetilde{S}_n\equiv-\frac{\ud^2}{\ud x^2}$, $\mathcal{D}(\widetilde{S}_n)\equiv\{g\in H^2(\mathbb{R}\setminus\{0\})\cap H^1(\mathbb{R})\,|\,g'(0^+)-g'(0^-)=-2\sqrt{n}\,g(0)\}$, in which case the self-adjoint extension $\widetilde{S}:=\bigoplus_{n\in\mathbb{N}}\widetilde{S}_n$ of $S$ is unbounded from below.

Certainly, if $S$ has \emph{finite} deficiency index, then any self-adjoint extension of $S$ is lower semi-bounded:


 \begin{lemma}\label{lem:II-finiteDefIndLowSemibdd}
  Let $S$ be a densely defined, symmetric, lower semi-bounded operator on the Hilbert space $\cH$ with finite deficiency index $d$. Then any self-adjoint extension of $S$ is lower semi-bounded.  
 \end{lemma}

 \begin{proof}
  Let $\widetilde{S}$ be a self-adjoint extension of $S$. Then $\mathcal{D}(\widetilde{S})=\mathcal{D}(\overline{S})+V$, where $V$ is a $d$-dimensional subspace of $\cH$: this is immediately seen from \eqref{eq:II-vNlike3}, in view of the fact that $\dim\ker(S^*-\overline{z}\mathbbm{1})=d$ for any $z\in\rho(\widetilde{S})$. The spectral projection $E^{(\widetilde{S})}((-\infty,\mathfrak{m}(S))$ of $\widetilde{S}$ (Sect.~\ref{sec:I_spectral_theorem}) must satisfy $\dim\ran\, E^{(\widetilde{S})}((-\infty,\mathfrak{m}(S))\leqslant d$, that is, the rank of $E^{(\widetilde{S})}((-\infty,\mathfrak{m}(S))$ cannot exceed $d$: for, otherwise, from $\mathcal{D}(\widetilde{S})=\mathcal{D}(\overline{S})+V$ and $\dim V=d$ one would deduce that there is a non-zero $f\in\mathcal{D}(\overline{S})\cap\ran\, E^{(\widetilde{S})}((-\infty,\mathfrak{m}(S))$, whence (Sect.~\ref{sec:I_funct_calc})
  \[
   \langle f,\overline{S}f\rangle\;=\;\langle f,\widetilde{S}f\rangle\;=\;\int_{(-\infty,\mathfrak{m}(S))}\lambda\,\ud\mu_f^{(\widetilde{S})}(\lambda)\;<\;\mathfrak{m}(S)\|f\|^2\,,
  \]
  a contradiction. Thus, $\dim\ran\, E^{(\widetilde{S})}((-\infty,\mathfrak{m}(S))\leqslant d$, which proves that $S$ is lower semi-bounded.  
 \end{proof}

 Of course, even when $S$ has \emph{infinite} deficiency index, it still admits semi-bounded self-adjoint extensions (one example is the Friedrichs extension). Within the general parametrisation \eqref{eq:defSB-new} from Theorem \ref{thm:VB-representaton-theorem}, the semi-bounded extensions are conveniently identified in terms of the Birman extension parameter $B_\star^{-1}$.

\begin{theorem}[Characterisation of semi-bounded extensions]\label{thm:semibdd_exts_operator_formulation}\index{theorem!Birman (semi-bounded self-adjoint extension)}
Let $S$ be a densely defined and positive symmetric operator with greatest lower bound $\mathfrak{m}(S)>0$ on the  Hilbert space $\cH$.
If, with respect to the notation of \eqref{eq:defSB} and \eqref{eq:Bto-1}, $S_B$ is a self-adjoint extension of $S$, and if $\alpha<\mathfrak{m}(S)$, then
\begin{equation}\label{eq:SBsmbb-iff-invBsmbb}
\begin{split}
\langle g,S_B g\rangle\;&\geqslant\;\alpha\,\|g\|^2\qquad\forall g\in\mathcal{D}(S_B) \\
& \Updownarrow \\
\langle v,B_\star^{-1} v\rangle\;\geqslant\;\alpha\|v\|^2+\:&\alpha^2\langle v,(S_\mathrm{F}-\alpha\mathbbm{1})^{-1} v\rangle\qquad\forall v\in\mathcal{D}(B_\star^{-1})\,.
\end{split}
\end{equation}
As an immediate consequence, $\mathfrak{m}(B_\star^{-1})\geqslant \mathfrak{m}(S_B)$ for any lower semi-bounded $S_B$.
In particular, positivity or strict positivity of the bottom of $S_B$ is equivalent to the same property for $B_\star^{-1}$, that is,
\begin{eqnarray}
  \mathfrak{m}(S_B)\;\geqslant \;0\quad&\Leftrightarrow\quad \mathfrak{m}(B_\star^{-1})\;\geqslant\; 0\,, \label{eq:positiveSBiffpositveB-1geq}\\
 \mathfrak{m}(S_B)\;> \;0\quad&\Leftrightarrow\quad \mathfrak{m}(B_\star^{-1})\;>\; 0\,. \label{eq:positiveSBiffpositveB-1gnoteq}
\end{eqnarray}
 Moreover, if $\mathfrak{m}(B_\star^{-1})>-\mathfrak{m}(S)$, then
 \begin{equation}\label{eq:bounds_mS_mB}
 \mathfrak{m}(B_\star^{-1})\;\geqslant\; \mathfrak{m}(S_B)\;\geqslant\;\frac{\mathfrak{m}(S) \,\mathfrak{m}(B_\star^{-1})}{\mathfrak{m}(S)+\mathfrak{m}(B_\star^{-1})}\,.
 \end{equation}
\end{theorem}


\begin{proof}
 One starts with the proof of \eqref{eq:SBsmbb-iff-invBsmbb}, assuming first that
 $\langle g,S_B g\rangle\geqslant\alpha\,\|g\|^2$ $\forall g\in\mathcal{D}(S_B)$, which is equivalent to $S_B[g]\geqslant\alpha\|g\|^2$ $\forall g\in\mathcal{D}[S_B]$ (Sect.~\ref{sec:I_forms}). 
 Take arbitrary $f\in\mathcal{D}(S_\mathrm{F})$ and $v\in\mathcal{D}(B_\star^{-1})$: in particular, $f\in\mathcal{D}[S_B]$ (because $S_\mathrm{F}\geqslant S_B$ and hence $\mathcal{D}(S_\mathrm{F})\subset\mathcal{D}[S_\mathrm{F}]\subset\mathcal{D}[S_B]$) and $v\in\mathcal{D}[S_B]\cap\ker S^*$ (Lemma \ref{lem:domB-1indomSBform}(i)), whence $g:=f+v\in\mathcal{D}[S_B]$.
 Moreover, $S_B[v]=\langle v,B_\star^{-1}v\rangle$ (Lemma \ref{lem:domB-1indomSBform}(ii)) and $S_B[f,v]=0$ (Theorem \ref{thm:II-S[f,u]-Krein-2}). Then $S_B[g]=S_B[f+v]=S_\mathrm{F}[f]+S_B[v]=\langle f,S_\mathrm{F}f\rangle+\langle v,B_\star^{-1} v\rangle$. Thus, the assumption that $S_B$ is bounded below by $\alpha$ is tantamount as
\[
\langle f,S_\mathrm{F} f\rangle+\langle v,B_\star^{-1} v\rangle\;\geqslant\;\alpha\,\big( \langle f,f\rangle+\langle f,v\rangle+\langle v,f\rangle+\langle v,v\rangle\big)
\]
for all such $f$'s and $v$'s, whence also, replacing $f\mapsto\lambda f$, $v\mapsto\mu v$,
\[
\begin{split}
\big(\langle f,S_\mathrm{F} f\rangle&-\alpha\|f\|^2\big)\,|\lambda|^2-\alpha\langle f,v\rangle\lambda\overline{\mu}-\alpha\langle v,f\rangle\overline{\lambda}\mu \\
&\qquad +\big(\langle v,B_\star^{-1} v\rangle-\alpha\|v\|^2\big)\,|\mu|^2\;\geqslant\;0\qquad\forall \lambda,\mu\in\mathbb{C}\,.
\end{split}
\]
Since $\alpha<\mathfrak{m}(S)$, and hence $\langle f,S_\mathrm{F} f\rangle-\alpha\|f\|^2> 0$, the last inequality holds true if and only if
\[\tag{*}\label{eq:II-proofbddbelow3}
\alpha^2|\langle f,v\rangle|^2\;\leqslant\;\big(\langle v,B_\star^{-1} v\rangle-\alpha\|v\|^2\big)\,\big(\langle f,S_\mathrm{F} f\rangle-\alpha\|f\|^2\big)
\]
for arbitrary $f\in\mathcal{D}(S_\mathrm{F})$ and $v\in\mathcal{D}(B_\star^{-1})$.
 By re-writing \eqref{eq:II-proofbddbelow3} as 
 \[
\langle v,B_\star^{-1} v\rangle-\alpha\|v\|^2 \;\geqslant\;\alpha^2\,\frac{|\langle f,v\rangle|^2}{\langle f,(S_\mathrm{F}-\alpha\mathbbm{1})f\rangle}\,,
\]
and by the fact that the above inequality is valid for arbitrary $f\in\mathcal{D}(S_\mathrm{F})$, and hence holds true also when the supremum over such $f$'s is taken, one finds
\[
\langle v,B_\star^{-1} v\rangle-\alpha\|v\|^2 \;\geqslant\;\alpha^2\langle v,(S_\mathrm{F}-\alpha\mathbbm{1})^{-1} v\rangle
\]
by means of a standard operator-theoretic argument applied to the bottom-positive operator $S_\mathrm{F}-\alpha\mathbbm{1}$ (see Lemma \ref{lem:A_A-1} below). This completes the proof of \eqref{eq:SBsmbb-iff-invBsmbb}.

  From \eqref{eq:SBsmbb-iff-invBsmbb} one deduces immediately both \eqref{eq:positiveSBiffpositveB-1geq}, by taking $\alpha=0$, and the implication ``$\mathfrak{m}(S_B)>0\Rightarrow \mathfrak{m}(B_\star^{-1})>0$'' in \eqref{eq:positiveSBiffpositveB-1gnoteq}, because $\mathfrak{m}(B_\star^{-1})\geqslant \mathfrak{m}(S_B)$. Conversely, if $\mathfrak{m}(B_\star^{-1})>0$, then it follows from \eqref{eq:Bto-1_original}-\eqref{eq:Bto-1} that  $B_\star^{-1}$ has a bounded inverse, that $U_0\equiv\ker S_B=\{0\}$, and that $B:\mathcal{D}(B)\equiv\widetilde{U}_1\subset U_1\to U_1$ is bounded; this, in turn, implies by Lemma \ref{lemma:Bbig}(ii) and by \eqref{eq:defB} that the operator $\mathscr{B}$ defined in \eqref{eq:defBbig} is bounded, which by Lemma \ref{lemma:Bbig}(i) means that $S_B$ has a bounded inverse (densely defined) on the whole $\cH_+$ and therefore (as $\cH=\cH_+\oplus U_0$) on the whole $\cH$. This fact excludes that $\mathfrak{m}(S_B)=0$, and since $\mathfrak{m}(B_\star^{-1})>0\Rightarrow \mathfrak{m}(S_B)\geqslant 0$ by \eqref{eq:positiveSBiffpositveB-1geq}, one finally concludes   $\mathfrak{m}(S_B)>0$, which completes the proof of \eqref{eq:positiveSBiffpositveB-1gnoteq}.

Last, it only remains to prove 
$\mathfrak{m}(S_B)\geqslant \mathfrak{m}(S)\mathfrak{m}(B_\star^{-1})(\mathfrak{m}(S)+\mathfrak{m}(B_\star^{-1}))^{-1}$ in \eqref{eq:bounds_mS_mB} (assuming $\mathfrak{m}(B_\star^{-1})>-\mathfrak{m}(S)$). 
In this case, for
\[
\alpha\;:=\;\frac{\mathfrak{m}(S) \,\mathfrak{m}(B_\star^{-1})}{\mathfrak{m}(S)+\mathfrak{m}(B_\star^{-1})}
\]
one has $\alpha<\mathfrak{m}(S)=\mathfrak{m}(S_\mathrm{F})$ and $\mathfrak{m}(B_\star^{-1})=\alpha\,\mathfrak{m}(S)(\mathfrak{m}(S)-\alpha)^{-1}$, whence $(\mathfrak{m}(S)-\alpha)^{-1}\geqslant (S_\mathrm{F}-\alpha)^{-1}$
and
\[
\begin{split}
\langle v,B_\star^{-1} v\rangle\;&\geqslant\;\mathfrak{m}(B_\star^{-1})\|v\|^2\;=\;\frac{\alpha\,\mathfrak{m}(S)}{\mathfrak{m}(S)-\alpha}\|v\|^2 \;=\;\alpha\|v\|^2+\frac{\alpha^2}{\mathfrak{m}(S_\mathrm{F})-\alpha}\|v\|^2  \\
&\geqslant\;\alpha\|v\|^2+\alpha^2\langle v,(S_\mathrm{F}-\alpha)^{-1}v\rangle\qquad\forall v\in\mathcal{D}(B_\star^{-1})\,.
\end{split}
\]
Owing to \eqref{eq:SBsmbb-iff-invBsmbb}, the latter inequality is equivalent to $\mathfrak{m}(S_B)\geqslant\alpha$, which completes the proof of \eqref{eq:bounds_mS_mB}.
\end{proof}

  \begin{lemma}\label{lem:A_A-1}
If $A$ is a self-adjoint operator on a Hilbert space $\cH$ with positive bottom ($\mathfrak{m}(A)>0$), then
\[
\sup_{f\in\mathcal{D}(A)}\frac{\;|\langle f,h\rangle|^2}{\langle f,Af\rangle}\;=\;\langle h,A^{-1}h\rangle\qquad\forall h\in\cH.
\]
\end{lemma}

\begin{proof}
Setting $g:=A^{\frac{1}{2}}f$ one has
\[
\sup_{f\in\mathcal{D}(A)}\frac{\;|\langle f,h\rangle|^2}{\langle f,Af\rangle}\;=\;\sup_{g\in\cH}\frac{\;|\langle A^{-\frac{1}{2}}g,h\rangle|^2}{\|g\|^2}\;=\;\sup_{\|g\|=1}|\langle g,A^{-\frac{1}{2}}h\rangle|^2
\]
and since $|\langle g,A^{-\frac{1}{2}}h\rangle|$ attains its maximum for $g=A^{-\frac{1}{2}}h/\|A^{-\frac{1}{2}}h\|$, the conclusion then follows.
\end{proof}

 \subsection{Semi-bounded extensions: quadratic form parametrisation}

 \begin{theorem}[Characterisation of semi-bounded extensions -- form version]\label{thm:semibdd_exts_form_formulation}\index{theorem!Birman (quadratic form of semi-bounded self-adjoint extension)}
 Let $S$ be a densely defined and positive symmetric operator with greatest lower bound $\mathfrak{m}(S)>0$ on the Hilbert space $\cH$.
 With respect to the notation of \eqref{eq:defSB} and \eqref{eq:Bto-1}, let $S_B$ be a \emph{lower semi-bounded} self-adjoint extension of $S$. Then
\begin{equation}\label{eq:D[SB]}
\mathcal{D}[B_\star^{-1}]\;=\; \mathcal{D}[S_B]\,\cap\,\ker S^*
\end{equation}
and
 \begin{eqnarray}
  \mathcal{D}[S_B]&=&\mathcal{D}[S_\mathrm{F}]\,\dotplus\,\mathcal{D}[B_\star^{-1}]\,, \label{eq:decomposition_of_form_domains-1} \\
  S_B[f+v,f'+v']&=&S_\mathrm{F}[f,f']\,+\,B_\star^{-1}[v,v'] \label{eq:decomposition_of_form_domains-2} \\
  & & \forall f,f'\in\mathcal{D}[S_\mathrm{F}],\;\forall v,v'\in\mathcal{D}[B_\star^{-1}]\,. \nonumber
 \end{eqnarray}
 As a consequence,
\begin{equation}\label{eq:extension_ordering}
S_{B_1}\,\geqslant\,S_{B_2}\qquad\Leftrightarrow\qquad {B_1}_\star^{-1}\,\geqslant\,{B_2}_\star^{-1}
\end{equation}
and
\begin{equation}
{B}_\star^{-1}\;\geqslant\;S_B\,.
\end{equation}
\end{theorem}

\begin{remark}
 Identity \eqref{eq:D[SB]} is the form version of the inclusion \eqref{eq:DB-1_into_DSB}, which instead holds for a generic (not necessarily semi-bounded) extension $S_B$. 
 As far as the analysis of quadratic forms of semi-bounded self-adjoint extensions is concerned, properties  \eqref{eq:D[SB]}-\eqref{eq:decomposition_of_form_domains-2} represent the fundamental technical improvement made by Birman \cite[Theorem 3]{Birman-1956} with respect to  Kre{\u\i}n's original theory, namely \eqref{eq:II-S[f,u]-Krein}-\eqref{eq:II-S[f,u]-Krein-3} from Theorem \ref{thm:II-S[f,u]-Krein-2}: here the novelty is the characterisation of the space $\mathcal{D}[\widetilde{S}]\cap\ker S^*$ in terms of the parameter $B$ (equivalently, $B_\star^{-1}$) of each semi-bounded extension $\widetilde{S}$. 
\end{remark}

\begin{proof}[Proof of Theorem \ref{thm:semibdd_exts_form_formulation}]
  Let $\alpha\in\mathbb{R}$ be such that $\alpha<\mathfrak{m}(S_B)$ and $|\alpha|>1$.
  In order to establish the inclusion $\mathcal{D}[B_\star^{-1}]\subset\mathcal{D}[S_B]\cap\ker S^*$ in \eqref{eq:D[SB]}, one exploits the fact that $\mathfrak{m}(B_\star^{-1})\geqslant \mathfrak{m}(S_B)>\alpha$ (Theorem \ref{thm:semibdd_exts_operator_formulation}) and therefore (Sect.~\ref{sec:I_forms}) $\mathcal{D}[B_\star^{-1}]$ is the completion of $\mathcal{D}(B_\star^{-1})$ in the norm associated with the scalar product
\[
\langle v_1,v_2\rangle_{(B_\star^{-1},\alpha)}\;:=\;\langle v_1,B_\star^{-1} v_2\rangle -\alpha\langle v_1,v_2\rangle\,,
\]
whereas $\mathcal{D}[S_B]$ is complete in the norm associated with the scalar product
\[
\langle v_1,v_2\rangle_{(S_B,\alpha)}\;:=\;S_B[v_1,v_2] -\alpha\langle v_1,v_2\rangle\,.
\]
 As $\langle v,B_\star^{-1} v\rangle=S_B[v]$ for all $v\in\mathcal{D}(B_\star^{-1})\subset\mathcal{D}[S_B]$ (Lemma \ref{lem:domB-1indomSBform}), 
\[
\|v\|_{(B_\star^{-1},\alpha)}^2\;=\;\langle v,B_\star^{-1} v\rangle -\alpha\|v\|^2\;=\;S_B[v]-\alpha\|v\|^2\;=\;\|v\|_{(S_B,\alpha)}^2\,.
\]
 Thus, the $\|\,\|_{(B_\star^{-1},\alpha)}$-completion of $\mathcal{D}(B_\star^{-1})$ 
does not exceed  $\mathcal{D}[S_B]$. On the other hand,
\[
\|v\|_{(B_\star^{-1},\alpha)}^2\;\geqslant\;(\mathfrak{m}(B_\star^{-1})-\alpha)\|v\|^2
\]
and the r.h.s.~above is obviously a norm with respect to which $\ker S^*$ is closed: therefore the $\|\,\|_{(B_\star^{-1},\alpha)}$-completion of $\mathcal{D}(B_\star^{-1})$ 
does not exceed  $\ker S^*$ either. This proves $\mathcal{D}[B_\star^{-1}]\subset\mathcal{D}[S_B]\cap\ker S^*$.

  For the opposite inclusion, recall first that the assumption $\mathfrak{m}(S_B)>\alpha$ implies
\[\tag{$\bullet$}\label{eq:II-proofbddform}
\alpha^2|\langle f,v\rangle|^2\;\leqslant\;\|f\|_{(S_B,\alpha)}^2\|v\|_{(S_B,\alpha)}^2 \qquad \forall f\in\mathcal{D}(S_\mathrm{F})\,,\;\forall v\in\mathcal{D}(B_\star^{-1})
\]
  (having used $\|v\|_{(B_\star^{-1},\alpha)}=\|v\|_{(S_B,\alpha)}$ $\forall v\in\mathcal{D}(B_\star^{-1})$), as seen already in the course of the proof of Theorem \ref{thm:semibdd_exts_operator_formulation}, when condition \eqref{eq:II-proofbddbelow3} therein was established. Observe that both $\mathcal{D}[B_\star^{-1}]$ and $\mathcal{D}[S_\mathrm{F}]$ are $\|\,\|_{(S_B,\alpha)}$-closed subspaces of $\mathcal{D}[S_B]$: the claim for $\mathcal{D}[B_\star^{-1}]$ has been just established in the first part of this theorem's proof, the claim for $\mathcal{D}[S_\mathrm{F}]$ follows from the properties $\mathcal{D}[S_\mathrm{F}]\subset\mathcal{D}[S_B]$, $S_B[f]=S_\mathrm{F}[f]$ $\forall f\in\mathcal{D}[S_\mathrm{F}]$, and $\mathfrak{m}(S_\mathrm{F})\geqslant \mathfrak{m}(S_B)>\alpha$ (Theorem \ref{thm:Friedrichs-ext}(vi)), because $\mathcal{D}[S_\mathrm{F}]$ is $\|\,\|_{(S_\mathrm{F},\alpha)}$-closed (Sect.~\ref{sec:I_forms}) and on it the $\|\,\|_{(S_\mathrm{F},\alpha)}$-norm and the $\|\,\|_{(S_B,\alpha)}$-norm are identical. Let now $u\in \mathcal{D}[S_B]\cap\ker S^*$ and let $(g_n)_{n\in\mathbb{N}}$ be a sequence in $\mathcal{D}(S_B)$ of approximants of $u$ in the $\|\,\|_{(S_B,\alpha)}$-norm. On account of Lemma \ref{lem:domB-1indomSBform}(i), $g_n=f_n+v_n$ for some $f_n\in\mathcal{D}(S_\mathrm{F})\subset\mathcal{D}[S_B]$ and $v_n\in\mathcal{D}(B_\star^{-1})\subset\mathcal{D}[S_B]$. This and \eqref{eq:II-proofbddform} above then imply
\[
\begin{split}
\|g_n&-g_m\|_{(S_B,\alpha)}^2\;\geqslant\;\|f_n-f_m\|_{(S_B,\alpha)}^2-2|\langle f_n-f_m,v_n-v_m\rangle_{(S_B,\alpha)}|+\|v_n-v_m\|_{(S_B,\alpha)}^2 \\
&\geqslant\;\|f_n-f_m\|_{(S_B,\alpha)}^2-\frac{2}{\alpha^2}|\|f_n-f_m\|_{(S_B,\alpha)}\|v_n-v_m\|_{(S_B,\alpha)}+\|v_n-v_m\|_{(S_B,\alpha)}^2 \\
&\geqslant\;(1-\alpha^{-2})\big(\|f_n-f_m\|_{(S_B,\alpha)}^2+\|v_n-v_m\|_{(B_\star^{-1},\alpha)}^2\big)\,.
\end{split} 
\]
 Therefore, since $1-\alpha^{-2}>0$, both $(f_n)_{n\in\mathbb{N}}$ and $(v_n)_{n\in\mathbb{N}}$ are Cauchy sequences, respectively, in $\mathcal{D}[S_\mathrm{F}]$ and $\mathcal{D}[B_\star^{-1}]$, with respect to the topology of the $\|\,\|_{(S_B,\alpha)}$-norm, with limits, say, $f_n\to f\in\mathcal{D}[S_\mathrm{F}]$ and $v_n\to v\in\mathcal{D}[B_\star^{-1}]$ as $n\to\infty$. Taking $n\to\infty$ in $g_n=f_n+v_n$ thus yields $u=f+v$. Having proved above that $\mathcal{D}[B_\star^{-1}]\subset\ker S^*$, one therefore concludes $f=u-v\in\ker S^*$, whence $f\in \mathcal{D}[S_\mathrm{F}]\cap\ker S^*=\{0\}$ (see \eqref{eq:II-def-SN-first3} from Theorem \ref{thm:II-KvNextension}(iv)). Then $u=v\in\mathcal{D}[B_\star^{-1}]$, and from the arbitrariness of $u$ one finally obtains $\mathcal{D}[B_\star^{-1}]\supset\mathcal{D}[S_B]\cap\ker S^*$. \eqref{eq:D[SB]} is thus established.

 Concerning  \eqref{eq:decomposition_of_form_domains-1}, the identity $\mathcal{D}[S_B]=\mathcal{D}[S_\mathrm{F}]\dotplus\mathcal{D}[B_\star^{-1}]$ is a direct consequence of \eqref{eq:D[SB]} and of \eqref{eq:II-S[f,u]-Krein}.

  Last, owing to the fact that $\mathcal{D}[B_\star^{-1}]$ is closed in $\mathcal{D}[S_B]$, identity \eqref{eq:SB-and-B-1} lifts to $S_B[v_1,v_2]=B_\star^{-1}[v_1,v_2]$ $\forall v_1,v_2\in\mathcal{D}[B_\star^{-1}]$. This and \eqref{eq:II-S[f,u]-Krein-3} then imply \eqref{eq:decomposition_of_form_domains-2}.
\end{proof}

 \begin{remark}
  Unlike the general Vi\v{s}ik-Birman representation of self-adjoint extensions, Theorem \ref{thm:VB-representaton-theorem}, which can be easily generalised to Theorem \ref{thm:VB-representaton-theorem-GENER} by replacing $S_\mathrm{F}$ with a distinguished self-adjoint extension $S_\mathrm{D}$ of $S$ such that $S_\mathrm{D}^{-1}\in\mathcal{B}(\cH)$, with no semi-boundedness assumption on $S$, in the characterisation of semi-bounded self-adjoint extensions provided by Theorems \ref{thm:semibdd_exts_operator_formulation}-\ref{thm:semibdd_exts_form_formulation} the role played by $S_\mathrm{F}$ is much more crucial (let alone the assumption $\mathfrak{m}(S)>0$) and it cannot be replaced by other distinguished self-adjoint extensions. Observe, in particular, that the property $S_\mathrm{F}\geqslant S_B$, whence in particular $\mathcal{D}(S_\mathrm{F})\subset\mathcal{D}[S_B]$, and the property $\mathfrak{m}(S)=\mathfrak{m}(S_\mathrm{F})$ were crucial throughout the reasoning of Lemma \ref{lem:domB-1indomSBform} and Theorems \ref{thm:semibdd_exts_operator_formulation}-\ref{thm:semibdd_exts_form_formulation}.  
 \end{remark}

  \subsection{Parametrisation of Friedrichs and Kre{\u\i}n-von Neumann extensions}\label{sec:II-F-N-paramerisation1}

\begin{proposition}[Parametrisation of $S_\mathrm{F}$ and $S_\mathrm{N}$]\label{prop:parametrisation_SF_SN} Let $S$ be a densely defined and positive symmetric operator with greatest lower bound $\mathfrak{m}(S)>0$ on the Hilbert space $\cH$.
 With respect to the notation of \eqref{eq:defSB} and \eqref{eq:Bto-1}, let $S_B$ be a \emph{positive} self-adjoint extension of $S$. 
\begin{enumerate}[(i)]
 \item $S_B$ is the \emph{Friedrichs extension}\index{Friedrichs extension} $S_\mathrm{F}$ when $\mathcal{D}[B_\star^{-1}]=\{0\}$, i.e., when $\mathcal{D}(B)=\ker S^*$ and $Bu=0$ $\forall u\in\ker S^*$. 
 \item $S_B$ is the \emph{Kre{\u\i}n-von Neumann extension}\index{Kre{\u\i}n-von Neumann extension} $S_\mathrm{N}$ when $\mathcal{D}(B_\star^{-1})=\mathcal{D}[B_\star^{-1}]=\ker S^*$ and $B_\star^{-1}u=0$ $\forall u\in\ker S^*$, i.e., when $\mathcal{D}(B)=\{0\}$.
\end{enumerate}
\end{proposition}

\begin{proof} (i) On account of \eqref{eq:decomposition_of_form_domains-1}-\eqref{eq:decomposition_of_form_domains-2}, $S_B=S_\mathrm{F}$ if and only if $\mathcal{D}[B_\star^{-1}]=\{0\}$. This is equivalent to $\mathcal{D}(B_\star^{-1})=\{0\}$. In turn, on account of \eqref{eq:Bto-1_original} and of $\ker S_\mathrm{F}=\{0\}$, the triviality of $\mathcal{D}(B_\star^{-1})$ implies $\mathrm{ran}\,B=\{0\}$, that is, $B$ is the zero operator on its domain. On the other hand, in the notation of \eqref{eq:U0-U1-H+}, the fact that $S_B=S_\mathrm{F}$ is equivalent to $U_0\equiv\ker S_\mathrm{F}=\{0\}$ and $U_1=\ker S^*$. As $B$ is the null self-adjoint operator in the Hilbert subspace $U_1$ (see \eqref{eq:defB} above), then $\mathcal{D}(B)=\ker S^*$ and $Bu=0$ $\forall u\in\ker S^*$. Conversely, such $B$, owing to the one-to-one correspondence $B\leftrightarrow S_B$ from Theorem \ref{thm:VB-representaton-theorem}, is necessarily the Birman parameter for $S_B=S_\mathrm{F}$.

 (ii) By comparing \eqref{eq:decomposition_of_form_domains-1}-\eqref{eq:decomposition_of_form_domains-2} with \eqref{eq:II-def-SN-first3}, $S_B=S_\mathrm{N}$ if and only if $\mathcal{D}[B_\star^{-1}]=\ker S^*$, in which case $B_\star^{-1}[u]=0$ $\forall u\in\ker S^*$. 
 On the other hand $\ker S_\mathrm{N}=\ker S^*$ (see \eqref{eq:II-def-SN-first2} above); therefore, in the notation of \eqref{eq:U0-U1-H+}, the fact that $S_B=S_\mathrm{N}$ is equivalent to $U_0\equiv\ker S_\mathrm{N}=\ker S^*$ and $U_1=\{0\}$, which through \eqref{eq:defB} implies $\mathcal{D}(B)=\{0\}$. Conversely, such $B$, owing to the one-to-one correspondence $B\leftrightarrow S_B$ from Theorem \ref{thm:VB-representaton-theorem}, is necessarily the Birman parameter for $S_B=S_\mathrm{N}$.
\end{proof}

  Summarising Proposition \ref{prop:parametrisation_SF_SN}, the distinguished extensions $S_\mathrm{F}$ and $S_\mathrm{N}$ are identified a priori in the Vi\v{s}ik-Birman parametrisation according to the scheme
  \begin{equation}\label{eq:B_for_Friedrichs}
\textrm{\qquad\;\, Friedrichs $(S_\mathrm{F})$:}\quad\qquad 
\begin{array}{c}
\mathcal{D}[B_\star^{-1}]=\{0\} \\
\textrm{``$B_\star^{-1}=\infty$''}
\end{array}\quad
\begin{array}{c}
\mathcal{D}(B)=\ker S^* \\
B=\mathbb{O}
\end{array}
\end{equation}
\begin{equation}\label{eq:B_for_Krein-vN}
\textrm{Kre{\u\i}n-von Neumann $(S_\mathrm{N})$:}\quad 
\begin{array}{c}
\mathcal{D}[B_\star^{-1}]=\ker S^* \\
B_\star^{-1}=\mathbb{O}
\end{array}\quad
\begin{array}{c}
\mathcal{D}(B)=\{0\} \\
\textrm{``$B=\infty$''}\,.
\end{array}
\end{equation}
  It is indeed customary to write formally ``$B_\star^{-1}=\infty$'' for $S_B=S_\mathrm{F}$ and ``$B=\infty$'' for $S_B=S_\mathrm{N}$, consistently with the extension ordering \eqref{eq:extension_ordering}.

  \begin{remark}
  With respect to the generalised representation
  \[
   \mathcal{D}(S_B)\;=\;\mathcal{D}(\overline{S})\,\dotplus\,(S_\mathrm{D}^{-1}+B)\widetilde{U}_1\,\dotplus\,U_0
  \]
  provided by Theorem \ref{thm:VB-representaton-theorem-GENER},
  there is no canonical choice any longer for the parameter $B$ to select the Friedrichs extension $S_\mathrm{F}$, if it exists. However, it remains true that the extension parameter
  \[
   \mathcal{D}(B)\,=\,\ker S^*\,,\qquad B\,=\,\mathbb{O}
  \]
  parametrises the distinguished extension $S_\mathrm{D}$.
\end{remark}

  \section{Kre{\u\i}n-Vi\v{s}ik-Birman self-adjoint extension theory re-parametrised}\label{sec:II-VBreparametrised}
 
 The Vi\v{s}ik-Birman\index{Kre{\u\i}n-Vi\v{s}ik-Birman theory} representation of the self-adjoint extensions of densely defined and positive symmetric operators (Theorem \ref{thm:VB-representaton-theorem}), or also its generalised version for densely defined and symmetric operators admitting a distinguished self-adjoint extensions with everywhere defined bounded inverse (Theorem  \ref{thm:VB-representaton-theorem-GENER}), was originally devised by Vi\v{s}ik \cite{Vishik-1949,Vishik-1952}, as mentioned already, in a framework of boundary value problems on domains, where meaningful extension conditions at the boundary and boundary forms are expressed directly and conveniently in terms of the extension parameter $B$. In other contexts, including the study of self-adjoint realisations of symmetric operators that represent formal quantum Hamiltonians, a somewhat more natural role is played by the Birman extension parameter $B_\star^{-1}$, instead of $B$, as it is rather through $B_\star^{-1}$ that specific spectral properties of each extension are monitored. In terms of $B_\star^{-1}$ a re-parametrisation of the extensions representation can be worked out, which is more frequently used in applications. This Section presents such an equivalent formulation of the theory. In retrospect, it is also worth referring to the corpus of results collected here as the Kre{\u\i}n-Vi\v{s}ik-Birman self-adjoint extension theory as a whole: in fact, it is already evident at this stage that Kre{\u\i}n's theorem on positive self-adjoint extensions (Theorem \ref{thm:II-Krein-final-thm}) is entirely contained in the Vi\v{s}ik-Birman classification provided by Theorems \ref{thm:semibdd_exts_operator_formulation} and \ref{thm:semibdd_exts_form_formulation}.

 Such a re-parametrisation from the original Vi\v{s}ik-Birman one is presented here following the recent analysis \cite{GMO-KVB2017}. There are various other precursors in the literature where the self-adjoint extension classification appears in the form of an explicit Vi\v{s}ik-Birman representation, or some strictly analogous expression, in which the extension parameter is (the analogue of) Birman's operator $B_\star^{-1}$, rather than $B$. Subsequent to the original works by Kre{\u\i}n, Vi\v{s}ik, and Birman, this was done by Grubb \cite{Grubb-1968} in 1968, by Faris \cite{faris-1975} in 1975, and by Alonso and Simon \cite{Alonso-Simon-1980,Alonso-Simon-1980-2} in 1980. (This does not include the separate development of abstract extension theories, started in those very years and active until current times, where the operator concept gradually began to be replaced by the notion of \emph{relation}\index{relation}, and led to the modern framework of boundary triplets\index{boundary triplets}.) The above-mentioned contributions all contain several \emph{partial} aspects of the complete scheme presented in this and in the following Section. In particular, with reference to the results discussed here, Grubb's analysis yields Theorem \ref{thm:VB-representaton-theorem_Tversion2} (see \cite[Theorem 13.6]{Grubb-DistributionsAndOperators-2009}), the bound \eqref{eq:bounds_mS_mB_Tversion} of Theorem \ref{thm:semibdd_exts_operator_formulation_Tversion} (see \cite[Theorem 13.17]{Grubb-DistributionsAndOperators-2009}), Theorem \ref{thm:VB-representaton-theorem_Grubbversion} (see \cite[Theorem II.2.1]{Grubb-1968}, as well as Kre{\u\i}n's decomposition formula for quadratic forms of Theorem \ref{thm:II-S[f,u]-Krein-2} (see \cite[Theorem 13.19]{Grubb-DistributionsAndOperators-2009}), whereas the Alonso-Simon analysis is restricted to the positive self-adjoint extension problem.

 The simple, but key point in the mentioned re-parametrisation is the following `inversion mechanism', that elucidates the link between $B$ and $B_\star^{-1}$.

\begin{proposition}\label{prop:Birmans_inversion}
Let $\mathcal{K}$ be a Hilbert space and let $\mathcal{S}(\mathcal{K})$ be the collection of the self-adjoint operators acting in Hilbert subspaces of $\mathcal{K}$. Given $T\in\mathcal{S}(\mathcal{K})$, let $V$ be the closed subspace of $\mathcal{K}$ which $T$ acts in, with domain $\mathcal{D}(T)\equiv\widetilde{V}$ dense in $V$, and let $W:=V^\perp$, i.e., $\mathcal{K}=V\oplus W$. Let $\phi(T)$ be the densely defined operator acting on the Hilbert subspace $\overline{\ran\,T}\oplus W$ of $\mathcal{K}$ defined by
\begin{equation}\label{eq:def-phi}
\begin{split}
\mathcal{D}(\phi(T))\;&:=\;\ran\,T\,\boxplus\,W\,, \\
\phi(T) Tv\;&:=\;v\qquad\forall v\in\mathcal{D}(T)\!\cap\!(\ker T)^\perp\,,\\
\phi(T) w\;&:=\;0\qquad\forall w\in W\,.
\end{split} 
\end{equation}
Then:
\begin{enumerate}[(i)]
 \item $\phi(T)\in\mathcal{S}(\mathcal{K})$;
 \item the map $\phi:\mathcal{S}(\mathcal{K})\to \mathcal{S}(\mathcal{K})$ is a bijection on $\mathcal{S}(\mathcal{K})$;
  \item $\phi^2$ is the identity map on $\mathcal{S}(\mathcal{K})$, that is, $\phi^{-1}=\phi$.
\end{enumerate}
\end{proposition}

\begin{remark}\label{rem:phi-inversion}
In short, $\phi$ provides a transformation of $T$ that is quite close to an inversion: it is the inverse of $T$ on its range, plus the zero operator on the orthogonal complement in $\mathcal{K}$ of the Hilbert subspace where $T$ acts on. In particular, if $T$ is densely defined in $\mathcal{K}$ itself and invertible, then $\phi(T)=T^{-1}$, that is, $\phi$ is precisely the inversion transformation.
\end{remark}

\begin{remark}\label{rem:B-1_phi_B}
By comparing \eqref{eq:Bto-1} and \eqref{eq:def-phi} in the special case  
$\mathcal{K}=\ker S^*$, $T=B$, $V=U_1$, and $W=U_0=\ker S_B$, one concludes that
\begin{equation}\label{eq:thetwoopparameters}
 B_\star^{-1}\;=\;\phi(B)\,,
\end{equation}
that is, the Birman operator $B_\star^{-1}$ is precisely the $\phi$-inversion of the Vi\v{s}ik operator $B$.
\end{remark}

\begin{proof}[Proof of Proposition \ref{prop:Birmans_inversion}]
The operator $(B|_{\mathcal{D}(T)\cap(\ker T)^\perp})^{-1}$ is self-adjoint in $\overline{\ran\,T}$ (Sect.~\ref{sec:I-symmetric-selfadj}), and its operator orthogonal sum with $\mathbb{O}$ on $W$, namely $\phi(T)$, is then self-adjoint in $\overline{\ran\,T}\oplus W$. For the rest of the proof, it obviously suffices to show that $\phi(\phi(T))=T$ for any $T\in\mathcal{S}(\mathcal{K})$. By construction $\phi(T)$ acts in the Hilbert space $V':=\overline{\ran T}\oplus W$ with domain $\widetilde{V}':=\ran T\boxplus W$ dense in $V'$. Setting $W':={V'}^\perp$, in view of the decomposition $\mathcal{K}=V'\oplus W'$ the operator $\phi(\phi(T))$ is therefore defined in the Hilbert subspace $\overline{\ran\,\phi(T)}\oplus W'$ of $\mathcal{K}$ according to
\[
\begin{split}
\mathcal{D}(\phi(\phi(T)))\;&:=\;\ran\,\phi(T)\,\boxplus\,W'\,, \\
\phi(\phi(T)) \phi(T)v'\;&:=\;v'\qquad\forall v'\in\mathcal{D}(\phi(T))\!\cap\!(\ker\phi(T))^\perp\,, \\
\phi(\phi(T)) w'\;&:=\;0\;\qquad\forall w'\in W'\,.
\end{split}
\]
Since $V=\overline{\ran\,T}\oplus\ker T$, and hence  $\mathcal{K}=\overline{\ran\,T}\oplus\ker T\oplus W$, then $W'=\ker T$. Thus,
\[
\mathcal{D}(\phi(\phi(T)))\;=\;\ran\,\phi(T)\,\boxplus\, W'\;=\;(\mathcal{D}(T)\cap\overline{\ran\,T})\,\boxplus\,\ker T \;=\;\mathcal{D}(T)\,.
\]
Now, $\phi(\phi(T)) w'=0=Tw'$ $\forall w'\in W'=\ker T$. It then remains to show that $\phi(\phi(T))$ and $T$ agree also on $\mathcal{D}(T)\cap\overline{\ran\,T})$. In fact, if $v$ is a vector in such a subspace, then $v=\phi(T) Tv$ and in view of $\phi(\phi(T)) \phi(T)Tv=Tv$ one has $\phi(\phi(T))v=Tv$. This completes the proof that 
$\phi(\phi(T))v=Tv$ for any $v\in\mathcal{D}(T)$ and since the two operators have also the same domain, the conclusion is $\phi(\phi(T))=T$.
\end{proof}

 A systematic application of Proposition \ref{prop:Birmans_inversion} allows to turn the classification results of Sect.~\ref{sec:II-VBparametrisation-orig} into the following equivalent formulations.

\begin{theorem}[Classification of self-adjoint extensions -- operator version]\label{thm:VB-representaton-theorem_Tversion}\index{theorem!Kre{\u\i}n-Vi\v{s}ik-Birman (self-adjoint extension)}
Let $S$ be a densely defined and positive symmetric operator with greatest lower bound $\mathfrak{m}(S)>0$ on the Hilbert space $\cH$. 
  There is a one-to-one correspondence between the  family of the self-adjoint extensions of $S$ in $\cH$ and the family of the self-adjoint operators in Hilbert subspaces of $\ker S^*$. If $T$ is any such operator, in the correspondence $T\leftrightarrow S_T$ each self-adjoint extension $S_T$ of $S$ is given by
\begin{equation}\label{eq:ST}
\begin{split}
S_T\;&=\;S^*\upharpoonright\mathcal{D}(S_T)\,, \\
\mathcal{D}(S_T)\;&=\;\left\{f+S_\mathrm{F}^{-1}(Tv+w)+v\left|
\begin{array}{c}
f\in\mathcal{D}(\overline{S})\,,\;v\in\mathcal{D}(T) \\
w\in\ker S^*\cap\mathcal{D}(T)^\perp
\end{array}
\right.\right\}.
\end{split}
\end{equation}
\end{theorem}

 \begin{theorem}[Generalised classification -- operator version]\label{thm:VB-representaton-theorem_Tversion2}\index{theorem!Kre{\u\i}n-Vi\v{s}ik-Birman (self-adjoint extension)}
 Let $S$ be a densely defined symmetric operator on the Hilbert space $\cH$, which admits a self-adjoint extension $S_\mathrm{D}$ that has everywhere defined bounded inverse on $\cH$. There is a one-to-one correspondence between the  family of the self-adjoint extensions of $S$ in $\cH$ and the family of the self-adjoint operators in Hilbert subspaces of $\ker S^*$. If $T$ is any such operator, in the correspondence $T\leftrightarrow S_T$ each self-adjoint extension $S_T$ of $S$ is given by
\begin{equation}\label{eq:ST-2}
\begin{split}
S_T\;&=\;S^*\upharpoonright\mathcal{D}(S_T) \\
\mathcal{D}(S_T)\;&=\;\left\{f+S_\mathrm{D}^{-1}(Tv+w)+v\left|
\begin{array}{c}
f\in\mathcal{D}(\overline{S})\,,\;v\in\mathcal{D}(T) \\
w\in\ker S^*\cap\mathcal{D}(T)^\perp
\end{array}
\right.\right\}.
\end{split}
\end{equation}
\end{theorem}

\begin{theorem}[Characterisation of semi-bounded extensions]\label{thm:semibdd_exts_operator_formulation_Tversion}\index{theorem!Kre{\u\i}n-Vi\v{s}ik-Birman (semi-bounded self-adjoint extension)}
Let $S$ be a densely defined and positive symmetric operator with greatest lower bound $\mathfrak{m}(S)>0$ on the Hilbert space $\cH$.
If, with respect to the notation of \eqref{eq:ST}, $S_T$ is a self-adjoint extension of $S$, and if $\alpha<\mathfrak{m}(S)$, then
\begin{equation}\label{eq:SBsmbb-iff-invBsmbb_Tversion}
\begin{split}
\langle g,S_T g\rangle\;&\geqslant\;\alpha\,\|g\|^2\qquad\forall g\in\mathcal{D}(S_T) \\
& \Updownarrow \\
\langle v,T v\rangle\;\geqslant\;\alpha\|v\|^2+\:&\alpha^2\langle v,(S_\mathrm{F}-\alpha\mathbbm{1})^{-1} v\rangle\qquad\forall v\in\mathcal{D}(T)\,.
\end{split}
\end{equation}
As an immediate consequence,
\begin{equation}\label{eq:II-mTgeqmST}
 \mathfrak{m}(T)\;\geqslant\;\mathfrak{m}(S_T)
\end{equation}
for any semi-bounded $S_T$.
In particular, positivity or strict positivity of the bottom of $S_T$ is equivalent to the same property for $T$, that is,
 \begin{equation}\label{eq:positiveSBiffpositveB-1_Tversion}
 \begin{split}
 \mathfrak{m}(S_T)\;\geqslant \;0\quad&\Leftrightarrow\quad \mathfrak{m}(T)\;\geqslant\; 0 \\
 \mathfrak{m}(S_T)\;> \;0\quad&\Leftrightarrow\quad \mathfrak{m}(T)\;>\; 0\,.
 \end{split}
 \end{equation}
Moreover, if $\mathfrak{m}(T)>-\mathfrak{m}(S)$, then
 \begin{equation}\label{eq:bounds_mS_mB_Tversion}
 \mathfrak{m}(T)\;\geqslant\; \mathfrak{m}(S_T)\;\geqslant\;\frac{\mathfrak{m}(S) \,\mathfrak{m}(T)}{\mathfrak{m}(S)+\mathfrak{m}(T)}\,.
 \end{equation}
\end{theorem}

\begin{theorem}[Characterisation of semi-bounded extensions -- form version]\label{thm:semibdd_exts_form_formulation_Tversion}\index{theorem!Kre{\u\i}n-Vi\v{s}ik-Birman (quadratic forms of semi-bounded self-adjoint extension)}
Let $S$ be a densely defined and positive symmetric operator with greatest lower bound $\mathfrak{m}(S)>0$ on the Hilbert space $\cH$. With respect to the notation of \eqref{eq:ST}, let $S_T$ be a  \emph{lower semi-bounded} (not necessarily positive) self-adjoint extension of $S$. Then
\begin{equation}\label{eq:D[SB]_Tversion}
\mathcal{D}[T]\;=\; \mathcal{D}[S_T]\,\cap\,\ker S^*
\end{equation}
and
 \begin{equation}\label{eq:decomposition_of_form_domains_Tversion}
 \begin{split}
 \mathcal{D}[S_T]\;&=\;\mathcal{D}[S_\mathrm{F}]\,\dotplus\,\mathcal{D}[T] \, ,\\
 S_T[f+v,f'+v']\;&=\;S_\mathrm{F}[f,f']\,+\,T[v,v'] \, ,\\
 &\forall f,f'\in\mathcal{D}[S_\mathrm{F}],\;\forall v,v'\in\mathcal{D}[T]\,.
 \end{split}
\end{equation}
As a consequence,
\begin{equation}\label{eq:extension_ordering_Tversion}
S_{T_1}\,\geqslant\,S_{T_2}\qquad\Leftrightarrow\qquad T_1\,\geqslant\,T_2
\end{equation}
and
\begin{equation}
T\;\geqslant\;S_T\,.
\end{equation}
\end{theorem}

\begin{proposition}[Parametrisation of $S_\mathrm{F}$ and $S_\mathrm{N}$]\label{prop:parametrisation_SF_SN_Tversion}
Let $S$ be a densely defined and positive symmetric operator with greatest lower bound $\mathfrak{m}(S)>0$ on the Hilbert space $\cH$.
With respect to the notation of \eqref{eq:ST}, let $S_T$ be a \emph{positive} self-adjoint extension of $S$. 
\begin{enumerate}[(i)]
 \item $S_T$ is the \emph{Friedrichs extension}\index{Friedrichs extension} when $\mathcal{D}[T]=\{0\}$ (``$\,T=\infty$'').
 \item $S_T$ is the \emph{Kre{\u\i}n-von Neumann extension}\index{Kre{\u\i}n-von Neumann extension} when $\mathcal{D}(T)=\mathcal{D}[T]=\ker S^*$ and $Tu=0$ $\forall u\in\ker S^*$ ($\,T=\mathbb{O}$).
\end{enumerate}
\end{proposition}

\begin{proof}[Proof of Theorems \ref{thm:VB-representaton-theorem_Tversion} and \ref{thm:VB-representaton-theorem_Tversion2}]
 In the following it will be shown how to deduce Theorem \ref{thm:VB-representaton-theorem_Tversion} from Theorem \ref{thm:VB-representaton-theorem}. In a completely analogous manner, Theorem \ref{thm:VB-representaton-theorem_Tversion2} is deduced from Theorem \ref{thm:VB-representaton-theorem-GENER}.

Let $S_B$ be a generic self-adjoint extension of $S$, pa\-ra\-me\-tri\-sed by $B$ according to Theorem \ref{thm:VB-representaton-theorem}, formula \eqref{eq:defSB}. Correspondingly, let $B_\star^{-1}$ be the Birman's operator \eqref{eq:Bto-1_original}-\eqref{eq:Bto-1}. Claim: $S_B$ is precisely of the form $S_T$ in \eqref{eq:ST} above, where $T=B_\star^{-1}$.

To prove that, consider a generic element  $g=f+(S_\mathrm{F}^{-1}+B)\widetilde{u}_1+u_0$  of $\mathcal{D}(S_B)$, as given by the decomposition \eqref{eq:defSB} for some $f\in\mathcal{D}(\overline{S})$, $\widetilde{u}_1\in\widetilde{U}_1=\mathcal{D}(B)$, and $u_0\in U_0=\ker S^*\cap\mathcal{D}(B)^\perp=\ker S_B$. One writes $\widetilde{u}_1=z+w$ for some $w\in\ker B$ and some $z\in \mathcal{D}(B)\cap\overline{\ran\,B}$ that are uniquely identified by  the decomposition $U_1=\overline{\mathcal{D}(B)}=\overline{\ran\,B}\oplus\ker B$, $\mathcal{D}(B)=(\mathcal{D}(B)\cap\overline{\ran\,B})\boxplus\ker B$. Owing to \eqref{eq:Bto-1}, $v:=B\widetilde{u}_1+u_0=B z+u_0\in\mathcal{D}(B_\star^{-1})$ and $B_\star^{-1} v=z$. Moreover, from
\[
\ker S^*\;=\;U_0\,\oplus\,U_1\;=\;\ker S_B\,\oplus\,\overline{\ran\,B}\,\oplus\,\ker B\;=\;\overline{\mathcal{D}(B_\star^{-1})}\,\oplus\,\ker B
\]
one deduces that $\ker B=\ker S^*\cap \mathcal{D}(B_\star^{-1})^\perp$.
Therefore,
\[
\begin{split}
g\;&=\;f+S_\mathrm{F}^{-1}\widetilde{u}_1+B\widetilde{u}_1+u_0\;=\;f+S_\mathrm{F}^{-1}(z+w)+v \\
&=\;f+S_\mathrm{F}^{-1}(B_\star^{-1} v+w)+v  \, ,\\
&\qquad\qquad v\in\mathcal{D}(B_\star^{-1})\,,\;w\in\ker S^*\cap \mathcal{D}(B_\star^{-1})^\perp\,,
\end{split}
\]
that is, $g$ is an element of $\mathcal{D}(S_T)$ defined in \eqref{eq:ST} above with $T=B_\star^{-1}$. It is straightforward to go through the same arguments and decompositions in reverse order to conclude that \emph{any} vector of the form $S_\mathrm{F}^{-1}(B_\star^{-1} v+w)+v$, where $ v\in\mathcal{D}(B_\star^{-1})$ and $w\in\ker S^*\cap \mathcal{D}(B_\star^{-1})^\perp$, can be re-written as $(S_\mathrm{F}^{-1}+B)\widetilde{u}_1+u_0$ for  $\widetilde{u}_1\in\widetilde{U}_1=\mathcal{D}(B)$, and $u_0\in U_0$ determined by
\[
\begin{split}
B_\star^{-1}v+w\;&=\;\widetilde{u}_1\,, \\
v\;&=\;B\widetilde{u}_1+u_0\,,
\end{split}
\]
which proves that any $g\in\mathcal{D}(S_T)$ is also an element of $\mathcal{D}(S_B)$. Thus,  \eqref{eq:defSB} and \eqref{eq:ST} define the same domain: $\mathcal{D}(S_B)=\mathcal{D}(S_T)$ for $T=B_\star^{-1}$. Since $S_B$ and $S_T$ are the restrictions to such a common domain of the same operator $S^*$, then $S_B=S_T$ for $T=B_\star^{-1}$, and the above claimed statement is proved.

As a consequence of this and of the one-to-one correspondence $S_B\leftrightarrow B$ of Theorem \ref{thm:VB-representaton-theorem}, the self-adjoint extensions of $S$ are \emph{all} of the form $S_T$ of \eqref{eq:ST} for some self-adjoint operator $T$ on a Hilbert subspace of $\ker S^*$.

What remains to be proved is that when $T$ runs in the family $\mathcal{S}(\ker S^*)$ of the self-adjoint operators on Hilbert subspaces of $\ker S^*$, the corresponding $S_T$'s give the whole family of self-adjoint extensions of $S$. This follows at once from \eqref{eq:thetwoopparameters}, since, owing to \eqref{eq:defSB} $B_\star^{-1}=\phi(B)=T$, and $\phi$ is a bijection in $\mathcal{S}(\ker S^*)$ (Proposition \ref{prop:Birmans_inversion}).
\end{proof}

\begin{proof}[Proof of Theorem \ref{thm:semibdd_exts_operator_formulation_Tversion}, Theorem \ref{thm:semibdd_exts_form_formulation_Tversion}, and Proposition \ref{prop:parametrisation_SF_SN_Tversion}]~ All the statements follow at once from their original versions, respectively Theorem \ref{thm:semibdd_exts_operator_formulation}, Theorem \ref{thm:semibdd_exts_form_formulation}, and Proposition \ref{prop:parametrisation_SF_SN}, and  from the fact that the extension parameter $T$ is precisely the parameter $B_\star^{-1}$ in Theorem \ref{thm:semibdd_exts_operator_formulation}, Theorem \ref{thm:semibdd_exts_form_formulation}, and Proposition \ref{prop:parametrisation_SF_SN}.
\end{proof}

\begin{remark}[Equivalence of Theorems \ref{thm:VB-representaton-theorem} and \ref{thm:VB-representaton-theorem_Tversion}]
Theorem \ref{thm:VB-representaton-theorem_Tversion}'s proof actually shows also that the original Vi\v{s}ik-Birman representation Theorem \ref{thm:VB-representaton-theorem} can be \emph{deduced} from Theorem \ref{thm:VB-representaton-theorem_Tversion} and that therefore the two theorems are equivalent -- and so too are Theorems \ref{thm:VB-representaton-theorem-GENER} and \ref{thm:VB-representaton-theorem_Tversion2}. Indeed, assuming the representation \eqref{eq:ST} for a generic self-adjoint extension $S_T$ of $S$, the same argument of Theorem \ref{thm:VB-representaton-theorem_Tversion}'s proof shows that in terms of the `$\phi$-inverse' $B:=\phi(T)$ of the parameter $T$ one can re-write $\mathcal{D}(S_T)$ in the form $\mathcal{D}(S_B)$ of \eqref{eq:defSB}, and therefore all self-adjoint extensions of $S$ have the form $S_B=S^*\upharpoonright\mathcal{D}(S_B)$ for some $B\in\mathcal{S}(\ker S^*)$. Moreover, since $\phi$ is a bijection on $\mathcal{S}(\ker S^*)$ (Proposition \ref{prop:Birmans_inversion}), one concludes that when $B$ runs over $\mathcal{S}(\ker S^*)$ the corresponding $S_B$ exhausts the whole family of self-adjoint extensions of $S$, thus obtaining Theorem \ref{thm:VB-representaton-theorem}.
\end{remark}

 With respect to self-adjoint extension representation now formulated as in Theorem \ref{thm:VB-representaton-theorem_Tversion}, yet another equivalent formulation is obtained.

\begin{theorem}[Classification of self-adjoint extensions -- operator version]\label{thm:VB-representaton-theorem_Grubbversion}\index{theorem!Kre{\u\i}n-Vi\v{s}ik-Birman (self-adjoint extension)}
Let $S$ be a densely defined and positive symmetric operator with greatest lower bound $\mathfrak{m}(S)>0$ on the Hilbert space $\cH$. 
  There is a one-to-one correspondence between the family of the self-adjoint extensions of $S$ in $\cH$ and the family of the self-adjoint operators in Hilbert subspaces of $\ker S^*$. 
 Let $T$ be any such operator, let $P_T:\cH\to\cH$ be the orthogonal projection onto $\overline{\mathcal{D}(T)}$, and let $P_*:\mathcal{D}(S^*)\to\mathcal{D}(S^*)$ be the (non-orthogonal, in general) projection onto $\ker S^*$ with respect to Kre{\u\i}n's decomposition formula $\mathcal{D}(S^*)=\mathcal{D}(S_\mathrm{F})\dotplus\ker S^*$ (Proposition \ref{prop:II-KVB-decomp-of-Sstar}). Then:
 \begin{enumerate}[(i)]
  \item In the correspondence $T\leftrightarrow S_T$, each self-adjoint extension $S_T$ of $S$ is given by
\begin{equation}\label{eq:ST_Grubb}
\begin{split}
S_T\;&=\;S^*\upharpoonright\mathcal{D}(S_T)\,, \\
\mathcal{D}(S_T)\;&=\;\left\{g\in\mathcal{D}(S^*)\left|\!
\begin{array}{c}
P_* g\in\mathcal{D}(T)\textrm{ and} \\
P_T S^*g=TP_{\!*} g
\end{array}\!
\right.\right\}.
\end{split}
\end{equation}
 \item The extension parameter $T$ in \eqref{eq:ST_Grubb} is precisely the same as in \eqref{eq:ST}, that is, the  self-adjoint extension representation given here is the same as the one given in Theorem \ref{thm:VB-representaton-theorem_Tversion}. In particular, given a self-adjoint extension $\widetilde{S}$ of $S$, its extension parameter $T$ (i.e., the operator $T$ for which $\widetilde{S}=S_T$) is the operator acting on the Hilbert space $\overline{P_*\mathcal{D}(\widetilde{S})}$ with domain $\mathcal{D}(T)=P_*\mathcal{D}(\widetilde{S})$ and action $TP_{\!*} g=P_T S_Tg$ $\forall g\in\mathcal{D}(\widetilde{S})$.
 \end{enumerate}
\end{theorem}


\begin{proof}
All one needs to prove is that the domain $\mathcal{D}(S_T)$ given by \eqref{eq:ST} can be re-written in the form \eqref{eq:ST_Grubb} \emph{with the same} $T$. If $g=f+S_\mathrm{F}^{-1}(Tv+w)+v$ is a generic element of the space $\mathcal{D}(S_T)$ according to \eqref{eq:ST}, then $P_*g=v$ (by Kre{\u\i}n's decomposition formula \eqref{eq:II-KreinDecomp}), $S^*g=\overline{S}f+Tv+w$, and $P_T S^*g=Tv$. Thus, $P_*g\in\mathcal{D}(T)$ and   $TP_*g=Tv=P_T S^*g$, which proves that $g$ belongs to the domain defined in  \eqref{eq:ST_Grubb}. For the converse, recall that for any $g\in\mathcal{D}(S^*)$ the Vi\v{s}ik-Birman decomposition formula \eqref{eq:II-VishBirDecomp} gives $g=f+S_\mathrm{F}^{-1}u+P_* g$ for some $f\in\mathcal{D}(\overline{S})$ and $u\in\ker S^*$. If now $g$ belongs to the domain defined in \eqref{eq:ST_Grubb}, then $v:=P_* g\in\mathcal{D}(T)\subset\ker S^*$ for some $T\in\mathcal{S}(\ker S^*)$, and the decomposition $\ker S^*=\overline{\mathcal{D}(T)}\oplus(\ker S^*\cap\mathcal{D}(T)^\perp)$ gives $u=P_Tu+w$ for some $w\in\ker S^*\cap\mathcal{D}(T)^\perp$. Since $P_T S^*g=P_T(\overline{S}f+u)=P_Tu$, but also $P_T S^*g=Tv$ (by \eqref{eq:ST_Grubb}), then $P_Tu=Tv$ and $u=Tv+w$. This proves that $g=f+S_\mathrm{F}^{-1}(Tv+w)+v$, which belongs to the domain $\mathcal{D}(S_T)$ given by \eqref{eq:ST}. Thus, \eqref{eq:ST} and \eqref{eq:ST_Grubb} define (for the same $T$) the same space $\mathcal{D}(S_T)$.
\end{proof}

\section{Invertibility, semi-boundedness, and negative spectrum in the Kre{\u\i}n-Vi\v{s}ik-Birman extension scheme}\markboth{Classical self-adjoint extension schemes}{Invertibility, semi-boundedness, and negative spectrum}\label{sec:II-spectralKVB}

 Given\index{Kre{\u\i}n-Vi\v{s}ik-Birman theory} a densely defined and lower semi-bounded symmetric operator on Hilbert space, say, with strictly positive lower bound, the possibility of monitoring invertibility, semi-boundedness, and negative spectrum of one of its self-adjoint extensions directly and explicitly in terms of Birman's extension parameter of the Vi\v{s}ik-Birman representation, provides a valuable advantage as compared to von Neumann's extension classification.

 Here the extension parametrisation $S_T\leftrightarrow T$ (Sect.~\ref{sec:II-VBreparametrised}) shall be exploited. A first important link between $S_T$ and $T$, which is straightforward although not made explicit in the original Vi\v{s}ik-Birman analysis, is the following.

\begin{theorem}[Invertibility]\label{thm:invertibility}
Let $S$ be a densely defined and positive symmetric operator with greatest lower bound $\mathfrak{m}(S)>0$ on the Hilbert space $\cH$, and let $S_T$ be a self-adjoint extension of $S$ according to the parametrisation \eqref{eq:ST} of Theorem \ref{thm:VB-representaton-theorem_Tversion}. Then
\begin{enumerate}[(i)]
 \item $\ker S_T=\ker T$, and therefore $S_T$ is injective $\Leftrightarrow$ $T$ is injective;
 \item $S_T$ is surjective $\Leftrightarrow$ $T$ is surjective;
  \item $S_T$ is invertible on the whole $\cH$ $\Leftrightarrow$ $T$ is invertible on the whole $\overline{\mathcal{D}(T)}$.
\end{enumerate}
\end{theorem}

\begin{proof} 
 If $v\in\ker T$, then according to \eqref{eq:ST} $v=f+S_\mathrm{F}^{-1}(Tv+w)+v\in\mathcal{D}(S_T)$ with $f=w=0$, whence $S_\mathrm{F}v=Tv=0$, that is, $v\in\ker S_T$. Conversely, if $g\in\ker S_T$, then $g=f+S_\mathrm{F}^{-1}(Tv+w)+v\in\mathcal{D}(S_T)$ for suitable $f,v,w$ according to \eqref{eq:ST} and $0=S_Tg=\overline{S}f+Tv+w\in\ran\overline{S}\boxplus\ran T\boxplus(\ker S^*\cap\mathcal{D}(T)^\perp)$: this implies $\overline{S}f=Tv=w=0$, which is the same as $f=Tv=w=0$, owing to the injectivity of $\overline{S}$. In particular, $g=v\in\ker T$. This completes the proof of (i). Concerning (ii), in the notation of \eqref{eq:ST} one has that  $\ran \,S_T=\ran\,\overline{S}\boxplus\ran\,T\boxplus(\ker S^*\cap\mathcal{D}(T)^\perp)$. Thus, $T$ is surjective $\Leftrightarrow$ $\ran\,T\boxplus(\ker S^*\cap\mathcal{D}(T)^\perp)=\overline{\ran\,T}\oplus(\ker S^*\cap\mathcal{D}(T)^\perp)=\overline{\mathcal{D}(T)}\oplus(\ker S^*\cap\mathcal{D}(T)^\perp)=\ker S^*$ $\Leftrightarrow$ $\ran\, S_T=\ran\,\overline{S}\oplus\ker S^*=\cH$ $\Leftrightarrow$ $S_T$ is surjective. Last, (iii) is an obvious consequence of (i) and (ii).
\end{proof}

\begin{remark}\label{rem:no_invertibility_control_with_SB}
Vi\v{s}ik-Birman's original parametrisation $S_B\leftrightarrow B$ for the extensions does \emph{not} allow to control invertibility of the extension in terms of the extension parameter, as opposed to the parametrisation $S_T\leftrightarrow T$. Compare indeed 
 \begin{eqnarray}
  \ker S_T\,&=&\, \ker T\,, \\
  \ker S_B \,&=&\, \ker S^*\cap\mathcal{D}(B)^\perp \label{eq:II-kerSBagain}
 \end{eqnarray}
 (the former following from Theorem \ref{thm:invertibility}(i), the latter from \eqref{eq:II-kerSB}). Clearly,  \eqref{eq:II-kerSBagain} shows that the injectivity of $S_B$ and the injectivity of $B$ are unrelated. Moreover, the identity $\ran\,S_B=\ran\,\overline{S}\oplus\mathcal{D}(B)$ (see \eqref{eq:II-ranSB} above) shows that the surjectivity of $S_B$ and the surjectivity of $B$ are unrelated too.
\end{remark}

  Theorem \ref{thm:invertibility} is immediately generalised to the setting of Theorem \ref{thm:VB-representaton-theorem_Tversion2}, namely when the given $S$ has a self-adjoint extension $S_\mathrm{D}$ invertible on the whole $\cH$. The proof can be repeated verbatim, replacing $S_\mathrm{F}$ wit $S_\mathrm{D}$.

  \begin{theorem}[Invertibility -- generalised version]\label{thm:invertibility-gen}
  Let $S$ be a densely defined symmetric operator on the Hilbert space $\cH$, which admits a self-adjoint extension $S_\mathrm{D}$ that has everywhere defined bounded inverse on $\cH$, and let $S_T$ be a self-adjoint extension of $S$ according to the parametrisation \eqref{eq:ST-2} of Theorem \ref{thm:VB-representaton-theorem_Tversion2}. Then
\begin{enumerate}[(i)]
 \item $\ker S_T=\ker T$, and therefore $S_T$ is injective $\Leftrightarrow$ $T$ is injective;
 \item $S_T$ is surjective $\Leftrightarrow$ $T$ is surjective;
  \item $S_T$ is invertible on the whole $\cH$ $\Leftrightarrow$ $T$ is invertible on the whole $\overline{\mathcal{D}(T)}$.
\end{enumerate}
\end{theorem}

 One further important link between $S_T$ and $T$, again not explicitly present in the Vi\v{s}ik-Birman discussion but analysable in that scheme, concerns the property of semi-boundedness, and in particular the \emph{semi-boundedness problem},\index{semi-boundedness problem} namely the problem of finding conditions under which the semi-boundedness of $S_T$ and of $T$ are equivalent (in general or under special circumstances).

 \begin{theorem}[Semi-boundedness]\label{thm:semi-boundedness_of_ST} 
 Let $S$ be a densely defined and positive symmetric operator with greatest lower bound $\mathfrak{m}(S)>0$ on the Hilbert space $\cH$, let $P_K:\cH\to\cH$ be the orthogonal projection onto $\ker S^*$, and for each $\alpha<\mathfrak{m}(S)$ let 
\begin{equation}\label{eq:Weyl_M}
M(\alpha)\;:=\;P_K(\alpha\mathbbm{1}+\alpha^2(S_\mathrm{F}-\alpha\mathbbm{1})^{-1})P_K\;=\;P_K(\alpha S_\mathrm{F}(S_\mathrm{F}-\alpha\mathbbm{1})^{-1})P_K\,.
\end{equation}
 Furthermore, let $S_T$ be a self-adjoint extension of $S$ according to the parametrisation \eqref{eq:ST} of Theorem \ref{thm:VB-representaton-theorem_Tversion}. Assume that $\mathfrak{m}(T)\in[-\infty,0)$, that is, $T$ is either unbounded below or with finite negative bottom (otherwise it is already known by \eqref{eq:positiveSBiffpositveB-1_Tversion} from Theorem \ref{thm:semibdd_exts_operator_formulation_Tversion} that
$\mathfrak{m}(T)\geqslant 0$ $\Leftrightarrow$ $\mathfrak{m}(S_T)\geqslant 0$).
Then the two conditions
\begin{enumerate}[(i)]
 \item $S_T$ is lower semi-bounded (on $\cH$)
 \item $T$ is lower semi-bounded (on $\overline{\mathcal{D}(T)}$)
\end{enumerate}
are equivalent if and only if $M(\alpha)$ ``diverges to $-\infty$ uniformly as $\alpha\to -\infty$'',\index{uniform divergence to $-\infty$} meaning that $\forall R>0$ $\exists\,\alpha_R<0$ such that $M(\alpha)\leqslant -R\mathbbm{1}$ for each $\alpha\leqslant\alpha_R$.
\end{theorem}

\begin{proof}
Since  (i) $\Rightarrow$ (ii) is always true (owing to \eqref{eq:SBsmbb-iff-invBsmbb_Tversion} from Theorem \ref{thm:semibdd_exts_operator_formulation_Tversion}), what must be proven is the equivalence between the implication (ii) $\Rightarrow$ (i) and the condition of uniform divergence to $-\infty$ for $M(\alpha)$. Assume (ii) $\Rightarrow$ (i), that is, assume that for arbitrary $-R<\mathfrak{m}(T)<0$ the condition $T\geqslant -R\mathbbm{1}$ implies  $S_T\geqslant\alpha_R\mathbbm{1}$ for some $\alpha_R<0$ and hence also $S_T\geqslant\alpha\mathbbm{1}$ $\forall\alpha\leqslant\alpha_R$ (if the lower bound $\alpha_R$ was non-negative, then $\mathfrak{m}(T)$ would be non-negative too, against the assumption). In turn, owing to \eqref{eq:SBsmbb-iff-invBsmbb_Tversion} and \eqref{eq:Weyl_M}, $S_T\geqslant\alpha\mathbbm{1}$ $\forall\alpha\leqslant\alpha_R$ is equivalent to $T\geqslant M(\alpha)$ $\forall\alpha\leqslant\alpha_R$. Then, for $T\geqslant -R\mathbbm{1}$ to imply $T\geqslant M(\alpha)$ $\forall\alpha\leqslant\alpha_R$, necessarily $M(\alpha)\leqslant -R\mathbbm{1}$ $\forall\alpha\leqslant\alpha_R$. Conversely, assume now that for arbitrary $R>0$ there exists $\alpha_R$ such that $M(\alpha)\leqslant -R\mathbbm{1}$ $\forall\alpha\leqslant\alpha_R$: one wants to deduce (ii) $\Rightarrow$ (i). To this aim, assume that $T$ is lower semi-bounded and apply the assumption for $R=-\mathfrak{m}(T)$: for the corresponding $\alpha_R$ one has $M(\alpha_R)\leqslant -R\mathbbm{1}=\mathfrak{m}(T)\mathbbm{1}\leqslant T$, which by \eqref{eq:SBsmbb-iff-invBsmbb_Tversion} implies $S_T\geqslant \alpha_R\mathbbm{1}$. 
\end{proof}

\begin{theorem}[Semi-boundedness problem for finite deficiency index]\label{cor:finite_deficiency_index} 
 Let $S$ be a densely defined and positive symmetric operator with greatest lower bound $\mathfrak{m}(S)>0$ on the Hilbert space $\cH$, and with \emph{finite} deficiency index. Let $S_T$ be a self-adjoint extension of $S$ according to the parametrisation \eqref{eq:ST} of Theorem \ref{thm:VB-representaton-theorem_Tversion}. Then the lower semi-boundedness of $S_T$ is \emph{equivalent} to the lower semi-boundedness of $T$. In particular (as stated already in Lemma \ref{lem:II-finiteDefIndLowSemibdd}), if $S$ is a lower semi-bounded symmetric operator with finite deficiency index, then any self-adjoint extension of $S$ is also lower semi-bounded.
\end{theorem}

\begin{proof}
 By assumption, $\dim\ker S^*<\infty$.
 The thesis follows from Theorem \ref{thm:semi-boundedness_of_ST} once one shows that $M(\alpha)$ diverges uniformly to $-\infty$.

 In fact, irrespectively of the finite or infinite dimensionality of $\ker S^*$, 
\begin{equation}\label{eq:M-non-uniform-divergence}
\lim_{\alpha\to-\infty}\langle u,M(\alpha)u\rangle\;=\;-\infty\qquad\forall u\in\ker S^*\setminus\{0\}\,. 
\end{equation}
Indeed, if $u\in\ker S^*\setminus\{0\}$, then necessarily $u\notin\mathcal{D}[S_\mathrm{F}]$ (owing to \eqref{eq:II-def-SN-first3} from Theorem \ref{thm:II-KvNextension}), whence (Sect.~\ref{sec:I_forms})
\[
\int_{[0,+\infty)}\lambda\;\ud\langle u,E^{(S_\mathrm{F})}(\lambda) u\rangle\;=\;+\infty\,,
\]
where $\ud E^{(S_\mathrm{F})}$ denotes the spectral measure of $S_\mathrm{F}$ (Sect.~\ref{sec:I_spectral_theorem}). Such a measure is only supported on $[0,+\infty)$ because $S_\mathrm{F}$ has the same positive greatest lower bound as $S$ (Theorem \ref{thm:Friedrichs-ext}(iii)). Therefore, since $\frac{\lambda\alpha}{\lambda-\alpha}\to-\lambda$ as $\alpha\to -\infty$,
\[
\langle u,M(\alpha)u\rangle\;=\;\int_{[0,+\infty)}\!\frac{\lambda\,\alpha}{\lambda-\alpha}\,\ud\langle u,E^{(S_\mathrm{F})}(\lambda) u\rangle\;\xrightarrow[]{\;\alpha\to -\infty\;}\;-\infty\,,
\]
which proves \eqref{eq:M-non-uniform-divergence}.

 If in addition $\dim\ker S^*<\infty$, then, as will be now shown, \eqref{eq:M-non-uniform-divergence} implies \emph{uniform} divergence to $-\infty$, in the sense of Theorem \ref{thm:semi-boundedness_of_ST}. For arbitrary $R>0$ decompose $u\in\ker S^*\setminus\{0\}$ as $u=f_R+v_R$ with
\[
f_R\;:=\;E^{(S_\mathrm{F})}([0,2R])u\,,\qquad v_R:=\;E^{(S_\mathrm{F})}((2R,+\infty))u\,.
\]
Observe that $f_R\in\mathcal{D}(S_\mathrm{F})$, because
\[
\int_{[0,+\infty)}\!\lambda^2\,\ud\langle f_R,E^{(S_\mathrm{F})}(\lambda) f_R\rangle\;=\;\int_{[0,2R]}\!\lambda^2\,\ud\langle f_R,E^{(S_\mathrm{F})}(\lambda) f_R\rangle\;\leqslant\;4R^2\|f_R\|^2\,,
\]
whereas necessarily $v_R\notin\mathcal{D}(S_\mathrm{F})$ because $u\notin\mathcal{D}(S_\mathrm{F})$. One has
\[\tag{a}\label{eq:II-findefind-a}
\begin{split}
\langle u&, M(\alpha) u \rangle \;=\;\int_{[0,+\infty)}\!\frac{\lambda\,\alpha}{\lambda-\alpha}\,\ud\langle u,E^{(S_\mathrm{F})}(\lambda) u\rangle \\
&=\;\int_{[0,2R]}\!\frac{\lambda\,\alpha}{\lambda-\alpha}\,\ud\langle f_R,E^{(S_\mathrm{F})}(\lambda) f_R\rangle\;+\;\int_{(2R,+\infty)}\!\frac{\lambda\,\alpha}{\lambda-\alpha}\,\ud\langle v_R,E^{(S_\mathrm{F})}(\lambda) v_R\rangle\,.
\end{split}
\]

In the second integral in the r.h.s~above $\lambda>2R$, whence $\frac{\lambda}{\lambda-R}<2$: therefore, choosing $\alpha<-2R$ implies $\alpha<-2R<-\frac{\lambda R}{\lambda-R}$, whence also $\frac{\lambda\alpha}{\lambda-\alpha}<-R$. Thus,
\[\tag{b}\label{eq:II-findefind-b}
\int_{(2R,+\infty)}\!\frac{\lambda\,\alpha}{\lambda-\alpha}\,\ud\langle v_R,E^{(S_\mathrm{F})}(\lambda) v_R\rangle\;<\;-R\,\|v_R\|^2\qquad (\alpha<-2R)\,.
\]

Next, concerning the first  integral in the r.h.s~of (a), assume for concreteness $\dim\ker S^*=d$ for some $d\in\mathbb{N}$. Obviously there is $d_R\in\mathbb{N}$, $d_R\leqslant d$, such that
\[
\dim E^{(S_\mathrm{F})} ([0,2R])\ker S^*\;=\;d_R\,,
\]
and let $\{\varphi_{R,1},\dots,\varphi_{R,d_R}\}$ be an orthonormal basis of the $d_R$-dimensional subspace  $E^{(S_\mathrm{F})} ([0,2R])\ker S^*\subset\mathcal{D}(S_\mathrm{F})$. Further, decompose $f_R=f_{R,1}+\cdots+f_{R,d_R}$ with $f_{R,j}:=\langle \varphi_{R,j},f_R\rangle \varphi_{R,j}$, $j=1,\dots,d_R$. Then
\[
\begin{split}
\int_{[0,2R]}\!\frac{\lambda\,\alpha}{\lambda-\alpha}\,\ud\langle &f_R,E^{(S_\mathrm{F})}(\lambda) f_R\rangle\;=\;\sum_{j=1}^{d_R}\int_{[0,2R]}\!\frac{\lambda\,\alpha}{\lambda-\alpha}\,\ud\langle f_{R,j},E^{(S_\mathrm{F})}(\lambda) f_{R,j}\rangle \\
&=\;\sum_{j=1}^{d_R}\;|\langle\varphi_{R,j},f_R\rangle|^2\int_{[0,2R]}\!\frac{\lambda\,\alpha}{\lambda-\alpha}\,\ud\langle \varphi_{R,j},E^{(S_\mathrm{F})}(\lambda) \varphi_{R,j}\rangle \\
&=\;\sum_{j=1}^{d_R}\;|\langle\varphi_{R,j},f_R\rangle|^2\langle \widetilde\varphi_{R,j},M(\alpha)\widetilde \varphi_{R,j}\rangle\,,
\end{split}
\]
where the $\widetilde\varphi_{R,j}$'s are vectors in $\ker S^{*}$ such that  $\varphi_{R,j}=E^{(S_\mathrm{F})}([0,2R]) \widetilde\varphi_{R,j}$ . Clearly, none of the $\widetilde\varphi_{R,j}$'s can be the zero vector: thus, for each considered $j$, owing to \eqref{eq:M-non-uniform-divergence}, $\langle \widetilde\varphi_{R,j},M(\alpha)\widetilde \varphi_{R,j}\rangle\to -\infty$ as $\alpha\to -\infty$. As there is only a \emph{finite} number $d_R\leqslant d$ of such terms, there necessarily exists a common threshold $\alpha_R<0$ such that
\[
\sup_{j\in\{1,\dots,d_R\}}\langle \widetilde\varphi_{R,j},M(\alpha)\widetilde \varphi_{R,j}\rangle\;\leqslant\;-R\qquad\forall\alpha\leqslant\alpha_R\,.
\]
Therefore,
\[\tag{c}\label{eq:II-findefind-c}
 \begin{split}
\int_{[0,2R]}\!\frac{\lambda\,\alpha}{\lambda-\alpha}\,\ud\langle f_R,E^{(S_\mathrm{F})}(\lambda) f_R\rangle\;&\leqslant\;-R\sum_{j=1}^{d_R}\;|\langle\varphi_{R,j},f_R\rangle|^2 \\
&=\;-R\|f_R\|^2\qquad (\alpha\leqslant\alpha_R)\,,
 \end{split}
\]
$\alpha_R$ only depending on $R$ (and on $d$), \emph{not} on $f_R$.

Plugging the bounds \eqref{eq:II-findefind-b} and \eqref{eq:II-findefind-c} into \eqref{eq:II-findefind-a} yields
\[
\langle u,M(\alpha)u\rangle\;<\;-R\|f_R\|^2-R\|v_R\|^2\;=\;-R\|u\|^2
\]
for $\alpha<\min\{-2R,\alpha_R\}$. From the arbitrariness of $u\in\ker S^*$ and of $R>0$ one concludes that $M(\alpha)\to-\infty$ uniformly as $\alpha\to-\infty$.

The final consequence stated in the theorem follows by applying the main statement to a shifted version $S'$ of $S$ that has strictly positive lower bound -- which does not alter the finiteness of the deficiency index (Sect.~\ref{sec:I-symmetric-selfadj} and \ref{sec:perturbation-spectra}): for any extension $S'_T$ of such $S'$ the extension parameter $T$ acts on a subspace of the finite-dimensional $\ker {S'}^*$, and is therefore lower semi-bounded; so too is then  $S'_T$.
\end{proof}

 \begin{corollary}[Semi-boundedness problem with compact inverse Friedrichs]\label{cor:compact_res_SF}
 Let $S$ be a densely defined and positive symmetric operator with greatest lower bound $\mathfrak{m}(S)>0$ on the Hilbert space $\cH$, and such that the inverse $S_\mathrm{F}^{-1}$ of its Friedrichs extension is compact. Let $S_T$ be a self-adjoint extension of $S$ according to the parametrisation \eqref{eq:ST} of Theorem \ref{thm:VB-representaton-theorem_Tversion}. Then the semi-boundedness of $S_T$ is equivalent to the semi-boundedness of $T$.
\end{corollary}

\begin{proof}
Since $S_\mathrm{F}^{-1}$ is compact, the spectrum of $S_\mathrm{F}$ only consists of a discrete set of eigenvalues, each of finite multiplicity (Riesz-Schauder theorem, Sect.~\ref{sec:I-spectrum}), whence the bound \eqref{eq:II-findefind-c} in the proof of Theorem \ref{cor:finite_deficiency_index}. The same conclusion then follows.
\end{proof}

\begin{remark}\label{rem:semiboundedness_problem} Concerning the \emph{semi-boundedness problem}\index{semi-boundedness problem} of finding conditions under which the semi-boundedness of $S_T$ and of $T$ are equivalent, the fact that the compactness of $S_\mathrm{F}^{-1}$ is a sufficient condition (that is, Corollary \ref{cor:compact_res_SF}) was noted originally by
Grubb \cite{Grubb-1974} and by
Gorba{\v{c}}uk and Miha{\u\i}lec \cite{Gorbachuk-Mihailets-1976} in the mid 1970's. 
More than a decade later the same property, and more generally the necessary and sufficient condition provided by Theorem \ref{thm:semi-boundedness_of_ST}, was proved \emph{with a boundary triplets language}\index{boundary triplets} by Derkach and Malamud \cite{Derkach-Malamud-1988-1991}.
In fact, it is easy to recognise that the operator-valued function $\alpha\mapsto M(\alpha)$ defined in \eqref{eq:Weyl_M} is the Weyl function\index{Weyl function} of a standard boundary triplet\index{boundary triplets} \cite[Example 14.12]{schmu_unbdd_sa}.
In \cite[Section 3]{Derkach-Malamud-1988-1991} one can also find examples in which such a condition is violated.  
\end{remark}

 \begin{corollary}[Semi-boundedness with finite-rank Birman parameter]
 Let $S$ be a densely defined and positive symmetric operator with greatest lower bound $\mathfrak{m}(S)>0$ on the Hilbert space $\cH$. All self-adjoint extensions $S_T$ of $S$, in the parametrisation \eqref{eq:ST} of Theorem \ref{thm:VB-representaton-theorem_Tversion}, whose extension parameter $T$ acts on a \emph{finite-dimensional} subspace of $\ker S^*$ are themselves lower semi-bounded. For the occurrence of unbounded below self-adjoint extensions of $S$ it is necessary (not sufficient) that $\dim\overline{\mathcal{D}(T)}=\infty$. 
\end{corollary}

\begin{proof} 
 Let $S_T$ be any extension as in the assumptions. $T$ is bounded (hence also semi-bounded) because the  Hilbert space $\overline{\mathcal{D}(T)}$ it acts on has finite dimension. Let $P_T:\cH\to\cH$ be the orthogonal projection onto $\overline{\mathcal{D}(T)}$ and set
\[
\widetilde{M}(\alpha)\;:=\;P_T(\alpha\mathbbm{1}+\alpha^2(S_\mathrm{F}-\alpha\mathbbm{1})^{-1})P_T\;=\;P_T\,M(\alpha)\,P_T\,,\qquad\alpha<\mathfrak{m}(S)\,.
\]
One can repeat for $\widetilde{M}(\alpha)$ the same arguments used in the proof of Theorem \ref{cor:finite_deficiency_index} to establish the uniform divergence of $M(\alpha)$ to $-\infty$, thus obtaining the same property for $\widetilde{M}(\alpha)$ on the finite-dimensional space $\overline{\mathcal{D}(T)}$ (the assumption $\dim\overline{\mathcal{D}(T)}=d<\infty$ implies $\dim E^{(S_\mathrm{F})}([0,2R])\overline{\mathcal{D}(T)}=d_R\leqslant d$, analogously to what argued in the proof of Theorem \ref{cor:finite_deficiency_index}). Therefore, $\exists\,\alpha<0$, with $|\alpha|$ sufficiently large, such that 
\[
\alpha\|v\|^2+\alpha^2\langle v,(S_\mathrm{F}-\alpha\mathbbm{1})^{-1}v\rangle\;<\;\mathfrak{m}(T)\|v\|^2\;\leqslant\;\langle v,Tv\rangle\qquad\forall v\in\mathcal{D}(T)\,.
\]
In turn, this implies $\mathfrak{m}(S_T)>\alpha$ on account of \eqref{eq:SBsmbb-iff-invBsmbb_Tversion} from Theorem \ref{thm:semibdd_exts_operator_formulation_Tversion}.
\end{proof}

\begin{remark} It is also worth remarking that, unless $S$ is essentially self-adjoint, in all other cases (i.e., whenever the deficiency index $\dim\ker S^*\geqslant 1$) there is no \emph{uniform} lower bound to the bottoms of the semi-bounded self-adjoint extensions of $S$:
\[
 \inf_{ \substack{ \widetilde{S} \\  \widetilde{S}=(\widetilde{S})^*\supset S \\ \mathfrak{m}(\widetilde{S})>-\infty } }\mathfrak{m}(\widetilde{S})\;=\;\inf_{ \substack{ T\in\mathcal{S}(\ker S^*) \\  \mathfrak{m}(S_T)>-\infty }}\mathfrak{m}(S_T)\;=\;-\infty\,.
\]
This is an immediate consequence of the bound $\mathfrak{m}(T)\geqslant \mathfrak{m}(S_T)$ given by \eqref{eq:SBsmbb-iff-invBsmbb_Tversion} (Theorem \ref{thm:semibdd_exts_operator_formulation_Tversion}), by taking extension parameters $T=-\gamma\mathbbm{1}$  for arbitrary $\gamma>0$.
\end{remark}

  Yet another fundamental link between any self-adjoint extension $S_T$ and its Birman extension parameter $T$ concerns the negative spectra of such two operators. This was established first by Birman \cite{Birman-1956}: Theorem \ref{thm:negative_spectrum_ST_T} and Corollary \ref{cor:negative_spectrum_ST_T_corollary} below are indeed a polished version of \cite[Theorem 4]{Birman-1956}.

  In fact, it was Kre{\u\i}n in his 1946 work \cite{Krein-1947} who made the first systematic analysis of the negative spectrum of the self-adjoint extensions of a densely defined and symmetric operator $S$ with strictly positive bottom: as mentioned at the beginning of Section \ref{sec:II-VBparametrisation-orig}, this was done under the restriction of finite deficiency index, and Birman then removed such a restriction. In particular, in \cite[Theorem 19]{Krein-1947} Kre{\u\i}n had demonstrated that if $\widetilde{S}$ is any self-adjoint extension of $S$, then $\widetilde{S}$ is lower semi-bounded and its negative spectrum consists of eigenvalues whose number is finite and equals (counting multiplicity) the number of negative critical values of the quadratic form of $\widetilde{S}$ restricted to $\mathcal{D}[\widetilde{S}]\cap\ker S^*$. Re-writing $\widetilde{S}=S_T$ in the current parametrisation \eqref{eq:ST}, one has $\mathcal{D}[\widetilde{S}]\cap\ker S^*=\mathcal{D}[T]$ (Theorem \ref{thm:semibdd_exts_form_formulation}): thus, Kre{\u\i}n's result in the case of finite deficiency index reads as the equality of the (finite) multiplicities of the negative spectra of $S_T$ and $T$. All this is extended to possibly infinite deficiency index as follows.

For convenience, set
\begin{equation}
\begin{split}
 \sigma_{\!-\!}(S_T)\;& :=\;\sigma(S_T)\cap (-\infty,0) \, , \\
 \sigma_{\!-\!}(T)\;& :=\;\sigma(T)\cap (-\infty,0)\,.
\end{split}
\end{equation}

\begin{theorem}[Negative spectrum]\label{thm:negative_spectrum_ST_T}
 Let $S$ be a densely defined and positive symmetric operator with greatest lower bound $\mathfrak{m}(S)>0$ on the  Hilbert space $\cH$, and let $S_T$ be a self-adjoint extension of $S$ according to the parametrisation \eqref{eq:ST} of Theorem \ref{thm:VB-representaton-theorem_Tversion}. Then $\sigma_{\!-\!}(S_T)$ consists of a bounded below set of finite-rank eigenvalues of $S_T$ whose only possible accumulation point is $0$ \emph{if and only if}  $\sigma_{\!-\!}(T)$ has the same property. When this is the case, and $\lambda_1\leqslant\lambda_2\leqslant\cdots<0$ and $t_1\leqslant t_2\leqslant\cdots<0$ are the ordered sequences of negative eigenvalues (counted with multiplicity) of $S_T$ and of $T$ respectively, then $\lambda_k\leqslant t_k$ for $k=1,2,\dots$. In particular, the two ground states $\lambda_1=\mathfrak{m}(S_T)$ of $S_T$ and $t_1=\mathfrak{m}(T)$ of $T$ satisfy $\lambda_1\leqslant t_1$, consistently with inequality \eqref{eq:II-mTgeqmST} from Theorem \ref{thm:semibdd_exts_operator_formulation_Tversion}.
\end{theorem}

\begin{corollary}\label{cor:negative_spectrum_ST_T_corollary}
Under the assumptions of Theorem \ref{thm:negative_spectrum_ST_T}: for some $N\in\mathbb{N}$, $\sigma_{\!-\!}(S_T)$ consists of $N$ eigenvalues if and only if $\sigma_{\!-\!}(T)$ consists of $N$ eigenvalues (counting multiplicity).
\end{corollary}

\begin{corollary}\label{cor:finite_negative_spectrum} Under the assumptions of Theorem \ref{thm:negative_spectrum_ST_T}:
\begin{enumerate}[(i)]
 \item if $S$ has finite deficiency index, then all self-adjoint extensions of $S$ have finite (possibly empty) negative spectrum, with finite-dimensional eigenvalues;
 \item if the extension parameter $T$ acts on a \emph{finite-dimensional} subspace of $\ker S^*$, then the negative spectrum $\sigma_-(S_T)$ of $S_T$ is finite, and consists of finite-dimensional eigenvalues.
\end{enumerate}
\end{corollary}

In preparation for the proof of Theorem \ref{thm:negative_spectrum_ST_T} and its corollaries, recall that $\mathcal{D}(T)\subset \mathcal{D}[S_T]$ and $\langle v,Tv\rangle=S_T[v]$ $\forall v\in\mathcal{D}(T)$ (Theorem \ref{thm:semibdd_exts_form_formulation_Tversion}), whence also
\begin{equation}\label{eq:spectral_integrals}
 \begin{split}
\int_{[\mathfrak{m}(T),+\infty)}\!t\,\langle v,\mathrm{d} E^{(T)}(t) v\rangle\;&=\;
\int_{[\mathfrak{m}(S_T),+\infty)}\!\lambda\,\langle v,\mathrm{d} E^{(S_T)}(\lambda) v\rangle \\
&\geqslant\;\int_{[\mathfrak{m}(S_T),0)}\!\lambda\,\langle v,\mathrm{d} E^{(S_T)}(\lambda) v\rangle\,,
 \end{split}
\end{equation}
where here and in the following, as usual (Sect. \ref{sec:I_spectral_theorem}) $\ud E^{(S_T)}$ and $\ud E^{(T)}$ are the spectral measures associated, respectively, with $S_T$ and $T$.

 It is also convenient to single out two useful facts (the first is straightforward).

\begin{lemma}\label{lemma:L1_L2}
If $V$ and $W$ are closed subspaces of the Hilbert space $\cH$ with $\dim V<\infty$ and $\dim W>\dim V$, then $W\cap V^\perp\neq\{0\}$.
\end{lemma}

\begin{lemma}\label{lem:gk_vk}
If $\varepsilon>0$ and, for some $N\in\mathbb{N}$, $g_1,\dots,g_N$ are linearly independent elements in $\mathcal{D}(S_T)\cap E^{(S_T)}((-\infty,-\varepsilon])\cH$, then the corresponding $v_1,\dots,v_N$ given by the decomposition \eqref{eq:ST}  $g_k=f_k+S_\mathrm{F}^{-1}(Tv_k+w_k)+v_k$, $k\in\{1,\dots,N\}$, are linearly independent in $\mathcal{D}(T)$.
\end{lemma}

\begin{proof}
Assume that $\sum_{k=1}^N c_k v_k=0$ for some $c_1,\dots,c_N\in\mathbb{C}$. One has $g:=\sum_{k=1}^N c_k g_k=\sum_{k=1}^N c_k (f_k+S_\mathrm{F}^{-1}(Tv_k+w_k))\in\mathcal{D}(S_\mathrm{F})$, whence $\langle g,S_T g\rangle=\langle g,S_\mathrm{F} g\rangle\geqslant \mathfrak{m}(S)\|g\|^2\geqslant 0$. On the other hand, 
\[
\begin{split}
\langle g,S_T g\rangle\;&=\;\int_{[\mathfrak{m}(S_T),+\infty)}\!\lambda\,\langle g,\mathrm{d} E^{(S_T)}(\lambda) g\rangle\;=\;\int_{[\mathfrak{m}(S_T),-\varepsilon]}\!\lambda\,\langle g,\mathrm{d} E^{(S_T)}(\lambda) g\rangle \\
&\leqslant\;-\varepsilon\int_{[\mathfrak{m}(S_T),-\varepsilon]}\!\langle g,\mathrm{d} E^{(S_T)}(\lambda) g\rangle\;\leqslant\;0\,,
\end{split}
\]
the second identity following from $g\in\mathcal{D}(S_T)\cap E^{(S_T)}((-\infty,-\varepsilon])\cH$. Thus, necessarily $\langle g,S_{\mathrm{F}} g\rangle=0$, and therefore $g=0$, whence by assumption $c_1=\cdots=c_N=0$.
\end{proof}

\begin{proof}[Proof of Theorem \ref{thm:negative_spectrum_ST_T}]
It is clear from \eqref{eq:positiveSBiffpositveB-1_Tversion} (Theorem \ref{thm:semibdd_exts_operator_formulation_Tversion}) that $\sigma_{\!-\!}(S_T)$ is empty if and only if so too is $\sigma_{\!-\!}(T)$, in which case the present theorem is trivially true. The proof must now cover the case of non-empty spectra.

Assume that  $\sigma_{\!-\!}(S_T)$ consists of a non-empty and bounded below set of finite-rank eigenvalues whose only possible accumulation point is $0$. In particular, $-\infty<\mathfrak{m}(S_T)<0$, whence (on account of \eqref{eq:II-mTgeqmST}-\eqref{eq:positiveSBiffpositveB-1_Tversion} from Theorem \ref{thm:semibdd_exts_operator_formulation_Tversion}),
$\mathfrak{m}(S_T)\leqslant \mathfrak{m}(T)<0$. In particular, $\sigma_{\!-\!}(T)$ is non-empty.
If, for contradiction, $\sigma_{\!-\!}(T)$ does not satisfy the same property of $\sigma_{\!-\!}(S_T)$, then there exists $\varepsilon>0$ such that
\[
 \dim E^{(T)}([\mathfrak{m}(T),-\varepsilon])\overline{\mathcal{D}(T)}\;=\;+\infty\,,
\]
whereas by assumption
\[
 \dim E^{(S_T)}([\mathfrak{m}(S_T),-{\textstyle\frac{1}{2}}\varepsilon])\cH\;<\;+\infty\,.
\]
By Lemma \ref{lemma:L1_L2} $\exists\,v\in E^{(T)}([\mathfrak{m}(T),-\varepsilon])\overline{\mathcal{D}(T)}$, $v\neq 0$, $v\perp E^{(S_T)}([\mathfrak{m}(S_T),-\frac{1}{2}\varepsilon])\cH$. As a consequence of this and of \eqref{eq:spectral_integrals},
\[
\begin{split}
-\varepsilon\|v\|^2\;&\geqslant\;\int_{[\mathfrak{m}(T),-\varepsilon)}\!t\,\langle v,\mathrm{d} E^{(T)}(t) v\rangle\;=\;\int_{[\mathfrak{m}(T),+\infty)}\!t\,\langle v,\mathrm{d} E^{(T)}(t) v\rangle \\
&\geqslant\;\int_{[\mathfrak{m}(S_T),0)}\!\lambda\,\langle v,\mathrm{d} E^{(S_T)}(\lambda) v\rangle\;=\;\int_{(-\frac{1}{2}\varepsilon,0)}\!\lambda\,\langle v,\mathrm{d} E^{(S_T)}(\lambda) v\rangle\;\geqslant\;-\frac{\varepsilon}{2}\|v\|^2\,,
\end{split}
\]
which is a contradiction because $v\neq 0$.

  For the converse, assume that $\sigma_{\!-\!}(T)$ consists of a non-empty and bounded below set of finite-rank eigenvalues of $T$ whose only possible accumulation point is $0$. In particular, $-\infty<\mathfrak{m}(T)<0$. If, for contradiction, $\sigma_{\!-\!}(S_T)$ does not satisfy the same property of $\sigma_{\!-\!}(T)$, then $\dim E^{(S_T)}((-\infty,-\varepsilon])\cH=+\infty$ for some $\varepsilon>0$. Therefore also
\[\tag{*}\label{eq:II-proofSigmaMINUSa}
\dim \big(E^{(S_T)}((-\infty,-\varepsilon])\cH\big)\cap\mathcal{D}(S_T)\;=\;+\infty \, ,
\]
because $\big(E^{(S_T)}((-\infty,-\varepsilon])\cH\big)\cap\mathcal{D}(S_T)$ is dense in $E^{(S_T)}((-\infty,-\varepsilon])\cH$.
%
Based on the decomposition \eqref{eq:ST} for generic $g\in\mathcal{D}(S_T)$ (namely, $g=f+S_\mathrm{F}^{-1}(Tv+w)+v$), set
\[
V_\varepsilon\;:=\;\left\{ 
v\in\mathcal{D}(T)\left|
\begin{array}{c}
g-v\in\mathcal{D}(S_\mathrm{F})\textrm{ for some} \\
g\in \big(E^{(S_T)}((-\infty,-\varepsilon])\cH\big)\cap\mathcal{D}(S_T) \\
\end{array}
\!\right.\right\}\,.
\]
In fact, owing to Lemma \ref{lem:gk_vk}, any $v\in V_\varepsilon$ identifies uniquely the corresponding $g\in \big(E^{(S_T)}((-\infty,-\varepsilon])\cH\big)\cap\mathcal{D}(S_T)$. Furthermore, Lemma \ref{lem:gk_vk} and \eqref{eq:II-proofSigmaMINUSa} yield $\dim V_\varepsilon=+\infty$.
On the other hand, let $\delta\in\mathbb{R}$ with $0<\delta<\min\{-\mathfrak{m}(T),\frac{\varepsilon \mathfrak{m}(S)}{2\mathfrak{m}(S)+\varepsilon}\}$: by assumption 
\[\tag{**}\label{eq:II-proofSigmaMINUSb}
 \dim E^{(T)}([\mathfrak{m}(T),-\delta])\overline{\mathcal{D}(T)}\;<\;+\infty\,.
\]
Lemma \ref{lemma:L1_L2} and \eqref{eq:II-proofSigmaMINUSa}-\eqref{eq:II-proofSigmaMINUSb} then imply the existence of a non-zero $v\in V_\varepsilon$ with $v\perp E^{(T)}([\mathfrak{m}(T),-\delta])\overline{\mathcal{D}(T)}$. For such $v$ one has
\[
\begin{split}
\langle v,Tv\rangle\;&=\;\int_{[\mathfrak{m}(T),+\infty)}\!t\,\langle v,\mathrm{d} E^{(T)}(t) v\rangle\;=\;\int_{(-\delta,+\infty)}\!t\,\langle v,\mathrm{d} E^{(T)}(t) v\rangle\;\geqslant\;-\delta\|v\|^2 \\
&\geqslant\;-\frac{\varepsilon \mathfrak{m}(S)}{2\mathfrak{m}(S)+\varepsilon}\,\|v\|^2\,,
\end{split}
\]
which can be re-written equivalently as
\[
\langle v,Tv\rangle+\frac{\varepsilon}{2}\,\|v\|^2\;\geqslant\;\frac{\:\varepsilon^2}{4}\,\frac{1}{\mathfrak{m}(S)+\frac{1}{2}\varepsilon}\,\|v\|^2\,.
\]
The last inequality implies
\[
\langle v,Tv\rangle+\frac{\varepsilon}{2}\,\|v\|^2\;\geqslant\;\frac{\:\varepsilon^2}{4}\langle v,(S_\mathrm{F}+\textstyle{\frac{1}{2}}\varepsilon)^{-1}v\rangle\,.
\]
If $g$ is the vector in $E^{(S_T)}(-\infty,-\varepsilon])\cH\cap\mathcal{D}(S_T)$ that corresponds to such $v\in V_\varepsilon$, by repeating the very same reasoning as in the proof of Theorem \ref{thm:semibdd_exts_operator_formulation} one sees that the latter condition is \emph{equivalent} to $\langle g,S_T g\rangle\geqslant-\frac{\varepsilon}{2}\|g\|^2$. However, 
this last finding is not compatible with the fact that 
\[
\langle g,S_T g\rangle\,=\int_{[\mathfrak{m}(S_T),+\infty)}\!\lambda\,\langle g,\mathrm{d} E^{(S_T)}(\lambda) g\rangle\,=\int_{[\mathfrak{m}(S_T),-\varepsilon)}\!\lambda\,\langle g,\mathrm{d} E^{(S_T)}(\lambda) g\rangle\,\leqslant\,-\varepsilon\|g\|^2\,,
\]
whence the contradiction.

This completes the proof of the equivalence of the considered condition for $\sigma_{\!-\!}(S_T)$ and $\sigma_{\!-\!}(T)$. When such a condition holds and the eigenvalues are labelled as in the statement of the theorem, the fact that $\lambda_k\leqslant t_k$ for $k=1,2,\dots$ is a consequence of the min-max principle (Sect.~\ref{sec:minmax}) for the self-adjoint operators $S_T$ and $T$, owing to the fact (Theorem \ref{thm:semibdd_exts_form_formulation_Tversion}) that $S_T\leqslant T$.
\end{proof}

\begin{proof}[Proof of Corollary \ref{cor:negative_spectrum_ST_T_corollary}]
Owing to Theorem \ref{thm:negative_spectrum_ST_T}, the property
\[
\sigma_{\!-\!}(S_T)=\{\textrm{eigenvalues }\lambda_1\leqslant\cdots\leqslant\lambda_N<0\}\qquad\textrm{for some $N\in\mathbb{N}$}
\]
is equivalent to
\[
\sigma_{\!-\!}(T)=\{\textrm{eigenvalues }t_1\leqslant\cdots\leqslant t_M<0\}\qquad\textrm{for some $M\in\mathbb{N}$}
\]
(with possibly repeated eigenvalues), and when this is the case $\lambda_1=\mathfrak{m}(S_T)\leqslant \mathfrak{m}(T)=t_1$. If $M>N$, then $\exists\,v\in (E^{(T)}([\mathfrak{m}(T),-\varepsilon])\overline{\mathcal{D}(T)})\cap(E^{(S_T)}([\mathfrak{m}(S_T),0))\cH)^\perp$, $v\neq 0$, for some $\varepsilon>0$ (in fact, $\forall\varepsilon\in(0,|t_M|]$), as a consequence of Lemma \ref{lemma:L1_L2}. Moreover, $v\in\mathcal{D}(T)$ because
\[
\int_{[\mathfrak{m}(T),+\infty)}\! t^2\langle v,\mathrm{d} E^{(T)}(t) v\rangle\;=\;\int_{[\mathfrak{m}(T),-\varepsilon]}\! t^2\langle v,\mathrm{d} E^{(T)}(t) v\rangle\;<\;+\infty\,,
\]
whence also $v\in\mathcal{D}[S_T]$ with $S_T[v]=\langle v,Tv\rangle$ (Theorem \ref{thm:semibdd_exts_form_formulation_Tversion}).
As a consequence of this and of \eqref{eq:spectral_integrals},
\[
\begin{split}
0\;>\;\int_{[\mathfrak{m}(T),-\varepsilon]}\!t\,\langle v,\mathrm{d} E^{(T)}(t) v\rangle\;&=\;\int_{[\mathfrak{m}(T),+\infty)}\!t\,\langle v,\mathrm{d} E^{(T)}(t) v\rangle \\
&\geqslant\;\int_{[\mathfrak{m}(S_T),0)}\!\lambda\,\langle v,\mathrm{d} E^{(S_T)}(\lambda) v\rangle\;=\;0\,,
\end{split}
\]
a contradiction. If instead $M<N$, it is convenient to use the fact that 
there is $\varepsilon>0$ (in fact, any  $\varepsilon\in(0,|\lambda_N|]$ does the job) for which 
Lemma \ref{lem:gk_vk} applied to the space $V_\varepsilon$ introduced in the proof of Theorem \ref{thm:negative_spectrum_ST_T} yields $\dim V_\varepsilon\geqslant N$. Then, on account of Lemma \ref{lemma:L1_L2},
$\exists\,v\in V_\varepsilon\cap(E^{(T)}([\mathfrak{m}(T),0))\overline{\mathcal{D}(T)})^\perp$, $v\neq 0$. In turn, as already observed in the proof of Theorem \ref{thm:negative_spectrum_ST_T}, such $v$ identifies uniquely a non-zero element $g\in E^{(S_T)}([\mathfrak{m}(S_T),-\varepsilon])\cH\cap\mathcal{D}(S_T) $ for which $g-v\in\mathcal{D}(S_\mathrm{F})$. For such $g$ and $v$, \eqref{eq:decomposition_of_form_domains_Tversion} yields $\langle g,S_T g\rangle\geqslant\langle v,Tv\rangle$. With these findings,
\[
\begin{split}
0\;&>\;\int_{[\mathfrak{m}(S_T),-\varepsilon]}\!\lambda\,\langle g,\mathrm{d} E^{(S_T)}(\lambda) g\rangle\;=\;\int_{[\mathfrak{m}(S_T),+\infty)}\!\lambda\,\langle g,\mathrm{d} E^{(S_T)}(\lambda) g\rangle\;=\;\langle g,S_T g\rangle \\
&\geqslant\; \langle v,Tv\rangle\;=\;\int_{[\mathfrak{m}(T),+\infty)}\!t\,\langle v,\mathrm{d} E^{(T)}(t) v\rangle\;\geqslant\;\int_{[\mathfrak{m}(T),0)}\!t\,\langle v,\mathrm{d} E^{(T)}(t) v\rangle\;=\;0\,,
\end{split}
\]
another contradiction. Thus, the conclusion is necessarily $M=N$.
\end{proof}

\begin{proof}[Proof of Corollary \ref{cor:finite_negative_spectrum}] In either case (i) and (ii) the extension parameter $T$ is self-adjoint on a finite-dimensional space, therefore its spectrum only consists of a finite number of (finite-dimensional) eigenvalues. This is true in particular for the negative spectrum of $T$. Then the conclusion follows from Corollary \ref{cor:negative_spectrum_ST_T_corollary}.
\end{proof}

 \section{Resolvents in the Kre{\u\i}n-Vi\v{s}ik-Birman extension scheme}\label{sec:II-resolventsKVB}

 The\index{Kre{\u\i}n-Vi\v{s}ik-Birman theory} \emph{resolvent difference} of two generic self-adjoint extensions of a densely defined (not necessarily lower semi-bounded) symmetric operator has profound relevance both in abstract terms and in a variety of applications and contexts. The problem of characterising such difference appears to have emerged as a central one already in the early 1940's -- and attracting constant interest ever after -- namely at a time when von Neumann's extension theory was already well grounded and Kre{\u\i}n's extension theory was getting shaped in its final form. It was both Kre{\u\i}n \cite{Krein-1944-KFormula11} and, independently, Na\u{\i}mark \cite{Naimark-1943-KFormula11} who in 1943-1944 established a resolvent difference formula, since then customarily referred to as the \emph{Kre{\u\i}n} or \emph{Kre{\u\i}n-Na\u{\i}mark resolvent formula}\index{Kre{\u\i}n-Na\u{\i}mark resolvent formula}, for any two self-adjoint extensions of a given densely defined (and closed) symmetric operator with deficiency indices (1,1). Shortly after,  in 1946, this was extended to the case of finite deficiency indices $(n,n)$ by Kre{\u\i}n \cite{Krein-1946-KFormulann}; a proof for this case is also provided in the classic monograph by Akhiezer and Glazman \cite[Section 84]{Akhiezer-Glazman-1961-1993}. Two decades later, in 1965, the final generalisation to possibly infinite deficiency indices was established by Saakjan \cite{Saakjan-1965}, whose rather compact proof was more recently reviewed and supplemented with additional results by Gesztesy, Makarov, and Tsekanovskii \cite{Gesztesy-Makarov-Tsekanovskii-1998}. By now the subject is standard material of specialised monographs and textbooks -- one may refer, for instance, to \cite[Section 14.6]{schmu_unbdd_sa} -- although the literature both on the Kre{\u\i}n-Na\u{\i}mark formula in specific applications and on Kre{\u\i}n-Na\u{\i}mark-like formulas of various abstract and applied interest is too extensive to be discussed exhaustively here. This is even more so because the subject has meanwhile been largely incorporated into and dealt with the general language of Gamma fields,\index{Gamma field} Weyl functions\index{Weyl function}, and boundary triplet theory\index{boundary triplets} -- see, e.g., \cite[Chapter 14]{schmu_unbdd_sa} or \cite{Behrndt-Hassi-deSnoo-boundaryBook} -- whose spirit diverts from the present discussion on classical extension theories.

 This Section is focussed on the more circumscribed topic of resolvent formulas (in fact, eventually, of Kre{\u\i}n-Na\u{\i}mark type) for generic self-adjoint extensions of a densely defined and lower semi-bounded symmetric operator, \emph{which can be derived within the Kre{\u\i}n-Vi\v{s}ik-Birman extension scheme}. The actual resolvent difference under consideration here is between the resolvents of the extension of interest and of the distinguished (e.g., Friedrichs) extension, and is expressed in terms of the corresponding extension parameter. In particular, Theorem \ref{thm:resolvent_formula_positive_extensions} below, obtained here from the general extension scheme of Sections \ref{sec:II-VBparametrisation-orig}-\ref{sec:II-VBreparametrised}, has a partial precursor for finite deficiency index already in Kre{\u\i}n's 1946 work \cite[Theorems 20 and 22]{Krein-1947}. It can be further tracked down to the independent derivations in Grubb's 1968 work \cite[Theorem II.1.4]{Grubb-1968} and, limited to positive invertible extensions, in Faris' 1975 monograph \cite[Theorem 15.1]{faris-1975}.

 \begin{theorem}[Resolvent formula for invertible extensions]\label{thm:resolvent_formula_positive_extensions}
  Let $S$ be a densely defined and positive symmetric operator with greatest lower bound $\mathfrak{m}(S)>0$ on the Hilbert space $\cH$.
  \begin{enumerate}[(i)]
   \item Let $S_T$, in the parametrisation \eqref{eq:ST} of Theorem \ref{thm:VB-representaton-theorem_Tversion}, be a self-adjoint extension of $S$ with everywhere defined and bounded inverse on $\cH$ and distinct from $S_{\mathrm{F}}$, and let $P_T:\cH\to\cH$ be the orthogonal projection onto the Hilbert subspace $\mathcal{K}\equiv\overline{\mathcal{D}(T)}$. Then the extension parameter $T$ is invertible with everywhere defined and bounded inverse on $\mathcal{K}$ and 
   \begin{equation}\label{eq:resolvent_formula_z=0}
S_T^{-1}\;=\;S_\mathrm{F}^{-1}\,+\,P_T \,T^{-1} P_T\,.
\end{equation}
     \item Let $S_B$, in the parametrisation \eqref{eq:defSB} of Theorem \ref{thm:VB-representaton-theorem}, be a self-adjoint extension of $S$ with everywhere defined and bounded inverse on $\cH$ and distinct from $S_{\mathrm{F}}$, and let $P_B:\cH\to\cH$ be the orthogonal projection onto the Hilbert subspace $\mathcal{K}\equiv\overline{\mathcal{D}(B)}$. Then 
     \begin{equation}\label{eq:resolvent_formula_z=0_B-version}
S_B^{-1}\;=\;S_\mathrm{F}^{-1}\,+\,P_B \,B \,P_B\,.
\end{equation}
     \end{enumerate}
\end{theorem}

  \begin{proof}
(i) The invertibility of $T$ follows by Theorem \ref{thm:invertibility}(iii).
Thus, \eqref{eq:resolvent_formula_z=0} is an identity between  bounded self-adjoint operators. A generic $h\in\cH=\ran\, S_T$ is written as $h=S_T g$ for some  
$g=f+S_\mathrm{F}^{-1}(Tv+w)+v=F+v$ and suitable $f\in\mathcal{D}(\overline{S})$, $v\in\mathcal{D}(T)$, $w=\ker S^*\cap\mathcal{D}(T)$, $F:=f+S_\mathrm{F}^{-1}(Tv+w)\in\mathcal{D}(S_\mathrm{F})$ (Theorem \ref{thm:VB-representaton-theorem_Tversion}). Then
\[
\langle h, S_T^{-1} h\rangle\;=\;\langle g, S_T g\rangle\;=\;\langle F, S_\mathrm{F} F\rangle +\langle v,S_\mathrm{F} F \rangle\;=\;\langle F, S_\mathrm{F} F\rangle + \langle v, Tv\rangle\,.
\]
On the other hand,
\[
\begin{split}
 \langle F, S_\mathrm{F} F\rangle\;&=\;\langle S_\mathrm{F} F, S_\mathrm{F}^{-1} S_\mathrm{F} F\rangle \;=\;\langle S_T g, S_\mathrm{F}^{-1} S_T g\rangle\;=\;\langle h, S_\mathrm{F}^{-1} h\rangle\,, \\
 \langle v, Tv\rangle\;&=\;\langle Tv, T^{-1} Tv\rangle\;=\;\langle P_T S_T g, T^{-1}P_T S_T g\rangle\;=\;\langle h, P_T T^{-1} P_T h\rangle\,,
\end{split}
\]
whence $\langle h, S_T^{-1} h\rangle=\langle h, S_\mathrm{F}^{-1} h\rangle+\langle h, P_T T^{-1} P_T h\rangle$.

 (ii) Any $h\in\cH=\ran\, S_B$ reads $h=S_B g$ for some  
$g=f+(S_\mathrm{F}^{-1}+B)\widetilde{u}_1+u_0$ and suitable $f\in\mathcal{D}(\overline{S})$, $\widetilde{u}_1\in\mathcal{D}(B)$, $u_0\in\ker S^*\cap \mathcal{D}(B)^\perp$ (Theorem \ref{thm:VB-representaton-theorem}), therefore $F:=f+S_\mathrm{F}^{-1}\widetilde{u}_1\in\mathcal{D}(S_\mathrm{F})$,  $h:=S_Bg=S_\mathrm{F} F=\overline{S}f+\widetilde{u}_1$, and $\widetilde{u}_1=P_B S_B g$. Then
\[
\begin{split}
\langle h, S_B^{-1} h\rangle\;&=\;\langle g, S_B g\rangle\;=\; \langle F, S_\mathrm{F} F\rangle + \langle  B\widetilde{u}_1+u_0, \overline{S}f+\widetilde{u}_1\rangle \\
&=\langle F, S_\mathrm{F} F\rangle + \langle \widetilde{u}_1, B\widetilde{u}_1\rangle \;=\;\langle h, S_\mathrm{F}^{-1} h\rangle\,.
\end{split}
\]
 On the other hand, $\langle F, S_\mathrm{F} F\rangle=\langle h, S_\mathrm{F}^{-1} h\rangle$, analogously to what argued in (i), and 
 \[
 \langle \widetilde{u}_1, B\widetilde{u}_1\rangle\;=\;\langle P_B S_B g, B P_B S_B g\rangle\;=\;\langle h, P_B B P_B h \rangle\,,
\]
 whence $\langle h, S_B^{-1} h\rangle=\langle h, S_\mathrm{F}^{-1} h\rangle+\langle h, P_B B P_B h \rangle$.
\end{proof}

\begin{corollary}\label{cor:resolvent_formula_positive_extensions}
 Let $S$ be a densely defined and positive symmetric operator with greatest lower bound $\mathfrak{m}(S)>0$ on the Hilbert space $\cH$.
Let $\widetilde{S}$ be a  self-adjoint extension of $S$ and let $\lambda<\mathfrak{m}(S)$ be such that $\widetilde{S}-\lambda\mathbbm{1}$ is invertible on the whole $\cH$ (for example, a semi-bounded  extension $\widetilde{S}$ and a real number $\lambda<\mathfrak{m}(\widetilde{S})$).
Let $T(\lambda)$ be the extension parameter, in the sense of the parametrisation \eqref{eq:ST} of Theorem \ref{thm:VB-representaton-theorem_Tversion}, of the operator $\widetilde{S}-\lambda\mathbbm{1}$ considered as a self-adjoint extension of the densely defined and bottom-positive symmetric operator $S(\lambda):=S-\lambda\mathbbm{1}$. Correspondingly, let $P(\lambda):\cH\to\cH$ be the orthogonal projection onto $\overline{\mathcal{D}(T(\lambda))}$. Then
%
%
\begin{equation}\label{eq:resolvent_formula_z}
(\widetilde{S}-\lambda\mathbbm{1})^{-1}\;=\;(S_\mathrm{F}-\lambda\mathbbm{1})^{-1}\,+\,P(\lambda) \,T(\lambda)^{-1} P(\lambda)\,.
\end{equation}
\end{corollary}

\begin{proof}
Since $\mathfrak{m}(S(\lambda))=\mathfrak{m}(S)-\lambda>0$, the assumptions of Theorem \ref{thm:resolvent_formula_positive_extensions} are matched and \eqref{eq:resolvent_formula_z=0} takes the form \eqref{eq:resolvent_formula_z} owing to the fact that the Friedrichs extension of $S(\lambda)$ is precisely $S_\mathrm{F}-\lambda\mathbbm{1}$ (Theorem \ref{thm:Friedrichs-ext}(vii)).
%
\end{proof}

 \begin{remark} The  resolvent difference $(\widetilde{S}-\lambda\mathbbm{1})^{-1}-(S_\mathrm{F}-\lambda\mathbbm{1})^{-1}$ considered in Corollary \ref{cor:resolvent_formula_positive_extensions} has only non-zero matrix elements, as \eqref{eq:resolvent_formula_z} shows, on a precise subspace of $\ker(S^*-\lambda\mathbbm{1})$. The bounded-operator-valued map $\lambda\mapsto P(\lambda)T(\lambda)^{-1}P(\lambda)$ remains here somewhat implicit, although of course $T(\lambda)$ and $P(\lambda)$ are unambiguously and constructively identified in terms of the given $\widetilde{S}-\lambda\mathbbm{1}$ (Theorem \ref{thm:VB-representaton-theorem_Grubbversion}(ii)).
\end{remark}

  Theorem \ref{thm:resolvent_formula_positive_extensions} has a direct counterpart in the setting of Theorem \ref{thm:VB-representaton-theorem_Tversion2}, namely when the given $S$ has a self-adjoint extension $S_\mathrm{D}$ invertible on the whole $\cH$. The proof is an immediate adaptation, replacing $S_\mathrm{F}$ with $S_\mathrm{D}$.

  \begin{theorem}\label{thm:res-gen}
  Let $S$ be a densely defined symmetric operator on the Hilbert space $\cH$, which admits a self-adjoint extension $S_\mathrm{D}$ that has everywhere defined bounded inverse on $\cH$. Let $S_T$ be a distinct self-adjoint extension of $S$, in the notation of the parametrisation \eqref{eq:ST-2} of Theorem \ref{thm:VB-representaton-theorem_Tversion2}, such that $S_T$ has everywhere defined and bounded inverse on $\cH$, and let $P_T:\cH\to\cH$ be the orthogonal projection onto the Hilbert subspace $\mathcal{K}\equiv\overline{\mathcal{D}(T)}$. Then the extension parameter $T$ is invertible with everywhere defined and bounded inverse on $\mathcal{K}$ and 
   \begin{equation}\label{eq:resolvent_formula_z0GEN}
S_T^{-1}\;=\;S_\mathrm{D}^{-1}\,+\,P_T \,T^{-1} P_T\,.
\end{equation} 
\end{theorem}

\begin{theorem}[Kre{\u\i}n's resolvent formula for deficiency index = 1]\label{thm:Kreins_resolvent_formula_1}
 Let $S$ be a densely defined and positive symmetric operator with greatest lower bound $\mathfrak{m}(S)>0$ on the Hilbert space $\cH$, and with deficiency index $\dim \ker S^*=1$. Let $\widetilde{S}$ be a self-adjoint extension of $S$ other than the Friedrichs extension $S_\mathrm{F}$.  
Let $v\in\ker S^*\!\setminus\!\{0\}$ and for each $\lambda\in(-\infty,\mathfrak{m}(S))\cap\rho(\widetilde{S})$ 
set
\begin{equation}\label{eq:v(z)}
v(\lambda)\;:=\;v + \lambda (S_\mathrm{F}-\lambda\mathbbm{1})^{-1} v\,.
\end{equation}
Then there exists an analytic function $\beta:(-\infty,\mathfrak{m}(S))\cap\rho(\widetilde{S})\to\mathbb{R}$, with $\beta(\lambda)> 0$, such that
\begin{equation}\label{eq:Kreins_resolvent_formula_1}
(\widetilde{S}-\lambda\mathbbm{1})^{-1}\;=\;(S_\mathrm{F}-\lambda\mathbbm{1})^{-1}\,+\,\beta(\lambda)\,|v(\lambda)\rangle\langle v(\lambda)|\,.
\end{equation}
$\beta(\lambda)$, $v(\lambda)$, and \eqref{eq:Kreins_resolvent_formula_1} admit an analytic continuation to $\rho(S_\mathrm{F})\cap\rho(\widetilde{S})$. 
\end{theorem}

\begin{proof} Clearly,  $\dim\ker(S^*-\lambda\mathbbm{1})=\dim \ker S^*=1$ (Sect.~\ref{sec:I-symmetric-selfadj}). Moreover, $\widetilde{S}$ is lower semi-bounded (Theorem \ref{cor:finite_deficiency_index}).

 As $\lambda<\mathfrak{m}(\widetilde{S})$, $\widetilde{S}-\lambda\mathbbm{1}$ is a bottom-positive self-adjoint extension of the densely defined and bottom-positive symmetric operator $S(\lambda):=S-\lambda\mathbbm{1}$.
Its extension parameter $T(\lambda)$, in the sense of the parametrisation \eqref{eq:ST} of Theorem \ref{thm:VB-representaton-theorem_Tversion}, is a bottom-positive self-adjoint operator on the one-dimensional space $\ker(S^*-\lambda\mathbbm{1})$: indeed, $\mathfrak{m}(T(\lambda))\geqslant \mathfrak{m}(\widetilde{S}-\lambda\mathbbm{1})>0$ (due to \eqref{eq:II-mTgeqmST} from Theorem \ref{thm:semibdd_exts_operator_formulation_Tversion}). Therefore, $T(\lambda)$ acts as the multiplication by a positive number $t(\lambda)$.

As is evident from \eqref{eq:v(z)}, $v(\lambda)\in\ker(S^*-\lambda\mathbbm{1})$. In addition, $v(\lambda)\neq 0$ for each admissible $\lambda$: this is obviously true if $\lambda=0$, and if it was not true  for $\lambda\neq 0$, then 
$\lambda (S_\mathrm{F}-\lambda\mathbbm{1})^{-1} v=-v\neq 0$, which would contradict $\mathcal{D}(S_\mathrm{F})\cap\ker S^*=\{0\}$ (Proposition \ref{prop:II-KVB-decomp-of-Sstar}). Thus, for every admissible $\lambda$ the operator $P(\lambda):=\|v(\lambda)\|^{-2}|v(\lambda)\rangle\langle v(\lambda)|$ is the orthogonal projection onto $\ker(S^*-\lambda\mathbbm{1})$.

 Now the resolvent formula \eqref{eq:resolvent_formula_z} takes precisely the form \eqref{eq:Kreins_resolvent_formula_1} with $\beta(\lambda):=\|v(\lambda)\|^{-2} \,t(\lambda)^{-1}$. In particular, $\beta(\lambda)>0$. Moreover,  $\lambda\mapsto(\widetilde{S}-\lambda\mathbbm{1})^{-1}$ and $\lambda\mapsto(S_\mathrm{F}-\lambda\mathbbm{1})^{-1}$ are analytic operator-valued functions on  $\rho(S_\mathrm{F})\cap\rho(\widetilde{S})$ (Sect.~\ref{sec:I-spectrum}) and so too is the vector-valued function $\lambda\mapsto v(\lambda)$ (because of \eqref{eq:v(z)}). Therefore, 
$\lambda\mapsto\beta(\lambda)$ is analytic on $\rho(S_\mathrm{F})\cap\rho(\widetilde{S})$, and real analytic on $(-\infty,\mathfrak{m}(S))\cap\rho(\widetilde{S})$. This completes the proof.
\end{proof}

 \section{Self-adjoint extensions with Friedrichs lower bound}

  As discussed already at the beginning of Section \ref{sec:II-KreinExtTheory}), the possible \emph{non-uniqueness} of self-adjoint extensions of a given lower semi-bounded and densely defined symmetric operator on Hilbert space, beside the certainly existing Friedrichs extension, all characterised by the property of having the same greatest lower bound of the considered symmetric operator itself, was known already since the 1936 work of  
  Freudenthal \cite{Freudenthal-1936}. As a standard pedagogical example (see, e.g., \cite[Section 1]{M-2020-Friedrichs}), the two self-adjoint negative Laplacians on a finite interval $[a,b]$, respectively with Dirichlet and with anti-periodic boundary conditions, both extend the negative Laplacian defined on smooth functions compactly supported in $[a,b]$, and both preserve the greatest lower bound: the Dirichlet Laplacian is the Friedrichs extension, but clearly Dirichlet and anti-periodic Laplacian are distinct operators, in particular they have different spectra.

  Self-adjoint extensions with the same greatest lower bound as the considered lower semi-bounded and densely defined symmetric operator $S$ are conveniently referred to as \emph{top extensions}\index{top extensions} of $S$. The sub-family of top extensions of $S$ is never empty, as it obviously contains the Friedrichs extension $S_\mathrm{F}$ (Theorem \ref{thm:Friedrichs-ext}(iii)). Moreover, according to Kre{\u\i}n's theory (Theorem \ref{thm:II-Krein-final-thm}), the top extensions of $S$ are themselves ordered by operator inclusion, hence in the sense of inclusion of their form domains, starting from the highest one, the Friedrichs extension. In addition, a careful application of the Vi\v{s}ik-Birman classification scheme (Sect.~\ref{sec:II-VBparametrisation-orig}-\ref{sec:II-VBreparametrised}) allows to parametrise the top extensions with a precise range of extension parameters, and to show the existence of a unique smallest top extension, defined in the following as the \emph{least-top extension} $S_\mathrm{LT}$ of $S$.

  The result of this analysis, recently performed in \cite{M-2020-Friedrichs}, is formulated as follows. Observe, as usual, that the generic lower semi-bounded case can be straightforwardly brought to the form covered by part (ii)-(iii) of Theorem \ref{thm:II-top-extensions} here below.

  \begin{theorem}[Extensions with Friedrichs lower bound]\label{thm:II-top-extensions}\index{top extensions}
   Let $S$ be a densely defined and lower semi-bounded symmetric operator on the Hilbert space $\cH$, and let $S_\mathrm{F}$ be its Friedrichs extension.
   \begin{enumerate}[(i)]
    \item Among all self-adjoint extensions $S^{\mathrm{\,top}}$ of $S$ with $\mathfrak{m}(S^{\mathrm{\,top}})=\mathfrak{m}(S)$, i.e., the \emph{top extensions} of $S$, there exist a unique lowest one, called \emph{least-top extension}\index{least-top extension}, and denoted as $S_\mathrm{LT}$. Any other top extension $S^{\mathrm{\,top}}$ satisfies
    \begin{equation}\label{eq:SLTStopSF}
     S_\mathrm{LT}\;\leqslant\;S^{\mathrm{\,top}}\;\leqslant\;S_\mathrm{F}\,.
    \end{equation}
   \end{enumerate}
   Assume further, for concreteness, that $S$ has greatest lower bound $\mathfrak{m}(S)>0$.
\begin{enumerate}
  \item[(ii)] Necessary and sufficient condition for $S$ to admit self-adjoint extensions other then the Friedrichs extensions and with the same lower bound $\mathfrak{m}(S)$, i.e, necessary and sufficient condition for $S_\mathrm{LT}\neq S_\mathrm{F}$, is that 
 \begin{equation}\label{eq:II-maincondSLT}
  \mathrm{ran}(S_\mathrm{F}-\mathfrak{m}(S)\mathbbm{1})^{\frac{1}{2}}\cap\ker S^*\;\neq\;\{0\}\,.
 \end{equation}
  \item[(iii)] When condition \eqref{eq:II-maincondSLT} is matched, the expression 
 \begin{equation}\label{eq:defq}
  \begin{split}
   \mathcal{D}[q]\;&:=\;\mathrm{ran}(S_\mathrm{F}-\mathfrak{m}(S)\mathbbm{1})^{\frac{1}{2}}\cap\ker S^* \, , \\
   q[v]\;&:=\;\mathfrak{m}(S)\|v\|^2+\mathfrak{m}(S)^2\big\| (S_\mathrm{F}-\mathfrak{m}(S)\mathbbm{1})^{-\frac{1}{2}}v\big\|^2
  \end{split}
 \end{equation}
 defines a symmetric, closed, and strictly positive quadratic form $q$ on the Hilbert subspace $\overline{\mathcal{D}[q]}\subset\cH$, and with respect to the extension parametrisation $S_T\leftrightarrow T$ given by \eqref{eq:ST} of Theorem \ref{thm:VB-representaton-theorem_Tversion} one has
 \begin{equation}\label{eq:II-SLT-T0}
  S_\mathrm{LT}\;\equiv\; S_{T_\circ}\,,
 \end{equation}
 where the extension parameter $T_\circ$ is the self-adjoint operator in $\overline{\mathcal{D}[q]}$ uniquely associated with $q$. 
 Within the same extension parametrisation,
 \begin{equation}\label{eq:II-Stop-T0}
  \mathfrak{m}(S_T)\;=\;\mathfrak{m}(S)\qquad\Leftrightarrow\qquad T\geqslant T_\circ\,.
 \end{equation}
%
 \end{enumerate}
 \end{theorem}

  \begin{proof}
   Upon shifting $S$ to $S+\lambda\mathbbm{1}$ with $\lambda>-\mathfrak{m}(S)$, then applying parts (ii) and (iii), and finally shifting backwards, one deduces part (i). 
   On account of \eqref{eq:II-Stop-T0} and \eqref{eq:extension_ordering_Tversion}, $S_\mathrm{LT}$ is the lowest top extension and all other top extensions are ordered as in \eqref{eq:SLTStopSF}. It then remains to prove parts (ii) and (iii) under the additional assumption $\mathfrak{m}(S)>0$.

   Concerning (ii), let $S_T$ be a self-adjoint extension of $S$, labelled by some $T\in\mathcal{S}(\ker S^*)$, with the property $\mathfrak{m}(S_T)=\mathfrak{m}(S_\mathrm{F})$ and $S_T\neq S_\mathrm{F}$. For each $n\in\mathbb{N}$ let $\mu_n:=\mathfrak{m}(S)-\frac{1}{n}$.
   Since $S_T\geqslant \mu_n\mathbbm{1}$, then
 \[
  \langle v,T v\rangle\;\geqslant\;\mu_n\|v\|^2+\:\mu_n^2\,\big\|(S_\mathrm{F}-\mu_n\mathbbm{1})^{-\frac{1}{2}} v\big\|^2 \qquad \forall v\in\mathcal{D}(T)
 \]
 (Theorem \ref{thm:semibdd_exts_operator_formulation_Tversion}), whence also, since $\mu_n\uparrow\mathfrak{m}(S)$,
 \[
  \limsup_{n\to\infty}\big\|(S_\mathrm{F}-\mu_n\mathbbm{1})^{-\frac{1}{2}} v\big\|^2\;<\;+\infty\,.
 \]
 In fact, for each $v$ the sequence of square norms $\|(S_\mathrm{F}-\mu_n\mathbbm{1})^{-\frac{1}{2}} v\|^2$ is monotone increasing. For, if $\mathfrak{m}(S)>\mu'>\mu$, then
 \[
  \begin{split}
   \big\|(S_\mathrm{F}-\mu'\mathbbm{1})^{-\frac{1}{2}} v\big\|^2&-\big\|(S_\mathrm{F}-\mu\mathbbm{1})^{-\frac{1}{2}} v\big\|^2 \\
   &=\;\int_{[\mathfrak{m}(S),+\infty)}\Big(\frac{1}{\lambda-\mu'}-\frac{1}{\lambda-\mu}\Big)\,\ud\langle v,E^{(S_\mathrm{F})}(\lambda),{v}\rangle\;\geqslant\;0\,,
  \end{split}
 \]
 where $\ud E^{(S_\mathrm{F})}$ is the spectral measure of $S_\mathrm{F}$. Therefore,
 \[
  \exists\lim_{n\to\infty}\big\|(S_\mathrm{F}-\mu_n\mathbbm{1})^{-\frac{1}{2}} v\big\|^2\;<\;+\infty\,.
 \]
 As the latter conclusion is tantamount as
 \[
  \exists\lim_{n\to\infty}\int_{[\mathfrak{m}(S),+\infty)}\frac{1}{\lambda-\mu_n}\,\ud\langle v,E^{(S_\mathrm{F})}(\lambda),{v}\rangle\;<\;+\infty\,,
 \]
 then by monotone convergence the function $\lambda\mapsto(\lambda-\mathfrak{m}(S))^{-1}$ is summable with respect to the scalar measure $\ud\langle v,E^{(S_\mathrm{F})}(\lambda),{v}\rangle$. Thus, $\|(S_\mathrm{F}-\mathfrak{m}(S)\mathbbm{1})^{-\frac{1}{2}} v\|^2<+\infty$, whence $v\in\mathrm{ran}(S_\mathrm{F}-\mathfrak{m}(S)\mathbbm{1})^{\frac{1}{2}}$. On the other hand, by assumption $S_T\neq S_\mathrm{F}$ and consequently $\mathcal{D}(T)$ is a non-trivial subspace of $\ker S^*$ (Proposition \ref{prop:parametrisation_SF_SN_Tversion}). Summarising,
 \[
  \mathcal{D}(T)\;\subset\;\mathrm{ran}(S_\mathrm{F}-\mathfrak{m}(S)\mathbbm{1})^{\frac{1}{2}}\cap\ker S^* \, ,
 \]
 and therefore $\mathrm{ran}(S_\mathrm{F}-\mathfrak{m}(S)\mathbbm{1})^{\frac{1}{2}}\cap\ker S^*$ is non-trivial. The necessity part of the statement (ii) is proved.

  Conversely, concerning the sufficiency part of (ii), and concerning part (iii), assume now that $\mathrm{ran}(S_\mathrm{F}-\mathfrak{m}(S)\mathbbm{1})^{\frac{1}{2}}\cap\ker S^*\neq\{0\}$.

  The fact that \eqref{eq:defq} defines a symmetric quadratic form with strictly positive lower bound is obvious. The closedness of $q$ is checked by showing (Sect.~\ref{sec:I_forms}) that for any sequence $(v_n)_{n\in\mathbb{N}}$ in $\mathcal{D}[q]$ and $v\in\overline{\mathcal{D}[q]}$ with $v_n\to v$ and $q[v_n-v_m]\to 0$ as $n,m\to\infty$ one actually has $v\in\mathcal{D}[q]$ and $q[v_n-v]\to 0$. 
  The above assumption on $(v_n)_{n\in\mathbb{N}}$ implies in particular $v_n \to v$ and $ (S_\mathrm{F}-\mathfrak{m}(S)\mathbbm{1})^{-\frac{1}{2}}v_n\to u$
for some $v\in\cH$. As $(S_\mathrm{F}-\mathfrak{m}(S)\mathbbm{1})^{-\frac{1}{2}}$ is self-adjoint and hence closed, then
\[
 \begin{split}
  v\;&\in\;\mathcal{D}( (S_\mathrm{F}-\mathfrak{m}(S)\mathbbm{1})^{-\frac{1}{2}})\;=\;\mathrm{ran}( (S_\mathrm{F}-\mathfrak{m}(S)\mathbbm{1})^{\frac{1}{2}})\,, \\
  u\;&=\;(S_\mathrm{F}-\mathfrak{m}(S)\mathbbm{1})^{-\frac{1}{2}}v\,.
 \end{split}
\]
A first conclusion, since $\ker S^*$ is closed in $\cH$ and hence $v\in\ker S^*$ as well, is that $v\in \mathrm{ran}( (S_\mathrm{F}-\mathfrak{m}(S)\mathbbm{1})^{\frac{1}{2}})\cap\ker S^*=\mathcal{D}[q]$. As a further conclusion, since $v_n\to v$ and  $(S_\mathrm{F}-\mathfrak{m}(S)\mathbbm{1})^{-\frac{1}{2}}v_n\to (S_\mathrm{F}-\mathfrak{m}(S)\mathbbm{1})^{-\frac{1}{2}}v$ in $\cH$, one has $q[v_n-v]\to 0$.

 As $q$ is densely defined in the Hilbert subspace $\overline{\mathcal{D}[q]}$, and it is symmetric, closed, and lower semi-bounded, then (Sect.~\ref{sec:I_forms}) $q$ uniquely identifies a self-adjoint operator $T_\circ$ on $\overline{\mathcal{D}[q]}$ defined by 
\[
 \begin{split}
 \mathcal{D}(T_\circ)\;&:=\;\big\{ v\in \mathcal{D}[q]\,|\,\exists z_v\in\cH\textrm{ with } \langle u,z_v\rangle=q[u,v]\;\forall u\in\mathcal{D}[q]\big\}\,, \\
 T_\circ v\;&:=\;z_v\,.
 \end{split}
\]
 Since $\mathcal{D}[q]\subset\ker S^*$ and $\ker S^*$ is closed in $\cH$, then $\overline{\mathcal{D}[q]}\subset\ker S^*$, thus proving that $T_\circ\in\mathcal{S}(\ker S^*)$. In particular, $\mathcal{D}(T_\circ)\varsupsetneq\{0\}$, because $\mathcal{D}[q]\varsupsetneq\{0\}$, whence $S_{T_\circ}\neq S_\mathrm{F}$ (the Friedrichs extension is parametrised instead by ``$T=\infty$'', in the sense $\mathcal{D}[T]=\{0\}$, as Proposition \ref{prop:parametrisation_SF_SN_Tversion} states). The existence of a non-Friedrichs top extension $S_\mathrm{LT}:=S_{T_\circ}$ is established, which completes the sufficiency statement of (ii).

 To complete the proof of part (iii), it remains to establish \eqref{eq:II-Stop-T0}. For one implication, let $T\in\mathcal{S}(\ker S^*)$ with $T\geqslant T_q$, meaning that $\mathcal{D}(T)\subset\mathcal{D}(T_q)$ and 
\[
  \langle v,Tv\rangle\;\geqslant\;\langle v,T_q v\rangle \;=\;q[v]\;=\;\mathfrak{m}(S)\|v\|^2+\mathfrak{m}(S)^2\big\| (S_\mathrm{F}-\mathfrak{m}(S)\mathbbm{1})^{-\frac{1}{2}}v\big\|^2\quad \forall v\in\mathcal{D}(T)\,.
\]
 For an arbitrary $\mu<\mathfrak{m}(S)$, reasoning as for the proof of the necessity part above, 
\[
 \big\| (S_\mathrm{F}-\mathfrak{m}(S)\mathbbm{1})^{-\frac{1}{2}}v\big\|^2\;>\;\big\| (S_\mathrm{F}-\mu\mathbbm{1})^{-\frac{1}{2}}v\big\|^2\,,
\]
whence 
\[
  \langle v,Tv\rangle\;>\;\mu\|v\|^2+\mu^2\big\| (S_\mathrm{F}-\mu\mathbbm{1})^{-\frac{1}{2}}v\big\|^2\qquad \forall v\in\mathcal{D}(T)\,,
\]
 and therefore also $S_T\geqslant\mu\mathbbm{1}$ (Theorem \ref{thm:semibdd_exts_operator_formulation_Tversion}).
 By the arbitrariness of $\mu$, $S_T\geqslant\mathfrak{m}(S)\mathbbm{1}$. Thus, $\mathfrak{m}(S_T)\geqslant\mathfrak{m}(S)$, whence necessarily $\mathfrak{m}(S_T)=\mathfrak{m}(S)$ (Theorem \ref{thm:Friedrichs-ext}(vi)). One implication in formula \eqref{eq:II-Stop-T0} is proved.

 For the converse implication in \eqref{eq:II-Stop-T0}, let $S_T$, for some $T\in\mathcal{S}(\ker S^*)$, be a self-adjoint extension of $S$ with $\mathfrak{m}(S_T)=\mathfrak{m}(S)$. It is also assumed that condition \eqref{eq:II-maincondSLT} is matched and therefore, on account of part (ii), $S_T\neq S_\mathrm{F}$. The extension parameter $T$ must satisfy
 \begin{equation*}\tag{a}\label{eq:coreq}
 \begin{split}
  \mathcal{D}(T)\;&\subset\;\mathrm{ran}(S_\mathrm{F}-\mathfrak{m}(S)\mathbbm{1})^{\frac{1}{2}}\cap\ker S^*\qquad\textrm{and} \\
  \langle v,Tv\rangle \;&\geqslant\; \mathfrak{m}(S)\|v\|^2+\mathfrak{m}(S)^2\big\| (S_\mathrm{F}-\mathfrak{m}(S)\mathbbm{1})^{-\frac{1}{2}}v\big\|^2\qquad\forall v\in\mathcal{D}(T)\,.
 \end{split}
 \end{equation*}
 This is a consequence of the reasoning employed for the proof of the necessity statement of (ii). Indeed, the inclusion for $\mathcal{D}(T)$ was explicitly proved therein, as well as 
  \[\tag{b}\label{eq:nowb}
  \langle v,T v\rangle\;\geqslant\;\mu_n\|v\|^2+\:\mu_n^2\,\big\|(S_\mathrm{F}-\mu_n\mathbbm{1})^{-\frac{1}{2}} v\big\|^2\qquad\forall n\in\mathbb{N}\,.
 \]
 Moreover, for each $v\in\mathcal{D}(T)$, 
 \[
  \lim_{n\to\infty}\big\|(S_\mathrm{F}-\mu_n\mathbbm{1})^{-\frac{1}{2}} v\big\|^2\;=\;\big\|(S_\mathrm{F}-\mathfrak{m}(S)\mathbbm{1})^{-\frac{1}{2}} v\big\|^2\,,
 \]
 because
  \[
  \begin{split}
   \big\|(S_\mathrm{F}-&\mathfrak{m}(S)\mathbbm{1})^{-\frac{1}{2}} v\big\|^2-\big\|(S_\mathrm{F}-\mu_n\mathbbm{1})^{-\frac{1}{2}} v\big\|^2 \\
   &=\;\int_{[\mathfrak{m}(S),+\infty)}\Big(\frac{1}{\lambda-\mathfrak{m}(S)}-\frac{1}{\lambda-\mu_n}\Big)\,\ud\nu_{v}(\lambda)\;\xrightarrow[]{n\to\infty}\;0
  \end{split}
 \]
 by dominated convergence. Therefore, the limit $n\to\infty$ in \eqref{eq:nowb} yields the second line of \eqref{eq:coreq}. In turn, \eqref{eq:coreq} reads $\mathcal{D}(T)\subset\mathcal{D}[q]=\mathcal{D}[T_\circ]$ and $\langle v,Tv\rangle\geqslant q[v]=T_\circ[v]$ $\forall v\in\mathcal{D}(T)$. Such a condition for the two lower semi-bounded self-adjoint operators $T$ and $T_\circ$ is equivalent to $T\geqslant T_\circ$ (Sect.~\ref{sec:I_forms}). Thus, also the converse implication in \eqref{eq:II-Stop-T0} and hence the whole part (iii) is proved. 
  \end{proof}

  \begin{remark}
  The proof of Theorem \ref{thm:II-top-extensions} shows that the identification and parametrisation of the top extensions\index{top extensions} of $S$, when for concreteness $\mathfrak{m}(S)>0$, crucially relies on the equivalence \eqref{eq:SBsmbb-iff-invBsmbb_Tversion} from Birman's Theorem \ref{thm:semibdd_exts_operator_formulation_Tversion}. An important caution is needed, though. $(S_\mathrm{F}-\mathfrak{m}(S)\mathbbm{1})^{-1}$ cannot be defined as a bounded operator \emph{on the whole} $\cH$, yet it makes sense on $\mathrm{ran}(S_\mathrm{F}-\mathfrak{m}(S)\mathbbm{1})$, and when the latter space has a non-trivial intersection with $\ker S^*$, then there are non-zero vectors $v\in \mathrm{ran}(S_\mathrm{F}-\mathfrak{m}(S)\mathbbm{1})\cap\ker S^*$ on which $\langle v,(S_\mathrm{F}-\mathfrak{m}(S)\mathbbm{1})^{-1}v\rangle$ is indeed well defined. This makes the right-hand side of \eqref{eq:SBsmbb-iff-invBsmbb_Tversion} meaningful also when $\mu=\mathfrak{m}(S)$.
  When on such $v$'s one can define an operator $T\in\mathcal{S}(\ker S^*)$ satisfying \eqref{eq:SBsmbb-iff-invBsmbb_Tversion} for $\alpha=\mathfrak{m}(S)$, that $T$ identifies a top extension $S_T$. Clearly, underlying \eqref{eq:SBsmbb-iff-invBsmbb_Tversion} is the quadratic form language, so the actual operator to possibly invert in some subspace of $\ker S^*$ is rather $(S_\mathrm{F}-\mathfrak{m}(S)\mathbbm{1})^{\frac{1}{2}}$, a positive self-adjoint operator with zero lower bound. In this respect, as $S_\mathrm{F}$ is self-adjoint on $\cH$, and so is $S_\mathrm{F}-\mathfrak{m}(S)\mathbbm{1}$ with lower bound zero, then upon decomposing 
\[
 \cH\;=\;\overline{\mathrm{ran}(S_\mathrm{F}-\mathfrak{m}(S)\mathbbm{1})}\oplus\ker(S_\mathrm{F}-\mathfrak{m}(S)\mathbbm{1})
\]
the negative powers $(S_\mathrm{F}-\mathfrak{m}(S)\mathbbm{1})^{-\delta}$, $\delta>0$, are naturally defined as self-adjoint operators on the Hilbert subspace $\overline{\mathrm{ran}(S_\mathrm{F}-\mathfrak{m}(S)\mathbbm{1})}$.
\end{remark}

Densely defined, lower semi-bounded, non-essentially self-adjoint operators may or may not admit self-adjoint extensions, other than the Friedrichs one, with the same Friedrichs lower bound (and if they do, there is an infinity of them); in other words, their least top extension may or may not coincide with their Friedrichs extension. Appendix \ref{sec:appendixFriedrichsTOP} redirects to pedagogical examples of this phenomenon.

%
%
%
%
%
%


%
%
%

\begin{partbacktext}
\part{Applications}

\end{partbacktext}
%
%
%
\chapter{Hydrogenoid spectra with central perturbations}
\label{chapter-Hydrogenoid} 

 In this first chapter of `modern applications', the self-adjoint realisations of hydrogenoid Hamiltonians with central, point-like perturbation and their spectral analysis, the subject is on the one hand classical, in that the self-adjoint extension problem was completed long ago by means of von Neumann's extension scheme, but on the other hand it displays a degree of present-day topicality as a playground for the application of the Kre{\u\i}n-Vi\v{s}ik-Birman scheme. This point of view has been recently studied in \cite{GM-hydrogenoid-2018}, which is the reference the whole chapter is modelled on.

 Re-doing the analysis within this scheme, indeed, shows in which respects the latter approach is cleaner and provides more directly and naturally the characterisation of the models and their spectral content, and moreover the discussion of this Chapter allows for a useful comparison between the von Neumann and the Kre{\u\i}n-Vi\v{s}ik-Birman approach.

\section{Hydrogenoid Hamiltonians with point-like perturbation at the centre}\label{sec:Hydro-intro}

 In the precise sense that will be formalised in a moment, the class of models examined in this Chapter are certain realistic types of \emph{singular}, \emph{point-like} perturbations of the familiar quantum Hamiltonian for the valence electron of hydrogenoid atoms,\index{Hydrogenoid Hamiltonian} namely the operator
\begin{equation}\label{eq:Hhydr}
H_{\mathrm{Hydr}} \; = \; -\frac{\hslash^2}{2m_{\mathrm{e}}} \Delta - \frac{Z e^2}{|x|}
\end{equation} 
on $L^2(\mathbb{R}^3)$ with domain of self-adjointness $H^2(\mathbb{R}^3)$, where $m_{\mathrm{e}}$ and $-e$ are, respectively, the electron's mass and charge ($e>0$), $Z$ is the atomic number of the nucleus, $\hslash$ is Planck's constant and $\Delta$ is the three-dimensional Laplacian.

 Beside the actual problem of characterising and classifying such perturbations as new self-adjoint operators, this Chapter presents also the corresponding spectral analysis, eventually discussing the deviations from the celebrated spectrum of the hydrogen atom \cite{Landau-Lifshitz-3,Cohen-Tannoudji-1977-2020,sakurai_napolitano_2017,weinberg_2015}
\begin{equation}\label{eq:spectrum_hydrogeonid}\index{Hydrogenoid spectrum}
 \begin{split}
 \sigma_{\mathrm{ess}}(H_{\mathrm{Hydr}})\;&=\;\sigma_{\mathrm{ac}}(H_{\mathrm{Hydr}})\;=\;[0,+\infty)\,,\quad\sigma_{\mathrm{sc}}(H_{\mathrm{Hydr}})\;=\;\emptyset\,, \\
 \sigma_{\mathrm{point}}(H_{\mathrm{Hydr}})\;&=\;\big\{E_n^{(\mathrm{H})}\,\big|\,n\in\mathbb{N} \big\}\,, \\
 E_n^{(\mathrm{H})}\;&\!:=\; -\frac{Z^2 e^4 m_{\mathrm{e}}}{\,2\hslash^2 n^2}     \;=\;-\frac{(Ze)^2}{\,2n^2 a_{\mathrm{B}}}  \;=\;-m_{\mathrm{e}}c^2\frac{\,(Z\alpha_{\mathrm{f}})^2}{2 n^2}\,,
 \end{split}
\end{equation}
where $c$ is the speed of light,
\begin{equation}
 a_{\mathrm{B}}\;=\;\frac{\hslash^2}{\,m_{\mathrm{e}} e^2}
\end{equation}
is the Bohr radius,\index{Bohr radius} and
\begin{equation}
 \alpha_{\mathrm{f}}\;=\;\frac{\:e^2}{\hslash c}\;\approx\;\frac{1}{137}
\end{equation}
is the fine structure constant.\index{fine structure constant} Throughout this Chapter, whenever physical constants are explicitly kept into account, CGS physical units will be adopted, thus in particular $4\pi\varepsilon_0=1$.

 It turns out that intimately related to this problem is the issue of the self-adjoint realisations of the \emph{radial} differential operator
\begin{equation}
 -\frac{\ud^2}{\ud r^2}+\frac{\nu}{r}\,,\qquad \nu\in\mathbb{R}
\end{equation}
on the Hilbert space $L^2(\mathbb{R}^+)$ of the half-line, and of the classification of all such realisations and the characterisation of their spectra.

 Subsections \ref{sec:III-darwin}-\ref{sec:III-radialproblem} present the needed preliminaries. The main results are then formulated and discussed in Subsection \ref{sec:IIImainresults}. The technical analysis is developed in Sections \ref{sec:Section_of_Classification} and \ref{sec:III-perturbations}.

\subsection{Fine structure and Darwin correction}\label{sec:III-darwin}\index{Darwin term}

 Standard calculations within first-order perturbation theory, made first by Sommerfeld even before the complete definition of quantum mechanics (see, e.g., the classical discussion \cite[\S 34]{LandauLifshitz-4}), show that the correction $\delta E_n^{(\mathrm{H})}$ to the $n$-th eigenvalue $E_n^{(\mathrm{H})}$ of \eqref{eq:spectrum_hydrogeonid} is given by
\begin{equation}\label{eq:1storderPert}
 \frac{\,\delta E_n^{(\mathrm{H})}}{E_n^{(\mathrm{H})}}\;=\;-\frac{\,(Z\alpha_{\mathrm{f}})^2}{n}\Big(\frac{1}{j+\frac{1}{2}}-\frac{3}{4n}\Big)\,,
\end{equation}
where $j$ is the quantum number of the total angular momentum, thus $j=\frac{1}{2}$ if $\ell=0$ and $j=\ell\pm\frac{1}{2}$ otherwise, in the standard notation that will be made explicit in a moment. (The net effect is therefore a partial removal of the degeneracy of $E_n^{(\mathrm{H})}$ in the spin of the electron and in the angular number $\ell$, a double degeneracy remaining for levels with the same $n$ and $\ell=j\pm\frac{1}{2}$, apart from the maximum possible value $j_{\mathrm{max}}=n-\frac{1}{2}$.)

 It is worth recalling from standard physical arguments (see, e.g., \cite[Chapter 6]{Thaller-Dirac-1992}) that the first-order perturbative scheme yielding \eqref{eq:1storderPert} corresponds to adding to $H_{\mathrm{Hydr}}$ corrections that arise in the non-relativistic limit from the Dirac operator for the considered atom.

 $H_{\mathrm{Hydr}}$ is indeed formally recovered as one of the two identical copies of the spinor Hamiltonian obtained from the Dirac operator as $c\to +\infty$, and the eigenvalues of the latter, once the rest energy $m_{\mathrm{e}}c^2$ is removed, converge to those of $H_{\mathrm{Hydr}}$, with three types of sub-leading corrections, to the first order in $1/c^{2}$:
\begin{itemize}
 \item the \emph{kinetic energy correction},\index{kinetic energy correction} interpreted in terms of the replacement of the relativistic with the non-relativistic energy, that classically amounts to the contribution
 \[
  \Big( \sqrt{c^2p^2-m_{\mathrm{e}}^2c^4}-m_{\mathrm{e}}c^2\Big)-\frac{\,p^2}{2m_{\mathrm{e}}}\;=\;-\frac{1}{8m_{\mathrm{e}}^2c^2}\,p^4+O(c^{-4})\,;
 \]
 \item the \emph{spin-orbit correction},\index{spin-orbit correction} interpreted in terms of the interaction of the magnetic moment of the electron with the magnetic field generated by the nucleus in the reference frame of the former, including also the effect of the Thomas precession;
 \item the \emph{Darwin term correction},\index{Darwin term} interpreted as an effective smearing out of the electrostatic interaction between the electron and nucleus due to the Zitterbewegung,\index{Zitterbewegung} the rapid quantum oscillations of the electron.
\end{itemize}
In fact, each such modified eigenvalue $E_n^{(\mathrm{H})}+\delta E_n^{(\mathrm{H})}$ is the first-order term of the expansion in powers of $1/c^2$ of $E_{n,j}-m_{\mathrm{e}}c^2$, where $E_{n,j}$ is the Dirac operator's eigenvalue given by the celebrated Sommerfeld \emph{fine structure formula}\index{Sommerfeld fine structure formula}
\begin{equation}
 E_{n,j}\;=\;m_{\mathrm{e}}c^2\Big(1+\frac{(Z\alpha_{\mathrm{f}})^2}{\:(n-j-\frac{1}{2}+\sqrt{\kappa^2-(Z\alpha_{\mathrm{f}})^2})^2}\Big)^{\!-\frac{1}{2}}\,.
\end{equation}

 On the Darwin correction,\index{Darwin term} in particular, the following is worth being recalled. This correction is induced by the interaction between the magnetic moment of the moving electron and the electric field $\mathbf{E}=-\nabla V$, where $V$ is the electric potential due to the charge distribution that generates $\mathbf{E}$. 
This effect, to the first order in perturbation theory, produces an additive term to the non-relativistic Hamiltonian, which formally reads \cite[\S 33]{LandauLifshitz-4} 
\begin{equation}\label{eq:HDarw1}
 H_{\mathrm{Darwin}}\;=\;-\frac{e \hslash^2}{8 m_{\mathrm{e}}^2 c^2}\,\mathrm{div}\mathbf{E}\;=\;\frac{e \hslash^2}{8 m_{\mathrm{e}}^3 c^2}\,\Delta V\,.
\end{equation}
For a hydrogenoid atom, $V(x)=-Ze^2/|x|$, whence $\Delta V=4\pi Z e\delta^{(3)}(x)$ (here $\delta^{(3)}$ denotes the three-dimensional Dirac delta distribution): the term \eqref{eq:HDarw1} is therefore to be regarded as a \emph{point-like perturbation} `supported' at the centre of the atom, whose nuclear charge creates the field $\mathbf{E}$. 
In this case one gives meaning to \eqref{eq:HDarw1} in the sense of the expectation 
\begin{equation}\label{HDarwexp}\index{Darwin term}
 \langle\psi,H_{\mathrm{Darwin}}\psi\rangle\;=\;\frac{\,\pi\, e^2\hslash^2 Z }{\,2 m_{\mathrm{e}}^2 c^2}\,|\psi(0)|^2\;=\;-E_n^{(\mathrm{H})}\,\frac{(Z\alpha_{\mathrm{f}})^2}{n}\Big(\frac{n a_{\mathrm{B}}}{Z}\Big)^{\!3}\,\pi\,|\psi(0)|^2\,.
\end{equation}

Unlike the semi-relativistic kinetic energy and spin-orbit corrections, the Darwin correction\index{Darwin term} only affects the $s$ orbitals ($\ell=0$, $j=\frac{1}{2}$), the wave functions of higher orbitals vanishing at $x=0$. Since the $s$-wave normalised eigenfunction $\psi_n^{(\mathrm{H})}$ corresponding to $E_n^{(\mathrm{H})}$ satisfies $|\psi_n^{(\mathrm{H})}\!(0)|^2=\frac{1}{\pi}(\frac{Z}{n a_B})^3$, \eqref{HDarwexp} implies 
\begin{equation}\label{eq:1storderPert-Darwin}\index{Darwin term}
 \Big(\frac{\,\delta E_n^{(\mathrm{H})}}{E_n^{(\mathrm{H})}}\Big)_{\mathrm{Darwin}}\;=\;\frac{\,(Z\alpha_{\mathrm{f}})^2}{n}\qquad\qquad (\ell=0)\,.
\end{equation}

\subsection{Point-like perturbations supported at the interaction centre}

The above classical considerations are one of the typical \emph{motivations} for the rigorous study of a `simplified fine structure', low-energy correction of the ideal (non-relativistic) hydrogenoid Hamiltonian \eqref{eq:Hhydr} that consists of a Darwin-like perturbation \index{Darwin term} only. In particular, one considers an additional interaction that is only present in the $s$-wave sector.

This amounts to constructing self-adjoint Hamiltonians with Coulomb plus point interaction centred at the origin, and this requires to go beyond the formal perturbative arguments that yielded the spectral correction \eqref{eq:1storderPert-Darwin}.

One natural approach, exploited first in the early 1980's works by Zorbas \cite{Zorbas-1980}, by   Albeverio, Gesztesy, H{\o}egh-Krohn, and Streit \cite{AGHKS-1983_Coul_plus_delta}, and by Bulla and Gesztesy \cite{Bulla-Gesztesy-1985}, is to regard such Hamiltonians as self-adjoint extensions of the densely defined, symmetric, semi-bounded from below operator
\begin{equation}\label{eq:Hring1}
  \mathring{H}_{\mathrm{Hydr}}\;=\;\Big(-\frac{\:\hslash^2}{2m_{\mathrm{e}}}\Delta-\frac{Z e^2}{|x|}\Big)\Big|_{C^\infty_0(\mathbb{R}^3\setminus \{0\})}\,.
\end{equation}

For clarity of presentation it is convenient to set $\nu:=-Ze^2$, in fact allowing $\nu$ to be positive or negative real, and to work in units $2m_{\mathrm{e}}=\hslash=e=1$. One then writes $H^{(\nu)}$ and $\mathring{H}^{(\nu)}$ for the operator $-\Delta+\frac{\nu}{|x|}$ defined, respectively, on the domain of self-adjointness $H^2(\mathbb{R}^3)$ or on the restriction domain $C^\infty_0(\mathbb{R}^3\setminus \{0\})$.

As was found in \cite{Zorbas-1980,AGHKS-1983_Coul_plus_delta,Bulla-Gesztesy-1985}, the self-adjoint extensions of $\mathring{H}^{(\nu)}$ on $L^2(\mathbb{R}^3)$ at fixed $\nu$ form a one-parameter family $\{H_\alpha^{(\nu)}|\alpha\in(-\infty,+\infty]\}$ of rank-one perturbations, in the resolvent sense, of the Hamiltonian $H^{(\nu)}$. This famous result is formulated in Theorem \ref{thm:3Dclass} below.

In the present Chapter such a result, and other findings, are \emph{re-obtained} through an alternative path, proposed first by Gallone and Michelangeli in 2018 \cite{GM-hydrogenoid-2018}. In the above-mentioned works \cite{Zorbas-1980,AGHKS-1983_Coul_plus_delta,Bulla-Gesztesy-1985} the standard self-adjoint extension theory a la von Neumann (Sect.~\ref{sec:II-vN-theory}) was applied, whereas here the construction and classification of the self-adjoint Hamiltonians for the considered model is based on the Kre{\u\i}n-Vi\v{s}ik-Birman extension scheme (Sect.~\ref{sec:II-VBreparametrised}-\ref{sec:II-resolventsKVB}).

 Last, for later usage it is convenient to recall the known explicit expression of the integral kernel of $(H^{(\nu)}-k^2\mathbbm{1})^{-1}$ \cite{Hostler-1967}:
\begin{equation}
 \begin{split}
  (H^{(\nu)}-k^2\mathbbm{1})^{-1}(x,y)\;=&\,\;\Gamma({\textstyle1+\frac{\ii\nu}{2k}})\,\frac{\,\mathscr{A}_{\nu,k}(x,y)}{4\pi|x-y|}\,,\qquad x,y\in\mathbb{R}^3\,,\;\; x\neq y \, , \\
  \mathscr{A}_{\nu,k}(x,y)\;:=&\;\Big(\frac{\ud}{\ud \xi}-\frac{\ud}{\ud \eta}\Big)\,\mathscr{M}_{-\frac{\ii\nu}{2k},\frac{1}{2}}(\xi)\,\mathscr{W}_{-\frac{\ii\nu}{2k},\frac{1}{2}}(\eta)\bigg|_{\substack{\xi=-\ii k z_- \\ \eta=-\ii k z_+}} \, , \\
  z_\pm\;:=\;|x|+|y|&\pm|x-y|\,,\qquad k^2\in\rho(H_\alpha^{(\nu)})\,,\quad \mathfrak{Im}k>0\,,
 \end{split}
\end{equation}
 where $\mathscr{M}_{a,b}$ and $\mathscr{W}_{a,b}$ are the Whittaker functions\index{Whittaker functions} \cite[Chapter 13]{Abramowitz-Stegun-1964}.

\subsection{Angular decomposition}\label{sec:angular_decomp}

As customary, the rotational symmetry of $H^{(\nu)}$ and $\mathring{H}^{(\nu)}$ is exploited by passing to polar coordinates $x\equiv(r,\Omega)\in\mathbb{R}^+\!\times\mathbb{S}^2$, $r:=|x|$, for $x\in\mathbb{R}^3$. This induces the standard isomorphism
\begin{equation}\label{eq:ang_decomp}
\begin{split}
 L^2(\mathbb{R}^3,\ud x)\;&\cong\;
 U^{-1}L^2(\mathbb{R}^+,\ud r)\otimes L^2(\mathbb{S}^2,\ud\Omega) \\
 &\cong\;\bigoplus_{\ell=0}^\infty \Big( U^{-1}L^2(\mathbb{R}^+,\ud r)\otimes\mathrm{span}\{Y_{\ell}^{-\ell},\dots,Y_{\ell}^\ell\}\Big)\,,
\end{split}
\end{equation}
where $U:L^2(\mathbb{R}^+,r^2\ud r)\to L^2(\mathbb{R}^+,\ud r)$ is the unitary $(Uf)(r)=rf(r)$, and the $Y^m_\ell$'s are the spherical harmonics\index{spherical harmonics} on $\mathbb{S}^2$, i.e., the common eigenfunctions of $\boldsymbol{L}^2$ and $\boldsymbol{L}_3$ of eigenvalue $\ell(\ell+1)$ and $m$ respectively,  $\boldsymbol{L}=x \times (- \ii \nabla)$ being the angular momentum operator.

 It is then straightforward to see that $\mathring{H}^{(\nu)}$ (and analogously $H^{(\nu)}$) is reduced with respect to the Hilbert space orthogonal sum \eqref{eq:ang_decomp} (Sect.~\ref{sec:I_invariant-reducing-ssp}); explicitly,
\begin{equation}\label{eq:ang_decomp_operator}
 \mathring{H}^{(\nu)}\;\cong\;\bigoplus_{\ell=0}^\infty \Big( U^{-1} h^{(\nu)}_\ell U\otimes\mathbbm{1} \Big)\,,
\end{equation}
where each $h^{(\nu)}_\ell$ is the operator on $L^2(\mathbb{R}^+,\ud r)$ defined by
\begin{equation}\label{eq:AngularOperator}
h_{\ell}^{(\nu)} \; := \; -\frac{\ud^2}{\ud r^2} +\frac{\,\ell(\ell+1)}{r^2}+\frac{\nu}{r} \,,\qquad \mathcal{D}\big(h_{\ell}^{(\nu)}\big) \; :=\; C^\infty_0(\mathbb{R}^+)\,.
\end{equation}

\subsection{The radial problem}\label{sec:III-radialproblem}

Owing to \eqref{eq:ang_decomp}-\eqref{eq:ang_decomp_operator}, the question of the self-adjoint extensions of $\mathring{H}^{(\nu)}$ on $L^2(\mathbb{R}^3,\ud x)$ can be answered after the same problem is controlled for each $h_{\ell}^{(\nu)}$ on $L^2(\mathbb{R}^+,\ud r)$. In particular (Sect.~\ref{sec:I_invariant-reducing-ssp} and \ref{sec:I-symmetric-selfadj}), orthogonal sums, with respect to the Hilbert space decomposition \eqref{eq:ang_decomp} of self-adjoint extensions of the $h_{\ell}^{(\nu)}$'s in each sector are actual self-adjoint extensions of $\mathring{H}^{(\nu)}$.

Based on Weyl limit-point limit-circle analysis\index{Weyl criterion!limit-point limit-circle}\index{Weyl limit-point/limit-circle}\index{limit-point/limit-circle}\index{theorem!Weyl (limit-point limit-circle criterion)} (Sect.~\ref{sec:WeylsCriterion}), all the block operators $h_{\ell}^{(\nu)}$ with $\ell\in\mathbb{N}$ are essentially self-adjoint, as they are in the limit-point case both  at infinity and at zero.

One could also argue (later this conclusion will be retrieved along a different path) that $h_{0}^{(\nu)}$ is still in the limit-point case at infinity, yet limit-circle at zero, thus having deficiency indices $(1,1)$ and hence (Theorem \ref{thm:vonN_thm_symm_exts}) admitting a one-parameter family of self-adjoint extensions.

The question of the self-adjoint realisations of $\mathring{H}^{(\nu)}$ is then boiled down to the self-adjointness problem for $h_{0}^{(\nu)}$.

This too is a problem studied since long. The first analysis dates back to Rellich \cite{Rellich-1944} (even though self-adjointness was not the driving notion back then) and is based on Green function methods to show that $-\frac{\ud^2}{\ud r^2} +\frac{\nu}{r}+\ii\mathbbm{1}$ is inverted by a bounded operator on Hilbert space when the appropriate boundary condition at the origin is selected. 
The study of the corresponding Friedrichs extension, also for the more general class of Sturm-Liouville operators, attracted an interest on its own, significantly in the works of Kalf \cite{Kalf-1978} and Rosenberger \cite{Rosenberger-1985}.
Some four decades after Rellich's work, Bulla and Gesztesy \cite{Bulla-Gesztesy-1985} (a concise summary of which may be found in \cite[Appendix D]{albeverio-solvable}) produced a `modern' classification based on the special version of von Neumann's extension theory for second order differential operators \cite[Chapt.~8]{Weidmann-LinearOperatosHilbert}, in which the extension parameter that labels each self-adjoint realisation governs a boundary condition at zero analogous to \eqref{eq:bc}. (It was already mentioned that the work \cite{Bulla-Gesztesy-1985} came a few years after Zorbas \cite{Zorbas-1980} and Albeverio, Gesztesy, H\o{}egh-Krohn, and Streit \cite{AGHKS-1983_Coul_plus_delta} had classified the self-adjoint realisations of the three-dimensional problem directly, i.e., without explicitly working out the reduction discusses in Sec.~\ref{sec:angular_decomp}.) More recently, Gesztesy and Zinchenko \cite{Gesztesy-Zinchenko-2006} extended the scope of \cite{Bulla-Gesztesy-1985} to more singular potentials than $r^{-1}$,
and Khalile and Pankrashkin \cite[Appendix C]{Khalile-Pankrashkin-2018} re-did the construction of the self-adjoint realisations for the radial problem (also for more general Whittaker operators\index{Whittaker operators}) by means of von Neumann and boundary triplet methods.\index{boundary triplets}



\subsection{Main results: radial problem, 3D problem, eigenvalue correction.}\label{sec:IIImainresults}

Here are the main results of the present Chapter. On the one hand, classical facts (namely Theorem \ref{thm:1Dclass} below for the radial problem and Theorem \ref{thm:3Dclass} for the singularly-perturbed hydrogenoid Hamiltonians) are reproduced through the alternative extension scheme of Kre{\u\i}n, Vi\v{s}ik, and Birman. Previously studied objects are characterised in an explicit, new form, specifically the Friedrichs realisation of the radial operator (Theorem \ref{thm:1Fried}) and the final formula for the central perturbation of the hydrogenoid spectra (Theorem \ref{thm:EV_corrections}).

Clearly, whereas the derivatives in \eqref{eq:Hring1} and \eqref{eq:AngularOperator} are \emph{classical}, the following formulas contain \emph{weak} derivatives.

%

The first step is the identification of the operator closure and of the Friedrichs realisation of the radial problem.

\begin{theorem}[Closure and Friedrichs extension of $h^{(\nu)}_0$]\label{thm:1Fried}
The operator $h_0^{(\nu)}$ is semi-bounded from below with deficiency index one. 
\begin{enumerate}[(i)]
\item One has
\begin{equation}
\begin{split}
 \mathcal{D}(\overline{h_0^{(\nu)}})\;&=\;H^2_0(\mathbb{R}^+)\;=\;\overline{C^\infty_0(\mathbb{R}^+)}^{\Vert \, \Vert_{H^2}} \, , \\
 \overline{h_0^{(\nu)}}f\; &= \; - f''+\frac{\nu}{r} f \, .
\end{split}
\end{equation}
\end{enumerate}
The Friedrichs extension $h_{0,\mathrm{F}}^{(\nu)}$ of $h_0^{(\nu)}$ has 
\begin{enumerate}
\item[(ii)] operator domain and action given by
 \begin{equation}\label{eq:DomainHF}
\begin{split}
\mathcal{D}(h_{0,\mathrm{F}}^{(\nu)})\;&=\;H^2(\mathbb{R}^+) \cap H^1_0(\mathbb{R}^+)\;=\;\{f \in H^2(\mathbb{R}^+)\,|\, \lim_{r \downarrow 0} f(r) = 0\} \, ,
\\
h_{0,\mathrm{F}}^{(\nu)} f \; &= \; - f''+\frac{\nu}{r} f \, ;
\end{split}
\end{equation}
\item[(iii)] quadratic form given by
\begin{equation}\label{eq:DomainHFform}
\begin{split}
\mathcal{D}[h_{0,\mathrm{F}}] & = H^1_0(\mathbb{R}^+) \, , \\
h_{0,\mathrm{F}}^{(\nu)}[f,h] \; &=\; \int_0^{+\infty} \Big(\overline{f'(r)} h'(r)+ \nu\, \frac{\overline{f(r)} h(r)}{r} \Big) \, \ud r \, ;
\end{split}
\end{equation}
\item[(iv)] resolvent with integral kernel
\begin{equation}\label{eq:DomainInverse}
\begin{split}
 \!\!\!\!\!\!\!\!\Big( h^{(\nu)}_{0,\mathrm{F}}&+\frac{\nu^2}{4 \kappa^2} \Big)^{-1}(r,\rho)\;= \\
 &=\;-\frac{\kappa \Gamma(1-\kappa)}{\nu} \begin{cases}
\mathscr{W}_{\kappa,\frac{1}{2}}( -\textstyle{\frac{\nu}{\kappa}} r) \mathscr{M}_{\kappa,\frac{1}{2}}(-\textstyle{\frac{\nu}{\kappa}} \rho) \qquad \text{if }0 < \rho < r \, ,\\
\mathscr{M}_{\kappa,\frac{1}{2}}(-\textstyle{\frac{\nu}{\kappa}} r)\mathscr{W}_{\kappa,\frac{1}{2}}( -\textstyle{\frac{\nu}{\kappa}} \rho) \qquad \text{if } 0 < r < \rho\,,
\end{cases}
\end{split}
\end{equation}
where
\begin{equation}
 \kappa \,\in \,(-\infty,0)\cup(0,1)\,,\qquad \mathrm{sign} \, \kappa \,=\, - \mathrm{sign} \, \nu\,,
\end{equation}
and where $\mathscr{W}_{a,b}(r)$ and $\mathscr{M}_{a,b}(r)$ are the Whittaker functions.\index{Whittaker functions}
\end{enumerate}
\end{theorem}

Next, using the Friedrichs extension as a \emph{reference extension} for the Kre{\u\i}n-Vi\v{s}ik-Birman scheme, all other self-adjoint realisations of the radial problem are classified, re-obtaining classical findings from the literature \cite{Rellich-1944,Bulla-Gesztesy-1985}.

\begin{theorem}[Self-adjoint realisations of $h^{(\nu)}_0$]
\label{thm:1Dclass}

\begin{enumerate}[(i)]
\item The self-adjoint extensions of $h_0^{(\nu)}$ form the family $(h_{0,\alpha}^{(\nu)})_{\alpha \in \mathbb{R} \cup \{\infty\}}$, where $\alpha=\infty$ labels the Friedrichs extension, and 
\begin{equation}\label{eq:dsa}
 \begin{split}
  \mathcal{D}(h^{(\nu)}_{0,\alpha})\;&=\;\left\{
  g\in L^2(\mathbb{R}^+)\left|
  \begin{array}{c}
   -g''+\textstyle{\frac{\nu}{r}g}\in L^2(\mathbb{R}^+) \\
   \textrm{and }\;g_1\;=\;4\pi\alpha\, g_0
  \end{array}\!
  \right.\right\} \, , \\
  h^{(\nu)}_{0,\alpha}\,g\;&=\;-g''+\frac{\nu}{r}\,g\,,
 \end{split}
\end{equation}
$g_0$ and $g_1$ being the existing limits
\begin{equation}\label{eq:g0g1limits-statements}
 \begin{split}
  g_0\;&:=\;\lim_{r\downarrow 0}g(r) \, , \\
  g_1\;&:=\;\lim_{r\downarrow 0}r^{-1}\big(g(r)-g_0(1+\nu r\ln r)\big)\,.
 \end{split}
\end{equation}
\item For given $\kappa \in (-\infty,0)\cup(0,1)$, $\mathrm{sign}\, \kappa = - \mathrm{sign}\, \nu$, one has
 \begin{eqnarray}
  \!\!\!\!\!\!\!\!\!\!\!\!\!\!\!\!\!\!\!\!\!\!\!\!\!\!\!\!\!\!\!\!\!\!\!\!\!\!\!\!\Big(h^{(\nu)}_{0,\alpha}+ \frac{\nu^2}{4 \kappa^2} \Big)^{-1} &=& \Big( h^{(\nu)}_{0,\infty}+\frac{\nu^2}{4 \kappa^2} \Big)^{-1}+ \frac{\Gamma(1-\kappa)^2}{4 \pi} \frac{1}{\alpha-\mathfrak{F}_{\nu,\kappa}} | \Phi_\kappa \rangle \langle \Phi_\kappa |\,, \label{eq:1Dresolvent} \\
  \mathfrak{F}_{\nu,\kappa}\;&:=&\;\frac{\nu}{4\pi}\big(\Psi(1-\kappa)+\ln(-{\textstyle\frac{\nu}{\kappa}})+(2\gamma-1)+{\textstyle\frac{1}{2 \kappa}}\big)\,, \label{eq:Fnuk}
 \end{eqnarray}
where $\Phi_\kappa(r)\;:=\;\mathscr{W}_{\kappa,\frac{1}{2}}(-\frac{\nu}{\kappa} r)$, $\gamma\approx 0.577$ is the Euler-Mascheroni constant,\index{Euler-Mascheroni constant} and $\Psi(z)=\Gamma'(z)/\Gamma(z)$ is the digamma function.\index{digamma function} 
\end{enumerate}
\end{theorem}

Consistently, when $\nu=0$ the boundary condition \eqref{eq:dsa} for the $\alpha$-extension takes the classical form $g'(0)=4\pi\alpha g(0)$, namely the well-known boundary condition for the generic self-adjoint Laplacian on half-line \cite{Kostrykin-Schrader-2006,GTV-2012,DM-2015-halfline}.


When the radial analysis is lifted back to the three-dimensional Hilbert space, one re-obtains, thus through an alternative path, the following classification result already available in the literature (see, e.g., \cite[Theorem I.2.1.2]{albeverio-solvable}).

\begin{theorem}[Self-adjoint realisations of $\mathring{H}^{(\nu)}$]
\label{thm:3Dclass}

\noindent The self-adjoint extensions of $\mathring{H}^{(\nu)}$ form the family $(H^{(\nu)}_\alpha)_{\alpha\in\mathbb{R}\cup\{\infty\}}$ characterised as follows. 
\begin{enumerate}[(i)]
 \item With respect to the canonical decomposition \eqref{eq:ang_decomp} of $L^2(\mathbb{R}^3)$, the extension $H^{(\nu)}_\alpha$ is reduced as
 \begin{equation}\label{eq:ang_decomp_Halpha}
 H^{(\nu)}_\alpha\;\cong\;\bigoplus_{\ell=0}^\infty \Big( U^{-1} h^{(\nu)}_{\ell,\alpha}\, U\otimes\mathbbm{1} \Big)\,,
\end{equation}
where $h^{(\nu)}_{0,\alpha}$ is characterised in Theorem \ref{thm:1Dclass} and $h^{(\nu)}_{\ell,\alpha}$, for $\ell\geqslant 1$, is the closure of $h^{(\nu)}_{\ell}$ introduced in \eqref{eq:AngularOperator}, namely the $L^2(\mathbb{R}^+)$-self-adjoint operator
\begin{equation}\label{eq:domhell1}
 \begin{split}
  \mathcal{D}(h^{(\nu)}_{\ell,\alpha})\;&=\;\{g\in L^2(\mathbb{R}^+)\,|\,-g''+\textstyle{\frac{\,\ell(\ell+1)}{r^2}g+\frac{\nu}{r}g}\in L^2(\mathbb{R}^+)\}  \, ,\\
  h^{(\nu)}_{\ell,\alpha}\,g\;&=\;-g''+\textstyle{\frac{\,\ell(\ell+1)}{r^2}g+\frac{\nu}{r}g}\,.
 \end{split}
\end{equation}
\item The choice $\alpha=\infty$ identifies the Friedrichs extension of $\mathring{H}^{(\nu)}$, which is precisely the self-adjoint hydrogenoid Hamiltonian
\begin{equation}\label{HnuFriedr}
 H^{(\nu)}\;=\;-\Delta+\frac{\nu}{\,|x|\,}\,,\qquad\mathcal{D}(H^{(\nu)})\;=\;H^2(\mathbb{R}^3)\,.
\end{equation}
 It is the only member of the family $(H^{(\nu)}_\alpha)_{\alpha\in\mathbb{R}\cup\{\infty\}}$ whose domain's functions have separately finite kinetic and finite potential energy, in the sense of energy forms.
 \item For given $\kappa \in (-\infty,0)\cup(0,1)$, $\mathrm{sign}\,\kappa=-\mathrm{sign}\,\nu$, one has
 \begin{equation}\label{eq:HnuResolvent}
  \Big(H^{(\nu)}_\alpha+\frac{\nu^2}{\,4\kappa^2}\,\mathbbm{1}\Big)^{\!-1}\;=\;\Big(H^{(\nu)}+\frac{\nu^2}{\,4\kappa^2}\,\mathbbm{1}\Big)^{\!-1} +\frac{1}{\,\alpha-\mathfrak{F}_{\nu,\kappa}\,}|\mathfrak{g}_{\nu,\kappa}\rangle\langle\mathfrak{g}_{\nu,\kappa}|\,,
 \end{equation}
 where
 \begin{equation}\label{eq:ourgnukappa}
 \mathfrak{g}_{\nu,\kappa}(x)\;:=\;\Gamma({\textstyle 1-\kappa})\,\frac{\,\,\mathscr{W}_{\kappa,\frac{1}{2}}(-\frac{\nu}{\kappa}|x|)}{4\pi|x|}
\end{equation}
and $\mathfrak{F}_{\nu,\kappa}$ is defined in \eqref{eq:Fnuk}.
  \item For given $\kappa \in (-\infty,0)\cup(0,1)$, $\mathrm{sign}\,\kappa=-\mathrm{sign}\,\nu$, one has
  \begin{equation}\label{eq:DomainHalpha}
   \begin{split}
    \mathcal{D}(H^{(\nu)}_\alpha)\;&=\;\Big\{ \psi=\varphi_\kappa+\frac{\varphi_\kappa(0)}{\,\alpha-\mathfrak{F}_{\nu,\kappa}\,}\,\mathfrak{g}_{\nu,\kappa}\Big|\,\varphi_\kappa\in H^2(\mathbb{R}^3)\Big\} \, , \\
    \Big(H^{(\nu)}_\alpha+\frac{\nu^2}{\,4\kappa^2}\,\mathbbm{1}\Big)\psi\;&=\;\Big(H^{(\nu)}+\frac{\nu^2}{\,4\kappa^2}\,\mathbbm{1}\Big)\varphi_\kappa\,,
   \end{split}
  \end{equation}
  the decomposition of each $\psi$ being unique.
\end{enumerate}
\end{theorem}

 It is worth observing that \eqref{eq:DomainHalpha} provides the typical decomposition of a generic element in $\mathcal{D}(H_\alpha^{(\nu)})$ into the \emph{regular} part $\varphi_\kappa \in H^2(\mathbb{R}^3)$ and the \emph{singular} part $\mathfrak{g}_{\nu,\kappa}\sim |x|^{-1}$ as $x\to 0$ with a precise boundary condition among the two.

 The uniqueness property of part (ii) above is another feature that emerges naturally within the Kre{\u\i}n-Vi\v{s}ik-Birman scheme, as will be argued in Corollary \ref{cor:Frie}. It gives the standard hydrogenoid Hamiltonian a somewhat physically distinguished status, in complete analogy with its semi-relativistic counterpart, the well-known distinguished realisation of the Dirac-Coulomb Hamiltonian (Theorem \ref{thm:recap}(ii)).

 The last result concerns the spectral analysis of each realisation $H_\alpha^{(\nu)}$. 

Since the  $H_\alpha^{(\nu)}$'s are rank-one perturbations, in the resolvent sense, of $H_{\alpha=\infty}^{(\nu)}\equiv H^{(\nu)}$,
then one deduces from \eqref{eq:spectrum_hydrogeonid} that 
\begin{equation}
 \sigma_{\mathrm{ess}}(H_\alpha^{(\nu)})\;=\;\sigma_{\mathrm{ac}}(H_\alpha^{(\nu)})\;=\;[0,+\infty)\,,\quad\sigma_{\mathrm{sc}}(H_\alpha^{(\nu)})\;=\;\emptyset\,,
\end{equation}
and only $\sigma_{\mathrm{point}}(H_\alpha^{(\nu)})$ differs from the corresponding $\sigma_{\mathrm{point}}(H^{(\nu)})$ (Sect.~\ref{sec:perturbation-spectra}).

Concerning the corrections to $\sigma_{\mathrm{point}}(H^{(\nu)})$ due to the central perturbation, one distinguishes among the two possible cases. If $\nu<0$, then the $n$-th eigenvalue $-\frac{\nu^2}{4n^2}$ in $\sigma_{\mathrm{point}}(H^{(\nu)})$ is $n^2$-fold degenerate, with partial $(2\ell+1)$-fold degeneracy in the sector of angular symmetry $\ell$ for all $\ell\in\{0,\dots,n-1\}$. All the eigenstates of $H^{(\nu)}$ with eigenvalue $-\frac{\nu^2}{4n^2}$ and with symmetry $\ell\geqslant 1$ are also eigenstates of any other realisation $H_\alpha^{(\nu)}$ with the same eigenvalue, because $H_\alpha^{(\nu)}$ is a perturbation of $H^{(\nu)}$ in the $s$-wave only. Thus, the effect of the central perturbation is a correction to the $\ell=0$ point spectrum of $H^{(\nu)}$, which consists of countably many non-degenerate eigenvalues $E_n:=-\frac{\nu^2}{4n^2}$, $n\in\mathbb{N}$.

If instead $\nu>0$, then a standard application of the Kato-Agmon-Simon Theorem\index{theorem!Kato-Agmon-Simon} (see e.g. \cite[Theorem XIII.58]{rs4}) gives $\sigma_{\mathrm{point}}(H^{(\nu)})=\emptyset$. Yet, if the central perturbation corresponds to an interaction that is attractive or at least not too much repulsive, then it can create one negative eigenvalue in the $\ell=0$ sector.

This is described in detail as follows.

\begin{theorem}[Eigenvalue corrections]
\label{thm:EV_corrections}

\noindent For given $\alpha\in\mathbb{R}\cup\{\infty\}$ and $\nu\in\mathbb{R}$, let $\sigma^{(0)}_{\mathrm{p}}(H^{(\nu)}_\alpha)$ be the point spectrum of the self-adjoint extension $H^{(\nu)}_\alpha$ with definite angular symmetry $\ell=0$ (`$s$-wave point spectrum'). Moreover, for $E<0$ let
 \begin{equation}\label{eq:Feigenvalues}
 \mathfrak{F}_\nu(E)\;:=\;\frac{\nu}{4\pi}\Big(\Psi\big(1+{\textstyle\frac{\nu}{2\sqrt{|E|}}}\big)+ \ln(2 \sqrt{|E|}) +2\gamma - 1 - {\textstyle\frac{\sqrt{|E|}}{\nu}} \Big)\,,
\end{equation}
 where $\gamma$ is the Euler-Mascheroni constant\index{Euler-Mascheroni constant} and $\Psi$ is the digamma function.\index{digamma function}
 \begin{enumerate}[(i)]
  \item If $\nu<0$, then the equation
  \begin{equation}\label{eq:theoremFEalpha}
   \mathfrak{F}_\nu(E)\;=\;\alpha
  \end{equation}
 admits countably many simple negative roots that form an increasing sequence $(E_n^{(\nu,\alpha)})_{n\in\mathbb{N}}$ accumulating at zero, and 
  \begin{equation}
  \sigma^{(0)}_{\mathrm{p}}(H^{(\nu)}_\alpha)\;=\;\big\{ E_n^{(\nu,\alpha)}\,\big|\, n\in\mathbb{N}\big\}\,.
 \end{equation}
 For the Friedrichs extension,
  \begin{equation}
   E_n^{(\nu,\alpha=\infty)}\;=\;E_n^{(\nu)}\;=\;-\frac{\nu^2}{4 n^2}\,,
  \end{equation}
 that is, the ordinary hydrogenoid eigenvalues.
 \item If $\nu>0$, then equation \eqref{eq:theoremFEalpha} has no negative roots if $\alpha\geqslant\alpha_\nu$, where
 \begin{equation}
  \alpha_\nu\;:=\;\frac{\nu}{4\pi}\,(\ln\nu+2\gamma-1)\,,
 \end{equation}
 and has one simple negative root $E_+^{(\nu,\alpha)}$ if $\alpha<\alpha_\nu$. Correspondingly,
 \begin{equation}
  \sigma^{(0)}_{\mathrm{p}}(H^{(\nu)}_\alpha)\;=\;
  \begin{cases}
   \emptyset & \textrm{ if }\;\alpha\geqslant\alpha_\nu\,, \\
   E_+^{(\nu,\alpha)}& \textrm{ if }\;\alpha<\alpha_\nu\,.
  \end{cases}
 \end{equation}
 \end{enumerate}
\end{theorem}

\begin{figure}
\includegraphics[width=6.2cm]{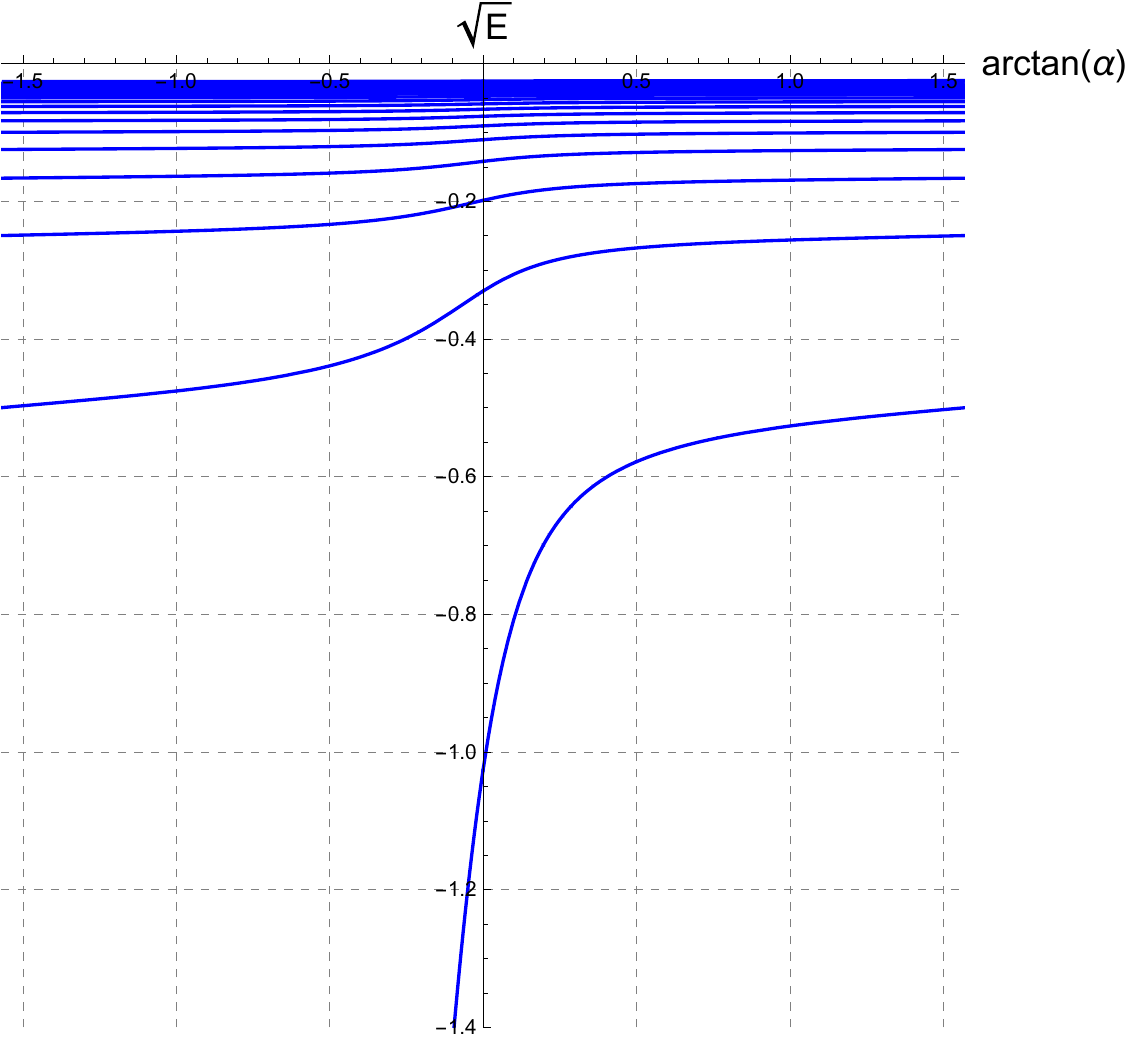}\quad
\includegraphics[width=5.8cm]{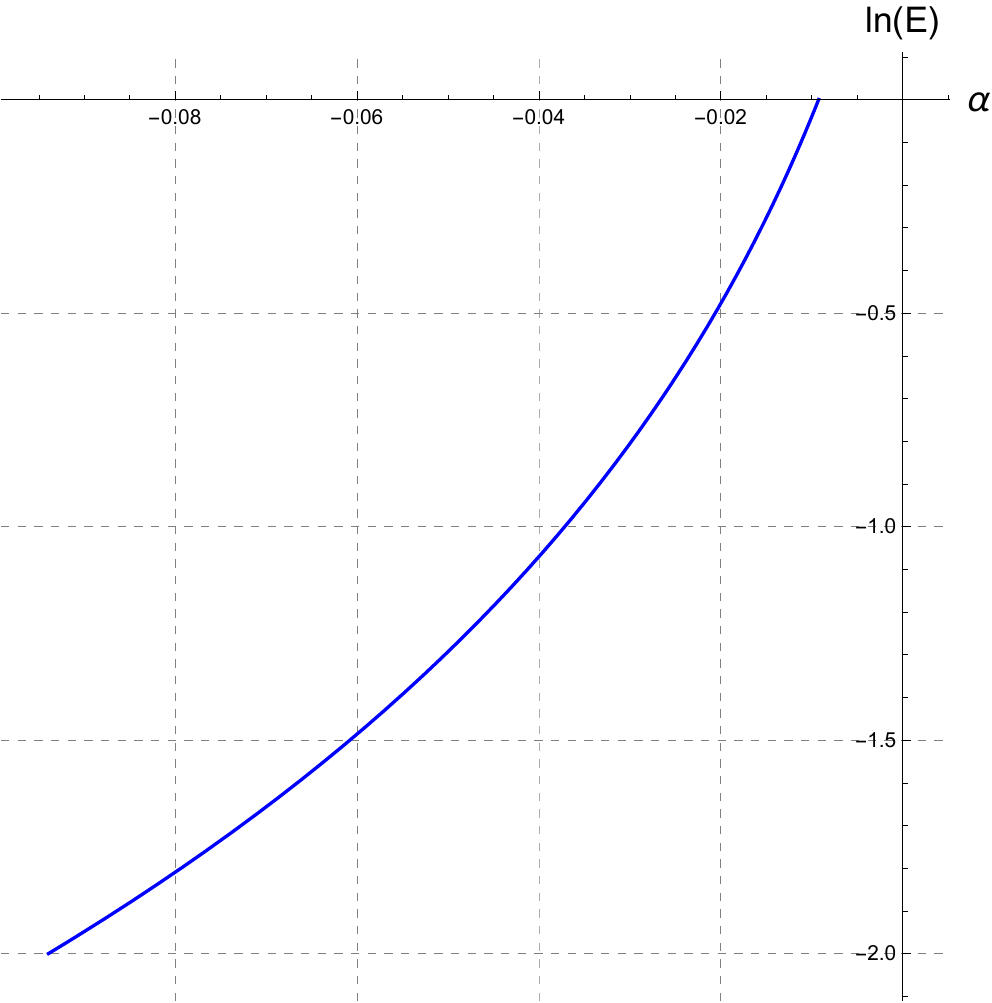}
\caption{Eigenvalues of the perturbed hydrogenoid Hamiltonian $H^{(\nu)}_\alpha$ for $\nu=-1$ (left) and $\nu=1$ (right). The scales of the energy $E$ and of the extension parameter $\alpha$ are modified to magnify the behaviour of the eigenvalues.} \label{fig:Eig}
\end{figure}

Figure  \ref{fig:Eig} displays the structure of the discrete spectrum described in Theorem \ref{thm:EV_corrections} above.

 Figure  \ref{fig:Eig} confirms (Sect.~\ref{sec:swaveEVproblem}) that when $ \nu < 0$ each $E_n^{(\nu,\alpha)}$ is smooth and strictly monotone in $\alpha$, with a typical \emph{fibred structure} of the union of all the discrete spectra $\sigma_{\text{disc}}(H^{(\nu)}_\alpha)$ 
\begin{equation}
(-\infty,0)\;=\bigcup_{\alpha \in (-\infty,+\infty]} \{ E_n^{(\nu,\alpha)}\, | \, n \in \mathbb{N}\} \;=\;\mathbb{R}\setminus \sigma_{\mathrm{ess}}(H_\alpha^{(\nu)})\,,
\end{equation}
(see Remark \ref{rem:spectra_fibre} below, as well as Remarks \ref{rem:DCfibredstructure} and \ref{cor:spectralanalysis} for an analogous phenomenon, respectively, for Dirac operators and for three-body Hamiltonians with zero-range integration),
and the correction $E_n^{(\nu,\alpha)}$ to the non-relativistic $E_n^{(\nu)}$ always \emph{decreases} the energy, with the intertwined relation $E_{n+1}^{(\nu,\alpha)}\geqslant E_n^{(\nu)}\geqslant E_n^{(\nu,\alpha)}$ (see Remark \ref{rem:EVdecreased}).

Analogously, when $\nu>0$,
\begin{equation}
(-\infty,0)\;=\bigcup_{\alpha \in (-\infty,\alpha_\nu)} \{ E_+^{(\nu,\alpha)}\} \;=\;\mathbb{R}\setminus \sigma_{\mathrm{ess}}(H_\alpha^{(\nu)})\,.
\end{equation}

\section{Hydrogenoid self-adjoint realisations}\label{sec:Section_of_Classification}

 This Section presents the construction that eventually yields Theorems \ref{thm:1Fried}, \ref{thm:1Dclass}, and \ref{thm:3Dclass}. The main focus are the self-adjoint extensions on $L^2(\mathbb{R}^+)$ of the radial operator  $h_{0}^{(\nu)}$. It is essentially equivalent, and technically more convenient, to study the self-adjoint extensions of the \emph{shifted} operator
\begin{equation}\label{eq:Symmetric}
 S\; :=\;-\frac{\ud^2}{\ud r^2} +\frac{\nu}{r}+\frac{\,\nu^2}{\,4\kappa^2}\,,\qquad\mathcal{D}(S)\;:=\;C^\infty_0(\mathbb{R}^+)\,,
\end{equation}
for generic 
\begin{equation}\label{eq:conditions_kappa}
 \kappa\,\in\,\mathbb{R}\,,\qquad\mathrm{sign}\,\kappa\;=\;-\mathrm{sign}\,\nu\,,\qquad0<|\kappa|<\textstyle\frac{1}{2}\,.
\end{equation}
Owing to \eqref{eq:spectrum_hydrogeonid}, $-\frac{\ud^2}{\ud r^2} +\frac{\nu}{r}\geqslant-\nu^2/4$, whence $S\geqslant{\textstyle\frac{1}{4}}\,\nu^2(\kappa^{-2}-1)$: thus, $S$ is densely defined and symmetric on $L^2(\mathbb{R}^+)$ with \emph{strictly positive bottom}. This feature will simplify the identification of the self-adjoint extensions of $S$, and allows for a direct application of the extension scheme from Section \ref{sec:II-VBreparametrised}: the corresponding extensions for $h_{0}^{(\nu)}$ are then obtained through a trivial shift.

It is also convenient to write
\begin{equation}
 \widetilde{S}\;: =\;-\frac{\ud^2}{\ud r^2} +\frac{\nu}{r}+\frac{\,\nu^2}{\,4\kappa^2} 
\end{equation}
when referring to the \emph{differential} action on functions in $L^2(\mathbb{R}^+)$, in the classical or the weak sense, with no reference to the operator domain.

 In view of Section \ref{sec:II-VBreparametrised}, an amount of preparatory steps are needed (Subsect.~\ref{sec:homo}-\ref{sec:SclosureSstar}), in which one identifies the subspaces $\mathcal{D}(\overline{S})$, $\ker S^*$, and $S_{\mathrm{F}}^{-1}\ker S^*$, $S_{\mathrm{F}}$ being the Friedrichs extension of $S$. 
In Subsection \ref{sec:SclosureSstar} $S_{\mathrm{F}}$ is characterised and Theorem \ref{thm:1Fried} is proved; in Subsection \ref{sec:KVB_for_S} the extensions of $S$ are classified and Theorem \ref{thm:1Dclass} is proved; last, in Subsection \ref{sec:3Dreconstr} Theorem~\ref{thm:3Dclass} is deduced from the previous results.

\subsection{The homogeneous radial problem}\label{sec:homo}

 The first subspace to be characterised is $\ker S^*$. The adjoint of $S$ is the maximal realisation\index{differential operator!maximal realisation} of the differential operator $\widetilde{S}$ (Sect.~\ref{sec:MinimalAndMaximalRealisations}) in $L^2(\mathbb{R}^+)$, i.e.,
\begin{equation}
 \begin{split}\label{eq:Sstar_maximal}
  \mathcal{D}(S^*)\;&=\;\left\{g\in L^2(\mathbb{R}^+)\left|\, \widetilde{S}g\in L^2(\mathbb{R}^+)\!\right.\right\}\,, \\
  S^*g\;&=\;\widetilde{S}g\;=\;-g''+\frac{\nu}{r}\,g+\frac{\nu^2}{4\kappa^2}\,g\,.
 \end{split}
\end{equation}
 Thus, $\ker S^*$ is formed by the square-integrable solutions to $\widetilde{S} u = 0$ on $\mathbb{R}^+$. Moreover, standard ODE arguments (see e.g. \cite[Theorems 5.2--5.4]{Wasow_asympt_expansions}) show that if $u$ solves $\widetilde{S} u = 0$, then it is smooth on $\mathbb{R}^+$, with possible singularity only at zero or infinity.

Through the change of variable $\rho:=-\frac{\nu}{\kappa}r$, $w(\rho):=u(r)$, where $-\frac{\nu}{\kappa}>0$ for every non-zero $\nu$ owing to \eqref{eq:conditions_kappa}, the differential problem becomes
\begin{equation}\label{eq:hypergeom}
 \Big(-\frac{\ud^2}{\ud \rho^2} -\frac{\kappa}{\rho}+\frac{1}{4}\Big)w\;=\;0\,,
\end{equation}
that is, a special case of the \emph{Whittaker equation}\index{Whittaker equation} $w''-(\frac{1}{4}-\frac{\kappa}{\rho}+(\frac{1}{4}-\mu^2)\frac{1}{\rho^2})w=0$ with parameter $\mu=\frac{1}{2}$ \cite[Eq.~(13.1.31)]{Abramowitz-Stegun-1964}.
The functions
\begin{eqnarray}
 \mathscr{M}_{\kappa,\frac{1}{2}}(\rho) &=& e^{-\frac{1}{2} \rho} \rho\, M_{1-\kappa,2}(\rho)\,, \\
 \mathscr{W}_{\kappa,\frac{1}{2}}(\rho) &=& e^{-\frac{1}{2} \rho} \rho \,U_{1-\kappa,2}(\rho) \label{eq:somespecialf}
\end{eqnarray}
form a pair $(\mathscr{M}_{\kappa,\frac{1}{2}},\mathscr{W}_{\kappa,\frac{1}{2}})$ of linearly independent solutions to \eqref{eq:hypergeom} \cite[Eq.~(13.1.32)-(13.1.33)]{Abramowitz-Stegun-1964}, where $M_{a,b}$ and $U_{a,b}$ are, respectively, the Kummer\index{Kummer functions} and the Tricomi\index{Tricomi functions} function \cite[Eq.~(13.1.2)-(13.1.3)]{Abramowitz-Stegun-1964}.

Owing to \cite[Eq.~(13.5.5), (13.5.7), (13.1.2), and (13.1.6)]{Abramowitz-Stegun-1964} as $\rho\downarrow 0$, and to \cite[Eq.~(13.1.4) and (13.1.8)]{Abramowitz-Stegun-1964} as $\rho\to +\infty$, one has the asymptotics
\begin{equation}\label{eq:AsymM0}
 \begin{split}
 \mathscr{M}_{\kappa,\frac{1}{2}}(\rho)\;&\stackrel{\rho\downarrow 0}{=}\; \rho -\frac{\kappa}{2} \rho^2 + \frac{1+2\kappa^2}{24} \rho^3 + O(\rho^4)\,, \\
 \mathscr{W}_{\kappa,\frac{1}{2}}(\rho)\;&\stackrel{\rho\downarrow 0}{=}\;\frac{1}{\Gamma(1-\kappa)} -\frac{ \kappa}{\Gamma(1-\kappa)} \,\rho \ln \rho \\
 &\qquad\quad +\frac{\,(2-4 \gamma) \kappa-2 \kappa\Psi(1-\kappa)  -1\,}{2 \Gamma(1-\kappa)}\, \rho+O(\rho^2 \ln \rho) \, ,
 \end{split}
\end{equation}
and 
\begin{equation}\label{eq:AsymMInf}
\begin{split}
 \mathscr{M}_{\kappa,\frac{1}{2}}(\rho)\;&\stackrel{\rho\to +\infty}{=}\;\frac{1}{\Gamma(1-\kappa)} e^{\rho/2} \rho^{-\kappa} (1+O(\rho^{-1}))\,, \\
 \mathscr{W}_{\kappa,\frac{1}{2}}(\rho)\;&\stackrel{\rho\to +\infty}{=}\;e^{-\rho/2} \rho^\kappa (1+O(\rho^{-1}))\,,
\end{split}
\end{equation}
where $\gamma$ is the Euler-Mascheroni constant\index{Euler-Mascheroni constant} and $\Psi$ is the digamma function.\index{digamma function} Since $0<|\kappa|<\frac{1}{2}$, the expressions \eqref{eq:AsymM0} and \eqref{eq:AsymMInf} make sense.

Therefore, only $\mathscr{W}_{\kappa,\frac{1}{2}}$ is square-integrable at infinity, whereas both $\mathscr{M}_{\kappa,\frac{1}{2}}$ and $\mathscr{W}_{\kappa,\frac{1}{2}}$ are square-integrable at zero. This implies that the square-integrable solutions to $\widetilde{S}u \; = \;0$ form a \emph{one}-dimensional space, that is, $\dim\ker S^*=1$. 

Explicitly, upon setting
\begin{equation}\label{eq:FPhi}
 \begin{split}
  F_\kappa(r)\;&:=\;\mathscr{M}_{\kappa,\frac{1}{2}}(\lambda r)\,, \\
  \Phi_\kappa(r)&:=\;\mathscr{W}_{\kappa,\frac{1}{2}}(\lambda r)\,,\qquad \lambda:=-{\textstyle\frac{\nu}{\kappa}}>0\,,
 \end{split}
\end{equation}
one has that
\begin{equation}\label{eq:kerSstar}
 \ker S^*\;=\;\mathrm{span}\{\Phi_\kappa\}\,,
\end{equation}
and $(F_\kappa,\Phi_\kappa)$ is a pair of linearly independent solutions to the original problem $\widetilde{S}u=0$.

\subsection{Inhomogeneous inverse radial problem}

Next, one focuses on the inhomogeneous problem $\widetilde{S}f=g$ in the unknown $f$ for given $g$. With respect to the fundamental system\index{fundamental system for an O.D.E.} $(F_\kappa,\Phi_\kappa)$ for $\widetilde{S}u=0$, the general solution is given by
\begin{equation}
 f\;=\;c_1 F_\kappa + c_2 \Phi_\kappa + f_{\mathrm{part}}
\end{equation}
for $c_1,c_2\in\mathbb{C}$ and some particular solution $f_{\mathrm{part}}$, i.e., $\widetilde{S}f_{\mathrm{part}}=g$.

The Wronskian\index{Wronskian}
\begin{equation}
W(\Phi_\kappa,F_\kappa)(r)\;:=\;\det \begin{pmatrix}
\Phi_\kappa(r) & F_\kappa(r) \\
\Phi_\kappa'(r) & F_\kappa'(r)
\end{pmatrix}
\end{equation}
relative to the pair $(F_\kappa,\Phi_\kappa)$ is actually constant in $r$, since it is evaluated on solutions to the homogeneous differential problem, with a value that can be computed by means of the asymptotics \eqref{eq:AsymM0} or \eqref{eq:AsymMInf} and amounts to
\begin{equation}\label{eq:value_of_W}
W(\Phi_\kappa,F_\kappa)\;=\;\frac{-\nu/\kappa}{\Gamma(1-\kappa)}\;=:\; W\,.
\end{equation}

A standard application of the method of variation of constants\index{variation of constants} \cite[Section 2.4]{Wasow_asympt_expansions} shows that one can take $f_{\mathrm{part}}$ to be
\begin{equation}
f_{\text{part}}(r) \;=\; \int_0^{+\infty} G(r,\rho) g(\rho) \, \ud \rho\,,
\end{equation}
where 
\begin{equation}\label{eq:Green}
G(r,\rho) \, := \,\frac{1}{W} \begin{cases}
\Phi_\kappa(r) F_\kappa(\rho) \, , \qquad \text{if }0 < \rho < r \, ,\\
F_\kappa(r) \Phi_\kappa(\rho) \, , \qquad \text{if } 0 < r < \rho \,.
\end{cases}
\end{equation}

The following property holds.

\begin{lemma}\label{lem:RGbddsa}
The integral operator $R_G$ on $L^2(\mathbb{R}^+,\ud r)$ with kernel $G(r,\rho)$ given by \eqref{eq:Green} is bounded and self-adjoint.
\end{lemma}

\begin{proof}
$R_G$ splits into the sum of four integral operators with kernels given by
\begin{equation*}
 \begin{split}
  G^{++}(r,\rho)\;&:=\;G(r,\rho)\,\mathbf{1}_{(1,+\infty)}(r)\,\mathbf{1}_{(1,+\infty)}(\rho)\,, \\
  G^{+-}(r,\rho)\;&:=\;G(r,\rho)\,\mathbf{1}_{(1,+\infty)}(r)\,\mathbf{1}_{(0,1)}(\rho)\,, \\
  G^{-+}(r,\rho)\;&:=\;G(r,\rho)\,\mathbf{1}_{(0,1)}(r)\,\mathbf{1}_{(1,+\infty)}(\rho)\,, \\
  G^{--}(r,\rho)\;&:=\;G(r,\rho)\,\mathbf{1}_{(0,1)}(r)\,\mathbf{1}_{(0,1)}(\rho)\,, 
 \end{split}
 \end{equation*}
where $\mathbf{1}_J$ denotes the characteristic function of the interval $J\subset\mathbb{R}^+$.
One can estimate each $G^{LM}(r,\rho)$, $L,M\in\{+,-\}$, by means of the short- and large-distance asymptotics \eqref{eq:AsymM0}-\eqref{eq:AsymMInf} for $F_\kappa$ and $\Phi_\kappa$. Calling $\lambda=-\frac{\nu}{\kappa}$ as in \eqref{eq:FPhi}, for example, 
\[
 \begin{split}
  |\Phi_\kappa(r) \,F_\kappa(\rho)\,\mathbf{1}_{(1,+\infty)}(r)\,\mathbf{1}_{(1,+\infty)}(\rho)|\;&\lesssim\; e^{-\frac{\lambda}{2}(r-\rho)} \,\left(\frac{r}{\rho}\right)^\kappa \, , \qquad\textrm{if }\,0<\rho<r\,, \\
  |F_\kappa(r)\, \Phi_\kappa(\rho)\,\mathbf{1}_{(1,+\infty)}(r)\,\mathbf{1}_{(1,+\infty)}(\rho)|\;&\lesssim\; e^{-\frac{\lambda}{2}(\rho-r)} \, \left(\frac{\rho}{r}\right)^\kappa \, , \qquad\;\textrm{if }\,0<r<\rho\,,
 \end{split}
\]
because $F_k$ diverges exponentially and $\Phi_\kappa$ vanishes exponentially as $r\to+\infty$. Thus,
\[
 |G^{++}(r,\rho)|\;\lesssim\;e^{-\frac{\lambda}{4}|r-\rho|}\,.
\]
With analogous reasoning one finds
\begin{equation*}\tag{*}\label{eq:matrix_estimates}
 \begin{split}
  |G^{++}(r,\rho)|\;&\lesssim\;e^{-\frac{\lambda}{4}|r-\rho|}\,\mathbf{1}_{(1,+\infty)}(r)\,\mathbf{1}_{(1,+\infty)}(\rho)\,, \\
  |G^{+-}(r,\rho)|\;&\lesssim\;e^{-\frac{\lambda}{4}r}\,\mathbf{1}_{(1,+\infty)}(r)\,\mathbf{1}_{(0,1)}(\rho)\,, \\
  |G^{-+}(r,\rho)|\;&\lesssim\;e^{-\frac{\lambda}{4}\rho}\,\mathbf{1}_{(0,1)}(r)\,\mathbf{1}_{(1,+\infty)}(\rho)\,, \\
  |G^{--}(r,\rho)|\;&\lesssim\;\mathbf{1}_{(0,1)}(r)\,\mathbf{1}_{(0,1)}(\rho)\,.
 \end{split}
\end{equation*}
The last three bounds in \eqref{eq:matrix_estimates} imply $G^{+-},G^{-+},G^{--}\in L^2(\mathbb{R}^+\times\mathbb{R}^+,\ud r\,\ud\rho)$ and therefore the corresponding integral operators are Hilbert-Schmidt operators on $L^2(\mathbb{R}^+)$ (Sect.~\ref{sec:bdd-compacts-unitaries-orthproj}). The first bound in \eqref{eq:matrix_estimates} allows to conclude, by an obvious Schur\index{Schur test} test, that also the integral operator with kernel $G^{++}(r,\rho)$ is bounded on $L^2(\mathbb{R}^+)$. This proves the overall boundedness of $R_G$. Its self-adjointness is then clear from \eqref{eq:Green}: the adjoint $R_G^*$ of $R_G$ has kernel $\overline{G(\rho,r)}$, but $G$ is real-valued and $G(\rho,r)=G(r,\rho)$, thus proving that $R_G^*=R_G$.
\end{proof}

\subsection{Distinguished extension and its inverse}

 
 The \emph{reference} self-adjoint extension of $S$ needed in the Kre\u{\i}n-Vi\v{s}ik-Birman scheme to classify all other extensions (Sect.~\ref{sec:II-VBreparametrised}) can be naturally chosen to be the Friedrichs extension $S_{\mathrm{F}}$, since the greatest lower bound of $S$ is strictly positive (Theorem \ref{thm:Friedrichs-ext}(ii)).

 Along this line one proves the following.
\begin{proposition}\label{eq:RGisSFinv}
 $R_G=S_{\mathrm{F}}^{-1}$.
\end{proposition}

This is checked in several steps. To begin with, one recognises that $R_G$ inverts a self-adjoint extension of $S$.

\begin{lemma}\label{eq:RGinvertsExtS}
 There exists a self-adjoint extension $\mathscr{S}$ of $S$ in $L^2(\mathbb{R}^+)$ which has everywhere defined and bounded inverse and such that $\mathscr{S}^{-1}=R_G$.
\end{lemma}

\begin{proof}
 $R_G$ is bounded and self-adjoint (Lemma \ref{lem:RGbddsa}), and by construction satisfies $\widetilde{S}\,R_G\,g=g$ $\forall g\in L^2(\mathbb{R}^+)$. Therefore, $R_G g=0$ for some $g\in L^2(\mathbb{R}^+)$ implies $g=0$, i.e., $R_G$ is injective. Then $R_G$ has dense range ($(\mathrm{ran}\,R_G)^\perp=\ker R_G$). 
 As a consequence (Sect.~\ref{sec:I-adjoint}), $(R_{G}^{-1})^*=(R_{G}^*)^{-1}=R_{G}^{-1}$, that is, $\mathscr{S}:=R_{G}^{-1}$ is self-adjoint. One thus has $R_G=\mathscr{S}^{-1}$ and from the identity $S^*R_G=\mathbbm{1}$ on $L^2(\mathbb{R}^+)$ one deduces that for any $f\in\mathcal{D}(\mathscr{S})$, say, $f=R_G g=\mathscr{S}^{-1} g$ for some $g\in L^2(\mathbb{R}^+)$, the identity $S^*f=\mathscr{S}f$ holds. This means that $S^*\supset\mathscr{S}$, whence also $\overline{S}=S^{**}\subset\mathscr{S}$, i.e., $\mathscr{S}$ is a self-adjoint extension of $S$.
\end{proof}

 The next preparatory step is the identification of the quadratic form of the Friedrichs extension (Theorem \ref{thm:Friedrichs-ext}(i)): the norm
 \begin{equation}\label{eq:NormF}
  \|f\|_{\mathrm{F}}^2\;:=\;\langle f,Sf\rangle+\langle f,f\rangle
 \end{equation}
 is equal, for $f\in C^\infty_0(\mathbb{R}^+)$, to
 \begin{equation}\label{eq:NormF2}
  \|f\|_{\mathrm{F}}^2\;=\;\|f'\|_{L^2}^2+\nu\|r^{-\frac{1}{2}}f\|_{L^2}^2+(\textstyle{\frac{\nu^2}{\,4\,\kappa^2}}+1)\|f\|_{L^2}^2\,,
 \end{equation}
 and  $\mathcal{D}[S_{\mathrm{F}}]$ is the closure of $\mathcal{D}(S)=C^\infty_0(\mathbb{R}^+)$ in the norm $\|\cdot\|_{\mathrm{F}}$. This yields at once the following.

\begin{lemma}\label{lem:Fform}
 The quadratic form of the Friedrichs extension of $S$ is given by
 \begin{equation}\label{eq:Fform}
  \begin{split}
  \mathcal{D}[S_{\mathrm{F}}]\;&=\;\big\{f\in L^2(\mathbb{R}^+)\,\big|\, \|f'\|_{L^2}^2+\nu\,\|r^{-\frac{1}{2}}f\|_{L^2}^2+\|f\|_{L^2}^2<+\infty\big\} \\
   S_{\mathrm{F}}[f,h]\;&=\;\int_0^{+\infty}\!\!\Big(\,\overline{f'(r)}h'(r)+\nu\,\frac{\,\overline{f(r)}h(r)}{r}+\frac{\nu^2}{\,4\,\kappa^2}\,\overline{f(r)}h(r)\Big)\ud r\,.
  \end{split}
 \end{equation}
 \end{lemma}

%

In fact, the Friedrichs form domain is a classical functional space. This follows from a standard application of the one-dimensional Hardy\index{Hardy inequality} inequality
\begin{equation}\label{eq:III-hardy1}
 \int_0^{+\infty}\frac{\;|f(x)|^2}{\;x^2}\,\ud x\;\leqslant\;2\int_0^{+\infty}|f'(x)|^2\,\ud x\qquad \forall\,f \in C^\infty_0(\mathbb{R}^+)\,.
\end{equation}

\begin{lemma}\label{lem:DomSFrid}
$\mathcal{D}[S_{\mathrm{F}}]\:=\:H^1_0(\mathbb{R}^+)\::=\:\overline{C^\infty_0(\mathbb{R}^+)}^{\Vert \, \Vert_{H^1}}$.
\end{lemma}

\begin{proof}
 The inequality\index{Hardy inequality} \eqref{eq:III-hardy1}, namely $\Vert r^{-1} f \Vert_{L^2} \leqslant 2 \,\Vert f' \Vert_{L^2}$, implies
\begin{equation*}
\Vert r^{-\frac{1}{2}} f \Vert^2_{L^2} \;\leqslant\; \frac{\varepsilon}{4} \,\Vert r^{-1} f \Vert^2_{L^2} +\frac{1}{\varepsilon}\, \Vert f \Vert^2_{L^2} \;\leqslant\; \varepsilon \Vert f' \Vert_{L^2}^2 + \varepsilon^{-1} \Vert f \Vert_{L^2}^2
\end{equation*}
for arbitrary $\varepsilon>0$. This and \eqref{eq:NormF2} imply on the one hand $\|f\|_{\mathrm{F}}\lesssim\|f\|_{H^1}$, and on the other hand
\[
 \|f\|_{\mathrm{F}}^2\;\geqslant\;(1-|\nu|\varepsilon)\|f'\|_{L^2}^2+(1+\textstyle{\frac{\nu^2}{\,4\,\kappa^2}}-|\nu|\varepsilon^{-1})\|f\|_{L^2}^2\,.
\]
The r.h.s.~above is equivalent to the $H^1$-norm provided that the coefficients of $\|f'\|_{L^2}^2$ and $\|f\|_{L^2}^2$ are strictly positive, which is the same as
\[
 \nu^2\;<\;\frac{|\nu|}{\varepsilon}\;<\;1+\frac{\nu^2}{\,4\,\kappa^2}\,.
\]
For given $\nu$ and $\kappa$, a choice of $\varepsilon>0$ satisfying the inequalities above is always possible, because $\nu^2<1+\frac{\nu^2}{\,4\,\kappa^2}$, or equivalently, $1+\nu^2(\frac{1}{4\kappa^2}-1)>0$, which is true owing to the assumption $0<|\kappa|<\frac{1}{2}$. One has thus shown that $\|f\|_{\mathrm{F}}\approx\|f\|_{H^1}$ in the sense of the equivalence of norms on $C^\infty_0(\mathbb{R}^+)$. Now, the $\|\cdot\|_{\mathrm{F}}$-completion of $C^\infty_0(\mathbb{R}^+)$ is by definition $\mathcal{D}[S_{\mathrm{F}}]$ (Theorem \ref{thm:Friedrichs-ext}(i)-(ii)), whereas the $\|\cdot\|_{H^1}$-completion is $H^1_0(\mathbb{R}^+)$: the Lemma is therefore proved.
\end{proof}

 Proceeding further, it is convenient for later purposes to highlight the following feature of $\mathrm{ran} \,R_G$.

\begin{lemma}\label{lem:ranRGinDrinv}
For every $g \in L^2(\mathbb{R}^+)$ one has
\begin{equation}\label{eq:Integral}
\int_0^{+\infty} \frac{| (R_G g)(r) |^2}{r^2} \ud r \;<\; +\infty\,,
\end{equation}
i.e.,
\begin{equation}
\mathrm{ran} \, R_G \subset \mathcal{D}(r^{-1})\,.
\end{equation}
\end{lemma}

\begin{proof}
Clearly, $\int_1^{+\infty} r^{-2} | (R_Gg)(r) |^2 \,\ud r \leqslant \Vert R_G \Vert^2_{\mathrm{op}} \Vert g \Vert^2_{L^2}$. It then remains to prove the finiteness of the integral in \eqref{eq:Integral} only for $r \in (0,1)$.
Owing to \eqref{eq:value_of_W} and \eqref{eq:Green},
\begin{equation*}\tag{*}\label{eq:HYDROG*here}
|(R_Gg)(r)|\;\lesssim\; |\Phi_\kappa(r)| \int_0^r |F_\kappa(\rho) g(\rho) | \, \ud \rho + |F_\kappa(r)| \int_0^{+\infty} |\Phi_\kappa(\rho) g(\rho)| \,\ud \rho\,.
\end{equation*}
 To further estimate the above quantity one exploits the asymptotics \eqref{eq:AsymM0}.
The first summand on the r.h.s.~above as a $O(r^{\frac{3}{2}})$-quantity as $r\downarrow 0$, because in this limit  $\Phi_\kappa$ is smooth and bounded, whereas $F_\kappa$ is smooth and vanishes as $O(r)$, and therefore
\begin{equation*}
\int_0^r |F_\kappa(\rho) g(\rho)| \,\ud \rho \;\leqslant\sup_{\rho \in [0,r]}|F_\kappa(\rho)|\, \Vert g \Vert_{L^2} \,r^{\frac{1}{2}}\,=\, O(r^{\frac{3}{2}}) \,.
\end{equation*} 
The second  summand on the r.h.s.~of \eqref{eq:HYDROG*here} is a $O(r)$-quantity as $r\downarrow 0$, because so is $F_\kappa(r)$ and because $\int_0^{+\infty}|\Phi_\kappa(\rho) g(\rho)|\, \ud \rho \leqslant \Vert \Phi_\kappa \Vert_{L^2} \Vert g \Vert_{L^2}$. Thus, $(R_Gg)(r)=O(r)$ as $r\downarrow 0$, whence the integrability of $r^{-2}|(R_Gg)(r)|^2$ at zero.
\end{proof}
 With the above preparation, one can finally prove that $R_G=S_{\mathrm{F}}^{-1}$.

\begin{proof}[Proof of Proposition \ref{eq:RGisSFinv}]
 $R_G=\mathscr{S}^{-1}$ for some $\mathscr{S}=\mathscr{S}^*\supset S$ (Lemma \ref{eq:RGinvertsExtS}), and one wants to conclude that $\mathscr{S}=S_{\mathrm{F}}$. This follows if one shows that $\mathcal{D}(\mathscr{S})\subset \mathcal{D}[S_{\mathrm{F}}]$, owing to the well-known property of $S_{\mathrm{F}}$ that distinguishes it from all other self-adjoint extensions of $S$ (Theorem \ref{thm:Friedrichs-ext}(v)).
 
 To this aim, given a generic $f=R_G g\in\mathrm{ran}\,R_G=\mathcal{D}(\mathscr{S})$ for some $g\in L^2(\mathbb{R}^+)$, one shows that $S_{\mathrm{F}}[f]:=S_{\mathrm{F}}[f,f]<+\infty$, the form of $S_{\mathrm{F}}$ being given by Lemma \ref{lem:Fform}. The fact that $\|f\|_{L^2}^2$ is finite is obvious, and the finiteness of $\|r^{-\frac{1}{2}}f\|_{L^2}^2$ follows by interpolation from Lemma \ref{lem:ranRGinDrinv}. One is thus left with proving that $\|f'\|_{L^2}^2<+\infty$, after which the conclusion then follows from \eqref{eq:Fform}.
 
 Now, $f\in\mathcal{D}(S^*)$ and therefore $-f''+\frac{\nu}{r} f+\frac{\nu^2}{\,4\,\kappa^2} f=g\in L^2(\mathbb{R}^+)$: this, and the already mentioned square-integrability of $f$ and $r^{-1}f$, yield $f''\in L^2(\mathbb{R}^+)$. It is then standard to deduce (see, e.g., \cite[Remark 4.21]{Grubb-DistributionsAndOperators-2009}) that $f'$ too belongs to $L^2(\mathbb{R}^+)$, thus concluding the proof.
\end{proof}

For later purposes one sets for convenience
\begin{equation}\label{eq:Psikappa}
 \Psi_\kappa\;:=\;S_{\mathrm{F}}^{-1}\Phi_\kappa\;=\;R_G\Phi_\kappa
\end{equation}
and proves the following.

\begin{lemma}\label{lem:RGPhi_atzero} The function $\Psi_\kappa$ satisfies the asymptotics
\begin{equation}\label{eq:RGPhi_atzero}
\Psi_\kappa(r) \;=\; \Gamma(1- \kappa) \Vert \Phi_\kappa \Vert_{L^2}^2\, r + O(r^2)\qquad\textrm{as } r \downarrow 0\,.
\end{equation}
\end{lemma}

\begin{proof}
Owing to \eqref{eq:value_of_W} and \eqref{eq:Green},
\begin{equation*}
(R_G \Phi_\kappa)(r) \;= \; -\frac{\kappa\,\Gamma(1-\kappa)}{\nu} \Big(\Phi_\kappa(r)\! \int_0^r F_\kappa(\rho) \Phi_\kappa(\rho) \,\ud \rho + F_\kappa(r)\! \int_r^{+\infty} \!\!\Phi_\kappa^2(\rho) \,\ud \rho\Big)\,.
\end{equation*}
As $r\downarrow 0$, \eqref{eq:AsymM0} and \eqref{eq:FPhi} imply that the first summand behaves as
\begin{equation*}
{\textstyle -\frac{\,\kappa\Gamma(1-\kappa)}{\nu}}\,\big({\textstyle\frac{1}{\Gamma(1-\kappa)}}+ O(r \ln r)\big)\! \int_0^r \!\big({\textstyle-\frac{\nu}{\kappa}} \rho+O(\rho^2)\big) \big({\textstyle\frac{1}{\Gamma(1-\kappa)}}+O(\rho\ln\rho) \big) \, \ud \rho\,,
\end{equation*}
which, after some simplifications, becomes
\begin{equation*}
\frac{1}{\,2\,\Gamma(1-\kappa)}\, r^2 + O(r^3 \ln r)\,.
\end{equation*}
The second summand turns out to be the leading term: indeed, as $r\downarrow 0$, $\int_r^\infty \Phi_\kappa^2\,\ud\rho = \| \Phi_\kappa \|_{L^2}^2 + O(r)$ and hence 
\begin{equation*}
-\frac{\kappa\,\Gamma(1-\kappa)}{\nu} \, F_\kappa(r)\! \int_r^{+\infty}\!\! \Phi_\kappa^2(\rho) \,\ud \rho\;=\;\Gamma(1- \kappa) \Vert \Phi_\kappa \Vert_{L^2}^2\, r + O(r^2)\,,
\end{equation*}
which completes the proof.
\end{proof}

\subsection{Operators $\overline{S}$, $S_{\mathrm{F}}$, and $S^*$}\label{sec:SclosureSstar}

The subspace $\mathcal{D}(S^*)$ implicitly characterised in \eqref{eq:Sstar_maximal} and the subspace $\mathcal{D}(S_{\mathrm{F}})$ have the structure (Proposition \ref{prop:II-KVB-decomp-of-Sstar})
\begin{eqnarray}
 \mathcal{D}(S^*)&=&\mathcal{D}(\overline{S})\dotplus S_{\mathrm{F}}^{-1}\ker S^*\dotplus\ker S^*\,, \\
 \mathcal{D}(S_{\mathrm{F}})&=&\mathcal{D}(\overline{S})\dotplus S_{\mathrm{F}}^{-1}\ker S^*\,,
\end{eqnarray}
which reads now, owing to \eqref{eq:kerSstar} and to \eqref{eq:Psikappa},
\begin{eqnarray}
 \mathcal{D}(S^*)&=&\big\{ g=f+ c_1 \Psi_\kappa +c_0\Phi_\kappa\,|\,f\in\mathcal{D}(\overline{S}),\,c_0,c_1\in\mathbb{C}\big\}\,,\label{eq:DSstar_firstversion} \\
 \mathcal{D}(S_{\mathrm{F}})&=&\mathcal{D}(\overline{S})\dotplus \mathrm{span}\{\Psi_\kappa\}\,. \label{eq:DSF_firstversion}
\end{eqnarray}

 Concerning $\mathcal{D}(\overline{S})$, this space is well studied in the literature, as $\overline{S}$ is the minimal realisation (Sect.~\ref{sec:MinimalAndMaximalRealisations})\index{differential operator!minimal realisation} of the classical differential operator $\widetilde{S}$. It is known, e.g., from \cite[Prop.~3.1(i)-(ii)]{DR-2017}, that the elements of $\mathcal{D}(\overline{S})$ display the following features.
 
\begin{lemma}\label{lem:Derez}
Let $f \in \mathcal{D}(\overline{S})$. Then the functions $f$ and $f'$
\begin{enumerate}[(i)]
 \item are continuous on $\mathbb{R}^+$ and vanish as $r\to +\infty$;
 \item vanish as $r\downarrow 0$ as
 \begin{equation}\label{eq:ffprimeDSclosure}
\begin{split}
f(r) = o(r^{\frac{3}{2}})\,, \qquad f'(r) = o (r^{\frac{1}{2}})\,.
\end{split}
\end{equation}
\end{enumerate}
\end{lemma}

 From this, one deduces a notable consequence.
\begin{lemma}\label{lem:DSclosure_is_H20}
 One has
 \begin{equation}\label{eq:DSclosure_is_H20}
  \mathcal{D}(\overline{S})\;=\;H^2_0(\mathbb{R}^+)\;=\;\overline{\,C^\infty_0(\mathbb{R}^+)\,}^{\|\,\|_{H^2}}\,.
 \end{equation}
\end{lemma}

\begin{proof}
 First, observe that
 \[\tag{i}
  \mathcal{D}(\overline{S})\;\subset\; H^2_0(\mathbb{R}^+)\,.
 \]
 Indeed, for any $f\in\mathcal{D}(\overline{S})$ one has $\widetilde{S}f=-f''+\frac{\nu}{r}f+\frac{\nu^2}{4\kappa^2}f\in L^2(\mathbb{R}^+)$, as well as $f\in L^2(\mathbb{R}^+)$ and $r^{-1}f\in L^2(\mathbb{R}^+)$, the latter following from \eqref{eq:ffprimeDSclosure}; therefore, $f''\in L^2(\mathbb{R}^+)$ and hence also $f'\in L^2(\mathbb{R}^+)$  (see, e.g., \cite[Remark 4.21]{Grubb-DistributionsAndOperators-2009}); thus, $f\in H^2(\mathbb{R}^+)$. Owing to \eqref{eq:ffprimeDSclosure} again, $f(0)=f'(0)=0$, whence $f\in H^2_0(\mathbb{R}^+)$. 
 
 One also has the inclusion
  \[\tag{ii}
  H^2_0(\mathbb{R}^+)\;\subset\;\mathcal{D}(S^*)\,.
 \]
 Indeed, for any $f\in H^2_0(\mathbb{R}^+)$ one has $f,f''\in L^2(\mathbb{R}^+)$, and $f\in C^1_0(\mathbb{R}^+)$ by Sobolev's Lemma,\index{Sobolev Lemma}\index{theorem!Sobolev (Lemma)} where $C^1_0(\mathbb{R}^+)$ is the space of the $C^1$-functions over $\mathbb{R}^+$ vanishing at zero together with their first derivative. Thus, $f(r)=o(r^{\frac{3}{2}})$ as $r\downarrow 0$, implying $r^{-1}f\in L^2(\mathbb{R}^+)$.
 Then $\widetilde{S}f=-f''+\frac{\nu}{r}f+\frac{\nu^2}{4\kappa^2}f\in L^2(\mathbb{R}^+)$, which by \eqref{eq:Sstar_maximal} means that $f\in \mathcal{D}(S^*)$.
 
 One has then the chain
 \[
 \begin{split}
  \mathcal{D}(\overline{S})\;&\subset\; H^2_0(\mathbb{R}^+)\;\subset\;\mathcal{D}(S^*)\;=\;\mathcal{D}(\overline{S})\dotplus\mathrm{span}\{\Psi_\kappa,\Phi_\kappa\} \\
  &\subset\; H^2_0(\mathbb{R}^+)\dotplus\mathrm{span}\{\Psi_\kappa,\Phi_\kappa\}\;\subset\;\mathcal{D}(S^*)\,,
 \end{split}
 \]
 where the first two inclusions are (i) and (ii) respectively, the identity that follows is an application of \eqref{eq:DSstar_firstversion}, then the next inclusion follows from (i) again and the sum remains direct because no non-zero element in $\mathrm{span}\{\Psi_\kappa,\Phi_\kappa\}$ belongs to $H^2_0(\mathbb{R}^+)$, and the last inclusion follows from (ii) and \eqref{eq:DSstar_firstversion}. Therefore,
 \[
  \mathcal{D}(\overline{S})\dotplus\mathrm{span}\{\Psi_\kappa,\Phi_\kappa\}\;=\;H^2_0(\mathbb{R}^+)\dotplus\mathrm{span}\{\Psi_\kappa,\Phi_\kappa\}\,,\quad\textrm{with }\mathcal{D}(\overline{S})\,\subset\, H^2_0(\mathbb{R}^+)\,,
 \]
 whence necessarily $\mathcal{D}(\overline{S})=H^2_0(\mathbb{R}^+)$.
 \end{proof}
 
 As a consequence, \eqref{eq:DSF_firstversion} now reads
 \begin{equation}\label{eq:DSF2}
  \mathcal{D}(S_{\mathrm{F}})\;=\;H^2_0(\mathbb{R}^+)\dotplus\mathrm{span}\{\Psi_\kappa\}\,,
 \end{equation}
  and one can additionally characterise $\mathcal{D}(S_{\mathrm{F}})$ as follows.
 
 \begin{lemma}\label{lem:DomOpFridS} One has
  \begin{equation}\label{eq:DSF_z}
   \begin{split}
    \mathcal{D}(S_{\mathrm{F}})\;&=\;H^2(\mathbb{R}^+)\cap H^1_0(\mathbb{R}^+) \\
    &=\;\big\{f\in H^2(\mathbb{R}^+)\,|\,f(r)=O(r)\;\textrm{as}\;r\downarrow 0\big\}\,.
   \end{split}
  \end{equation}
 \end{lemma}

  \begin{proof}
  Based on \eqref{eq:DSF_firstversion} and \eqref{eq:DSclosure_is_H20}, let $\phi=f+c\,\Psi_k\in \mathcal{D}(S_{\mathrm{F}})$ for generic $f\in H^2_0(\mathbb{R}^+)$ and $c\in\mathbb{C}$. From $-\Psi_\kappa''+\frac{\nu}{r}\Psi_k+\frac{\nu^2}{4\kappa^2}\Psi_\kappa=S_{\mathrm{F}}\Psi_\kappa=\Phi_\kappa\in L^2(\mathbb{R}^+)$  (see \eqref{eq:Psikappa} above) and from Lemma \ref{lem:ranRGinDrinv} one deduces that $\Psi_\kappa''\in L^2(\mathbb{R}^+)$ and hence $\Psi_\kappa\in H^2(\mathbb{R}^+)$, which proves that $\mathcal{D}(S_{\mathrm{F}})\subset H^2(\mathbb{R}^+)$. Moreover, $\mathcal{D}(S_{\mathrm{F}})\subset\mathcal{D}[S_{\mathrm{F}}]=H^1_0(\mathbb{R}^+)$, owing to Lemma \ref{lem:DomSFrid}, whence the conclusion $\mathcal{D}(S_{\mathrm{F}})\subset H^2(\mathbb{R}^+)\cap H^1_0(\mathbb{R}^+)$.   
  For the converse inclusion, any $\phi\in H^2(\mathbb{R}^+)\cap H^1_0(\mathbb{R}^+)$ is re-written as $\phi=f+\frac{\phi'(0)}{\Psi_\kappa'(0)}\Psi_\kappa$ with $f:=\phi-\frac{\phi'(0)}{\Psi_\kappa'(0)}\Psi_\kappa$ (it is clear from the proof of Lemma \ref{lem:RGPhi_atzero} that $\Psi_\kappa'(0)=-\Gamma(1-\kappa)\|\Phi_\kappa\|_{L^2}^2\neq 0$). By linearity $f\in H^2(\mathbb{R}^+)$, by the assumptions on $\phi$ and \eqref{eq:RGPhi_atzero} $f(0)=0$, and by construction $f'(0)=0$. Thus, $f\in H^2_0(\mathbb{R}^+)$. Then $\phi\in\mathcal{D}(S_{\mathrm{F}})$ owing to \eqref{eq:DSF2}.
  \end{proof}

 In turn, one can now re-write \eqref{eq:DSstar_firstversion} as
 \begin{equation}\label{eq:DSstar_secondversion}
  \begin{split}
   \mathcal{D}(S^*)\;&=\;H^2_0(\mathbb{R}^+)\dotplus\mathrm{span}\{\Psi_\kappa,\Phi_\kappa\} \\
   &=\;\big(H^2(\mathbb{R}^+)\cap H^1_0(\mathbb{R}^+)\big)\dotplus\mathrm{span}\{\Phi_\kappa\}\,.
  \end{split}
 \end{equation}
 
 This finally leads to the proof of Theorem \ref{thm:1Fried}.

\begin{proof}[Proof of Theorem \ref{thm:1Fried}]
Since $S-h_0^{(\nu)}$ is bounded, both $h_0^{(\nu)}$ and $S$ have deficiency index one (Section \ref{sec:perturbation-spectra}). Parts (i) and (ii) follow at once, respectively from Lemma~\ref{lem:DomOpFridS} and Lemma~\ref{lem:DomSFrid}, since the shift does not modify the domains. Part (iii) follows from
\[
\Big(h_{0,\mathrm{F}}^{-1}+\frac{\nu^2}{4 \kappa^2} \Big)^{-1} \;=\; S_{\mathrm{F}}^{-1} \;=\;R_G\,,
\]
and from the expression \eqref{eq:Green} for the kernel of $R_G$, using the definitions \eqref{eq:FPhi} and 
\eqref{eq:value_of_W}.
\end{proof}

\subsection{Kre{\u\i}n-Vi\v{s}ik-Birman classification of the extensions}\label{sec:KVB_for_S}

 According to Theorem \ref{thm:VB-representaton-theorem_Tversion}, specialised to the present case of deficiency index one, the self-adjoint extensions of $S$ correspond to those restrictions of $S^*$ to subspaces of $\mathcal{D}(S^*)$ that, in terms of formula \eqref{eq:DSstar_firstversion}, are identified by the condition
\begin{equation}\label{eq:c1betac0}
 c_1\;=\;\beta c_0\qquad\textrm{for some }\,\beta\in\mathbb{R}\cup\{\infty\}\,.
\end{equation}
The extension parametrised by $\beta=\infty$ has the domain \eqref{eq:DSF2} and is therefore the Friedrichs extension. The multiplication by $\beta$ on the one-dimensional space $\ker S^*=\mathrm{span}\{\Psi_\kappa\}$ is the actual Birman extension parameter\index{Birman extension parameter} of the self-adjoint realisation labelled by $\beta$.

\begin{remark}
 If one replaces the restriction condition \eqref{eq:c1betac0} with the same expression where now $\beta$ is allowed to be a generic complex number, this gives all possible \emph{closed} extensions of $S$ between $\overline{S}$ and $S^*$, as follows by a straightforward application of Grubb's extension theory (see, e.g., \cite[Chapter 13]{Grubb-DistributionsAndOperators-2009}), namely the natural generalisation of the Kre{\u\i}n-Vi\v{s}ik-Birman theory for closed extensions. A recent application of Grubb's theory to operators of point interactions, including $(-\Delta)|_{C^\infty_0(\mathbb{R}^3\setminus\{0\})}$ in $L^2(\mathbb{R}^3)$, from the point of view of Friedrichs systems, is presented in \cite{EM-FriedrichsDelta2017}. 
\end{remark}

 Denote with $S_\beta$ the extension selected by \eqref{eq:c1betac0} for given $\beta$. Owing to \eqref{eq:DSstar_firstversion} and \eqref{eq:c1betac0}, a generic $g\in\mathcal{D}(S_\beta)$ decomposes as
\begin{equation}\label{eq:SelfAdjointFunction-0}
 g\;=\;f+\beta c_0\Psi_\kappa+c_0\Phi_\kappa
\end{equation}
for unique $f\in H^2_0(\mathbb{R}^+)$ and $c_0\in\mathbb{C}$. The asymptotics \eqref{eq:AsymM0}, \eqref{eq:RGPhi_atzero}, and \eqref{eq:ffprimeDSclosure} imply
\begin{equation}\label{eq:SelfAdjointFunction}
 \begin{split}
  g(r)\;&=\;\frac{c_0}{\Gamma(1-\kappa)}+\frac{c_0\, \nu}{\Gamma(1-\kappa)}\, r \ln r \\
  &\quad + \Big(c_0 \, \nu \, \frac{\,2\Psi(1-\kappa)+2 \ln(-\frac{\nu}{\kappa})+(4 \gamma-2) +\kappa^{-1}\,}{2\, \Gamma(1- \kappa)}+c_0\, \beta \,\Gamma(1 - \kappa)  \Vert \Phi_\kappa \Vert^2\Big)\,r \\
  &\quad + o(r^{\frac{3}{2}})\qquad\textrm{as }\,r\downarrow 0\,.
 \end{split}
\end{equation}
The $O(1)$-term and $O(r\ln r)$-term in \eqref{eq:SelfAdjointFunction} come from $\Phi_\kappa$, and so does the first $O(r)$-term; the second $O(r)$-term comes instead from $\Psi_\kappa$; the $o(r^{\frac{3}{2}})$-remainder comes from $f$.

The analogous asymptotics for a generic function $g\in\mathcal{D}(S^*)$ is
\begin{equation}\label{eq:AdjointFunction}
g(r)\;=\; C_0\Big(\frac{1}{\Gamma(1-\kappa)}+\frac{\nu}{\Gamma(1-\kappa)} r \ln r\Big) + C_1 r + o(r^{\frac{3}{2}})\qquad\textrm{as }\,r\downarrow 0
\end{equation}
for some $C_0,C_1\in\mathbb{C}$, as follows again from \eqref{eq:AsymM0}, \eqref{eq:RGPhi_atzero}, and \eqref{eq:ffprimeDSclosure} applied to \eqref{eq:DSstar_firstversion}.
Comparing \eqref{eq:SelfAdjointFunction} with \eqref{eq:AdjointFunction} one concludes the following.

\begin{proposition}[Classification of extensions at $\ell=0$: shift-dependent formulation]\label{prop:classificationThm}

\noindent
The self-adjoint extensions of $S$ form a family $\{S_\beta\,|\,\beta\in\mathbb{R}\cup\{\infty\}\}$. The extension with $\beta=\infty$ is the Friedrichs extension $S_{\mathrm{F}}$. For $\beta\in\mathbb{R}$, the extension $S_\beta$ is the restriction of $S^*$ to the domain $\mathcal{D}(S_\beta)$ that consists of all functions in $\mathcal{D}(S^*)$ for which the coefficient $C_0$ of the leading term $\frac{1}{\Gamma(1-\kappa)}+\frac{\nu}{\Gamma(1-\kappa)} r \ln r$ and the coefficient $C_1$ of the next $O(r)$-sub-leading term, as $r\downarrow 0$, are constrained by the relation
\begin{equation}\label{eq:bc}
 \frac{C_1}{C_0} \; = \; c_{\nu,\kappa}\, \beta+ d_{\nu,\kappa}\,,
\end{equation}
where
\begin{equation}\label{eq:c-d-coeff}
 \begin{split}
  c_{\nu,\kappa} \;& := \; \Gamma(1 - \kappa)\, \|\Phi_\kappa \|_{L^2}^2\,,\\
  d_{\nu,\kappa} \; &:= \; \nu\,\frac{\, 2 \Psi (1- \kappa)+2 \ln(-\frac{\nu}{\kappa})+2(2\gamma - 1)   +\kappa^{-1}\,}{2\, \Gamma(1- \kappa)}\,.
 \end{split}
\end{equation}
Equivalently,
\begin{equation}\label{eq:structural_classification}
 \begin{split}
  S_\beta\;&=\;S^*\upharpoonright\mathcal{D}(S_\beta)\,, \\
  \mathcal{D}(S_\beta)\;&=\;\big\{g=f+\beta c_0\Psi_\kappa+c_0\Phi_\kappa\,\big|\,f\in H^2_0(\mathbb{R}^+),\, c_0\in\mathbb{C} \big\}\,.
 \end{split}
\end{equation}
\end{proposition}

Within the Kre{\u\i}n-Vi\v{s}ik-Birman extension scheme an equivalent classification in terms of quadratic forms is available. In the present setting, Theorem \ref{thm:semibdd_exts_form_formulation_Tversion} yields at once the following.

\begin{proposition}[Shift-dependent classification at $\ell=0$: form version]\label{prop:form_Sbeta}

\noindent The self-adjoint extensions of $S$ form a family $\{S_\beta\,|\,\beta\in\mathbb{R}\cup\{\infty\}\}$. The extension with $\beta=\infty$ is the Friedrichs extension $S_{\mathrm{F}}$. For $\beta\in\mathbb{R}$, the extension $S_\beta$ has quadratic form
 \begin{equation}\label{eq:formextensions}
 \begin{split}
  \mathcal{D}[S_\beta]\;&=\;\mathcal{D}[S_{\mathrm{F}}]\dotplus\mathrm{span}\{\Phi_\kappa\}\,, \\
  S_\beta[\phi_\kappa+c_\kappa\Phi_\kappa]\;&=\;S_{\mathrm{F}}[\phi_\kappa]+\beta|c_\kappa|^2\|\Phi_\kappa\|_{L^2}^2
 \end{split}
\end{equation}
for generic $\phi_\kappa\in\mathcal{D}[S_{\mathrm{F}}]$ and $c_\kappa\in\mathbb{C}$. 
\end{proposition}

Thus, the classification provided by Proposition \ref{prop:classificationThm} identifies each extension \emph{directly from the short distance behaviour of the elements of its domain}, and the self-adjointness condition \eqref{eq:bc} is a constrained \emph{boundary condition} as $r\downarrow 0$ (see Remark \ref{rem:previous_literature} below for further comments).
This turns out to be particularly informative for practical purposes, including the next goal of classification of the discrete spectra of the $S_\beta$'s.

The Friedrichs extension, $\beta=\infty$, is read out from \eqref{eq:bc} as $C_0=0$ and $C_1=c_{\nu,k}$, upon interpreting $C_0\beta=1$. In this case, as expected, \eqref{eq:structural_classification} takes the form of \eqref{eq:DSF2} and \eqref{eq:formextensions} is interpreted as $\mathcal{D}[S_{\beta=\infty}]=\mathcal{D}[S_{\mathrm{F}}]$. Moreover, the following feature of $S_{\mathrm{F}}$ is now obvious from \eqref{eq:structural_classification} and from the short-distance asymptotics of $\Phi_\kappa$ and $\Psi_\kappa$ given by \eqref{eq:AsymM0} and \eqref{eq:RGPhi_atzero} above.

\begin{corollary}\label{cor:Frie}
 The Friedrichs extension $S_{\mathrm{F}}$ is the \emph{only} member of the family $\{S_\beta\,|\,\beta\in\mathbb{R}\cup\{\infty\}\}$ with operator domain contained in $\mathcal{D}[r^{-1}]$, i.e., it is the only self-adjoint extension whose domain's functions have finite expectation of the potential (and hence also of the kinetic) energy. 
\end{corollary}

Another immediate consequence of the extension parametrisation \eqref{eq:structural_classification}, as an application of the Kre{\u\i}n resolvent formula \index{Kre{\u\i}n-Na\u{\i}mark resolvent formula} for deficiency index one (Theorem \ref{thm:Kreins_resolvent_formula_1}), is the following.

\begin{corollary}\label{cor:resolventSbeta}
The self-adjoint extension $S_\beta$  is invertible if and only if $\beta\neq 0$, in which case
\begin{equation}
S_\beta^{-1}\; = \; S_{\mathrm{F}}^{-1}+\frac{1}{\beta} \frac{1}{\Vert \Phi_\kappa \Vert^2} |\Phi_\kappa \rangle \langle \Phi_\kappa|\,.
\end{equation}
\end{corollary}

\begin{remark}\label{rem:previous_literature}
 As mentioned in Section \ref{sec:III-radialproblem}, the present boundary-condition-driven classification of the self-adjoint realisations of the differential operator $\widetilde{S}$ on the half-line has several precursors in the literature \cite{Rellich-1944,Bulla-Gesztesy-1985}. In fact, the analysis of radial Schr\"{o}dinger operators with Coulomb potentials, and more generally of Whittaker operators\index{Whittaker operators} $-\frac{\ud^2}{\ud r^2}+(\frac{1}{4}-\mu^2)\frac{1}{\,r^2}-\frac{\kappa}{r}$ on half-line, is also quite active in the present days \cite{Gesztesy-Zinchenko-2006,Bruneau-Derezinski-Georgescu-2011,Derezinski-Richard-2017,DR-2017}.
 The very `spirit' of the structural formula \eqref{eq:structural_classification} is to link, through the extension parameter $\beta$, the \emph{regular} (in this context: rapidly vanishing) behaviour at the origin of the component $f+\beta c_0\Psi_\kappa$ with the \emph{singular} (non-vanishing) behaviour of the component $c_0\Phi_\kappa$ of a generic $g\in\mathcal{D}(S_\beta)$, and the boundary condition of self-adjointness \eqref{eq:bc} is a convenient re-phrasing of that. The regular/singular jargon becomes even more stringent when lifting this analysis to the three dimensional case, as remarked after Theorem \ref{thm:3Dclass}.
\end{remark}

The $\beta$-parametrisation in Propositions \ref{prop:classificationThm} and \ref{prop:form_Sbeta} is shift-\emph{dependent} and it is convenient now to re-scale $\beta$ so as to re-parametrise the extensions in a shift-\emph{independent} way. To this aim, for $g\in\mathcal{D}(S^*)$ set
\begin{equation}\label{eq:g0g1limits}
 \begin{split}
  g_0\;&:=\;\frac{C_0}{\,\Gamma(1-\kappa)}\;=\;\lim_{r\downarrow 0}g(r)\,, \\
  g_1\;&:=\;C_1\;=\;\lim_{r\downarrow 0}r^{-1}\big(g(r)-g_0(1+\nu r\ln r)\big)\,,
 \end{split}
\end{equation}
so that \eqref{eq:AdjointFunction} reads
\begin{equation}
 g\;=\;g_0(1 + \nu\,r\ln r)+g_1 r +o(r^{\frac{3}{2}})\qquad\textrm{as }r\downarrow 0\,,
\end{equation}
and define also
\begin{equation}\label{eq:alphabeta}
 \alpha\;:=\;\frac{1}{4\pi}\,\Gamma(1-\kappa)\,( c_{\nu,\kappa}\, \beta+ d_{\nu,\kappa} )\,.
\end{equation}
Then, as is obvious from \eqref{eq:bc}-\eqref{eq:c-d-coeff},
\begin{equation}\label{eq:DSbeta-alpha}
 \mathcal{D}(S_\beta)\;=\;\{g\in\mathcal{D}(S^*)\,|\,g_1=4\pi\alpha g_0\}\,.
\end{equation}
Moreover, an easy computation applying \eqref{eq:alphabeta} yields
 \begin{equation}\label{eq:beta-alpha-computation}
  \frac{1}{\beta} \frac{1}{\Vert \Phi_\kappa \Vert^2}\;=\;\frac{{\,\Gamma(1-\kappa)^2}}{4\pi}\frac{1}{ \alpha -\frac{\nu}{4\pi}\big(\Psi(1-\kappa)+\ln(-\frac{\nu}{\kappa})+(2\gamma-1)+\frac{1}{2 \kappa}\big)}\,.
 \end{equation}

This leads directly to the proof of the main result for the radial problem.

\begin{proof}[Proof of Theorem \ref{thm:1Dclass}] Removing the shift from $S$ to $h_0^{(\nu)}$ does not alter the domain of the corresponding self-adjoint extensions or adjoints, and modifies trivially their action. Thus, part (i) follows from Proposition \ref{prop:classificationThm} and from formulas \eqref{eq:g0g1limits} and   \eqref{eq:DSbeta-alpha} for $\mathcal{D}(S_\beta)$, using the expression \eqref{eq:Sstar_maximal} for $\mathcal{D}(S^*)$, whereas part (ii) follows from Corollary \ref{cor:resolventSbeta} with $S_\beta^{-1}=\big( h_{0,\alpha}^{(\nu)}+\frac{\nu^2}{4 \kappa^2} \big)^{-1}$ and $S_{\mathrm{F}}^{-1}=\big( h_{0,\mathrm{F}}^{(\nu)}+\frac{\nu^2}{4 \kappa^2} \big)^{-1}$, together with the identity \eqref{eq:beta-alpha-computation}. So far one has worked with $0<|\kappa|<\frac{1}{2}$: thanks to the uniqueness of the analytic continuation, this determines unambiguously the resolvent at any point in the resolvent set. One can then extend all our previous formulas to the whole regime $(-\infty,0)\cup(0,1)$ for which the expression $\Gamma(1-\kappa)$ still makes sense.
\end{proof}

\subsection{Reconstruction of the 3D hydrogenoid extensions}\label{sec:3Dreconstr}

Finally, the previous conclusions can now be re-phrased in terms of self-adjoint realisations of the hydrogenoid-type operator
\begin{equation}
 \mathring{H}^{(\nu)}\;=\;-\Delta+\frac{\nu}{\,|x|\,}\,,\qquad\mathcal{D}(\mathring{H}^{(\nu)})\;=\;C^\infty_0(\mathbb{R}^3\!\setminus\!\{0\})
\end{equation}
(see \eqref{eq:Hring1} above) on $L^2(\mathbb{R}^3)$.
The self-adjoint extensions of the shifted operator $\mathring{H}^{(\nu)}+\eta{\mathbbm{1}}$, $\eta:=\frac{\nu^2}{4\kappa^2}$, in the sector of angular symmetry $\ell=0$ of $L^2(\mathbb{R}^3)$ are precisely those found in Proposition \ref{prop:classificationThm}.

\begin{proof}[Proof of Theorem \ref{thm:3Dclass}]~

\noindent
\underline{Part (i)}. Formula \eqref{eq:ang_decomp_Halpha} is obvious from \eqref{eq:ang_decomp}-\eqref{eq:ang_decomp_operator} and from Theorem \ref{thm:1Dclass}.

\noindent
 \underline{Part (ii)}. Obviously the unique self-adjoint extension of $\mathring{H}^{(\nu)}$, hence necessarily the Friedrichs extension, in the sectors with angular symmetry $\ell\geqslant 1$ is the projection onto such sectors of the operator \eqref{HnuFriedr}, owing to part (i) of this theorem. In the sector $\ell=0$ the operator \eqref{HnuFriedr} acts as $(-\frac{\ud^2}{\ud r^2}+\frac{\nu}{r})\otimes\mathbbm{1}$ and it remains to recognise that its radial domain consists of those $f$'s in $H^2(\mathbb{R}^+)$ that vanish as $f(r)=O(r)$ as $r\downarrow 0$, because this is precisely $\mathcal{D}(S_{\mathrm{F}})$. This is standard: spherically symmetric elements of $H^2(\mathbb{R}^3)$ are functions $F(|x|)$ for $F\in L^2(\mathbb{R}^+,r^2\,\ud r)$ such that $\Delta_x F(|\cdot|)\in L^2(\mathbb{R}^3,\ud x)$ and hence $\frac{1}{r^2}\frac{\ud}{\ud r}(r^2\frac{\ud}{\ud r} F)\in L^2(\mathbb{R}^+,r^2\ud r)$; on the other hand $F=\frac{f}{r}$ for $f\in L^2(\mathbb{R}^+,\ud r)$, whence $\frac{1}{r^2}\frac{\ud}{\ud r}(r^2\frac{\ud}{\ud r} F)=\frac{\,f''}{r}$, and the square-integrability of $\Delta_x F(|\cdot)|$ reads $f''\in L^2(\mathbb{R}^+,\ud r)$; therefore, $f\in H^2(\mathbb{R}^+)$ and $f(r)=rF(r)=O(r)$ as $r\downarrow 0$. Last, the feature mentioned in the statement which identifies uniquely the Friedrichs extension follow from Proposition \ref{prop:classificationThm} and Corollary \ref{cor:Frie}, thanks to the equivalence $\alpha=\infty$ $\Leftrightarrow$ $\beta=\infty$.
 
 \underline{Part (iii)}. Owing to parts (i) and (ii) one only has to establish \eqref{eq:HnuResolvent} over the sector $\ell=0$. In this sector, \emph{radially},
 \[
  (h^{(\nu)}_{0,\alpha}+{\textstyle\frac{\nu^2}{\,4\kappa^2}}\,\mathbbm{1})^{-1}\;=\;(h^{(\nu)}_{0,\infty}+{\textstyle\frac{\nu^2}{\,4\kappa^2}}\,\mathbbm{1})^{-1}+\frac{1}{\beta} \frac{1}{\Vert \Phi_\kappa \Vert^2} |\Phi_\kappa \rangle \langle \Phi_\kappa|
 \]
 owing to Corollary \ref{cor:resolventSbeta}. 
 Formula \eqref{eq:ourgnukappa} reads 
 \[
  \mathfrak{g}_{\nu,k}(x)\;=\;
  \frac{\,\Gamma(1-\kappa)}{4\pi}\,\frac{\Phi_\kappa(|x|)}{|x|}\;=\;\frac{\,\Gamma(1-\kappa)}{\sqrt{4\pi}}\,\frac{\Phi_\kappa(|x|)}{|x|}\otimes Y_0^0\,,
 \]
 and therefore the projection $|\mathfrak{g}_{\nu,k}\rangle\langle\mathfrak{g}_{\nu,k}|$ acting on $L^2(\mathbb{R}^3)$ acts radially in the $\ell=0$ sector as the projection $\frac{\,\Gamma(1-\kappa)^2}{4\pi}|\Phi_\kappa\rangle\langle\Phi_\kappa|$.
 This proves that 
 \[
  \Big(H^{(\nu)}_\alpha+\frac{\nu^2}{\,4\kappa^2}\,\mathbbm{1}\Big)^{\!-1}\;=\;\Big(H^{(\nu)}+\frac{\nu^2}{\,4\kappa^2}\,\mathbbm{1}\Big)^{\!-1} +\frac{1}{\beta} \frac{1}{\Vert \Phi_\kappa \Vert^2}\,\frac{4\pi}{\,\Gamma(1-\kappa)^2}\,|\mathfrak{g}_{\nu,k}\rangle\langle\mathfrak{g}_{\nu,k}|\,.
 \]
 Combining the formula above with \eqref{eq:beta-alpha-computation} finally yields the resolvent formula \eqref{eq:HnuResolvent}.
 
  \underline{Part (iv)}. This is a standard consequence of part (ii) -- see, e.g., the argument in the proof of \cite[Theorems I.1.1.3 and I.2.1.2]{albeverio-solvable}.  
\end{proof}

\section{Perturbation of the discrete spectra}\label{sec:III-perturbations}

 In the remaining part of the present Chapter Theorem \ref{thm:EV_corrections} is proved together with a few additional observations. 

 It is to be stressed once again that a different path is followed here, based on the radial analysis of extensions developed in Section \ref{sec:Section_of_Classification}, as compared to the classical approach \cite{Zorbas-1980,AGHKS-1983_Coul_plus_delta,Bulla-Gesztesy-1985} that determines the eigenvalues as poles of the resolvent \eqref{eq:HnuResolvent} (see Remark \ref{rem:swavepoles} below). This completes the present approach entirely based on the  Kre{\u\i}n-Vi\v{s}ik-Birman extension theory.

\subsection{The $s$-wave eigenvalue problem}\label{sec:swaveEVproblem}

For fixed $\alpha\in\mathbb{R}$ and $\nu\in\mathbb{R}$ let $\Psi\in\mathcal{D}(H^{(\nu)}_\alpha)$ and $E<0$ satisfy $H^{(\nu)}_\alpha\Psi=E\Psi$ with $\Psi$ belonging to the $L^2$-sector with angular symmetry $\ell=0$. 

In view of \eqref{eq:ang_decomp},
\begin{equation}
 \Psi(x)\;=\;\frac{\,g(|x|)\,}{\,\sqrt{4\pi}\,|x|\,}
\end{equation}
for some $g\in\mathcal{D}(S_\beta)\subset L^2(\mathbb{R}^+)$ such that $S_\beta g=(E+\frac{\nu^2}{4\kappa^2})g$, where $\beta$ is given by \eqref{eq:alphabeta} for the chosen $\alpha$ and $\nu$, and a chosen $\kappa\in(0,1)$ (see \eqref{eq:conditions_kappa} above). Thus,
\begin{equation}\label{eq:radEVproblem1}
\Big(\!-\frac{\ud^2}{\ud r^2} +\frac{\nu}{r} \Big) \,g\;=\; E g\,.
\end{equation}

Passing to re-scaled energy $\mathscr{E}$, radial variable $\rho$, coupling $\vartheta$, and unknown $h$ defined by
\begin{equation}
 \begin{split}
  \mathscr{E}\;&:=\;\frac{\,4\kappa^2 E}{\nu^2}+1\,,\qquad\qquad\qquad\qquad\!\! \rho\;:=\;-r\,\frac{\,\nu\sqrt{1-\mathscr{E}}}{\kappa}\;=\;2 r \sqrt{|E|\,} \, , \\
  \vartheta\;&:=\;\frac{\kappa}{\,\sqrt{1-\mathscr{E}}\,}\;=\;\frac{-\nu}{\,2\sqrt{|E|}}\,,\qquad\; u(\rho)\;:=\;g(r)\,,
 \end{split}
\end{equation}
the eigenvalue problem \eqref{eq:radEVproblem1} takes the form
\begin{equation}\label{eq:radEVproblem2}
 \Big(\!-\frac{\ud^2}{\ud \rho^2} -\frac{\vartheta}{\rho}+\frac{1}{4} \Big) \,u\;=\;0\,,
\end{equation}
namely a Whittaker equation \index{Whittaker equation} of the same type \eqref{eq:hypergeom} above, whose only square-integrable solutions on $\mathbb{R}^+$, analogously to what argued in Section \ref{sec:homo}, are the multiples of the Whittaker function\index{Whittaker functions}
\begin{equation}
u(\rho)\;=\;\mathscr{W}_{\vartheta,\frac{1}{2}}(\rho) \;=\;e^{-\frac{1}{2}\rho}\, \rho\, U_{1-\vartheta,2}(\rho)
\end{equation}
(the special functions $\mathscr{W}_{\vartheta,\frac{1}{2}}$ and $U_{1-\vartheta,2}$ were introduced already in \eqref{eq:somespecialf}).
Therefore, up to multiples, the solution to \eqref{eq:radEVproblem1} is
\begin{equation}
g(r)\;=\; 
\mathscr{W}_{\vartheta,\frac{1}{2}}(2r\sqrt{|E|})\,.
\end{equation}

By means of the expansion \eqref{eq:AsymM0} and of the identity $\nu=-2\sqrt{|E|}\,\vartheta$ one finds
\begin{equation}
 \begin{split}
  g_0\;&=\;\frac{1}{\,\Gamma(1+\frac{\nu}{2\sqrt{|E|}})} \, ,\\
	g_1\; &=\; \nu \, \frac{\, \psi\big(1+\frac{\nu}{2\sqrt{|E|}}\big)+ \ln(2 \sqrt{|E|}) +2\gamma - 1 - {\textstyle\frac{\sqrt{|E|}}{\nu}}}{\, \Gamma\big(1+\frac{\nu}{2\sqrt{|E|}}\big)}\,,
 \end{split}
\end{equation}
and such two constants must satisfy the condition $g_1=4\pi\alpha g_0$, as prescribed by Theorem \ref{thm:1Dclass} , because the considered eigenfunction $\Psi$ belongs to $\mathcal{D}(H_\alpha^{(\nu)})$.
It is thus proved that $E$ is an eigenvalue for $H_\alpha^{(\nu)}$ if and only if
\begin{equation}\label{eq:FEalpha}
 \mathfrak{F}_\nu(E)\;=\;\alpha
\end{equation}
with $\mathfrak{F}_\nu$ defined in \eqref{eq:Feigenvalues}.

When $\nu<0$ the function $(-\infty,0)\ni E\mapsto \mathfrak{F}_\nu(E)$ has vertical asymptotes corresponding to non-positive arguments $1+\frac{\nu}{2\sqrt{|E|}}=-q$, $q\in\mathbb{N}_0$, of the digamma function \index{digamma function} $\Psi$, i.e., at the points $E=E_n^{(\nu)}$ defined by
\begin{equation}\label{eq:EnHydr}
 E_n^{(\nu)}\;:=\;-\frac{\nu^2}{4 n^2}\,,\qquad n\,:=\,q+1\,\in\,\mathbb{N}\,.
\end{equation}
The sequence $(E_n)_{n\in\mathbb{N}}$ is increasing and converges to zero. Within each interval $(E_n,E_{n+1})$ the function $E\mapsto \mathfrak{F}_\nu(E)$ is smooth and strictly monotone increasing, and moreover
\[
 \lim_{E\to -\infty}\,\mathfrak{F}_\nu(E)\;=\;-\infty\,.
\]
Thus, for any $\alpha\in\mathbb{R}$ the equation \eqref{eq:FEalpha} does admit countably many negative simple roots, which form the increasing sequence $(E_n^{(\nu,\alpha)})_{n\in\mathbb{N}}$ and accumulate at zero. Therefore, the $s$-wave point spectrum of $H^{(\nu)}_\alpha$ consists precisely of the $E_n^{(\nu,\alpha)}$'s. In the extremal case $\alpha=\infty$ one has $E_n^{(\nu,\alpha=\infty)}=E_n^{(\nu)}$: indeed, the $s$-wave point spectrum of the Friedrichs extension $H^{(\nu)}$ is the ordinary non-relativistic hydrogenoid $s$-wave spectrum, as given by \eqref{eq:EnHydr}.

When $\nu>0$ the function $(-\infty,0)\ni E\mapsto \mathfrak{F}_\nu(E)$ is smooth and strictly monotone increasing, with
\[
 \lim_{E\to-\infty}\,\mathfrak{F}_\nu(E)=-\infty\,,\qquad  \lim_{E\uparrow 0}\mathfrak{F}_\nu(E)\;=\;\frac{\nu}{4\pi}\,(\ln\nu+2\gamma-1)\;=:\;\alpha_\nu\,,
\]
the latter limit following from \eqref{eq:Feigenvalues} owing to the asymptotics \cite[Eq.~(6.3.18)]{Abramowitz-Stegun-1964} that here reads
\[
 \psi\big(1+{\textstyle\frac{\nu}{2\sqrt{|E|}}}\big)\;\stackrel{E\to 0}{=} \;\ln{\textstyle\frac{\nu}{2\sqrt{|E|}}}+O(\sqrt{|E|})\,.
\]
Thus, the equation \eqref{eq:FEalpha} has no negative roots if $\alpha\geqslant\alpha_\nu$ and one negative root if $\alpha<\alpha_\nu$.

This completes the proof of Theorem \ref{thm:EV_corrections}.

\begin{figure}
\includegraphics[width=6cm]{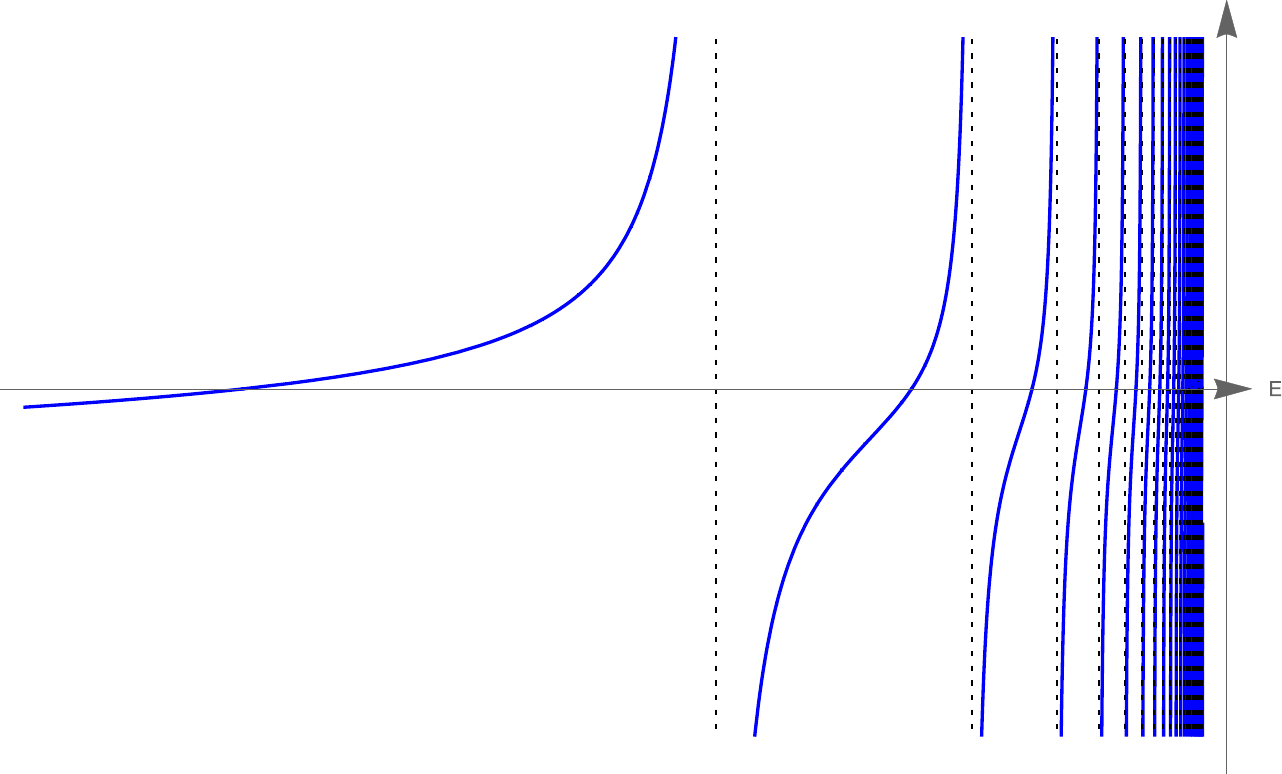}\quad
\includegraphics[width=6cm]{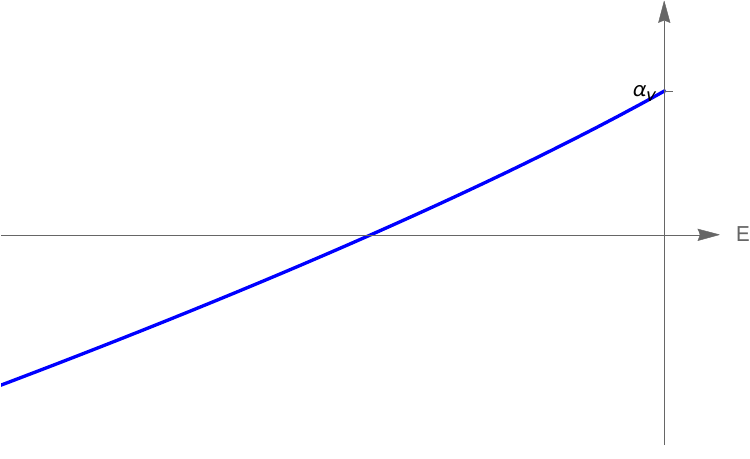}
\caption{Behaviour of the function $(-\infty,0)\ni E\mapsto \mathfrak{F}_\nu(E)$ for $\nu<0$ (left) and $\nu>0$ (right). Dashed lines in the figure represent vertical asymptotes.} \label{fig:Fnu}
\end{figure}

\subsection{Further remarks}\label{sec:EVfurtherremarks}

\begin{remark}\label{rem:EVdecreased}
 The result of Theorem \ref{thm:EV_corrections} when $\nu<0$ confirms that $H^{(\nu)}\geqslant H^{(\nu)}_\alpha$, namely that the Friedrichs extension is larger (in the sense of self-adjoint operator ordering) than any other extension. In particular, $E_{n+1}^{(\nu,\alpha)}\geqslant E_n^{(\nu)}\geqslant E_n^{(\nu,\alpha)}$. 
\end{remark}

\begin{remark}\label{rem:spectra_fibre}
 As is clear from the behaviour of the roots to  $\mathfrak{F}_\nu(E)=\alpha$ (Fig.~\ref{fig:Fnu})
 \begin{equation}
  \bigcup_{\alpha}\sigma_{\mathrm{p}}^{(0)}(H^{(\nu)}_\alpha)\;=\;(-\infty,0)\,.
 \end{equation}
 In this sense the spectra $\sigma_{\mathrm{p}}^{(0)}(H^{(\nu)}_\alpha)$ fibre, as $\alpha$ runs over $\mathbb{R}\cup\{\infty\}$, the whole negative real line.
\end{remark}

\begin{remark}
 When $\nu<0$ one has
 \[
  \lim_{\alpha\to-\infty}E_1^{(\nu,\alpha)}\;=\;-\infty\,,\qquad \lim_{\alpha\to-\infty}E_{n+1}^{(\nu,\alpha)}\;=\;E_{n}^{(\nu)}\,,\quad n=2,3,\dots
 \]
 Both limits are obvious from the behaviour of the function $\mathfrak{F}_\nu(E)$ (Fig.~\ref{fig:Fnu}); the former in particular is a consequence of general facts of the Kre{\u\i}n-Vi\v{s}ik-Birman theory, in the following sense. That there exists only one eigenvalue of $H^{(\nu)}_\alpha$ below the bottom $E_1^{(\nu)}$ of the Friedrichs extension $H^{(\nu)}$ is a consequence of $\mathring{H}^{(\nu)}$ having deficiency index one and of Corollary \ref{cor:negative_spectrum_ST_T_corollary}. Moreover, such eigenvalue, which is precisely $E_1^{(\nu,\alpha)}$, must satisfy 
 \[
  E_1^{(\nu,\alpha)}\;\leqslant\;\beta\;=\;\frac{1}{\,c_{\nu,\kappa}}\Big(\frac{4\pi\alpha}{\Gamma(1-\kappa)}-d_{\nu,\kappa}\Big)
 \]
 for any fixed $\kappa\in(0,1)$, as a consequence of \eqref{eq:alphabeta} and of Theorem \ref{thm:negative_spectrum_ST_T}. Thus, the limit $\alpha\to-\infty$ in the above inequality reproduces the limit for $E_1^{(\nu,\alpha)}$.
\end{remark}

 \begin{remark}
  When $\nu>0$ Theorem \ref{thm:EV_corrections} implies that $ H^{(\nu)}_\alpha\geqslant\mathbb{O}$ if and only if $\alpha\geqslant\alpha_\nu$. This fact too can be understood in terms of a general property of the Kre{\u\i}n-Vi\v{s}ik-Birman theory (Theorem \ref{thm:semibdd_exts_operator_formulation_Tversion}), which in the present setting reads
  \[
   H^{(\nu)}_\alpha+\frac{\nu^2}{4\kappa^2}\mathbbm{1}\;\geqslant\;\mathbb{O}\qquad\Leftrightarrow\qquad\beta\,\geqslant\, 0
  \]
  for any $\kappa<0$. The limit $\kappa\to -\infty$ and \eqref{eq:alphabeta} then yields
  \[
   \begin{split}
    H^{(\nu)}_\alpha\;\geqslant\;\mathbb{O}\qquad\Leftrightarrow\qquad \alpha\,& \geqslant\,\lim_{\kappa\to-\infty}\,\frac{\Gamma(1-\kappa)}{4\pi}\, d_{\nu,\kappa}\\
    &=\,\lim_{\kappa\to-\infty}\,\frac{\nu}{4\pi}\,\big(\psi(1-\kappa)+\ln({\textstyle-\frac{\nu}{\kappa}})+2\gamma-1+{\textstyle\frac{1}{2\kappa}}\big) \\
    &=\: \frac{\nu}{4\pi}\,(\ln\nu+2\gamma-1)\;=\;\alpha_\nu\,,
   \end{split}
  \]
 the limit above following again from the asymptotics \cite[Eq.~(6.3.18)]{Abramowitz-Stegun-1964}.  
 \end{remark}

\begin{remark}\label{rem:swavepoles}
 Having identified the eigenvalues of $H^{(\nu)}_\alpha$ as the roots of $\mathfrak{F}_\nu(E)=\alpha$ is clearly consistent with the fact that such eigenvalues are all the poles  of the resolvent $(H^{(\nu)}_\alpha-z\mathbbm{1})^{-1}$, $z=-\frac{\nu^2}{4\kappa^2}$, determined in \eqref{eq:HnuResolvent}, i.e., the values $E=-\frac{\nu^2}{4\kappa^2}$ with $\kappa$ determined by $\mathfrak{F}_{\nu,\kappa}=\alpha$, which is precisely another way of writing  $\mathfrak{F}_\nu(E)=\alpha$.
 \end{remark}

\chapter{Dirac-Coulomb Hamiltonians for heavy nuclei}
\label{chapter-Dirac-Coulomb} 

 The Dirac electron bound to a nucleus which the electron is bound to was among the first non-trivial and central models in the early days of quantum mechanics. The striking effectiveness of the Sommerfeld fine structure formula in providing the energy levels of the system gave this model the highest profile. Whereas the essential self-adjointness of the minimal operator could be established perturbatively for sufficiently small magnitudes of the Coulomb coupling, it had been known for decades, in explicit form at least since the 1970's, that for sufficiently large couplings a multitude of distinct self-adjoint realisations is available. Almost the entirety of the interest was then focussed on the existence of a distinguished Dirac-Coulomb self-adjoint Hamiltonian, characterised by being the only one with finite expectations, separately, of the kinetic part and of the potential part. Only in very recent times were the other self-adjoint realisations identified, first within the von Neumann extension scheme, and subsequently within the Kre{\u\i}n-Vi\v{s}ik-Birman scheme. This Chapter, modelled on the recent works \cite{MG_DiracCoulomb2017,GM-2017-DC-EV} presents this second approach and develops the spectral theory for each of the self-adjoint extensions.

\section{One-body Dirac-Coulomb models in sub-critical and critical regimes}\label{sec:IV-DCintroetc}

In quantum mechanics the Hilbert space for a relativistic electron or positron (or more generally a relativistic spin-$\frac{1}{2}$ particle) is
\begin{equation}
 \cH\;:=\;L^2(\mathbb{R}^3)\otimes\mathbb{C}^4\;\cong\;L^2(\mathbb{R}^3,\mathbb{C}^4)
\end{equation}
and the (formal) Hamiltonian of the free particle is
\begin{equation}
 H_0\;:=\;-\ii c \hslash \,\bm{\alpha}\cdot \bm{\nabla}+\beta m_{\mathrm{e}} c^2
\end{equation}
acting on $\cH$, where $\hslash$ is Planck's (reduced) constant, $c$ is the speed of light, $m_{\mathrm{e}}$ is the mass of the electron, and $\bm{\alpha}\equiv(\alpha_1,\alpha_2,\alpha_3)$ and $\beta$ are the $4\times 4$ matrices 
\begin{equation}
 \beta\;=\;\begin{pmatrix} 
            \mathbbm{1} & \mathbb{O} \\
            \mathbb{O} & -\mathbbm{1}
           \end{pmatrix}\,,\qquad
 \alpha_j\;=\;\begin{pmatrix}
               \mathbb{O} & \sigma_j \\
               \sigma_j & \mathbb{O}
              \end{pmatrix}\,,\qquad j\in\{1,2,3\}\,,
\end{equation}
having denoted here by $\mathbbm{1}$ and $\mathbb{O}$, respectively, the identity and the zero $2\times 2$ matrix, and by $\sigma_j$, as customary, the Pauli matrices\index{Pauli matrices}
\begin{equation}
 \sigma_1\;=\;\begin{pmatrix}
               0 & 1 \\ 1 & 0
              \end{pmatrix}\,,\qquad
 \sigma_2\;=\;\begin{pmatrix}
               0 & -\ii \\ \ii & 0
              \end{pmatrix}\,,\qquad
 \sigma_3\;=\;\begin{pmatrix}
               1 & 0 \\ 0 & -1
              \end{pmatrix}\,.
\end{equation}
Explicitly, the scalar product between any two elements $\psi\equiv(\psi_1,\psi_2,\psi_3,\psi_4)$ and $\phi\equiv(\phi_1,\phi_2,\phi_3,\phi_4)$ in $\cH$ is given by
\begin{equation}
 \langle\psi,\phi\rangle_\cH\;=\;\sum_{j=1}^4\int_{\mathbb{R}^3}\overline{\psi_j(x)}\,\phi_j(x)\,\ud x\,,
\end{equation}
and $H_0$ is the first order matrix-valued differential operator
\begin{equation}
 H_0\;=\;\begin{pmatrix}
          m_{\mathrm{e}} c^2\mathbbm{1} & -\ii\hslash c \,\bm{\sigma}\cdot\bm{\nabla} \\
          -\ii\hslash c \,\bm{\sigma}\cdot\bm{\nabla} &  -m_{\mathrm{e}}c^2\mathbbm{1}
         \end{pmatrix}
\end{equation}
(where $\bm{\sigma}\equiv(\sigma_1,\sigma_2,\sigma_3)$), known as the free Dirac operator.

Throughout this Chapter, whenever physical constants are explicitly kept into account, CGS physical units will be adopted, thus in particular $4\pi\varepsilon_0=1$.

The properties of $H_0$ are well known \cite{Thaller-Dirac-1992}. $H_0$ is essentially self-adjoint on $C^\infty_0(\mathbb{R}^3\!\setminus\!\{0\})\otimes\mathbb{C}^4$ with domain of self-adjointness
\begin{equation}
 H^1(\mathbb{R}^3)\otimes\mathbb{C}^4\;\cong\; H^1(\mathbb{R}^3,\mathbb{C}^4)\,,
\end{equation}
 and the spectrum of its closure (as a self-adjoint operator on $\cH$) is purely absolutely continuous and given by
\begin{equation}
 \sigma(H_0)\;=\;\sigma_{\mathrm{ac}}(H_0)\;=\;(-\infty,-m_{\mathrm{e}}c^2]\cup[m_{\mathrm{e}}c^2,+\infty)\,.
\end{equation}
In fact, $H_0$ is unitarily equivalent to
\begin{equation}
 \widetilde{H_0}\;:=\;\begin{pmatrix}
                       \mathbbm{1}\sqrt{-c^2\Delta+m_{\mathrm{e}}^2c^4} & 0 \\
                       0 & -\mathbbm{1}\sqrt{-c^2\Delta+m_{\mathrm{e}}^2c^4}
                      \end{pmatrix}.
\end{equation}

When the particle is subject to the external scalar field due to the Coulomb interaction with a nucleus of atomic number $Z$ placed in the origin of $\mathbb{R}^3$, this is accounted for by the Dirac-Coulomb Hamiltonian\index{minimal Hamiltonian} given by
\begin{equation}\label{eq:IV-Hformal}
 H\;:=\;-\ii c \hslash \,\bm{\alpha}\cdot \bm{\nabla}+\beta m_{\mathrm{e}} c^2-\frac{e^2 Z}{\hslash\,|x|}\mathbbm{1}\;=\;H_0-\frac{\,cZ\alpha_\mathrm{f}\,}{|x|}\mathbbm{1}\,,
\end{equation}
where now $\mathbbm{1}$ is the $4\times 4$ identity matrix (no confusion should arise here and henceforth on the symbol $\mathbbm{1}$, being its meaning of identity self-explanatory from the context), $e$ is the elementary charge, and 
\begin{equation}
 \alpha_\mathrm{f}\;=\;\frac{e^2}{\hslash c}\;\approx\;\frac{1}{137}
\end{equation}
is the fine structure constant\index{fine structure constant} (recall that in the current units $4\pi\varepsilon_0=1$).

The operator $H$ can be naturally minimally defined as a symmetric operator on the dense $C^\infty_0(\mathbb{R}^3\!\setminus\!\{0\})\otimes\mathbb{C}^4$ of $\cH$.

However, the possibility that this yields an unambiguous (i.e., a unique self-adjoint) realisation of the Hamiltonian depends on the magnitude of the coupling $Z\alpha_\mathrm{f}$, hence of the nuclear charge $Z$. It is indeed well known \cite{Thaller-Dirac-1992} that the formal operator \eqref{eq:IV-Hformal} is essentially self-adjoint on $C^\infty_0(\mathbb{R}^3\!\setminus\!\{0\},\mathbb{C}^4)$ \emph{only} when $Z\alpha_\mathrm{f}\leqslant\frac{\sqrt{3}}{\,2}$, implying $Z\leqslant 118$ (and in particular the domain of self-adjointness when $Z\alpha_{\mathrm{f}}<\frac{\sqrt{3}}{2}$ is $\mathcal{D}(H)=H^1(\mathbb{R}^3,\mathbb{C}^4)$), in which case the spectrum consists of the essential part $\sigma_{\mathrm{ess}}(H)=(-\infty,-m_{\mathrm{e}}c^2]\cup[m_{\mathrm{e}}c^2,+\infty)$ plus a discrete spectrum in the `gap' $(-m_{\mathrm{e}}c^2,m_{\mathrm{e}}c^2)$ consisting of eigenvalues $E_{n,\kappa}$ given by the celebrated Sommerfeld fine structure formula\index{Sommerfeld fine structure formula}

\begin{equation}\label{eq:Somm-form}
E_{n,\kappa}\;=\;m_{\mathrm{e}}c^2\Big(1+\frac{(Z\alpha_\mathrm{f})^2}{\big(n+\sqrt{\kappa^2-(Z\alpha_\mathrm{f})^2}\,\big)^{\!2}}\Big)^{\!-\frac{1}{2}},\quad n\in\mathbb{N}_0,\,\kappa\in\mathbb{Z}\!\setminus\!\{0\}\,.
\end{equation}

Although the above regime of $Z$ covers all currently known elements (the last one to be discovered, the Oganesson\index{Oganesson} ${}^{294}_{118}$Og, thus $Z=118$, was first synthesised in 2002 and formally named in 2016), the problem of the self-adjoint realisation of the Dirac-Coulomb Hamiltonian above the threshold $Z\alpha_\mathrm{f}=\frac{\sqrt{3}}{\,2}$ has been topical since long and so is still today. Even the observation that the problem only arises due to the idealisation of point-like nuclei (and also because one neglects the anomalous magnetic moment of the electron) does not diminish its relevance, given the extreme experimental precision, for example, of the Sommerfeld fine structure formula\index{Sommerfeld fine structure formula} for the eigenvalues of $H$ when $Z\leqslant 118$.

From the mathematical side, the study of the self-adjoint extensions of the Dirac-Coulomb Hamiltonian has a long and active history until modern days \cite{Evans-1970,Weidmann-1971,Rejto-1971,Schmincke-1972,Schmincke-1972-distinguished,Gustafson-Rejto-1973,Wust-1975,Kalf-Schmincke-Wust-1975,Nenciu-1976,Wust-1977,Chernoff-1977,Klaus-Wust-1978,Landgren-Rejto-JMP1979,Landgren-Rejto-Martin-JMP1980,Landgren-Rejto-1981,Burnap-Brysk-Zweifel-NuovoCimento1981,Arai-Yamada-RIMS-1982,Arai-Yamada-RIMS-1983,Kato-1983,Berthier-Georgescu-JFA1987,Thaller-Dirac-1992,Xia-1999,Georgescu-Mantoiu-JOT2001,Esteban-Loss-JMP2007,Voronov-Gitman-Tyutin-TMP2007,Arrizabalaga-JMP2011,Arrizabalaga-Duoandikoetxea-Vega_2012_JMP2013,Hogreve-2013_JPhysA,Esteban-Lewin-Sere-2017_DC-minmax-levels}.
A recent, concise survey of this vast literature is presented in \cite{Gallone-AQM2017}. Theorem \ref{thm:recap} here below summarises all such, by now classical, facts.
For the clarity of presentation,  natural units $c=\hslash=m_{\mathrm{e}}=e=1$ are adopted henceforth (unless when explicitly declared otherwise), so as to get rid of mathematically inessential parameters, the coupling constant of relevance thus becoming $\nu\equiv - Z\alpha_\mathrm{f}$.

\begin{theorem}[Self-adjoint extensions of the minimal Dirac-Coulomb]\label{thm:recap}
On the Hilbert space $\cH=L^2(\mathbb{R}^3,\mathbb{C}^4)$ consider, for fixed $\nu\in\mathbb{R}$, the operator
\begin{equation}\label{eq:IV-minimalH}
 \begin{split}
  H\;&=\;H_0+\frac{\nu}{\,|x|\,}\mathbbm{1}\,,\qquad H_0\;=\;-\ii\,\bm{\alpha}\cdot \bm{\nabla}+\beta\,, \\
  \mathcal{D}(H)\;&=\;\mathcal{D}(H_0)\;=\;C^\infty_0(\mathbb{R}^3\!\setminus\!\{0\},\mathbb{C}^4)\,.
 \end{split}
\end{equation}
\begin{enumerate}
 \item[(i)] \emph{(Sub-critical regime.)} If $|\nu|\leqslant\frac{\sqrt{3}}{2}$, then $H$ is essentially self-adjoint and if $|\nu| < \frac{\sqrt{3}}{2}$, then $\mathcal{D}(\overline{H})=H^1(\mathbb{R}^3,\mathbb{C}^4)$.
 \item[(ii)] \emph{(Critical regime.)} If $\frac{\sqrt{3}}{2}<|\nu|<1$, then $H$ admits an infinity of self-adjoint extensions, among which there is a `distinguished' one,\index{distinguished Dirac-Coulomb Hamiltonian} $H_{\mathrm{D}}$, uniquely characterised by the properties
 \begin{equation}
  \mathcal{D}(H_{\mathrm{D}})\,\subset\,\mathcal{D}\big(|H_0|^{\frac{1}{2}}\big)\quad\mathrm{or}\quad  \mathcal{D}(H_{\mathrm{D}})\,\subset\,\mathcal{D}\big(|x|^{-\frac{1}{2}}\big)\,,
 \end{equation}
 that is, $H_{\mathrm{D}}$ is the unique extension whose operator domain is in one, and hence also in the other, among the kinetic energy form domain $\mathcal{D}[H_0]=\mathcal{D}(|H_0|^{\frac{1}{2}})$ and the potential energy form domain $\mathcal{D}[|x|^{-1}]=\mathcal{D}(|x|^{-\frac{1}{2}})$. Moreover, $0\notin\sigma(H_{\mathrm{D}})$.
 \item[(iii)] \emph{(Super-critical regime.)} If $|\nu|\geqslant 1$, then $H$ admits an infinity of self-adjoint extensions, without a distinguished one in the sense of the operator $H_{\mathrm{D}}$ in the critical regime. In fact, when $|\nu|>1$ every self-adjoint extension of $H$ has infinitely many eigenfunctions not belonging to $\mathcal{D}(|x|^{-\frac{1}{2}})$.
\end{enumerate}
In either regime, the spectrum of any self-adjoint extension $\widetilde{H}$ of $H$ is such that
\begin{equation}
 \begin{split}
  \sigma_\mathrm{ess}(\widetilde{H})\;&=\;\sigma(\overline{H_0})\;=\;(-\infty,-1]\cup[1,+\infty) \, ,\\
  \sigma_\mathrm{disc}(\widetilde{H})\;&\subset\;(-1,1)\,.
 \end{split}
\end{equation}
\end{theorem}

It is worth remarking that for Coulomb-like matrix-valued interactions $V(x)$ that are \emph{not} of the form $\nu|x|^{-1}\mathbbm{1}$ but still satisfy  $|V(x)|\leqslant\;\nu|x|^{-1}$, the sub-critical regime described in Theorem \ref{thm:recap}(i) only ranges up to $|\nu|<\frac{1}{2}$, and counterexamples are well known of operators $H_0+V$ with $|V(x)|\leqslant\;(\frac{1}{2}+\varepsilon)|x|^{-1}$ for arbitrary $\varepsilon>0$ and failing to be essentially self-adjoint on $C^\infty_0(\mathbb{R}^3\!\setminus\!\{0\},\mathbb{C}^4)$ \cite{Arai-Yamada-RIMS-1983}.


In this Chapter the primary focus is the critical regime, $|\nu|\in(\frac{\sqrt{3}}{2},1)$. This is a regime of ultra-heavy nuclei, in fact nuclei of elements that one expects to discover in the next future. It is the first regime where the Kato-Rellich-like perturbative arguments (Sect.~\ref{sec:perturbation-spectra}),\index{theorem!Kato-Rellich} applicable for small $\nu$'s, cease to work. It is also regarded as a physically meaningful regime, because as long as $|\nu|<1$ the Sommerfeld fine structure formula\index{Sommerfeld fine structure formula}\index{Sommerfeld fine structure formula} still provides, formally, bound states for real energy levels, which only become complex when $|\nu|>1$, thus predicting an instability of the atom ('the \emph{$Z=137$ catastrophe}`).\index{Z=137 catastrophe}

In fact, the critical regime for the Dirac-Coulomb operator is already intensively studied, with a special focus on the `distinguished' self-adjoint extension $H_{\mathrm{D}}$\index{distinguished Dirac-Coulomb Hamiltonian} \cite{Evans-1970,Weidmann-1971,Schmincke-1972-distinguished,Wust-1975,Nenciu-1976,Wust-1977,Klaus-Wust-1978,Landgren-Rejto-JMP1979,Arai-Yamada-RIMS-1983,Kato-1983,Xia-1999,Esteban-Loss-JMP2007,Voronov-Gitman-Tyutin-TMP2007,Arrizabalaga-JMP2011,Arrizabalaga-Duoandikoetxea-Vega_2012_JMP2013,Hogreve-2013_JPhysA,Esteban-Lewin-Sere-2017_DC-minmax-levels}, which, as seen in Theorem \ref{thm:recap}(ii), has the characteristic feature of having finite expectations separately of the kinetic and the potential part, and moreover is invertible with everywhere-defined bounded inverse. Conceptually, and qualitatively, this has very much in common with the analogous, scaling-critical problem of the self-adjoint realisation of the (formal) non-relativistic and pseudo-relativistic Schr\"{o}dinger operators
\[
 -\Delta+\frac{\nu}{\,|x|^2}\qquad\mathrm{or}\qquad \sqrt{1-\Delta}+\frac{\nu}{\,|x|\,}\qquad\mathrm{on}\;L^2(\mathbb{R}^3)
\]
when $\nu$ is out of the perturbative regime, 
an issue that is both standard textbook material (such as the Weyl limit-point limit-circle analysis reviewed in Section \ref{sec:WeylsCriterion}) and object of recent research \cite{LeYaouanc-Oliver-Raynal-JMP1997,Bruneau-Derezinski-Georgescu-2011,Fall-Felli-JFA2014}.

Beyond the characterisation of the distinguished Dirac-Coulomb Hamiltonian, the next goal is the classification of all other self-adjoint extensions of the minimal operator\index{minimal Hamiltonian} \eqref{eq:IV-minimalH}, and their spectral analysis, with special reference to those that display physically realistic features.

 The extensions of $H$ are more efficiently and naturally identified upon recognising the reduced structure of $H$ with respect to a natural orthogonal decomposition of the Hilbert space in sectors of definite angular symmetry (in the usual sense reviewed in Section \ref{sec:I_invariant-reducing-ssp}). This is the customary canonical decomposition of $H$ into \emph{partial wave operators} \cite[Section 4.6]{Thaller-Dirac-1992}, which is a consequence of its spherical symmetry.

 By expressing $x\equiv(x_1,x_2,x_3)\in\mathbb{R}^3$ in polar coordinates $x=(r,\Omega)\in\mathbb{R}^+\!\times\mathbb{S}^2$, $r:=|x|$, the map $\psi(x)\mapsto r\psi(x_1(r,\Omega),x_2(r,\Omega),x_3(r,\Omega))$ induces a unitary isomorphism
\[
 L^2(\mathbb{R}^3,\mathbb{C}^4,\ud x)\;\xrightarrow[]{\cong}\;L^2(\mathbb{R}^+,\ud r)\otimes L^2(\mathbb{S}^2,\mathbb{C}^4,\ud\Omega)\,.
\]
In terms of the observables
\[
\begin{split}
 \bm{L}\,&=\,\bm{x}\times(-\ii\bm{\nabla})\,,\qquad\qquad\quad\;\bm{S}\,=\,-\frac{1}{4}\bm{\alpha}\times\bm{\alpha}\,, \\
 \bm{J}\,&=\,\bm{L}+\bm{S}\equiv(J_1,J_2,J_3)\,,\qquad\!\! K\,=\,\beta(2\bm{L}\cdot\bm{S}+\mathbbm{1})\,,
\end{split}
\]
one further decomposes
\begin{equation}
 L^2(\mathbb{S}^2,\mathbb{C}^4,\ud\Omega)\;\cong\;\bigoplus_{j\in\frac{1}{2}+\mathbb{N}_0}\;\;\;\bigoplus_{m_j=-j}^j\;\bigoplus_{\kappa_j=\pm(j+\frac{1}{2})}\mathcal{K}_{m_j,\kappa_j}\,,
\end{equation}
where 
\begin{equation}
 \mathcal{K}_{m_j,\kappa_j}\;:=\;\mathrm{span}\{\Psi^+_{m_j,\kappa_j},\Psi^-_{m_j,\kappa_j}\}\;\cong\;\mathbb{C}^2
\end{equation}
and $\Psi^+_{m_j,\kappa_j}$ and $\Psi^-_{m_j,\kappa_j}$ are two orthonormal vectors in $\mathbb{C}^4$, and simultaneous eigenvectors of the observables
\[
 J^2\!\upharpoonright\! L^2(\mathbb{S}^2,\mathbb{C}^4,\ud\Omega)\,,\qquad J_3\!\upharpoonright\! L^2(\mathbb{S}^2,\mathbb{C}^4,\ud\Omega)\,,\qquad K\!\upharpoonright\! L^2(\mathbb{S}^2,\mathbb{C}^4,\ud\Omega)\,,
\]
namely, the \emph{total angular momentum} and the \emph{spin-orbit} operators, with eigenvalues, respectively, $j(j+1)$, $m_j$, and $\kappa_j$. It then turns out that each subspace
\begin{equation}\label{eq:IV-def_space_H_mj_kj}
 \cH_{m_j,\kappa_j}\;:=\;L^2(\mathbb{R}^+,\ud r)\otimes\mathcal{K}_{m_j,\kappa_j}\;\cong\;L^2(\mathbb{R}^+,\mathbb{C}^2,\ud r)
\end{equation}
is a reducing subspace (Sect.~\ref{sec:I_invariant-reducing-ssp}) for $H$, which, through the overall isomorphism
\begin{equation}\label{eq:IV-overalliso}
U\;:\;L^2(\mathbb{R}^3,\mathbb{C}^4,\ud x)\;\xrightarrow[]{\cong}\;\bigoplus_{j\in\frac{1}{2}+\mathbb{N}_0}\;\;\;\bigoplus_{m_j=-j}^j\;\bigoplus_{\kappa_j=\pm(j+\frac{1}{2})}\cH_{m_j,\kappa_j}\,,
\end{equation}
is therefore unitarily equivalent to the operator orthogonal sum\index{operator orthogonal sum}
\begin{equation}\label{eq:IV-Dirac_operator_decomposition}
 UHU^*\;=\;\bigoplus_{j\in\frac{1}{2}+\mathbb{N}}\;\;\;\bigoplus_{m_j=-j}^j\;\bigoplus_{\kappa_j=\pm(j+\frac{1}{2})}\;h_{m_j,\kappa_j}\,,
\end{equation}
where
\begin{equation}\label{eq:IV-def_operator_h_mj_kj}
\begin{split}
  h_{m_j,\kappa_j}\;&:=\;\begin{pmatrix}
                   1+\frac{\nu}{r} & -\frac{\ud}{\ud r}+\frac{\kappa_j}{r} \\
                   \frac{\ud}{\ud r}+\frac{\kappa_j}{r} & -1+\frac{\nu}{r}
                  \end{pmatrix}, \\
 \qquad\mathcal{D}(h_{m_j,\kappa_j})\;&:=\;C^\infty_0(\mathbb{R}^+)\otimes \mathcal{K}_{m_j,\kappa_j}\;\cong\;C^\infty_0(\mathbb{R}^+,\mathbb{C}^2)\,.
\end{split}
\end{equation}
Thus, \eqref{eq:IV-def_operator_h_mj_kj} defines a densely defined and symmetric operator on the Hilbert space \eqref{eq:IV-def_space_H_mj_kj} and the overall problem of the self-adjoint realisation of $H$ is reduced to the same problem in each reducing subspace.

In particular, it is of physical relevance to consider each operator
\begin{equation}
 h_{m_j}\;:=\;h_{m_j,\kappa_j=j+\frac{1}{2}}\oplus h_{m_j,\kappa_j=-(j+\frac{1}{2})}
\end{equation}
acting block-diagonal-wise, with the two different spin-orbit components,\index{spin-orbit components} on the Hilbert eigenspace $L^2(\mathbb{R}^+,\mathbb{C}^4,\ud r)$ of $(j,m_j)$-eigenvalue for $J^2$ and $J_3$.

 Now, an application of Weyl limit-point limit-circle arguments\index{Weyl criterion!limit-point limit-circle}\index{Weyl limit-point/limit-circle}\index{limit-point/limit-circle}\index{theorem!Weyl (limit-point limit-circle criterion)} (Sect.~\ref{sec:WeylsCriterion}), whose proof is deferred to Sect.~\ref{sec:IVdefindex} (see, e.g., \cite[Chapter 6.B]{Weidmann-book1987} or \cite[Section 2]{Gallone-AQM2017} for previous discussions), shows that for sufficiently small $|\nu|$ (corresponding to the sub-critical regime) each block $h_{m_j,\kappa_j}$ is essentially self-adjoint and hence so too is $H$, whereas the larger $|\nu|$ in the critical and super-critical regime, the larger is the finite number of blocks that loose essential self-adjointness and reach deficiency indices $(1,1)$.

 \begin{proposition}\label{prop:deficiency_indices}~
  \begin{enumerate}[(i)]
   \item The operator $h_{m_j,\kappa_j}$ is essentially self-adjoint on its domain with respect to the Hilbert space $\cH_{m_j,\kappa_j}$ if and only if
 \begin{equation}
  \nu^2\;\leqslant\;\kappa_j^2-\textstyle{\frac{1}{4}}\,,
 \end{equation}
 and it has deficiency indices $(1,1)$ otherwise. In particular, $H$ is essentially self-adjoint when $|\nu|<\frac{\sqrt{3}}{2}$.
   \item Assume that $\sqrt{n^2-\frac{1}{4}}<|\nu|\leqslant\sqrt{(n+1)^2-\frac{1}{4}}$ for some $n\in\mathbb{N}$. Then $H$ has equal deficiency indices $d_+(H)=d_-(H)=2n(n+1)$. 
  \end{enumerate}
 \end{proposition}

%
%
%
%

%
 Thus, in the regime $|\nu|\in(\frac{\sqrt{3}}{2},1)$ of interest, only the operators of the decomposition \eqref{eq:IV-Dirac_operator_decomposition} with $\kappa_j^2=1$, i.e.,
 \begin{equation}\label{eq:IV-4op}
  h_{\frac{1}{2},1}\,,\quad h_{-\frac{1}{2},1}\,,\quad h_{\frac{1}{2},-1}\,,\quad h_{-\frac{1}{2},-1}\,,
 \end{equation}
 have deficiency indices $(1,1)$, all others being essentially self-adjoint.
 As a consequence, the operator
\begin{equation}
 h_{\frac{1}{2},1}\oplus h_{\frac{1}{2},-1}\oplus h_{-\frac{1}{2},1}\oplus h_{-\frac{1}{2},-1}\,,
\end{equation}
and hence $H$ itself, has deficiency indices $(4,4)$. 
This means (Theorem \ref{thm:vonN_thm_symm_exts}) that there is a 16-real-parameter family of self-adjoint extensions of $H$. 
From the operator-theoretic point of view, the self-adjoint extension problem for $h_{\frac{1}{2},1}$ is the very same as for the other three operators (and in fact $h_{\frac{1}{2},1}$ and $h_{-\frac{1}{2},1}$ have the same formal action on $L^2(\mathbb{R}^+,\mathbb{C}^2)$, and so have $h_{\frac{1}{2},-1}$ and $h_{-\frac{1}{2},-1}$). 

There is room for extensions only on the sector $j=\frac{1}{2}$ of lowest total angular momentum $J^2$. In practice, as the analysis of Sections \ref{sec:mainresults}-\ref{sec:IV-proofs} shows, each self-adjoint realisation is characterised by certain boundary condition at the origin, where the Coulomb singularity is placed. For higher $j$'s the large angular momentum makes the Coulomb singularity lesser and lesser relevant, and on such sectors $H$ is already essentially self-adjoint.

  Physically, the relevant class of extensions is rather the \emph{one}-parameter sub-family consisting of the orthogonal sum of self-adjoint extensions for each of the four block operators \eqref{eq:IV-4op}, plus of course the sum of the closures of all the already essentially self-adjoint blocks, where each of the non-trivial self-adjoint extensions of a single block is characterised by the same boundary condition of self-adjointness at the origin.

  There are of course many more self-adjoint extensions: those that preserve the operator orthogonal sum structure in blocks of definite angular and spin-orbit symmetry, but are characterised by a different behaviour around the origin for each symmetry sector, and even more generally those that are not reduced by the decomposition \eqref{eq:IV-overalliso} and hence are characterised by boundary conditions at the origin that couple the short-distance behaviour of components of the wave function in different sectors of symmetry. Such non-uniform (in $m_j$ and $\kappa_j$) or non-local behaviours are considered to be unphysical. (In Chapter \ref{chapter-Grushin}, specifically in Sections \ref{sec:genextscrHa} and \ref{sec:uniformlyfirbredext}, a completely analogous selection of physically relevant self-adjoint realisations out of a much larger family of extensions is performed for a different model, again discarding those extensions that are not in the form of an operator orthogonal sum in each of the sectors of relevance and do not display the same magnitude of the boundary condition sector-wise.)

  In this Chapter the self-adjoint extension problem for the each minimally defined block operator $h_{m,\kappa}$, with $m=\pm\frac{1}{2}$ and $\kappa=\pm 1$, is discussed and solved (Sect.~\ref{sec:mainresults}-\ref{sec:IV-proofs}) within the Kre{\u\i}n-Vi\v{s}ik-Birman scheme and the spectra of each self-adjoint realisations are characterised (Sect.~\ref{sec:Somm}-\ref{sec:IV-betaspectrum}).

\section{Self-adjoint realisations of Dirac-Coulomb blocks of definite angular and spin-orbit symmetry}\markboth{Dirac-Coulomb Hamiltonians for heavy nuclei}{Self-adjoint realisations of blocks of definite angular and spin-orbit symmetry}\label{sec:mainresults}

 Both classical extension schemes discussed in Chapter \ref{chaper-extension-schemes} are applicable to the minimally defined block operators \eqref{eq:IV-4op}, the one a la von Neumann because its applicability is universal, and the one a la Kre{\u\i}n-Vi\v{s}ik-Birman because the existence of a distinguished self-adjoint extension that is invertible with everywhere-defined bounded inverse provides the required \emph{reference extension} needed in Theorem \ref{thm:VB-representaton-theorem_Tversion2}.

 Quite strikingly, despite the operator of interest is one of the most relevant from the very early days of the almost one century old modern quantum theory, and despite all the tools for identifying its self-adjoint realisations had been available for decades, the classification of the family of Dirac-Coulomb extensions is only relatively recent, by Voronov, Gitman, and Tyutin \cite{Voronov-Gitman-Tyutin-TMP2007} in 2007 and by Hogreve \cite{Hogreve-2013_JPhysA} in 2013, both within the von Neumann extension scheme, then subsequently by Gallone and Michelangeli \cite{MG_DiracCoulomb2017,GM-2017-DC-EV} in 2017 within the Kre{\u\i}n-Vi\v{s}ik-Birman scheme, then right after by Cassano and Pizzichillo \cite{Cassano-Pizzichillo-2017} within a `brute force' scheme that retains features of both and consists of monitoring explicitly, with standard differential equation techniques, all possible self-adjoint behaviours at the origin for sub-domains of the adjoint operator, and finally again by Cassano and Pizzichillo \cite{Cassano-Pizzichillo-DCtriplets-2019} who re-phrased their previous classification \cite{Cassano-Pizzichillo-2017} in a boundary triplet language. (Notably, \cite{Cassano-Pizzichillo-2017,Cassano-Pizzichillo-DCtriplets-2019} cover the more general case of matrix-valued interaction $V(x)\sim|x|^{-1}$ that include an explicit electric, scalar, and anomalous magnetic potential.)

 Such a `delay' is most plausibly due to the manifest success that the Sommerfeld fine structure formula\index{Sommerfeld fine structure formula} had in predicting energy levels with an excellent matching with the experimental values, and in fact providing the discrete spectrum of the distinguished (canonical) Dirac-Coulomb Hamiltonian:\index{distinguished Dirac-Coulomb Hamiltonian} in this sense, the quest for alternative, physically distinct self-adjoint realisations became less central. Yet, modern experimental manipulations of cold atoms, specifically in this context photoionisation microscopy with excitation
of a quasi-bound Stark state in Hydrogen atoms (see, e.g., \cite{Stodolna-et-al-PRL2013}), give a compelling indication that central, point-like perturbations of the standard Hamiltonian are indeed feasible, which does put the question of the other Dirac-Coulomb self-adjoint realisations back at the centre of the mathematical investigation.

  With the extension classification obtained within the von Neumann scheme well available in the above-mentioned works \cite{Voronov-Gitman-Tyutin-TMP2007,Hogreve-2013_JPhysA}, this chapter focuses on the application of the Kre{\u\i}n-Vi\v{s}ik-Birman scheme, as in this context it turns out to be both considerably less laborious and more naturally and directly informative, characterising the self-adjoint realisation of the Dirac-Coulomb Hamiltonian straightforwardly in terms of local boundary conditions at the Coulomb singularity.

  The problem is thus the identification, by means of Theorem \ref{thm:VB-representaton-theorem_Tversion2}, of the self-adjoint extensions of the operator $h_{m_j,\kappa_j}$ defined in  \eqref{eq:IV-def_operator_h_mj_kj} on the Hilbert space $\cH_{m_j,\kappa_j}$ defined in \eqref{eq:IV-def_space_H_mj_kj}, using as reference extension precisely the distinguished extension of Theorem \ref{thm:recap}(ii).

 In abstract terms one considers the Hilbert space $L^2(\mathbb{R}^+,\mathbb{C}^2)$ with scalar product
\begin{equation}\label{eq:IV-scalar_product}
 \begin{split}
  \langle \psi,\phi\rangle_{L^2(\mathbb{R}^+,\mathbb{C}^2)}\,&=\,\int_0^{+\infty}\langle\psi(r),\phi(r)\rangle_{\mathbb{C}^2}\,\ud r \,=\,\sum_{\alpha=\pm}\int_0^{+\infty}\overline{\psi^\alpha(r)}\,\phi^\alpha(r)\,\ud r\, , \\
  &\quad\psi\equiv\begin{pmatrix}
                   \psi^+\! \\ \psi^-\!
                  \end{pmatrix},\,
        \phi\equiv\begin{pmatrix}
                   \phi^+\! \\ \phi^-\!
                  \end{pmatrix}\,\in\,L^2(\mathbb{R}^+,\mathbb{C}^2)\,,
 \end{split} 
\end{equation}
 and the (minimally defined) operator
\begin{equation}\label{eq:IV-def_operator_S}
\begin{split}
  h\;&:=\;\begin{pmatrix}
                   1+\frac{\nu}{r} & -\frac{\ud}{\ud r}+\frac{\kappa}{r} \\
                   \frac{\ud}{\ud r}+\frac{\kappa}{r} & -1+\frac{\nu}{r}
                  \end{pmatrix}, \\
 \qquad\mathcal{D}(h)\;&:=\;C^\infty_0(\mathbb{R}^+,\mathbb{C}^2)\,, \\
 &\qquad\quad (\kappa=\pm 1)\,.
\end{split}
\end{equation}
$h$ is non-semi-bounded, densely defined, and symmetric, and following from Proposition \ref{prop:deficiency_indices} it has deficiency indices $(1,1)$. It is convenient to use a separate notation for the  differential action (namely, the formal operator) associated with $h$ and thus write
\begin{equation}\label{eq:IV-tildeS}
 \widetilde{h}
 \begin{pmatrix}
  f^+\\ f^-
 \end{pmatrix}\;:=\;\begin{pmatrix}
                   1+\frac{\nu}{r} & -\frac{\ud}{\ud r}+\frac{\kappa}{r} \\
                   \frac{\ud}{\ud r}+\frac{\kappa}{r} & -1+\frac{\nu}{r}
                  \end{pmatrix}\begin{pmatrix}
  f^+ \\ f^-
 \end{pmatrix}.
\end{equation}
Since $\widetilde{h}$ has real smooth coefficients, and is formally self-adjoint, then (Sect.~\ref{sec:MinimalAndMaximalRealisations}) the operator closure $\overline{h}$ and the adjoint $h^*$ of $h$ are, respectively, the minimal and the maximal realisation of $\widetilde{h}$, that is, they both act as $\widetilde{h}$ respectively on
\begin{equation}\label{DSclosureDS*}
 \begin{split}
  \mathcal{D}(\overline{h})\;&=\;\overline{C^\infty_0(\mathbb{R}^+,\mathbb{C}^2)}^{\|\cdot\|_{\Gamma(h)}}\,, \\
  \mathcal{D}(h^*)\;&=\;\{\psi\in L^2(\mathbb{R}^+,\mathbb{C}^2)\,|\,\widetilde{h}\psi \in L^2(\mathbb{R}^+,\mathbb{C}^2)\}\,,
 \end{split}
\end{equation}
where $\|\cdot\|_{\Gamma(h)}$ is the graph norm (Sect.~\ref{sec:I-preliminaries}) associated with $h$. 
 
It is also often convenient to re-express $\nu$ in terms of the new parameter
\begin{equation}\label{eq:IV-defB}
 B\;:=\;\sqrt{\kappa^2-\nu^2}\,.
\end{equation}
In the critical regime $|\nu|\in(\frac{\sqrt{3}}{2},1)$, and for the blocks $\kappa=\pm 1$, one has $B\in(0,\frac{1}{2})$.

 In order to apply Theorem \ref{thm:VB-representaton-theorem_Tversion2} one needs the intermediate results of Propositions \ref{prop:kerS*}, \ref{prop:SD}, and \ref{prop:Sclosure} below (whose proofs are deferred to Section \ref{sec:IV-proofs}).

First one needs a characterisation of $\ker h^*$.

\begin{proposition}\label{prop:kerS*}
 Let $h$ be the operator on $L^2(\mathbb{R}^+,\mathbb{C}^2,\ud r)$ defined in \eqref{eq:IV-def_operator_S} with $|\nu|\in(\frac{\sqrt{3}}{2},1)$. The operator $h^*$ has a one-dimensional kernel, spanned by the function
 \begin{equation}\label{eq:IV-vector_Phi}
  \Phi\:\equiv\:\begin{pmatrix}
            \Phi^+\! \\ \Phi^-
           \!\end{pmatrix}
 \end{equation}
 with
 \begin{equation}\label{eq:IV-vector_Phi_components}
  \Phi^{\pm}(r)\;:=\;e^{-r}r^{-B}\Big( {\frac{\pm(\kappa+\nu)+B}{\kappa+\nu}}\,U_{-B,1-2B}(2r)-{\frac{2rB}{\kappa+\nu}}\,U_{1-B,2-2B}(2r)\Big)\,,
 \end{equation}
where
$U_{a,b}(r)$ is the Tricomi function\index{Tricomi functions} \cite[Sec.~13.1.3]{Abramowitz-Stegun-1964}. $\Phi$ is analytic on $(0,+\infty)$ with asymptotics
\begin{equation}\label{eq:IV-Phi_asymptotics}
 \begin{split}
  \Phi(r)\;&=\;r^{-B}\,\textstyle{\frac{\Gamma(2B)}{\Gamma(B)}}
  \begin{pmatrix}
   \;\frac{\kappa+\nu+B}{\kappa+\nu} \\
   -\frac{\kappa+\nu-B}{\kappa+\nu}
  \end{pmatrix}+
  \begin{pmatrix}
   q^+\! \\ q^-\!
  \end{pmatrix}r^B+O(r^{1-B})\quad\textrm{as }\;r\downarrow 0 \, ,\\
   \Phi(r)\;&=\;2^B\begin{pmatrix}
               1 \\ -1
              \end{pmatrix}r^{-B}e^{-r}
              (1+O(r^{-1}))\quad\,\textrm{as }\;r\to +\infty\,,
 \end{split}
\end{equation}
where $q^\pm$ are both non-zero and explicitly given by \eqref{eq:IV-def_qpm} below.
\end{proposition}

Next, one needs to identify a `distinguished', reference extension $h_{\mathrm{D}}$ of $h$ which be self-adjoint and with everywhere defined inverse, and to characterise the action of $h_{\mathrm{D}}^{-1}$ on $\ker h^*$.

\begin{proposition}\label{prop:SD} Let $h$ be the operator on $L^2(\mathbb{R}^+,\mathbb{C}^2,\ud r)$ defined in \eqref{eq:IV-def_operator_S} with $|\nu|\in(\frac{\sqrt{3}}{2},1)$.
\begin{enumerate}
 \item[(i)]  There exists a self-adjoint realisation $h_{\mathrm{D}}$ of $h$ with the property that
 \begin{equation}\label{eq:IV-SD_uniqueness_properties}
  \mathcal{D}(h_{\mathrm{D}})\subset H^{\frac{1}{2}}(\mathbb{R}^+,\mathbb{C}^2)\quad\textrm{or}\quad \mathcal{D}(h_{\mathrm{D}})\subset\mathcal{D}[r^{-1}]\,,
 \end{equation}
 where the latter is the form domain of the multiplication operator by $r^{-1}$ on each component of $L^2(\mathbb{R}^+,\mathbb{C}^2)$ (the space of `finite potential energy'). $h_{\mathrm{D}}$ is the only self-adjoint realisation of $h$ satisfying \eqref{eq:IV-SD_uniqueness_properties}.
 \item[(ii)] $h_{\mathrm{D}}$ is invertible on $L^2(\mathbb{R}^+,\mathbb{C}^2)$ with everywhere defined and bounded inverse. The explicit integral kernel of $h_{\mathrm{D}}$ is given by \eqref{eq:IV-Green}.
 \item[(iii)] In terms of the spaces $\mathcal{D}(\overline{h})$ and $\ker h^*$ one has
 \begin{equation}\label{eq:IV-decomp_DSD}
  \mathcal{D}(h_{\mathrm{D}})\;=\;\mathcal{D}(\overline{h})\dotplus h_{\mathrm{D}}^{-1}\ker h^*\,.
 \end{equation}
 Moreover,
 \begin{equation}\label{eq:IV-Krein_decomp_formula}
  \begin{split}
  \mathcal{D}(h^*)\;&=\; \mathcal{D}(h_{\mathrm{D}})\dotplus\ker h^*\,,\\
  &=\;\mathcal{D}(\overline{h})\dotplus h_{\mathrm{D}}^{-1}\ker h^*\dotplus\ker h^*\,.
  \end{split}
 \end{equation}
 \item[(iv)] The vector $h_{\mathrm{D}}^{-1}\Phi$, where $\Phi\in\ker h^*$ is given by \eqref{eq:IV-vector_Phi}-\eqref{eq:IV-vector_Phi_components}, satisfies the asymptotics
 \begin{equation}\label{SDPhi_vanishing}
  (h_{\mathrm{D}}^{-1}\Phi)(r)\;=\;\begin{pmatrix} p^+\! \\ p^-\!\end{pmatrix} r^B+o(r^{\frac{1}{2}})\qquad \textrm{as }\;r\downarrow 0\,,
 \end{equation}
\end{enumerate}
where  $p^{\pm}$ are both non-zero and explicitly given in \eqref{ed:defppm} below.
\end{proposition}

 \begin{remark}
  In Proposition \ref{prop:SD}(iv) a slightly more elaborate argument improves the estimate of the reminder in \eqref{SDPhi_vanishing} to $O(r^{1-B})$; however, this is not needed in the analysis that follows.
 \end{remark}


Last, an amount of information is needed on the domain of the operator closure $\overline{h}$ of $h$. 

\begin{proposition}\label{prop:Sclosure} Let $h$ be the operator on $L^2(\mathbb{R}^+,\mathbb{C}^2,\ud r)$ defined in \eqref{eq:IV-def_operator_S} with $|\nu|\in(\frac{\sqrt{3}}{2},1)$, then
\begin{equation}\label{eq:DhclosureequalH10}
	\mathcal{D}(\overline{h}) \;=\; H^1_0(\mathbb{R}^+,\mathbb{C}^2) \, .
\end{equation} 
In particular, any $f\in\mathcal{D}(\overline{h})$ satisfies the short-distance behaviour
\begin{equation}\label{eq:IV-f_vanishing}
 f(r)\;=\;o(r^{\frac{1}{2}})\qquad \textrm{as } \quad r\downarrow 0\,.
\end{equation}
\end{proposition}

With Propositions \ref{prop:kerS*}, \ref{prop:SD}, and \ref{prop:Sclosure} at hand, one arrives at the general self-adjoint extension classification as follows.

\begin{theorem}[Classification of the self-adjoint Dirac-Coulomb realisations]\label{thm:classification_structure} The self-adjoint extensions of the operator $h$ in $L^2(\mathbb{R}^+,\mathbb{C}^2,\ud r)$ defined in \eqref{eq:IV-def_operator_S} with $|\nu|\in(\frac{\sqrt{3}}{2},1)$ constitute a one-parameter family $(h_\beta)_{\beta\in\mathbb{R}\cup{\{\infty\}}}$ of restrictions of the adjoint operator $h^*$ determined in \eqref{DSclosureDS*}, each of which is given by
\begin{equation}\label{eq:IV-Sbeta}
 \begin{split}
  h_\beta\;&:=\;h^*\upharpoonright\mathcal{D}(h_\beta) \, , \\
  \mathcal{D}(h_\beta)\;&:=\;\left\{g=f+c(\beta h_{\mathrm{D}}^{-1}\Phi+\Phi)\left|\!
  \begin{array}{c}
   f\in H^1_0(\mathbb{R}^+,\mathbb{C}^2) \\
   c\in\mathbb{C}
  \end{array}\right.\!\!\right\}.
 \end{split}
\end{equation}
Here $h_{\mathrm{D}}$ is the distinguished self-adjoint extension of $h$ identified in Proposition \ref{prop:SD} and $\Phi$ is the spanning element of $\ker h^*$ identified in Proposition \ref{prop:kerS*}. In this parametrisation the distinguished extension $h_{\mathrm{D}}$ corresponds to $\beta=\infty$. For each $g\in\mathcal{D}(h_\beta)$ the function $f\in H^1_0(\mathbb{R}^+,\mathbb{C}^2)$ and the constant $c\in\mathbb{C}$ are uniquely determined. 
\end{theorem}

 \begin{proof} 
 As mentioned already, this is a direct application of Theorem \ref{thm:VB-representaton-theorem_Tversion2}.
One extension is the distinguished extension $h_{\mathrm{D}}$, with domain $\mathcal{D}(h_{\mathrm{D}})=\mathcal{D}(\overline{h})\dotplus h_{\mathrm{D}}^{-1}\ker h^*$ (Proposition \ref{prop:SD}(iii)), which is of the form \eqref{eq:IV-Sbeta} for $\beta=\infty$: with respect to the general formula \eqref{eq:ST-2}, this is the extension that corresponds to the Birman extension parameter\index{Birman extension parameter} $T$ defined on $\{0\}\subset\ker h^*$. Since $\dim\ker h^*=1$ (Proposition \ref{prop:kerS*}), for all other extensions of $h$ the corresponding Birman extension parameter, having to be self-adjoint on the whole one-dimensional $\mathrm{span}\{\Phi\}$, is necessarily the multiplication operator by a real scalar $\beta$. Then \eqref{eq:ST-2} takes precisely the form \eqref{eq:IV-Sbeta}. The uniqueness of the decomposition of $g\in\mathcal{D}(h_\beta)$ into $g\in\mathcal{D}(h_\beta)$ is a direct consequence of the direct sum decomposition \eqref{eq:IV-Krein_decomp_formula} of Proposition \ref{prop:SD}(iii) and the explicit characterisation of $\mathcal{D}(\overline{h})$ from Proposition \ref{prop:Sclosure}.
\end{proof}

 Paired to the parametrisation \eqref{eq:IV-Sbeta}, one can re-express the above classification in terms of local boundary conditions at the centre of the Coulomb singularity.

\begin{theorem}[Self-adjoint Dirac-Coulomb realisations via local boundary conditions]\label{thm:classification_bc}
 
\noindent
Let $h$ be the operator in $L^2(\mathbb{R}^+,\mathbb{C}^2,\ud r)$ defined in \eqref{eq:IV-def_operator_S} with $|\nu|\in(\frac{\sqrt{3}}{2},1)$.
\begin{enumerate}
 \item[(i)] Any function $g\equiv\begin{pmatrix} g^+ \\ g^-\end{pmatrix}\in\mathcal{D}(h^*)$ satisfies the short-distance asymptotics
 \begin{equation}\label{eq:IV-coeff_a_b}
 \begin{split}
  \lim_{r\downarrow 0} \,r^B g(r)\;&=:\;g_0 \\
  \lim_{r\downarrow 0} \,r^{-B}(g(r)-g_0r^{-B})\;&=:\;g_1
 \end{split}
 \end{equation}
 for some existing limiting values $g_0,g_1\in\mathbb{C}^2$.
 In particular,
 \begin{equation}\label{eq:IV-coeff_a_b_BIS}
  g(r)\;=\;g_0\, r^{-B}+g_1r^B+o(r^{\frac{1}{2}})\qquad\textrm{as }\;r\downarrow 0\,.
 \end{equation}
 \item[(ii)] The self-adjoint extensions of $h$ constitute a one-parameter family $(h_{\beta})_{\beta\in\mathbb{R}\cup{\{\infty\}}}$ of restrictions of the adjoint operator $h^*$, each of which is given by
\begin{equation}\label{eq:IV-Sbeta_bc}
 \begin{split}
  h_{\beta}\;&:=\;h^*\upharpoonright\mathcal{D}(h_{\beta}) \, , \\
  \mathcal{D}(h_{\beta})\;&:=\;\Big\{g\in\mathcal{D}(h^*)\,\Big|\,\frac{g_1^+}{g_0^+}=c_{\nu,\kappa} \beta+d_{\nu,\kappa}\Big\}\,,
 \end{split}
\end{equation}
where
\begin{equation}\label{eq:IV-defcd}
 \begin{split}
  c_{\nu,\kappa}\;&:=\;p^+\Big(\frac{\Gamma(2B)}{\Gamma(B)}\,\frac{\kappa+\nu+B}{\kappa+\nu}\Big)^{\!-1} \, , \\
  d_{\nu,\kappa}\;&:=\;q^+\Big(\frac{\Gamma(2B)}{\Gamma(B)}\,\frac{\kappa+\nu+B}{\kappa+\nu}\Big)^{\!-1},
 \end{split}
\end{equation}
and
\begin{eqnarray}
 q^\pm &:=&\frac{4^B(-B\pm(\kappa+\nu))\Gamma(-2B)}{(\kappa+\nu)\Gamma(-B)}\,, \label{eq:IV-def_qpm} \\
 p^{\pm}&:=& q^{\pm}\,\frac{(\kappa+\nu)\cos(B\pi)}{4^B B}\|\Phi\|_{L^2(\mathbb{R}^+,\mathbb{C}^2)}^2 \,. \label{ed:defppm}
\end{eqnarray}
This extension parametrisation is precisely the same as that of Theorem \ref{thm:classification_structure}.
\end{enumerate}
\end{theorem}

 \begin{proof} (i) It was determined in Propositions \ref{prop:kerS*}, \ref{prop:SD}, and \ref{prop:Sclosure} that a generic $g\in\mathcal{D}(h^*)$ decomposes with respect to \eqref{eq:IV-Krein_decomp_formula} as
\begin{equation}\label{eq:IV-decomp_g_DS*}
\begin{pmatrix} g^+ \\ g^-\end{pmatrix}\;=\;\begin{pmatrix} f^+ \\ f^-\end{pmatrix}+a\, h_{\mathrm{D}}^{-1}\begin{pmatrix} \Phi^+ \\ \Phi^-\end{pmatrix}+\frac{b}{\gamma}\begin{pmatrix} \Phi^+ \\ \Phi^-\end{pmatrix}\quad \gamma\,:=\,{\frac{\Gamma(2B)}{\Gamma(B)}\,\frac{\kappa+\nu+B}{\kappa+\nu}}
\end{equation}
for some $a,b\in\mathbb{C}$, and moreover, as $r\downarrow 0$,
\[
 \begin{split}
  f(r)\;&=\;o(r^{\frac{1}{2}})\,, \\
  (h_{\mathrm{D}}^{-1}\Phi)(r)\;&=\; \begin{pmatrix} p^+\! \\ p^-\!\end{pmatrix} r^B+o(r^{\frac{1}{2}})\,, \\
  r^B\Phi(r)\;&=\;\begin{pmatrix} 1 \\ -\frac{\kappa+\nu-B}{\kappa+\nu+B}\end{pmatrix}
  \gamma+\begin{pmatrix}
    q^+\! \\q^-\!
   \end{pmatrix} r^{2B}+o(r^{\frac{1}{2}+B})
 \end{split}
\]
(see, respectively, \eqref{eq:IV-f_vanishing}, \eqref{SDPhi_vanishing}, and \eqref{eq:IV-Phi_asymptotics} above). Therefore, the limit in the first component yields $r^Bg^+\!(r)\xrightarrow[]{r\downarrow 0}b$, and also
\[
 \begin{split}
  r^{-B}&(g^+\!(r)-br^{-B})\;=\;r^{-B}\big(f^+\!(r)+a(h_{\mathrm{D}}^{-1}\Phi)^+\!(r)+b\gamma^{-1}\Phi^+\!(r)-br^{-B}\big) \\
  &=\;a\,p^+ +b\,q^+\gamma^{-1} + o(r^{\frac{1}{2}-B})\,,
 \end{split}
\]
that is, $r^{-B}(g^+\!(r)-br^{-B})\xrightarrow[]{r\downarrow 0}a\,p^+ +b\,q^+\gamma^{-1}$. Thus, \eqref{eq:IV-coeff_a_b} and \eqref{eq:IV-coeff_a_b_BIS} follow by setting
\[\tag{*}\label{eq:IV-proof2}
 g_0^+\;:=\;b\,,\qquad g_1^+\;:=\;a p^+ + b q^+\gamma^{-1}
\]
%
(an analogous argument holds for the lower components).

(ii) Necessary and sufficient condition for $g\in\mathcal{D}(h^*)$ to belong to the domain $\mathcal{D}(h_\beta)$ of the extension $h_\beta$ determined by \eqref{eq:IV-Sbeta} of Theorem \ref{thm:classification_structure} is that in the decomposition \eqref{eq:IV-decomp_g_DS*} above the coefficients $a$ and $b$ satisfy $a=\beta b\gamma^{-1}$. Owing to \eqref{eq:IV-proof2} and \eqref{eq:IV-defcd}, the latter condition reads $g_1^+/g_0^+=c_{\nu,\kappa} \beta+d_{\nu,\kappa}$.
\end{proof}

  \begin{remark}
The proof of Theorem \ref{thm:classification_bc} shows that the decomposition of $g\in\mathcal{D}(h_\beta)$ determined by \eqref{eq:IV-Sbeta}, and hence $c$ and $f$, are explicitly given by
 \begin{equation}\label{eq:IV-comp_of_g}
 \begin{split}
  c\;&=\;(\textstyle{\frac{\Gamma(B)}{\Gamma(2B)}\,\frac{\kappa+\nu}{\kappa+\nu+B}})\cdot\displaystyle{\lim_{r\downarrow 0}} \;r^B g^+\!(r) \, ,\\
  f\;&=\;g-c(\beta h_{\mathrm{D}}^{-1}\Phi+\Phi)\,.
 \end{split}
 \end{equation}
 Indeed, in the notation of \eqref{eq:IV-decomp_g_DS*} therein, $b=\gamma c$. In fact, the same argument shows that the first equation in \eqref{eq:IV-comp_of_g} determines the component $c\Phi\in\ker h^*$ of a generic $g\in\mathcal{D}(h^*)$, and hence defines the (non-orthogonal) projection $\mathcal{D}(h^*)\to\ker h^*$, $g\mapsto c\Phi$ induced by the decomposition formula \eqref{eq:IV-Krein_decomp_formula}.
When $\beta\neq 0$, one has equivalently 
\begin{equation}\label{eq:IV-comp_of_g_alt}
 c\;=\;\beta^{-1}\!\!\int_0^{+\infty}\!\langle\Phi(r),(\widetilde{h}g)(r)\rangle_{\mathbb{C}^2}\,\ud r \,.
\end{equation}
Indeed $\widetilde{h}g=h_\beta g=h^*g=\overline{h}f+c\beta\Phi$ and $\ran\overline{h}\perp\ker h^*$, whence $\langle\Phi,\widetilde{h}g\rangle_{L^2(\mathbb{R}^3,\mathbb{C}^2)}=c\beta$. 
\end{remark}

 \begin{remark}
 As is typical when the operator to be self-adjointly realised is a differential operator, one interprets \eqref{eq:IV-Krein_decomp_formula} as the canonical decomposition of an element $g\in\mathcal{D}(h^*)$ into a `regular' and a `singular' part
 \begin{equation}
  \begin{split}
   g_{\mathrm{reg}}\;&:=\;f+a\, h_{\mathrm{D}}^{-1}\Phi\in\mathcal{D}(h_{\mathrm{D}}) \, ,\\
   g_{\mathrm{sing}}\;&:=\;\frac{b}{\gamma}\,\Phi\in\ker h^*\,,
  \end{split}
 \end{equation}
 where $a,b\in\mathbb{C}$ and $f\in\mathcal{D}(\overline{h})$ are determined by $g$ and $\gamma=\frac{\Gamma(2B)}{\Gamma(B)}\,\frac{\kappa+\nu-B}{\kappa+\nu}$. Indeed $\mathcal{D}(h_{\mathrm{D}})$ has a higher regularity then $\ker h^*$: functions in the former space vanish at zero, as follows from \eqref{SDPhi_vanishing}-\eqref{eq:IV-f_vanishing}, whereas $\Phi$ diverges at zero, as seen in \eqref{eq:IV-Phi_asymptotics}. In this language, $r^{-B}g_{\mathrm{reg}}^+(r)\xrightarrow[]{r\downarrow 0}a p^+$ and $r^{B}g_{\mathrm{sing}}^+(r)\xrightarrow[]{r\downarrow 0}b$, and the self-adjointness condition \eqref{eq:IV-Sbeta_bc} that selects, among the elements in $\mathcal{D}(h^*)$, only those in $\mathcal{D}(h_{\beta})$ reads
 \begin{equation}\label{eq:IV-bc-reg-sing}
  \Big(\frac{\gamma}{p^+}\lim_{r\downarrow 0}\;r^{-B}g_{\mathrm{reg}}^+(r)\Big) \;=\; (c_{\nu,\kappa} \beta+d_{\nu,\kappa})\,\Big(\lim_{r\downarrow 0}\;r^{B}g_{\mathrm{sing}}^+(r)\Big)\,,
 \end{equation}
 that is, the ratio between $\gamma(p^+\!)^{-1}$ times the coefficient of the leading vanishing term of $g_{\mathrm{reg}}^+$ and the coefficient of the leading divergent term of $g_{\mathrm{sing}}^+$ is indexed by the real extension parameter $\beta$.
\end{remark}

 One can further supplement the above picture with an additional amount of information concerning the invertibility, the resolvent, and the spectral gap of each realisation $h_\beta$, all features that are directly read out from the Kre{\u\i}n-Vi\v{s}ik-Birman scheme (Sect.~\ref{sec:II-spectralKVB}). The simple proof of this last result is deferred to Section \ref{sec:resolvent}.

\begin{theorem}[Invertibility, resolvent, and estimate on the spectral gap]\label{thm:DC-invertibility-resolvent-gap}

 The elements of the family $(h_\beta)_{\beta\in\mathbb{R}\cup{\{\infty\}}}$ of the self-adjoint extensions of the operator $h$ in $L^2(\mathbb{R}^+,\mathbb{C}^2)$ defined in \eqref{eq:IV-def_operator_S} with $|\nu|\in(\frac{\sqrt{3}}{2},1)$, labelled according to the parametrisation of Theorem \ref{thm:classification_structure}, have the following properties.
\begin{enumerate}
 \item[(i)] $h_\beta$ is invertible on the whole $L^2(\mathbb{R}^+,\mathbb{C}^2)$ if and only if $\beta\neq 0$.
 \item[(ii)] For each invertible extension $h_\beta$,
 \begin{equation}\label{eq:IV-Sbeta-1}
  h_\beta^{-1}\;=\;h_{\mathrm{D}}^{-1}+\frac{1}{\,\beta\|\Phi\|_{L^2}^{2}}\:|\Phi\rangle\langle\Phi|\,.
 \end{equation}
 \item[(iii)] For each invertible extension $h_\beta$,
 \begin{equation}\label{eq:IV-sigmaess}
  \sigma_{\mathrm{ess}}(h_\beta)\;=\;\sigma_{\mathrm{ess}}(h_{\mathrm{D}})\;=\;(-\infty,-1]\cup[1,+\infty)\,,
 \end{equation}
 and the gap in the spectrum $\sigma(h_\beta)$ around $E=0$ is at least the interval $(-E(\beta),E(\beta))$, where
 \begin{equation}\label{eq:estbetagap}
  E(\beta)\;:=\;\frac{|\beta|}{\,|\beta| \|h_{\mathrm{D}}^{-1}\|_{\mathrm{op}}+1\,}\,.
 \end{equation}
\end{enumerate}
\end{theorem}

 \section{Extension mechanism on each symmetry block at criticality}\label{sec:IV-proofs}

  Propositions \ref{prop:deficiency_indices}, \ref{prop:kerS*}, \ref{prop:SD}, and \ref{prop:Sclosure}, as well as Theorem \ref{thm:DC-invertibility-resolvent-gap} are now proved, following the general scheme of Sections \ref{sec:II-VBreparametrised}-\ref{sec:II-resolventsKVB}. Throughout this Section, $h$ is the operator on $L^2(\mathbb{R}^+,\mathbb{C}^2)$ defined in \eqref{eq:IV-def_operator_S} with $|\nu|\in(\frac{\sqrt{3}}{2},1)$ (critical regime).

  \subsection{Deficiency index computation}\label{sec:IVdefindex}

 \begin{proof}[Proof of Proposition \ref{prop:deficiency_indices}]
  (i) The essential self-adjointness or the magnitude of the deficiency indices of $h_{m_j,\kappa_j}$ remain unaltered by considering instead 
  \[
   h_{m_j,\kappa_j}-\sigma_3\;=\;\begin{pmatrix}
                   \frac{\nu}{r} & -\frac{\ud}{\ud r}+\frac{\kappa_j}{r} \\
                   \frac{\ud}{\ud r}+\frac{\kappa_j}{r} & \frac{\nu}{r}
                  \end{pmatrix}
  \]
  (on the same domain $C^\infty_0(\mathbb{R}^+,\mathbb{C}^2)$ as for $h_{m_j,\kappa_j}$), as this amounts to perturbing $h_{m_j,\kappa_j}$ with a bounded and hence infinitesimally smaller operator (Sect.~\ref{sec:perturbation-spectra}).

  In turn, in view of the general scheme of Section \ref{sec:WeylsCriterion}, the square-integrability or lack thereof of the solutions to the differential problem
  \[
   \begin{pmatrix}
                   \frac{\nu}{r} & -\frac{\ud}{\ud r}+\frac{\kappa_j}{r} \\
                   \frac{\ud}{\ud r}+\frac{\kappa_j}{r} & \frac{\nu}{r}
                  \end{pmatrix}u\;=\;\lambda u
  \]
  on half-line can be monitored by simply taking $\lambda=0$, in which case two linearly independent solutions are 
  \[\tag{*}\label{eq:IVlplcproof}
   \begin{pmatrix}  \sqrt{\kappa_j^2-\nu^2}-\kappa_j \\
\nu \end{pmatrix} r^{ \sqrt{\kappa_j^2-\nu^2}}\qquad\textrm{ and }\qquad \begin{pmatrix} - \sqrt{\kappa_j^2-\nu^2}-\kappa_j \\
\nu \end{pmatrix} r^{- \sqrt{\kappa_j^2-\nu^2}}
  \]
  if $\kappa_j^2\neq\nu^2$, and
  \[\tag{**}\label{eq:IVlplcproof2}
   \begin{pmatrix} 1 \\ \displaystyle-\frac{\kappa_j}{\nu} \end{pmatrix}\qquad\textrm{ and }\qquad \begin{pmatrix} \ln r \\ \displaystyle-\frac{1}{\nu}-\frac{\kappa_j}{\nu}\ln r \end{pmatrix}
  \]
  if $\kappa_j^2=\nu^2$.

  Based on this information, one applies the Weyl criterion\index{Weyl criterion!limit-point limit-circle}\index{Weyl limit-point/limit-circle}\index{limit-point/limit-circle}\index{theorem!Weyl (limit-point limit-circle criterion)} (Sect.~\ref{sec:WeylsCriterion}) as follows 
  When $\kappa_j^2=\nu^2$ both solutions \eqref{eq:IVlplcproof2} are square-integrable near zero and fail to be so at infinity, meaning that $h_{m_j,\kappa_j}$ is in the limit-circle case at the origin and in the limit-point case at infinity, and therefore has deficiency indices $(1,1)$ (Sect.~\ref{sec:WeylsCriterion}).

  When instead $\kappa_j^2\neq\nu^2$, at infinity one or both solutions \eqref{eq:IVlplcproof} are not square-integrable (due the exponential divergence, or to oscillatory behaviour), meaning that $h_{m_j,\kappa_j}$ is in the limit-point case at infinity; near zero, if $\nu^2>\kappa_j^2-\frac{1}{4}$, then both solutions \eqref{eq:IVlplcproof} are square-integrable and $h_{m_j,\kappa_j}$ is in the limit-circle case at the origin and therefore has deficiency indices $(1,1)$; last, if $\nu^2\leqslant \kappa_j^2-\frac{1}{4}$, then one of \eqref{eq:IVlplcproof} fails to be square-integrable, whereas the other is, and $h_{m_j,\kappa_j}$ is in the limit-point case at the origin, and therefore is essentially self-adjoint.

  This completes the proof of the essential self-adjointness of $h_{m_j,\kappa_j}$ if and only if $\nu^2\leqslant \kappa_j^2-\frac{1}{4}$, and hence establishes part (i). Additionally, it shows that when $\nu^2>\kappa_j^2-\frac{1}{4}$ the operator $h_{m_j,\kappa_j}$ has deficiency indices $(1,1)$.

  (ii) Under the condition $n^2-\frac{1}{4}<\nu^2\leqslant(n+1)^2-\frac{1}{4}$ each block operator $h_{m_j,\kappa_j}$ with $|\kappa_j|\in\{1,\dots,n\}$ has deficiency indices $(1,1)$, because then $\kappa_j^2-\frac{1}{4}\leqslant n^2-\frac{1}{4}<\nu^2$, all other blocks being essentially self-adjoint, because $|\kappa_j|\in\{n+1,n+2,\dots\}$ implies $\kappa_j^2-\frac{1}{4}\geqslant (n+1)^2-\frac{1}{4}\geqslant\nu^2$. Either conclusion is based on the above findings from part (i). Each non-self-adjoint block contributes with one unit to the total deficiency indices $d_+(H)=d_-(H)=:d(H)$ of $H$ (Sect.~\ref{sec:I-symmetric-selfadj}), so it remains to count the multiplicity of such blocks.

  One sees from \eqref{eq:IV-Dirac_operator_decomposition} that, since $\kappa_j=\pm(j+\frac{1}{2})$ for some $j\in\frac{1}{2}+\mathbb{N}_0$, the contributing indices $|\kappa_j|\in\{1,\dots,n\}$ are labelled by the $n$ semi-integers $j\in\{\frac{1}{2},\dots,n-\frac{1}{2}\}$. For each such $j$ one has blocks $h_{m_j,\kappa_j}$ labelled by $m_j\in\{-j,-j+1,\dots,j\}$, which give $2j+1$ possibilities, and by $\kappa_j=\pm(j+\frac{1}{2})$, which give $2$ possibilities. Therefore,
  \[
   \begin{split}
       d(H)\;&=\;\sum_{j=\frac{1}{2}}^{n-\frac{1}{2}}\;\;\sum_{m_j=-j}^j\;\sum_{\kappa_j=\pm(j+\frac{1}{2})} 1\;=\;\sum_{j=\frac{1}{2}}^{n-\frac{1}{2}}\;2(2j+1) \\
       &=\;\sum_{j=1}^{n}2\big(2(j-{\textstyle\frac{1}{2}})+1\big)\;=\;4\sum_{j=1}^{n}j\;=\;2n(n+1)\,.
   \end{split}
  \]
  Part (ii) is thus proved.
 \end{proof}

 \subsection{The homogeneous problem: kernel of $h^*$.}\label{sec:kernel}

 As emerges in general from the proof of Theorem \ref{thm:VB-representaton-theorem-GENER}, the fact (Theorem \ref{thm:recap}(ii)) that the minimal operator $H$ defined in \eqref{eq:IV-minimalH} admits an infinity of self-adjoint extensions implies that $H^*$ has a non-trivial kernel. On the other hand, from the operator orthogonal sum\index{operator orthogonal sum} \eqref{eq:IV-Dirac_operator_decomposition} that expresses $H$ (up to isomorphism), and from the standard fact that in an operator orthogonal sum the adjoint of the sum is the sum of the adjoints (Sect.~\ref{sec:I_invariant-reducing-ssp}), one must conclude that some block operators $h_{m_j,\kappa_j}$ -- in fact, precisely the non-self-adjoint ones -- have adjoint with non-trivial kernel. Thus, the operator $h$ under consideration has non-trivial adjoint. Its kernel is determined by the $L^2$-solutions $u$ to the homogeneous differential problem $\widetilde{h}u=0$, where $\widetilde{h}$ is the differential operator \eqref{eq:IV-tildeS} (and owing to \eqref{DSclosureDS*}).

 The tools for such ordinary differential equations are standard, and applied here below. It is worth remarking that along the alternative extension scheme of $h$ a la von Neumann, one has to solve the differential problem $\widetilde{h}u=z u$ for non-real $z\in\mathbb{C}$, say, $z=\pm\ii$, which requires a somewhat more extended discussion -- see, e.g., \cite[Section 3]{Voronov-Gitman-Tyutin-TMP2007} or \cite[Sections 4-6]{Hogreve-2013_JPhysA}.

 As usual, one writes the spinorial unknown $u$ as $u(r)=\begin{pmatrix} u^+\!(r) \\ u^-\!(r) \end{pmatrix}$, $r\in\mathbb{R}^+$. Upon transforming $u$ into $\varphi$ with
\begin{equation}\label{eq:IV-u-varphi}
 \varphi(r)\;:=\;{\textstyle\frac{1}{2}}(\mathbf{A}u)({\textstyle\frac{r}{2}})\,e^{r/2}\,,\qquad \mathbf{A}:=\frac{1}{\sqrt{2}}\begin{pmatrix} 1 & 1 \\ 1 & -1 \end{pmatrix},
\end{equation}
the differential system $\widetilde{h}u=0$ takes the form
\begin{equation}\label{eq:IV-diff_varphi}
 \begin{cases}
  \,(\varphi^+\!)'\,=\,\varphi^+\!-\frac{\kappa-\nu}{r}\,\varphi^- \, ,\\
  \,(\varphi^-\!)'\,=\,-\frac{\kappa+\nu}{r}\,\varphi^+\,.
 \end{cases}
\end{equation}
Therefore, $\varphi^-$ is a solution to
\begin{equation}\label{eq:IV-confl}
 r(\varphi^-\!)''+(1-r)(\varphi^-\!)'-\textstyle{\frac{\nu^2-\kappa^2}{r}}\,\varphi^-\!\;=\;0\,,
\end{equation}
equivalently,
\begin{equation}\label{eq:IV-xi-phi}
 \xi(r)\,:=\,r^B\varphi^-\!(r)
\end{equation}
is a solution to
\begin{equation}\label{eq:IV-ODE_hyper}
 r\xi''+(1-2B-r)\xi'+B\,\xi\;=\;0\,.
\end{equation}
The second order ordinary differential equation \eqref{eq:IV-ODE_hyper} is the confluent hypergeometric equation\index{confluent hypergeometric equation} \cite[(13.1.1) and (13.1.11)]{Abramowitz-Stegun-1964}, two linearly independent solutions of which are the confluent hypergeometric functions\index{confluent hypergeometric functions} of first and second kind, that is, respectively, the Kummer function\index{Kummer functions} $M_{a,b}(r)$ \cite[(13.1.2)]{Abramowitz-Stegun-1964} and the Tricomi function\index{Tricomi functions} $U_{a,b}(r)$ \cite[(13.1.3)]{Abramowitz-Stegun-1964}, with $a=-B$ and $b=1-2B$.

The solutions
\begin{equation}
 \xi_0(r):=M_{-B,1-2B}(r)\,,\qquad \xi_\infty(r):=U_{-B,1-2B}(r)
\end{equation}
to \eqref{eq:IV-ODE_hyper} determine, via \eqref{eq:IV-xi-phi} and the second of \eqref{eq:IV-diff_varphi}, two linearly independent solutions $\varphi_0=\begin{pmatrix}\varphi_0^+\! \\ \varphi_0^-\! \end{pmatrix}$ and $\varphi_\infty=\begin{pmatrix}\varphi_\infty^+\! \\ \varphi_\infty^-\! \end{pmatrix}$ to \eqref{eq:IV-diff_varphi}. Using the properties \cite[(13.4.8) and (13.4.21)]{Abramowitz-Stegun-1964}
\[
M_{a,b}'(r)\,=\,\frac{a}{b}\,M_{a+1,b+1}(r)\,,\qquad U_{a,b}'(r)\,=\,-a\,U_{a+1,b+1}(r)\,,
\]
and the inverse transformation of \eqref{eq:IV-u-varphi}, that is, $u(r)=2\,e^{-r/2}(\mathbf{A}^{-1}\varphi)(2r)$, where $\mathbf{A}^{-1}=\mathbf{A}$, yields the following two linearly independent solutions to the original problem $\widetilde{h}u=0$:
\begin{equation}\label{eq:IV-solutions_u0_uinf}
\begin{split}
\!\!\!\!\!\!u_0(r)&:=\frac{1}{e^{r}r^{B}}\!\begin{pmatrix}
 \textstyle{\frac{\kappa+\nu+B}{\kappa+\nu}}\,M_{-B,1-2B}(2r)+\textstyle{\frac{2rB}{(\kappa+\nu)(1-2B)}}\,M_{1-B,2-2B}(2r) \\
 \textstyle{-\frac{\kappa+\nu-B}{\kappa+\nu}}\,M_{-B,1-2B}(2r)+\textstyle{\frac{2rB}{(\kappa+\nu)(1-2B)}}\,M_{1-B,2-2B}(2r)
                               \end{pmatrix}, \\
 \!\!\!\!\!\!\!\!u_\infty(r)&:=\frac{1}{e^{r}r^{B}}\!\begin{pmatrix}
 \textstyle{\frac{\kappa+\nu+B}{\kappa+\nu}}\,U_{-B,1-2B}(2r)-\textstyle{\frac{2rB}{\kappa+\nu}}\,U_{1-B,2-2B}(2r) \\
 \textstyle{-\frac{\kappa+\nu-B}{\kappa+\nu}}\,U_{-B,1-2B}(2r)-\textstyle{\frac{2rB}{\kappa+\nu}}\,U_{1-B,2-2B}(2r)
                               \end{pmatrix}
\end{split}
\end{equation}
(in fact, an irrelevant common pre-factor $2^{-B}$ has been neglected). Both $u_0$ and $u_\infty$ are real-valued and smooth on $\mathbb{R}^+$.

Because of the asymptotics \cite[(13.1.2), (13.5.1), and (13.5.5)]{Abramowitz-Stegun-1964}
\begin{equation}\label{eq:asy1}
 \begin{split}
  M_{a,b}(r)\;&\stackrel{r\to+\infty}{=} \;\frac{\,e^r\,r^{a-b}\,}{\Gamma(a)}\,(1+O(r^{-1}))\,, \\
  M_{a,b}(r)\;&\stackrel{r\downarrow 0}{=}\;1+O(r)\qquad\textrm{with } b\notin\mathbb{N}\,,
 \end{split}
\end{equation}
and \cite[(13.1.2), (13.1.3), (13.5.2), (13.5.8), and (13.5.10)]{Abramowitz-Stegun-1964}
\begin{equation}\label{eq:asy2}
 \begin{split}
  U_{a,b}(r)\;&\stackrel{r\to+\infty}{=}\;r^{-a}(1+O(r^{-1}))\,, \\
  U_{a,b}(r)\;&\stackrel{r\downarrow 0}{=}\;\frac{\Gamma(1-b)}{\Gamma(1+a-b)}+\frac{\Gamma(b-1)}{\Gamma(a)}\,r^{1-b}+O(r)\quad \textrm{with }b\in(0,1)\,,
   \\
  U_{a,b}(r)\;&\stackrel{r\downarrow 0}{=}\;\frac{\Gamma(b-1)}{\Gamma(a)}\,r^{-(b-1)}+O(1)\qquad\qquad\qquad\;\textrm{with }b\in(1,2)\,,
 \end{split}
\end{equation}
one deduces that both $u_0$ and $u_\infty$ are square-integrable around $r=0$, whereas only $u_\infty$ is square-integrable at infinity, and moreover
\begin{equation}\label{eq:IV-asymptotics_for_u0}
 \begin{split}
  u_0(r)\;&\stackrel{r\downarrow 0}{=}\;\begin{pmatrix}
               \frac{\kappa+\nu+B}{\kappa+\nu} \\ -\frac{\kappa+\nu-B}{\kappa+\nu}
              \end{pmatrix} r^{-B} +O(r^{1-B})\,,\\
  u_0(r)\;&\stackrel{r\to+\infty}{=}\;-\textstyle{\frac{2^B(1-2B)}{\Gamma(-B)(\kappa+\nu)}}\begin{pmatrix}
               1 \\ 1
              \end{pmatrix}r^Be^r (1+O(r^{-1}))\,,
 \end{split}
\end{equation}
and
\begin{equation}\label{eq:IV-asymptotics_for_uinfty}
 \begin{split}
  u_\infty(r)\;&\stackrel{r\downarrow 0}{=}\;\textstyle{\frac{\Gamma(2B)}{\Gamma(B)}}\begin{pmatrix}
               \frac{\kappa+\nu-B}{\kappa+\nu} \\ -\frac{\kappa+\nu+B}{\kappa+\nu}
              \end{pmatrix} r^{-B} +\begin{pmatrix} q^+\! \\ q^-\! \end{pmatrix} r^B +O(r^{1-B})\,,\\
  u_\infty(r)\;&\stackrel{r\to+\infty}{=}\;2^B\begin{pmatrix}
               1 \\ -1
              \end{pmatrix}r^{-B}e^{-r}
              (1+O(r^{-1}))\,,
 \end{split}
\end{equation}
where
\begin{equation}
 q^\pm\;:=\;\frac{4^B(-B\pm(\kappa+\nu))\Gamma(-2B)}{(\kappa+\nu)\Gamma(-B)}\,.
\end{equation}
Observe that $q^\pm\neq 0$.

Therefore, there is only a \emph{one}-dimensional space of solutions to $\widetilde{h}u=0$ which are square-integrable, and hence $\ker h^*$ is one-dimensional. For convenience, one chooses as the spanning vector the function $\Phi:=u_\infty$. Then \eqref{eq:IV-asymptotics_for_uinfty} implies \eqref{eq:IV-Phi_asymptotics} and Proposition \ref{prop:kerS*} is proved.

\subsection{Distinguished extension $h_{\mathrm{D}}$}\label{sec:distinguished}
 The characterisation of the distinguished extension\index{distinguished Dirac-Coulomb Hamiltonian} of $h$ declared in Proposition \ref{prop:SD} goes through the differential problem $\widetilde{h}f=g$ in the unknown $f$ for given $g\in L^2(\mathbb{R}^+,\mathbb{C})$, so as to determine the domain of invertibility of the differential operator $\widetilde{h}$. The strategy here is an adaptation to the present first order differential operator with Coulomb singularity of the analogous problem for homogeneous Schr\"{o}dinger operators of Bessel type\index{Bessel operator} on half-line, a subject that is well classical \cite{Bruneau-Derezinski-Georgescu-2011,Derezinski-Georgescu-2021} and which is also encountered, in a different context, in Sections \ref{sec:homo}, \ref{sec:ess-self-adj} and \ref{sec:zero_mode}.

In order to set up the problem conveniently, it is beneficial first to replace the pair $(u_0,u_\infty)$ of linearly independent solutions \eqref{eq:IV-solutions_u0_uinf} to $\widetilde{h}u=0$ to the new pair $(v_0,v_\infty)$ given by
\begin{equation}\label{eq:IV-v0_vinf}
 \begin{split}
  v_0\;&:=u_\infty-{ \frac{\Gamma(2B)}{\Gamma(B)}}\,u_0\,, \\
  v_\infty\;&:=\;u_\infty \,.
 \end{split}
\end{equation}
This preserves the linear independence of $v_0$ and $v_\infty$ with the virtue of having two solutions with different power-law in the asymptotics as $r\downarrow 0$: from \eqref{eq:IV-v0_vinf} and \eqref{eq:IV-asymptotics_for_u0}-\eqref{eq:IV-asymptotics_for_uinfty} one finds 
\begin{equation}\label{eq:IV-asymptotics_for_v0vinf}
\begin{array}{l}
 \begin{split}
 v_0(r)\;&=\;\begin{pmatrix} q^+\! \\ q^-\! \end{pmatrix} r^B+O(r^{1-B})\\
 v_\infty(r)\;&=\;{\frac{\Gamma(2B)}{\Gamma(B)}}\begin{pmatrix}
               \frac{\kappa+\nu+B}{\kappa+\nu} \\ -\frac{\kappa+\nu-B}{\kappa+\nu}
              \end{pmatrix} r^{-B} +O(r^B)
 \end{split}
\end{array}\qquad\textrm{as } r\downarrow 0\,,
\end{equation}
where $q^\pm$ is given by \eqref{eq:IV-def_qpm}. At large distances, $v_0$ and $v_\infty$ have exponential 
asymptotics as $u_0$ and $u_\infty,$ namely
\begin{equation}\label{eq:IV-asymptotics_for_v0vinf_large_distances}
\begin{array}{l}
 \begin{split}
 v_0(r)\;&=\;-{\frac{1}{2} \,\frac{2^{B} B}{(\kappa+\nu) \cos(B \pi)}} \begin{pmatrix}
               1 \\ 1
              \end{pmatrix}r^B e^r(1+O(r^{-1}))\\
 v_\infty(r)\;&=\;2^B\begin{pmatrix}
               1 \\ -1
              \end{pmatrix}r^{-B} e^{-r} 
              (1+O(r^{-1}))
 \end{split}
\end{array}\;\textrm{as } r\to+\infty\,.
\end{equation}

 With respect to the fundamental system\index{fundamental system for an O.D.E.} $(v_0,v_\infty)$, the general solution to the inhomogeneous problem $\widetilde{h}f=g$ has the form
\begin{equation}\label{eq:IV-ODE-decomp}
 f\;=\;A_0v_0+A_\infty v_\infty+f_\mathrm{part}\,,
\end{equation}
where $A_0$ and $A_\infty$ run over $\mathbb{C}$ and $f_\mathrm{part}$ is a particular solution, namely, $\widetilde{h}f_\mathrm{part}=g$.

 $f_\mathrm{part}$ is now determined through standard variation of constants\index{variation of constants} \cite[Section 2.4]{Wasow_asympt_expansions}. First, one re-writes $\widetilde{h}f=g$ in normal form as
\begin{equation}\label{eq:IV-normal_form}
 y'+\mathbf{V}(r)y\,=\,g\,,\qquad y\,:=\,-\ii\sigma_2 f\,,
\end{equation}
where
\begin{equation}\label{eq:IV-normal_form_potential}
 \mathbf{V}(r)\,:=\frac{1}{r}
 \begin{pmatrix}
 -\kappa & \nu \\
 -\nu & \kappa 
 \end{pmatrix}+
 \begin{pmatrix}
  0 & 1 \\ 1 & 0
 \end{pmatrix}\,.
\end{equation}
 One also introduces the Wronskian\index{Wronskian}
\begin{equation}
 \mathbb{R}^+\ni r\mapsto W_r(v_0,v_\infty)\;:=\;\det\begin{pmatrix}
                             v_0^+(r) & v_\infty^+(r) \\
                             v_0^-(r) & v_\infty^-(r)
                            \end{pmatrix}.
\end{equation}
This is precisely the Wronskian\index{Wronskian} $ W_r(-\ii\sigma_2v_0,-\ii\sigma_2v_\infty)$ of two fundamental solutions of the problem written in normal form, because $\det(-\ii\sigma_2)=1$ and hence $ W_r(-\ii\sigma_2v_0,-\ii\sigma_2v_\infty)=W_r(v_0,v_\infty)$. Moreover, since $\mathbf{V}(r)$ is trace-less for any $r\in\mathbb{R}^+$, Liouville's theorem\index{theorem!Liouville} implies that $W_r(\ii\sigma_2 v_0,\ii\sigma_2 v_\infty)$ is constant, and so is also $W_r(v_0,v_\infty)$. Therefore,
\begin{equation}\label{eq:IV-computation_of_wronskian}
\begin{split}
 W_r(v_0,v_\infty)\;&=\;\lim_{r\downarrow 0}W_r(v_0,v_\infty)\;=\; {\frac{4^B B}{(\kappa+\nu)\cos(B \pi)}} \;=:\;W_0^\infty\,.
\end{split}
\end{equation}
The limit in \eqref{eq:IV-computation_of_wronskian} above follows straightforwardly from the asymptotics \eqref{eq:IV-asymptotics_for_v0vinf} and from the expression \eqref{eq:IV-def_qpm} for $q^\pm$. Clearly, $W_0^\infty\neq 0$. Then through variation of constants for the differential problem \eqref{eq:IV-normal_form}, and with the further transformation $f=-\ii\sigma_2y$, one finds eventually
\begin{equation}
 f_\mathrm{part}(r)\;=\;\int_0^{+\infty}\!\!G(r,\rho)\,g(\rho)\,\ud \rho\,,
\end{equation}
where
\begin{equation}\label{eq:IV-Green}
 G(r,\rho)\;:=\; \frac{1}{\:W_0^\infty}
 \begin{cases}
 \begin{pmatrix}
                    v^+_\infty(r) v^+_0(\rho) & v^+_\infty(r) v^-_0(\rho) \\
                    v^-_\infty(r) v^+_0(\rho) & v^-_\infty(r) v^-_0(\rho)
                   \end{pmatrix}, & \textrm{if } 0<\rho<r\,, \\
                   \\
  \begin{pmatrix}
                    v^+_0(r) v^+_\infty(\rho) & v^+_0(r) v^-_\infty(\rho) \\
                    v^-_0(r) v^+_\infty(\rho) & v^-_0(r) v^-_\infty(\rho)
                   \end{pmatrix}, & \textrm{if } 0<r<\rho\,.
 \end{cases}
\end{equation}

Next, one observes the following.

\begin{lemma}\label{lem:IV-RG_bounded}
 The integral operator $R_G$ on $L^2(\mathbb{R}^+,\mathbb{C},\ud r)$ with kernel $G(r,\rho)$ given by \eqref{eq:IV-Green} is bounded and self-adjoint.
\end{lemma}

\begin{proof}
 For each $r,\rho\in\mathbb{R}^+$, $G(r,\rho)$ is the sum of the four terms
 \begin{equation}\label{eq:IV-four_term_G}
 \begin{split}
  G^{++}(r,\rho)\;&:=\;G(r,\rho)\,\mathbf{1}_{(1,+\infty)}(r)\,\mathbf{1}_{(1,+\infty)}(\rho)\,, \\
  G^{+-}(r,\rho)\;&:=\;G(r,\rho)\,\mathbf{1}_{(1,+\infty)}(r)\,\mathbf{1}_{(0,1)}(\rho)\,, \\
  G^{-+}(r,\rho)\;&:=\;G(r,\rho)\,\mathbf{1}_{(0,1)}(r)\,\mathbf{1}_{(1,+\infty)}(\rho)\,, \\
  G^{--}(r,\rho)\;&:=\;G(r,\rho)\,\mathbf{1}_{(0,1)}(r)\,\mathbf{1}_{(0,1)}(\rho)\,, 
 \end{split}
 \end{equation}
where $\mathbf{1}_J$ denotes the characteristic function of the interval $J\subset\mathbb{R}^+$, and correspondingly $R_G$ splits into the sum of four integral operators with kernel given by \eqref{eq:IV-four_term_G}.

Now, for each entry of $G^{LM}(r,\rho)$, with $L,M\in\{+,-\}$, a point-wise estimate in $(r,\rho)$ can be derived from the short and large distance asymptotics for $v_0$ and $v_\infty$. For example, the entry $G^{++}_{11}(r,\rho)$ in the first row and first column of $G^{++}(r,\rho)$ is controlled as
\[
 \begin{split}
  |v^+_\infty(r) \,v^+_0(\rho)\,\mathbf{1}_{(1,+\infty)}(r)\,\mathbf{1}_{(1,+\infty)}(\rho)|\;&\lesssim\; e^{-r} \,e^\rho \, (\rho/r)^B\,,\qquad\textrm{if }\,0<\rho<r\,, \\
  |v^+_0(r)\, v^+_\infty(\rho)\,\mathbf{1}_{(1,+\infty)}(r)\,\mathbf{1}_{(1,+\infty)}(\rho)|\;&\lesssim\; e^r\,e^{-\rho}\, (r/\rho)^B\,, \qquad\textrm{if }\,0<r<\rho\,,
 \end{split}
\]
because $v_0$ diverges exponentially and $v_\infty$ vanishes exponentially as $r\to+\infty$, \eqref{eq:IV-asymptotics_for_v0vinf_large_distances}; thus,
\[
 |G^{++}_{11}(r,\rho)|\;\lesssim\;e^{-|r-\rho|}\,(\rho r)^B \,.
\]
In fact, the asymptotics for $v_0$ and $v_\infty$ are the same for both components, so one can also conclude that 
\[
 \|G^{++}(r,\rho)\|_{M_2(\mathbb{C})}\;\lesssim\;e^{-|r-\rho|}\, (\rho r)^B\, ,
\]
where $\|\cdot\|_{M_2(\mathbb{C})}$ denotes the matrix norm. The estimate of the other kernels is perfectly analogous, and one finds
\begin{equation}\label{eq:IV-matrix_estimates}
 \begin{split}
  \|G^{++}(r,\rho)\|_{M_2(\mathbb{C})}\;&\lesssim\;(r \rho)^B e^{-|r-\rho|}\,\mathbf{1}_{(1,+\infty)}(r)\,\mathbf{1}_{(1,+\infty)}(\rho)\,, \\
  \|G^{+-}(r,\rho)\|_{M_2(\mathbb{C})}\;&\lesssim\;r^Be^{-\rho}\,\mathbf{1}_{(1,+\infty)}(r)\,\mathbf{1}_{(0,1)}(\rho)\,, \\
  \|G^{-+}(r,\rho)\|_{M_2(\mathbb{C})}\;&\lesssim\;e^{-r}\rho^B\,\mathbf{1}_{(0,1)}(r)\,\mathbf{1}_{(1,+\infty)}(\rho)\,, \\
  \|G^{--}(r,\rho)\|_{M_2(\mathbb{C})}\;&\lesssim\;(r^B\rho^{-B}+r^{-B}\rho^B)\,\mathbf{1}_{(0,1)}(r)\,\mathbf{1}_{(0,1)}(\rho)\,.
 \end{split}
\end{equation}

The last three estimates in \eqref{eq:IV-matrix_estimates} show at once that the kernels $G^{+-}(r,\rho)$, $G^{-+}(r,\rho)$, and $G^{--}(r,\rho)$ are in $L^2(\mathbb{R}^+\times\mathbb{R}^+,M_2(\mathbb{C}),\ud r\,\ud\rho)$ and therefore the corresponding integral operators are Hilbert-Schmidt operators (Sect. \ref{sec:bdd-compacts-unitaries-orthproj}), hence bounded, on $L^2(\mathbb{R}^+,\mathbb{C}^2,\ud r)$. The first estimate in \eqref{eq:IV-matrix_estimates} allows to conclude, by an obvious Schur test,\index{Schur test} that also the integral operator with kernel $G^{++}(r,\rho)$ is bounded on $L^2(\mathbb{R}^+,\mathbb{C}^2,\ud r)$. This proves the overall boundedness of $R_G$.

The self-adjointness of $R_G$ is clear from \eqref{eq:IV-Green}: the adjoint $R_G^*$ of $R_G$ has kernel $\overline{G(\rho,r)^T}$, but $G$ is real-valued and $G(\rho,r)=G(r,\rho)$, thus proving that $R_G^*=R_G$.
\end{proof}

The integral operator $R_G$ has a relevant mapping property that is more directly read out from the following alternative representation.
If $f_\mathrm{part}=R_G\,g$, then
\begin{equation}\label{eq:IV-fpart_alternative_repr}
 f_\mathrm{part}(r)\;=\;\Theta_\infty^{(g)}(r)\,v_0(r)+\Theta_0^{(g)}(r)\,v_\infty(r)\,,
\end{equation}
where
\begin{equation}\label{eq:IV-fpart_alternative_repr-Thetas}
 \begin{split}
  \Theta_0^{(g)}(r)\;&:=\;\frac{1}{\:W_0^\infty}\int_0^r\langle{\,\overline{v_0(\rho)}}\,,\,g(\rho)\,\rangle_{\mathbb{C}^2}\,\ud \rho\,, \\
    \Theta_\infty^{(g)}(r)\;&:=\;\frac{1}{\:W_0^\infty}\int_r^{+\infty}\!\langle{\,\overline{v_\infty(\rho)}}\,,\,g(\rho)\,\rangle_{\mathbb{C}^2}\,\ud \rho\,,
 \end{split}
\end{equation}
and $W_0^\infty(\neq 0)$ is the constant computed in \eqref{eq:IV-computation_of_wronskian}. Indeed, from \eqref{eq:IV-Green},
\[
\begin{split}
 f_\mathrm{part}(r)\;&=\;\int_0^{+\infty}\!G(r,\rho)\,g(\rho)\,\ud \rho \\
 &=\;\frac{1}{W_0^{\infty}}\begin{pmatrix} 
 v_\infty^+(r) \\ v_\infty^-(r)
 \end{pmatrix}\int_0^r\big(v_0^+(\rho)g^+(\rho)+v_0^-(\rho)g^-(\rho)\big)\,\ud \rho \\
 &\qquad +\frac{1}{W_0^{\infty}}\begin{pmatrix} 
 v_0^+(r) \\ v_0^-(r)
 \end{pmatrix}\int_r^{+\infty}\!\!\big(v_\infty^+(\rho)g^+(\rho)+v_\infty^-(\rho)g^-(\rho)\big)\,\ud \rho\,,
\end{split}
\]
that is, \eqref{eq:IV-fpart_alternative_repr}.

\begin{lemma}\label{RG_maps_in_energyform}
For every $g\in L^2(\mathbb{R}^+,\mathbb{C}^2)$ one has
\begin{equation}\label{eq:IV-Rgg_finite_pot_energy}
 \int_0^{+\infty}\frac{\|(R_G\,g)(r)\|^2_{\mathbb{C}^2}}{r}\,\ud r\;<\;+\infty\,,
\end{equation}
i.e.,
\begin{equation}
 \mathrm{ran}\,R_G\;\subset\;\mathcal{D}[r^{-1}]\,.
\end{equation}
\end{lemma}

\begin{proof}
Clearly, $\int_1^{+\infty}r^{-1}\|(R_G\,g)(r)\|^2_{\mathbb{C}^2}\,\ud r\leqslant\|R_G\|_{\mathrm{op}}^{2}\|g\|^2_{L^2(\mathbb{R}^+,\mathbb{C}^+)}$. It then remains to prove the finiteness of the integral in \eqref{eq:IV-Rgg_finite_pot_energy} only for $r \in (0,1)$.
One represents $f=R_G\,g\in\mathrm{ran}\,R_G$ as in \eqref{eq:IV-fpart_alternative_repr}-\eqref{eq:IV-fpart_alternative_repr-Thetas}. For $r\in(0,1)$ one has
\[
 \begin{split}
   |\Theta_0^{(g)}(r)|\;&\leqslant\;|W_0^\infty|^{-1}\,\|v_0\,\mathbf{1}_{(0,r)}\|_{L^2(\mathbb{R}^+,\mathbb{C}^2)}\,\|g\|_{L^2(\mathbb{R}^+,\mathbb{C}^2)}\;\leqslant\;C_{g,\nu}\,r^{B+\frac{1}{2}}\,,  \\
   |\Theta_\infty^{(g)}(r)|\;&\leqslant\;|W_0^\infty|^{-1}\,\|v_\infty\,\mathbf{1}_{(r,\infty)}\|_{L^2(\mathbb{R}^+,\mathbb{C}^2)}\,\|g\|_{L^2(\mathbb{R}^+,\mathbb{C}^2)}\;\leqslant\;C_{g,\nu}
 \end{split}
\]
for some constant $C_{g,\nu}>0$ depending on $g$ and $\nu$ only, having used the short distance asymptotics \eqref{eq:IV-asymptotics_for_v0vinf} for $v_0$ and $v_\infty$.
Combining now the above bounds again with \eqref{eq:IV-asymptotics_for_v0vinf} one sees that on the interval $(0,1)$ the functions $r\mapsto \Theta_0^{(g)}(r)\,v_\infty(r)$ and $r\mapsto \Theta_\infty^{(g)}(r)\,v_0(r)$ are continuous and vanish when $r\to 0$, respectively, as $r^{1/2}$ and $r^B$, which makes the integral $\int_0^{1}r^{-1}\|(R_G\,g)(r)\|^2_{\mathbb{C}^2}\,\ud r$ finite.
\end{proof}

Combining Lemmas \ref{lem:IV-RG_bounded} and \ref{RG_maps_in_energyform} together, Proposition \ref{prop:SD} can be now proved.

\begin{proof}[Proof of Proposition \ref{prop:SD}]

Concerning parts (i) and (ii), the integral operator $R_G$ on $L^2(\mathbb{R}^+,\mathbb{C})$ with kernel given by \eqref{eq:IV-Green} is bounded and self-adjoint (Lemma \ref{lem:IV-RG_bounded}), and by construction satisfies $\widetilde{h}\,R_G\,g=g$ $\forall g\in L^2(\mathbb{R}^+,\mathbb{C})$. 
Therefore, $R_G g=0$ for some $g\in L^2(\mathbb{R}^+)$ implies $g=0$, i.e., $R_G$ is injective. Then $R_G$ has dense range ($(\mathrm{ran}\,R_G)^\perp=\ker R_G$). 
 As a consequence (Sect.~\ref{sec:I-adjoint}), $(R_{G}^{-1})^*=(R_{G}^*)^{-1}=R_{G}^{-1}$, that is, $\mathscr{S}:=R_{G}^{-1}$ is self-adjoint. One thus has $R_G=\mathscr{S}^{-1}$ and from the identity $h^*R_G=\mathbbm{1}$ on $L^2(\mathbb{R}^+)$ one deduces that for any $f\in\mathcal{D}(\mathscr{S})$, say, $f=R_G g=\mathscr{S}^{-1} g$ for some $g\in L^2(\mathbb{R}^+)$, the identity $h^*f=\mathscr{S}f$ holds. This means that $h^*\supset\mathscr{S}$, whence also $\overline{S}=S^{**}\subset\mathscr{S}$, i.e., $\mathscr{S}$ is a self-adjoint extension of $S$.

%

Because of Lemma \ref{RG_maps_in_energyform}, the space $\mathcal{D}(\mathscr{S})=\mathrm{ran}\,R_G$ is contained in the potential energy form domain $\mathcal{D}[r^{-1}]$: owing to Theorem \ref{thm:recap}(ii) then $\mathscr{S}$ must be the reduction to the subspace $\cH_{\pm \frac{1}{2},\pm 1}$ of the distinguished\index{distinguished Dirac-Coulomb Hamiltonian} self-adjoint extension of the Dirac-Coulomb operator $H$: it shall be denoted with $h_{\mathrm{D}}$. As such, $h_{\mathrm{D}}$ is the unique self-adjoint realisation of $h$ satisfying the property \eqref{eq:IV-SD_uniqueness_properties}, it is invertible, and its kernel is precisely given by \eqref{eq:IV-Green}.


 Concerning part (iii), the decompositions \eqref{eq:IV-decomp_DSD} and \eqref{eq:IV-Krein_decomp_formula} are canonical (Proposition \ref{prop:II-KVB-decomp-of-Sstar}), once a self-adjoint extension $h_{\mathrm{D}}$ of $h$ is available with everywhere defined and bounded inverse.

 Last, concerning part (iv), one recalls from the previous discussion that $\Phi=u_\infty=v_\infty$ and $h_{\mathrm{D}}^{-1}\Phi=R_Gv_\infty$. A closed expression for the latter function is given by \eqref{eq:IV-fpart_alternative_repr} above, which now reads
\[
 (h_{\mathrm{D}}^{-1}\Phi)(r)\;=\;\Theta_\infty^{(v_\infty)}(r)\,v_0(r)+\Theta_0^{(v_\infty)}(r)\,v_\infty(r)\,.
\]
From \eqref{eq:IV-fpart_alternative_repr-Thetas} and \eqref{eq:IV-asymptotics_for_v0vinf} one deduces
\[
 \begin{split}
  |\Theta_0^{(v_\infty)}(r)|\;&\leqslant\;|W_0^\infty|^{-1}\int_0^r\big|\langle{\,\overline{v_0(\rho)}}\,,\,v_\infty(\rho)\,\rangle_{\mathbb{C}^2}\big|\,\ud \rho \\
  &\lesssim\;\int_0^r(\rho^B+O(\rho^{1-B}))(\rho^{-B}+O(\rho^{B}))\,\ud \rho \\
  &=\;r+o(r)\quad\textrm{as }\;r\downarrow 0
 \end{split}
\]
and 
 \[
 \begin{split}
  \Theta_\infty^{(v_\infty)}(r)\;&=\;\frac{1}{\,W_0^\infty}\int_r^{+\infty}\langle{\,\overline{v_\infty(\rho)}}\,,\,v_\infty(\rho)\,\rangle_{\mathbb{C}^2}\,\ud \rho \\
  &=\;\frac{1}{\,W_0^\infty}\,\|v_\infty\|_{L^2(\mathbb{R}^+,\mathbb{C}^2)}^2(1+o(1))\quad\textrm{as }\;r\downarrow 0\,.
 \end{split}
\]
Therefore, using again the short distance asymptotics \eqref{eq:IV-asymptotics_for_v0vinf},
\[\tag{*}\label{eq:IV-anotherstar}
 \begin{split}
  (h_{\mathrm{D}}^{-1}\Phi)(r)\;=\;\frac{\|v_\infty\|_{L^2(\mathbb{R}^+,\mathbb{C}^2)}^2}{\,W_0^\infty}\,
  \begin{pmatrix}
   q^+\! \\ q^-\!
  \end{pmatrix}
  r^{B}+o(r^{B})
 \end{split}
\]
where $q^{\pm}$ is given by \eqref{eq:IV-def_qpm}.
Upon setting
\begin{equation}
 p^{\pm}\;:=\; q^{\pm}\,(W_0^\infty)^{-1}\|v_\infty\|_{L^2(\mathbb{R}^+,\mathbb{C}^2)}^2 
\end{equation}
 one then obtains the leading term of \eqref{SDPhi_vanishing}. The remainder is in fact smaller than $o(r^B)$. This can be seen by comparing the above asymptotics for $h_{\mathrm{D}}^{-1}\Phi$ with the expression \eqref{eq:IV-decomp_a0ainf_b0binf} established in Lemma \ref{lem:IV-decomp_a0ainf_b0binf} (which is applicable, because $h_{\mathrm{D}}^{-1}\Phi\in\mathcal{D}(h_{\mathrm{D}})\subset\mathcal{D}(h^*)$), namely
\[
 h_{\mathrm{D}}^{-1}\Phi\;=\;a_0^{(h_{\mathrm{D}}^{-1}\Phi)}\,v_0+a_\infty^{(h_{\mathrm{D}}^{-1}\Phi)}\,v_\infty+b_\infty^{(h_{\mathrm{D}}^{-1}\Phi)}\,v_0+ b_0^{(h_{\mathrm{D}}^{-1}\Phi)}\,v_\infty\,.
\]
For the latter, one has the asymptotics
\[\tag{**}\label{eq:IV-anotherstarstar}
\begin{split}
 \!\!(h_{\mathrm{D}}^{-1}\Phi)(r)\;&=\;\mathbf{c}_0 \,a_0^{(h_{\mathrm{D}}^{-1}\Phi)}\,(r^B+O(r^{1-B}))+\mathbf{c}_\infty\, a_\infty^{(h_{\mathrm{D}}^{-1}\Phi)}\,(r^{-B}+O(r^B)) \\
 &\qquad +o(r^{\frac{1}{2}})\quad\textrm{as }\;r\downarrow 0\,.
\end{split}
\]
as follows from \eqref{eq:IV-asymptotics_for_v0vinf} and \eqref{eq:IV-bvbg_asympt} for some non-zero constants $\mathbf{c}_0,\mathbf{c}_\infty\in\mathbb{C}^2$. In order for \eqref{eq:IV-anotherstar} and \eqref{eq:IV-anotherstarstar} to be compatible, necessarily $a_\infty^{(h_{\mathrm{D}}^{-1}\Phi)}=0$. This implies that after the leading order $r^B$ there comes a remainder $o(r^{\frac{1}{2}})$, thus yielding \eqref{SDPhi_vanishing} and completing the proof of part (iv).
\end{proof}

\subsection{Operator closure $\overline{h}$}\label{sec:closure}

 In order to establish Proposition \ref{prop:Sclosure} an actually stronger result is now proved for the characterisation of $\mathcal{D}(\overline{h})$, namely Proposition \ref{prop:characterisationDS} below. 
 

 To this aim, it is convenient to exploit yet another representation of $\mathcal{D}(h^*)$. 

\begin{lemma}\label{lem:IV-decomp_a0ainf_b0binf}
 For each $g\in\mathcal{D}(h^*)$ there exist, uniquely determined, complex constants $a_0^{(g)},a_\infty^{(g)}$ and functions
 \begin{equation}\label{eq:IV-b0binf}
  \begin{split}
   b_0^{(g)}(r)\;&:=\;\frac{1}{W_0^\infty}\int_0^r\langle\, \overline{v_0(\rho)}\,,\,(h^*\!g)(\rho)\,\rangle_{\mathbb{C}^2}\,\ud\rho\,, \\
   b_\infty^{(g)}(r)\;&:=\;-\frac{1}{W_0^\infty}\int_0^r\langle \,\overline{v_\infty(\rho)}\,,\,(h^*\!g)(\rho)\,\rangle_{\mathbb{C}^2}\,\ud\rho
  \end{split}
 \end{equation}
 on $\mathbb{R}^+$ such that
 \begin{equation}\label{eq:IV-decomp_a0ainf_b0binf}
  g\;=\;a_0^{(g)}\,v_0+a_\infty^{(g)}\,v_\infty+b_\infty^{(g)}\,v_0+ b_0^{(g)}\,v_\infty\,,
 \end{equation}
 where $v_0$ and $v_\infty$ are the two linearly independent solutions \eqref{eq:IV-v0_vinf} to the homogeneous  problem $\widetilde{h}v=0$ (recall that they are real and smooth on $\mathbb{R}^+$) and $W_0^\infty$ is the constant computed in \eqref{eq:IV-computation_of_wronskian}. Moreover, both $ b_0^{(g)}(r)$ and $b_\infty^{(g)}(r)$ vanish as $r\downarrow 0$, and
 \begin{equation}\label{eq:IV-bvbg_asympt}
  b_\infty^{(g)}(r)\,v_0(r)+ b_0^{(g)}(r)\,v_\infty(r)\;=\;o(r^{1/2})\qquad\textrm{as }\;r\downarrow 0\,.
 \end{equation}
\end{lemma}

\begin{proof}
Let $\psi:=h^*g=\widetilde{h}g$. Then, as already argued in \eqref{eq:IV-ODE-decomp} and \eqref{eq:IV-fpart_alternative_repr}-\eqref{eq:IV-fpart_alternative_repr-Thetas}, $g$ is expressed in terms of $\psi$ as
\[\tag{*}\label{eq:IV-new-star}
 g\;=\;A_0\,v_0+A_\infty\,v_\infty+\Theta^{(\psi)}_\infty\,v_0+ \Theta_0^{(\psi)}\,v_\infty
\]
for some $A_0,A_\infty\in\mathbb{C}$ that are now uniquely identified by $g$. From \eqref{eq:IV-fpart_alternative_repr-Thetas} and \eqref{eq:IV-b0binf} one sees that
\[
 \begin{split}
  \Theta_\infty^{(\psi)}(r)\;&=\;b_\infty^{(g)}(r)\,, \\
  \Theta_0^{(\psi)}(r)\;&=\;-\frac{1}{\:W_0^\infty}\int_0^r\langle{\,\overline{v_0(\rho)}}\,,\,(h^*\!g)(\rho)\,\rangle_{\mathbb{C}^2}\,\ud \rho \\
  &= \;b_0^{(g)}(r)-\frac{1}{\:W_0^\infty}\int_0^{+\infty}\!\!\langle{\,\overline{v_0(\rho)}}\,,\,(h^*\!g)(\rho)\,\rangle_{\mathbb{C}^2}\,\ud \rho\,.
 \end{split}
\]
Then \eqref{eq:IV-new-star} implies \eqref{eq:IV-decomp_a0ainf_b0binf} at once, upon setting 
\[
 \begin{split}
  a_0^{(g)}\;&:=\;A_0-\frac{1}{\:W_0^\infty}\int_0^{+\infty}\!\!\langle{\,\overline{v_0(\rho)}}\,,\,(h^*\!g)(\rho)\,\rangle_{\mathbb{C}^2}\,\ud \rho\,, \\
  a_\infty^{(g)}\;&:=\;A_\infty\,.
 \end{split}
\]
Observe that the constant added above to $A_0$ is finite and bounded by
\[
 |W_0^\infty|^{-1}\|v_0\|_{L^2(\mathbb{R}^+,\mathbb{C}^2)}\|h^*\!g\|_{L^2(\mathbb{R}^+,\mathbb{C}^2)}\,.
\]
As for the proof of \eqref{eq:IV-bvbg_asympt}, by means of the short distance asymptotics \eqref{eq:IV-asymptotics_for_v0vinf} for $v_0$ and $v_\infty$ one finds
\[
 \begin{split}
  |b_\infty^{(g)}(r)\,v_0(r)|\;&\lesssim\;r^{-B}\!\int_0^{r}\rho^B\|g(\rho)\|_{\mathbb{C}^2}\,\ud \rho\;\leqslant\;\int_0^{r}\|g(\rho)\|_{\mathbb{C}^2}\,\ud \rho \\
  &\;\leqslant\;r^{\frac{1}{2}}\,\|g\|_{L^2([0,r],\mathbb{C}^2)}\;=\;o(r^{\frac{1}{2}})
 \end{split}
\]
and 
\[
 \begin{split}
  |b_0^{(g)}(r)\,v_\infty(r)|\;&\lesssim\;r^{B}\!\int_0^{r}\rho^{-B}\|g(\rho)\|_{\mathbb{C}^2}\,\ud \rho\;\leqslant\;r^B\|\rho^{-B}\|_{L^2[0,r]}\|g\|_{L^2([0,r],\mathbb{C}^2)} \\
  &\;\lesssim\;r^{\frac{1}{2}}\,\|g\|_{L^2([0,r],\mathbb{C}^2)}\;=\;o(r^{\frac{1}{2}})\,,
 \end{split}
\]
and \eqref{eq:IV-bvbg_asympt} then follows.
\end{proof}

The next preparatory step is to introduce, for later convenience, the \emph{generalised Wronskian}\index{Wronskian} of any two square-integrable functions,
\begin{equation}\label{eq:IV-def_wronskian}
 \mathbb{R}^+\ni r\mapsto W_r(\psi,\phi)\;:=\;\det
 \begin{pmatrix}
  \psi^{+}(r) & \phi^+(r) \\
  \psi^{-}(r) & \phi^-(r)
 \end{pmatrix},\quad\psi,\phi\in L^2(\mathbb{R}^+,\mathbb{C}^2)\,,
\end{equation}
and the \emph{boundary form}\index{boundary form} for any two functions in $\mathcal{D}(h^*)$,
\begin{equation}\label{eq:IV-boundary_form}
 \omega(g,f)\;:=\;\langle h^*g,f\rangle-\langle g,h^*f\rangle\,,\qquad f,g\in\mathcal{D}(h^*)\,.
\end{equation}
The boundary form is anti-symmetric, i.e.,
\begin{equation}\label{eq:IV-omega_antisymm}
 \omega(f,g)\;=\;-\overline{\omega(f,g)}\,,
\end{equation}
and it is related to the Wronskian\index{Wronskian}  by
\begin{equation}\label{eq:IV-omegaWr}
 \omega(g,f)\;=\;-\lim_{r\downarrow 0}W_r(\overline{g},f)\,.
\end{equation}
Indeed, using $\widetilde{h}=-\ii\sigma_2\,{\textstyle\frac{\ud}{\ud r}}+\mathbf{V}(r)$ from \eqref{eq:IV-normal_form}-\eqref{eq:IV-normal_form_potential}, one has
\[
\begin{split}
 \omega(g,f)\;&=\;\int_0^{+\infty}\!\!\ud r\,\big( \langle (\widetilde{h}g)(r),f(r)\rangle_{\mathbb{C}^2}-\langle g(r),(\widetilde{h}f)(r) \rangle_{\mathbb{C}^2} \big) \\
 &=\;\int_0^{+\infty}\!\!\ud r\,\big( \langle -\ii\sigma_2g'(r),f(r)\rangle_{\mathbb{C}^2}-\langle g(r),-\ii\sigma_2 f'(r) \rangle_{\mathbb{C}^2} \big) \\
 &=\;\lim_{r\downarrow 0}\,\big(\,\overline{g^-(r)}\,f^+(r)-\overline{g^+(r)}\,f^-(r)\big)\;=\;-\lim_{r\downarrow 0}W_r(\overline{g},f)\,.
\end{split}
\]

It is also convenient to introduce the (two-dimensional) space of solutions to the differential problem $\widetilde{h}v=0$,
\begin{equation}
 \mathcal{L}\;:=\;\{v:\mathbb{R}^+\to\mathbb{C}^2\,|\,\widetilde{h}v=0\}\;=\;\mathrm{span}\{v_0,v_\infty\}\,,
\end{equation}
As well known, $r\mapsto W_r(u,v)$ is constant whenever $u,v\in\mathcal{L}$, and this constant is zero if and only if $u$ and $v$ are linearly dependent. It will be important also to keep into account that any $v\in\mathcal{L}$ is square-integrable around $r=0$, as determined in \eqref{eq:IV-asymptotics_for_v0vinf}.

\begin{lemma}\label{lem:IV-Lv_vanishes_DS}
For given $v\in\mathcal{L}$, 
\begin{equation}\label{eq:IV-def_Lv}
 \begin{split}
  & L_v:\mathcal{D}(h^*)\to\mathbb{C} \\
  & \;\;\quad\qquad g\longmapsto L_v(g)\;:=\;\lim_{r\downarrow 0}W_r(\overline{v},g)
 \end{split}
\end{equation}
defines a linear functional on $\mathcal{D}(h^*)$ which vanishes on $\mathcal{D}(\overline{h})$.
\end{lemma}

\begin{proof}
The linearity of $L_v$ is obvious, and the finiteness of $L_v(g)$ for $g\in\mathcal{D}(h^*)$ is checked as follows.  

Decompose $g=a_0^{(g)}\,v_0+a_\infty^{(g)}\,v_\infty+b_\infty^{(g)}\,v_0+ b_0^{(g)}\,v_\infty$ as in \eqref{eq:IV-decomp_a0ainf_b0binf} and $v=c_0v_0+c_\infty v_\infty$ in the basis of $\mathcal{L}$. Owing to \eqref{eq:IV-def_Lv}, it suffices to control the finiteness of $L_{v_0}(g)$ and $L_{v_\infty}(g)$. By linearity,
\[
 L_{v_0}(g)\;=\;a_0^{(g)}\,L_{v_0}(v_0)+a_\infty^{(g)}\,L_{v_0}(v_\infty)+L_{v_0}(b_\infty^{(g)}\,v_0+ b_0^{(g)}\,v_\infty)\,.
\]
Moreover,
\[
 \begin{split}
  L_{v_0}(v_0)\;&=\;\lim_{r\downarrow 0}W_r(v_0,v_0)\;=\;0\,, \\
  L_{v_0}(v_\infty)\;&=\;W_0^{\infty}\,, \\
  L_{v_0}(b_\infty^{(g)}\,v_0)\;&=\;\lim_{r\downarrow 0}\,W_r(v_0,b_\infty^{(g)}\,v_0)\;=\;\lim_{r\downarrow 0}\,b_\infty^{(g)}(r)W_r(v_0,v_0)\;=\;0\,, \\
  L_{v_0}(b_0^{(g)}\,v_\infty)\;&=\;\lim_{r\downarrow 0}\,W_r(v_0,b_0^{(g)}\,v_\infty)\;=\;\lim_{r\downarrow 0}\,b_0^{(g)}(r)W_r(v_0,v_\infty)\;=\;0\,.
 \end{split}
\]
The conclusion is $L_{v_0}(g)=a_\infty^{(g)}W_0^{\infty}$. Analogously, $L_{v_\infty}(g)=-a_0^{(g)}W_0^{\infty}$, and this establishes the finiteness of $L_v(g)$.

 Next, it is now proved that if $f\in\mathcal{D}(\overline{h})$, then $L_v(f)=0$. Let $\chi\in C^\infty_0([0,+\infty))$ be such that $\chi(r)=1$ for $r\in[0,\frac{1}{2}]$ and $\chi(r)=0$ for $r\in[1,+\infty)$. One has that $v\chi\in\mathcal{D}(h^*)$, indeed $v\chi\in L^2(\mathbb{R}^+,\mathbb{C}^2)$ and
\[
 \begin{split}
  \widetilde{h}(v\chi)\;&=\;(-\ii\sigma_2\,{\textstyle\frac{\ud}{\ud r}}+\mathbf{V}(r))v\chi\;=\;\chi(-\ii\sigma_2\,{\textstyle\frac{\ud}{\ud r}}+\mathbf{V}(r))v-\ii\sigma_2v\chi' \\
  &=\;(\widetilde{h}v)\chi-\ii\sigma_2v\chi'\;=\;-\ii\sigma_2v\chi'\;\in\;L^2(\mathbb{R}^+,\mathbb{C}^2)\,,
 \end{split}
\]
where the identities $\widetilde{h}=-\ii\sigma_2\,{\textstyle\frac{\ud}{\ud r}}+\mathbf{V}(r)$ and $\widetilde{h}v=0$ were used. Moreover, because of the behaviour of $\chi$ around $r=0$, the Wronskians $W_r(\overline{v\chi},g)$ and $W_r(\overline{v},g)$ are asymptotically equal as $r\downarrow 0$, that is, $ L_{v\chi}=L_{v}$
As a consequence of this latter fact and  of \eqref{eq:IV-omegaWr},
\[
 \begin{split}
  L_{v}(f)\;&=\;L_{v\chi}(f)\;=\;\lim_{r\downarrow 0}W_r(\overline{v\chi},f)\;=\;-\omega(v\chi,f) \\
  &=\;\langle v\chi,h^*f\rangle-\langle h^*(v\chi),f\rangle\;=\;\langle v\chi,\overline{h}f\rangle-\langle v\chi,\overline{h}f\rangle\;=\;0\,,
 \end{split}
\]
which completes the proof.
\end{proof}

 The preparations above allow to characterise the operator closure's domain $\mathcal{D}(\overline{h})$ in an efficient way.

\begin{proposition}\label{prop:characterisationDS}
 Let $f\in\mathcal{D}(h^*)$. The following conditions are equivalent:
 \begin{enumerate}
  \item[(i)] $f\in\mathcal{D}(\overline{h})$;
  \item[(ii)] $\omega(f,g)=0$ for all $g\in\mathcal{D}(h^*)$;
  \item[(iii)] $L_v(f)=0$ for all $v\in\mathcal{L}$;
  \item[(iv)] with respect to the decomposition \eqref{eq:IV-decomp_a0ainf_b0binf} for $f$, $a_0^{(f)}=a_\infty^{(f)}=0$.
 \end{enumerate}
\end{proposition}

\begin{proof} The implication (i)$\Rightarrow$(ii) follows at once from
\[
 \omega(f,g)\;=\;\langle h^*f,g\rangle-\langle f,h^*g\rangle\;=\;\langle\overline{h}f,g\rangle-\langle\overline{h}f,g\rangle\;=\;0\,.
\]
For the converse implication (ii)$\Rightarrow$(i), one observes that
\[
 0\;=\;\omega(f,g)\;=\;\langle h^*f,g\rangle-\langle f,h^*g\rangle\qquad\forall g\in\mathcal{D}(h^*)
\]
is equivalent to $\langle h^*f,g\rangle=\langle f,h^*g\rangle$ $\forall g\in\mathcal{D}(h^*)$, which implies that $f\in\mathcal{D}(h^{**})=\mathcal{D}(\overline{h})$.

The implication (i)$\Rightarrow$(iii) is given by Lemma \ref{lem:IV-Lv_vanishes_DS}. For the converse, assume that $L_v(f)=0$ for all $v\in\mathcal{L}$: the goal is now to show that $\omega(f,g)=0$ for all $g\in\mathcal{D}(h^*)$: indeed, since the equivalence (i)$\Leftrightarrow$(ii) is already established, one then concludes that $f\in\mathcal{D}(\overline{h})$, and hence (iii)$\Rightarrow$(i). Owing to the decomposition \eqref{eq:IV-decomp_a0ainf_b0binf} for $g$,
\[
 \omega(f,g)\;=\;a_0^{(g)}\omega(f,v_0)+a_\infty^{(g)}\omega(f,v_\infty)+\omega(f,b_\infty^{(g)}\,v_0)+\omega(f,b_0^{(g)}\,v_\infty)\,.
\]
One has
\[
 \overline{\omega(f,v_0)}\;=\;-\omega(v_0,f)\;=\;\lim_{r\downarrow 0}W_r(\overline{v_0},f)\;=\;L_{v_0}(f)\;=\;0\,,
\]
having used \eqref{eq:IV-omega_antisymm} in the first step, \eqref{eq:IV-omegaWr} in the second, \eqref{eq:IV-def_Lv} in the third, and the assumption $L_v(f)=0$ for all $v\in\mathcal{L}$ in the last step. Analogously,
\[
 \overline{\omega(f,v_\infty)}\;=\;-\omega(v_\infty,f)\;=\;\lim_{r\downarrow 0}W_r(\overline{v_\infty},f)\;=\;L_{v_\infty}(f)\;=\;0\,.
\]
Therefore, $\omega(f,v_0)=\omega(f,v_\infty)=0$, and one is left with
\[
\begin{split}
 \overline{\omega(f,g)}\;&=\;\overline{\omega(f,b_\infty^{(g)}\,v_0)}+\overline{\omega(f,b_0^{(g)}\,v_\infty)} \;=\;-\omega(b_\infty^{(g)}\,v_0,f)-\omega(b_0^{(g)}\,v_\infty,f) \\
 &=\;\lim_{r\downarrow 0}\Big( W_r\big(\,\overline{b_\infty^{(g)}\,v_0}\,,f\big)+W_r\big(\overline{b_0^{(g)}\,v_\infty}\,,f\big)\Big) \\
 &=\;\lim_{r\downarrow 0}\Big( \,\overline{b_\infty^{(g)}(r)}\,W_r(\overline{v_0},f)+\overline{b_0^{(g)}(r)}\,W_r(\overline{v_\infty},f)\Big)\,.
\end{split}
\]
As $r\downarrow 0$, $W_r(\overline{v_0},f)\to L_{v_0}(f)=0$ and $W_r(\overline{v_\infty},f)\to  L_{v_\infty}(f)=0$, and also (as seen in Lemma \ref{lem:IV-decomp_a0ainf_b0binf}) $b_\infty^{(g)}(r)\to 0$ and $b_0^{(g)}(r)\to 0$, whence $\omega(f,g)=0$. This completes the proof of the implication (iii)$\Rightarrow$(i).

Last, in order to establish the equivalence (i)$\Leftrightarrow$(iv), decompose $f$ as in \eqref{eq:IV-decomp_a0ainf_b0binf}, namely,
\[
  f\;=\;a_0^{(f)}\,v_0+a_\infty^{(f)}\,v_\infty+b_\infty^{(f)}\,v_0+ b_0^{(f)}\,v_\infty\,.
\]
 A simple computation gives
\[
 \begin{split}
  L_{v_0}(f)\;&=\;\lim_{r\downarrow 0}W_r(\overline{v_0},f)\;=\;a_0^{(f)}\,\lim_{r\downarrow 0}W_r(\overline{v_0},v_0)+a_\infty^{(f)}\,\lim_{r\downarrow 0}W_r(\overline{v_0},v_\infty) \\
  &\qquad\qquad +\lim_{r\downarrow 0}b_\infty^{(f)}(r)\,W_r(\overline{v_0},v_0)+\lim_{r\downarrow 0}b_0^{(f)}(r)\,W_r(\overline{v_0},v_\infty) \\
  &\;=\;a_\infty^{(f)}\,W_0^{\infty}\,.
 \end{split}
\]
Indeed, $W_r(\overline{v_0},v_0)=W_r(v_0,v_0)=0$, and  $W_r(\overline{v_0},v_\infty)=W_r(v_0,v_\infty)\to W_0^{\infty}$, $b_\infty^{(f)}(r)\to 0$, and $b_0^{(f)}(r)\to 0$ as $r\downarrow 0$. Similarly,
\[
 L_{v_\infty}(f)\;=\;-a_0^{(f)}\,W_0^{\infty}\,.
\]
Because of the already proved equivalence (i)$\Leftrightarrow$(iii), one then concludes that $f\in\mathcal{D}(\overline{h})$ if and only if $L_{v_0}(f)=L_{v_\infty}(f)=0$, which from the above computation is tantamount as $a_0^{(f)}=a_\infty^{(f)}=0$. This completes the proof.
\end{proof}

\begin{proof}[Proof of Proposition \ref{prop:Sclosure}]
Owing to Proposition \ref{prop:characterisationDS}(iv), a generic $f\in\mathcal{D}(\overline{h})$ admits the unique representation $f=b_\infty^{(f)}\,v_0+ b_0^{(f)}\,v_\infty$. The asymptotics \eqref{eq:IV-bvbg_asympt} from Lemma \ref{lem:IV-decomp_a0ainf_b0binf} then implies $f(r)=o(r^{1/2})$ as $r \downarrow 0$, thus establishing \eqref{eq:IV-f_vanishing}.

Concerning the identity \eqref{eq:DhclosureequalH10}, for the inclusion $\mathcal{D}(\overline{h}) \subset H^1_0(\mathbb{R}^+,\mathbb{C}^2)$ one observes that, for arbitrary $f\in\mathcal{D}(\overline{h})$, the function $g:=\overline{h}f\in L^2(\mathbb{R}^+,\mathbb{C}^2)$ satisfies  
\[
	\begin{split}
		\frac{\;\ud f^-}{\ud r} \;&=\; - g^+ + \frac{1}{r} f^- + \left( 1 + \frac{\nu}{r} \right) f^+ \, ,\\
		\frac{\;\ud f^+}{\ud r} \;&=\; g^+-\frac{1}{r} f^+ - \left(\frac{\nu}{r}-1 \right) f^- \, .
	\end{split}
\]
Proposition \ref{prop:characterisationDS} and the short-distance asymptotics of $f$ ensure that all the terms in the r.h.s.~are square-integrable, whence $\frac{\; \ud f^{\pm}}{\ud r} \in L^2(\mathbb{R}^+)$. Thus, $f \in H^1(\mathbb{R}^+,\mathbb{C}^2)$, and moreover $f\in H^1_0(\mathbb{R}^+,\mathbb{C}^2)=\{u \in H^1(\mathbb{R}^+,\mathbb{C}^2) \, |\, u(0)=0\}$, as a consequence of \eqref{eq:IV-f_vanishing}.

For the opposite inclusion, let $f \in H^1_0(\mathbb{R}^+,\mathbb{C}^2)$. Then $f(0)=0$, $f' \in L^2(\mathbb{R}^+,\mathbb{C}^2)$, and 
\[
	|f(r)|\;\leq\; \int_0^r |f'(\rho)| \, \ud \rho \;\leq\; r^{\frac{1}{2}} \Vert f'\Vert_{L^2([0,r],\mathbb{C}^2)} \;=\; o(r^{\frac{1}{2}})\qquad\textrm{as }r\downarrow 0 \, .
\]
Together with the square-integrability of $f$ and $f'$, the latter inequality implies that $\widetilde{h} f \in L^2(\mathbb{R}^+,\mathbb{C}^2)$ and therefore, by definition of maximal domain, $f \in \mathcal{D}(h^*)$. Then (Lemma \ref{lem:IV-decomp_a0ainf_b0binf}) $f$ undergoes the decomposition \eqref{eq:IV-decomp_a0ainf_b0binf} which, due to the short-scale asymptotics of $v_0$ and $v_\infty$ in \eqref{eq:IV-asymptotics_for_v0vinf}, is compatible with $f(r)=o(r^{\frac{1}{2}})$ if and only if $a_0^{(f)}=a_\infty^{(f)}=0$. From Proposition \ref{prop:characterisationDS} one concludes that $f \in \mathcal{D}(\overline{h})$, thus proving the inclusion $H^1_0(\mathbb{R}^+,\mathbb{C}^2)\subset\mathcal{D}(\overline{h})$. 
\end{proof}

 \begin{remark}
  It is worth observing that \cite[Eq.~(67)]{Voronov-Gitman-Tyutin-TMP2007} expresses the vanishing rate of elements of what is here the space $\mathcal{D}(\overline{h})$ only as $O(r^{\frac{1}{2}})$, as $r\downarrow 0$, whereas it is proved here that the actual vanishing rate is $o(r^{\frac{1}{2}})$. In fact, the $o(r^{\frac{1}{2}})$-rate was previously mentioned, but not proved, already in \cite[Section 2]{Burnap-Brysk-Zweifel-NuovoCimento1981}.
 \end{remark}

 \begin{remark}
  Continuing the above-mentioned parallel between the present analysis for $\mathcal{D}(\overline{h})$ and the analogous problem for a homogeneous Schr\"{o}dinger operator of Bessel type\index{Bessel operator} on half-line, say, $\mathfrak{b}=-\frac{\ud^2}{\ud r^2}+\nu r^{-2}$, $\nu>-\frac{1}{4}$, it is worth remarking the following analogy with the classical literature on the latter subject:  in \cite[Proposition 4.17]{Bruneau-Derezinski-Georgescu-2011} the family of self-adjoint realisations of $\mathfrak{b}$ is shown to be a collection $(h_{\theta})_{\theta\in[0,2\pi)}$ where $\mathcal{D}(h_\theta)$ is formed by elements that as $r\downarrow 0$ have the form
\[
 f+c(r^{\frac{1}{2}-m}\cos\theta+r^{\frac{1}{2}+m}\sin\theta)
\]
for some $c\in\mathbb{C}$ and some function $f$ with $f(r)\sim r^{-3/2}$ as $r\downarrow 0$, where $m:=\sqrt{\nu+\frac{1}{4}}$\,. In fact, one would say in the present language that also in that case it is possible to identify a `distinguished' extension in the regime $\nu>-\frac{1}{4}$, the one with $\theta=\frac{\pi}{2}$, which has the property that $\mathcal{D}(h_{\theta=\pi/2})\subset\mathcal{D}[r^{-2}]$. 
 \end{remark}

\subsection{Resolvents and spectral gap}\label{sec:resolvent}

%
%

\begin{proof}[Proof of Theorem \ref{thm:DC-invertibility-resolvent-gap}]
 Part (i) is an immediate consequence of Theorem \ref{thm:invertibility-gen}, since the Birman extension parameter\index{Birman extension parameter} in the present case is the multiplication by $\beta$. (This is consistent with the representation formula \eqref{eq:IV-Sbeta}, which clearly implies that when $\beta=0$ the extension $S_{\beta=0}$ has a kernel.)

 Part (ii) is an immediate consequence of Theorem \ref{thm:res-gen}(i), because the orthogonal projection $P_T$ has in the present case the expression $P_T=\|\Phi\|_{L^2(\mathbb{R}^+,\mathbb{C}^2)}^{-2}|\Phi\rangle\langle\Phi|$.

 Concerning part (iii), \eqref{eq:IV-sigmaess} is a consequence (Sect.~\ref{sec:perturbation-spectra}) of the fact that the resolvent difference \eqref{eq:IV-Sbeta-1} between the $\beta$-extension and the distinguished extension is compact. Moreover, using \eqref{eq:IV-Sbeta-1} one re-writes $h_\beta f = Ef$ as
\begin{equation*}
f=E \,h_\beta^{-1} f=E\,\Big(h_{D}^{-1} +\frac{1}{\;\beta\|\Phi\|_{L^2(\mathbb{R}^+,\mathbb{C}^2)}^2}\:|\Phi \rangle \langle \Phi |\Big)f\,.
\end{equation*}
This equation is surely solved by $f=0$ and, if $E\in(-E(\beta),E(\beta))$, then the operator on the r.h.s.~is a contraction. Thus $f=0$ is the only function which satisfies the eigenvalue equation $h_\beta f = Ef$ and therefore there cannot be eigenvalues in such a regime of $E$.
\end{proof}

 \section{Sommerfeld formula and distinguished extension}\label{sec:Somm}

 Prior to determining, in the forthcoming Section \ref{sec:IV-betaspectrum}, the spectrum of the generic self-adjoint extension $h_\beta$ characterised in Theorems \ref{thm:classification_structure} and \ref{thm:classification_bc}, it is instructive to revisit the classical derivation of the Sommerfeld fine structure formula\index{Sommerfeld fine structure formula} in order to spot at which precise point the reasoning selects the spectrum of one particular self-adjoint realisation, which turns out to be precisely the distinguished\index{distinguished Dirac-Coulomb Hamiltonian} $h_\mathrm{D}$ (and actually also one mirror version of it), and instead is not able to identify the spectra (hence modified Sommerfeld formulas) for the other self-adjoint realisations.

  As a matter of fact, formula \eqref{eq:Somm-form} predates the identification of the extension family $(h_\beta)_{\beta\in\mathbb{R}\cup\{\infty\}}$, has a remarkable agreement with the experimental measurement of the energy eigenvalues of the Hydrogen atom, and reproduces, in the \emph{non-relativistic limit},\index{non-relativistic limit} the discrete spectrum of the Schr\"{o}dinger-Coulomb model for Hydrogen\index{Hydrogenoid spectrum} (see \eqref{eq:spectrum_hydrogeonid} from Chapter \ref{chapter-Hydrogenoid}), that is (taking in \eqref{eq:Somm-form} $|\kappa|=1$ and setting $E_n\equiv E_{n,\kappa}$),
  \[
 E_n-m_{\mathrm{e}} c^2\;=\;m_{\mathrm{e}} c^2\left(\bigg(1+\frac{(\frac{Z e^2}{\hslash c})^2}{\Big(n+\sqrt{1-(\frac{Z e^2}{\hslash c})^2}\Big)^2}\bigg)^{\!-\frac{1}{2}}-1\right)\;\xrightarrow[]{\:c\to +\infty\:}\;-\frac{Z^2 e^4 m_{\mathrm{e}}}{\,2\hslash^2(n+1)^2}\,.
\]

  Moreover, \eqref{eq:Somm-form} yields real eigenvalues not only in the sub-critical regime $|\nu|\leqslant\frac{\sqrt{3}}{2}$ (Theorem \ref{thm:recap}), but also in the critical regime $|\nu|\in(\frac{\sqrt{3}}{2},1)$ where an infinite multiplicity of distinct self-adjoint realisations is known to exist (with the physical $\nu$ given by $\nu=- Z\alpha_\mathrm{f}<0$), and only produces complex eigenvalues, thus indicating an instability of the atom, in the super-critical regime $|\nu|>1$ (the already-mentioned `$Z=137$ catastrophe'). \index{Z=137 catastrophe} At criticality it is therefore meaningful to inquire which discrete spectrum is now such formula referring to, and how it need be corrected for the spectra of the other self-adjoint Dirac-Coulomb realisations.

  Eventually, one wants to solve the eigenvalue problem
\begin{equation}\label{eq:EigenvalueEq}
h_\beta  \psi\;=\;E\psi\,,\qquad\psi\in\mathcal{D}(h_\beta)\,,\qquad E\in(-1,1)
\end{equation}
 for a generic extension $h_\beta$ provided by Theorem \ref{thm:classification_structure} in the regime $|\nu|\in(\frac{\sqrt{3}}{2},1)$, the associated differential problem to which being
 \begin{equation}\label{eq:diffprobnow}
  \widetilde{h}\psi\;=\;=E\psi\,,\qquad E\in(-1,1)\,.
 \end{equation}
 In this Section the focus is on the classical approaches to \eqref{eq:diffprobnow} and why such methods produce the fine structure formula\index{Sommerfeld fine structure formula} without detecting different spectra for different self-adjoint realisations. In the following Section the actual problem \eqref{eq:EigenvalueEq} is finally discussed and solved.

 Such classical approaches lie either within the framework of ordinary differential equations, with truncation of asymptotic series, or within the framework of supersymmetric quantum mechanics. It actually turns out that there are no explicit alternatives: indeed, in the differential equation approach (Sect.~\ref{sec:ODEmethods}) the only alternative to truncating series is to deal with eigenfunctions expressed by infinite series, and imposing the eigenfunction with eigenvalue $E$ to belong to some domain $\mathcal{D}(h_\beta)$ does not produce a closed formula for $E$ any longer; on the other hand, in the supersymmetric approach (Sect.~\ref{sec:susy}) the first order differential eigenvalue problem is studied by an auxiliary second order differential problem whose solutions only exhibit the boundary condition typical of the distinguished (or also of the `mirror' distinguished) extension, with no access to different boundary conditions.

 For concreteness of presentation, throughout this Section it is assumed that $\kappa=1$ and $\nu>0$ (with straightforward counterpart reasoning when $\kappa=-1$ and/or $\nu<0$): one has to keep in mind that depending on the sign of $\nu$ the actual form for the fine structure formula \index{Sommerfeld fine structure formula} is \eqref{eq:EVEnk1}, as determined in Corollary \ref{cor:eigenvalues_distinguished} from Section \ref{sec:IV-betaspectrum} below.

  \subsection{Eigenvalue problem by truncation of asymptotic series}\label{sec:ODEmethods}

The historically first approach (see, e.g., \cite[Section 14]{Bethe-Slapeter-1957}) for the determination of the eigenvalues of the Dirac-Coulomb Hamiltonian is based on methods of ordinary differential equations.

By direct inspection it is seen that the two linearly independent solutions to the differential problem \eqref{eq:diffprobnow} 
have large-$r$ asymptotics $e^{r\sqrt{1-E^2}}$ and $e^{-r\sqrt{1-E^2}}$, only the second one being square-integrable and hence admissible. This suggests the natural re-scaling $\psi\mapsto U\psi=:\phi$ defined by
\begin{equation}\label{eq:rescaling_map_U}
 (U\psi)(\rho)\;:=\;{\frac{1}{\sqrt{2}(1-E^2)^{\frac{1}{4}}}}\,\exp\big({\textstyle\frac{\rho}{2\sqrt{1-E^2}}}\big)\,\psi\big({\textstyle{\frac{\rho}{2\sqrt{1-E^2}}}}\big)\,,
\end{equation}
which induces the unitary operator $U: L^2(\mathbb{R}^+, \mathbb{C}^2,\ud  r) \to L^2(\mathbb{R}^+,\mathbb{C}^2,e^{-\rho} \, \ud \rho)$ and yields the unitarily equivalent problem
\begin{equation}\label{eq:EigenvalueEq1}
 U(h_\beta -E\mathbbm{1})U^{-1}\phi\;=\;0\,,\qquad\quad \phi\;:=\;U\psi\;\in\; U\mathcal{D}(h_\beta)\,,
\end{equation}
where
\begin{equation}\label{eq:EigenvalueEq2}
U(h_\beta -E\mathbbm{1})U^{-1}\;=\; 2\sqrt{1-E^2} \begin{pmatrix}
\frac{1}{2} \sqrt{\frac{1-E}{1+E}} + \frac{\nu}{\rho} & \frac{1}{2}-\frac{\ud}{\ud\rho} + \frac{1}{\rho} \\
-\frac{1}{2}+\frac{\ud}{\ud\rho}+\frac{1}{\rho} & -\frac{1}{2} \sqrt{\frac{1+E}{1-E}}+ \frac{\nu}{\rho}
\end{pmatrix}.
\end{equation}

The operator \eqref{eq:EigenvalueEq2} has a pole of order one at $\rho=0$, implying that the differential equation \eqref{eq:EigenvalueEq1} can be recast as 
\begin{equation}\label{eq:DiffEq}
\rho \,\phi' \;=\; A(\rho)\, \phi
\end{equation}
with
\begin{equation}\label{eq:VectorFieldMatrix}
A(\rho)\;:=\;\begin{pmatrix}
-1 & -\nu \\
\nu & 1
\end{pmatrix}+
\frac{1}{2}
\begin{pmatrix}
1 &  \sqrt{\frac{1+E}{1-E}} \\
\sqrt{\frac{1-E}{1+E}} & 1
\end{pmatrix}\rho\,.
\end{equation}
In particular it is explicitly checked that $\rho\mapsto A(\rho)$ is holomorphic.

It turns out that the differential problem \eqref{eq:DiffEq}-\eqref{eq:VectorFieldMatrix} is suited for the following standard result in the theory of ordinary differential equations (see, e.g., \cite[Theorems 5.1 and 5.4]{Wasow_asympt_expansions}).


\begin{proposition}\label{prop:Wasow}
Let $z\mapsto B(z)$ be a matrix-valued function whose entries are holomorphic at $z=0$ and whose Taylor series 
$B(z)=\sum_{j=0}^\infty B_j z^j$, say, of radius of convergence $r_B$, has the zero-th component $B_0$ diagonal and with eigenvalues that do not differ by integers. Then there exists a holomorphic matrix-valued function $z\mapsto P(z)$ whose Taylor series $P(z)=\sum_{j=0}^\infty P_j z^j$ converges for $|z|<r_B$ and has zero-th component $P_0=\mathbbm{1}$, such that the transformation
\begin{equation}\label{eq:TransformationDiffEq}
y(z)\;=\;P(z) f(z)
\end{equation}
reduces the differential equation
\begin{equation}\label{eq:GenDiffEq}
z y'(z)\;=\;B(z) y(z)
\end{equation}
to the form
\begin{equation}\label{eq:NormalizedDiffEq}
z f'(z) \;=\;B_0 f(z).
\end{equation}
\end{proposition}

Proposition \ref{prop:Wasow} is indeed applicable to \eqref{eq:DiffEq}-\eqref{eq:VectorFieldMatrix} for any $\nu\in(0,1)\!\setminus\!\{\frac{\sqrt{3}}{2}\}$ because in this case the matrix $A_0=A(0)$ is diagonalisable and its two distinct eigenvalues $\pm B = \pm \sqrt{1-\nu^2}$ do not differ by an integer (indeed, $2B \notin \mathbb{Z}$). (For the purpose of the discussion of this Section, we do not need to cover the exceptional case $\nu=\frac{\sqrt{3}}{2}$ which presents particular features -- see, e.g., \cite{Esteban-Lewin-Sere-2017_DC-minmax-levels}.)

 The (more relevant) critical regime $\nu\in(\frac{\sqrt{3}}{2},1)$ is discussed here first: the argument for the sub-critical values $\nu\in(0,\frac{\sqrt{3}}{2})$ is even simpler and is presented at the end of this Subsection.

Proposition \ref{prop:Wasow} implies at once that the \emph{general solution} to \eqref{eq:DiffEq}-\eqref{eq:VectorFieldMatrix} has the form
\begin{equation}
\phi(\rho)\;=\;G P(\rho) \left(\begin{matrix} \rho^B & 0 \\ 0 & \rho^{-B} \end{matrix}\right) \phi_0
\end{equation}
for some holomorphic matrix-valued $P(\rho)$ and some vector $\phi_0\in \mathbb{C}^2$, where $G$ is the matrix that diagonalises $A_0$. Component-wise,
\begin{align}
\label{eq:Series1}\phi^{+}(\rho)&\;=\; \sum_{j=0}^\infty a_j^{(B)} \rho^{B+j} + \sum_{j=0}^\infty a_j^{(-B)} \rho^{-B+j}\\
\label{eq:Series2}\phi^{-}(\rho)&\;=\; \sum_{j=0}^\infty b_j^{(B)} \rho^{B+j} + \sum_{j=0}^\infty b_j^{(-B)} \rho^{-B+j}
\end{align}
for suitable coefficients $a_j^{(B)},b_j^{(B)},a_j^{(-B)},b_j^{(-B)}\in\mathbb{C}$, $j\in\mathbb{N}_0$, that must satisfy the consistency relations obtained by plugging \eqref{eq:Series1}-\eqref{eq:Series2} into \eqref{eq:DiffEq}. In doing so, one recognises that $\rho^{B+j}$-powers and $\rho^{-B+j}$-powers never get multiplied among themselves, and moreover each type of powers only gets multiplied by $a_j$ or $b_j$ coefficient of the same type; the net result, when equating to zero the coefficients of each power in the identity $\rho \phi'(\rho)-A(\rho)\phi(\rho)=0$ is the \emph{double set} of recursive equations
\begin{align}
\label{eq:Eq_First}{\textstyle \frac{1}{2} \sqrt{\frac{1-E}{1+E}}} \,a_j^{(\pm B)} + \nu\, a_{j+1}^{(\pm B)} + {\textstyle \frac{1}{2} }\,b_j^{(\pm B)} + (-j\mp B) b_{j+1}^{(\pm B)} &\;=\;0 \, ,\\
\label{eq:Eq_Second}{\textstyle -\frac{1}{2}} a_j^{(\pm B)} +(j\pm B+2) \,a_{j+1}^{(\pm B)} - {\textstyle \frac{1}{2} \sqrt{\frac{1+E}{1-E}}}\, b_j^{(\pm B)} + \nu \,b_{j+1}^{(\pm B)} &\;=\;0 \, ,\\
\label{eq:Eq_Zeroeth}\nu\,a_0^{(\pm B)}-(\pm B-1)\,b_0^{(\pm B)} &\;=\;0\,,
\end{align}
that is, the upper signs for the $B$-part and the lower signs for the $-B$-part of \eqref{eq:Series1}-\eqref{eq:Series2}.

The above recursive relations are conveniently re-written in a more manageable form upon introducing $\alpha_j^{(\pm B)}$ and $\beta_j^{(\pm B)}$ through
\begin{equation}\label{eq:change}
 a_j^{(\pm B)}\;=\; \sqrt{1+E} \,(\alpha_j^{(\pm B)}+\beta_j^{(\pm B)})\,,\qquad b_j^{(\pm B)}\;=\;\sqrt{1-E} \, (\alpha_j^{(\pm B)}-\beta_j^{(\pm B)})\,,
\end{equation}
which yields
\begin{eqnarray}
\textstyle \big(\frac{\nu}{\sqrt{1-E^2}} +1 \big)\alpha_{j}^{(\pm B)}+\big( \frac{ E \nu}{\sqrt{1-E^2}} +j\pm B \big) \beta_{j}^{(\pm B)}\;&=&\;0  \label{eq:recursive01}\\
\textstyle\alpha_j^{(\pm B)}+\big(\frac{E \nu}{\sqrt{1-E^2}} -j-1\mp B \big)\, \alpha_{j+1}^{(\pm B)} + \big(\frac{\nu}{\sqrt{1- E^2}} - 1 \big) \,\beta_{j+1}^{(\pm B)} \;&=&\;0 \label{eq:recursive02}\\
\textstyle\big(\frac{\nu E}{\sqrt{1-E^2}} \mp B\big)\,\alpha_0^{(\pm B)}-\big(\frac{\nu}{\sqrt{1-E^2}} -1\big) \,\beta_0^{(\pm B)} \;&=& \;0\,. \label{eq:recursive00}
\end{eqnarray}
Now, plugging \eqref{eq:recursive01} into \eqref{eq:recursive02} yields 
\begin{equation}\label{eq:alpha_beta}
\alpha_{j+1}^{(\pm B)}\;=\;\frac{\frac{E \nu}{\sqrt{1-E^2}} + j \pm B +1 }{(j\pm B+1)^2-B^2}\,\alpha_j^{(\pm B)}\,.
\end{equation}

From \eqref{eq:alpha_beta} one sees that, unless $\alpha^{(\pm B)}_{j_0}=0$ for some $j_0$, in which case $\alpha^{(\pm B)}_{j}=0$ for all $j\geqslant j_0$, one has
\begin{equation}
\frac{\alpha_{j+1}^{(\pm B)}}{\alpha_j^{(\pm B)}}\;=\; j^{-1}+O(j^{-2})\qquad\textrm{as } j \to +\infty\,,
\end{equation}
implying that $\sum_j \alpha_j^{(\pm B)} \rho^{j}$ grows faster than $e^{\rho/2}$ at infinity and hence fails to belong to 
$L^2(\mathbb{R}^+, \mathbb{C}, e^{-\rho} \, \ud \rho)$. Through the transformation \eqref{eq:change} this implies that 
\begin{itemize}
 \item at least one among $\sum_j a_j^{(B)} \rho^{B+j}$ and $\sum_j b_j^{(B)} \rho^{B+j}$,
 \item and at least one among $\sum_j a_j^{(-B)} \rho^{-B+j}$ and $\sum_j b_j^{(-B)} \rho^{-B+j}$
\end{itemize}
are series that diverge faster than $e^{\rho/2}$. This poses the issue of admissibility (in particular, of the square-integrability) of the spinor-valued function $\phi$ given by \eqref{eq:Series1}-\eqref{eq:Series2}, for which the only possible affirmative answers are the following three.

 \medskip

\textbf{\underline{First case}:} $\phi\in L^2(\mathbb{R}^+, \mathbb{C}^2, e^{-\rho} \, \ud \rho)$ because the $B$-series in  \eqref{eq:Series1} and the $B$-series in \eqref{eq:Series2} are actually truncated (i.e., polynomials), whereas the $(-B)$-series in  \eqref{eq:Series1} and the $(-B)$-series in \eqref{eq:Series2} vanish identically. This is obtained by imposing that $\alpha^{(B)}_{n+1}=0$ for some $n\in\mathbb{N}_0$ and that all the $a^{(-B)}_j$'s and $b^{(-B)}_j$'s vanish. Then \eqref{eq:alpha_beta} constrains $E$ to attain one of the values 
\begin{equation}\label{eq:Truncation}
E_n= - \Big(1+\frac{\nu^2}{(n+\sqrt{1-\nu^2})^2}\Big)^{-\frac{1}{2}} \qquad n\in\mathbb{N}\,.
\end{equation}
From \eqref{eq:recursive01} it is seen that the vanishing of $\alpha_{n+1}$ implies the vanishing of $\beta_j$ for all $j \geqslant n+2$ while, from \eqref{eq:recursive02}, one sees that $\beta_{n+1} \neq 0$. By direct inspection in \eqref{eq:recursive00} one sees that also $E_{n=0}$ given by \eqref{eq:Truncation} is an eigenvalue for which $\beta_0 \neq 0 $ and $\alpha_0 =0$ (it is crucial in this step that $\nu > 0$). Hence, for each value $E_n$, the corresponding $\phi$ has the form
\begin{equation}\label{eq:SommerfeldFormulaSeries}
\phi_{n}(\rho)\;=\;\rho^Be^{-\rho\sqrt{1-E_n^2}} \, \sum_{j=0}^{n+1}\begin{pmatrix} a_j^{(B)} \\ b_j^{(B)}\end{pmatrix}\rho^{j}\,,
\end{equation}
and through the inverse transformation $\psi=U^{-1}\phi$ of \eqref{eq:EigenvalueEq1} it is immediately recognised that $\psi$ satisfies the boundary condition \eqref{eq:IV-Sbeta_bc} with $\beta=\infty$. This leads to the discrete spectrum of the \emph{distinguished}\index{distinguished Dirac-Coulomb Hamiltonian} extension $h_{\mathrm{D}}$: formula \eqref{eq:Truncation} is precisely the fine structure formula\index{Sommerfeld fine structure formula} for $\nu>0$.

 \medskip

\textbf{\underline{Second case}:} $\phi\in L^2(\mathbb{R}^+, \mathbb{C}^2, e^{-\rho} \, \ud \rho)$ because the $(-B)$-series in  \eqref{eq:Series1} and the $(-B)$-series in \eqref{eq:Series2} are finite polynomials, whereas the $B$-series in  \eqref{eq:Series1} and the $B$-series in \eqref{eq:Series2} vanish identically. This is obtained by imposing that $\alpha^{(-B)}_{n+1}=0$ for some $n\in\mathbb{N}_0$ and that all the $a^{(B)}_j$'s and $b^{(B)}_j$'s vanish. In this case \eqref{eq:alpha_beta} constrains $E$ to attain one of the values 
\begin{equation}\label{eq:anti-dist}
\begin{split}
 E_n&\;=\;- \displaystyle\Big(1+\frac{\nu^2}{(n-\sqrt{1-\nu^2})^2} \Big)^{-\frac{1}{2}} \qquad n\in\mathbb{N}\,,\\
 E_0&\;=\; B \,,
\end{split}
\end{equation}
the value $E_0$ being obtained by direct inspection in \eqref{eq:recursive00} analogously to what is done for the analogous point in the previous case) and for each such value, the corresponding $\phi$ has the form
\begin{equation}
\phi_{n}(\rho)\;=\;  \rho^{-B}e^{-\rho\sqrt{1-E_n^2}} \, \sum_{j=0}^{n+1}\begin{pmatrix} a_j^{(-B)} \\ b_j^{(-B)}\end{pmatrix}\rho^{j}
\,.
\end{equation}
Through the inverse transformation $\psi=U^{-1}\phi$ of \eqref{eq:EigenvalueEq1} it is immediately recognised that $\psi$ satisfies the boundary condition \eqref{eq:IV-Sbeta_bc} with
\begin{equation}\label{eq:mdbeta}
\beta\;=\; -\frac{d_{\nu,1}}{c_{\nu,1}}\,.
\end{equation}
This is another self-adjoint realisation of the Dirac-Coulomb Hamiltonian, different from $h_{\mathrm{D}}$, which arises in this second case, that mirrors the first case for the distinguished extension. One refers to this realisation as the \emph{mirror distinguished} extension $h_{\mathrm{MD}}$. This analysis thus produces the discrete spectrum of $h_{\mathrm{MD}}$, the eigenvalue formula \eqref{eq:anti-dist} providing the modification of the Sommerfeld fine structure formula\index{Sommerfeld fine structure formula} for this Dirac-Coulomb Hamiltonian.

It is crucial to observe at this point that the \emph{two eigenvalue formulas \eqref{eq:Truncation} and \eqref{eq:anti-dist} do not have any value in common}. As a consequence, even if combining together the truncation of the first case (in the $B$-series) and the truncation of the second case (in the $(-B)$-series) would produce a function $\phi$ that belongs to $L^2(\mathbb{R}^+, \mathbb{C}, e^{-\rho} \, \ud \rho)$, such $\phi$ could not correspond to any definite value $E$, i.e., $\phi$ could not be a solution to \eqref{eq:EigenvalueEq1}.

Truncation in \eqref{eq:Series1}-\eqref{eq:Series2} produces admissible solutions only of the form of truncated series of $B$-type or truncated series of $(-B)$-type. This explains why the only remaining case is the following.

 \medskip

\textbf{\underline{Third case}:}  $\phi$ has the form \eqref{eq:Series1}-\eqref{eq:Series2} where \emph{both} component $\phi^+$ and $\phi^-$ contain two series that diverge faster than $e^{\rho/2}$ at infinity, whose sum however produces a compensation such that $\phi$ belongs to $L^2(\mathbb{R}^+, \mathbb{C}^2, e^{-\rho} \, \ud \rho)$. This yields then an admissible eigenfunction $\psi=U^{-1}\phi$ with eigenvalue $E$. Matching the coefficients of the expansion
\[
 \phi(\rho)\;=\;\rho^{-B}\begin{pmatrix} a_0^{(-B)} \\ b_0^{(-B)}  \end{pmatrix} +\rho^B\begin{pmatrix} a_0^{(B)} \\ b_0^{(B)}  \end{pmatrix} +\cdots\qquad\textrm{as } \rho \downarrow 0\,,
\]
through the transformation $\psi=U^{-1}\phi$, to the general boundary condition \eqref{eq:IV-Sbeta_bc} indicates which domain $\mathcal{D}(h_\beta)$ the vector $\psi$ belongs to.

Clearly, since in the third case above no truncation occurs in \eqref{eq:Series1}-\eqref{eq:Series2}, the recursive formulas for the coefficients are now of no use and it is not possible to infer from them any closed formula for the eigenvalues of the realisation $h_\beta$, $\beta\notin\{-\frac{d_{\nu,1}}{c_{\nu,1}},\infty\}$. In this sense, as announced at the beginning of this Section, the ordinary differential equation methods discussed here only select the discrete spectrum (and a closed eigenvalue formula) for the distinguished extension $h_{\mathrm{D}}$ and for the mirror distinguished extension $h_{\mathrm{MD}}$.

 Last, one observes that in the sub-critical regime $\nu\in(0,\frac{\sqrt{3}}{2})$, i.e., $B\in (\frac{1}{2},1)$, the argument that led to the general form \eqref{eq:Series1}-\eqref{eq:Series2} is precisely the same, but of course in this regime $\rho^{-B}$ fails to be square-integrable near the origin, meaning that the whole $(-B)$-series in \eqref{eq:Series1}-\eqref{eq:Series2} must vanish identically. The only admissible solution is then that obtained with a truncation as in the first case, which leads again, as should be, to the Sommerfeld fine structure formula\index{Sommerfeld fine structure formula} \eqref{eq:Truncation}.

  \subsection{Eigenvalue problem by supersymmetric methods}\label{sec:susy}

A second, by now classical \cite{Sukumar_1985,Grosse-1987,Cooper-Khare-Mustoa-Wipf-1988,Panahi-Bakhshi_2011}, approach to determine the fine structure formula\index{Sommerfeld fine structure formula} exploits the supersymmetric structure of the eigenvalue problem \eqref{eq:EigenvalueEq}.

By means of the bounded and invertible linear transformation $A:L^2(\mathbb{R}^+,\mathbb{C}^2)\to L^2(\mathbb{R}^+,\mathbb{C}^2)$ defined by
\begin{equation}
 A\,\xi\;:=\;
 \begin{pmatrix}
  -(1+B) & \nu \\
  \nu & -(1+B)
 \end{pmatrix}
 \begin{pmatrix}
  \xi^+ \\ \xi^-
 \end{pmatrix}
\end{equation}
it is convenient to turn the problem \eqref{eq:EigenvalueEq} into the form
\begin{equation}\label{eq:EigenvalueEq_v2}
\begin{split}
 0\;&=\;\sigma_2 \,A^{-1}\,\sigma_2\,(h_\beta-E\mathbbm{1})\,A\,\phi \\
 &=\;\left[\begin{pmatrix}
      0 & -\frac{\ud}{\ud r}+\frac{B}{r}+\frac{\nu E}{B} \\
      \frac{\ud}{\ud r}+\frac{B}{r}+\frac{\nu E}{B} & 0
     \end{pmatrix}
    -\begin{pmatrix}
      \frac{E}{B}-1 & 0 \\
      0 & \frac{E}{B}+1 
     \end{pmatrix}\right]\phi\,,
\end{split}
\end{equation} 
having set
\begin{equation}\label{eq:psiAphi}
 \phi\;:=\;A^{-1}\psi\,. 
\end{equation}

Next, in terms of the differential operators
\begin{equation}
 D^{\pm}\;:=\;\pm\frac{\ud}{\ud r}+\frac{B}{r}+\frac{\nu E}{B}
\end{equation}
acting on scalar functions, and of the differential operators
\begin{equation}
 Q\;:=\;\begin{pmatrix}
         \mathbb{O} & D^- \\
         D^+ & \mathbb{O}
        \end{pmatrix}\qquad\textrm{and}\qquad
 \mathsf{H}\;:=\;Q^2\;=\;\begin{pmatrix}
         D^-D^+ & \mathbb{O} \\
         \mathbb{O} & D^+D^-
        \end{pmatrix}
\end{equation}
acting on spinor functions, equation \eqref{eq:EigenvalueEq_v2} reads
\begin{equation}\label{eq:ev_Qmatrixform}
 Q\phi\;=\;\begin{pmatrix}
      \frac{E}{B}-1 & 0 \\
      0 & \frac{E}{B}+1 
     \end{pmatrix}\phi\,,
\end{equation}
whence
\begin{equation}\label{eq:EigenvalueEq_v3}
 \mathsf{H}\phi\;=\;Q^2\phi\;=\;Q\begin{pmatrix}
      \frac{E}{B}-1 & 0 \\
      0 & \frac{E}{B}+1 
     \end{pmatrix}\phi\;=
     \begin{pmatrix}
      \frac{E}{B}+1 & 0 \\
      0 & \frac{E}{B}-1 
     \end{pmatrix}Q\phi\;=\;{\textstyle(\frac{E^2}{B^2}-1)}\,\phi\,,
\end{equation}
equivalently,
\begin{equation}\label{eq:EigenvalueEq_v4}
 \begin{split}
  D^+D^-\phi^-\;&=\;{\textstyle(\frac{E^2}{B^2}-1)}\,\phi^-\,, \\
  D^-D^+\phi^+\;&=\;{\textstyle(\frac{E^2}{B^2}-1)}\,\phi^+\,.
 \end{split}
\end{equation}

Equation \eqref{eq:EigenvalueEq_v3} or \eqref{eq:EigenvalueEq_v4} are the \emph{supersymmetric}\index{supersymmetric operator} form of \eqref{eq:EigenvalueEq}. The structure is indeed the same as for a generic triple $(\mathscr{H},\mathscr{P},\mathscr{Q})$, where (see, e.g., \cite[Section 6.3]{Cycon-F-K-S-Schroedinger_ops} and \cite[Section 5.1]{Thaller-Dirac-1992}), for some densely defined operator $D$ on $L^2(\mathbb{R}^+)$,
\begin{equation}
 \mathscr{Q}\;:=\;
 \begin{pmatrix}
  \mathbb{O} & D^* \\
  D & \mathbb{O}
 \end{pmatrix},\quad
  \mathscr{P}\;:=\;
 \begin{pmatrix}
  \mathbbm{1} & \mathbb{O} \\
  \mathbb{O} & -\mathbbm{1}
 \end{pmatrix},\quad
 \mathscr{H}\;:=\;\mathscr{Q}^2\;=\;
 \begin{pmatrix}
  D^*D & \mathbb{O} \\
  \mathbb{O} & DD^*
 \end{pmatrix}
\end{equation}
are self-adjoint operators on $L^2(\mathbb{R}^+)\oplus L^2(\mathbb{R}^+)\cong L^2(\mathbb{R}^+,\mathbb{C}^2)$ with the properties that
\begin{equation}
\mathscr{P}^2=\mathbbm{1}\,,\quad\mathscr{P}\mathcal{D}(\mathscr{H})=\mathcal{D}(\mathscr{H})\,,\quad\mathscr{P}\mathcal{D}(\mathscr{Q})=\mathcal{D}(\mathscr{Q})\,,\quad\{\mathscr{Q},\mathscr{P}\}=\mathbb{O}\,.
\end{equation}
 In the supersymmetric jargon, $\mathscr{P}$ is an involution (the \emph{grading operator}),\index{grading operator} $\mathscr{Q}$ is a \emph{supercharge}\index{supercharge} with respect to such involution, and $\mathscr{H}$ is a Hamiltonian \emph{with supersymmetry}.\index{supersymmetric operator} Moreover, standard spectral arguments show that the two spectra $\sigma(D^*D)$ and $\sigma(DD^*)$ with respect to $L^2(\mathbb{R}^+)$ lie both in $[0,+\infty)$ and coincide, and in particular the eigenvalues are the same, but for possibly the value zero.

 Even if the domain of $D^{\pm}$ in $L^2(\mathbb{R}^+)$ was left undeclared, it is clear that the two operators are formally adjoint to each other. The fact that the eigenvalues of $D^+D^-$ and $D^-D^+$ relative to square-integrable eigenfunctions are non-negative follows from a trivial integration by parts; the fact that those such eigenvalues that are strictly positive are the same for both $D^+D^-$ and $D^-D^+$ is also an immediate algebraic consequence, for $D^-D^+f=\lambda f$ for $\lambda\neq 0$ implies that $D^+f\neq 0$ and $D^+D^-(D^+f)=\lambda(D^+f)$, the same then holding also when roles of $D^+$ and $D^-$ are exchanged.

The solutions $(E,\psi)$ to the problem \eqref{eq:EigenvalueEq}, with chosen realisation $h_\beta$, can be read out from \eqref{eq:EigenvalueEq_v3}-\eqref{eq:EigenvalueEq_v4}.

 One starts with the `ground state' solution, referring here to the lowest possible eigenvalue of $\mathsf{H}$, namely the value zero, and hence, because of \eqref{eq:EigenvalueEq_v3}, the smallest possible $|E|$ for the eigenvalue $E$ of the considered realisation $h_\beta$. First of all, the ground state energy $E_0$ must satisfy $E_0^2=B^2$, as follows from \eqref{eq:EigenvalueEq_v3}.

Out of the two possibilities, one is then to take $D^-\phi^-=0$ in \eqref{eq:EigenvalueEq_v4}, with $E=E_0$ to be determined, which is an ordinary differential equation whose solutions are the multiples of
\[
 \phi^-(r)\;=\;r^{B}\,e^{\frac{\nu E_0}{B}r}\,.
\]

For such $\phi^-$ to be square-integrable, $\nu E_0<0$, thus $E_0=-B$ since $\nu>0$. Correspondingly, the second equation in \eqref{eq:EigenvalueEq_v4} is $D^-D^+\phi^+=0$ for some $\phi^+\in L^2(\mathbb{R}^+)$. This is equivalent to $D^+\phi^+=0$, thanks to the fact that $D^-$ is the formal adjoint of $D^+$. The latter ordinary differential equation is solved by the multiples of $r^{-B}\,e^{-\frac{\nu E_0}{B}r}$, which is not square-integrable at infinity, whence $\phi^+=0$. Alternatively, one may argue that the corresponding $\phi^+$ to the above $\phi^-$ is read out directly from \eqref{eq:ev_Qmatrixform}: it must be (a multiple of)
\[
 ({\textstyle \frac{E_0}{B}-1})^{-1}(D^-\phi^-)(r)\;=\;({\textstyle \frac{E_0}{B}-1})^{-1}({\textstyle \frac{\ud}{\ud r}+\frac{B}{r}+\frac{\nu E_0}{B}})(r^{B}\,e^{\frac{\nu E_0}{B}r})
\]
and it must be square-integrable, which forces $\phi^+$ to be necessarily null, for the above function fails to be square-integrable at the origin.

 This way, one thus finds a solution $(E,\phi)$ to the problem \eqref{eq:EigenvalueEq_v4} with smallest possible $|E|$ and square-integrable $\phi$, namely the pair $(E_0,\phi_0)$ (up to multiples of $\phi_0$) given by
\begin{equation}\label{eq:gs_1_phi}
 E_0\;=\;-B\,,\qquad \phi_0(r)\;=\;r^{B}\,e^{\frac{\nu E_0}{B}r}
 \begin{pmatrix} 
 0 \\ 1
 \end{pmatrix}.
\end{equation}
Through \eqref{eq:IV-defB} and the transformation \eqref{eq:psiAphi}, and in view of the general boundary condition of self-adjointness \eqref{eq:IV-Sbeta_bc}, one sees that \eqref{eq:gs_1_phi} corresponds to the pair $(E_0,\psi_0)$ given by
\begin{equation}
 E_0\;=\;-\Big(1+\frac{\nu^2}{1-\nu^2}\Big)^{-\frac{1}{2}}\,,\qquad \psi_0(r)\;=\;r^{B}\,e^{\frac{\nu E_0}{B}r}
 \begin{pmatrix} 
 \nu \\ -(1+B)
 \end{pmatrix}\in\mathcal{D}(h_{\mathrm{D}})\,,
\end{equation}
which is the ground state solution to the initial eigenvalue problem \eqref{eq:EigenvalueEq} for $\beta=\infty$, and hence for the \emph{distinguished} Dirac-Coulomb Hamiltonian.\index{distinguished Dirac-Coulomb Hamiltonian}

By a completely analogous reasoning, the other possibility is to look for ground state solutions to \eqref{eq:EigenvalueEq_v4} with $D^+\phi^+=0$, and $E=E_0$ to be determined, an ordinary differential equation solved by the multiples of
\[
 \phi^+(r)\;=\;r^{-B}\,e^{-\frac{\nu E_0}{B}r}\,,
\]
and such $\phi^+$ is only square-integrable if $E_0=B>0$. Correspondingly, the first equation in \eqref{eq:EigenvalueEq_v4} is $D^+D^-\phi^-=0$, equivalently, $D^-\phi^-=0$, which is solved by multiples of $r^{B}\,e^{\frac{\nu E_0}{B}r}$; the latter function failing to be square-integrable at infinity, one thus ends up with the solution $(E_0,\phi_0)$ (up to multiples of $\phi_0$) given by
\begin{equation}\label{eq:gs_2_phi}
 E_0\;=\;B\,,\qquad \phi_0(r)\;=\;r^{-B}\,e^{-\frac{\nu E_0}{B}r}
 \begin{pmatrix} 
 1 \\ 0
 \end{pmatrix}.
\end{equation}
Thus, again using \eqref{eq:psiAphi}, and comparing the expansion
\[
 r^{-B}\,e^{-\frac{\nu E_0}{B}r}\;=\;r^{-B}-{\textstyle\frac{\nu E_0}{B}}\,r^{1-B}+o(r^{1-B})\qquad\textrm{as }r\downarrow 0
\]
with the general classification general boundary condition of self-adjointness \eqref{eq:IV-Sbeta_bc}, whence now $g_0^+=1$, $g_1^+=0$, $c_{\nu,1}\beta+d_{\nu,1}=0$, one sees that another ground state solution to \eqref{eq:EigenvalueEq} is the pair $(E_0,\psi_0)$ given by
\begin{equation}\label{eq:E0MD}
 \begin{split}
  E_0\;&=\;\Big(1+\frac{\nu^2}{1-\nu^2}\Big)^{-\frac{1}{2}}\,, \\
  \psi_0(r)\;&=\;r^{-B}\,e^{-\frac{\nu E_0}{B}r}
 \begin{pmatrix} 
 -(1+B) \\ \nu 
 \end{pmatrix}\;\in\;\mathcal{D}(h_{\mathrm{MD}})\,,
 \end{split}
\end{equation}
and this is the ground state solution for the \emph{mirror distinguished} ($\beta=-d_{\nu,1}/c_{\nu,1}$) self-adjoint realisation $h_{\mathrm{MD}}$ already introduced in Subsection \ref{sec:ODEmethods}, formula \eqref{eq:mdbeta}. 

Significantly, no other realisations can be monitored through the supersymmetric scheme above, but those with $\beta=\infty$ or $\beta=-d_{\nu,1}/c_{\nu,1}$.

The  excited states too are determined within the supersymmetric scheme. Let
\begin{equation}\label{eq:Bplus}
 D_n^{\pm}\;:=\;\pm\frac{\ud}{\ud r}+\frac{B_n}{r}+\frac{\nu E}{B_n}\,,\qquad B_n:=B+n\,,\qquad n\in\mathbb{N}_0\,.
\end{equation}
Clearly, $B=B_0$, $D^\pm=D^\pm_0$. $D_n^+$ and $D_n^-$ are formally adjoint. From
\[
 \begin{split}
  D^\pm_n D^\mp_n\;&=\;-\frac{\ud^2}{\ud r^2}+\frac{B_n(B_n\mp 1)}{r^2}+\frac{2\nu E}{r}+\frac{\nu^2 E^2}{B_n^2}
 \end{split}
\]
one deduces
\begin{equation}\label{eq:Bshift}
 \begin{split}
  & D^\pm_n D^\mp_n f\;=\;\big({\textstyle E^2(1+\frac{\nu^2}{B_n^2}})-1\big)f \\
  & \qquad\Leftrightarrow\qquad-f''+{\textstyle \frac{B_n(B_n\mp 1)}{r^2}}f+{\textstyle\frac{2\nu E}{r}}f-E^2f\;=\;0\,.
   \end{split}
\end{equation}
 This shows that the equation in \eqref{eq:Bshift} with the lower signs is the same as the equation with the upper signs and with $B_n$ replaced by $B_{n+1}$. This is the basis for an iterative argument, as follows.

As a first step, as a consequence of \eqref{eq:Bshift}, the equation $D^-D^+\phi^+=(\frac{E^2}{B^2}-1)\phi^+$ of the problem \eqref{eq:EigenvalueEq_v3} is equivalent to $D_1^+D_1^-\phi^+=(E^2(1+\frac{\nu^2}{(B+1)^2})-1)\phi^+$, which can be regarded as the first scalar equation of
\begin{equation}\label{eq:susyproblem_step1}
\begin{pmatrix}
         D_1^-D_1^+ & \mathbb{O} \\
         \mathbb{O} & D_1^+D_1^-
        \end{pmatrix}
\begin{pmatrix}
 \xi_1^+ \\ \xi_1^-
\end{pmatrix}\;=\;\big({\textstyle E^2(1+\frac{\nu^2}{(B+1)^2})-1}\big)
\begin{pmatrix}
 \xi_1^+ \\ \xi_1^-
\end{pmatrix},\qquad \xi_1^-:=\phi^+\,.
\end{equation}
The ground state solution $(E_1,\xi_1^{(\textrm{gs})})$ to the new supersymmetric problem \eqref{eq:susyproblem_step1} is obtained in complete analogy to the argument that led to \eqref{eq:gs_1_phi}, whence
\begin{equation}\label{eq:Eqxi1}
 E_1\;=\;-\big({\textstyle 1+\frac{\nu^2}{(B+1)^2}} \big)^{\!-\frac{1}{2}}\,,\qquad \xi_1^{(\textrm{gs})}(r)\;=\;r^{B+1}e^{\frac{\nu E_1}{B+1}r}
 \begin{pmatrix} 0 \\ 1 \end{pmatrix}.
\end{equation}
(The other solution that one would find in complete analogy to the argument that led to \eqref{eq:gs_2_phi} is not square-integrable.)
In turn, using $\phi^+=\xi_1^-$, \eqref{eq:Eqxi1} corresponds to a solution $\phi_1^+$ to the equation $D^-D^+\phi^+=(\frac{E_1^2}{B^2}-1)\phi^+$, and hence to a solution $(E_1,\phi_1)$ to the original problem  \eqref{eq:ev_Qmatrixform}-\eqref{eq:EigenvalueEq_v3},
given by
\begin{equation}
 \begin{split}
  E_1\;&=\;-\big({\textstyle 1+\frac{\nu^2}{(B+1)^2}} \big)^{\!-\frac{1}{2}}\;<\;E_0\;<\;0\,, \\
  \phi_1^+(r)\;&=\;r^{B+1}e^{\frac{\nu E_1}{B+1}r}\,, \\
  \phi_1^-(r)\;&=\;({\textstyle \frac{E_1}{B}+1})^{-1}(D^+\phi_1^+)(r)\,.
 \end{split}
\end{equation}
Clearly, $(D^+\phi_1^+)(r)\sim r^B$ as $r\downarrow 0$, and all together $\psi_1:=A\phi_1\in \mathcal{D}(h_{\mathrm{D}})$: thus, $(E_1,\psi_1)$ gives the first excited state for the eigenvalue problem \eqref{eq:EigenvalueEq} for the \emph{distinguished} realisation $h_{\mathrm{D}}$.

The procedure is repeated for the iterated supersymmetric problems
\begin{equation}\label{eq:susyproblem_stepn}
\begin{split}
\begin{pmatrix}
         \mathbb{O} & D_{n-1}^-   \\
         D_{n+1}^+ & \mathbb{O}
        \end{pmatrix}\begin{pmatrix}
 \xi_{n-1}^+ \\ \xi_{n-1}^-
\end{pmatrix} \;&=\;\begin{pmatrix}
        E\sqrt{1+\frac{\nu^2}{B^2_{n-1}}}-1  & \mathbb{O}   \\
        \mathbb{O} & E\sqrt{1+\frac{\nu^2}{B^2_{n-1}}}+1
        \end{pmatrix}\begin{pmatrix}
 \xi_{n-1}^+ \\ \xi_{n-1}^-
\end{pmatrix}, \\
\begin{pmatrix}
         D_n^-D_n^+ & \mathbb{O} \\
         \mathbb{O} & D_n^+D_n^-
        \end{pmatrix}
\begin{pmatrix}
 \xi_n^+ \\ \xi_n^-
\end{pmatrix}\;&=\;\big({\textstyle E^2(1+\frac{\nu^2}{B_n^2})-1}\big)
\begin{pmatrix}
 \xi_n^+ \\ \xi_n^-
\end{pmatrix},\qquad \xi_n^-=\xi_{n-1}^+\,.
\end{split}
\end{equation}
The admissible ground state solution $(E_n,\xi_n^{(\textrm{gs})})$ for the second equation in \eqref{eq:susyproblem_stepn} is
\begin{equation}
 E_n\;=\;-\big({\textstyle 1+\frac{\nu^2}{(B+n)^2}} \big)^{\!-\frac{1}{2}}\,,\qquad \xi_n^{(\textrm{gs})}(r)\;=\;r^{B+n}e^{\frac{\nu E_n}{B+n}r}
 \begin{pmatrix} 0 \\ 1 \end{pmatrix};
\end{equation}
then, by the first equation in \eqref{eq:susyproblem_stepn} and the preceding iterations, the pair $(E_n,\phi_n)$ with
\begin{equation}
\phi_n\;:=\;\begin{pmatrix}
  D_{n-1}^+ \big(r^{B+n}e^{\frac{\nu E_n}{B+n}r}\big)\\
  D_0^+D_1^+\cdots D_{n-1}^+\big(r^{B+n}e^{\frac{\nu E_n}{B+n}r}\big)
\end{pmatrix}
\end{equation}
gives the $n$-th excited state solution to the original problem  \eqref{eq:ev_Qmatrixform}-\eqref{eq:EigenvalueEq_v3}. One immediately recognises that $\phi_n^\pm(r)\sim r^B$ as $r\downarrow 0$, whence $\psi_n:=A\phi_n\in \mathcal{D}(h_{\mathrm{D}})$: thus, $(E_n,\psi_n)$ gives the $n$-th excited state for the eigenvalue problem \eqref{eq:EigenvalueEq} for the \emph{distinguished} realisation $h_{\mathrm{D}}$.

With the analysis above one reproduces all energy levels of Sommerfeld's formula
\begin{equation}
 E_n\;=\;-\big({\textstyle 1+\frac{\nu^2}{(n+\sqrt{1-\nu^2})^2}} \big)^{\!-\frac{1}{2}}\,,\qquad n\in\mathbb{N}_0
\end{equation}
and recognises that they all correspond to bound states for the distinguished realisation $h_{\mathrm{D}}$ of the Dirac-Coulomb Hamiltonian.\index{distinguished Dirac-Coulomb Hamiltonian}

By a completely symmetric iterative analysis which starts using
\begin{equation}
B_n:=B-n, \qquad n \in \mathbb{N}_0
\end{equation} 
instead of \eqref{eq:Bplus} and the same definitions for $D^\pm_n$ one sees that also the pairs $(E_n,\psi_n)$, with $\psi_n:=A\phi_n$ and
\begin{equation}\label{eq:as}
 \begin{split}
E_n\;&:=\;-\big({\textstyle 1+\frac{\nu^2}{(n-\sqrt{1-\nu^2})^2}} \big)^{\!-\frac{1}{2}}\,, \\
\phi_n\;&:=\;\begin{pmatrix}
  D_0^+D_1^+\cdots D_{n-1}^+\big(r^{-B+n}e^{\frac{\nu E_n}{-B+n}r}\big)\\
  D_{n-1}^+ \big(r^{-B+n}e^{\frac{\nu E_n}{-B+n}r}\big)
\end{pmatrix},
 \end{split}
\end{equation}
provide a complete set of solutions to the eigenvalue problem \eqref{eq:EigenvalueEq} for the \emph{mirror distinguished} realisation $h_{\mathrm{MD}}$ ($\psi_n\in\mathcal{D}(h_\beta)$ for $\beta=-d_{\nu,1}/c_{\nu,1}$).

\section{Discrete spectra for critical Dirac-Coulomb Hamiltonians}\label{sec:IV-betaspectrum}

\begin{theorem}\label{thm:spectrum-beta}
 Let $h$ be the operator on $L^2(\mathbb{R}^+,\mathbb{C}^2,\ud r)$ defined in \eqref{eq:IV-def_operator_S} with $|\nu|\in(\frac{\sqrt{3}}{2},1)$ and $\kappa=\pm 1$, and let $(h_\beta)_{\beta\in\mathbb{R}\cup\{\infty\}}$ be the family of self-adjoint realisations of $h$, as found in Theorems \ref{thm:classification_structure} and \ref{thm:classification_bc}. The discrete spectrum of a generic realisation $h_\beta$ consists of the countable collection 
 \begin{equation}
  \sigma_{\mathrm{disc}}(h_\beta)\;=\;\big\{E_n^{(\beta)}\,|\,n\in\mathbb{N}_0\,, n\geqslant n_0\big\}\;\subset\;(-1,1)
 \end{equation}
 of eigenvalues $E_n^{(\beta)}$ which are all the possible roots, enumerated in decreasing order when $\nu>0$ and in increasing order when $\nu<0$, of the transcendental equation
 \begin{equation}\label{eq:fEn_formula}
  \mathfrak{F}_{\nu,\kappa}(E_n^{(\beta)})\;=\;c_{\nu,\kappa} \,\beta + d_{\nu,\kappa}\,,
 \end{equation}
 where the constants $c_{\nu,\kappa}$ and $d_{\nu,\kappa}$ are given by \eqref{eq:IV-defcd}, and
 \begin{equation}\label{eq:Fnu}
 \begin{split}
  \mathfrak{F}_{\nu,\kappa}(E)\;&:=\; \big(2 \sqrt{1-E^2}\big)^{2\sqrt{1-\nu^2}}\;\frac{\Gamma(-2\sqrt{1-\nu^2})}{\Gamma(2\sqrt{1-\nu^2})}\;\frac{\nu\sqrt{\frac{1-E}{1+E}}+\kappa-\sqrt{1-\nu^2}}{\nu\sqrt{\frac{1-E}{1+E}}+\kappa+\sqrt{1-\nu^2}}\;\times \\
  &\qquad\qquad\qquad\times\frac{\Gamma\big(\frac{\nu E }{\sqrt{1-E^2}}+\sqrt{1-\nu^2}\big)}{\Gamma\big(\frac{\nu E}{\sqrt{1-E^2}}-\sqrt{1-\nu^2}\big)}\,.
 \end{split}
 \end{equation}
 The starting index of the enumeration is $n_0=0$ if $\kappa$ and $\nu$ have the same sign, and $n_0=1$ otherwise.
\end{theorem}

Equation \eqref{eq:fEn_formula}  provides the implicit formula for the eigenvalues of the generic extension $h_\beta$. 
A formula of the eigenfunctions corresponding to the eigenvalues $E^{(\beta)}_n$ is found in the course of the proof -- see \eqref{eq:psi_eigenf} below.

In particular, equation \eqref{eq:fEn_formula} contains the finite structure formula\index{Sommerfeld fine structure formula} 
for the distinguished extension $h_{\mathrm{D}}$ ($\beta=\infty$). 

\begin{corollary}\label{cor:eigenvalues_distinguished}
 Under the assumptions of Theorem \ref{thm:spectrum-beta}, let $h_{\mathrm{D}}$ be the distinguished (i.e., $\beta=\infty$) self-adjoint extension of $h$. Then the eigenvalues $(E_n)_{n=n_0}^\infty$ of $h_{\mathrm{D}}$ are given by
 \begin{equation}\label{eq:EVEnk1}
  E_n\;=\;-\,\mathrm{sign}(\nu)\,\Big(1+\frac{\nu^2}{(n+\sqrt{1-\nu^2})^2}\Big)^{\!-\frac{1}{2}}\,,
 \end{equation}
 the starting index of the enumeration being $n_0=0$ if $\kappa$ and  $\nu$ have the same sign, and $n_0=1$ otherwise.
\end{corollary}

 \begin{proof}[Proof of Theorem \ref{thm:spectrum-beta}]
 One starts with the differential problem \eqref{eq:diffprobnow}, re-written in the form \eqref{eq:EigenvalueEq1}-\eqref{eq:EigenvalueEq2}.

For a solution $\phi$ to \eqref{eq:EigenvalueEq1} with given $E\in(-1,1)$ it is convenient to introduce, in analogy to \eqref{eq:change}, the two scalar functions $u_1$ and $u_2$ such that 
\begin{equation}\label{eq:def_u1u2}
 \begin{split}
  \phi^+&\;=\;\sqrt{1+E} \, (u_1+u_2) \, ,\\
  \phi^-&\;=\; \sqrt{1-E} \, (u_1-u_2)\,.
 \end{split}
\end{equation}
Plugging \eqref{eq:def_u1u2} into \eqref{eq:EigenvalueEq1}-\eqref{eq:EigenvalueEq2} yields
\begin{equation}\label{eq:u1u2eq}
 \begin{split}
  \textstyle u_2'+\Big(\frac{\kappa}{\rho}+\frac{\nu}{\rho\,\sqrt{1-E^2}\,} \Big) u_1 +\frac{\nu E }{\rho\,\sqrt{1-E^2}\,} u_2 \;&=\; 0 \, , \\
   \textstyle -u_1'+\Big(1+\frac{\nu E }{\rho\,\sqrt{1-E^2}\,} \Big) u_1+ \Big(\frac{\nu}{\rho\,\sqrt{1-E^2}\,}-\frac{\kappa}{\rho} \Big) u_2 \;&=\;0\,,
 \end{split}
\end{equation}
and solving for $u_1$ in the first equation above and plugging it into the second equation gives a second order differential equation for $u_2$ which, re-written for the scalar function $v:=\rho^B u_2$, takes the form 
\begin{equation}\label{eq:hypergeom-v}
 \rho \,v''+(1-2B-\rho) \,v'-\Big(\frac{\nu E}{\sqrt{1-E^2}}-B \Big) v \; = \; 0\,.
\end{equation}

Equation \eqref{eq:hypergeom-v} is a confluent hypergeometric equation\index{confluent hypergeometric equation}, analogously to \eqref{eq:IV-confl} above, thus with two linearly independent solutions given by the Kummer function\index{Kummer functions} $M_{a,b}$ and the Tricomi function\index{Tricomi functions} $U_{a,b}$ already considered for \eqref{eq:IV-confl}, now with parameters 
\begin{equation}\label{eq:parameters_ab}
 \textstyle a\;=\;\frac{\nu E}{\sqrt{1-E^2}}-B\,,\qquad b\;=\;1-2B\,.
\end{equation}
 As observed already, only $U_{a,b}$ belongs to $L^2(\mathbb{R}^+,\mathbb{C}, e^{-\rho}\ud\rho)$: indeed, \eqref{eq:asy1}-\eqref{eq:asy2} show that
\[
 \begin{array}{l}
  M_{a,b}(\rho)\;=\;e^r\,\frac{\,r^{a-b}}{\Gamma(a)}(1+O(r^{-1})) \\
  \,\,U_{a,b}(\rho)\;=\;r^{-a}(1+O(r^{-1}))
 \end{array}\qquad \textrm{as } r\to +\infty\,.
\]
 With $u_2=\rho^{-B} v=\rho^{-B} U_{a,b}(\rho)$, and with $u_1$ determined by \eqref{eq:u1u2eq} and the property
\[
 U'_{a,b}(\rho)\;=\;-a\, U_{a+1,b+1}(\rho)\,,
\]
 one reconstructs the solution $\phi$ by means of \eqref{eq:def_u1u2} and finds
\begin{equation}
 \phi^{\pm}(\rho)\;=\;\frac{\rho^{-B}}{\,\kappa+\frac{\nu}{\sqrt{1-E^2}}} \textstyle \big( \big(B\pm\nu\sqrt{\frac{1-E}{1+E}}\pm \kappa \big) \,U_{a,b}(\rho)+a\, \rho\, U_{a+1,b+1}(\rho) \big)\,.
\end{equation}

Correspondingly, the solution $\psi=U^{-1}\phi$ to the differential problem \eqref{eq:diffprobnow}, where $U:L^2(\mathbb{R}^+,\mathbb{C}^2,\ud r)\to L^2(\mathbb{R}^+,\mathbb{C}^2,e^{-\rho}\ud \rho)$ is the unitary map \eqref{eq:rescaling_map_U}, takes the form
\begin{equation}\label{eq:psi_eigenf}
\begin{split}
 \psi^{\pm}(r)\;&=\;\frac{(2r\sqrt{1-E^2})^{-B}\,e^{-r\sqrt{1-E^2}}}{\,\kappa+\frac{\nu}{\sqrt{1-E^2}}} \Big( \textstyle \sqrt{1\pm E}\,\big(B\pm\nu\sqrt{\frac{1-E}{1+E}}\pm \kappa \big) \,U_{a,b}(2r\sqrt{1-E^2}) \\
 &\qquad\qquad\qquad\qquad\qquad\quad+2ar\sqrt{1-E^2}\, U_{a+1,b+1}(2r\sqrt{1-E^2}) \Big)\,.
\end{split}
\end{equation}
From the above expression one deduces the asymptotics
\begin{equation}\label{eq:psiplus}
\begin{split}
\!\!\!\!\!\!\!\psi^+(r) \;&=\; \textstyle\frac{\Gamma(1-b)}{\Gamma(1+a-b)} \big(B+\nu\sqrt{\frac{1-E}{1+E}}+ \kappa \big) r^{-B} \\
&\qquad + \frac{\Gamma(b-1)}{\Gamma(a)} (2\sqrt{1-E^2})^{2B} \big(\nu \sqrt{\frac{1-E}{1+E}} + \kappa - B \big) r^B  \\
&\qquad + o(r^{1/2})\qquad\textrm{as }r\downarrow 0\,.
\end{split}
\end{equation}

Since $\widetilde{h}\psi=E\psi\in L^2(\mathbb{R}^+,\mathbb{C}^2,\ud r)$, then $\psi\in\mathcal{D}(h^*)$, owing to \eqref{DSclosureDS*}. By comparing 
\eqref{eq:psiplus} above with the general short-distance asymptotics \eqref{eq:IV-coeff_a_b_BIS} for a function in $\mathcal{D}(h^*)$ one reads out the coefficients
\begin{equation}\label{eq:g0g1_final}
 \begin{split}
  g_0^+ \;&=\; \textstyle \frac{\Gamma(2 B)}{\Gamma(\frac{\nu E}{\sqrt{1-E^2}}+B)} \big(\nu\sqrt{\frac{1-E}{1+E}} + \kappa+B \big)\,, \\
g_1^+ \;&=\; \textstyle(2 \sqrt{1-E^2})^{2B}\,\frac{\Gamma(-2 B)}{\Gamma(\frac{\nu E }{\sqrt{1-E^2}} - B)} \, \big(\nu \sqrt{\frac{1-E}{1+E}}  + \kappa - B \big)
 \end{split}
\end{equation}
 of the small-$r$ expansion $\psi(r)=g_0 r^{-B}+g_1 r^B+o(r^{\frac{1}{2}})$.

 One can now apply to such $\psi$ the classification formulas from Theorem \ref{thm:classification_bc}. Upon setting
  \begin{equation}\label{eq:setting_of_F}
  \begin{split}
   \mathfrak{F}_{\nu,\kappa}(E)\;:=&\;\frac{g_1^+}{g_0^+} \\
   =&\; (2 \sqrt{1-E^2})^{2B}\;\frac{\Gamma(-2B)}{\Gamma(2B)}\;\frac{\Gamma(\frac{\nu E }{\sqrt{1-E^2}}+B)}{\Gamma(\frac{\nu E}{\sqrt{1-E^2}}-B)}\;\frac{\nu\sqrt{\frac{1-E}{1+E}}+\kappa-B}{\nu\sqrt{\frac{1-E}{1+E}}+\kappa+B}\,,
  \end{split}
 \end{equation}
 one deduces from \eqref{eq:g0g1_final} and \eqref{eq:IV-Sbeta_bc} that the function $\psi\in\mathcal{D}(h^*)$ determined so far actually belongs to $\mathcal{D}(h_\beta)$, and therefore is a solution to $h_\beta\psi=E\psi$, if and only if $E$ satisfies
\begin{equation}\label{eq:fE_beta_equation}
 \mathfrak{F}_{\nu,\kappa}(E)\;=\;c_{\nu,\kappa}\, \beta + d_{\nu,\kappa}\,,
\end{equation}
which then proves \eqref{eq:fEn_formula}.

It is straightforward to deduce from the properties of the $\Gamma$-function that the map $(-1,1)\ni E\mapsto \mathfrak{F}_{\nu,\kappa}(E)$ has the following features. First, $\mathfrak{F}_{\nu,\kappa}$ has vertical asymptotes corresponding to the roots of 
\begin{equation}\label{eq:roots}
 \frac{{ \textstyle\Gamma\big(\frac{\nu E }{\sqrt{1-E^2}}+B}\big)}{{ \textstyle\Gamma\big(\frac{\nu E }{\sqrt{1-E^2}}-B}\big)}\times\frac{{\textstyle \nu\sqrt{\frac{1-E}{1+E}}+\kappa-B}}{{\textstyle{\nu\sqrt{\frac{1-E}{1+E}}+\kappa+B}}}\;=\;\infty\,.
\end{equation}
 Second, such roots (see also the proof of Corollary \ref{cor:eigenvalues_distinguished} below) are indeed countably many, and the corresponding asymptotes are located at the points $E=E_n$, with $E_n$ given by formula \eqref{eq:EVEnk1}. Therefore, the asymptotes accumulate at $E=-1$ for $\nu>0$ and at $E=1$ for $\nu<0$. When $\nu>0$, in each interval $(E_{n+1},E_{n})$, as well as in the interval $(E_{n_0},1)$, $\mathfrak{F}_{\nu,\kappa}$ is smooth and strictly monotone decreasing; the value $\mathfrak{F}_\nu(1)$ is finite and negative. When $\nu<0$ one has conversely that in each interval $(E_{n},E_{n+1})$, as well as in the interval $(-1,E_{n_0})$, $\mathfrak{F}_{\nu,\kappa}$ is smooth and strictly monotone increasing.

Thus, the range of $\mathfrak{F}_{\nu,\kappa}$ is the whole real line, which makes the equation \eqref{eq:fE_beta_equation} always solvable for any $\beta$, again with a countable collection of roots. This completes the proof.
\end{proof}

\begin{figure}[t!]
\begin{center}
\includegraphics[scale=0.35]{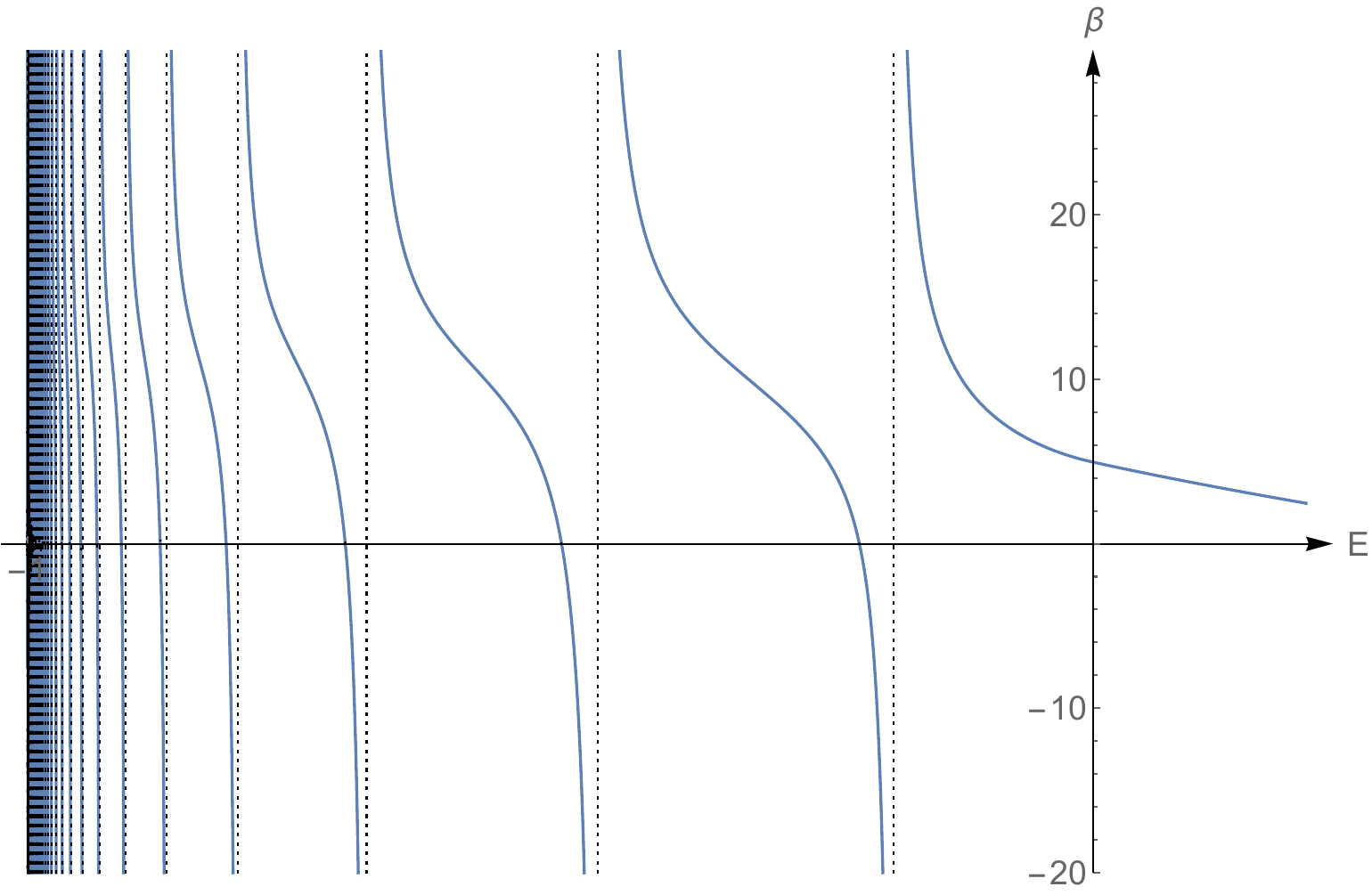}
\end{center}
\caption{Plot of $\mathfrak{F}_{\nu,\kappa}(E)$ for $\kappa=1$ and $\nu=0.9$ for $E \in (-1,0.3)$.}\label{fig:mathfrakF}
\end{figure}

The behaviour of $E\mapsto \mathfrak{F}_{\nu,\kappa}(E)$ discussed in the proof above is illustrated in Figure \ref{fig:mathfrakF}  for $\kappa=1$ and $\nu>0$. Observe that in this case the points $E_n$ where the vertical asymptotes are located at are all negative and $E_n\to -1$ as $n\to +\infty$. For
$\beta\in(-\infty,\mathfrak{F}_{\nu,\kappa}(1))\cup(d_{\nu,\kappa},+\infty)$ all such roots are strictly negative, whereas for $\beta\in (\mathfrak{F}_{\nu,\kappa}(1),d_{\nu,\kappa})$ the lowest root (and only that one) is strictly positive. As expected, $\mathfrak{F}_{\nu,\kappa}(0)=d_{\nu,\kappa}$, as one can easily see by comparing the value $\mathfrak{F}_{\nu,\kappa}(0)$ obtained from \eqref{eq:setting_of_F} with the quantity $d_{\nu,\kappa}$ given by \eqref{eq:IV-defcd} and \eqref{eq:IV-def_qpm}.

%

\begin{proof}[Proof of Corollary \ref{cor:eigenvalues_distinguished}]
 The goal is to determine the roots of $\mathfrak{F}_{\nu,\kappa}(E)=\infty$, equivalently, the roots of  equation \eqref{eq:roots}. For each of the four factors
 \[
  \begin{split}
   P_\nu(E)\;&:=\; \textstyle\Gamma\big(\frac{\nu E }{\sqrt{1-E^2}}+B\big)\,, \\
   Q_{\nu,\kappa}(E)\;&:=\; \textstyle \nu\sqrt{\frac{1-E}{1+E}}+\kappa-B\,, \\
   R_{\nu,\kappa}(E)\;&:=\;\textstyle\nu\sqrt{\frac{1-E}{1+E}}+\kappa+B\,, \\
   S_{\nu}(E)\;&:= \;\textstyle\Gamma\big(\frac{\nu E }{\sqrt{1-E^2}}-B\big)
  \end{split}
 \]

 on the l.h.s.~of \eqref{eq:roots} it is straightforward to find the following.
 \begin{itemize}
  \item $P_\nu(E)=\infty$ for $\frac{\nu E }{\sqrt{1-E^2}}+B=-n$, $n\in\mathbb{N}_0$, and hence for $E=-\mathrm{sign}(\nu)\,\mathcal{E}_n$ with 
  \begin{equation}\label{eq:ProvaCorollario}
   \mathcal{E}_n\;:=\;\Big({ 1+\frac{\nu^2}{(n+\sqrt{1-\nu^2})^2} \Big)^{\!-\frac{1}{2}}}\,.
  \end{equation}
   \item $Q_{\nu,\kappa}(E)=0$ for 
  \[
   \begin{array}{ll}
    E=-B\,, & \quad\textrm{if $\kappa=-1$, and $\nu>0$}, \\
    E=B\,,  & \quad\textrm{if $\kappa=1$ and $\nu<0$}, \\
    \textrm{no value of $E$} & \quad \textrm{otherwise\,.}
   \end{array}
  \]
  \item $R_{\nu,\kappa}(E)=0$ for 
  \[
   \begin{array}{ll}
    E=-B\,, & \quad\textrm{if $\kappa=1$ and $\nu<0$}, \\
    E=B\,, & \quad\textrm{if $\kappa=-1$, and $\nu>0$}, \\
    \textrm{no value of $E$} & \quad \textrm{otherwise\,.}
   \end{array}
  \]
    \item $S_\nu(E)=\infty$ for $\frac{\nu E }{\sqrt{1-E^2}}-B=-n$, $n\in\mathbb{N}_0$, and hence for $E=\mathrm{sign}(\nu)\,\mathcal{E}_{-n}$ with $\mathcal{E}_n$ defined in \eqref{eq:ProvaCorollario}.
 \end{itemize}

 Therefore, for the problem $\mathfrak{F}_{\nu,\kappa}(E)=\infty$, which is equivalent to 
 \begin{equation*}
  Z_{\nu,\kappa}(E)\;:=\;\frac{P_\nu(E)}{S_\nu(E)} \,\frac{Q_{\nu,\kappa}(E)}{R_{\nu,\kappa}(E)}\;=\;\infty\,,
 \end{equation*}
  one can distinguish the following cases. 
  
For all $\kappa$ and $\nu$, then $Z_{\nu,\kappa}(E)=\infty$ at least for $E=-\mathrm{sign}(\nu) \mathcal{E}_n$ with $n\geqslant 1$ (which makes $P_\nu$ diverge, keeping $Q_{\nu,\kappa}$, $R_{\nu,\kappa}$, and $S_\nu$ finite); the remaining possibilities $E=\pm B$ have to be discussed separately. 

If $\kappa$ and $\nu$ have the same sign, then $\lim_{E \to \pm B} Z_{\nu,\kappa}(E)$ is either zero or infinity because only one among $P_\nu$ and $S_\nu$ diverges, $Q_{\nu,\kappa}$ and $R_{\nu,\kappa}$ remaining finite. Explicitly,
\[
\begin{split}
	\lim_{E \to \mp B} Z_{\nu,\kappa}(E)\;&=\;\infty \, ,\qquad \textrm{if } \nu \gtrless 0\,, \\
		\lim_{E \to \pm B} Z_{\nu,\kappa}(E)\;&=\;0 \, ,\qquad\;\: \textrm{if } \nu \gtrless 0\,.
\end{split}
\]
Thus, the value $E=-\mathrm{sgn}(\nu)B$ is admissible and $E=\mathrm{sgn}(\nu)B$ is to be discarded. This proves formula \eqref{eq:EVEnk1} for the case $\kappa$ and $\nu$ with the same sign.

If instead $\kappa$ and $\nu$ have opposite sign, then $\lim_{E\to \pm B} Z_{\nu,\kappa}(E)$ must be either determined resolving the indeterminate form $P_\nu\cdot Q_{\nu,\kappa}=\infty\cdot 0$ ($R_{\nu,\kappa}$ and $S_\nu$ being finite) or resolving the indeterminate form $S_\nu\cdot R_{\nu,\kappa}=\infty\cdot 0$ ($P_\nu$ and $Q_{\nu,\kappa}$ being finite). Owing to the asymptotics $\Gamma(x) \sim x^{-1}$ as $x \to 0$ all these limits are finite and non-zero, which makes the values $\pm B$ not admissible. This discussion proves formula \eqref{eq:EVEnk1} for the case in which $\kappa$ and $\nu$ have opposite sign.
\end{proof}

\begin{figure}[t!]
\begin{center}
\includegraphics[scale=0.43]{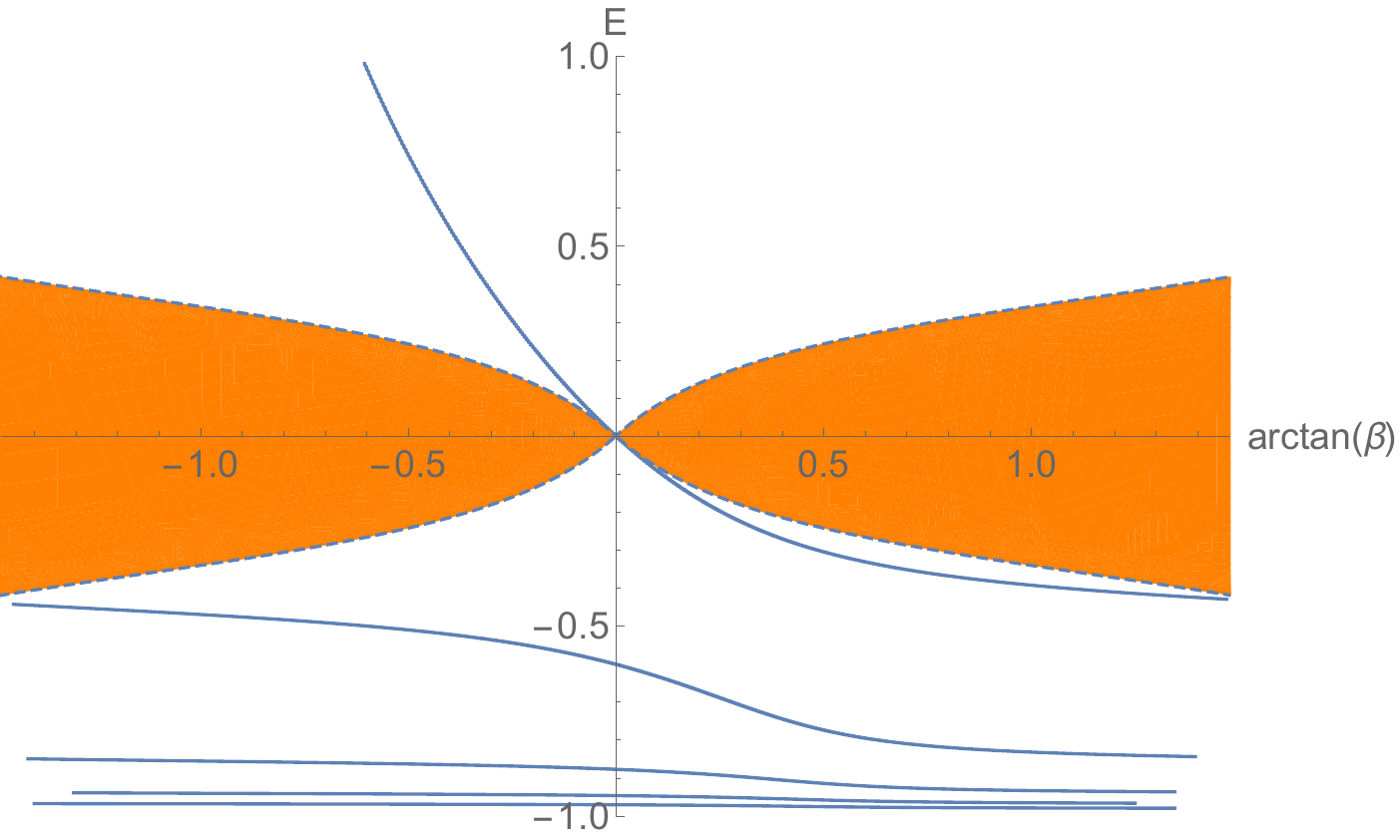}
\end{center}
\caption{Numerical computation of the eigenvalues $E^{(\beta)}_n$ as functions of $\beta$, for $\kappa=1$ and $\nu=0.9$. The shaded area is the region $|E|<E(\beta)$, with $E(\beta)$ given by \eqref{eq:estbetagap}, and indicates the estimated gap in the spectrum around zero, according to Theorem \ref{thm:spectrum-beta}(vi).}\label{fig:Evs}
\end{figure}

The first five eigenvalues $E^{(\beta)}_0,\dots,E^{(\beta)}_4$ for generic $\beta$ are plotted in Figure \ref{fig:Evs} for the concrete case $\kappa=1$, $\nu>0$. This is obtained by computing numerically the intersection points of the curve $E\mapsto \mathfrak{F}_{\nu,\kappa}(E)$ with horizontal lines corresponding to various values of $c_{\nu,\kappa}\,\beta+d_{\nu,\kappa}$. In this case when $\beta>0$ all eigenvalues are strictly negative (and accumulate to $-1$),  whereas for a region of negative $\beta$'s the first eigenvalue is positive. As expected, $E^{(\beta)}_0=0$ only for $\beta=0$: this corresponds to the sole non-invertible extension.

%

 \begin{remark}\label{rem:DCfibredstructure}
  It follows from the detailed discussion of the behaviour of  $\mathfrak{F}_{\nu,\kappa}(E)$ (in particular, of the vertical asymptotes of $\mathfrak{F}_{\nu,\kappa}(E)$) that each $E^{(\beta)}_n$ is smooth and strictly monotone in $\beta$, and it moves with continuity from $\beta=(+\infty)^-$ to $\beta=(-\infty)^+$. This results in a typical \emph{fibred structure} of the union of all the discrete spectra $\sigma_{\mathrm{disc}}(h_\beta)$, with
\begin{equation}
 \bigcup_{\beta\in(-\infty,+\infty]}\big\{E_n^{(\beta)}\,|\,n\in\mathbb{N}_0\,, n\geqslant n_0\big\}\;=\;(-1,1)\,.
\end{equation}
This is a common phenomenon for the discrete spectra of one-parameter families of self-adjoint extensions of a given densely defined symmetric operator, where each extension is a rank-one perturbation, in the resolvent sense, of a reference extension: the complement of the essential spectrum, which is the same for all the extensions, is fibred by the union of all discrete spectra. (This spectral phenomenon occurs also in the models examined in Chapter \ref{chapter-Hydrogenoid} (Remark \ref{rem:spectra_fibre}) and in Chapter \ref{cor:spectralanalysis} (Remark \ref{cor:spectralanalysis})).
 \end{remark}

 \begin{remark}
  In Figure \ref{fig:Evs} the width of the spectral gap of $h_\beta$ is highlighted (for the choice $\kappa=1$, $\nu>0$). This shows that the estimate \eqref{eq:estbetagap} for such gap (Theorem \ref{thm:DC-invertibility-resolvent-gap}(iii)), becomes more and more accurate, and in fact exact, for $\beta\to 0$ and $\beta\to +\infty$, and it is reasonably informative in between. Owing to Corollary \ref{cor:eigenvalues_distinguished} one can now write
\begin{equation}\label{eq:normahdm1}
 \|h_{\mathrm{D}}^{-1}\|\;=\;B^{-1}\;=\;(1-\nu^2)^{-\frac{1}{2}}\,,
\end{equation}
   which, plugged into \eqref{eq:estbetagap}, gives the the value
  \begin{equation}
  E(\beta)\;=\;-\frac{\beta}{1+ \beta (1-\nu^2)^{-\frac{1}{2}}}
 \end{equation}
  for the interval $(-E(\beta),E(\beta))$ entirely contained in the gap of $\sigma(h_\beta)$ around the zero energy.
 \end{remark}


%
%
%
\chapter{Quantum particle on Grushin structures}
\label{chapter-Grushin} 

   The study of a quantum particle on degenerate Riemannian manifolds, and the problem of the purely geometric confinement away from the singularity locus of the metric, as opposite to the dynamical transmission across the singularity, has recently attracted a considerable amount of attention in relation to Grushin structures (defined, e.g., in \cite{Calin-Chang-SubRiemannianGeometry}, as well as in Section \ref{sec:V-GruplaneGrucyl}) and to the induced confining effective potentials on cylinder, cone, and plane (as in the works by Nenciu and Nenciu \cite{Nenciu-Nenciu-2009}, Boscain and Laurent \cite{Boscain-Laurent-2013}, Boscain, Prandi, and Seri \cite{Boscain-Prandi-Seri-2014-CPDE2016}, Prandi, Rizzi, and Seri \cite{Prandi-Rizzi-Seri-2016}, Boscain and Prandi \cite{Boscain-Prandi-JDE-2016}, Franceschi, Prandi, and Rizzi \cite{Franceschi-Prandi-Rizzi-2017}, Gallone, Michelangeli, and Pozzoli \cite{GMP-Grushin-2018,GMP-Grushin2-2020,GMP-heat-2021}, Boscain and Neel \cite{Boscain-Neel-2019}, Pozzoli \cite{PozzoliGru-2020volume}, Beschastnyi, Boscain, and Pozzoli \cite{Boscain-Beschastnnyi-Pozzoli-2020}, Gallone and Michelangeli \cite{GM-Grushin3-2020}), as well as, more generally, on two-step two-dimensional almost-Riemannian structures (Boscain and Laurent \cite{Boscain-Laurent-2013}, Beschastnyi, Boscain, and Pozzoli \cite{Boscain-Beschastnnyi-Pozzoli-2020}, Beschastnyi \cite{IB-2021}), or also generalisations to almost-Riemannian structures in any dimension and of any step, and even to sub-Riemannian geometries, provided that certain geometrical assumptions on the singular set are taken
   (Franceschi, Prandi, and Rizzi \cite{Franceschi-Prandi-Rizzi-2017}, Prandi, Rizzi, and Seri \cite{Prandi-Rizzi-Seri-2016}). On a related note, a satisfactory interpretation of the heat-confinement in the Grushin cylinder is known in terms of Brownian motions (Boscain and Neel \cite{Boscain-Neel-2019}) and random walks (Agrachev, Boscain, Neel, and Rizzi \cite{Agra-Bosca-Neel-Rizzi-2018}).

   Underlying such analyses there is a natural problem of control of essential self-adjointness or lack thereof, whence also a natural problem of identification, classification, and analysis of self-adjoint extensions, for the minimally defined Laplace-Beltrami operator on manifold.

   Such questions are discussed in this Chapter for the planar, and, more extensively, the cylindric realisation of a Grushin-type structure, where both the self-adjointness characterisation (Sect.~\ref{sec:constant-fibre-integral}-\ref{sec:proof_xy_Euclidean}) and the spectral analysis (Sect.~\ref{sec:V-spectralanalysis}-\ref{sec:V-scattering}) are carried on by means of the Kre{\u\i}n-Vi\v{s}ik-Birman extension scheme. The whole chapter is modelled on the recent works \cite{GMP-Grushin-2018,PozzoliGru-2020volume,GMP-Grushin2-2020,GM-Grushin3-2020,GMP-heat-2021}.

\section{Grushin-type plane and Grushin-type cylinder}\label{sec:V-GruplaneGrucyl}

  The quantum models studied in this Chapter are built upon two concrete geometric settings that are representative of the general class of those two-dimensional incomplete Riemannian manifolds customarily referred to as \emph{Grushin-type manifolds}\index{Grushin!manifold} \cite[Chapter 11]{Calin-Chang-SubRiemannianGeometry}: the \emph{Grushin(-type) plane}\index{Grushin!plane} and the \emph{Grushin(-type) cylinder}.\index{Grushin!cylinder}

  The \emph{Grushin plane}\index{Grushin!plane} is the Riemannian manifold $M_1$, and more generally a \emph{Grushin(-type) plane} is a Riemannian manifold $M_\alpha\equiv(M,g_\alpha)$, for some $\alpha\in\mathbb{R}$,
  where
    \begin{equation}\label{eq:Mgalpha}
 	M^\pm \; := \; \mathbb{R}^\pm_x \times \mathbb{R}_y \, \qquad \mathcal{Z} \;:=\; \{0\} \times \mathbb{R}_y \, , \qquad M\;:=\; M^+ \cup M^- 
\end{equation}
and 
 \begin{equation}\label{eq:galphaeverywhere}
    g_\alpha\;:=\;\ud x\otimes\ud x+\frac{1}{\;|x|^{2\alpha}}\,\ud y\otimes\ud y\,.
 \end{equation}
  In its \emph{one-side} version it is the Grushin-type half-plane (left or right).
  A straightforward computation \cite{Agrachev-Boscain-Sigalotti-2008,Boscain-Laurent-2013,Pozzoli_MSc2018} shows that the Gaussian (sectional) curvature\index{Gaussian (sectional) curvature} $K_\alpha$ of $M_\alpha$ is
\begin{equation}\label{eq:curvature}
 K_\alpha(x,y)\;=\;-\frac{\,\alpha(\alpha+1)\,}{x^2}\,,
\end{equation}
hence $M_\alpha$ is a hyperbolic manifold whenever $\alpha>0$.
 Each $M_\alpha$ is clearly parallelizable, a global orthonormal frame being
\begin{equation}\label{eq:frame}
\{X_1,X_2^{(\alpha)}\}\;=\;\left\{ 
\begin{pmatrix}
 1 \\ 0
\end{pmatrix},\;
\begin{pmatrix}
 0 \\ |x|^{\alpha}
\end{pmatrix}
\right\}\equiv\;\Big\{\frac{\partial}{\partial x},|x|^{\alpha}\frac{\partial}{\partial y}\Big\}.
\end{equation} 
 One has the Lie bracket\index{Lie bracket}
\begin{equation}
 [X_1,X_2^{(\alpha)}]\;=\;\begin{pmatrix}
 0 \\ \alpha|x|^{\alpha-1}
\end{pmatrix},
\end{equation}
 and moreover, when $\alpha\in\mathbb{N}$ the fields $X_1,X_2^{(\alpha)}$ are smooth and define an \emph{almost-Riemannian structure}\index{almost-Riemannian structure} on $\mathbb{R}^2=M^+\cup\mathcal{Z}\cup M^-$, for a rigorous definition of which one may refer to \cite[Section 1]{Agrachev-Boscain-Sigalotti-2008} or \cite[Section 7.1]{Prandi-Rizzi-Seri-2016}: indeed the Lie bracket generating condition\index{Lie bracket generating condition} 
\begin{equation}
 \dim\mathrm{Lie}_{(x,y)}\,\mathrm{span}\{X_1,X_2^{(\alpha)}\}\;=\;2\qquad\forall(x,y)\in\mathbb{R}^2
\end{equation}
is satisfied in this case. For $\alpha\in\mathbb{R}\setminus\mathbb{N}$ the field $X_2^{(\alpha)}$ is \emph{not} smooth and in addition the structure is \emph{not} Lie-bracket-generating.

  It is worth adding that the notion of almost Riemannian structure informally speaking refers to a smooth $d$-dimensional  manifold $\mathcal{M}$ equipped with a family of smooth vector fields $X_1,...,X_d$ satisfying the Lie bracket generating condition: if $\mathcal{Z} \subset\mathcal{M}$ is the embedded hyper-surface of points where the $X_j$'s are not linearly independent, on $\mathcal{M}\setminus \mathcal{Z}$ the fields $X_1,\dots,X_d$ define a Riemannian structure which however becomes singular on $\mathcal{Z}$.

  In complete analogy, one defines the (two-sided) \emph{Grushin cylinder}\index{Grushin!cylinder} $M_1$ and, more generally, a \emph{Grushin(-type) cylinder} as that Riemannian manifold $M_\alpha\equiv(M,g_\alpha)$, for some $\alpha\in\mathbb{R}$, with
  \begin{equation}\label{eq:Mgalpha-cylinder}
 M^{\pm}\;:=\;\mathbb{R}^{\pm}_x\times\mathbb{S}^1_y\,,\qquad\mathcal{Z}\;:=\;\{0\}\times\mathbb{S}^1_y\,,\qquad M\;:=\;M^+\cup M^-
\end{equation}
 and $g_\alpha$ given by \eqref{eq:galphaeverywhere}, now with $(x,y)\in M$. The value $\alpha=0$ selects the Euclidean cylinder (with the circular section at $x=0$ punctured off). Analogously to the planar setting, one computes the curvature \eqref{eq:curvature} and moreover when $\alpha\in\mathbb{N}$ the fields $X_1,X_2^{(\alpha)}$ define an \emph{almost-Riemannian structure} on $\mathbb{R}\times\mathbb{S}^1=M^+\cup\mathcal{Z}\cup M^-$, owing to their smoothness and the validity of the Lie bracket generating condition 
\begin{equation}
 \dim\mathrm{Lie}_{(x,y)}\,\mathrm{span}\{X_1,X_2^{(\alpha)}\}\;=\;2\qquad\forall(x,y)\in\mathbb{R}\times\mathbb{S}^1\,.
\end{equation}

 In either case, plane or cylinder, to each $M_\alpha$ one naturally associates the Riemannian volume form\index{Riemannian volume form}
\begin{equation}\label{eq:volumeform}
 \mu_\alpha\;:=\;\mathrm{vol}_{g_\alpha}\;=\;\sqrt{\det g_\alpha}\,\ud x\wedge\ud y\;=\;|x|^{-\alpha}\,\ud x\wedge\ud y\,.
\end{equation}
 By means of \eqref{eq:frame} and \eqref{eq:volumeform} one computes
\begin{equation}
 \begin{split}
  X_1^2\;&=\;\frac{\partial^2}{\partial x^2}\,,\qquad \quad \, \,\,\,\,\,\, \mathrm{div}_{\mu_\alpha}X_1\;=\;-\frac{\alpha}{|x|}\,, \\
  (X_2^{(\alpha)})^2\;&=\;|x|^{2\alpha}\frac{\partial^2}{\partial y^2}\,,\quad \mathrm{div}_{\mu_\alpha}X_2^{(\alpha)}\;=\;0\,,
 \end{split}
\end{equation}
whence
\begin{equation}\label{eq:Deltamualpha}
\begin{split}
 \Delta_{\mu_\alpha}\;&=\;\mathrm{div}_{\mu_\alpha}\circ\nabla \\
 &=\;X_1^2+X_2^2+(\mathrm{div}_{\mu_\alpha}X_1)X_1+(\mathrm{div}_{\mu_\alpha}X_2^{(\alpha)})X_2^{(\alpha)} \\
 &=\;\frac{\partial^2}{\partial x^2}+|x|^{2\alpha}\frac{\partial^2}{\partial y^2}-\frac{\alpha}{|x|}\,\frac{\partial}{\partial x}\,,
\end{split}
\end{equation}
which is the (Riemannian) Laplace-Beltrami operator\index{Laplace-Beltrami operator} on $M_\alpha$.

 \section{Geodesic incompleteness}

 It is straightforward to observe that the Grushin-type plane (respectively, Grushin-type cylinder) $M_\alpha$ is geodesically incomplete\index{geodesic incompleteness} for $\alpha>0$. Indeed, 
 $M_\alpha$ is obviously incomplete as a metric space, which can be seen by the non-convergent Cauchy sequence of points $(n^{-1},y_0)\in M$ as $n\to\infty$ for fixed $y_0\in\mathbb{R}$ (respectively $y_0\in\mathbb{S}^1$), and the conclusion then follows from a standard Hopf-Rinow theorem\index{theorem!Hopf-Rinow} 
\cite[Theorem 2.8, Chapter 7]{DoCarmo-Riemannian}.

\begin{figure}
\begin{center}
\includegraphics[width=5cm]{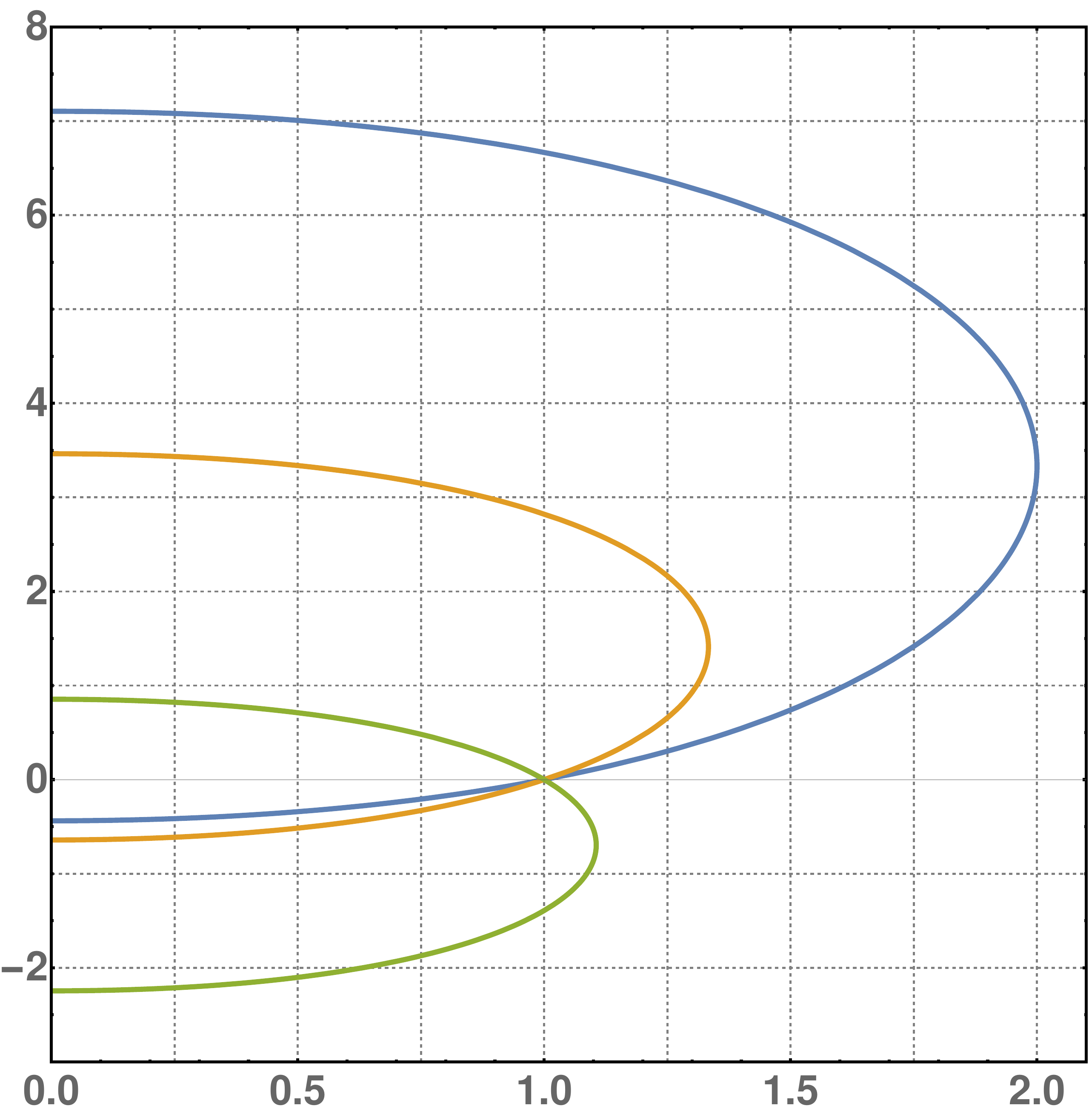}\quad\includegraphics[width=5cm]{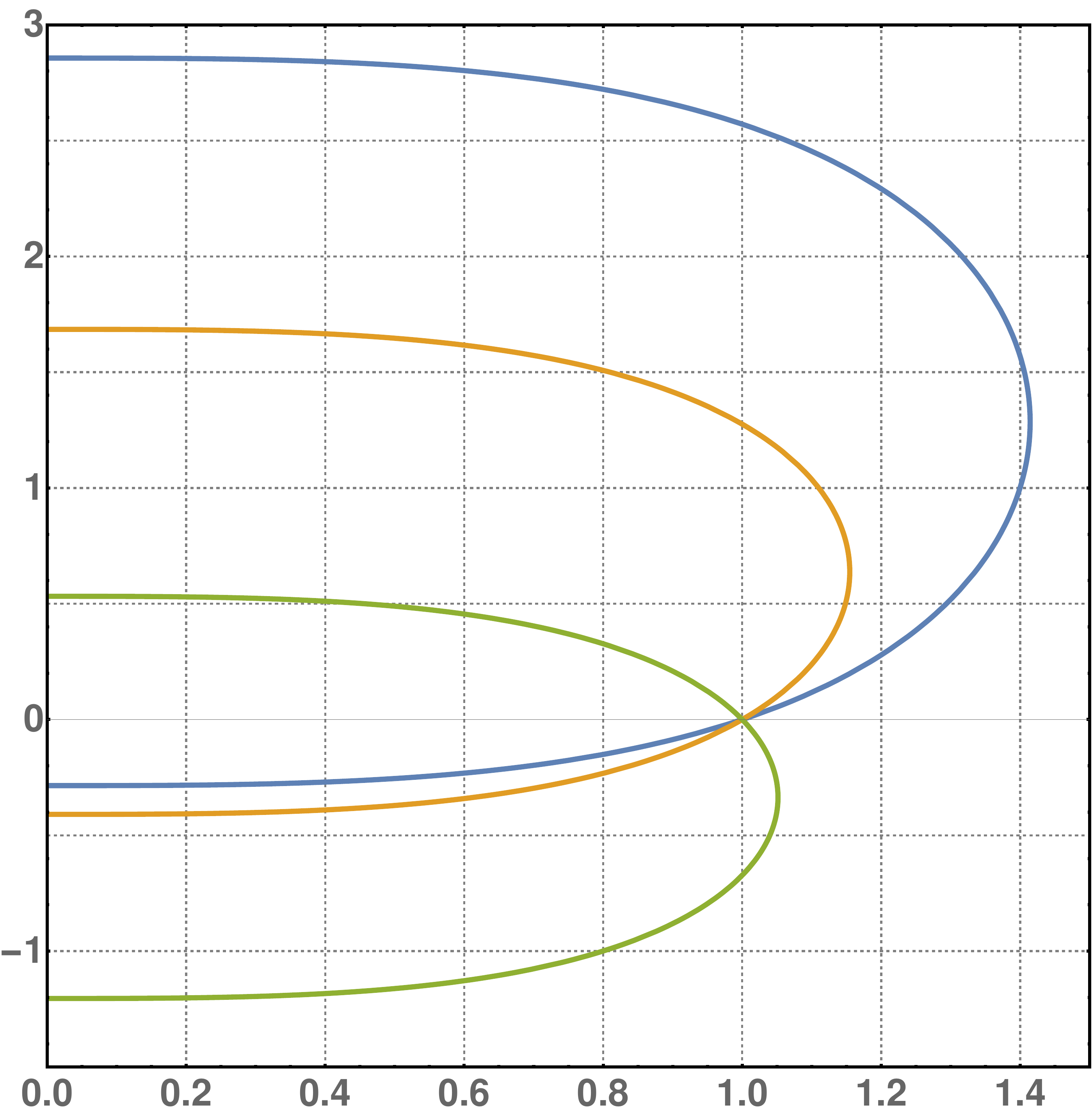}
\caption{Geodesics in Grushin-type (half-)plane obtained from \eqref{eq:system_for_geodesics} for various choices of $\theta$. Left: $\alpha=\frac{1}{2}$. Right: $\alpha=1$.} \label{fig:geodesics}
\end{center}
\end{figure}

 In fact, a \emph{stronger} property holds: not only is it possible to find \emph{one} geodesic curve that passes through a given arbitrary point $(x_0,y_0)\in M$ and reaches the boundary $\partial M$ in finite time in the past or in the future -- which is in fact geodesic incompleteness, and in the present case is trivially seen by considering the geodesic $y=y_0$ in the discussion that follows -- but furthermore it can be proved that \emph{all} geodesics passing through any $(x_0,y_0)$ at $t=0$ intercept the $y$-axis at finite times $t_\pm$ with $t_-<0<t_+$, with the sole exception of the geodesic line $y=y_0$ along which the boundary is reached only in one direction of time (Fig.~\ref{fig:geodesics}).

 \begin{proposition}\label{thm:incompleteness}
  Let $\alpha>0$ and let $M_\alpha\equiv(M,g_\alpha)$ denote either the Grushin-type plane or the Grushin-type cylinder defined above. \emph{All} geodesics passing through a generic point $(x_0,y_0)\in M$ escape from $M_\alpha$.
 \end{proposition}

 The proof for the special case $\alpha=1$ can be found, e.g., in \cite[Sect.~11.2]{Calin-Chang-SubRiemannianGeometry} or \cite[Sect.~3.1]{Boscain-Laurent-2013}; the general case is discussed here below following \cite[Theorem 2.2]{GMP-Grushin-2018}.


 \begin{proof}[Proof of Proposition \ref{thm:incompleteness}]
 The following proof is for the Grushin-type right half-plane; the argument for the two-sided plane or for the cylinder is virtually identical. Recall that, as a consequence of Pontryagin's maximum principle\index{theorem!Pontryagin (maximum principle)}\index{maximum principle (Pontryagin)} (see, e.g., \cite[Sect.~3.4]{Agrachev-Barilari-Boscain-subriemannian} or \cite[Sect.~2.2]{Boscain-Laurent-2013}), the geodesics on $M_\alpha$ are projections onto $M$ of solutions to the Hamilton equations associated with the Hamiltonian that, with respect to the orthonormal frame \eqref{eq:frame}, reads
\begin{equation}\label{eq:Hamiltonian_for_geodesics}
 h_\alpha(x,y;P_x,P_y)\;:=\;\frac{1}{2}\big(\langle X_1,P\rangle_{\mathbb{R}^2}^2+\langle X_2^{(\alpha)},P\rangle_{\mathbb{R}^2}^2\big) \;=\;\frac{1}{2}\big(P_x^2+x^{2\alpha} P_y^2\big)\,,
 \end{equation}
where $P:=(P_x,P_y)\in T^*_{(x,y)}M$ is the vector of the momenta associated with the coordinates $(x,y)$. The corresponding Hamiltonian system is therefore
\begin{equation}\label{eq:Hamiltoniansystem}
 \begin{split}
  \dot{x}\;&=\;P_x\,,\qquad\qquad \dot{P}_x\;=\;-\alpha x^{2\alpha-1}\,, \\
  \dot{y}\;&=\;x^{2\alpha} P_y\,,\qquad\;\;\; \dot{P}_y\;=\;0\,.
 \end{split}
\end{equation}
The local existence and uniqueness of a solution to \eqref{eq:Hamiltoniansystem} with prescribed values of $(x,y)$ and $(P_x,P_y)$ at $t=0$ is standard.

One deduces from \eqref{eq:Hamiltoniansystem} that the geodesic passing at $t=0$ through the point $(x(0),y(0))=(1,0)\in M$ with direction $(\dot{x}(0),\dot{y}(0))=(\cos\theta,\sin\theta)$, for fixed $\theta\in[0,2\pi)$, is the solution $(x(t),y(t))$ to
\begin{equation}\label{eq:system_for_geodesics}
 \begin{cases}
  \;\ddot{x}+\alpha\,\sin^2\theta\,x^{2\alpha-1}\,=\,0\,,\quad & x(0)=1\,,\quad \dot{x}(0)=\cos\theta \,, \\
  \;y(t)\,=\,\sin\theta\int_0^t x(\tau)^{2\alpha}\,\ud \tau\,.
 \end{cases}
\end{equation}

The exceptional cases $\theta=0$ and $\theta=\pi$ yield, respectively, the geodesics $(1+t,0)$ and $(1-t,0)$, both reaching the boundary $\partial M$, respectively at the instants $t=-1$ and $t=1$.

Generically, there are two instants $t_\pm$ with $t_-<0<t_+$ such that $x(t_\pm)=0$ (Fig.~\ref{fig:geodesics}).
This is seen as customary by exploiting the conservation of $h_\alpha$ along each geodesic, that is, the conservation of the quantity
\begin{equation}\label{eq:Exxdot}
E(x,\dot x) \; := \; \frac{1}{2}\big(\dot{x}^2+ \sin^2 \theta \,x^{2 \alpha}\big)\;=\;\frac{1}{2}
\end{equation}
computed from \eqref{eq:Hamiltonian_for_geodesics}-\eqref{eq:Hamiltoniansystem} with the initial values $(x(0),y(0))=(1,0)$, $(\dot{x}(0),\dot{y}(0))$ $=(\cos\theta,\sin\theta)$, or equivalently computed from \eqref{eq:system_for_geodesics}. Suitably interpreting the sign of $\dot{x}$ and integrating the identity
\begin{equation}
 \ud t\;=\;\pm \big(2 E(x(0),\dot{x}(0))-\sin^2 \theta\, x^{2\alpha}\big)^{-\frac{1}{2}}\ud x
\end{equation}
obtained from \eqref{eq:Exxdot}, one computes the time $T(x_{\mathrm{in}}\to x_{\mathrm{fin}})$ needed for $x(t)$ to reach a final value $x_{\mathrm{fin}}$ from and initial value $x_{\mathrm{in}}$ along a geodesic $\gamma$.
Thus,
\begin{itemize}
 \item if $\cos\theta\leqslant 0$, then $\dot{x}(t)=\cos^2\theta-\alpha\sin^2\int_0^t x(\tau)^{2\alpha-1}\ud\tau<0$, and therefore
 \[
  \begin{split}
   T\;&=\;-\int_{1}^0\frac{ \ud x}{\sqrt{1-\sin^2 \theta \, x^{2\alpha}}} \;\leqslant\; \int_0^1 \frac{\ud x}{\sqrt{1-x^{2\alpha}}} \\
   &=\;\frac{1}{\alpha}\int_0^1  \frac{ \ud s}{\,s^{1-\frac{1}{\alpha}} \sqrt{1- s}\sqrt{1+s}}\;<\;+\infty\,;
  \end{split}
 \]
 \item if $\cos \theta > 0$ and $\sin \theta \neq 0$, then $\dot{x}$ changes sign at the critical point $x=x_c:=|\sin\theta|^{-\frac{1}{\alpha}}$ and the above computation is modified as
\begin{equation*}
\begin{split}
T\;&=\; \int_1^{x_c} \frac{\ud x}{\sqrt{1-\sin^2 \theta \,x^{2 \alpha}}} - \int_{x_c}^0 \frac{\ud x}{\sqrt{1-\sin^2 \theta \,x^{2 \alpha}}}\\
&= \;\int_0^{|\sin\theta|^{-\frac{1}{\alpha}}}\!\!\!\frac{\ud x}{\sqrt{1-\sin^2\theta\,x^{2\alpha}}\,}\;=\; \frac{1}{|\sin \theta|^{\frac{1}{\alpha}}} \int_0^1 \frac{\ud s}{\sqrt{1-s^{2\alpha}}} \;<\;+\infty\,.
\end{split}
\end{equation*}
 \end{itemize}

This argument shows the finiteness of the above-mentioned positive instant $t_+$ of reach of $\partial M$, and the finiteness of $t_-$ follows by the same argument reverting the sign of $t$ in the equations.

Clearly the choice of the initial point $(1,0)$ is non-restrictive and \emph{any} other point $(x_0,y_0)\in M$ can be treated the same way, thanks to the translational invariance of the metric along the $y$-direction.
 \end{proof}

\section{Geometric quantum confinement and transmission protocols}\index{geometric quantum confinement}

 The Grushin-type plane (respectively, Grushin-type cylinder) $M_\alpha$ naturally leads to the introduction of the Hilbert space
\begin{equation}\label{eq:Halphaspace}
 \cH_\alpha\;:=\;L^2(M,\ud\mu_\alpha)\,,
\end{equation}
understood as the completion of $C^\infty_c(M)$ (the space of smooth and compactly supported functions on $M$) with respect to the scalar product
\begin{equation}
 \langle \psi,\varphi\rangle_{\alpha}\;:=\;\iint_{M}\overline{\psi(x,y)}\,\varphi(x,y)\,\frac{1}{|x|^{\alpha}}\,\ud x\,\ud y\,.
\end{equation}
Quantum mechanically, $\cH_\alpha$ is the Hilbert space for a Schr\"{o}dinger particle constrained on $M_\alpha$.

 One can also naturally associate to the Laplace-Beltrami operator\index{Laplace-Beltrami operator} $\Delta_{\mu_\alpha}$ (see \eqref{eq:Deltamualpha} above) a minimal domain dense in $\cH_\alpha$, the smooth functions compactly supported away from $\partial M$, thus defining 
the \emph{minimal free Hamiltonian}\index{minimal free Hamiltonian}
\begin{equation}\label{eq:V-Halpha}
 H_\alpha\;:=\;-\Delta_{\mu_\alpha}\,,\qquad\mathcal{D}(H_\alpha)\;:=\;C^\infty_c(M)\,.
\end{equation}
 The Green identity shows that $H_\alpha$ is a (densely defined) symmetric and non-negative operator acting on $\cH_\alpha$. $H_\alpha$ is the candidate quantum Hamiltonian of kinetic energy, or `free' quantum Hamiltonian, namely the Hamiltonian governing the evolution of a quantum particle constrained on $M_\alpha$ which evolves free from interactions with external electric or magnetic fields, and only moves under the effect of the underlying geometry of $M_\alpha$, in complete analogy to the free motion along geodesics of a classical particle constrained on $M_\alpha$. Obviously, the above quantum picture makes sense for \emph{self-adjoint} realisations of $H_\alpha$.

 The problem of realising $H_\alpha$ self-adjointly on $\cH_\alpha$ is clearly well-posed, as the operator is lower semi-bounded, has therefore equal deficiency indices (Sect.~\ref{sec:I-symmetric-selfadj}), and hence admits self-adjoint extensions (Theorem \ref{thm:vonN_thm_selfadj_exts}). The alternative between $H_\alpha$ having only one trivial self-adjoint extension, its closure, or an infinity of inequivalent ones, is interpreted as the circumstance that the Grushin plane (respectively, cylinder) induces the \emph{geometric quantum confinement}\index{geometric quantum confinement} of the particle constrained on $M_\alpha$, or instead a quantum \emph{transmission protocol}\index{transmission protocol}\index{Grushin transmission protocol} across the metric's singularity locus $\mathcal{Z}$.

 Such an interpretation stems naturally from the fact that, with respect to the orthogonal direct sum decomposition
 \begin{equation}\label{eq:V-Hspacedirectsum}
  \cH_\alpha\;=\;L^2(M,\ud\mu_\alpha)\;\cong\;L^2(M^-,\ud\mu_\alpha)\oplus L^2(M^+,\ud\mu_\alpha)\,,
 \end{equation}
 the operator $H_\alpha$ is reduced (Sect.~\ref{sec:I_invariant-reducing-ssp}) as
 \begin{equation}
  \begin{split}
     H_\alpha\;=&\; H_\alpha^-\oplus H_\alpha^+\,, \\
     H_\alpha^\pm\;:=&\;-\Delta_{\mu_\alpha}\,,\quad\mathcal{D}(H_\alpha^\pm)\;:=\;C^\infty_c(M^\pm)\,,
  \end{split}
 \end{equation}
 whence also (Sect.~\ref{sec:I_invariant-reducing-ssp})
 \begin{equation}\label{eq:V-Halphadirectsum}
   \overline{H_\alpha}\;=\; \overline{H_\alpha^-}\oplus \overline{H_\alpha^+}\,.
 \end{equation}
 Thus, if (depending on the metric's singularity and hence on $\alpha$) $H_\alpha$ is essentially self-adjoint on $\cH_\alpha$, and hence so too are $H_\alpha^\pm$ on $L^2(M^\pm,\ud\mu_\alpha)$, then \eqref{eq:V-Halphadirectsum} expresses the orthogonal sum of self-adjoint operators (each in the respective Hilbert space), and
  \begin{equation}
  e^{-\ii t \overline{H_\alpha}}\;=\;e^{-\ii t \overline{H_\alpha^-}}\oplus e^{-\ii t \overline{H_\alpha^+}}\,,\qquad \forall\,t\in\mathbb{R}\,.
 \end{equation}
 Such a decomposition of the Schr\"{o}dinger flow implies, in particular, that starting from an initial datum $\psi_0\in\mathcal{D}(\overline{H_\alpha})$ with support, say, only within $M^+$, the \emph{unique} solution $\psi\in C^1(\mathbb{R}_t,L^2(M,\ud\mu_\alpha))$ to the Cauchy problem
 \begin{equation}
  \begin{cases}
   \;\ii\frac{\ud}{\ud t} \psi \!\!&=\;\overline{H_\alpha}\,\psi \, , \\
   \;\psi|_{t=0}\!\!&=\;\psi_0
  \end{cases}
 \end{equation}
 remains for all times supported (`confined', or `trapped') in $M^+$. The quantum particle in this case never crosses the $y$-axis towards the other half of the manifold, stays permanently away from $\mathcal{Z}$, and no quantum information escapes from $M^+$. \emph{For all times} the quantum particle's wave-function remains in $\mathcal{D}(\overline{H_\alpha^+})$ and need not be characterised by non-trivial boundary conditions at $\mathcal{Z}$. It is just the geometry of $M_\alpha$ that produces this effect (the essential self-adjointness of $H_\alpha$) and therefore one speaks of purely \emph{geometric quantum confinement}\index{geometric quantum confinement}.

 On the opposite, when $H_\alpha$ admits non-trivial self-adjoint extensions, these are characterised by suitable, non-trivial boundary conditions of self-adjointess at $\mathcal{Z}$. Such conditions encode an interaction located at the metric's singularity, allowing for a non-vanishing behaviour of functions in the extension's domain in the vicinity of $\mathcal{Z}$. The quantum particle is thus allowed to reach (to `touch') the boundary of $M_\alpha$ from one side or both, and the purely geometric quantum confinement is lost.

 Among the non-trivial self-adjoint extensions of $H_\alpha$ there are those that preserve the decomposition structure \eqref{eq:V-Halphadirectsum}: for them, the overall Schr\"{o}dinger flow is still reduced to the left and to the right, meaning that no crossing of probability density takes place at $\mathcal{Z}$ (the $L^2$-norm of the wave function at later times remains constant separately on each half), yet with a non-trivial interaction of the particle with at least one side of the boundary.

 Then there are also self-adjoint extensions of $H_\alpha$ that are \emph{not} reduced with respect to \eqref{eq:V-Hspacedirectsum}: for each of them the boundary conditions of self-adjointness connect the behaviour of a generic function of the extension's domain from the two sides of the metric's singularity and thus prescribe an actual protocol of transmission\index{protocol of transmission}\index{Grushin transmission protocol} between the two halves.

 Notably, a sharp contrast emerges between the quantum and the classical picture. Classically, a Grushin plane or cylinder is always geodesically incomplete (Proposition \ref{thm:incompleteness}), meaning that a free particle always touches $\mathcal{Z}$ in finite time. Quantum-mechanically, instead, it turns out (Theorems \ref{thm:V-plane-conf-noconf} and \ref{thm:Halpha_esa_or_not}) that there are regimes of $\alpha$ for which $H_\alpha$ is essentially self-adjoint, thus meaning that, upon taking its closure, it is unambiguously realised as a self-adjoint free particle Hamiltonian describing dynamics where the particle remains always separated from the boundary of the manifold.

 \section{Constant-fibre direct integral for the Grushin plane}\label{sec:constant-fibre-integral}

 The self-adjointness problem for the minimal free Hamiltonian\index{minimal free Hamiltonian} $H_\alpha$ 
 on the Hilbert space $\cH_\alpha=L^2(M,\ud\mu_\alpha)$ of Grushin-type \emph{plane} reveals a cleaner and more intrinsic and manageable structure 
 in a suitable unitarily equivalent version that is presented in this Section.

  In the discussion that follows a somewhat more general Grushin-type metric of relevance in almost-Riemannian geometry can be accommodated. Mimicking the construction of Section \ref{sec:V-GruplaneGrucyl}, one considers the manifold $M_f\equiv(M,g_f)$ by replacing \eqref{eq:Mgalpha} with
\begin{equation}\label{eq:Mf}
 g_f\;:=\;\ud x\otimes\ud x+f^2(x)\,\ud y\otimes\ud y
\end{equation}
for some measurable function $f$ on $\mathbb{R}$ satisfying
\begin{equation}\label{eq:assumtpions_f}
 \begin{array}{ll} 
  \mathrm{(i)} &   f(x)>0\;\;\;\forall x\neq 0 \\
  \mathrm{(ii)} &  f(x)\geqslant \kappa \;\;\;\textrm{in a neighbourhood of $x=0$ for some $\kappa>0$} \\
  \mathrm{(iii)} &  \textrm{$f\in C^\infty(\mathbb{R}\!\setminus\!\{0\})$} \\
  \mathrm{(iv)} & 2f(x)f''(x)-f'(x)^2\geqslant 0 \;\;\;\textrm{in a neighbourhood of $x=0$}\,.
 \end{array}
\end{equation}
The ordinary Grushin-type metric is recovered when $f(x)=|x|^{-\alpha}$ and $\alpha\geqslant 0$ (compare \eqref{eq:Mf} with \eqref{eq:galphaeverywhere}): in this regime of $\alpha$ all four assumptions \eqref{eq:assumtpions_f} are satisfied. The smoothness condition (iii) is merely required to match the actual definition of Riemannian manifold; apart from that, only $C^2$-regularity is relevant for the present analysis. Condition (iv) is also going to be weakened.

%
%
%

 In this setting one has a a generalised Grushin plane with global orthonormal frame 
\begin{equation}\label{eq:frame-f}
\{X_1,X_2^{(f)}\}\;=\;\left\{ 
\begin{pmatrix}
 1 \\ 0
\end{pmatrix},\;
\begin{pmatrix}
 0 \\ 1/f(x)
\end{pmatrix}
\right\}\equiv\;\Big\{\frac{\partial}{\partial x},\frac{1}{f(x)}\frac{\partial}{\partial y}\Big\}\,,
\end{equation}
and a computation analogous to \eqref{eq:volumeform}-\eqref{eq:Deltamualpha} shows that the associated Laplace-Beltrami\index{Laplace-Beltrami operator} operator $\Delta_f\equiv\Delta_{\mu_{g_f}}$ ($\mu_{g_f}\equiv\mathrm{vol}_{g_f}=f(x)\,\ud x\wedge\ud y$) is given by
\begin{equation}\label{eq:unaltra}
 \Delta_{f}\;=\;\frac{\partial^2}{\partial x^2}+\frac{1}{f^2(x)}\,\frac{\partial^2}{\partial y^2}+\frac{f'(x)}{f(x)}\,\frac{\partial}{\partial x}\,.
\end{equation}
 In complete analogy to $\cH_\alpha$ and $H_\alpha$, the \emph{minimal free Hamiltonian}\index{minimal free Hamiltonian} 
\begin{equation}\label{Hf}
 H_f\;:=\;-\Delta_{f}\,,\qquad\mathcal{D}(H_f)\;:=\;C^\infty_c(M)
\end{equation}
 is a densely defined, symmetric, non-negative operator in the Hilbert space
 \begin{equation}
  \cH_f\;:=\;L^2(M,\ud\mu_{g_f})\,.
 \end{equation}

%
%

Through the unitary transformation
\begin{equation}
 U_f^+:L^2(\mathbb{R}^+\times\mathbb{R},f(x)\ud x\ud y)\stackrel{\cong}{\longrightarrow}L^2(\mathbb{R}^+\times\mathbb{R},\ud x\ud y)\,,\qquad \psi\mapsto f^{1/2}\psi
\end{equation}
a simple computation shows that
\begin{equation}\label{eq:diff_OP}
 \begin{split}
 U_f^+H_f^+(U_f^+)^{-1}\;&=\;-\frac{\partial^2}{\partial x^2}-\frac{1}{f^2}\frac{\partial^2}{\partial y^2}+\frac{2ff''-f'^2}{4f^2} \\
 \mathcal{D}(U_f^+ H_f^+ (U_f^+)^{-1})\;&=\;C^\infty_c(\mathbb{R}^+_x\times\mathbb{R}_y)\,.
 \end{split}
\end{equation}
The further unitary $\mathcal{F}_2^+:L^2(\mathbb{R}^+\times\mathbb{R},\ud x\ud y)\stackrel{\cong}{\longrightarrow}L^2(\mathbb{R}^+\times\mathbb{R},\ud x\ud \xi)$, consisting of the Fourier transform in the $y$-variable only, produces the operator on $ L^2(\mathbb{R}^+\times\mathbb{R},\ud x\ud \xi)$
\begin{equation}
 \mathscr{H}_f^+\;:=\;\mathcal{F}_2^+U_f^+H_f^+(U_f^+)^{-1}(\mathcal{F}_2^+)^{-1}
\end{equation}
whose domain and action are given by
\begin{equation}\label{eq:Hscrf}
 \begin{split}
  \mathscr{H}_f^+\;&=\;-\frac{\partial^2}{\partial x^2}+\frac{\xi^2}{f^2}+\frac{2ff''-f'^2}{4f^2} \\
  \mathcal{D}(\mathscr{H}_f^+)\;&=\;\{\psi\in L^2(\mathbb{R}^+\times\mathbb{R},\ud x\ud \xi)\,|\,\psi\in\mathcal{F}_2^+C^\infty_c(\mathbb{R}^+_x\times\mathbb{R}_y)\}\,.
 \end{split}
\end{equation}
Thus, for each $\psi\in\mathcal{D}(\mathscr{H}_f^+)$ the functions $\psi(\cdot,\xi)$ are compactly supported in $x$ inside $(0,+\infty)$ for every $\xi$, whereas the functions $\psi(x,\cdot)$ are some special case of Schwartz functions for every $x$.
The choice $f(x)=x^{-\alpha}$, $\alpha>0$, yields the operator
\begin{equation}
  \begin{split}
  \mathscr{H}_\alpha^+\;&=\;-\frac{\partial^2}{\partial x^2}+\xi^2 x^{2\alpha}+\frac{\,\alpha(2+\alpha)\,}{4x^2} \\
  \mathcal{D}(\mathscr{H}_\alpha^+)\;&=\;\{\psi\in L^2(\mathbb{R}^+\times\mathbb{R},\ud x\ud \xi)\,|\,\psi=\mathcal{F}_2C^\infty_c(\mathbb{R}^+_x\times\mathbb{R}_y)\}\,.
 \end{split}
\end{equation}

 Obvious two-sided counterpart formulas hold for 
 \begin{equation}
  \begin{split}
   U_f\;&:=\;U_f^-\oplus U_f^+\,, \\
   \mathcal{F}_2\;&:=\;  \mathcal{F}_2^- \oplus  \mathcal{F}_2^+\,, \\
    \mathscr{H}_f\;&:=\; \mathcal{F}_2 U_f H_f U_f^{-1} \mathcal{F}_2^{-1} \;=\;\mathscr{H}_f^-\oplus\mathscr{H}_f^+\,. \\
  \end{split}
 \end{equation}
 The self-adjointness problem for $H_f$, resp.~$H_f^\pm$, is tantamount as the self-adjointness problem for $\mathscr{H}_f$, resp.~$\mathscr{H}_f^\pm$. For the latter control there is clearly no loss of generality in considering the right half plane only, which will be done henceforth.

 The adjoint of $\mathscr{H}_f^+$ is easily characterised as follows.

\begin{lemma}\label{lemma:Hfstar}
 The adjoint of $\mathscr{H}_f^+$ is the operator
\begin{equation}\label{eq:Hfstarplane}
 \begin{split}
  (\mathscr{H}_f^+)^*\;&=\;-\frac{\partial^2}{\partial x^2}+\frac{\xi^2}{f^2}+\frac{2ff''-f'^2}{4f^2} \, ,\\
  \mathcal{D}((\mathscr{H}_f^+)^*)\;&=\;
  \left\{\!\!
  \begin{array}{c}
   \psi\in L^2(\mathbb{R}^+\times\mathbb{R},\ud x\ud \xi)\;\;\textrm{such that} \\
   \big(-\frac{\partial^2}{\partial x^2}+\frac{\xi^2}{f^2}+\frac{2ff''-f'^2}{4f^2}\big)\psi\in L^2(\mathbb{R}^+\times\mathbb{R},\ud x\ud \xi)
  \end{array}
  \!\!\right\}.
 \end{split}
\end{equation}
\end{lemma}

\begin{proof}
 $\mathscr{H}_f^+$ is unitarily equivalent, via Fourier transform in the second variable, to the minimally defined differential operator \eqref{eq:diff_OP}, whose adjoint is the maximally defined realisation of the same differential action (Sect.~\ref{sec:MinimalAndMaximalRealisations}), thus with domain consisting of the elements $F$'s such that both $F$ and $(-\frac{\partial^2}{\partial x^2}-\frac{1}{f^2}\frac{\partial^2}{\partial y^2}+\frac{2ff''-f'^2}{4f^2})F$ belong to $L^2(\mathbb{R}^+\times\mathbb{R},\ud x\ud y)$. Fourier-transforming such adjoint then yields \eqref{eq:Hfstarplane}. 
\end{proof}

Whereas obviously $L^2(\mathbb{R}^+_x\times\mathbb{R}_\xi,\ud x\ud \xi)\cong L^2(\mathbb{R}^+,\ud x)\otimes  L^2(\mathbb{R},\ud\xi)$, the operator $\mathscr{H}_f$ is not a product with respect to the above factorisation, it rather reads as the sum of two products
\begin{equation}
 \mathscr{H}_f^+\;=\;\Big(-\frac{\partial^2}{\partial x^2}+\frac{2ff''-f'^2}{4f^2}\Big)\otimes\mathbbm{1}_\xi+\frac{1}{f^2}\mathbbm{1}_x\otimes \xi^2
\end{equation}
each of which with the same domain as $\mathscr{H}_f$ itself. The second summand is manifestly essentially self-adjoint on $L^2(\mathbb{R}^+,\ud x)\otimes  L^2(\mathbb{R},\ud\xi)$, whereas the self-adjointness of the first summand boils down to the analysis of the factor acting on $L^2(\mathbb{R}^+,\ud x)$ only; yet, there is no general guarantee that the sum of the two preserves the essential self-adjointness.

It is more natural to regard $\mathscr{H}_f$ with respect to the \emph{constant-fibre direct integral}\index{constant-fibre!direct integral of Hilbert spaces} structure (Sect.~\ref{sec:I-preliminaries})
\begin{equation}\label{L^2integralDecomp}
 \begin{split}
  \cH^+\;:=\;L^2(\mathbb{R}^+\times\mathbb{R},\ud x\ud \xi)\;&\cong\;L^2\big(\mathbb{R},\ud \xi\,;L^2(\mathbb{R}^+,\ud x)\big) \\
  &\equiv\;\int_{\mathbb{R}}^{\oplus}\ud \xi \,L^2(\mathbb{R}^+,\ud x)\,,
 \end{split}
\end{equation}
thus thinking of $L^2(\mathbb{R}^+_x\times\mathbb{R}_\xi,\ud x\ud \xi)$ as $L^2(\mathbb{R}^+,\ud x)$-valued square-integrable functions of $\xi\in\mathbb{R}$. The space $\mathfrak{h}^+:=L^2(\mathbb{R}^+,\ud x)$ is the (constant) \emph{fibre}\index{fibre!of constant-fibre direct integral} of the direct integral and the scalar products satisfy
\begin{equation}\label{eq:scalar_products_fibred}
 \langle \psi,\varphi\rangle_{\cH^+}\;=\;\int_{\mathbb{R}}\langle \psi(\cdot,\xi),\varphi(\cdot,\xi)\rangle_{\mathfrak{h}^+}\,\ud \xi\,.
\end{equation}

For each $\xi\in\mathbb{R}$ it is natural to introduce the operator
\begin{equation}\label{eq:Afplaneintro}
 A_f^+(\xi)\;:=\;-\frac{\ud^2}{\ud x^2}+\frac{\xi^2}{f^2}+\frac{2ff''-f'^2}{4f^2}\,,\qquad\mathcal{D}(A_f^+(\xi))\;:=\;C^\infty_c(\mathbb{R}^+)
\end{equation}
acting on the fibre Hilbert space $\mathfrak{h}$. When $f(x)=x^{-\alpha}$ one just writes
\begin{equation}\label{eq:Aalphaplaneintro}
 A_\alpha^+(\xi)\;:=\;-\frac{\ud^2}{\ud x^2}+\xi^2 x^{2\alpha}+\frac{\,\alpha(2+\alpha)\,}{4x^2}\,,\quad\,\mathcal{D}(A_\alpha^+(\xi))\;:=\;C^\infty_c(\mathbb{R}^+)\,.
\end{equation}
By construction, each $A_f^+(\xi)$ is a densely defined symmetric operator on $\mathfrak{h}^+$, with the \emph{same} domain irrespective of $\xi$, and it is also positive because of the assumptions \eqref{eq:assumtpions_f}(iv) on $f$. As such, the $A_f^+(\xi)$'s are all closable and positive (Sect.~\ref{sec:I-symmetric-selfadj}), with the same closure's domain, dense in $\mathfrak{h}$. Arguing as for Lemma \ref{lemma:Hfstar} (Sect.~\ref{sec:MinimalAndMaximalRealisations}) one has
\begin{equation}\label{eq:Afstarplane}
 \begin{split}
  A_f^+(\xi)^*\;&=\;-\frac{\ud^2}{\ud x^2}+\frac{\xi^2}{f^2}+\frac{2ff''-f'^2}{4f^2} \\
  \mathcal{D}(A_f^+(\xi)^*)\;&=\;
  \left\{\!\!
  \begin{array}{c}
   \psi\in L^2(\mathbb{R}^+,\ud x)\;\;\textrm{such that} \\
   \big(-\frac{\ud^2}{\ud x^2}+\frac{\xi^2}{f^2}+\frac{2ff''-f'^2}{4f^2}\big)\psi\in L^2(\mathbb{R}^+,\ud x)
  \end{array}
  \!\!\right\}.
 \end{split}
\end{equation}

Next, with respect to the decomposition \eqref{L^2integralDecomp} one defines the auxiliary operator $B_f$ in the Hilbert space $\cH$ by
\begin{equation}\label{ed:defB}
 \begin{split}
  \mathcal{D}(B_f)\;&:=\;\left\{\psi\in\cH\,\left|\! 
  \begin{array}{l}
   \mathrm{(i)}\quad\psi(\cdot,\xi)\in\mathcal{D}(\overline{A_f^+(\xi)})\textrm{ for almost every }\xi \\
   \mathrm{(ii)}\,\displaystyle\int_\mathcal{\mathbb{R}}\big\|\overline{A_f^+(\xi)}\psi(\cdot,\xi)\big\|_{\mathfrak{h}}^2\,\ud\xi<+\infty
  \end{array}
  \!\!\right.\right\} \, , \\
  (B_f\psi)(x,\xi)\;&:=\;\big(\overline{A_f^+(\xi)}\,\psi(\cdot,\xi)\big)(x)\,.
 \end{split}
\end{equation}
 It is customary to use, for the \emph{whole} \eqref{ed:defB}, the symbol
\begin{equation}\label{eq:Bfdecomposable}
 B_f\;=\;\int_{\mathbb{R}}^{\oplus}\overline{A_f^+(\xi)}\,\ud\xi\,.
\end{equation}

 \begin{remark}
  The fact that the
$\overline{A_f^+(\xi)}$'s have all the \emph{same} dense domain in $\mathfrak{h}$ guarantees that the decomposition \eqref{eq:Bfdecomposable} of $B_f$ is unique and hence unambiguous: if one also had $B_f=\int_{\mathbb{R}}^{\oplus}B_f(\xi)\ud\,\xi$ for a map $\xi\mapsto B_f(\xi)$ with $\mathcal{D}(B_f(\xi))=\mathcal{D}(\overline{A_f^+(\xi)})=\mathcal{D}$, a common dense domain in $\mathfrak{h}$, then necessarily $\overline{A_f^+(\xi)}=B_f(\xi)$ for almost every $\xi\in\mathbb{R}$.
 \end{remark}

\begin{remark}\label{rem:no-integral-decomp}
As suggestive as it would be, it is however important to observe that the operator of interest, $\mathscr{H}_f^+$, is \emph{not} decomposable in the form $\int_\mathbb{R}^{\oplus}A_f^+(\xi)\,\ud \xi$. Indeed, the analogue of condition (i) in \eqref{ed:defB} would be satisfied, but condition (ii) would be not. More precisely, by definition an element $\psi\in\mathcal{D}(\int_\mathbb{R}^{\oplus}A_f^+(\xi)\,\ud \xi)$ does satisfy the property $\psi(\cdot,\xi)\in\mathcal{D}(A_f^+(\xi))=C^\infty_c(\mathbb{R}^+)$ for every $\xi$, as is the case for the elements of $\mathcal{D}(\mathscr{H}_f^+)$. Moreover, it satisfies
\begin{equation}\label{eq:Afpsi}
 \begin{split}
 +\infty\;&>\int_\mathcal{\mathbb{R}}\big\|A_f^+(\xi)\psi(\cdot,\xi)\big\|_{\mathfrak{h}}^2\,\ud\xi \\
 &=\iint_{\mathbb{R}^+\times\mathbb{R}}\Big|\Big(-\frac{\partial^2}{\partial x^2}+\frac{\xi^2}{f^2}+\frac{2ff''-f'^2}{4f^2}\Big)\psi(x,\xi)\Big|^2\ud x\,\ud\xi\,,
 \end{split}
\end{equation}
and \eqref{eq:Afpsi} does not necessarily imply that for every $x$ the function $\psi(x,\cdot)$ is the Fourier transform of a $C^\infty_0(\mathbb{R})$-function as it has to be for an element of $\mathcal{D}(\mathscr{H}_f^+)$. Condition \eqref{eq:Afpsi} is surely satisfied by other functions besides all those in $\mathcal{D}(\mathscr{H}_f^+)$. Thus, one has the (proper) inclusion
\begin{equation}\label{eq:BextendsHf}
 B_f\;\varsupsetneq\; \mathscr{H}_f^+\,.
\end{equation}
\end{remark}

The operator $B_f$ is not just an extension of $\mathscr{H}_f^+$, it is a \emph{closed} symmetric extension.

\begin{proposition}\label{prop:BsymmetricAndClosed}~
\begin{enumerate}[(i)]
 \item $B_f$ is symmetric.
 \item $B_f$ is closed.
\end{enumerate}
\end{proposition}

\begin{proof}
 Symmetry is immediately checked by means of \eqref{eq:scalar_products_fibred}, thanks to the symmetry of each $\overline{A_f^+(\xi)}$. Concerning the closedness, let $(\psi_n)_{n\in\mathbb{N}}$, $\psi$, and $\Psi$ be, respectively, a sequence and two functions in $\mathcal{D}(B_f)$ such that $\psi_n\to\psi$ and $B\psi_n\to\Psi$ in $\cH^+$ as $n\to +\infty$. Thus,
 \[
  \begin{split}
   \int_{\mathbb{R}}\|\psi_n(\cdot,\xi)-\psi(\cdot,\xi)\|_{\mathfrak{h}^+}^2\,\ud \xi\;&\xrightarrow[]{\;n\to +\infty\;}0\,, \\
   \int_{\mathbb{R}}\big\|\overline{A_f^+(\xi)}\,\psi_n(\cdot,\xi)-\Psi(\cdot,\xi)\|_{\mathfrak{h}^+}^2\,\ud \xi\;&\xrightarrow[]{\;n\to +\infty\;}0\,,
  \end{split}
 \]
 which implies that, \emph{up to extracting a subsequence}, and for \emph{almost} every $\xi$, $\psi_n(\cdot,\xi)\to\psi(\cdot,\xi)$ and $\overline{A_f^+(\xi)}\,\psi_n(\cdot,\xi)\to\Psi(\cdot,\xi)$ in $\mathfrak{h}^+$ as $n\to +\infty$. Owing to the closedness of $\overline{A_f^+(\xi)}$, one must conclude (Sect.~\ref{sec:I-bdd-closable-closed}) that
 \[
  \psi(\cdot,\xi)\in\mathcal{D}(\overline{A_f^+(\xi)})\qquad\textrm{and}\qquad \overline{A_f^+(\xi)}\,\psi(\cdot,\xi)\;=\;\Psi(\cdot,\xi)
 \]
 for almost every $\xi$. Therefore,
\[
 \int_\mathcal{\mathbb{R}}\big\|\overline{A_f^+(\xi)}\psi(\cdot,\xi)\big\|_{\mathfrak{h}^+}^2\,\ud\xi\;=\;\|\Psi\|_{\cH}^2\;<\;+\infty.
\]
 Both conditions (i) and (ii) of \eqref{ed:defB} are satisfied, which proves that $\psi\in\mathcal{D}(B_f)$ and $B_f\psi=\Psi$, that is (Sect. \ref{sec:I-bdd-closable-closed}), the closedness of $B_f$. 
\end{proof}

The convenience of the auxiliary operator $B_f$ relies on the important circumstance that the self-adjointess of $B_f$ can be characterised in terms of the same property in each fibre. One direction of this fact is a general well-known property \cite[Theorem XIII.85(i)]{rs4} that for completeness is re-stated and proved here in the particular case under consideration.

\begin{proposition}\label{prop:RSpropBA}
 If $A_f^+(\xi)$ is essentially self-adjoint for almost every $\xi\in\mathbb{R}$, then $B_f$ is self-adjoint.
\end{proposition}

 \begin{proof}
  $B_f$ is densely defined, symmetric, and closed (Proposition \ref{prop:BsymmetricAndClosed}), hence to establish its self-adjointness it suffices to show (Sect.~\ref{sec:I-symmetric-selfadj}) that $\mathrm{ran}(B_f\pm\ii\mathbbm{1})=\cH^+$, the Hilbert space \eqref{L^2integralDecomp}.
  By symmetry of $A_f^+(\xi)$, $(\overline{A_f^+(\xi)}+\ii\mathbbm{1})^{-1}$ is everywhere defined and bounded on the fibre $\mathfrak{h}^+=L^2(\mathbb{R}^+,\ud x)$ and with norm bounded by 1 uniformly in $\xi$ (Sect.~\ref{sec:I-symmetric-selfadj}). Therefore, the definition (in the sense of \eqref{eq:Bfdecomposable})
  \[
   C_f^+\;:=\;\int_{\mathbb{R}}^{\oplus}(\overline{A_f^+(\xi)}+\ii\mathbbm{1})^{-1}\,\ud\xi\,,
  \]
  namely
  \begin{equation*}
 \begin{split}
  \mathcal{D}(C_f^+)\;&:=\;\left\{\psi\in\cH\,\left|\! 
  \begin{array}{l}
   \mathrm{(i)}\quad\psi(\cdot,\xi)\in \mathfrak{h}\textrm{ for almost every }\xi \\
   \mathrm{(ii)}\,\displaystyle\int_\mathcal{\mathbb{R}}\big\|(\overline{A_f^+(\xi)}+\ii\mathbbm{1})^{-1}\psi(\cdot,\xi)\big\|_{\mathfrak{h}^+}^2\,\ud\xi<+\infty
  \end{array}
  \!\!\right.\right\}, \\
  (C_f^+\psi)(x,\xi)\;&:=\;(\overline{A_f^+(\xi)}+\ii\mathbbm{1})^{-1}\,\psi(\cdot,\xi)\big)(x)\,,
 \end{split}
\end{equation*}
  identifies an everywhere defined and bounded operator on $\cH^+$. Now, for arbitrary $\Phi\in\cH^+$ set $\Psi:=C_f^+\Phi$. Then $\Psi(\cdot,\xi)\in\mathrm{ran}(\overline{A_f^+(\xi)}+\ii\mathbbm{1})^{-1}=\mathcal{D}(\overline{A_f^+(\xi)})$ and
  \[
   \big\|\overline{A_f^+(\xi)}\Psi(\cdot,\xi)\big\|_{\mathfrak{h}^+}\;=\;\big\|\overline{A_f^+(\xi)}(\overline{A_f^+(\xi)}+\ii\mathbbm{1})^{-1}\,\Phi(\cdot,\xi)\big)\big\|_{\mathfrak{h}^+}\;\leqslant\;\|\Phi(\cdot,\xi)\|_{\mathfrak{h}^+}\,.
  \]
  As $\Phi\in\cH$, $\xi\mapsto\|\Phi(\cdot,\xi)\|_{\mathfrak{h}^+}$ is square-integrable on $\mathbb{R}$, and the above estimate then shows that $\xi\mapsto \|\overline{A_f^+(\xi)}\Psi(\cdot,\xi)\|_{\mathfrak{h}^+}$ is square-integrable too. Therefore, in view of \eqref{ed:defB}, $\Psi\in\mathcal{D}(B_f)$. When applying $B_f+\ii\mathbbm{1}$ to $\Psi$ one finds, fibre-wise for almost every $\xi$,
  \[
   (\overline{A_f^+(\xi)}+\ii\mathbbm{1})\Psi(\cdot,\xi)\;=\;(\overline{A_f^+(\xi)}+\ii\mathbbm{1})(\overline{A_f^+(\xi)}+\ii\mathbbm{1})^{-1}\,\Phi(\cdot,\xi)\;=\;\Phi(\cdot,\xi)\,,
  \]
  that is, $\Phi=(B_f+\ii\mathbbm{1})\Psi\in\mathrm{ran}(B_f)$. This proves that $\mathrm{ran}(B_f+\ii\mathbbm{1})=\cH^+$. The proof that $\mathrm{ran}(B_f-\ii\mathbbm{1})=\cH^+$ is completely analogous.  
 \end{proof}

 Concerning the opposite statement, one has the following.

\begin{proposition}\label{prop:Bselfadj-implies-Axiselfadj}
 If $B_f$ is self-adjoint, then $A_f^+(\xi)$ is essentially self-adjoint for almost every $\xi\in\mathbb{R}$.
\end{proposition}

\begin{proof}
 It follows by assumption that for any $\varphi\in\cH^+$ there exists $\psi_\varphi\in\mathcal{D}(B_f)$ with $\varphi=(B_f+\ii)\psi_\varphi$. Thus, as an identity in $\mathfrak{h}^+$,
 \[
  \varphi(\cdot,\xi)\;=\;\big(\overline{A_f^+(\xi)}+i\mathbbm{1}\big)\,\psi_\varphi(\cdot,\xi)\qquad\textrm{for almost every }\,\xi\,.
 \]
 In particular, running $\varphi$ over all the $C^\infty_c(\mathbb{R}^+_x\times\mathbb{R}_\xi)$-functions and fixing $\xi_0\in\mathbb{R}$, obviously $\varphi(\cdot,\xi_0)$ spans the whole space of $C^\infty_c(\mathbb{R}^+_x)$-functions, which is a dense of $\mathfrak{h}^+$: with this choice the above identity implies that $\mathrm{ran}(\overline{A_f^+(\xi_0)}+i\mathbbm{1}\big)$ is dense in $\mathfrak{h}^+$ and hence $A_f^+(\xi_0)$ is essentially self-adjoint. 
\end{proof}

 The combination of Propositions \ref{prop:RSpropBA} and \ref{prop:Bselfadj-implies-Axiselfadj} finally allows for a complete control of the essential self-adjointess of $\mathscr{H}_\alpha$, and hence also of $H_\alpha$ via unitary equivalence. This control is deferred to Section \ref{sec:ess-self-adj} for a more direct comparison with the same problem on the cylinder.

 \section{Constant-fibre orthogonal sum for the Grushin cylinder}\label{sec:constant-fibre-sum}

 Also the self-adjointness problem for the minimal free Hamiltonian\index{minimal free Hamiltonian} $H_\alpha$ 
 on the Hilbert space $\cH_\alpha=L^2(M,\ud\mu_\alpha)$ of Grushin-type \emph{cylinder} is intrinsically cleaner and more manageable in a suitable unitarily equivalent version. In this Section such a unitary equivalence is worked out in analogy to Section \ref{sec:constant-fibre-integral}, yet with an obvious but more explicit discussion of both the one-sided and the two-sided (left/right) problem, in order to prepare the ground for the subsequent identification and classification of non-trivial self-adjoint extensions.

 In view of the left-right orthogonal decomposition 
 \begin{equation*}
  \cH_\alpha\;=\;L^2(M,\ud\mu_\alpha)\;\cong\;L^2(M^-,\ud\mu_\alpha)\oplus L^2(M^+,\ud\mu_\alpha)
 \end{equation*}
 (see \eqref{eq:V-Hspacedirectsum} above), one switches from $L^2(M^\pm,\ud\mu_\alpha)$ to the new Hilbert spaces
\begin{equation}\label{eq:global-unitary-pm}
  \cH^\pm\;:=\;\mathcal{F}_2^{\pm} U_\alpha^{\pm}L^2(M^\pm,\ud\mu_\alpha)
\end{equation}
where $U_\alpha^{\pm}$ and $\mathcal{F}_2^{\pm}$ are unitary transformations defined, respectively, as
\begin{equation}\label{eq:unit1}
\begin{split}
 U_\alpha^\pm:L^2(\mathbb{R}^\pm\times\mathbb{S}^1,|x|^{-\alpha}\ud x\ud y)&\stackrel{\cong}{\longrightarrow}L^2(\mathbb{R}^\pm\times\mathbb{S}^1,\ud x\ud y) \, ,\\
 f &\; \mapsto\;\phi\;:=\;  |x|^{-\frac{\alpha}{2}}f\,,
\end{split}
\end{equation}
and
\begin{equation}\label{eq:defF2}
 \begin{split}
  \mathcal{F}_2^{\pm}:L^2(\mathbb{R}^\pm\times\mathbb{S}^1,\ud x\ud y)&\stackrel{\cong}{\longrightarrow}
 L^2(\mathbb{R}^\pm,\ud x)\otimes\ell^2(\mathbb{Z})\,, \\
  \phi &\;\mapsto\;\psi\;\equiv\;(\psi_k)_{k\in\mathbb{Z}}\,, \\
  e_k(y)\;:=\;\frac{e^{\ii k y}}{\sqrt{2\pi}}\,,&\qquad \psi_k(x)\,:=\int_0^{2\pi}\overline{e_k(y)}\,\phi(x,y)\,\ud y\,,\qquad x\in\mathbb{R}^{\pm}
 \end{split}
\end{equation}
(thus, $\phi(x,y)=\sum_{k\in\mathbb{Z}}\psi_k(x)e_k(y)$ in the $L^2$-convergent sense).

Up to canonical isomorphisms,
\begin{equation}\label{eq:EquivalentHSpaces}
	\mathcal{H}^\pm \;= \; \bigoplus_{k \in \mathbb{Z}} L^2(\mathbb{R}^\pm, \ud x) \; \cong  \; \ell^2(\mathbb{Z},L^2(\mathbb{R}^\pm)) \; \cong \; L^2(\mathbb{R}^\pm) \otimes \ell^2(\mathbb{Z})\,,
\end{equation}
therefore $\mathcal{H}^+$ and $\mathcal{H}^-$ display a natural \emph{constant-fibre orthogonal sum}\index{constant-fibre!direct sum of Hilbert spaces} structure
\begin{equation}\label{L^2directDecomp}
  \cH^\pm\;=\;\bigoplus_{k\in\mathbb{Z}} \;\mathfrak{h}^\pm\,,\qquad \mathfrak{h}^\pm\;:=\;L^2(\mathbb{R}^\pm,\ud x)
\end{equation}
with one-sided \emph{constant fibre}\index{fibre!of constant-fibre direct sum} $\mathfrak{h}_\pm$ and scalar product 
\begin{equation}
 \big\langle (\psi_k)_{k\in\mathbb{Z}} , (\widetilde{\psi}_k)_{k\in\mathbb{Z}} \big\rangle_{\cH^{\pm}}=\;\sum_{k\in\mathbb{Z}}\,\int_{\mathbb{R}^\pm}\overline{\psi_k(x)}\,\widetilde{\psi}_k(x)\,\ud x\;\equiv\;\sum_{k\in\mathbb{Z}}\,\langle \psi_k,\widetilde{\psi}_k\rangle_{\mathfrak{h}^\pm}\,.
\end{equation}

By means of the transformations \eqref{eq:unit1}-\eqref{eq:defF2} one defines
\begin{equation}\label{eq:tildeHalpha}
 \mathsf{H}_\alpha^\pm\;:=\;U_\alpha^\pm \,H_\alpha^\pm \,(U_\alpha^\pm)^{-1}
\end{equation}
acting on $L^2(\mathbb{R}^\pm\times\mathbb{S}^1,\ud x\ud y)$, i.e., 
\begin{equation}\label{eq:explicit-tildeHalpha}
 \begin{split}
  \mathcal{D}(\mathsf{H}_\alpha^\pm)\;&=\;C^\infty_c(\mathbb{R}^\pm_x\times\mathbb{S}^1_y)\,, \\
  \mathsf{H}_\alpha^\pm\phi\;&=\;\Big(-\frac{\partial^2}{\partial x^2}- |x|^{2\alpha}\frac{\partial^2}{\partial y^2}+\frac{\,\alpha(2+\alpha)\,}{4x^2}\Big)\phi\,,
 \end{split}
\end{equation}
 and one also defines
\begin{equation}\label{eq:unitary_transf_pm}
 \mathscr{H}_\alpha^\pm\;:=\;\mathcal{F}^{\pm}_2\, U_\alpha^\pm \,H_\alpha^\pm \,(U_\alpha^\pm)^{-1}(\mathcal{F}_2^{\pm})^{-1}\;=\;\mathcal{F}^{\pm}_2\,\mathsf{H}_\alpha^\pm(\mathcal{F}_2^{\pm})^{-1}
\end{equation}
acting on $\cH^{\pm}$, i.e.,
\begin{equation}\label{eq:actiondomainHalpha}
  \begin{split}
  \mathcal{D}(\mathscr{H}_\alpha^\pm)\;&=\;\Big\{\psi\equiv(\psi_k)_{k\in\mathbb{Z}}\in \bigoplus_{k\in\mathbb{Z}} L^2(\mathbb{R}^\pm,\ud x)\,\Big|\,\psi\in\mathcal{F}_2^{\pm}C^\infty_c(\mathbb{R}^\pm_x\times\mathbb{S}^1_y)\Big\}\,, \\
    \mathscr{H}_\alpha^\pm\psi\;&=\;\Big(\Big(-\frac{\ud^2}{\ud x^2}+k^2 |x|^{2\alpha}+\frac{\,\alpha(2+\alpha)\,}{4x^2}\Big)\psi_k\Big)_{k\in\mathbb{Z}} \,.
 \end{split}
\end{equation}
In particular, for each $\psi^\pm\in\mathcal{D}(\mathscr{H}_\alpha^\pm)$ the components $\psi^\pm_k(\cdot)$ are compactly supported in $x$ inside $\mathbb{R}^\pm$ for every $k\in\mathbb{Z}$, and moreover
\begin{equation}\label{eq:forcondii}
 \begin{split}
  \sum_{k\in\mathbb{Z}}\Big\|&\Big(-\frac{\ud^2}{\ud x^2}+k^2 |x|^{2\alpha}+\frac{\,\alpha(2+\alpha)\,}{4x^2}\Big)\psi^\pm_k \Big\|_{L^2(\mathbb{R}^\pm,\ud x)}^2 \\
  &=\;\|\mathscr{H}_\alpha^\pm\psi^\pm\|_{\cH^\pm}^2\;=\;\|(\mathcal{F}_2^{\pm})^{-1}\mathscr{H}_\alpha^\pm\mathcal{F}^{\pm}_2\phi^\pm\|_{L^2(\mathbb{R}^\pm_x\times\mathbb{S}^1_y)}^2 \\
  &=\;\Big\| \Big(-\frac{\partial^2}{\partial x^2}- |x|^{2\alpha}\frac{\partial^2}{\partial y^2}+\frac{\,\alpha(2+\alpha)\,}{4x^2}\Big)\phi^\pm  \Big\|_{L^2(\mathbb{R}^\pm_x\times\mathbb{S}^1_y)}^2\;<\;+\infty\,,
 \end{split}
\end{equation}
 having used that $\phi^\pm=\mathcal{F}_2^\pm\psi\in C^\infty_c(\mathbb{R}^\pm_x\times\mathbb{S}^1_y)$.

 The obvious two-sided counterparts of the above formulas are
 \begin{eqnarray}
  U_\alpha \;&:=&\; U_\alpha^-\oplus U_\alpha^+\,, \label{eq:V-two-sided-1}\\
  \mathcal{F}_2 \;&:=&\;  \mathcal{F}_2^-\oplus \mathcal{F}_2^+\,, \\
  \cH \;&:=&\; \cH^-\oplus\cH^+\;=\;\mathcal{F}_2U_\alpha L^2(M,\ud\mu_\alpha)\,, \label{eq:Hxispace} \\
  \cH \;&\cong&\;\bigoplus_{k \in \mathbb{Z}} L^2(\mathbb{R}, \ud x) \; \cong  \; \ell^2(\mathbb{Z},L^2(\mathbb{R})) \; \cong \; L^2(\mathbb{R}^\pm) \otimes \ell^2(\mathbb{Z})\,, \label{eq:V-two-sided-4} \\
  \cH\;&=&\;\bigoplus_{k\in\mathbb{Z}} \;\mathfrak{h}\,,\qquad \mathfrak{h}\;:=\;L^2(\mathbb{R},\ud x)\;\cong\;\mathfrak{h}^-\oplus \mathfrak{h}^+\,, \label{eq:Hoplusfrakh} \\
  \mathsf{H}_\alpha \;&:=&\; U_\alpha \,H_\alpha \,(U_\alpha^\pm)^{-1}\,,  \label{eq:V-two-sided-6} \\
  \mathscr{H}_\alpha \;&:=&\; \mathcal{F}_2\, U_\alpha \,H_\alpha \,U_\alpha^{-1}\mathcal{F}_2^{-1}\;=\;\mathcal{F}_2\,\mathsf{H}_\alpha \mathcal{F}_2^{-1}\,. \label{eq:V-two-sided-7}
 \end{eqnarray}

  Owing to the above unitary equivalences, the self-adjointness problem for $H_\alpha^\pm$ in $L^2(M^\pm,\ud\mu_\alpha)$ is tantamount as the self-adjointness problem for $\mathscr{H}_\alpha^\pm$ in $\cH^\pm$, and the same holds for $H_\alpha$ with respect to its unitarily equivalent version $\mathscr{H}_\alpha$. Furthermore, when non-trivial self-adjoint extensions exist for $H_\alpha^\pm$ (resp., $H_\alpha$), they can be equivalently (and in practice more conveniently) identified as self-adjoint extensions of $\mathscr{H}_\alpha^\pm$ (resp., $\mathscr{H}_\alpha$).

  The efficiency of this approach is due to the crucial circumstance that the \emph{adjoint} of $\mathscr{H}_\alpha$ turns out to be reduced with respect to the orthogonal sum \eqref{eq:Hoplusfrakh}, as will be clear from Lemma \ref{lem:Halphaadj-decomposable} below, and as such its self-adjoint restrictions can be more explicitly discussed in terms of the self-adjointness realisations of each $k$-component. Beside, it turns out that whereas $\mathscr{H}_\alpha$, if not essentially self-adjoint, has infinite deficiency index, the problem on each fibre can be formulated for operators with deficiency index equal to 2, thus making the analysis on fibre completely manageable.

  Explicitly (see Section \ref{sec:MinimalAndMaximalRealisations} and, for analogy, Lemma \ref{lemma:Hfstar}), the adjoint of $\mathsf{H}_\alpha$ is the maximal realisation of the same differential operator, that is,
\begin{equation}\label{eq:HHalphaadjoint}
 \begin{split}
  \mathcal{D}((\mathsf{H}_\alpha^\pm)^*)\;&=\;\left\{
  \begin{array}{c}
   \phi\in L^2(\mathbb{R}^\pm\times\mathbb{S}^1,\ud x \ud y)\textrm{ such that} \\
   \Big(-\frac{\partial^2}{\partial x^2}- |x|^{2\alpha}\frac{\partial^2}{\partial y^2}+\frac{\,\alpha(2+\alpha)\,}{4x^2}\Big)\phi\in L^2(\mathbb{R}^\pm\times\mathbb{S}^1,\ud x \ud y)
  \end{array}
  \right\}, \\
  (\mathsf{H}_\alpha^\pm)\phi\;&=\;\Big(-\frac{\partial^2}{\partial x^2}- |x|^{2\alpha}\frac{\partial^2}{\partial y^2}+\frac{\,\alpha(2+\alpha)\,}{4x^2}\Big)\phi\,.
 \end{split}
\end{equation}
This, and the unitary equivalence \eqref{eq:unitary_transf_pm}, yield at once
\begin{equation}\label{eq:Hfstar}
  \begin{split}
  \mathcal{D}((\mathscr{H}_\alpha^\pm)^*)\;&=\;
  \left\{\!\!
  \begin{array}{c}
   \psi\equiv(\psi_k)_{k\in\mathbb{Z}}\in \bigoplus_{k\in\mathbb{Z}} L^2(\mathbb{R}^\pm,\ud x)\;\;\textrm{such that} \\ 
   \;\displaystyle\sum_{k\in\mathbb{Z}}\Big\|\Big(-\frac{\ud^2}{\ud x^2}+k^2 |x|^{2\alpha}+\frac{\,\alpha(2+\alpha)\,}{4x^2}\Big)\psi_k \Big\|_{L^2(\mathbb{R}^\pm,\ud x)}^2\;<\;+\infty
  \end{array}
  \!\!\right\} , \\
   (\mathscr{H}_\alpha^\pm)^*\psi\;&=\;\Big(\Big(-\frac{\ud^2}{\ud x^2}+k^2 |x|^{2\alpha}+\frac{\,\alpha(2+\alpha)\,}{4x^2}\Big)\psi_k\Big)_{k\in\mathbb{Z}} \,.
 \end{split}
\end{equation}
Clearly, $\frac{\ud^2}{\ud x^2}$ is a weak derivative in \eqref{eq:HHalphaadjoint}-\eqref{eq:Hfstar} and a classical derivative in \eqref{eq:actiondomainHalpha}.
Furthermore, with respect to the decomposition \eqref{eq:Hxispace},
\begin{equation}\label{eq:Hfstar_sum}
 (\mathscr{H}_\alpha)^*\;=\;(\mathscr{H}_\alpha^-)^*\oplus(\mathscr{H}_\alpha^+)^*\,.
\end{equation}
 In \eqref{eq:HHalphaadjoint}-\eqref{eq:Hfstar_sum}, and systematically in the following, it is tacitly assumed that the adjoint is taken every time with respect to the corresponding underlying Hilbert space.

 In turn, formula \eqref{eq:Hfstar} suggests to define on each one-sided fibre $\mathfrak{h}_\pm$, hence for each $k\in\mathbb{Z}$, the operators
\begin{equation}\label{eq:Axi}
 A_\alpha^\pm(k)\;:=\;-\frac{\ud^2}{\ud x^2}+k^2 x^{2\alpha}+\frac{\,\alpha(2+\alpha)\,}{4x^2}\,,\quad\,\mathcal{D}(A_\alpha(k))\;:=\;C^\infty_c(\mathbb{R}^+)\,,
\end{equation}
as well as the two-sided counterpart
\begin{equation}\label{eq:Axibilateral}
 \begin{split}
  \mathcal{D}(A_\alpha(k))\;&:=\;C^\infty_c(\mathbb{R}^-)\boxplus C^\infty_c(\mathbb{R}^+)\,, \\
  A_\alpha(k)\;&:=\;A_\alpha^-(k)\oplus A_\alpha^+(k)\,.
 \end{split}
\end{equation}
 By construction, each $A_{\alpha}(k)$ is a densely defined and symmetric operator on $\mathfrak{h}$ (in fact, also non-negative, as seen in Lemma \ref{lem:V-Aalphak-positive} below), with the \emph{same} domain irrespective of $k$. As such (Sect.~\ref{sec:I-symmetric-selfadj}), all the $A_{\alpha}(k)$'s are closable and each $\overline{A_{\alpha}(k)}$ is non-negative in $\mathfrak{h}$. Analogously to \eqref{eq:Hfstar} (Sect.~\ref{sec:MinimalAndMaximalRealisations}), 
\begin{equation}\label{eq:Afstar}
 \begin{split}
  \mathcal{D}(A_{\alpha}^\pm(k)^*)\;&=\;
  \left\{\!\!
  \begin{array}{c}
   g^\pm\in L^2(\mathbb{R}^\pm,\ud x)\;\;\textrm{such that} \\
   \big(-\frac{\ud^2}{\ud x^2}+k^2 |x|^{2\alpha}+\frac{\,\alpha(2+\alpha)\,}{4x^2}\big)g^\pm\in L^2(\mathbb{R}^\pm,\ud x)
  \end{array}
  \!\!\right\}, \\
   A_{\alpha}^\pm(k)^*g^\pm\;&=\;\Big(-\frac{\ud^2}{\ud x^2}+k^2 |x|^{2\alpha}+\frac{\,\alpha(2+\alpha)\,}{4x^2}\Big) g^\pm\,,
 \end{split}
\end{equation}
and
\begin{equation}\label{eq:Afstar_sum}
 A_{\alpha}(k)^*\;=\;A_{\alpha}^-(k)^*\oplus A_{\alpha}^+(k)^*\,.
\end{equation}

  \begin{remark}\label{rem:V-DAastar-samek}
   \eqref{eq:Afstar} shows that $\mathcal{D}(A_{\alpha}(k)^*)=\mathcal{D}(A_{\alpha}(k')^*)$ for non-zero $k,k'\in\mathbb{Z}$.
  \end{remark}

   \begin{remark}\label{rem:Halphanotsum}
In complete analogy to Remark \ref{rem:no-integral-decomp}, it is crucial to observe that $\mathscr{H}_\alpha$ is \emph{not} reduced as $\bigoplus_{k\in\mathbb{Z}} A_\alpha(k)$ (in the general sense of Section \ref{sec:I_invariant-reducing-ssp}), and in fact
\begin{equation}\label{eq:Halphanotsum}
 \mathscr{H}_\alpha\;\varsubsetneq\;\bigoplus_{k\in\mathbb{Z}} A_\alpha(k)\,.
\end{equation}
Indeed, as seen in \eqref{eq:forcondii},
\[
 \sum_{k}\|A_\alpha(k)\psi_k\|_{\mathfrak{h}}^2\;=\; \textstyle\Big\| \Big(-\frac{\partial^2}{\partial x^2}- |x|^{2\alpha}\frac{\partial^2}{\partial y^2}+\frac{\,\alpha(2+\alpha)\,}{4x^2}\Big)\phi  \Big\|_{L^2(\mathbb{R}^\pm_x\times\mathbb{S}^1_y)}^2\,,
\]
where $\psi=\mathcal{F}_2\phi$,
the finiteness of which is guaranteed by $\phi\in C^\infty_c(\mathbb{R}^\pm_x\times\mathbb{S}^1_y)$ when $\psi\in\mathcal{D}(\mathscr{H}_\alpha)$, but of course is also guaranteed by a much larger class of $\phi$'s that are still smooth and compactly supported in $x$, but not smooth in $y$ -- thus, $\psi$'s that do not belong to $\mathcal{D}(\mathscr{H}_\alpha)$. 
\end{remark}

  Although $\mathscr{H}_\alpha$ is strictly contained in $\bigoplus_{k\in\mathbb{Z}} A_\alpha(k)$ (Remark \ref{rem:Halphanotsum}), the two operators have actually the \emph{same} adjoint and closure.

\begin{lemma}\label{lem:Halphaadj-decomposable}
 One has
\begin{equation}\label{eq:Halphaadj-decomposable}
 \mathscr{H}_\alpha^*\;=\;\bigoplus_{k\in\mathbb{Z}} \,A_\alpha(k)^*
\end{equation}
and 
\begin{equation}\label{eq:Halphaclosure-decomposable}
 \overline{\mathscr{H}_\alpha}\;=\;\bigoplus_{k\in\mathbb{Z}} \,\overline{A_\alpha(k)}\,,
\end{equation}
i.e.,
\begin{equation}
   \begin{split}
  \mathcal{D}(\mathscr{H}_\alpha^*)\;&:=\;\left\{\psi\equiv(\psi_k)_{k\in\mathbb{Z}}\in\cH\,\left|\! 
  \begin{array}{l}
   \mathrm{(i)}\quad\psi_k\in\mathcal{D}(A_\alpha(k)^*) \;\;\forall k\in\mathbb{Z} \\
   \mathrm{(ii)}\,\displaystyle\sum_{k\in\mathbb{Z}}\big\|A_\alpha(k)^*\psi_k\big\|_{\mathfrak{h}}^2<+\infty
  \end{array}
  \!\!\right.\right\},  \\
  \mathscr{H}_\alpha^*\psi\;&:=\;\big(A_\alpha(k)^*\,\psi_k\big)_{k\in\mathbb{Z}}\,,
 \end{split}
\end{equation}
and 
\begin{equation}
   \begin{split}
  \mathcal{D}(\overline{\mathscr{H}_\alpha})\;&:=\;\left\{\psi\equiv(\psi_k)_{k\in\mathbb{Z}}\in\cH\,\left|\! 
  \begin{array}{l}
   \mathrm{(i)}\quad\psi_k\in\mathcal{D}(\overline{A_\alpha(k)}) \;\;\forall k\in\mathbb{Z} \\
   \mathrm{(ii)}\,\displaystyle\sum_{k\in\mathbb{Z}}\big\|\overline{A_\alpha(k)}\psi_k\big\|_{\mathfrak{h}}^2<+\infty
  \end{array}
  \!\!\right.\right\},  \\
  \overline{\mathscr{H}_\alpha}\psi\;&:=\;\big(\overline{A_\alpha(k)}\,\psi_k\big)_{k\in\mathbb{Z}}\,.
 \end{split}
\end{equation}
Analogously,
 \begin{equation}
 (\mathscr{H}_\alpha^\pm)^*\;=\;\bigoplus_{k\in\mathbb{Z}} \,A_\alpha^\pm(k)^*\,,\qquad \overline{\mathscr{H}_\alpha^\pm}\;=\;\bigoplus_{k\in\mathbb{Z}} \,\overline{A_\alpha^\pm(k)}\,.
\end{equation}
Moreover,
 \begin{equation}\label{eq:Halphaadj-sumkernel}
  \ker \mathscr{H}_\alpha^*\;=\;\bigoplus_{k\in\mathbb{Z}}\,\ker A_\alpha(k)^*\,.
 \end{equation}
\end{lemma}

\begin{proof}
 On the one hand, $\mathscr{H}_\alpha^*\supset (\bigoplus_{k\in\mathbb{Z}} A_\alpha(k))^*=\bigoplus_{k\in\mathbb{Z}} A_\alpha(k)^*$ (owing to \eqref{eq:Halphanotsum} and the general reducing properties discussed in Section \ref{sec:I_invariant-reducing-ssp}). For the opposite inclusion, namely $\mathscr{H}_\alpha^*\subset \bigoplus_{k\in\mathbb{Z}} A_\alpha(k)^*$, recall that for $\xi\equiv(\xi_k)_{k\in\mathbb{Z}}\in\mathcal{D}(\mathscr{H}_\alpha)$ one has $\xi_k\in C^\infty_c(\mathbb{R}\setminus\{0\})=\mathcal{D}(A_\alpha(k))$ $\forall k\in\mathbb{Z}$. Therefore, if $\psi\equiv(\psi_k)_{k\in\mathbb{Z}}\in\mathcal{D}(\mathscr{H}_\alpha^*)$, then there exists $\eta\equiv(\eta_k)_{k\in\mathbb{Z}}\in\cH$ such that
 \[
  \sum_{k\in\mathbb{Z}}\langle\eta_k,\xi_k\rangle_{\mathfrak{h}}\;=\;\langle \eta,\xi\rangle_{\cH}\;=\;\langle\psi,\mathscr{H}_\alpha\,\xi\rangle_{\cH}\;=\;\sum_{k\in\mathbb{Z}}\langle\psi_k,A_\alpha(k)\,\xi_k\rangle_{\mathfrak{h}}\qquad\forall \xi\in\mathcal{D}(\mathscr{H}_\alpha)\,.
 \]
 By localising $\xi$ separately in each fibre $\mathfrak{h}_k$ one then deduces that for each $k\in\mathbb{Z}$ $\psi_k\in\mathcal{D}(A_\alpha(k)^*)$ and $\eta_k=A_\alpha(k)^*\psi_k$, whence also $\sum_{k\in\mathbb{Z}}\big\|A_\alpha(k)^*\psi_k\big\|_{\mathfrak{h}}^2=\|\eta\|^2_\cH<+\infty$. This means that
 \[
  \psi\in\mathcal{D}(\bigoplus_{k\in\mathbb{Z}}A_\alpha(k)^*)\qquad\textrm{and}\qquad \mathscr{H}_\alpha^*\psi=(A_\alpha(k)^*\psi_k)_{k\in\mathbb{Z}}=(\bigoplus_{k\in\mathbb{Z}} A_\alpha(k)^*)\psi\,,
 \]
 thus proving \eqref{eq:Halphaadj-decomposable}. Taking further the adjoint on both sides of \eqref{eq:Halphaadj-decomposable} yields \eqref{eq:Halphaclosure-decomposable} (Sect.~\ref{sec:I_invariant-reducing-ssp}), whence also \eqref{eq:Halphaadj-sumkernel} (Sect.~\ref{sec:I_invariant-reducing-ssp}). 
\end{proof}

  Lemma \ref{lem:Halphaadj-decomposable} has an immediate consequence on the problem of essential self-adjointness of $\mathscr{H}_\alpha$, and hence $H_\alpha$.

  \begin{corollary}\label{cor:V-esa-via-fibres}
   The operator $\mathscr{H}_\alpha$, and hence $H_\alpha$, is essentially self-adjoint if and only if so too are all operators $A_\alpha(k)$, $k\in\mathbb{Z}$, each in the respective Hilbert space.
  \end{corollary}

  \begin{proof}
   An obvious consequence of the unitary equivalence $H_\alpha\cong\mathscr{H}_\alpha$ and of the comparison between \eqref{eq:Halphaadj-decomposable} and \eqref{eq:Halphaclosure-decomposable}.
  \end{proof}

  Another evident, but crucial technical point for constructing self-adjoint restrictions of $\mathscr{H}_\alpha^\pm$ and $\mathscr{H}_\alpha$ is the following.

\begin{proposition}\label{prop:BextendsHalpha}
 Let $\{B(k)\,|\,k\in\mathbb{Z}\}$ be a collection of operators on the fibre space $\mathfrak{h}$ (resp., $\mathfrak{h}^\pm$) such that, for each $k$, $B(k)$ is a self-adjoint extension of $A_\alpha(k)$ (resp., $A_\alpha^\pm(k)$), and let
 \begin{equation}\label{eq:B_direct_integral}
 B\;=\;\bigoplus_{k\in\mathbb{Z}} \,B(k)\,.
\end{equation}
 Then $B$ is a self-adjoint extension of $\mathscr{H}_\alpha$ (resp., $\mathscr{H}_\alpha^\pm$).
\end{proposition}


\begin{proof}
$B$ is an actual extension of $\mathscr{H}_\alpha$, because 
\[
 \mathscr{H}_\alpha\;\subset\;\bigoplus_{k\in\mathbb{Z}} \,A_\alpha(k)\;\subset\;\bigoplus_{k\in\mathbb{Z}} \,B(k)\,.
\]
It is straightforward to see that $B$ is symmetric, so in order to establish the self-adjointness of $B$ one only needs to prove that $\mathrm{ran}(B\pm\ii\mathbbm{1})=\cH$ (Sect.~\ref{sec:I-symmetric-selfadj}). 
For generic $\eta\equiv(\eta_k)_{k\in\mathbb{Z}}\in\cH$ set then $\psi_k:=(B(k)+\ii\mathbbm{1})^{-1}\eta_k$ $\forall k\in\mathbb{Z}$. By construction $\psi_k\in\mathcal{D}(B(k))$,  $\|\psi_k\|_{\mathfrak{h}}\leqslant\|\eta_k\|_{\mathfrak{h}}$, and $\|B(k)\psi_k\|_{\mathfrak{h}}\leqslant\|\eta_k\|_{\mathfrak{h}}$, whence also $\sum_{k\in\mathbb{Z}}\|\psi_k\|_{\mathfrak{h}}^2<+\infty$ and  $\sum_{k\in\mathbb{Z}}\|B(k)\psi_k\|_{\mathfrak{h}}^2<+\infty$. Therefore, $\psi\equiv(\psi_k)_{k\in\mathbb{Z}}\in\mathcal{D}(B)$. Moreover, $(B+\ii\mathbbm{1})\psi=((B(k)+\ii\mathbbm{1})\psi_k)_{k\in\mathbb{Z}}=(\eta_k)_{k\in\mathbb{Z}}=\eta$. This proves that $\mathrm{ran}(B +\ii\mathbbm{1})=\cH$. Analogously, $\mathrm{ran}(B -\ii\mathbbm{1})=\cH$.
\end{proof}

Proposition \ref{prop:BextendsHalpha} provides a \emph{mechanism} for constructing self-adjoint operators $B$ of the form \eqref{eq:B_direct_integral} by re-assembling, fibre by fibre in the momentum number $k$ conjugate to $y$, self-adjoint extensions of the fibre operators $A_\alpha(k)$;  by further exploiting the canonical unitary equivalence
\begin{equation}
 B\;\stackrel{\cong}{\longmapsto}\; \mathcal{F}_2^{-1} U_\alpha^{-1}\;B\:\mathcal{F}_2 U_\alpha\,,
\end{equation}
this yields the actual self-adjoint extensions of $H_\alpha$. With self-explanatory meaning, one refers to such extensions as \emph{momentum-fibred extensions}, or simply \emph{fibred extensions}.\index{fibred extensions}

Thus, fibred extensions have the distinctive feature of being characterised, in position-momentum coordinates $(x,k)$, by boundary conditions on the elements $\psi$ of their domain which connect the behaviour of \emph{each} mode $\psi_k(x)$ as $x\to 0^+$ and $x\to 0^-$, with no jump between different modes. In other words, such extensions are \emph{local} in momentum, whence their primary physical and conceptual relevance.

Evidently, $\mathscr{H}_\alpha$ (and hence $H_\alpha$) admits plenty of extensions that are \emph{non-local} in momentum, namely with boundary condition as $x\to 0^\pm$ that mixes different $k$-modes. 


It is also clear that a generic fibred extension of $\mathscr{H}_\alpha$ may or may not be reduced into a `left' and `right' component by the Hilbert space direct sum \eqref{eq:Hxispace}, whereas $\mathscr{H}_\alpha$ itself certainly is. Indeed, at the level of each fibre, the extension $B(k)$ may or may not be reduced by the sum $\mathfrak{h}=\mathfrak{h}^-\oplus\mathfrak{h}^+$ as is instead $A_\alpha(k)$ by construction (see \eqref{eq:Axibilateral} above). In fact, the decoupling between left and right half-cylinder may hold for \emph{all} modes $k\in\mathbb{Z}$ or only for some $k$'s. In the former case, the resulting extension of $\mathscr{H}_\alpha$ is in fact a mere `juxtaposition' of two separate extensions for $\mathscr{H}_\alpha^{\pm}$ in the left/right half-cylinder.

 \section{Confinement mechanisms on plane and cylinder}\label{sec:ess-self-adj}

 For a Grushin-type plane one has the following.
  
\begin{theorem}[Quantum confinement vs transmission in Grushin-type plane]\label{thm:V-plane-conf-noconf} For $\alpha\in\mathbb{R}$ let $M_\alpha\equiv(M,\mu_\alpha)$ be the Grushin-type plane introduced in \eqref{eq:Mgalpha} and let $H_\alpha$ be the operator on $\cH_\alpha=L^2(M,\ud\mu_\alpha)$ introduced in \eqref{eq:V-Halpha}.
 \begin{enumerate}[(i)]
  \item If $\alpha\in(-\infty,-1)\cup[1,+\infty)$, then $H_\alpha$ is essentially self-adjoint and therefore the Grushin plane $M_\alpha$ induces geometric quantum confinement.
  \item If $\alpha\in[-1,1)$, then $H_\alpha$ is not essentially self-adjoint and has infinite deficiency index.
 \end{enumerate}
\end{theorem}

 For a Grushin-type cylinder one has the following.

 \begin{theorem}[Quantum confinement vs transmission in Grushin-type cylinder]\label{thm:Halpha_esa_or_not}  
 For $\alpha\in\mathbb{R}$ let $M_\alpha\equiv(M,\mu_\alpha)$ be the Grushin-type cylinder introduced in \eqref{eq:Mgalpha-cylinder} and let $H_\alpha$ be the operator on $\cH_\alpha=L^2(M,\ud\mu_\alpha)$ introduced in \eqref{eq:V-Halpha}.
   \begin{enumerate}[(i)]
   \item If $\alpha\in(-\infty,-3]\cup[1,+\infty)$, then the operator $H_\alpha$ is essentially self-adjoint, and therefore the Grushin cylinder $M_\alpha$ induces geometric quantum confinement.
   \item If $\alpha\in(-3,-1]$, then the operator $H_\alpha$ is not essentially self-adjoint and it has deficiency index $2$.
   \item If $\alpha\in(-1,1)$, then the operator $H_\alpha$ is not essentially self-adjoint and it has infinite deficiency index.
  \end{enumerate}
 \end{theorem}

 Theorems \ref{thm:V-plane-conf-noconf} and \ref{thm:Halpha_esa_or_not} are qualitatively similar, yet with important differences in the range of essential self-adjointness and in the magnitude of the deficiency index. In either case their proof can be boiled down to the self-adjointness problem for the fibre operator $A_\alpha(\xi)$, or $A_\alpha(k)$ (introduced, respectively, in \eqref{eq:Aalphaplaneintro} and \eqref{eq:Axi}), based on the unitary equivalence re-formulation discussed in Sections \ref{sec:constant-fibre-integral} and \ref{sec:constant-fibre-sum}.

 Besides, instrumental for the proof of Theorem \ref{thm:V-plane-conf-noconf}, one can deal with the generalised metric \eqref{eq:Mf} and thus establish the following counterpart.

\begin{theorem}\label{thm:generalisation_to_f}
 Let $f$ be a measurable function satisfying assumptions \eqref{eq:assumtpions_f} and let $H_f$ be the corresponding operator defined in \eqref{Hf}. 
 \begin{enumerate}[(i)]
  \item If, point-wise for every $x\neq 0$, 
  \begin{equation}\label{eq:fcondition_conf}
   2 f f''-f'^2\;\geqslant\;\frac{3}{x^2}\,f^2\,,
  \end{equation}
  then $H_f$ is essentially self-adjoint with respect to the Hilbert space $\cH_f$, and therefore the generalised Grushin plane $M_f$ induces geometric quantum confinement.
  \item If, point-wise for every $x\neq 0$, 
  \begin{equation}\label{eq:fcondition_noconf}
   2 f f''-f'^2\;\leqslant\;\frac{3-\varepsilon}{x^2}\,f^2\qquad\textrm{for some }\varepsilon>0\,,
  \end{equation}
  then $H_f$ is not essentially self-adjoint and has infinite deficiency index.
 \end{enumerate}
 \end{theorem}

  The proof of Theorems \ref{thm:V-plane-conf-noconf}-\ref{thm:generalisation_to_f} is presented in this Section. The cylindric case (Theorem \ref{thm:Halpha_esa_or_not}) was repeatedly obtained in recent works \cite{Boscain-Laurent-2013,Boscain-Prandi-JDE-2016,Prandi-Rizzi-Seri-2016,Franceschi-Prandi-Rizzi-2017}, which implicitly also contain the tools for the analysis of the planar case (Theorem \ref{thm:V-plane-conf-noconf}). The subsequent analysis of \cite{Pozzoli_MSc2018,GMP-Grushin-2018,PozzoliGru-2020volume,GMP-Grushin2-2020} has clarified and generalised the fibred structure of the problem, with a full treatment of plane and cylinder.

 \begin{remark}
 The lack of geometric quantum confinement in $M_\alpha$ for $\alpha\in(0,1)$ is compatible with the quantum confinement in \emph{regular} almost-Riemannian structures proved recently in \cite[Theorem 7.1]{Prandi-Rizzi-Seri-2016}. Recall, as observed already in Section \ref{sec:V-GruplaneGrucyl}, that $M_\alpha$ does not carry an almost-Riemannian structure when $\alpha\in(0,1)$.
\end{remark}

 \begin{remark}
   Conditions \eqref{eq:fcondition_conf}-\eqref{eq:fcondition_noconf} are homogeneous in $f$, thus the same conclusion holds for $f(x)=\lambda x^{-\alpha}$, $\lambda>0$: this amounts to dilate the $y$-axis, in practice leaving the metric unchanged. 
  \end{remark}

 For the time being, $A_f^+(\xi)$, or $A_\alpha^+(\xi)$ in particular, shall be used for both the operators \eqref{eq:Aalphaplaneintro} and \eqref{eq:Axi}, initially considered in their one-sided version, on the fibre Hilbert space $L^2(\mathbb{R}^+,\ud x)$. In either case, $A_f^+(\xi)$ is a Schr\"{o}dinger operator of the form
\begin{equation}
 A_f^+(\xi)\;=\;-\frac{\ud^2}{\ud x^2}+W_{\xi,f}\,,\qquad W_{\xi,f}(x)\;:=\;\frac{\xi^2}{f^2}+\frac{2ff''-f'^2}{4f^2}\,,
\end{equation}
 the potential $W_{\xi,f}$ being everywhere real-valued and continuous, and positive around the origin, thanks to assumptions \eqref{eq:assumtpions_f} (i) and (iii). As such, one refers to just \emph{the} deficiency index of $A_f(\xi)$, and the essential self-adjointness of $A_f(\xi)$ is studied by classical methods, specifically with Weyl's limit-point/limit-circle analysis (Sect.~\ref{sec:WeylsCriterion}).\index{Weyl criterion!limit-point limit-circle}\index{Weyl limit-point/limit-circle}\index{limit-point/limit-circle}\index{theorem!Weyl (limit-point limit-circle criterion)}
  
  \begin{proposition}\label{prop:Axiselfadjointness}
 Let $\xi\in\mathbb{R}$ and let $f$ satisfy assumptions \eqref{eq:assumtpions_f}.
 \begin{enumerate}[(i)]
  \item If $2ff''-f'^2\geqslant 3 x^{-2} f^2$, then $A_f^+(\xi)$ is essentially self-adjoint.
  \item If $2ff''-f'^2\leqslant (3-\varepsilon) x^{-2} f^2$ for some $\varepsilon>0$, then $A_f^+(\xi)$ is not essentially self-adjoint and has unit deficiency index. 
 \end{enumerate}
\end{proposition}

 \begin{proof}
  $W_{\xi,f}$ being everywhere real-valued and continuous, and positive near the origin,  $A_f^+(\xi)$ is always in the limit-point \emph{at infinity} (it suffices to take $M(x)=x^2$ in the analysis of Section \ref{sec:WeylsCriterion}), and only the behaviour at \emph{at zero} remains to be investigated. Here one has two possibilities (Sect.~\ref{sec:WeylsCriterion}):
\begin{itemize}
 \item if $2ff''-f'^2\geqslant 3 x^{-2} f^2$, then $W_{\xi,\alpha}(x)\geqslant\frac{3}{4x^2}$, in which case $A_f^+(\xi)$ is in the limit-point at zero;
 \item if instead  $2ff''-f'^2\leqslant (3-\varepsilon) x^{-2} f^2$ for some $\varepsilon>0$, since $f^{-2}\leqslant\kappa^{-2}$ near $x=0$ (owing to \eqref{eq:assumtpions_f}(ii)), then $W_{\xi,\alpha}(x)\leqslant\kappa^{-2}\xi^2+(3-\varepsilon)/(4x^2)$, whence also, for some $\xi$-dependent $\widetilde\varepsilon\in(0,\varepsilon)$, $W_{\xi,\alpha}(x)\leqslant(3-\widetilde{\varepsilon})/(4x^2)$: in this case $A_f(\xi)^+$ is in the limit-circle at zero.
\end{itemize}
 The Weyl criterion\index{Weyl criterion!limit-point limit-circle}\index{Weyl limit-point/limit-circle}\index{limit-point/limit-circle}\index{theorem!Weyl (limit-point limit-circle criterion)} (Sect.~\ref{sec:WeylsCriterion}) then leads to the conclusion.
 \end{proof}

 Theorem \ref{thm:generalisation_to_f}'s proof then unfolds as follows.

   \begin{proof}[Proof of Theorem \ref{thm:generalisation_to_f}]
    For part (i), assume $2ff''-f'^2\geqslant 3 x^{-2} f^2$.
    The operator $A_f(\xi)$ is essentially self-adjoint (Proposition \ref{prop:Axiselfadjointness}(i)), therefore $B_f$ is self-adjoint (Proposition \ref{prop:RSpropBA}). It turns out that in this case 
\begin{equation*}\tag{a}\label{eq:Hfstar_in_Bf}
 (\mathscr{H}_f^+)^*\;\subset\;B_f\,.
\end{equation*}
 To establish \eqref{eq:Hfstar_in_Bf} it suffices to show that $\mathcal{D}((\mathscr{H}_f^+)^*)\subset\mathcal{D}(B_f)$, as the differential action of the two operators is the same (Lemma \ref{lemma:Hfstar}). To this aim, for generic $F\in\mathcal{D}((\mathscr{H}_f^+)^*)$ formula \eqref{eq:Hfstarplane} gives
\[\tag{b}\label{eq:Vproofb}
\begin{split}
 +\infty\;&>\;\Big\|\Big(-\frac{\partial^2}{\partial x^2}+\frac{\xi^2}{f^2}+\frac{2ff''-f'^2}{4f^2}\Big)F\Big\|^2_{\cH^+} \\
 &=\;\int_{\mathbb{R}}\ud\xi\, \Big\|\Big(-\frac{\ud^2}{\ud x^2}+\frac{\xi^2}{f^2}+\frac{2ff''-f'^2}{4f^2}\Big)F(\cdot,\xi)\Big\|^2_{\mathfrak{h}^+} \,,
\end{split}
\]
whence
\[
 \Big\|\Big(-\frac{\ud^2}{\ud x^2}+\frac{\xi^2}{f^2}+\frac{2ff''-f'^2}{4f^2}\Big)F(\cdot,\xi)\Big\|^2_{\mathfrak{h^+}}\;<\;+\infty\quad\textrm{for almost every }\xi\in\mathbb{R}\,.
\]
The latter formula, owing to \eqref{eq:Afstarplane}, can be re-written as
\[
 F(\cdot,\xi)\,\in\,\mathcal{D}(A_f^+(\xi)^*)\quad\textrm{for almost every }\xi\in\mathbb{R}\,,
\]
and since in the present regime $A_f^+(\xi)^*=\overline{A_f^+(\xi)}$, one can also write
\[\tag{c}\label{eq:Vc}
 F(\cdot,\xi)\,\in\,\mathcal{D}(\overline{A_f^+(\xi)})\quad\textrm{for almost every }\xi\in\mathbb{R}\,.
\]
Since \eqref{eq:Vproofb} and \eqref{eq:Vc} imply that $F\in\mathcal{D}(B_f)$, property \eqref{eq:Hfstar_in_Bf} is proved. Now one combines the inclusion $B_f\supset \overline{\mathscr{H}_f^+}$, that follows from \eqref{eq:BextendsHf} and from the closedness of $B_f$, with the inclusion $B_f\subset \overline{\mathscr{H}_f^+}$, that follows from \eqref{eq:Hfstar_in_Bf} by taking the adjoint, because $\overline{\mathscr{H}_f^+}=\mathscr{H}_f^{**}\supset B_f^*=B_f$, having used the self-adjointness of $B_f$ valid in the present regime. The conclusion is $\overline{\mathscr{H}_f^+}=B_f$, that is, $\mathscr{H}_f^+$ is essentially self-adjoint. The same obviously holds for $\mathscr{H}_f^-$ and $\mathscr{H}_f=\mathscr{H}_f^-\oplus \mathscr{H}_f^+$ as well.

 For part (ii), by unitary equivalence the first claim to prove is the lack of essential self-adjointness in $\cH^+=L^2(\mathbb{R}^+\times\mathbb{R},\ud x\,\ud\xi)$ of the operator $\mathscr{H}_f^+$ defined in \eqref{eq:Hscrf}.
 By assumption now $2ff''-f'^2\leqslant (3-\varepsilon) x^{-2} f^2$ for some $\varepsilon>0$. 
 In this case, for no $\xi\in\mathbb{R}$ is $A_f^+(\xi)$ essentially self-adjoint (Proposition \ref{prop:Axiselfadjointness}(ii)), and therefore $B_f$ is not self-adjoint (Proposition \ref{prop:Bselfadj-implies-Axiselfadj}). 
 On the other hand, $B_f$ is a closed symmetric extension of $\mathscr{H}_f^+$, owing to \eqref{eq:BextendsHf} and to Proposition \ref{prop:BsymmetricAndClosed}, whence $\overline{\mathscr{H}_f^+}\subset B_f$.
 Now, if $\mathscr{H}_f^+$ was essentially self-adjoint, it could not be $\overline{\mathscr{H}_f^+}= B_f$, because this would violate the lack of self-adjointness of $B_f$. But it could not happen either that $B_f$ is a \emph{proper} extension of $\overline{\mathscr{H}_f^+}$, because self-adjoint operators are maximally symmetric. Therefore, $\mathscr{H}_f^+$ is not essentially self-adjoint, nor are $\mathscr{H}_f^-$ and $\mathscr{H}_f=\mathscr{H}_f^-\oplus \mathscr{H}_f^+$. In this case, the subspaces $\ker((\mathscr{H}_f^+)^*\pm\ii\mathbbm{1})$ are non-trivial, and the last claim to prove is
\begin{equation*}\tag{d}\label{eq:infinite_def_space}
 \dim\ker((\mathscr{H}_f^+)^*\pm\ii\mathbbm{1})\;=\;\infty\,.
\end{equation*}
 Since by assumption each $\overline{A_f^+(\xi)}$ is not self-adjoint, there exists $\varphi_{\xi}\in\mathcal{D}(A_f^+(\xi)^*)$ with $\|\varphi_{\xi}\|_{\mathfrak{h}^+}=1$ such that
\begin{equation*}\tag{e}\label{eq:eigenf_adjoint}
 (A_f(\xi)^+)^*\,\varphi_{\xi}\;=\;\ii \varphi_{\xi}\,.
\end{equation*}
From this, it is possible to deduce that for any compact interval $J\subset\mathbb{R}$, with $\mathbf{1}_J$ characteristic function of $J$, the function $\Phi_J$ defined by $\Phi_J(x,\xi):=\varphi_{\xi}(x)\mathbf{1}_J(\xi)$ satisfies
\begin{equation*}\tag{f}\label{eq:Phi_eigenfunction}
 \Phi_J\,\in\,\mathcal{D}((\mathscr{H}_f^+)^*)\,,\qquad (\mathscr{H}_f^+)^*\,\Phi_J\;=\;\ii\,\Phi_J\,.
\end{equation*}
 Indeed, the fact that $\Phi_J\in\cH^+$ follows from $\|\Phi_J\|_{\cH^+}^2=\int_\mathbb{R}\ud\xi\,\mathbf{1}_J(\xi)\,\|\varphi_{\xi}\|_{\mathfrak{h}^+}^2=|J|$, where $|J|$ denotes the Lebesgue measure of $J$. Moreover, for any $\psi\in\mathcal{D}(\mathscr{H}_f^+)$,
\[
 \begin{split}
  \langle\Phi_J,\mathscr{H}_f^+ \psi\rangle_{\cH^+}\;&=\;\iint_{\mathbb{R}^+\times\mathbb{R}}\ud x\,\ud\xi\,\overline{\varphi_{\xi}(x)}\mathbf{1}_J(\xi)\,A_f^+(\xi)\psi(x,\xi) \\
  &=\int_J\ud\xi\,\langle \varphi_\xi,A_f^+(\xi)\psi(\cdot,\xi)\rangle_{\mathfrak{h}^+}\;=\;\int_J\ud\xi\,\langle A_f^+(\xi)^*\varphi_\xi,\psi(\cdot,\xi)\rangle_{\mathfrak{h}^+} \\
  &=\;-\ii\int_J\ud\xi\,\langle \varphi_\xi,\psi(\cdot,\xi)\rangle_{\mathfrak{h}^+}\;=\;\langle\,\ii\,\Phi_J, \psi\rangle_{\cH^+}\,,
 \end{split}
\]
where \eqref{eq:eigenf_adjoint} was used in the fourth identity, and this establishes precisely \eqref{eq:Phi_eigenfunction}. By the arbitrariness of $J$, and the obvious orthogonality $\Phi_J\perp\Phi_{J'}$ whenever $J\cap J'=\varnothing$, the above reasoning produces an
infinity of linearly independent eigenfunctions of $(\mathscr{H}_f^+)^*$ with eigenvalue $\ii$, and the same can be clearly repeated for the eigenvalue $-\ii$. This establishes \eqref{eq:infinite_def_space}.
\end{proof}

   When, in particular, $f(x)=|x|^{-\alpha}$, one has 
   \begin{eqnarray}
      \frac{2ff''-f'^2}{4f^2}\;&=&\;\frac{\,\alpha(2+\alpha)\,}{4 x^2}\,, \label{eq:V-specialchoice1}\\
      \frac{2ff''-f'^2}{4f^2}-\frac{3}{ 4x^2}\;&=&\;\frac{\,(\alpha-1)(\alpha+3)\,}{4 x^2}\,, \label{eq:V-specialchoice2}  \\
       W_{\xi,f}(x)\;&\equiv&\;W_{\xi,\alpha}(x)\;:=\;\xi^2 x^{2\alpha}+\frac{\,\alpha(2+\alpha)\,}{4x^2}\,. \label{eq:V-Wxialphapotential}
   \end{eqnarray}
   With this choice, assumptions \eqref{eq:assumtpions_f}(i) and (iii) are automatically satisfied; for assumption \eqref{eq:assumtpions_f}(ii) to be matched, one must take $\alpha\geqslant 0$; and for the validity of assumption  \eqref{eq:assumtpions_f}(iv) one must take $\alpha(2+\alpha)\geqslant 0$, that is, $\alpha\in(-\infty,-2]\cup [0,+\infty)$. Thus, in the regime $\alpha\geqslant 0$ Theorem \ref{thm:generalisation_to_f} is indeed applicable: condition \eqref{eq:fcondition_conf} reads $(\alpha-1)(\alpha+3)\geqslant 0$, compatibly with $\alpha\geqslant 0$, that is, $\alpha\geqslant 1$, and this is a regime of essential self-adjointness for $\mathscr{H}_\alpha$; conversely, condition \eqref{eq:fcondition_noconf} reads $(\alpha-1)(\alpha+3)< 0$, compatibly with $\alpha\geqslant 0$, that is, $\alpha\in[0,1)$, and this is a regime of infinite deficiency index. This shows that Theorem \ref{thm:generalisation_to_f} implies Theorem \ref{thm:V-plane-conf-noconf} for all $\alpha\geqslant 0$.

  For the proof of Theorem \ref{thm:V-plane-conf-noconf} in the remaining regime $\alpha<0$ the reasoning requires an easy adaptation, as discussed below, still within the constant-fibre direct integral scheme. Prior to that, it is informative to highlight that the one-sided fibre operator $A_\alpha^+(\xi)$ is non-negative irrespective of $\alpha$ and hence, having equal deficiency indices (Sect.~\ref{sec:I-symmetric-selfadj}), it is either essentially self-adjoint or it admits self-adjoint extensions. This follows from the standard one-dimensional inequality by Hardy\label{index:Hardy inequality}
\begin{equation}\label{eq:hardy1}
 \int_0^{+\infty}\frac{\;|h(x)|^2}{\;x^2}\,\ud x\;\leqslant\;2\int_0^{+\infty}|h'(x)|^2\,\ud x\qquad \forall\,h \in C^\infty_0(\mathbb{R}^+)\,.
\end{equation}

  \begin{lemma}\label{lem:V-Aalphak-positive}
   For any $\alpha\in\mathbb{R}$ and $\xi\in\mathbb{R}$ the operator $A_\alpha^+(\xi)$ is non-negative.
  \end{lemma}

  \begin{proof}
   The claim is obvious for $\alpha\in(-\infty,-2]\cup[0,+\infty)$ because in this range $W_{\xi,\alpha}(x)>0$ for all $x>0$. On the other hand, the Hardy inequality \eqref{eq:hardy1} implies
  \[
   \begin{split}
   & \langle h,A_\alpha^+(\xi) h\rangle_{L^2} \\
   &\qquad =\;|\xi|^2\int_{\mathbb{R}^+} x^{2\alpha}|h(x)|^2\,\ud x+\int_{\mathbb{R}^+}|h'(x)|^2\,\ud x+\frac{\alpha(2+\alpha)}{4}\int_{\mathbb{R}^+}\frac{\;|h(x)|^2}{\;x^2}\,\ud x \\
   &\qquad \geqslant\; \Big(1-\frac{|\alpha(2+\alpha)|}{2}\Big)\int_{\mathbb{R}^+}|h'(x)|^2\,\ud x\;\geqslant\;0 \\
   & \qquad\qquad\qquad \forall h\in\mathcal{D}(A_\alpha^+(\xi))=C^\infty_0(\mathbb{R}^+)\,,\quad \forall\alpha\in[-\sqrt{3}-1,\sqrt{3}-1]\,.
   \end{split}
  \]
    Therefore, the range $\alpha\in(-2,0)$ is covered too.
  \end{proof}

  
  It is then appropriate to only speak of \emph{the} deficiency index of $A_\alpha^+(\xi)$. Proposition \ref{prop:Axiselfadjointness} is replaced by the following.

    \begin{proposition}\label{prop:Axiselfadjointness-general}
 Let $\xi\in\mathbb{R}$ and let $\alpha\in\mathbb{R}$.
 \begin{enumerate}[(i)]
  \item If $\alpha\in (-\infty,-3]\cup[1,+\infty)$, then $A_\alpha^+(\xi)$ is essentially self-adjoint irrespective of $\xi$.
  \item If $\alpha\in(-1,1)$, then, irrespective of $\xi$, $A_\alpha^+(\xi)$ is not essentially self-adjoint and has unit deficiency index.
  \item If $\alpha=-1$, then $A_{-1}^+(\xi)$ is essentially self-adjoint for $|\xi|\geqslant 1$ and has unit deficiency index for $|\xi|<1$.
  \item If $\alpha\in(-3,-1)$ then $A_\alpha^+(\xi)$ is essentially self-adjoint for $\xi\neq 0$ and has unit deficiency index for $\xi=0$.
 \end{enumerate}
\end{proposition}

 \begin{proof}
  This is another application of the Weyl limit-point limit-circle criterion\index{Weyl criterion!limit-point limit-circle}\index{Weyl limit-point/limit-circle}\index{limit-point/limit-circle}\index{theorem!Weyl (limit-point limit-circle criterion)} (Sect.~\ref{sec:WeylsCriterion}). 
  Since, irrespective of $\alpha,\xi\in\mathbb{R}$, $W_{\xi,\alpha}(x)\geqslant-x^2$ near infinity, $A_\alpha^+(\xi)$ is in the limit point case at infinity.

  Near zero, one sees from \eqref{eq:V-specialchoice2}-\eqref{eq:V-Wxialphapotential} that $W_{\xi,\alpha}(x)\geqslant \frac{3}{4x^2}$ (for all $x>0$, and in particular around the origin) for  $(\alpha-1)(\alpha+3)\geqslant 0$, namely $\alpha\in (-\infty,-3]\cup[1,+\infty)$. In this regime of $\alpha$ the operator $A_\alpha^+(\xi)$ is in the limit-point case at zero and hence it is essentially self-adjoint. This proves part (i).

  In the remaining regime $\alpha\in(-3,1)$ one easily finds the following:
  \begin{enumerate}[(a)]
   \item if $\alpha\in(-3,-1)$ and $\xi\neq 0$, then $W_{\xi,\alpha}(x)> \frac{3}{4x^2}$ near zero, therefore $A_\alpha^+(\xi)$ is in the limit-point case at zero and hence it is essentially self-adjoint;
   \item if $\alpha\in(-3,-1)$ and $\xi=0$, then $|W_{\xi,\alpha}(x)|< \frac{3}{4x^2}$ near zero, therefore $A_\alpha^+(\xi)$ is in the limit-circle case at zero and hence it has unit deficiency index; 
   \item if $\alpha=-1$, then 
     \[
   W_{\xi,-1}(x)\;=\;\frac{\xi^2-\frac{1}{4}}{x^2}\,; 
  \]
   therefore, when $|\xi|<1$ one has $|W_{\xi,-1}(x)|<\frac{3}{4x^2}$ for all $x>0$, thus $A_\alpha^+(\xi)$ is in the limit-circle case at zero and hence it has unit deficiency index, whereas when instead $|\xi|\geqslant 1$ one has $W_{\xi,-1}(x)\geqslant\frac{3}{4x^2}$ for all $x>0$, thus $A_\alpha^+(\xi)$ is in the limit-point case at zero and hence it is essentially self-adjoint; in fact, $A_{-1}(\xi)$ is precisely the minimally defined \emph{Bessel operator}\index{Bessel operator}\index{operator!Bessel}, for which it is well-known (see, e.g., \cite{Derezinski-Georgescu-2021}) that it is essentially self-adjoint for $|\xi|\geqslant 1$ and has unit deficiency index for $|\xi|<1$;
   \item if $\alpha\in(-1,0)$, then $|W_{\xi,-1}(x)|<\frac{3}{4x^2}$ near zero, and if $\alpha\in[0,1)$, then $0<W_{\xi,-1}(x)<\frac{3}{4x^2}$ for all $x>0$: in either case $A_\alpha^+(\xi)$ is in the limit-circle case at zero and hence it has unit deficiency index.
  \end{enumerate}

  Items (a)-(b) above prove part (iv), item (c) proves part (iii), item (d) proves part (ii).
 \end{proof}

  \begin{proof}[Proof of Theorem \ref{thm:V-plane-conf-noconf}]
   As previously argued, the remaining range to consider is $\alpha<0$. For completeness of discussion, the whole range $\alpha\in\mathbb{R}$ is included here. In fact, the proof consists of inserting the information of Proposition \ref{prop:Axiselfadjointness-general} into the very same scheme of the proof of Theorem \ref{thm:generalisation_to_f}. Actually, Proposition \ref{prop:Axiselfadjointness-general} implies that $A_\alpha^+(\xi)$ is essentially self-adjoint, for all $\xi\in\mathbb{R}$ or at most for $\xi\neq 0$, whenever $\alpha\in(-\infty,-1)\cup[1,+\infty)$ (essential self-adjointness holds with only a zero-measure exception of $\xi$). The same holds obviously on the whole two-sided fibre for $A_\alpha(\xi)=A_\alpha^-(\xi)\oplus A_\alpha^+(\xi)$, as a direct sum of two one-sided essentially self-adjoint operators. Therefore (Proposition \ref{prop:RSpropBA}), the concrete analogue of $B_f$, namely
   \[
    B_\alpha\;:=\;\int_{\mathbb{R}}^{\oplus}\overline{A_\alpha^+(\xi)}\,\ud\xi
   \]
   is self-adjoint in the Hilbert space $\cH$. The proof now proceeds precisely as the proof of Theorem \ref{thm:generalisation_to_f}(i). Part (i) of Theorem \ref{thm:V-plane-conf-noconf} is established.

   On the other hand, Proposition \ref{prop:Axiselfadjointness-general} shows that when $\alpha\in(-1,1)$ the one-sided fibre operator $A_\alpha^+(\xi)$ has deficiency index 1, hence its two-sided version $A_\alpha(\xi)=A_\alpha^-(\xi)\oplus A_\alpha^+(\xi)$ has deficiency index 2, irrespective of $\xi\in\mathbb{R}$. Besides, when $\alpha=-1$, $A_\alpha(\xi)$ has deficiency index 2 for all $\xi\in(-1,1)$. Thus, for the whole range $\alpha\in[-1,1)$ $B_\alpha$ is not essentially self-adjoint (Proposition \ref{prop:Bselfadj-implies-Axiselfadj}). The proof now proceeds precisely as the proof of Theorem \ref{thm:generalisation_to_f}(ii): the lack of essential self-adjointness of $B_\alpha$ prevents $\mathscr{H}_\alpha^+$ to be so, and the very same localisation-in-$\xi$ argument shows that the deficiency index is infinite. Part (ii) of Theorem \ref{thm:V-plane-conf-noconf} is established.
  \end{proof}

  \begin{proof}[Proof of Theorem \ref{thm:Halpha_esa_or_not}]
   Owing to Corollary \ref{cor:V-esa-via-fibres}, $\mathscr{H}_\alpha$ (and hence $H_\alpha$) is essentially self-adjoint if and only if so too are all the one-sided fibre operators $A_\alpha^\pm(k)$, and Proposition \ref{prop:Axiselfadjointness-general}(i) ensures that this is the case only when $\alpha\in(-\infty,-3]\cup[1,+\infty)$. Part (i) is thus proved.
   
   When $\alpha\in(-3,-1]$, $A_\alpha(0)=A_\alpha^-(0)\oplus A_\alpha^+(0)$ has deficiency index 2, whereas all other $A_\alpha(k)$ with $k\in\mathbb{Z}\setminus\{0\}$ are essentially self-adjoint (Proposition \ref{prop:Axiselfadjointness-general}(iii)-(iv)). This information, and the comparison between \eqref{eq:Halphaadj-decomposable}-\eqref{eq:Halphaclosure-decomposable} of Lemma \ref{lem:Halphaadj-decomposable} shows that $\mathscr{H}_\alpha$ has deficiency index 2. Part (ii) is proved.
   
   Last, when $\alpha\in(-1,1)$, $A_\alpha(k)$ has deficiency index 2 for all $k\in\mathbb{Z}$ (Proposition \ref{prop:Axiselfadjointness-general}(ii): with this information, \eqref{eq:Halphaadj-decomposable}-\eqref{eq:Halphaclosure-decomposable} now imply that $\mathscr{H}_\alpha$ has infinite deficiency index, thus proving part (iii) as well.   
  \end{proof}

  In retrospect, the different ranges of essential self-adjointness and magnitude of deficiency index for the minimal (Laplace-Beltrami) Hamiltonian $H_\alpha$ on Grushin-type plane and on Grushin-type cylinder are due to the following circumstance \cite{PozzoliGru-2020volume}.
  \begin{itemize}
   \item In the constant-fibre direct integral structure for the plane a single mode $\xi$ (or, in fact, a measure-zero collection of modes) of non essential self-adjointness for the fibre operator $A_\alpha(\xi)$ does not affect the overall self-adjointness of
   \[
    \int_\mathbb{R} \overline{A_\alpha(\xi)}\,\ud \xi\,.
   \]
   This yields essential self-adjointness for $H_\alpha$ in the whole regime $\alpha\in(-\infty,-1)\cup[1,+\infty)$ even though for $\alpha\in(-3,-1)$ the zero mode fibre operator $A_\alpha(0)$ has non-trivial deficiency index. In the remaining regime $\alpha\in[-1,1)$, $A_\alpha(\xi)$ contributes with deficiency index 2 in all modes $\xi\in\mathbb{R}$ when $\alpha\in(-1,1)$, and in all modes $\xi\in(-1,1)$ when $\alpha=-1$, thereby yielding infinite deficiency index for $H_\alpha$.
   \item In the constant-fibre orthogonal sum structure for the cylinder, each $k$-mode operator $A_\alpha(k)$ need be essentially self-adjoint, for the overall self-adjointness of
   \[
    \bigoplus_{k\in\mathbb{Z}}A_\alpha(k)\,,
   \]
   and this is the case only for $\alpha\in(-\infty,-3]\cup[1,+\infty)$ because of the lack of essential self-adjointness in the zero-mode when $\alpha\in(-3,-1]$. Since in the regime $\alpha\in(-3,-1]$ $A_\alpha(k)$ only fails to be essentially self-adjoint when $k=0$, in such a regime the overall contribution to the deficiency index of $H_\alpha$ is indeed 2. In the remaining range $\alpha\in(-1,1)$ the deficiency index is 2 for each mode $k\in\mathbb{Z}$ and therefore the overall deficiency index of $H_\alpha$ is infinite.
  \end{itemize}

%
%

 \section{Related settings on almost Riemannian manifolds}\label{sec:related-and-plane}\index{almost-Riemannian structure}



  Beside the above concrete cylindric and planar settings, the deep connection between geometry and self-adjointness is investigated for the problem of geometric confinement on more general almost-Riemannian structures.

   This includes \emph{two-step two-dimensional almost-Riemannian structures}\index{almost-Riemannian structure} \cite{Agrachev-Boscain-Sigalotti-2008}, characterised by an orthonormal frame for the metric in the vicinity of the singularity locus $\mathcal{Z}$ of the form 
  \begin{equation}
   X_1(x,y)\;=\;\dfrac{\partial}{\partial x}\,, \qquad X_2(x,y)\;=\;xe^{\phi(x,y)}\dfrac{\partial}{\partial y}
  \end{equation}
  (to be compared to \eqref{eq:frame} with $\alpha=1$) for some smooth function $\phi$ vanishing at $x=0$. The essential self-adjointness of the corresponding minimally defined Laplace-Beltrami\index{Laplace-Beltrami operator} in the case of compactified $\mathcal{Z}$ was established in \cite{Boscain-Laurent-2013}.

   From a related perspective, the already observed circumstance that Grushin-type cylinders or planes are, classically, geodesically incomplete, but can induce, quantum-mechanically, geometric confinement (a condition that occurs more generally for regular almost-Riemannian manifold\index{almost-Riemannian structure}  with compact singular set), poses an intriguing problem as far as semi-classical analysis is concerned. Indeed, reinstating Planck's constant ($\varepsilon\equiv\hslash$) in the Schr\"{o}dinger equation
   \begin{equation}\label{eq:Schreps}
    \ii\partial_t\psi+\varepsilon^2\Delta_{\mu_\alpha}\psi\;=\;0\,,\qquad\varepsilon>0
   \end{equation}
   (in the regime of $\alpha$ in which the minimally defined $\Delta_{\mu_\alpha}$ is unambiguously realised self-adjointly), semi-classics show, informally speaking, that as $\varepsilon\downarrow 0$ solutions get concentrated and evolves around geodesics. Therefore, the above-mentioned classical/quantum discrepancy makes the semi-classical analysis necessarily break down in the limit.

   Such a discordance between classical and quantum picture can be at least partially resolved by appealing to different quantisation procedures on the considered Riemannian manifold, in practice considering corrections of the Laplace-Beltrami operator\index{Laplace-Beltrami operator} that have a suitable interpretation of free kinetic energy, much in the original spirit of \cite{DeWitt-1957}. Most of coordinate-invariant quantisation procedures (including path integral quantisation, covariant Weyl quantisation, geometric quantisation, and finite-dim\-en\-sional approximation to Wiener measures) modify $\Delta_{\mu_\alpha}$ with a term that depends on the scalar curvature $R_\alpha$ (which, in two dimensions, is twice the Gaussian curvature $K_\alpha$). This produces a replacement in \eqref{eq:Schreps}, in two dimensions, of $-\Delta_{\mu_\alpha}$ with the \emph{curvature Laplacian}\index{curvature Laplacian}
   \begin{equation}
    -\Delta_{\mu_\alpha}+cK_\alpha
   \end{equation}
   for suitable $c> 0$. In the recent work \cite{Boscain-Beschastnnyi-Pozzoli-2020} it was indeed shown, for generic two-step two-dimensional almost-Riemannian manifolds with compact singular set, that irrespective of $c\in(0,\frac{1}{2})$ the above correction washes essential self-adjointness out, yielding a quantum picture where the Schr\"{o}dinger particle does reach the singularity much as the classical particle does. (At the expenses of some further technicalities, the whole regime $c>0$ can be covered as well.)

   For concreteness, in the Grushin cylinder the effect of the curvature correction is evidently understood as a compensation between $K=-\frac{2}{x^2}$ (see \eqref{eq:curvature} above) and the singular term $\frac{3}{4x^2}$ of the (unitary equivalent) Laplace-Beltrami\index{Laplace-Beltrami operator} operator. Still, the classical/quantum discrepancy discussed so far remains unexplained in more general settings.

   Concerning, instead, the heat flow\index{heat flow}, a satisfactory interpretation of the heat-confinement in the Grushin cylinder is known in terms of Brownian motions \cite{Boscain-Neel-2019} and random walks \cite{Agra-Bosca-Neel-Rizzi-2018}: roughly speaking, random particles are lost in the infinite area strip around $\mathcal{Z}$: the latter, in practice, acts as a barrier. Clearly, whereas curvature Laplacians are meaningful in the above context of inducing a non-confining (transmitting) Schr\"{o}dinger flow on two-step two-dimensional almost-Riemannian manifolds (including the Grushin cylinders), thus making quantum and classical picture more alike and well connected by semi-classics, this has no direct meaning instead in application to the heat flow on Riemannian or almost Riemannian manifolds. Indeed, as long as one regards the heat equation on manifold as a limit of a space-time discretised random walk, the stochastic process' generator \emph{is} the Laplace-Beltrami operator.\index{Laplace-Beltrami operator}
   

  Generalisations of \cite{Boscain-Laurent-2013} have been established in \cite{Franceschi-Prandi-Rizzi-2017,Prandi-Rizzi-Seri-2016} from two-step two-dimensional almost-Riemannian structures\index{almost-Riemannian structure}  to any dimension, any step, and even to sub-Riemannian geometries, provided that certain geometrical assumptions on the singular set are taken. The main difficulty is the treatment of the \emph{tangency} (or \emph{characteristic}) \emph{points}:\index{characteristic (tangency) points} these are points belonging to the singularity of the metric structure where the vector distribution is tangent to the singularity. They are never present in Grushin cylinder or two-step almost-Riemannian structures,\index{almost-Riemannian structure}  but may appear for example in three-step structures, such as, for instance,
\begin{equation}\label{ex:parabola}
X_1(x,y)\;=\;\dfrac{\partial}{\partial x}\,, \qquad X_2(x,y)\;=\;(y-x^2) \dfrac{\partial}{\partial y}\, ,\quad (x,y)\in \mathbb{R}^2 \,,
\end{equation}
 where the singularity is the parabola $y=x^2$ and the origin $(0,0)$ is a tangency point. Virtually nothing is known on the heat or the quantum confinement on such singular structures, including the simplest example \eqref{ex:parabola} (see \cite{Franceschi-Prandi-Rizzi-2015} for further remarks). First preliminary results in this respect were recently obtained in \cite{IB-2021}, where the interpretation of almost-Riemannian structures as special Lie manifolds permits to study some closure properties of singular perturbations of the Laplace-Beltrami operator\index{Laplace-Beltrami operator} even in the presence of tangency points. This opens new perspectives of treating several types of different singularities in sub-Riemannian geometry within the same unifying theory.

  \section{Extensions on one-sided fibre}\label{sec:fibre-extensions}

 From this point on, and for the rest of this Chapter, only the range $\alpha\in[0,1)$ for the metric on Grushin-type cylinders will be analysed.

 This is the most relevant regime of lack of essential self-adjointness (Theorem \ref{thm:Halpha_esa_or_not}) for the minimal free Hamiltonian\index{minimal free Hamiltonian} $H_\alpha$ introduced in \eqref{eq:V-Halpha}, and hence for the existence of a multiplicity of distinct transmission protocols\index{transmission protocol} (self-adjoint extensions). Indeed, it is both a regime of infinite deficiency index and of singularity of the Grushin metric.

 The ultimate goal (Theorem \ref{thm:H_alpha_fibred_extensions}) is to identify, classify, and characterise, in the spirit of the discussion made at the end of Section \ref{sec:constant-fibre-sum}, the most relevant family of physically meaningful self-adjoint extensions of $H_\alpha$ on the Grushin Hilbert space $\cH_\alpha$. This is efficiently done in the unitarily equivalent framework of the minimal operator $\mathscr{H}_\alpha$ on the Hilbert space $\cH$ (see \eqref{eq:V-two-sided-1}-\eqref{eq:V-two-sided-7} above). Indeed, although $\mathscr{H}_\alpha$ is not reduced with respect to the natural constant-fibre orthogonal decomposition \eqref{eq:V-two-sided-4}-\eqref{eq:Hoplusfrakh} of $\cH$ (Remark \ref{rem:Halphanotsum}), its adjoint $\mathscr{H}_\alpha^*$ is (Lemma \ref{lem:Halphaadj-decomposable}) which, as already mentioned (Sect.~\ref{sec:constant-fibre-sum}), naturally links the problem of self-adjoint \emph{restrictions} of $\mathscr{H}_\alpha^*$ to the knowledge of the self-adjoint realisations of each fibre operator $A_\alpha(k)$. In turn, the self-adjointness problem for the two-sided $A_\alpha(k)$ is solved once one solves its one-sided version for the operators $A_\alpha^\pm(k)$.

 The object of this Section is therefore the classification the self-adjoint extensions of the (right) one-sided fibre operators $A_\alpha(k)^+$ defined in \eqref{eq:Axi} for $\alpha\in[0,1)$ and $k\in\mathbb{Z}$, with respect to the one-sided fibre Hilbert space $L^2(\mathbb{R}^+)$. For simplicity of notation, the superscript `$+$' will be dropped throughout this Section, just writing $A_\alpha(k)$ for $A_\alpha(k)^+$, and $\langle \cdot,\cdot\rangle_{L^2}$ and $\|\cdot\|_{L^2}$ for scalar products and norms taken in $L^2(\mathbb{R}^+)$, with analogous notation for the Sobolev norms. Obviously, the whole discussion can be repeated verbatim for $A_\alpha(k)^-$ in $L^2(\mathbb{R}^-)$, with completely analogous conclusions.

 In fact, the entirety of the methods of the following discussion are adjustable and applicable as well to the other regime $\alpha\in(-3,0)$ of lack of self-adjointness for $A_\alpha(k)$ (Proposition \ref{prop:Axiselfadjointness-general}). What is specific of $\alpha\in[0,1)$ and will be used explicitly is the (strict) positivity property
 \begin{equation}\label{eq:Axibottom}
 \langle h,A_\alpha(k)h\rangle_{L^2}\;\geqslant\;M_{\alpha,k}\|h\|_{L^2}^2\qquad\forall h\in \mathcal{D}(A_\alpha(k))
\end{equation}
 that follows from \eqref{eq:Axi} and from
 \begin{equation}
   M_{\alpha,k}\;:=\;\min_{x\in\mathbb{R}^+}\Big( k^2 x^{2\alpha}+\frac{\,\alpha(2+\alpha)\,}{4x^2}\Big)\;=\;(1+\alpha)\big(\textstyle{\frac{2+\alpha}{4}}\big)^{\frac{\alpha}{1+\alpha}}|k|^{\frac{2}{1+\alpha}} \, ,
 \end{equation}
 valid for all $k\in\mathbb{Z}$ and $\alpha\geqslant 0$. Observe that when $k=0$, in particular,
\begin{equation}\label{eq:Axibottom-zero}
 \inf_{h\in\mathcal{D}(A_\alpha(0))\setminus\{0\}}\frac{\langle h,A_\alpha(0)h\rangle_{L^2}}{\|h\|_{L^2}^2}\;=\;0\,.
\end{equation}
 (Actually, \eqref{eq:Axibottom} is a refinement of the already available information $A_\alpha(k)\geqslant\mathbb{O}$ obtained by means of Hardy inequality\index{Hardy inequality} in the course of the proof of Proposition \ref{prop:Axiselfadjointness-general}.) Property \eqref{eq:Axibottom} leads to a determination of the deficiency space of $A_\alpha(k)$ that is specific of the regime $\alpha\in[0,1)$ (Sect.~\ref{subsec:homog_problem}): the computation would be different for the other non essential self-adjoint regime $\alpha\in(-3,0)$.

 Proposition \ref{prop:Axiselfadjointness-general} states that in the regime of interest $\alpha\in[0,1)$ the operator $A_\alpha(k)$ has unit deficiency index irrespective of $k\in\mathbb{Z}$, hence admits a one-(real-)parameter family of self-adjoint extensions. The extension problem for the special case $\alpha=0$ is a classical one (see, e.g., \cite{GTV-2012,DM-2015-halfline}), since $A_0(k)$ is the minimally defined, shifted Laplacian $-\frac{\ud^2}{\ud x^2}+ k^2$ on $L^2(\mathbb{R}^+)$. The discussion that follows is therefore focussed on the regime $\alpha\in(0,1)$: the final results, in the limit $\alpha\downarrow 0$, do match with the known formulas for extensions of the minimal Laplacian on half-line.

 The fact that $A_\alpha(k)$ is non-negative makes the Kre{\u\i}n-Vi\v{s}ik-Birman extension scheme applicable. In particular (see \eqref{eq:Axibottom}), the fact that $A_\alpha(k)$ has \emph{strictly positive} lower bound for $k\in\mathbb{Z}\setminus\{0\}$ allows one to directly apply the Kre{\u\i}n-Vi\v{s}ik-Birman scheme in the form discussed in Section \ref{sec:II-VBreparametrised}. The final result for this analysis is Theorem \ref{thm:fibre-thm} below. The mode $k=0$ is then discussed upon shifting $A_\alpha(0)$ to $A_\alpha(0)+\mathbbm{1}$ and re-doing the analysis: the final result in this case is Theorem \ref{thm:fibre-thm_zero_mode}.

For convenience of notation, set
\begin{equation}
 C_\alpha\;:=\;\frac{\,\alpha(2+\alpha)}{4}\,,
\end{equation}
 so that $C_\alpha\in[0,\frac{3}{4})$, and 
\begin{equation}\label{eq:Saxi}
 S_{\alpha,k}\;:=\; -\frac{\ud^2}{\ud x^2}+ k^2 x^{2\alpha}+\frac{C_\alpha}{x^2}\,,
\end{equation}
namely the differential operator (with no domain specification) representing the action of both $A_\alpha(k)$ and $A_\alpha(k)^*$, where the derivative is classical or weak depending on the context.

 Within the Kre{\u\i}n-Vi\v{s}ik-Birman  scheme (Sect.~\ref{sec:II-VBreparametrised}) one needs to characterise $\overline{A_\alpha(k)}$, $A_{\alpha,\mathrm{F}}(k)$ (the Friedrichs extension), and $\ker A_\alpha(k)^*$ in order to classify the self-adjoint extensions of $A_\alpha(k)$. In turn, in order to characterise the operator closure, the  Friedrichs extension, as well as any other self-adjoint extension, it suffices to indicate the corresponding domains, since all such operators are restrictions of the adjoint $A_\alpha(k)^*$, and as such they all act with the action of the differential operator $S_{\alpha,k}$.

 The main result in this context is the following.

\begin{theorem}\label{thm:fibre-thm}
 Let $\alpha\in[0,1)$ and $ k\in\mathbb{Z}\!\setminus\!\{0\}$.
 \begin{enumerate}[(i)]
  \item The operator closure of $A_\alpha(k)$ has domain
  \begin{equation}\label{eq:thm_Aclosure}
   \mathcal{D}(\overline{A_\alpha(k)})\;=\; H^2_0(\mathbb{R}^+)\cap L^2(\mathbb{R}^+,\langle x\rangle^{4\alpha}\,\ud x)\,.
  \end{equation}
  \item The adjoint of $A_\alpha(k)$ has domain
  \begin{equation}
   \begin{split}
    \mathcal{D}(A_\alpha(k)^*)\;&=\;\left\{\!\!
  \begin{array}{c}
   g\in L^2(\mathbb{R}^+)\;\;\textrm{such that} \\
   \big(-\frac{\ud^2}{\ud x^2}+ k^2 x^{2\alpha}+\frac{\,\alpha(2+\alpha)\,}{4x^2}\big)g\in L^2(\mathbb{R}^+)
  \end{array}
  \!\!\right\} \\
   &=\;\mathcal{D}(\overline{A_\alpha(k)})\dotplus\mathrm{span}\{\Psi_{\alpha,k}\}\dotplus\mathrm{span}\{\Phi_{\alpha,k}\}\,,
   \end{split} 
  \end{equation}
   where $\Phi_{\alpha,k}$ and $\Psi_{\alpha,k}$ are two smooth functions on $\mathbb{R}^+$ explicitly defined, in terms of modified Bessel functions\index{Bessel functions}, respectively by formula \eqref{eq:Phi_and_F} and by formulas \eqref{eq:V-value_of_W}, \eqref{eq:V-Green}, \eqref{eq:newRGalphaforus}, and \eqref{eq:defPsi} below. Moreover,
   \begin{equation}\label{eq:kerAxistar-in-thm}
  \ker A_\alpha(k)^*\;=\;\mathrm{span}\{\Phi_{\alpha,k}\}\,.
 \end{equation}
   \item The Friedrichs extension of $A_\alpha(k)$ has operator domain
   \begin{equation}\label{eq:thmAFoperator}
    \begin{split}
     \mathcal{D}(A_{\alpha,\mathrm{F}}(k))\;&=\;\left\{\!\!
  \begin{array}{c}
   g\in \mathcal{D}(A_\alpha(k)^*)\;\;\textrm{such that} \\
   g(x)\,\stackrel{x\downarrow 0}{=}\,g_1x^{1+\frac{\alpha}{2}}+o(x^{\frac{3}{2}})\,,\; g_1\in\mathbb{C}
  \end{array}
  \!\!\right\}  \\
     &=\;\mathcal{D}(\overline{A_\alpha(k)})\dotplus\mathrm{span}\{\Psi_{\alpha,k}\}
    \end{split}
   \end{equation}
    and form domain
   \begin{equation}\label{eq:thmAFform}
    \mathcal{D}[A_{\alpha,\mathrm{F}}(k)]\;=\;H^1_0(\mathbb{R}^+)\cap L^2(\mathbb{R}^+,\langle x\rangle^{2\alpha}\,\ud x)\,.
   \end{equation}
    Moreover, $A_{\alpha,\mathrm{F}}(k)$ is the only self-adjoint extension of $A_\alpha(k)$ whose operator domain is entirely contained in $\mathcal{D}(x^{-1})$, namely the self-adjointness domain of the operator of multiplication by $x^{-1}$. 
   \item  The self-adjoint extensions of $A_\alpha(k)$ in $L^2(\mathbb{R}^+)$ form the family
   \[
\{ A_\alpha^{[\gamma]}(k)\,|\,\gamma\in\mathbb{R}\cup\{\infty\}\}\,.    
   \]
 The extension with $\gamma=\infty$ is the Friedrichs extension, and for generic $\gamma\in\mathbb{R}$ one has
 \begin{equation}\label{eq:V-gammaextonesided}
  \mathcal{D}(A_\alpha^{[\gamma]}(k))\,=\,
  \left\{\!\!
  \begin{array}{c}
   g\in \mathcal{D}(A_\alpha(k)^*)\;\;\textrm{such that} \\
   g(x)\,\stackrel{x\downarrow 0}{=}\,g_0 x^{-\frac{\alpha}{2}}+\gamma\, g_0\,x^{1+\frac{\alpha}{2}}+o(x^{\frac{3}{2}})\,, \\
   g_0\in\mathbb{C}
  \end{array}
  \!\!\right\} \, .
  \end{equation}
 \end{enumerate}
\end{theorem}

Concerning the spaces indicated in \eqref{eq:thm_Aclosure} and \eqref{eq:thmAFform}, recall that by definition and by a standard Sobolev embedding
\begin{equation}\label{eq:H10}
 \begin{split}
  H^1_0(\mathbb{R}^+)\;&=\;\overline{C^\infty_c(\mathbb{R}^+)}^{\|\,\|_{H^1}} \\
  &=\;\{\varphi\in L^2(\mathbb{R}^+)\,|\,\varphi'\in L^2(\mathbb{R}^+)\textrm{ and }\varphi(0)=0\}\,,
 \end{split}
\end{equation}
and
\begin{equation}\label{eq:H20}
 \begin{split}
  H^2_0(\mathbb{R}^+)\;&=\;\overline{C^\infty_c(\mathbb{R}^+)}^{\|\,\|_{H^2}} \\
  &=\;\{\varphi\in L^2(\mathbb{R}^+)\,|\,\varphi',\varphi''\in L^2(\mathbb{R}^+)\textrm{ and }\varphi(0)=\varphi'(0)=0\}\,.
 \end{split}
\end{equation}

The proof of Theorem \ref{thm:fibre-thm} requires an amount of preparatory material that is presented in Sections \ref{subsec:homog_problem}-\ref{subsec:distinguished} and will be finally completed in Section \ref{subsec:proof_of_fibrethm}.

\subsection{Homogeneous differential problem: kernel of $A_\alpha(k)^*$}\label{subsec:homog_problem}

 In this Subsection the kernel of the adjoint $A_\alpha(k)^*$ is characterised.

To this aim, one makes use of the modified Bessel functions\index{Bessel functions} $K_\nu$ and $I_\nu$ \cite[Section 9.6.1]{Abramowitz-Stegun-1964}, that are two explicit, linearly independent, smooth solutions to the modified Bessel equation\index{Bessel equation}
\begin{equation}\label{eq:modifiedBessEq}
z^2 w''+z w'-(z^2+\nu^2) w \;=\; 0\,,\qquad z\in\mathbb{R}^+
\end{equation}
with parameter $\nu\in\mathbb{C}$. In particular, in terms of $K_{\frac{1}{2}}$ and $I_{\frac{1}{2}}$ one defines the functions
\begin{equation}\label{eq:Phi_and_F}
\begin{split}
\Phi_{\alpha,k}(x)\;&:=\; \sqrt{x}\,K_{\frac{1}{2}}\big({\textstyle\frac{|k|}{1+\alpha}}\,x^{1+\alpha}\big)\,, \\
F_{\alpha,k}(x)\;&:=\; \sqrt{x}\,I_{\frac{1}{2}}\big({\textstyle\frac{|k|}{1+\alpha}}\,x^{1+\alpha}\big)\,.
\end{split}
\end{equation}
Explicitly, as can be deduced from \cite[Eq.~(10.2.4), (10.2.13), and (10.2.14)]{Abramowitz-Stegun-1964},
\begin{equation}\label{eq:Phi_and_F_explicit}
\begin{split}
\Phi_{\alpha,k}(x)\;&:=\; {\textstyle\sqrt{\frac{\pi(1+\alpha)}{2|k|}}}\,x^{-\alpha/2}\,e^{-\frac{|k|}{1+\alpha}x^{1+\alpha}}\,, \\
F_{\alpha,k}(x)\;&:=\; {\textstyle\sqrt{\frac{2(1+\alpha)}{\pi|k|}}}\,x^{-\alpha/2}\,\sinh\big({\textstyle\frac{|k|}{1+\alpha}}\,x^{1+\alpha}\big)\,.
\end{split}
\end{equation}
From \eqref{eq:Phi_and_F_explicit} one obtains the short-distance asymptotics
\begin{equation}\label{eq:Asymtotics_0}
 \begin{split}
 \Phi_{\alpha,k}(x)\;&\stackrel{x\downarrow 0}{=}\; {\textstyle\sqrt{\frac{\pi (1+\alpha)}{2 |k|}}}\, x^{-\frac{\alpha}{2}} -{\textstyle\sqrt{\frac{\pi \, |k|}{2(1+\alpha)}}}\, x^{1+\frac{\alpha}{2}}+{\textstyle\sqrt{\frac{\pi |k|^3}{8(1+\alpha)^3}}}\, x^{2+\frac{3}{2}\alpha}+O(x^{3+\frac{5}{2}\alpha})\,, \\
 F_{\alpha,k}(x)\;&\stackrel{x\downarrow 0}{=}\;{\textstyle\sqrt{\frac{2 |k|}{(1+\alpha) \pi}}}\, x^{1+\frac{\alpha}{2} }+O(x^{3+\frac{5}{2}\alpha})\,,
 \end{split}
\end{equation}
and the large-distance asymptotics
\begin{equation}\label{eq:Asymtotics_Inf}
\begin{split}
 \Phi_{\alpha,k}(x)\;&\stackrel{x\to +\infty}{=}\;{\textstyle\sqrt{\frac{\pi (1+\alpha)}{2 |k|}}}\, e^{-\frac{|k| x^{1+\alpha}}{1+\alpha}} x^{-\frac{\alpha}{2}}(1+O(x^{-(1+\alpha)}))\,, \\
 F_{\alpha,k}(x)\;&\stackrel{x\to +\infty}{=}\;{\textstyle\sqrt{\frac{ 1+\alpha}{2 \pi |k|}}}\, e^{\frac{|k| x^{1+\alpha}}{1+\alpha}} x^{-\frac{\alpha}{2}}(1+O(x^{-(1+\alpha)}))\,,
\end{split}
\end{equation}
as well as the norm
\begin{equation}\label{eq:Phinorm}
\| \Phi_{\alpha,k} \|_{L^2}^2\;=\;\pi\,(1+\alpha)^{\frac{1-\alpha}{1+\alpha}}\,\Gamma\big({\textstyle\frac{1-\alpha}{1+\alpha}}\big)\, (2|k|)^{-\frac{2}{1+\alpha}}\,.
\end{equation}

\begin{lemma}\label{lem:kerAxistar}
 Let $\alpha\in(0,1)$ and $ k\in\mathbb{Z} \!\setminus\! \{0\}$. One has
 \begin{equation}\label{eq:kerAxistar}
  \ker A_\alpha(k)^*\;=\;\mathrm{span}\{\Phi_{\alpha,k}\}\,.
 \end{equation}
\end{lemma}

\begin{proof}
 Owing to \eqref{eq:Afstar}, a generic $h\in\ker A_\alpha(k)^*$ belongs to $L^2(\mathbb{R}^+)$ and satisfies
 \[\tag{i}
  S_{\alpha,k}\,h\;=\;-h''+ k^2 x^{2\alpha}h+C_\alpha\,x^{-2}h\;=\;0\,.
 \]
 Setting
 \[\tag{ii}
  z\;:=\;\frac{|k|}{1+\alpha}\,x^{1+\alpha}\,,\qquad w(z)\;:=\;\frac{h(x)}{\sqrt{x}}\,,\qquad\nu\;:=\;\frac{\sqrt{1+4C_\alpha}}{2(1+\alpha)}\;=\;\frac{1}{2}\,,
 \]
 the ordinary differential equation (i) takes precisely the form \eqref{eq:modifiedBessEq} with the considered $\nu$. The two linearly independent solutions $K_{\frac{1}{2}}$ and $I_{\frac{1}{2}}$ to \eqref{eq:modifiedBessEq} yield, through the transformation (ii) above, the two linearly independent solutions \eqref{eq:Phi_and_F} to (i). In fact, only $\Phi_{\alpha,k}$ is square-integrable, whereas $F_{\alpha,k}$ fails to be so at infinity (as is seen from \eqref{eq:Phinorm}-\eqref{eq:Asymtotics_Inf}). Formula \eqref{eq:kerAxistar} is thus proved.
\end{proof}

\subsection{Non-homogeneous inverse differential problem}\label{sec:non-homogeneous_problem}

 The next focus is the non-homogeneous problem
\begin{equation}
 S_{\alpha,k}\,u\;=\;g
\end{equation}
in the unknown $u$ for given $g$. With respect to the fundamental system\index{fundamental system for an O.D.E.} $\{F_{\alpha,k},\Phi_{\alpha,k}\}$ given by \eqref{eq:Phi_and_F}, of solutions for the problem $S_{\alpha,k}\,u=0$, the general solution is given by
\begin{equation}\label{eq:ODE_general_sol}
 u\;=\;c_1 F_{\alpha,k} + c_2 \Phi_{\alpha,k} + u_{\mathrm{part}}
\end{equation}
for $c_1,c_2\in\mathbb{C}$ and some particular solution $u_{\mathrm{part}}$, i.e., $S_{\alpha,k}\,u_{\mathrm{part}}=g$.

The Wronskian
\begin{equation}
W(\Phi_{\alpha,k},F_{\alpha,k})(r)\;:=\;\det \begin{pmatrix}
\Phi_{\alpha,k}(r) & F_{\alpha,k}(r) \\
\Phi_{\alpha,k}'(r) & F_{\alpha,k}'(r)
\end{pmatrix}
\end{equation}
relative to the fundamental system\index{fundamental system for an O.D.E.} $\{F_{\alpha,k},\Phi_{\alpha,k}\}$ is clearly constant in $r$, since it is evaluated on solutions to the homogeneous differential problem, with a value that can be computed by means of the asymptotics \eqref{eq:Asymtotics_0} or \eqref{eq:Asymtotics_Inf} and amounts to
\begin{equation}\label{eq:V-value_of_W}
W(\Phi_{\alpha,k},F_{\alpha,k})\;=\;1+\alpha\;=:\; W\,.
\end{equation}

A standard application of the method of variation of constants\index{variation of constants} \cite[Section 2.4]{Wasow_asympt_expansions} shows that one can take $u_{\mathrm{part}}$ to be
\begin{equation}\label{eq:upart}
u_{\text{part}}(r) \;=\; \int_0^{+\infty} G_{\alpha,k}(r,\rho) g(\rho) \, \ud \rho\,,
\end{equation}
where 
\begin{equation}\label{eq:V-Green}
G_{\alpha,k}(r,\rho) \, := \,\frac{1}{W} \begin{cases}
\Phi_{\alpha,k}(r) F_{\alpha,k}(\rho)\,, \qquad \text{if }0 < \rho < r\,,\\
F_{\alpha,k}(r) \Phi_{\alpha,k}(\rho)\,, \qquad \text{if } 0 < r < \rho \,.
\end{cases}
\end{equation}

For $a\in\mathbb{R}$ and $ k\in\mathbb{Z} \!\setminus\! \{0\}$, let $R_{G_{\alpha,k}}^{(a)}$ be the integral operator acting on functions $g$ on $\mathbb{R}^+$ as
\begin{equation}
 \begin{split}
  \big(R_{G_{\alpha,k}}^{(a)}g\big)(x)\;&:=\;\int_0^{+\infty} \mathscr{G}_{\alpha,k}^{(a)}(x,\rho)\,g(\rho)\,\ud \rho\,, \\
  \mathscr{G}_{\alpha,k}^{(a)}(x,\rho)\;&:=\;x^a\,k^2\, G_{\alpha,k}(x,\rho)\,,
 \end{split}
\end{equation}
and let
\begin{equation}\label{eq:RGalphaforus}
 R_{G_{\alpha,k}}\;:=\;|k|^{-2}\,R_{G_{\alpha,k}}^{(0)}\,,
\end{equation}
whence
\begin{equation}\label{eq:newRGalphaforus}
 ( R_{G_{\alpha,k}}g)(x)\;=\;\;\int_0^{+\infty}G_{\alpha,k}(r,\rho)\,g(\rho)\,\ud \rho\,.
\end{equation}

The following property holds.

\begin{lemma}\label{lem:V-RGbddsa}
Let $\alpha\in(0,1)$ and $ k\in\mathbb{Z} \!\setminus\! \{0\}$.
\begin{enumerate}
 \item[(i)] For each $a\in(-\frac{1-\alpha}{2},2\alpha]$, $R_{G_{\alpha,k}}^{(a)}$ can be realised as an everywhere defined, bounded operator on $L^2(\mathbb{R}^+, \ud x)$, which is also self-adjoint if $a=0$.
 \item[(ii)] When $a=2\alpha$, the operator $R_{G_{\alpha,k}}^{(2\alpha)}$ is bounded \emph{uniformly} in $k$.
\end{enumerate}
\end{lemma}

\begin{remark}
 For the purposes of the present Section, Lemma \ref{lem:V-RGbddsa} is overabundant, in that the \emph{uniformity} in $k$ of the norm of $R_{G_{\alpha,k}}^{(2\alpha)}$ is not needed. Instead, this information will be crucial in Subsect.~\ref{sec:control-of-tildephi}. It is to prove the boundedness claim (i) in a form that implies the $k$-uniformity of claim (ii) that one has to go through a somewhat lengthy proof.
\end{remark}

For the proof of Lemma \ref{lem:V-RGbddsa} it is convenient to re-write, by means of \eqref{eq:Phi_and_F_explicit} and \eqref{eq:V-value_of_W}, for any $ k\in\mathbb{Z} \!\setminus\! \{0\}$,
\begin{equation}
 \!\!\!\mathscr{G}_{\alpha,k}^{(a)}(x,\rho)\;=\;
 \begin{cases}
  \;|k|\,x^{a-\frac{\alpha}{2}}\,\rho^{-\frac{\alpha}{2}}\,e^{-\frac{|k|}{1+\alpha}x^{1+\alpha}}\,\sinh\big(\frac{|k|}{1+\alpha}\rho^{1+\alpha}\big)\,, & \textrm{ if }0<\rho<x\,, \\
  \;|k|\,x^{a-\frac{\alpha}{2}}\,\rho^{-\frac{\alpha}{2}}\,e^{-\frac{|k|}{1+\alpha}\rho^{1+\alpha}}\,\sinh\big(\frac{|k|}{1+\alpha}x^{1+\alpha}\big)\,, & \textrm{ if }0<x<\rho\,.
 \end{cases}
\end{equation}
It is also convenient to use the obvious bound
\begin{equation}\label{eq:GleqGtilde}
 \mathscr{G}_{\alpha,k}^{(a)}(x,\rho)\;\leqslant\;\widetilde{\mathscr{G}_{\alpha,k}^{(a)}}(x,\rho)\;
\end{equation}
with
\begin{equation}\label{eq:defGtilde}
 \widetilde{\mathscr{G}_{\alpha,k}^{(a)}}(x,\rho)\;:=\;\begin{cases}
  \;|k|\,x^{a-\frac{\alpha}{2}}\,\rho^{-\frac{\alpha}{2}}\,e^{-\frac{|k|}{1+\alpha}x^{1+\alpha}}\,e^{\frac{|k|}{1+\alpha}\rho^{1+\alpha}}\,, & \textrm{ if }0<\rho<x\,, \\
  \;|k|\,x^{a-\frac{\alpha}{2}}\,\rho^{-\frac{\alpha}{2}}\,e^{-\frac{|k|}{1+\alpha}\rho^{1+\alpha}}\,e^{\frac{|k|}{1+\alpha}x^{1+\alpha}}\,, & \textrm{ if }0<x<\rho\,.
 \end{cases}
\end{equation}

\begin{proof}[Proof of Lemma \ref{lem:V-RGbddsa}]
$R_{G_{\alpha,k}}^{(a)}$ splits into the sum of four integral operators with non-negative kernels given by
\[
\begin{split}
	  \mathscr{G}^{++}_{\alpha,k,a}(x,\rho)\;&:=\;\mathscr{G}_{\alpha,k}^{(a)}(x,\rho)\,\mathbf{1}_{(M,+\infty)}(x)\,\mathbf{1}_{(M,+\infty)}(\rho)\,, \\
	  \mathscr{G}^{+-}_{\alpha,k,a}(x,\rho)\;&:=\;\mathscr{G}_{\alpha,k}^{(a)}(x,\rho)\,\mathbf{1}_{(M,+\infty)}(x)\,\mathbf{1}_{(0,M)}(\rho)\,, \\
  \mathscr{G}^{-+}_{\alpha,k,a}(x,\rho)\;&:=\;\mathscr{G}_{\alpha,k}^{(a)}(x,\rho)\,\mathbf{1}_{(0,M)}(x)\,\mathbf{1}_{(M,+\infty)}(\rho)\,, \\
  \mathscr{G}^{--}_{\alpha,k,a}(x,\rho)\;&:=\;\mathscr{G}_{\alpha,k}^{(a)}(x,\rho)\,\mathbf{1}_{(0,M)}(x)\,\mathbf{1}_{(0,M)}(\rho)
 \end{split}
\]
for some cut-off $M>0$.

The $(-,-)$ operator is a Hilbert-Schmidt operator on $L^2(\mathbb{R}^+)$. Indeed, owing to \eqref{eq:GleqGtilde}-\eqref{eq:defGtilde},
\[
 \begin{split}
  \mathscr{G}^{--}_{\alpha,k,a}(x,\rho)\;&\leqslant\;|k| x^{a-\frac{\alpha}{2}}\,\rho^{-\frac{\alpha}{2}}\,e^{-\frac{|k|}{1+\alpha}|x^{1+\alpha}-\rho^{1+\alpha}|}\,\mathbf{1}_{(0,M)}(x)\,\mathbf{1}_{(0,M)}(\rho) \\
  &\leqslant\;|k| x^{a-\frac{\alpha}{2}}\,\rho^{-\frac{\alpha}{2}}\,\mathbf{1}_{(0,M)}(x)\,\mathbf{1}_{(0,M)}(\rho)\,,
 \end{split}
\]
whence, for $a>-\frac{1}{2}(1-\alpha)$,
\[
 \begin{split}
  \iint_{\mathbb{R}^+\times\mathbb{R}^+}\,\ud x\,\ud\rho\,\big|\mathscr{G}^{--}_{\alpha,k,a}(x,\rho) \big|^2\;&\leqslant\;k^2\!\int_0^M\ud x\,x^{2a-\alpha}\!\int_0^M\ud\rho\,\rho^{-\alpha} \\
  &=\;\frac{k^2\,M^{2(a+1-\alpha)}}{(2a+1-\alpha)(1-\alpha)}\,.
 \end{split}
\]

Also the $(-,+)$ operator is a Hilbert-Schmidt operator on $L^2(\mathbb{R}^+)$. Indeed,
\[
  \mathscr{G}^{-+}_{\alpha,k,a}(x,\rho)\;\leqslant\;|k|\,e^{\frac{|k|}{1+\alpha}M^{1+\alpha}} x^{a-\frac{\alpha}{2}}\,\rho^{-\frac{\alpha}{2}}\,e^{-\frac{|k|}{1+\alpha}\rho^{1+\alpha}}\,\mathbf{1}_{(0,M)}(x)\,\mathbf{1}_{(M,+\infty)}(\rho)\,,
\]
whence, for $a>-\frac{1}{2}(1-\alpha)$,
\[
 \begin{split}
  \iint_{\mathbb{R}^+\times\mathbb{R}^+}&\,\ud x\,\ud\rho\,\big|\mathscr{G}^{-+}_{\alpha,k,a}(x,\rho) \big|^2 \\
  &\leqslant\;k^2\,e^{\frac{2|k|}{1+\alpha}M^{1+\alpha}}\!\int_0^M\ud x\,x^{2a-\alpha}\!\int_M^{+\infty}\!\ud\rho\,\rho^{-\alpha}\,e^{-\frac{2|k|}{1+\alpha}\rho^{1+\alpha}} \\
  &\leqslant\;k^2\,M^{-2\alpha}\,e^{\frac{2|k|}{1+\alpha}M^{1+\alpha}}\!\int_0^M\ud x\,x^{2a-\alpha}\!\int_M^{+\infty}\!\ud\rho\,\rho^{\alpha}\,e^{-\frac{2|k|}{1+\alpha}\rho^{1+\alpha}} \\
  &=\;\frac{|k|}{2}\,M^{-2\alpha}\,e^{\frac{2|k|}{1+\alpha}M^{1+\alpha}}\!\int_0^M\ud x\,x^{2a-\alpha}\!\int_{\frac{2|k|}{1+\alpha}M^{1+\alpha}}^{+\infty}\ud y\,e^{-y} \\
  &=\;\frac{|k|\,M^{2a+1-3\alpha}}{2(2a+1-\alpha)}\,.
 \end{split}
\]

With analogous reasoning,
\[
  \mathscr{G}^{+-}_{\alpha,k,a}(x,\rho)\;\leqslant\;|k|\,e^{\frac{|k|}{1+\alpha}M^{1+\alpha}}\rho^{-\frac{\alpha}{2}} x^{a-\frac{\alpha}{2}}\,e^{-\frac{|k|}{1+\alpha}x^{1+\alpha}}\,\mathbf{1}_{(M,+\infty)}(x)\,\mathbf{1}_{(0,M)}(\rho)\,,
\]
whence
\[
 \begin{split}
   \iint_{\mathbb{R}^+\times\mathbb{R}^+}&\,\ud x\,\ud\rho\,\big|\mathscr{G}^{+-}_{\alpha,k,a}(x,\rho) \big|^2 \;\leqslant\;k^2\,e^{\frac{2|k|}{1+\alpha}M^{1+\alpha}}\!\int_0^M\ud\rho\,\rho^{-\alpha}\!\int_M^{+\infty}\!\ud x\,x^{2a-\alpha}\,e^{-\frac{2|k|}{1+\alpha}x^{1+\alpha}} \\
   &=\;\frac{\,k^2M^{1-\alpha}}{1-\alpha}\,e^{\frac{2|k|}{1+\alpha}M^{1+\alpha}}\!\int_M^{+\infty}\!\ud x\,x^{2a-\alpha}\,e^{-\frac{2|k|}{1+\alpha}x^{1+\alpha}}\,.
 \end{split} 
\]
In turn, integrating by parts, and for $a\leqslant\frac{1}{2}+\frac{3}{2}\alpha$,
\[
 \begin{split}
  \int_M^{+\infty}\!\ud x&\,x^{2a-\alpha}\,e^{-\frac{2|k|}{1+\alpha}x^{1+\alpha}} \\
  &=\;\frac{M^{2a-2\alpha}}{2|k|}\,e^{-\frac{2|k|}{1+\alpha}M^{1+\alpha}}+\frac{a-\alpha}{|k|}\int_{M}^{+\infty}\!\ud x\,x^{2a-1-3\alpha}\,x^{\alpha}\,e^{-\frac{2|k|}{1+\alpha}x^{1+\alpha}} \\
  &\leqslant\;\frac{M^{2a-2\alpha}}{2|k|}\,e^{-\frac{2|k|}{1+\alpha}M^{1+\alpha}}+\frac{\,(a-\alpha)M^{2a-1-3\alpha}}{2k^2}\!\int_{\frac{2|k|}{1+\alpha}M^{1+\alpha}}^{+\infty}\ud y\,e^{-y} \\
  &=\;e^{-\frac{2|k|}{1+\alpha}M^{1+\alpha}}\Big(\frac{M^{2a-2\alpha}}{2|k|}+\frac{\,(a-\alpha)M^{2a-1-3\alpha}}{2k^2}\Big)\,.
 \end{split}
\]
Thus,
\[
  \iint_{\mathbb{R}^+\times\mathbb{R}^+}\,\ud x\,\ud\rho\,\big|\mathscr{G}^{+-}_{\alpha,k,a}(x,\rho) \big|^2 \;\leqslant\;\frac{1}{\,2(1-\alpha)}\,\big(2|k| M^{2a+1-3\alpha}+(a-\alpha)M^{2(a-2\alpha)}\big)\,,
\]
which shows that the $(+,-)$ operator is a Hilbert-Schmidt operator on $L^2(\mathbb{R}^+)$.

Last, it will be now shown, by means of a standard Schur test,\index{Schur test} that the norm of the $(+,+)$ operator is bounded by $\sqrt{AB}$, where
\[
 \begin{split}
  A\;&:=\;\sup_{x\in(M,+\infty)}\int_M^{+\infty}\!\ud\rho\,\,\mathscr{G}_{\alpha,k}^{(a)}(x,\rho)\,, \\
  B\;&:=\;\sup_{\rho\in(M,+\infty)}\int_M^{+\infty}\!\ud x\,\,\mathscr{G}_{\alpha,k}^{(a)}(x,\rho)\,.
 \end{split}
\]

Owing to \eqref{eq:GleqGtilde}-\eqref{eq:defGtilde},
\[
 \begin{split}
  A\;&\leqslant\;A_1+A_2\,, \\
  B\;&\leqslant\;B_1+B_2\,,
 \end{split}
\]
with
\[
 \begin{split}
  A_1\;&:=\;\sup_{x\in(M,+\infty)}|k|\,x^{a-\frac{\alpha}{2}}\,e^{-\frac{|k|}{1+\alpha}x^{1+\alpha}}\!\int_M^{x}\ud\rho\,\rho^{-\frac{\alpha}{2}}\,e^{\frac{|k|}{1+\alpha}\rho^{1+\alpha}}\,, \\
  A_2\;&:=\;\sup_{x\in(M,+\infty)}|k|\,x^{a-\frac{\alpha}{2}}\,e^{\frac{|k|}{1+\alpha}x^{1+\alpha}}\!\int_x^{+\infty}\!\ud\rho\,\rho^{-\frac{\alpha}{2}}\,e^{-\frac{|k|}{1+\alpha}\rho^{1+\alpha}}\,, \\
  B_1\;&:=\;\sup_{\rho\in(M,+\infty)}|k|\,\rho^{-\frac{\alpha}{2}}\,e^{-\frac{|k|}{1+\alpha}\rho^{1+\alpha}}\!\int_M^{\rho}\ud x\,x^{a-\frac{\alpha}{2}}\,e^{\frac{|k|}{1+\alpha}x^{1+\alpha}}\,, \\
  B_2\;&:=\;\sup_{\rho\in(M,+\infty)}|k|\,\rho^{-\frac{\alpha}{2}}\,e^{\frac{|k|}{1+\alpha}\rho^{1+\alpha}}\!\int_{\rho}^{+\infty}\ud x\,x^{a-\frac{\alpha}{2}}\,e^{-\frac{|k|}{1+\alpha}x^{1+\alpha}}\,.
 \end{split}
\]

Concerning $A_1$, integration by parts yields
\[
 \begin{split}
  |k|\!\int_M^{x}\ud\rho\,\rho^{-\frac{\alpha}{2}}\,e^{\frac{|k|}{1+\alpha}\rho^{1+\alpha}}\;&=\;x^{-\frac{3}{2}\alpha}\,e^{\frac{|k|}{1+\alpha}x^{1+\alpha}}-M^{-\frac{3}{2}\alpha}\,e^{\frac{|k|}{1+\alpha}M^{1+\alpha}} \\
  &\qquad\qquad\quad+\frac{3\alpha}{2}\!\int_M^x\ud\rho\,\rho^{-(1+\frac{3}{2}\alpha)}\,e^{\frac{|k|}{1+\alpha}\rho^{1+\alpha}} \, ,
 \end{split}
\]
and choosing $M\geqslant M_\circ$, where
\[
 M_\circ\;:=\;\Big(\frac{2+3\alpha}{2|k|}\Big)^{\frac{1}{1+\alpha}}
\]
is the point of absolute minimum of the function $\rho\mapsto\rho^{-(1+\frac{3}{2}\alpha)}\,e^{\frac{|k|}{1+\alpha}\rho^{1+\alpha}}$, yields
\[
 \begin{split}
  |k|\!\int_M^{x}\ud\rho\,\rho^{-\frac{\alpha}{2}}\,e^{\frac{|k|}{1+\alpha}\rho^{1+\alpha}}\;&\leqslant\;x^{-\frac{3}{2}\alpha}\,e^{\frac{|k|}{1+\alpha}x^{1+\alpha}}+\frac{3\alpha}{2}\,x^{-(1+\frac{3}{2}\alpha)}\,e^{\frac{|k|}{1+\alpha}x^{1+\alpha}}\!\int_0^x\ud\rho \\
 &=\;{\textstyle\big(1+\frac{3}{2}\alpha\big)}x^{-\frac{3}{2}\alpha}\,e^{\frac{|k|}{1+\alpha}x^{1+\alpha}}\,.
 \end{split}
\]
Therefore,
\[
 A_1\;\leqslant\;\sup_{x\in(M,+\infty)}{\textstyle\big(1+\frac{3}{2}\alpha\big)}x^{a-2\alpha}\;=\;{\textstyle\big(1+\frac{3}{2}\alpha\big)}M^{a-2\alpha}\,,
\]
the last identity being valid for $a\leqslant 2\alpha$.

Concerning $A_2$,
\[
\begin{split}
 |k|\!\int_x^{+\infty}\!\ud\rho&\,\rho^{-\frac{\alpha}{2}}\,e^{-\frac{|k|}{1+\alpha}\rho^{1+\alpha}}\;\leqslant\;|k|\,x^{-\frac{3}{2}\alpha}\!\int_x^{+\infty}\!\ud\rho\,\rho^{\alpha}\,e^{-\frac{|k|}{1+\alpha}\rho^{1+\alpha}} \\
 &=\;x^{-\frac{3}{2}\alpha}\!\int_{\frac{|k|}{1+\alpha}x^{1+\alpha}}^{+\infty}\ud y\,e^{-y}\;=\;x^{-\frac{3}{2}\alpha}\,e^{-\frac{|k|}{1+\alpha}x^{1+\alpha}}\,,
\end{split}
\]
whence, when  $a\leqslant 2\alpha$,
\[
 A_2\;\leqslant\;\sup_{x\in(M,+\infty)}x^{a-2\alpha}\;=\;M^{a-2\alpha}\,.
\]

Concerning $B_1$,
\[
 \begin{split}
  |k|\!\int_M^{\rho}\ud x&\,x^{a-\frac{\alpha}{2}}\,e^{\frac{|k|}{1+\alpha}x^{1+\alpha}}\;=\;|k|\!\int_M^{\rho}\ud x\,x^{a-\frac{3}{2}\alpha}\,x^{\alpha}\,e^{\frac{|k|}{1+\alpha}x^{1+\alpha}} \\
  &\leqslant\;|k|\int_M^{\rho}\ud x\,x^{\alpha}\,e^{\frac{|k|}{1+\alpha}x^{1+\alpha}} \times
  \begin{cases}
   \;\rho^{a-\frac{3}{2}\alpha}\,, & \textrm{ if } a\geqslant\frac{3}{2}\alpha \\
   \;M^{a-\frac{3}{2}\alpha}\,, & \textrm{ if } a<\frac{3}{2}\alpha
  \end{cases} \\
  &\leqslant\;\int_0^{\frac{|k|}{1+\alpha}\rho^{1+\alpha}}\!\ud y\;e^y\times
  \begin{cases}
   \;\rho^{a-\frac{3}{2}\alpha}\,, & \textrm{ if } a\geqslant\frac{3}{2}\alpha \\
   \;M^{a-\frac{3}{2}\alpha}\,, & \textrm{ if } a<\frac{3}{2}\alpha
  \end{cases} \\
  &\leqslant\;e^{\frac{|k|}{1+\alpha}\rho^{1+\alpha}}\times
  \begin{cases}
   \;\rho^{a-\frac{3}{2}\alpha}\,, & \textrm{ if } a\geqslant\frac{3}{2}\alpha \\
   \;M^{a-\frac{3}{2}\alpha}\,, & \textrm{ if } a<\frac{3}{2}\alpha\,,
  \end{cases} 
  \end{split}
\]
whence
\[
 \begin{split}
  B_1\;&\leqslant\;\sup_{\rho\in(M,+\infty)}\begin{cases}
   \;\rho^{a-2\alpha}\,, & \textrm{ if } a\geqslant\frac{3}{2}\alpha\,, \\
   \;\rho^{-\frac{\alpha}{2}}\,M^{a-\frac{3}{2}\alpha}\,, & \textrm{ if } a<\frac{3}{2}\alpha\,.
  \end{cases} 
 \end{split}
\]
In either case, as long as  $a\leqslant 2\alpha$,
\[
 B_1\;\leqslant\;M^{a-2\alpha}\,.
\]

Concerning $B_2$, one splits the analysis between $a\leqslant\frac{3}{2}\alpha$ and $a>\frac{3}{2}\alpha$. In the former case,
\[
 \begin{split}
  |k|\!\int_{\rho}^{+\infty}\ud x&\,x^{a-\frac{\alpha}{2}}\,e^{-\frac{|k|}{1+\alpha}x^{1+\alpha}}\;\leqslant\;\rho^{a-\frac{3}{2}\alpha}|k|\!\int_{\rho}^{+\infty}\ud x\,x^{\alpha}\,e^{-\frac{|k|}{1+\alpha}x^{1+\alpha}} \\
  &=\;\rho^{a-\frac{3}{2}\alpha}\!\int_{\frac{|k|}{1+\alpha}}^{+\infty}\ud y\,e^{-y}\;=\;\rho^{a-\frac{3}{2}\alpha}\,e^{-\frac{|k|}{1+\alpha}\rho^{1+\alpha}}\,,
 \end{split}
\]
whence, as long as $a\leqslant 2\alpha$,
\[
 B_2\;\leqslant\sup_{\rho\in(M,+\infty)}\rho^{a-2\alpha}\;\leqslant\;M^{a-2\alpha}\,.
\]

When instead $a>\frac{3}{2}\alpha$, then, integrating by parts and using $a\leqslant1+\frac{5}{2}\alpha$,
\[
 \begin{split}
  |k|\!\int_{\rho}^{+\infty}\ud x&\,x^{a-\frac{\alpha}{2}}\,e^{-\frac{|k|}{1+\alpha}x^{1+\alpha}} \\
  &=\;\rho^{a-\frac{3}{2}\alpha}\,e^{-\frac{|k|}{1+\alpha}\rho^{1+\alpha}}+\big(a-{\textstyle\frac{3\alpha}{2}}\big)\!\int_{\rho}^{+\infty}\!\ud x\,x^{a-\frac{3}{2}\alpha-1}\,e^{-\frac{|k|}{1+\alpha}x^{1+\alpha}} \\
  &\leqslant\;\rho^{a-\frac{3}{2}\alpha}\,e^{-\frac{|k|}{1+\alpha}\rho^{1+\alpha}}+\big(a-{\textstyle\frac{3\alpha}{2}}\big)\rho^{a-\frac{5}{2}\alpha-1}\!\int_{\rho}^{+\infty}\!\ud x\,x^{\alpha}\,e^{-\frac{|k|}{1+\alpha}x^{1+\alpha}} \\
  &=\;\rho^{a-\frac{3}{2}\alpha}\,e^{-\frac{|k|}{1+\alpha}\rho^{1+\alpha}}+\big(a-{\textstyle\frac{3\alpha}{2}}\big)\rho^{a-\frac{5}{2}\alpha-1}|k|^{-1}\!\int_{\frac{|k|}{1+\alpha}\rho^{1+\alpha}}^{+\infty}\ud y\,e^{-y} \\
  &=\;e^{-\frac{|k|}{1+\alpha}\rho^{1+\alpha}}\big(\rho^{a-\frac{3}{2}\alpha}+(a-{\textstyle\frac{3\alpha}{2}})\,|k|^{-1}\rho^{a-\frac{5}{2}\alpha-1}\big)\,,
 \end{split}
\]
whence
\[
 \begin{split}
  B_2\;&\leqslant\sup_{\rho\in(M,+\infty)}\big(\rho^{a-2\alpha}+(a-{\textstyle\frac{3\alpha}{2}})\,|k|^{-1}\rho^{a-3\alpha-1}\big) \\
  &\leqslant\;M^{a-2\alpha}\big(1+(a-{\textstyle\frac{3\alpha}{2}})\,(|k|M^{1+\alpha})^{-1}\big)\,.
 \end{split}
\]

This completes the proof of the boundedness, via a Schur test,\index{Schur test} of the $(+,+)$ operator.

Summarising, with the above choice of the cut-off $M\geqslant M_\circ$, and intersecting all the above restrictions of $a$ in terms of $\alpha$, that is, $-\frac{1}{2}(1-\alpha)\leqslant a\leqslant 2\alpha$, it is established that there is an overall constant $Z_{a,\alpha}>0$ such that
\[
 \begin{split}
  \big\|R_{G_{\alpha,k}}^{(a)}\big\|^2_{\mathrm{op}}\;&\leqslant\;Z_{a,\alpha}\Big(k^2 M^{2(a+1-\alpha)}+|k|\,M^{2a+1-3\alpha}+M^{2a-4\alpha} \\
  &\qquad\qquad\quad +M^{2a-4\alpha}(|k|\,M^{1+\alpha})^{-1}\Big)\,.
 \end{split}
\]

This yields the statement of boundedness of part (i). The self-adjointness of $R_{G_{\alpha,k}}=|k|^{-2}\,R_{G_{\alpha,k}}^{(0)}$ is clear from \eqref{eq:V-Green}: the adjoint $R_{G_{\alpha,k}}^*$ has kernel $\overline{G_{\alpha,k}(\rho,r)}$, but $G$ is real-valued and $G_{\alpha,k}(\rho,r)=G_{\alpha,k}(r,\rho)$, whence indeed $R_{G_{\alpha,k}}^*=R_{G_{\alpha,k}}$. Thus, part (i) is proved.

As for part (ii), for the cut-off one chooses $M= M_\circ$ when $a=2\alpha$. In this case,
\[
 \begin{split}
  |k|\,M^{1+\alpha}\;&=\;{\textstyle 1+\frac{3}{2}\alpha}\,, \\
  |k|\,M^{2a+1-3\alpha}\;&=\;|k|\,M^{1+\alpha}\;=\;{\textstyle 1+\frac{3}{2}\alpha}\,,\\
  k^2 M^{2(a+1-\alpha)}\;&=\;\big(|k|\,M^{1+\alpha}\big)^2\;=\;\big( {\textstyle 1+\frac{3}{2}\alpha}\big)^2\,,
 \end{split}
\]
implying that there is an updated constant $\widetilde{Z}_{a,\alpha}>0$ such that 
\[
 \big\|R_{G_{\alpha,k}}^{(2\alpha)}\big\|_{\mathrm{op}}\;\leqslant\;\widetilde{Z}_{a,\alpha}
\]
uniformly in $k$. Thus, also part (ii) is proved.
\end{proof}

A relevant consequence of Lemma \ref{lem:V-RGbddsa} is the following large-distance decaying behaviour of a generic function of the form $R_{G_{\alpha,k}}u$.

\begin{corollary}\label{cor:RGtoWeightedL2} Let $\alpha\in(0,1)$ and $ k\in\mathbb{Z}\!\setminus\!\{0\}$. Then
\begin{equation}
	\mathrm{ran}\,R_{G_{\alpha,k}} \;\subset\; L^2(\mathbb{R}^+,\langle x\rangle^{4\alpha} \ud x)\,.
\end{equation}
\end{corollary}

\begin{proof}
By Lemma \ref{lem:V-RGbddsa} one knows that both $x^{2\alpha} R_{G_{\alpha,k}}$ and $R_{G_{\alpha,k}}$ are bounded in $L^2(\mathbb{R}^+,\ud x)$. Therefore, for any $u\in L^2(\mathbb{R}^+,\ud x)$ one has that both $R_{G_{\alpha,k}} u$ and $x^{2\alpha}R_{G_{\alpha,k}} u$ must belong to $L^2(\mathbb{R}^+,\ud x)$, whence indeed $R_{G_{\alpha,k}} u\in L^2(\mathbb{R}^+,(1+x^{4\alpha}) \ud x)$.
\end{proof}

Moreover, one recognises that $R_{G_{\alpha,k}}$ inverts a self-adjoint extension of $A_\alpha(k)$.

\begin{lemma}\label{eq:V-RGinvertsExtS}
 Let $\alpha\in(0,1)$ and $ k\in\mathbb{Z}\!\setminus\! \{0\}$. There exists a self-adjoint extension $\mathscr{A}_\alpha(k)$ of $A_\alpha(k)$ in $L^2(\mathbb{R}^+)$ which has everywhere defined and bounded inverse and such that $\mathscr{A}_{\alpha}(k)^{-1}=R_{G_{\alpha,k}}$.
\end{lemma}

\begin{proof}
 $R_{G_{\alpha,k}}$ is bounded and self-adjoint (Lemma \ref{lem:V-RGbddsa}), and by construction satisfies $S_{\alpha,k}\,R_{G_{\alpha,k}}\,g=g$ $\forall g\in L^2(\mathbb{R}^+)$. Therefore, $R_{G_{\alpha,k}} g=0$ for some $g\in L^2(\mathbb{R}^+)$ implies $g=0$, i.e., $R_{G_{\alpha,k}}$ is injective. Then $R_{G_{\alpha,k}}$ has dense range ($(\mathrm{ran}\,R_{G_{\alpha,k}})^\perp=\ker R_{G_{\alpha,k}}$). As a consequence (Sect.~\ref{sec:I-adjoint}), $(R_{G_{\alpha,k}}^{-1})^*=(R_{G_{\alpha,k}}^*)^{-1}=R_{G_{\alpha,k}}^{-1}$, that is, $\mathscr{A}_\alpha(k):=R_{G_{\alpha,k}}^{-1}$ is self-adjoint. One thus has $R_{G_{\alpha,k}}=\mathscr{A}_{\alpha}(k)^{-1}$ and from the identity $A_\alpha(k)^*R_{G_{\alpha,k}}=\mathbbm{1}$ on $L^2(\mathbb{R}^+)$ one deduces that for any $h\in\mathcal{D}(\mathscr{A}_\alpha(k))$, say, $h=R_{G_{\alpha,k}} g=\mathscr{A}_\alpha(k)^{-1} g$ for some $g\in L^2(\mathbb{R}^+)$, the identity $A_\alpha(k)^*h=\mathscr{A}_\alpha(k)h$ holds. This means that $A_\alpha(k)^*\supset\mathscr{A}_\alpha(k)$, whence also $\overline{A_\alpha(k)}=A_\alpha(k)^{**}\subset\mathscr{A}_{\alpha}(k)$, i.e., $\mathscr{A}_{\alpha}(k)$ is a self-adjoint extension of $A_\alpha(k)$.
\end{proof}

 To conclude this Subsection one proves the following useful asymptotics for the function
\begin{equation}\label{eq:defPsi}
 \Psi_{\alpha,k}\::=\;R_{G_{\alpha,k}}\Phi_{\alpha,k}\,.
\end{equation}

\begin{lemma}\label{lem:Psi_asymptotics}
 Let $\alpha\in(0,1)$ and $ k\in\mathbb{Z}\!\setminus\! \{0\}$. Then
 \begin{equation}\label{eq:Psi_asymptotics}
  \Psi_{\alpha,k}(x)\;\stackrel{x\downarrow 0}{=}\;{\textstyle\sqrt{\frac{2|k|}{\,\pi(1+\alpha)^3}}}\;\|\Phi_{\alpha,k}\|_{L^2}^2\:x^{1+\frac{\alpha}{2}}+o(x^{\frac{3}{2}})\,.
 \end{equation}
\end{lemma}

\begin{proof}
Owing to \eqref{eq:V-Green} and \eqref{eq:newRGalphaforus},
\[
	\Psi_{\alpha,k}(x)\;=\;\frac{1}{W} \Big( \Phi_{\alpha,k}(x)\!\int_0^x \! F_{\alpha,k}(\rho) \Phi_{\alpha,k}(\rho) \ud \rho + F_{\alpha,k}(x)\!\int_x^{+\infty} \!\! \Phi_{\alpha,k}(\rho)^2 \ud \rho \Big)\,.
\]
By means of \eqref{eq:Asymtotics_0} one then finds
\[
	\Phi_{\alpha,k}(x)\!\int_0^x \! F_{\alpha,k}(\rho) \Phi_{\alpha,k}(\rho) \ud \rho \;\stackrel{x\downarrow 0}{=}\; {\textstyle \sqrt{\frac{\pi(1+\alpha)}{8 |k|}} }x^{-\frac{\alpha}{2} + 2}+o(x^3)\;\stackrel{x\downarrow 0}{=}\;o(x^{\frac{3}{2}})
\]
(having explicitly used that $\alpha \in (0,1)$), and
\[
 \begin{split}
  F_{\alpha,k}(x)\!\int_x^{+\infty} \!\! \Phi_{\alpha,k}(\rho)^2 \ud \rho\:&\stackrel{x\downarrow 0}{=}\: F_{\alpha,k}(x)\Big(\|\Phi_{\alpha,k}\|_{L^2}^2-\int_0^x\!\Phi_{\alpha,k}(\rho)^2 \ud \rho\Big) \\
  &\stackrel{x\downarrow 0}{=}\:\;{\textstyle\sqrt{\frac{2|k|}{\,\pi(1+\alpha)}}}\;\|\Phi_{\alpha,k}\|_{L^2}^2\:x^{1+\frac{\alpha}{2}}+O(x^{2-\frac{\alpha}{2}})\,.
 \end{split}
\]
The latter quantity is leading, and using the expression \eqref{eq:V-value_of_W} for $W$ yields \eqref{eq:Psi_asymptotics}.
\end{proof}

In fact, using \eqref{eq:V-value_of_W},  \eqref{eq:V-Green}, and \eqref{eq:newRGalphaforus} as in the proof above, and using the explicit expression \eqref{eq:Phi_and_F_explicit} for $\Phi_{\alpha,k}$ and $F_{\alpha,k}$, one finds
\begin{equation}
 \begin{split}
  \!\!\!\!\!\!\!\!\Psi_{\alpha,k}(x)\;=\;{\textstyle\sqrt{\frac{\pi(1+\alpha)}{2|k|^3}}}&\bigg(x^{-\frac{\alpha}{2}}e^{-\frac{|k|}{1+\alpha}x^{1+\alpha}}\!\int_0^x\ud\rho\,\rho^{-\alpha}\sinh{\textstyle(\frac{|k|}{1+\alpha}\rho^{1+\alpha})}\,e^{-\frac{|k|}{1+\alpha}\rho^{1+\alpha}} \\
  &\quad +x^{-\frac{\alpha}{2}}\sinh{\textstyle(\frac{|k|}{1+\alpha}x^{1+\alpha})}\!\int_x^{+\infty}\!\!\ud\rho\,\rho^{-\alpha}\,e^{-\frac{2|k|}{1+\alpha}\rho^{1+\alpha}}\bigg),
 \end{split}
\end{equation}
or also, with a change of variable $\rho\mapsto|k|^{\frac{1}{1+\alpha}}\rho$,
 \begin{equation}\label{eq:explicitPsika}
 \begin{split}
  \Psi_{\alpha,k}(x)\;&=\;{\textstyle\sqrt{\frac{\pi(1+\alpha)}{2}}}|k|^{-\frac{5+\alpha}{2(1+\alpha)}}\times \\
  &\qquad \times\bigg(x^{-\frac{\alpha}{2}}e^{-\frac{|k|}{1+\alpha}x^{1+\alpha}}\!\int_0^{x|k|^{\frac{1}{1+\alpha}}}\ud\rho\,\rho^{-\alpha}\sinh{\textstyle(\frac{\rho^{1+\alpha}}{1+\alpha})}\,e^{-\frac{\,\rho^{1+\alpha}}{1+\alpha}} \\
  &\qquad\qquad\quad +x^{-\frac{\alpha}{2}}\sinh{\textstyle(\frac{|k|}{1+\alpha}x^{1+\alpha})}\!\int_{x|k|^{\frac{1}{1+\alpha}}}^{+\infty}\!\!\ud\rho\,\rho^{-\alpha}\,e^{-\frac{2\rho^{1+\alpha}}{1+\alpha}}\bigg)\,.
 \end{split}
\end{equation}
 Such explicit expressions for $\Psi_{\alpha,k}$ will not be needed until Subsect.~\ref{sec:q0q1}.

\subsection{Operator closure $\overline{A_\alpha(k)}$}

The next fundamental ingredient for the Kre\u{\i}n-Vi\v{s}ik-Birman extension scheme is the characterisation of the operator closure $\overline{A_\alpha(k)}$ of $A_\alpha(k)$. 


\begin{proposition}\label{prop:domAclosure}
 Let $\alpha\in(0,1)$ and $ k\in\mathbb{Z}\!\setminus\! \{0\}$. Then
 \begin{equation}\label{eq:DomAclosure-Grushin}
  \mathcal{D}(\overline{A_\alpha(k)})\;=\;H^2_0(\mathbb{R}^+)\cap L^2(\mathbb{R}^+,\langle x\rangle^{4\alpha}\,\ud x)\,.
 \end{equation}
\end{proposition}

Here $H^2_0(\mathbb{R}^+)$ is the space \eqref{eq:H20} and, by definition,
\begin{equation}
 \mathcal{D}(\overline{A_\alpha(k)})\;=\;\overline{C^\infty_c(\mathbb{R}^+)}^{\|\,\|_{\Gamma(A_\alpha(k))}}\,,
\end{equation}
where $\|\cdot\|_{\Gamma(A_\alpha(k))}$ is the graph norm (Sect.~\ref{sec:I-preliminaries})
\begin{equation}
\begin{split}
 \|\varphi\|_{\Gamma(A_\alpha(k))}^2\;&=\;\|-\varphi''+ k^2 x^{2\alpha}\varphi+C_\alpha x^{-2}\varphi\|_{L^2(\mathbb{R}^+)}^2+\|\varphi\|_{L^2(\mathbb{R}^+)}^2 \\
 &\qquad\forall\varphi\in\mathcal{D}(A_\alpha(k))=C^\infty_c(\mathbb{R}^+)\,.
\end{split}
\end{equation}

 Proposition \ref{prop:domAclosure} is proved in several steps.

First, in the same spirit of Lemma \ref{lem:IV-decomp_a0ainf_b0binf},
one produces a useful representation of $\mathcal{D}(A_\alpha(k)^*)$ based on the differential nature \eqref{eq:Afstar}  of the adjoint $A_\alpha(k)^*$.

\begin{lemma}\label{lem:odedecomp}
 Let $\alpha\in(0,1)$ and $ k\in\mathbb{Z}\!\setminus\! \{0\}$.
 \begin{enumerate}[(i)]
  \item For each $g \in \mathcal{D}(A_\alpha(k)^*)$ there exist uniquely determined constants $a_0^{(g)}, a_\infty^{(g)} \in \mathbb{C}$ and functions
\begin{equation}\label{eq:b0binfty}
\begin{split}
b_0^{(g)}(x) \;&:=\; \frac{1}{W} \int_0^x F_{\alpha,k}(\rho) (A_\alpha(k)^* g)(\rho) \, \ud \rho\,, \\
b_\infty^{(g)} (x) \;&:= \; - \frac{1}{W} \int_0^x \Phi_{\alpha,k}(\rho) (A_\alpha(k)^* g)(\rho) \, \ud \rho
\end{split}
\end{equation}
on $\mathbb{R}^+$ such that
\begin{equation}\label{eq:G-ODE-Decomposition}
g \;=\; a_0^{(g)} F_{\alpha,k} + a_\infty^{(g)} \Phi_{\alpha,k} + b_\infty^{(g)} F_{\alpha,k} + b_0^{(g)} \Phi_{\alpha,k}
\end{equation}
with $\Phi_{\alpha,k}$ and $F_{\alpha,k}$ defined in \eqref{eq:Phi_and_F} and $W=-(1+\alpha)$ as in \eqref{eq:V-value_of_W}.
\item The functions $b_0^{(g)}$ and $b_\infty^{(g)}$ satisfy the properties
\begin{eqnarray}
 & & b_0^{(g)},b_\infty^{(g)}\in AC(\mathbb{R}^+)\,, \label{eq:b_areAC} \\
 & & b_0^{(g)}(x)\:\stackrel{x\downarrow 0}{=}\: o(1)\,,\quad b_\infty^{(g)}(x)\:\stackrel{x\downarrow 0}{=}\: o(1)\,, \label{eq:b_vanishes}\\
 & & b_\infty^{(g)}(x) F_{\alpha,k}(x) + b_0^{(g)} (x) \Phi_{\alpha,k}(x) \:\stackrel{x\downarrow 0}{=}\: o(x^{\frac{3}{2}})\,.\label{eq:sdasimpt}
\end{eqnarray}
 \end{enumerate}
\end{lemma}

\begin{proof}
(i) Let $h:=A_\alpha(k)^*g=S_{\alpha,k}\, g$. As already observed at the beginning of Section \ref{sec:non-homogeneous_problem}, $g$ can be expressed in terms of $h$ by the standard representation
\[
	g=A_0 F_{\alpha,k} + A_\infty \Phi_{\alpha,k}+ \Theta^{(h)}_\infty F_{\alpha,k} + \Theta^{(h)}_0 \Phi_{\alpha,k}
\]
for some constants $A_0, A_\infty \in \mathbb{C}$ determined by $h$ and some $h$-dependent functions explicitly given, as follows from \eqref{eq:ODE_general_sol}, \eqref{eq:upart}, and \eqref{eq:V-Green}, by
\[
	\begin{split}
		\Theta_0^{(h)}(x)\;&:=\; \frac{1}{W} \int_0^x F_{\alpha,k}(\rho) h(\rho)\, \ud \rho\,,\\
		\Theta_\infty^{(h)} (x) \;&:=\; \frac{1}{W} \int_x^{+\infty} \!\!\Phi_{\alpha,k}(\rho) h(\rho) \, \ud \rho\,.
	\end{split}
\]
Comparing the latter formulas with \eqref{eq:b0binfty}-\eqref{eq:G-ODE-Decomposition}, one deduces 
\[
	\begin{split}
		\Theta^{(h)}_0(x)\;&=\;b_0^{(g)}(x)\,,\\
		\Theta^{(h)}_\infty(x)\;&=\;\frac{1}{W} \int_x^{+\infty}\!\! \Phi_{\alpha,k}(\rho) (A_\alpha(k)^* g)(\rho) \, \ud \rho \\
		&=\;W^{-1} \,\langle \Phi_{\alpha,k}, A_\alpha(k)^* g \rangle_{L^2(\mathbb{R}^+)}+b_\infty^{(g)}(x)\,.
	\end{split}
\]
So \eqref{eq:G-ODE-Decomposition} is proved upon setting
\[
\begin{split}
	a_0^{(g)}\;&:=\;A_0+W^{-1} \,\langle \Phi_{\alpha,k}, A_\alpha(k)^* g \rangle_{L^2(\mathbb{R}^+)}\,, \\
	a_\infty^{(g)} \;&:=\; A_\infty\,.
	\end{split}
\]

(ii) Since $\Phi_{\alpha,k}$, $F_{\alpha,k}$ and $A_\alpha(k)^*g$ are all square-integrable on the interval $[0,x]$, the integrand functions in \eqref{eq:b0binfty} are $L^1$-functions on $[0,x]$: this proves \eqref{eq:b_areAC} and justifies the simple estimates
\[
 \begin{split}
  |b_0^{(g)}(x)| \;&\lesssim\; \Vert F_{\alpha,k} \Vert_{L^2((0,x))} \Vert A_\alpha^*(k) g \Vert_{L^2((0,x))} \:\stackrel{x\downarrow 0}{=}\: o(1)\,, \\
  |b_\infty^{(g)}(x)| \;&\lesssim\; \Vert \Phi_{\alpha,k} \Vert_{L^2((0,x))} \Vert A_\alpha^*(k) g \Vert_{L^2((0,x))}\:\stackrel{x\downarrow 0}{=}\: o(1)\,,
 \end{split}
\]
so \eqref{eq:b_vanishes} is proved too. Last, one finds
\[
 \begin{split}
  |b_\infty^{(g)}(x) F_{\alpha,k}(x) | \;&\lesssim\; x^{1+\frac{\alpha}{2}} \Big(\int_0^x \!\rho^{-\alpha} \, \ud \rho \Big)^{\frac{1}{2}} \Vert h \Vert_{L^2((0,x))} \;\lesssim\; x^{\frac{3}{2}} o(1) \;=\; o(x^{\frac{3}{2}})\,, \\
  |b_0^{(g)}(x) \Phi_{\alpha,k}(x)| \;&\lesssim\; x^{-\frac{\alpha}{2}}\! \int_0^x \rho^{1+\frac{\alpha}{2}} |h(\rho)| \ud \rho \;\leqslant\; x \Vert h \Vert_{L^2((0,x))} x^{\frac{1}{2}} \;=\; o(x^{\frac{3}{2}})\,,
 \end{split}
\]
and \eqref{eq:sdasimpt} follows.
\end{proof}

\begin{remark}
 As is evident from the proof of Lemma \ref{lem:odedecomp}, the decomposition \eqref{eq:G-ODE-Decomposition} is valid for a generic solution $g$ to $S_{\alpha,k}\, g=h$, irrespective of whether $g$ belongs to $\mathcal{D}(A_\alpha(k)^*)$ or not (i.e., irrespective of whether $h$ is square-integrable or not), thus \eqref{eq:G-ODE-Decomposition} is a consequence of  general facts of the theory of linear ordinary differential equations. It is only in Lemma \ref{lem:odedecomp}(ii) that the property $g\in\mathcal{D}(A_\alpha(k)^*)$ (i.e., $h\in L^2(\mathbb{R}^+)$) was explicitly used. 
\end{remark}

Next, proceeding towards the proof of Proposition \ref{prop:domAclosure}, it is convenient to introduce, for any two functions in $\mathcal{D}(A_\alpha(k)^*)$, the \emph{generalised Wronskian}\index{Wronskian}
\begin{equation}
\mathbb{R}^+ \ni x \mapsto W_x(g,h) \;:=\; \det \begin{pmatrix}
g(x) & h(x) \\
g'(x) & h'(x)
\end{pmatrix}, \qquad g,h \in \mathcal{D}(A_\alpha(k)^*)
\end{equation}
and the \emph{boundary form}\index{boundary form}
\begin{equation}\label{eq:boundaryform}
\omega(g,h)\;:=\; \langle A_\alpha(k)^* g, h \rangle_{L^2}-\langle g, A_\alpha(k)^* h \rangle_{L^2}, \qquad g,h \in \mathcal{D}(A_\alpha(k)^*).
\end{equation}
The boundary form is anti-symmetric, i.e.,
\begin{equation}
\omega(h,g)=-\overline{\omega(g,h)},
\end{equation}
and it is related to the Wronskian by
\begin{equation}\label{eq:omega-W}
\omega(g,h)\;=\;- \lim_{x \downarrow 0} W_x (\overline{g},h) \, .
\end{equation}
Indeed,
\[
	\begin{split}
		\omega(g,h)\;&=\; \int_0^{+\infty} (\overline{A_\alpha(k)^* g})(\rho) \,h(\rho) \, \ud \rho - \int_0^{+\infty} \overline{g(\rho)}\, (A_\alpha(k)^* h)(\rho) \ud \rho  \\
		&=\;\lim_{x \downarrow 0}\Big( \int_x^{+\infty} (\overline{- g''(\rho)}\, h(\rho) \,\ud \rho + \int_x^{+\infty} \overline{g(\rho)}\, h''(\rho)\, \ud \rho \Big) \\
		&=\; \lim_{x \downarrow 0} \Big(\overline{g'(x)} h(x) - \overline{g(x)} h'(x) \Big)\;=\;-\lim_{x \downarrow 0} W_x(\overline{g},h) \, .
	\end{split}
\]

It is also convenient to refer to the two-dimensional space of solutions to the differential problem $S_{\alpha,k}\,u = 0$ as
\begin{equation}
\mathcal{L}\;:=\; \{ u: \mathbb{R}^+ \to \mathbb{C} \,| \, S_{\alpha,k}\,u = 0 \} \;=\; \mathrm{span} \, \{\Phi_{\alpha,k},F_{\alpha,k}\}\,,
\end{equation}
where the second identity follows from what argued in the proof of Lemma \ref{lem:kerAxistar}.
As well known, $x \mapsto W_x(u,v)$ is constant whenever $u,v \in \mathcal{L}$, and this constant is zero if and only if $u$ and $v$ are linearly dependent. Clearly, any $u \in \mathcal{L}$ is square-integrable around $x=0$, as follows from the asymptotics \eqref{eq:Asymtotics_0}.

\begin{lemma}\label{lem:LV}
Let $\alpha\in(0,1)$ and $ k\in\mathbb{Z} \!\setminus\! \{0\}$.
For given $u \in \mathcal{L}$,
\begin{equation}\label{eq:LinearLV}
\begin{split}
L_u : \mathcal{D}(&A_\alpha(k)^*) \to \mathbb{C}\,, \\
&g \mapsto L_u(g)\;:=\; \lim_{x \downarrow 0} W_x(\overline{u},g)
\end{split}
\end{equation}
defines a linear functional on $\mathcal{D}(A^*_\alpha(k))$ which vanishes on $\mathcal{D}(\overline{A_\alpha(k)})$.
\end{lemma}

\begin{proof}
The linearity of $L_u$ is obvious. Its finiteness of $L_u(g)$ is checked as follows. One decomposes (according to \eqref{eq:G-ODE-Decomposition} and using the basis of $\mathcal{L}$)
\[
 \begin{split}
  g\;&=\;a_0^{(g)} F_{\alpha,k}+a_\infty^{(g)} \Phi_{\alpha,k}+b_\infty^{(g)} F_{\alpha,k} + b_0^{(g)} \Phi_{\alpha,k}\,, \\
  u\;&=\;c_0 F_{\alpha,k}+c_\infty \Phi_{\alpha,k} \, .
 \end{split}
\]
Owing to \eqref{eq:LinearLV} it suffices to control the finiteness of $L_{F_{\alpha,k}}(g)$ and $L_{\Phi_{\alpha,k}}(g)$. By linearity
\[\tag{i}
\begin{split}
 	L_{F_{\alpha,k}}(g) \;&=\; a_0^{(g)} L_{F_{\alpha,k}}(F_{\alpha,k})+a_\infty^{(g)} L_{F_{\alpha,k}}(\Phi_{\alpha,k})+L_{F_{\alpha,k}}(b_\infty^{(g)} F_{\alpha,k}+b_0^{(g)} \Phi_{\alpha,k})\,, \\
 	L_{\Phi_{\alpha,k}}(g) \;&=\; a_0^{(g)} L_{\Phi_{\alpha,k}}(F_{\alpha,k})+a_\infty^{(g)} L_{\Phi_{\alpha,k}}(\Phi_{\alpha,k})+L_{\Phi_{\alpha,k}}(b_\infty^{(g)} F_{\alpha,k}+b_0^{(g)} \Phi_{\alpha,k}) \, .
\end{split}
\]
Moreover, obviously,
\[\tag{ii}
 \begin{split}
  L_{F_{\alpha,k}}(F_{\alpha,k})\;&=\;L_{\Phi_{\alpha,k}}(\Phi_{\alpha,k})\;=\;0\,, \\
     L_{F_{\alpha,k}}(\Phi_{\alpha,k})\;&=\;-W\;=\;-L_{\Phi_{\alpha,k}}(F_{\alpha,k})\,.
 \end{split}
\]
 The following properties are then claimed:
\[\tag{iii}
  L_{F_{\alpha,k}}\big(b_\infty^{(g)} F_{\alpha,k} + b_0^{(g)} \Phi_{\alpha,k}\big)\;=\;0\;=\;L_{\Phi_{\alpha,k}}\big(b_\infty^{(g)} F_{\alpha,k} + b_0^{(g)} \Phi_{\alpha,k}\big)\,.
\]
Plugging (ii) and (iii) into (i) the finiteness
\[
 L_{F_{\alpha,k}}(g)\;=\;-Wa_\infty^{(g)}\,,\qquad L_{\Phi_{\alpha,k}}(g)\;=\;Wa_0^{(g)}
\]
follows.

To prove (iii) one computes
\[
 \begin{split}
  &\det\begin{pmatrix}
       F_{\alpha,k} & b_\infty^{(g)} F_{\alpha,k} + b_0^{(g)} \Phi_{\alpha,k} \\
       F_{\alpha,k}' & (b_\infty^{(g)} F_{\alpha,k} + b_0^{(g)} \Phi_{\alpha,k})'
      \end{pmatrix}= \\
  &\qquad=\;F_{\alpha,k}^2(b_\infty^{(g)})'+F_{\alpha,k}(b_0^{(g)})'\Phi_{\alpha,k}+F_{\alpha,k}b_\infty^{(g)}F_{\alpha,k}'-F_{\alpha,k}'b_0^{(g)} \Phi_{\alpha,k} \\
  &\qquad=\;F_{\alpha,k}b_\infty^{(g)}F_{\alpha,k}'-F_{\alpha,k}'b_0^{(g)} \Phi_{\alpha,k}\,,
 \end{split}
\]
having used the cancellation 
\[
 F_{\alpha,k}^2(b_\infty^{(g)})'+F_{\alpha,k}(b_0^{(g)})'\Phi_{\alpha,k}\;=\;0\,,
\]
that follows from \eqref{eq:b0binfty}. Therefore, by means of the asymptotics \eqref{eq:Asymtotics_0} and \eqref{eq:b_vanishes} as $x\downarrow 0$, namely,
\[
 \begin{split}
  F_{\alpha,k}(x)\;&=\;O(x^{1+\frac{\alpha}{2}})\,,\qquad F_{\alpha,k}'(x)=O(x^{\frac{\alpha}{2}})\,,\qquad\Phi_{\alpha,k}(x)\;=\;O(x^{-\frac{\alpha}{2}})\,, \\
  b_0^{(g)}(x)\;&=\;o(1)\,,\qquad b_\infty^{(g)}(x)\;=\;o(1)\,,
 \end{split}
\]
 one concludes 
\[
  L_{F_{\alpha,k}}\big(b_\infty^{(g)} F_{\alpha,k} + b_0^{(g)} \Phi_{\alpha,k}\big)\;=\;\lim_{x\downarrow 0}\big(F_{\alpha,k}b_\infty^{(g)}F_{\alpha,k}'-F_{\alpha,k}'b_0^{(g)} \Phi_{\alpha,k} \big)\;=\;0\,.
\]
The proof of the second identity in (iii) is completely analogous.

  It remains to demonstrate that $L_u (\varphi) =0$ for $\varphi \in \mathcal{D}(\overline{A_\alpha(k)})$ and $u\in\mathcal{L}$. Although $u$ does not necessarily belong to $\mathcal{D}(A_\alpha(k)^*)$ (it might fail to be square-integrable at infinity), the function $\chi u$ surely does for $\chi \in C^\infty_0([0,+\infty))$ with $\chi(x)=1$ on $x \in[0,\frac{1}{2}]$ and $\chi(x)=0$ on $x \in[1,+\infty)$. This fact follows from \eqref{eq:Afstar} observing that $\chi u\in L^2(\mathbb{R}^+)$ and also
\[
	S_{\alpha,k}(u \chi)\;=\;\chi S_{\alpha,k}\,u- 2u'\chi'-u \chi''\;=\;-2 u'\chi' - u \chi'' \in L^2(\mathbb{R}^+)\,.
\]
The choice of $\chi$ guarantees that the Wronskians $W_x(\overline{u \chi}, g)$ and $W_x(\overline{u},g)$ coincide in a neighbourhood of $x=0$, that is, $L_{u \chi}=L_u$. Therefore, by means of \eqref{eq:boundaryform}, \eqref{eq:omega-W}, and \eqref{eq:LinearLV} one deduces
\[
	\begin{split}
		L_u(\varphi)\;&=\; L_{u \chi}(\varphi)\;=\;\lim_{x \downarrow 0} W_x(\overline{u \chi},\varphi)\;=\;-\omega(u \chi, \varphi)\\
		&=\;\langle u\chi, A_\alpha(k)^* \varphi \rangle - \langle A_\alpha(k)^*u \chi, \varphi \rangle \;=\; \langle u \chi, \overline{A_\alpha(k)} \varphi \rangle - \langle u \chi, \overline{A_\alpha(k)} \varphi \rangle \;=\;0\,,
	\end{split}
\]
which completes the proof.
\end{proof}

With this preparatory material at hand, the space $\mathcal{D}(\overline{A_\alpha(k)})$ can be characterised in the form here below.

\begin{lemma}\label{prop:EquivalentClosure}
Let $\alpha\in(0,1)$, $ k\in\mathbb{Z}\!\setminus\! \{0\}$, and $\varphi \in \mathcal{D}(A_\alpha(k)^*)$. The following conditions are equivalent:
\begin{enumerate}[(i)]
\item $\varphi \in \mathcal{D}(\overline{A_\alpha(k)})$,
\item $\omega(\varphi,g)=0$ for all $g \in \mathcal{D}(A_\alpha(k)^*)$,
\item $L_u(\varphi)=0$ for all $u \in \mathcal{L}$,
\item in the decomposition \eqref{eq:G-ODE-Decomposition} of $\varphi$ one has $a_0^{(\varphi)}=a_\infty^{(\varphi)}=0$.
\end{enumerate}
\end{lemma}

\begin{proof}
The implication (i) $\Rightarrow$ (ii) follows at once from
\[
	\omega(\varphi,g)\;=\; \langle A_\alpha(k)^* \varphi, g \rangle - \langle \varphi, A_\alpha(k)^* g \rangle \;=\; \langle \overline{A_\alpha(k)} \varphi, g \rangle - \langle \overline{A_\alpha(k)} \varphi,g \rangle \;=\; 0\,.
\]
For the converse implication (i) $\Leftarrow$ (ii), one observes that the property
\[
	0\;=\;\omega(\varphi,g)\;=\; \langle A_\alpha(k)^* \varphi, g \rangle - \langle \varphi, A_\alpha(k)^* g \rangle \qquad \forall g \in \mathcal{D}(A_\alpha(k)^*)
\]
is equivalent to $\langle A_\alpha(k)^* \varphi,g \rangle = \langle \varphi, A_\alpha(k)^* g \rangle$ $ \forall g \in \mathcal{D}(S^*)$, which implies that $\varphi \in \mathcal{D}(A_\alpha(k)^{**})=\mathcal{D}(\overline{A_\alpha(k)})$.

The implication (i) $\Rightarrow$ (iii) is given by Lemma \ref{lem:LV}. Concerning the implication (iii) $\Rightarrow$ (ii): now $L_u(\varphi)=0$ for all $u \in \mathcal{L}$ and one wants to prove that for such $\varphi$ one has $\omega(\varphi,g)=0$ for all $g \in \mathcal{D}(A_\alpha(k)^*)$. Owing to the decomposition \eqref{eq:G-ODE-Decomposition} for $g$,
\[
	\omega(\varphi,g)\;=\; a_0^{(g)} \omega(\varphi,F_{\alpha,k})+a_\infty^{(g)} \omega(\varphi,\Phi_{\alpha,k})+\omega(\varphi, b_\infty^{(g)} F_{\alpha,k})+\omega(\varphi,b_0^{(g)} \Phi_{\alpha,k}) \, .
\]
The first two summands on the r.h.s.~above are zero: indeed,
\[
	\overline{\omega(\varphi,F_{\alpha,k})} \;=\; - \omega(F_{\alpha,k},\varphi)\;=\;\lim_{x \downarrow 0} W_x(\overline{F_{\alpha,k}},\varphi)\;=\;L_{F_{\alpha,k}}(\varphi)\;=\;0\,
\]
having used in the last step the assumption that $L_u(\varphi)=0$ for all $u \in \mathcal{L}$, and analogously, $\overline{\omega(\varphi,\Phi_{\alpha,k})}=L_{\Phi_{\alpha,k}}(\varphi)=0$.
Therefore, 
\[
	\begin{split}
		\overline{\omega(\varphi,g)} \;&=\; \overline{\omega(\varphi, b_\infty^{(g)} F_{\alpha,k})}+\overline{\omega(\varphi,b_0^{(g)} \Phi_{\alpha,k})} \\
		&=\;-\omega(b_\infty^{(g)} F_{\alpha,k},\varphi)-\omega(b_0^{(g)} \Phi_{\alpha,k}, \varphi )\\
		&=\;\lim_{x \downarrow 0} \big( W_x(b_\infty^{(g)} F_{\alpha,k},\varphi)+W_x(b_0^{(g)} \Phi_{\alpha,k}, \varphi )\big)\\
		&=\;\lim_{x \downarrow 0} \big( b_\infty^{(g)}\,W_x( F_{\alpha,k},\varphi)+b_0^{(g)} \,W_x(\Phi_{\alpha,k}, \varphi )\big)\\	
		&=\;b_\infty^{(g)} L_{F_{\alpha,k}}(\varphi)+b_0^{(g)} L_{\Phi_{\alpha,k}}(\varphi)\;=\;0\,,
	\end{split}
\]
having used again the assumption (ii) in the last step (observe also the helpful cancellation $(b_\infty^{(g)})'F_{\alpha,k}\varphi+(b_0^{(g)})'\Phi_{\alpha,k}\varphi=0$ occurred in computing the determinants in the fourth step).

Properties (i), (ii), and (iii) are thus equivalent.
Last, one establishes the equivalence (i) $\Leftrightarrow$ (iv). Representing $\varphi$ according to \eqref{eq:G-ODE-Decomposition} as 
\[
 \varphi\;=\;a_0^{(\varphi)} F_{\alpha,k} + a_\infty^{(\varphi)} \Phi_{\alpha,k} + b_\infty^{(\varphi)} F_{\alpha,k} + b_0^{(\varphi)} \Phi_{\alpha,k}\,,
\]
and using the identities $W_x(F_{\alpha,k},F_{\alpha,k})=0$ and $W_x(F_{\alpha,k},\Phi_{\alpha,k})=-W$,
one finds
\[
  L_{F_{\alpha,k}}(\varphi)\;=\;\lim_{x\downarrow 0} W_x(F_{\alpha,k},\varphi)\;=\;-Wa_\infty^{(\varphi)}+\lim_{x\downarrow 0}W_x(F_{\alpha,k},b_\infty^{(\varphi)} F_{\alpha,k} + b_0^{(\varphi)} \Phi_{\alpha,k})\,.
\]
The determinant in the latter Wronskian has the very same form of the determinant computed in the proof of Lemma \ref{lem:LV}: using the same cancellation $F_{\alpha,k}^2(b_\infty^{(\varphi)})'+F_{\alpha,k}(b_0^{(\varphi)})'\Phi_{\alpha,k}=0$ and the usual short-distance asymptotics one finds
\[
 L_{F_{\alpha,k}}(\varphi)\;=\;-Wa_\infty^{(\varphi)}\,.
\]
In a completely analogous fashion,
\[
 L_{\Phi_{\alpha,k}}(\varphi)\;=\;Wa_0^{(\varphi)}\,.
\]
Therefore, $\varphi \in \mathcal{D}(\overline{A_\alpha(k)})$ if and only if $L_u(\varphi)=0$ for all $u \in \mathcal{L}$ (because (i) $\Leftrightarrow$ (iii)), and the latter property is equivalent to $a_0^{(\varphi)}=a_\infty^{(\varphi)}=0$.
\end{proof}

 It is now possible to characterise the short-distance behaviour of the functions in $\mathcal{D}(\overline{A_\alpha(k)})$ and of their derivative.

\begin{lemma}\label{lem:BehaviourZeroClosure}
Let $\alpha\in(0,1)$ and $ k\in\mathbb{Z}\setminus \{0\}$. If 
$\varphi \in \mathcal{D}(\overline{A_\alpha(k)})$, then $\varphi(x)=o(x^{\frac{3}{2}})$ and $\varphi'(x)=o(x^{\frac{1}{2}})$ as $x \downarrow 0$.
\end{lemma}

\begin{proof}
Owing to Lemma \ref{prop:EquivalentClosure},
\[
 \varphi\;=\;b_\infty^{(\varphi)} F_{\alpha,k} + b_0^{(\varphi)} \Phi_{\alpha,k}\,.
\]
Thus, $\varphi=o(x^{\frac{3}{2}})$ follows from  \eqref{eq:sdasimpt} of Lemma \ref{lem:odedecomp}. Moreover,
 \[
 \varphi'\;=\;\big(b_\infty^{(\varphi)} F_{\alpha,k} + b_0^{(\varphi)} \Phi_{\alpha,k}\big)'\;=\;b_\infty^{(\varphi)} F_{\alpha,k}' + b_0^{(\varphi)} \Phi_{\alpha,k}'\,,
\]
thanks to the cancellation $(b_\infty^{(\varphi)})'F_{\alpha,k}+(b_0^{(\varphi)})'\Phi_{\alpha,k}=0$ that follows from \eqref{eq:b0binfty}.
From the short-distance asymptotics \eqref{eq:Asymtotics_0} one has
\[
 \begin{split}
  F_{\alpha,k}(x)\;&=\;O(x^{1+\frac{\alpha}{2}})\,,\qquad F_{\alpha,k}'(x)=O(x^{\frac{\alpha}{2}})\,, \\
  \Phi_{\alpha,k}(x)\;&=\;O(x^{-\frac{\alpha}{2}})\,,\qquad\! \Phi_{\alpha,k}(x)'\;=\;O(x^{-(1+\frac{\alpha}{2})})\,,
 \end{split}
\]
whence
\[
 \begin{split}
  |b_\infty^{(\varphi)}(x) F_{\alpha,k}'(x)|\;&\lesssim\;x^{\frac{\alpha}{2}}\,\|A_\alpha(k)^*\varphi\|_{L^2((0,x))}\Big(\int_0^x |\rho^{-\frac{\alpha}{2}}|^2\ud \rho\Big)^{\!\frac{1}{2}} \\
  &\lesssim\;x^{\frac{1}{2}}\,\|\overline{A_\alpha(k)}\varphi\|_{L^2((0,x))}\;=\;o(x^{\frac{1}{2}})\,,
 \end{split}
\]
and also
\[
 \begin{split}
  |b_0^{(\varphi)}(x) \Phi_{\alpha,k}'(x)|\;&\lesssim\;\frac{1}{\:x^{1+\frac{\alpha}{2}}}\,\|A_\alpha(k)^*\varphi\|_{L^2((0,x))}\Big(\int_0^x |\rho^{1+\frac{\alpha}{2}}|^2\ud \rho\Big)^{\!\frac{1}{2}} \\
  &\lesssim\;x^{\frac{1}{2}}\,\|\overline{A_\alpha(k)}\varphi\|_{L^2((0,x))}\;=\;o(x^{\frac{1}{2}})\,.
 \end{split}
\]
The proof is thus completed.
\end{proof}

  With the above preparations, Proposition \ref{prop:domAclosure} can now be proved.

\begin{proof}[Proof of Proposition \ref{prop:domAclosure}]
 First one proves the inclusion
 \[\tag{*}
 H^2_0(\mathbb{R}^+)\cap L^2(\mathbb{R}^+,\langle x\rangle^{4\alpha}\,\ud x)\;\subset\;\mathcal{D}(\overline{A_\alpha(k)})\,.
 \]
 If $\varphi$ belongs to the space on the l.h.s.~of (*), then $\varphi''\in L^2(\mathbb{R})$, $x^{2\alpha}\varphi\in L^2(\mathbb{R})$, and $\varphi(x)=o({x^{\frac{3}{2}}})$ as $x\downarrow 0$, whence also $x^{-2}\varphi\in L^2(\mathbb{R})$. As a consequence, $-\varphi''+ k^2 x^{2\alpha}\varphi+C_\alpha x^{-2}\varphi\in L^2(\mathbb{R})$, i.e., owing to \eqref{eq:Afstar}, $\varphi\in\mathcal{D}(A_\alpha(k)^*)$. Representing now $\varphi$ according to \eqref{eq:G-ODE-Decomposition} as
 \[
  \varphi\;=\;a_0^{(\varphi)} F_{\alpha,k} + a_\infty^{(\varphi)} \Phi_{\alpha,k} + b_\infty^{(\varphi)} F_{\alpha,k} + b_0^{(\varphi)} \Phi_{\alpha,k}\,,
 \]
 one deduces that $a_0^{(\varphi)}=a_\infty^{(\varphi)}=0$, for otherwise the behaviour \eqref{eq:Asymtotics_0} of $\Phi_{\alpha,k}$ and $F_{\alpha,k}$ as $x\downarrow 0$ would be incompatible with $\varphi(x)=o({x^{\frac{3}{2}}})$. Instead, the component $b_\infty^{(\varphi)} F_{\alpha,k} + b_0^{(\varphi)} \Phi_{\alpha,k}$ displays the $o({x^{\frac{3}{2}}})$-behaviour, as seen from \eqref{eq:sdasimpt}. Lemma \ref{prop:EquivalentClosure} then implies $\varphi\in\mathcal{D}(\overline{A_\alpha(k)})$, which proves (*).

 Next, one proves the opposite inclusion
 \[\tag{**}
 H^2_0(\mathbb{R}^+)\cap L^2(\mathbb{R}^+,\langle x\rangle^{4\alpha}\,\ud x)\;\supset\;\mathcal{D}(\overline{A_\alpha(k)})\,.
 \]
 Owing to Lemma \ref{eq:V-RGinvertsExtS} there exists a self-adjoint extension $\mathscr{A}_\alpha(k)$ of $\overline{A_\alpha(k)}$ with $\mathcal{D}(\mathscr{A}_\alpha(k))=\mathrm{ran}R_{G_{\alpha,k}}$, and owing to Corollary \ref{cor:RGtoWeightedL2} $\mathrm{ran}R_{G_{\alpha,k}}\subset L^2(\mathbb{R}^+,\langle x\rangle^{4\alpha}\,\ud x)$. Therefore, $\mathcal{D}(\overline{A_\alpha(k)})\subset L^2(\mathbb{R}^+,\langle x\rangle^{4\alpha}\,\ud x)$. It remains to prove that $\mathcal{D}(\overline{A_\alpha(k)})\subset H^2_0(\mathbb{R}^+)$. For $\varphi\in \mathcal{D}(\overline{A_\alpha(k)})\subset\mathcal{D}(A_\alpha(k)^*)$ formula \eqref{eq:Afstar} prescribes that $g:=-\varphi''+ k^2 x^{2\alpha}\varphi+C_\alpha x^{-2}\varphi\in L^2(\mathbb{R})$. As proved right above, $x^{2\alpha}\varphi\in L^2(\mathbb{R})$, whereas the property $x^{-2}\varphi\in L^2(\mathbb{R}^+)$ follows from Lemma \ref{lem:BehaviourZeroClosure}. Then by linearity $\varphi''\in L^2(\mathbb{R}^+)$, which also implies $\varphi\in H^2(\mathbb{R}^+)$ by standard arguments \cite[Remark 4.21]{Grubb-DistributionsAndOperators-2009}. Lemma \ref{lem:BehaviourZeroClosure} ensures that $\varphi(0)=\varphi'(0)=0$, and one concludes (see \eqref{eq:H20} above) that $\varphi\in H^2_0(\mathbb{R}^+)$. This completes the proof of (**).
 \end{proof}

\subsection{Distinguished extension and induced classification}\label{subsec:distinguished}

In the Kre\u{\i}n-Vi\v{s}ik-Birman extension scheme one characterises all self-adjoint extensions of $A_\alpha(k)$ in terms of a \emph{reference} extension with everywhere defined bounded inverse: the Friedrichs extension $A_{\alpha,\mathrm{F}}(k)$ is surely so, since the bottom of $A_\alpha(k)$ is strictly positive, as seen in \eqref{eq:Axibottom} above.

In fact, $A_{\alpha,\mathrm{F}}(k)$ will be identified at a later stage, meanwhile implementing the the classification scheme with respect to the distinguished extension $\mathscr{A}_\alpha(k)$ determined in Lemma \ref{eq:V-RGinvertsExtS}. All this is only going to be temporary, and will allow one to recognise that actually $\mathscr{A}_\alpha(k)=A_{\alpha,\mathrm{F}}(k)$.

When $\mathscr{A}_\alpha(k)$ is taken as a reference, the other self-adjoint extensions of $A_\alpha(k)$ constitute a one-real-parameter-family $\{ A_\alpha^{[\beta]}(k)\,|\,\beta\in\mathbb{R}\}$ (because the deficiency index is 1, Proposition \ref{prop:Axiselfadjointness-general}(ii)), each member of which, according to Theorem \ref{thm:VB-representaton-theorem_Tversion}, is given by
\begin{equation}\label{eq:temp_fibre_classif-prelim}
 \begin{split}
  \mathcal{D}(A_\alpha^{[\beta]}(k))\;&:=\;
  \left\{\!\!
  \begin{array}{c}
   g=\varphi+c\beta\mathscr{A}_\alpha(k)^{-1}\Phi_{\alpha,k}+c\,\Phi_{\alpha,k} \\
   \textrm{with } \varphi\in\mathcal{D}(\overline{A_\alpha(k)})\,,\;c\in\mathbb{C}
  \end{array}
  \!\!\right\}, \\
  A_\alpha^{[\beta]}(k)g\;&:=\;A_\alpha(k)^*g\;=\;\overline{A_\alpha(k)}\,\varphi+c\beta\Phi_{\alpha,k}\,.
 \end{split}
\end{equation}
 Moreover (Proposition \ref{prop:II-KVB-decomp-of-Sstar}),
\begin{equation}\label{eq:Dadjoint-prelim}
 \begin{split}
  \mathcal{D}(A_\alpha(k)^*)\;&=\;\mathcal{D}(\overline{A_\alpha(k)})\dotplus\mathscr{A}_\alpha(k)^{-1}\mathrm{span}\{\Phi_{\alpha,k}\}\dotplus\mathrm{span}\{\Phi_{\alpha,k}\}\,, \\
  \mathcal{D}(\mathscr{A}_\alpha(k))\;&=\;\mathcal{D}(\overline{A_\alpha(k)})\dotplus\mathscr{A}_\alpha(k)^{-1}\mathrm{span}\{\Phi_{\alpha,k}\}\,.
 \end{split}
\end{equation}

Owing to Lemma \ref{eq:V-RGinvertsExtS} and to \eqref{eq:defPsi} one can re-write \eqref{eq:temp_fibre_classif-prelim} and \eqref{eq:Dadjoint-prelim} as 
\begin{equation}\label{eq:temp_fibre_classif}
 \begin{split}
  \mathcal{D}(A_\alpha^{[\beta]}(k))\;&=\;\big\{\,g=\varphi+c(\beta\, \Psi_{\alpha,k}+\Phi_{\alpha,k})\,\big|\,\varphi\in\mathcal{D}(\overline{A_\alpha(k)})\,,\;c\in\mathbb{C}\big\}\,, \\
  A_\alpha^{[\beta]}(k)g\;&=\;A_\alpha(k)^*g\;=\;\overline{A_\alpha(k)}\,\varphi+c\,\beta\,\Phi_{\alpha,k}
 \end{split}
\end{equation}
and 
\begin{equation}\label{eq:Dadjoint}
 \begin{split}
  \mathcal{D}(A_\alpha(k)^*)\;&=\;\mathcal{D}(\overline{A_\alpha(k)})\dotplus\mathrm{span}\{\Psi_{\alpha,k}\}\dotplus\mathrm{span}\{\Phi_{\alpha,k}\}\,, \\
  \mathcal{D}(\mathscr{A}_\alpha(k))\;&=\;\mathcal{D}(\overline{A_\alpha(k)})\dotplus\mathrm{span}\{\Psi_{\alpha,k}\}\,.
 \end{split}
\end{equation}

By comparing \eqref{eq:Dadjoint} with the short-range asymptotics for $\Phi_{\alpha,k}$ (formula \eqref{eq:Asymtotics_0} above), for $\Psi_{\alpha,k}$ (Lemma \ref{lem:Psi_asymptotics}), and for the elements of $\mathcal{D}(\overline{A_\alpha(k)})$ (Lemma \ref{lem:BehaviourZeroClosure}), it is immediate to deduce that for a function
\begin{equation}\label{eq:g_in_Dstar}
 g\;=\;\varphi+c_1\Psi_{\alpha,k}+c_0\Phi_{\alpha,k}\;\in\;\mathcal{D}(A_\alpha(k)^*)
\end{equation}
(with $\varphi\in \mathcal{D}(\overline{A_\alpha(k)})$ and $c_0,c_1\in\mathbb{C}$)
the limits
\begin{equation}\label{eq:limitsg0g1}
 \begin{split}
  g_0\;&:=\;\lim_{x\downarrow 0}\,x^{\frac{\alpha}{2}}g(x)\;=\;c_0{\textstyle\sqrt{\frac{\pi(1+\alpha)}{2|k|}}}\,, \\
  g_1\;&:=\;\lim_{x\downarrow 0}\,x^{-(1+\frac{\alpha}{2})}(g(x)-g_0 x^{-\frac{\alpha}{2}}) \\
  &\,\,=\;c_1{\textstyle\sqrt{\frac{2|k|}{\pi(1+\alpha)^3}}}\,\|\Phi_{\alpha,k}\|_{L^2}^2-c_0{\textstyle\sqrt{\frac{\pi|k|}{2(1+\alpha)}}} 
 \end{split}
\end{equation}
exist and are finite, and one has the asymptotics
\begin{equation}\label{eq:gatzero}
 g(x)\;\stackrel{x\downarrow 0}{=}\;  g_0x^{-\frac{\alpha}{2}}+g_1x^{1+\frac{\alpha}{2}}+o(x^{\frac{3}{2}})\,.
\end{equation}

In turn, by comparing \eqref{eq:g_in_Dstar} with \eqref{eq:temp_fibre_classif} one sees that for given $\beta$ the domain of the extension $A_\alpha^{[\beta]}(k)$ consists of all those $g$'s in $\mathcal{D}(A_\alpha(k)^*)$ that, decomposed as in \eqref{eq:g_in_Dstar}, satisfy the condition
\begin{equation}\label{eq:V-c1betac0}
 c_1\;=\;\beta\,c_0\,.
\end{equation}
Moreover, replacing $c_0$ and $c_1$ of the expression \eqref{eq:g_in_Dstar} with $g_0$ and $g_1$ according to \eqref{eq:limitsg0g1}, the self-adjointness condition \eqref{eq:V-c1betac0} takes the form
\begin{equation}\label{eq:g1gammag0}
 g_1\;=\;\gamma \,g_0\,,\qquad \gamma\;:=\;\frac{|k|}{1+\alpha}\Big(\frac{\, 2 \|\Phi_{\alpha,k}\|_{L^2}^2 \,}{\pi(1+\alpha)}\, \beta-1\Big)\,.
\end{equation}
 One can therefore equivalently parametrise each extension with the new real parameter $\gamma$ and write $A_\alpha^{[\gamma]}(k)$ in place of $A_\alpha^{[\beta]}(k)$, with $\beta$ and $\gamma$ linked by \eqref{eq:g1gammag0}.

  The following is thus proved.

\begin{proposition}\label{prop:temp-class}
 Let $\alpha\in(0,1)$ and $ k\in\mathbb{Z}\setminus\{0\}$. The self-adjoint extensions of $A_\alpha(k)$ in $L^2(\mathbb{R}^+)$ form the family $\{ A_\alpha^{[\gamma]}(k)\,|\,\gamma\in\mathbb{R}\cup\{\infty\}\}$. The extension with $\gamma=\infty$ is the reference extension $\mathscr{A}_\alpha(k)=R_{G_{\alpha,k}}^{-1}$, where $R_{G_{\alpha,k}}$ is the operator defined by \eqref{eq:newRGalphaforus}. For generic $\gamma\in\mathbb{R}$ one has
 \begin{equation}\label{eq:tempclass}
 \begin{split}
  A_\alpha^{[\gamma]}(k)\;&=\;A_\alpha(k)^*\Big|_{\mathcal{D}(A_\alpha^{[\gamma]}(k))}\,, \\
  \mathcal{D}(A_\alpha^{[\gamma]}(k))\;&=\;\{g\in\mathcal{D}(A_\alpha(k)^*)\,|\,g_1=\gamma g_0\}\,,
 \end{split}
 \end{equation}
 where, for each $g$, the constants $g_0$ and $g_1$ are defined by the limits \eqref{eq:limitsg0g1}.
\end{proposition}

Although the above classification is not yet in the desired final form, it allows one to make now an important identification.

\begin{proposition}\label{eq:V-RGisSFinv}
 Let $\alpha\in(0,1)$ and $ k\in\mathbb{Z}\setminus\{0\}$. Then  $\mathscr{A}_\alpha(k)=A_{\alpha,\mathrm{F}}(k)$, and hence $R_{G_{\alpha,k}}=A_{\alpha,\mathrm{F}}(k)^{-1}$ and $\Psi_{\alpha,k}=(A_{\alpha,\mathrm{F}}(k))^{-1}\Phi_{\alpha,k}$.
\end{proposition}

For the proof of Proposition \ref{eq:V-RGisSFinv} it is convenient to recall the following.

\begin{lemma}\label{lem:V-Fform}
 Let $\alpha\in(0,1)$ and $ k\in\mathbb{Z}\setminus\{0\}$. The quadratic form of the Friedrichs extension of $A_\alpha(k)$ is given by
 \begin{equation}\label{eq:V-Fform}
  \begin{split}
  \mathcal{D}[A_{\alpha,\mathrm{F}}(k)]\;&=\;\big\{g\in L^2(\mathbb{R}^+)\,\big|\, \|g'\|_{L^2}^2+\|x^{\alpha} g\|_{L^2}^2+\|x^{-1}g\|_{L^2}^2<+\infty\big\}\,, \\
   A_{\alpha,\mathrm{F}}(k)[g,h]\;&=\;\int_0^{+\infty}\!\!\Big(\,\overline{g'(x)}h'(x)+ k^2x^{2\alpha}\,\overline{g(x)}h(x)+C_\alpha\frac{\,\overline{g(x)}h(x)}{\,x^2}\Big)\ud x\,.
  \end{split}
 \end{equation}
 \end{lemma}

\begin{proof}
 A standard application of the general Friedrichs construction (Theorem \ref{thm:Friedrichs-ext}(i)): indeed, $\mathcal{D}[A_{\alpha,\mathrm{F}}(k)]$ is the closure of $\mathcal{D}(A_\alpha(k))=C^\infty_c(\mathbb{R}^+)$ in the norm
 \[
  \begin{split}
   \|g\|_{\mathrm{F}}^2\;:=&\;\langle g,A_\alpha(k) g\rangle_{L^2}+\langle g,g\rangle_{L^2} \\
   =&\; \|g'\|_{L^2}^2+ k^2\|x^{\alpha} g\|_{L^2}^2+C_\alpha\|x^{-1}g\|_{L^2}^2+\|g\|_{L^2}^2\,.
  \end{split}
 \]
 Then \eqref{eq:V-Fform} follows at once from the above formula, since $ k^2> 0$ and $C_\alpha>0$.
\end{proof}

\begin{proof}[Proof of Proposition \ref{eq:V-RGisSFinv}]
%
%
%

 Let $g\in\mathcal{D}(A_\alpha^{[\gamma]}(k))$ for some $\gamma\in\mathbb{R}$. 
 The short-distance expansion \eqref{eq:gatzero}, combined with the self-adjointness condition \eqref{eq:tempclass}, yields
 \[
  x^{-1}g(x)\;\stackrel{x\downarrow 0}{=}\;g_0\, x^{-(1+\frac{\alpha}{2})}+\gamma\,g_0\, x^{\frac{\alpha}{2}}+o(x^{\frac{1}{2}})\,.
 \]
 Therefore, in general (namely whenever $g_0\neq 0$) $x^{-1}g$ is \emph{not} square-integrable at zero. When this is the case, formula \eqref{eq:V-Fform} prevents $g$ from belonging to $\mathcal{D}[A_{\alpha,\mathrm{F}}(k)]$.
 This shows that \emph{no} extension $ A_\alpha^{[\gamma]}(k)$, $\gamma\in\mathbb{R}$, has operator domain entirely contained in $\mathcal{D}[A_{\alpha,\mathrm{F}}(k)]$. The latter statement does not cover  $\mathscr{A}_\alpha(k)$ ($\gamma=\infty$). Now, $A_{\alpha,\mathrm{F}}(k)$ can be none of the $ A_\alpha^{[\gamma]}(k)$'s, $\gamma\in\mathbb{R}$, because the Friedrichs extension has indeed  operator domain inside $\mathcal{D}[A_{\alpha,\mathrm{F}}(k)]$ -- in fact, it is the unique extension with such property (Theorem \ref{thm:Friedrichs-ext}(v)).
 Necessarily the conclusion is that $A_{\alpha,\mathrm{F}}(k)$ and $\mathscr{A}_\alpha(k)$ are the same.
%
%
%
%
%
\end{proof}

A straightforward consequence of Proposition \ref{eq:V-RGisSFinv} (and of its proof) is the following.

\begin{corollary}\label{cor:AF_in_x-1}
Let $\alpha\in(0,1)$ and $ k\in\mathbb{Z}\setminus\{0\}$.  
 The Friedrichs extension $A_{\alpha,\mathrm{F}}(k)$ of $A_\alpha(k)$ is the only self-adjoint extension whose operator domain is contained in $\mathcal{D}(x^{-1})$. 
\end{corollary}

\subsection{One-sided extensions for non-zero modes}\label{subsec:proof_of_fibrethm}

 Based on the preceding discussion, Theorem \ref{thm:fibre-thm} can be now demonstrated. The case $\alpha=0$, as mentioned, is classical (see, e.g., \cite[Section 6.2.2.1]{GTV-2012}): the operator closure has domain $H^2_0(\mathbb{R}^+)$, the adjoint has domain $H^2(\mathbb{R}^+)$, the Friedrichs extension is the Dirichlet Laplacian and has form domain $H^1_0(\mathbb{R}^+)$, and formula \eqref{eq:V-gammaextonesided} for $\alpha=0$ does match the expression, already available in the literature, of a generic self-adjoint extension.

 Concerning the proof of Theorem \ref{thm:fibre-thm} when $\alpha\in(0,1)$, part (i) is precisely Proposition \ref{prop:domAclosure}. Part (ii) follows from \eqref{eq:Afstar} and \eqref{eq:Dadjoint} concerning the operator domain, and from Lemma \ref{lem:kerAxistar} concerning the kernel.

Part (iv), the actual classification of extensions, is the rephrasing of Proposition \ref{prop:temp-class}, using the fact that the reference extension is $\mathscr{A}_\alpha(k)=A_{\alpha,\mathrm{F}}(k)$ (Proposition \ref{eq:V-RGisSFinv}), and plugging the self-adjointness condition $g_1=\gamma g_0$ into the general asymptotics \eqref{eq:gatzero}.

In part (iii), formula \eqref{eq:thmAFoperator} for the operator domain follows from \eqref{eq:Dadjoint} (with $\mathscr{A}_\alpha(k)=A_{\alpha,\mathrm{F}}(k)$) and from the short-range asymptotics for $\Psi_{\alpha,k}$ (Lemma \ref{lem:Psi_asymptotics}), and for the elements of $\mathcal{D}(\overline{A_\alpha(k)})$ (Lemma \ref{lem:BehaviourZeroClosure}) -- which is the same as taking formally $\gamma=\infty$ in the general asymptotics. The distinctive property of $A_{\alpha,\mathrm{F}}(k)$ with respect to the space $\mathcal{D}(x^{-1})$ is given by Corollary \ref{cor:AF_in_x-1}.

Thus, it remains to prove \eqref{eq:thmAFform} for the form domain of $A_{\alpha,\mathrm{F}}(k)$. The inclusion $\mathcal{D}[A_{\alpha,\mathrm{F}}(k)]\subset H^1_0(\mathbb{R}^+)\cap L^2(\mathbb{R}^+,\langle x\rangle^{2\alpha}\,\ud x)$ follows directly from Lemma \ref{lem:V-Fform}, as \eqref{eq:V-Fform} prescribes that if $g\in\mathcal{D}[A_{\alpha,\mathrm{F}}(k)]$, then $g',x^\alpha g,x^{-1}g\in L^2(\mathbb{R}^+)$, and the latter condition implies necessarily $g(0)=0$. Conversely, if $g\in H^1_0(\mathbb{R}^+)$ \emph{and} $g\in L^2(\mathbb{R}^+,\langle x\rangle^{2\alpha}\,\ud x)$, then $g(x)\stackrel{x\downarrow 0}{=}o(x^{\frac{1}{2}})$ and all three norms $\|g'\|_{L^2}$, $\|x^{\alpha} g\|_{L^2}$, and $\|x^{-1}g\|_{L^2}$ are finite. Owing to \eqref{eq:V-Fform}, $g\in \mathcal{D}[A_{\alpha,\mathrm{F}}(k)]$.

The proof of Theorem \ref{thm:fibre-thm} is thus completed.

\subsection{One-sided extensions for the zero mode}\label{sec:zero_mode}

 The analysis of Subsections \ref{subsec:homog_problem}-\ref{subsec:distinguished} is to be modified when $k=0$, because the greatest lower bound of $A_\alpha(0)$ is zero (see \eqref{eq:Axibottom-zero} above). The conceptual scheme is the same, but applied now to the \emph{shifted} operator $A_\alpha(0)+\mathbbm{1}$ that by construction has strictly positive greatest lower bound.

%
 
 Thus, one replaces the expression \eqref{eq:Dadjoint-prelim}, namely
\[
 \mathcal{D}(A_\alpha(k)^*)\;=\;\mathcal{D}(\overline{A_\alpha(k)})\dotplus(A_{\alpha,\mathrm{F}}(k))^{-1}\ker A_\alpha(k)^*\dotplus\ker A_\alpha(k)^*\,, \quad k\in\mathbb{Z}\setminus\{0\}\,,
\]
 with
\begin{equation}
 \begin{split}
  \mathcal{D}(A_\alpha(0)^*&+\mathbbm{1})\;=\;\mathcal{D}(\overline{A_\alpha(0)}+\mathbbm{1})\dotplus \\
  &\dotplus(A_{\alpha,\mathrm{F}}(0)+\mathbbm{1})^{-1}\ker (A_\alpha(0)^*+\mathbbm{1})\dotplus\ker (A_\alpha(0)^*+\mathbbm{1})
 \end{split}
\end{equation}
 (that follows again from Proposition \ref{prop:II-KVB-decomp-of-Sstar}). Obviously, $\mathcal{D}(A_\alpha(0)^*+\mathbbm{1})=\mathcal{D}(A_\alpha(0)^*)$ and $\mathcal{D}(\overline{A_\alpha(0)}+\mathbbm{1})=\mathcal{D}(\overline{A_\alpha(0)})$, and analogously the domain of each extension is insensitive to the shift by $\mathbbm{1}$. 

In fact, the zero mode fibre operator $A_\alpha(0)$ is a classical Bessel operator\index{Bessel operator} and its self-adjoint realisations are known in the literature, obtained by other means and from a different perspective (see, e.g., \cite{Bruneau-Derezinski-Georgescu-2011,Derezinski-Georgescu-2021}): an amount of details will be therefore omitted here.

 To begin with, concerning the homogeneous problem 
\begin{equation}\label{eq:homokzero}
 0\;=\;(S_{\alpha,0}+\mathbbm{1})h\;=\;-h''+C_\alpha x^{-2} h+h\,,
\end{equation}
 one sees that setting
\[
 w(z)\;:=\;\frac{h(x)}{\sqrt{x}}\,,\qquad \nu\;:=\;\sqrt{\frac{1+4C_\alpha}{4}}\;=\;\frac{1+\alpha}{2}\,,
\]
\eqref{eq:homokzero} takes the form of the modified Bessel equation\index{Bessel equation}
\begin{equation}
 x^2w''+x w'-(z^2+\nu^2)w\;=\;0\,,\qquad x\in\mathbb{R}^+\,.
\end{equation}
From the two linearly independent solutions $K_\nu$ and $I_\nu$ to the latter \cite[Section 9.6]{Abramowitz-Stegun-1964} one therefore has that
\begin{equation}\label{eq:Phi_and_F_mode_0}
\begin{split}
	\Phi_{\alpha,0}(x)\;&:=\;\sqrt{x}\, K_{\frac{1+\alpha}{2}}(x)\,, \\
	F_{\alpha,0}(x)\;&:=\; \sqrt{x}\, I_{\frac{1+\alpha}{2}}(x)
\end{split}
\end{equation}
are two linearly independent solutions to \eqref{eq:homokzero}. In fact, only $\Phi_{\alpha,0}$ is square-integrable, as is seen from the short-distance asymptotics \cite[Eq.~(9.6.2) and (9.6.10)]{Abramowitz-Stegun-1964} 
\begin{equation}\label{eq:AsymPhi00}
\begin{split}
	\Phi_{\alpha,0}\;&\overset{x \downarrow 0}{=}\;2^{\frac{\alpha-1}{2}} \Gamma\left({\textstyle\frac{1+\alpha}{2}}\right) x^{-\frac{\alpha}{2}}-
	 \textstyle{\frac{\Gamma\left({\textstyle\frac{1-\alpha}{2}} \right)}{2^{\frac{1+\alpha}{2}}(1+\alpha)}}  x^{1+\frac{\alpha}{2}}
	  + O(x^{2-\frac{\alpha}{2}})\,,	\\
	F_{\alpha,0}(x) \;&\overset{x \downarrow 0}{=}\; \textstyle{\big(2^{\frac{1+\alpha}{2}}\Gamma\left(\frac{3+\alpha}{2}\right)\big)}^{-1} x^{1+\frac{\alpha}{2}} + O(x^{3+\frac{\alpha}{2}})\,,
\end{split}
\end{equation}
and from the large-distance asymptotics \cite[Eq.~(9.7.1) and (9.7.2)]{Abramowitz-Stegun-1964} 
\begin{equation}\label{eq:Asym_Phi_F_0_Infty}
\begin{split}
	\Phi_{\alpha,0}(x)\;&\overset{x \to +\infty}{=}\; \textstyle{\sqrt{\frac{\pi}{2}}}\, e^{-x}\, (1+O(x^{-1}))\,, \\
		F_{\alpha,0}(x) \;&\overset{x \to +\infty}{=}\; \textstyle{\frac{1}{\sqrt{2 \pi}}}\, e^{x} \,(1+O(x^{-1})) \, .
\end{split}
\end{equation}

Thus, in analogy to Lemma \ref{lem:kerAxistar}, one finds:
\begin{lemma}\label{lem:kerAxistarzero}
 For $\alpha\in(0,1)$,
 \begin{equation}\label{eq:kerAxistarzero}
  \ker (A_\alpha(0)^*+\mathbbm{1})\;=\;\mathrm{span}\{\Phi_{\alpha,0}\}\,.
 \end{equation}
\end{lemma}

Next, concerning the non-homogeneous problem
\begin{equation}\label{eq:Inhomog_ODE_mode_0}
	S_{\alpha,0}u + u \; = \; g
\end{equation}
in the unknown $u$ for given $g$, the Wronskian relative to the fundamental system\index{fundamental system for an O.D.E.} $\{\Phi_{\alpha,0}, F_{\alpha,0}\}$ is constant in $r$ and explicitly given by
\begin{equation}
	W(\Phi_{\alpha,0},F_{\alpha,0})\; =\; \det \begin{pmatrix}
		\Phi_{\alpha,0}(r) & F_{\alpha,0}(r) \\
		\Phi_{\alpha,0}'(r) & F_{\alpha,0}'(r)	
	\end{pmatrix}
	\; =\; 1\,,
\end{equation}
as one computes based on the asymptotics \eqref{eq:AsymPhi00} or \eqref{eq:Asym_Phi_F_0_Infty}. By standard variation of constants,\index{variation of constants} a particular solution to \eqref{eq:Inhomog_ODE_mode_0} is
\begin{equation}
	u_{\text{part}} (r) \; = \; \int_0^{+\infty} G_{\alpha,0}(r,\rho) g(\rho) \, \ud \rho
\end{equation}
with
\begin{equation}\label{eq:GreenMode0}
	G_{\alpha,0}(r,\rho) \; := \; \begin{cases}
		\Phi_{\alpha,0}(r) F_{\alpha,0}(\rho)\,, \qquad \text{if\; $0<\rho<r$}\,, \\
		F_{\alpha,0}(r) \Phi_{\alpha,0}(\rho)\,, \qquad \text{if\; $0< r < \rho$}\,.
	\end{cases}
\end{equation}
With the same arguments used for Lemma  \ref{lem:V-RGbddsa}, using now the asymptotics \eqref{eq:AsymPhi00}-\eqref{eq:Asym_Phi_F_0_Infty}, one finds the following analogue (an explicit proof of which can be found also in \cite[Lemma 4.4]{Bruneau-Derezinski-Georgescu-2011}).

\begin{lemma}
Let $\alpha \in (0,1)$. Let $R_{G_{\alpha,0}}$ be the operator associated to the integral kernel \eqref{eq:GreenMode0}. $R_{G_{\alpha,0}}$ can be realised as an everywhere defined, bounded, and self-adjoint operator on $L^2(\mathbb{R}^+,\ud r)$.
\end{lemma}

Analogously to \eqref{eq:defPsi} one sets 
\begin{equation}\label{eq:defPsi_zero_mode}
	\Psi_{\alpha,0}(x)\;:=\; R_{G_{\alpha,0}} \Phi_{\alpha,0}\,.
\end{equation}
The proof of Lemma \ref{lem:Psi_asymptotics} can be then repeated verbatim, with $\Phi_{\alpha,0}$ and $F_{\alpha,0}$ in place of $\Phi_{\alpha,k}$ and $F_{\alpha,k}$, so as to obtain:

\begin{lemma}\label{lem:Psi_asymptotics_mode_zero}
 For $\alpha\in(0,1)$,
 \begin{equation}\label{eq:Psi_Asymptotics_mode_zero}
	\Psi_{\alpha,0}(x)\;\overset{x \downarrow 0}{=}\; \textstyle{\big( 2^{\frac{1+\alpha}{2}}\Gamma\left(\frac{3+\alpha}{2}\right)\!\big)}^{\!-1}\, \Vert \Phi_{\alpha,0} \Vert_{L^2}^2\, x^{1+\frac{\alpha}{2}} + o(x^{\frac{3}{2}})\,.
\end{equation}
\end{lemma}

Concerning $\overline{A_\alpha(0)}$, it suffices to import from the literature the following analogue of Lemma \ref{lem:BehaviourZeroClosure}.

\begin{lemma}\label{lem:BehaviourZeroClosure_zero_mode}
Let $\alpha \in (0,1)$. Then $\mathcal{D}(\overline{A_\alpha(0)})=H^2_0(\mathbb{R}^+)$. In particular, every $\varphi \in \mathcal{D}(\overline{A_\alpha(0)})$ satisfies $\varphi(x)=o(x^{\frac{3}{2}})$ and $\varphi'(x)=o(x^{\frac{1}{2}})$ as $x \downarrow 0$.
\end{lemma}

\begin{proof}
A direct consequence of \cite[Theorem 4.1]{Derezinski-Georgescu-2021}: in the notation therein $\overline{A_\alpha(0)}$ is the operator $L_\delta^\mathrm{min}$ with $\delta-\frac{1}{4}=C_\alpha$ (the present $\delta$ replaces the notation $\alpha$ from \cite{Derezinski-Georgescu-2021} so as not to clash with the current meaning of $\alpha$), that is $\delta=(\frac{1+\alpha}{2})^2$; the requirement $\mathfrak{Re}\sqrt{\delta}<1$ needed for the applicability of \cite[Theorem 4.1]{Derezinski-Georgescu-2021} is therefore satisfied, since $\alpha\in(0,1)$.
\end{proof}

As a further step, repeating the argument used for Lemma \ref{eq:V-RGinvertsExtS}, one concludes that $R_{G_{\alpha,0}}^{-1}$ is a self-adjoint extension of $A_\alpha(0)+\mathbbm{1}$ with everywhere defined and bounded inverse, whose domain clearly contains $\Psi_{\alpha,0}$. Such a reference extension induces a classification of all other self-adjoint extensions in complete analogy to what discussed in Subsect.~\ref{subsec:distinguished}. Thus, \eqref{eq:temp_fibre_classif} and \eqref{eq:Dadjoint} are valid in the identical form also when $k=0$, and the short-range asymptotics for $\Phi_{\alpha,0}$ (formula \eqref{eq:AsymPhi00}), for $\Psi_{\alpha,0}$ (Lemma \ref{lem:Psi_asymptotics_mode_zero}), and for the elements of $\mathcal{D}(\overline{A_\alpha(0)})$ (Lemma \ref{lem:BehaviourZeroClosure_zero_mode}) imply that for a generic
\begin{equation}\label{eq:g_in_Dstar_zero_mode}
 g\;=\;\varphi+c_1\Psi_{\alpha,0}+c_0\Phi_{\alpha,0}\;\in\;\mathcal{D}(A_\alpha(0)^*)
\end{equation}
(with $\varphi\in \mathcal{D}(\overline{A_\alpha(0)})$ and $c_0,c_1\in\mathbb{C}$)
the limits
\begin{equation}\label{eq:limitsg0g1_zero_mode}
\begin{split}
	g_0 \; :=& \; \lim_{x \downarrow 0}  x^{\frac{\alpha}{2}} g(x) \; = \;  c_0\,\textstyle{2^{-\frac{1-\alpha}{2}} \Gamma\left( {\textstyle \frac{1+\alpha}{2}}\right)}\,, \\
	g_1\;  :=& \;\lim_{x \downarrow 0} x^{-(1+\frac{\alpha}{2})} (g(x)-g_0x^{-\frac{\alpha}{2}}) \\
	=& \; c_1 \textstyle{\big(2^{\frac{1+\alpha}{2}}\Gamma({\textstyle\frac{3+\alpha}{2}})\big)}^{\!-1} \Vert \Phi_{\alpha,0} \Vert^2_{L^2(\mathbb{R}^+)} -
	c_0 \textstyle{\big(2^{\frac{1+\alpha}{2}}(1+\alpha)\big)}^{\!-1}\Gamma({\textstyle\frac{1-\alpha}{2}})
\end{split}
\end{equation}
exist and are finite, and one has the asymptotics
\begin{equation}\label{eq:gatzero_zero_mode}
 g(x)\;\stackrel{x\downarrow 0}{=}\;  g_0x^{-\frac{\alpha}{2}}+g_1x^{1+\frac{\alpha}{2}}+o(x^{\frac{3}{2}})\,.
\end{equation}
Then, analogously to \eqref{eq:V-c1betac0}-\eqref{eq:g1gammag0}, the condition of self-adjointness reads again as $c_1=\beta c_0$ for some $\beta\in\mathbb{R}$, or equivalently as 
\begin{equation}\label{eq:g1gammag0_zero_mode}
 g_1\;=\;\gamma g_0\,,\qquad \gamma\;:=\;{\frac{\|\Phi_{\alpha,0}\|_{L^2}^2}{2^\alpha\Gamma(\frac{1+\alpha}{2})\Gamma(\frac{3+\alpha}{2})}\Big(\beta-\frac{\Gamma(\frac{1-\alpha}{2})\Gamma(\frac{3+\alpha}{2})}{(1+\alpha)\|\Phi_{\alpha,0}\|_{L^2}^2}\Big)}.
\end{equation}
This yields an obvious analogue of the `temporary' classification of Proposition \ref{prop:temp-class}, where if $A_\alpha^{[\gamma]}(0)+\mathbbm{1}$ is a self-adjoint extension of $A_\alpha(0)+\mathbbm{1}$, so is $A_\alpha^{[\gamma]}(0)$ for $A_\alpha(0)$, with $\mathcal{D}(A_\alpha^{[\gamma]}(0)+\mathbbm{1})=\mathcal{D}(A_\alpha^{[\gamma]}(0))$.

In fact, based on the very same argument of Lemma \ref{lem:V-Fform}, repeated now for the characterisation of the form domain of $A_{\alpha,\mathrm{F}}(0)$, one can also reproduce the argument of Proposition \ref{eq:V-RGisSFinv}, establishing the following analogue.

\begin{proposition}\label{eq:RGisSFinv_zero_mode}
 For $\alpha\in(0,1)$, one has $A_{\alpha,\mathrm{F}}(0)+\mathbbm{1}=R_{G_{\alpha,0}}^{-1}$ and $\Psi_{\alpha,0}=(A_{\alpha,\mathrm{F}}(0)+\mathbbm{1})^{-1}\Phi_{\alpha,0}$.
\end{proposition}

Notably, the following useful characterisation of the domain of the Friedrichs extension of $A_\alpha(0)$ is available in the literature.

\begin{proposition}
 For $\alpha\in(0,1)$,
 \begin{equation}\label{eq:Friedrichs_mode_0}
	\mathcal{D}(A_{\alpha,\mathrm{F}}(0)) = \mathcal{D}(\overline{A_\alpha(0)}) + \mathrm{span}\{ x^{1+\frac{\alpha}{2}} P \}\,,
\end{equation}
where $P\in C^\infty_c([0,+\infty))$ with $P(0)=1$. 
\end{proposition}

\begin{proof}
 In the notation of \cite{Bruneau-Derezinski-Georgescu-2011}, the Friedrichs extension is the operator $H_m^{\theta}$ with $m^2-\frac{1}{4}=C_\alpha$, hence $m=\frac{1+\alpha}{2}\in(0,1)$, and with $\theta=\frac{\pi}{2}$ (\cite[Proposition 4.19]{Bruneau-Derezinski-Georgescu-2011}), whereas $\overline{A_\alpha(0)}$ is the operator $L_m^\mathrm{min}$. In turn, such $H_m^{\theta}$ is recognised to be the operator $L_m^{u_{\theta}}$, where $u_\theta$ is the function that for $\theta=\frac{\pi}{2}$ has the form $u_{\pi/2}(x)=x^{1+\alpha/2}$ (\cite[Proposition 4.17(1)]{Bruneau-Derezinski-Georgescu-2011}). With this correspondence, the formula $\mathcal{D}(L_m^{u_{\theta}})=\mathcal{D}(L_m^\mathrm{min})+\mathrm{span}\{u_\theta P\}$ (\cite[Proposition A.5]{Bruneau-Derezinski-Georgescu-2011}) then yields precisely \eqref{eq:Friedrichs_mode_0}. 
\end{proof}

With all the ingredients collected so far, and based on a straightforward adaptation of the arguments of Subsect.~\ref{subsec:proof_of_fibrethm}, the above `temporary' classification then takes the following final form.

\begin{theorem}\label{thm:fibre-thm_zero_mode}
 Let $\alpha\in[0,1)$.
 \begin{enumerate}[(i)]
  \item The adjoint of $A_\alpha(0)$ has domain
  \begin{equation}\label{eq:Dadjoint_kequalzero}
   \begin{split}
    \mathcal{D}(A_\alpha(0)^*)\;&=\;\left\{\!\!
  \begin{array}{c}
   g\in L^2(\mathbb{R}^+)\;\;\textrm{such that} \\
   \big(-\frac{\ud^2}{\ud x^2}+\frac{\,\alpha(2+\alpha)\,}{4x^2}\big)g\in L^2(\mathbb{R}^+)
  \end{array}
  \!\!\right\} \\
   &=\;H^2_0(\mathbb{R}^+)\dotplus\mathrm{span}\{\Psi_{\alpha,0}\}\dotplus\mathrm{span}\{\Phi_{\alpha,0}\}\,,
   \end{split} 
  \end{equation}
   where $\Phi_{\alpha,0}$ and $\Psi_{\alpha,0}$ are two smooth functions on $\mathbb{R}^+$ explicitly defined, in terms of modified Bessel functions, respectively by formulas \eqref{eq:Phi_and_F_mode_0}, \eqref{eq:GreenMode0}, and \eqref{eq:defPsi_zero_mode}. Moreover,
   \begin{equation}\label{eq:kerAxistar-in-thm_zero_mode}
  \ker (A_\alpha(0)^*+\mathbbm{1})\;=\;\mathrm{span}\{\Phi_{\alpha,0}\}\,.
 \end{equation}
   \item The self-adjoint extensions of $A_\alpha(0)$ in $L^2(\mathbb{R}^+)$ form the family
   \[
\{ A_\alpha^{[\gamma]}(0)\,|\,\gamma\in\mathbb{R}\cup\{\infty\}\}\,.    
   \]
 The extension with $\gamma=\infty$ is the Friedrichs extension $A_{\alpha,\mathrm{F}}(0)$, whose domain is given by \eqref{eq:Friedrichs_mode_0}, and moreover $(A_{\alpha,\mathrm{F}}(0)+\mathbbm{1})^{-1}=R_{G_{\alpha,0}}$, the everywhere defined and bounded operator with integral kernel given by \eqref{eq:GreenMode0}. For generic $\gamma\in\mathbb{R}$ one has
 \begin{equation}
  \mathcal{D}(A_\alpha^{[\gamma]}(0))\,=\,
  \left\{\!\!
  \begin{array}{c}
   g\in\mathcal{D}(A_\alpha(0)^*) \;\;\textrm{such that}\\
   g(x)\,\stackrel{x\downarrow 0}{=}\,g_0 x^{-\frac{\alpha}{2}}+\gamma\, g_0\,x^{1+\frac{\alpha}{2}}+o(x^{\frac{3}{2}}) \\
   \textrm{with } g_0\in\mathbb{C}
  \end{array}
  \!\!\right\}.
  \end{equation}
 \end{enumerate}
\end{theorem}

\section{Extensions on two-sided fibre}\label{sec:bilateralfibreext}

 For the problem of the self-adjoint extensions of the two-sided (`bilateral') operator
\begin{equation}
 \begin{split}
  \mathcal{D}(A_\alpha(k))\;&=\;C^\infty_c(\mathbb{R}^-)\boxplus C^\infty_c(\mathbb{R}^+)\,, \\
  A_\alpha(k)\;&=\;A_\alpha^-(k)\oplus A_\alpha^+(k)\,,
 \end{split}
\end{equation}
 with respect to the fibre Hilbert space 
\begin{equation}\label{eq:againdecomp}
 \mathfrak{h}\;=\;L^2(\mathbb{R})\;\cong\;L^2(\mathbb{R}^-)\oplus L^2(\mathbb{R}^+)\,,
\end{equation}
 (see \eqref{eq:Hoplusfrakh} and \eqref{eq:Axibilateral} above) one has to `double' the analysis of Section \ref{sec:fibre-extensions}.

Each $g\in L^2(\mathbb{R})$ is canonically expressed as
\begin{equation}\label{eq:gdecomp}
 g\;=\;g^-\oplus g^+\;\equiv\;
 \begin{pmatrix}
  g^- \\ g^+
 \end{pmatrix}
\,,\qquad g^\pm(x)\,:=\,g(x)\;\;\textrm{ for }x\in\mathbb{R}^\pm\,,
\end{equation}
and 
\begin{equation}
 A_\alpha(k)g\;=\;S_{\alpha,k}g^-\oplus S_{\alpha,k}g^+\,,\qquad S_{\alpha,k}\;:=\; -\frac{\ud^2}{\ud x^2}+ k^2 |x|^{2\alpha}+\frac{C_\alpha}{x^2}\,.
\end{equation}

As $A_\alpha^\pm(k)$ has deficiency index 1 in $L^2(\mathbb{R}^\pm)$, $A_\alpha(k)$ has deficiency index 2 in $L^2(\mathbb{R})$, and therefore has a richer variety of extensions.

Among them, one has extensions of form
\begin{equation}\label{eq:fibre_uncoupled_ext}
 B_\alpha^-(k) \oplus B_\alpha^+(k)\,,
\end{equation}
where $B_\alpha^\pm(k)$ is a self-adjoint extension of $A_\alpha^\pm(k)$ in $L^2(\mathbb{R}^\pm)$, namely a member of the family described in Theorem \ref{thm:fibre-thm}. Extensions of type \eqref{eq:fibre_uncoupled_ext} are \emph{reduced} (Sect.~\ref{sec:I_invariant-reducing-ssp}) with respect to the decomposition \eqref{eq:againdecomp}: they are \emph{decoupled} self-adjoint realisations of the differential operator $S_{\alpha,k}$, with no constraint between the behaviour as $x\to 0^+$ and $x\to 0^-$. The Friedrichs extension is of this type (Lemma \ref{lem:sumofFriedrichs}):
\begin{equation}\label{eq:biFriedrichs}
 A_{\alpha,\mathrm{F}}(k)\;=\;A_{\alpha,\mathrm{F}}^-(k)\oplus A_{\alpha,\mathrm{F}}^+(k)\,,
\end{equation}
where $A_{\alpha,\mathrm{F}}^\pm$ is the Friedrichs extension of $A_\alpha^\pm(k)$ in $L^2(\mathbb{R}^\pm)$ as given by Theorem \ref{thm:fibre-thm}(iii)).

Generic extensions do not necessarily have the reduced structure \eqref{eq:fibre_uncoupled_ext} and may be characterised by \emph{coupled} bilateral boundary conditions. 


Following the same steps of Section \ref{sec:fibre-extensions} one investigates now self-adjoint \emph{restrictions} of $A_\alpha(k)^*=A_\alpha^-(k)^*\oplus A_\alpha^+(k)^*$ (see \eqref{eq:Afstar_sum} above).

 In passing from the one-sided to the two-sided analysis it is convenient to introduce unique expressions for the functions of relevance, $\Phi_{\alpha,k}$ and $\Psi_{\alpha,k}$, valid for the left and the right side:
\begin{equation}\label{eq:PhiPsitilde}
 \widetilde{\Phi}_{\alpha,k}(x)\;:=\;\Phi_{\alpha,k}(|x|)\,,\qquad\widetilde{\Psi}_{\alpha,k}(x)\;:=\;\Psi_{\alpha,k}(|x|)\,,
\end{equation}
understanding $\widetilde{\Phi}_{\alpha,k}$ and $\widetilde{\Psi}_{\alpha,k}$ both as functions on $\mathbb{R}^-$ and on $\mathbb{R}^+$, depending on the context. Such functions are defined in \eqref{eq:Phi_and_F_explicit} and \eqref{eq:defPsi} when $k\neq 0$, and in \eqref{eq:Phi_and_F_mode_0} and \eqref{eq:defPsi_zero_mode} when $k=0$.

 The case $k\in\mathbb{Z}\setminus\{0\}$ shall be discussed first. One deduces at once, respectively from Proposition \ref{prop:domAclosure}, Lemma \ref{lem:kerAxistar}, formula \eqref{eq:defPsi}, and Proposition \ref{eq:V-RGisSFinv}, that
\begin{eqnarray}
 \mathcal{D}(\overline{A_\alpha(k)})\!&=&\!\big( H^2_0(\mathbb{R}^-)\boxplus H^2_0(\mathbb{R}^+)\big)\cap L^2(\mathbb{R},\langle x\rangle^{4\alpha}\,\ud x)\,, \label{eq:Dclosureclosure}\\
 \ker A_\alpha(k)^* \!&=&\!\mathrm{span}\{\widetilde{\Phi}_{\alpha,k}\}\oplus\mathrm{span}\{\widetilde{\Phi}_{\alpha,k}\}\,, \label{eq:bilateralkernel} \\
 A_{\alpha,\mathrm{F}}(k)^{-1}\ker A_\alpha(k)^* \!&=&\!\mathrm{span}\{\widetilde{\Psi}_{\alpha,k}\}\oplus\mathrm{span}\{\widetilde{\Psi}_{\alpha,k}\}\,, \label{eq:bilateralAFinvKer}
\end{eqnarray}
whence also (Proposition \ref{prop:II-KVB-decomp-of-Sstar}) 
\begin{equation}\label{eq:bilateralDAxistar}
 \begin{split}
  \mathcal{D}(A_\alpha(k)^*)\;&=\;\big( H^2_0(\mathbb{R}^-)\boxplus H^2_0(\mathbb{R}^+)\big)\cap L^2(\mathbb{R},\langle x\rangle^{4\alpha}\,\ud x) \\
  &\qquad \dotplus\mathrm{span}\{\widetilde{\Psi}_{\alpha,k}\}\oplus\mathrm{span}\{\widetilde{\Psi}_{\alpha,k}\} \\
  &\qquad \dotplus\mathrm{span}\{\widetilde{\Phi}_{\alpha,k}\}\oplus\mathrm{span}\{\widetilde{\Phi}_{\alpha,k}\}\,,
 \end{split}
\end{equation}
namely, the analogue of \eqref{eq:Dadjoint}.

In the notation of \eqref{eq:gdecomp}, a generic $g\in \mathcal{D}(A_\alpha(k)^*)$ has therefore the short-range asymptotics
\begin{equation}\label{eq:gbilateralasympt}
 g(x)\;\equiv\;\begin{pmatrix} g^-(x) \\ g^+(x) \end{pmatrix}\;\stackrel{x\to 0}{=} \;\begin{pmatrix} g_0^- \\ g_0^+ \end{pmatrix} |x|^{-\frac{\alpha}{2}}+\begin{pmatrix} g_1^- \\ g_1^+ \end{pmatrix}|x|^{1+\frac{\alpha}{2}}+o(|x|^{\frac{3}{2}})
\end{equation}
for suitable $g_0^\pm,g_1^\pm\in\mathbb{C}$ given by the (thereby existing) limits
 \begin{equation}\label{eq:bilimitsg0g1}
  \begin{split}
   g_0^\pm\;&=\;\lim_{x\to 0^\pm} |x|^{\frac{\alpha}{2}}g^\pm(x)\,, \\
   g_1^\pm\;&=\;\lim_{x\to 0^\pm} |x|^{-(1+\frac{\alpha}{2})}\big(g^\pm(x)-g_0^\pm |x|^{-\frac{\alpha}{2}}\big)\,.
  \end{split}
 \end{equation}
 Formula \eqref{eq:gbilateralasympt} follows from \eqref{eq:bilateralDAxistar} and the usual short-range asymptotics for $\Phi_{\alpha,k}$, $\Psi_{\alpha,k}$, and $\mathcal{D}(\overline{A_\alpha(k)})$.


Now, the Kre\u{\i}n-Vi\v{s}ik-Birman extension theory provides a one-to-one correspondence between self-adjoint extensions of $A_\alpha(k)$
and self-adjoint operators $T$ in Hilbert subspaces of $\ker A_\alpha(k)^*$: denoting by $A_\alpha^{(T)}\!(k)$ each such extension, and by $\mathcal{K}\subset \ker A_\alpha(k)^*$ the Hilbert subspace where $T$ acts in, $A_\alpha^{(T)}\!(k)$ is the restriction of $A_\alpha(k)^*$ to the domain (Theorem \ref{thm:VB-representaton-theorem_Tversion})
\begin{equation}\label{eq:Birman_formula}
 \mathcal{D}\big(A_\alpha^{(T)}\!(k)\big)\;=\;
 \left\{
 \begin{array}{c}
  g=\varphi+A_{\alpha,\mathrm{F}}(k)^{-1}(Tv+w)+v \\
  \textrm{with} \\
  \varphi\in\big( H^2_0(\mathbb{R}^-)\boxplus H^2_0(\mathbb{R}^+)\big)\cap L^2(\mathbb{R},\langle x\rangle^{4\alpha}\,\ud x)\,, \\
  v\in\mathcal{K}\,,\quad w\in \mathrm{span}\{\widetilde{\Phi}_{\alpha,k}\}\oplus\mathrm{span}\{\widetilde{\Phi}_{\alpha,k}\}\,,\quad w\perp v
 \end{array}
 \right\}.
\end{equation}
Clearly, $\dim\mathcal{K}$ can be equal to 0, 1, or 2.  The former case corresponds to taking formally `$T=\infty$' on $\mathcal{D}(T)=\{0\}$, and reproduces the Friedrichs extension. The other two cases produce the rest of the family of extensions.

 The preceding discussion has an immediate counterpart when $k=0$, based on the findings of Section \ref{sec:zero_mode}. The above formulas are valid for $k=0$ too, except for \eqref{eq:Dclosureclosure}, that need be replaced with
\begin{equation}\label{eq:V-Closure-DirectSumFibre}
 \mathcal{D}(\overline{A_\alpha(0)})\;=\;H^2_0(\mathbb{R}^-)\oplus H^2_0(\mathbb{R}^+)
\end{equation}
(on account of Lemma \ref{lem:BehaviourZeroClosure_zero_mode}), and except for \eqref{eq:bilateralDAxistar}, that consequently reads now
\begin{equation}\label{eq:bilateralDAxistar_zero_mode}
 \begin{split}
  \mathcal{D}(A_\alpha(0)^*)\;&=\;H^2_0(\mathbb{R}^-)\boxplus H^2_0(\mathbb{R}^+) \\
  &\qquad \dotplus\mathrm{span}\{\widetilde{\Psi}_{\alpha,0}\}\oplus\mathrm{span}\{\widetilde{\Psi}_{\alpha,0}\} \\
  &\qquad \dotplus\mathrm{span}\{\widetilde{\Phi}_{\alpha,0}\}\oplus\mathrm{span}\{\widetilde{\Phi}_{\alpha,0}\}\,.
 \end{split}
\end{equation}
Thus, when $k=0$ formula \eqref{eq:Birman_formula} takes the form
\begin{equation}\label{eq:Birman_formula_zero_mode}
 \mathcal{D}\big(A_\alpha^{(T)}\!(0)\big)\;=\;
 \left\{
 \begin{array}{c}
  g=\varphi+(A_{\alpha,\mathrm{F}}(0)+\mathbbm{1})^{-1}(Tv+w)+v \\
  \textrm{with} \\
  \varphi\in H^2_0(\mathbb{R}^-)\boxplus H^2_0(\mathbb{R}^+)\,, \\
  v\in\mathcal{K}\,,\quad w\in \mathrm{span}\{\widetilde{\Phi}_{\alpha,0}\}\oplus\mathrm{span}\{\widetilde{\Phi}_{\alpha,0}\}\,,\quad w\perp v
 \end{array}
 \right\},
\end{equation}
where now $\mathcal{K}$ is a Hilbert subspace of $\ker(A_\alpha(0)^*+\mathbbm{1})$ and $T$ is a self-adjoint operator in $\mathcal{K}$.
 
 The classification of two-sided extensions of $A_\alpha(k)$ is then formulated as follows.

\begin{theorem}\label{thm:bifibre-extensions}
 Let $\alpha\in[0,1)$ and $k\in\mathbb{Z}$. Each self-adjoint extension $B_\alpha(k)$ of $A_\alpha(k)$ acts as 
 \begin{equation}
  B_\alpha(k) g\;=\;S_{\alpha,k} \,g^- \oplus S_{\alpha,k} \,g^+
 \end{equation}
 on a generic $g$ of its domain, written in the notation of \eqref{eq:gdecomp} and \eqref{eq:gbilateralasympt}-\eqref{eq:bilimitsg0g1}. The family of self-adjoint extensions of $A_\alpha(k)$ is formed by the following sub-families.
 \begin{itemize}
  \item \underline{Friedrichs extension}. 
  
  It is the operator \eqref{eq:biFriedrichs}. Its domain consists of those functions in $\mathcal{D}(A_\alpha(k)^*)$ whose asymptotics \eqref{eq:gbilateralasympt} has $g_0^\pm=0$.
    \item \underline{Family $\mathrm{I_R}$}.
  
  It is the family $\{A_{\alpha,\mathrm{R}}^{[\gamma]}(k)\,|\,\gamma\in\mathbb{R}\}$ defined, with respect to the  asymptotics \eqref{eq:gbilateralasympt}, by
  \begin{equation}\label{eq:bifibre-AR}
   \mathcal{D}(A_{\alpha,\mathrm{R}}^{[\gamma]}(k))\;=\;\{g\in\mathcal{D}(A_\alpha(k)^*)\,|\,g_0^-=0\,,\;g_1^+=\gamma g_0^+\}\,.
  \end{equation}
   \item \underline{Family $\mathrm{I_L}$}.
  
  It is the family $\{A_{\alpha,\mathrm{L}}^{[\gamma]}(k)\,|\,\gamma\in\mathbb{R}\}$ defined, with respect to the  asymptotics \eqref{eq:gbilateralasympt}, by
  \begin{equation}\label{eq:bifibre-AL}
   \mathcal{D}(A_{\alpha,\mathrm{L}}^{[\gamma]}(k))\;=\;\{g\in\mathcal{D}(A_\alpha(k)^*)\,|\,g_1^-=\gamma g_0^-\}\,,\;g_0^+=0\,.
   \end{equation}
   \item \underline{Family $\mathrm{II}_a$} with $a\in\mathbb{C}$.
   
   It is the family $\{A_{\alpha,a}^{[\gamma]}(k)\,|\,\gamma\in\mathbb{R}\}$ defined, with respect to the  asymptotics \eqref{eq:gbilateralasympt}, by
  \begin{equation}\label{eq:bifibre-Aag}
   \mathcal{D}(A_{\alpha,a}^{[\gamma]}(k))\;=\;\left\{g\in\mathcal{D}(A_\alpha(k)^*)\left|\!
   \begin{array}{c}
    g_0^+=a\,g_0^- \\
    g_1^-+\overline{a}\,g_1^+=\gamma\, g_0^-
   \end{array}\!\!
   \right.\right\}.
  \end{equation}
   \item \underline{Family $\mathrm{III}$}.
   
   It is the family $\{A_{\alpha}^{[\Gamma]}(k)\,|\,\Gamma\equiv(\gamma_1,\gamma_2,\gamma_3,\gamma_4)\in\mathbb{R}^4\}$ defined, with respect to the  asymptotics \eqref{eq:gbilateralasympt}, by
     \begin{equation}\label{eq:bifibre-AG}
   \mathcal{D}(A_{\alpha}^{[\Gamma]}(k))\;=\;\left\{g\in\mathcal{D}(A_\alpha(k)^*)\left|\!
   \begin{array}{c}
    g_1^-=\gamma_1 g_0^-+\zeta g_0^+ \\
    g_1^+=\overline{\zeta} g_0^-+\gamma_4 g_0^+ \\
    \zeta:=\gamma_2+\ii\gamma_3
   \end{array}\!\!
   \right.\right\}.
  \end{equation}
 \end{itemize}
 The families $\mathrm{I_R}$, $\mathrm{I_L}$, $\mathrm{II}_a$ for all $a\in\mathbb{C}\setminus\{0\}$, and $\mathrm{III}$ are mutually disjoint and, together with the Friedrichs extension, exhaust the family of self-adjoint extensions of $A_\alpha(k)$. 
\end{theorem}

 Each self-adjoint extension being of the form $A_\alpha^{(T)}\!(k)$ for some self-adjoint $T$ in a Hilbert subspace $\mathcal{K}\subset\ker A_\alpha(k)^*$ if $k\in\mathbb{Z}\setminus\{0\}$, or $\mathcal{K}\subset(\ker A_\alpha(0)^*+\mathbbm{1})$ if $k=0$ (see \eqref{eq:Birman_formula} and \eqref{eq:Birman_formula_zero_mode} above), the proof of Theorem \ref{thm:bifibre-extensions} eventually shows the correspondence between each extension family and the choice of $\mathcal{K}$ as summarised by Table \ref{tab:extensions}. Thus, extensions of type $\mathrm{I_R}$, $\mathrm{I_L}$, and  $\mathrm{II}_a$ correspond to $\dim\mathcal{K}=1$, type-$\mathrm{III}$ extensions correspond to to $\dim\mathcal{K}=2$, and the Friedrichs extension is the case with $\dim\mathcal{K}=0$.

\begin{table}[t!]
\begin{center}
\begin{tabular}{|c|c|c|c|c|}
 \hline
 \begin{tabular}{c} family of \\ extensions \end{tabular} & space $\mathcal{K}$ & \begin{tabular}{c} boundary \\ conditions \end{tabular} & parameters & notes \\
 \hline
 \hline
 Friedrichs & $\{0\} \oplus \{0\}$ & $g_0^{\pm}=0$ &  & \begin{tabular}{c} bilateral \\ confining \end{tabular}\\
 \hline
  $\mathrm{I_R}$ & $\{0\}\oplus \mathrm{span}\{\widetilde{\Phi}_{\alpha,k}\}$ & $\begin{array}{c} g_0^-= 0 \\ g_1^+=\gamma g_0^+ \end{array}$ & $\gamma\in\mathbb{R}$ & \begin{tabular}{c} left \\ confining \end{tabular} \\
 \hline
   $\mathrm{I_L}$ & $\mathrm{span}\{\widetilde{\Phi}_{\alpha,k}\}\oplus \{0\}$ & $\begin{array}{c} g_1^-=\gamma g_0^- \\ g_0^+= 0 \end{array}$ & $\gamma\in\mathbb{R}$ & \begin{tabular}{c} right \\ confining \end{tabular} \\
 \hline
  \begin{tabular}{c} $\mathrm{II}_a$ \\ $a\in\mathbb{C}$ \end{tabular} & $\mathrm{span}\{\widetilde{\Phi}_{\alpha,k}\oplus a \widetilde{\Phi}_{\alpha,k}\}$ & $\begin{array}{c}
    g_0^+=a\,g_0^- \\
    g_1^-+\overline{a}\,g_1^+=\gamma\, g_0^-
   \end{array}$ & $\gamma\in\mathbb{R}$ & \begin{tabular}{c} bridging \\ for $a=1$ \\ and $\gamma=0$ \end{tabular} \\
  \hline
    $\mathrm{III}$ & $\mathrm{span}\{\widetilde{\Phi}_{\alpha,k}\}\oplus \mathrm{span}\{\widetilde{\Phi}_{\alpha,k}\}$ & $\begin{array}{c}
    g_1^-=\gamma_1 g_0^-+\zeta g_0^+ \\
    g_1^+=\overline{\zeta} g_0^-+\gamma_4 g_0^+ \\
    \zeta:=\gamma_2+\ii\gamma_3
   \end{array}$ & \begin{tabular}{c} $\gamma_j\in\mathbb{R}$ \\ $j=1,2,3,4$ \end{tabular} &  \\
 \hline
\end{tabular}
\medskip
\caption{Summary of all possible boundary conditions of self-adjointness for the bilateral-fibre extensions of $A_\alpha(k)$}\label{tab:extensions}
\end{center}
\end{table}

\begin{proof}[Proof of Theorem \ref{thm:bifibre-extensions}]
 Consider first the modes $k\in\mathbb{Z}\setminus\{0\}$. The classification formula \eqref{eq:Birman_formula} includes zero-, one-, and two-dimensional spaces $\mathcal{K}$.

 The choice $\mathcal{K}=\{0\}\oplus \{0\}$ yields the extension with domain
 \[
  \mathcal{D}(\overline{A_\alpha(k)})\dotplus A_{\alpha,\mathrm{F}}(k)^{-1}\ker A_\alpha(k)^*\;=\;\mathcal{D}(A_{\alpha,\mathrm{F}}(k))\,,
 \]
namely the Friedrichs extension. Formula \eqref{eq:thmAFoperator} of Theorem \ref{thm:fibre-thm}, applied on both sides $\mathbb{R}^+$ and $\mathbb{R}^-$, then implies $g_0^+=0=g_0^-$.

 The choice $\mathcal{K}=\{0\}\oplus \mathrm{span}\{\widetilde{\Phi}_{\alpha,k}\}$ yields the extensions in the domain of which a function $g=\varphi+A_{\alpha,\mathrm{F}}(k)^{-1}(Tv+w)+v$ is decoupled into a component $g^-$ in the domain of $A_{\alpha,\mathrm{F}}^-(k)$ (the Friedrichs extension of $A_{\alpha}^-(k)$) and a component $g^+$ in the domain of a self-adjoint extension of $A_\alpha^+(k)$ in $L^2(\mathbb{R}^+)$. This identifies a family $\{A_{\alpha,\mathrm{R}}^{[\gamma]}(k)\,|\,\gamma\in\mathbb{R}\}$ of extensions with
 \[
  A_{\alpha,\mathrm{R}}^{[\gamma]}(k)\;=\;A_{\alpha,\mathrm{F}}^-(k)\oplus A_{\alpha}^{+,[\gamma]}(k)\,,
 \]
 where $A_{\alpha}^{+,[\gamma]}(k)$ denotes here the generic extension of $A_\alpha^+(k)$, according to the classification of Theorem \ref{thm:fibre-thm}(iv), for which therefore $g_1^+=\gamma g_0^+$. The symmetric choice $\mathcal{K}=\mathrm{span}\{\widetilde{\Phi}_{\alpha,k}\}\oplus \{0\}$ is treated in a completely analogous way.

 The next one-dimensional choice is $\mathcal{K}=\mathrm{span}\{\widetilde{\Phi}_{\alpha,k}\oplus a \widetilde{\Phi}_{\alpha,k}\}$ for some $a\in\mathbb{C}$. The case $a=0$ can be excluded, as it yields type-$\mathrm{I_L}$ extensions already discussed above. Formula \eqref{eq:Birman_formula} is now to be specialised with
 \[
  v\in\mathcal{K}\,,\qquad w\in \mathcal{K}^\perp\cap 
  \big(\mathrm{span}\{\widetilde{\Phi}_{\alpha,k}\}\oplus\mathrm{span}\{\widetilde{\Phi}_{\alpha,k}\}\big)=\mathrm{span}\{\widetilde{\Phi}_{\alpha,k}\oplus (-\overline{a}^{-1}) \widetilde{\Phi}_{\alpha,k}\}\,.
 \]
 The generic self-adjoint operator $T$ on $\mathcal{K}$ is now the multiplication by some $\tau\in\mathbb{R}$.
 Then \eqref{eq:Birman_formula} reads
 \[
 \begin{split}
  g\;&=\;\varphi+A_{\alpha,\mathrm{F}}(k)^{-1}\left( \tau c_0 \begin{pmatrix} \widetilde{\Phi}_{\alpha,k} \\ a \widetilde{\Phi}_{\alpha,k} \end{pmatrix} + \widetilde{c}_0 \begin{pmatrix} \widetilde{\Phi}_{\alpha,k} \\ -\overline{a}^{-1}\,  \widetilde{\Phi}_{\alpha,k}\end{pmatrix} \right)+c_0 \begin{pmatrix} \widetilde{\Phi}_{\alpha,k} \\ a \widetilde{\Phi}_{\alpha,k} \end{pmatrix} \\
  &=\;\varphi+\begin{pmatrix} (\tau c_0+\widetilde{c}_0)\widetilde{\Psi}_{\alpha,k} \\ (\tau c_0 a-\widetilde{c}_0 \,\overline{a}^{-1})\widetilde{\Psi}_{\alpha,k}  \end{pmatrix}+c_0 \begin{pmatrix} \widetilde{\Phi}_{\alpha,k} \\ a \widetilde{\Phi}_{\alpha,k} \end{pmatrix}
 \end{split}
 \]
 for generic coefficients $c_0,\widetilde{c}_0\in\mathbb{C}$.
 From the expression above one finds that the limits \eqref{eq:bilimitsg0g1}, computed with the short-range asymptotics \eqref{eq:Asymtotics_0} and \eqref{eq:Psi_asymptotics} (and Lemma \ref{lem:BehaviourZeroClosure}), amount to
 \[
  \begin{split}
   g_0^-\;&=\;c_0\textstyle\sqrt{\frac{\pi(1+\alpha)}{2|k|}}\,, \\
   g_0^+\;&=\;c_0 \,a\textstyle\sqrt{\frac{\pi(1+\alpha)}{2|k|}}\,, \\
   g_1^-\;&=\;\textstyle(\tau c_0+\widetilde{c}_0)\sqrt{\frac{2|k|}{\pi(1+\alpha)^3}}\|\Phi_{\alpha,k}\|_{L^2(\mathbb{R}^+)}^2-c_0\sqrt{\frac{\pi|k|}{2(1+\alpha)}}\,, \\
   g_1^+\;&=\;\textstyle(\tau c_0 a-\widetilde{c}_0\,\overline{a}^{-1})\sqrt{\frac{2|k|}{\pi(1+\alpha)^3}}\|\Phi_{\alpha,k}\|_{L^2(\mathbb{R}^+)}^2-c_0 a \sqrt{\frac{\pi|k|}{2(1+\alpha)}}\,.
  \end{split}
 \]
 The first two equations above yield $g_0^+=ag_0^-$. The last two yield
 \[
  \begin{split}
  g_1^- +\overline{a}\,g_1^+\;&=\;\textstyle c_0(1+|a|^2)\Big(\tau\sqrt{\frac{2|k|}{\pi(1+\alpha)^3}}\|\Phi_{\alpha,k}\|_{L^2(\mathbb{R}^+)}^2-\sqrt{\frac{\pi|k|}{2(1+\alpha)}} \Big) \\
  &=\;g_0^-\textstyle(1+|a|^2) \frac{|k|}{1+\alpha} \Big( \frac{\,2 \|\Phi_{\alpha,k}\|_{L^2(\mathbb{R}^+)}^2 \,}{\pi (1+\alpha)} \, \tau -1 \Big)\,,
  \end{split} 
 \]
 having replaced $c_0=g_0^-\sqrt{\frac{2|k|}{\pi(1+\alpha)}}$. One can also write
 \[
  g_1^- +\overline{a}\,g_1^+\;=\;\gamma\, g_0^-
 \]
 after re-parametrising the extension parameter as
 \begin{equation}\label{eq:IkTheorem51}
  \gamma\;:=\;(1+|a|^2) \frac{|k|}{1+\alpha} \Big( \frac{\,2 \|\Phi_{\alpha,k}\|_{L^2(\mathbb{R}^+)}^2 \,}{\pi (1+\alpha)} \, \tau -1 \Big)\,\in\,\mathbb{R}\,.
 \end{equation}
 
 This completes the identification of the extensions $A_{\alpha,a}^{[\gamma]}(k)$.

 The remaining choice for $\mathcal{K}$ is $\mathcal{K}=\mathrm{span}\{\widetilde{\Phi}_{\alpha,k}\}\oplus \mathrm{span}\{\widetilde{\Phi}_{\alpha,k}\}$, namely the whole $\ker A_\alpha(k)^*$. In this case formula \eqref{eq:Birman_formula} only has $v$-vectors and no $w$-vectors, and the self-adjoint $T$ is represented by a generic $2\times 2$ Hermitian matrix
 \[
  T\;=\;
  \begin{pmatrix}
   \tau_1 & \tau_2 +\ii\tau_3 \\
   \tau_2 -\ii\tau_3 &\tau_4
  \end{pmatrix},\qquad\tau_1,\tau_2,\tau_3,\tau_4\in\mathbb{R}\,.
 \]
 Then \eqref{eq:Birman_formula} reads
 \[
  \begin{split}
   g\;&=\;\varphi+A_{\alpha,\mathrm{F}}(k)^{-1}T\begin{pmatrix} c_0^-\widetilde{\Phi}_{\alpha,k} \\ c_0^+\widetilde{\Phi}_{\alpha,k} \end{pmatrix}+\begin{pmatrix} c_0^-\widetilde{\Phi}_{\alpha,k} \\ c_0^+\widetilde{\Phi}_{\alpha,k} \end{pmatrix} \\
   &=\;\varphi+\begin{pmatrix} (\tau_1 c_0^-+(\tau_2+\ii\tau_3)c_0^+) \widetilde{\Psi}_{\alpha,k} \\ ((\tau_2-\ii\tau_3)c_0^-+\tau_4c_0^+) \widetilde{\Psi}_{\alpha,k}\end{pmatrix}+\begin{pmatrix} c_0^-\widetilde{\Phi}_{\alpha,k} \\ c_0^+\widetilde{\Phi}_{\alpha,k} \end{pmatrix}
  \end{split}
 \]
 for generic coefficients $c_0^\pm\in\mathbb{C}$.
 From the expression above one finds that the limits \eqref{eq:bilimitsg0g1}, computed with the short-range asymptotics \eqref{eq:Asymtotics_0} and \eqref{eq:Psi_asymptotics} (and Lemma \ref{lem:BehaviourZeroClosure}), amount to
 \[
  \begin{split}
   g_0^\pm\;&=\;c_0^\pm\textstyle\sqrt{\frac{\pi(1+\alpha)}{2|k|}}\,, \\
   g_1^-\;&=\;\textstyle c_0^-\Big(\tau_1\sqrt{\frac{2|k|}{\pi(1+\alpha)^3}}\|\Phi_{\alpha,k}\|_{L^2(\mathbb{R}^+)}^2-\sqrt{\frac{\pi|k|}{2(1+\alpha)}}\Big)+c_0^+(\tau_2+\ii\tau_3)\sqrt{\frac{2|k|}{\pi(1+\alpha)^3}}\|\Phi_{\alpha,k}\|_{L^2(\mathbb{R}^+)}^2\,, \\
   g_1^+\;&=\;\textstyle c_0^-(\tau_2-\ii\tau_3)\sqrt{\frac{2|k|}{\pi(1+\alpha)^3}}\|\Phi_{\alpha,k}\|_{L^2(\mathbb{R}^+)}^2+c_0^+\Big(\tau_4\sqrt{\frac{2|k|}{\pi(1+\alpha)^3}}\|\Phi_{\alpha,k}\|_{L^2(\mathbb{R}^+)}^2+\sqrt{\frac{\pi|k|}{2(1+\alpha)}}\Big).
  \end{split}
 \]
 Replacing $c_0^\pm=g_0^\pm\sqrt{\frac{2|k|}{\pi(1+\alpha)}}$ in the last two equations above, and re-defining the extension parameters as
 \begin{equation}\label{eq:IIkTheorem51}
  \begin{split}
   \gamma_1\;&:=\;  \frac{|k|}{1+\alpha} \Big( \frac{\,2 \|\Phi_{\alpha,k}\|_{L^2(\mathbb{R}^+)}^2 \,}{\pi (1+\alpha)} \, \tau_1 -1 \Big)\,, \\
   \gamma_2+\ii\gamma_3\;&:=\; (\tau_2+\ii\tau_3)\frac{2|k|}{\pi(1+\alpha)^2}\|\Phi_{\alpha,k}\|_{L^2(\mathbb{R}^+)}^2\,, \\
   \gamma_4\;&:=\;  \frac{|k|}{1+\alpha} \Big( \frac{\,2 \|\Phi_{\alpha,k}\|_{L^2(\mathbb{R}^+)}^2 \,}{\pi (1+\alpha)} \, \tau_4 -1 \Big),,  \end{split}
 \end{equation}
 yields precisely the boundary condition that characterises the extension $A_{\alpha}^{[\Gamma]}(k)$ with $\Gamma=(\gamma_1,\gamma_2,\gamma_3,\gamma_4)$.

 Last, the above reasonings can be now repeated when $k=0$, based on the classification formula \eqref{eq:Birman_formula_zero_mode}. The only modifications needed are the replacement of $A_{\alpha,\mathrm{F}}(k)^{-1}$ with $(A_{\alpha,\mathrm{F}}(0)+\mathbbm{1})^{-1}$, and the use, instead of the short-range asymptotics given by \eqref{eq:Asymtotics_0}, \eqref{eq:Psi_asymptotics}, and Lemma \ref{lem:BehaviourZeroClosure} valid for $k\neq 0$, of the short-range asymptotics given by \eqref{eq:AsymPhi00}, \eqref{eq:Psi_Asymptotics_mode_zero}, and Lemma \ref{lem:BehaviourZeroClosure_zero_mode} valid for $k=0$.
 
 The net result concerning the extensions of type $\mathrm{II}_a$, namely the extensions $A^{[\gamma]}_{\alpha,a}(0)$, is that \eqref{eq:IkTheorem51} is replaced by 
 \begin{equation}\label{eq:I0Theorem51}
  	\gamma \;:=\;  {\frac{(1+|a|^2)\Gamma\left( \frac{1-\alpha}{2}\right)}{2^\alpha (1+\alpha)\Gamma\left( \frac{1+\alpha}{2}\right)} \Big( \frac{(1+\alpha)\Vert \Phi_{\alpha,0} \Vert_{L^2(\mathbb{R}^+)}^2}{\Gamma\left( \frac{3+\alpha}{2}\right)\Gamma\left( \frac{1-\alpha}{2}\right)}  \, \tau - 1 \Big)}\;\in\;\mathbb{R}\,.
 \end{equation}

 Analogously, concerning the extensions of type $\mathrm{III}$, namely the extensions $A^{[\Gamma]}_{\alpha}(0)$, \eqref{eq:IIkTheorem51} is now replaced by 
  \begin{equation}\label{eq:II0Theorem51}
  \begin{split}
   \gamma_1\;&:=\;  \frac{\Gamma(\frac{1-\alpha}{2})}{2^\alpha (1+\alpha)\Gamma(\frac{1+\alpha}{2})} \Big( \frac{\, (1+\alpha) \|\Phi_{\alpha,0}\|_{L^2(\mathbb{R}^+)}^2 \,}{\Gamma(\frac{3+\alpha}{2}) \Gamma (\frac{1-\alpha}{2}) } \, \tau_1 -1 \Big)\,, \\
   \gamma_2+\ii\gamma_3\;&:=\; (\tau_2+\ii\tau_3)\frac{\| \Phi_{\alpha,0} \|_{L^2(\mathbb{R}^+)}^2}{2^\alpha \Gamma(\frac{3+\alpha}2) \Gamma(\frac{1+\alpha}2)}\,, \\
   \gamma_4\;&:=\;  \frac{\Gamma(\frac{1-\alpha}{2})}{2^\alpha (1+\alpha)\Gamma(\frac{1+\alpha}{2})} \Big( \frac{\, (1+\alpha) \|\Phi_{\alpha,0}\|_{L^2(\mathbb{R}^+)}^2 \,}{\Gamma(\frac{3+\alpha}{2}) \Gamma (\frac{1-\alpha}{2}) } \, \tau_4 -1 \Big)\,.  \end{split}
 \end{equation}
 
 The proof is now completed.
\end{proof}


Whereas Theorem \ref{thm:bifibre-extensions} expresses the various conditions of self-adjointness in terms of the representation \eqref{eq:gdecomp} and \eqref{eq:gbilateralasympt}-\eqref{eq:bilimitsg0g1} of a generic $g\in\mathcal{D}(A_\alpha(k)^*)$, that is, in terms of the short-range behaviour of $g$, for the forthcoming analysis it is convenient to re-formulate the above classification in two further equivalent forms.

The first one refers to the representation \eqref{eq:gdecomp}, \eqref{eq:bilateralDAxistar}, and \eqref{eq:bilateralDAxistar_zero_mode}
of $g\in\mathcal{D}(A_\alpha(k)^*)$,  that is,
 \begin{equation}\label{eq:gkrepresentationc0c1}
 g\;=\;\begin{pmatrix}\widetilde{\varphi}^- \\ \widetilde{\varphi}^+\end{pmatrix}+\begin{pmatrix} c_{1}^-\widetilde{\Psi}_{\alpha,k} \\ c_{1}^+\widetilde{\Psi}_{\alpha,k} \end{pmatrix}+\begin{pmatrix} c_{0}^-\widetilde{\Phi}_{\alpha,k} \\ c_{0}^+\widetilde{\Phi}_{\alpha,k} \end{pmatrix}
\end{equation}
with $\widetilde{\varphi}^\pm\in\mathcal{D}(\overline{A^\pm_\alpha(k)})$ and $c_0^\pm,c_1^\pm\in\mathbb{C}$. Then the proof of Theorem \ref{thm:bifibre-extensions} demonstrates also the following.

\begin{theorem}\label{thm:bifibre-extensionsc0c1}
 Let $\alpha\in[0,1)$ and $k\in\mathbb{Z}$. The family of self-adjoint extensions of $A_\alpha(k)$ is formed by the following sub-families.
 \begin{itemize}
  \item \underline{Friedrichs extension}.  It is the operator \eqref{eq:biFriedrichs}. Its domain consists of those functions in $\mathcal{D}(A_\alpha(k)^*)$ whose representation \eqref{eq:gkrepresentationc0c1} has $c_0^\pm=0$.
    \item \underline{Family $\mathrm{I_R}$}.   It is the family $\{A_{\alpha,\mathrm{R}}^{[\gamma]}(k)\,|\,\gamma\in\mathbb{R}\}$ defined, with respect to the  representation \eqref{eq:gkrepresentationc0c1}, by
  \[
   \mathcal{D}(A_{\alpha,\mathrm{R}}^{[\gamma]}(k))\;=\;\{g\in\mathcal{D}(A_\alpha(k)^*)\,|\,c_0^-=0\,,\;c_1^+=\beta c_0^+\}\,,
  \]
  where $\beta$ and $\gamma$ are related by \eqref{eq:g1gammag0} for $k\neq 0$ and \eqref{eq:g1gammag0_zero_mode} for $k=0$.
   \item \underline{Family $\mathrm{I_L}$}.   It is the family $\{A_{\alpha,\mathrm{L}}^{[\gamma]}(k)\,|\,\gamma\in\mathbb{R}\}$ defined, with respect to the  representation \eqref{eq:gkrepresentationc0c1}, by
  \[
   \mathcal{D}(A_{\alpha,\mathrm{L}}^{[\gamma]}(k))\;=\;\{g\in\mathcal{D}(A_\alpha(k)^*)\,|\,c_1^-=\beta c_0^-\,,\;c_0^+=0\}\,,
  \]
  where $\beta$ and $\gamma$ are related by \eqref{eq:g1gammag0} for $k\neq 0$ and \eqref{eq:g1gammag0_zero_mode} for $k=0$.
   \item \underline{Family $\mathrm{II}_a$} with $a\in\mathbb{C}$.   It is the family $\{A_{\alpha,a}^{[\gamma]}(k)\,|\,\gamma\in\mathbb{R}\}$ defined by
%
   \[
    \mathcal{D}(A_{\alpha,a}^{[\gamma]}(k))\;=\;\left\{\!
    \begin{array}{c}
     g\in\mathcal{D}(A_\alpha(k)^*)\textrm{ with \eqref{eq:gkrepresentationc0c1} of the form } \\
      g=\begin{pmatrix}\widetilde{\varphi}^- \\ \widetilde{\varphi}^+\end{pmatrix}+\begin{pmatrix} (\tau c_0+\widetilde{c}_0)\widetilde{\Psi}_{\alpha,k} \\ (\tau c_0 a-\widetilde{c}_0 \,\overline{a}^{-1})\widetilde{\Psi}_{\alpha,k}  \end{pmatrix}+c_0 \begin{pmatrix} \widetilde{\Phi}_{\alpha,k} \\ a \widetilde{\Phi}_{\alpha,k} \end{pmatrix}
    \end{array}
    \!\right\},
   \]
  where $\tau$ and $\gamma$ are related by \eqref{eq:IkTheorem51} if $k\neq 0$, and by \eqref{eq:I0Theorem51} if $k=0$.
   \item \underline{Family $\mathrm{III}$}.   It is the family $\{A_{\alpha}^{[\Gamma]}(k)\,|\,\Gamma\equiv(\gamma_1,\gamma_2,\gamma_3,\gamma_4)\in\mathbb{R}^4\}$ defined by
   \[
     \mathcal{D}(A_{\alpha}^{[\Gamma]}(k))\;=\;\left\{\!
     \begin{array}{c}
       g\in\mathcal{D}(A_\alpha(k)^*) \textrm{ satisfying \eqref{eq:gkrepresentationc0c1} with } \\
       \begin{pmatrix}
		c_{1}^- \\ c_1^+
	\end{pmatrix} = \begin{pmatrix}
		\tau_1 & \tau_2 + \ii \tau_3 \\
		\tau_2 - \ii \tau_3 & \tau_4
	\end{pmatrix} \begin{pmatrix}
		c_0^- \\
		c_0^+
	\end{pmatrix}
     \end{array}
     \!\right\},
   \]
   where $(\tau_1,\tau_2,\tau_3,\tau_4)$ and $(\gamma_1,\gamma_2,\gamma_3,\gamma_4)$ are related by \eqref{eq:IIkTheorem51} if $k\neq 0$ and \eqref{eq:II0Theorem51} if $k=0$.   
 \end{itemize}
\end{theorem}

The second alternative for the self-adjointness conditions is in fact a very close re-phrasing of Theorem \ref{thm:bifibre-extensions}, with the same short-range parameters $g_0^\pm$ and $g_1^\pm$ and the same classification parameters $\gamma$ or $\Gamma$, except that it is referred to the following representation of $g$, which is valid identically for any $x\in\mathbb{R}\setminus\{0\}$, and not just as $|x|\to 0$.

To this aim, and also for later convenience, a cut-off function $P$ is introduced, belonging to $C^\infty_c(\mathbb{R})$ and such that
\begin{equation}\label{eq:Pcutoff}
 P(x)\;=\;
 \begin{cases}
  \;1 &\textrm{ if }\;|x|<1\,, \\
  \;0 &\textrm{ if }\;|x|>2\,.
 \end{cases}
\end{equation}
In fact, in the following Theorem it is enough that $P$ be smooth, compactly supported, and with $P(0)=1$; the general assumption \eqref{eq:Pcutoff} is kept nevertheless for later use.

\begin{theorem}\label{prop:g_with_Pweight}
 Let $\alpha\in[0,1)$ and let $k\in\mathbb{Z}$. Then for any $g\in\mathcal{D}(A_\alpha(k)^*)$ there exist a unique $\varphi\in\mathcal{D}(\overline{A_{\alpha}(k)})$ (in particular, $\varphi^\pm\in H^2_0(\mathbb{R}^\pm)$) and uniquely determined coefficients 
 $g_0^\pm,g_1^\pm\in\mathbb{C}$ such that
 \begin{equation}\label{eq:gwithPweight}
  g(x)\;=\;\varphi(x)+g_0\,|x|^{-\frac{\alpha}{2}}\,P(x)+g_1\,|x|^{1+\frac{\alpha}{2}}\,P(x)\qquad\forall x\in\mathbb{R}\setminus\{0\}
 \end{equation}
 in the usual notation
 \begin{equation*}
  \varphi(x)\equiv\begin{pmatrix} \varphi^-(x) \\ \varphi^+(x) \end{pmatrix},\quad g_0\equiv\begin{pmatrix} g_0^- \\ g_0^+\end{pmatrix}, \quad g_1\equiv\begin{pmatrix} g_1^- \\ g_1^+\end{pmatrix}.
 \end{equation*}
 Here $g_0^\pm$ and $g_1^\pm$ are precisely the same as in the asymptotics \eqref{eq:gbilateralasympt}-\eqref{eq:bilimitsg0g1}. Therefore, the same classification of Theorem \ref{thm:bifibre-extensions} in terms of $g_0^\pm$ and $g_1^\pm$ is applicable. 
\end{theorem}

\begin{proof}
 For $k\in\mathbb{Z}\setminus\{0\}$ decompose $g\in\mathcal{D}(A_\alpha(k)^*)$ as $g^\pm=\widetilde{\varphi}^\pm+c_1^\pm\widetilde{\Psi}_{\alpha,k}+c_0^\pm\widetilde{\Phi}_{\alpha,k}$ with respect to the decomposition \eqref{eq:gkrepresentationc0c1}. Only the component $g^+$ shall be discussed, dropping temporarily the `$+$' superscript: the discussion for $g^-$ is completely analogous. Thus, $g=\widetilde{\varphi}+c_1\widetilde{\Psi}_{\alpha,k}+c_0\widetilde{\Phi}_{\alpha,k}$ for all $x>0$ and for uniquely determined $\widetilde{\varphi}\in\mathcal{D}(\overline{A_{\alpha}(k)})$ and $c_0,c_1\in\mathbb{C}$. Define
 \[
  \begin{split}
   L_{0,k}(x)\,&:=\,\Big({\textstyle\sqrt{\frac{\pi(1+\alpha)}{2|k|}}-\sqrt{\frac{\pi|k|}{2(1+\alpha)}}\,|x|^{1+\alpha}} \Big)\,P(x)\,, \\
   L_{1,k}(x)\,&:=\,{\textstyle\sqrt{\frac{2|k|}{\pi(1+\alpha)^3}}}\,\|\Phi_{\alpha,k}\|_{L^2(\mathbb{R}^+)}^2 \,P(x)
  \end{split}
 \]
 and re-write
 \[
 \begin{split}
  g\;&=\;\widetilde{\varphi}+c_1(\widetilde{\Psi}_{\alpha,k}-|x|^{1+\frac{\alpha}{2}}L_{1,k})+c_0(\widetilde{\Phi}_{\alpha,k}-|x|^{-\frac{\alpha}{2}}L_{0,k})+c_1\,|x|^{1+\frac{\alpha}{2}}L_{1,k}+c_0\,|x|^{-\frac{\alpha}{2}}L_{0,k} \\
  &=\;\varphi+\Big( c_1{\textstyle\sqrt{\frac{2|k|}{\pi(1+\alpha)^3}}}\,\|\Phi_{\alpha,k}\|_{L^2(\mathbb{R}^+)}^2-c_0{\textstyle\sqrt{\frac{\pi|k|}{2(1+\alpha)}}}\Big)\,|x|^{1+\frac{\alpha}{2}}P+c_0{\textstyle\sqrt{\frac{\pi(1+\alpha)}{2|k|}}}\,|x|^{-\frac{\alpha}{2}} P\,,
 \end{split}
 \]
 having set 
 \[
  \varphi\;:=\;\widetilde{\varphi}+c_1(\widetilde{\Psi}_{\alpha,k}-|x|^{1+\frac{\alpha}{2}}L_{1,k})+c_0(\widetilde{\Phi}_{\alpha,k}-|x|^{-\frac{\alpha}{2}}L_{0,k})\,.
 \]
 Because of the relation \eqref{eq:limitsg0g1} between $c_0,c_1$ and $g_0,g_1$, one also has
 \[
  g\;=\;\varphi+g_0\,|x|^{-\frac{\alpha}{2}}P+g_1\,|x|^{1+\frac{\alpha}{2}}P\,.
 \]
 One can now argue that $\varphi\in\mathcal{D}(\overline{A_{\alpha}(k)})$. To this aim, first one observes that both $|x|^{-\frac{\alpha}{2}}L_{0,k}$ and $|x|^{1+\frac{\alpha}{2}}L_{1,k}$ belong to $\mathcal{D}(A_\alpha(k)^*)$. The latter statement, owing to \eqref{eq:Afstar} and \eqref{eq:Saxi}, is proved by checking the square-integrability of $S_{\alpha,k}(|x|^{-\frac{\alpha}{2}}L_{0,k})$ and of $S_{\alpha,k}(|x|^{1+\frac{\alpha}{2}}L_{1,k})$. Since $P$ localises $L_{0,k}$ and $L_{1,k}$ around $x=0$, square-integrability must only be checked \emph{locally}. It is then routine to see that
 \[
  \begin{array}{lllll}
   -(|x|^{-\frac{\alpha}{2}}L_{0,k})''\,   +k^2 |x|^{2\alpha}(|x|^{-\frac{\alpha}{2}}L_{0,k})\,   +C_\alpha x^{-2}(|x|^{-\frac{\alpha}{2}}L_{0,k})\,, \\
   -(|x|^{1+\frac{\alpha}{2}}L_{1,k})''\,   +k^2 |x|^{2\alpha}(|x|^{1+\frac{\alpha}{2}}L_{1,k})\,   +C_\alpha x^{-2}(|x|^{1+\frac{\alpha}{2}}L_{1,k})
  \end{array}
 \]
 are both square-integrable around $x=0$. As a consequence, both $(\widetilde{\Psi}_{\alpha,k}-|x|^{1+\frac{\alpha}{2}}L_{1,k})$ and $(\widetilde{\Phi}_{\alpha,k}-|x|^{-\frac{\alpha}{2}}L_{0,k})$ are elements of $\mathcal{D}(A_\alpha(k)^*)$. Therefore, owing to the representation \eqref{eq:Dadjoint}-\eqref{eq:limitsg0g1}, in order to check that such two functions also belong to $\mathcal{D}(\overline{A_{\alpha}(k)})$, it suffices to verify the limits 
 \[
 \begin{split}
  \lim_{x\to 0}|x|^{\frac{\alpha}{2}}(\widetilde{\Psi}_{\alpha,k}-|x|^{1+\frac{\alpha}{2}}L_{1,k})\;=\;\lim_{x\to 0}|x|^{\frac{\alpha}{2}}(\widetilde{\Phi}_{\alpha,k}-|x|^{-\frac{\alpha}{2}}L_{0,k})\;&=\;0\,, \\
  \lim_{x\to 0}|x|^{-(1+\frac{\alpha}{2})}(\widetilde{\Psi}_{\alpha,k}-|x|^{1+\frac{\alpha}{2}}L_{1,k})\;=\;\lim_{x\to 0}|x|^{-(1+\frac{\alpha}{2})}(\widetilde{\Phi}_{\alpha,k}-|x|^{-\frac{\alpha}{2}}L_{0,k})\;&=\;0\,.
 \end{split}
 \]
 This is straightforward to check, thanks to the short-distance asymptotics that were chosen for $L_{0,k}$ and $L_{1,k}$ precisely so as to suitably match with the short-distance asymptotics  \eqref{eq:Asymtotics_0} of $\widetilde{\Phi}_{\alpha,k}$ and \eqref{eq:Psi_asymptotics} of $\widetilde{\Psi}_{\alpha,k}$. This finally shows that $\varphi\in\mathcal{D}(\overline{A_{\alpha}(k)})$ and establishes \eqref{eq:gwithPweight}. Of course, if conversely a function $g$ of the form \eqref{eq:gwithPweight} is given with $\varphi\in\mathcal{D}(\overline{A_{\alpha}(k)})$, unfolding the above arguments one sees that $g\in\mathcal{D}(A_\alpha(k)^*)$.

 If instead $k=0$, the same argument can be then repeated decomposing now $g\in\mathcal{D}(A_\alpha(0)^*)$ as $g^\pm=\widetilde{\varphi}^\pm+c_1^\pm\widetilde{\Psi}_{\alpha,0}+c_0^\pm\widetilde{\Phi}_{\alpha,0}$ according to the decomposition \eqref{eq:bilateralDAxistar_zero_mode}, and using now the short-range asymptotics \eqref{eq:AsymPhi00}, \eqref{eq:Psi_Asymptotics_mode_zero}, and Lemma \ref{lem:BehaviourZeroClosure_zero_mode} valid for $k=0$. The straightforward details are omitted. 
\end{proof}

\section{General extensions of $\mathscr{H}_\alpha$}\label{sec:genextscrHa}

 The fibre analysis of Section \ref{sec:bilateralfibreext} allows now to study of the self-adjoint extensions, in the Hilbert space \eqref{eq:Hxispace}, namely
\begin{equation}\label{eq:Hspaceonceagain}
 \cH\;\cong\;\bigoplus_{k\in\mathbb{Z}}\mathfrak{h}_k\;\cong\;\ell^2(\mathbb{Z},\mathfrak{h})\,,\qquad \mathfrak{h}_k\;\equiv\;\mathfrak{h}\;\cong\;L^2(\mathbb{R}^-)\oplus L^2(\mathbb{R}^+)\,,
\end{equation}
of the minimal operator $\mathscr{H}_\alpha$ introduced in \eqref{eq:actiondomainHalpha} and \eqref{eq:V-two-sided-7}.

It was already observed that it is the peculiar orthogonal sum structure of $\mathscr{H}_\alpha^*=\bigoplus_{k\in\mathbb{Z}}A_\alpha(k)^*$ (Lemma \ref{lem:Halphaadj-decomposable}) that makes the self-adjoint extension problem for $\mathscr{H}_\alpha$ (i.e., the self-adjoint restriction problem for $\mathscr{H}_\alpha^*$) efficiently solvable. This is done, to begin with, in Theorem \ref{thm:Halphageneralext}, after an amount of simple preliminaries (Lemmas \ref{lem:HalphaFrie-decomposable}-\ref{lem:gkkrepr}). Then, physically meaningful extensions are identified with the subsequent analysis of Section \ref{sec:uniformlyfirbredext}.

 Since $\mathscr{H}_\alpha\subset\bigoplus_{k\in\mathbb{Z}} A_\alpha(k)$ (see \eqref{eq:Halphanotsum}), and each $A_\alpha(k)$ is non-negative (see \eqref{eq:Axibottom}), $\mathscr{H}_\alpha$ itself is non-negative, and therefore has Friedrichs extension $\mathscr{H}_{\alpha,\mathrm{F}}$ (Theorem \ref{thm:Friedrichs-ext}).

 \begin{lemma}\label{lem:HalphaFrie-decomposable}
  Let $\alpha\in[0,1)$. One has
  \begin{equation}\label{eq:HalphaFriedrichs_unif-fibred}
   \mathscr{H}_{\alpha,\mathrm{F}}\;=\;\bigoplus_{k\in\mathbb{Z}}\,A_{\alpha,\mathrm{F}}(k)\,.
  \end{equation}
 \end{lemma}

 \begin{proof}
  One applies Lemma \ref{lem:sumofFriedrichs} to $\overline{\mathscr{H}_\alpha}=\bigoplus_{k\in\mathbb{Z}} \,\overline{A_\alpha(k)}$.  
 \end{proof}

There is an obvious peculiarity of the mode $k=0$ that is to be dealt with separately. Indeed, \eqref{eq:Axibottom} and \eqref{eq:Axibottom-zero} (and \eqref{eq:II-mSFequalmS}) show that the bottom of the spectrum of $A_{\alpha,\mathrm{F}}(k)$ is strictly positive when $k\in\mathbb{Z}\setminus\{0\}$, explicitly
\begin{equation}\label{eq:AFbottom}
 A_{\alpha,\mathrm{F}}(k)\;\geqslant\;(1+\alpha)\big(\textstyle{\frac{2+\alpha}{4}}\big)^{\frac{\alpha}{1+\alpha}}\,\mathbbm{1}_k\,,\qquad k\in\mathbb{Z}\setminus\{0\}\,,\quad\alpha\geqslant 0\,,
\end{equation}
 and instead amounts precisely to zero when $k=0$. It is then convenient to consider a positive shift of $\mathscr{H}_\alpha$ in the zero mode only, the new operator $\mathscr{H}_\alpha+\mathbbm{1}_0$ where $\mathbbm{1}_0$ acts, with respect to the decomposition \eqref{eq:Hspaceonceagain}, as the identity in the $0$-th fibre and as the zero operator in all other fibres. Clearly, $\mathscr{H}_\alpha$ and $\mathscr{H}_\alpha+\mathbbm{1}_0$ have precisely the same domain, and the same holds for the respective adjoints and the respective Friedrichs extensions.

\begin{lemma}\label{lem:psikinspaces}
 Let $\alpha\in[0,1)$. Let $(\psi_k)_{k\in\mathbb{Z}}\in\cH\cong\ell^2(\mathbb{Z},\mathfrak{h})$. Then:
 \begin{enumerate}[(i)]
  \item $(\psi_k)_{k\in\mathbb{Z}}\in\ker( \mathscr{H}_\alpha+\mathbbm{1}_0)^*$ if and only if
  \begin{equation}
   \psi_k\;=\;c_{0,k}^-\widetilde{\Phi}_{\alpha,k}\oplus c_{0,k}^+\widetilde{\Phi}_{\alpha,k}
   \qquad\quad\forall k\in\mathbb{Z}
  \end{equation}
  for coefficients $c_{0,k}^\pm\in\mathbb{C}$ such that
  \begin{equation}
   \sum_{k\in \mathbb{Z}\setminus\{0\}}|k|^{-\frac{2}{1+\alpha}}|c_{0,k}^\pm|^2\:<\:+\infty\,.
  \end{equation}
  Thus, there is a natural Hilbert space isomorphism 
 \begin{equation}
  \ker( \mathscr{H}_\alpha+\mathbbm{1}_0)^*\;\cong\;\ell^2\big(\mathbb{Z},\mathbb{C}^2,\|\Phi_{\alpha,k}\|^2_{L^2}\big)\;=\;\ell^2(\mathbb{Z},\mathbb{C}^2,\mu_k)
 \end{equation}
 (the latter `$=$' sign being meant set-theoretically) with
 \begin{equation}\label{eq:DefMeasureMuK}
  \mu_k\,:=\,
  \begin{cases}
   |k|^{-\frac{2}{1+\alpha}}\,,
   & k\neq 0\,, \\
   1\,, & k=0\,,
  \end{cases}
 \end{equation}
 where $\ell^2(\mathbb{Z},\mathbb{C}^2,\mu_k)$ is the Hilbert space of sequences $\Big(\begin{pmatrix} c_k^- \\ c_k^+\end{pmatrix}\Big)_{k\in\mathbb{Z}}$ with obvious (component-wise) vector space structure and with scalar product
 \begin{equation}
  \left\langle \Big(\begin{pmatrix} c_k^- \\ c_k^+\end{pmatrix}\Big)_{k\in\mathbb{Z}},\Big(\begin{pmatrix} d_k^- \\ d_k^+\end{pmatrix}\Big)_{k\in\mathbb{Z}}\right\rangle_{\!\ell^2(\mathbb{Z},\mathbb{C}^2,\mu_k)}=\:\;\sum_{k\in\mathbb{Z}}\mu_k\big(\,\overline{c_k^-}\,d_k^-+\overline{c_k^+}\,d_k^+\big)\,.
 \end{equation}
  \item $(\psi_k)_{k\in\mathbb{Z}}\in(\mathscr{H}_{\alpha,\mathrm{F}}+\mathbbm{1}_0)^{-1}\ker( \mathscr{H}_\alpha+\mathbbm{1}_0)^*$ if and only if
  \begin{equation}
   \psi_k\;=\;c_{1,k}^-\widetilde{\Psi}_{\alpha,k}\oplus c_{1,k}^+\widetilde{\Psi}_{\alpha,k}
   \qquad\quad\forall k\in\mathbb{Z}
  \end{equation}
  for coefficients $c_{1,k}^\pm\in\mathbb{C}$ such that
  \begin{equation}\label{eq:Regularityg1}
   \sum_{k\in \mathbb{Z}\setminus\{0\}}|k|^{-\frac{2}{1+\alpha}}|c_{1,k}^\pm|^2\:<\:+\infty\,.
  \end{equation}
 \end{enumerate}
\end{lemma}

\begin{proof}
 Part (i) follows from $\ker \mathscr{H}_\alpha^*=\bigoplus_{k\in\mathbb{Z}}\ker A(k)^*$ (Lemma \ref{lem:Halphaadj-decomposable}, formula \eqref{eq:Halphaadj-sumkernel}), from $\ker (A(k)^*+\delta_{k,0}\mathbbm{1}_0)=\mathrm{span}\{\widetilde{\Phi}_{\alpha,k}\}\oplus\mathrm{span}\{\widetilde{\Phi}_{\alpha,k}\}$ (Lemmas \ref{lem:kerAxistar} and \ref{lem:kerAxistarzero}, and formula \eqref{eq:bilateralkernel}), and from $\|\Phi_{\alpha,k}\|_{L^2(\mathbb{R}^+)}^2\sim |k|^{-\frac{2}{1+\alpha}}$ for $k\neq 0$ (formula \eqref{eq:Phinorm}). Part (ii) follows from the identity
 \[
  \begin{split}
  (\mathscr{H}_{\alpha,\mathrm{F}}+\mathbbm{1}_0)^{-1}\ker( \mathscr{H}_\alpha+\mathbbm{1}_0)^*\;&=\;\bigoplus_{k\in\mathbb{Z}\setminus\{0\}}\!(A_{\alpha,\mathrm{F}}(k))^{-1}\ker A_\alpha(k)^*  \\
  &\qquad\qquad \oplus (A_\alpha(0)+\mathbbm{1})^{-1}\ker(A_\alpha(0)^*+\mathbbm{1})\,,
  \end{split}
  \]
 which is a consequence of Lemma \ref{lem:Halphaadj-decomposable} (eq.~\eqref{eq:Halphaadj-sumkernel}) and Lemma \ref{lem:HalphaFrie-decomposable}, from the identity
 \[
  (A_{\alpha,\mathrm{F}}(k)+\delta_{k,0}\mathbbm{1})^{-1}\ker (A_\alpha(k)^*+\delta_{k,0}\mathbbm{1})\;=\;\mathrm{span}\{\widetilde{\Psi}_{\alpha,k}\}\oplus\mathrm{span}\{\widetilde{\Psi}_{\alpha,k}\}\,,
 \]
 which is a consequence of Lemmas \ref{lem:kerAxistar} and \ref{lem:kerAxistarzero}, and of Propositions  \ref{eq:V-RGisSFinv} and \ref{eq:RGisSFinv_zero_mode}, from the consequent identity
 \[
  \sum_{k\in\mathbb{Z}\setminus\{0\}}\left\|A_\alpha(k)^*\!\begin{pmatrix} c_{1,k}^-\widetilde{\Psi}_{\alpha,k} \\ c_{1,k}^+\widetilde{\Psi}_{\alpha,k}\end{pmatrix}\right\|_{\mathfrak{h}}^2\;=\;
  \sum_{k\in\mathbb{Z}\setminus\{0\}}\left\|\begin{pmatrix} c_{1,k}^-\widetilde{\Phi}_{\alpha,k} \\ c_{1,k}^+\widetilde{\Phi}_{\alpha,k}\end{pmatrix}\right\|_{\mathfrak{h}}^2\,,
 \]
 and again from the normalisation $\|\Phi_{\alpha,k}\|_{L^2(\mathbb{R}^+)}^2\sim |k|^{-\frac{2}{1+\alpha}}$. 
\end{proof}

 There is a natural follow-up to Lemma \ref{lem:psikinspaces}, concerning the fibre-wise structure of the domain of $\mathscr{H}_\alpha^*$. 
 Recall, to this aim, the general `canonical' representation of $\mathcal{D}(\mathscr{H}_\alpha^*)$ (Proposition \ref{prop:II-KVB-decomp-of-Sstar}):
 \begin{equation}\label{eq:repreDHstar}
  \begin{split}
  \mathcal{D}&(\mathscr{H}_\alpha^*)\;=\;\mathcal{D}((\mathscr{H}_\alpha+\mathbbm{1}_0)^*) \\
  &=\;\mathcal{D}(\overline{\mathscr{H}_\alpha+\mathbbm{1}_0})\dotplus(\mathscr{H}_{\alpha,\mathrm{F}}+\mathbbm{1}_0)^{-1}\ker( \mathscr{H}_\alpha+\mathbbm{1}_0)^*\dotplus \ker( \mathscr{H}_\alpha+\mathbbm{1}_0)^* \\
  &=\;\mathcal{D}(\overline{\mathscr{H}_\alpha})\dotplus(\mathscr{H}_{\alpha,\mathrm{F}}+\mathbbm{1}_0)^{-1}\ker( \mathscr{H}_\alpha^*+\mathbbm{1}_0)\dotplus\ker( \mathscr{H}_\alpha^*+\mathbbm{1}_0)\,.
  \end{split}
 \end{equation}

\begin{lemma}\label{lem:gkkrepr}
 Let $\alpha\in[0,1)$. Let $(g_k)_{k\in\mathbb{Z}}\in\cH\cong\ell^2(\mathbb{Z},\mathfrak{h})$. Then $(g_k)_{k\in\mathbb{Z}}\in\mathcal{D}(\mathscr{H}_\alpha^*)$ if and only if
 \begin{equation}\label{eq:gkrepresentation}
 g_k\;=\;\begin{pmatrix}\widetilde{\varphi}_k^- \\ \widetilde{\varphi}_k^+\end{pmatrix}+\begin{pmatrix} c_{1,k}^-\widetilde{\Psi}_{\alpha,k} \\ c_{1,k}^+\widetilde{\Psi}_{\alpha,k} \end{pmatrix}+\begin{pmatrix} c_{0,k}^-\widetilde{\Phi}_{\alpha,k} \\ c_{0,k}^+\widetilde{\Phi}_{\alpha,k} \end{pmatrix}\qquad\quad\forall k\in\mathbb{Z}
\end{equation}
with
\begin{equation}\label{eq:pileupcond1}
  (\widetilde{\varphi}_k)_{k\in\mathbb{Z}}\:\in\:\mathcal{D}(\overline{\mathscr{H}_\alpha})\,,\qquad  \widetilde{\varphi}_k\,\equiv\,\begin{pmatrix}\widetilde{\varphi}_k^- \\ \widetilde{\varphi}_k^+\end{pmatrix}\\
\end{equation}
and
\begin{eqnarray}\label{eq:pileupcond2}
 & &  \sum_{k\in \mathbb{Z}\setminus\{0\}}|k|^{-\frac{2}{1+\alpha}}|c_{0,k}^\pm|^2\:<\:+\infty\,, \\
 & &  \sum_{k\in \mathbb{Z}\setminus\{0\}}|k|^{-\frac{2}{1+\alpha}}|c_{1,k}^\pm|^2\:<\:+\infty\,. \label{eq:pileupcond3}
\end{eqnarray} 
\end{lemma}

\begin{proof} From the representation \eqref{eq:repreDHstar} one deduces at once that in order for $(g_k)_{k\in\mathbb{Z}}$ to belong to  $\mathcal{D}(\mathscr{H}_\alpha^*)$ it is necessary and sufficient that 
 \[
  (g_k)_{k\in\mathbb{Z}}\;=\;(\widetilde{\varphi}_k)_{k\in\mathbb{Z}}+(\psi_k)_{k\in\mathbb{Z}}+(\xi_k)_{k\in\mathbb{Z}}
 \]
 for some $(\widetilde{\varphi}_k)_{k\in\mathbb{Z}}\in\mathcal{D}(\overline{\mathscr{H}_\alpha})$, $(\psi_k)_{k\in\mathbb{Z}}\in (\mathscr{H}_{\alpha,\mathrm{F}}+\mathbbm{1}_0)^{-1}\ker( \mathscr{H}_\alpha^*+\mathbbm{1}_0)$, and $(\xi_k)_{k\in\mathbb{Z}}\in \ker( \mathscr{H}_\alpha^*+\mathbbm{1}_0)$. The conclusion then follows from Lemma \ref{lem:psikinspaces}. 
\end{proof}

\begin{remark} 
 It was already known, from the identity $\mathscr{H}_\alpha^*=\bigoplus_{k\in\mathbb{Z}}A_\alpha(k)^*$ and from the analysis of $A_\alpha(k)^*$ made in Section \ref{sec:bilateralfibreext} (formulas \eqref{eq:bilateralDAxistar} and \eqref{eq:bilateralDAxistar_zero_mode}), that an element in $\mathcal{D}(\mathscr{H}_\alpha^*)$ must have the form $(g_k)_{k\in\mathbb{Z}}$ with $g_k$ satisfying \eqref{eq:gkrepresentation} for some $\widetilde{\varphi}_k\in\mathcal{D}(\overline{A_\alpha(k)})$ and some $c_{0,k}^\pm,c_{1,k}^\pm\in\mathbb{C}$. However, a generic collection $(g_k)_{k\in\mathbb{Z}}$ in $\ell^2(\mathbb{Z},\mathfrak{h})$ of $g_k$'s satisfying \eqref{eq:gkrepresentation} does not necessarily belong to $\mathcal{D}(\mathscr{H}_\alpha^*)$, in particular the corresponding collection 
 $(\widetilde{\varphi}_k)_{k\in\mathbb{Z}\setminus\{0\}}$ does not necessarily belong to $\mathcal{D}(\overline{\mathscr{H}_\alpha})$. Only under the conditions prescribed by Lemma \ref{lem:gkkrepr} 
  can one pile up such $g_k$'s so as to obtain an actual element in $\mathcal{D}(\mathscr{H}_\alpha^*)$ (in fact, \eqref{eq:pileupcond1}-\eqref{eq:pileupcond3} impose some kind of \emph{uniformity} in $k$ of $\widetilde{\varphi}_k$, $c_{0,k}^\pm$, $c_{1,k}^\pm$). 
\end{remark}

\begin{remark}\label{rem:regularitydeficiencyspace}
Lemmas \ref{lem:psikinspaces}(i) and \ref{lem:gkkrepr} characterise $\ker(\mathscr{H}_\alpha^*+\mathbbm{1}_0)$, the deficiency space for $\mathscr{H}_\alpha+\mathbbm{1}_0$, which is isomorphic to the deficiency space of the original operator $H_\alpha$. By exploiting the same unitary equivalence \eqref{eq:unitary_transf_pm}, it was determined in \cite{Posilicano-2014-sum-trace-maps} that the deficiency space of $H_\alpha^+$ is isomorphic to $H^{-\frac{1}{2}\frac{1-\alpha}{1+\alpha}}(\mathbb{S}^1)$ -- more precisely, isomorphic to $H^{\frac{1}{2}\frac{1-\alpha}{1+\alpha}}(\mathbb{S}^1)$ or equivalently to $H^{-\frac{1}{2}\frac{1-\alpha}{1+\alpha}}(\mathbb{S}^1)$ depending on the different explicit isomorphisms (namely the different `coordinate systems', or also the different `boundary triplets') between the trace space and the deficiency space. The present analysis is thus completely consistent with those findings: indeed, $\mathcal{F}_2:H^{-\frac{1}{2}\frac{1-\alpha}{1+\alpha}}(\mathbb{S}^1)\stackrel{\cong}{\longrightarrow}\ell^2(\mathbb{Z},\mathbb{C}^2, \mu_k)$.
\end{remark}

After the above preparations, the subsequent discussion takes two separate directions. One, which is completed here in the remaining part of the present Section, is the characterisation of the \emph{whole family} of self-adjoint extensions of $\mathscr{H}_\alpha$ in $\cH$, an information that surely deserves interest per se. Another, which is the object of the next Section, is the study of the \emph{distinguished family} of extensions of $\mathscr{H}_\alpha$ produced by Proposition \ref{prop:BextendsHalpha}. In fact, for the latter a clean and explicit description can be further obtained when going back to the physical variables $(x,y)$ -- and this turns out to be indeed the physically relevant sub-family of self-adjoint Hamiltonians on the Grushin cylinder.

\begin{theorem}\label{thm:Halphageneralext}
 Let $\alpha\in[0,1)$. There is a one-to-one correspondence $S\leftrightarrow \mathscr{H}_\alpha^{(S)}$ between the self-adjoint extensions $\mathscr{H}_\alpha^{(S)}$ of $\mathscr{H}_\alpha$ and the self-adjoint operators $S$ defined in Hilbert subspaces of $\ker (\mathscr{H}_\alpha^*+\mathbbm{1}_0)\cong\ell^2(\mathbb{Z},\mathbb{C}^2,\mu_k)$. If $S$ is any such operator, the corresponding extension $\mathscr{H}_\alpha^{(S)}$ is given by
 \begin{equation}\label{eq:Halphageneralext}
  \begin{split}
   \mathcal{D}(\mathscr{H}_\alpha^{(S)})\;&=\;\left\{
   \begin{array}{c}
    \psi=\widetilde{\varphi}+ (\mathscr{H}_{\alpha,\mathrm{F}}+\mathbbm{1}_0)^{-1}(Sv+w)+v \\
    \textrm{such that} \\
    \widetilde{\varphi}\in\mathcal{D}(\overline{\mathscr{H}_\alpha})\,,\quad v\in\mathcal{D}(S)\,, \\
    w\in\ker (\mathscr{H}_\alpha^*+\mathbbm{1}_0)\cap \mathcal{D}(S)^\perp
   \end{array}\right\} \, , \\
   (\mathscr{H}_\alpha^{(S)}+\mathbbm{1}_0)\psi\;&=\;(\overline{\mathscr{H}_\alpha}+\mathbbm{1}_0)\widetilde{\varphi}+Sv+w\,.
  \end{split}
 \end{equation}
\end{theorem}

\begin{proof}
 A direct application of Theorem \ref{thm:VB-representaton-theorem_Tversion}. The second formula in \eqref{eq:Halphageneralext} follows from the first as $(\mathscr{H}_\alpha^{(S)}+\mathbbm{1}_0)=(\mathscr{H}_\alpha^*+\mathbbm{1}_0)\upharpoonright\mathcal{D}(\mathscr{H}_\alpha^{(S)})$. 
\end{proof}

Theorem \ref{thm:Halphageneralext} encompasses a huge variety of extensions, since $\mathscr{H}_\alpha$ has infinite deficiency index (Theorem \ref{thm:Halpha_esa_or_not}(iii)). The self-adjointness condition \eqref{eq:Halphageneralext} for $\mathscr{H}_\alpha^{(S)}$, which is in fact a \emph{restriction condition} on the domain $\mathscr{H}_\alpha^*$, selects, in terms of the representation \eqref{eq:repreDHstar}, and among the generic elements 
\[
 \psi\;=\;\widetilde{\varphi}+ (\mathscr{H}_{\alpha,\mathrm{F}}+\mathbbm{1}_0)^{-1}\eta+\xi 
\]
of $\mathcal{D}(\mathscr{H}_\alpha^*)$, only those for which the vectors $\xi,\eta\in\ker (\mathscr{H}_\alpha^*+\mathbbm{1}_0)$ (customarily referred to as the \emph{charges}\index{charge} of $\psi$, see e.g.~\cite{MO-2016} and references therein) satisfy
\[ 
\begin{split}
 \xi\;&=\;v\;\in\;\mathcal{D}(S)\,, \\
 \eta\;&=\;Sv+w\,,\quad w\in\ker (\mathscr{H}_\alpha^*+\mathbbm{1}_0)\cap \mathcal{D}(S)^\perp\,.
\end{split}
\]
In this respect, the above condition produces in general a complicated, non-fibre-preserving mixing of the charge $\eta$ with respect to the charge $\xi$: such a mixing is encoded in the auxiliary operator $S$.

For physically relevant extensions the above mixing is absent instead, and the restriction condition of self-adjointness operates \emph{independently in each fibre}, namely in each momentum mode $k$. This is the case when
\begin{equation}\label{eq:fibredS}
 S\;=\;\bigoplus_{k\in\mathbb{Z}}S(k)\qquad \textrm{ on } \qquad \ker (\mathscr{H}_\alpha^*+\mathbbm{1}_0)\;=\;\bigoplus_{k}\,\ker(A_\alpha(k)^*+\delta_{k,0}\mathbbm{1})
\end{equation}
for operators $S(k)$'s each of which is self-adjoint on a (zero-, one-, two-dimensional) subspace $\mathcal{K}$ of the two-dimensional space $\ker(A_\alpha(k)^*+\delta_{k,0}\mathbbm{1})$. Extensions \eqref{eq:Halphageneralext} where $S$ is of the type \eqref{eq:fibredS} are \emph{fibred}\index{fibred extensions} in the sense that the self-adjointness condition is compatible with the fibre structure.

Explicitly, if $\mathscr{H}_\alpha^{(S)}$ is a fibred extension of $\mathscr{H}_\alpha$, then a generic element $(g_k)_{k\in\mathbb{Z}}$ of $\mathcal{D}(\mathscr{H}_\alpha^{(S)})$ is such that 
\begin{equation}\label{eq:Birman_formula-fibred}
 g_k\;=\;\widetilde{\varphi}_k+(A_{\alpha,\mathrm{F}}(k)+\delta_{k,0}\mathbbm{1})^{-1}(S(k) v_k+w_k)+v_k\,,\qquad k\in\mathbb{Z}\,,
\end{equation}
for some $\widetilde{\varphi}_k\in\mathcal{D}(\overline{A_\alpha(k)})$, $v_k\in\mathcal{D}(S(k))$, $w_k\in \ker (A_\alpha(k)^*+\delta_{k,0}\mathbbm{1})\cap\mathcal{D}(S(k))^\perp$.  Comparing \eqref{eq:Birman_formula-fibred} with \eqref{eq:Birman_formula} and \eqref{eq:Birman_formula_zero_mode} one immediately sees that the component $g_k$ belongs to the domain of the extension $A_\alpha^{(S(k))}(k)$ of $A_\alpha(k)$ (following the notation of \eqref{eq:Birman_formula} and \eqref{eq:Birman_formula_zero_mode}) with respect to the Hilbert space $\mathfrak{h}$.
Thus, fibred extensions of $\mathscr{H}_\alpha$ are precisely of the form $\bigoplus_{k\in\mathbb{Z}}B(k)$, where each $B(k)$ is a self-adjoint extension of $A_\alpha(k)$ in $\mathfrak{h}$, namely the extensions produced through the mechanism discussed in Proposition \ref{prop:BextendsHalpha}.

\section{Uniformly fibred extensions of $\mathscr{H}_\alpha$}\label{sec:uniformlyfirbredext}\index{fibred extensions!uniformly}

 The most relevant and physically meaningful sub-class of self-adjoint extensions of $\mathscr{H}_\alpha$ are those customarily referred to as \emph{uniformly fibred extensions}. For them, a more explicit and convenient characterisation is obtained, Theorem \ref{thm:classificationUF} below, as compared to the general classification of Theorem \ref{thm:Halphageneralext}.

\subsection{Generalities and classification theorem}

 The focus now is on extensions that on the one hand are \emph{fibred}\index{fibred extensions}, in the sense discussed in the end of Section \ref{sec:genextscrHa}, hence reduced as the direct sum of self-adjoint extensions of $A_\alpha(k)$ on each fibre, and therefore with conditions of self-adjointness that do not couple different fibres, and which \emph{in addition} display the following kind of uniformity.

 Recall that a generic fibred extension\index{fibred extensions} acts on each fibre as a generic self-adjoint realisation of $A_\alpha(k)$ that belongs to one of the families of the classification of Theorem \ref{thm:bifibre-extensions}, and is therefore parametrised (apart when it is $A_{\alpha,\mathrm{F}}(k)$) by one real parameter or four real parameters. Such extension types and extension parameters may differ fibre by fibre, say, parameter $\gamma^{(k_1)}$ for an extension of type $\mathrm{I_R}$ or $\mathrm{I_L}$ or $\mathrm{II}_{a_k}$ on the $k_1$-th fibre, and parameters $\gamma_1^{(k_2)},\dots,\gamma_4^{(k_2)}$ for an extension of type $\mathrm{III}$ on the $k_2$-th fibre.

\emph{Uniformly} fibred extensions\index{fibred extensions!uniformly} are those for which fibre by fibre the type of extension of $A_\alpha(k)$ is the same, and all have the same extension parameter(s) $\gamma$ (and $a$), or $\gamma_1,\dots,\gamma_4$.

By definition, uniformly fibred extensions can be therefore grouped into sub-families in complete analogy to those of Theorem \ref{thm:bifibre-extensions}:
\begin{itemize}
  \item \underline{Friedrichs extension}: the operator $\mathscr{H}_{\alpha,\mathrm{F}}=\bigoplus_{k\in\mathbb{Z}}A_{\alpha,\mathrm{F}}(k)$ (see Lemma \ref{lem:HalphaFrie-decomposable}); 
  \item \underline{Family $\mathrm{I_R}$}: operators of the form
  \begin{equation}\label{eq:HalphaR_unif-fibred}
   \mathscr{H}_{\alpha,\mathrm{R}}^{[\gamma]}\;:=\;\bigoplus_{k\in\mathbb{Z}}A_{\alpha,\mathrm{R}}^{[\gamma]}(k)
  \end{equation}
  for some $\gamma\in\mathbb{R}$;
  \item \underline{Family $\mathrm{I_L}$}: operators of the form
  \begin{equation}\label{eq:HalphaL_unif-fibred}
   \mathscr{H}_{\alpha,\mathrm{L}}^{[\gamma]}\;:=\;\bigoplus_{k\in\mathbb{Z}}A_{\alpha,\mathrm{L}}^{[\gamma]}(k)
  \end{equation}
 for some $\gamma\in\mathbb{R}$;
  \item \underline{Family $\mathrm{II}_a$} for given $a\in\mathbb{C}$: operators of the form
  \begin{equation}\label{eq:Halpha-IIa_unif-fibred}
   \mathscr{H}_{\alpha,a}^{[\gamma]}\;:=\;\bigoplus_{k\in\mathbb{Z}}A_{\alpha,a}^{[\gamma]}(k)
  \end{equation}
 for some $\gamma\in\mathbb{R}$;
  \item \underline{Family $\mathrm{III}$}: operators of the form
  \begin{equation}\label{eq:Halpha-III_unif-fibred}
   \mathscr{H}_{\alpha}^{[\Gamma]}\;:=\;\bigoplus_{k\in\mathbb{Z}}A_{\alpha}^{[\Gamma]}(k)
  \end{equation}
 for some $\Gamma\equiv(\gamma_1,\gamma_2,\gamma_3,\gamma_4)\in\mathbb{R}^4$.
 \end{itemize}

Physically, uniformly fibred extensions\index{fibred extensions!uniformly} have surely a special status in that the boundary condition experienced as $x\to 0$ by the quantum particle governed by any such Hamiltonian has both the same form and the same `magnitude' (hence the same $\gamma$-parameter, or $\gamma_j$-parameters) irrespective of the transversal momentum, namely the quantum number $k$.

 Mathematically, uniformly fibred extensions have a completely explicit description not only in mixed position-momentum variables $(x,k)$, namely extensions of $\mathscr{H}_\alpha$, but also in the original physical coordinates $(x,y)$, namely extensions of the symmetric operator $\mathsf{H}_\alpha=\mathcal{F}_2^{-1}\mathscr{H}_\alpha\mathcal{F}_2$ acting on $L^2(\mathbb{R}\times\mathbb{S}^1,\ud x\ud y)$, explicitly described in \eqref{eq:explicit-tildeHalpha} and \eqref{eq:V-two-sided-6}.

This is the content of the main result of this Section.

\begin{theorem}\label{thm:classificationUF}
 Let $\alpha\in[0,1)$. The densely defined, symmetric operator
 \[
  \begin{split}
   \mathsf{H}_\alpha\;&=\;\mathcal{F}_2^{-1}\mathscr{H}_\alpha\mathcal{F}_2\;=\;-\frac{\partial^2}{\partial x^2}- |x|^{2\alpha}\frac{\partial^2}{\partial y^2}+\frac{\,\alpha(2+\alpha)\,}{4x^2}\,, \\
  \mathcal{D}(\mathsf{H}_\alpha^\pm)\;&=\;C^\infty_c(\mathbb{R}^\pm_x\times\mathbb{S}^1_y)
  \end{split}
 \]
 admits, among others, the following families of self-adjoint extensions in $L^2(\mathbb{R}\times\mathbb{S}^1,\ud x \ud y):$  
 \begin{itemize}
  \item \underline{Friedrichs extension}: $\mathsf{H}_{\alpha,\mathrm{F}}$, where $\mathsf{H}_{\alpha,\mathrm{F}}=\mathcal{F}_2^{-1}\mathscr{H}_{\alpha,\mathrm{F}}\mathcal{F}_2$;
  \item \underline{Family $\mathrm{I_R}$}: $\{\mathsf{H}_{\alpha,\mathrm{R}}^{[\gamma]}\,|\,\gamma\in\mathbb{R}\}$, where $\mathsf{H}_{\alpha,\mathrm{R}}^{[\gamma]}=\mathcal{F}_2^{-1}\mathscr{H}_{\alpha,\mathrm{R}}^{[\gamma]}\,\mathcal{F}_2$;
  \item \underline{Family $\mathrm{I_L}$}: $\{\mathsf{H}_{\alpha,\mathrm{L}}^{[\gamma]}\,|\,\gamma\in\mathbb{R}\}$, where $\mathsf{H}_{\alpha,\mathrm{L}}^{[\gamma]}=\mathcal{F}_2^{-1}\mathscr{H}_{\alpha,\mathrm{L}}^{[\gamma]}\,\mathcal{F}_2$;
  \item \underline{Family $\mathrm{II}_a$} with $a\in\mathbb{C}$: $\{\mathsf{H}_{\alpha,a}^{[\gamma]}\,|\,\gamma\in\mathbb{R}\}$, where $\mathsf{H}_{\alpha,a}^{[\gamma]}=\mathcal{F}_2^{-1} \mathscr{H}_{\alpha,a}^{[\gamma]}\,\mathcal{F}_2$;
  \item \underline{Family $\mathrm{III}$}: $\{\mathsf{H}_{\alpha}^{[\Gamma]}\,|\,\Gamma\equiv(\gamma_1,\gamma_2,\gamma_3,\gamma_4)\in\mathbb{R}^4\}$, where $\mathsf{H}_{\alpha}^{[\Gamma]}=\mathcal{F}_2^{-1} \mathscr{H}_{\alpha}^{[\Gamma]}\,\mathcal{F}_2$.
 \end{itemize}
 Each element from any such family is characterised by being the \emph{restriction} of the adjoint operator
  \begin{equation}\label{eq:HHalphaadjointagain}
  \begin{split}
    \mathcal{D}(\mathsf{H}_\alpha^*)\;&=\;\left\{\!
  \begin{array}{c}
   \phi\in L^2(\mathbb{R}\times\mathbb{S}^1,\ud x \ud y)\textrm{ such that} \\
   \Big(-\frac{\partial^2}{\partial x^2}- |x|^{2\alpha}\frac{\partial^2}{\partial y^2}+\frac{\,\alpha(2+\alpha)\,}{4x^2}\Big)\phi^\pm\in L^2(\mathbb{R}^\pm\times\mathbb{S}^1,\ud x \ud y)
  \end{array}\!
  \right\}, \\
  (\mathsf{H}_\alpha^\pm)^*\phi^\pm\;&=\;-\frac{\partial^2\phi^\pm}{\partial x^2}- |x|^{2\alpha}\frac{\partial^2\phi^\pm}{\partial y^2}+\frac{\,\alpha(2+\alpha)\,}{4x^2}\phi^\pm
  \end{split}
\end{equation}
 to the functions
 \[
  \phi\;=\;\begin{pmatrix} \phi^- \\ \phi^+ \end{pmatrix},\qquad \phi^\pm\;\in\;L^2(\mathbb{R}^\pm\times\mathbb{S}^1,\ud x \ud y)
 \]
 for which the limits
  \begin{eqnarray}
   \phi_0^\pm(y)\!&=&\!\lim_{x\to 0^\pm} |x|^{\frac{\alpha}{2}}\,\phi^\pm(x,y)\,, \label{eq:limitphi0} \\
   \phi_1^\pm(y)\!&=&\!\lim_{x\to 0^\pm} |x|^{-(1+\frac{\alpha}{2})}\big(\phi^\pm(x,y)-\phi^\pm_0(y)|x|^{-\frac{\alpha}{2}}\big) \label{eq:limitphi1} \\
   &=&\!\pm(1+\alpha)^{-1}\lim_{x\to 0^\pm} |x|^{-\alpha}\partial_x\big(|x|^{\frac{\alpha}{2}}\phi^\pm(x,y)\big) \nonumber
  \end{eqnarray}
 exist and are finite for almost every $y\in\mathbb{S}^1$, and satisfy the following boundary conditions, depending on the considered type of extension, for almost every $y\in\mathbb{R}$:
 \begin{eqnarray}
  \phi_0^\pm(y)\,=\,0\,, \qquad \quad\;\;& & \textrm{if }\;  \phi\in\mathcal{D}(\mathsf{H}_{\alpha,\mathrm{F}})\,, \label{eq:DHalpha_cond3_Friedrichs-NOWEIGHTS}\\
  \begin{cases}
   \;\phi_0^-(y)= 0\,,  \\
   \;\phi_1^+(y)=\gamma \phi_0^+(y)\,,
  \end{cases} & & \textrm{if }\;  \phi\in\mathcal{D}(\mathsf{H}_{\alpha,\mathrm{R}}^{[\gamma]})\,, \\
   \begin{cases}
   \;\phi_1^-(y)=\gamma \phi_0^-(y)\,, \\
   \;\phi_0^+(y)= 0 \,,
  \end{cases} & & \textrm{if }\;  \phi\in\mathcal{D}(\mathsf{H}_{\alpha,\mathrm{L}}^{[\gamma]}) \,, \label{eq:DHalpha_cond3_L-NOWEIGHTS}\\
     \begin{cases}
   \;\phi_0^+(y)=a\,\phi_0^-(y)\,, \\
   \;\phi_1^-(y)+\overline{a}\,\phi_1^+(y)=\gamma \phi_0^-(y)\,,
  \end{cases} & & \textrm{if }\;  \phi\in\mathcal{D}(\mathsf{H}_{\alpha,a}^{[\gamma]})\,, \label{eq:DHalpha_cond3_IIa-NOWEIGHTS} \\
   \begin{cases}
   \;\phi_1^-(y)=\gamma_1 \phi_0^-(y)+(\gamma_2+\ii\gamma_3) \phi_0^+(y)\,, \\
   \;\phi_1^+(y)=(\gamma_2-\ii\gamma_3) \phi_0^-(y)+\gamma_4 \phi_0^+(y)\,,
  \end{cases} & & \textrm{if }\;  \phi\in\mathcal{D}(\mathsf{H}_{\alpha}^{[\Gamma]})\,. \label{eq:DHalpha_cond3_III-NOWEIGHTS}
 \end{eqnarray} 
 Moreover,
 \begin{equation}
  \phi_0^\pm \in H^{s_{0,\pm}}(\mathbb{S}^1, \ud y)\qquad\textrm{ and }\qquad \phi_1^\pm\in H^{s_{1,\pm}}(\mathbb{S}^1,\ud y)
 \end{equation}
 with
 \begin{itemize}
 	\item $s_{1,\pm}=\frac{1}{2}\frac{1-\alpha}{1+\alpha}$\qquad\qquad\qquad\qquad\qquad\; for the Friedrichs extension,
 	\item $s_{1,-}=\frac{1}{2}\frac{1-\alpha}{1+\alpha}$, $s_{0,+}=s_{1,+}=\frac{1}{2}\frac{3+\alpha}{1+\alpha}$ \quad\;\; for extensions of type $\mathrm{I}_{\mathrm{R}}$,
 	\item  $s_{1,+}=\frac{1}{2}\frac{1-\alpha}{1+\alpha}$, $s_{0,-}=s_{1,-}=\frac{1}{2}\frac{3+\alpha}{1+\alpha}$ \quad\;\; for extensions of type $\mathrm{I}_{\mathrm{L}}$,
 	\item $s_{1,\pm}=s_{0,\pm}=\frac{1}{2}\frac{1-\alpha}{1+\alpha}$ \qquad\qquad\qquad \;\;\;\,\, for extensions of type $\mathrm{II}_a$,
 	\item $s_{1,\pm}=s_{0,\pm}=\frac{1}{2}\frac{3+\alpha}{1+\alpha}$ \qquad\qquad\qquad \;\;\;\,\, for extensions of type $\mathrm{III}$.
 \end{itemize}
\end{theorem}

\subsection{General strategy}\label{sec:genstrategy}

The proof of Theorem \ref{thm:classificationUF} requires a detailed analysis, concisely explained here below. All the preparation is developed in Subsections \ref{sec:genstrategy} through \ref{sec:q0q1}, and the proof is discussed in Subsection \ref{sec:proofclassifthm}.

The trivial part is of course the reconstruction of each uniformly fibred extension of $\mathscr{H}_\alpha$ through a direct sum of self-adjoint extensions of the $A_\alpha(k)$'s. Instead, the difficult part is to extract the appropriate information so as to export the boundary conditions of self-adjointness from the mixed position-momentum variables $(x,k)$ to the physical coordinates $(x,y)$. The inverse Fourier transform $\mathcal{F}_2^{-1}$ is indeed a non-local operation, and in order to `add up' the boundary conditions initially available $k$ by $k$, one needs suitable \emph{uniformity} controls in $k$.

Let $\mathscr{H}_\alpha^{\mathrm{u.f.}}$ be a uniformly fibred extension of $\mathscr{H}_\alpha$. A generic element $(g_k)_{k\in\mathbb{Z}}\in\mathcal{D}(\mathscr{H}_\alpha^{\mathrm{u.f.}})$ can be represented as in \eqref{eq:gkrepresentation} with the `summability' conditions \eqref{eq:pileupcond1}-\eqref{eq:pileupcond2} that guarantee $(g_k)_{k\in\mathbb{Z}}$ to belong to $\mathcal{D}(\mathscr{H}_\alpha^*)$ (Lemma \ref{lem:gkkrepr}), plus additional constraints among the coefficients $c_{0,k}^\pm$ and $c_{1,k}^\pm$ that guarantee that $\mathcal{D}(\mathscr{H}_\alpha^{\mathrm{u.f.}})$ is indeed a domain of self-adjointness. Actually, the latter requirement imposes \emph{stronger} summability conditions on the $c_{0,k}^\pm$'s and $c_{1,k}^\pm$'s, as is discussed in Subsection \ref{subsec:g0g1Sobolev}.

However, the above-mentioned representation \eqref{eq:gkrepresentation} for the elements of $\mathcal{D}(\mathscr{H}_\alpha^{\mathrm{u.f.}})$ is problematic when one has to describe $\mathcal{F}_2^{-1}\mathcal{D}(\mathscr{H}_\alpha^{\mathrm{u.f.}})$, namely the same domain in $(x,y)$-coordinates (it is immediate from \eqref{eq:V-two-sided-7} that $\mathcal{F}_2^{-1}\mathcal{D}(\mathscr{H}_\alpha^{\mathrm{u.f.}})$ is the domain of the self-adjoint extension $\mathcal{F}_2^{-1}\mathscr{H}_\alpha^{\mathrm{u.f.}}\mathcal{F}_2$ of $\mathsf{H}_\alpha=\mathcal{F}_2^{-1}\mathscr{H}_\alpha\mathcal{F}_2$).

More precisely, when applying $\mathcal{F}_2^{-1}$ to \eqref{eq:gkrepresentation}, one looses control on the self-adjointness constraint that now becomes a rather implicit condition between the $(x,y)$-functions
\begin{equation}\label{eq:badF2-1}
 \mathcal{F}_2^{-1} \left(\begin{pmatrix} c_{1,k}^-\widetilde{\Psi}_{\alpha,k} \\ c_{1,k}^+\widetilde{\Psi}_{\alpha,k} \end{pmatrix}\right)_{\!k\in\mathbb{Z}}\,,\qquad\qquad \mathcal{F}_2^{-1} \left(\begin{pmatrix} c_{0,k}^-\widetilde{\Phi}_{\alpha,k} \\ c_{0,k}^+\widetilde{\Phi}_{\alpha,k} \end{pmatrix}\right)_{\!k\in\mathbb{Z}}.
\end{equation}
Recall indeed from \eqref{eq:defF2} that
\[
 (\mathcal{F}_2^+)^{-1}(( c_{1,k}^+\widetilde{\Psi}_{\alpha,k}))_{k\in\mathbb{Z}}\;=\;\frac{1}{\sqrt{2\pi}}\sum_{k\in\mathbb{Z}}c_{1,k}\widetilde{\Psi}_{\alpha,k}(x) e^{\ii k y}\,,
\]
and similarly for the other components: on such functions of $x$ and $y$ it is not evident if differentiating or taking the limit $x\to 0$ term by term in the series in $k$ is actually justified -- and it is precisely in terms of such operations that the final boundary conditions are going to be expressed.

From another perspective, the known regularity and asymptotic properties of $\widetilde{\Psi}_{\alpha,k}$ (and, analogously, $\widetilde{\Phi}_{\alpha,k}$) may well provide the above information on the function  $(\mathcal{F}_2^+)^{-1}((\widetilde{\Psi}_{\alpha,k}))_{k\in\mathbb{Z}}$, but it is not evident how to read out useful information from $(\mathcal{F}_2^+)^{-1}(( c_{1,k}^+\widetilde{\Psi}_{\alpha,k}))_{k\in\mathbb{Z}}$ so as to finally express the boundary conditions of self-adjointness in terms of limits as $x\to 0$ of the functions in the domain and on their derivatives.

As taking the inverse Fourier transform directly on \eqref{eq:gkrepresentation} appears not to be informative in practice, it is convenient to follow a second route inspired to the alternative representation \eqref{eq:gwithPweight} (Theorem \ref{prop:g_with_Pweight}).

Now the generic element $(g_k)_{k\in\mathbb{Z}}\in\mathcal{D}(\mathscr{H}_\alpha^{\mathrm{u.f.}})$ is represented for each $k$ as
 \begin{equation}\label{eq:gwithPweight_k}
  g_k\;=\;\begin{pmatrix} \varphi_k^- \\ \varphi_k^+ \end{pmatrix}+\begin{pmatrix} g_{0,k}^- \\ g_{0,k}^+\end{pmatrix}|x|^{-\frac{\alpha}{2}}\,P+\begin{pmatrix} g_{1,k}^- \\ g_{1,k}^+\end{pmatrix}|x|^{1+\frac{\alpha}{2}}\,P\,,
 \end{equation}
where each $\varphi_k\in\mathcal{D}(\overline{A_\alpha(k)})$ and $P$ is the short-scale cut-off \eqref{eq:Pcutoff}.

The evident advantage of \eqref{eq:gwithPweight_k}, as compared to \eqref{eq:gkrepresentation}, is that computing
\begin{equation}
 \phi\;:=\;\mathcal{F}_2^{-1}(g_k)_{k\in\mathbb{Z}}
\end{equation}
and using the linearity of $\mathcal{F}_2^{-1}$ yields \emph{formally}
\begin{equation}\label{eq:afterF2-1}
 \phi(x,y)\;=\;\varphi(x,y)+g_1(y)|x|^{1+\frac{\alpha}{2}}P(x)+g_0(y)|x|^{-\frac{\alpha}{2}}P(x)
\end{equation}
with
\begin{eqnarray}
 \varphi\!\!&:=&\!\!\mathcal{F}_2^{-1}(\varphi_k)_{k\in\mathbb{Z}}\,, \label{eq:F2-1phi}\\
 g_0\!\!&:=&\!\!\mathcal{F}_2^{-1}(g_{0,k})_{k\in\mathbb{Z}}\,, \label{e1:F2-1g0}\\
 g_1\!\!&:=&\!\!\mathcal{F}_2^{-1}(g_{1,k})_{k\in\mathbb{Z}} \label{eq:F2-1g1}\,.
\end{eqnarray}

In \eqref{eq:afterF2-1} the function $\varphi$ is expected to retain the regularity in $x$ and the fast vanishing properties, as $x\to 0$, of each $\varphi_k$, and hence $\varphi$ is expected to be a sub-leading term when taking $\lim_{x\to 0}\phi(x,y)$ and $\lim_{x\to 0}\partial_x\phi(x,y)$; on the other hand, the regularity and short-distance behaviour in $x$ of the other two summands on the r.h.s.~of \eqref{eq:afterF2-1} are immediately read out, unlike the situation with the functions \eqref{eq:badF2-1}. Moreover, and most importantly, since $\mathscr{H}_\alpha^{\mathrm{u.f.}}$ is a \emph{uniformly fibred} extension, the boundary condition of self-adjointness in \eqref{eq:gwithPweight_k} (namely a condition among those listed in the third column of Table \ref{tab:extensions}) takes the same form, with the same extension parameter, irrespective of $k$, and therefore is immediately exported, in the same form and with the same extension parameter, between $g_0(y)$ and $g_1(y)$ for almost every $y\in\mathbb{S}^1$.

The above reasoning paves the way to a classification of the family of uniformly fibred extensions of $H_\alpha$ in terms of explicit boundary conditions as $x\to 0$.

Clearly, so far \eqref{eq:afterF2-1} is only formal: one must guarantee that \eqref{eq:F2-1phi}-\eqref{eq:F2-1g1} are actually well-posed and define square-integrable functions in the corresponding variables, with the desired properties. This is in fact the price to pay for the present strategy, whereas for the functions \eqref{eq:badF2-1} it was clear a priori that $\mathcal{F}_2^{-1}$ is applicable, thanks to Lemma \ref{eq:repreDHstar}.

As will be further commented at a later stage (Sect.~\ref{sec:singular_decomposition_adjoint}), such a strategy leads to the following somewhat awkward circumstance:  whereas Lemma \ref{eq:repreDHstar} guarantees that applying $\mathcal{F}_2^{-1}$ on $(g_k)_{k\in\mathbb{Z}}$ represented as in \eqref{eq:gkrepresentation} yields three distinct functions, each of which belongs to $\mathcal{F}_2^{-1}\mathcal{D}(\mathscr{H}_\alpha^*)=\mathcal{D}(\mathsf{H}_\alpha^*)$, the three summands on the r.h.s.~of \eqref{eq:afterF2-1} will be proved to belong to $L^2(\mathbb{R}\times\mathbb{S}^1,\ud x\ud y)$, \emph{none} of which being however in $\mathcal{D}(\mathsf{H}_\alpha^*)$ in general! -- only their sum is, due to cancellations of singularities. This explains why the analysis is going to be onerous.

\subsection{Integrability and Sobolev regularity of $g_0$ and $g_1$}\label{subsec:g0g1Sobolev}

Following the programme outlined in the previous Subsection, it will be now shown that \eqref{e1:F2-1g0} and \eqref{eq:F2-1g1} indeed define functions in $L^2(\mathbb{S}^1,\ud y)$ with suitable regularity.

\begin{proposition}\label{prop:g0g1Sobolev}
 Let $\alpha\in[0,1)$ and let $(g_k)_{k\in\mathbb{Z}}\in\mathcal{D}(\mathscr{H}_\alpha^{\mathrm{u.f.}})$, where $\mathscr{H}_\alpha^{\mathrm{u.f.}}$ is one of the operators \eqref{eq:HalphaFriedrichs_unif-fibred} or \eqref{eq:HalphaR_unif-fibred}-\eqref{eq:Halpha-III_unif-fibred}, for given parameters $\gamma\in\mathbb{R}$, $a\in\mathbb{C}$, $\Gamma\in\mathbb{R}^4$, depending on the type. With respect to the representation \eqref{eq:gwithPweight_k} of each $g_k$, one has the following. 
\begin{enumerate}[(i)]
	\item If $\mathscr{H}_{\alpha}^{\mathrm{u.f.}}$ is the Friedrichs extension, then
	\begin{equation}\label{eq:g0g1Sobolev_typeF}
		\sum_{k \in \mathbb{Z}} |k|^{\frac{1-\alpha}{1+\alpha}}|g_{1,k}^\pm|^2 \; < \;+\infty \,,\qquad g_{0,k}^\pm\;=\;0\,.
	\end{equation}
	\item If $\mathscr{H}_{\alpha}^{\mathrm{u.f.}}$ is of type $\mathrm{I}_{\mathrm{R}}$, then
	\begin{equation}\label{eq:g0g1Sobolev_typeIR}
		\begin{split}
		\sum_{k \in \mathbb{Z}} |k|^{\frac{1-\alpha}{1+\alpha}}|g_{1,k}^-|^2  \; < \; +\infty& \,, \qquad g_{0,k}^- \;=\;0\,, \\
		\sum_{k \in \mathbb{Z}}  |k|^{\frac{3+\alpha}{1+\alpha}}|g_{1,k}^+|^2 \; < \; + \infty& \,, \qquad \sum_{k \in \mathbb{Z}}  |k|^{\frac{3+\alpha}{1+\alpha}}|g_{0,k}^+|^2 \; < \; + \infty\,.
		\end{split}
	\end{equation}
	\item If $\mathscr{H}_{\alpha}^{\mathrm{u.f.}}$ is of type $\mathrm{I}_{\mathrm{L}}$, then
	\begin{equation}\label{eq:g0g1Sobolev_typeIL}
		\begin{split}
		\sum_{k \in \mathbb{Z}} |k|^{\frac{1-\alpha}{1+\alpha}}|g_{1,k}^+|^2  \; < \; +\infty& \,, \qquad g_{0,k}^+ \;=\;0\,, \\
		\sum_{k \in \mathbb{Z}} |k|^{\frac{3+\alpha}{1+\alpha}} |g_{1,k}^-|^2  \; < \; + \infty& \,, \qquad \sum_{k \in \mathbb{Z}} |k|^{\frac{3+\alpha}{1+\alpha}}|g_{0,k}^-|^2  \; < \; + \infty\,.
		\end{split}
	\end{equation}
	\item If $\mathscr{H}_{\alpha}^{\mathrm{u.f.}}$ is of type $\mathrm{II}_a$, then
	\begin{equation}\label{eq:g0g1Sobolev_typeIIa}
		\begin{split}
		&\sum_{k \in \mathbb{Z}} |k|^{\frac{1-\alpha}{1+\alpha}}|g_{1,k}^\pm|^2  \; < \; +\infty \,, \qquad \sum_{k \in \mathbb{Z}}  |k|^{\frac{3+\alpha}{1+\alpha}} |g_{0,k}^\pm|^2\; < \; + \infty\,, \\
		&\sum_{k \in \mathbb{Z}} |k|^{\frac{3+\alpha}{1+\alpha}}|g_{1,k}^-+ \overline{a} g_{1,k}^+|^2  \; < \; + \infty\,. 
		\end{split}
	\end{equation}
	\item If $\mathscr{H}_{\alpha}^{\mathrm{u.f.}}$ is of type $\mathrm{III}$, then
	\begin{equation}\label{eq:g0g1Sobolev_typeIII}
	   \sum_{k\in\mathbb{Z}}|k|^{\frac{3+\alpha}{1+\alpha}}|g^\pm_{0,k}|^2\;<\;+\infty\,,\qquad \sum_{k\in\mathbb{Z}}|k|^{\frac{3+\alpha}{1+\alpha}}|g^\pm_{1,k}|^2\;<\;+\infty\,.
	\end{equation}
\end{enumerate} 
\end{proposition}

\begin{corollary}\label{cor:(g0)k_(g1)k_in_Hs}
 Under the assumptions of Proposition \ref{prop:g0g1Sobolev}, $(g^\pm_{0,k})_{k\in\mathbb{Z}}$ and $(g^\pm_{1,k})_{k\in\mathbb{Z}}$ belong $\ell^2(\mathbb{Z})$. Hence, \eqref{e1:F2-1g0} and \eqref{eq:F2-1g1} define functions $y\mapsto g_0^\pm(y)$ and $y\mapsto g_1^\pm(y)$ that belong to $L^2(\mathbb{S}^1,\ud y)$. In particular, the summability properties \eqref{eq:g0g1Sobolev_typeF}-\eqref{eq:g0g1Sobolev_typeIII} imply that $g_0^\pm \in H^{s_{0,\pm}}(\mathbb{S}^1, \ud y)$ and $g_1^\pm\in H^{s_{1,\pm}}(\mathbb{S}^1,\ud y)$, where the order of such Sobolev spaces is, respectively,
 \begin{enumerate}[(i)]
 	\item $s_{1,\pm}=\frac{1}{2}\frac{1-\alpha}{1+\alpha}$\qquad\qquad\qquad\qquad\qquad\; for the Friedrichs extension,
 	\item $s_{1,-}=\frac{1}{2}\frac{1-\alpha}{1+\alpha}$, $s_{0,+}=s_{1,+}=\frac{1}{2}\frac{3+\alpha}{1+\alpha}$ \quad\;\, for extensions of type $\mathrm{I}_{\mathrm{R}}$,
 	\item  $s_{1,+}=\frac{1}{2}\frac{1-\alpha}{1+\alpha}$, $s_{0,-}=s_{1,-}=\frac{1}{2}\frac{3+\alpha}{1+\alpha}$ \quad\;\, for extensions of type $\mathrm{I}_{\mathrm{L}}$,
 	\item $s_{1,\pm}=s_{0,\pm}=\frac{1}{2}\frac{1-\alpha}{1+\alpha}$ \qquad\qquad\qquad \;\;\;\,\, for extensions of type $\mathrm{II}_a$,
 	\item $s_{1,\pm}=s_{0,\pm}=\frac{1}{2}\frac{3+\alpha}{1+\alpha}$ \qquad\qquad\qquad \;\;\,\; for extensions of type $\mathrm{III}$.
 \end{enumerate}
 \end{corollary}

\begin{proof}[Proof of Proposition \ref{prop:g0g1Sobolev}]
For each case, the proof is organised in two stages. First, one considers each family of extensions on fibre Hilbert space, as characterised by Theorem \ref{thm:bifibre-extensionsc0c1} in terms of certain self-adjointness constraints between the coefficients $c_0^\pm$ and $c_1^\pm$ of the representation \eqref{eq:repreDHstar}-\eqref{eq:gkrepresentation} of the elements of $\mathcal{D}(\mathscr{H}_\alpha^*)$, and one shows that owing to such constraints the a priori summability \eqref{eq:pileupcond2}-\eqref{eq:pileupcond3} of the $c_0^\pm$'s and $c_1^\pm$'s is actually enhanced (see also Remark \ref{rem:enhanced_summability} below). Then, one exports the resulting summability of the $c_0^\pm$'s and $c_1^\pm$'s on to the $g_0^\pm$'s and $g_1^\pm$'s by means of the relations
\begin{eqnarray}
 g_{0,k}^\pm & = &c_{0,k}^\pm \textstyle{\sqrt{\frac{\pi(1+\alpha)}{2 |k|}}} \label{eq:Relationg0c0} \\
 g_{1,k}^\pm & = & c_{1,k}^\pm \textstyle{\sqrt{\frac{2 |k|}{\pi (1+\alpha)^3}}} \Vert \Phi_{\alpha,k} \Vert_{L^2(\mathbb{R}^+)}^2 - c_{0,k}^\pm \textstyle{\sqrt{\frac{\pi |k|}{2(1+\alpha)}}}\label{eq:Relationg1c1}
\end{eqnarray}
valid for $k\neq 0$ (see \eqref{eq:limitsg0g1} above). Obviously, it suffices to prove the final summability properties for $k\in\mathbb{Z}\setminus\{0\}$. Recall from \eqref{eq:Phinorm} that
\[
 \Vert \Phi_{\alpha,k} \Vert_{L^2(\mathbb{R}^+)}^2\;\sim\; |k|^{-\frac{2}{1+\alpha}}\,,
\]
namely for some multiplicative constant depending only on $\alpha$.


(i) Theorem \ref{thm:bifibre-extensionsc0c1} states that for this case $c_{0,k}^\pm =0$. This, together with  \eqref{eq:pileupcond2} and \eqref{eq:Relationg1c1}, yields
\[
	+ \infty \; > \sum_{k\in\mathbb{Z}\setminus\{0\}} |k|^{-\frac{2}{1+\alpha}} |c_{1,k}^\pm|^2 \;\sim\sum_{k\in\mathbb{Z}\setminus\{0\}} |k|^{\frac{1-\alpha}{1+\alpha}} |g_{1,k}^\pm |^2\,.
\]

(ii) Theorem \ref{thm:bifibre-extensionsc0c1} states that for this case $c_{0,k}^+=0$ and $c_{1,k}^+ =\beta_k c_{0,k}^+$ with $\beta_k$ given for $k\neq 0$ by 
\[
\gamma \; = \; \textstyle\frac{|k|}{1+\alpha}\Big(\frac{\, 2 \|\Phi_{\alpha,k}\|_{L^2}^2 \,}{\pi(1+\alpha)}\, \beta_k-1\Big)
\]
(see \eqref{eq:g1gammag0} above), that is, $\beta_k\sim |k|^{\frac{2}{1+\alpha}}$ at the leading order in $k$. (Here the operator of multiplication by $\beta_k$ is what was denoted in abstract by $S(k)$ in the discussion following Theorem \eqref{thm:Halphageneralext} -- see \eqref{eq:fibredS} above.) This, together with  \eqref{eq:pileupcond3} and \eqref{eq:Relationg0c0} yields
\[
 +\infty\;>\sum_{k\in\mathbb{Z}\setminus\{0\}} |k|^{-\frac{2}{1+\alpha}} |c_{1,k}^+|^2\;=\sum_{k\in\mathbb{Z}\setminus\{0\}} |k|^{-\frac{2}{1+\alpha}} |\beta_k c_{0,k}^+|^2\;\sim\sum_{k\in\mathbb{Z}\setminus\{0\}} |k|^{\frac{3+\alpha}{1+\alpha}}|g_{0,k}^+|^2\,.
\]
From this, one also obtains
\[
 \sum_{k\in\mathbb{Z}\setminus\{0\}} |k|^{\frac{3+\alpha}{1+\alpha}}|g_{1,k}^+|^2\;<\;+\infty\,,
\]
owing to the self-adjointness condition in the form $g_{1,k}^+=\gamma g_{0,k}^+$ (Theorem \ref{thm:bifibre-extensions}). As for the summability of the $c_{1,k}^-$, one proceeds precisely as in case (i).

(iii) The reasoning for this case is completely analogous as for case (ii), upon exchanging the `$+$' coefficients with the `$-$' coefficients.

(iv) Theorem \ref{thm:bifibre-extensionsc0c1} states for this case 
\[
	\begin{split}
		c_{0,k}^- \;&=\; c_{0,k} \, , \qquad\qquad\! c_{1,k}^- \; = \; \tau_k c_{0,k} + \widetilde{c}_{0,k} \, , \\
		c_{0,k}^+ \;& = \; a  c_{0,k}\,, \qquad\quad\; c_{1,k}^+ \;=\; \tau_k a c_{0,k} - \overline{a}^{-1} \widetilde{c}_{0,k}\,,
	\end{split}
\]
with $\tau_k$ given for $k \neq 0$ by
\[
			  \gamma\;=\;{(1+|a|^2) \frac{|k|}{1+\alpha}} \Big( \frac{\,2 \|\Phi_{\alpha,k}\|_{L^2(\mathbb{R}^+)}^2 \,}{\pi (1+\alpha)} \, \tau_k -1 \Big)\,,
\]
(see \eqref{eq:IkTheorem51} above), that is, $\tau_k \sim |k|^{\frac{2}{1+\alpha}}$ at the leading order in $k$. This, together with the a priori bounds \eqref{eq:pileupcond3}, and with \eqref{eq:Relationg0c0}, yields
\[
\begin{split}
 + \infty  \; &> \; \sum_{k\in\mathbb{Z}\setminus\{0\}} |k|^{-\frac{2}{1+\alpha}} |c_{1,k}^- + \overline{a} c_{1,k}^+ |^2 \;=\; \sum_{k\in\mathbb{Z}\setminus\{0\}} |k|^{-\frac{2}{1+\alpha}} |(1+|a|^2) \tau_k c_{0,k} |^2 \\
 &\sim \; \sum_{k\in\mathbb{Z}\setminus\{0\}}|k|^{\frac{3+\alpha}{1+\alpha}} |g_{0,k}^-|^2 \, .
\end{split}
\]
From this, and self-adjointness conditions $g_{0,k}^+= a g_{0,k}^-$ and $g_{1,k}^-+\overline{a} g_{1,k}^+= \gamma g_{0,k}^-$  (Theorem \ref{thm:bifibre-extensions}), one obtains the last two conditions in \eqref{eq:g0g1Sobolev_typeIIa}. As for establishing the first condition in \eqref{eq:g0g1Sobolev_typeIIa}, one has
\[
 \begin{split}
  \sum_{k\in\mathbb{Z}\setminus\{0\}} &|k|^{\frac{1-\alpha}{1+\alpha}}\,|g_{1,k}^{\pm}|^2 \\
  &\leqslant\; \sum_{k\in\mathbb{Z}\setminus\{0\}}|k|^{\frac{1-\alpha}{1+\alpha}}\,|c_{1,k}^{\pm}|^2{\textstyle{\frac{4|k|}{\pi(1+\alpha^3)}}}\|\Phi_{\alpha,k}\|_{L^2(\mathbb{R}^+)}^4+ \!\!\sum_{k\in\mathbb{Z}\setminus\{0\}}|k|^{\frac{1-\alpha}{1+\alpha}}\,|c_{0,k}^{\pm}|^2\textstyle{\frac{\pi|k|}{(1+\alpha)}} \\
  &\sim\;\sum_{k\in\mathbb{Z}\setminus\{0\}}|k|^{-\frac{2}{1+\alpha}}|c_{1,k}^{\pm}|^2+\!\!\sum_{k\in\mathbb{Z}\setminus\{0\}}|k|^{\frac{3+\alpha}{1+\alpha}}|g_{0,k}^{\pm}|^2\;<\;+\infty\,,
 \end{split}
\]
having used \eqref{eq:Relationg1c1} for the first step, \eqref{eq:Relationg0c0} for the second step, and the a priori bounds \eqref{eq:pileupcond3} as well as the already proved second condition in \eqref{eq:g0g1Sobolev_typeIIa} for the last step.

(v) Theorem \ref{thm:bifibre-extensionsc0c1} states for this case
\[
	\begin{pmatrix}
		c_{1,k}^- \\
		c_{1,k}^+
	\end{pmatrix} \; = \; \begin{pmatrix}
		\tau_{1,k} & \tau_{2,k} + \ii \tau_{3,k} \\
		\tau_{2,k} - \ii \tau_{3,k} & \tau_{4,k}
	\end{pmatrix}
	\begin{pmatrix}
		c_{0,k}^- \\
		c_{0,k}^+
	\end{pmatrix}
\]
with
 \begin{equation*}
  \begin{split}
   \gamma_1\;&=\; \textstyle \frac{|k|}{1+\alpha} \Big( \frac{\,2 \|\Phi_{\alpha,k}\|_{L^2(\mathbb{R}^+)}^2 \,}{\pi (1+\alpha)} \, \tau_{1,k} -1 \Big)\,, \\
   \gamma_2+\ii\gamma_3\;&=\; (\tau_{2,k}+\ii\tau_{3,k})\textstyle\frac{2|k|}{\pi(1+\alpha)^2}\|\Phi_{\alpha,k}\|_{L^2(\mathbb{R}^+)}^2\,, \\
   \gamma_4\;&=\; \textstyle \frac{|k|}{1+\alpha} \Big( \frac{\,2 \|\Phi_{\alpha,k}\|_{L^2}^2 \,}{\pi (1+\alpha)} \, \tau_{4,k} -1 \Big)  \end{split}
 \end{equation*}
(see \eqref{eq:IIkTheorem51} above). Thus,
\[
\tau_{1,k}\;\sim\;|k|^{\frac{2}{1+\alpha}}\,,\qquad   \tau_{2,k}\pm\ii\tau_{3,k}\;\sim\;|k|^{\frac{1-\alpha}{1+\alpha}} \,,\qquad \tau_{4,k}\;\sim\;|k|^{\frac{2}{1+\alpha}}\,,
\]
and 
\[
	\begin{pmatrix}
		c_{1,k}^- \\
		c_{1,k}^+
	\end{pmatrix} \; \sim \; \begin{pmatrix}
		|k|^{\frac{2}{1+\alpha}} & |k|^{\frac{1-\alpha}{1+\alpha}} \\
		|k|^{\frac{1-\alpha}{1+\alpha}} & |k|^{\frac{2}{1+\alpha}}
	\end{pmatrix}
	\begin{pmatrix}
		c_{0,k}^- \\
		c_{0,k}^+
	\end{pmatrix}\;\sim\;
	 \begin{pmatrix}
		|k|^{\frac{5+\alpha}{2(1+\alpha)}} & |k|^{\frac{3-\alpha}{2(1+\alpha)}} \\
		|k|^{\frac{3-\alpha}{2(1+\alpha)}} & |k|^{\frac{5+\alpha}{2(1+\alpha)}}
	\end{pmatrix}
	\begin{pmatrix}
		g_{0,k}^- \\
		g_{0,k}^+
	\end{pmatrix}
\]
at the leading order in $k$, having used \eqref{eq:Relationg0c0} in the last asymptotics. As the above matrix has determinant of leading order $|k|^{\frac{5+\alpha}{1+\alpha}}$, a standard inversion formula yields
\[
 \begin{pmatrix}
		g_{0,k}^- \\
		g_{0,k}^+
	\end{pmatrix}\;\sim\;|k|^{-\frac{5+\alpha}{2(1+\alpha)}}
	 \begin{pmatrix}
		1 & -|k|^{-1} \\
		-|k|^{-1} & 1
	\end{pmatrix}\begin{pmatrix}
		c_{1,k}^- \\
		c_{1,k}^+
	\end{pmatrix},
\]
whence
\[
 |g_{0,k}^-|^2+|g_{0,k}^+|^2\;\lesssim\;|k|^{-\frac{5+\alpha}{1+\alpha}}\big(|c_{0,k}^-|^2+|c_{0,k}^+|^2\big)
\]
at the leading order in $k$. Therefore,
\[
 \begin{split}
   \sum_{k\in\mathbb{Z}\setminus\{0\}}|k|^{\frac{3+\alpha}{1+\alpha}}\, |g_{0,k}^\pm|^2\;\lesssim \sum_{k\in\mathbb{Z}\setminus\{0\}}|k|^{-\frac{2}{1+\alpha}}\, \big(|c_{0,k}^-|^2+|c_{0,k}^+|^2\big)\;<\;+\infty\,,
 \end{split}
\]
having used the a priori bound  \eqref{eq:pileupcond3} for the last step. This establishes the first condition in \eqref{eq:g0g1Sobolev_typeIII}. The second condition follows at once from the first by means of the self-adjointness constraints 
\[
 \begin{split}
  g_{1,k}^-\;&=\;\gamma_1 g_{0,k}^-+(\gamma_2+\ii\gamma_3)g_{0,k}^+\,, \\
  g_{1,k}^+\;&=\;(\gamma_2-\ii\gamma_3)g_{0,k}^-+\gamma_4 g_{0,k}^+
 \end{split}
\]
from Theorem \ref{thm:bifibre-extensions}.
\end{proof}


\begin{remark}[Enhanced summability]\label{rem:enhanced_summability}  Let $(g_k)_{k\in\mathbb{Z}}\in\mathcal{D}(\mathscr{H}_\alpha^*)$. As established in Lemma \ref{lem:gkkrepr}, the coefficients $c_{0,k}$ given by the representation \eqref{eq:repreDHstar}-\eqref{eq:gkrepresentation} of $g_k$ satisfy
 \[
  \sum_{k\in\mathbb{Z}\setminus\{0\}}|k|^{-\frac{2}{1+\alpha}}|c_{0,k}^\pm|^2\,\;<\;+\infty\,.
 \]
 If \emph{in addition}  $(g_k)_{k\in\mathbb{Z}}\in\mathcal{D}(\mathscr{H}_\alpha^{\mathrm{u.f.}})$ for some uniformly-fibred extension of $\mathscr{H}_\alpha$, then Prop.~\ref{prop:g0g1Sobolev} above shows that the coefficients $g_{0,k}$ given by the representation \eqref{eq:gwithPweight_k} of $g_k$ satisfy
 \[
  \sum_{k\in\mathbb{Z}\setminus\{0\}}|k|^{\frac{3+\alpha}{1+\alpha}}|g_{0,k}^\pm|^2\,\;<\;+\infty
 \]
 (this covers also the case when the $g_{0,k}^+$'s or the $g_{0,k}^-$'s are all zero, depending on the considered type of extension). The latter condition, owing to \eqref{eq:Relationg0c0} and hence $g_{0,k}^\pm\sim |k|^{-\frac{1}{2}}c_{0,k}^\pm$, implies
  \[
  \sum_{k\in\mathbb{Z}\setminus\{0\}}|k|^{\frac{2}{1+\alpha}}|c_{0,k}^\pm|^2\,\;<\;+\infty\,.
 \]
 Thus, the condition of belonging to $\mathcal{D}(\mathscr{H}_\alpha^{\mathrm{u.f.}})$, instead of merely to $\mathcal{D}(\mathscr{H}_\alpha^*)$, enhances the summability of the sequence $(c_{0,k}^\pm)_{k\in\mathbb{Z}}$. 
%
\end{remark}

\subsection{Decomposition of the adjoint into singular terms}\label{sec:singular_decomposition_adjoint}

As alluded to at the end of Subsection \ref{sec:genstrategy}, it is shown here that the decomposition induced by \eqref{eq:gwithPweight_k} of a generic element in the domain of a uniformly fibred extension $\mathscr{H}_\alpha^{\mathrm{u.f.}}$, namely
\begin{equation}\label{eq:singular_decomposition_adjoint}
 (g_k)_{k\in\mathbb{Z}}\;=\;(\varphi_k)_{k\in\mathbb{Z}}+\big(g_{1,k}|x|^{1+\frac{\alpha}{2}}P\big)_{k\in\mathbb{Z}}+\big(g_{0,k}|x|^{-\frac{\alpha}{2}}P\big)_{k\in\mathbb{Z}}\,,
\end{equation}
unavoidably displays an annoying form of singularity, which affects the subsequent analysis, in the following precise sense.

\begin{lemma}\label{lem:singular_decomposition_adjoint}
Let $\alpha\in[0,1)$ and let $\mathscr{H}_\alpha^{\mathrm{u.f.}}$ be a uniformly fibred self-adjoint extension. There exists $ (g_k)_{k\in\mathbb{Z}}\in\mathcal{D}(\mathscr{H}_\alpha^{\mathrm{u.f.}})$ such that, with respect to the decomposition \eqref{eq:singular_decomposition_adjoint},
\[
 \begin{split}
   \big(g_{1,k}|x|^{1+\frac{\alpha}{2}}P\big)_{k\in\mathbb{Z}}\;&\notin\;\mathcal{D}(\mathscr{H}_\alpha^*)\,, \\
   \big(g_{0,k}|x|^{-\frac{\alpha}{2}}P\big)_{k\in\mathbb{Z}}\;&\notin\;\mathcal{D}(\mathscr{H}_\alpha^*)\,,
 \end{split}
\]
with the obvious exception of those objects above that are prescribed to be identically zero for all elements of the domain of the considered uniformly fibred extension.
\end{lemma}


 Clearly, the fact that
 \begin{equation}\label{eq:varhisquareintegrable}
  (\varphi_k)_{k\in\mathbb{Z}}\in\ell^2(\mathbb{Z},L^2(\mathbb{R},\ud x))
 \end{equation}
 follows at once by difference from \eqref{eq:singular_decomposition_adjoint}, because owing to Corollary \ref{cor:(g0)k_(g1)k_in_Hs} both $(g_{1,k}|x|^{1+\frac{\alpha}{2}}P)_{k\in\mathbb{Z}}$ and $(g_{0,k}|x|^{-\frac{\alpha}{2}}P)_{k\in\mathbb{Z}}$ belong to $\ell^2(\mathbb{Z},L^2(\mathbb{R},\ud x))$. However, whereas in \eqref{eq:gwithPweight_k} and \eqref{eq:singular_decomposition_adjoint} each $\varphi_k$ belongs to $\mathcal{D}(\overline{A_{\alpha}(k)})$, their collection $(\varphi_k)_{k\in\mathbb{Z}}$ may fail to belong to $\mathcal{D}(\overline{\mathscr{H}_\alpha})$ because it may even fail to belong to $\mathcal{D}(\mathscr{H}_\alpha^*)$!

In preparation for the proof of Lemma \ref{lem:singular_decomposition_adjoint}, a simple computation shows that
\[
 \begin{split}
   A_\alpha^\pm(k)^*\big(|x|^{-\frac{\alpha}{2}}P\big)\;&=\; \alpha|x|^{-(1+\frac{\alpha}{2})}P'-|x|^{-\frac{\alpha}{2}}P''+k^2|x|^{\frac{3\alpha}{2}} P\,, \\
  A_\alpha^\pm(k)^*\big(|x|^{1+\frac{\alpha}{2}}P\big)\;&=\;-(2+\alpha)|x|^{\frac{\alpha}{2}}P'-|x|^{1+\frac{\alpha}{2}}P''+k^2|x|^{1+\frac{5\alpha}{2}} P
 \end{split}
\]
for any $k\in\mathbb{Z}$ and $x\gtrless 0$ depending on the `+' or the `$-$' case. In particular, as the cut-off function $P$ is constantly equal to one when $|x|<1$, 
\begin{equation}\label{eq:action_on_short_x}
 \begin{split}
 \mathbf{1}_{I^{\pm}}(x)A_\alpha^\pm(k)^*\big(|x|^{-\frac{\alpha}{2}}P\big)\;&=\;\mathbf{1}_{I^{\pm}}(x)k^2|x|^{\frac{3\alpha}{2}}\,, \\
 \mathbf{1}_{I^{\pm}}(x) A_\alpha^\pm(k)^*\big(|x|^{1+\frac{\alpha}{2}}P\big)\;&=\;\mathbf{1}_{I^{\pm}}(x) k^2|x|^{1+\frac{5\alpha}{2}}\,,
 \end{split}
\end{equation}
where $I^-:=(-1,0)$ and $I^+:=(0,1)$. One can now see that this implies
\begin{eqnarray}
 \big\|(\mathscr{H}_\alpha^\pm)^*\big(g_{0,k}^{\pm}|x|^{-\frac{\alpha}{2}}P\big)_{k\in\mathbb{Z}}\big\|^2_{\cH^\pm}\!&\geqslant&\!\sum_{k\in\mathbb{Z}}k^4|g_{0,k}^\pm|^2\,, \label{eq:Hastar-shortdist-g0}\\
 \big\|(\mathscr{H}_\alpha^\pm)^*\big(g_{1,k}^{\pm}|x|^{1+\frac{\alpha}{2}}P\big)_{k\in\mathbb{Z}}\big\|^2_{\cH^\pm}\!&\geqslant&\!\sum_{k\in\mathbb{Z}}k^4|g_{1,k}^\pm|^2\,.  \label{eq:Hastar-shortdist-g1}
\end{eqnarray}
Indeed,
\[
 \begin{split}
  \big\|(\mathscr{H}_\alpha^+)^*\big(g_{1,k}^{+}x^{1+\frac{\alpha}{2}}P\big)_{k\in\mathbb{Z}}\big\|_{\cH^+}^2\;&=\;\sum_{k\in\mathbb{Z}}\big\|A_\alpha^+(k)^*\big(g_{1,k}^+x^{1+\frac{\alpha}{2}}P\big)\big\|^2_{L^2(\mathbb{R}^+,\ud x)} \\
  &\geqslant\;\sum_{k\in\mathbb{Z}}\big\|g_{1,k}^+ k^2 x^{1+\frac{5\alpha}{2}}\big\|^2_{L^2((0,1),\ud x)} \\
  &=\;(3+5\alpha)^{-1}\sum_{k\in\mathbb{Z}}k^4|g_{1,k}^+|^2\,,
 \end{split}
\]
where \eqref{eq:Halphaadj-decomposable} was used in the first step and \eqref{eq:action_on_short_x} in the second; all other cases for \eqref{eq:Hastar-shortdist-g0}-\eqref{eq:Hastar-shortdist-g1} are obtained in a completely analogous way.

%
%
%
%
 
\begin{proof}[Proof of Lemma \ref{lem:singular_decomposition_adjoint}]
 The proof consists of a discussion, case by case, of all possible types of uniformly fibred extensions.\index{fibred extensions!uniformly} For arbitrary $\varepsilon>0$ let
\[
 \begin{split}
  a_k(\varepsilon)\;&:=\;
  \begin{cases}
   \;|k|^{\frac{1}{1+\alpha}-\frac{1}{2}(1+\varepsilon)}\,, & \;\textrm{ if }k\in\mathbb{Z}\setminus\{0\}\,, \\
   \qquad 0\,, & \;\textrm{ if }k=0\,,
  \end{cases} \\
  b_k(\varepsilon)\;&:=\;
  \begin{cases}
   \;|k|^{-\frac{1}{1+\alpha}-\frac{1}{2}(1+\varepsilon)}\,, & \textrm{if }k\in\mathbb{Z}\setminus\{0\} \,,\\
   \qquad 0\,, & \textrm{if }k=0\,.
  \end{cases}
 \end{split}
\]

(i) Friedrichs extension $\mathscr{H}_{\alpha,\mathrm{F}}=\bigoplus_{k\in\mathbb{Z}}A_{\alpha,\mathrm{F}}(k)$.
For this case one chooses $(g_k)_{k\in\mathbb{Z}}$ with
\[
 g_k\;:=\;\begin{pmatrix}
  a_k(\varepsilon)\,\widetilde{\Psi}_{\alpha,k} \\
  a_k(\varepsilon)\,\widetilde{\Psi}_{\alpha,k}
 \end{pmatrix}.
\]
With respect to the representation \eqref{eq:gkrepresentation}, $c_{0,k}^\pm=0$ and $c_{1,k}^\pm=a_k(\varepsilon)$. Therefore,
\[
 \sum_{k\in\mathbb{Z}}|k|^{-\frac{2}{1+\alpha}}|c_{1,k}^\pm|^2\;=\;\sum_{k\in\mathbb{Z}\setminus\{0\}}|k|^{-1-\varepsilon}\;<\;+\infty
\]
and, owing to Lemma \ref{lem:gkkrepr}, $(g_k)_{k\in\mathbb{Z}}\in\mathcal{D}(\mathscr{H}_\alpha^*)$.
Moreover, by construction $g_k$ satisfies the conditions of self-adjointness characterising $\mathcal{D}(A_{\alpha,\mathrm{F}}(k))$ stated in Theorem \ref{thm:bifibre-extensionsc0c1}; thus, $(g_k)_{k\in\mathbb{Z}}\in\mathcal{D}(\mathscr{H}_{\alpha,\mathrm{F}})$. Expressing now $(g_k)_{k\in\mathbb{Z}}$ in the representation \eqref{eq:singular_decomposition_adjoint}, formulas \eqref{eq:Relationg0c0}-\eqref{eq:Relationg1c1} yield
\[
 g_{0,k}^\pm\;=\;0\,,\qquad g_{1,k}^{\pm}\;\sim\;|k|^{-\frac{1}{2}(\frac{2}{1+\alpha}+\varepsilon)}\quad(k\neq 0)\,,
\]
whence
\[
 \sum_{k\in\mathbb{Z}\setminus\{0\}}k^4|g_{1,k}^\pm|^2\;\sim\sum_{k\in\mathbb{Z}\setminus\{0\}}|k|^{\frac{2+4\alpha}{1+\alpha}-\varepsilon}\;=\;+\infty\quad\Leftrightarrow\quad\varepsilon\in(0,{\textstyle\frac{3+5\alpha}{1+\alpha}}]\,.
\]
Thus, for $\varepsilon\in(0,{\textstyle\frac{3+5\alpha}{1+\alpha}}]$, one deduces from \eqref{eq:Hastar-shortdist-g1} that $(g_{1,k}|x|^{1+\frac{\alpha}{2}}P)_{k\in\mathbb{Z}}\notin\mathcal{D}(\mathscr{H}_\alpha^*)$.

(ii) Extensions of type $\mathrm{I}_{\mathrm{R}}$: for $\gamma\in\mathbb{R}$ consider $\mathscr{H}_{\alpha,\mathrm{R}}^{[\gamma]}=\bigoplus_{k\in\mathbb{Z}}A_{\alpha,\mathrm{R}}^{[\gamma]}(k)$. For this case one chooses $(g_k)_{k\in\mathbb{Z}}$ with
\[
 g_k\;:=\;\begin{pmatrix}
		a_k(\varepsilon) \widetilde{\Psi}_{\alpha,k}\\
		\beta_k b_k(\varepsilon) \widetilde{\Psi}_{\alpha,k}+b_k(\varepsilon) \widetilde{\Phi}_{\alpha,k}
	\end{pmatrix}
\]
and $\beta_k$ given by
\[
 \gamma \; = \; \textstyle\frac{|k|}{1+\alpha}\Big(\frac{\, 2 \|\Phi_{\alpha,k}\|_{L^2(\mathbb{R}^+)}^2 \,}{\pi(1+\alpha)}\, \beta_k-1\Big) \, .
\]
From \eqref{eq:Phinorm}, $\|\Phi_{\alpha,k}\|_{L^2(\mathbb{R}^+)}^2\sim|k|^{-\frac{2}{1+\alpha}}$ (for some multiplicative $\alpha$-dependent constant), whence $\beta_k\sim |k|^{\frac{2}{1+\alpha}}$ at the leading order in $k\in\mathbb{Z}\setminus\{0\}$. With respect to the representation \eqref{eq:gkrepresentation},
\[
 \begin{array}{rclcrcl}
  c_{0,k}^- \! &=& \! 0\,, & \quad & c_{1,k}^- \! &=& \! a_k(\varepsilon)\;=\;|k|^{\frac{1}{1+\alpha}-\frac{1}{2}(1+\varepsilon)}\,, \\
  c_{0,k}^+ \! &=& \! b_k(\varepsilon)\;=\;|k|^{-\frac{1}{1+\alpha}-\frac{1}{2}(1+\varepsilon)}\,, & \; & c_{1,k}^+ \! &=& \! \beta_k b_k(\varepsilon)\;\sim\;|k|^{\frac{1}{1+\alpha}-\frac{1}{2}(1+\varepsilon)}\,,
 \end{array}
\]
at the leading order in $k\in\mathbb{Z}\setminus\{0\}$, whereas all the above coefficients vanish for $k=0$. Therefore,
\[
 \begin{split}
  \sum_{k\in\mathbb{Z}}|k|^{-\frac{2}{1+\alpha}}|c_{0,k}^+|^2\;&=\sum_{k\in\mathbb{Z}\setminus\{0\}}|k|^{-\frac{4}{1+\alpha}-1-\varepsilon}\;<\;+\infty\,, \\
  \sum_{k\in\mathbb{Z}}|k|^{-\frac{2}{1+\alpha}}|c_{1,k}^\pm|^2\;&= \sum_{k\in\mathbb{Z}\setminus\{0\}}|k|^{-1-\varepsilon}\;<\;+\infty\,,
 \end{split}
\]
which implies, owing to Lemma \ref{lem:gkkrepr}, that $(g_k)_{k\in\mathbb{Z}}\in\mathcal{D}(\mathscr{H}_\alpha^*)$.
Moreover, by construction $g_k$ satisfies the conditions of self-adjointness characterising $\mathcal{D}(A_{\alpha,\mathrm{R}}^{[\gamma]}(k))$ stated in Theorem \ref{thm:bifibre-extensionsc0c1}; thus, $(g_k)_{k\in\mathbb{Z}}\in\mathcal{D}(\mathscr{H}_{\alpha,\mathrm{R}}^{[\gamma]})$. Expressing now $(g_k)_{k\in\mathbb{Z}}$ in the representation \eqref{eq:singular_decomposition_adjoint}, formulas \eqref{eq:Relationg0c0}-\eqref{eq:Relationg1c1} yield
\[
 \begin{array}{rclcrcl}
  g_{0,k}^- \! &=& \! 0\,, & \quad & g_{1,k}^- \! &\sim& \! |k|^{-\frac{1}{2}(\frac{2}{1+\alpha}+\varepsilon)} \,, \\
  g_{0,k}^+ \! &\sim& \!|k|^{-\frac{1}{2}(\frac{4+2\alpha}{1+\alpha}+\varepsilon)}\,, & \; & g_{1,k}^+ \! &\sim& \!|k|^{-\frac{1}{2}(\frac{4+2\alpha}{1+\alpha}+\varepsilon)}\,,
 \end{array}
\]
for $k\in\mathbb{Z}\setminus\{0\}$, up to multiplicative pre-factors depending on $\alpha$ and $\gamma$ only, all the above coefficients vanishing for $k=0$. From this one obtains
\[
 \begin{split}
  \sum_{k\in\mathbb{Z}}k^4|g_{0,k}^+|^2\;&\sim\sum_{k\in\mathbb{Z}\setminus\{0\}}|k|^{\frac{2\alpha}{1+\alpha}-\varepsilon}\;=\;+\infty\quad\Leftrightarrow\quad\varepsilon\in(0,{\textstyle\frac{1+3\alpha}{1+\alpha}}]\,, \\
 \sum_{k\in\mathbb{Z}}k^4|g_{1,k}^+|^2\;&\sim\sum_{k\in\mathbb{Z}\setminus\{0\}}|k|^{\frac{2\alpha}{1+\alpha}-\varepsilon}\;=\;+\infty\quad\Leftrightarrow\quad\varepsilon\in(0,{\textstyle\frac{1+3\alpha}{1+\alpha}}]\,, \\
  \sum_{k\in\mathbb{Z}}k^4|g_{1,k}^-|^2\;&\sim\sum_{k\in\mathbb{Z}\setminus\{0\}}|k|^{\frac{2+4\alpha}{1+\alpha}-\varepsilon}\;=\;+\infty\quad\Leftrightarrow\quad\varepsilon\in(0,{\textstyle\frac{3+5\alpha}{1+\alpha}}]\,.
 \end{split}
\]
Thus, for $\varepsilon\in(0,{\textstyle\frac{1+3\alpha}{1+\alpha}}]$, one deduces from \eqref{eq:Hastar-shortdist-g0}-\eqref{eq:Hastar-shortdist-g1} that $(g_{0,k}|x|^{-\frac{\alpha}{2}}P)_{k\in\mathbb{Z}}\notin\mathcal{D}(\mathscr{H}_\alpha^*)$ and $(g_{1,k}|x|^{1+\frac{\alpha}{2}}P)_{k\in\mathbb{Z}}\notin\mathcal{D}(\mathscr{H}_\alpha^*)$.

(iii)  Extensions of type $\mathrm{I}_{\mathrm{L}}$: for $\gamma\in\mathbb{R}$ consider $\mathscr{H}_{\alpha,\mathrm{L}}^{[\gamma]}=\bigoplus_{k\in\mathbb{Z}}A_{\alpha,\mathrm{L}}^{[\gamma]}(k)$. For this case one chooses $(g_k)_{k\in\mathbb{Z}}$ with
\[
 g_k\;:=\;\begin{pmatrix}
		\beta_k b_k(\varepsilon) \widetilde{\Psi}_{\alpha,k}+b_k(\varepsilon) \widetilde{\Phi}_{\alpha,k} \\
		a_k(\varepsilon) \widetilde{\Psi}_{\alpha,k}
	\end{pmatrix},
\]
with the same $\beta_k$ as in case (ii). With the obvious exchange of `+' and `-' components, the reasoning is the same as in case (ii).

(iv) Extensions of type $\mathrm{II}_a$ for given $a\in\mathbb{C}\setminus\{0\}$: for $\gamma\in\mathbb{R}$ consider $ \mathscr{H}_{\alpha,a}^{[\gamma]}=\bigoplus_{k\in\mathbb{Z}}A_{\alpha,a}^{[\gamma]}(k)$.
For this case one chooses $(g_k)_{k\in\mathbb{Z}}$ with
\[
 g_k\;:=\;\begin{pmatrix}
		\big(\tau_k b_k(\varepsilon) + a_k(\varepsilon)\big) \widetilde{\Psi}_{\alpha,k}+b_{k}(\varepsilon)\widetilde{\Phi}_{\alpha,k} \\
		\big(\tau_k a b_k(\varepsilon)- \overline{a}^{-1} a_k(\varepsilon)\big)\widetilde{\Psi}_{\alpha,k}+a b_{k}(\varepsilon)\widetilde{\Phi}_{\alpha,k}
	\end{pmatrix}
\]
and $\tau_k$ given by 
\[
			  \gamma\;=\;\textstyle{(1+|a|^2) \frac{|k|}{1+\alpha}} \Big( \frac{\,2 \|\Phi_{\alpha,k}\|_{L^2(\mathbb{R}^+)}^2 \,}{\pi (1+\alpha)} \, \tau_k -1 \Big)\,.
\]
In particular, $\tau_k\sim |k|^{\frac{2}{1+\alpha}}$ at the leading order in $k\in\mathbb{Z}\setminus\{0\}$. With respect to the representation \eqref{eq:gkrepresentation},
\[
  c_{0,k}^\pm\;\sim\;|k|^{-\frac{1}{1+\alpha}-\frac{1}{2}(1+\varepsilon)} \,,\qquad c_{1,k}^\pm\;\sim\;|k|^{\frac{1}{1+\alpha}-\frac{1}{2}(1+\varepsilon)}
\]
at the leading order in $k\in\mathbb{Z}\setminus\{0\}$, whereas all the above coefficients vanish for $k=0$. Therefore,
\[
 \begin{split}
  \sum_{k\in\mathbb{Z}}|k|^{-\frac{2}{1+\alpha}}|c_{0,k}^\pm|^2\;&=\sum_{k\in\mathbb{Z}\setminus\{0\}}|k|^{-\frac{4}{1+\alpha}-1-\varepsilon}\;<\;+\infty\,, \\
  \sum_{k\in\mathbb{Z}}|k|^{-\frac{2}{1+\alpha}}|c_{1,k}^\pm|^2\;&= \sum_{k\in\mathbb{Z}\setminus\{0\}}|k|^{-1-\varepsilon}\;<\;+\infty\,, 
 \end{split}
\]
which implies, owing to Lemma \ref{lem:gkkrepr}, that $(g_k)_{k\in\mathbb{Z}}\in\mathcal{D}(\mathscr{H}_\alpha^*)$.
Moreover, by construction $g_k$ satisfies the conditions of self-adjointness characterising $\mathcal{D}(A_{\alpha,a}^{[\gamma]}(k))$ stated in Theorem \ref{thm:bifibre-extensionsc0c1}; thus, $(g_k)_{k\in\mathbb{Z}}\in\mathcal{D}(\mathscr{H}_{\alpha,a}^{[\gamma]})$. Expressing now $(g_k)_{k\in\mathbb{Z}}$ in the representation \eqref{eq:singular_decomposition_adjoint}, formulas \eqref{eq:Relationg0c0}-\eqref{eq:Relationg1c1} yield
\[
 g_{0,k}^\pm\;\sim\;|k|^{-\frac{1}{2}(\frac{4+2\alpha}{1+\alpha}-\varepsilon)}\,,\qquad g_{1,k}^\pm\;\sim\;|k|^{-\frac{1}{2}(\frac{2}{1+\alpha}+\varepsilon)}
\]
at the leading order in $k\in\mathbb{Z}\setminus\{0\}$, all the above coefficients vanishing for $k=0$. From this one obtains
\[
 \begin{split}
  \sum_{k\in\mathbb{Z}}k^4|g_{0,k}^\pm|^2\;&\sim\sum_{k\in\mathbb{Z}\setminus\{0\}}|k|^{\frac{2\alpha}{1+\alpha}-\varepsilon}\;=\;+\infty\quad\Leftrightarrow\quad\varepsilon\in(0,{\textstyle\frac{1+3\alpha}{1+\alpha}}]\,, \\
  \sum_{k\in\mathbb{Z}}k^4|g_{1,k}^\pm|^2\;&\sim\sum_{k\in\mathbb{Z}\setminus\{0\}}|k|^{\frac{2+4\alpha}{1+\alpha}-\varepsilon}\;=\;+\infty\quad\Leftrightarrow\quad\varepsilon\in(0,{\textstyle\frac{3+5\alpha}{1+\alpha}}]\,.
 \end{split}
\]
Thus, for $\varepsilon\in(0,{\textstyle\frac{1+3\alpha}{1+\alpha}}]$, one deduces from \eqref{eq:Hastar-shortdist-g0}-\eqref{eq:Hastar-shortdist-g1} that $(g_{0,k}|x|^{-\frac{\alpha}{2}}P)_{k\in\mathbb{Z}}\notin\mathcal{D}(\mathscr{H}_\alpha^*)$ and $(g_{1,k}|x|^{1+\frac{\alpha}{2}}P)_{k\in\mathbb{Z}}\notin\mathcal{D}(\mathscr{H}_\alpha^*)$.

(v) Extensions of type $\mathrm{III}$: for $\Gamma\in\mathbb{R}^4$ consider $  \mathscr{H}_{\alpha}^{[\Gamma]}=\bigoplus_{k\in\mathbb{Z}}A_{\alpha}^{[\Gamma]}(k)$.
For this case one chooses $(g_k)_{k\in\mathbb{Z}}$ with
\[
 g_k\;:=\;\begin{pmatrix}
		\big(\tau_{1,k}+\tau_{2,k}+ \ii \tau_{3,k}\big)b_{k}(\varepsilon) \widetilde{\Psi}_{\alpha,k}+b_{k}(\varepsilon)\widetilde{\Phi}_{\alpha,k} \\
		\big(\tau_{2,k}- \ii \tau_{3,k}+\tau_{4,k}\big)b_{k}(\varepsilon)\widetilde{\Psi}_{\alpha,k}+b_{k}(\varepsilon)\widetilde{\Phi}_{\alpha,k}
	\end{pmatrix}
\]
and $(\tau_{1,k},\tau_{2,k},\tau_{3,k},\tau_{4,k})$ given by
\[
  \begin{split}
   \gamma_1\;&=\; \textstyle \frac{|k|}{1+\alpha} \Big( \frac{\,2 \|\Phi_{\alpha,k}\|_{L^2(\mathbb{R}^+)}^2 \,}{\pi (1+\alpha)} \, \tau_{1,k} -1 \Big)\,, \\
   \gamma_2+\ii\gamma_3\;&=\; (\tau_{2,k}+\ii\tau_{3,k})\textstyle\frac{2|k|}{\pi(1+\alpha)^2}\|\Phi_{\alpha,k}\|_{L^2(\mathbb{R}^+)}^2\,, \\
   \gamma_4\;&=\; \textstyle \frac{|k|}{1+\alpha} \Big( \frac{\,2 \|\Phi_{\alpha,k}\|_{L^2(\mathbb{R}^+)}^2 \,}{\pi (1+\alpha)} \, \tau_{4,k} -1 \Big).  \end{split}
\]
In particular, 
\[
\tau_{1,k}\;\sim\;|k|^{\frac{2}{1+\alpha}}\,,\qquad   \tau_{2,k}\pm\ii\tau_{3,k}\;\sim\;|k|^{\frac{1-\alpha}{1+\alpha}} \,,\qquad \tau_{4,k}\;\sim\;|k|^{\frac{2}{1+\alpha}}\,,
\]
at the leading order in $k\in\mathbb{Z}\setminus\{0\}$. With respect to the representation \eqref{eq:gkrepresentation},
\[
  c_{0,k}^\pm\;\sim\;|k|^{-\frac{1}{1+\alpha}-\frac{1}{2}(1+\varepsilon)} \,,\qquad c_{1,k}^\pm\;\sim\;|k|^{\frac{1}{1+\alpha}-\frac{1}{2}(1+\varepsilon)}
\]
at the leading order in $k\in\mathbb{Z}\setminus\{0\}$, whereas all the above coefficients vanish for $k=0$. 
From this point one repeats verbatim the reasoning of part (iv).
%
\end{proof}

\subsection{Detecting short-scale asymptotics and regularity}\label{subsec:auxiliaryLemma}

 As observed with \eqref{eq:varhisquareintegrable}, $\mathcal{F}_2^{-1}$ is applicable to $(\varphi_k)_{k\in\mathbb{Z}}$ and thus \eqref{eq:F2-1phi} defines a function $\varphi\in L^2(\mathbb{R}\times\mathbb{S}^1,\ud x\ud y)$. The next step in the strategy outlined in Subsect.~\ref{sec:genstrategy} is to show convenient short-scale asymptotics as $x\to 0$ for $\varphi(x,y)$ and $\partial_x\varphi(x,y)$.

 Evidently, the possibility that $\varphi\notin\mathcal{F}_2^{-1}\mathcal{D}(\mathscr{H}_\alpha^*)=\mathcal{D}(\mathsf{H}_\alpha^*)$ (Lemma \ref{lem:singular_decomposition_adjoint}) complicates this analysis: no regularity or short-scale asymptotics of the elements of $\mathcal{D}(\mathsf{H}_\alpha^*)$ can be claimed a priori for $\varphi$.

 For the above purposes the following auxiliary result will be used.

 \begin{lemma}\label{lem:grand_auxiliary_lemma}
  Let $\alpha\in[0,1)$ and let $R:(0,1)\times\mathbb{S}^1\to\mathbb{C}$ be a function such that
  \begin{enumerate}
   \item[(a)] $\big\| x^{-(\frac{3}{2}+\frac{\alpha}{2})}R\,\big\|_{L^2((0,1)\times\mathbb{S}^1,\ud x\ud y)}\;<\;+\infty$\,,
   \item[(b)] $\big\| \partial_x^2R\,\big\|_{L^2((0,1)\times\mathbb{S}^1,\ud x\ud y)}\;<\;+\infty$\,.
  \end{enumerate}
 Then for almost every $y\in\mathbb{S}^1$ the function $(0,1)\ni x\mapsto R(x,y)$ belongs to $H^2_0((0,1])$ and as such it satisfies the following properties:
 \begin{enumerate}
  \item[(i)] $R(\cdot,y)\in C^1(0,1)$,
  \item[(ii)] $R(x,y)\stackrel{x\downarrow 0}{=}o(x^{{\frac{3}{2}}})$,
  \item[(iii)] $\partial_x R(x,y)\stackrel{x\downarrow 0}{=}o(x^{{\frac{1}{2}}})$.
 \end{enumerate}  
 \end{lemma}

 \begin{remark}
  $H^2_0((0,1])$ in the statement of Lemma \ref{lem:grand_auxiliary_lemma} is the closure of $C^\infty_0((0,1])$ in the $H^2$-norm. The edge $x=1$ is included so as to mean that there is no vanishing constraint at $x=1$ for the elements of $H^2_0((0,1])$ and their derivatives: only vanishing as $x\downarrow 0$ emerges, in the form of conditions (ii) and (iii).
 \end{remark}

 \begin{proof}[Proof of Lemma \ref{lem:grand_auxiliary_lemma}]
  Assumption (a) in Lemma \ref{lem:grand_auxiliary_lemma} implies that $R(\cdot,y)\in L^2((0,1))$, and hence together with (b) it implies that $R(\cdot,y)\in H^2((0,1))$ for a.e.~$y\in\mathbb{S}^1$.
  Therefore $R(\cdot,y)=a_y+b_yx+r_y(x)$ for a.e.~$y\in\mathbb{S}^1$, for some $a_y,b_y\in\mathbb{C}$ and $r_y\in H^2_0((0,1])$. For compatibility with assumption (a), necessarily $a_y=b_y=0$, whence $R(\cdot,y)\in H^2_0((0,1])$ for a.e.~$y\in\mathbb{S}^1$.  
 \end{proof}

 The application of Lemma \ref{lem:grand_auxiliary_lemma} to the present context proceeds as follows.

 First of all, since the goal is to characterise for fixed $y\in\mathbb{S}^1$ the behaviour and the regularity of $x\mapsto\varphi(x,y)$ as $x\to 0$ from \emph{each side} of the singular point $x=0$, it suffices to analyse the case $x>0$; then completely analogous conclusions are obtained for $x<0$. Lemma \ref{lem:grand_auxiliary_lemma} is thus meant to be applied to the restriction $R(x,y)=\varphi(x,y)\mathbf{1}_{(0,1)}(x)$.

 In fact, since in general $\varphi\in L^2(\mathbb{R}\times\mathbb{S}^1,\ud x\ud y)\setminus\mathcal{D}(\mathsf{H}_\alpha^*)$, it is not possible to check the assumptions (a) and (b) of Lemma \ref{lem:grand_auxiliary_lemma} directly for $\varphi$. One has to opt instead for splitting $\varphi$ into a component in $\mathcal{D}(\overline{\mathsf{H}_\alpha})$ plus a remainder, the explicit form of which will allow to apply Lemma \ref{lem:grand_auxiliary_lemma}.

 This idea is implicit in the very choice of $(\varphi_k)_{k\in\mathbb{Z}}$ made in \eqref{eq:gwithPweight_k}. Recall that for given $(g_k)_{k\in\mathbb{Z}}$ one could represent
 \[
  g_k^{\pm}\;=\;\varphi_k^{\pm}+g_{1,k}^{\pm}|x|^{1+\frac{\alpha}{2}}P+g_{0,k}^{\pm}|x|^{-\frac{\alpha}{2}}P\,,
 \]
 and also
 \[
  g_k^{\pm}\;=\;\widetilde{\varphi}_k^{\pm}+c_{1,k}^{\pm}\widetilde{\Psi}_{\alpha,k}+c_{0,k}^{\pm}\widetilde{\Phi}_{\alpha,k}\,,
 \]
 where
  \begin{equation}\label{eq:phitildeinDclosure}
  (\widetilde{\varphi}_k^{\pm})_{k\in\mathbb{Z}}\;\in\;\mathcal{D}\bigg(\bigoplus_{k\in\mathbb{Z}}\overline{A_\alpha^\pm(k)}\bigg)\;=\;\mathcal{D}\big(\overline{\mathscr{H}_\alpha^\pm}\big)
 \end{equation}
 Moreover, as argued in the proof of Theorem \ref{prop:g_with_Pweight}, for each $k\in\mathbb{Z}\setminus\{0\}$ one can split
 \begin{equation}\label{eq:phi=phitilde+q}
  \varphi_k^{\pm}\;=\;\widetilde{\varphi}_k^{\pm}+\vartheta_k^{\pm}\,,
 \end{equation}
 while keeping
 \begin{equation}\label{eq:varthetazeromode}
  \widetilde{\varphi}_0^{\pm}\;\equiv\; \varphi_0^{\pm}\qquad\textrm{ and }\qquad \vartheta_0^{\pm}\;\equiv\; 0\qquad\textrm{ when }\qquad k=0\,,
 \end{equation}
 where
 \begin{equation}\label{eq:q=q0+q1}
  \vartheta_k^{\pm}\;=\; \vartheta_{0,k}^{\pm}+\vartheta_{1,k}^{\pm} 
 \end{equation}
 with
 \begin{eqnarray}
  & & \vartheta_{0,k}^{\pm}\;:=\;c_{0,k}^\pm\Big(\widetilde{\Phi}_{\alpha,k}-{\textstyle\sqrt{\frac{\pi(1+\alpha)}{2|k|}}|x|^{-\frac{\alpha}{2}}P+\sqrt{\frac{\pi|k|}{2(1+\alpha)}}\,|x|^{1+\frac{\alpha}{2}} P}\Big) \, , \label{eq:defq0q1decomp-q0} \\
  & & \vartheta_{1,k}^{\pm}\;:=\;c_{1,k}^\pm\Big(\widetilde{\Psi}_{\alpha,k}-{\textstyle\sqrt{\frac{2|k|}{\pi(1+\alpha)^3}}}\,\|\Phi_{\alpha,k}\|_{L^2(\mathbb{R}^+)}^2\,|x|^{1+\frac{\alpha}{2}}P\Big) \, , \label{eq:defq0q1decomp-q1}
 \end{eqnarray}
 and 
  \begin{equation}\label{eq:regularityoftheta01}
  \vartheta_{0,k}^{\pm},\vartheta_{1,k}^{\pm}\;\in\;\mathcal{D}\big(\overline{A^\pm_\alpha(k)}\big)\;=\;H^2_0(\mathbb{R}^\pm)\cap L^2(\mathbb{R}^\pm,\langle x\rangle^{4\alpha}\,\ud x)\,.
 \end{equation}
 It is important to remember, for later convenience, that the zero mode is all cast into $\widetilde{\varphi}_0^{\pm}\equiv \varphi_0^{\pm}$, hence $(\vartheta_k)_{k\in\mathbb{Z}}\equiv(\vartheta_k)_{k\in\mathbb{Z}\setminus\{0\}}$.

 The decomposition \eqref{eq:phi=phitilde+q}-\eqref{eq:defq0q1decomp-q1} induces the splitting
 \begin{equation}\label{eq:fftk}
  (\varphi_k)_{k\in\mathbb{Z}}\;=\;(\widetilde{\varphi}_k)_{k\in\mathbb{Z}}+(\vartheta_k)_{k\in\mathbb{Z}}
 \end{equation}
 as an identity in $\ell^2(\mathbb{Z},L^2(\mathbb{R}^+,\ud x))$, where $(\vartheta_k)_{k\in\mathbb{Z}}$ does not necessarily belong to $\mathcal{D}(\mathscr{H}_\alpha^*)$, as $(\varphi_k)_{k\in\mathbb{Z}}$ does not either (Lemma \ref{lem:singular_decomposition_adjoint}).
 In turn, owing to \eqref{eq:varhisquareintegrable} and \eqref{eq:phitildeinDclosure}, the identity \eqref{eq:fftk} yields the splitting
 \begin{equation}\label{eq:splittingphiphitildetheta}
  \varphi(x,y)\;=\;\widetilde{\varphi}(x,y)+\vartheta(x,y)\,,\qquad(x,y)\in\mathbb{R}\times\mathbb{S}^1\,,
 \end{equation}
 with
 \begin{eqnarray}
    & & \widetilde{\varphi}\;:=\;\mathcal{F}_2^{-1}(\widetilde{\varphi}_k)_{k\in\mathbb{Z}}\;\in\;\mathcal{F}_2^{-1}\mathcal{D}\big(\overline{\mathscr{H}_\alpha^\pm}\big)\;=\;\mathcal{D}(\overline{\mathsf{H}_\alpha})\,, \\
    & & \vartheta\, \;:=\; \mathcal{F}_2^{-1} (\vartheta_k)_{k\in\mathbb{Z}}\;\in\;L^2(\mathbb{R}\times\mathbb{S}^1,\ud x\ud y)\,. \label{eq:deffunctionvartheta}
 \end{eqnarray}
  Here $\vartheta$ may fail to belong to $\mathcal{D}(\mathsf{H}_\alpha^*)$, precisely as $\varphi$.

 The explicit information that $\widetilde{\varphi}\in\mathcal{D}(\overline{\mathsf{H}_\alpha})$ and the explicit expression for $\vartheta$ will finally allow one to apply Lemma \ref{lem:grand_auxiliary_lemma} separately to each of them. This is the object of Subsections \ref{sec:control-of-tildephi} and \ref{sec:q0q1}.

\subsection{Control of $\widetilde{\varphi}$}\label{sec:control-of-tildephi}

 The focus here is the regularity and the behaviour as $x\to 0^\pm$ of the functions belonging to $\mathcal{D}\big(\overline{\mathsf{H}_\alpha^\pm}\big)$.

Clearly, from \eqref{eq:explicit-tildeHalpha},
\begin{equation}\label{eq:DHaclosedclosure}
 \mathcal{D}\big(\overline{\mathsf{H}_\alpha^\pm}\big)\;=\;\overline{C^\infty_c(\mathbb{R}^\pm_x\times\mathbb{S}^1_y)}^{\|\,\|_{\Gamma(\mathsf{H}_\alpha)}}\,,
\end{equation}
 $\|h\|_{\Gamma(\mathsf{H}_\alpha)}:=\big(\|h\|_{L^2(\mathbb{R}^\pm_x\times\mathbb{S}^1_y)}^2+\|\mathsf{H}_\alpha^\pm h\|_{L^2(\mathbb{R}^\pm_x\times\mathbb{S}^1_y)}^2\big)^{\frac{1}{2}}$ being the graph norm (Sect.~\ref{sec:I-preliminaries}).

 Recall also, from $\overline{\mathsf{H}_\alpha^\pm}\subset(\mathsf{H}_\alpha^\pm)^*$ and from \eqref{eq:HHalphaadjoint}, that 
\begin{equation}\label{eq:actionofHaclosed}
 \overline{\mathsf{H}_\alpha^\pm}\,\widetilde{\varphi}^\pm\;=\;\Big(-\frac{\partial^2}{\partial x^2}- |x|^{2\alpha}\frac{\partial^2}{\partial y^2}+\frac{\,C_\alpha}{|x|^2}\Big)\widetilde{\varphi}^\pm\qquad\forall\widetilde{\varphi}^\pm\in\mathcal{D}\big(\overline{\mathsf{H}_\alpha^\pm}\big)\,.
\end{equation}

The main result here is the following.

\begin{proposition}\label{prop:Hclosurecontrol}
 Let $\alpha\in[0,1)$. There exists a constant $K_\alpha>0$ such that for any $\widetilde{\varphi}^\pm\in\mathcal{D}(\overline{\mathsf{H}_\alpha^\pm})$ one has
 \begin{equation}\label{eq:Hclosurecontrol}
  \big\|\,|x|^{-2}\widetilde{\varphi}^\pm\,\big\|_{L^2(\mathbb{R}^\pm_x\times\mathbb{S}^1_y)}+\big\|\partial_x^2\widetilde{\varphi}^\pm\big\|_{L^2(\mathbb{R}^\pm_x\times\mathbb{S}^1_y)}\;\leqslant\;K_\alpha\,\big\|\overline{\mathsf{H}_\alpha^\pm}\,\widetilde{\varphi}^\pm\big\|_{L^2(\mathbb{R}^\pm_x\times\mathbb{S}^1_y)}\,.
 \end{equation}
 When $\alpha\uparrow 1$, then $K_\alpha\to +\infty$. As a consequence, $\widetilde{\varphi}^\pm$ satisfies the assumptions of Lemma \ref{lem:grand_auxiliary_lemma} and therefore, for almost every $y\in\mathbb{S}^1$,
 \begin{enumerate}
  \item[(i)] the function $x\mapsto\widetilde{\varphi}^\pm(x,y)$ belongs to $C^1(0,1)$,
  \item[(ii)] $\widetilde{\varphi}^\pm(x,y)=o(|x|^{\frac{3}{2}})$ as $x\to 0^\pm$,  
  \item[(iii)] $\partial_x\widetilde{\varphi}^\pm(x,y)=o(|x|^{\frac{1}{2}})$ as $x\to 0^\pm$.
 \end{enumerate} 
\end{proposition}

As one only needs information on the limit separately from each side of the singularity, it is enough to consider the `+' case: the same conclusions will apply also to the `-' case. Thus, in the remaining part of this Subsection,  it is convenient to simply write $\widetilde{\varphi}$ for $\widetilde{\varphi}^+\in\mathcal{D}(\overline{\mathsf{H}_\alpha^+})$.

The proof of Proposition \ref{prop:Hclosurecontrol} relies on two technical estimates. The first is an iterated version of the standard one-dimensional Hardy inequality\index{Hardy inequality} \eqref{eq:hardy1}, namely
\begin{equation}
\Vert r^{-1} h \Vert_{L^2(\mathbb{R}^+,\ud r)} \;\leqslant\; 2 \,\Vert h' \Vert_{L^2(\mathbb{R}^+,\ud r)}\qquad \forall\,h \in C^\infty_0(\mathbb{R}^+)\,.
\end{equation}

\begin{lemma}[Rellich inequality]\label{lem:doubleHardy}
For any $h\in C^\infty_c(\mathbb{R}^+)$ one has
\begin{equation}\label{eq:hardy2}
 \|r^{-2}h\|_{L^2(\mathbb{R}^+,\ud r)}\;\leqslant\;\frac{4}{3}\,\|h''\|_{L^2(\mathbb{R}^+,\ud r)}\,.
 \end{equation} 
\end{lemma}

\begin{corollary}\label{cor:doubleHardy}
 Let $\widetilde{\varphi}\in C^\infty_c(\mathbb{R}^+_x\times \mathbb{S}^1_y)$. Then
 \begin{equation}
  \|x^{-2}\widetilde{\varphi}\|_{L^2(\mathbb{R}^+_x\times\mathbb{S}^1_y)}\;\leqslant\;\frac{4}{3}\|\partial_x^2 \widetilde{\varphi}\|_{L^2(\mathbb{R}^+_x\times\mathbb{S}^1_y)}\,.
 \end{equation}
\end{corollary}

 For completeness of presentation, here is a direct, fast proof of Lemma \ref{lem:doubleHardy}, which is simpler than the general demonstration (see, e.g., \cite{Rellich-ICM1954} or \cite[Chapter 6]{Balinsky-Evans-Lewis-BOOK-Hardy}) in the $d$-dimensional case.

\begin{proof}[Proof of Lemma \ref{lem:doubleHardy}]
 As $h\in C^\infty_c(\mathbb{R}^+)$, all the considered integrals are finite, because the integrand functions are supported away from zero, and moreover integration by parts produces no boundary terms. One has
 \[
  \begin{split}
   \|r^{-2}h\|_{L^2}^2\;&=\;\int_0^{+\infty}\frac{|h(r)|^2}{r^4}\,\ud r\;=\;-\frac{1}{3}\int_0^{+\infty}\Big(\frac{1}{r^3}\Big)'\,\overline{h(r)}\,h(r)\,\ud r \\
   &=\;\frac{1}{3}\int_0^{+\infty}\frac{1}{r^3}\,\big(\overline{h(r)}\,h(r)\big)'\,\ud r\;=\;\frac{2}{3}\,\mathfrak{Re}\int_0^{+\infty}\frac{\overline{h(r)}\,h'(r)}{r^3}\,\ud r\,,
  \end{split}
 \]
 and in turn, by means of a weighted Cauchy-Schwarz inequality and Hardy's inequality,
 \[
  \begin{split}   
  \Big|\int_0^{+\infty}\frac{\overline{h(r)}\,h'(r)}{r^3}\,\ud r\Big|\;&\leqslant\;{\textstyle\frac{1}{2}}a\|r^{-2}h\|_{L^2}^2+{\textstyle\frac{1}{2}} a^{-1}\|r^{-1}h'\|_{L^2}^2 \\
  &\leqslant\;{\textstyle\frac{1}{2}}a\|r^{-2}h\|_{L^2}^2+2 a^{-1}\|h''\|_{L^2}^2
  \end{split}
 \]
 for some $a>0$.
 Thus,
 \[
  \|r^{-2}h\|_{L^2}^2\;\leqslant\;{\textstyle\frac{1}{3}}a\,\|r^{-2}h\|_{L^2}^2+{\textstyle\frac{4}{3}}a^{-1}\,\|h''\|_{L^2}^2\,,
 \]
 whence
 \[
   \|r^{-2}h\|_{L^2}^2\;\leqslant\;\frac{4}{a(3-a)}\,\|h''\|_{L^2}^2\,.
 \]
 Optimising over $a\in(0,3)$ yields $a=\frac{3}{2}$, which corresponds to $\|r^{-2}h\|_{L^2}^2\leqslant\frac{16}{9}\|h''\|_{L^2}^2$. This is precisely \eqref{eq:hardy2}.
 \end{proof}

The second needed estimate is meant to control the term $x^{2\alpha}\partial_y^2$ of $\overline{\mathsf{H}_\alpha}$ and reads as follows.

\begin{lemma}\label{lem:boundedness_x2alphad2y}
 Let $\alpha\in[0,1)$. There exists a constant $D_\alpha>0$ such that for any $\widetilde{\varphi}\in\mathcal{D}(\overline{\mathsf{H}_\alpha^+})$ one has
 \begin{equation}\label{eq:boundedness_x2alphad2y}
  \|x^{2\alpha}\partial_y^2\widetilde{\varphi}\|_{L^2(\mathbb{R}^+_x\times\mathbb{S}^1_y)}\;\leqslant\;D_\alpha\big\|\overline{\mathsf{H}_\alpha^+}\,\widetilde{\varphi}\big\|_{L^2(\mathbb{R}^+_x\times\mathbb{S}^1_y)}\,.
 \end{equation}
\end{lemma}

\begin{proof}
 It is enough to prove \eqref{eq:boundedness_x2alphad2y} for any $\widetilde{\varphi}\in C^\infty_c(\mathbb{R}^+_x\times\mathbb{S}^1_y)$; then the general inequality is merely obtained by closure, owing to \eqref{eq:DHaclosedclosure}. To this aim, let $(\widetilde{\varphi}_k)_{k\in\mathbb{Z}}:=\mathcal{F}_2^+\widetilde{\varphi}\in\cH^+\cong\ell^2(\mathbb{Z},L^2(\mathbb{R}^+,\ud x))$. One has
 \begin{equation*}
  \begin{split}
   \big\|x^{2\alpha}\partial_y^2 \widetilde{\varphi}\big\|^2_{L^2(\mathbb{R}^+_x\times\mathbb{S}^1_y)}\;&=\;\sum_{k\in\mathbb{Z}}\|x^{2\alpha}k^2\,\widetilde{\varphi}_k\|_{L^2(\mathbb{R}^+)}^2 \\
   &=\;\sum_{k\in\mathbb{Z}\setminus\{0\}}\|x^{2\alpha}k^2\,R_{G_{\alpha,k}}A_{\alpha,\mathrm{F}}(k)\widetilde{\varphi}_k\|_{L^2(\mathbb{R}^+)}^2 \\
   &\leqslant\;\sum_{k\in\mathbb{Z}\setminus\{0\}}\|x^{2\alpha}k^2\,R_{G_{\alpha,k}}\big\|_{\mathrm{op}}^2\big\|\overline{A_{\alpha}^+(k)}\widetilde{\varphi}_k\|_{L^2(\mathbb{R}^+)}^2\,,
  \end{split}
 \end{equation*}
where Plancherel's formula was used in the first identity and Proposition \ref{eq:V-RGisSFinv} in the second identity. Owing to Lemma \ref{lem:V-RGbddsa}(ii), $\|x^{2\alpha}k^2\,R_{G_{\alpha,k}}\|_{\mathrm{op}}\leqslant D_\alpha$ uniformly in $k$ for some $D_\alpha>0$. Based on this fact, and on Lemma \ref{lem:Halphaadj-decomposable} (formula \eqref{eq:Halphaclosure-decomposable}), one then has
\[
 \begin{split}
   \big\|x^{2\alpha}\partial_y^2 \widetilde{\varphi}\big\|^2_{L^2(\mathbb{R}^+_x\times\mathbb{S}^1_y)}\;&\leqslant\; D_\alpha^2\sum_{k\in\mathbb{Z}}\big\|\overline{A_{\alpha}^+(k)}\widetilde{\varphi}_k\|_{L^2(\mathbb{R}^+)}^2\;=\;D_\alpha^2\,\big\|\overline{\mathscr{H}_\alpha^+}(\widetilde{\varphi}_k)_{k\in\mathbb{Z}}\big\|_{\cH}^2 \\
   &=\;D_\alpha^2\,\big\|\overline{\mathsf{H}_\alpha^+}\,\widetilde{\varphi}\big\|_{L^2(\mathbb{R}^+_x\times\mathbb{S}^1_y)}^2\,,
 \end{split}
\]
which completes the proof.
\end{proof}

Based upon the above estimates, Proposition \ref{prop:Hclosurecontrol} can now be proved.

\begin{proof}[Proof of Proposition \ref{prop:Hclosurecontrol}]
 Again, it suffices to establish \eqref{eq:Hclosurecontrol} when $\widetilde{\varphi}\in C^\infty_c(\mathbb{R}^+_x\times\mathbb{S}^1_y)$, and then conclude by density from \eqref{eq:DHaclosedclosure}.
 
 One has
 \begin{equation*}
 \begin{split}
  \|\partial_x^2\widetilde{\varphi}\|_{L^2(\mathbb{R}_x\times\mathbb{S}^1_y)}\;&\leqslant\;\big\|\overline{\mathsf{H}_\alpha^+}\,\widetilde{\varphi}\big\|_{L^2(\mathbb{R}_x\times\mathbb{S}^1_y)}+\big\|x^{2\alpha}\partial_y^2 \widetilde{\varphi}\big\|_{L^2(\mathbb{R}^+_x\times\mathbb{S}^1_y)}+C_\alpha\|x^{-2}\widetilde{\varphi}\|_{L^2(\mathbb{R}_x^+\times\mathbb{S}^1_y)} \\
  &\leqslant\;\big\|\overline{\mathsf{H}_\alpha^+}\,\widetilde{\varphi}\big\|_{L^2(\mathbb{R}_x^+\times\mathbb{S}^1_y)}+D_\alpha\big\|\overline{\mathsf{H}_\alpha^+}\,\widetilde{\varphi}\big\|_{L^2(\mathbb{R}^+_x\times\mathbb{S}^1_y)}+\frac{4C_\alpha}{3}\|\partial_x^2\widetilde{\varphi}\|_{L^2(\mathbb{R}_x^+\times\mathbb{S}^1_y)}\,,
 \end{split}
 \end{equation*}
 where the first inequality is a triangular inequality based on \eqref{eq:actionofHaclosed}, whereas the second inequality follows directly from Corollary \ref{cor:doubleHardy} and Lemma \ref{lem:boundedness_x2alphad2y}.
 
 Therefore,
 \[
  \|\partial_x^2\widetilde{\varphi}\|_{L^2(\mathbb{R}_x^+\times\mathbb{S}^1_y)}\;\leqslant\;\frac{1+D_\alpha}{\,1-\frac{4}{3}C_\alpha}\,\big\|\overline{\mathsf{H}_\alpha^+}\,\widetilde{\varphi}\big\|_{L^2(\mathbb{R}_x^+\times\mathbb{S}^1_y)}\,.
 \]
 As $C_\alpha=\frac{1}{4}\alpha(2+\alpha)$, the constant $K_\alpha:=(1+D_\alpha)(1-\frac{4}{3}C_\alpha)^{-1}$ is strictly positive for any $\alpha$ of interest, namely, $\alpha\in(0,1)$. Moreover, $K_\alpha\to +\infty$ as $\alpha\uparrow 1$ (indeed, tracing back the constant $D_\alpha$ through the proof of Lemma 3.3 where it was imported from in Lemma \ref{lem:boundedness_x2alphad2y}, it is easy to see that $D_\alpha$ does not diverge when $\alpha\uparrow 1$). The proof is thus completed.  
\end{proof}

\subsection{Control of $\vartheta$}\label{sec:q0q1}

 As a counterpart to Subsection \ref{sec:control-of-tildephi}, the needed short-scale behaviour of the function $\vartheta\in L^2(\mathbb{R}\times\mathbb{S}^1,\ud x\ud y)$ defined in  \eqref{eq:deffunctionvartheta} is established here.

 Recall that $\vartheta^\pm$ may well fail to belong to $\mathcal{D}(\overline{\mathsf{H}_\alpha^\pm})$ and therefore cannot be controlled by means of Proposition \ref{prop:Hclosurecontrol}: a separate analysis is needed, based on the explicit expression and homogeneity properties of $\vartheta$.

 The main result here is the following.

 \begin{proposition}\label{prop:varthetacontrol}
   Let $\alpha\in[0,1)$. For almost every $y\in\mathbb{S}^1$,
 \begin{enumerate}
  \item[(i)] the function $x\mapsto\vartheta^\pm(x,y)$ belongs to $C^1(0,1)$,
  \item[(ii)] $\vartheta^\pm(x,y)=o(|x|^{\frac{3}{2}})$ as $x\to 0^\pm$,  
  \item[(iii)] $\partial_x\vartheta^\pm(x,y)=o(|x|^{\frac{1}{2}})$ as $x\to 0^\pm$.
 \end{enumerate} 
 \end{proposition}

 In preparation for the proof of this result, in terms of the functions
\begin{equation}\label{eq:funtionsh0h1}
 \begin{split}
  h_{0,k}\;&:=\; {\textstyle{\sqrt{\frac{2}{\pi(1+\alpha)}}}}\,|k|^{\frac{1}{2(1+\alpha)}}\Big(\Phi_{\alpha,k}-{\textstyle\sqrt{\frac{\pi(1+\alpha)}{2|k|}}\,x^{-\frac{\alpha}{2}}+\sqrt{\frac{\pi|k|}{2(1+\alpha)}}\,x^{1+\frac{\alpha}{2}}}\Big)\,,\\
  h_{1,k}\;&:=\;{\textstyle{\sqrt{\frac{2}{\pi(1+\alpha)}}}}\,|k|^{\frac{5}{2(1+\alpha)}}\Big(\Psi_{\alpha,k}-{\textstyle\sqrt{\frac{2|k|}{\pi(1+\alpha)^3}}}\,\|\Phi_{\alpha,k}\|_{L^2(\mathbb{R}^+)}^2\,|x|^{1+\frac{\alpha}{2}}\Big)
 \end{split}
\end{equation}
 defined on $\mathbb{R}^+$ for each $k\in\mathbb{Z}\setminus\{0\}$, one sees from \eqref{eq:defq0q1decomp-q0}-\eqref{eq:defq0q1decomp-q1} that 
 \begin{equation}\label{eq:q0q1scaling}
  \begin{split}
   \vartheta_{0,k}^\pm(x)\;&=\;c_{0,k}^{\pm}{\textstyle\sqrt{\frac{\pi(1+\alpha)}{2}}}\,|k|^{-\frac{1}{2(1+\alpha)}}\,h_{0,k}(|x|)\,,\qquad 0<\pm x<1\,, \\
   \vartheta_{1,k}^\pm(x)\;&=\;c_{1,k}^{\pm}{\textstyle\sqrt{\frac{\pi(1+\alpha)}{2}}}\,|k|^{-\frac{5}{2(1+\alpha)}}\,h_{1,k}(|x|)\,,\qquad 0<\pm x<1\,.
  \end{split}
 \end{equation}
 Clearly the above identities are not valid when $|x|>1$.

\begin{lemma}\label{lem:homogeneity}
 Let $\alpha\in[0,1)$ and $k\in\mathbb{Z}\setminus\{0\}$. For $x\in\mathbb{R}^+$ one has
\begin{equation}\label{eq:q0q1scaling2}
  h_{0,k}(x)\;:=\;w_0\big(|k|x^{1+\alpha}\big)\,,\qquad h_{1,k}(x)\;:=\;w_1\big(|k|x^{1+\alpha}\big)
\end{equation}
with
\begin{equation}\label{eq:q0q1scalingw0}
 w_0(x)\;:=\;x^{-\frac{\alpha}{2(1+\alpha)}}\big(e^{-\frac{x}{1+\alpha}}-1+{\textstyle\frac{x}{1+\alpha}}\big)
\end{equation}
and 
\begin{equation}\label{eq:q0q1scalingw1}
 \begin{split}
  w_1(x)\;&:=\;x^{-\frac{\alpha}{2(1+\alpha)}}\,e^{-\frac{x}{1+\alpha}}\int_0^{x^{\frac{1}{1+\alpha}}}\ud\rho\,\rho^{-\alpha}\sinh{\textstyle(\frac{\rho^{1+\alpha}}{1+\alpha})}\,e^{-\frac{\,\rho^{1+\alpha}}{1+\alpha}} \\
  &\qquad +x^{-\frac{\alpha}{2(1+\alpha)}}\,\sinh{\textstyle(\frac{x}{1+\alpha})}\int_{x^{\frac{1}{1+\alpha}}}^{+\infty}\!\!\ud\rho\,\rho^{-\alpha}\,e^{-\frac{2\rho^{1+\alpha}}{1+\alpha}} \\
  &\qquad-2^{-\frac{1-\alpha}{1+\alpha}}\,(1+\alpha)^{-\frac{1+3\alpha}{1+\alpha}}\,\Gamma\big({\textstyle\frac{1-\alpha}{1+\alpha}}\big)\,x^{\frac{2+\alpha}{2(1+\alpha)}}\,.
 \end{split}
\end{equation} 
\end{lemma}

\begin{proof}
 Plugging the explicit expression \eqref{eq:Phi_and_F_explicit} for $\Phi_{\alpha,k}$ into the first formula in \eqref{eq:funtionsh0h1}  one finds
 \[
  h_{0,k}(x)\;=\;\big(|k|^{\frac{1}{1+\alpha}}x)^{-\frac{\alpha}{2}}\Big(e^{-\frac{|k|}{1+\alpha}x^{1+\alpha}}-1+{\textstyle\frac{|k|x^{1+\alpha}}{1+\alpha}}\Big)\;=\;w_0\big(|k|x^{1+\alpha}\big)
 \]
 with $w_0$ defined by \eqref{eq:q0q1scalingw0}. Analogously, inserting the expression \eqref{eq:explicitPsika} for $\Psi_{\alpha,k}$ and the expression \eqref{eq:Phinorm} for $\| \Phi_{\alpha,k} \|_{L^2(\mathbb{R}^+)}^2$ into the second formula in \eqref{eq:funtionsh0h1}, one obtains
 \[
 \begin{split}
  h_{1,k}^{\pm}(x)\;&=\; \big(|k|^{\frac{1}{1+\alpha}}x\big)^{-\frac{\alpha}{2}}e^{-\frac{|k|x^{1+\alpha}}{1+\alpha}}\!\int_0^{x|k|^{\frac{1}{1+\alpha}}}\ud\rho\,\rho^{-\alpha}\sinh{\textstyle(\frac{\rho^{1+\alpha}}{1+\alpha})}\,e^{-\frac{\,\rho^{1+\alpha}}{1+\alpha}} \\
  &\qquad\quad +\big(|k|^{\frac{1}{1+\alpha}}x\big)^{-\frac{\alpha}{2}}\sinh{\textstyle\big(\frac{|k|x^{1+\alpha}}{1+\alpha}\big)}\!\int_{x|k|^{\frac{1}{1+\alpha}}}^{+\infty}\!\!\ud\rho\,\rho^{-\alpha}\,e^{-\frac{2\rho^{1+\alpha}}{1+\alpha}} \\
  &\qquad\quad-2^{-\frac{1-\alpha}{1+\alpha}}\,(1+\alpha)^{-\frac{1+3\alpha}{1+\alpha}}\,\Gamma\big({\textstyle\frac{1-\alpha}{1+\alpha}}\big)\,\big(|k|^{\frac{1}{1+\alpha}}x\big)^{1+\frac{\alpha}{2}}  \\
  &=\;w_1\big(|k|x^{1+\alpha}\big)
 \end{split}
 \]
 with $w_1$ defined by \eqref{eq:q0q1scalingw1}.
\end{proof}

\begin{lemma}\label{lem:scaling_on_norms_hjk}
 Let $\alpha\in[0,1)$ and $k\in\mathbb{Z}\setminus\{0\}$. The functions $h_{0,k}$ and $h_{1,k}$ defined in \eqref{eq:funtionsh0h1} satisfy
 \begin{eqnarray}
  \big\|x^{-2}h_{j,k}\big\|_{L^2((0,1))}^2 &\leqslant&|k|^{\frac{3}{1+\alpha}}\,\big\|x^{-2}h_{j,1}\big\|_{L^2(\mathbb{R}^+)}^2\,, \label{eq:scalingx-2h} \\
  \big\|h_{j,k}''\big\|_{L^2((0,1))}^2&\leqslant& |k|^{\frac{3}{1+\alpha}}\,\big\|h_{j,1}''\big\|_{L^2(\mathbb{R}^+)}^2 \label{eq:scalingd2h}
 \end{eqnarray}
 for $j\in\{0,1\}$. 
\end{lemma}

\begin{proof}
 By means of the homogeneity properties \eqref{eq:q0q1scaling2} one finds
 \[
  \begin{split}
    \big\|x^{-2}h_{j,k}\big\|_{L^2((0,1))}^2\;&=\;\int_0^1\big|x^{-2} w_j\big(|k|x^{1+\alpha}\big)\big|^2\,\ud x \\
    &=\;|k|^{\frac{3}{1+\alpha}}\int_0^{|k|^{\frac{1}{1+\alpha}}}|x^{-2}w_j(x^{1+\alpha})|^2\,\ud x \\
    &\leqslant\;|k|^{\frac{3}{1+\alpha}}\int_0^{+\infty}|x^{-2}h_{j,1}(x)|^2\,\ud x
  \end{split}
 \]
 and 
\[
 \begin{split}
  \big\|&h_{j,k}''\big\|_{L^2((0,1))}^2\;=\;\int_0^1\Big| \frac{\ud^2}{\ud x^2}\,w_j\big(|k|x^{1+\alpha}\big) \Big|^2\,\ud x \\
  &=\;\int_0^1\big| (1+\alpha)^2|k|^2 x^{2\alpha}w_j''\big(|k|x^{1+\alpha}\big)+\alpha(1+\alpha)|k|x^{-(1-\alpha)}w_j'\big(|k|x^{1+\alpha}\big) \big|^2\,\ud x \\
  &=\;|k|^{\frac{3}{1+\alpha}}\int_0^{|k|^{\frac{1}{1+\alpha}}}\big| (1+\alpha)^2 x^{2\alpha}w_j''\big(x^{1+\alpha}\big)+\alpha(1+\alpha)x^{-(1-\alpha)}w_j'\big(x^{1+\alpha}\big) \big|^2\,\ud x \\
  &=\;|k|^{\frac{3}{1+\alpha}}\int_0^{|k|^{\frac{1}{1+\alpha}}}\Big| \frac{\ud^2}{\ud x^2}\,w_j\big(x^{1+\alpha}\big) \Big|^2\,\ud x\;\leqslant\;|k|^{\frac{3}{1+\alpha}}\int_0^{+\infty}|h_{j,1}''(x)|^2\,\ud x\,,
 \end{split}
\]
which proves, respectively, \eqref{eq:scalingx-2h} and \eqref{eq:scalingd2h}.
\end{proof}

\begin{lemma}\label{lem:finitenessnormsh0h1}
 Let $\alpha\in[0,1)$. The functions $h_{0,1}$ and $h_{1,1}$ defined in \eqref{eq:funtionsh0h1} satisfy
  \begin{eqnarray}
  \big\|x^{-2}h_{j,1}\big\|_{L^2(\mathbb{R}^+)}^2&<&+\infty \,,\label{eq:x-2h-finite}\\
  \big\|h_{j,1}''\big\|_{L^2(\mathbb{R}^+)}^2&<&+\infty \label{eq:D2h-finite}
 \end{eqnarray}
 for $j\in\{0,1\}$. 
\end{lemma}

\begin{proof}
As $h_{0,1}$ (resp., $h_{1,1}$) only agrees with $\vartheta^+_{0,1}$ (resp., $\vartheta^+_{1,1}$) over the interval $(0,1)$, apart from a $\alpha$-dependent pre-factor, one cannot deduce \eqref{eq:x-2h-finite}-\eqref{eq:D2h-finite} from \eqref{eq:regularityoftheta01}, because the considered norms are over the whole $\mathbb{R}^+$. However, the reasoning made in the proof of Theorem \ref{prop:g_with_Pweight}, which led to \eqref{eq:regularityoftheta01}, can be essentially repeated here. Clearly, both $h_{0,1}$ and $h_{1,1}$ are $C^\infty(\mathbb{R}^+)$-functions; therefore, the finiteness of the norms in \eqref{eq:x-2h-finite}-\eqref{eq:D2h-finite} is only to be checked as $x\downarrow 0$ and $x\to+\infty$. In fact, for
\[
 h_{0,1}\;=\;x^{-\frac{\alpha}{2}}\big(e^{-\frac{x^{1+\alpha}}{1+\alpha}}-1+{\textstyle\frac{x^{1+\alpha}}{1+\alpha}}\big)
\]
one can perform a straightforward computation and find
\[
 \begin{split}
  & \;\,h_{0,1}(x)\;\stackrel{x\downarrow 0}{=}\;x^{2+\frac{3}{2}\alpha}(1+O(x^{1+\alpha}))\,, \\
  & h_{0,1}(x)\stackrel{x\to +\infty}{=}{\textstyle\frac{1}{1+\alpha}}x^{1-\frac{\alpha}{2}}(1+O(x^{-1})) \,,
 \end{split}
\]
and 
\[
 \begin{split}
  &h_{0,1}''(x)(x)\;\stackrel{x\downarrow 0}{=}\;x^{\frac{3}{2}\alpha}\big({\textstyle\frac{9}{8}-\frac{1}{8(1+\alpha)^2}}\big)(1+O(x^{1+\alpha}))\,, \\
  &\quad h_{0,1}''(x)\stackrel{x\to +\infty}{=}{\textstyle\frac{\alpha(2+\alpha)}{4(1+\alpha)}}x^{-(1+\frac{\alpha}{2})}(1+o(1))\,.
  \end{split}
\]
Such asymptotics imply \eqref{eq:x-2h-finite}-\eqref{eq:D2h-finite} when $j=0$, as $\alpha\in(0,1)$. Concerning
\[
  h_{1,1}\;=\;{\textstyle{\sqrt{\frac{2}{\pi(1+\alpha)}}}}\,\Psi_{\alpha,1}-{\textstyle\frac{2}{\pi(1+\alpha)^2}}\,\|\Phi_{\alpha,1}\|_{L^2(\mathbb{R}^+)}^2\,x^{1+\frac{\alpha}{2}}\,,
\]
the square-integrability of $x^{-2}h_{1,1}$ is controlled analogously to the proof of Theorem \ref{prop:g_with_Pweight}: the short-distance asymptotics \eqref{eq:Psi_asymptotics} for $\Psi_{\alpha,1}$ gives a convenient compensation in $h_{1,1}$ as $x\downarrow 0$, whereas at infinity the control can be simply made term by term, as $\Psi_{\alpha,1}\in L^2(\mathbb{R}^+)$. Thus, \eqref{eq:x-2h-finite} is also proved for $j=1$. Next, one considers
\[
 h_{1,1}''\;=\;{\textstyle{\sqrt{\frac{2}{\pi(1+\alpha)}}}}\,\Psi_{\alpha,1}''-{\textstyle\frac{2}{\pi(1+\alpha)^2}}\,\|\Phi_{\alpha,1}\|_{L^2(\mathbb{R}^+)}^2\,{\textstyle\frac{\alpha(2+\alpha)}{2}}\,x^{-(1-\frac{\alpha}{2})}\,.
\]
As $\Psi_{\alpha,1}=R_{G_{\alpha,1}}\Phi_{\alpha,1}$ and $R_{G_{\alpha,1}}=(A_{\alpha,\mathrm{F}}^+(1))^{-1}$ (see \eqref{eq:defPsi} and Proposition \ref{eq:V-RGisSFinv} above), then
\[
 \begin{split}
  \Psi_{\alpha,1}''\;&=\;-\big({\textstyle -\frac{\ud^2}{\ud x^2}}+x^{2\alpha}+{\textstyle\frac{\alpha(2+\alpha)}{2}}x^{-2}\big)R_{G_{\alpha,1}}\Phi_{\alpha,1}+\big(x^{2\alpha}+{\textstyle\frac{\alpha(2+\alpha)}{2}}x^{-2}\big)\Psi_{\alpha,1} \\
  &=\;-\Phi_{\alpha,1}+\big(x^{2\alpha}+{\textstyle\frac{\alpha(2+\alpha)}{2}}x^{-2}\big)\Psi_{\alpha,1}\,,
 \end{split}
\]
whence
\[
 \begin{split}
  h_{1,1}''\;&=\;-{\textstyle{\sqrt{\frac{2}{\pi(1+\alpha)}}}}\,\Phi_{1,\alpha}+{\textstyle{\sqrt{\frac{2}{\pi(1+\alpha)}}}}\big(x^{2\alpha}+{\textstyle\frac{\alpha(2+\alpha)}{2}}x^{-2}\big)\Psi_{\alpha,1} \\ 
 &\qquad\qquad  -{\textstyle\frac{2}{\pi(1+\alpha)^2}}\,\|\Phi_{\alpha,1}\|_{L^2(\mathbb{R}^+)}^2\,{\textstyle\frac{\alpha(2+\alpha)}{2}}\,x^{-(1-\frac{\alpha}{2})} \\
 &=\;-{\textstyle{\sqrt{\frac{2}{\pi(1+\alpha)}}}}\,\Phi_{1,\alpha}+{\textstyle{\sqrt{\frac{2}{\pi(1+\alpha)}}}}\,x^{2\alpha}\,\Psi_{\alpha,1}+{\textstyle\frac{\alpha(2+\alpha)}{2}}\,x^{-2}\,h_{1,1}\,.
 \end{split}
\]
Each of the three summands on the r.h.s.~above belongs to $L^2(\mathbb{R}^+)$: in particular, the second so does because $\Psi_{\alpha,1}\in\mathrm{ran}\,R_{G_{\alpha,k}}\subset L^2(\mathbb{R}^+,\langle x\rangle^{4\alpha} \ud x)$ (Corollary \ref{cor:RGtoWeightedL2}). This proves \eqref{eq:D2h-finite} for $j=1$.
\end{proof}

From \eqref{eq:q0q1scaling}, and from Lemmas \ref{lem:scaling_on_norms_hjk} and \ref{lem:finitenessnormsh0h1}, one immediately deduces:

\begin{corollary}\label{cor:varthetasummability}
 Let $\alpha\in[0,1)$ and $k\in\mathbb{Z}\setminus\{0\}$. Then
 \begin{equation}\label{eq:varthetasummability0}
  \begin{split}
   \big\|x^{-2}\vartheta_{0,k}^\pm\big\|_{L^2(I^\pm)}^2\;&\lesssim\;|c_{0,k}^\pm|^2\,|k|^{\frac{2}{1+\alpha}}\,, \\
   \big\|(\vartheta_{0,k}^\pm)''\big\|_{L^2(I^\pm)}^2\;&\lesssim\;|c_{0,k}^\pm|^2\,|k|^{\frac{2}{1+\alpha}}\,,
  \end{split}
 \end{equation}
 and 
  \begin{equation}\label{eq:varthetasummability1}
  \begin{split}
   \big\|x^{-2}\vartheta_{1,k}^\pm\big\|_{L^2(I^\pm)}^2\;&\lesssim\;|c_{1,k}^\pm|^2\,|k|^{-\frac{2}{1+\alpha}}\,, \\
   \big\|(\vartheta_{1,k}^\pm)''\big\|_{L^2(I^\pm)}^2\;&\lesssim\;|c_{1,k}^\pm|^2\,|k|^{-\frac{2}{1+\alpha}}\,,
  \end{split}
 \end{equation}
 with $I^+=(0,1)$ and $I^-=(-1,0)$, where the constants in the above inequalities only depend on $\alpha$. 
\end{corollary}

In fact, \eqref{eq:varthetasummability0}-\eqref{eq:varthetasummability1} are trivially true also for $k=0$: recall indeed (see \eqref{eq:varthetazeromode} above) that $\vartheta_0\equiv 0$.

 \begin{proof}[Proof of Proposition \ref{prop:varthetacontrol}]
  It clearly suffices to discuss the proof for the `+' component $\vartheta^+=\mathcal{F}_2^{-1}(\vartheta^+_k)_{k\in\mathbb{Z}}$. Recall also that $\vartheta_0^+\equiv 0$.

  Now, owing to Corollary \ref{cor:varthetasummability},
  \[
   \begin{split}
    \big\|x^{-2}(\vartheta^+_{0,k})_{k\in\mathbb{Z}}\big\|^2_{\ell^2(\mathbb{Z},L^2((0,1),\ud x))}\;&\lesssim\;\sum_{k\in\mathbb{Z}\setminus\{0\}}|c_{0,k}^\pm|^2\,|k|^{\frac{2}{1+\alpha}}\,, \\
    \big\|((\vartheta_{0,k}^\pm)'')_{k\in\mathbb{Z}}\big\|^2_{\ell^2(\mathbb{Z},L^2((0,1),\ud x))}\;&\lesssim\;\sum_{k\in\mathbb{Z}\setminus\{0\}}|c_{0,k}^\pm|^2\,|k|^{\frac{2}{1+\alpha}}\,.
   \end{split}
  \]
  The series on the r.h.s.~above are \emph{convergent}, because of the enhanced summability of the $c_{0,k}$'s, due to the fact that the initially considered $(g_k)_{k\in\mathbb{Z}}$ belongs to the domain of a uniformly fibred extension (as observed already in Remark \ref{rem:enhanced_summability}).

  As a first consequence, $(\vartheta^+_{0,k})_{k\in\mathbb{Z}}$ belongs to $\ell^2(\mathbb{Z},L^2((0,1),\ud x))$, and so too does $(\vartheta^+_{1,k})_{k\in\mathbb{Z}}$ by difference from $(\vartheta^+_{k})_{k\in\mathbb{Z}}$: therefore, the inverse Fourier transform can be separately applied to
  \[
   \vartheta^+\;=\;\mathcal{F}_2^{-1}(\vartheta^+_k)_{k\in\mathbb{Z}}\;=\;\mathcal{F}_2^{-1}(\vartheta^+_{0,k})_{k\in\mathbb{Z}}+\mathcal{F}_2^{-1}(\vartheta^+_{1,k})_{k\in\mathbb{Z}}\,.
  \]

  As a further consequence, the above estimates imply, by means of Plancherel's formula, 
  \[
   \begin{split}
    \big\|x^{-2}\mathcal{F}_2^{-1}(\vartheta^+_{0,k})_{k\in\mathbb{Z}}\big\|^2_{L^2((0,1)\times\mathbb{S}^1,\ud x \ud y)}\;=\;\big\|x^{-2}(\vartheta^+_{0,k})_{k\in\mathbb{Z}}\big\|^2_{\ell^2(\mathbb{Z},L^2((0,1),\ud x))}\;&<\;+\infty \,,\\
    \big\|\partial^2_{x}\mathcal{F}_2^{-1}(\vartheta^+_{0,k})_{k\in\mathbb{Z}}\big\|^2_{L^2((0,1)\times\mathbb{S}^1,\ud x \ud y)}\;=\;\big\|(\partial_x^2\vartheta^+_{0,k})_{k\in\mathbb{Z}}\big\|^2_{\ell^2(\mathbb{Z},L^2((0,1),\ud x))}\;&<\;+\infty\,.
   \end{split}
  \]

  Analogously, Corollary \ref{cor:varthetasummability} also implies
  \[
   \begin{split}
    \big\|x^{-2}(\vartheta^+_{1,k})_{k\in\mathbb{Z}}\big\|^2_{\ell^2(\mathbb{Z},L^2((0,1),\ud x))}\;&\lesssim\;\sum_{k\in\mathbb{Z}\setminus\{0\}}|c_{1,k}^\pm|^2\,|k|^{-\frac{2}{1+\alpha}} \,,\\
    \big\|((\vartheta_{1,k}^\pm)'')_{k\in\mathbb{Z}}\big\|^2_{\ell^2(\mathbb{Z},L^2((0,1),\ud x))}\;&\lesssim\;\sum_{k\in\mathbb{Z}\setminus\{0\}}|c_{1,k}^\pm|^2\,|k|^{-\frac{2}{1+\alpha}}\,,
   \end{split}
  \]
  and the series on the r.h.s.~above are \emph{finite} because of the general summability for elements in $\mathcal{D}(\mathscr{H}_{\alpha}^*)$ established in Lemma \ref{lem:gkkrepr}, formula \eqref{eq:pileupcond3}. Thus, for almost every $y\in\mathbb{S}^1$,
  \[
   \begin{split}
    \big\|x^{-2}\mathcal{F}_2^{-1}(\vartheta^+_{1,k})_{k\in\mathbb{Z}}\big\|^2_{L^2((0,1)\times\mathbb{S}^1,\ud x \ud y)}\;&<\;+\infty\,, \\
    \big\|\partial^2_{x}\mathcal{F}_2^{-1}(\vartheta^+_{1,k})_{k\in\mathbb{Z}}\big\|^2_{L^2((0,1)\times\mathbb{S}^1,\ud x \ud y)}\;&<\;+\infty\,.
   \end{split}
  \]
  
  Summarising, 
  \[
   \|x^{-2}\vartheta^+\|_{L^2((0,1)\times\mathbb{S}^1,\ud x \ud y)}+ \|\partial_x^2\vartheta^+\|_{L^2((0,1)\times\mathbb{S}^1,\ud x \ud y)}\;<\;+\infty\,.
  \]
  Therefore, $\vartheta^+$ satisfies the assumptions (a) and (b) of Lemma \ref{lem:grand_auxiliary_lemma} (for, obviously, $|x^{-(\frac{3}{2}+\frac{\alpha}{2})}\vartheta^+(x,y)|\leqslant|x^{-2}\vartheta^+(x,y)|$ when $x\in(0,1)$, since $\alpha\in(0,1)$). The thesis then follows by applying Lemma \ref{lem:grand_auxiliary_lemma}.  
 \end{proof}

 \subsection{Proof of the uniformly fibred classification}\label{sec:proofclassifthm}

 \begin{proof}[Proof of Theorem \ref{thm:classificationUF}]
 The goal is to characterise the domain
 \[
  \mathcal{D}(\mathcal{F}_2^{-1}\mathscr{H}_\alpha^{\mathrm{u.f.}}\mathcal{F}_2)
 \]
of the various uniformly fibred extensions of $\mathsf{H}_\alpha=\mathcal{F}_2^{-1}\mathscr{H}_\alpha\mathcal{F}_2$.

The expression \eqref{eq:HHalphaadjointagain} for $\mathsf{H}_\alpha^*$ provided in the statement of the theorem was already found in \eqref{eq:HHalphaadjoint}.

Next, consider a generic $\phi=\mathcal{F}_2^{-1}(g_k)_{k\in\mathbb{Z}}\in\mathcal{D}(\mathcal{F}_2^{-1}\mathscr{H}_\alpha^{\mathrm{u.f.}}\mathcal{F}_2)$, where $(g_k)_{k\in\mathbb{Z}}\in\mathcal{D}(\mathscr{H}_\alpha^{\mathrm{u.f.}})$. Owing to the definitions \eqref{eq:afterF2-1}-\eqref{eq:F2-1g1} and to Corollary \ref{cor:(g0)k_(g1)k_in_Hs}, 
\begin{equation}\label{eq:nowsafespliting}
 \phi(x,y)\;=\;\varphi(x,y)+g_1(y)|x|^{1+\frac{\alpha}{2}}P(x)+g_0(y)|x|^{-\frac{\alpha}{2}}P(x)\,,
\end{equation}
where $P$ is a smooth cut-off which is identically equal to one for $|x|<1$ and zero for $|x|>2$, and $g_0,g_1\in L^2(\mathbb{S}^1)$ with further Sobolev regularity as specified therein.

Moreover, upon splitting $\varphi=\widetilde{\varphi}+\vartheta$ as in \eqref{eq:splittingphiphitildetheta}, and using Proposition \ref{prop:Hclosurecontrol} for $\widetilde{\varphi}$ and Proposition \ref{prop:varthetacontrol} for $\vartheta$, one deduces that for almost every $y\in\mathbb{S}^1$
 \begin{itemize}
  \item the function $x\mapsto\varphi^\pm(x,y)$ belongs to $C^1(0,1)$,
  \item $\varphi^\pm(x,y)=o(|x|^{\frac{3}{2}})$ as $x\to 0^\pm$,  
  \item $\partial_x\varphi^\pm(x,y)=o(|x|^{\frac{1}{2}})$ as $x\to 0^\pm$.
 \end{itemize} 
Plugging this information into \eqref{eq:nowsafespliting} yields
\[
 \begin{split}
  \lim_{x\to 0^\pm} |x|^{\frac{\alpha}{2}}\,\phi^\pm(x,y)\;&=\;g_0^\pm(y)\,, \\
  \lim_{x\to 0^\pm} |x|^{-(1+\frac{\alpha}{2})}\big(\phi^\pm(x,y)-g_0^\pm(y)|x|^{-\frac{\alpha}{2}}\big)\;&=\;  g_1^\pm(y)+\lim_{x\to 0^\pm} |x|^{-(1+\frac{\alpha}{2})}\varphi^\pm(x,y)\\
  &=\;g_1^\pm(y)\,,
 \end{split}
\]
namely
\begin{equation}\label{eq:g0g1f0f1}
 g_0\;=\;\phi_0\,,\qquad g_1\;=\;\phi_1\,,
\end{equation}
proving also that the limits \eqref{eq:limitphi0}, as well as the limits of the first line of \eqref{eq:limitphi1}, do exist. Also, the Sobolev regularity stated for $\phi_0$ and $\phi_1$ follows directly from Corollary \ref{cor:(g0)k_(g1)k_in_Hs}.

The second identity in \eqref{eq:limitphi1} is obtained as follows. By means of \eqref{eq:nowsafespliting} one computes
 \[
  \begin{split}
   \pm(1+\alpha)^{-1}&\lim_{x\to 0^\pm} |x|^{-\alpha}\partial_x\big(|x|^{\frac{\alpha}{2}}\phi^\pm(x,y)\big)\;=\; \\
   &=\;\pm(1+\alpha)^{-1}\lim_{x\to 0^\pm} |x|^{-\alpha}\partial_x\big(|x|^{\frac{\alpha}{2}}\varphi^\pm(x,y)+g_1^\pm(y)|x|^{1+\alpha}+g_0^\pm(y)\big) \\
   &=\;g_1^\pm(y)\pm(1+\alpha)^{-1}\lim_{x\to 0^\pm} |x|^{-\alpha}\partial_x\big(|x|^{\frac{\alpha}{2}}\varphi^\pm(x,y)\big)\,.
  \end{split}
 \]
 On the other hand,
 \[
   \lim_{x\to 0^\pm} |x|^{-\alpha}\partial_x \big(|x|^{\frac{\alpha}{2}}\varphi^\pm(x,y)\big)= \lim_{x\to 0^\pm} \big({\textstyle\frac{\alpha}{2}|x|^{-(1+\frac{\alpha}{2})}}\varphi^\pm(x,y)+|x|^{-\frac{\alpha}{2}}\partial_x\varphi^\pm(x,y)\big)=0\,,
 \]
 having used the properties  $\varphi^\pm(x,y)=o(|x|^{\frac{3}{2}})$ and $\partial_x\varphi^\pm(x,y)=o(|x|^{\frac{1}{2}})$ as $x\to 0^\pm$. This yields the second identity in \eqref{eq:limitphi1}.

It remains to show that for each type of extension, the stated boundary conditions of self-adjointness do hold for $\phi_0$ and $\phi_1$. As, by  \eqref{e1:F2-1g0}-\eqref{eq:F2-1g1} and by \eqref{eq:g0g1f0f1}
\[
 \begin{split}
  \phi_0^{\pm}(y)\;&=\;\frac{1}{\sqrt{2\pi}}\sum_{k\in\mathbb{Z}} e^{\ii k y}g_{0,k}^{\pm}\,, \\
  \phi_1^{\pm}(y)\;&=\;\frac{1}{\sqrt{2\pi}}\sum_{k\in\mathbb{Z}} e^{\ii k y}g_{1,k}^{\pm}\,,
 \end{split}
\]
the above series being $L^2(\mathbb{S}^1)$-convergent,
and since for each \emph{uniformly fibred} extension $\mathscr{H}_\alpha^{\mathrm{u.f.}}$ the boundary conditions are expressed by \emph{the same linear combinations} of the $g_{0,k}^{\pm}$'s and $g_{1,k}^{\pm}$'s for each $k$, then now the boundary conditions of self-adjointness in terms of $\phi_0$ and $\phi_1$ are immediately read out from those of the classification Theorem \ref{thm:bifibre-extensions} for bilateral-fibre extensions (see also Table \ref{tab:extensions}) in terms of $g_{0,k}^{\pm}$ and $g_{1,k}^{\pm}$.
\end{proof}

\section{Classification of local transmission protocols on cylinder}\label{sec:proof_xy_Euclidean}

 Translating Theorem \ref{thm:classificationUF} through the unitary equivalence introduced in Section \ref{sec:constant-fibre-sum}, it is finally possible to characterise the physically relevant sub-family of self-adjoint realisations of the minimal free Hamiltonian\index{minimal free Hamiltonian} $H_\alpha$ on the Grushin-type cylinder Hilbert space $\cH_\alpha$ (as defined in \eqref{eq:Halphaspace} and \eqref{eq:V-Halpha}) in the regime $\alpha\in[0,1)$ of lack of geometric quantum confinement, local singularity of the metric, and infinity of deficiency index (Theorem \ref{thm:Halpha_esa_or_not}).

  The first important property is that $H_\alpha^*$ is decomposed with respect to
  \begin{equation}\label{eq:V-Hspacedirectsum2}
  \cH_\alpha\;=\;L^2(M,\ud\mu_\alpha)\;\cong\;L^2(M^-,\ud\mu_\alpha)\oplus L^2(M^+,\ud\mu_\alpha)
 \end{equation}
 (see \eqref{eq:V-Hspacedirectsum} above).

 \begin{proposition}\label{prop:adjoint_on_M} 
  Let $\alpha\geqslant 0$. One has
   \begin{equation}
    H_\alpha^*\;=\;(H_\alpha^-)^*\oplus(H_\alpha^+)^* \, ,
   \end{equation}
    where $(H_\alpha^\pm)^*$, the adjoint of $H_\alpha^\pm$ in $L^2(M^\pm,\ud\mu_\alpha)$, is the differential operator whose domain and action are given by
   \begin{equation}
   \begin{split}
     \mathcal{D}((H_\alpha^\pm)^*)\;&=\;\big\{f^\pm\in L^2(M^\pm,\ud\mu_\alpha)\,\big|\,-\Delta_{\mu_\alpha}f^\pm\in L^2(M^\pm,\ud\mu_\alpha)\big\} \, , \\
     (H_\alpha^\pm)^*\,f^\pm\;&=\;-\Delta_{\mu_\alpha}f^\pm\,.
   \end{split}
   \end{equation}
 \end{proposition}

 \begin{proof}
  The thesis follows directly from the analogous statement \eqref{eq:HHalphaadjointagain} in Theorem \ref{thm:classificationUF} for $(\mathsf{H}_\alpha^\pm)^*$, by exploiting the unitary equivalence 
  \begin{equation*}
  \begin{split}
   H_\alpha^\pm\;&=\;(U_\alpha^\pm)^{-1}\mathsf{H}_\alpha^\pm\,U_\alpha^\pm\,, \\
   (H_\alpha^\pm)^*\;&=\;(U_\alpha^\pm)^{-1}(\mathsf{H}_\alpha^\pm)^*\,U_\alpha^\pm
  \end{split}
  \end{equation*}
  (se \eqref{eq:tildeHalpha} and \eqref{eq:V-two-sided-6} above), 
  where $\phi^\pm=U_\alpha^\pm f^\pm=|x|^{-\frac{\alpha}{2}}f^\pm$ (see \eqref{eq:unit1} and \eqref{eq:V-two-sided-1}). Tacitly one also uses the property that the adjoint of the direct sum is the direct sum of the adjoints (Sect.~\ref{sec:I_invariant-reducing-ssp}).  
  \end{proof}

 Next, the special sub-class of self-adjoint restrictions of $(H_\alpha^\pm)^*$, hence extensions of $H_\alpha$, is described, characterised by local boundary conditions.

    \begin{theorem}\label{thm:H_alpha_fibred_extensions}
 Let $\alpha\in[0,1)$. The operator $H_\alpha$ admits, among others, the following families of self-adjoint extensions in $L^2(M,\ud\mu_\alpha)$:
 \begin{itemize}
  \item \underline{Friedrichs extension}: $H_{\alpha,\mathrm{F}}$;
  \item \underline{Family $\mathrm{I_R}$}: $\{H_{\alpha,\mathrm{R}}^{[\gamma]}\,|\,\gamma\in\mathbb{R}\}$;
  \item \underline{Family $\mathrm{I_L}$}: $\{H_{\alpha,\mathrm{L}}^{[\gamma]}\,|\,\gamma\in\mathbb{R}\}$;
  \item \underline{Family $\mathrm{II}_a$} with $a\in\mathbb{C}$: $\{H_{\alpha,a}^{[\gamma]}\,|\,\gamma\in\mathbb{R}\}$;
  \item \underline{Family $\mathrm{III}$}: $\{H_{\alpha}^{[\Gamma]}\,|\,\Gamma\equiv(\gamma_1,\gamma_2,\gamma_3,\gamma_4)\in\mathbb{R}^4\}$.
 \end{itemize}
 Each operator belonging to any such family is a restriction of $H_\alpha^*$, and hence its differential action is precisely $-\Delta_{\mu_\alpha}$. The domain of each of the above extensions is characterised as the space of the functions $f\in L^2(M,\ud\mu_\alpha)$ satisfying the following properties.
  \begin{enumerate}
  \item[(i)] \underline{Integrability and regularity}:
  \begin{equation}\label{eq:DHalpha_cond1}
  \sum_{\pm}\;\iint_{\mathbb{R}_x^\pm\times\mathbb{S}^1_y}\big|(\Delta_{\mu_\alpha}f^\pm)(x,y)\big|^2\,\ud\mu_\alpha(x,y)\;<\;+\infty\,.
 \end{equation}
  \item[(ii)] \underline{Boundary condition}: The limits
 \begin{eqnarray}
  f_0^\pm(y)&:=&\lim_{x\to 0^\pm}f^\pm(x,y)\,, \label{eq:DHalpha_cond2_limits-1}\\
  f_1^\pm(y)&:=&\pm(1+\alpha)^{-1}\lim_{x\to 0^\pm}\Big(\frac{1}{\:|x|^\alpha}\,\frac{\partial f(x,y)}{\partial x}\Big) \label{eq:DHalpha_cond2_limits-2}
  \end{eqnarray}
 exist and are finite for almost every $y\in\mathbb{S}^1$, and depending on the considered type of extension, and for almost  every $y\in\mathbb{R}$, they satisfy
 \begin{eqnarray}
  f_0^\pm(y)\,=\,0\,, \qquad \quad\;\;& & \textrm{if }\;  f\in\mathcal{D}(H_{\alpha,\mathrm{F}})\,, \label{eq:DHalpha_cond3_Friedrichs}\\
  \begin{cases}
   \;f_0^-(y)= 0 \,, \\
   \;f_1^+(y)=\gamma f_0^+(y)\,,
  \end{cases} & & \textrm{if }\;  f\in\mathcal{D}(H_{\alpha,\mathrm{R}}^{[\gamma]})\,, \\
   \begin{cases}
   \;f_1^-(y)=\gamma f_0^-(y)\,, \\
   \;f_0^+(y)= 0 \,,
  \end{cases} & & \textrm{if }\;  f\in\mathcal{D}(H_{\alpha,\mathrm{L}}^{[\gamma]}) \,, \label{eq:DHalpha_cond3_L}\\
     \begin{cases}
   \;f_0^+(y)=a\,f_0^-(y)\,, \\
   \;f_1^-(y)+\overline{a}\,f_1^+(y)=\gamma f_0^-(y)\,,
  \end{cases} & & \textrm{if }\;  f\in\mathcal{D}(H_{\alpha,a}^{[\gamma]})\,, \label{eq:DHalpha_cond3_IIa} \\
   \begin{cases}
   \;f_1^-(y)=\gamma_1 f_0^-(y)+(\gamma_2+\ii\gamma_3) f_0^+(y)\,, \\
   \;f_1^+(y)=(\gamma_2-\ii\gamma_3) f_0^-(y)+\gamma_4 f_0^+(y)\,,
  \end{cases} & & \textrm{if }\;  f\in\mathcal{D}(H_{\alpha}^{[\Gamma]})\,. \label{eq:DHalpha_cond3_III}
 \end{eqnarray} 
 \end{enumerate} 
     Moreover,
 \begin{equation}\label{eq:traceregularity}
  f_0^\pm \in H^{s_{0,\pm}}(\mathbb{S}^1, \ud y)\qquad\textrm{ and }\qquad f_1^\pm\in H^{s_{1,\pm}}(\mathbb{S}^1,\ud y)
 \end{equation}
 with
 \begin{itemize}
 	\item $s_{1,\pm}=\frac{1}{2}\frac{1-\alpha}{1+\alpha}$\qquad\qquad\qquad\qquad\qquad\; for the Friedrichs extension,
 	\item $s_{1,-}=\frac{1}{2}\frac{1-\alpha}{1+\alpha}$, $s_{0,+}=s_{1,+}=\frac{1}{2}\frac{3+\alpha}{1+\alpha}$ \quad\;\; for extensions of type $\mathrm{I}_{\mathrm{R}}$,
 	\item  $s_{1,+}=\frac{1}{2}\frac{1-\alpha}{1+\alpha}$, $s_{0,-}=s_{1,-}=\frac{1}{2}\frac{3+\alpha}{1+\alpha}$ \quad\;\; for extensions of type $\mathrm{I}_{\mathrm{L}}$,
 	\item $s_{1,\pm}=s_{0,\pm}=\frac{1}{2}\frac{1-\alpha}{1+\alpha}$ \qquad\qquad\qquad \;\;\;\;\,\,for extensions of type $\mathrm{II}_a$,
 	\item $s_{1,\pm}=s_{0,\pm}=\frac{1}{2}\frac{3+\alpha}{1+\alpha}$ \qquad\qquad\qquad \;\;\;\,\, for extensions of type $\mathrm{III}$.
 \end{itemize}
\end{theorem}

\begin{proof}
 Also in this case, the proof is a matter of exporting the classification of Theorem \ref{thm:classificationUF} for the uniformly fibred self-adjoint extensions of $\mathsf{H}_\alpha$, via unitary equivalence, to the corresponding extensions of
 \[
  H_\alpha\;=\;U_\alpha^{-1}\mathsf{H}_\alpha\,U_\alpha\,.
 \]

 One then defines
 \[
  \begin{split}
   H_{\alpha,\mathrm{F}}\;&:=\;U_\alpha^{-1}\,\mathsf{H}_{\alpha,\mathrm{F}}\,U_\alpha\,, \\
   H_{\alpha,\mathrm{R}}^{[\gamma]}\;&:=\;U_\alpha^{-1}\,\mathsf{H}_{\alpha,\mathrm{R}}^{[\gamma]}\,U_\alpha\,, \\
   H_{\alpha,\mathrm{L}}^{[\gamma]}\;&:=\;U_\alpha^{-1}\,\mathsf{H}_{\alpha,\mathrm{L}}^{[\gamma]}\,U_\alpha\,, \\
   H_{\alpha,a}^{[\gamma]}\;&:=\;U_\alpha^{-1}\,\mathsf{H}_{\alpha,a}^{[\gamma]}\,U_\alpha\,, \\
   H_{\alpha}^{[\Gamma]}\;&:=\;U_\alpha^{-1}\,\mathsf{H}_{\alpha}^{[\Gamma]}\,U_\alpha\,.
  \end{split}
 \]
 By construction, the above operators are self-adjoint and extend $H_\alpha$. They are restrictions of $H_\alpha^*$ and as such, in view of Proposition \ref{prop:adjoint_on_M}, each element in their domain satisfy the integrability and regularity condition \eqref{eq:DHalpha_cond1}.

 A generic function $f$ in the domain of one of the above extensions is by construction, owing to \eqref{eq:unit1}, of the form
 \[
  f\;=\;|x|^{\frac{\alpha}{2}}\phi
 \]
 for some $\phi$ in the domain of the corresponding unitarily equivalent operator. This and  \eqref{eq:limitphi0}-\eqref{eq:limitphi1} then yield
 \[
  \begin{split}
   \phi_0^\pm(y)\;&=\;\lim_{x\to 0^\pm}f(x,y)\;=:\;f_0^\pm(y) \\
   \phi_1^\pm(y)\;&=\;\pm(1+\alpha)^{-1}\lim_{x\to 0^\pm}|x|^{-\alpha}\partial_xf(x,y)\;=:\;f_1^\pm(y)\,.
  \end{split}
 \]
 One thus sees that for a.e.~$y\in\mathbb{S}^1$ the limits \eqref{eq:DHalpha_cond2_limits-1}-\eqref{eq:DHalpha_cond2_limits-2} do exists, and are finite for a.e.~$y$ because both $\phi_0$ and $\phi_1$ belong to $L^2(\mathbb{S}^1)$.

 In fact, the additional Sobolev regularity of $f_0$ and $f_1$ is the same as for $\phi_0$ and $\phi_1$, and it is immediately imported from Theorem \ref{thm:classificationUF}.

 The very same applies to the expression of the boundary conditions of self-adjointness for each family of extensions: \eqref{eq:DHalpha_cond3_Friedrichs-NOWEIGHTS}-\eqref{eq:DHalpha_cond3_III-NOWEIGHTS} immediately imply \eqref{eq:DHalpha_cond3_Friedrichs}-\eqref{eq:DHalpha_cond3_III}.
\end{proof}

  Requirement \eqref{eq:DHalpha_cond1} in Theorem \ref{thm:H_alpha_fibred_extensions} amounts to saying that all the considered extensions are contained in $H_\alpha^*$ (Proposition \ref{prop:adjoint_on_M}). Each of the requirements \eqref{eq:DHalpha_cond3_Friedrichs}-\eqref{eq:DHalpha_cond3_III} then expresses the corresponding condition of self-adjointness.

 The common feature of all such extensions is that their boundary conditions as $x\to 0$ have the \emph{same} form uniformly in $y\in\mathbb{R}$. In this precise sense, those are \emph{local} extensions.

 It is also clear that the Friedrichs extension, as well as type-$\mathrm{I_R}$ and type-$\mathrm{I_L}$ extensions, are reduced with respect to the Hilbert space decomposition \eqref{eq:V-Hspacedirectsum2}: each such operator is the orthogonal sum of two self-adjoint operators, respectively on $L^2(M^+,\ud\mu_\alpha)$ and $L^2(M^-,\ud\mu_\alpha)$, characterised by independent boundary conditions at the singularity locus $\mathcal{Z}$ from the right and from the left. On the contrary, type-$\mathrm{II}_a$ (with $a\neq 0$) and type-$\mathrm{III}$ extensions are not reduced \emph{in general}: the boundary condition couples the behaviour as $x\to 0^+$ and $x\to 0^-$.

 The physical picture emerging from Theorem \ref{thm:H_alpha_fibred_extensions} is then the following.
\begin{itemize}
 \item The Friedrichs extension $H_{\alpha,\mathrm{F}}$ models quantum confinement on each half of the Grushin cylinder, with no interaction of the particle with the boundary and no dynamical transmission between the two halves. 
 \item Type-$\mathrm{I_R}$ and type-$\mathrm{I_L}$ extensions model systems with no dynamical transmission across $\mathcal{Z}$, but with possible non-trivial interaction of the quantum particle with the boundary respectively from the right or from the left, with confinement on the opposite side. For instance, a quantum particle governed by $H_{\alpha,\mathrm{R}}^{[\gamma]}$ may `touch' the boundary from the right, but not from the left, and moreover it cannot trespass the singularity region.
 \item Type-$\mathrm{II}_a$ and type-$\mathrm{III}$ extensions model in general, dynamical transmission through the boundary.
\end{itemize}

One further observation on Theorem \ref{thm:H_alpha_fibred_extensions} (see also Remark \ref{rem:regularitydeficiencyspace} for a more explicit comment on this point) concerns the regularity \eqref{eq:traceregularity} of the boundary functions $f_0$ and $f_1$ in terms of which the various conditions of self-adjointness are expressed. In fact, \eqref{eq:DHalpha_cond2_limits-1}-\eqref{eq:DHalpha_cond2_limits-2} define the \emph{trace maps}\index{trace map}
\[
\begin{split}
& \gamma_0^\pm:\mathcal{D}(\widetilde{H}_\alpha)\cap L^2(M^\pm,\ud\mu_\alpha)\to H^{s_0,\pm}(\mathbb{S}^1)\,,  \\
& \gamma_1^\pm:\mathcal{D}(\widetilde{H}_\alpha)\cap L^2(M^\pm,\ud\mu_\alpha)\to H^{s_1,\pm}(\mathbb{S}^1)
\end{split}
\]
(actually, concrete examples of what one customarily refers to as \emph{abstract trace maps} -- see, e.g., \cite[Section 2]{Posilicano-2014-sum-trace-maps}), where $\widetilde{H}_\alpha$ stands for one of the considered extensions of $H_\alpha$. In fact, property \eqref{eq:traceregularity} is completely consistent with the abstract analysis, recently developed in \cite{Posilicano-2014-sum-trace-maps}, of the trace space, and hence also, isomorphically speaking, of the deficiency space, of the operator $H_\alpha$.

Of course, also in the language of \cite{Posilicano-2014-sum-trace-maps}, namely the framework of direct sum of trace maps, the hard part of the job that remains to be done for the classification of the (local) extensions of $H_\alpha$, and which is completed here, is the passage from the `natural' direct sum setting, namely the description of the restrictions of the direct sum operator $\mathscr{H}_\alpha^*=\bigoplus_{k\in\mathbb{Z}}(A_\alpha(k))^*$, to the original Grushin setting, namely the corresponding descriptions of the restrictions of $H_\alpha^*$.

\section{Spectral analysis of uniformly fibred extensions}\label{sec:V-spectralanalysis}

 The Kre{\u\i}n-Vi\v{s}ik-Birman scheme provides general tools (Sect.~\ref{sec:II-spectralKVB}) that can be applied in particular to the self-adjoint extensions classified in Theorem \ref{thm:H_alpha_fibred_extensions}, in order to extract relevant spectral information. This yields the following main results.

 First, it is of relevance for concrete models of Grushin-type transmission to characterise those Laplace-Beltrami\index{Laplace-Beltrami operator} self-adjoint realisations $\widetilde{H}_\alpha$ that are \emph{non-negative} (as the original $H_\alpha$ from \eqref{eq:V-Halpha}), and hence generate the \emph{semi-group} $(e^{-t \widetilde{H}_\alpha})_{t\geqslant 0}$ for the flow of the heat equation\index{heat flow}
\begin{equation}\label{eq:heat}
 \partial_t f\;=\;\Delta_\alpha f\,.
\end{equation}
Among non-negative generators, it then becomes of interest to select those that in addition are \emph{Markovian} and hence generate \emph{Markovian semi-groups}\index{Markovian semi-group}, defined by the property
\begin{equation}
 0\leqslant f\leqslant 1\quad(x,y)\textrm{-a.e.}\quad\Rightarrow\quad 0\leqslant e^{-t \widetilde{H}_\alpha}f\leqslant 1\quad(x,y)\textrm{-a.e.},\;\;\;\forall t\geqslant 0\,. 
\end{equation}
Each such Markovian extension $\widetilde{H}_\alpha$ therefore generates a Markov process's\index{Markov process} $(X_t)_{t\geqslant 0}$, for which it is relevant to inquire the possible stochastic completeness and recurrence. The former property, in particular, expresses the circumstance that the process $(X_t)_{t\geqslant 0}$ has infinite lifespan almost surely, which is interpreted as the fact that along the evolution \eqref{eq:heat} the heat is not absorbed by $\mathcal{Z}$. Such a programme (see, e.g., \cite{Fukushima-Oshima-Takeda}) was carried out in \cite{Boscain-Prandi-JDE-2016} for certain distinguished non-negative realisations of $\Delta_\alpha$.
 
 In this respect, the following holds.
 
\begin{theorem}[Positive extensions]\label{thm:positivity}
Let $\alpha\in[0,1)$. With respect to the self-adjoint extensions of the minimal operator $H_\alpha$ classified in Theorem \ref{thm:H_alpha_fibred_extensions}, and in terms of the extension parameters $\gamma$ and $\Gamma$ introduced therein,
\begin{itemize}
		\item the Friedrichs extension $H_{\alpha,\mathrm{F}}$ is non-negative;
		\item extensions in the family $\mathrm{I_R}$, $\mathrm{I_L}$, and $\mathrm{II}_a$, $a\in\mathbb{C}$, are non-negative if and only if $\gamma\geqslant 0$;
		\item extensions in the family $\mathrm{III}$ are non-negative if and only if so is the matrix
		\[
		 \widetilde{\Gamma}\;:=\;\begin{pmatrix}
		  \gamma_1 & \gamma_2+\ii\gamma_3 \\
		  \gamma_2-\ii\gamma_3 & \gamma_4
		 \end{pmatrix},
		\]
                i.e., if and only if $\gamma_1 + \gamma_4 > 0$ and $\gamma_1 \gamma_4\geqslant \gamma_2^2 + \gamma_3^2$.
\end{itemize}
\end{theorem}

   Next, the spectrum of each self-adjoint Hamiltonian of the family described by Theorem \ref{thm:H_alpha_fibred_extensions} can be characterised. The structure of each spectrum turns out to consist of a common essential spectrum, the non-negative half-line, in which an infinity of eigenvalues are embedded, plus a finite negative discrete spectrum.

  Using the convenient notation
	\begin{equation}
		\lfloor x \rfloor \;:=\; \begin{cases}
			\;n & \textrm{if $x \in (n,n+1]$ for some }n\in\mathbb{N}_0\,, \\
			\;0 & \textrm{if $x \leqslant 0$}\,,
		\end{cases}
	\end{equation}
  the result is the following.

\begin{theorem}[Spectral analysis of $H_\alpha$]\label{thm:MainSpectral} 
 Let $\alpha\in[0,1)$. With respect to the self-adjoint extensions of the minimal operator $H_\alpha$ classified in Theorem \ref{thm:H_alpha_fibred_extensions}, and in terms of the extension parameters $\gamma$ and $\Gamma$ introduced therein, each such extension has finite negative discrete spectrum, and essential spectrum equal to $[0,+\infty)$. In particular, any such operator is lower semi-bounded. The essential spectrum contains in each case a (countable) infinity of embedded eigenvalues, with no accumulation, each of finite multiplicity. The number $\mathcal{N}_-(\widetilde{H}_\alpha)$ of negative eigenvalues for each considered extension $\widetilde{H}_\alpha$, counted with their multiplicity, is computed as follows.
\begin{equation}
 \mathcal{N}_-\big(H_{\alpha,\mathrm{F}}\big)\;=\;0\,.
\end{equation}
\begin{equation}
 \mathcal{N}_-\big(H_{\alpha,\mathrm{R}}^{[\gamma]}\big)\;=\;\mathcal{N}_-\big(H_{\alpha,\mathrm{L}}^{[\gamma]}\big)\;=\;2 \lfloor (1+\alpha) |\gamma| \rfloor +1\,,\qquad\gamma<0\,.
\end{equation}
\begin{equation}
 \mathcal{N}_-\big(H_{\alpha,a}^{[\gamma]}\big)\;=\;2\left\lfloor \frac{1+\alpha}{1+|a|^2} |\gamma| \right\rfloor	 +1\,,\qquad \gamma<0\,,\;a\in\mathbb{C}\,.
\end{equation}
  \begin{equation}\label{eq:multnegspec-III}
   \begin{split}
    \mathcal{N}_-\big(H_{\alpha}^{[\Gamma]}\big)\;&=\;\textstyle 2\bigg\lfloor -(1+\alpha)\big(\gamma_1+\gamma_4+\sqrt{(\gamma_1-\gamma_4)^2+4 (\gamma_2^2 + \gamma_3^2)}\,\big)\bigg\rfloor \\
    & \quad + \textstyle 2\bigg\lfloor -(1+\alpha)\big(\gamma_1+\gamma_4-\sqrt{(\gamma_1-\gamma_4)^2+4 (\gamma_2^2 + \gamma_3^2)}\,\big)\bigg\rfloor \\
    & \quad + n_0(\Gamma)
   \end{split}
  \end{equation}
   with
   \begin{equation}
    n_0(\Gamma)\,:=\,\begin{cases}
   \;2\,, & \textrm{if}\quad\gamma_1 \, \gamma_4>\gamma_2^2 + \gamma_3^2 \quad \text{and} \quad \gamma_1+\gamma_4<0\,, \\
   \;1\,, & \textrm{if}\quad \gamma_1 \, \gamma_4<\gamma_2^2 + \gamma_3^2\quad\textrm{or}\quad
   \begin{cases}
    \;\gamma_1 \, \gamma_4 =\gamma_2^2 + \gamma_3^2\,, \\
    \;\gamma_1+\gamma_4<0\,,
   \end{cases} \\
   \;0\,, & \textrm{if}\quad \gamma_1 \, \gamma_4\geqslant\gamma_2^2 + \gamma_3^2 \quad \text{and} \quad \gamma_1+\gamma_4>0\,.
  \end{cases}
   \end{equation}
\end{theorem}

For those Hamiltonian admitting negative bound states and hence a negative lowest-energy eigenstate (the ground state\index{ground state}), the ground state's energy and wave function are now characterised. 
The latter is expressed in terms of the special function $K_{\frac{1+\alpha}{2}}$, smooth on $\mathbb{R}^+$, where $K_\nu$ denotes the modified Bessel function \cite[Section 9.6.1]{Abramowitz-Stegun-1964}.


\begin{theorem}[Ground-states]\label{thm:GroundStateCH}
Let $\alpha\in[0,1)$, $\gamma\in\mathbb{R}$, $a\in\mathbb{C}$, $\Gamma\equiv(\gamma_1,\gamma_2,\gamma_3,\gamma_4)\in\mathbb{R}^4$.
\begin{enumerate}
 \item[(i)] When $H_{\alpha,\mathrm{R}}^{[\gamma]}$ (respectively, $H_{\alpha,\mathrm{L}}^{[\gamma]}$) has negative spectrum, i.e., when $\gamma<0$, it has a unique ground state with energy $ E_0\big(H_{\alpha,\mathrm{R}}^{[\gamma]}\big)$ (respectively, $ E_0\big(H_{\alpha,\mathrm{L}}^{[\gamma]}\big)$) given by
 \begin{equation}\label{eq:gsenergyR}
  E_0\big(H_{\alpha,\mathrm{R}}^{[\gamma]}\big)\;=\;E_0\big(H_{\alpha,\mathrm{L}}^{[\gamma]}\big)\;=\; -\left(2\,\frac{\Gamma(\frac{1+\alpha}{2})}{\,\Gamma(-\frac{1+\alpha}{2})}\, \gamma \right)^{\frac{2}{1+\alpha}} 
 \end{equation}
 and non-normalised eigenfunction $\Phi_{\alpha}^{(\mathrm{I_R})}$ (respectively, $\Phi_{\alpha}^{(\mathrm{I_L})}$) given by
 \begin{equation}\label{eq:gs--R}
 \begin{split}
  \Phi_{\alpha}^{(\mathrm{I_R})}(x,y)\;&=\;
  \begin{cases}
   \qquad 0\,, & x<0\,, \\
   \;x^{\frac{1+\alpha}{2}}K_{\frac{1+\alpha}{2}}(x\sqrt{E})\,, & x>0\,,
  \end{cases} \\
 \Phi_{\alpha}^{(\mathrm{I_L})}(x,y)\;&=\;
  \begin{cases}
   \;(-x)^{\frac{1+\alpha}{2}}K_{\frac{1+\alpha}{2}}(-x\sqrt{E})\,, & x<0\,, \\
   \qquad 0\,, & x>0\,,
  \end{cases}
 \end{split}
\end{equation}
  where for short $E\equiv -E_0\big(H_{\alpha,\mathrm{R}}^{[\gamma]}\big)=-E_0\big(H_{\alpha,\mathrm{L}}^{[\gamma]}\big)$.
  \item[(ii)] When $H_{\alpha,a}^{[\gamma]}$ has negative spectrum, i.e., when $\gamma<0$, it has a unique ground state with energy $E_0\big(H_{\alpha,a}^{[\gamma]}\big)$ given by
 \begin{equation}
  E_0\big(H_{\alpha,a}^{[\gamma]}\big)\;=\;-\left(\frac{2\,\Gamma(\frac{1+\alpha}{2})}{(1+|a|^2)\Gamma(-\frac{1+\alpha}{2})} \,\gamma \right)^{\frac{2}{1+\alpha}}
 \end{equation}
 and non-normalised eigenfunction $\Phi_{\alpha}^{(\mathrm{II}_a)}$ given by
 \begin{equation}
  \Phi_{\alpha}^{(\mathrm{II}_a)}(x,y)\;=\;
  \begin{cases}
   \; (-x)^{\frac{1+\alpha}{2}}K_{\frac{1+\alpha}{2}}(-x\sqrt{E})\,, & x<0\,, \\
   \;a x^{\frac{1+\alpha}{2}}K_{\frac{1+\alpha}{2}}(x\sqrt{E})\,, & x>0\,,
  \end{cases}
\end{equation}
  where for short $E\equiv -E_0\big(H_{\alpha,a}^{[\gamma]}\big)$.
 \item[(iii)] When $H_{\alpha}^{[\Gamma]}$ has negative spectrum (see Theorems \ref{thm:positivity}-\ref{thm:MainSpectral}), its ground state energy $E_0\big(H_{\alpha}^{[\Gamma]}\big)$ is given by
 \begin{equation}
  E_0\big(H_{\alpha}^{[\Gamma]}\big)\;=\;-\left(\frac{\,2^\alpha\,\Gamma(\frac{1+\alpha}{2})}{\Gamma(-\frac{1+\alpha}{2})} \Big( \gamma_1+\gamma_4 - \sqrt{(\gamma_1-\gamma_4)^2 +4 (\gamma_2^2+\gamma_3^2)} \,\Big)\right)^{\frac{2}{1+\alpha}}.
 \end{equation}
  The ground state has at most two-fold degeneracy. It is non-degenerate if and only if, under the negativity condition, i.e., $\gamma_1+\gamma_4<\sqrt{(\gamma_1-\gamma_4)^2 +4 (\gamma_2^2+\gamma_3^2)}$, additionally one has $\gamma_1\neq\gamma_4$ or $\gamma_2^2+\gamma_3^2>0$, in which case the non-normalised  eigenfunction $\Phi_{\alpha}^{(\mathrm{III})}$ is given by
\begin{equation}
\begin{split}
 & \Phi_{\alpha}^{(\mathrm{III})}(x,y)\;=\;\begin{cases}
  \!\!\begin{array}{l}
   \big({\textstyle \gamma_1-\gamma_4 - \sqrt{(\gamma_1-\gamma_4)^2 +4 (\gamma_2^2+\gamma_3^2)}}\big)\,\times \\
   \qquad \times \,(-x)^{\frac{1+\alpha}{2}}K_{\frac{1+\alpha}{2}}(-x\sqrt{E})\,,
  \end{array}
  & x<0\,, \\
  \;2(\gamma_2-\ii\gamma_3)\,x^{\frac{1+\alpha}{2}}\,K_{\frac{1+\alpha}{2}}(x\sqrt{E})\,, & x>0\,,
 \end{cases}
\end{split}
\end{equation}
where for short $E\equiv -E_0\big(H_{\alpha}^{[\Gamma]}\big)$.
 The ground state is two-fold degenerate if and only if $\gamma_1=\gamma_4<0$ and $\gamma_2=\gamma_3=0$, in which case its eigenspace is spanned by the non-normalised eigenfunctions $\Phi_{\alpha,c_1,c_2}^{(\mathrm{III})}$ given by
  \begin{equation}
  \Phi_{\alpha,c_1,c_2}^{(\mathrm{III})}(x,y)\;=\;
  \begin{cases}
   \; c_1 (-x)^{\frac{1+\alpha}{2}}K_{\frac{1+\alpha}{2}}(-x\sqrt{E})\,, & x<0\,, \\
   \; c_2\, x^{\frac{1+\alpha}{2}}K_{\frac{1+\alpha}{2}}(x\sqrt{E})\,, & x>0\,,
  \end{cases}
  \end{equation}
 where $c_1,c_2\in\mathbb{C}$, and again $E\equiv -E_0\big(H_{\alpha}^{[\Gamma]}\big)$.
\end{enumerate}
\end{theorem}

 The Friedrichs extension $H_{\alpha,\mathrm{F}}$ is not covered by Theorem \ref{thm:GroundStateCH} because it has no negative spectrum, its spectrum being $[0,+\infty)$ and purely essential. It is of interest, nevertheless, to get information on the first (lowest) energy eigenstate of $H_{\alpha,\mathrm{F}}$ embedded in the essential spectrum (as described in Theorem \ref{thm:MainSpectral}).

  \begin{proposition}\label{prop:Friedrichs-groundstate}
  For given $\alpha\in[0,1)$, the (lowest) energy eigenstate of $H_{\alpha,\mathrm{F}}$ has eigenvalue $E_0(H_{\alpha,\mathrm{F}})$ embedded in $[0,+\infty)$, which is two-fold degenerate, and is estimated as
 \begin{equation}\label{eq:estimate-Fgroundstate}
	(1+\alpha)\left({\textstyle\frac{2+\alpha}{4} }\right)^{\frac{\alpha}{1+\alpha}}  \; \leqslant \;    E_0(H_{\alpha,\mathrm{F}}) \; \leqslant \;  \frac{2^{\frac{1-\alpha}{1+\alpha}} (1+\alpha)^{\frac{1+3\alpha}{1+\alpha}}}{\alpha^{\frac{\alpha}{1+\alpha}} \Gamma(\frac{3+\alpha}{1+\alpha})}\,.
\end{equation}
  In particular, $E_0(H_{\alpha,\mathrm{F}})$ is strictly positive, and the lowest spectral point $0$ in the spectrum of $H_{\alpha,\mathrm{F}}$ is not an eigenvalue.  
 \end{proposition}

\begin{figure}[t!]
	\includegraphics[width=6.5cm]{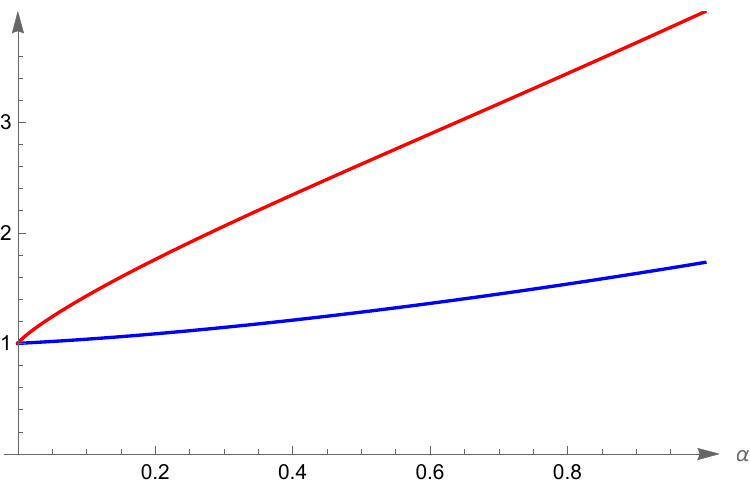}
	\caption{Lower and upper bounds \eqref{eq:estimate-Fgroundstate}}
	\label{fig:variational-estimate}
\end{figure}

  Figure \ref{fig:variational-estimate} shows the effectiveness and narrowness of the bounds \eqref{eq:estimate-Fgroundstate} for all admissible values of $\alpha$.

\begin{figure}[t!]
	\includegraphics[width=0.46\textwidth]{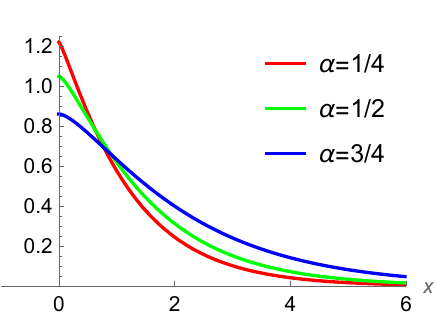} $\quad$  
\includegraphics[width=0.46\textwidth]{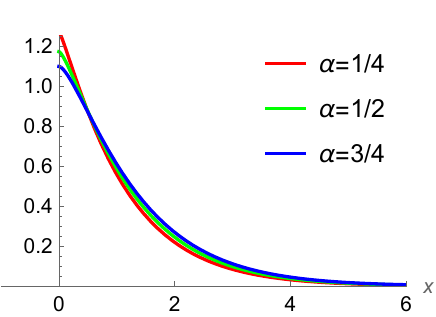}
	\caption{$x$-profile of the (constant in $y$) normalised ground state wave function of the Hamiltonian $H_{\alpha,\mathrm{R}}^{[\gamma]}$ for various $\alpha$. Left: case $\gamma=-1$, thus with different ground state energies depending on $\alpha$ according to \eqref{eq:gs--R}. Right: case with ground state energy $E_0\big(H_{\alpha,\mathrm{R}}^{[\gamma]}\big)=-1$, thus with different $\gamma$ for each $\alpha$, according to \eqref{eq:gsenergyR}. The smaller the parameter $\alpha$, the more pronounced the localisation of the wave function around the metric's singularity.}
	\label{fig:3}
\end{figure}

The structure of the ground state wave functions described in Theorem \ref{thm:GroundStateCH} is qualitatively the same for each model and consists of a behaviour $|x|^{\frac{1+\alpha}{2}}K_{\frac{1+\alpha}{2}}(|x|)$, on both half-lines when applicable, while being constant in $y$. The latter feature, as emerges from Subsections \ref{subsec:Spectrum0}-\ref{subsec:SpectrumK}, is due to the compactness of the $y$-variable in $M$: the lowest energy level of the Hamiltonian is a contribution from the ``zero-th'' mode of functions in $y$, the constant-in-$y$ functions. The function $|x|^{\frac{1+\alpha}{2}}K_{\frac{1+\alpha}{2}}(|x|)$ is localised around $x=0$ with exponential fall off at infinity, thus expressing the localisation of the ground states around the Grushin singularity of the manifold. The $\sqrt{E}$-factor in the argument of the Bessel function does not affect such a conclusion. In fact, irrespective of $\alpha$, given any negative number $-E$ there is one model out of each family of Theorem \ref{thm:H_alpha_fibred_extensions} with ground state energy level precisely equal to $-E$: indeed, it is always possible to choose the extension parameters $\gamma=\gamma^{(\mathrm{I})}$ for $\mathrm{I_R}$ or $\mathrm{I_L}$,  $\gamma=\gamma^{(\mathrm{II}_a)}$ for $\mathrm{II}_a$, and $\Gamma$ for $\mathrm{III}$ such that
\[
 \textstyle\gamma^{(\mathrm{I})}\;=\;\frac{1}{\,1+|a|^2}\,\gamma^{(\mathrm{II}_a)}\;=\;2^{\alpha-1}\big(\gamma_1+\gamma_4 - \sqrt{(\gamma_1-\gamma_4)^2 +4 (\gamma_2^2+\gamma_3^2)}\big)\;=:\;\theta\;<0\,,
\]
in which case Theorem \ref{thm:GroundStateCH} implies 
\[
 -E\;\equiv\;E_0\big(H_{\alpha,\mathrm{R}}^{[ \gamma^{(\mathrm{I})}]}\big)\;=\;E_0\big(H_{\alpha,a}^{[\gamma^{(\mathrm{II}_a)}]}\big)\;=\;E_0\big(H_{\alpha}^{[\Gamma]}\big)\;=\;-\textstyle\left(\frac{\,2\,\theta\,\Gamma(\frac{1+\alpha}{2})}{\,\Gamma(-\frac{1+\alpha}{2})}\right)^{\frac{2}{1+\alpha}}.
\]
In particular (Figure \ref{fig:3}), at larger $\alpha$'s the de-localisation away from $x=0$ becomes more pronounced, as the Grushin metric becomes more singular.

\begin{figure}[!t]
\captionsetup[subfigure]{labelformat=empty} 
  \centering
  \subfloat[][$\begin{array}{c}\textrm{excited state of mode $k=\pm 1$} \\ \textrm{for $H_{\alpha,a}^{[\gamma]}$ with $a=1,\gamma=1$}\end{array}$]
  {\includegraphics[width=0.45\textwidth]{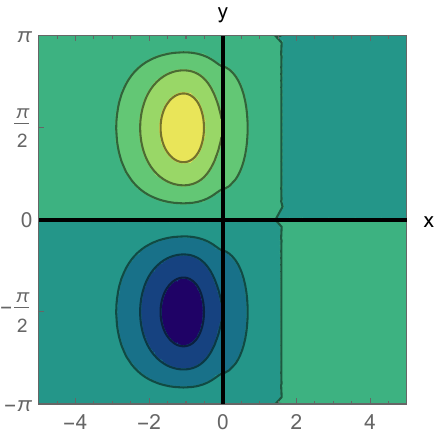} }
  \subfloat[][$\begin{array}{c}\textrm{excited state of mode $k=\pm 1$} \\ \textrm{for $H_{\alpha,a}^{[\gamma]}$ with $a=-2,\gamma=-1$}\end{array}$]
  {\includegraphics[width=0.45\textwidth]{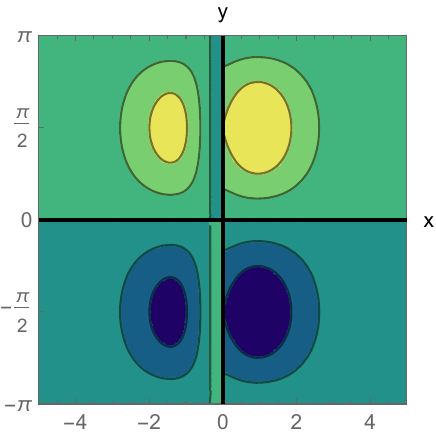} } \\
  \vspace{-0.4cm}
  \subfloat[][$\begin{array}{c}\textrm{eigenstate of mode $k=\pm 1$} \\ \textrm{for $H_{\alpha,\mathrm{F}}$}\end{array}$]
  {\includegraphics[width=0.45\textwidth]{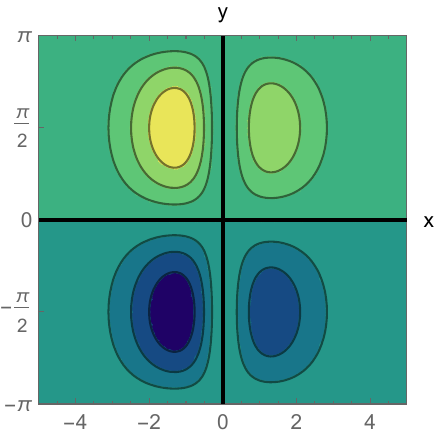} } 
  \subfloat[][$\begin{array}{c}\textrm{excited state of mode $k=\pm 1$} \\ \textrm{for $H_{\alpha,a}^{[\gamma]}$ with $a=1,\gamma=0$  (bridging)}\end{array}$]
  {\includegraphics[width=0.45\textwidth]{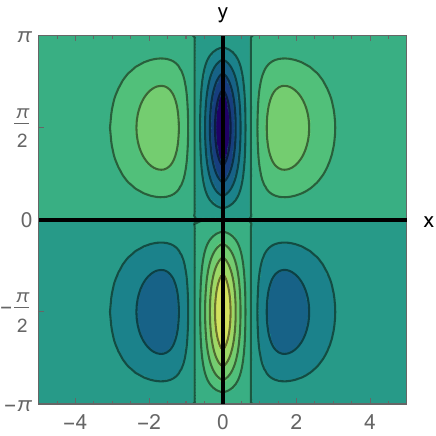} }
  \caption{Contour plot of bound states for various Hamiltonians of Theorem \ref{thm:H_alpha_fibred_extensions}, all with $\alpha=\frac{1}{2}$. The considered bound states, whose energy is embedded in the spectral interval $[0,+\infty)$, are produced by the mode $k=\pm 1$ of the fibred representation of the Hamiltonian described in Section \ref{sec:constant-fibre-sum} and Theorem \ref{thm:fibredHalpha-spectrum}: they thus display the simplest admissible $y$-oscillation ($\sin y$).}
  \label{fig:contourplots}
\end{figure}

Above the ground state described in Theorem \ref{thm:GroundStateCH}, each Hamiltonian $H_{\alpha,\mathrm{F}}$, $H_{\alpha,\mathrm{R}}^{[\gamma]}$, $H_{\alpha,\mathrm{L}}^{[\gamma]}$, $H_{\alpha,a}^{[\gamma]}$, $H_{\alpha}^{[\Gamma]}$ exhibits an infinite multitude of eigenvalues, all of finite multiplicity, a finite number of them negative, all the others embedded in the essential spectrum $[0,+\infty)$ (Theorem \ref{thm:MainSpectral}). Unlike the ground state wave function, all such excited states do display oscillatory behaviour in the $y$-variable.

Figure \ref{fig:contourplots} displays the contour plots on $(-L,L)_x\times [0,2\pi]_y$, for conveniently large $L>0$, of eigenstate wave functions of various types of Hamiltonians, in all cases with energy level given by the `$k=\pm 1$ mode' (in the precise sense of Theorem \ref{thm:fibredHalpha-spectrum} below), meaning that the $y$-oscillation is of the form $\sin y$. At one extreme of the range of possible behaviours there is the Friedrichs extension, whose bound states are well confined \emph{away} from the Grushin singularity $x=0$. Intermediate behaviours are those with some degree of discontinuity at $x=0$, in which case the transmission protocol governed by the corresponding Hamiltonian is affected by partial absorption at the Grushin singularity. At the other extreme, the distinguished protocol governed by $H_{\alpha,a}^{[\gamma]}$ with $a=1$ and $\gamma=0$ results instead in a smooth behaviour of the eigenstate wave functions (like the one displayed in Figure \ref{fig:contourplots}) around $x=0$, a signature of complete communication between the two half-cylinders.

   Theorems \ref{thm:positivity}-\ref{thm:GroundStateCH} and Proposition \ref{prop:Friedrichs-groundstate} are proved in Section \ref{subsec:V-scatt-proofs} after performing in Sections \ref{subsec:Spectrum0}-\ref{subsec:SpectrumK} the spectral analysis of the self-adjoint operators of the families identified in the two-sided fibre extension analysis (Theorems \ref{thm:bifibre-extensions}-\ref{prop:g_with_Pweight}), namely
   \begin{equation}\label{eq:families-ext-Aak}
 \begin{split}
  & A_{\alpha,\mathrm{F}}(k)\,,\\
  & \{A_{\alpha,\mathrm{R}}^{[\gamma]}(k)\,|\,\gamma\in\mathbb{R}\}\,, \\
  & \{A_{\alpha,\mathrm{L}}^{[\gamma]}(k)\,|\,\gamma\in\mathbb{R}\}\,, \\
  & \{A_{\alpha,a}^{[\gamma]}(k)\,|\,\gamma\in\mathbb{R}\}\qquad\textrm{for fixed $a\in\mathbb{C}$}\,, \\
  & \{A_{\alpha}^{[\Gamma]}(k)\,|\,\Gamma\equiv(\gamma_1,\gamma_2,\gamma_3,\gamma_4)\in\mathbb{R}^4\}\,,
 \end{split}
\end{equation} 
and re-assembling the result on each fibre by means of the orthogonal sums \eqref{eq:HalphaFriedrichs_unif-fibred} and \eqref{eq:HalphaR_unif-fibred}-\eqref{eq:Halpha-III_unif-fibred}.

  Recall from Section \ref{sec:bilateralfibreext} that the final parametrisation of the extensions \eqref{eq:families-ext-Aak} in terms of $\gamma\in\mathbb{R}$ or $\Gamma\in\mathbb{R}^4$ emerged as a convenient re-labelling (yielding the more transparent boundary conditions of Table \ref{tab:extensions}) of such extensions initially identified with the intrinsic Birman extension parameter \index{Birman extension parameter} of the Kre{\u\i}n-Vi\v{s}ik-Birman classification scheme (Theorem \ref{thm:VB-representaton-theorem_Tversion}). For the present purposes, it is convenient to reinstate the Birman parametrisation, as it is in terms of the spectral properties of the Birman parameter of each such self-adjoint realisation that the spectral properties of the labelled operator can be characterised (Sect.~\ref{sec:II-spectralKVB}).

  The Birman parameters for the extensions \eqref{eq:families-ext-Aak} are easily read out from the proof of Theorem  \ref{thm:bifibre-extensions} (they also appear in the formulation of Theorem \ref{thm:bifibre-extensionsc0c1}). They are self-adjoint operators acting on the one- and two-dimensional spaces $\mathbb{C}$ and $\mathbb{C}^2$, and hence, respectively, with the multiplications by a real number and by a hermitian $2\times 2$ matrix, explicitly given as follows.
 \begin{itemize}
	\item The Birman parameter for the sub-families $\{A_{\alpha,\mathrm{R}}^{[\gamma]}(k)\,|\,\gamma\in\mathbb{R}\}$ and $\{A_{\alpha,\mathrm{L}}^{[\gamma]}(k)\,|\,\gamma\in\mathbb{R}\}$ is the $\mathbb{C}\to\mathbb{C}$ multiplication by the real number $\beta_k$, that is linked to the parameter $\gamma$ by the formula
	\begin{equation}\label{eq:GammaIRIL}
	\beta_k\;=\; \frac{2^{\frac{1-\alpha}{1+\alpha}}(1+\alpha)^{\frac{2\alpha}{1+\alpha}}|k|^{\frac{2}{1+\alpha}}}{\Gamma(\frac{1-\alpha}{1+\alpha})} \left( \frac{1+\alpha}{|k|} \gamma + 1 \right)\qquad (k\neq 0),
	\end{equation}
	and the extension $A_{\alpha,\mathrm{R}}^{[\gamma]}(k)$ (resp., $A_{\alpha,\mathrm{L}}^{[\gamma]}(k)$) is identified by $\beta_k$.
	\item The Birman parameter for the sub-family $\{A_{\alpha,a}^{[\gamma]}(k)\,|\,\gamma\in\mathbb{R}\}$, at fixed $a\in\mathbb{C}$, is the $\mathbb{C}\to\mathbb{C}$ multiplication by the real number $\tau_k$, that is linked to the parameter $\gamma$ by the formula
	\begin{equation}\label{eq:GammaTauIIa}
	\tau_k \;=\; \frac{2^{\frac{1-\alpha}{1+\alpha}}(1+\alpha)^{\frac{2\alpha}{1+\alpha}}|k|^{\frac{2}{1+\alpha}}}{\Gamma(\frac{1-\alpha}{1+\alpha})} \left(1+\frac{(1+\alpha) \gamma}{(1+|a|^2) |k|} \right)\qquad (k\neq 0),
	\end{equation}
	and the extension $A_{\alpha,a}^{[\gamma]}(k)$ is identified by $\tau_k$.
	\item The Birman parameter for the sub-family $\{A_{\alpha}^{[\Gamma]}(k)\,|\,\Gamma\in\mathbb{R}^4\}$ is the $\mathbb{C}^2\to\mathbb{C}^2$ linear map induced by the hermitian matrix
	\begin{equation}\label{eq:IIkTheorem51new}
	\begin{split}
	 			T_k\;&\equiv\;\begin{pmatrix}
			\tau_{1,k} & \tau_{2,k} + \ii \tau_{3,k} \\
			\tau_{2,k} - \ii \tau_{3,k} & \tau_{4,k}
			\end{pmatrix} \\
			&=\; \frac{\,2^{\frac{1-\alpha}{1+\alpha}}(1+\alpha)^{\frac{2\alpha}{1+\alpha}}|k|^{\frac{2}{1+\alpha}}}{\Gamma(\frac{1-\alpha}{1+\alpha})} \begin{pmatrix}
				1+ \frac{1+\alpha}{|k|} \gamma_1 & \frac{1+\alpha}{|k|}(\gamma_2 + \ii \gamma_3) \\
				\frac{1+\alpha}{|k|}(\gamma_2 - \ii \gamma_3) & 1+ \frac{1+\alpha}{|k|} \gamma_4
			\end{pmatrix} \\
			&\qquad (k\neq 0),
	\end{split}
	\end{equation}
	and the extension $\{A_{\alpha}^{[\Gamma]}(k)$ is identified by $T_k$.
\end{itemize}

 \subsection{Spectral analysis of the zero mode}\label{subsec:Spectrum0}

 The mode $k=0$ in the collection \eqref{eq:families-ext-Aak} requires a separate analysis, essentially due to the fact that for the self-adjoint extensions of $A_\alpha(0)$ explicit expressions for the Birman extension parameter were not computed.
 
 There is nothing conceptually deep preventing one from identifying the Birman parameter also when $k=0$. Simply, unlike the $A_\alpha(k)$'s when $k\in\mathbb{Z}\setminus\{0\}$, the non-negative expectations $\langle g,A_\alpha(0)g \rangle_{\mathfrak{h}}$ include  zero at their bottom (see \eqref{eq:Axibottom}-\eqref{eq:Axibottom-zero}), therefore the Friedrichs extension of $A_\alpha(0)$ does not admit an everywhere defined and bounded inverse on $L^2(\mathbb{R})$. This makes the standard Kre\u{\i}n-Vi\v{s}ik-Birman scheme not applicable to the pivot extension $A_{\alpha,\mathrm{F}}(0)$, and indeed the extension analysis for $A_\alpha(0)$ goes through the auxiliary shifted operator $A_\alpha(0)+\mathbbm{1}$ (Sect.~\ref{sec:zero_mode}).

  The spectral picture for the zero mode on fibre is the following.

\begin{proposition}\label{prop:SpectrumK0} Let $\alpha\in[0,1)$. 
\begin{enumerate}
 \item[(i)] The essential spectrum of any self-adjoint extension \eqref{eq:families-ext-Aak} of $A_\alpha(0)$ satisfy
\begin{equation}\label{eq:sess0}
 \begin{split}
  &\sigma_{\mathrm{ess}}(A_{\alpha,\mathrm{F}}(0))\;=\;\sigma_{\mathrm{ess}}\big(A_{\alpha,\mathrm{R}}^{[\gamma]}(0)\big)\;=\;\sigma_{\mathrm{ess}}\big(A_{\alpha,\mathrm{L}}^{[\gamma]}(0)\big) \\
  &\quad =\;\sigma_{\mathrm{ess}}\big(A_{\alpha,a}^{[\gamma]}(0)\big)\;=\;\sigma_{\mathrm{ess}}\big(A_{\alpha}^{[\Gamma]}(0)\big)\;=\;[0,+\infty)
 \end{split}
\end{equation}
for any $\gamma\in\mathbb{R}$, $a\in\mathbb{C}$, $\Gamma\in\mathbb{R}^4$. In all cases, the essential spectrum does not contain embedded eigenvalues.
\item[(ii)] The discrete spectrum of any such extension can be therefore only strictly negative, and moreover it is empty for $A_{\alpha,\mathrm{F}}(0)$, consists of at most one negative non-degenerate eigenvalue for $A_{\alpha,\mathrm{R}}^{[\gamma]}(0)$, $A_{\alpha,\mathrm{L}}^{[\gamma]}(0)$, and $A_{\alpha,a}^{[\gamma]}(0)$, and consists of at most two negative eigenvalues for $A_{\alpha}^{[\Gamma]}(0)$, counted with multiplicity.
\item[(iii)] $A_{\alpha,\mathrm{R}}^{[\gamma]}(0)$ and $A_{\alpha,\mathrm{L}}^{[\gamma]}(0)$ admit one negative eigenvalue, denoted respectively as $E_0\big(A_{\alpha,\mathrm{R}}^{[\gamma]}(0)\big)$ and $E_0\big(A_{\alpha,\mathrm{L}}^{[\gamma]}(0)\big)$, if and only if $\gamma<0$, in which case
\begin{equation}\label{eq:EigenvalueIR}
E_0\big(A_{\alpha,\mathrm{R}}^{[\gamma]}(0)\big)\;=\;E_0\big(A_{\alpha,\mathrm{L}}^{[\gamma]}(0)\big)\;=\; -\left(2\,\frac{\Gamma(\frac{1+\alpha}{2})}{\,\Gamma(-\frac{1+\alpha}{2})}\, \gamma \right)^{\frac{2}{1+\alpha}} \, .
\end{equation}
The corresponding (non-normalised) eigenfunctions are, respectively,
\begin{equation}\label{eq:eigenf-IRIL}
 \begin{split}
  g_{\alpha}^{(\mathrm{I_R})}(x)\;&:=\;
  \begin{cases}
   \qquad 0\,, & x<0\,, \\
   \;\sqrt{x}\,K_{\frac{1+\alpha}{2}}(x\sqrt{E})\,, & x>0\,,
  \end{cases} \\
  g_{\alpha}^{(\mathrm{I_L})}(x)\;&:=\;
  \begin{cases}
   \;\sqrt{-x}\,K_{\frac{1+\alpha}{2}}(-x\sqrt{E})\,, & x<0\,, \\
   \qquad 0\,, & x>0\,,
  \end{cases}
 \end{split}
\end{equation}
 where for short $E\equiv -E_0\big(A_{\alpha,\mathrm{R}}^{[\gamma]}(0)\big)=-E_0\big(A_{\alpha,\mathrm{L}}^{[\gamma]}(0)\big)$.
\item[(iv)] For given $a\in\mathbb{C}$, $A_{\alpha,a}^{[\gamma]}(0)$ admits one negative eigenvalue, denoted as $E_0\big(A_{\alpha,a}^{[\gamma]}(0)\big)$, if and only if $\gamma<0$, in which case
\begin{equation}\label{eq:EigenvalueIIa}
E_0\big(A_{\alpha,a}^{[\gamma]}(0)\big)\;=\; -\left(\frac{2\,\Gamma(\frac{1+\alpha}{2})}{(1+|a|^2)\Gamma(-\frac{1+\alpha}{2})} \,\gamma \right)^{\frac{2}{1+\alpha}} \, .
\end{equation}
The corresponding (non-normalised) eigenfunction is
\begin{equation}\label{eq:eigenf-IIa}
  g_{\alpha}^{(\mathrm{II}_a)}(x)\;:=\;
  \begin{cases}
   \; \sqrt{-x}\,K_{\frac{1+\alpha}{2}}(-x\sqrt{E})\,, & x<0\,, \\
   \;a \sqrt{x}\,K_{\frac{1+\alpha}{2}}(x\sqrt{E})\,, & x>0\,,
  \end{cases}
\end{equation}
 where for short $E\equiv -E_0\big(A_{\alpha,a}^{[\gamma]}(0)\big)$.
\item[(v)] $A_{\alpha}^{[\Gamma]}(0)$ admits at most two negative eigenvalues: exactly two, given by
\begin{equation}\label{eq:III0-evs}
 \begin{split}
  E_0\big(A_{\alpha}^{[\Gamma]}(0)\big)\;&:=\;-\left(\frac{\,2^\alpha\,\Gamma(\frac{1+\alpha}{2})}{\Gamma(-\frac{1+\alpha}{2})} \Big( \gamma_1+\gamma_4 - \sqrt{(\gamma_1-\gamma_4)^2 +4 (\gamma_2^2+\gamma_3^2)} \,\Big)\right)^{\frac{2}{1+\alpha}}, \\
  E_1\big(A_{\alpha}^{[\Gamma]}(0)\big)\;&:=\; -\left(\frac{\,2^\alpha\,\Gamma(\frac{1+\alpha}{2})}{\Gamma(-\frac{1+\alpha}{2})}\Big( \gamma_1+\gamma_4 + \sqrt{(\gamma_1-\gamma_4)^2 +4 (\gamma_2^2+\gamma_3^2)} \,\Big)\right)^{\frac{2}{1+\alpha}},
 \end{split}
\end{equation}
with $E_0\big(A_{\alpha}^{[\Gamma]}(0)\big)\leqslant E_1\big(A_{\alpha}^{[\Gamma]}(0)\big)$, 
 if and only if
		\begin{equation}\label{eq:III0-2neg}
			\gamma_1+\gamma_4 \; < \; 0 \qquad \text{and} \qquad \gamma_1 \, \gamma_4 \; > \; \gamma_2^2 + \gamma_3^2\,,
		\end{equation}
 only one negative eigenvalue, the quantity $E_0\big(A_{\alpha}^{[\Gamma]}(0)\big)$ above, if and only if
\begin{equation}\label{eq:III0-only1neg}
  \gamma_1 \, \gamma_4 \; < \; \gamma_2^2 + \gamma_3^2\qquad\textrm{or}\qquad
  \begin{cases}
   \;\gamma_1 \, \gamma_4 \; = \; \gamma_2^2 + \gamma_3^2\,, \\
   \;\gamma_1+\gamma_4\;<\;0\,,
  \end{cases}
\end{equation}
or no negative eigenvalue at all, if and only if
\begin{equation}\label{eq:III0-none-neg}
  \gamma_1+\gamma_4 - \sqrt{(\gamma_1-\gamma_4)^2 +4 (\gamma_2^2+\gamma_3^2)} \;\geqslant \;0\,.
\end{equation}
The lowest negative eigenvalue $E_0\big(A_{\alpha}^{[\Gamma]}(0)\big)$, if existing, is non-degenerate when additionally $\gamma_1\neq\gamma_4$ or $\gamma_2^2+\gamma_3^2>0$, in which case its (non-normalised) eigenfunction is
\begin{equation}\label{eq:eigenf-III-nondegen}
 g_{\alpha}^{(\mathrm{III})}(x)\;:=\;\begin{cases}
  \!\!
  \begin{array}{l}
   \big({\textstyle \gamma_1-\gamma_4 - \sqrt{(\gamma_1-\gamma_4)^2 +4 (\gamma_2^2+\gamma_3^2)}}\big)\,\times \\
   \qquad \times\,\sqrt{-x}\,K_{\frac{1+\alpha}{2}}(-x\sqrt{E})\,,
  \end{array} 
  & x<0\,, \\
  \;2(\gamma_2-\ii\gamma_3)\,\sqrt{x}\,K_{\frac{1+\alpha}{2}}(x\sqrt{E})\,, & x>0\,,
 \end{cases}
\end{equation}
 where for short $E\equiv -E_0\big(A_{\alpha}^{[\Gamma]}(0)\big)$.
 Instead, the negative $E_0\big(A_{\alpha}^{[\Gamma]}(0)\big)$, if existing, is two-fold degenerate when  additionally $\gamma_1=\gamma_4$ and $\gamma_2=\gamma_3=0$, in which case its two-dimensional eigenspace is spanned by the (non-normalised) eigenfunctions
 \begin{equation}\label{eq:eigenf-III-degen}
 \begin{split}
  g_{\alpha,+}^{(\mathrm{III})}(x)\;&:=\;
  \begin{cases}
   \qquad 0\,, & x<0\,, \\
   \;\sqrt{x}\,K_{\frac{1+\alpha}{2}}(x\sqrt{E})\,, & x>0\,,
  \end{cases} \\
  g_{\alpha,-}^{(\mathrm{III})}(x)\;&:=\;
  \begin{cases}
   \;\sqrt{-x}\,K_{\frac{1+\alpha}{2}}(-x\sqrt{E})\,, & x<0\,. \\
   \qquad 0\,, & x>0\,.
  \end{cases}
 \end{split}
\end{equation}
\end{enumerate}
\end{proposition}

As each self-adjoint extension of $A_\alpha(0)$ is a restriction of the adjoint \eqref{eq:Afstar}, the eigenvalue problem takes in all cases the form 
\begin{equation}\label{eq:evproblem0}
 \Big(-\frac{\ud^2}{\ud x^2}+\frac{\,\alpha(2+\alpha)\,}{4x^2}+E\Big) g^\pm\;=\;0\,,
\end{equation}
where $-E$ is the eigenvalue and $g\equiv\begin{pmatrix} g^- \\ g^+ \end{pmatrix}\in\mathfrak{h}$ is the corresponding eigenfunction. This yields two independent ODEs, one on $\mathbb{R}^-$ and one on $\mathbb{R}^+$, yet identical in form: it then suffices to only solve, say, the one with $x>0$. Negative eigenvalues are found by restricting to $E>0$.

Upon re-scaling $\xi:=x\sqrt{E}$, $w(x\sqrt{E}):= x^{-\frac{1}{2}} g^+(x)$, the positive half-line version of \eqref{eq:evproblem0} becomes
\begin{equation}
	\xi^2 \frac{\ud^2 w}{\ud \xi^2}+ \xi \frac{\ud w}{\ud \xi} -(\xi^2 + \nu^2) w \; = \; 0\qquad\quad (\textstyle{\nu:=\frac{1+\alpha}{2}})\,,
\end{equation}
namely a modified Bessel equation,\index{Bessel equation} the two linearly independent solution of which are the modified Bessel functions\index{Bessel functions} $K_\nu$ and $I_\nu$ \cite[Section 9.6.1]{Abramowitz-Stegun-1964}. These are smooth functions over $\mathbb{R}^+$ satisfying the asymptotics
\begin{equation}\label{eq:k0asymptx0}
 \begin{split}
 I_{\frac{1+\alpha}{2}}(\xi)\;&\stackrel{\xi\downarrow 0}{=}\;\frac{1}{2^{\frac{1+\alpha}{2}} \Gamma({\textstyle\frac{3+\alpha}{2}})} \xi^{\frac{1+\alpha}{2}} + O(\xi^{\frac{5+\alpha}{2}})\,, \\
 K_{\frac{1+\alpha}{2}}(\xi) \;&\stackrel{\xi\downarrow 0}{=}\;2^{\frac{\alpha-1}{2}} \Gamma({\textstyle\frac{1+\alpha}{2}}) \,	\xi^{-\frac{1+\alpha}{2}} + 2^{-\frac{3+\alpha}{2}} \Gamma({\textstyle\frac{1+\alpha}{2}}) \, \xi^{\frac{1+\alpha}{2}} + O(\xi^{\frac{3-\alpha}{2}})
 \end{split}
\end{equation}
(following from \cite[Eq.~(9.6.19)]{Abramowitz-Stegun-1964}) and 
\begin{equation}
\begin{split}
 I_{\frac{1+\alpha}{2}}(\xi)\;&\stackrel{\xi\to +\infty}{=}\;\frac{e^\xi}{\sqrt{2 \pi \xi}\,} (1+O(\xi^{-1}))\,, \\
K_{\frac{1+\alpha}{2}}(\xi)\;&\stackrel{\xi\to +\infty}{=}\;{\textstyle\sqrt{\frac{\pi}{2 \xi}}}\, e^{-\xi}(1+O(\xi^{-1}))
\end{split}
\end{equation}
(following from \cite[Eq.~(9.7.1)-(9.7.2)]{Abramowitz-Stegun-1964}).

One therefore deduces that the square-integrable solutions to \eqref{eq:evproblem0} when $x>0$ form a one-dimensional space spanned by
\begin{equation}\label{eq:k0gensol}
	u_E(x) \; := \; \sqrt{x}\, K_{\frac{1+\alpha}{2}}( x\sqrt{E})\,.
\end{equation}
Thus, the general solution to \eqref{eq:evproblem0} belonging to $\mathfrak{h}\cong L^2(\mathbb{R}^-)\oplus  L^2(\mathbb{R}^+)$ has the form
\begin{equation}\label{eq:GenericGE}
	g_E(x) \; \equiv \; \begin{pmatrix}
		B u_E(-x) \\
		C u_E(x) 
	\end{pmatrix},\qquad B,C \in \mathbb{C}\,.
\end{equation}

Now, \eqref{eq:k0asymptx0} and \eqref{eq:k0gensol} yield
\begin{equation}
 \begin{split}
  & u_E(|x|) \;\stackrel{x\to 0}{=}\; 2^{-\frac{1-\alpha}{2}} \Gamma({\textstyle{\frac{1+\alpha}{2}}}) E^{-\frac{1+\alpha}{4}} |x|^{-\frac{\alpha}{2}} \\
  &\qquad\qquad\quad + 2^{-\frac{3+\alpha}{2}} \Gamma(\textstyle{-\frac{1+\alpha}{2}}) E^{\frac{1+\alpha}{4}} |x|^{1+\frac{\alpha}{2}}+ O(|x|^{2-\frac{\alpha}{2}}) \, .
 \end{split}
\end{equation}
By means of such asymptotics, the boundary values \eqref{eq:bilimitsg0g1} for the function \eqref{eq:GenericGE} are computed as
\begin{equation}\label{eq:gE_boudaryvalues}
	\begin{split}
		g_{E,0}^- \;&= \; B \, 2^{-\frac{1-\alpha}{2}} \Gamma({\textstyle{\frac{1+\alpha}{2}}}) E^{-\frac{1+\alpha}{4}}\,, \\
		g_{E,0}^+ \;& = \; C \, 2^{-\frac{1-\alpha}{2}} \Gamma({\textstyle{\frac{1+\alpha}{2}}}) E^{-\frac{1+\alpha}{4}}\,, \\
		g_{E,1}^- \;& = \; B \, 2^{-\frac{3+\alpha}{2}} \Gamma(\textstyle{-\frac{1+\alpha}{2}}) E^{\frac{1+\alpha}{4}}\,, \\
		g_{E,1}^+ \;& = \; C \, 2^{-\frac{3+\alpha}{2}} \Gamma(\textstyle{-\frac{1+\alpha}{2}}) E^{\frac{1+\alpha}{4}}\,.
	\end{split}
\end{equation}

\begin{proof}[Proof of Proposition \ref{prop:SpectrumK0}]~

Part (i) follows from the fact that $A_\alpha(0)$ has finite deficiency index (equal to 2) and therefore all its self-adjoint extension have the same essential spectrum, say, of the Friedrichs extension. Recall indeed that self-adjoint extensions of the same densely defined, operator, and lower semi-bounded operator, when the deficiency index is finite, have the respective resolvents that differ by a finite rank operator (Corollary \ref{cor:resolvent_formula_positive_extensions} or Section \ref{sec:perturbation-spectra}), and that, in turn, two self-adjoint operators with compact resolvent difference have the same essential spectrum (Sect.~\ref{sec:perturbation-spectra}). For the latter, indeed $\sigma_{\mathrm{ess}}(A_{\alpha,\mathrm{F}}(0))=[0,+\infty)$, as is easily seen from the existence of a singular Weyl sequence (Sect.~\ref{sec:I-parts-of-spectrum}) relative to any spectral point $\lambda\geqslant 0$ (the Schr\"{o}dinger operator in \eqref{eq:evproblem0} behaves like the free operator at large distances). Moreover, \eqref{eq:evproblem0} being the eigenvalue problem for a Schr\"{o}dinger operator with potential $V(x)=c|x|^{-2}$, $c>0$, it has no $L^2$-solution with $E\geqslant 0$, which proves the absence of embedded eigenvalues.

Concerning part (ii), $A_{\alpha,\mathrm{F}}(0)$ has the same lower bound zero as the original $A_\alpha(0)$ (Theorem \ref{thm:Friedrichs-ext}(iii)), and therefore does not have negative spectrum. As for the other extensions, the multiplicity of their negative spectrum is computed explicitly in the proof of the next claims of the present Proposition. In fact, it can be quantified a priori by observing that the $A_{\alpha,\mathrm{R}}^{[\gamma]}(0)$'s, $A_{\alpha,\mathrm{L}}^{[\gamma]}(0)$'s, and $A_{\alpha,a}^{[\gamma]}(0)$'s form one-parameter sub-families of extensions, meaning that their Birman parameter acts on a one-dimensional space, whereas the $A_{\alpha}^{[\Gamma]}(0)$'s form a four-parameter sub-family, meaning that their Birman parameter acts on a two-dimensional space. As the (finite) multiplicity of the negative spectrum of an extension and of its Birman parameter are the same (Corollary \ref{cor:negative_spectrum_ST_T_corollary}), the conclusion then follows.

For the remaining parts (iii)-(v), set for convenience
\begin{equation*}
\mu_0 \;:=\; - 2^{1+\alpha}\frac{\Gamma(\frac{1+\alpha}{2})}{\Gamma(-\frac{1+\alpha}{2})}\;>\;0
\end{equation*}
(indeed, $\Gamma(-\frac{1+\alpha}{2}) < 0$ and $\Gamma(\frac{1+\alpha}{2})>0$, as $\alpha \in [0,1)$).

The general square-integrable eigenfunction \eqref{eq:GenericGE} belongs to the domain of $A_{\alpha,\mathrm{R}}^{[\gamma]}(0)$ if and only if its boundary values \eqref{eq:gE_boudaryvalues} satisfy the conditions \eqref{eq:bifibre-AR}, that in this case read $B=0$, as expected, and
\[
	E^{\frac{1+\alpha}{2}} \;=\; - \mu_0 \, \gamma
\] 
irrespective of $C\in\mathbb{C}$. $E$ being strictly positive, this only makes sense when $\gamma<0$, in which case the expression for the unique negative eigenvalue
\[
 E_0\big(A_{\alpha,\mathrm{R}}^{[\gamma]}(0)\big)\;=\;-E\;=\;-(-\mu_0\gamma)^{\frac{2}{1+\alpha}}
\]
of $A_{\alpha,\mathrm{R}}^{[\gamma]}(0)$ takes the form \eqref{eq:EigenvalueIR}. Owing to \eqref{eq:GenericGE}, the corresponding (non-normalised) eigenfunction has the form \eqref{eq:eigenf-IRIL}. The reasoning for $A_{\alpha,\mathrm{L}}^{[\gamma]}(0)$ is completely analogous. Part (iii) is thus proved.

Along the same line, the eigenfunction \eqref{eq:GenericGE} belongs to $\mathcal{D}\big(A_{\alpha,a}^{[\gamma]}(0)\big)$ if and only if its boundary values \eqref{eq:gE_boudaryvalues} satisfy the conditions \eqref{eq:bifibre-Aag}, that in this case read
\[
	\begin{split}
		C \; &= \; a B\,, \\
		E^{\frac{1+\alpha}{2}} \;&= \; -\frac{\mu_0}{1+|a|^2} \gamma\,.
	\end{split}
\]
As $E>0$, one must have $\gamma<0$, in which case the expression for the only negative eigenvalue $E_0\big(A_{\alpha,a}^{[\gamma]}(0)\big)=-E$ of $A_{\alpha,a}^{[\gamma]}(0)$ takes the form \eqref{eq:EigenvalueIIa}. Owing to \eqref{eq:GenericGE}, the corresponding (non-normalised) eigenfunction has the form \eqref{eq:eigenf-IIa}. This proves part (iv).

Last, the eigenfunction \eqref{eq:GenericGE} belongs to $\mathcal{D}\big(A_{\alpha}^{[\Gamma]}(0)\big)$ if and only if \eqref{eq:gE_boudaryvalues} matches \eqref{eq:bifibre-AG}, i.e.,
\[
	\begin{pmatrix}
		g_{E,1}^- \\
		g_{E,1}^+ 
	\end{pmatrix}
	\; = \;
	\begin{pmatrix}
		\gamma_1 & \gamma_2 + \ii \gamma_3 \\
		\gamma_2 - \ii \gamma_3 & \gamma_4
	\end{pmatrix}
	\begin{pmatrix}
		g_{E,0}^- \\
		g_{E,0}^+
	\end{pmatrix}.
\]
The latter is equivalent to
\[
	2^{1+\alpha}\,\frac{\Gamma(\frac{1+\alpha}{2})}{\,\Gamma(-\frac{1+\alpha}{2})} \begin{pmatrix}
		\gamma_1 & \gamma_2+ \ii \gamma_3 \\
		\gamma_2 - \ii \gamma_3 & \gamma_4
	\end{pmatrix} \begin{pmatrix}
		B \\ C
	\end{pmatrix}\;=\;E^{\frac{1+\alpha}{2}}\begin{pmatrix}
		B \\
		C 
	\end{pmatrix} ,
\]
meaning that $E^{\frac{1+\alpha}{2}}$ is an eigenvalue of
\[
	T \; := \; -\mu_0 \begin{pmatrix}
		\gamma_1 & \gamma_2+ \ii \gamma_3 \\
		\gamma_2 - \ii \gamma_3 & \gamma_4
	\end{pmatrix}  \, .
\]
Thus, $-E$ (with $E>0$) is a negative eigenvalue of $A_\alpha^{[\Gamma]}(0)$ if and only of $E^{\frac{1+\alpha}{2}}$ is a positive eigenvalue of the matrix $T$, and equating the the eigenvector of $T$ to $\begin{pmatrix} B \\ C \end{pmatrix}$ yields the condition on the constants $B$ and $C$ in \eqref{eq:GenericGE} for the corresponding eigenfunction of $A_\alpha^{[\Gamma]}(0)$.

Elementary arguments show that the eigenvalues of $T$ are two and real (due to hermiticity) and 
\begin{itemize}
 \item \emph{both positive} if and only if $\mathrm{Tr}(T)>0$ and $\det(T)>0$,
 \item \emph{only one positive} if and only if one of the following two possibilities occurs: $\det(T)< 0$ (corresponding to two distinct eigenvalues of opposite sign), or $\det(T)=0$ and $\mathrm{Tr}(T)>0$ (corresponding to a positive and a zero eigenvalue).
\end{itemize}
Computing
\[
 \mathrm{Tr}(T)\;=\;-\mu_0(\gamma_1+\gamma_4)\,,\qquad\det(T)\;=\;-\mu_0^2(\gamma_2^2+\gamma_3^2-\gamma_1\gamma_4) \, ,
\]
one then concludes that the conditions for $T$ to have, respectively, exactly one, or two positive eigenvalues, and hence for $A_{\alpha}^{[\Gamma]}(0)$ to have, respectively, exactly one, or two negative eigenvalues, take respectively the form \eqref{eq:III0-only1neg} and \eqref{eq:III0-2neg}.

Explicitly, the eigenvalues of $T$ are 
\[
	\lambda^\pm\;:=\;-\frac{\,\mu_0}{2}  \Big( \gamma_1+\gamma_4 \pm \sqrt{(\gamma_1-\gamma_4)^2 +4 (\gamma_2^2+\gamma_3^2)} \,\Big),
\]
with $\lambda^-\geqslant\lambda^+$, and the quantities
\[
 \begin{split}
  E_0\big(A_{\alpha}^{[\Gamma]}(0)\big)\;&:=\;-E^-=-(\lambda^-)^{\frac{2}{1+\alpha}}  \, ,\\
  E_1\big(A_{\alpha}^{[\Gamma]}(1)\big)\;&:=\;-E^+=-(\lambda^+)^{\frac{2}{1+\alpha}}
 \end{split}
\]
(hence, the expressions \eqref{eq:III0-evs}), defined when applicable depending on the conditions \eqref{eq:III0-2neg}-\eqref{eq:III0-only1neg}, are the corresponding negative eigenvalues of $A_{\alpha}^{[\Gamma]}(0)$ (only the first, or both, when applicable). The condition $\lambda^-\leqslant 0$, hence \eqref{eq:III0-none-neg}, clearly identifies the case when $A_{\alpha}^{[\Gamma]}(0)$ does not have negative eigenvalues at all.

The eigenvalues of $T$ are distinct ($\lambda^->\lambda^+$) when $\gamma_1\neq\gamma_4$ or when at least one among $\gamma_2,\gamma_3$ is non-zero: with distinct eigenvalues, the largest has eigenvector (proportional to)
\[
 \begin{pmatrix}
  B \\ C
 \end{pmatrix} \;\equiv\;
 \begin{pmatrix}
  \gamma_1-\gamma_4 - \sqrt{(\gamma_1-\gamma_4)^2 +4 (\gamma_2^2+\gamma_3^2)} \\
  2(\gamma_2-\ii\gamma_3)
 \end{pmatrix},
\]
implying that the ground state eigenfunction \eqref{eq:GenericGE} of $A_{\alpha}^{[\Gamma]}(0)$ relative to the non-degenerate lowest negative eigenvalue $ E_0\big(A_{\alpha}^{[\Gamma]}(0)\big)$ has the form \eqref{eq:eigenf-III-nondegen}. Instead, $T$ has coincident eigenvalues ($\lambda^-=\lambda^+$) when $\gamma_1=\gamma_4$ and $\gamma_2=\gamma_3=0$, i.e., when $T$ is a multiple of the identity. In this case any vector $\begin{pmatrix}
  B \\ C
 \end{pmatrix}\in\mathbb{C}^2$ is eigenvector for $T$, implying that the $A_{\alpha}^{[\Gamma]}(0)$ has a two-dimensional eigenspace relative to the two-fold degenerate negative eigenvalue  $ E_0\big(A_{\alpha}^{[\Gamma]}(0)\big)=-\mu_0\gamma_1$, spanned by the eigenfunctions \eqref{eq:eigenf-III-degen}.

The proof of part (v) is thus completed.
\end{proof}

 \subsection{Spectral analysis of non-zero modes}\label{subsec:SpectrumK}

  For non-zero modes, Proposition \ref{prop:SpectrumK0} has the following counterpart.

\begin{proposition}\label{prop:NegativeEigenvaluesKDiffZero} Let $\alpha\in[0,1)$, $k\in\mathbb{Z}\setminus\{0\}$, $a\in\mathbb{C}$.
\begin{enumerate}
 \item[(i)] The spectrum of any self-adjoint extension \eqref{eq:families-ext-Aak} of $A_\alpha(k)$ is purely discrete. $A_{\alpha,\mathrm{F}}(k)$ has no negative eigenvalue, all other extensions have at most finitely many.
 \item[(ii)] $A_{\alpha,\mathrm{R}}^{[\gamma]}(k)$ and $A_{\alpha,\mathrm{L}}^{[\gamma]}(k)$ have at most one negative eigenvalue, denoted, respectively, by $E_0\big(A_{\alpha,\mathrm{R}}^{[\gamma]}(k)\big)$ and $E_0\big(A_{\alpha,\mathrm{L}}^{[\gamma]}(k)\big)$. Such negative eigenvalue exists if and only if $\gamma<-|k|/(1+\alpha)$, in which case it is non-degenerate with
 	\begin{equation}\label{eq:negEV-k_extIRIL}
 	\begin{split}
 	 & \left.
	\begin{array}{l}
	 E_0\big(A_{\alpha,\mathrm{R}}^{[\gamma]}(k)\big) \\
	 E_0\big(A_{\alpha,\mathrm{L}}^{[\gamma]}(k)\big)
	\end{array}\!\!
	\right\}\;\leqslant\; \frac{\,2^{\frac{1-\alpha}{1+\alpha}}(1+\alpha)^{\frac{2\alpha}{1+\alpha}}|k|^{\frac{2}{1+\alpha}}}{\Gamma(\frac{1-\alpha}{1+\alpha})} \big({\textstyle\frac{1+\alpha}{|k|}}\gamma+1 \big) \\
	&\qquad\qquad\qquad\quad\quad\;\;(\gamma<-{\textstyle\frac{|k|}{1+\alpha}})\,.
 	\end{split}
	\end{equation}
 \item[(iii)] $A_{\alpha,a}^{[\gamma]}(k)$ has at most one negative eigenvalue, denoted by $E_0\big(A_{\alpha,a}^{[\gamma]}(k)\big)$. Such negative eigenvalue exists if and only if $\gamma<-|k|(1+|a|^2)/(1+\alpha)$, in which case
 \begin{equation}\label{eq:negEV-k_extIIa}
 \begin{split}
  E_0\big(A_{\alpha,a}^{[\gamma]}(k)\big)\;&\leqslant\;\frac{\,2^{\frac{1-\alpha}{1+\alpha}}(1+\alpha)^{\frac{2\alpha}{1+\alpha}}|k|^{\frac{2}{1+\alpha}}}{\Gamma(\frac{1-\alpha}{1+\alpha})}\big({\textstyle\frac{1+\alpha}{(1+|a|^2)|k|}}\gamma+1 \big) \\
  &\qquad\! (\gamma<{-\textstyle\frac{(1+|a|^2) |k|}{1+\alpha}})\,.
 \end{split}
 \end{equation}
 \item[(iv)] $A_{\alpha}^{[\Gamma]}(k)$ has at most two negative eigenvalues: exactly two if and only if
 \begin{equation}\label{eq:negEV-k_extIII_2ev}
 |k|\;<\;-\textstyle(1+\alpha)\big(\gamma_1+\gamma_4+\sqrt{(\gamma_1-\gamma_4)^2+4 (\gamma_2^2 + \gamma_3^2)}\,\big)\,,
 \end{equation}
 only one if and only if
 \begin{equation}\label{eq:negEV-k_extIII_1ev}
 \begin{split}
  & -\textstyle(1+\alpha)\big(\gamma_1+\gamma_4+\sqrt{(\gamma_1-\gamma_4)^2+4 (\gamma_2^2 + \gamma_3^2)}\,\big)\;\leqslant \\
  & \qquad\qquad \leqslant\; |k|\;<\;-\textstyle(1+\alpha)\big(\gamma_1+\gamma_4-\sqrt{(\gamma_1-\gamma_4)^2+4 (\gamma_2^2 + \gamma_3^2)}\,\big)\,,
 \end{split}
 \end{equation}
 and none if and only if
 \begin{equation}\label{eq:negEV-k_extIII_0ev}
  |k|\;\geqslant\;-\textstyle(1+\alpha)\big(\gamma_1+\gamma_4-\sqrt{(\gamma_1-\gamma_4)^2+4 (\gamma_2^2 + \gamma_3^2)}\,\big)\,.
 \end{equation}
In the first two cases the lowest negative eigenvalue $ E_0\big(A_{\alpha}^{[\Gamma]}(k)\big)$ satisfies
\begin{equation}\label{eq:negEV-k_extIIIbound}
\begin{split}
 E_0\big(A_{\alpha}^{[\Gamma]}(k)\big)\;&\leqslant\; \frac{\,2^{\frac{1-\alpha}{1+\alpha}}(1+\alpha)^{\frac{2\alpha}{1+\alpha}}|k|^{\frac{2}{1+\alpha}}}{\Gamma(\frac{1-\alpha}{1+\alpha})}\,\times \\
 & \times\left(1+\frac{1+\alpha}{|k|}\Big(\gamma_1+\gamma_4-\sqrt{(\gamma_1-\gamma_4)^2+4 (\gamma_2 + \gamma_3^2)}\,\Big)\right).
\end{split}
\end{equation}
\end{enumerate}
\end{proposition}

 For the proof of Proposition \ref{prop:NegativeEigenvaluesKDiffZero} the following additional information on the fibre operators is needed.

\begin{lemma}\label{lem:OrderRelationOps}
	Let $\widetilde{\mathscr{H}}_\alpha=\bigoplus_{k\in\mathbb{Z}}\widetilde{A}_\alpha(k)$ be any of the operators \eqref{eq:HalphaFriedrichs_unif-fibred} or \eqref{eq:HalphaR_unif-fibred}-\eqref{eq:Halpha-III_unif-fibred}. If $|k| > |k'|$, then $\mathcal{D}(\widetilde{A}_\alpha(k))\subset \mathcal{D}(\widetilde{A}_\alpha(k'))$ and
	\[
	 \langle h,\widetilde{A}_\alpha(k)h\rangle_{L^2(\mathbb{R})}\;>\;\langle h,\widetilde{A}_\alpha(k')h\rangle_{L^2(\mathbb{R})}\qquad \forall h\in\mathcal{D}(\widetilde{A}_\alpha(k))\setminus \{0\}\,.
	\]
%
%
\end{lemma}

\begin{proof}
 For each non-zero $h$ the inequality among expectations is an obvious computation:
 \[
  \langle h,\widetilde{A}_\alpha(k)h\rangle_{L^2(\mathbb{R})}-\langle h,\widetilde{A}_\alpha(k')h\rangle_{L^2(\mathbb{R})}\;=\;\langle h, |x|^{2\alpha}(|k|^2-|k'|^2) h \rangle_{L^2(\mathbb{R})}\;>\;0\,.
 \]
 So one has to prove the inclusion of domains. The actual identity $\mathcal{D}(\widetilde{A}_\alpha(k))= \mathcal{D}(\widetilde{A}_\alpha(k'))$ for $k,k'\in\mathbb{Z}\setminus\{0\}$ was already observed in Remark \ref{rem:V-DAastar-samek}. When instead $|k|>0$ the inclusion $\mathcal{D}(\widetilde{A}_\alpha(k))\subset \mathcal{D}(\widetilde{A}_\alpha(0))$ follows from Theorem \ref{prop:g_with_Pweight}, that provides a representation of $\mathcal{D}(\widetilde{A}_\alpha(0))$ and $\mathcal{D}(\widetilde{A}_\alpha(k))$ in terms of $\mathcal{D}(\overline{A_\alpha(0)})$ and $\mathcal{D}(\overline{A_\alpha(k)})$, and from the inclusion $\mathcal{D}(\overline{A_\alpha(0)})\subset\mathcal{D}(\overline{A_\alpha(k)})$, as one may see from comparing \eqref{eq:DomAclosure-Grushin} and \eqref{eq:V-Closure-DirectSumFibre} above.
\end{proof}

\begin{proof}[Proof of Proposition \ref{prop:NegativeEigenvaluesKDiffZero}] (i) Any self-adjoint extension of $A_\alpha(k)$ is a suitable restriction of the adjoint \eqref{eq:Afstar}-\eqref{eq:Afstar_sum}, and therefore acts as the differential operator (see \eqref{eq:Saxi} above)
\[
 -\frac{\ud^2}{\ud x^2}+k^2 |x|^{2\alpha}+\frac{\,\alpha(2+\alpha)\,}{4x^2}
\]
on $L^2(\mathbb{R}^\pm)$. On each half-line this is a Schr\"{o}dinger operator of the form $-\frac{\ud^2}{\ud x^2}+V(x)$ with $V(x)\to+\infty$ as $|x|\to+\infty$ and it is well known (see, e.g., \cite[Section 5.5]{Titchmarsh-EigenfExpans-1962}) that such operators have purely discrete spectrum. In turn, the spectrum of each extension on the direct sum $L^2(\mathbb{R}^-)\oplus L^2(\mathbb{R}^+)$ is the union of the spectra relative to each half-line, hence is itself purely discrete. In particular, $A_{\alpha,\mathrm{F}}(k)$ is strictly positive (see \eqref{eq:AFbottom} above) and therefore has no negative spectrum. All other extensions may produce negative bound states: their number is finite because the original operator $A_\alpha(k)$ has finite deficiency index (Proposition \ref{prop:Axiselfadjointness-general}). More precisely (Corollary \ref{cor:negative_spectrum_ST_T_corollary}), the number of negative eigenvalue of each extension is the same as the number of negative bound states of the corresponding Birman extension parameter -- the operators listed in \eqref{eq:GammaIRIL}-\eqref{eq:IIkTheorem51new} -- and therefore amounts up to one for $A_{\alpha,\mathrm{R}}^{[\gamma]}(k)$, $A_{\alpha,\mathrm{L}}^{[\gamma]}(k)$, and $A_{\alpha,a}^{[\gamma]}(k)$, and up to two for $A_{\alpha}^{[\Gamma]}(k)$.

(ii) $A_{\alpha,\mathrm{R}}^{[\gamma]}(k)$ and $A_{\alpha,\mathrm{L}}^{[\gamma]}(k)$ are associated to the one-dimensional Birman parameter \eqref{eq:GammaIRIL}, which admits negative spectrum, in the precise number of one negative eigenvalue, if and only if $\beta_k<0$. Such condition is equivalent to $\gamma<-|k|/(1+\alpha)$. Moreover (Theorem \ref{thm:negative_spectrum_ST_T}), the lowest negative eigenvalue of the Birman parameter is an upper bound of the lowest negative eigenvalue of the corresponding extension. The conditions $E_0\big(A_{\alpha,\mathrm{R}}^{[\gamma]}(k)\big)\leqslant\beta_k<0$ and $E_0\big(A_{\alpha,\mathrm{L}}^{[\gamma]}(k)\big)\leqslant\beta_k<0$ thus yield \eqref{eq:negEV-k_extIRIL}.

(iii) The reasoning is precisely the same as for part (ii), now with respect to the Birman parameter \eqref{eq:GammaTauIIa}. The condition $E_0\big(A_{\alpha,a}^{[\gamma]}(k)\big)\leqslant\tau_k<0$ yields \eqref{eq:negEV-k_extIIa}.

(iv) Now the Birman parameter is the two-dimensional hermitian matrix $T_k$ given by \eqref{eq:IIkTheorem51new}. One can drop out the positive multiplicative pre-factor
\[
 \mu_k\;:=\;\frac{\,2^{\frac{1-\alpha}{1+\alpha}}(1+\alpha)^{\frac{2\alpha}{1+\alpha}}|k|^{\frac{2}{1+\alpha}}}{\Gamma(\frac{1-\alpha}{1+\alpha})}\;>\;0\,,
\]
and the number of negative eigenvalues for $T_k$ is equivalent to the number of negative eigenvalues for 
	\[
			\widetilde{T}_k\;:=\; \begin{pmatrix}
				1+ \frac{1+\alpha}{|k|} \gamma_1 & \frac{1+\alpha}{|k|}(\gamma_2 + \ii \gamma_3) \\
				\frac{1+\alpha}{|k|}(\gamma_2 - \ii \gamma_3) & 1+ \frac{1+\alpha}{|k|} \gamma_4
			\end{pmatrix}.
	\]
The latter has indeed two real eigenvalues (due to hermiticity), explicitly,
\[
  \lambda_k^\pm\;:=\;1+\frac{1+\alpha}{|k|}\Big(\gamma_1+\gamma_4\pm\sqrt{(\gamma_1-\gamma_4)^2+4 (\gamma_2^2 + \gamma_3^2)}\,\Big).
\]

Three possibilities can occur: $\lambda_k^-\leqslant\lambda_k^+<0$, $\lambda_k^-<0\leqslant\lambda_k^+$, or $0\leqslant\lambda_k^-\leqslant\lambda_k^+$. The first corresponds to the condition
\[
 |k|\;<\;-\textstyle(1+\alpha)\big(\gamma_1+\gamma_4+\sqrt{(\gamma_1-\gamma_4)^2+4 (\gamma_2^2 + \gamma_3^2)}\,\big)\,,
\]
in which case $T_k$ has two negative eigenvalues (namely $\mu_k\lambda^\pm_k$), and so too does $A_{\alpha}^{[\Gamma]}(k)$ with the upper bound $\mu_k\lambda^-_k\geqslant E_0\big(A_{\alpha}^{[\Gamma]}(k)\big)$.

The second possibility corresponds to the condition
\[
 \begin{split}
  & -\textstyle(1+\alpha)\big(\gamma_1+\gamma_4+\sqrt{(\gamma_1-\gamma_4)^2+4 (\gamma_2^2 + \gamma_3^2)}\,\big)\;\leqslant \\
  & \qquad\qquad \leqslant\; |k|\;<\;-\textstyle(1+\alpha)\big(\gamma_1+\gamma_4-\sqrt{(\gamma_1-\gamma_4)^2+4 (\gamma_2^2 + \gamma_3^2)}\,\big)\,,
 \end{split}
\]
in which case $T_k$ has only one negative eigenvalue (namely $\mu_k\lambda^-_k$), and so too does $A_{\alpha}^{[\Gamma]}(k)$, again with $\mu_k\lambda^-_k\geqslant E_0\big(A_{\alpha}^{[\Gamma]}(k)\big)$.

The third possibility is
\[
 |k|\;\geqslant\;-\textstyle(1+\alpha)\big(\gamma_1+\gamma_4-\sqrt{(\gamma_1-\gamma_4)^2+4 (\gamma_2^2 + \gamma_3^2)}\,\big)\,,
\]
for an infinite number of $k$'s it is never empty, and corresponds to the fact that $T_k$ has no negative eigenvalues, hence $A_{\alpha}^{[\Gamma]}(k)$ has neither.

Part (iv) is thus proved.
\end{proof}

  \subsection{Reconstruction of the spectral content of fibred extensions}\label{subsec:V-scatt-proofs}

 The previous results on each fibre can be now assembled together to produce the spectral analysis of the fibred operators \eqref{eq:HalphaFriedrichs_unif-fibred} and \eqref{eq:HalphaR_unif-fibred}-\eqref{eq:Halpha-III_unif-fibred}.

\begin{theorem}\label{thm:fibredHalpha-spectrum}
 Let $\alpha\in[0,1)$, $\gamma\in\mathbb{R}$, $a\in\mathbb{C}$, $\Gamma\equiv(\gamma_1,\gamma_2,\gamma_3,\gamma_4)\in\mathbb{R}^4$. The spectra of the self-adjoint extensions of type \eqref{eq:HalphaFriedrichs_unif-fibred} or \eqref{eq:HalphaR_unif-fibred}-\eqref{eq:Halpha-III_unif-fibred} of the operator $\mathscr{H}_\alpha$ defined in \eqref{eq:unitary_transf_pm}-\eqref{eq:actiondomainHalpha} and \eqref{eq:V-two-sided-7} have the following properties.
 \begin{enumerate}
  \item[(i)] Essential spectrum:
  \begin{equation}
 \begin{split}
  &\sigma_{\mathrm{ess}}(\mathscr{H}_{\alpha,\mathrm{F}})\;=\;\sigma_{\mathrm{ess}}\big(\mathscr{H}_{\alpha,\mathrm{R}}^{[\gamma]}\big)\;=\;\sigma_{\mathrm{ess}}\big(\mathscr{H}_{\alpha,\mathrm{L}}^{[\gamma]}\big) \\
  &\quad =\;\sigma_{\mathrm{ess}}\big(\mathscr{H}_{\alpha,a}^{[\gamma]}\big)\;=\;\sigma_{\mathrm{ess}}\big(\mathscr{H}_{\alpha}^{[\Gamma]}\big)\;=\;[0,+\infty)\,.
 \end{split}
\end{equation}
 \item[(ii)] Discrete spectrum: the discrete spectrum of each such operator is only negative and consists of finitely many eigenvalues. In particular, any such operator is lower semi-bounded.
 \item[(iii)] For each considered extension, the essential spectrum $[0,+\infty)$ contains (countably) infinite embedded eigenvalues, each of finite multiplicity. There is no accumulation of embedded eigenvalues.
 \item[(iv)] $\mathscr{H}_{\alpha,\mathrm{F}}$ has no negative eigenvalue, thus no discrete spectrum.
 \item[(v)] Both $\mathscr{H}_{\alpha,\mathrm{R}}^{[\gamma]}$ and $\mathscr{H}_{\alpha,\mathrm{L}}^{[\gamma]}$ have negative spectrum if and only if $\gamma<0$, in which case their negative spectrum is finite, discrete, and consists respectively of $\mathcal{N}_-\big(\mathscr{H}_{\alpha,\mathrm{R}}^{[\gamma]}\big)$ and  $\mathcal{N}_-\big(\mathscr{H}_{\alpha,\mathrm{L}}^{[\gamma]}\big)$ negative eigenvalues, counted with multiplicity, with
 \begin{equation}\label{eq:multnegspec-IRIL}
  \mathcal{N}_-\big(\mathscr{H}_{\alpha,\mathrm{R}}^{[\gamma]}\big)\;=\;\mathcal{N}_-\big(\mathscr{H}_{\alpha,\mathrm{L}}^{[\gamma]}\big)\;=\;2\lfloor (1+\alpha)|\gamma|\rfloor +1\qquad(\gamma<0)\,.
 \end{equation}
 \item[(vi)] $\mathscr{H}_{\alpha,a}^{[\gamma]}$ has negative spectrum if and only if $\gamma<0$, in which case its negative spectrum is finite, discrete, and consists of $\mathcal{N}_-\big(\mathscr{H}_{\alpha,a}^{[\gamma]}\big)$ negative eigenvalues, counted with multiplicity, with 
 \begin{equation}\label{eq:multnegspec-IIa}
  \mathcal{N}_-\big(\mathscr{H}_{\alpha,a}^{[\gamma]}\big)\;=\;2\Big\lfloor \frac{1+\alpha}{\,1+|a|^2\,}|\gamma|\Big\rfloor +1\qquad(\gamma<0)\,.
 \end{equation}
  \item[(vii)] $\mathscr{H}_{\alpha}^{[\Gamma]}$ has negative spectrum if and only if $\gamma_1+\gamma_4 - \sqrt{(\gamma_1-\gamma_4)^2 +4 (\gamma_2^2+\gamma_3^2)}<0$, in which case its negative spectrum is finite, discrete, and consists of $\mathcal{N}_-\big(\mathscr{H}_{\alpha}^{[\Gamma]}\big)$ negative eigenvalues, counted with multiplicity, with 
  \begin{equation}\label{eq:multnegspec-IIImathscrHaG}
   \begin{split}
    \mathcal{N}_-\big(\mathscr{H}_{\alpha}^{[\Gamma]}\big)\;&=\;\textstyle 2\bigg\lfloor -(1+\alpha)\big(\gamma_1+\gamma_4+\sqrt{(\gamma_1-\gamma_4)^2+4 (\gamma_2^2 + \gamma_3^2)}\,\big)\bigg\rfloor \\
    & \quad + \textstyle 2\bigg\lfloor -(1+\alpha)\big(\gamma_1+\gamma_4-\sqrt{(\gamma_1-\gamma_4)^2+4 (\gamma_2^2 + \gamma_3^2)}\,\big)\bigg\rfloor \\
    & \quad + n_0(\Gamma)
   \end{split}
  \end{equation}
   with
   \begin{equation}
    n_0(\Gamma)\,:=\,\begin{cases}
   \;2\,, & \textrm{if}\quad\gamma_1 \, \gamma_4>\gamma_2^2 + \gamma_3^2 \quad \text{and} \quad \gamma_1+\gamma_4<0\,, \\
   \;1\,, & \textrm{if}\quad \gamma_1 \, \gamma_4<\gamma_2^2 + \gamma_3^2\quad\textrm{or}\quad
   \begin{cases}
    \;\gamma_1 \, \gamma_4 =\gamma_2^2 + \gamma_3^2\,, \\
    \;\gamma_1+\gamma_4<0\,,
   \end{cases} \\
   \;0\,, & \textrm{if}\quad \gamma_1 \, \gamma_4\geqslant\gamma_2^2 + \gamma_3^2 \quad \text{and} \quad \gamma_1+\gamma_4>0\,.
  \end{cases}
   \end{equation}
 \end{enumerate}
\end{theorem}

\begin{corollary}\label{cor:positivity}
 The operator $\mathscr{H}_{\alpha,\mathrm{F}}$ is non-negative. The operators $\mathscr{H}_{\alpha,\mathrm{R}}^{[\gamma]}$, $\mathscr{H}_{\alpha,\mathrm{L}}^{[\gamma]}$, and $\mathscr{H}_{\alpha,a}^{[\gamma]}$ are non-negative if and only if $\gamma\geqslant 0$. The operator $\mathscr{H}_{\alpha}^{[\Gamma]}$ is non-negative if and only if 
 $\gamma_1\gamma_4\geqslant\gamma_2^2 + \gamma_3^2$ and $\gamma_1+\gamma_4>0$.
\end{corollary}

\begin{proof}[Proof of Theorem \ref{thm:fibredHalpha-spectrum}]
 In the following $\widetilde{\mathscr{H}}_\alpha=\bigoplus_{k\in\mathbb{Z}}\widetilde{A}_\alpha(k)$ refers to any of the operators \eqref{eq:HalphaFriedrichs_unif-fibred} or \eqref{eq:HalphaR_unif-fibred}-\eqref{eq:Halpha-III_unif-fibred}.

 Obviously, $\lambda\in\mathbb{R}$ is an eigenvalue of $\widetilde{\mathscr{H}}_\alpha$, and hence $\widetilde{\mathscr{H}}_\alpha \psi=\lambda \psi$ for some non-zero $\psi\equiv(\psi_k)_{k\in\mathbb{Z}}\in\cH$ (the fibred Hilbert space \eqref{eq:Hxispace}), if and only if $\widetilde{A}_\alpha(k) \psi_k=\lambda \psi_k$ for all the $k$'s in $\mathbb{Z}$ for which $\psi_k$ is non-zero in the fibre Hilbert space $\mathfrak{h}$.

 The first focus is on the negative eigenvalues $\lambda$ of $\widetilde{\mathscr{H}}_\alpha$. They necessarily come as negative eigenvalues of some of the $\widetilde{A}_\alpha(k)$'s. Owing to Propositions \ref{prop:SpectrumK0}-\ref{prop:NegativeEigenvaluesKDiffZero}, each $\widetilde{A}_\alpha(k)$ contributes with a finite number of negative eigenvalues (counting multiplicity). Moreover, and most importantly, only a finite number of $\widetilde{A}_\alpha(k)$'s have negative eigenvalues.

 Here is the proof of the latter claim for $\widetilde{\mathscr{H}}_\alpha\equiv\mathscr{H}_{\alpha,\mathrm{R}}^{[\gamma]}$; the analogous control of the other extensions is postponed to the second part of the proof. Thus, when $\widetilde{A}_\alpha(k)\equiv A_{\alpha,\mathrm{R}}^{[\gamma]}(k)$ and $k\in\mathbb{Z}\setminus\{0\}$, Proposition \ref{prop:NegativeEigenvaluesKDiffZero}(ii) states that this operator admits exactly one negative, non-degenerate eigenvalue if and only if $|k|<-(1+\alpha)\gamma$. This is never the case when $\gamma\geqslant 0$, whereas when $\gamma<0$ the number of $k$-modes with negative eigenvalue is precisely $2\lfloor (1+\alpha)|\gamma|\rfloor +1$, having now added to the counting also the negative eigenvalue of $A_{\alpha,\mathrm{R}}^{[\gamma]}(0)$ when $\gamma<0$ (Proposition \ref{prop:SpectrumK0}(iii)). Therefore, $\mathscr{H}_{\alpha,\mathrm{R}}^{[\gamma]}$ with $\gamma<0$ has exactly $2\lfloor (1+\alpha)|\gamma|\rfloor +1$ negative eigenvalues, counted with multiplicity, namely a \emph{finite} number of them. As already mentioned, the analogous quantification for the other types of extensions is deferred to the the second part of the proof.

 The conclusion so far is that $\widetilde{\mathscr{H}}_\alpha$ has only a finite number of negative eigenvalues, counted with multiplicity, which therefore all belong to $\sigma_{\mathrm{disc}}\big(\widetilde{\mathscr{H}}_\alpha\big)$. All other eigenvalues of $\widetilde{\mathscr{H}}_\alpha$ are non-negative, hence embedded in
 \[
  [0,+\infty)\;=\;\sigma_{\mathrm{ess}}\big(\widetilde{A}_\alpha(0)\big)\;=\;\bigcup_{k\in\mathbb{Z}}\sigma_{\mathrm{ess}}\big(\widetilde{A}_\alpha(k)\big)\;\subset\;\sigma_{\mathrm{ess}}\big(\widetilde{\mathscr{H}}_\alpha\big)
 \]
 (having used \eqref{eq:sess0} from Proposition \ref{prop:SpectrumK0} for the first identity, Proposition \ref{prop:NegativeEigenvaluesKDiffZero}(i) for the second, and
 \[
  \sigma_{\mathrm{ess}}(\widetilde{\mathscr{H}}_\alpha) \;\supset\; \bigcup_{k \in \mathbb{Z}}\sigma_{\mathrm{ess}}(\widetilde{A}_\alpha(k))
 \]
 from Section \ref{sec:I-parts-of-spectrum}), and therefore such eigenvalues do not belong to $\sigma_{\mathrm{disc}}\big(\widetilde{\mathscr{H}}_\alpha\big)$. The latter set is therefore only negative and finite. This proves part (ii).

 As argued above, $[0,+\infty)\subset\sigma_{\mathrm{ess}}\big(\widetilde{\mathscr{H}}_\alpha\big)$. Should there be $\lambda<0$ in $\sigma_{\mathrm{ess}}\big(\widetilde{\mathscr{H}}_\alpha\big)$, such $\lambda$ would either be an eigenvalue of infinite multiplicity or an accumulation point of $\sigma\big(\widetilde{\mathscr{H}}_\alpha\big)$. The first option is excluded, because as observed already negative eigenvalues of $\widetilde{\mathscr{H}}_\alpha$ can only be negative eigenvalues of a finite number of the $\widetilde{A}_\alpha(k)$'s, hence all with finite multiplicity. The second option is excluded as well, owing to
 \[
  \sigma(\widetilde{\mathscr{H}}_\alpha) \;=\;\overline{\,\bigcup_{k \in \mathbb{Z}} \sigma(\widetilde{A}_\alpha(k))}
 \]
 from Section \ref{sec:I-parts-of-spectrum}, which implies that a negative accumulation point of $\sigma\big(\widetilde{\mathscr{H}}_\alpha\big)$ must be the limit of negative points in the $\sigma\big(\widetilde{A}_\alpha(k)\big)$'s: the negative spectra of the $\widetilde{A}_\alpha(k)$'s are actually non-empty only for finitely many $k$'s, and each such negative spectrum is finite. Then, necessarily $\sigma_{\mathrm{ess}}\big(\widetilde{\mathscr{H}}_\alpha\big)=[0,+\infty)$. This proves part (i).

 Concerning the non-$\sigma_{\mathrm{disc}}\big(\widetilde{\mathscr{H}}_\alpha\big)$ eigenvalues embedded in $[0,+\infty)$, once again each of them must be an eigenvalue for some of the $\widetilde{A}_\alpha(k)$'s. As each $\widetilde{A}_\alpha(k)$ with $k\neq 0$ has infinitely many positive eigenvalues (Proposition \ref{prop:NegativeEigenvaluesKDiffZero}(i)), the number of embedded eigenvalues for $\widetilde{\mathscr{H}}_\alpha$, counting multiplicity, is (countably) infinite as well. Now, the only possibility for any such eigenvalue $\lambda\geqslant 0$ to have itself infinite multiplicity is that $\lambda$ is \emph{simultaneously} an eigenvalue for infinitely many distinct $\widetilde{A}_\alpha(k)$. This cannot be the case, though. Indeed, each $\widetilde{A}_\alpha(k)$ is lower semi-bounded and
 \[
  \mathfrak{m}\big(\widetilde{A}_\alpha(k)\big)\;\geqslant\;\frac{\,\mathfrak{m}\big(A_\alpha(k)\big)\cdot \mathfrak{m}(\mathcal{B}_k)}{\,\mathfrak{m}\big(A_\alpha(k)\big)+\mathfrak{m}(\mathcal{B}_k)\,}
 \]
 (see \eqref{eq:bounds_mS_mB_Tversion} from Theorem \ref{thm:semibdd_exts_operator_formulation_Tversion}).
  In the above formula $\mathcal{B}_k$ indicates the Birman extension parameter\index{Birman extension parameter} of $\widetilde{A}_\alpha(k)$, namely one of the operators \eqref{eq:GammaIRIL}-\eqref{eq:IIkTheorem51new}, $\mathfrak{m}(\mathcal{B}_k)$ denotes the bottom of its spectrum, and $\mathfrak{m}\big(A_\alpha(k)\big)$ can be non-restrictively taken to be the lower bound \eqref{eq:Axibottom}. For sufficiently large $|k|$,
 \[
  \mathfrak{m}\big(A_\alpha(k)\big)\,\sim\,|k|^{\frac{2}{1+\alpha}}\,,\qquad \mathfrak{m}(\mathcal{B}_k)\,\sim\,|k|^{\frac{1-\alpha}{1+\alpha}}\,,
 \]
 whence $ \mathfrak{m}\big(\widetilde{A}_\alpha(k)\big)\gtrsim\,|k|^{\frac{1-\alpha}{1+\alpha}}$. This shows that the lowest eigenvalue of $\widetilde{A}_\alpha(k)$ grows with $|k|$, thus making impossible for a fixed $\lambda\geqslant 0$ to be simultaneous eigenvalue of infinitely many $\widetilde{A}_\alpha(k)$'s. The conclusion is that $\lambda$ has finite multiplicity. The same reasoning excludes that the embedded eigenvalues accumulate to some limit points. This completes the proof of part (iii).

 What remains to prove now is the precise quantification of the multiplicity of the finite negative discrete spectrum of the operator $\widetilde{\mathscr{H}}_\alpha$ for each possible type $\mathscr{H}_{\alpha,\mathrm{F}},\mathscr{H}_{\alpha,\mathrm{R}}^{[\gamma]},\mathscr{H}_{\alpha,\mathrm{L}}^{[\gamma]},\mathscr{H}_{\alpha,a}^{[\gamma]},\mathscr{H}_{\alpha}^{[\Gamma]}$, mimicking the same reasoning made above for $\mathscr{H}_{\alpha,\mathrm{R}}^{[\gamma]}$. Clearly $\mathscr{H}_{\alpha,\mathrm{F}}$ has no negative spectrum because the $A_{\alpha,\mathrm{F}}(k)$'s have neither (Propositions \ref{prop:SpectrumK0}(ii) and \ref{prop:NegativeEigenvaluesKDiffZero}(i)). And the multiplicity \eqref{eq:multnegspec-IRIL} of the negative spectrum of $\mathscr{H}_{\alpha,\mathrm{R}}^{[\gamma]}$ has been already computed above -- the same clearly applies to $\mathscr{H}_{\alpha,\mathrm{L}}^{[\gamma]}$. Thus, parts (iv) and (v) are proved.

 Concerning $\mathscr{H}_{\alpha,a}^{[\gamma]}$, Proposition \ref{prop:NegativeEigenvaluesKDiffZero}(iii) states that $A_{\alpha,a}^{[\gamma]}$ with $k\in\mathbb{Z}\setminus\{0\}$ admits exactly one negative, non-degenerate eigenvalue if and only if $|k|<-\frac{1+\alpha}{1+|a|^2}\gamma$. This is never the case when $\gamma\geqslant 0$, whereas when $\gamma<0$ the number of $k$-modes with negative eigenvalue is precisely $2\lfloor \frac{1+\alpha}{\,1+|a|^2\,}|\gamma|\rfloor +1$, having now added to the counting also the negative eigenvalue of $A_{\alpha,a}^{[\gamma]}(0)$ when $\gamma<0$ (Proposition \ref{prop:SpectrumK0}(iv)). Therefore, $\mathscr{H}_{\alpha,a}^{[\gamma]}$ with $\gamma<0$ has exactly $2\lfloor \frac{1+\alpha}{\,1+|a|^2\,}|\gamma|\rfloor +1$ negative eigenvalues, counted with multiplicity. Part (vi) is proved.

 Last, concerning $\mathscr{H}_{\alpha}^{[\Gamma]}$, the counting goes as follows. In the non-zero modes, the $A_{\alpha}^{[\Gamma]}(k)$'s with \emph{exactly two} negative eigenvalues are those for which (see \eqref{eq:negEV-k_extIII_2ev} from Proposition \ref{prop:NegativeEigenvaluesKDiffZero})
 \begin{equation*}
 |k|\;<\;-\textstyle(1+\alpha)\big(\gamma_1+\gamma_4+\sqrt{(\gamma_1-\gamma_4)^2+4 (\gamma_2^2 + \gamma_3^2)}\,\big)\,,
 \end{equation*}
 therefore their number amounts to
 \[
  N_2\;:=\;\textstyle 2\bigg\lfloor -(1+\alpha)\big(\gamma_1+\gamma_4+\sqrt{(\gamma_1-\gamma_4)^2+4 (\gamma_2^2 + \gamma_3^2)}\,\big)\bigg\rfloor\,.
 \]
 The $A_{\alpha}^{[\Gamma]}(k)$'s instead with \emph{only one} negative eigenvalue are those for which (see \eqref{eq:negEV-k_extIII_1ev} above)
 \begin{equation*}
 \begin{split}
  & -\textstyle(1+\alpha)\big(\gamma_1+\gamma_4+\sqrt{(\gamma_1-\gamma_4)^2+4 (\gamma_2^2 + \gamma_3^2)}\,\big)\;\leqslant \\
  & \qquad\qquad \leqslant\; |k|\;<\;-\textstyle(1+\alpha)\big(\gamma_1+\gamma_4-\sqrt{(\gamma_1-\gamma_4)^2+4 (\gamma_2^2 + \gamma_3^2)}\,\big)\,,
 \end{split}
 \end{equation*}
 therefore their number amounts to
 \[
  \begin{split}
     & N_1\;:=\;\textstyle 2\bigg\lfloor -(1+\alpha)\big(\gamma_1+\gamma_4-\sqrt{(\gamma_1-\gamma_4)^2+4 (\gamma_2^2 + \gamma_3^2)}\,\big)\bigg\rfloor \\
     &\qquad \qquad \textstyle -2\bigg\lfloor -(1+\alpha)\big(\gamma_1+\gamma_4+\sqrt{(\gamma_1-\gamma_4)^2+4 (\gamma_2^2 + \gamma_3^2)}\,\big)\bigg\rfloor\,.
  \end{split}
 \]
 All other $A_{\alpha}^{[\Gamma]}(k)$'s do not have negative eigenvalues. Thus, the contribution in terms of number of negative eigenvalues of $\mathscr{H}_{\alpha}^{[\Gamma]}$ from its non-zero modes is $2N_2+N_1$, counting multiplicity. To this quantity one has to add the number $n_0$ of negative eigenvalues from the zero-mode, namely from $A_{\alpha}^{[\Gamma]}(0)$: owing to Proposition \ref{prop:SpectrumK0}(v),
 \[
  n_0\;=\;
  \begin{cases}
   \;2\,, & \textrm{if}\quad\gamma_1 \, \gamma_4>\gamma_2^2 + \gamma_3^2 \quad \text{and} \quad \gamma_1+\gamma_4<0\,, \\
   \;1\,, & \textrm{if}\quad \gamma_1 \, \gamma_4<\gamma_2^2 + \gamma_3^2\quad\textrm{or}\quad
   \begin{cases}
    \;\gamma_1 \, \gamma_4 =\gamma_2^2 + \gamma_3^2\,, \\
    \;\gamma_1+\gamma_4<0\,,
   \end{cases} \\
   \;0\,, & \textrm{otherwise, $\quad$i.e., if}\quad  \gamma_1+\gamma_4 - \sqrt{(\gamma_1-\gamma_4)^2 +4 (\gamma_2^2+\gamma_3^2)} \;\geqslant \;0\,.
  \end{cases}
 \]
 The conclusion is that the number $\mathcal{N}_-\big(\mathscr{H}_{\alpha}^{[\Gamma]}\big)$ of negative eigenvalues of $\mathscr{H}_{\alpha}^{[\Gamma]}$, counting multiplicity, is precisely $2N_2+N_1+n_0$, which yields the expression \eqref{eq:multnegspec-IIImathscrHaG}. Moreover, \emph{none} of the fibre operators $A_{\alpha}^{[\Gamma]}(k)$, $k\in\mathbb{Z}$, admits negative eigenvalues if and only if (see \eqref{eq:III0-none-neg} and \eqref{eq:negEV-k_extIII_0ev} above)
 \[
  \begin{split}
   &\gamma_1+\gamma_4 - \textstyle\sqrt{(\gamma_1-\gamma_4)^2 +4 (\gamma_2^2+\gamma_3^2)} \;\geqslant \;0 \quad\textrm{and}\\
   &|k|\;\geqslant\;-\textstyle(1+\alpha)\big(\gamma_1+\gamma_4-\sqrt{(\gamma_1-\gamma_4)^2+4 (\gamma_2^2 + \gamma_3^2)}\,\big)\quad \forall k\in\mathbb{Z}\setminus\{0\}\,,
  \end{split}
 \]
 which is equivalent to just $\gamma_1+\gamma_4 - \sqrt{(\gamma_1-\gamma_4)^2 +4 (\gamma_2^2+\gamma_3^2)}\geqslant0$. The latter condition characterises the absence of negative spectrum for $\mathscr{H}_{\alpha,a}^{[\gamma]}$, can be also interpreted as the non-negativity of the lowest eigenvalue of the hermitian matrix
 \[
  \widetilde{\Gamma}\;:=\;
  \begin{pmatrix}
   \gamma_1 & \gamma_2+\ii\gamma_3 \\
   \gamma_2-\ii\gamma_3 & \gamma_4
  \end{pmatrix},
 \]
 and is therefore equivalent to $\mathrm{Tr}(\widetilde{\Gamma})>0$ and $\det(\widetilde{\Gamma})\geqslant 0$, i.e., to $\gamma_1+\gamma_4>0$ and $\gamma_1\gamma_4\geqslant\gamma_2^2 + \gamma_3^2$. Part (vii) is thus proved.
\end{proof}

 Last, by inverting the unitary transformation of \eqref{eq:V-two-sided-7}, thus exploiting the identities
 \begin{equation}\label{eq:inverseUniary_all}
  \begin{split}
   H_{\alpha,\mathrm{F}} \;&=\; (U_\alpha)^{-1}(\mathcal{F}_2)^{-1} \mathscr{H}_{\alpha,\mathrm{F}}\,\mathcal{F}_2\,U_\alpha\,, \\
 H_{\alpha,\mathrm{R}}^{[\gamma]}  \;&=\;(U_\alpha)^{-1}(\mathcal{F}_2)^{-1} \mathscr{H}_{\alpha,\mathrm{R}}^{[\gamma]}\,\mathcal{F}_2\,U_\alpha\,,\\
 H_{\alpha,\mathrm{L}}^{[\gamma]}  \;&=\;(U_\alpha)^{-1}(\mathcal{F}_2)^{-1} \mathscr{H}_{\alpha,\mathrm{L}}^{[\gamma]}\,\mathcal{F}_2\,U_\alpha\,, \\
 H_{\alpha,a}^{[\gamma]}  \;&=\;(U_\alpha)^{-1}(\mathcal{F}_2)^{-1}\mathscr{H}_{\alpha,a}^{[\gamma]}\,\mathcal{F}_2\,U_\alpha\,,\\
 H_{\alpha}^{[\Gamma]}  \;&=\;(U_\alpha)^{-1}(\mathcal{F}_2)^{-1}\mathscr{H}_{\alpha}^{[\Gamma]}\,\mathcal{F}_2\,U_\alpha\,,
  \end{split}
 \end{equation}
%
 one can now translate the information obtained so far on the uniformly fibred self-adjoint extensions of $\mathscr{H}_\alpha$ to the corresponding self-adjoint extensions of $H_\alpha$.

\begin{proof}[Proof of Theorem \ref{thm:positivity}]
 The non-negativity is preserved between the pairs of unitarily equivalent operators in  \eqref{eq:inverseUniary_all}, therefore the theorem follows at once from Corollary \ref{cor:positivity}.
\end{proof}

\begin{proof}[Proof of Theorem \ref{thm:MainSpectral}]
 Discrete and essential spectrum are preserved between the pairs of unitarily equivalent operators in  \eqref{eq:inverseUniary_all}, as well as the multiplicity and degeneracy of eigenvalues, therefore the theorem follows at once from Theorem \ref{thm:fibredHalpha-spectrum}.
\end{proof}

\begin{proof}[Proof of Theorem \ref{thm:GroundStateCH}]
 Consider the regime when an operator of the type \eqref{eq:HalphaFriedrichs_unif-fibred} or \eqref{eq:HalphaR_unif-fibred}-\eqref{eq:Halpha-III_unif-fibred}, collectively denoted as $\widetilde{\mathscr{H}}_\alpha=\bigoplus_{k\in\mathbb{Z}}\widetilde{A}_\alpha(k)$, has negative spectrum (Theorem \ref{thm:fibredHalpha-spectrum}). Build $\psi\equiv(\psi_k)_{k\in\mathbb{Z}}\in\cH$ whose components are all zero but for $\psi_0$, taken to be the eigenfunction in $\mathfrak{h}$ relative to the zero-mode lowest negative eigenvalue $E_0( \widetilde{A}_\alpha(0))$ (Proposition \ref{prop:SpectrumK0}). Then $\widetilde{\mathscr{H}}_\alpha\,\psi=E_0( \widetilde{A}_\alpha(0))\psi$. Moreover, as a consequence of the strict ordering established in Lemma \ref{lem:OrderRelationOps}, $E_0( \widetilde{A}_\alpha(0))$ is the ground state energy of $\widetilde{\mathscr{H}}_\alpha$ with ground state vector $\psi$. The degeneracy of $E_0( \widetilde{A}_\alpha(0))$ as lowest eigenvalue of $\widetilde{\mathscr{H}}_\alpha$ is the same as the degeneracy as lowest eigenvalue of $\widetilde{A}_\alpha(0)$. With this reasoning and Proposition \ref{prop:SpectrumK0} one characterises the ground state of $\widetilde{\mathscr{H}}_\alpha$ for each of the types  \eqref{eq:HalphaR_unif-fibred}-\eqref{eq:Halpha-III_unif-fibred}. Let now $\widetilde{H}_\alpha$ be the unitarily equivalent counterpart of $\widetilde{\mathscr{H}}_\alpha$ through the transformations in \eqref{eq:inverseUniary_all}, hence one of the self-adjoint operators acting on $L^2(M,\mu_\alpha)$ and classified in Theorem \ref{thm:H_alpha_fibred_extensions} -- apart from the Friedrichs extension that clearly has no negative spectrum. The lowest negative eigenvalue and its degeneracy are preserved by unitary equivalence, hence can be immediately read out from Proposition \ref{prop:SpectrumK0}, through the above reasoning. The corresponding eigenfunction is transformed from $\cH$ to $L^2(M,\ud\mu_\alpha)$ by inverting \eqref{eq:global-unitary-pm}. Thus, in view of \eqref{eq:defF2} $(\mathcal{F}_2^{-1}\psi)(x,y)= \frac{1}{\sqrt{2\pi}}\,\psi_0(x)$, and in view of \eqref{eq:unit1} $(U_\alpha^{-1}\mathcal{F}_2^{-1}\psi)(x,y)= \frac{1}{\sqrt{2\pi}}\,|x|^{\frac{\alpha}{2}}\,\psi_0(x)$. As the original $\psi$ has only zero-mode, the transformed ground state function $U_\alpha^{-1}\mathcal{F}_2^{-1}\psi$ is \emph{constant} in $y$. Applying such a scheme to the explicit eigenfunctions $g_{\alpha}^{(\mathrm{I_R})}$, $g_{\alpha}^{(\mathrm{I_L})}$, $g_{\alpha}^{(\mathrm{II}_a)}$, $g_{\alpha}^{(\mathrm{III})}$, $g_{\alpha,+}^{(\mathrm{III})}$, $g_{\alpha,-}^{(\mathrm{III})}$ identified in Proposition \ref{prop:SpectrumK0} yields at once the corresponding ground state functions for $H_{\alpha,\mathrm{R}}^{[\gamma]}$, $H_{\alpha,\mathrm{L}}^{[\gamma]}$, $H_{\alpha,a}^{[\gamma]}$, $H_{\alpha}^{[\Gamma]}$. 
\end{proof}

 \begin{proof}[Proof of Proposition \ref{prop:Friedrichs-groundstate}]
 Clearly, due to unitary equivalence, it suffices to reason in terms of $\mathscr{H}_{\alpha,\mathrm{F}}$. As argued already for the proof of Theorem \ref{thm:fibredHalpha-spectrum}, the lowest eigenvalue of $\mathscr{H}_{\alpha,\mathrm{F}}$ must be an eigenvalue for some of the fibre operators $A_{\alpha,\mathrm{F}}(k)$. $A_{\alpha,\mathrm{F}}(0)$ has no eigenvalues (Proposition \ref{prop:SpectrumK0}(ii)). The $A_{\alpha,\mathrm{F}}(k)$'s with $k\neq 0$ have indeed purely discrete positive spectrum (Proposition \ref{prop:NegativeEigenvaluesKDiffZero}(i)), and the lowest contribution to the eigenvalues of $\mathscr{H}_{\alpha,\mathrm{F}}$ only comes from the lowest, non-degenerate eigenvalue of $A_{\alpha,\mathrm{F}}(1)$ and the lowest, non-degenerate eigenvalue of $A_{\alpha,\mathrm{F}}(-1)$, owing to the operator ordering established in Lemma \ref{lem:OrderRelationOps}. Such eigenvalues are equal (because the two fibre operators are unitarily equivalent through the left$\leftrightarrow$right symmetry), but of course the two eigenfunctions in the fibre correspond to two linearly independent eigenfunctions for $\mathscr{H}_{\alpha,\mathrm{F}}$. Therefore $E_0(H_{\alpha,\mathrm{F}})$ is two-fold degenerate. Next, one knows from \eqref{eq:AFbottom} that 
 \begin{equation*}
 \mathfrak{m}(A_{\alpha,\mathrm{F}}(1))\;\geqslant\;(1+\alpha)\big(\textstyle{\frac{2+\alpha}{4}}\big)^{\frac{\alpha}{1+\alpha}}\,.
\end{equation*}
 Then the inequality $E_0(H_{\alpha,\mathrm{F}})\geqslant \mathfrak{m}(A_{\alpha,\mathrm{F}}(1))$ yields the lower bound in \eqref{eq:estimate-Fgroundstate}. For the upper bound one performs a variational argument. Consider the family $(h_b)_{b>0}$ of trial functions in $\mathfrak{h}\cong L^2(\mathbb{R})$ defined by
 \[
  h_b(x)\;:=\;|x|^{1+\frac{\alpha}{2}}e^{-b|x|^{1+\alpha}}\,\mathbf{1}_{\mathbb{R}^+}(x)\,,
 \]
 where $\mathbf{1}_{\mathbb{R}^+}$ is the characteristic function of $(0,+\infty)$. Clearly, $h_b\in\mathcal{D}(A_{\alpha,\mathrm{F}}(1))$ $\forall b>0$ (Theorem \ref{thm:bifibre-extensions}). A direct computation yields
 \[
  \frac{\;\langle h_b, A_{\alpha,\mathrm{F}}(1) h_b \rangle_{L^2(\mathbb{R})}}{\langle h_b,h_b \rangle_{L^2(\mathbb{R})}}\,=\,\frac{2^{\frac{1-\alpha}{1+\alpha}}}{ \Gamma (\frac{3+\alpha}{1+\alpha})}\frac{1+b^2(1+\alpha)^2}{b^{\frac{2 \alpha}{1+\alpha}}}\,,
 \]
 whence
 \[
 \begin{split}
   E_0(H_{\alpha,\mathrm{F}})\;&\leqslant\;\min_{b>0}\,\frac{\;\langle h_b, A_{\alpha,\mathrm{F}}(1) h_b \rangle_{L^2(\mathbb{R})}}{\langle h_b,h_b \rangle_{L^2(\mathbb{R})}}\;=\;\frac{2^{\frac{1-\alpha}{1+\alpha}} (1+\alpha)^{\frac{1+3\alpha}{1+\alpha}}}{\alpha^{\frac{\alpha}{1+\alpha}} \Gamma(\frac{3+\alpha}{1+\alpha})}\,.
 \end{split}
 \]
 This provides the upper bound in \eqref{eq:estimate-Fgroundstate}. 
 \end{proof}

\section{Scattering in transmission protocols}\label{sec:V-scattering}

  \subsection{Scattering on Grushin-type cylinder}

 The Hamiltonian $H_{\alpha,a}^{[\gamma]}$ with $a=1$ and $\gamma=0$ imposes the local behaviour (see  \eqref{eq:DHalpha_cond2_limits-1}-\eqref{eq:DHalpha_cond2_limits-2} and \eqref{eq:DHalpha_cond3_IIa} above) 
\begin{equation}\label{eq:bridging_conditions}
 \begin{split}
  \lim_{x\to 0^-}f(x,y)\;&=\;\lim_{x\to 0^+}f(x,y)\,, \\
  \lim_{x\to 0^-}\Big(\frac{1}{\:|x|^\alpha}\,\frac{\partial f(x,y)}{\partial x}\Big)\;&=\;\lim_{x\to 0^+}\Big(\frac{1}{\:|x|^\alpha}\,\frac{\partial f(x,y)}{\partial x}\Big)\,,
 \end{split}
\end{equation}
which quantum-mechanically is interpreted as the continuity of the spatial probability density of the particle in the region around $\mathcal{Z}$ and of the momentum in the direction normal to $\mathcal{Z}$, defined with respect to the weight $|x|^{-\alpha}$ induced by the metric.

This occurrence corresponds to the `optimal' transmission across the boundary, with the dynamics developing the best `bridging' between left and right half-cylinder. Such a Hamiltonian is indeed referred to as the \emph{bridging}\index{bridging extension} realisation of the free Hamiltonian on Grushin cylinder, an extension identified first in \cite[Proposition 3.11]{Boscain-Prandi-JDE-2016} (clearly, within the present scheme one is able to recover and study it as a distinguished element of the general classification of Theorem \ref{thm:H_alpha_fibred_extensions}).

 In order to clarify the peculiarity of the bridging transmission protocol, one is led to the last object of the present Chapter, namely the scattering over Grushin-type cylinders.

Intuitively speaking, far away from the Grushin singularity $\mathcal{Z}$ the metric tends to become flat and the action $-\Delta_\alpha$ of each free Hamiltonians considered so far tends to resemble that of the free Laplacian $-\Delta$, plus the correction due to the $(|x|^{-1}\partial_x)$-term, on wave functions $f(x,y)$ that are constant in $y$. This suggests that at very large distances a particle evolves free from the effects of the underlying geometry, and one can speak of scattering states of energy $E>0$. The precise shape of the wave function $f_{\mathrm{scatt}}$ of such a scattering state, at this informal level, can be easily guessed to be of the form
\begin{equation}\label{eq:fscatt}
 f_{\mathrm{scatt}}(x,y)\;\sim\;|x|^{\frac{\alpha}{2}}e^{\pm\ii x\sqrt{E}}\qquad \textrm{as }|x|\to +\infty\,.
\end{equation}
Indeed, $-\Delta_\alpha f_{\mathrm{scatt}}\sim Ef_{\mathrm{scatt}}+\frac{\alpha(2+\alpha)}{4|x|^2}f_{\mathrm{scatt}}$, that is, up to a very small $O(|x|^{-2})$-correction, $f_{\mathrm{scatt}}$ is a generalised eigenfunction of $-\Delta_\alpha$ with eigenvalue $E$. In fact, this is fully justified on a mathematically rigorous level (Sect.~\ref{sec:scattering-fibre}), upon studying the scattering of the unitarily equivalent self-adjoint operators \eqref{eq:Halpha-IIa_unif-fibred}-\eqref{eq:Halpha-III_unif-fibred}: scattering states are constant in $y$ and of the form \eqref{eq:fscatt}.

 One can then examine, as in the standard stationary scattering analysis, the scattering of a flux of particles injected into the Grushin cylinder at large distances and shot towards the singularity $\mathcal{Z}$ with given energy $E>0$: by monitoring the spatial density of the transmitted flux and the reflected flux, normalised with respect to the density of the incident flux, one quantifies in the usual way the \emph{transmission coefficient}\index{transmission coefficient} and \emph{reflection coefficient}\index{reflection coefficient} for the scattering (see, e.g., \cite[Chapter 8.C]{Cohen-Tannoudji-1977-2020} or \cite[Chapter 6]{sakurai_napolitano_2017}).

Obviously, no scattering across the singularity occurs for Friedrichs, or type-$\mathrm{I}_\mathrm{R}$, or type-$\mathrm{I}_\mathrm{L}$ quantum protocols. The focus here is on type-$\mathrm{II}_a$ scattering, as it includes in particular the bridging protocol; the conclusions for type-$\mathrm{III}$ scattering are qualitatively analogous. The proof is then deferred to Subsection \ref{sec:scattering-fibre}.

\begin{theorem}[Scattering]\label{thm:scattering} Let $\alpha\in[0,1)$, $a\in\mathbb{C}$, $\gamma\in\mathbb{R}$.
 The transmission coefficient $T_{\alpha,a,\gamma}(E)$ and the reflection coefficient $R_{\alpha,a,\gamma}(E)$ at given energy $E>0$, as defined above for the transmission protocol governed by the Hamiltonian $H_{\alpha,a}^{[\gamma]}$, are given by
  \begin{equation}\label{eq:TR-IIa-thm}
  \begin{split}
   T_{\alpha,a,\gamma}(E)\;&=\;\left|\frac{\,E^{\frac{1+\alpha}{2}}(1+e^{\ii\pi\alpha})\,\Gamma(\frac{1-\alpha}{2})\,\overline{a}}{\,E^{\frac{1+\alpha}{2}}\Gamma(\frac{1-\alpha}{2})(1+|a|^2)+\ii\,\gamma\,2^{1+\alpha}e^{\ii\frac{\pi}{2}\alpha}\Gamma(\frac{3+\alpha}{2})} \right|^2 \,, \\
   R_{\alpha,a,\gamma}(E)\;&=\;\left|\frac{\,E^{\frac{1+\alpha}{2}}\Gamma(\frac{1-\alpha}{2})\,(1-|a|^2\,e^{\ii\pi\alpha})+\ii\,\gamma\,2^{1+\alpha}e^{\ii\frac{\pi}{2}\alpha}\Gamma(\frac{3+\alpha}{2})}{\,E^{\frac{1+\alpha}{2}}\Gamma(\frac{1-\alpha}{2})(1+|a|^2)+\ii\,\gamma\,2^{1+\alpha}e^{\ii\frac{\pi}{2}\alpha}\Gamma(\frac{3+\alpha}{2})}\right|^2\,.
  \end{split}
 \end{equation}
  They satisfy 
  \begin{equation}\label{eq:TpEi1-thm}
  T_{\alpha,a,\gamma}(E)+R_{\alpha,a,\gamma}(E)\;=\;1\,,
 \end{equation}
 and when $\gamma=0$ they are independent of $E$.
 The scattering is reflection-less ($R_{\alpha,a,\gamma}(E)=0$) when
 \begin{equation}\label{eq:Etransm-thm}
   E\;=\;\bigg(\frac{\,2^{1+\alpha}\,\gamma\,\Gamma(\frac{3+\alpha}{2})\sin\frac{\pi}{2}\alpha\,}{\,\Gamma(\frac{1-\alpha}{2})(1-\cos\pi\alpha)\,}\bigg)^{\!\frac{2}{1+\alpha}},	
 \end{equation}
 provided that $\alpha\in(0,1)$, $|a|=1$, and $\gamma>0$. In the high energy limit the scattering is independent of the extension parameter $\gamma$ and one has
 \begin{equation}\label{eq:highenergyscatt-thm}
  \begin{split}
   \lim_{E\to+\infty}T_{\alpha,a,\gamma}(E)\;&=\;\frac{\,2\,|a|^2(1+\cos\pi\alpha)}{(1+|a|^2)^2}\,, \\
   \lim_{E\to+\infty}R_{\alpha,a,\gamma}(E)\;&=\;\frac{\,1+|a|^4-2|a|^2\cos\pi\alpha}{(1+|a|^2)^2}\,,
  \end{split}
 \end{equation}
 whereas in the low energy limit, for $\gamma\neq 0$,
  \begin{equation}\label{eq:lowenergyscatt-thm}
  \begin{split}
   \lim_{E\downarrow 0}T_{\alpha,a,\gamma}(E)\;&=\;0\,, \\
   \lim_{E\downarrow 0}R_{\alpha,a,\gamma}(E)\;&=\;1\,.
  \end{split}
 \end{equation}
\end{theorem}

\begin{figure}[!h]
\captionsetup[subfigure]{labelformat=empty} 
  \centering
  \subfloat[][$\alpha=\frac{2}{3},a=\ii,\gamma=\frac{1}{2}$]
  {\includegraphics[width=0.45\textwidth]{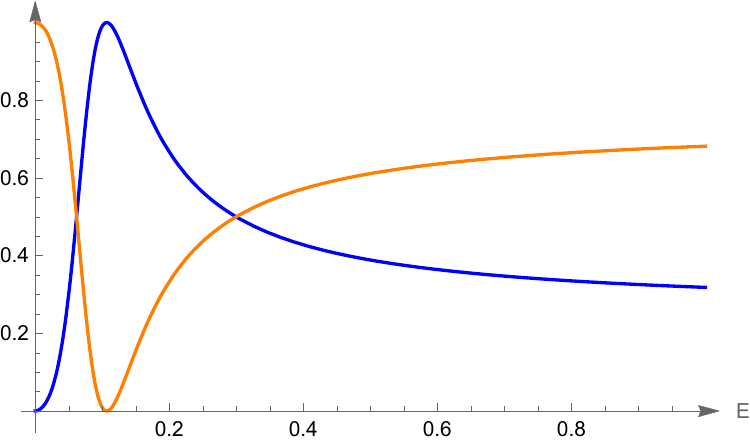} }
  \subfloat[][$\alpha=\frac{2}{5},a=\frac{\ii}{2},\gamma=-\frac{1}{5}$]
  {\includegraphics[width=0.45\textwidth]{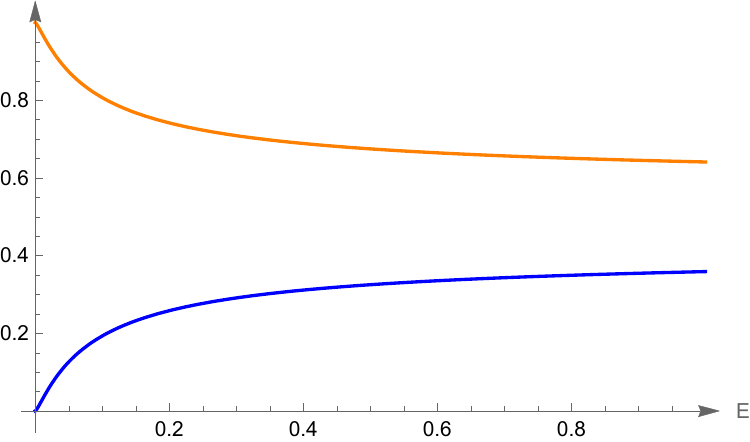} } \\
  \caption{Plot of the coefficients $ T_{\alpha,a,\gamma}(E)$ (blue curve) and $ R_{\alpha,a,\gamma}(E)$ (orange curve) according to formula \eqref{eq:TR-IIa-thm}. In the first case, for a special value of $E$ the scattering is reflection-less.}
  \label{fig:genericTR}
\end{figure}

 Figure \ref{fig:genericTR} displays two representative behaviours of $ T_{\alpha,a,\gamma}(E)$ and $ R_{\alpha,a,\gamma}(E)$.
 
Specialising the general results of Theorems \ref{thm:H_alpha_fibred_extensions} and \ref{thm:scattering} for the bridging protocol ($a=1$, $\gamma=0$), one may summarise its distinguished status as follows:
\begin{itemize}
 \item \emph{no spatial filter:} the bridging Hamiltonian, as well as all type-$\mathrm{II}_a$ protocols with $a=1$, imposes the local continuity of the wave function at $\mathcal{Z}$ (first identity in \eqref{eq:bridging_conditions}), thus a transmission with no jump in the particle's probability density from one side to the other of the singularity;
 \item \emph{no energy filter}: in the scattering governed by the bridging Hamiltonian, as well as by all type-$\mathrm{II}_a$ protocols with $\gamma=0$, the fraction of transmitted (and reflected) flux does not depend on the incident energy,  
 \begin{equation}
 \begin{split}
   T^{\mathrm{bridg.}}(E)\;:=\;T_{\alpha,1,0}(E)\;&=\;\frac{1}{2}\,(1+\cos\pi\alpha)\,, \\
   R^{\mathrm{bridg.}}(E)\;:=\; R_{\alpha,1,0}(E)\;&=\;\frac{1}{2}\,(1-\cos\pi\alpha)\,,
  \end{split}
\end{equation}
 meaning that the singularity does not act as a filter in the energy.  
\end{itemize}

 \begin{remark}
  At the upper edge of the considered range for the parameter $\alpha$, based on \eqref{eq:TR-IIa-thm} above one finds
	\begin{equation}
		\begin{split}
			&\lim_{\alpha \uparrow 1} T_{\alpha,a,\gamma}(E) \;=\; 0 \, , \\
			&\lim_{\alpha \uparrow 1} R_{\alpha,a,\gamma}(E) \;=\; 1 \, .
		\end{split}
	\end{equation}
Thus, the more singular the metric, the less transmitting the type-$\mathrm{II}_a$ protocol, up to the threshold $\alpha=1$ corresponding to the regime of geometric quantum confinement, where indeed the scattering becomes transmission-less (complete reflection).
 \end{remark}

 \begin{remark}
  The occurrence of the scattering only in the mode $k=0$ is inherently connected with the ``shape'' of the Grushin cylinder as $|x| \to +\infty$. If $\alpha \in(0,1)$, the expression \eqref{eq:galphaeverywhere} of the metric indicates that the larger $|x|$ the more the cylinder shrinks in the transversal direction, up to closing to a single point at infinity: this forces the incident particle incoming from infinity to only have zero angular momentum. When $\alpha=0$ the Grushin cylinder is an actual infinite cylinder in the Euclidean metric: thus, if one was to replace the singular interaction, supported on the circle $\{0\}\times\mathbb{S}^1_y$, of the models considered here, with a localised potential around $x=0$ and with $y$ dependence, one would then be able to engineer a flow of incoming particles with non-zero angular momentum, thus spiralling around the cylinder's axis and scattering through different sectors. In the present model, instead, the \emph{local}  boundary conditions at the singularity locus are $y$-\emph{independent}, also when $\alpha=0$: this makes the analysis independent in each sector. 
 \end{remark}

 \begin{remark}
  Concerning the possible existence of the special energy value \eqref{eq:Etransm-thm} at which the scattering is reflection-less, this is a phenomenon one is familiar with already from toy models such as the one-dimensional scattering over a finite rectangular barrier, and occurs when the incoming wave at that energy can ``conspire'' in the most efficient way with the boundary conditions at the scattering centre, so as to have zero reflection. It is worth observing that as $\alpha\to 1^-$, namely when the magnitude of the singularity of the metric reaches the threshold beyond which there is only geometric quantum confinement, thus no scattering, the reflection-less energy \eqref{eq:Etransm-thm} decreases with $\alpha$ until it vanishes: this means that at the $\alpha=1$ threshold the reflection-less scattering disappears, consistently with the fact that the whole scattering is inhibited. 
 \end{remark}

 \subsection{Scattering on fibre}\label{sec:scattering-fibre}

  Owing to Propositions \ref{prop:SpectrumK0} and \ref{prop:NegativeEigenvaluesKDiffZero}, inquiring the scattering on fibre is meaningful only for the mode $k=0$, where indeed the self-adjoint realisations of the fibre operator have all essential spectrum $[0,+\infty)$ with no embedded eigenvalues -- in particular, absolutely continuous.

As all the zero-mode self-adjoint operators \eqref{eq:families-ext-Aak} act with the differential action \eqref{eq:Afstar}, one is concerned in each case with the one-dimensional scattering on the potential
\begin{equation}
 V(x)\;:=\;\frac{\alpha(2+\alpha)}{4|x|^2}
\end{equation}
and with the additional ``internal'' interaction governed by one of the boundary conditions
 \eqref{eq:bifibre-AR}-\eqref{eq:bifibre-AG} at $x=0$ -- understanding the protocols $\mathrm{I}_\mathrm{R}$ and $\mathrm{I}_\mathrm{L}$ as describing an interaction with the potential $V$ and with the barrier at $x=0$ only on the corresponding half-line.

In all cases the quest for scattering states leads to determining the generalised eigenfunctions relative to an energy $E>0$ as suitable solutions $g\equiv\begin{pmatrix} g^- \\ g^+\end{pmatrix}$ to the ordinary differential equation
\begin{equation}\label{eq:ODEpm}
 -g''+V g\;=\;E g
\end{equation}
on each half-line when applicable.

In complete analogy to the reasoning for \eqref{eq:evproblem0}-\eqref{eq:GenericGE} one now solves the problem
\begin{equation}\label{eq:ODEpos}
 u''(x)+\big(E-{\textstyle\frac{\alpha(2+\alpha)}{4|x|^2}}\big)u(x)\;=\;0\,,\qquad x>0\,,
\end{equation}
in terms of modified Bessel functions,\index{Bessel functions} switching now for convenience to the \emph{Hankel functions}\index{Hankel functions} (Bessel functions of third kind) $H_\nu^{(1)}$ and $H_\nu^{(2)}$, where
\begin{equation}
 \begin{split}
  H_\nu^{(1)}(z)\;&=\;\frac{\ii}{\sin\nu\pi}\big(e^{-\ii\pi\nu}J_\nu(z)-J_{-\nu}(z)\big)\,, \\
  H_\nu^{(2)}(z)\;&=\;-\frac{\ii}{\sin\nu\pi}\big(e^{\ii\pi\nu}J_\nu(z)-J_{-\nu}(z)\big)\,,
 \end{split}
\end{equation}
and $J_\nu$ is the ordinary Bessel function of first kind \cite[Eq.~(9.1.3)-(9.1.4)]{Abramowitz-Stegun-1964}. Here $\nu=\frac{1+\alpha}{2}$. One thus writes the two linearly independent solutions  $u_E^{(1)}$ and $u_E^{(2)}$ to \eqref{eq:ODEpos} as
\begin{equation}\label{eq:v1v2soll}
 \begin{split}
  u_E^{(1)}(x)\;:=\;\sqrt{x}\, H_{\frac{1+\alpha}{2}}^{(1)}(x\sqrt{E})\,, \\
  u_E^{(2)}(x)\;:=\;\sqrt{x}\, H_{\frac{1+\alpha}{2}}^{(2)}(x\sqrt{E})\,. 
 \end{split}
\end{equation}
In turn, the general solution to \eqref{eq:ODEpm} on $\mathbb{R}$ has the form
\begin{equation}\label{eq:generalgEscatt}
 \;g_E(x)\;\equiv\;\begin{pmatrix} g_E^-(x) \\ g_E^+(x) \end{pmatrix}\;=\;
 \begin{pmatrix}
  A_1^- u_E^{(1)}(-x)+A_2^-u_E^{(2)}(-x) \\
  A_1^+ u_E^{(1)}(x)+A_2^+u_E^{(2)}(x)
 \end{pmatrix},\qquad A_1^\pm,A_2^\pm\in\mathbb{C}\,.
\end{equation}

From \eqref{eq:v1v2soll} and from the asymptotics at short \cite[Eq.~(9.1.10)]{Abramowitz-Stegun-1964} and large \cite[Eq.~(9.2.3)-(9.2.4)]{Abramowitz-Stegun-1964} distances for the Hankel functions one computes
\begin{equation}\label{eq:Hankel-small}
 \begin{split}
 u_E^{(1)}(x)\;&\stackrel{x\downarrow 0}{=}\;-\frac{\ii\,2^{\frac{1+\alpha}{2}}}{\,E^{\frac{1+\alpha}{4}}\Gamma(\frac{1-\alpha}{2})\sin(\frac{1+\alpha}{2}\pi)}\,x^{-\frac{\alpha}{2}} \\
  &\qquad\qquad+\frac{\ii\,E^{\frac{1+\alpha}{4}}e^{-\ii\pi\frac{1+\alpha}{2}}}{2^{\frac{1+\alpha}{2}}\Gamma(\frac{3+\alpha}{2})\sin(\frac{1+\alpha}{2}\pi)}\,x^{1+\frac{\alpha}{2}}+O(x^{2-\frac{\alpha}{2}})\,, \\
  u_E^{(2)}(x)\;&\stackrel{x\downarrow 0}{=}\;\frac{\ii\,2^{\frac{1+\alpha}{2}}}{\,E^{\frac{1+\alpha}{4}}\Gamma(\frac{1-\alpha}{2})\sin(\frac{1+\alpha}{2}\pi)}\,x^{-\frac{\alpha}{2}} \\
  &\qquad\qquad-\frac{\ii\,E^{\frac{1+\alpha}{4}}e^{\ii\pi\frac{1+\alpha}{2}}}{2^{\frac{1+\alpha}{2}}\Gamma(\frac{3+\alpha}{2})\sin(\frac{1+\alpha}{2}\pi)}\,x^{1+\frac{\alpha}{2}}+O(x^{2-\frac{\alpha}{2}})\,,
 \end{split}
\end{equation}
and 
\begin{equation}\label{eq:Hankel-large}
 \begin{split}
  u_E^{(1)}(x)\;&\stackrel{x\to+\infty}{=}\;\sqrt{\frac{2}{\pi}}\,E^{-\frac{1}{4}}\,e^{i(x\sqrt{E}-\frac{\pi}{4}(2+\alpha))}\,(1+O(x^{-1}))\,, \\
  u_E^{(2)}(x)\;&\stackrel{x\to+\infty}{=}\;\sqrt{\frac{2}{\pi}}\,E^{-\frac{1}{4}}\,e^{-i(x\sqrt{E}-\frac{\pi}{4}(2+\alpha))}\,(1+O(x^{-1}))\,.
 \end{split}
\end{equation}
Incidentally, \eqref{eq:Hankel-large} explains the convenience of choosing $H_{\frac{1+\alpha}{2}}^{(1)},H_{\frac{1+\alpha}{2}}^{(2)}$ to express a pair of linearly independent solutions to \eqref{eq:ODEpos}: at large distances, $u_E^{(1)}$ and $u_E^{(2)}$ behave as \emph{plane waves}, with an evident advantage for their interpretation in the scattering arguments that follow.

The local behaviour of the general solution \eqref{eq:generalgEscatt} around $x=0$, and concretely speaking the boundary values \eqref{eq:bilimitsg0g1}, are given by
\begin{equation}\label{eq:boundarygpmscatt}
 \begin{split}
  g_{E,0}^\pm\;&=\;\frac{\ii\,2^{\frac{1+\alpha}{2}}}{\,E^{\frac{1+\alpha}{4}}\Gamma(\frac{1-\alpha}{2})\sin(\frac{1+\alpha}{2}\pi)}\,(A_2^\pm-A_1^\pm)\,,  \\
  g_{E,1}^\pm\;&=\;\frac{E^{\frac{1+\alpha}{4}}e^{-\ii\pi\frac{\alpha}{2}}}{2^{\frac{1+\alpha}{2}}\Gamma(\frac{3+\alpha}{2})\sin(\frac{1+\alpha}{2}\pi)}\,(A_1^\pm + e^{\ii\pi\alpha}A_2^\pm)\,,
 \end{split}
\end{equation}
as is easy to compute from \eqref{eq:Hankel-small} and \eqref{eq:bilimitsg0g1}.

 Focussing, in particular, on the $\mathrm{II}_a$-type scattering, one finds the following.

 \begin{proposition}\label{prop:scatteringIIa} Let $\alpha\in[0,1)$, $a\in\mathbb{C}$, and $\gamma\in\mathbb{R}$.
 \begin{enumerate}
  \item[(i)] The one-dimensional Schr\"{o}dinger scattering governed by the Hamiltonian $A_{\alpha,a}^{[\gamma]}(0)$ has transmission and reflection coefficients equal respectively to
   \begin{equation}\label{eq:TR-IIa}
  \begin{split}
   T_{\alpha,a,\gamma}(E)\;&=\;\left|\frac{\,E^{\frac{1+\alpha}{2}}(1+e^{\ii\pi\alpha})\,\Gamma(\frac{1-\alpha}{2})\,\overline{a}}{\,E^{\frac{1+\alpha}{2}}\Gamma(\frac{1-\alpha}{2})(1+|a|^2)+\ii\,\gamma\,2^{1+\alpha}e^{\ii\frac{\pi}{2}\alpha}\Gamma(\frac{3+\alpha}{2})} \right|^2 \,, \\
   R_{\alpha,a,\gamma}(E)\;&=\;\left|\frac{\,E^{\frac{1+\alpha}{2}}\Gamma(\frac{1-\alpha}{2})\,(1-|a|^2\,e^{\ii\pi\alpha})+\ii\,\gamma\,2^{1+\alpha}e^{\ii\frac{\pi}{2}\alpha}\Gamma(\frac{3+\alpha}{2})}{\,E^{\frac{1+\alpha}{2}}\Gamma(\frac{1-\alpha}{2})(1+|a|^2)+\ii\,\gamma\,2^{1+\alpha}e^{\ii\frac{\pi}{2}\alpha}\Gamma(\frac{3+\alpha}{2})}\right|^2\,,
  \end{split}
 \end{equation}
 and
 \begin{equation}\label{eq:TpEi1}
  T_{\alpha,a,\gamma}(E)+R_{\alpha,a,\gamma}(E)\;=\;1\,.
 \end{equation}
 \item[(ii)] Reflection and transmission coefficients are independent of energy $E$ for all extensions with $\gamma=0$.
 \item[(iii)] The scattering is reflection-less ($R_{\alpha,a,\gamma}(E)=0$) when
 \begin{equation}\label{eq:Etransm}
   E\;=\;\bigg(\frac{\,2^{1+\alpha}\,\gamma\,\Gamma(\frac{3+\alpha}{2})\sin\frac{\pi}{2}\alpha\,}{\,\Gamma(\frac{1-\alpha}{2})(1-\cos\pi\alpha)\,}\bigg)^{\!\frac{2}{1+\alpha}},	
 \end{equation}
 provided that $\alpha\in(0,1)$, $|a|=1$, and $\gamma>0$. 
 \item[(iv)] In the high energy limit the scattering is independent of the extension parameter $\gamma$ and one has
 \begin{equation}\label{eq:highenergyscatt}
  \begin{split}
   \lim_{E\to+\infty}T_{\alpha,a,\gamma}(E)\;&=\;\frac{\,2\,|a|^2(1+\cos\pi\alpha)}{(1+|a|^2)^2}\,, \\
   \lim_{E\to+\infty}R_{\alpha,a,\gamma}(E)\;&=\;\frac{\,1+|a|^4-2|a|^2\cos\pi\alpha}{(1+|a|^2)^2}\,.
  \end{split}
 \end{equation}
 \item[(v)] In the low energy limit, for $\gamma\neq 0$,
  \begin{equation}\label{eq:lowenergyscatt}
  \begin{split}
   \lim_{E\downarrow 0}T_{\alpha,a,\gamma}(E)\;&=\;0\,, \\
   \lim_{E\downarrow 0}R_{\alpha,a,\gamma}(E)\;&=\;1\,.
  \end{split}
 \end{equation}
 \end{enumerate} 
 \end{proposition}

\begin{proof} The generalised eigenfunctions $g_E$ for $A_{\alpha,a}^{[\gamma]}$ at energy $E>0$ have the form \eqref{eq:generalgEscatt} with parameters $A_1^\pm,A_2^\pm$ such that the corresponding boundary values \eqref{eq:boundarygpmscatt} satisfy
 \[
  \begin{cases}
   \qquad g_{E,0}^+\,=\,a\,g_{E,0}^-\,, \\
   \;g_{E,1}^-+\overline{a}g_{E,1}^+\,=\,\gamma\,g_{E,0}^-
  \end{cases}
 \]
 (see \eqref{eq:bifibre-Aag} above), whence
 \[\tag{a}\label{eq:taaac1}
  \begin{cases}
   \qquad A_2^+-A_1^+\,=\,a(A_2^--A_1^-) \,,\\
   \;\displaystyle\frac{\,E^{\frac{1+\alpha}{2}}e^{-\ii\pi\frac{\alpha}{2}}\Gamma(\frac{1-\alpha}{2})}{2^{1+\alpha}\Gamma(\frac{3+\alpha}{2})}\,(A_1^-+\overline{a}\,A_1^+ + e^{\ii\pi\alpha}(A_2^-+\overline{a}\,A_2^+)\,= \\
   \;\qquad\qquad\qquad \;=\,\ii\,\gamma(A_2^--A_1^-)\,.
  \end{cases}
 \]

 By standard scattering arguments, and in view of the large distance asymptotics \eqref{eq:Hankel-large}, the occurrence of an incident \emph{plane wave} $e^{-\ii x\sqrt{E}}$ from $+\infty$ towards the origin (with ``momentum'' $\sqrt{E}$), producing after the scattering a transmitted plane wave $e^{-\ii x\sqrt{E}}$ towards $-\infty$ and a reflected plane wave $e^{\ii x\sqrt{E}}$ back towards $+\infty$, corresponds to the choice
 \[\tag{b}\label{eq:taaac2}
  A_2^+\,=\,1\,,\qquad A_2^-\,=\,0\,.
 \]
 Indeed, in this case \eqref{eq:generalgEscatt} takes the asymptotic form
 \[
  g_E(x)\;\sim\;
   {\textstyle\sqrt{\frac{2}{\pi}}}\,E^{-\frac{1}{4}}e^{\ii\frac{\pi}{4}(2+\alpha)}\begin{pmatrix}
   A_1^-\,e^{-\ii\frac{\pi}{2}(2+\alpha)}\,e^{-\ii x\sqrt{E}} \\
   A_1^+\,e^{-\ii\frac{\pi}{2}(2+\alpha)}\,e^{\ii x\sqrt{E}}+e^{-\ii x\sqrt{E}}
  \end{pmatrix}\qquad\textrm{as }|x|\to +\infty\,,
 \]
 with conventionally unit amplitude for the source incident plane wave from $+\infty$ (up to the overall multiplicative pre-factor $\sqrt{\frac{2}{\pi}}\,E^{-\frac{1}{4}}e^{\ii\frac{\pi}{4}(2+\alpha)}$), and clearly no incident plane wave from $-\infty$. The transmission and reflection coefficients for the scattering are then equal respectively to
 \[
   T^{(\leftarrow)}_{\alpha,a,\gamma}(E)\;=\;|A_1^-|^2\,,\qquad R^{(\leftarrow)}_{\alpha,a,\gamma}(E)\;=\;|A_1^+|^2\,.
 \]

 Solving \eqref{eq:taaac1} with the condition \eqref{eq:taaac2} yields
 \[
  \begin{split}
   A_1^-\;&=\;-\frac{\,E^{\frac{1+\alpha}{2}}(1+e^{\ii\pi\alpha})\,\Gamma(\frac{1-\alpha}{2})\,\overline{a}}{\,E^{\frac{1+\alpha}{2}}\Gamma(\frac{1-\alpha}{2})(1+|a|^2)+\ii\,\gamma\,2^{1+\alpha}e^{\ii\frac{\pi}{2}\alpha}\Gamma(\frac{3+\alpha}{2})} \,,\\
   A_1^+\;&=\;\frac{\,E^{\frac{1+\alpha}{2}}\Gamma(\frac{1-\alpha}{2})\,(1-|a|^2\,e^{\ii\pi\alpha})+\ii\,\gamma\,2^{1+\alpha}e^{\ii\frac{\pi}{2}\alpha}\Gamma(\frac{3+\alpha}{2})}{\,E^{\frac{1+\alpha}{2}}\Gamma(\frac{1-\alpha}{2})(1+|a|^2)+\ii\,\gamma\,2^{1+\alpha}e^{\ii\frac{\pi}{2}\alpha}\Gamma(\frac{3+\alpha}{2})}\,.
  \end{split}
 \]

 Mimicking the reasoning above, the occurrence of an incident plane wave $e^{\ii x\sqrt{E}}$ from $-\infty$ towards the origin, producing after the scattering a transmitted plane wave $e^{\ii x\sqrt{E}}$ towards $+\infty$ and a reflected plane wave $e^{-\ii x\sqrt{E}}$ back towards $-\infty$, corresponds to the choice
 \[\tag{c}\label{eq:taaac3}
  A_2^-\,=\,1\,,\qquad A_2^+\,=\,0\,,
 \]
 in which case \eqref{eq:generalgEscatt} becomes
 \[
  g_E(x)\;\sim\;
   {\textstyle\sqrt{\frac{2}{\pi}}}\,E^{-\frac{1}{4}}e^{\ii\frac{\pi}{4}(2+\alpha)}\begin{pmatrix}
   A_1^-\,e^{-\ii\frac{\pi}{2}(2+\alpha)}\,e^{-\ii x\sqrt{E}}+e^{\ii x\sqrt{E}} \\
   A_1^+\,e^{-\ii\frac{\pi}{2}(2+\alpha)}\,e^{\ii x\sqrt{E}}
  \end{pmatrix}\qquad\textrm{as }|x|\to +\infty\,.
 \]
 The transmission and reflection coefficients for the scattering are now, respectively,
 \[
   T^{(\rightarrow)}_{\alpha,a,\gamma}(E)\;=\;|A_1^+|^2\,,\qquad R^{(\rightarrow)}_{\alpha,a,\gamma}(E)\;=\;|A_1^-|^2\,.
 \]
  Solving \eqref{eq:taaac1} with the new condition \eqref{eq:taaac3} yields
 \[
  \begin{split}
   A_1^-\;&=\;\frac{\,E^{\frac{1+\alpha}{2}}\Gamma(\frac{1-\alpha}{2})\,(|a|^2-\,e^{\ii\pi\alpha})+\ii\,\gamma\,2^{1+\alpha}e^{\ii\frac{\pi}{2}\alpha}\Gamma(\frac{3+\alpha}{2})}{\,E^{\frac{1+\alpha}{2}}\Gamma(\frac{1-\alpha}{2})(1+|a|^2)+\ii\,\gamma\,2^{1+\alpha}e^{\ii\frac{\pi}{2}\alpha}\Gamma(\frac{3+\alpha}{2})}\,, \\
   A_1^+\;&=\;-\frac{\,E^{\frac{1+\alpha}{2}}(1+e^{\ii\pi\alpha})\,\Gamma(\frac{1-\alpha}{2})\,a}{\,E^{\frac{1+\alpha}{2}}\Gamma(\frac{1-\alpha}{2})(1+|a|^2)+\ii\,\gamma\,2^{1+\alpha}e^{\ii\frac{\pi}{2}\alpha}\Gamma(\frac{3+\alpha}{2})}\,.
  \end{split}
 \]
 An easy computation shows that 
 \[
  \begin{split}
   T^{(\rightarrow)}_{\alpha,a,\gamma}(E)\;=\;T^{(\leftarrow)}_{\alpha,a,\gamma}(E)\;&=:\;T_{\alpha,a,\gamma}(E)\,, \\
   R^{(\rightarrow)}_{\alpha,a,\gamma}(E)\;=\;R^{(\leftarrow)}_{\alpha,a,\gamma}(E)\;&=:\;R_{\alpha,a,\gamma}(E)\,,
  \end{split}
 \]
 whence the final expressions \eqref{eq:TR-IIa}, as well as the validity of \eqref{eq:TpEi1} (see also Remark \ref{rem:Jcont} below). This proves part (i).

 All the claims of part (ii), (iv), and (v) then follow straightforwardly from  \eqref{eq:TR-IIa}. Concerning part (iii), one sees from \eqref{eq:TR-IIa} that the scattering is reflection-less when
 \begin{equation*}\tag{d}\label{eq:taaac4}
   E\;=\;\bigg(\frac{\,\ii\,\gamma\,2^{1+\alpha}\,e^{\ii\frac{\pi}{2}\alpha}\,\Gamma(\frac{3+\alpha}{2})}{(|a|^2 e^{\ii\pi\alpha}-1)\Gamma(\frac{1-\alpha}{2})}\bigg)^{\!\frac{2}{1+\alpha}},
 \end{equation*}
 provided that
 \[
  0\;<\;\frac{\ii\gamma\,e^{\ii\frac{\pi}{2}\alpha}}{|a|^2e^{\ii\pi\alpha}-1}\;=\;\frac{\,(|a|^2+1)\gamma\sin\frac{\pi}{2}\alpha+\ii(|a|^2-1)\gamma\cos\frac{\pi}{2}\alpha\,}{(|a|^2\cos\pi\alpha-1)^2+|a|^4\sin^2\pi\alpha}\,,
 \]
 a condition that is only satisfied when $\alpha\in(0,1)$, \emph{and} $|a|=1$, \emph{and} $\gamma>0$. If this is the case, the quantity computed above becomes
 \[
  \frac{\gamma\sin\frac{\pi}{2}\alpha}{\,1-\cos\pi\alpha\,}\,,
 \]
 and the expression \eqref{eq:taaac4} takes finally the form \eqref{eq:Etransm}. 
\end{proof}

 \begin{remark}\label{rem:Jcont}
  The identity $ T_{\alpha,a,\gamma}(E)+R_{\alpha,a,\gamma}(E)=1$ was checked directly in the proof above. In fact, it could have been claimed a priori, as one does with the scattering governed by a Schr\"{o}dinger operator $H_{\mathrm{Schr}}=-\frac{\ud^2}{\ud x^2}+V$ with smooth and fast decaying real potential $V$, but with a subtle difference in the reasoning. For  $H_{\mathrm{Schr}}$, the problem $H_{\mathrm{Schr}}g=Eg$ with $E>0$ is an ODE on $\mathbb{R}$ whose (two linearly independent) solutions are continuous, and with continuous derivative, over the whole $\mathbb{R}$. For any such $g$, the probability current
  \begin{equation}\label{eq:current}
   J_g(x)\;:=\;-\ii\big(\overline{g(x)}g'(x)-g(x)\overline{g'(x)}\big)\;=\;2\,\mathfrak{Im}\big(\overline{g(x)}g'(x)\big)
  \end{equation}
  is actually conserved $\forall x\in\mathbb{R}$ because
  \[
   J_g'(x)\;=\;2\,\mathfrak{Im}\big(\overline{g(x)}g''(x)\big)\;=\;2\,\mathfrak{Im}\big((V(x)-E)|g(x)|^2\big)\;=\;0
  \]
  (having used the identity $-g''+Vg=Eg$ and the reality of $V$). In the present setting, one could also reason by saying that on each open half-line $\mathbb{R}^\pm$ the ODE $H_{\mathrm{Schr}}g=Eg$ gives rise to a conserved current, and the left and right currents do coincide because, taking the limit $x\to 0^\pm$ in \eqref{eq:current}, one exploits the continuity of $g$ and $g'$ at $x=0$. In turn, the conserved current allows one to conclude the following: for a solution $g$ to $H_{\mathrm{Schr}}g=Eg$ with asymptotics
  \begin{equation}\label{eq:rtsol}
   g(x)\;\sim\;
   \begin{cases}
    \;e^{\ii x\sqrt{E}}+r\,e^{-\ii x\sqrt{E}} & \textrm{as } x\to -\infty \\
    \;\qquad t\,e^{\ii x\sqrt{E}} & \textrm{as } x\to +\infty
   \end{cases}
  \end{equation}
  (such $g$ does exists, as $V$ has fast decrease) a simple calculation yields
  \begin{equation}
   \lim_{x\to-\infty}J_g(x)\;=\;2\sqrt{E}\,(1-|r|^2)\,,\qquad  \lim_{x\to+\infty}J_g(x)\;=\;2\sqrt{E}\,|t|^2\,,
  \end{equation}
  whence $|t|^2+|r|^2=1$ because $J_g(x)$ is constant. This gives the conservation of the incident flux into the sum of reflected and transmitted flux. In the present case, \eqref{eq:ODEpos} is a differential problem consisting of \emph{two} separate ODEs on each half-line, to which one superimposes the boundary condition at $x=0$ characteristic for the operator $A_{\alpha,a}^{[\gamma]}$. Precisely as in the ordinary Schr\"{o}dinger case, one has probability currents
    \begin{equation}\label{eq:currents}
   J^\pm_g(x)\;:=\;2\,\mathfrak{Im}\big(\overline{g^\pm(x)}(g^\pm)'(x)\big)
  \end{equation}
  on $\mathbb{R}^\pm$, each of which is conserved, and such that for a solution of type \eqref{eq:rtsol}
    \begin{equation}
   \lim_{x\to-\infty}J^-_g(x)\;=\;2\sqrt{E}\,(1-|r|^2)\,,\qquad  \lim_{x\to+\infty}J^+_g(x)\;=\;2\sqrt{E}\,|t|^2\,.
  \end{equation}
  However, the overall conservation $J^-=J^+$, which would lead again to $|t|^2+|r|^2=1$, cannot follow from the continuity of $g$ and $g'$ at $x=0$ as for the ordinary Schr\"{o}dinger case: such functions actually diverge as $|x|\to 0$, more precisely
  \begin{equation}\label{eq:asintg}
   \begin{split}
       g(x)\;&=\;
   \begin{cases}
    \;g_0^-(-x)^{-\frac{\alpha}{2}}+g_1^-(-x)^{1+\frac{\alpha}{2}}+o(x^{\frac{3}{2}}) & \textrm{as } x\uparrow 0^-\,, \\
    \;g_0^+x^{-\frac{\alpha}{2}}+g_1^+ x^{1+\frac{\alpha}{2}}+o(x^{\frac{3}{2}}) & \textrm{as } x\downarrow 0^+\,,
   \end{cases} \\
   g'(x)\;&=\;
   \begin{cases}
    \;\frac{\alpha}{2}\,g_0^-(-x)^{-(1+\frac{\alpha}{2})}-(1+\frac{\alpha}{2})\,g_1^-(-x)^{\frac{\alpha}{2}}+o(x^{\frac{1}{2}}) & \textrm{as } x\uparrow 0^-\,, \\
    \;-\frac{\alpha}{2}\,g_0^+ x^{-(1+\frac{\alpha}{2})}+(1+\frac{\alpha}{2})\,g_1^+ x^{\frac{\alpha}{2}}+o(x^{\frac{1}{2}}) & \textrm{as } x\downarrow 0^+\,,
   \end{cases}
   \end{split}
  \end{equation}
  with
   \begin{equation}\label{eq:againcond}
  \begin{cases}
   \qquad g_{0}^+\,=\,a\,g_{0}^- \,,\\
   \;g_{1}^-+\overline{a}g_{1}^+\,=\,\gamma\,g_{0}^-\,.
  \end{cases}
 \end{equation}
    A new check is therefore needed: from \eqref{eq:currents} and \eqref{eq:asintg} one computes
  \begin{equation}
   J^\pm\;:=\;\lim_{x\to 0^\pm}J^\pm(x)\;=\;\pm \,2\,(1+\alpha)\,\mathfrak{Im}\big(\overline{g_0^\pm}\,g_1^\pm \big)
  \end{equation}
 and using \eqref{eq:againcond} one finds
  \begin{equation}
   \begin{split}
    J^-\;&=\;-2(1+\alpha)\,\mathfrak{Im}\Big(\frac{1}{\overline{a}}\overline{g_0^+}\cdot\Big(\gamma\,\frac{1}{a}g_0^+-\overline{a}g_1^+\Big)\Big) \\
    &=\;2(1+\alpha)\,\mathfrak{Im}\big(\overline{g_0^+}\,g_1^+ \big)\;=\;J^+\,.
   \end{split}
  \end{equation}
 This finally establishes the a priori information that $ T_{\alpha,a,\gamma}(E)+R_{\alpha,a,\gamma}(E)=1$.  
   \end{remark}

 \begin{proof}[Proof of Theorem \ref{thm:scattering}]
  One exploits once again the unitary equivalence \eqref{eq:inverseUniary_all} and studies the scattering governed by the Hamiltonian $\mathscr{H}_{\alpha,a}^{[\gamma]}$. Based on the fibred structure \eqref{eq:Halpha-IIa_unif-fibred} of the latter operator, it is clear that the scattering takes place independently in each $k$-channel. And owing to the analysis of Propositions \ref{prop:SpectrumK0} and \ref{prop:NegativeEigenvaluesKDiffZero} one sees that the $k=0$ channel is the only meaningful one. In this case the scattering was studied in Proposition \ref{prop:scatteringIIa}. In particular, the free incident flux on the $k=0$ fibre at energy $E>0$ is described by the plane wave $g_E(x)\sim e^{\pm\ii x\sqrt{E}}$ as $|x|\to +\infty$: this corresponds to a scattering state $\psi\equiv(\psi_k)_{k\in\mathbb{Z}}$ for $\mathscr{H}_{\alpha,a}^{[\gamma]}$ where $\psi_0\equiv g_E$ and $\psi_k\equiv 0$ if $k\neq 0$. One thus sees, by inverting the transformations \eqref{eq:unit1}-\eqref{eq:defF2}, that the large distance behaviour of the incident flux for the scattering governed by $H_{\alpha,a}^{[\gamma]}$ on the Grushin manifold $M_\alpha$ has precisely the form \eqref{eq:fscatt}. All the claimed properties are invariant under unitary transformations and then follows at once from Proposition \ref{prop:scatteringIIa}.
  \end{proof}


%
%
%
\chapter{Models of zero-range interaction for the bosonic trimer at unitarity}
\label{chapter-bosonic_trimer} 

 Models of particles coupled by a non-trivial interaction whose spatial range is zero, hence a \emph{contact} interaction,\index{contact interaction} represent one of the deepest modern applications in quantum mechanics of self-adjoint extension theory (see, e.g., \cite{Albeverio-HK-Streit-1977-enforms,Thomas-L-E-PhysRevD-1984-multiparticle,albeverio-solvable,Koshmanenko-1993,albeverio-makarov-1997-attractors,albeverio-lakaev-makarov-1998,albverio-kurasov-2000-sing_pert_diff_ops,Dimock-Rajeev-2004,Albeverio-Figari-2018}). This is also a field that combines a long history of well-established results with really hard problems that are still open. This is even more so in view of the rather intriguing circumstance that recent experimental advances have made the subject highly topical in theoretical and experimental cold atom physics, in parallel with the purely mathematical investigation, a scenario where formal physical heuristics produce beautiful results and conjectures that still need be demonstrated in full mathematical rigour. Self-adjoint extension theory kicks in to identify physically meaningful self-adjoint realisations of formal Hamiltonians, and then to study their spectral properties. Mathematically this is often particularly challenging because in this context one is \emph{not} dealing with ordinary Schr\"{o}dinger operators with usual structure of kinetic plus potential part: indeed, the `interaction potential' is rather to be thought of as some sort of delta-like profile (with the important warning, as specified later, that in two and three dimensions these are \emph{not} and \emph{cannot} be Dirac deltas), thus a large amount of techniques that are specific from Schr\"{o}dinger operator theory are not directly applicable. While giving multiple references to a growing and active literature, this chapter is focussed on a specific class of zero-range models, in fact on one of the most relevant and instructive `building blocks', namely a system of three identical bosons with non-trivial contact interaction, keeping the recent work \cite{M2020-BosonicTrimerZeroRange} as main reference. There is also a pedagogical motivation for choosing such bosonic playground: as these models are naturally unbounded from below, it turns out that a clever combination of both von Neumann and Kre{\u\i}n-Vi\v{s}ik-Birman extension schemes is required.

\section{Introduction and background}\label{sec:BOSCHAPTintro}

The focus of this chapter is a class of models for a three-dimensional quantum system of this kind: three non-relativistic, identical bosons are coupled among themselves by means of an isotropic two-body interaction of zero spatial range and, for the main part of our analysis, with infinite scattering length. The interaction does not couple the spins. The notation $\xx$ or $\pp$ will be used throughout to emphasise three-dimensional position or momentum variables.


It is fair to say that the system under consideration has undergone various phases of interests over the decades, both in the physical and in the mathematical literature, until the present days. Originally, and also without imposing the bosonic symmetry, it emerged as the typical picture for interacting nucleons in early nuclear physics, at the scale of which the inter-particle interaction may well be considered of zero range as compared to the atomic scales. Instead, in more recent times it has been a system of interest in cold atom physics, given the modern experimental advances in inducing effective zero-range interactions in a Bose gas or in heteronuclear gaseous mixtures by means of sophisticated Feschbach-resonance techniques.\index{Feschbach resonance}

As the perspective here is mainly mathematical, even if driven by strong physical inspiration, it is worth stressing an important and long lasting difference between the approaches.

\emph{Physical} investigations of the quantum three-body problem with zero-range interaction have always had as primary interest the characterisation of the \emph{bound states} of the system. To this aim, at least in the more modern literature (given its vastness, one may refer to the recent reviews \cite{Braaten-Hammer-2006,Naidon-Endo-Review_Efimov_Physics-2017}), the eigenvalue problem is invariably set up in terms of the free Hamiltonian (all in all particles subject to a zero-range interaction are meant to move as free bodies except when they come on top of each other), with the constraint that the three-body eigenfunction must display the `physical' short-range asymptotics 
\begin{equation}\label{eq:preBP}
 \psi(\xx_1,\xx_2,\xx_3)\;\sim\;\frac{1}{|\xx_i-\xx_j|}-\frac{1}{a}\qquad \textrm{as }|\xx_i-\xx_j|\to 0\,,
\end{equation}
where $a$ is the $s$-wave scattering length in each two-body channel. The behaviour \eqref{eq:preBP} was identified by Bethe and Peierls in 1935 \cite{Bethe_Peierls-1935} as the actual leading behaviour of eigenfunctions with `contact' interaction. Next, solutions are obtained, with an ad hoc analysis applicable to the eigenvalue problem only, and not to the generality of states in the domain of the underlying Hamiltonian, by reducing the three-body \emph{eigenfunction} to a convenient triple of two-body channel \emph{Faddeev components},\index{Faddeev components} in a combination of which that encodes the possible bosonic or fermionic symmetry, where each Faddeev component is a function of one pair of internal Jacobi coordinates. In the case of three identical bosons,
\begin{equation}
 \psi(\xx_1,\xx_2,\xx_3)\;=\;\chi(\xx_{12},\xx_{12,3})+\chi(\xx_{23},\xx_{23,1})+\chi(\xx_{31},\xx_{31,2})\,,
\end{equation}
where
\begin{equation}
 \xx_{ij}\;=\;\xx_j-\xx_i\,,\qquad \xx_{ij,k}\;=\;\xx_k-\frac{\xx_i+\xx_j}{2}\,.
\end{equation}
Based on the Faddeev equations formalism for the three-body system \cite{Faddeev-1963-eng-1965-3body,Fedorov-Jensen-1993}, the original problem is thus boiled down to a single Faddeev component $\chi$. At this level the problem is conveniently separable upon switching from Jacobi to hyper-radial coordinates\index{hyper-radial coordinates} and expanding $\chi$ into definite angular momentum terms, and in each sector of definite angular symmetry the problem becomes tractable analytically and numerically.

The above line of reasoning, in fact encompassing a multitude of similar variants, is due to an original idea of Landau, elaborated in the mid 1950's by Skornyakov and Ter-Martirosyan \cite{TMS-1956} in a famous study of the three-body quantum system with \emph{zero-range} interactions. (Actually, \cite{TMS-1956} predates by a couple of years Faddeev's first work \cite{Faddeev-scattering-1960} on the three-body scattering theory, and makes use of Green's function methods. Then in \cite{Faddeev-scattering-1960} Faddeev showed that the equation identified by Skornyakov and Ter-Martirosyan for solving the three-body eigenvalue problem could be recovered in the formal limit of zero interaction range from the ordinary scheme of Faddeev equations.)

In atomic physics the approach sketched above is the basis of what one has customarily referred to since then as \emph{zero-range methods} \index{zero-range methods}\cite{Demkov-Ostrovskii-book}. The same approach resurfaced in the early 1970's by Efimov \cite{Efimov-1971,Efimov-1973} in his famous work on quantum three-body systems with \emph{finite-range} two-body interactions (with important precursors such as Macek \cite{Macek-1968} in the usage of hyper-radial equations for three-body energy levels). Efimov's analysis established a reference for the subsequent literature on cold-atom few-body systems.

The catch here is that such a physical scheme is solid when the inter-particle interactions are realised, say, by potentials $V_{ij}$ that are sufficiently regular and have short range, thereby making the underlying three-body Hamiltonian 
\begin{equation}
 -\frac{1}{2m_1}\Delta_{\xx_1}-\frac{1}{2m_2}\Delta_{\xx_2}-\frac{1}{2m_3}\Delta_{\xx_3}+V_{12}(\xx_1-\xx_2)+V_{23}(\xx_2-\xx_3)+V_{13}(\xx_1-\xx_3)
\end{equation}
(in units $\hslash=1$) unambiguously realised as a self-adjoint operator on the three-body Hilbert space, and thus giving rise to a well-posed set of Faddeev equations. At zero range, instead, the model is formally thought of as 
\begin{equation}\label{eq:fistFormalHamilt}
 -\frac{1}{2m_1}\Delta_{\xx_1}-\frac{1}{2m_2}\Delta_{\xx_2}-\frac{1}{2m_3}\Delta_{\xx_3}+\mu_{12}\delta(\xx_1-\xx_2)+\mu_{23}\delta(\xx_2-\xx_3)+\mu_{13}\delta(\xx_1-\xx_3)
\end{equation}
for some coupling constants $\mu_{ij}$, where the $\delta$-functions are symbolically used to mean some sort of non-trivial interaction only supported at the configurations $\xx_j=\xx_k$: as \eqref{eq:fistFormalHamilt} is not an ordinary Schr\"{o}dinger operator, for it Faddeev components of the three-body eigenfunctions and the corresponding Faddeev equations do not make sense strictly speaking, but for a formal limit of zero interaction range.

In short, physical zero-range methods\index{zero-range methods} determine eigenfunctions and eigenvalues of a \emph{formal Hamiltonian that otherwise remains unspecified}.

The signature of a possible remaining ambiguity of the physical approach is the emergence of an unphysical continuum of eigenvalues, an occurrence that depends on the masses, the attractive or repulsive nature of the interaction, and the bosonic or fermionic exchange symmetry in \eqref{eq:fistFormalHamilt}. When this happens, an (infinite) discrete set of bound states is selected by imposing an additional restriction to the admissible eigenfunctions. Such a restriction may be suitably interpreted as a \emph{three-body short-range boundary condition}.\index{contact condition!three-body} This occurrence was initially observed by Skornyakov \cite{Skornyakov-1959} right after his joint work \cite{TMS-1956} with Ter-Martirosyan, and was analysed by Danilov \cite{Danilov-1961} who selected the admissible solutions in the spirit of the additional experimental three-body parameter proposed at the same time by Gribov \cite{Gribov-1959}. That choice was soon after justified on more rigorous operator-theoretic grounds by Faddeev and Minlos \cite{Minlos-Faddeev-1961-1,Minlos-Faddeev-1961-2}. (It is actually remarkable that such  Russian key contributions all span a fistful of years, from the work \cite{TMS-1956} by Skornyakov and Ter-Martirosyan in 1956 to the period 1959-1961 with the works by Skornyakov \cite{Skornyakov-1959}, Gribov \cite{Gribov-1959}, Danilov \cite{Danilov-1961}, Faddeev \cite{Faddeev-scattering-1960}, and Faddeev and Minlos \cite{Minlos-Faddeev-1961-1,Minlos-Faddeev-1961-2}.) The possible necessity of an additional three-body parameter and its physical interpretation have become by now a standard picture in the physical literature of cold atoms in the zero-range regime \cite[Section 4]{Naidon-Endo-Review_Efimov_Physics-2017}.

\emph{Mathematical} investigations of the quantum three-body problem with zero-range interaction, on the other hand, have pursued over the decades a different programme: to \emph{characterise first the Hamiltonian} of the system, as an explicitly declared self-adjoint operator on Hilbert space, through its operator or form domain of self-adjointness and its action on each function of the domain, \emph{and only after to analyse the spectral properties}.

This conceptual scheme was brought up first in the already mentioned seminal works by Faddeev and Minlos \cite{Minlos-Faddeev-1961-1,Minlos-Faddeev-1961-2}, which were deeply mathematical in nature. There, rigorous Hamiltonians of contact interaction were proposed as suitable self-adjoint extensions of the symmetric operator
\begin{equation}\label{eq:freeHamiltRestricted}
 \Big(-\frac{1}{m_1}\Delta_{\xx_1}-\frac{1}{m_2}\Delta_{\xx_2}-\frac{1}{m_2}\Delta_{\xx_2}\Big)\Big|_{C^\infty_c((\mathbb{R}^3_{\xx_1}\times\mathbb{R}^3_{\xx_2}\times\mathbb{R}^3_{\xx_3})\setminus\Gamma)}\,,
\end{equation}
namely the free three-body Hamiltonian restricted on smooth functions that are compactly supported away from the \emph{coincidence manifold}\index{coincidence manifold}
\begin{equation}
 \Gamma\;:=\;\bigcup_{i,j}\Gamma_{ij}\,,\qquad \Gamma_{ij}\;:=\;\{(\xx_1,\xx_2,\xx_3)\,|\,\xx_i=\xx_j\}\,.
\end{equation}
The motivation is that any such extension encodes by construction a singular interaction only supported at the points of $\Gamma$. (Such a scheme lied on the very same footing as the analogous rigorous construction of \emph{two-body} zero-range interaction Hamiltonians, initially proposed in 1960 by Berezin and Faddeev \cite{Berezin-Faddeev-1961}.) In order for the analysis to produce physically meaningful results, the actual extensions of \eqref{eq:freeHamiltRestricted} to be considered are only those defined on domains of self-adjointness consisting of wave-functions that display the Bethe-Peierls short-range asymptotics\index{contact condition!Bethe-Peierls} \eqref{eq:preBP}.

All this has been then specialised among various lines, including:
\begin{itemize}
 \item a more operator-theoretic line in the Faddeev-Minlos spirit, developed from the mid 1980's to the recent years by Minlos (also in collaboration with Mel$'$nikov, Mogilner, and Shermatov) \cite{Minlos-1987,Minlos-Shermatov-1989,mogilner-shermatov-PLA-1990,Menlikov-Minlos-1991,Menlikov-Minlos-1991-bis,Minlos-TS-1994,Shermatov-2003,Minlos-2011-preprint_May_2010,Minlos-2010-bis,Minlos-2012-preprint_30sett2011,Minlos-2014-I_RusMathSurv,Minlos-2014-II_preprint-2012}, with also recent contributions by Yoshitomi \cite{Yoshitomi_MathSlov2017}, by Michelangeli and Ottolini \cite{MO-2016,MO-2017}, and by Becker, Michelangeli, and Ottolini \cite{BMO-2017};
 \item a line exploiting quadratic forms methods, initiated at the end of the 1980's by Dell'Antonio, Figari, and Teta and mainly developed in the following decades by an Italian community \cite{Teta-1989,dft-Nparticles-delta,DFT-proc1995,Finco-Teta-2012,CDFMT-2012,michelangeli-schmidbauer-2013,Correggi-Finco-Teta-2015_N+1,CDFMT-2015,Basti-Teta-2015,MP-2015-2p2,Basti-Figari-Teta-Rendiconti2018}, 
 with also recent contributions by Moser and Seiringer \cite{Moser-Seiringer-2017,Moser-Seiringer-2018-2p2};
 \item a side line by Pavlov and his school \cite{Kuperin-Makarov-Merk-Motovilov-Pavlov-1989-JMP1990,Makarov-Melezhik-Motovilov-1995}, retaining the same ideas, but aimed at rigorously constructing variants of the formal Hamiltonian \eqref{eq:fistFormalHamilt} for particles with spin, and spin-spin contact interactions;
 \item a further line where (three-dimensional) three-body Hamiltonians with zero-range interactions are constructed as rigorous limits, in the resolvent sense, of ordinary Schr\"{o}dinger operators with potentials that scale up to a delta-like profile -- an idea discussed first by Albeverio, H\o{}egh-Krohn, and Wu \cite{Albe-HK-Wu-1981} in the early 1980's, and subsequently by Dimock and Rajeev \cite{Dimock-Rajeev-2004} in the planar analogue (more recently, one- and two-dimensional counterpart results have been established in \cite{BastiEtAl2018-1d-resLim,Griesemer-Hofacker-Linden-2019,Griesemer-Michael-2021}); closely related to this is a ground state Dirichlet form approach developed by Albeverio, H\o{}egh-Krohn, and Streit \cite{Albeverio-HK-Streit-1977-enforms}.
\end{itemize}

For what exposed so far, it is clear that the physical and the mathematical branches of the literature on the quantum three-body problem on point interaction, albeit very deeply cross-intersecting, are not immediately transparent to each other. The rigorous definition of the self-adjoint Hamiltonian is much more laborious than the formal diagonalisation made by physicists, and unavoidably requires the analysis of technical features of the Hamiltonian other than the observable energy levels. Besides, the Hamiltonians of interest not having the form of a Schr\"{o}dinger operator, the mathematical analysis faces the lack of various powerful tools from Schr\"{o}dinger operator theory.

Furthermore, the implementation of the Bethe-Peierls asymptotics\index{contact condition!Bethe-Peierls} \eqref{eq:preBP}, a crucial step of the mathematical modelling, yields various technical difficulties.

First, \eqref{eq:preBP} is a \emph{point-wise} asymptotics and need be understood as an expansion in a precise \emph{functional} sense in order to be meaningfully implemented in the operator-theoretic construction of the Hamiltonian.

Next, there is an arbitrariness in the modelling as to prescribing the Bethe-Peierls condition for \emph{all} the functions of the desired domain of self-adjointness, or possibly just for a meaningful \emph{subspace}, e.g., the eigenfunctions only.

In addition, once a realisation of the minimal operator \eqref{eq:freeHamiltRestricted} is found that fulfills the Bethe-Peierls condition,\index{contact condition!Bethe-Peierls} a possibility that one encounters is that this is only a symmetric operator with a multiplicity of self-adjoint extensions, so that another parameter must be introduced to label each extension beside the given scattering length $a$ entering \eqref{eq:preBP}, in complete analogy to the three-body parameter of the physicists.

Another possibility is that after implementing the Bethe-Peierls asymptotics, the resulting candidate Hamiltonian, be it already self-adjoint or not, is unbounded from below (beside being obviously unbounded from above, as is the initial operator \eqref{eq:freeHamiltRestricted}). That multi-particle quantum models of zero-range interaction may be as such is known since when Thomas in 1935 \cite{Thomas1935}, modelling the tritium as if the range of the interaction was exactly zero, showed that the scattering of the proton over the two neutrons would result in an infinity of bound states accumulating at minus infinity (the \emph{Thomas collapse}\index{Thomas effect (collapse onto the centre)}, in the sense of `fall of the particles to the centre'), and this is well familiar in modern cold atom theoretical physics. Yet, this complicates the mathematical treatment, for instance making the quadratic form approach unsuited.

Related to that, one is then also concerned with producing meaningful regularisations of those models obtained along the conceptual path described above, where the spectral instability is removed and yet certain relevant features of the effective Hamiltonian are retained.

The present chapter presents a comprehensive and up-to-date discussion of all such instances, and in particular a systematic discussion of the technical procedures for the rigorous construction of self-adjoint Hamiltonians of physical relevance. This also allows one to clarify certain steps of the operator-theoretic construction that are notoriously subtle for the correct identification of a domain of self-adjointness. The main reference for this material is the recent work \cite{M2020-BosonicTrimerZeroRange}.

The main results in this context, Theorems \ref{thm:generalclassification}, \ref{thm:globalTMSext}, \ref{thm:H0beta}, \ref{thm:spectralanalysis}, and \ref{thm:regularised-models}, present respectively:
\begin{itemize}
 \item the general classification of all self-adjoint realisations of the minimal operator \eqref{eq:freeHamiltRestricted} (a vast class that of course includes also physically non-relevant operators, i.e., realisations characterised by non-local boundary conditions), 
 \item the characterisation of all those extensions displaying the physical short-scale structure for the functions of their domains,
 \item the rigorous construction of a class of canonical models with the physical short-scale structure, and their spectral analysis,
 \item the counterpart for a class of regularised models where the instability is cured at an effective level.
\end{itemize}

The bosonic trimer with zero-range interaction has a natural parameter to be declared in the first place, the \emph{scattering length} \index{scattering length} $a$ of the two-body interaction. It is the above-mentioned parameter governing the short-scale asymptotics \eqref{eq:preBP}. Whereas throughout our general discussion on physically relevant extensions $a$ will be kept generic, for a sharper presentation the final construction of the canonical models is done in the regime $a=\infty$. In physics this is referred to as the \emph{unitary regime}\index{unitary regime}, and many-body systems with two-body interaction of infinite scattering length are customarily called \emph{unitary gases}\index{unitary gases} \cite{Castin-Werner-2011_-_review} (for the connection with the optical theorem in which the choice $a=\infty$ maximises the scattering amplitude, and the fact that in turn the optical theorem is a consequence of the unitarity of the quantum evolution). The unitary regime is surely the physically most relevant one, for its applications in cold atom physics and for an amount of universality properties that it displays at the spectral level: this regime will be therefore selected for a large part of the present analysis.

On a more technical level, a synergy of the self-adjoint extension theories discussed in Chapter \ref{chaper-extension-schemes} is needed along the discussion. As the minimal operator \eqref{eq:freeHamiltRestricted} is non-negative, it is natural to apply to it the Kre{\u\i}n-Vi\v{s}ik-Birman scheme for semi-bounded symmetric operators (Sect.~\ref{sec:II-VBreparametrised}). This produces a vast multiplicity of self-adjoint extensions, the largest part of which are characterised by boundary conditions at the coincidence configuration set $\Gamma$ that are unphysical for being non-local or for not satisfying the short-scale Bethe-Peierls two-body asymptotics. However, latter condition turns out to be naturally implementable on the general expression of \emph{symmetric} extensions of the minimal operator \eqref{eq:freeHamiltRestricted}, but this procedure selects operators that may fail to be self-adjoint and, most importantly, are not lower semi-bounded. To them only the extension scheme a la von Neumann (Sect.~\ref{sec:II-vN-theory}) is then applicable.

Once mathematically well-posed (i.e., self-adjoint) and physically meaningful Hamiltonians are constructed, it is fairly manageable to express their quadratic forms, as is done in the sequel. Of course, as in several previous works, one can revert the order and study first a given quadratic form, typically selected by a physically grounded educated guess, proving that it actually represent a self-adjoint operator. What escapes such approach is the systematic classification of all extensions of interest: the standard classification theorems, indeed, are essentially formulated as operator classifications.

To conclude, there are surely various interesting directions along which it would be desirable to continue this study. To mention some of the most attractive ones, a more explicit theoretic dictionary between this mathematical approach and the physical zero-range methods,\index{zero-range methods} the characterisation of the quantum dynamics under the considered Hamiltonians, and an extension of such models to many-body systems with zero-range interaction.

\section{Admissible Hamiltonians}\label{sec:generalextscheme}

\subsection{The minimal operator}

In order to discuss realisations of the formal Hamiltonian \eqref{eq:fistFormalHamilt} as self-adjoint extensions of \eqref{eq:freeHamiltRestricted}, one factors out the translation invariance by introducing the centre of mass and the internal coordinates
\begin{equation}
 \yy_{\mathrm{c.m.}}\;:=\;\frac{\xx_1+\xx_2+\xx_3}{3}\,,\qquad \yy_1\;:=\;\xx_1-\xx_3\,,\qquad \yy_2\;:=\;\xx_2-\xx_3\,,
\end{equation}
and by re-writing
\begin{equation}
  -\frac{1}{2m}\Delta_{\xx_1}-\frac{1}{2m}\Delta_{\xx_2}-\frac{1}{2m}\Delta_{\xx_3}\;=\;-\frac{1}{6m}\Delta_{\yy_{\mathrm{c.m.}}}+\frac{1}{m}\mathring{H}\,,
\end{equation}
where
\begin{equation}\label{eq:Hring-initial}
 \mathring{H}\;:=\;-\Delta_{\yy_1}-\Delta_{\yy_2}-\nabla_{\yy_1}\cdot\nabla_{\yy_2}\,.
\end{equation}

In absolute coordinates, as opposed to the internal ones, three-body wave-functions $\Psi(\xx_1,\xx_2,\xx_3)$ display bosonic symmetry, that is, they are invariant under exchange of any pair of variables, hence under the corresponding transformation of the internal coordinates, according to the scheme
\begin{equation}\label{eq:bosonic_transformations}
 \begin{array}{lcl}
  \begin{cases}
   \xx_1\leftrightarrow\xx_2 \\
   \xx_3\textrm{ fixed}
  \end{cases} & \Leftrightarrow\;\;\; &
  \begin{cases}
   \yy_1\leftrightarrow\yy_2 \\
   \yy_2-\yy_1\leftrightarrow-(\yy_2-\yy_1)
  \end{cases} \\ \\
   \begin{cases}
   \xx_1\leftrightarrow\xx_3 \\
   \xx_2\textrm{ fixed}
  \end{cases} & \Leftrightarrow\;\;\; &
  \begin{cases}
   \yy_1\leftrightarrow -\yy_1 \\
   \yy_2\leftrightarrow \yy_2-\yy_1 
  \end{cases} \\ \\
   \begin{cases}
   \xx_2\leftrightarrow\xx_3 \\
   \xx_1\textrm{ fixed}
  \end{cases} & \Leftrightarrow\;\;\; &
  \begin{cases}
   \yy_1\leftrightarrow -(\yy_2-\yy_1) \\
   \yy_2\leftrightarrow -\yy_2\,. 
  \end{cases}
 \end{array}
\end{equation}
Transformations \eqref{eq:bosonic_transformations} clearly preserve the centre-of-mass variable.
Therefore, bosonic symmetry selects, within the Hilbert space of the internal coordinates
\begin{equation}\label{eq:Hinternal}
 \cH\;:=\;L^2(\mathbb{R}^3\times\mathbb{R}^3,\ud\yy_1\ud\yy_2)\,,
\end{equation}
the \emph{bosonic sector}\index{bosonic sector}, namely the Hilbert subspace
\begin{equation}\label{eq:Hbosonic}
 \begin{split}
 \cH_\mathrm{b}\;\equiv\;\;& L^2_\mathrm{b}(\mathbb{R}^3\times\mathbb{R}^3,\ud\yy_1\ud\yy_2) \\
  :=\;&
 \left\{
 \begin{array}{c}
  \psi\in L^2(\mathbb{R}^3\times\mathbb{R}^3,\ud\yy_1\ud\yy_2)\,\textrm{ such that} \\
  \psi(\yy_1,\yy_2)=\psi(\yy_2,\yy_1)=\psi(-\yy_1,\yy_2-\yy_1)
 \end{array}
 \right\}
 \end{split}
\end{equation}
in the sense of almost-everywhere identities between square-integrable functions.

The meaningful problem is then to characterise the self-adjoint extensions, with respect to $\cH_\mathrm{b}$ of the densely defined, closed, and symmetric operator
\begin{equation}\label{eq:domHring-initial}
\begin{split} 
 \mathcal{D}(\mathring{H})\;&:=\;\cH_\mathrm{b}\:\cap\: H^2_0\big( (\mathbb{R}^3_{\yy_1}\times\mathbb{R}^3_{\yy_2})\setminus\Gamma\big) \\
 \mathring{H}\;&:=\;-\Delta_{\yy_1}-\Delta_{\yy_2}-\nabla_{\yy_1}\cdot\nabla_{\yy_2}\,,
\end{split}
\end{equation}
where $\Gamma$ is the \emph{coincidence manifold}\index{coincidence manifold}
\begin{equation}
\Gamma\;:=\;\bigcup_{j=1}^3\Gamma_j\qquad \textrm{with}\qquad
\begin{cases}\label{eq:hyperplanes}
 \Gamma_1\;:=\;\{\yy_2=\mathbf{0}\} \, , \\
 \Gamma_2\;:=\;\{\yy_1=\mathbf{0}\} \, ,\\
 \Gamma_3\;:=\;\{\yy_1=\yy_2\} \, , 
\end{cases}
\end{equation}
in the sense of hyperplanes in $\mathbb{R}^3\times\mathbb{R}^3$,
and 
\begin{equation}
 H^2_0\big( (\mathbb{R}^3_{\yy_1}\times\mathbb{R}^3_{\yy_2})\setminus\Gamma\big)\;:=\;\overline{C^\infty_c\big( (\mathbb{R}^3_{\yy_1}\times\mathbb{R}^3_{\yy_2})\setminus\Gamma\big)}^{\|\,\|_{H^2}}\,.
\end{equation}
In the notation \eqref{eq:hyperplanes}, the hyperplane $\Gamma_j$ is the set of configurations where the two particles different from the $j$-th one coincide.

In short, $\mathring{H}$ is the operator closure of $-\Delta_{\yy_1}-\Delta_{\yy_2}-\nabla_{\yy_1}\cdot\nabla_{\yy_2}$ initially defined on the bosonic smooth functions on $\mathbb{R}^3\times\mathbb{R}^3$ which are compactly supported away from the coincidence manifold $\Gamma$.

As the reasonings that will follow are somewhat more informative in the momentum representation, one shall often switch to the  variables $\pp_1,\pp_2$ that are Fourier conjugate to $\yy_1,\yy_2$ (yet, all the present considerations can be straightforwardly re-phrased in position coordinates). One deduces from \eqref{eq:Hbosonic} that, for any $\psi\in\cH$,
\begin{equation}\label{eq:bosonicmomentum}
 \psi\in\cH_{\mathrm{b}}\qquad\Leftrightarrow\qquad\widehat{\psi}(\pp_1,\pp_2)\;=\;\widehat{\psi}(\pp_2,\pp_1)\;=\;\widehat{\psi}(\pp_1,-\pp_1-\pp_2)
\end{equation}
for almost every $\pp_1,\pp_2$.
Moreover, 
\begin{equation}\label{eq:psiy0}
 \psi|_{\Gamma_1}(\yy_1)\;=\;\psi(\yy_1,\mathbf{0})\;=\;\frac{1}{\;(2\pi)^{3}}\iint_{\mathbb{R}^3\times\mathbb{R}^3}\ud\pp_1\ud\pp_2\,e^{\ii \yy_1 \cdot\pp_1}\widehat{\psi}(\pp_1,\pp_2)\,,
\end{equation}
from which, using the identity
\[
 \frac{1}{\;(2\pi)^{3}}\int_{\mathbb{R}^3}\ud\yy_1\,e^{\ii\yy_1\cdot(\qq_1-\pp_1)}=\delta(\qq_1-\pp_1)
\]
(where henceforth $\delta$ denotes the usual three-dimensional $\delta$-distribution),
one deduces
\begin{equation}\label{eq:tracemomentum}
 \widehat{\psi|_{\Gamma_1}}(\pp_1)\;=\;\frac{1}{\;(2\pi)^{\frac{3}{2}}}\int_{\mathbb{R}^3}\widehat{\psi}(\pp_1,\pp_2)\,\ud\pp_2\,,
\end{equation}
and analogous expressions for $\widehat{\psi|_{\Gamma_2}}$ and $\widehat{\psi|_{\Gamma_3}}$.
This includes also the possibility that the evaluation of $\psi$ at a coincidence hyperplane makes \eqref{eq:psiy0}-\eqref{eq:tracemomentum} infinite for (almost) every value of the remaining variable.

By a standard trace theorem (see, e.g.,  \cite[Lemma 16.1]{Trtar-SobSpaces_interp}), if $\psi\in H^2(\mathbb{R}^3\times\mathbb{R}^3)$, then its evaluation $\psi|_{\Gamma_j}$ at the $j$-th coincidence hyperplane is a function in $H^{\frac{1}{2}}(\mathbb{R}^3)$, hence not necessarily continuous. Thus, when $f\in\mathcal{D}(\mathring{H})$ the vanishing ``$f|_{\Gamma_j}=0$'' in $H^{\frac{1}{2}}(\mathbb{R}^3)$ is to be understood by duality as 
\begin{equation}\label{eq:vanishingduality}
 0\;=\;\langle \eta, f|_{\Gamma_j}\rangle_{H^{-\frac{1}{2}},H^{\frac{1}{2}}}\;=\;\int_{\mathbb{R}^3}\overline{\widehat{\eta}(\pp)}\;\widehat{f|_{\Gamma_j}}(\pp)\,\ud\pp\qquad\forall\eta\in H^{-\frac{1}{2}}(\mathbb{R}^3)\,.
\end{equation}
This means that for any $f\in\mathcal{D}(\mathring{H})$ the vanishing at $\Gamma_1$, $\Gamma_2$, or $\Gamma_3$ corresponds, respectively, to
\begin{equation}\label{eq:triplevanishing}
 \begin{split}
  \iint_{\mathbb{R}^3\times\mathbb{R}^3}\widehat{f}(\pp_1,\pp_2)\,\widehat{\eta}(\pp_1)\,\ud\pp_1\ud\pp_2\;&=\;0\,,
 \\
 \iint_{\mathbb{R}^3\times\mathbb{R}^3}\widehat{f}(\pp_1,\pp_2)\,\widehat{\eta}(\pp_2)\,\ud\pp_1\ud\pp_2\;&=\;0\,, \\
 \iint_{\mathbb{R}^3\times\mathbb{R}^3}\widehat{f}(\pp_1,\pp_2)\,\widehat{\eta}(-\pp_1-\pp_2)\,\ud\pp_1\ud\pp_2\;&=\;0\,,
 \end{split}
\end{equation}
for each $\eta\in H^{-\frac{1}{2}}(\mathbb{R}^3)$,
as one may conclude combining \eqref{eq:bosonicmomentum}, \eqref{eq:tracemomentum} (and its counterparts by symmetry), and \eqref{eq:vanishingduality}.

The following is therefore proved.


\begin{lemma}\label{lem:Hminimal} The definition \eqref{eq:domHring-initial} is equivalent to
\begin{equation}\label{eq:Hringshort}
 \begin{split}
   \mathcal{D}(\mathring{H})\;&=\;\left\{ 
   \begin{array}{c}
   f\in\cH_{\mathrm{b}}\cap H^2(\mathbb{R}^3\times\mathbb{R}^3) \\
   \textrm{$f$ satisfies \eqref{eq:triplevanishing} }\forall\eta\in H^{-\frac{1}{2}}(\mathbb{R}^3)
   \end{array}
   \right\} \\
  \widehat{\mathring{H} f}(\pp_1,\pp_2)\;&=\;(\pp_1^2+\pp_2^2+\pp_1\cdot\pp_2)\widehat{f}(\pp_1,\pp_2)\,.
 \end{split}
\end{equation}
\end{lemma}

A convenient shorthand shall be
\begin{equation}
 H_\mathrm{b}^s(\mathbb{R}^3\times\mathbb{R}^3)\;:=\;\cH_{\mathrm{b}}\cap H^s(\mathbb{R}^3\times\mathbb{R}^3)\,,\qquad s\geqslant 0\,,
\end{equation}
for Sobolev spaces with bosonic symmetry.

 It is also convenient to observe that for any $\lambda>0$
\begin{equation}\label{eq:lambda-equiv-1}
 \pp_1^2+\pp_2^2+\pp_1\cdot\pp_2+\lambda\;\sim\;\pp_1^2+\pp_2^2+1\,,
\end{equation}
in the sense that each quantity controls the other from above and from below.

\subsection{Friedrichs extension}

It is clear that $\mathring{H}$ is lower semi-bounded, with lower bound $\mathfrak{m}(\mathring{H})=0$. As such, it has a distinguished extension, the Friedrichs extension $\mathring{H}_{\mathrm{F}}$ (Theorem \ref{thm:Friedrichs-ext}).

\begin{lemma}\label{lem:Friedrichs}
 The Friedrichs extension $\mathring{H}_{\mathrm{F}}$ of $\mathring{H}$ is the self-adjoint operator acting as
 \begin{equation}\label{eq:HFspace}
  (\mathring{H}_{\mathrm{F}} \phi)(\yy_1,\yy_2) \;=\;-\Delta_{\yy_1}\phi(\yy_1,\yy_2)-\Delta_{\yy_2}\phi(\yy_1,\yy_2)-\nabla_{\yy_1}\cdot\nabla_{\yy_2}\phi(\yy_1,\yy_2)\,,
 \end{equation}
 or equivalently
 \begin{equation}\label{eq:HFmomentum}
   (\widehat{\mathring{H}_{\mathrm{F}} \phi})(\pp_1,\pp_2)\;=\;(\pp_1^2+\pp_2^2+\pp_1\cdot\pp_2)\widehat{\phi}(\pp_1,\pp_2)\,,
 \end{equation}
 defined on the domain
 \begin{equation}\label{eq:HFdomain}
  \mathcal{D}(\mathring{H}_{\mathrm{F}})\;=\;H_\mathrm{b}^2(\mathbb{R}^3\times\mathbb{R}^3)\,.
 \end{equation}
 Its quadratic form is
 \begin{equation}\label{eq:HFform}
 \begin{split}
  \mathcal{D}[\mathring{H}_{\mathrm{F}}]\;&=\;H_\mathrm{b}^1(\mathbb{R}^3\times\mathbb{R}^3) \\
  \mathring{H}_{\mathrm{F}}[\phi]\;&=\;\frac{1}{2}\iint_{\mathbb{R}^3\times\mathbb{R}^3}\Big(\big|(\nabla_{\yy_1}+\nabla_{\yy_2})\phi\big|^2+\big|\nabla_{\yy_1}\phi\big|^2+\big|\nabla_{\yy_2}\phi\big|^2\Big)\,\ud\yy_1\,\ud\yy_2\,.
 \end{split}
 \end{equation}
\end{lemma}

\begin{proof}
The \emph{form} domain $\mathcal{D}[\mathring{H}]$ of $\mathring{H}$ is the completion of $\mathcal{D}(\mathring{H})=\cH_\mathrm{b}\cap H^2_0( (\mathbb{R}^3_{\yy_1}\times\mathbb{R}^3_{\yy_2})\setminus\Gamma)$ in the norm $\|f\|_{\mathrm{F}}:=(\langle f,\mathring{H}f\rangle+\|f\|_{\cH}^2)^{\frac{1}{2}}$, and $\|f\|_{\mathrm{F}}\approx\|f\|_{H^1}$ owing to \eqref{eq:lambda-equiv-1}. Reasoning as done in collaboration with Ottolini in \cite[Lemma 3(ii)]{MO-2016}, the above-mentioned completion is precisely $H^1_{\mathrm{b}}(\mathbb{R}^3\times\mathbb{R}^3)$. Thus, $\mathcal{D}[\mathring{H}]=H^1_{\mathrm{b}}(\mathbb{R}^3\times\mathbb{R}^3)$. Since \eqref{eq:HFspace}-\eqref{eq:HFdomain} obviously defines a self-adjoint extension of $\mathring{H}$ with domain entirely contained in $\mathcal{D}[\mathring{H}]$, necessarily such operator is the Friedrichs extension of $\mathring{H}$. The explicit formula for the evaluation of the quadratic form follows from the identity $\pp_1^2+\pp_2^2+\pp_1\cdot\pp_2=\frac{1}{2}(\pp_1+\pp_2)^2+\frac{1}{2}\pp_1^2+\frac{1}{2}\pp_2^2$.
\end{proof}

As $\mathring{H}$ is bounded from below, its self-adjoint realisations may be identified by means of the Kre{\u\i}n-Vi\v{s}ik-Birman extension scheme for semi-bounded symmetric operators (see Sect.~\ref{sec:II-VBreparametrised}). 

In such scheme each extension is conveniently parametrised with respect to a reference extension that has everywhere-defined bounded inverse. Now, the Friedrichs extension $\mathring{H}_{\mathrm{F}}$ has zero at the bottom of its spectrum and hence is not everywhere invertible in $\cH_\mathrm{b}$. One then searches for self-adjoint realisations of the shifted operator $\mathring{H}+\lambda\mathbbm{1}$, for some $\lambda>0$, since obviously
\begin{equation}
 \mathcal{D}(\mathring{H}+\lambda\mathbbm{1})\;=\;\mathcal{D}(\mathring{H})\,,\qquad (\mathring{H}+\lambda\mathbbm{1})_{\mathrm{F}}=\mathring{H}_{\mathrm{F}}+\lambda\mathbbm{1}\qquad \forall\lambda>0\,,
\end{equation}
and the latter operator is indeed everywhere invertible in $\cH_\mathrm{b}$. Once the self-adjoint extensions of $\mathring{H}+\lambda\mathbbm{1}$ are identified, the corresponding ones for $\mathring{H}$ are then read out from the former by removing the shift.

The data needed for the classification of the self-adjoint extensions of  $\mathring{H}+\lambda\mathbbm{1}$, according to the Kre{\u\i}n-Vi\v{s}ik-Birman theory, are the deficiency subspace $\ker(\mathring{H}^*+\lambda\mathbbm{1})$ and the action of $(\mathring{H}_{\mathrm{F}}+\lambda\mathbbm{1})^{-1}$ on such space. These data will be identified in the following.

First of all, obviously,
\begin{equation}\label{eq:HFinverse}
 ((\mathring{H}_{\mathrm{F}}+\lambda\mathbbm{1})^{-1}\psi)\,{\textrm{\large $\widehat{\,}$\normalsize}}\,(\pp_1,\pp_2)\;=\;(\pp_1^2+\pp_2^2+\pp_1\cdot\pp_2+\lambda)^{-1}\widehat{\psi}(\pp_1,\pp_2)
\end{equation}
for every $\psi\in\cH_\mathrm{b}$ and $\lambda>0$. Next, the adjoint will be characterised.

\subsection{Adjoint}

For given $\xi\in H^{-\frac{1}{2}}(\mathbb{R}^3)$ and $\lambda>0$ let $u_\xi^\lambda$ be the function defined by
\begin{equation}\label{eq:uxi}
 \widehat{u_\xi^\lambda}(\pp_1,\pp_2)\;:=\;\frac{\widehat{\xi}(\pp_1)+\widehat{\xi}(\pp_2)+\widehat{\xi}(-\pp_1-\pp_2)}{\pp_1^2+\pp_2^2+\pp_1\cdot\pp_2+\lambda}\,.
\end{equation}

 Recall from \eqref{eq:Hinternal}-\eqref{eq:Hbosonic} that $\cH=L^2(\mathbb{R}^3\times\mathbb{R}^3,\ud\yy_1\ud\yy_2)$ and $\cH_{\mathrm{b}}=L^2_{\mathrm{b}}(\mathbb{R}^3\times\mathbb{R}^3,\ud\yy_1\ud\yy_2)$.

\begin{lemma}\label{lem:uxiproperties}~

\begin{enumerate}[(i)]
 \item For every $\lambda>0$ there exists a constant $c_\lambda>0$ such that for every $\xi\in H^{-\frac{1}{2}}(\mathbb{R}^3)$ one has
 \begin{equation}\label{eq:equivalencenorms}
  c_\lambda^{-1}\|\xi\|_{H^{-\frac{1}{2}}(\mathbb{R}^3)}\;\leqslant\;\|u_\xi^\lambda\|_{\cH}\;\leqslant c_\lambda\|\xi\|_{H^{-\frac{1}{2}}(\mathbb{R}^3)}\,.
 \end{equation}
  \item For every $\xi\in H^{-\frac{1}{2}}(\mathbb{R}^3)$, $u_\xi^\lambda\in\cH_{\mathrm{b}}$.
 \item If $u_\xi^\lambda=u_\eta^\lambda$ for some $\xi,\eta\in H^{-\frac{1}{2}}(\mathbb{R}^3)$ and $\lambda>0$, then $\xi=\eta$.
 \item For $\xi\in H^{-\frac{1}{2}}(\mathbb{R}^3)$ and $\lambda,\mu>0$ one has $u_\xi^\lambda-u^\mu_\xi\in H^2_\mathrm{b}(\mathbb{R}^3\times\mathbb{R}^3)$.
\end{enumerate} 
\end{lemma}

\begin{proof}
 Part (i) can be proved by easily mimicking the very same argument of \cite[Lemma B.2]{CDFMT-2015}. Part (ii) follows from (i) and from the invariance of \eqref{eq:uxi} under the transformations \eqref{eq:bosonic_transformations}. Part (iii) follows from (i), owing to the linearity $\xi\mapsto u_\xi$. Part (iv) follows from the identity
 \[
  \widehat{u_\xi^\lambda}(\pp_1,\pp_2)-\widehat{u_\xi^\mu}(\pp_1,\pp_2)\;=\;\frac{(\mu-\lambda)\big(\widehat{\xi}(\pp_1)+\widehat{\xi}(\pp_2)+\widehat{\xi}(-\pp_1-\pp_2)\big)}{\,(\pp_1^2+\pp_2^2+\pp_1\cdot\pp_2+\lambda)\,(\pp_1^2+\pp_2^2+\pp_1\cdot\pp_2+\mu)\,}
 \]
 and from \eqref{eq:lambda-equiv-1} and \eqref{eq:equivalencenorms}.
\end{proof}

\begin{lemma}\label{lem:Hstaretc} Let $\lambda>0$.
\begin{enumerate}[(i)]
  \item One has
 \begin{equation}\label{eq:kerHstarlambda}
  \ker(\mathring{H}^*+\lambda\mathbbm{1})\;=\;\big\{u_\xi^\lambda\,|\,\xi\in H^{-\frac{1}{2}}(\mathbb{R}^3)\big\}\,. 
 \end{equation}
 \item One has
 \begin{equation}\label{eq:DHFdecomposed}
 \begin{split}
 \mathcal{D}(\mathring{H}_{\mathrm{F}})\;&=\; H_\mathrm{b}^2(\mathbb{R}^3\times\mathbb{R}^3) \\
  &=\;\left\{
   \phi\in \cH_{\mathrm{b}}\,\left|
   \begin{array}{c}
    \displaystyle\widehat{\phi}\,=\,\widehat{f^\lambda}+\frac{\widehat{u_\eta^\lambda}}{\pp_1^2+\pp_2^2+\pp_1\cdot\pp_2+\lambda} \\
    \textrm{for } f^\lambda\in\mathcal{D}(\mathring{H})\,,\;\eta\in H^{-\frac{1}{2}}(\mathbb{R}^3)
   \end{array}
   \right.\right\}.
   \end{split}
 \end{equation}
 \item One has
 \begin{equation}\label{eq:DHstardecomposed}
 \begin{split}
   \mathcal{D}(\mathring{H}^*)\;&=\left\{
   g\in \cH_{\mathrm{b}}\,\left|
   \begin{array}{c}
    \displaystyle\widehat{g}\,=\,\widehat{\phi^\lambda}+\widehat{u_\xi^\lambda} \\
    \textrm{for } \phi^\lambda\in H_\mathrm{b}^2(\mathbb{R}^3\times\mathbb{R}^3) \,,\;\xi\in H^{-\frac{1}{2}}(\mathbb{R}^3)
   \end{array}
   \!\!\right.\right\} \\
   &=\left\{
   g\in \cH_{\mathrm{b}}\,\left|
   \begin{array}{c}
    \displaystyle\widehat{g}\,=\,\widehat{f^\lambda}+\frac{\widehat{u_\eta^\lambda}}{\pp_1^2+\pp_2^2+\pp_1\cdot\pp_2+\lambda}+\widehat{u_\xi^\lambda} \\
    \textrm{for } f^\lambda\in\mathcal{D}(\mathring{H})\,,\;\xi,\eta\in H^{-\frac{1}{2}}(\mathbb{R}^3)
   \end{array}
   \right.\right\}
 \end{split}
 \end{equation}
 and 
 \begin{equation}
  ((\mathring{H}^*+\lambda\mathbbm{1})g)\,{\textrm{\large $\widehat{\,}$\normalsize}}\,\;=\;(\pp_1^2+\pp_2^2+\pp_1\cdot\pp_2+\lambda)\widehat{\phi^\lambda}
 \end{equation}
 with
 \begin{equation}
  \widehat{\phi^\lambda}\;:=\;\widehat{f^\lambda}+\frac{\widehat{u_\eta^\lambda}}{\pp_1^2+\pp_2^2+\pp_1\cdot\pp_2+\lambda}\,,
 \end{equation}
 or equivalently
 \begin{equation}\label{eq:Hstarg}
  \begin{split}
   \widehat{(\mathring{H}^* g)}(\pp_1,\pp_2)\;&=\;(\pp_1^2+\pp_2^2+\pp_1\cdot \pp_2) \widehat{g}(\pp_1,\pp_2) \\
   &\qquad -\big(\widehat{\xi}(\pp_1)+\widehat{\xi}(\pp_2)+\widehat{\xi}(-\pp_1-\pp_2) \big)\,.
  \end{split}
 \end{equation}
\end{enumerate}
The decompositions in \eqref{eq:DHFdecomposed} and \eqref{eq:DHstardecomposed} are unique, in the sense that for each $\phi\in\mathcal{D}(\mathring{H}_{\mathrm{F}})$ there exist unique $f^\lambda,\eta$ satisfying the decomposition in \eqref{eq:DHFdecomposed}, and for each $g\in \mathcal{D}(\mathring{H}^*)$ there exist unique $f^\lambda,\eta,\xi$ satisfying the decomposition in \eqref{eq:DHstardecomposed}.
\end{lemma}

 \begin{proof}
  Any $u\in\ker(\mathring{H}^*+\lambda\mathbbm{1})=\mathrm{ran}(\mathring{H}+\lambda\mathbbm{1})^\perp$ is characterised by
  \[
   \iint_{\mathbb{R}^3}\widehat{u}(\pp_1,\pp_2)\,(\pp_1^2+\pp_2^2+\pp_1\cdot \pp_2+\lambda)\,\widehat{f}(\pp_1,\pp_2)\ud\pp_1\ud\pp_2\;=\;0\qquad\forall f\in\mathcal{D}(\mathring{H})\,.
  \]
  Combining this with \eqref{eq:triplevanishing}, one deduces that $\widehat{u}(\pp_1,\pp_2)\,(\pp_1^2+\pp_2^2+\pp_1\cdot \pp_2+\lambda)$ must be a linear combination of $\widehat{\xi}(\pp_1)$, $\widehat{\xi}(\pp_2)$, and $\widehat{\xi}(-\pp_1-\pp_2)$ for a generic $\xi\in H^{-\frac{1}{2}}(\mathbb{R}^3)$; as $u\in\cH_\mathrm{b}$, this combination must be the sum (up to an overall multiplicative pre-factor). Thus, $\widehat{u}(\pp_1,\pp_2)\,(\pp_1^2+\pp_2^2+\pp_1\cdot \pp_2+\lambda)=\widehat{\xi}(\pp_1)+\widehat{\xi}(\pp_2)+\widehat{\xi}(-\pp_1-\pp_2)$ (having re-absorbed the common pre-factor in $\xi$), which proves part (i). Parts (ii) and (iii) then follow from part (i) and from \eqref{eq:HFinverse} as an application of the von Neumann formulas\index{von Neumann's formula}
  \[
   \begin{split}
    \mathcal{D}(\mathring{H}_{\mathrm{F}})\;&=\;\mathcal{D}(\mathring{H})\dotplus (\mathring{H}_{\mathrm{F}}+\lambda\mathbbm{1})^{-1}\ker(\mathring{H}^*+\lambda\mathbbm{1}) \\
    \mathcal{D}(\mathring{H}^*)\;&=\;\mathcal{D}(\mathring{H})\dotplus (\mathring{H}_{\mathrm{F}}+\lambda\mathbbm{1})^{-1}\ker(\mathring{H}^*+\lambda\mathbbm{1})\dotplus \ker(\mathring{H}^*+\lambda\mathbbm{1})
   \end{split}
  \]
 (namely, \eqref{eq:II-vNlike1}-\eqref{eq:II-vNlike3} from Proposition \ref{prop:II-vNlike-decomp}).  
 \end{proof}

 \begin{remark}
 \eqref{eq:kerHstarlambda}-\eqref{eq:DHstardecomposed} show that functions in $\mathcal{D}(\mathring{H}^*)$ have a `regular' $H^2$-com\-po\-nent and a `singular' $L^2$-component, with no constraint among the two. The regular part is the domain of $\mathring{H}_{\mathrm{F}}$, the singular part is the kernel of $\mathring{H}^*+\lambda\mathbbm{1}$.  Because of the possible singularity of a generic $g\in\mathcal{D}(\mathring{H}^*)$, the action on $g$ of the differential operator $-\Delta_{\yy_1}-\Delta_{\yy_2}-\nabla_{\yy_1}\cdot\nabla_{\yy_2}$ produces in general a non-$L^2$ output. More precisely, \eqref{eq:Hstarg} shows that one has to subtract from $(-\Delta_{\yy_1}-\Delta_{\yy_2}-\nabla_{\yy_1}\cdot\nabla_{\yy_2})g$ the distribution
 \begin{equation}\label{eq:livingonhyperplanes}
  (2\pi)^{\frac{3}{2}}\big(\xi(\yy_1)\delta(\yy_2)+\delta(\yy_1)\xi(\yy_2)+\delta(\yy_1-\yy_2)\xi(-\yy_2)\big)
 \end{equation}
 (that is, the inverse Fourier transform of the second summand in \eqref{eq:Hstarg}), a distribution supported at the coincidence manifold $\Gamma$, in order to obtain the $L^2$-function $\mathring{H}^* g$.
 \end{remark}

 \begin{remark}
 In position coordinates $(\yy_1,\yy_2)$, each of the two functions $u_\eta^\lambda$ and $u_\xi^\lambda$ appearing in the expression \eqref{eq:DHstardecomposed} of a generic element $g\in\mathcal{D}(\mathring{H}^*)$ is obtained by taking the \emph{convolution} of the Green function $\mathcal{G}_\lambda$ relative to $-\Delta_{\yy_1}-\Delta_{\yy_2}-\nabla_{\yy_1}\cdot\nabla_{\yy_2}+\lambda$ with a distribution of the form \eqref{eq:livingonhyperplanes} for the two considered labelling functions $\eta,\xi$. This structure, and the fact that in \eqref{eq:livingonhyperplanes} $\xi$ (and $\eta$) is interpreted as a function on the union of the coincidence hyperplanes, is formally analogous to the familiar picture in electrostatics, where $u_\xi^\lambda$ is the `potential' relative to the `charge' $\xi$. For this reason, as has been customary since long in this context \cite{dft-Nparticles-delta}, one retains the nomenclature that $\xi$ and $\eta$ are the \emph{charges} for the function $g$. In this respect, by \emph{charges}\index{charge} one means functions in $H^{-\frac{1}{2}}(\mathbb{R}^3)$.  
 \end{remark}

 For $g\in\mathcal{D}(\mathring{H}^*)$, there is a unique charge $\xi$ at each parameter $\lambda>0$ satisfying the decomposition \eqref{eq:DHstardecomposed}, but a priori $\xi$ might be $\lambda$-dependent. It will be now shown that this cannot be the case, and one can speak of \emph{the} charge $\xi$ of $g$ tout court (understanding $\xi$, as usual, as the charge of the singular part of $g$, not to be confused with the charge $\eta$ appearing in the regular part of $g$).

 \begin{lemma}\label{lem:chargexiofg}
  Let $\lambda,\lambda'>0$ and $g\in\mathcal{D}(\mathring{H}^*)$. If, according to \eqref{eq:DHstardecomposed},
  \[
   \widehat{g}\;=\;\widehat{\phi^\lambda}+\widehat{u_\xi^\lambda} \;=\;\widehat{\phi^{\lambda'}}+\widehat{u_{\xi'}^{\lambda'}}
  \]
  for some $\phi^\lambda,\phi^{\lambda'}\in H_\mathrm{b}^2(\mathbb{R}^3\times\mathbb{R}^3)$ and $\xi,\xi'\in H^{-\frac{1}{2}}(\mathbb{R}^3)$, then $\xi=\xi'$.  
 \end{lemma}

 \begin{proof}
  Since $u_{\xi}^{\lambda'}-u_\xi^\lambda\in H^2_\mathrm{b}(\mathbb{R}^3\times\mathbb{R}^3)$ (Lemma \ref{lem:uxiproperties}(iv)), then
  \[
   F^{\lambda'}\;:=\;\phi^\lambda-\big(u_{\xi}^{\lambda'}-u_{\xi}^{\lambda} \big)
  \]
 defines a function in $H^2_\mathrm{b}(\mathbb{R}^3\times\mathbb{R}^3)$, and
 \[
  \widehat{g}\;=\;\widehat{\phi^\lambda}+\widehat{u_\xi^\lambda} \;=\;\widehat{F^{\lambda'}}+\widehat{u_{\xi}^{\lambda'}}\,.
 \]
 Comparing the latter identity with $\widehat{g}=\widehat{\phi^{\lambda'}}+\widehat{u_{\xi'}^{\lambda'}}$, the uniqueness of the decomposition \eqref{eq:DHstardecomposed} of $g$ with parameter $\lambda'$ implies 
 \[
  F^{\lambda'}\;=\;\phi^{\lambda'}\,,\qquad u_{\xi}^{\lambda'}\;=\;u_{\xi'}^{\lambda'}\,.
 \]
 In turn, the latter identity implies $\xi=\xi'$ (Lemma \ref{lem:uxiproperties}(iii)).  
 \end{proof}

 \subsection{Deficiency subspace}\label{sec:deficiencysubspace}
 
 Lemma \ref{lem:Hstaretc} shows that $\mathring{H}$ has \emph{infinite} deficiency index\index{deficiency indices}, as is the dimensionality of the deficiency subspace $\ker(\mathring{H}^*+\lambda\mathbbm{1})$ independently of $\lambda>0$.

 In the self-adjoint extension problem under study, it is convenient to use a unitarily isomorphic version of  $\ker(\mathring{H}^*+\lambda\mathbbm{1})$, as will be now characterised.

 To this aim, for $\eta\in H^{-\frac{1}{2}}(\mathbb{R}^3)$ and $\lambda>0$ one defines (for a.e.~$\pp$)
 \begin{equation}\label{eq:Wlambda}
  (\widehat{W_\lambda\eta})(\pp)\;:=\;\frac{3\pi^2}{\sqrt{\frac{3}{4}\pp^2+\lambda}}\,\widehat{\eta}(\pp)+6\int_{\mathbb{R}^3}\frac{\widehat{\eta}(\qq)}{(\pp^2+\qq^2+\pp\cdot \qq+\lambda)^2}\,\ud\qq\,.
 \end{equation}

 \begin{lemma}\label{lem:Wlambdaproperties} Let $\lambda>0$.
 
 \begin{enumerate}[(i)]
  \item For generic $\xi,\eta\in H^{-\frac{1}{2}}(\mathbb{R}^3)$ one has $W_\lambda\eta\in H^{\frac{1}{2}}(\mathbb{R}^3)$ and
   \begin{equation}\label{eq:scalar_products}
 \langle u_\xi,u_\eta\rangle_{\cH}\;=\;\langle \xi,W_\lambda\eta\rangle_{H^{-\frac{1}{2}}(\mathbb{R}^3),H^{\frac{1}{2}}(\mathbb{R}^3)}\,.
 \end{equation}
  \item Formula \eqref{eq:Wlambda} defines a positive, bounded, linear bijection 
\[
	W_\lambda: H^{-\frac{1}{2}}(\mathbb{R}^3)\to H^{\frac{1}{2}}(\mathbb{R}^3) \, .
\]
 \end{enumerate}
 \end{lemma}

 \begin{proof}
  By suitably exploiting symmetry in exchanging the integration variables,
  \[
   \begin{split}
    &\langle u_\xi,u_\eta\rangle_{\cH}\;= \\
    &=\;\iint_{\mathbb{R}^3\times\mathbb{R}^3}\ud\pp_1\ud\pp_2\,\frac{\overline{\widehat{\xi}(\pp_1)}+\overline{\widehat{\xi}(\pp_2)}+\overline{\widehat{\xi}(-\pp_1-\pp_2)}}{\pp_1^2+\pp_2^2+\pp_1\cdot\pp_2+\lambda}\,\frac{\widehat{\eta}(\pp_1)+\widehat{\eta}(\pp_2)+\widehat{\eta}(-\pp_1-\pp_2)}{\pp_1^2+\pp_2^2+\pp_1\cdot\pp_2+\lambda} \\
    &=\;3\iint_{\mathbb{R}^3\times\mathbb{R}^3}\ud\pp_1\ud\pp_2\,\frac{\overline{\widehat{\xi}(\pp_1)}\,\widehat{\eta}(\pp_1)}{(\pp_1^2+\pp_2^2+\pp_1\cdot\pp_2+\lambda)^2} \\
    &\qquad\qquad+6\iint_{\mathbb{R}^3\times\mathbb{R}^3}\ud\pp_1\ud\pp_2\,\frac{\overline{\widehat{\xi}(\pp_1)}\,\widehat{\eta}(\pp_2)}{(\pp_1^2+\pp_2^2+\pp_1\cdot\pp_2+\lambda)^2}\,.
   \end{split}
  \]
  In the first summand on the r.h.s.~above one computes
  \[
   \int_{\mathbb{R}^3}\frac{\ud\pp_2}{(\pp_1^2+\pp_2^2+\pp_1\cdot\pp_2+\lambda)^2}\;=\;\frac{\pi^2}{\,\sqrt{\frac{3}{4}\pp_1^2+\lambda}\,}\,,
  \]
  which eventually yields, in view of the definition \eqref{eq:Wlambda},
  \[
   \langle u_\xi,u_\eta\rangle_{\cH}\;=\;\int_{\mathbb{R}^3}\overline{\widehat{\xi}(\pp)}\, (\widehat{W_\lambda\eta})(\pp)\,\ud\pp\,.
  \]
    From the latter identity and \eqref{eq:equivalencenorms} one deduces
  \[
\begin{split}
\|W_\lambda\eta\|_{H^{\frac{1}{2}}}\;&=\;\sup_{\|\xi\|_{H^{-\frac{1}{2}}}=1}\Big|\int_{\mathbb{R}^3}\overline{\widehat{\xi}(\pp)}\,\widehat{(W_\lambda\eta)}(\pp)\,\ud \pp\,\Big|\;=\;\sup_{\|\xi\|_{H^{-\frac{1}{2}}}=1}\big|\langle u_\xi,u_\eta\rangle_{\cH}\big| \\
&\leqslant\;\sup_{\|\xi\|_{H^{-\frac{1}{2}}}=1}\|u_\xi\|_{\cH}\|u_\eta\|_{\cH}\;\leqslant \;\textrm{const}\cdot\|\eta\|_{H^{-\frac{1}{2}}}\qquad\forall\eta\in H^{-\frac{1}{2}}(\mathbb{R}^3)\,,
\end{split}
\]
  which shows that $W_\lambda H^{-\frac{1}{2}}(\mathbb{R}^3)\subset H^{\frac{1}{2}}(\mathbb{R}^3)$ and that \eqref{eq:scalar_products} holds true. This completes the proof of part (i).

Concerning part (ii), the reasoning above has shown that the map $W_\lambda:H^{-\frac{1}{2}}(\mathbb{R}^3)\to H^{\frac{1}{2}}(\mathbb{R}^3)$ is bounded. By \eqref{eq:equivalencenorms} and \eqref{eq:scalar_products},
  \[
   \langle \eta,W_\lambda\eta\rangle_{H^{-\frac{1}{2}},H^{\frac{1}{2}}}\;=\;\|u_\eta\|^2_{\cH}\;\geqslant\;c_\lambda^{-2}\,\|\eta\|_{H^{-\frac{1}{2}}}^2\,,
  \]
  which implies coercivity
\[
 \|W_\lambda\eta\|_{H^{\frac{1}{2}}}\;\geqslant\;c_\lambda^{-2}\,\|\eta\|_{H^{-\frac{1}{2}}}\,.
\]
This demonstrates that $W_\lambda$ is a positive, injective $H^{-\frac{1}{2}}\to H^{\frac{1}{2}}$ map. $W_\lambda$ is thus invertible on $\mathrm{ran}\,W_\lambda$ and by boundedness $\mathrm{ran}\,W_\lambda$ is closed in $H^{\frac{1}{2}}(\mathbb{R}^3)$. It only remains to show that $\mathrm{ran}\,W_\lambda$ is also dense in $H^{\frac{1}{2}}(\mathbb{R}^3)$ to conclude that $W_\lambda^{-1}$ is everywhere defined and bounded. Now, testing by duality an \emph{arbitrary} $\xi\in H^{-\frac{1}{2}}(\mathbb{R}^3)$ against  $\mathrm{ran}\,W_\lambda\subset H^{\frac{1}{2}}(\mathbb{R}^3)$ one sees that
\[
 \begin{split}
  &\langle \xi,W_\lambda\eta\rangle_{H^{-\frac{1}{2}},H^{\frac{1}{2}}}\;=\;0\qquad\forall\eta\in H^{-\frac{1}{2}}(\mathbb{R}^3) \\
  &\Rightarrow\;\langle u_\xi,u_\eta\rangle_{\cH}\;=\;0\qquad\quad\:\,\forall u_\eta\in\ker (\mathring{H}^*+\lambda\mathbbm{1}) \\
  &\Rightarrow\;u_\xi\,=\,0 \\
  &\Rightarrow\;\xi\,=\,0\,,
 \end{split}
\]
and this implies that $\mathrm{ran}\,W_\lambda$ is dense in $H^{\frac{1}{2}}(\mathbb{R}^3)$. Part (ii) is proved.  
 \end{proof}

 As a direct consequence of Lemma \ref{lem:Wlambdaproperties}, the expression
\begin{equation}\label{eq:W-scalar-product}
\langle \xi,\eta\rangle_{H^{-\frac{1}{2}}_{W_\lambda}}\;:=\;\langle \xi,W_\lambda\,\eta\rangle_{H^{-\frac{1}{2}},H^{\frac{1}{2}}}\;=\;\langle u_\xi,u_\eta\rangle_{\cH}
\end{equation}
defines a scalar product in $H^{-\frac{1}{2}}(\mathbb{R}^3)$. It is \emph{equivalent} to the standard scalar product of $H^{-\frac{1}{2}}(\mathbb{R}^3)$, as follows by combining \eqref{eq:W-scalar-product} with \eqref{eq:equivalencenorms}.

One shall denote by $H^{-\frac{1}{2}}_{W_\lambda}(\mathbb{R}^3)$ the Hilbert space consisting of the $H^{-\frac{1}{2}}(\mathbb{R}^3)$-functions and equipped with the scalar product \eqref{eq:W-scalar-product}.
Then the map
\begin{equation}\label{eq:isomorphism_Ulambda}
\begin{split}
U_\lambda\,:\,\ker (\mathring{H}^*+\lambda\mathbbm{1})\;&\;\xrightarrow[]{\;\;\;\cong\;\;\;}\;H^{-\frac{1}{2}}_{W_\lambda}(\mathbb{R}^3)\,,\qquad u_\xi \longmapsto \,\xi
\end{split}
\end{equation}
is an isomorphism between Hilbert spaces, with $\ker (\mathring{H}^*+\lambda\mathbbm{1})$ equipped with  the standard scalar product inherited from $\cH$.

\subsection{Extensions classification}

Consistently with the notation of Section \ref{sec:II-VBreparametrised}, $\mathcal{S}(\mathcal{K})$ is the collection of all self-adjoint operators $A:\mathcal{D}(A)\subset\mathcal{K}'\to\mathcal{K}'$ acting in Hilbert subspaces $\mathcal{K}'$ of a given Hilbert space $\mathcal{K}$.
The Kre{\u\i}n-Vi\v{s}ik-Birman theory determines that, given $\lambda>0$ and hence the deficiency subspace $\ker (\mathring{H}^*+\lambda\mathbbm{1})$, the self-adjoint extensions of $\mathring{H}$ are in an explicit one-to-one correspondence with the elements in $\mathcal{S}(\ker (\mathring{H}^*+\lambda\mathbbm{1}))$.
Equivalently, by unitary isomorphism (Subsect.~\ref{sec:deficiencysubspace}), each self-adjoint extension of $\mathring{H}$ is labelled by an element of $\mathcal{S}(H^{-\frac{1}{2}}_{W_\lambda}(\mathbb{R}^3))$. In practice, the latter viewpoint is going to be more informative.

The extension classification takes the following form.

\begin{theorem}\label{thm:generalclassification}
 Let $\lambda>0$.
 \begin{enumerate}[(i)]
  \item The self-adjoint extensions of $\mathring{H}$ in $\cH_\mathrm{b}$ constitute the family
  \begin{equation}\label{eq:family}
   \big\{\mathring{H}_{\mathcal{A}_\lambda}\,\big|\,\mathcal{A}_\lambda\in \mathcal{S}\big(H^{-\frac{1}{2}}_{W_\lambda}(\mathbb{R}^3)\big)\big\}\,,
  \end{equation}
  where
  \begin{equation}\label{eq:domDHAshort}
   \mathcal{D}(\mathring{H}_{\mathcal{A}_\lambda})\;=\;\left\{ g\in\mathcal{D}(\mathring{H}^*)\left|\!
   \begin{array}{c}
   \eta=\mathcal{A}_\lambda\xi+\chi \\
   \xi\in\mathcal{D}(\mathcal{A}_\lambda) \\
   \chi\in \mathcal{D}(\mathcal{A}_\lambda)^{\perp_{\lambda}}\cap H^{-\frac{1}{2}}_{W_\lambda}(\mathbb{R}^3)
   \end{array}
   \!\!\right.\right\}
  \end{equation}
  or equivalently
  \begin{equation}\label{eq:domDHA}
   \mathcal{D}(\mathring{H}_{\mathcal{A}_\lambda})\;=\;\left\{
   g\in\cH_{\mathrm{b}}\left|
   \begin{array}{c}
    \widehat{g}\,=\,\widehat{\phi^\lambda}+\widehat{u_\xi^\lambda}\;\textrm{ with} \\
    \widehat{\phi^\lambda}\,=\,\widehat{f^\lambda}+\displaystyle\frac{\widehat{u_\eta^\lambda}}{\pp_1^2+\pp_2^2+\pp_1\cdot\pp_2+\lambda}\,, \\
    f^\lambda\in\mathcal{D}(\mathring{H})\,, \\
    \eta=\mathcal{A}_\lambda\xi+\chi\,,\quad \xi\in\mathcal{D}(\mathcal{A}_\lambda)\,, \\
    \chi\in \mathcal{D}(\mathcal{A}_\lambda)^{\perp_{\lambda}}\cap H^{-\frac{1}{2}}_{W_\lambda}(\mathbb{R}^3)
   \end{array}
   \!\!\right.\right\},
  \end{equation}
 and 
 \begin{equation}\label{eq:domDHA-actionDHA}
  \big((\mathring{H}_{\mathcal{A}_\lambda}+\lambda\mathbbm{1})g\big)\,{\textrm{\large $\widehat{\,}$\normalsize}}\,\;=\;(\pp_1^2+\pp_2^2+\pp_1\cdot\pp_2+\lambda)\,\widehat{\phi^\lambda}\,.
 \end{equation}
 In \eqref{eq:domDHAshort}-\eqref{eq:domDHA} above $\perp_{\lambda}$ refers to the orthogonality in the $H^{-\frac{1}{2}}_{W_\lambda}$-scalar product. The Friedrichs extension $\mathring{H}_{\mathrm{F}}$, namely the operator \eqref{eq:HFspace}-\eqref{eq:HFdomain}, corresponds to the formal choice `$\mathcal{A}_\lambda=\infty$' on  $\mathcal{D}(\mathcal{A}_\lambda)=\{0\}$.
  \item An extension $\mathring{H}_{\mathcal{A}_\lambda}$ is lower semi-bounded with
  \[
   \mathring{H}_{\mathcal{A}_\lambda}\;\geqslant\;-\Lambda\mathbbm{1}\qquad\textrm{for some $\Lambda>0$}
  \]
  if and only if, $\forall\xi\in\mathcal{D}(\mathcal{A}_\lambda)$,
  \[
  \begin{split}
   \langle\xi,\mathcal{A}_\lambda\xi&\rangle_{H^{-\frac{1}{2}}_{W_\lambda}}\;\geqslant \\
   &\geqslant\;(\lambda-\Lambda)\|\xi\|^2_{H^{-\frac{1}{2}}_{W_\lambda}} +(\lambda-\Lambda)^2\langle\xi,(\mathring{H}_{\mathrm{F}}+\Lambda\mathbf{1})^{-1}\xi\rangle_{H^{-\frac{1}{2}}_{W_\lambda}}\,.
  \end{split}
  \]
  In particular,
   \begin{equation}\label{eq:positiveSBiffpositveB-1_Tversion-CHAPTBOS}
 \begin{split}
 \mathfrak{m}(\mathring{H}_{\mathcal{A}_\lambda})\;\geqslant \;-\lambda\quad&\Leftrightarrow\quad \mathfrak{m}(\mathcal{A}_\lambda)\;\geqslant\; 0 \, , \\
 \mathfrak{m}(\mathring{H}_{\mathcal{A}_\lambda})\;> \;-\lambda\quad&\Leftrightarrow\quad \mathfrak{m}(\mathcal{A}_\lambda)\;>\; 0\,.
 \end{split}
 \end{equation}
  Moreover, if $\mathfrak{m}(\mathcal{A}_\lambda)>-\lambda$, then
 \begin{equation}\label{eq:bounds_mS_mB_Tversion-CHAPTBOS}
 \mathfrak{m}(\mathcal{A}_\lambda)\;\geqslant\; \mathfrak{m}(\mathring{H}_{\mathcal{A}_\lambda})+\lambda\;\geqslant\;\frac{\lambda \,\mathfrak{m}(\mathcal{A}_\lambda)}{\,\lambda+\mathfrak{m}(\mathcal{A}_\lambda)}\,.
  \end{equation}
 \item The quadratic form of any lower semi-bounded extension $\mathring{H}_{\mathcal{A}_\lambda}$ is given by
 \begin{equation}\label{eq:decomposition_of_form_domains_Tversion-CHAPTBOS}
 \begin{split}
 \mathcal{D}[\mathring{H}_{\mathcal{A}_\lambda}]\;&=\;\mathcal{D}[\mathring{H}_{\mathrm{F}}]\,\dotplus\,U_\lambda^{-1}\mathcal{D}[\mathcal{A}_\lambda] \, , \\
 \mathring{H}_{\mathcal{A}_\lambda}[\phi^\lambda+u_\xi^\lambda]\;&=\;\mathring{H}_{\mathrm{F}}[\phi^\lambda] +\lambda\Big(\|\phi^\lambda\big\|^2_{\cH}-\big\|\phi^\lambda+u_\xi^\lambda\big\|^2_{\cH} \Big)+\mathcal{A}_\lambda[\xi] \\
 &\forall \phi^\lambda\in\mathcal{D}[\mathring{H}_{\mathrm{F}}]=H_\mathrm{b}^1(\mathbb{R}^3\times\mathbb{R}^3)\,,\;\forall \xi\in\mathcal{D}[\mathcal{A}_\lambda]\,,
 \end{split}
\end{equation}
 and the lower semi-bounded extensions are ordered in the sense of quadratic forms according to the analogous ordering of the labelling operators, that is,
\begin{equation}\label{eq:extension_ordering_Tversion-CHAPTBOS}
\mathring{H}_{\mathcal{A}_\lambda^{(1)}}\,\geqslant\,\mathring{H}_{\mathcal{A}_\lambda^{(2)}}\qquad\Leftrightarrow\qquad \mathcal{A}_\lambda^{(1)}\,\geqslant\,\mathcal{A}_\lambda^{(2)}\,.
\end{equation}
\end{enumerate}
\end{theorem}

Recall (Sect.~\ref{sec:I-symmetric-selfadj}) that the symbol $\mathfrak{m}$ in \eqref{eq:positiveSBiffpositveB-1_Tversion-CHAPTBOS}-\eqref{eq:bounds_mS_mB_Tversion-CHAPTBOS} denotes the bottom of the spectrum of the considered operator.

Theorem \ref{thm:generalclassification} is a direct application of the general extension scheme a la Kre{\u\i}n-Vi\v{s}ik-Birman (Theorems \ref{thm:VB-representaton-theorem_Tversion}-\ref{thm:semibdd_exts_form_formulation_Tversion}) to the minimal operator $\mathring{H}+\lambda\mathbbm{1}$, given the data provided by Lemmas \ref{lem:Hminimal}, \ref{lem:Friedrichs}, and \ref{lem:Hstaretc}, and exploiting the Hilbert space isomorphism \eqref{eq:isomorphism_Ulambda} in order to re-phrase the classification formulas in terms of the unitarily isomorphic version $H^{-\frac{1}{2}}_{W_\lambda}(\mathbb{R}^3)$ of the deficiency subspace.

 It is customary to refer to each $\mathcal{A}_\lambda$ as the \emph{(Vi\v{s}ik-)Birman extension parameter} or \emph{labelling operator} for the extension $H_{\mathcal{A}_\lambda}$. (Strictly speaking, as discussed in Section \ref{sec:II-VBparametrisation-orig}, the present $\mathcal{A}_\lambda$ is a unitarily equivalent version of the extension parameter in the sense of Birman \index{Birman extension parameter} \cite{Birman-1956,KM-2015-Birman}, and the inverse of $\mathcal{A}_\lambda$ on $\mathrm{ran}\,\mathcal{A}_\lambda$ is a unitarily equivalent version of the extension parameter in the sense of Vi\v{s}ik \index{Vi\v{s}ik extension parameter} \cite{Vishik-1952}.)

\begin{remark}\label{rem:restrictions}
 The domain of each $\mathring{H}_{\mathcal{A}_\lambda}$, as indicated by \eqref{eq:domDHAshort}, is a suitable restriction of the domain of $\mathring{H}^*$ obtained by selecting only those functions $g$ whose charges $\eta_g$ and $\xi_g$ are constrained by the self-adjointness condition
\begin{equation}
  \eta_g\;=\;\mathcal{A}_\lambda\xi_g+\chi_g
\end{equation}
for some additional charge $\chi_g\in\mathcal{D}(\mathcal{A}_\lambda)^{\perp_\lambda}$, whence
\begin{equation}\label{eq:constraintetaxi}
 ( \eta_g-\mathcal{A}_\lambda\xi_g )\;\in\;\mathcal{D}(\mathcal{A}_\lambda)^{\perp_\lambda}\cap H^{-\frac{1}{2}}(\mathbb{R}^3)\,.
\end{equation}
\end{remark}

\begin{remark}\label{rem:samedomains}
 Fixed a self-adjoint extension $\mathscr{H}$ of $\mathring{H}$ and representing it as $\mathscr{H}=\mathring{H}_{\mathcal{A}_\lambda^{(\mathscr{H})}}$ for suitable labelling operators $\mathcal{A}_\lambda^{(\mathscr{H})}$ for each $\lambda>0$ according to Theorem \ref{thm:generalclassification}, one has
 \begin{equation}
  \mathcal{D}\big(\mathcal{A}_{\lambda}^{(\mathscr{H})}\big)\;=\;\mathcal{D}\big(\mathcal{A}_{\lambda'}^{(\mathscr{H})}\big)\qquad\forall\lambda,\lambda'>0\,.
 \end{equation}
 That is, the explicit action of each labelling operator changes with $\lambda$, but the domain stays fixed. This is an obvious consequence of the identity
 \begin{equation}
   \mathcal{D}\big(\mathcal{A}_{\lambda}^{(\mathscr{H})}\big)\;=\;\left\{\xi\in H^{-\frac{1}{2}}(\mathbb{R}^3)\,\left|\!
   \begin{array}{c}
    \xi\textrm{ is the singular-part charge of $g$} \\
    \textrm{ for some }g\in\mathcal{D}(\mathscr{H})
   \end{array}
   \!\!\right.\right\}
 \end{equation}
 that follows from the uniqueness of the charge $\xi$ for each $g$ (Lemma \ref{lem:chargexiofg}). 
\end{remark}

\section{Two-body short-scale singularity}\label{sec:two-body-short-scale-sing}

The domain of each self-adjoint extension $\mathring{H}_{\mathcal{A}_\lambda}$ of $\mathring{H}$ is a suitable restriction of the domain of $\mathring{H}^*$. Each restriction of self-adjointness must be a constraint of the form \eqref{eq:constraintetaxi} on the charges $\eta$ and $\xi$ (Remark \ref{rem:restrictions}). As such two charges characterise respectively the regular ($\phi^\lambda$) and the singular ($u_\xi^\lambda$) part of a generic $g\in\mathcal{D}(\mathring{H}^*)$, indirectly this constraint is a condition linking $\phi^\lambda$ and $u_\xi^\lambda$ (which otherwise would be independent). In practice this amounts to selecting those $g$'s from $\mathcal{D}(\mathring{H}^*)$ which display an admissible type of short-scale asymptotics as $|\yy_1|\to 0$, of $|\yy_2|\to 0$, or $|\yy_2-\yy_1|\to 0$, that is, when \emph{two} of the three particles of the trimer come on top of each other. This procedure will eventually identify the \emph{physically meaningful} self-adjoint extensions of $\mathring{H}$.

\subsection{Short-scale structure}

For the functions $\psi\in L^2(\mathbb{R}^3\times\mathbb{R}^3,\ud\yy_1\ud\yy_2)$ of interest, the following is a convenient and informative way to monitor the behaviour of $\psi(\yy_1,\yy_2)$ as $|\yy_2|\to 0$ at fixed $\yy_1$. 

 Writing $\yy_2\in\mathbb{R}^3$ in spherical coordinates as $\yy_2\equiv |\yy_2|\Omega_{\yy_2}$, with $\Omega_{\yy_2}\in\mathbb{S}^2$, and for $\rho>0$ and almost every $\yy_1\in\mathbb{R}^3$, one defines
\begin{equation}
 \psi_{\mathrm{av}}(\yy_1;\rho)\;:=\;\frac{1}{4\pi}\int_{\mathbb{S}^2}\psi(\yy_1,\rho\Omega)\,\ud\Omega\,.
\end{equation}
Thus, the function $\yy_1\mapsto\psi_{\mathrm{av}}(\yy_1;\rho)$ is the spherical average of the function $\yy_1\mapsto\psi(\yy_1,\yy_2)$ over the sphere with $|\yy_2|=\rho$.

For later purposes, one is concerned with certain meaningful behaviours of $\psi_{\mathrm{av}}(\yy_1;\rho)$ as $\rho\to 0$ at fixed $\yy_1$, namely when it either approaches a finite value or instead diverges as $\rho^{-1}$. With no pretension of full generality, let us adopt the following characterisation: a measurable function $\varphi:[0,+\infty)\to\mathbb{C}$ is said to display \emph{$\mathcal{Z}$-behaviour} (at zero) \index{Z@$\mathcal{Z}$-behaviour (at zero)} when $\varphi\in L^2(\mathbb{R}^+,\rho^2\ud \rho)$, $\varphi$ is continuous in a neighbourhood $(0,\varepsilon_\varphi)$ for some $\varepsilon_\varphi>0$,  and
\begin{equation}\label{eq:Z}
 \frac{\int_0^{+\infty}\!\ud\rho\,\frac{\,\sin\rho-\rho\cos\rho}{\rho}\,\varphi(\frac{\rho}{R})}{\varphi(\frac{1}{R})}\;\xrightarrow[]{\,R\to +\infty\,}\;\frac{\pi}{2}\,c_\varphi
\end{equation}
for some constant $c_\varphi\in\mathbb{C}$. In terms of the even extension $\phi(\rho):=\varphi(|\rho|)$, $\rho\in\mathbb{R}$, \eqref{eq:Z} is equivalent to
\begin{equation}\label{eq:Z2}
 \frac{\,\int_{-R}^{R}\widehat{\phi}(s)\,\ud s-2 R\,\widehat{\phi}(R)\,}{\,\int_\mathbb{R}e^{\ii s/R}\,\widehat{\phi}(s)\,\ud s\,}\;\xrightarrow[]{\,R\to +\infty\,}\;c_\varphi\,.
\end{equation}
The request that $\varphi\in L^2(\mathbb{R}^+,\rho^2\ud \rho)$ is made precisely with the function $\rho\mapsto\psi_{\mathrm{av}}(\yy_1;\rho)$ in mind, of course.

Observe that \eqref{eq:Z}-\eqref{eq:Z2} is just a convenient way to characterise the behaviour of $\varphi(\rho)$ as $\rho\to 0$. This is clear if one interprets \emph{separately} the two summands that emerge from the above expressions (but in general one does want to include the possible effect of \emph{compensation} between them). Thus, for instance, if $\phi$ is a Schwartz function with $\phi(0)=\varphi(0)\neq 0$, standard Riemann-Lebesgue and Fourier transform arguments yield
\[
 \begin{split}
  \int_0^{+\infty}\!\ud\rho\,\frac{\sin\rho}{\rho}\,\varphi({\textstyle\frac{\rho}{R}})\;&\xrightarrow[]{\,R\to +\infty\,}\;\varphi(0)\int_0^{+\infty}\!\ud\rho\,\frac{\sin\rho}{\rho}\;=\;\frac{\pi}{2}\,\varphi(0)\,, \\
  \int_0^{+\infty}\!\ud\rho\,\cos\rho\,\varphi({\textstyle\frac{\rho}{R}})\;&=\;\sqrt{\frac{\pi}{2}}\,R\,\widehat{\phi}(R)\;\xrightarrow[]{\,R\to +\infty\,}\;0\,,
 \end{split} 
\]
therefore in this case \eqref{eq:Z} is satisfied with $c_\varphi=1$. More generally, the asymptotic finiteness of the quantities
\[
 \frac{\,\int_{-R}^{R}\widehat{\phi}(s)\,\ud s\,}{\,\int_\mathbb{R}e^{\ii s/R}\,\widehat{\phi}(s)\,\ud s\,}\,,\qquad\frac{\,R\,\widehat{\phi}(R)\,}{\phi(\frac{1}{R})}
\]
as $R\to +\infty$ encodes a prescription on $\phi(\rho)$ as $\rho$ vanishes, including when $\phi$ (hence $\varphi$) is singular at $\rho=0$. In fact, \eqref{eq:Z}-\eqref{eq:Z2} encode in general a possible compensation among the above two summands. For instance, for the function $\varphi=\rho^{-1}\mathbf{1}_{(0,1)}$ one finds
\[
\begin{split}
\frac{\int_0^{+\infty}\!\ud\rho\,\frac{\,\sin\rho-\rho\cos\rho}{\rho}\,\varphi(\frac{\rho}{R})}{\varphi(\frac{1}{R})}\,&=\,\int_0^R\ud\rho\,\frac{\,\sin\rho-\rho\cos\rho}{\rho^2}\,\\
&=\,\Big[-\frac{\sin\rho}{\rho}\,\Big]_{0}^R\,\xrightarrow[]{\,R\to +\infty\,}\,1\,,
\end{split}
\]
meaning that in this case \eqref{eq:Z} is satisfied with $c_\varphi=\frac{2}{\pi}$.

Clearly, the $\mathcal{Z}$-behaviour is not the most general behaviour of $\rho\mapsto\psi_{\mathrm{av}}(\yy_1;\rho)$ when $\psi\in L^2(\mathbb{R}^3\times\mathbb{R}^3,\ud\yy_1\ud\yy_2)$ or even, for later applications, when $\psi$ belongs to the domain of self-adjoint operator of interest. It is generic enough, though, to comprise both functions $\varphi$ with sufficient regularity at $\rho=0$ and integrability over $[0,+\infty)$, and functions with enough integrability and local $\rho^{-1}$-singularity.

\begin{lemma}\label{lem:shortscalegeneric}
  Let $\psi\in\cH_\mathrm{b}$ such that for almost every $\yy_1$ the function $\rho\mapsto\psi_{\mathrm{av}}(\yy_1;\rho)$ has $\mathcal{Z}$-behaviour, for concreteness uniformly in $\yy_1$ (thus, with the same constant in the limit \eqref{eq:Z}).
 For $R>0$ and a.e.~$\pp_1$ let
 \begin{equation}\label{eq:ApsiR}
  \widehat{A}_{\psi,R}(\pp_1)\;:=\;\frac{1}{\;(2\pi)^{\frac{3}{2}}}\int_{\!\substack{ \\ \\ \pp_2\in\mathbb{R}^3 \\ |\pp_2|<R}}\widehat{\psi}(\pp_1,\pp_2)\,\ud \pp_2\,.
 \end{equation}
 Then, for a.e.~$\yy_1$, and for some constant $c_\psi\in\mathbb{C}$,
 \begin{equation}\label{eq:AR}
  A_{\psi,R}(\yy_1)\;=\; c_\psi\,\psi_{\mathrm{av}}(\yy_1;{\textstyle\frac{1}{R})}\,(1+o(1))\qquad \textrm{as }\;R\to +\infty\,.
 \end{equation}
\end{lemma}

\begin{remark}
 In the assumption of the Lemma $\psi$ may be singular at $\yy_2=0$, in which case both sides of \eqref{eq:AR} diverge with $R$. If instead $\psi$ is suitably regular at $\yy_2=0$, then the r.h.s.~converges to $\psi(\yy_1,\mathbf{0})$, consistently with \eqref{eq:psiy0}-\eqref{eq:tracemomentum} above. 
\end{remark}

\begin{proof}[Proof of Lemma \ref{lem:shortscalegeneric}]
 One has
 \[
  \begin{split}
    A_{\psi,R}(\yy_1)\;&=\;\frac{1}{\;(2\pi)^3}\iint_{\mathbb{R}^3\times\mathbb{R}^3}\ud\pp_1\ud\pp_2\,e^{\ii\pp_1\cdot\yy_1}\,\mathbf{1}_{\{|\pp_2|<R\}}(\pp_2)\,\widehat{\psi}(\pp_1,\pp_2) \\
    &=\;\iint_{\mathbb{R}^3\times\mathbb{R}^3}\ud\zz_1\ud\zz_2\,\delta(\zz_1+\yy_1)\,\delta_R(\zz_2)\,\psi(\zz_1,\zz_2) \\
    &=\;\int_{\mathbb{R}^3}\ud\zz_2\,\delta_R(\zz_2)\,\psi(-\yy_1,\zz_2)\;=\;\int_{\mathbb{R}^3}\ud\zz_2\,\delta_R(\zz_2)\,\psi(\yy_1,-\zz_2)\,,
  \end{split}
 \]
 where
 \[
  \begin{split}
   \delta_R(\zz_2)\;:=&\;\,\Big(\frac{\mathbf{1}_{\{|\pp_2|<R\}}}{(2\pi)^{\frac{3}{2}}}\Big)^{\!\vee}(\zz_2)\;=\;\frac{1}{\;(2\pi)^{3}}\int_{\!\substack{ \\ \\ \pp_2\in\mathbb{R}^3 \\ |\pp_2|<R}}e^{\ii \pp_2\cdot\zz_2}\,\ud\pp_2 \\
   =&\;\,\frac{2\pi}{\;(2\pi)^{3}}\int_0^R\ud r\,r^2\int_{-1}^1\,\ud t\,e^{\ii r |\zz_2| t} \\ 
   =&\;\,\frac{2 R^3}{\;(2\pi)^{2}}\,\frac{\,\sin R|\zz_2|-R|\zz_2|\cos R|\zz_2|\,}{(R|\zz_2|)^3}\,.
  \end{split}
 \]
 In fact, $\delta_R$ is a smooth, approximate delta-distribution in three dimensions. Thus,
 \[
  \begin{split}
   A_{\psi,R}(\yy_1)\;&=\;\frac{2}{\;(2\pi)^2}\int_{\mathbb{R}^3}\ud\zz_2\,\frac{\,\sin|\zz_2|-|\zz_2|\cos |\zz_2|\,}{|\zz_2|^3}\,\psi(\yy_1,{\textstyle\frac{1}{R}}\zz_2) \\
   &=\;\frac{2}{\;(2\pi)^2}\int_0^{+\infty}\!\ud\rho\,\frac{\,\sin\rho-\rho\cos\rho}{\rho}\Big(\int_{\mathbb{S}^2}\ud\Omega\,\psi(\yy_1,{\textstyle\frac{1}{R}}\rho\,\Omega)\Big) \\
   &=\;\frac{2}{\pi}\int_0^{+\infty}\!\ud\rho\,\frac{\,\sin\rho-\rho\cos\rho}{\rho}\,\psi_{\mathrm{av}}(\yy_1,\textstyle\frac{\rho}{R})\,.
  \end{split}  
 \]
 By assumption (see \eqref{eq:Z} above),
 \[
  \lim_{R\to +\infty}\frac{\int_0^{+\infty}\!\ud\rho\,\frac{\,\sin\rho-\rho\cos\rho}{\rho}\,\psi_{\mathrm{av}}(\yy_1,\textstyle\frac{\rho}{R})}{\psi_{\mathrm{av}}(\yy_1,\textstyle\frac{1}{R})}\;=\;\frac{\pi}{2}\,c_\psi
 \]
 for some constant $c_\psi\in\mathbb{C}$. Therefore,
 \[
  A_{\psi,R}(\yy_1)\,=\,\frac{2}{\pi}\int_0^{+\infty}\!\ud\rho\,\frac{\,\sin\rho-\rho\cos\rho}{\rho}\,\psi_{\mathrm{av}}(\yy_1,\textstyle\frac{\rho}{R})\,\stackrel{R\to +\infty}{=}  \,c_\psi\,\psi_{\mathrm{av}}(\yy_1,\textstyle\frac{1}{R})\,(1+o(1))\,,
 \]
which completes the proof. 
\end{proof}

\begin{remark}
 Should, more realistically, the function $\rho\mapsto\psi_{\mathrm{av}}(\yy_1,\rho)$ in the above proof display $\mathcal{Z}$-behaviour non-uniformly in $\yy_1$, the counterpart of the $c_\psi$-constant would be a function $c_\psi(\yy_1)$. In this respect one is not really interested in pushing such generality forward: the $\mathcal{Z}$-behaviour was merely introduced to visualise, in meaningful concrete cases, the correspondence between the two expressions \eqref{eq:ApsiR} and \eqref{eq:AR} with explicit dependence on the cut-off parameter $R$. 
\end{remark}

\subsection{The $T_\lambda$ operator}\label{sec:Tlambdaoperator}

In practice, the computation of \eqref{eq:ApsiR} for elements of $\mathcal{D}(\mathring{H}^*)$ produces a quantity that for convenience is analysed separately in this Subsection, before resuming the discussion in the following Subsect.~\ref{sec:largemomentumasympt}.

For $\xi\in H^{-\frac{1}{2}}(\mathbb{R}^3)$ and $\lambda>0$ one defines (for a.e.~$\pp$)
\begin{equation}\label{eq:Tlambda}
  (\widehat{T_\lambda\xi})(\pp)\;:=\;2\pi^2\sqrt{\frac{3}{4}\pp^2+\lambda}\;\widehat{\xi}(\pp)-2\int_{\mathbb{R}^3}\frac{\widehat{\xi}(\qq)}{\,\pp^2+\qq^2+\pp\cdot \qq+\lambda\,}\,\ud\qq\,.
 \end{equation}

At least for $\xi\in H^{-\frac{1}{2}+\varepsilon}(\mathbb{R}^3)$, $\varepsilon>0$, \eqref{eq:Tlambda} does produce an almost-everywhere finite quantity, for
\[
\Big|\int_{\mathbb{R}^3}\frac{\widehat{\xi}(\qq)}{\,\pp^2+\qq^2+\pp\cdot \qq+\lambda\,}\,\ud\qq\Big|\;\lesssim\;\|\xi\|_{H^{-\frac{1}{2}+\varepsilon}}\Big(\int_{\mathbb{R}^3}\frac{(\qq^2+1)^{\frac{1}{2}-\varepsilon}}{(\pp^2+\qq^2+1)^2}\,\ud q\Big)^{\frac{1}{2}}\;<\;+\infty\,.
\]
Instead, the example $\widehat{\xi}_0(\qq):=\mathbf{1}_{\{|\qq|\geqslant 2\}}(|\qq|\log|\qq|)^{-1}$ shows that \eqref{eq:Tlambda} may be \emph{infinite} for generic $H^{-\frac{1}{2}}$-functions.

The map $\xi\mapsto T_\lambda\xi$ is central in the present analysis. It commutes with the rotations in $\mathbb{R}^3$ and therefore, upon densely defining it over $H^s(\mathbb{R}^3)$, $s\geqslant-\frac{1}{2}$, one has
\begin{equation}\label{eq:decompTTell}
 T_\lambda\;=\;\bigoplus_{\ell=0}^\infty T_\lambda^{(\ell)}
\end{equation}
in the sense of direct sum of operators on Hilbert space (Sect.~\ref{sec:I-preliminaries} and \ref{sec:I_invariant-reducing-ssp}) with respect to the canonical decomposition
\begin{equation}\label{eq:bigdecomp}
 \begin{split}
 H^s(\mathbb{R}^3)\;&\cong\;\bigoplus_{\ell=0}^\infty \Big( L^2(\mathbb{R}^+,(1+p^2)^s p^2\ud p) \otimes \mathrm{span}\big\{ Y_{\ell,n}\,|\,n=-\ell,\dots,\ell \big\} \Big)  \\
 &\equiv\;\bigoplus_{\ell=0}^\infty \,H^s_{\ell}(\mathbb{R}^3)\,.
 \end{split}
\end{equation}
Here the $Y_{\ell,n}$'s form the usual orthonormal basis of $L^2(\mathbb{S}^2)$ of spherical harmonics \index{spherical harmonics} and each $\xi\in H^s(\mathbb{R}^3)$ decomposes with respect to \eqref{eq:bigdecomp} as
\begin{equation}\label{eq:xihatangularexpansion}
 \begin{split}
  \widehat{\xi}(\pp)\;&=\;\sum_{\ell=0}^\infty\sum_{n=-\ell}^\ell f_{\ell,n}^{(\xi)}(|\pp|) Y_{\ell,n}(\Omega_{\pp})\;=\;\sum_{\ell=0}^\infty\widehat{\xi^{(\ell)}}(\pp) \, , \\
  \widehat{\xi^{(\ell)}}(\pp)\;&\!:=\;\sum_{n=-\ell}^\ell f_{\ell,n}^{(\xi)}(|\pp|) Y_{\ell,n}(\Omega_{\pp})
 \end{split}
\end{equation}
in polar coordinates $\pp\equiv|\pp|\Omega_\pp$. Explicitly, 
\begin{equation}
 \langle\xi,\eta\rangle_{H^s}\;=\;\sum_{\ell=0}^\infty\langle\xi^{(\ell)},\eta^{(\ell)}\rangle_{H^s_\ell}\;=\;\sum_{\ell=0}^\infty\sum_{n=-\ell}^\ell\int_{\mathbb{R}^+}\overline{f_{\ell,n}^{(\xi)}(p)}\,f_{\ell,n}^{(\eta)}(p)\,(1+p^2)^s p^2\ud p 
\end{equation}
%
%
%
%
%
and 
\begin{equation}\label{eq:TlambdaTlambdaell}
 \widehat{T_\lambda\xi}\;=\;\sum_{\ell=0}^\infty \widehat{T_\lambda^{(\ell)}\xi^{(\ell)}}\,.
\end{equation}

Let $P_\ell$ be the Legendre polynomial \index{Legendre polynomials} of order $\ell=0,1,2,\dots$, namely
\begin{equation}\label{def_Legendre}
P_\ell(t)\equiv\frac{1}{2^\ell \ell!}\,\frac{\ud^\ell}{\ud t^\ell}\,(t^2-1)^\ell\,.
\end{equation}

\begin{lemma}[Decomposition properties of $T_\lambda$]\label{lem:Tlambdadecomposition}
 Let $\lambda>0$ and $\xi^{(\ell)},\eta^{(\ell)}\in H^s_\ell(\mathbb{R}^3)$. 
 Then, with respect to the representation \eqref{eq:xihatangularexpansion},
 \begin{enumerate}[(i)]
  \item $T_\lambda^{(\ell)}$ acts trivially (i.e., as the identity) on the angular components of $\widehat{\xi}^{(\ell)}$, and acts as
  \begin{equation}\label{eq:fellsector}
   f_{\ell,n}^{(\xi)}(p)\:\mapsto\:2\pi^2\sqrt{{\textstyle\frac{3}{4}}p^2+\lambda}\,f_{\ell,n}^{(\xi)}(p)-4\pi\!\!\int_{\mathbb{R}^+}\!\ud q\,q^2  f_{\ell,n}^{(\xi)}(q)\!\int_{-1}^1\frac{P_\ell(t)\,\ud t}{p^2+q^2+p \,q \,t+\lambda}
  \end{equation}
 on each radial component;
  \item one has
  \begin{equation}\label{eq:xiTxipre}
    \int_{\mathbb{R}^3} \overline{\,\widehat{\xi}(\pp)}\, \big(\widehat{T_\lambda\eta}\big)(\pp)\,\ud\pp\;=\;\sum_{\ell=0}^\infty \int_{\mathbb{R}^3} \overline{\,\widehat{\xi^{(\ell)}}(\pp)}\, \big(\widehat{T_\lambda^{(\ell)}\eta^{(\ell)}}\big)(\pp)\,\ud\pp
  \end{equation}
 and
 \begin{equation}\label{eq:xiTxi}
   \begin{split}
   \!\!\!\!\!\!\!\!\!\!\!\!\int_{\mathbb{R}^3}& \overline{\,\widehat{\xi^{(\ell)}}(\pp)}\, \big(\widehat{T_\lambda^{(\ell)}\eta^{(\ell)}}\big)(\pp)\,\ud\pp\;=\;2\pi^2\!\int_{\mathbb{R}^+}\!\ud p\,p^2\,\overline{f_{\ell,n}^{(\xi)}(p)}f_{\ell,n}^{(\eta)}(p)\sqrt{{\textstyle\frac{3}{4}}p^2+\lambda} \\
   & -4\pi\!\iint_{\mathbb{R}^+\times\mathbb{R}^+}\ud p\,\ud q\,p^2q^2\,\overline{f_{\ell,n}^{(\xi)}(p)}\,f_{\ell,n}^{(\eta)}(q)\!\int_{-1}^1\ud t\,\frac{P_\ell(t)}{p^2+q^2+p \,q \,t+\lambda}\,.
    \end{split}
  \end{equation}
 \end{enumerate}
\end{lemma}

\begin{proof}
   The triviality of the action of $T_\lambda^{(\ell)}$ on the angular components is due to the invariance of $T_\lambda$ under rotations. All other formulas are then straightforwardly derived from \eqref{eq:Tlambda} by exploiting the following standard expansion in Legendre polynomials and the addition formula for spherical harmonics:\index{spherical harmonics} 
\begin{equation}\label{expansion-Leg-poly}
\begin{split}
&\frac{1}{\pp^2+\qq^2+\pp\cdot\qq+\lambda} \;=\;\sum_{\ell=0}^\infty\frac{2\ell+1}{2}\!\int_{-1}^1\ud t\,\frac{P_\ell(t)\,P_\ell(\cos(\theta_{\pp,\qq}))}{\pp^2+\qq^2+|\pp|\,|\qq|\,+\lambda} \\
&\quad =\; \sum_{\ell=0}^\infty 2\pi\int_{-1}^1\ud t\,\frac{P_\ell(t)}{\pp^2+\qq^2+|\pp|\,|\qq|\,t+\lambda}\sum_{r=-\ell}^\ell\overline{Y_{\ell r}(\Omega_{\pp})}\,Y_{\ell r}(\Omega_{\qq})
\end{split}
\end{equation}
    (see, e.g., \cite[Eq.~(8.814)]{Gradshteyn-tables-of-integrals-etc}).
  \end{proof}

\begin{lemma}[Mapping properties of $T_\lambda$]\label{lem:Tlambdaproperties} Let $\lambda>0$.

\begin{enumerate}[(i)]
 \item For each $s\geqslant 1$ \eqref{eq:Tlambda} defines an operator
 \[
  T_\lambda:\mathcal{D}(T_\lambda)\subset L^2(\mathbb{R}^3)\to L^2(\mathbb{R}^3)\,,\qquad \mathcal{D}(T_\lambda)\;:=\;H^s(\mathbb{R}^3)
 \]
 that is densely defined and symmetric in $L^2(\mathbb{R}^3)$.
 \item One has
 \begin{equation}\label{eq:Tlambdamapping}
  \| T_\lambda \xi\|_{H^{s-1}}\;\lesssim\;\|\xi\|_{H^s}\qquad \forall\xi\in H^s(\mathbb{R}^3)\,,\qquad s\in\Big(-\frac{1}{2},\frac{3}{2}\Big)\,,
 \end{equation}
 i.e., \eqref{eq:Tlambda} defines a bounded operator $T_\lambda:H^s(\mathbb{R}^3)\to H^{s-1}(\mathbb{R}^3)$ for every $s\in(-\frac{1}{2},\frac{3}{2})$.
 \item One has
 \begin{equation}\label{eq:TlambdamappingN}
  \| T_\lambda^{(\ell)} \xi\|_{H^{s-1}}\;\lesssim\;\|\xi\|_{H^s}\qquad \forall\xi\in H^s_\ell(\mathbb{R}^3)\,,\qquad s\in\Big[-\frac{1}{2},\frac{3}{2}\Big]\,,\quad\ell\in\mathbb{N}\,,
 \end{equation}
 i.e., \eqref{eq:Tlambda} defines a bounded operator $T_\lambda:H^s_\ell(\mathbb{R}^3)\to H^{s-1}_\ell(\mathbb{R}^3)$ for every $s\in[-\frac{1}{2},\frac{3}{2}]$, provided that $\ell\in\mathbb{N}$. In the sector $\ell=0$ \eqref{eq:TlambdamappingN} fails in general at the endpoints in $s$ and only \eqref{eq:Tlambdamapping} is valid.
 \item For any other $\lambda'>0$ one has
 \begin{equation}
  \|(T_{\lambda'}-T_\lambda)\xi\|_{H^{\frac{1}{2}}}\;\lesssim\;|\lambda'-\lambda|\,\|\xi\|_{H^{-\frac{1}{2}}}\,.
 \end{equation}
 \item For $s\geqslant\frac{1}{2}$ and $\xi,\eta\in H^s(\mathbb{R}^3)$
 one has
 \begin{equation}\label{eq:Tlambdaexchange}
 \int_{\mathbb{R}^3} \overline{\,\widehat{\xi}(\pp)}\, \big(\widehat{T_\lambda\eta}\big)(\pp)\,\ud\pp\;=\;\int_{\mathbb{R}^3} \overline{\,\widehat{T_\lambda\xi}(\pp)}\, \widehat{\eta}(\pp)\,\ud\pp\quad\qquad(s\geqslant\textstyle\frac{1}{2})
 \end{equation}
 and the quantity above is real and finite.
\end{enumerate}
\end{lemma}

\begin{proof}
 All claims (i)-(iii) are obvious for the multiplicative part of $T_\lambda$, namely the first summand on the r.h.s.~of \eqref{eq:Tlambda}, and need only be proved for the integral part of $T_\lambda$. The latter, apart from an irrelevant multiplicative pre-factor, is the same as the multiplicative part of the `fermionic' counterpart of $T_\lambda$, namely the analogous operator emerging in the analysis of a trimer consisting of two identical fermions and a third different particle. All the claimed properties were already demonstrated in that case by Michelangeli and Ottolini in \cite[Propositions 3 and 4, Corollary 2]{MO-2016}.

 Concerning part (iv),
 \[
 \begin{split}
  ((T_{\lambda'}-T_\lambda)&\xi)\,{\textrm{\large $\widehat{\,}$\normalsize}}\,(\pp)\;=\; \frac{2\pi^2(\lambda'-\lambda)}{\:\sqrt{\frac{3}{4}\pp^2+\lambda'}+\sqrt{\frac{3}{4}\pp^2+\lambda}\:}\,\widehat{\xi}(\pp)\\
  & +2(\lambda'-\lambda)\int_{\mathbb{R}^3}\,\frac{\widehat{\xi}(\qq)}{\,(\pp^2+\qq^2+\pp\cdot\qq+\lambda)\,(\pp^2+\qq^2+\pp\cdot\qq+\lambda')\,}\,\ud\qq\,,
 \end{split}
 \]
 whence
 \[
   \big|(T_{\lambda'}-T_\lambda)\xi)\,{\textrm{\large $\widehat{\,}$\normalsize}}\,(\pp)\big|\;\lesssim\;|\lambda'-\lambda|\,\bigg( \frac{2\pi^2\,|\widehat{\xi}(\pp)|}{\,\sqrt{\frac{3}{4}\pp^2+\lambda}\,}+\int_{\mathbb{R}^3}\,\frac{|\widehat{\xi}(\qq)|}{\,(\pp^2+\qq^2+1)^2\,}\,\ud\qq\bigg)\,.
 \]
 Thus, $\xi\mapsto(T_{\lambda'}-T_\lambda)\xi$ has the same behaviour as $W_\lambda$, and hence the same $H^{-\frac{1}{2}}\to H^{\frac{1}{2}}$ boundedness.

 Concerning (v), the only non-trivial piece of the claim regards the integral part of $T_\lambda$, namely the identity
  \begin{equation*}
 \begin{split}
  \int_{\mathbb{R}^3}&\ud\pp\,\overline{\widehat{\xi}(\pp)}\,\Big(\int_{\mathbb{R}^3}\ud\qq\,\frac{\widehat{\eta}(\qq)}{\,\pp^2+\qq^2+\pp\cdot\qq+\lambda\,}\Big) \\
   &=\;\int_{\mathbb{R}^3}\ud\qq\,\Big(\int_{\mathbb{R}^3}\ud\pp\,\frac{\overline{\widehat{\xi}(\pp)}}{\,\pp^2+\qq^2+\pp\cdot\qq+\lambda\,}\Big)\,\widehat{\eta}(\qq)\,.
 \end{split}
 \end{equation*}
 The exchange of integration order above is indeed legitimate, as the assumptions on $\xi,\eta$ guarantee the applicability of Fubini-Tonelli theorem. More precisely,
 \[
 \begin{split}
  &\bigg|\int_{\mathbb{R}^3} \ud\pp\,\overline{\widehat{\xi}(\pp)}\,\Big(\int_{\mathbb{R}^3}\ud\qq\,\frac{\widehat{\eta}(\qq)}{\,\pp^2+\qq^2+\pp\cdot\qq+\lambda\,}\Big)\bigg| \\
  &\;\lesssim\;	 \int_{\mathbb{R}^3}\ud\pp\,|\widehat{\xi}(\pp)|\,\Big(\int_{\mathbb{R}^3}\ud\qq\,\frac{|\widehat{\eta}(\qq)|}{\,\pp^2+\qq^2+1\,}\Big) \;\leqslant\;\|\xi\|_{H^{\frac{1}{2}}}\bigg\|\int_{\mathbb{R}^3}\ud\qq\,\frac{|\widehat{\eta}(\qq)|}{\,\pp^2+\qq^2+1\,} \bigg\|_{H^{-\frac{1}{2}}}\\
  &\;\lesssim\;\|\xi\|_{H^{\frac{1}{2}}}\|\eta\|_{H^{\frac{1}{2}}}\;\leqslant\;\|\xi\|_{H^{s}}\|\eta\|_{H^{s}}\;<\;+\infty\qquad\forall s\geqslant\frac{1}{2}\,,
 \end{split}
 \]
 where \eqref{eq:lambda-equiv-1} was applied in the first inequality, and \eqref{eq:Tlambdamapping} in the third (estimate \eqref{eq:Tlambdamapping} refers to the whole $T_\lambda$, but as commented above in the course of its proof it is actually established by demonstrating the only non-trivial piece of the estimate, namely the one involving the integral part of $T_\lambda$). 
\end{proof}


\begin{remark}\label{rem:Tl-failstomap}
 $T_\lambda$ fails to map $H^{\frac{3}{2}}(\mathbb{R}^3)$ into $H^{\frac{1}{2}}(\mathbb{R}^3)$ as is the case, for instance, for the action of $T_\lambda$ on the class of spherically symmetric functions in $\mathcal{F}^{-1} C^\infty_c(\mathbb{R}^3_\pp)$. Indeed, if $\xi$ has symmetry $\ell=0$ and $\widehat{\xi}\in C^\infty_c(\mathbb{R}^3_\pp)$, then the contribution from the integral part of $(\widehat{T_\lambda\xi})(\pp)$ is of the order of (see \eqref{eq:fellsector} above)
 \[
 \begin{split}
  \int_{\mathrm{supp}\,f^{(\xi)}}&\ud q\,q^2\,f^{(\xi)}(q)\int_{-1}^1\frac{\ud t}{\,|\pp|^2+q^2+|\pp|qt+\lambda\,} \\
  &=\;\frac{1}{|\pp|}\int_{\mathrm{supp}\,f^{(\xi)}}\ud q\,q\,f^{(\xi)}(q)\,\log\Big(1+\frac{2|\pp|q}{\,|\pp|^2+q^2-|\pp|qt+\lambda\,}\Big)\,,
 \end{split}
 \]
 which, both in the limit $|\pp|\to 0$ and $|\pp|\to +\infty$ is of the order of
 \[
  \int_{\mathrm{supp}\,f^{(\xi)}}\ud q\,\frac{q^2}{\,|\pp|^2+q^2-|\pp|qt+\lambda\,}\,f^{(\xi)}(q)\;\sim\;\frac{1}{\,\pp^2+1\,}\,.
 \]
 The contribution from the multiplicative part of $(\widehat{T_\lambda\xi})(\pp)$ is obviously a compactly supported function, the conclusion therefore is $(\widehat{T_\lambda\xi})(\pp)\sim (\pp^2+1)^{-1}$, and the latter is a $H^{\frac{1}{2}-\varepsilon}$-function $\forall\varepsilon>0$ not belonging to $H^{\frac{1}{2}}(\mathbb{R}^3)$. 
\end{remark}

\begin{remark}\label{rem:whensymmetric}
 Parts (i) and (v) of Lemma \ref{lem:Tlambdaproperties} present two regimes of validity of the identity \eqref{eq:Tlambdaexchange} when $\xi,\eta\in H^s(\mathbb{R}^3)$ for $s\geqslant\frac{1}{2}$. In the regime $s\geqslant 1$, each side of the \eqref{eq:Tlambdaexchange} is a product of two $L^2$-functions and such identity amounts to the symmetry of $T_\lambda$ in $L^2(\mathbb{R}^3)$ with domain $H^s(\mathbb{R}^3)$. For $\frac{1}{2}\leqslant s<1$, instead, $T_\lambda$ does not make sense any longer as an operator on $L^2(\mathbb{R}^3)$, and yet \eqref{lem:Tlambdaproperties} still expresses the symmetry of the action of $T_\lambda$ on $H^s$-functions, and hence also the reality of the considered integrals. 
\end{remark}

 Additional relevant properties $T_\lambda$ are discussed in Subsect.~\ref{sec:Tlambdaestimates} below.

\subsection{Large momentum asymptotics}\label{sec:largemomentumasympt}

\begin{lemma}\label{lem:largepasympt-star}
 Let $g\in\mathcal{D}(\mathring{H}^*)$ and let $\lambda>0$. Then, decomposing $\widehat{g}=\widehat{\phi^\lambda}+\widehat{u_\xi^\lambda}$ with $\widehat{\phi^\lambda}=\widehat{f^\lambda}+(\pp_1^2+\pp_2^2+\pp_1\cdot\pp_2+\lambda)^{-1}\,\widehat{u_\eta^\lambda}$ as demonstrated in Lemma \ref{lem:Hstaretc}, in the limit $R\to +\infty$ one has the asymptotics
 \begin{equation}\label{eq:g-largep2-star}
  \int_{\!\substack{ \\ \\ \pp_2\in\mathbb{R}^3 \\ |\pp_2|<R}}\widehat{g}(\pp_1,\pp_2)\,\ud\pp_2\;=\;4\pi R\,\widehat{\xi}(\pp_1)+\Big({\textstyle\frac{1}{3}}(\widehat{W_\lambda\eta})(\pp_1)-(\widehat{T_\lambda\xi})(\pp_1)\Big)+o(1)
 \end{equation}
 as well as the identity
 \begin{equation}\label{eq:phi-largep2-star}
  \int_{\mathbb{R}^3}\widehat{\phi^\lambda}(\pp_1,\pp_2)\,\ud\pp_2\;=\;{\textstyle\frac{1}{3}}(\widehat{W_\lambda\eta})(\pp_1)\,.
 \end{equation} 
\end{lemma}

An immediate corollary of Lemma \ref{lem:largepasympt-star}, obtained by means of Lemma \ref{lem:shortscalegeneric} taking $R=|\yy_2|^{-1}\to +\infty$, is the following.

\begin{corollary}\label{cor:largepasympt-star}
 Under the assumptions of Lemmas \ref{lem:largepasympt-star} and \ref{lem:shortscalegeneric} one has
 \begin{equation}\label{eq:g-largep2-star-yversion}
 \!\!\!(2\pi)^{\frac{3}{2}} c_g\,g_{\mathrm{av}}(\yy_1;|\yy_2|)\,\stackrel{|\yy_2|\to 0}{=}\,\frac{4\pi}{|\yy_2|}\xi(\yy_1)+\Big( {\textstyle\frac{1}{3}}(W_\lambda\eta)(\yy_1)-(T_\lambda\xi)(\yy_1)\Big) + o(1)  
 \end{equation}
 for some constant $c_g\in\mathbb{C}$, and 
 \begin{equation}\label{eq:phi-largep2-star-yversion}
   \phi^\lambda(\yy_1,\mathbf{0})\;=\;\frac{1}{\,3\,(2\pi)^{\frac{3}{2}}}\,(W_\lambda\eta)(\yy_1)
 \end{equation}
 for a.e.~$\yy_1$. 
\end{corollary}

\begin{remark}\label{rem:gexpansionfinite}
 Not for all $g\in \mathcal{D}(\mathring{H}^*)$ are \eqref{eq:g-largep2-star} and \eqref{eq:g-largep2-star-yversion} \emph{finite} quantities, but surely they are if the charge $\xi$ of $g$ has at $H^{-\frac{1}{2}+\varepsilon}$-regularity for some $\varepsilon>0$ (as argued right after the definition \eqref{eq:Tlambda}). 
\end{remark}

\begin{proof}[Proof of Lemma \ref{lem:largepasympt-star}]
 For what observed in Remark \ref{rem:gexpansionfinite}, one tacitly restricts the computations to those $\xi$'s making the following integrals finite (e.g., all $\xi$'s with $H^{-\frac{1}{2}+\varepsilon}$-regularity), for otherwise the corresponding identities to prove are all identities between infinites.
 
 One has
 \[
  \begin{split}
    &\int_{\!\substack{ \\ \\ \pp_2\in\mathbb{R}^3 \\ |\pp_2|<R}}\widehat{u_\xi^\lambda}(\pp_1,\pp_2)\ud\pp_2\;=\;\widehat{\xi}(\pp_1) \int_{\!\substack{ \\ \\ \pp_2\in\mathbb{R}^3 \\ |\pp_2|<R}}\frac{\ud \pp_2}{\,\pp_1^2+\pp_2^2+\pp_1\cdot\pp_2+\lambda\,} \\
    &\quad +\int_{\!\substack{ \\ \\ \pp_2\in\mathbb{R}^3 \\ |\pp_2|<R}}\frac{\widehat{\xi}(\pp_2)}{\,\pp_1^2+\pp_2^2+\pp_1\cdot\pp_2+\lambda\,}\,\ud\pp_2+\int_{\!\substack{ \\ \\ \pp_2\in\mathbb{R}^3 \\ |\pp_2|<R}}\frac{\widehat{\xi}(-\pp_1-\pp_2)}{\,\pp_1^2+\pp_2^2+\pp_1\cdot\pp_2+\lambda\,}\,\ud\pp_2\,.
  \end{split}
 \]
 Both last two summands on the r.h.s.~above converge as $R\to +\infty$ to
 \[
  \int_{\mathbb{R}^3}\frac{\widehat{\xi}(\pp_2)}{\,\pp_1^2+\pp_2^2+\pp_1\cdot\pp_2+\lambda\,}\,\ud\pp_2
 \]
 (for the third one this follows after an obvious change of the integration variable). Moreover,
 \[
\begin{split}
\int_{\!\substack{ \\ \\ \pp_2\in\mathbb{R}^3 \\ |\pp_2|<R}}&\frac{\ud \pp_2}{\,\pp_1^2+\pp_2^2+\pp_1\cdot\pp_2+\lambda\,}\;=\;2\pi\int_0^R\ud r\,r^2\int_{-1}^1\frac{\ud t}{\pp_1^2+r^2+|\pp_1| r t+\lambda} \\
&=\;\frac{2\pi}{|\pp_1|}\int_0^R r\log\frac{r^2+\pp_1^2+|\pp_1|r+\lambda}{r^2+\pp_1^2-|\pp_1|r+\lambda}\,\ud r \\
&=\;2\pi R\,\Big(1+\frac{R}{2|\pp_1|}\log\frac{R^2+\pp_1^2+|\pp_1|R+\lambda}{R^2+\pp_1^2-|\pp_1| R+\lambda}\Big) \\
&\qquad\quad+2\pi\sqrt{\frac{3}{4} \pp_1^2+\lambda}\,\Big(\!\arctan\frac{|\pp_1|-2R}{2\sqrt{\frac{3}{4} \pp_1^2+\lambda}}-\arctan\frac{|\pp_1|+2R}{2\sqrt{\frac{3}{4} \pp_1^2+\lambda}}\,\Big) \\
&\qquad\quad+\pi\frac{\pp_1^2+\lambda}{4\sqrt{\frac{3}{4} \pp_1^2+\lambda}}\,\log\frac{R^2+\pp_1^2+|\pp_1|R+\lambda}{R^2+\pp_1^2-|\pp_1| R+\lambda} \\
&=\;4\pi R-2\pi^2\sqrt{\frac{3}{4} \pp_1^2+\lambda}+o(1)\qquad\textrm{as }\;R\to +\infty\,.
\end{split}
\]
Thus,
\[
\begin{split}
 &\int_{\!\substack{ \\ \\ \pp_2\in\mathbb{R}^3 \\ |\pp_2|<R}}\widehat{u_\xi^\lambda}(\pp_1,\pp_2)\ud\pp_2 \\
 &=\;4\pi R\,\widehat{\xi}(\pp_1)-2\pi^2\sqrt{\frac{3}{4} \pp_1^2+\lambda}\,\widehat{\xi}(\pp_1)+2 \int_{\mathbb{R}^3}\frac{\widehat{\xi}(\pp_2)}{\,\pp_1^2+\pp_2^2+\pp_1\cdot\pp_2+\lambda\,}\,\ud\pp_2+o(1) \\
 &=\;4\pi R\,\widehat{\xi}(\pp_1)-(\widehat{T_\lambda}\xi)(\pp_1)+o(1)\,.
\end{split}
\]

Next, one computes (using $\int_{\mathbb{R}^3}f^\lambda(\pp_1,\pp_2)\ud\pp_2=0$)
\[
 \begin{split}
 &\int_{\mathbb{R}^3}\widehat{\phi^\lambda}(\pp_1,\pp_2)\,\ud\pp_2\;=\;\int_{\mathbb{R}^3}\frac{\widehat{u}_\eta(\pp_1,\pp_2)}{\,\pp_1^2+\pp_2^2+\pp_1\cdot\pp_2+\lambda\,} \\
 &=\;\widehat{\eta}(\pp_1) \int_{\mathbb{R}^3}\frac{\ud \pp_2}{\,(\pp_1^2+\pp_2^2+\pp_1\cdot\pp_2+\lambda)^2} \\
    &\quad +\int_{\mathbb{R}^3}\frac{\widehat{\eta}(\pp_2)}{\,(\pp_1^2+\pp_2^2+\pp_1\cdot\pp_2+\lambda)^2}\,\ud\pp_2+\int_{\mathbb{R}^3}\frac{\widehat{\eta}(-\pp_1-\pp_2)}{\,(\pp_1^2+\pp_2^2+\pp_1\cdot\pp_2+\lambda)^2}\,\ud\pp_2\,.
 \end{split}
\]
By an obvious change of variable one sees that the last two summands are the same. Moreover, 
\[
 \int_{\mathbb{R}^3}\frac{\ud \pp_2}{\,(\pp_1^2+\pp_2^2+\pp_1\cdot\pp_2+\lambda)^2}\;=\;\frac{\pi^2}{\,\sqrt{\frac{3}{4} \pp_1^2+\lambda}\,}\,.
\]
Therefore,
\[
\begin{split}
 \int_{\mathbb{R}^3}\widehat{\phi^\lambda}(\pp_1,\pp_2)\,\ud\pp_2\;&=\;\frac{\pi^2}{\,\sqrt{\frac{3}{4} \pp_1^2+\lambda}\,}+2\int_{\mathbb{R}^3}\frac{\widehat{\eta}(\pp_2)}{\,(\pp_1^2+\pp_2^2+\pp_1\cdot\pp_2+\lambda)^2}\,\ud\pp_2 \\
 &=\;\frac{1}{3}\,(\widehat{W_\lambda\eta})(\pp_1)
\end{split}
\]
This proves \eqref{eq:phi-largep2-star}, and combining this with the above results for $\widehat{u_\xi^\lambda}$ one proves \eqref{eq:g-largep2-star}. 
\end{proof}

By exploiting the bosonic symmetry and repeating the above arguments with respect the other coincidence hyperplanes, one finally obtains the following picture:
\begin{itemize}
 \item a function $f\in \mathcal{D}(\mathring{H})$ vanishes by definition in a neighbourhood of the coincidence manifold \index{coincidence manifold} $\Gamma$;
 \item at each hyperplane, away from the configuration of triple coincidence, any function $\phi\in\mathcal{D}(\mathring{H}_{\mathrm{F}})$ is finite almost everywhere, as shown by \eqref{eq:phi-largep2-star-yversion};
 \item a generic $g\in\mathcal{D}(\mathring{H}^*)$ displays the $|\yy|^{-1}$ singularity, as shown by \eqref{eq:g-largep2-star-yversion}.
\end{itemize}
Actually, \eqref{eq:g-largep2-star-yversion} and  \eqref{eq:phi-largep2-star-yversion} express the short-scale behaviour counterpart of the large momentum asymptotics \eqref{eq:g-largep2-star} and \eqref{eq:phi-largep2-star}, respectively -- and always with the caveat that the asymptotics for $g$ have finite coefficients only for a subclass of charges $\xi$ (which includes all charges with $H^{-\frac{1}{2}+\varepsilon}$-regularity).

The leading singularity of $g$ is of order $|\yy|^{-1}$ in the relative variable with respect to the considered coincidence hyperplane. Explicitly, in terms of the charges $\xi$ and $\eta$ of $g$,
\begin{equation}\label{eq:shortscalegeneric2}
 g_{\mathrm{av}}(\yy_1;|\yy_2|)\,\stackrel{|\yy_2|\to 0}{=}\,c_g^{-1}\sqrt{\frac{2}{\pi}\,}\,\Big(\frac{\xi(\yy_1)}{|\yy_2|}+\omega_{\xi,\eta}(\yy_1)\Big)+o(1)\,,
\end{equation}
\begin{equation}
 \omega_{\xi,\eta}(\yy_1)\;:=\; {\textstyle\frac{1}{4\pi}}\big({\textstyle\frac{1}{3}}(W_\lambda\eta)(\yy_1)-(T_\lambda\xi)(\yy_1)\big)\,,
\end{equation}
point-wise almost-everywhere in $\yy_1$. Analogous expressions hold with respect to the other coincidence hyperplanes, with the same $\xi$ and $\omega_{\xi,\eta}$.

 $\xi$ and $\omega_{\xi,\eta}$  are interpreted in \eqref{eq:shortscalegeneric2} as functions supported on the coincidence hyperplane ($\{\yy_2=\mathbf{0}\}$ in this case). The leading singularity's coefficient $\xi$ has some $H^{-\frac{1}{2}}$-regularity (in fact, more than that). The next-to-leading singularity's coefficient $\omega_{\xi,\eta}$, in general, is not even $H^{-\frac{1}{2}}$-regular, owing to the mapping properties of $W_\lambda$ (Lemma \ref{lem:Wlambdaproperties}) and $T_\lambda$ (Lemma \ref{lem:Tlambdaproperties}). But if the charge $\xi$ is absent in $g$, and hence $g\in H^2(\mathbb{R}^3\times\mathbb{R}^3,\ud\yy_1\ud\yy_2)$, then $\omega_{\xi,\eta}$ has the same regularity of $W_\lambda\eta$, namely the very $H^{\frac{1}{2}}(\mathbb{R}^3)$-regularity prescribed by the trace theorem.

When a self-adjoint extension $\mathring{H}_{\mathcal{A}_\lambda}$ is considered, and hence the subspace $\mathcal{D}(\mathring{H}_{\mathcal{A}_\lambda})$ is selected out of $\mathcal{D}(\mathring{H}^*)$ by means of the constraint \eqref{eq:constraintetaxi} on the charges $\xi$ and $\eta$, in practice one makes a choice in the class of leading coefficients $\xi$ and sub-leading coefficients $\omega_{\xi,\eta}$ of the short-scale expansion \eqref{eq:shortscalegeneric2}  which amounts to taking
\begin{equation}\label{eq:selection}
 \begin{split}
  \xi \;&\in\;\mathcal{D}(\mathcal{A}_\lambda)\,, \\
  4\pi\,\omega_{\xi,\eta}\;&=\; {\textstyle\frac{1}{3}} W_\lambda(\mathcal{A}_\lambda\xi+\chi)-T_\lambda\xi\quad\textrm{for some }\chi\in\mathcal{D}(\mathcal{A}_\lambda)^{\perp_\lambda}\cap H^{-\frac{1}{2}}(\mathbb{R}^3)\,.
 \end{split}
\end{equation}


\section{Ter-Martirosyan-Skornyakov extensions}\label{sec:TMSextension-section}

\subsection{TMS and BP asymptotics}\label{sec:TMS-BP-asympt}

Given $g\in\cH_\mathrm{b}$ such that $\int_{|\pp_2|<R}\widehat{g}(\pp_1,\pp_2)\ud\pp_2<+\infty$ for all $R>0$, one says that $g$ satisfies the \emph{Ter-Martirosyan-Skornyakov (TMS) condition} \index{contact condition!Ter-Martirosyan-Skornyakov} with parameter $a\in(\mathbb{R}\setminus\{0\})\cup\{\infty\}$ if there exists a function $\xi_\circ$ such that
 \begin{equation}\label{eq:g-TMS-generic}
  \int_{\!\substack{ \\ \\ \pp_2\in\mathbb{R}^3 \\ |\pp_2|<R}}\widehat{g}(\pp_1,\pp_2)\,\ud\pp_2\,\stackrel{R\to +\infty}{=}\,4\pi \Big(R-\frac{1}{a}\Big)\widehat{\xi_\circ}(\pp_1)+o(1)\,.
 \end{equation}
An operator $K$ on $\cH_\mathrm{b}$ for which all $g$'s of $\mathcal{D}(K)$ with $\int_{|\pp_2|<R}\widehat{g}(\pp_1,\pp_2)\ud\pp_2<+\infty$ $\forall R>0$ satisfy \eqref{eq:g-TMS-generic} shall be called a \emph{Ter-Martirosyan-Skornyakov operator}.\index{Ter-Martirosyan-Skornyakov!operator} (For the time being this definition is kept deliberately general: $K$ may or may not be densely defined, symmetric, self-adjoint, etc., and nothing is said about the class $\xi_\circ$ belongs to.)

For those $g$'s of $\mathcal{D}(K)$ for which it is possible to repeat the arguments of Lemma \ref{lem:shortscalegeneric}, \eqref{eq:g-TMS-generic} amounts to
 \begin{equation}\label{eq:g-TMS-BP-generic}
 g_{\mathrm{av}}(\yy_1;|\yy_2|)\,\stackrel{|\yy_2|\to 0}{=}\,c_g^{-1}\sqrt{\frac{2}{\pi}\,}\,\Big(\frac{1}{|\yy_2|}-\frac{1}{a}\Big)\,\xi_\circ(\yy_1) + o(1)  
 \end{equation}
   point-wise almost everywhere in $\yy_1\in\mathbb{R}^3$. (The numerical pre-factors appearing on the r.h.s.~of \eqref{eq:g-TMS-generic} and \eqref{eq:g-TMS-BP-generic} are merely prepared for the forthcoming application to the analysis of the extensions of $\mathring{H}$.)

The case $a=0$ in \eqref{eq:g-TMS-generic}-\eqref{eq:g-TMS-BP-generic} would correspond to $\xi_\circ\equiv 0$ and hence to the fact that the quantity $\int_{\mathbb{R}^3}\widehat{g}(\pp_1,\pp_2)\ud\pp_2$ is finite for almost every $\pp_1$. As the TMS condition is meant to pinpoint an actual \emph{singularity} of $g$ at each coincidence hyperplane, one conventionally excludes $a=0$ from the above definition.

 In fact, \eqref{eq:g-TMS-generic}-\eqref{eq:g-TMS-BP-generic} describe a short-scale structure of $g$ in the vicinity of each coincidence hyperplane (but away from the triple coincidence point) which in spatial coordinates has precisely the form of the Bethe-Peierls contact condition \index{contact condition!Bethe-Peierls} \eqref{eq:preBP} expected on physical grounds for the eigenfunctions of a quantum trimer with zero-range interaction: in this interpretation, $a$ is the two-body $s$-wave scattering length \index{scattering length} of the interaction.

 One refers to \eqref{eq:g-TMS-BP-generic} too as the Bethe-Peierls (BP) condition and one equivalently says that in the TMS condition \eqref{eq:g-TMS-generic}, resp., the BP condition \eqref{eq:g-TMS-BP-generic}, the quantity
  \begin{equation}\label{eq:a-alpha}
  \alpha\;:=\;-\frac{4\pi}{a}\;\in\;\mathbb{R}
 \end{equation}
  is the `inverse (negative) scattering length' (in suitable units).

 This indicates that within the huge variety of self-adjoint extensions of $\mathring{H}$ (Theorem \ref{thm:generalclassification}), the physically meaningful ones are those displaying the TMS condition for functions of their domain.

 De facto some arbitrariness in the modelling still remains, as will be elaborated further on in due time (Subsect.~\ref{sec:definite-ell-general} and \ref{sec:variants}), for one could deem an extension `physically meaningful'
 \begin{itemize}
  \item in the restrictive sense that \emph{all} functions in the domain of the extension satisfy the TMS asymptotics (meaning, all functions $g$ for which, at any $R>0$, the quantity $\int_{|\pp_2|<R}\widehat{g}(\pp_1,\pp_2)\ud\pp_2$ is finite),
  \item in the milder sense that only \emph{some} relevant functions do, for instance declaring the physical asymptotics for functions with given symmetry, or for certain \emph{eigenfunctions} of the extension.
 \end{itemize}

  Let us examine first the possibility that \emph{at least one} function in the domain of a self-adjoint extension of $\mathring{H}$ satisfies the TMS condition.

 \begin{lemma}\label{eq:oneTMSfunction}
  Let $\mathscr{H}$ be a self-adjoint extension of $\mathring{H}$ and let $\alpha\in\mathbb{R}$. Assume that there exists $g\in\mathcal{D}(\mathscr{H})$ satisfying the TMS condition \eqref{eq:g-TMS-generic} for the given $\alpha$ and for some function $\xi_\circ$. One has the following.
  \begin{enumerate}[(i)]
   \item $\xi_\circ$ must coincide with the charge $\xi$ of (the singular part of) $g$ (Lemma \ref{lem:chargexiofg}):
   \begin{equation}\label{eq:xixicirc}
     \xi\;=\;\xi_\circ \,.
   \end{equation}
   \item \emph{For every} shift parameter $\lambda>0$ with respect to which the canonical representation \eqref{eq:domDHA} of $g$ is written, the charges $\xi\in\mathcal{D}(\mathcal{A}_\lambda)$ and $\chi\in\mathcal{D}(\mathcal{A}_\lambda)^{\perp_\lambda}\cap H^{-\frac{1}{2}}(\mathbb{R}^3)$ of $g$ must satisfy
   \begin{eqnarray}
     {\textstyle\frac{1}{3}} W_\lambda(\mathcal{A}_\lambda\xi+\chi)\;&=&\;T_\lambda\xi+\alpha\xi \, , \label{eq:TMSoncharges} \\
     T_\lambda\xi+\alpha\xi&\in& H^{\frac{1}{2}}(\mathbb{R}^3)\,. \label{eq:Ta12}
   \end{eqnarray}
   \item \emph{For every} shift parameter $\lambda>0$, $g$ and its regular part $\phi^\lambda$ must satisfy 
 \begin{eqnarray}
  \int_{\!\substack{ \\ \\ \pp_2\in\mathbb{R}^3 \\ |\pp_2|<R}}\widehat{g}(\pp_1,\pp_2)\,\ud\pp_2 \;&=&\; (4\pi R+\alpha)\,\widehat{\xi}(\pp_1)+o(1)\,, \label{eq:g-largep2-TMS0} \\
  \int_{\mathbb{R}^3}\widehat{\phi^\lambda}(\pp_1,\pp_2)\,\ud\pp_2 \;&=&\; (\widehat{T_\lambda\xi})(\pp_1)+\alpha\,\widehat{\xi}(\pp_1) \label{eq:phi-largep2-TMS0}\,, \\
  \phi^\lambda(\yy_1,\mathbf{0}) \;&=&\; \frac{1}{\,(2\pi)^{\frac{3}{2}}}\,\Big((T_\lambda\xi)(\yy_1)+\alpha\,\xi(\yy_1)\Big) \label{eq:phi-largep2-star-yversion-TMS0}\,,
 \end{eqnarray}
 \eqref{eq:g-largep2-TMS0}, \eqref{eq:phi-largep2-TMS0}, \eqref{eq:phi-largep2-star-yversion-TMS0} being equivalent.
 \end{enumerate}
 \end{lemma}

 \begin{proof} Let $\lambda>0$ and write
  \[
   \widehat{g}\,=\,\widehat{f^\lambda}+\displaystyle\frac{\widehat{u_{\mathcal{A}_\lambda\xi+\chi}^\lambda}}{\pp_1^2+\pp_2^2+\pp_1\cdot\pp_2+\lambda}+\widehat{u_\xi^\lambda}
  \]
 according to \eqref{eq:domDHA}. 
 Comparing the asymptotics \eqref{eq:g-largep2-star} valid for such $g$ with the asymptotics \eqref{eq:g-TMS-generic} assumed in the hypothesis, one deduces
 \[
  \begin{split}
   \xi\;&=\;\xi_\circ \\
   {\textstyle\frac{1}{3}} W_\lambda(\mathcal{A}_\lambda\xi+\chi)-T_\lambda\xi\; &=\;\alpha\,\xi_\circ\,.
  \end{split}
 \]
 From the arbitrariness of $\lambda$ one concludes that the identity \eqref{eq:TMSoncharges} holds true \emph{irrespective of $\lambda$}. Moreover, $T_\lambda\xi+\alpha\xi$ must make sense as a function in $\mathrm{ran}\,W_\lambda=H^{\frac{1}{2}}(\mathbb{R}^3)$ (Lemma \ref{lem:Wlambdaproperties}) irrespective of $\lambda$. This completes the proof of parts (i) and (ii).
  Plugging \eqref{eq:TMSoncharges} into \eqref{eq:g-largep2-star} and \eqref{eq:phi-largep2-star} yields \eqref{eq:g-largep2-TMS0} and \eqref{eq:phi-largep2-TMS0}, and in fact one can be derived one from the other, by comparison with the corresponding general identities \eqref{eq:g-largep2-star} and \eqref{eq:phi-largep2-star}. In turn, \eqref{eq:phi-largep2-TMS0} and \eqref{eq:phi-largep2-star-yversion-TMS0} correspond to each other via Fourier transform. Parts (ii) and (iii) too is proved.
 \end{proof}

 \begin{remark}
  Owing to Lemma \ref{eq:oneTMSfunction}(ii), \eqref{eq:phi-largep2-star-yversion-TMS0} must be an identity in $H^{\frac{1}{2}}(\mathbb{R}^3)$. This is indeed consistent with the $H^2(\mathbb{R}^3\times\mathbb{R}^3)\to H^{\frac{1}{2}}(\mathbb{R}^3)$ trace properties.
 \end{remark}


 \begin{remark}
  The `lesson' from Lemma \ref{eq:oneTMSfunction} is that TMS condition \emph{and} self-adjoint\-ness of the extension impose strong restrictions. A function $g$ fulfilling the TMS condition \emph{inside the domain of a self-adjoint extension of $\mathring{H}$} must satisfy the restrictions \eqref{eq:xixicirc}-\eqref{eq:Ta12}  $\forall\lambda>0$. 
 \end{remark}

 \subsection{Generalities on TMS extensions}\label{sec:generalitiesTMSext}
 
 There are two relevant types of operators related with $\mathring{H}$ and compatible with the emergence of large-momenta / short-scale asymptotics of Ter-Martirosyan-Skornyakov / Bethe-Peierls type:\index{contact condition!Bethe-Peierls}\index{contact condition!Ter-Martirosyan-Skornyakov}
 
 \noindent \emph{an operator $\mathscr{H}$ on $\cH_\mathrm{b}$ such that $\mathring{H}\subset\mathscr{H}=\mathscr{H}^*$ (respectively, $\mathring{H}\subset\mathscr{H}\subset\mathscr{H}^*$)  and such that \emph{every} $g\in\mathcal{D}(\mathscr{H})$ satisfies the TMS condition \eqref{eq:g-largep2-TMS0} with the same given $\alpha$ shall be called a Ter-Martirosyan-Skornyakov self-adjoint extension (respectively, Ter-Martirosyan-Skornyakov symmetric extension) of $\mathring{H}$ with parameter $\alpha$}.

 \begin{remark}\label{rem:TMSsym}
 While the above definition in the \emph{self-adjoint} case is self-explanatory, based on the preceding analysis, as the self-adjoint extensions of $\mathring{H}$ are classified in Theorem \ref{thm:generalclassification} and the circumstance that $g\in\mathcal{D}(\mathscr{H})$ satisfies the TMS condition is analysed in Lemma \ref{eq:oneTMSfunction}, a clarification is in order for the \emph{symmetric} case.
 In fact, formulas \eqref{eq:domDHA}-\eqref{eq:domDHA-actionDHA} above make sense also when $\mathcal{A}_\lambda$ is simply symmetric (not necessarily self-adjoint) in $H^{-\frac{1}{2}}_{W_\lambda}(\mathbb{R}^3)$, in which case the operator $H_{\mathcal{A}_\lambda}$ thus defined is evidently still an extension of $\mathring{H}$. Let us show that $H_{\mathcal{A}_\lambda}$ is also symmetric. For generic $g\in\mathcal{D}(\mathscr{H})$,
 \[
  \begin{split}
   &\langle g , (H_{\mathcal{A}_\lambda} +\lambda\mathbbm{1})g\rangle_{\cH_\mathrm{b}}\;=\;\big\langle \phi^\lambda+ u_\xi^\lambda , (H_{\mathcal{A}_\lambda} +\lambda\mathbbm{1}) (\phi^\lambda+ u_\xi^\lambda)\big\rangle_{\cH_\mathrm{b}} \\
   &=\;\big\langle \phi^\lambda , (H_{F} +\lambda\mathbbm{1}) \phi^\lambda\big\rangle_{\cH_\mathrm{b}}+ \big\langle u_\xi^\lambda , (H_{F} +\lambda\mathbbm{1}) \phi^\lambda\big\rangle_{\cH_\mathrm{b}} \\
   &=\;\big\langle \phi^\lambda , (H_{F} +\lambda\mathbbm{1}) \phi^\lambda\big\rangle_{\cH_\mathrm{b}}+ \big\langle u_\xi^\lambda , u_{\mathcal{A}_\lambda\xi+\chi}^\lambda\big\rangle_{\cH_\mathrm{b}} \\
    &=\;\big\langle \phi^\lambda , (H_{F} +\lambda\mathbbm{1}) \phi^\lambda\big\rangle_{\cH_\mathrm{b}}+ \langle \xi,\mathcal{A}_\lambda\xi+\chi\rangle_{H^{-\frac{1}{2}}_{W_\lambda}} \\
    &=\;\big\langle \phi^\lambda , (H_{F} +\lambda\mathbbm{1}) \phi^\lambda\big\rangle_{\cH_\mathrm{b}}+ \langle \xi,\mathcal{A}_\lambda\xi\rangle_{H^{-\frac{1}{2}}_{W_\lambda}}\;\in\;\mathbb{R}\,.
  \end{split}
 \]
 (Here $\langle  u_\xi^\lambda , (\mathring{H}+\lambda\mathbbm{1})f^\lambda\rangle_{\cH_\mathrm{b}}=0$ was used in the third step, \eqref{eq:W-scalar-product} in the fourth, and $\chi\perp_\lambda\xi$ in the fifth.) For the reality of the above expression it indeed suffices $\mathcal{A}_\lambda$ to be symmetric in $H^{-\frac{1}{2}}_{W_\lambda}(\mathbb{R}^3)$. The proof of Lemma \ref{eq:oneTMSfunction} can be just repeated for the symmetric $\mathcal{A}_\lambda$ and the same conclusions hold for the symmetric extension $\mathscr{H}$ considered now.
 \end{remark}

 TMS \emph{symmetric} extensions \index{Ter-Martirosyan-Skornyakov!symmetric extension} of $\mathring{H}$ will play a crucial role in Section \ref{sec:lzero}. For the time being, let us focus on TMS \emph{self-adjoint} extensions, \index{Ter-Martirosyan-Skornyakov!self-adjoint extension} and comment on their symmetric counterpart at the end of this Subsection (Remark \ref{rem:remonsym}).

 The requirement that a self-adjoint extension of $\mathring{H}$ \emph{as a whole} be a Ter-Martirosyan-Skornyakov operator imposes a precise choice of the corresponding Birman operators $\mathcal{A}_\lambda$.\index{Birman extension parameter}

 \begin{lemma}\label{lem:generalTMSext}
  Let $\lambda>0$ and $\alpha\in\mathbb{R}$. Let $\mathcal{A}_\lambda\in\mathcal{K}(H^{-\frac{1}{2}}_{W_\lambda}(\mathbb{R}^3))$ and let $\mathring{H}_{\mathcal{A}_\lambda}$ be the corresponding self-adjoint extension of $\mathring{H}$.
  The following two conditions are equivalent.
  \begin{enumerate}[(i)]
   \item Every $g\in\mathcal{D}(\mathring{H}_{\mathcal{A}_\lambda})$ satisfies the TMS condition \eqref{eq:TMSoncharges} with the given $\alpha$.
   \item $\mathcal{D}(\mathcal{A}_\lambda)$ is dense in $H^{-\frac{1}{2}}(\mathbb{R}^3)$,
      $(T_\lambda+\alpha\mathbbm{1})\mathcal{D}(\mathcal{A}_\lambda)\subset H^{\frac{1}{2}}(\mathbb{R}^3)$, and 
   \begin{equation}\label{eq:Alambdaexpression}
     \mathcal{A}_\lambda\;=\;3 W_\lambda^{-1}(T_\lambda+\alpha\mathbbm{1})\,.
   \end{equation}
  \end{enumerate}
  \end{lemma}

 \begin{proof}
  The implication (ii) $\Rightarrow$ (i) is obvious from Lemma \ref{lem:largepasympt-star}. Conversely, if \eqref{eq:TMSoncharges} is to be satisfied by every $g\in\mathcal{D}(\mathring{H}_{\mathcal{A}_\lambda})$, then owing to Lemma \ref{eq:oneTMSfunction}
  \[
   \begin{cases}
    T_\lambda\xi+\alpha\,\xi\in H^{\frac{1}{2}}(\mathbb{R}^3) \\
    \chi=3W_\lambda^{-1}(T_\lambda\xi+\alpha\,\xi)-\mathcal{A}_\lambda\xi
   \end{cases}
   \quad \forall\xi\in\mathcal{D}(\mathcal{A}_\lambda)\,,\;\forall\chi\in\mathcal{D}(\mathcal{A}_\lambda)^{\perp_\lambda}\cap H^{-\frac{1}{2}}(\mathbb{R}^3)\,.
  \]
  The first condition means precisely $(T_\lambda+\alpha\mathbbm{1})\mathcal{D}(\mathcal{A}_\lambda)\subset H^{\frac{1}{2}}(\mathbb{R}^3)$, and the second condition can only be satisfied if $\mathcal{D}(\mathcal{A}_\lambda)^{\perp_\lambda}\cap H^{-\frac{1}{2}}(\mathbb{R}^3)$ is trivial, namely when $\mathcal{D}(\mathcal{A}_\lambda)$ is dense in $H^{-\frac{1}{2}}(\mathbb{R}^3)$ and $\mathcal{A}_\lambda=3 W_\lambda^{-1}(T_\lambda+\alpha\mathbbm{1})$.
  \end{proof}

 \begin{theorem}\label{thm:globalTMSext}
  Let $\alpha\in\mathbb{R}$ and let $\mathscr{H}$ be an operator on $\cH_{\mathrm{b}}$. The following two possibilities are equivalent.
  \begin{enumerate}[(i)]
   \item $\mathscr{H}$ is a Ter-Martirosyan-Skornyakov self-adjoint extension of $\mathring{H}$ with inverse scattering length $\alpha$.
   \item There exists a subspace $\mathcal{D}\subset H^{-\frac{1}{2}}(\mathbb{R}^3)$ such that, for one and hence for all $\lambda>0$,
   \begin{itemize}
   \item[1.] $\mathcal{D}$ is dense in $H^{-\frac{1}{2}}(\mathbb{R}^3)$,
   \item[2.] $(T_\lambda+\alpha\mathbbm{1})\mathcal{D}\subset H^{\frac{1}{2}}(\mathbb{R}^3)$,
   \item[3.] the operator
   \begin{equation}\label{eq:Alfinally}
    \begin{split}
     \mathcal{A}_\lambda\;&:=\;3 W_\lambda^{-1}(T_\lambda+\alpha\mathbbm{1}) \, ,\\
     \mathcal{D}(\mathcal{A}_\lambda)\;&:=\;\mathcal{D}
    \end{split}
   \end{equation}
   is self-adjoint in $H^{-\frac{1}{2}}_{W_\lambda}(\mathbb{R}^3)$,
   \item[4.] $\mathscr{H}=\mathring{H}_{\mathcal{A}_\lambda}$.
   \end{itemize}   
  \end{enumerate}
   When (i) or (ii) are matched, for one and hence for all $\lambda>0$ one has
   \begin{equation}\label{eq:HAl-tms-dom}
  \mathcal{D}(\mathscr{H})\;=\;
  \left\{g=\phi^\lambda+u_\xi^\lambda\left|\!
  \begin{array}{c}
   \phi^\lambda\in H^2_\mathrm{b}(\mathbb{R}^3\times\mathbb{R}^3)\,,\;\xi\in\mathcal{D}\,, \\
   \displaystyle\int_{\mathbb{R}^3}\widehat{\phi^\lambda}(\pp_1,\pp_2)\,\ud\pp_2\,=\,(\widehat{T_\lambda\xi})(\pp_1)+\alpha\,\widehat{\xi}(\pp_1)
  \end{array}
  \!\!\!\right.\right\} \, ,
  \end{equation}
  where for each $\lambda$ the above decomposition of $g$ in terms of $\phi^\lambda$ and $\xi$ is unique, and 
  \begin{equation}\label{eq:HAl-tms-act}
   (\mathscr{H}+\lambda\mathbbm{1})g\;=\;(\mathring{H}_{\mathrm{F}}+\lambda\mathbbm{1})\phi^\lambda\,.
  \end{equation}	
   \end{theorem}

   \begin{proof}
    Assume that every $g\in\mathcal{D}(\mathscr{H})$ satisfies the TMS asymptotics \eqref{eq:g-largep2-TMS0} with the same $\alpha$.  Applying Lemma \ref{lem:generalTMSext} one obtains all four conditions 1.~through 4.~listed in part (ii), for every $\lambda>0$, except that $\mathcal{D}$ is replaced by $\mathcal{D}(\mathcal{A}_\lambda)$ for each considered $\lambda$. But all such $\mathcal{D}(\mathcal{A}_\lambda)$'s are in fact the same subspace (Remark \ref{rem:samedomains}). 
%
    The proof of (i) $\Rightarrow$ (ii) is completed.
    
    Conversely, assume that (ii) holds true for \emph{one} $\lambda_\circ>0$. Applying Lemma \ref{lem:generalTMSext} one deduces that $\mathscr{H}$ is a Ter-Martirosyan-Skornyakov self-adjoint extension \index{Ter-Martirosyan-Skornyakov!self-adjoint extension} of $\mathring{H}$ with inverse scattering length $\alpha$. Since one knows already that (i) $\Rightarrow$ (ii), then condition (ii) holds true for \emph{any} other $\lambda>0$ as well. This establishes the full implication (ii) $\Rightarrow$ (i).
    
    Under condition (i), or equivalently (ii), \eqref{eq:HAl-tms-dom}-\eqref{eq:HAl-tms-act} then follow from \eqref{eq:domDHA}-\eqref{eq:domDHA-actionDHA} of Theorem \ref{thm:generalclassification}(i) and from \eqref{eq:phi-largep2-TMS0} and \eqref{eq:Alfinally}. 
   \end{proof}

  It is worth stressing that formulas \eqref{eq:HAl-tms-dom}-\eqref{eq:HAl-tms-act} alone, considered for some subspace $\mathcal{D}$ of $H^{-\frac{1}{2}}(\mathbb{R}^3)$, evidently define an extension $\mathscr{H}$ of $\mathring{H}$; however, they do not necessarily make $\mathscr{H}$ a self-adjoint extension. It is convenient for later purposes to formulate this point in the form of a separate corollary.

  \begin{corollary}\label{cor:globalTMSext}
   Let $\alpha\in\mathbb{R}$, $\lambda>0$, let $\mathcal{D}$ be a subspace of $H^{-\frac{1}{2}}(\mathbb{R}^3)$, and let $\mathscr{H}$ be the operator defined by \eqref{eq:HAl-tms-dom}-\eqref{eq:HAl-tms-act}. Then $\mathscr{H}$ is self-adjoint in $\cH_\mathrm{b}$ if and only if $\mathcal{D}$ is dense in $H^{-\frac{1}{2}}(\mathbb{R}^3)$, $(T_\lambda+\alpha\mathbbm{1})\mathcal{D}\subset H^{\frac{1}{2}}(\mathbb{R}^3)$, and the operator \eqref{eq:Alfinally} is self-adjoint in $H^{-\frac{1}{2}}_{W_\lambda}(\mathbb{R}^3)$.   
  \end{corollary}

  Thus, the quest of Ter-Martirosyan-Skornyakov self-adjoint extensions \index{Ter-Martirosyan-Skornyakov!self-adjoint extension} of $\mathscr{H}$ in $\cH_{\mathrm{b}}$ is boiled down to the self-adjointness problem of $W_\lambda^{-1}(T_\lambda+\alpha\mathbbm{1})$ in $H^{-\frac{1}{2}}_{W_\lambda}(\mathbb{R}^3)$ with domain $\mathcal{D}$, hence in practice to the \emph{problem of finding a domain of self-adjointness for the formal action $\xi\mapsto W_\lambda^{-1}(T_\lambda+\alpha\mathbbm{1})\xi$}. This task actually constitutes the hard part of the rigorous modelling of physically meaningful Hamiltonians of zero-range interactions for the considered bosonic trimer.

  For an operator satisfying either condition of Theorem \ref{thm:globalTMSext} one shall use the natural notation $\mathscr{H}_\alpha$, so as to emphasise the only relevant parameter of the considered Ter-Martirosyan-Skornyakov (self-adjoint) extension of $\mathring{H}$. This must be done keeping in mind that in principle for the same $\alpha\in\mathbb{R}$ there could be distinct operators of the form $\mathscr{H}_\alpha$, that is, distinct domains of self-adjointness for $W_\lambda^{-1}(T_\lambda+\alpha\mathbbm{1})$ in $H^{-\frac{1}{2}}_{W_\lambda}(\mathbb{R}^3)$ (Corollary \ref{cor:globalTMSext}), in analogy with the familiar existence of a variety of distinct domains of self-adjointness in $L^2(0,1)$ for the same differential operator $-\frac{\ud^2}{\ud x^2}$.

  \begin{remark}\label{rem:remonsym}
   The reasonings that led to Theorem \ref{thm:globalTMSext} have an obvious counterpart for Ter-Martirosyan-Skornyakov \emph{symmetric} extensions \index{Ter-Martirosyan-Skornyakov!symmetric extension} of $\mathring{H}$. 
   \begin{enumerate}[(i)]
    \item Lemma \ref{lem:generalTMSext} is equally valid when $\mathcal{A}_\lambda$ is only assumed to be symmetric in $H^{-\frac{1}{2}}_{W_\lambda}(\mathbb{R}^3)$ and the corresponding $\mathring{H}_{\mathcal{A}_\lambda}$ is a Ter-Martirosyan-Skornyakov symmetric extension of $\mathring{H}$, based on the observations made in Remark \ref{rem:TMSsym}. 
    \item By means of such a `symmetric version' of Lemma \ref{lem:generalTMSext}, the proof of Theorem \ref{thm:globalTMSext} can be straightforwardly adjusted so as to establish that:
    
    \noindent \emph{$\mathscr{H}$ is a Ter-Martirosyan-Skornyakov symmetric extension of $\mathring{H}$ with inverse scattering length $\alpha\in\mathbb{R}$ if and only if there exists a subspace $\mathcal{D}\subset H^{-\frac{1}{2}}(\mathbb{R}^3)$ such that, for one and hence for all $\lambda>0$, $\mathcal{D}$ is dense in $H^{-\frac{1}{2}}(\mathbb{R}^3)$, $(T_\lambda+\alpha\mathbbm{1})\mathcal{D}\subset H^{\frac{1}{2}}(\mathbb{R}^3)$, the operator $\mathcal{A}_\lambda:=3 W_\lambda^{-1}(T_\lambda+\alpha\mathbbm{1})$ is symmetric in $H^{-\frac{1}{2}}_{W_\lambda}(\mathbb{R}^3)$ on the domain $\mathcal{D}$, and $\mathscr{H}=\mathring{H}_{\mathcal{A}_\lambda}$.}    
%
   \end{enumerate}
  \end{remark}

  \subsection{Symmetry and self-adjointness of the TMS parameter}\index{Ter-Martirosyan-Skornyakov!extension parameter}

  As emerged in Subsect.~\ref{sec:generalitiesTMSext}, the operator \eqref{eq:Alfinally} is the correct Birman operator\index{Birman extension parameter} labelling symmetric or self-adjoint TMS extensions of $\mathring{H}$ in terms of the general parametrisation provided by Theorem \ref{thm:generalclassification} (and Remarks \ref{rem:TMSsym} and \ref{rem:remonsym}).

  The symmetry or self-adjointness, in the respective Hilbert spaces, of the auxiliary operators $\mathcal{A}_\lambda$ and $T_\lambda$ on the domain $\mathcal{D}$ are closely related (albeit deceptively, in a sense), as will be now discussed.

  To avoid ambiguities, the standard notation $T^*$, $\overline{T}$, etc., is reserved for the adjoint of $T$, its operator closure, and so on, with respect to the underlying $L^2$-space (as done for $\mathring{H}^*$ as an operator on $\cH_{\mathrm{b}}$), which in this context shall be $L^2(\mathbb{R}^3)$, and one writes instead $\mathcal{A}_\lambda^\star$, $\overline{\mathcal{A}_\lambda}^\lambda$, $\perp_\lambda$, etc., with reference to $H^{-\frac{1}{2}}_{W_\lambda}(\mathbb{R}^3)$.

  \begin{lemma}\label{lem:symsym}
   Let $\lambda>0$, $\alpha\in\mathbb{R}$, and let $\mathcal{D}$ be a dense subspace of $L^2(\mathbb{R}^3)$ such that $(T_\lambda+\alpha\mathbbm{1})\mathcal{D}\subset H^{\frac{1}{2}}(\mathbb{R}^3)$. Consider both $\mathcal{A}_\lambda:=3W_\lambda^{-1}(T_\lambda+\alpha\mathbbm{1})$ and $T_\lambda$ as operators with domain $\mathcal{D}$. Then
   \[
    \mathcal{A}_\lambda\;\subset\;\mathcal{A}_\lambda^\star\qquad \Leftrightarrow \qquad T_\lambda\;\subset\;T_\lambda^*\,,
   \]
  that is, the symmetry of $\mathcal{A}_\lambda$ in $H^{-\frac{1}{2}}_{W_\lambda}(\mathbb{R}^3)$ is equivalent to the symmetry of $T_\lambda$ in $L^2(\mathbb{R}^3)$.  
  \end{lemma}

  \begin{proof}
   $\mathcal{D}$ is dense in $L^2(\mathbb{R}^3)$ and hence in $H^{-\frac{1}{2}}(\mathbb{R}^3)\cong H^{-\frac{1}{2}}_{W_\lambda}(\mathbb{R}^3)$.  
   Owing to \eqref{eq:W-scalar-product},
   \[
    \langle \xi, \mathcal{A}_\lambda\xi\rangle_{H^{-\frac{1}{2}}_{W_\lambda}}\;=\;3\langle\xi,(T_\lambda+\alpha\mathbbm{1})\xi\rangle_{L^2}\qquad\forall\xi\in\mathcal{D}\,.
   \]
  Therefore, the reality of the l.h.s.~is equivalent to the reality of the r.h.s.
  \end{proof}

  \begin{lemma}\label{lem:two-selfadj-problems}
   Let $\lambda>0$, $\alpha\in\mathbb{R}$, and let $\mathcal{D}$ be a dense subspace of $L^2(\mathbb{R}^3)$ such that $(T_\lambda+\alpha\mathbbm{1})\mathcal{D}\subset H^{\frac{1}{2}}(\mathbb{R}^3)$. Consider both $\mathcal{A}_\lambda:=3W_\lambda^{-1}(T_\lambda+\alpha\mathbbm{1})$ and $T_\lambda$ as operators with domain $\mathcal{D}$. Assume that $T_\lambda\,=\,T_\lambda^*$. Then,
   \begin{enumerate}[(i)]
    \item $\mathcal{D}(\mathcal{A}_\lambda^\star)\cap L^2(\mathbb{R}^3)\,=\,\mathcal{D}(\mathcal{A}_\lambda)=\mathcal{D}$;
    \item $\mathcal{A}_\lambda\,=\,\mathcal{A}_\lambda^\star$ if and only if $\mathcal{D}(\mathcal{A}_\lambda^\star)\subset L^2(\mathbb{R}^3)$.
   \end{enumerate}
  \end{lemma}

  \begin{proof} Clearly (ii) follows from (i). Concerning (i), the inclusion $\mathcal{D}(\mathcal{A}_\lambda^\star)\cap L^2(\mathbb{R}^3)\supset\mathcal{D}(\mathcal{A}_\lambda)$ is obvious. Let now $\eta\in\mathcal{D}(\mathcal{A}_\lambda^\star)\cap L^2(\mathbb{R}^3)$. Then, for some $c_\eta>0$,
   \[
   \Big| \langle\eta,\mathcal{A}_\lambda\xi\rangle_{H^{-\frac{1}{2}}_{W_\lambda}}\Big|\;\leqslant\; c_\eta\,\|\xi\|_{L^2}^2\qquad\forall\xi\in\mathcal{D}=\mathcal{D}(\mathcal{A}_\lambda)\,.
  \]
    Equivalently, owing to \eqref{eq:W-scalar-product},
  \[
   \big| \langle\eta,(T_\lambda+\alpha\mathbbm{1})\xi\rangle_{L^2}\big|\;\leqslant\; {\textstyle\frac{1}{3}}c_\eta\,\|\xi\|_{L^2}^2\qquad\forall\xi\in\mathcal{D}=\mathcal{D}(T_\lambda)\,.
  \]
   Therefore, $\eta\in\mathcal{D}(T_\lambda^*)=\mathcal{D}(T_\lambda)=\mathcal{D}(\mathcal{A}_\lambda)$.  
  \end{proof}

 \subsection{TMS extensions in sectors of definite angular momentum}\label{sec:definite-ell-general}

 As the maps $\xi\mapsto T_\lambda\xi$, $\xi\mapsto W_\lambda\xi$, $\xi\mapsto W_\lambda^{-1}\xi$ all commute with the rotations in $\mathbb{R}^3$ (Subsect.~\ref{sec:Tlambdaoperator}), and so too does therefore the map $\xi\mapsto W_\lambda^{-1}T_\lambda$, then the TMS parameter \index{Ter-Martirosyan-Skornyakov!extension parameter} $\mathcal{A}_\lambda=3W_\lambda^{-1}(T_\lambda+\alpha\mathbbm{1})$ is naturally reduced in each sector of definite angular momentum.

 More precisely, with respect to the decomposition \eqref{eq:bigdecomp}-\eqref{eq:xihatangularexpansion}, and following the same reasoning therein, one then has
\begin{equation}\label{eq:WlambdaWlambdaell}
 \widehat{W_\lambda\xi}\;=\;\sum_{\ell=0}^\infty \widehat{W_\lambda^{(\ell)}\xi^{(\ell)}}\,,
\end{equation}
where each $W_\lambda^{(\ell)}$ is non-trivial only radially. Moreover,
 \begin{equation}
  \begin{split}
   \langle \xi,\eta\rangle_{H^{-\frac{1}{2}}_{W_\lambda}}\;&=\; \int_{\mathbb{R}^3} \overline{\,\widehat{\xi}(\pp)}\, \big(\widehat{W_\lambda\eta}\big)(\pp)\,\ud\pp \\
   &=\;\sum_{\ell=0}^{\infty}\int_{\mathbb{R}^3} \overline{\,\widehat{\xi^{(\ell)}}(\pp)}\, \big(\widehat{W_\lambda^{(\ell)}\eta^{(\ell)}}\big)(\pp)\,\ud\pp\,,
  \end{split}
 \end{equation}
where
 \begin{equation}\label{eq:Wellsp}
   \begin{split}
   &\int_{\mathbb{R}^3} \overline{\,\widehat{\xi^{(\ell)}}(\pp)}\, \big(\widehat{W_\lambda^{(\ell)}\eta^{(\ell)}}\big)(\pp)\,\ud\pp \\
   &\;=\;\sum_{n=-\ell}^{\ell}\bigg(\int_{\mathbb{R}^+}\!\ud p\,p^2\,\frac{3\pi^2}{\sqrt{{\textstyle\frac{3}{4}}p^2+\lambda}\,}\,\overline{f_{\ell,n}^{(\xi)}(p)}\,f_{\ell,n}^{(\eta)}(p) \\
   &\quad +12\pi\!\iint_{\mathbb{R}^+\times\mathbb{R}^+}\ud p\,\ud q\,p^2q^2\,\overline{f_{\ell,n}^{(\xi)}(p)}\,f_{\ell,n}^{(\eta)}(q)\!\int_{-1}^1\ud t\,\frac{P_\ell(t)}{(p^2+q^2+p\,q \,t+\lambda)^2}\bigg) \\
   &\;\equiv\;\sum_{n=-\ell}^{\ell}\big\langle f_{\ell,n}^{(\xi)},f_{\ell,n}^{(\eta)} \big\rangle_{W_\lambda^{(\ell)}}\,,
    \end{split}
  \end{equation}
in complete analogy to \eqref{eq:xiTxipre}-\eqref{eq:xiTxi}.

As the scalar product $\langle\cdot,W_\lambda\cdot\rangle_{H^{-\frac{1}{2}},H^{\frac{1}{2}}}$ is equivalent to the ordinary $H^{-\frac{1}{2}}$-scalar product, so is the scalar product $\langle\cdot,\cdot\rangle_{W_\lambda^{(\ell)}}$ defined by \eqref{eq:Wellsp} equivalent to the ordinary scalar product in $ L^2(\mathbb{R}^+,(1+p^2)^{-\frac{1}{2}} p^2\ud p) $. The latter is therefore a Hilbert space also when equipped with $\langle\cdot,\cdot\rangle_{W_\lambda^{(\ell)}}$, in which case one shall denote it with $L^2_{W_\lambda^{(\ell)}}(\mathbb{R}^+)$. One thus has the canonical Hilbert space isomorphism
\begin{equation}\label{eq:RpRpW}
 L^2(\mathbb{R}^+,(1+p^2)^{-\frac{1}{2}} p^2\ud p)\;\cong\;L^2_{W_\lambda^{(\ell)}}(\mathbb{R}^+)\,.
\end{equation}
By means of \eqref{eq:RpRpW} one re-writes \eqref{eq:bigdecomp} as
\begin{equation}\label{eq:bigdecompW}
 \begin{split}
 H^{-\frac{1}{2}}_{W_\lambda}(\mathbb{R}^3)\;&\cong\;H^{-\frac{1}{2}}(\mathbb{R}^3)\; \\
 &\cong\;\bigoplus_{\ell=0}^\infty \Big( L^2_{W_\lambda^{(\ell)}}(\mathbb{R}^+) \otimes \mathrm{span}\big\{ Y_{\ell,n}\,|\,n=-\ell,\dots,\ell \big\} \Big)  \\
 &\equiv\;\bigoplus_{\ell=0}^\infty \,H^{-\frac{1}{2}}_{W_\lambda,\ell}(\mathbb{R}^3) \,.
 \end{split}
\end{equation}
The expansion \eqref{eq:xihatangularexpansion} of a generic $\xi\in H^{-\frac{1}{2}}(\mathbb{R}^3)$ is equivalently referred to the ordinary decomposition or the $\lambda$-decomposition of the space \eqref{eq:bigdecompW}.

 With respect to \eqref{eq:bigdecompW} $\mathcal{A}_\lambda$ is reduced as
 \begin{equation}\label{eq:Alamdbareducedell}
  \mathcal{A}_{\lambda}\;=\;\bigoplus_{\ell=0}^{\infty}\,\mathcal{A}_{\lambda}^{(\ell)}\;=\;\bigoplus_{\ell=0}^{\infty}\,3 W_\lambda^{-1}(T_\lambda^{(\ell)}+\alpha\mathbbm{1})\,.
 \end{equation}
 The problem of finding a domain $\mathcal{D}$ of symmetry or of self-adjointness for $\mathcal{A}_\lambda$ with respect to $H^{-\frac{1}{2}}_{W_\lambda}(\mathbb{R}^3)$ is tantamount as finding a domain $\mathcal{D}_\ell$ of symmetry or of self-adjointness for $\mathcal{A}_\lambda^{(\ell)}$ with respect to $H^{-\frac{1}{2}}_{W_\lambda,\ell}(\mathbb{R}^3)$ for each $\ell\in\mathbb{N}_0$. This is the object of Section \ref{sec:higherell} and \ref{sec:lzero}.

 The subspace of $\mathcal{D}(\mathcal{A}_\lambda)$ consisting of elements $g$ with charge $\xi\in\mathcal{D}_\ell$ is sometimes referred to as the \emph{charge domain}\index{charge domain} of the TMS (symmetric or self-adjoint) extension of $\mathring{H}$ in the $\ell$-th sector of definite angular momentum.

\section{Sectors of higher angular momenta}\label{sec:higherell}

 In the modelling of the bosonic trimer with zero-range interaction, all the relevant physics is expected in the sector of \emph{zero} angular momentum, since in each two-body channel particles undergo a low-energy, and hence essentially an $s$-wave scattering.

 In this respect, the \emph{characterisation} of the quantum Hamiltonian is somewhat artificial in the sectors of higher (non-zero) angular momentum, as in practice experimental observations do not involve states in such sectors. For instance one could simply consider a Hamiltonian where for each $\ell\in\mathbb{N}$ the Birman parameter\index{Birman extension parameter} of formula \eqref{eq:domDHA} has domain $\mathcal{D}_\ell=\{0\}$ and value `$\mathcal{A}_\lambda=\infty$' on it, namely the Friedrichs extension of $\mathring{H}$ in those sectors, whereas only the $\ell=0$ is defined non-trivially. This would model a total absence of interaction at higher angular momenta.

 A more typical choice is to define a model that in all $\ell$-sectors, not only $\ell=0$, is characterised by the physical TMS asymptotics with inverse scattering length, and to do so by making a somewhat canonical construction for $\ell\neq 0$, and a non-trivial one for $\ell=0$. Such programme is presented in this Section for non-zero $\ell$. The analysis of $\ell=0$ is deferred to Section \ref{sec:lzero}.
 

 \subsection{$T_\lambda$-estimates}\label{sec:Tlambdaestimates}
 
 Let us import here a set of useful estimates established in the already mentioned work \cite{CDFMT-2012} by Correggi, Dell'Antonio, Finco, Michelangeli, and Teta.

 For given $\lambda>0$ and $\ell\in\mathbb{N}$ one introduces the shorthands 
 \begin{equation}
   \begin{split}
    \Phi_\lambda[f,g]\;&:=\;2\pi^2\!\int_{\mathbb{R}^+}\!\ud p\,p^2\,\sqrt{{\textstyle\frac{3}{4}}p^2+\lambda}\,\overline{f(p)}\,g(p) \, ,\\
    \Psi_{\lambda,\ell}[f,g]\;&:=\;2\pi\!\iint_{\mathbb{R}^+\times\mathbb{R}^+}\ud p\,\ud q\,p^2q^2\,\overline{f(p)}\,g(q)\!\int_{-1}^1\ud t\,\frac{P_\ell(t)}{p^2+q^2+p \,q \,t+\lambda} \, ,
   \end{split}
 \end{equation}
 and 
  \begin{equation}
    \begin{split}
   \Phi_\lambda[f]\;&:=\;\Phi_\lambda[f,f] \, ,\\
   \Psi_{\lambda,\ell}[f]\;&:=\;\Psi_{\lambda,\ell}[f,f] \, ,
  \end{split}
  \end{equation}
 so that \eqref{eq:xiTxipre}-\eqref{eq:xiTxi} now read
  \begin{equation}\label{eq:xiTxi-updated}
   \int_{\mathbb{R}^3} \overline{\,\widehat{\xi}(\pp)}\, \big(\widehat{T_\lambda\eta}\big)(\pp)\,\ud\pp\;=\;\sum_{\ell=0}^\infty\sum_{n=-\ell}^\ell\Big( \Phi_\lambda\big[f_{\ell,n}^{(\xi)},f_{\ell,n}^{(\eta)}\big]-2\Psi_{\lambda,\ell}\big[f_{\ell,n}^{(\xi)},f_{\ell,n}^{(\eta)}\big]\Big)\,.
  \end{equation}

  \begin{lemma}
   Let $\lambda>0$ and $\ell\in\mathbb{N}$. Let $f:\mathbb{R}\to\mathbb{C}$ make the quantities below finite.
   \begin{enumerate}[(i)]
    \item One has
    \begin{equation}\label{eq:Psilambdaordering}
     \begin{split}
    0\;\leqslant\;\Psi_{\lambda,\ell}[f]\;\leqslant\;\Psi_{0,\ell}[f] & \qquad\textrm{for even $\ell$} \, , \\
    \Psi_{0,\ell}[f]\;\leqslant\;\Psi_{\lambda,\ell}[f]\;\leqslant\; 0 & \qquad\textrm{for odd $\ell$}.
   \end{split}
    \end{equation}
    \item One has 
    \begin{equation}\label{eq:PhiPsilambdazero}
     \begin{split}
      \Phi_0[f]\;&=\;\pi^2\sqrt{3}\int_\mathbb{R}\ud s\,|f^\sharp(s)|^2 \, ,\\
      \Psi_{0,\ell}[f]\;&=\;\int_\mathbb{R}\ud s\,S_\ell(s)\,|f^\sharp(s)|^2 \, ,
     \end{split}
    \end{equation}
    where
    \begin{equation}
     f^\sharp(s)\;:=\;\frac{1}{\sqrt{2\pi}}\int_{\mathbb{R}}\,\ud x\,e^{-\ii s x}\,e^{2x}f(e^x)
    \end{equation} 
    and 
    \begin{equation}\label{eq:Sell}
     S_\ell(s)\;:=\;2\pi^2\int_{-1}^1\ud t\,P_\ell(t)\,\frac{\sinh(s\arccos\frac{t}{2})}{\sin(\arccos\frac{t}{2})\,\sinh\pi s}\,.
    \end{equation}
    \item $S_\ell:\mathbb{R}\to\mathbb{R}$ is a smooth even function, strictly monotone on $\mathbb{R}^\pm$, and such that
    \begin{equation}\label{eq:Slzeroordering}
     \begin{split}
      0\;\leqslant\;S_{\ell+2}(s)\;\leqslant\; S_{\ell}(s)\;\leqslant\;S_{\ell}(0) & \qquad\textrm{for even $\ell$} \, ,\\
      S_{\ell}(0)\;\leqslant\;S_{\ell}(s)\;\leqslant\;S_{\ell+2}(s)\;\leqslant\;0   & \qquad\textrm{for odd $\ell$} \, .
     \end{split}
    \end{equation}
   \end{enumerate}      
  \end{lemma}

  \begin{proof}
   Part (i) follows from \cite[Lemma 3.2]{CDFMT-2012}. Part (ii) from \cite[Lemma 3.3]{CDFMT-2012}. Part (iii) from \cite[Lemma 3.5]{CDFMT-2012}.   
  \end{proof}

  The values of $S_\ell(s)$ that will be relevant in the present analysis are
  \begin{equation}\label{eq:Sellzerospecialvalues}
   \begin{split}
    S_0(0)\;&=\;\frac{\,2\pi^3}{3}\;>\;0 \, ,\\
    S_1(0)\;&=\;-8\pi\Big(1-\frac{\pi}{2\sqrt{3}}\Big)\;<\;0 \, ,\\
    S_2(0)\;&=\;\frac{\,\pi^2}{3}(5\pi-9\sqrt{3})\;>\;0\,,
   \end{split}
  \end{equation}
 as one easily computes from \eqref{eq:Sell} and \eqref{def_Legendre}.

  By means of the estimates above, one obtains the following important bounds.

  \begin{lemma}\label{lem:xiTxi-equiv-H12}
   Let $\lambda>0$ and let $\xi\in H^{\frac{1}{2}}(\mathbb{R}^3)$.
   Then
   \begin{eqnarray}
    \int_{\mathbb{R}^3} \overline{\,\widehat{\xi}(\pp)}\, \big(\widehat{T_\lambda\xi}\big)(\pp)\,\ud\pp&\leqslant&\kappa^+\cdot 2\pi^2\!\int_{\mathbb{R}^3}\sqrt{\frac{3}{4}\pp^2+\lambda}\,|\widehat{\xi}(\pp)|^2\,\ud\pp \, ,	\label{eq:xiTxiFromAbove} \\
    \int_{\mathbb{R}^3} \overline{\,\widehat{\xi}(\pp)}\, \big(\widehat{T_\lambda\xi}\big)(\pp)\,\ud\pp&\geqslant&\kappa^-\cdot 2\pi^2\!\int_{\mathbb{R}^3}\sqrt{\frac{3}{4}\pp^2+\lambda}\,|\widehat{\xi}(\pp)|^2\,\ud\pp \, ,	\label{eq:xiTxiFromBelow}
   \end{eqnarray}
  where
  \begin{equation}
  \begin{split}
   \kappa^+\;&=\;\frac{16}{\pi\sqrt{3}}-\frac{5}{3}  \, ,\\
    \kappa^-\;&=\;
   \begin{cases}
    -\displaystyle\Big(\frac{4\pi}{3\sqrt{3}}-1\Big) & \textrm{ if $\xi$ is non-trivial on $H^{\frac{1}{2}}_{\ell=0}(\mathbb{R}^3)$} \, ,\\
    \quad 7-\frac{10\pi}{3\sqrt{3}} & \textrm{ if }\;\xi\in\bigoplus_{\ell=1}^\infty H^{\frac{1}{2}}_\ell(\mathbb{R}^3)\,.
   \end{cases}
  \end{split}
  \end{equation}
 In particular, if $\xi\perp  H^{\frac{1}{2}}_{\ell=0}(\mathbb{R}^3)$, then $\kappa^->0$ and
 \begin{equation}
  \int_{\mathbb{R}^3} \overline{\,\widehat{\xi}(\pp)}\, \big(\widehat{T_\lambda\xi}\big)(\pp)\,\ud\pp\;\approx\;\|\xi\|_{H^{\frac{1}{2}}}^2\qquad (\ell\neq 0)
 \end{equation}
  in the sense of equivalence of norms (with $\lambda$-dependent multiplicative constants).
  \end{lemma}

  \begin{proof}
   Expanding $\xi$ as in \eqref{eq:xihatangularexpansion} and using \eqref{eq:xiTxi-updated} one has   
    \[\tag{*}\label{eq:xtxisums}
   \int_{\mathbb{R}^3} \overline{\,\widehat{\xi}(\pp)}\, \big(\widehat{T_\lambda\xi}\big)(\pp)\,\ud\pp\;=\;\sum_{\ell=0}^\infty\sum_{n=-\ell}^\ell\Big( \Phi_\lambda\big[f_{\ell,n}^{(\xi)}\big]-2\Psi_{\lambda,\ell}\big[f_{\ell,n}^{(\xi)}\big]\Big)\,.
  \]
   Owing to \eqref{eq:Psilambdaordering}, \eqref{eq:PhiPsilambdazero}, \eqref{eq:Slzeroordering}, and \eqref{eq:Sellzerospecialvalues},
   \[
    \begin{split}
     &\sum_{\ell=0}^\infty\sum_{n=-\ell}^\ell\Psi_{\lambda,\ell}\big[f_{\ell,n}^{(\xi)}\big]\;\;\geqslant\;\sum_{\substack{  \ell\in\mathbb{N}_0 \\ \ell\textrm{ odd} }}\sum_{n=-\ell}^\ell\Psi_{\lambda,\ell}\big[f_{\ell,n}^{(\xi)}\big]\;\geqslant\;\sum_{\substack{  \ell\in\mathbb{N}_0 \\ \ell\textrm{ odd} }}\sum_{n=-\ell}^\ell\Psi_{0,\ell}\big[f_{\ell,n}^{(\xi)}\big] \\
     &\;=\;\sum_{\substack{  \ell\in\mathbb{N}_0 \\ \ell\textrm{ odd} }}\sum_{n=-\ell}^\ell\int_\mathbb{R}\ud s\,S_\ell(s)\,\big|\big(f_{\ell,n}^{(\xi)}\big)^\sharp(s)\big|^2\;\geqslant\;S_1(0)\sum_{\substack{  \ell\in\mathbb{N}_0 \\ \ell\textrm{ odd} }}\sum_{n=-\ell}^\ell\int_\mathbb{R}\ud s\,\big|\big(f_{\ell,n}^{(\xi)}\big)^\sharp(s)\big|^2 \\
     &=\;-\frac{\,8\pi(1-\frac{\pi}{2\sqrt{3}})\,}{\pi^2\sqrt{3}}\sum_{\substack{  \ell\in\mathbb{N}_0 \\ \ell\textrm{ odd} }}\sum_{n=-\ell}^\ell\Phi_0\big[f_{\ell,n}^{(\xi)}\big]\;\geqslant\;-\frac{4}{3}\Big(\frac{2\sqrt{3}}{\pi}-1\Big)\sum_{\ell=0}^\infty\sum_{n=-\ell}^\ell\Phi_0\big[f_{\ell,n}^{(\xi)}\big] \\
     &\geqslant\;-\frac{4}{3}\Big(\frac{2\sqrt{3}}{\pi}-1\Big)\sum_{\ell=0}^\infty\sum_{n=-\ell}^\ell\Phi_\lambda\big[f_{\ell,n}^{(\xi)}\big]\,.
    \end{split}
   \]
    Plugging this into \eqref{eq:xtxisums} yields \eqref{eq:xiTxiFromAbove}. Analogously,
      \[
    \begin{split}
     &\sum_{\ell=0}^\infty\sum_{n=-\ell}^\ell\Psi_{\lambda,\ell}\big[f_{\ell,n}^{(\xi)}\big]\;\;\leqslant\;\sum_{\substack{  \ell\in\mathbb{N}_0 \\ \ell\textrm{ even} }}\sum_{n=-\ell}^\ell\Psi_{\lambda,\ell}\big[f_{\ell,n}^{(\xi)}\big]\;\leqslant\;\sum_{\substack{  \ell\in\mathbb{N}_0 \\ \ell\textrm{ even} }}\sum_{n=-\ell}^\ell\Psi_{0,\ell}\big[f_{\ell,n}^{(\xi)}\big] \\
     &\;=\;\sum_{\substack{  \ell\in\mathbb{N}_0 \\ \ell\textrm{ even} }}\sum_{n=-\ell}^\ell\int_\mathbb{R}\ud s\,S_\ell(s)\,\big|\big(f_{\ell,n}^{(\xi)}\big)^\sharp(s)\big|^2\;\leqslant\;S_{0}(0)\sum_{\substack{  \ell\in\mathbb{N}_0 \\ \ell\textrm{ even} }}\sum_{n=-\ell}^\ell\int_\mathbb{R}\ud s\,\big|\big(f_{\ell,n}^{(\xi)}\big)^\sharp(s)\big|^2 \\
     &=\;\frac{\,\frac{2\pi^3}{3}\,}{\pi^2\sqrt{3}}\sum_{\substack{  \ell\in\mathbb{N}_0 \\ \ell\textrm{ even} }}\sum_{n=-\ell}^\ell\Phi_0\big[f_{\ell,n}^{(\xi)}\big]\;\leqslant\;\frac{2\pi}{3\sqrt{3}}\sum_{\ell=0}^\infty\sum_{n=-\ell}^\ell\Phi_0\big[f_{\ell,n}^{(\xi)}\big]\;\leqslant\;\frac{2\pi}{3\sqrt{3}}\sum_{\ell=0}^\infty\sum_{n=-\ell}^\ell\Phi_\lambda\big[f_{\ell,n}^{(\xi)}\big]\,,
    \end{split}
   \]
    which combined with \eqref{eq:xtxisums} yields \eqref{eq:xiTxiFromBelow} in the general case. 
    In the particular case when $\xi$ has no $\ell=0$ component the previous computation becomes
    \[
     \begin{split}
       &\sum_{\ell=1}^\infty\sum_{n=-\ell}^\ell\Psi_{\lambda,\ell}\big[f_{\ell,n}^{(\xi)}\big]\;\;\leqslant\;S_{2}(0)\sum_{\substack{  \ell\in\mathbb{N} \\ \ell\textrm{ even} }}\sum_{n=-\ell}^\ell\int_\mathbb{R}\ud s\,\big|\big(f_{\ell,n}^{(\xi)}\big)^\sharp(s)\big|^2 \\
       &=\;\frac{\,\frac{\:\pi^2}{3}(5\pi-9\sqrt{3})\,}{\pi^2\sqrt{3}}\sum_{\substack{  \ell\in\mathbb{N} \\ \ell\textrm{ even} }}\sum_{n=-\ell}^\ell\Phi_0\big[f_{\ell,n}^{(\xi)}\big]\;\leqslant\;\frac{\,5\pi-9\sqrt{3}\,}{3\sqrt{3}}\sum_{\ell=1}^\infty\sum_{n=-\ell}^\ell\Phi_\lambda\big[f_{\ell,n}^{(\xi)}\big]
     \end{split}
    \]
    and plugging the latter estimate into \eqref{eq:xtxisums}, where now the $\ell=0$ summands are absent, one obtains \eqref{eq:xiTxiFromBelow} for this case.
    \end{proof}

  \subsection{Self-adjointness for $\ell\geqslant 1$}\label{sec:selfadj-ellnotzero}

  Here a domain of self-adjointness for the TMS parameter \index{Ter-Martirosyan-Skornyakov!extension parameter} $\mathcal{A}_\lambda$ in $H^{-\frac{1}{2}}_{W_\lambda,\ell}(\mathbb{R}^3)$ will be characterised when $\ell\in\mathbb{N}$. It is convenient to realise first $\mathcal{A}_\lambda$ as a symmetric operator and then construct canonically a self-adjoint realisation of it.

  Set
  \begin{equation}\label{eq:lambdaalpha}
    \lambda_\alpha\;:=\;
    \begin{cases}
     \;\;0 & \textrm{ if }\;\alpha\geqslant 0 \, ,\\
     \alpha^2/(2\pi^2\kappa^-)^2& \textrm{ if }\;\alpha<0\qquad \big(\kappa^-=7-\frac{10\pi}{3\sqrt{3}}\big)\,.
    \end{cases}
  \end{equation}

  \begin{lemma}\label{lem:Atildenot0}
   For $\lambda>0$, $\alpha\in\mathbb{R}$, and $\ell\in\mathbb{N}$, let
  \begin{equation}\label{eq:Dtildeell}
   \widetilde{\mathcal{D}}_\ell\;:=\; H_\ell^{\frac{3}{2}}(\mathbb{R}^3)
  \end{equation}
  and 
  \begin{equation}\label{eq:Alambdatildenot0}
   \begin{split}
       \widetilde{\mathcal{A}_{\lambda}^{(\ell)}}\;&:=\;3W_\lambda^{-1}\big(T_\lambda^{(\ell)}+\alpha\mathbbm{1}\big) \, , \\
       \mathcal{D}\big( \widetilde{\mathcal{A}_{\lambda}^{(\ell)}}\big)\;&:=\; \widetilde{\mathcal{D}}_\ell\,.
   \end{split}
  \end{equation}
  One has the following.
  \begin{enumerate}[(i)]
   \item $ \widetilde{\mathcal{A}_{\lambda}^{(\ell)}}$ is a densely defined symmetric operator in $H^{-\frac{1}{2}}_{W_\lambda,\ell}(\mathbb{R}^3)$.
   \item If $\lambda>\lambda_\alpha$, then $\mathfrak{m}\big(\widetilde{\mathcal{A}_{\lambda}^{(\ell)}}\big)>0$, i.e., $\widetilde{\mathcal{A}_{\lambda}^{(\ell)}}$ has strictly positive lower bound.
  \end{enumerate}
  \end{lemma}

  \begin{proof}
   (i) Obviously $\widetilde{\mathcal{D}}_\ell$ is dense in $H^{-\frac{1}{2}}_{W_\lambda,\ell}(\mathbb{R}^3)$. Moreover, $(T_\lambda^{(\ell)}+\alpha\mathbbm{1}\big)\widetilde{\mathcal{D}}_\ell\subset H^{\frac{1}{2}}_\ell(\mathbb{R}^3)$ (Lemma \ref{lem:Tlambdaproperties}(iii)) and $H^{\frac{1}{2}}_\ell(\mathbb{R}^3)=\mathrm{ran}\,W_{\lambda}^{(\ell)}$ (Lemma \ref{lem:Wlambdaproperties}(ii)), therefore \eqref{eq:Alambdatildenot0} is a well-posed definition for a densely defined operator in $H^{-\frac{1}{2}}_{W_\lambda,\ell}(\mathbb{R}^3)$. The map $\widetilde{\mathcal{D}}_\ell\ni\xi\mapsto T_\lambda^{(\ell)}\xi$ is densely defined and symmetric in $L^2(\mathbb{R}^3)$ (Lemma \ref{lem:Tlambdaproperties}(i)). All assumptions of Lemma \ref{lem:symsym} are then satisfied in the $\ell$-th sector: one then concludes that $ \widetilde{\mathcal{A}_{\lambda}^{(\ell)}}$ is symmetric in $H^{-\frac{1}{2}}_{W_\lambda,\ell}(\mathbb{R}^3)$.   
   
   (ii) When $\xi\in\widetilde{\mathcal{D}}_\ell$,
   \[
    \begin{split}
     \frac{1}{3}\big\langle\xi, \widetilde{\mathcal{A}_{\lambda}^{(\ell)}}\xi\big\rangle_{H^{-\frac{1}{2}}_{W_\lambda}}\;&=\;\big\langle\xi,(T_\lambda^{(\ell)}+\alpha\mathbbm{1}\big)\big\rangle_{L^2} \\
     &\geqslant\;2\pi^2\kappa^-\!\int_{\mathbb{R}^3}\sqrt{\frac{3}{4}\pp^2+\lambda}\,|\widehat{\xi}(\pp)|^2\,\ud\pp+\alpha\|\xi\|_{L^2}^2 \\
     &\geqslant\;(2\pi^2\kappa^-\sqrt{\lambda}+\alpha)\|\xi\|_{L^2}^2\;\geqslant\;c_\lambda(2\pi^2\kappa^-\sqrt{\lambda}+\alpha)\|\xi\|^2_{H^{-\frac{1}{2}}_{W_\lambda}}
    \end{split}
   \]
   for some $c_\lambda>0$, having used \eqref{eq:W-scalar-product} in the first step, \eqref{eq:xiTxiFromBelow} in the second, and the isomorphism $H^{-\frac{1}{2}}(\mathbb{R}^3)\cong H^{-\frac{1}{2}}_{W_\lambda}(\mathbb{R}^3)$ in the last. Thus, $\mathfrak{m}(\widetilde{\mathcal{A}_{\lambda}^{(\ell)}})\geqslant 3c_\lambda(2\pi^2\kappa^-\sqrt{\lambda}+\alpha)=6\pi^2c_\lambda\kappa^-(\sqrt{\lambda}-\sqrt{\lambda_\alpha})$ and the thesis follows.   
  \end{proof}

  Being densely defined, symmetric, and lower semi-bounded,  $\widetilde{\mathcal{A}_{\lambda}^{(\ell)}}$ has its Friedrichs self-adjoint extension (Theorem \ref{thm:Friedrichs-ext}). That will be the actual TMS parameter \index{Ter-Martirosyan-Skornyakov!extension parameter} $\mathcal{A}_{\lambda}^{(\ell)}$. 
%
  
  \begin{proposition}\label{prop:Alambdaellnot0}
   Let $\alpha\in\mathbb{R}$, $\lambda>\lambda_\alpha$, and $\ell\in\mathbb{N}$. Define 
   \begin{equation}\label{eq:domainDell}
    \mathcal{D}_\ell\;:=\;\big\{\xi\in H_\ell^{\frac{1}{2}}(\mathbb{R}^3)\,\big|\,T_\lambda^{(\ell)}\xi\in H_\ell^{\frac{1}{2}}(\mathbb{R}^3)\big\}\,.
   \end{equation}
   The operator 
   \begin{equation}\label{AFop-ellnot0}
    \begin{split}
     \mathcal{D}\big(\mathcal{A}_{\lambda}^{(\ell)}\big)\;&:=\;\mathcal{D}_\ell \, ,\\
     \mathcal{A}_{\lambda}^{(\ell)}\;&:=\;3 W_\lambda^{-1}\big(T_\lambda^{(\ell)}+\alpha\mathbbm{1}\big)\,.
    \end{split}
   \end{equation}
   is the Friedrichs extension of $\widetilde{\mathcal{A}_{\lambda}^{(\ell)}}$ with respect to $H^{-\frac{1}{2}}_{W_\lambda,\ell}(\mathbb{R}^3)$ and therefore is self-adjoint in such space.  
   Its sesquilinear form is
   \begin{equation}\label{AFform-ellnot0}
    \begin{split}
     \mathcal{D}\big[\mathcal{A}_{\lambda}^{(\ell)}\big]\;&=\;H_\ell^{\frac{1}{2}}(\mathbb{R}^3) \, , \\
     \mathcal{A}_{\lambda}^{(\ell)}[\eta,\xi]\;&=\;3\big\langle\eta,\big(T_\lambda^{(\ell)}+\alpha\mathbbm{1}\big)\xi \big\rangle_{H^{\frac{1}{2}},H^{-\frac{1}{2}}}\,.
    \end{split}
   \end{equation}
  \end{proposition}

  \begin{proof}
   The definition \eqref{AFop-ellnot0} is well-posed, as $\big(T_\lambda^{(\ell)}+\alpha\mathbbm{1}\big)\mathcal{D}_\ell\subset H^{\frac{1}{2}}_\ell(\mathbb{R}^3)=\mathrm{ran} W_\lambda$.  
   Lemma \ref{lem:xiTxi-equiv-H12} and the fact that $\widetilde{\mathcal{A}_{\lambda}^{(\ell)}}$ has strictly positive lower bound (Lemma \ref{lem:Atildenot0}(ii)) imply that the map
   \[
    \xi\;\mapsto\;\|\xi\|_{\mathcal{A}}\;:=\;\big\langle \xi,\widetilde{\mathcal{A}_{\lambda}^{(\ell)}}\xi\big\rangle_{H^{-\frac{1}{2}}_{W_\lambda}}^{\frac{1}{2}}\;=\;\Big( 3\big\langle\xi,\big(T_\lambda^{(\ell)}+\alpha\mathbbm{1}\big)\xi\big\rangle_{L^2}\Big)^{\frac{1}{2}}
   \]
   is a norm, and is actually equivalent to the $H^{\frac{1}{2}}$-norm. Let us temporarily denote by $\mathcal{A}_{\mathrm{F}}$ the Friedrichs extension of $\widetilde{\mathcal{A}_{\lambda}^{(\ell)}}$ with respect to $H^{-\frac{1}{2}}_{W_\lambda,\ell}(\mathbb{R}^3)$.

   As prescribed by the Friedrichs construction (Theorem \ref{thm:Friedrichs-ext}(i)), $\mathcal{A}_{\mathrm{F}}$ has form domain
   \[
    \mathcal{D}[\mathcal{A}_{\mathrm{F}}]\;=\;\overline{\mathcal{D}\big( \widetilde{\mathcal{A}_{\lambda}^{(\ell)}}\big)}^{\|\,\|_{\mathcal{A}	}}\;=\;\overline{H^{\frac{3}{2}}(\mathbb{R}^3)}^{\|\,\|_{H^{\frac{1}{2}}}}\;=\;H^{\frac{1}{2}}(\mathbb{R}^3) \, ,
   \]
   and for $\xi,\eta\in H_\ell^{\frac{1}{2}}(\mathbb{R}^3)$
   \[
    \mathcal{A}_{\mathrm{F}}[\eta,\xi]\;=\;\lim_{n\to\infty}\big\langle\eta_n, \widetilde{\mathcal{A}_{\lambda}^{(\ell)}}\xi_n\big\rangle_{H^{-\frac{1}{2}}_{W_\lambda}}\;=\;3\lim_{n\to\infty}\big\langle\eta_n,\big(T_\lambda^{(\ell)}+\alpha\mathbbm{1}\big)\xi_n \big\rangle_{L^2} \, ,
   \]
  for any two sequences $(\xi_n)_n$ and $(\eta_n)_n$ in $H_\ell^{\frac{3}{2}}(\mathbb{R}^3)$ such that $\xi_n\to\xi$ and $\eta_n\to\eta$ in the $\|\,\|_{\mathcal{A}}$-norm, namely in $H_\ell^{\frac{1}{2}}(\mathbb{R}^3)$. Now, interpreting
  \[
   \begin{split}
    \big\langle\eta_n,\big(T_\lambda^{(\ell)}+\alpha\mathbbm{1}\big)\xi_n \big\rangle_{L^2}\;=\;\big\langle\eta_n,\big(T_\lambda^{(\ell)}+\alpha\mathbbm{1}\big)\xi_n \big\rangle_{H^{\frac{1}{2}},H^{-\frac{1}{2}}}
   \end{split}
  \]
  and using the fact that $T_\lambda^{(\ell)}+\alpha\mathbbm{1}$ is a bounded $H_\ell^{\frac{1}{2}}\to H_\ell^{-\frac{1}{2}}$ map (Lemma \ref{lem:Tlambdaproperties}(ii)), one sees that $(T_\lambda^{(\ell)}+\alpha\mathbbm{1}\big)\xi_n\to (T_\lambda^{(\ell)}+\alpha\mathbbm{1}\big)\xi$ in $H_\ell^{-\frac{1}{2}}(\mathbb{R}^3)$ and therefore
  \[
   \mathcal{A}_{\mathrm{F}}[\eta,\xi]\;=\;3\lim_{n\to\infty}\big\langle\eta_n,\big(T_\lambda^{(\ell)}+\alpha\mathbbm{1}\big)\xi_n \big\rangle_{H^{\frac{1}{2}},H^{-\frac{1}{2}}}\;=\;3\big\langle\eta,\big(T_\lambda^{(\ell)}+\alpha\mathbbm{1}\big)\xi \big\rangle_{H^{\frac{1}{2}},H^{-\frac{1}{2}}}\,.
  \]
 Formula \eqref{AFform-ellnot0} is thus proved.

 Next, the operator $\mathcal{A}_{\mathrm{F}}$ is derived from its quadratic form in the usual manner, that is (Sect.~\ref{sec:I_forms}),
 \[
  \begin{split}
   \mathcal{D}(\mathcal{A}_{\mathrm{F}})\;&=\;\left\{
   \xi\in\mathcal{D}[\mathcal{A}_{\mathrm{F}}]\,\left|\!
   \begin{array}{c}
    \exists\,\zeta_\xi\in H^{-\frac{1}{2}}_{W_\lambda,\ell}(\mathbb{R}^3)\textrm{ such that } \\
    \langle\eta,\zeta_\xi\rangle_{H^{-\frac{1}{2}}_{W_\lambda}}\,=\,\mathcal{A}_{\mathrm{F}}[\eta,\xi]\;\;\;\forall\eta\in\mathcal{D}[\mathcal{A}_{\mathrm{F}}]
   \end{array}
   \!\!\right.\right\} \\
   \mathcal{A}_{\mathrm{F}}\,\xi\;&=\;\zeta_\xi
  \end{split}
 \]
 This means, owing to \eqref{eq:W-scalar-product} and \eqref{AFform-ellnot0}, that $\xi\in \mathcal{D}(\mathcal{A}_{\mathrm{F}})$ if and only if $\xi$ is a $H_\ell^{\frac{1}{2}}$-function with
 \[
  \big\langle\eta, W_\lambda^{(\ell)}\zeta_\xi-3\big(T_\lambda^{(\ell)}+\alpha\mathbbm{1}\big)\xi\big\rangle_{H^{\frac{1}{2}},H^{-\frac{1}{2}}}\;=\;0\qquad \forall \eta\in H_\ell^{\frac{1}{2}}(\mathbb{R}^3)\,.
 \]
 for some $\zeta_\xi\in H_\ell^{-\frac{1}{2}}(\mathbb{R}^3)$, and therefore equivalently
 \[
  \xi\in H_\ell^{\frac{1}{2}}(\mathbb{R}^3)\qquad\textrm{ and }\qquad 3\big(T_\lambda^{(\ell)}+\alpha\mathbbm{1}\big)\xi\;=\;W_\lambda^{(\ell)}\zeta_\xi\,.
 \]
 The second condition, owing to the $H^{-\frac{1}{2}}_\ell\to H^{\frac{1}{2}}_\ell$ bijectivity  of $W_\lambda^{(\ell)}$ (Lemma \ref{lem:Wlambdaproperties}(ii)), is tantamount as $\big(T_\lambda^{(\ell)}+\alpha\mathbbm{1}\big)\xi\in H_\ell^{\frac{1}{2}}(\mathbb{R}^3)$, and moreover $\zeta_\xi=3 W_\lambda^{-1}\big(T_\lambda^{(\ell)}+\alpha\mathbbm{1}\big)\xi$. Formula \eqref{AFop-ellnot0} is proved.   
  \end{proof}

 It is instructive to remark that whereas on the domain $\mathcal{D}_\ell$ the operator $\mathcal{A}_\lambda^{(\ell)}=3W_\lambda^{-1}\big(T_\lambda^{(\ell)}+\alpha\mathbbm{1}\big)$ is self-adjoint with respect to $H^{-\frac{1}{2}}_{W_\lambda,\ell}(\mathbb{R}^3)$, and therefore $T_\lambda^{(\ell)}$ on the same domain is symmetric with respect to $L^2(\mathbb{R}^3)$ (Lemma \ref{lem:symsym}), \emph{however} $T_\lambda^{(\ell)}$ is \emph{not} self-adjoint in $L^2_\ell(\mathbb{R}^3)$.

 \begin{lemma}\label{lem:exampleMinloswrong}
   Let $\lambda>0$, and $\ell\in\mathbb{N}$. The operator
   \begin{equation}
    \begin{split}
     \mathcal{D}\big(\mathsf{T}^{(\ell)}_\lambda\big)\;&:=\;\mathcal{D}_\ell\;=\;\big\{\xi\in H_\ell^{\frac{1}{2}}(\mathbb{R}^3)\,\big|\, T_{\lambda}^{(\ell)}\xi\in H_\ell^{\frac{1}{2}}(\mathbb{R}^3)\big\}  \, ,\\
     \mathsf{T}^{(\ell)}_\lambda\xi\;&:=\; T_\lambda\xi\qquad\forall\xi\in\mathcal{D}\big(\mathsf{T}^{(\ell)}_\lambda\big)
    \end{split}
   \end{equation}
 is densely defined and symmetric with respect to the Hilbert space $L^2_\ell(\mathbb{R}^3)$. However, it is not self-adjoint.
  \end{lemma}

  \begin{proof}
   It was already argued prior to stating the Lemma that $\mathsf{T}^{(\ell)}_\lambda$ is densely defined and symmetric in $L^2_\ell(\mathbb{R}^3)$. In fact, the symmetry property
   \[
    \big\langle\eta,\mathsf{T}^{(\ell)}_\lambda\xi \big\rangle_{L^2}\;=\;\big\langle\mathsf{T}^{(\ell)}_\lambda\eta,\xi \big\rangle_{L^2}\qquad\forall\xi,\eta\in \mathcal{D}_\ell
   \]
  also follows directly from Lemma \ref{lem:Tlambdaproperties}(v), because $\xi,\eta\in H^{\frac{1}{2}}(\mathbb{R}^3)$ and $\mathsf{T}^{(\ell)}_\lambda\xi,\mathsf{T}^{(\ell)}_\lambda\eta\in  H^{\frac{1}{2}}(\mathbb{R}^3)\subset L^2(\mathbb{R}^3)$.

  With respect to $L^2_\ell(\mathbb{R}^3)$ the quadratic form
   \[
   \begin{split}
    \mathcal{D}\big(q^{(\ell)}_\lambda\big)\;&:=\;H^{\frac{1}{2}}_\ell(\mathbb{R}^3) \, , \\
    q^{(\ell)}_\lambda[\xi]\;&:=\;\big\langle\xi,T_\lambda^{(\ell)}\xi\big\rangle_{H^{\frac{1}{2}},H^{-\frac{1}{2}}}\;\approx\;\|\xi\|_{H^{\frac{1}{2}}}^2
   \end{split}
   \]
  is densely defined, coercive and hence lower semi-bounded with strictly positive lower bound (as follows from Lemma \ref{lem:xiTxi-equiv-H12}), and closed (because $ \mathcal{D}\big(q^{(\ell)}_\lambda\big)$ is obviously closed with respect to the norm induced by the form, namely the $H^{\frac{1}{2}}$-norm). As such, $q^{(\ell)}$ is the quadratic form of the self-adjoint operator
    \[
  \begin{split}
   \mathcal{D}(\mathsf{Q}_\lambda^{(\ell)})\;&:=\;\left\{
   \xi\in\mathcal{D}\big(q^{(\ell)}_\lambda\big)\,\left|\!
   \begin{array}{c}
    \exists\,\zeta_\xi\in L^2_\ell(\mathbb{R}^3)\textrm{ such that } \\
    \langle\eta,\zeta_\xi\rangle_{L^2}\,=\,q^{(\ell)}_\lambda[\eta,\xi]\;\;\;\forall\eta\in\mathcal{D}\big(q^{(\ell)}_\lambda\big)
   \end{array}
   \!\!\right.\right\}  \, ,\\
   \mathsf{Q}_\lambda^{(\ell)}\,\xi\;&:=\;\zeta_\xi\,.
  \end{split}
 \]
  Equivalently, $\xi\in \mathcal{D}\big(\mathsf{Q}_\lambda^{(\ell)}\big)$ if and only if $\xi$ is an $H^{\frac{1}{2}}_\ell$-function such that
  \[
   \big\langle\eta,\zeta_\xi-T_{\lambda}^{(\ell)}\xi\big\rangle_{H^{\frac{1}{2}},H^{-\frac{1}{2}}}\;=\;0\qquad\forall\eta\in H_\ell^{\frac{1}{2}}(\mathbb{R}^3)
  \]
 for some $\zeta_\xi\in L^2_\ell(\mathbb{R}^3)$, and therefore, equivalently,
 \[
  \xi\in H_\ell^{\frac{1}{2}}(\mathbb{R}^3)\qquad\textrm{ and }\qquad T_{\lambda}^{(\ell)}\xi\,=\,\zeta_\xi\,.
 \]
 The second condition above is tantamount as $ T_{\lambda}^{(\ell)}\xi\in L^2_\ell(\mathbb{R}^3)$. In conclusion,
    \[
  \begin{split}
   \mathcal{D}\big(\mathsf{Q}_\lambda^{(\ell)}\big)\;&=\;\big\{\xi\in H_\ell^{\frac{1}{2}}(\mathbb{R}^3)\,\big|\, T_{\lambda}^{(\ell)}\xi\in L^2_\ell(\mathbb{R}^3)\big\}\, ,\\
   \mathsf{Q}_\lambda^{(\ell)}\,\xi\;&=\;T_{\lambda}^{(\ell)}\xi\,.
  \end{split}
 \]

 At this point it is clear that
 \[
  \mathsf{T}^{(\ell)}_\lambda\;\subset\;\mathsf{Q}_\lambda^{(\ell)}\;=\;\big(\mathsf{Q}_\lambda^{(\ell)}\big)^*\,.
 \]
 The lack of self-adjointness of $\mathsf{T}^{(\ell)}_\lambda$ is then evident from the strict inclusion $\mathcal{D}\big(\mathsf{T}^{(\ell)}_\lambda\big)\varsubsetneq\mathcal{D}\big(\mathsf{Q}_\lambda^{(\ell)}\big)$.  
\end{proof}

\section{Sector of zero angular momentum}\label{sec:lzero}

  The problem of finding a domain $\mathcal{D}_0$ of self-adjointness in the Hilbert space $H^{-\frac{1}{2}}_{W_\lambda,\ell=0}(\mathbb{R}^3)$ for the operator $3W_\lambda^{-1}\big(T_\lambda^{(\ell=0)}+\alpha\mathbbm{1}\big)$ (in the following the full `$\ell=0$' superscript will be often shortened), is more subtle than the analogous problem for $\ell\in\mathbb{N}$ (Subsect.~\ref{sec:selfadj-ellnotzero}), and so too is the quest for a domain $\widetilde{\mathcal{D}_0}$ of sole symmetry.

  This is related with the fact that no Sobolev space $H^s_{\ell=0}(\mathbb{R}^3)$ is entirely mapped by $T_\lambda$ into $H^{\frac{1}{2}}(\mathbb{R}^3)$ (Remark \ref{rem:Tl-failstomap}), so $\widetilde{\mathcal{D}_0}$ cannot be a standard Sobolev space (as opposite to when $\ell\neq 0$: Lemma \ref{lem:Atildenot0}). A related difficulty, that emerges indirectly from the discussion of Lemma \ref{lem:xiTxi-equiv-H12}, is the fact that when $\ell=0$ the map
  \[
   \xi\;\longmapsto\;\int_{\mathbb{R}^3} \overline{\,\widehat{\xi}(\pp)}\, \big(\widehat{T_\lambda\xi}\big)(\pp)\,\ud\pp
  \]
  does not induce any longer an equivalent $H^{\frac{1}{2}}$-norm (see Remark \ref{rem:noequivH12norm} below). In fact, as will be seen in the following, any reasonable choice of a domain $\widetilde{\mathcal{D}_0}$ of symmetry for  $W_\lambda^{-1}T_\lambda^{(\ell=0)}$ makes it an unbounded below operator, unlike the lower semi-boundedness of $\mathcal{A}_\lambda^{(\ell)}$ when $\ell\neq 0$ (Lemma \ref{lem:Atildenot0}(ii)).

  These difficulties require an improved analysis that is presented in this Section. They are also the source of various past mistakes leading to ill-posed models: Section \ref{sec:illposed} discusses such a perspective.

  Here the same conceptual path as in Section \ref{sec:higherell} will be followed. First one discusses the symmetric case (symmetric realisation of $W_\lambda^{-1}T_\lambda^{(\ell=0)}$ and hence Ter-Martirosyan-Skornyakov symmetric extension of $\mathring{H}$), then the self-adjoint case  (self-adjoint $W_\lambda^{-1}T_\lambda^{(\ell=0)}$ and hence self-adjoint TMS extension). For each two steps, an amount of technical preparation is needed.

  The choice made here is to discuss explicitly only a special scenario, in fact the physically most relevant one, beside being mathematically more manageable: zero-range interaction with infinite scattering length, hence $\alpha=0$. This is the regime of \emph{unitarity} \index{unitary regime} already mentioned in Section \ref{sec:BOSCHAPTintro}. 

  \subsection{Mellin-like transformations}

  For fixed $\lambda>0$, to each charge of interest $\xi\in H_{\ell=0}^{s}(\mathbb{R}^3)$, written according to \eqref{eq:xihatangularexpansion} as
  \begin{equation}\label{eq:0xi}
   \widehat{\xi}(\pp)\;=\;\frac{1}{\sqrt{4\pi}}f(|\pp|)\,,\qquad \pp\equiv|\pp|\Omega_{\pp}\,,\qquad f\in L^2(\mathbb{R}^+,(1+p^2)^sp^2\ud p)\,,
  \end{equation}
   one associates an odd, measurable function $\theta:\mathbb{R}\to\mathbb{C}$ defined by
  \begin{equation}\label{ftheta-1}
   \begin{split}
    \theta(x)\;&:=\;
    \begin{cases}
     \lambda f\big(\frac{2\sqrt{\lambda}}{\sqrt{3}}\sinh x\big)\sinh x\cosh x & \textrm{ if }x\geqslant 0 \, , \\
     -\theta(-x) & \textrm{ if }x< 0 \, ,
    \end{cases} \\
    x\;&:=\;\log\bigg(\sqrt{\frac{3p^2}{4\lambda}}+\sqrt{\frac{3p^2}{4\lambda}+1}\bigg)\,,\qquad p\,:=\,|\pp|\,.
   \end{split}
  \end{equation}
  The inverse transformation is
  \begin{equation}\label{ftheta-2}
   \begin{split}
    f(p)\;&=\;\frac{\,\theta\Big(\log\Big(\sqrt{\frac{3p^2}{4\lambda}}+\sqrt{\frac{3p^2}{4\lambda}+1}\Big)\Big)\,}{\sqrt{\frac{3}{4}p^2\,}\,\sqrt{\frac{3}{4}p^2+\lambda}} \, , \\
    p\;&=\;\frac{2\sqrt{\lambda}}{\sqrt{3}}\,\sinh x\qquad \textrm{ for }x\geqslant 0\,.
   \end{split}
  \end{equation}
   The above change of variable $p\leftrightarrow x$ is a homeomorphism on $\mathbb{R}^+$, with also
   \begin{equation}\label{ftheta-3-eq:pxchangevar}
    \sinh x\;=\;\sqrt{\frac{3}{4}p^2\,}\,,\quad\sqrt{\lambda}\cosh x\;=\;\sqrt{\frac{3}{4}p^2+\lambda\,}\,,\quad\ud p\;=\;\frac{2\sqrt{\lambda}}{\sqrt{3}}\,\cosh x\,\ud x\,.
   \end{equation}
 It is for later convenience that the function $\theta$ initially set on $\mathbb{R}^+$ is extended by odd parity over the whole real line. 

 One refers to the function $\theta$ defined by \eqref{eq:0xi}-\eqref{ftheta-1} as the \emph{re-scaled radial component associated with the charge $\xi$ and with parameter $\lambda$}. When such a correspondence need be emphasised, one writes $\theta^{(\xi)}$.

   Using \eqref{ftheta-1}-\eqref{ftheta-3-eq:pxchangevar} and the fact that $1+p^2\sim\frac{3}{4}p^2+\lambda$, in the sense that each side is controlled from above and from below by the other with suitable overall $\lambda$-dependent multiplicative constants, a straightforward computation gives
   \begin{equation}\label{eq:mellinnorms}
    \big\|\xi\big\|_{H^s(\mathbb{R}^3)}^2\;\approx\; 
    \big\|(\cosh x)^{s-\frac{1}{2}}\theta\big\|_{L^2(\mathbb{R})}^2
   \end{equation}
   in the sense of equivalence of norms (with $\lambda$-dependent pre-factors).

  Here further definitions and properties will be introduced which are going to be useful in the course of the present discussion.

   The function $\theta$ having odd parity on $\mathbb{R}$, one has the identities
   \begin{equation}\label{eq:h-identities}
    \begin{split}
     \int_{\mathbb{R}^+}\ud x\,\theta(x)&\Big(\log \frac{\,2\cosh(x+y)-1\,}{\,2\cosh(x+y)+1\,}+\log \frac{\,2\cosh(x-y)+1\,}{\,2\cosh(x-y)-1\,}\Big) \\
     &=\;\int_{\mathbb{R}}\ud x\,\theta(x) \log \frac{\,2\cosh(x+y)-1\,}{\,2\cosh(x+y)+1\,}\;\\
     &=\;\int_{\mathbb{R}}\ud x\,\theta(x) \log \frac{\,2\cosh(x-y)+1\,}{\,2\cosh(x-y)-1\,}\,.
    \end{split}
   \end{equation}

  Moreover (see, e.g., \cite[I.1.9.(50)]{Erdelyi-Tables1}),
  \begin{equation}\label{eq:logfourier}
   \Big(\log \frac{\,2\cosh x+1\,}{\,2\cosh x-1\,}\Big){\!\!\textrm{\huge ${\,}^{\widehat{\,}}$\normalsize}}\,(s)\;=\;\sqrt{2\pi}\,\frac{\sinh\frac{\pi}{6}s}{\,s\,\cosh \frac{\pi}{2}s}\,.
  \end{equation}
  By means of \eqref{eq:logfourier}, taking the Fourier transform in the following convolution yields
   \begin{equation}\label{eq:fourierconvolution}
     \bigg(\int_{\mathbb{R}}\ud x\,\theta(y) \log \frac{\,2\cosh(x-y)+1\,}{\,2\cosh(x-y)-1\,}\bigg){\!\!\textrm{\huge ${\,}^{\widehat{\,}}$\normalsize}}\,(s)\;=\;2\pi\,\widehat{\theta}(s)\,\frac{\sinh\frac{\pi}{6}s}{\,s\,\cosh \frac{\pi}{2}s}\,.
   \end{equation}
   Here and in the following the $s$-dependence in $\widehat{\theta}(s)$ is only symbolic, to indicate that the object $\widehat{\theta}$ is a distribution on test functions of $s\in\mathbb{R}$. Of course in special cases $\widehat{\theta}(s)$ may well be an ordinary function.

\begin{figure}[t!]
\begin{center}
\includegraphics[width=8cm]{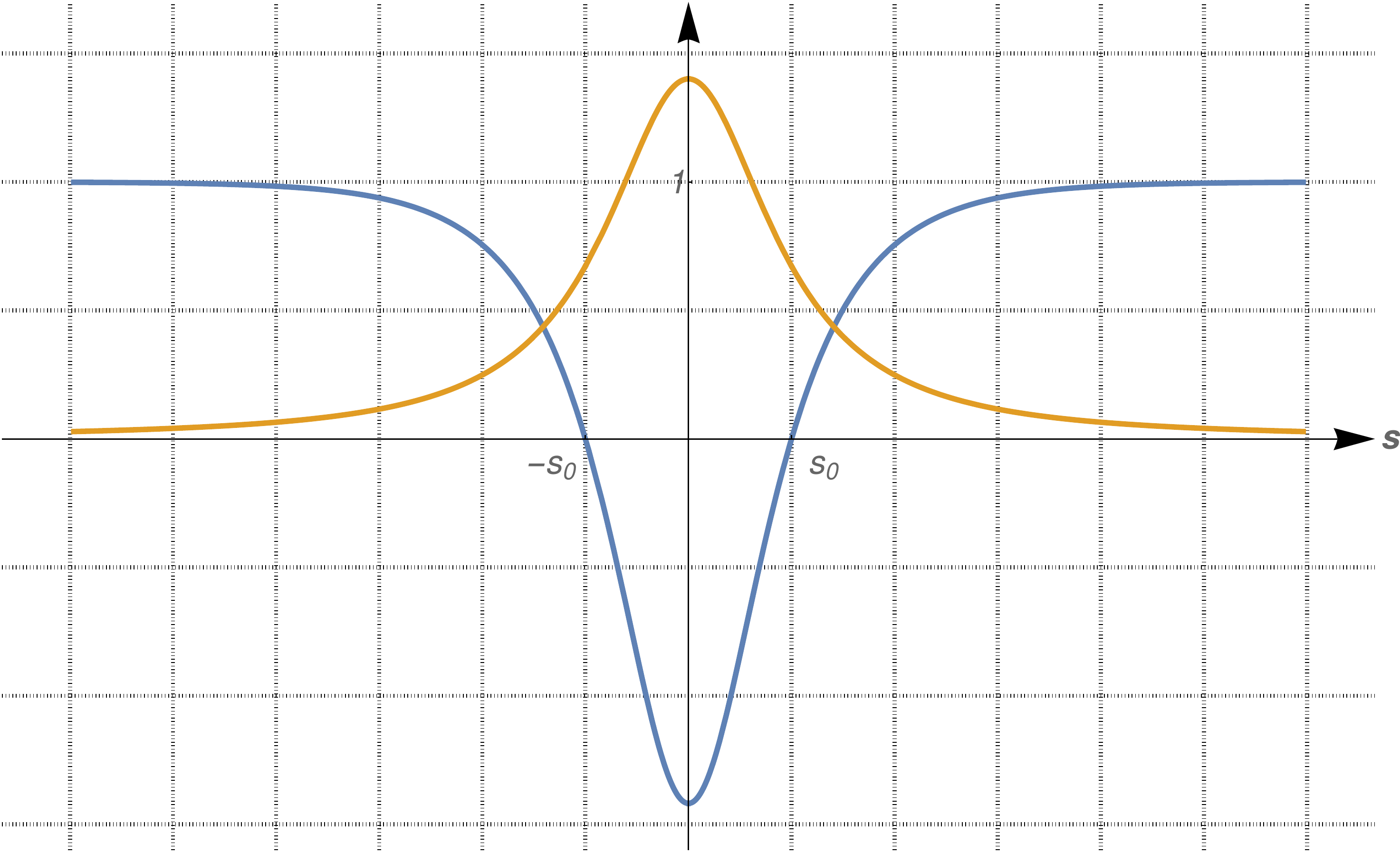}
\caption{Plot of the functions $\widehat{\gamma}(s)$ (blue) and $\widehat{\gamma}_+(s)$ (orange), defined respectively in \eqref{eq:gamma-distribution} and \eqref{eq:gammaplus}.}\label{fig:gammagammaplus}
\end{center}
\end{figure}

  Let $\gamma$ be the distribution on $\mathbb{R}$ defined by
  \begin{equation}\label{eq:gamma-distribution}
   \widehat{\gamma}(s)\;:=\;1-\frac{8}{\sqrt{3}}\,\frac{\sinh\frac{\pi}{6}s}{\,s\,\cosh \frac{\pi}{2}s}\,.
  \end{equation}
  The function $\mathbb{R}\ni s\mapsto\widehat{\gamma}(s)$ is smooth, even, strictly monotone increasing (resp., decreasing) for $s>0$ (resp., $s<0$), with values in $[1-\frac{4\pi}{3\sqrt{3}},1)$, asymptotically approaching $1$ as $s\to\pm\infty$, and with absolute minimum $ \widehat{\gamma}(0)=-(\frac{4\pi}{3\sqrt{3}}-1)$. The equation $\widehat{\gamma}(s)=0$ has thus simple roots $s=\pm s_0$, with $s_0\approx 1.0062$. One also defines
  \begin{equation}\label{eq:gammaplus}
   \widehat{\gamma}_+(s)\;:=\;\frac{1}{\,(s-s_0)(s+s_0)\,}\, \widehat{\gamma}(s)\,.	
  \end{equation}
  $\widehat{\gamma}_+$ is therefore strictly positive, smooth, even, monotone to zero decreasing for $s>0$ with $s^{-2}$ decay, and with absolute maximum  $\widehat{\gamma}_+(0)=s_0^{-2}(\frac{4\pi}{3\sqrt{3}}-1)$ (Figure \ref{fig:gammagammaplus}).

   Further quantities of interest involving $\xi$ are conveniently expressed in terms of the auxiliary function $\theta$ or its (one-dimensional) Fourier transform $\widehat{\theta}$.

   \begin{lemma}\label{lem:xithetaidentities}
    Let $\lambda>0$, $s\in\mathbb{R}$, and $\xi$ be as in \eqref{eq:0xi}. One has the identities  
    \begin{equation}\label{eq:Tlxi-theta}
     \big(\widehat{T_\lambda^{(0)}\xi}\big)(\pp)\;=\;\frac{1}{\sqrt{4\pi}\,|\pp|}\,\frac{4\pi^2}{\sqrt{3}\,}\Big(\theta(x)-\frac{4}{\pi\sqrt{3}}\int_{\mathbb{R}}\ud y\,\theta(y) \log \frac{\,2\cosh(x-y)+1\,}{\,2\cosh(x-y)-1\,}\Big),
    \end{equation}
        \begin{equation}\label{eq:Tlambda0xi-with-theta}
     \begin{split}
      \big\|T_\lambda^{(0)}\xi\big\|_{H^{s}(\mathbb{R}^3)}^2\;&\approx\;\int_{\mathbb{R}}\ud x\,(\cosh x)^{1+2s}\,\bigg|\,\theta(x)-\frac{4}{\pi\sqrt{3}}\int_{\mathbb{R}}\ud y\,\theta(y) \log \frac{\,2\cosh(x-y)+1\,}{\,2\cosh(x-y)-1\,}\bigg|^2,
     \end{split}
    \end{equation}
   and 
    \begin{equation}\label{eq:xiTxi-with-theta}
     \int_{\mathbb{R}^3} \overline{\,\widehat{\xi}(\pp)}\, \big(\widehat{T_\lambda^{(0)}\xi}\big)(\pp)\,\ud\pp\;=\;\frac{\,8\pi^2}{3\sqrt{3}}\int_{\mathbb{R}}\ud s\, \widehat{\gamma}(s)\,|\widehat{\theta}(s)|^2\,,
    \end{equation}
    with $x$ and $\theta$ given by \eqref{ftheta-1}, and $\gamma$ given by \eqref{eq:gamma-distribution}.
    In \eqref{eq:Tlxi-theta} it is understood that $x\geqslant 0$, and \eqref{eq:Tlambda0xi-with-theta} is meant as an equivalence of norms (with $\lambda$-dependent multiplicative constant).    
   \end{lemma}

 \begin{proof}
  Specialising formula \eqref{eq:fellsector} of Lemma \ref{lem:Tlambdadecomposition}(i) with the Legendre polynomial $P_0\equiv 1$ gives
  \[
   \big(\widehat{T_\lambda^{(0)}\xi}\big)(\pp)\;=\;\frac{1}{\sqrt{4\pi}\,}\,\frac{1}{p}\Big(2\pi^2pf(p)\sqrt{\frac{3}{4}p^2+\lambda\,}\,-\,4\pi\int_{\mathbb{R}^+}\!\ud q\,q f(q)\,\log\frac{\,p^2+q^2+pq+\lambda\,}{p^2+q^2-pq+\lambda}\Big)\,.
  \]
  With $p=\frac{2\sqrt{\lambda}}{\sqrt{3}}\sinh x$ and $q=\frac{2\sqrt{\lambda}}{\sqrt{3}}\sinh y$ one has 
  \[\tag{*}\label{logppqq}
   \begin{split}
    \frac{\,p^2+q^2+pq+\lambda\,}{p^2+q^2-pq+\lambda}\;&=\;\frac{\,\sinh^2 x+\sinh^2y+\sinh x\,\sinh y+\frac{3}{4}\,}{\,\sinh^2 x+\sinh^2y-\sinh x\,\sinh y+\frac{3}{4}\,} \\
    &=\;\frac{\,2\cosh(x+y)-1\,}{\,2\cosh(x+y)+1\,}\,\frac{\,2\cosh(x-y)+1\,}{\,2\cosh(x-y)-1\,}\,.
   \end{split}
  \]
 Using the latter identity and \eqref{ftheta-2} one then finds
  \[
   \sqrt{4\pi}\,|\pp|\,\big(\widehat{T_\lambda^{(0)}\xi}\big)(\pp)\;=\;\frac{4\pi^2}{\sqrt{3}\,}\Big(\theta(x)-\frac{4}{\pi\sqrt{3}}\int_{\mathbb{R}^+}\!\ud y\,\theta(y)\big(A_x(y)+B_x(y)\big)\Big)\,,
  \]
  where
  \[
   A_x(y)\;:=\;\log \frac{\,2\cosh(x+y)-1\,}{\,2\cosh(x+y)+1\,}\,,\qquad B_x(y)\;:=\;\log \frac{\,2\cosh(x-y)+1\,}{\,2\cosh(x-y)-1\,}\,.
  \]
  Combining this with \eqref{eq:h-identities} yields \eqref{eq:Tlxi-theta}. 
  
  Next, by means of \eqref{ftheta-3-eq:pxchangevar} and \eqref{eq:Tlxi-theta} one finds
   \[
    \begin{split}
      &\big\|T_\lambda^{(0)}\xi\big\|_{H^{s}(\mathbb{R}^3)}^2 \\
      &\approx\;\int_{\mathbb{R}^+}\ud p\,p^2\,\Big(\frac{3}{4}p^2+\lambda\Big)^s\,\bigg|\frac{1}{p}\,\Big(\theta(x)-\frac{4}{\pi\sqrt{3}}\int_{\mathbb{R}}\ud y\,\theta(y) \log \frac{\,2\cosh(x-y)+1\,}{\,2\cosh(x-y)-1\,}\Big)\bigg|^2 \\
      &=\frac{\,2\lambda^{\frac{1}{2}+s}}{\sqrt{3}}\int_{\mathbb{R}^+}\ud x\,(\cosh x)^{1+2s}\,\bigg|\,\theta(x)-\frac{4}{\pi\sqrt{3}}\int_{\mathbb{R}}\ud y\,\theta(y) \log \frac{\,2\cosh(x-y)+1\,}{\,2\cosh(x-y)-1\,}\bigg|^2.
    \end{split}
   \]
   Owing to the odd parity of $\theta$ and to \eqref{eq:h-identities}, the integrand function above is invariant under change of variable $x\mapsto -x$, therefore the last line can be re-written as
   \[
    \frac{\,\lambda^{\frac{1}{2}+s}}{\sqrt{3}}\int_{\mathbb{R}}\ud x\,(\cosh x)^{1+2s}\,\bigg|\,\theta(x)-\frac{4}{\pi\sqrt{3}}\int_{\mathbb{R}}\ud y\,\theta(y) \log \frac{\,2\cosh(x-y)+1\,}{\,2\cosh(x-y)-1\,}\bigg|^2 \, .
   \]
  This gives \eqref{eq:Tlambda0xi-with-theta}.

   Concerning \eqref{eq:xiTxi-with-theta}, specialising formulas \eqref{eq:xiTxipre}-\eqref{eq:xiTxi} of Lemma \ref{lem:Tlambdadecomposition}(ii) with the Legendre polynomial $P_0\equiv 1$ gives
    \begin{equation*}
   \begin{split}
    &\frac{1}{2\pi^2}\int_{\mathbb{R}^3} \overline{\,\widehat{\xi}(\pp)}\, \big(\widehat{T_\lambda^{(0)}\xi}\big)(\pp)\,\ud\pp\;=\;\int_{\mathbb{R}^+}\!\ud p\,\sqrt{{\textstyle\frac{3}{4}}p^2+\lambda}\,|p f(p)|^2 \\
   &\qquad -\frac{2}{\pi}\iint_{\mathbb{R}^+\times\mathbb{R}^+}\ud p\,\ud q\, (p \overline{f(p)})\,(q f(q))\,\log\frac{\,p^2+q^2+pq+\lambda\,}{p^2+q^2-pq+\lambda}
    \end{split}
  \end{equation*}
   The first summand on the r.h.s.~above can be re-written as
   \[
    \int_{\mathbb{R}^+}\!\ud p\,\sqrt{{\textstyle\frac{3}{4}}p^2+\lambda}\,|p f(p)|^2\;=\;\frac{8}{\,3\sqrt{3}}\int_{\mathbb{R}^+}\ud x\,|\theta(x)|^2\;=\;\frac{4}{\,3\sqrt{3}}\int_{\mathbb{R}}\ud x\,|\theta(x)|^2\,,
   \]
  having used \eqref{ftheta-2}-\eqref{ftheta-3-eq:pxchangevar} in the first step and the odd parity of $\theta$ in the second. Analogously, and using also \eqref{logppqq} and \eqref{eq:h-identities}, the second summand becomes
  \[
   \begin{split}
    & \frac{2}{\pi}\iint_{\mathbb{R}^+\times\mathbb{R}^+}\ud p\,\ud q\, (p \overline{f(p)})\,(q f(q))\,\log\frac{\,p^2+q^2+pq+\lambda\,}{p^2+q^2-pq+\lambda} \\
    &=\;\frac{32}{9\pi}\iint_{\mathbb{R}^+\times\mathbb{R}^+}\ud x\,\ud y\,\overline{\theta(x)}\,\theta(y)\Big(\log \frac{\,2\cosh(x+y)-1\,}{\,2\cosh(x+y)+1\,}+\log \frac{\,2\cosh(x-y)+1\,}{\,2\cosh(x-y)-1\,}\Big) \\
     &=\;\frac{16}{9\pi}\iint_{\mathbb{R}\times\mathbb{R}}\ud x\,\ud y\,\overline{\theta(x)}\,\theta(y)\log \frac{\,2\cosh(x-y)+1\,}{\,2\cosh(x-y)-1\,}\,.
   \end{split}
  \]
  Thus,
      \begin{equation*}
   \begin{split}
    \frac{1}{2\pi^2}\int_{\mathbb{R}^3} \overline{\,\widehat{\xi}(\pp)}\, &\big(\widehat{T_\lambda^{(0)}\xi}\big)(\pp)\,\ud\pp\;=\;\frac{4}{\,3\sqrt{3}}\bigg(\int_{\mathbb{R}}\ud x\,|\theta(x)|^2-\\
   &\qquad -\frac{4}{\pi\sqrt{3}}\iint_{\mathbb{R}\times\mathbb{R}}\ud x\,\ud y\,\overline{\theta(x)}\,\theta(y)\log \frac{\,2\cosh(x-y)+1\,}{\,2\cosh(x-y)-1\,}\bigg)\,.
    \end{split}
  \end{equation*}
  Applying Parseval's identity in both summands of the above r.h.s., and using \eqref{eq:fourierconvolution} in the second summand, one gets
  \[
   \frac{1}{2\pi^2}\int_{\mathbb{R}^3} \overline{\,\widehat{\xi}(\pp)}\, \big(\widehat{T_\lambda^{(0)}\xi}\big)(\pp)\,\ud\pp\;=\;\frac{4}{\,3\sqrt{3}}\bigg(\int_{\mathbb{R}}\ud s\,|\widehat{\theta}(s)|^2-\frac{8}{\sqrt{3}}\int_{\mathbb{R}}\ud s\,|\widehat{\theta}(s)|^2\frac{\sinh\frac{\pi}{6}s}{\,s\,\cosh \frac{\pi}{2}s}\bigg).
  \]
  This, and the definition \eqref{eq:gamma-distribution}, finally prove \eqref{eq:xiTxi-with-theta}.   
  \end{proof}

  \begin{remark}\label{rem:noequivH12norm}
   As the function $\widehat{\gamma}$ attains both positive and negative values, formula \eqref{eq:xiTxi-with-theta} shows that the pairing $\int_{\mathbb{R}^3} \overline{\,\widehat{\xi}(\pp)}\, \big(\widehat{T_\lambda^{(0)}\xi}\big)(\pp)\,\ud\pp$ is \emph{not equivalent} to the $L^2$-norm of the associated re-scaled radial function $\theta^{(\xi)}$, and therefore (owing to \eqref{eq:mellinnorms}) does not induce an equivalent $H^{\frac{1}{2}}$-norm, in contrast with the analogous properties in the sectors with $\ell\neq 0$ (Lemma \ref{lem:xiTxi-equiv-H12}).   
  \end{remark}

  \begin{lemma}\label{lem:xiWxi-zero}
   Let $\lambda>0$ and let $\xi_1,\xi_2$ be spherically symmetric functions with re-scaled radial components $\theta_1,\theta_2$ respectively, according to the definition \eqref{eq:0xi}-\eqref{ftheta-1}. Then 
   \begin{equation}\label{eq:xiWxi-zero}
    \begin{split}
      &\int_{\mathbb{R}^3} \overline{\,\widehat{\xi}_1(\pp)}\, \big(\widehat{W_\lambda^{(0)}\xi_2}\big)(\pp)\,\ud\pp\;=\;\frac{4\pi^2}{\,\lambda\sqrt{3}\,}\int_{\mathbb{R}}\frac{\,\overline{\theta_1(x)}\,\theta_2(x)}{\,(\cosh x)^2}\,\ud x \\
      &\qquad +\frac{32\pi}{\lambda}\iint_{\mathbb{R}\times\mathbb{R}}\frac{\overline{\theta_1(x)}\,\theta_2(y)}{\,(2\cosh(x+y)+1)\,(2\cosh(x-y)-1)\,}\,\ud x\,\ud y
    \end{split}
   \end{equation}
   and also
   \begin{equation}\label{eq:xiWxi-zero-FOURIER}
    \begin{split}
      &\int_{\mathbb{R}^3} \overline{\,\widehat{\xi}_1(\pp)}\, \big(\widehat{W_\lambda^{(0)}\xi_2}\big)(\pp)\,\ud\pp\;=\;\frac{2\pi^2}{\,\lambda\sqrt{3}\,}\int_{\mathbb{R}}\Big(\frac{s}{\,\sinh\frac{\pi}{2}s}*\overline{\widehat{\theta}_1}\Big)(s)\,\widehat{\theta}_2(s)\,\ud s \\
      &\qquad +\frac{16\pi}{3\lambda}\iint_{\mathbb{R}\times\mathbb{R}}\overline{\widehat{\theta}_1(s)}\,\widehat{\theta}_2(t)\,\frac{\sinh\frac{\pi}{6}(s+t)}{\sinh \frac{\pi}{2}(s+t)}\,\frac{\sinh\frac{\pi}{3}(s-t)}{\sinh \frac{\pi}{2}(s-t)}\,\ud s\,\ud t\,.
    \end{split}
   \end{equation}
   In the double integrals above the order of integration is not specified, tacitly understanding that $\theta_1
   \theta_2$ (or $\widehat{\theta}_1\widehat{\theta}_2$) is sufficiently integrable, depending on the applications.
  \end{lemma}

  \begin{proof}
   Specialising formula \eqref{eq:Wellsp} with the Legendre polynomial $P_0\equiv 1$ gives
   \begin{equation*}
   \begin{split}
   &\int_{\mathbb{R}^3} \overline{\,\widehat{\xi_1}(\pp)}\, \big(\widehat{W_\lambda^{(0)}\xi_2}\big)(\pp)\,\ud\pp \\
   &\;=\;\int_{\mathbb{R}^+}\!\ud p\,\frac{3\pi^2}{\sqrt{{\textstyle\frac{3}{4}}p^2+\lambda}\,}\,(p\overline{f_1(p)})\,(pf_2(p)) \\
   &\qquad +12\pi\!\iint_{\mathbb{R}^+\times\mathbb{R}^+}\ud p\,\ud q\,(p\overline{f_1(p)})\,(q f_2(q))\,\frac{2 p q}{\,(p^2+q^2+\lambda)^2-(p q)^2}\,.
    \end{split}
  \end{equation*}
    The first summand on the r.h.s.~above can be re-written as
   \[
   \begin{split}
    \int_{\mathbb{R}^+}\!\ud p\,\frac{3\pi^2}{\sqrt{{\textstyle\frac{3}{4}}p^2+\lambda}\,}\,(p\overline{f_1(p)})\,(pf_2(p))\;&=\;\frac{8\pi^2}{\,\lambda\sqrt{3}\,}\int_{\mathbb{R}^+}\frac{\,\overline{\theta_1(x)}\,\theta_2(x)}{\,(\cosh x)^2}\,\ud x \\
    &=\;\frac{4\pi^2}{\,\lambda\sqrt{3}\,}\int_{\mathbb{R}}\frac{\,\overline{\theta_1(x)}\,\theta_2(x)}{\,(\cosh x)^2}\,\ud x\,,
   \end{split}
   \]
  having used \eqref{ftheta-2}-\eqref{ftheta-3-eq:pxchangevar} in the first step and the odd parity of $\theta_1$ and $\theta_2$ in the second. Next, with $p=\frac{2\sqrt{\lambda}}{\sqrt{3}}\sinh x$ and $q=\frac{2\sqrt{\lambda}}{\sqrt{3}}\sinh y$, one re-writes
  \[
   \begin{split}
    \frac{2 p q}{\,(p^2+q^2+\lambda)^2-(p q)^2}\;&=\;\frac{1}{\,p^2+q^2-pq+\lambda\,}-\frac{1}{\,p^2+q^2+pq+\lambda\,} \\
     &=\;\frac{3}{\lambda}\big( a(x,y)-b(x,y)\big)\,,
   \end{split}
  \]
   where
   \[
    \begin{split}
     a(x,y)\;& :=\;\frac{1}{\,(2\cosh(x+y)+1)\,(2\cosh(x-y)-1)\,} \, ,\\
     b(x,y)\;& :=\;\frac{1}{\,(2\cosh(x+y)-1)\,(2\cosh(x-y)+1)\,} \, .
    \end{split}
   \]
  This and \eqref{ftheta-2}-\eqref{ftheta-3-eq:pxchangevar} then imply
  \[
   \begin{split}
    & 12\pi\!\iint_{\mathbb{R}^+\times\mathbb{R}^+}\ud p\,\ud q\,(p\overline{f_1(p)})\,(q f_2(q))\,\frac{2 p q}{\,(p^2+q^2+\lambda)^2-(p q)^2} \\
    &=\;\frac{64\pi}{\lambda}\iint_{\mathbb{R}^+\times\mathbb{R}^+}\ud x\,\ud y\,\overline{\theta_1(x)}\,\theta_2(y)\,\big( a(x,y)-b(x,y)\big) \\
    &=\;\frac{32\pi}{\lambda}\iint_{\mathbb{R}\times\mathbb{R}}\ud x\,\ud y\,\overline{\theta_1(x)}\,\theta_2(y)\,a(x,y) \\
    &=\;\frac{32\pi}{\lambda}\iint_{\mathbb{R}\times\mathbb{R}}\frac{\overline{\theta_1(x)}\,\theta_2(y)}{\,(2\cosh(x+y)+1)\,(2\cosh(x-y)-1)\,}\,\ud x\,\ud y \, ,
   \end{split}
  \]
  the second identity being due to the odd parity of $\theta_1$ and $\theta_2$ and to the obvious relations $a(-x,-y)=a(x,y)$, $a(-x,y)=b(x,y)=a(x,-y)$. Adding up the two summands worked out above yields finally  \eqref{eq:xiWxi-zero}.

  Concerning \eqref{eq:xiWxi-zero-FOURIER},
  \[
    \Big(\frac{1}{(\cosh x)^2}\Big){}^{\textrm{\LARGE $\widehat{\,}$\normalsize}}\;(s)\;=\;\sqrt{\frac{\pi}{2}}\,\frac{s}{\,\sinh\frac{\pi}{2}s}\,,
  \]
  whence
  \[
   \frac{4\pi^2}{\,\lambda\sqrt{3}\,}\int_{\mathbb{R}}\frac{\,\overline{\theta_1(x)}\,\theta_2(x)}{\,(\cosh x)^2}\,\ud x\;=\;\frac{2\pi^2}{\,\lambda\sqrt{3}\,}\int_{\mathbb{R}}\Big(\frac{s}{\,\sinh\frac{\pi}{2}s}*\overline{\widehat{\theta}_1}\Big)(s)\,\widehat{\theta}_2(s)\,\ud s
  \]
%
%
  Moreover (see, e.g., \cite[I.1.9.(6)]{Erdelyi-Tables1}),
  \[
   \begin{split}
    \Big(\frac{1}{\,2\cosh x -1}\Big){}^{\textrm{\LARGE $\widehat{\,}$\normalsize}}\;(s)\;&=\;\sqrt{\frac{2\pi}{3}}\,\frac{\sinh\frac{2\pi}{3}s}{\sinh \pi s} \, , \\
    \Big(\frac{1}{\,2\cosh x +1}\Big){}^{\textrm{\LARGE $\widehat{\,}$\normalsize}}\;(s)\;&=\;\sqrt{\frac{2\pi}{3}}\,\frac{\sinh\frac{\pi}{3}s}{\sinh \pi s}\,,
   \end{split}
  \]
  whence
  \[
   \begin{split}
    &\frac{1}{2\pi}\iint_{\mathbb{R}\times\mathbb{R}}e^{-\ii s x}\,e^{-\ii t y}\frac{1}{\,(2\cosh(x+y)+1)\,(2\cosh(x-y)-1)\,}\,\ud x\,\ud y \\
    &=\;\frac{1}{4\pi}\iint_{\mathbb{R}\times\mathbb{R}}\,e^{-\ii\frac{s+t}{2} x}\,e^{-\ii\frac{s-t}{2} x}\,\frac{1}{\,2\cosh x +1}\,\frac{1}{\,2\cosh x -1}\,\ud x\,\ud y \\
    &=\;\frac{1}{6}\,\frac{\sinh\frac{\pi}{6}(s+t)}{\sinh \frac{\pi}{2}(s+t)}\,\frac{\sinh\frac{\pi}{3}(s-t)}{\sinh \frac{\pi}{2}(s-t)}\,,
   \end{split}
  \]
  and
  \[
   \begin{split}
    &\frac{32\pi}{\lambda}\iint_{\mathbb{R}\times\mathbb{R}}\frac{\overline{\theta_1(x)}\,\theta_2(y)}{\,(2\cosh(x+y)+1)\,(2\cosh(x-y)-1)\,}\,\ud x\,\ud y \\
    &=\;\frac{16\pi}{3\lambda}\iint_{\mathbb{R}\times\mathbb{R}}\overline{\widehat{\theta}_1(s)}\,\widehat{\theta}_2(t)\,\frac{\sinh\frac{\pi}{6}(s+t)}{\sinh \frac{\pi}{2}(s+t)}\,\frac{\sinh\frac{\pi}{3}(s-t)}{\sinh \frac{\pi}{2}(s-t)}\,\ud s\,\ud t\,.
   \end{split}
  \]
  Adding up all together yields \eqref{eq:xiWxi-zero-FOURIER}. 
  \end{proof}

  \subsection{Radial Ter-Martirosyan-Skornyakov equation}
  
  The next technical tool is the solution formula for the equation
  \begin{equation}\label{eq:TMSeq0}
   T_\lambda^{(0)}\xi\;=\;\eta
  \end{equation}
  in the unknown $\xi$ and with datum $\eta$, spherically symmetric on $\mathbb{R}^3$. \eqref{eq:TMSeq0} is customarily referred to as the \emph{Ter-Martirosyan-Skornyakov equation}\index{Ter-Martirosyan-Skornyakov!equation} for the sector of zero angular momentum. It appeared for the first time in \cite[Eq.~(12)]{TMS-1956}, the already-mentioned work by Ter-Martirosyan and Skornyakov, whence the name.

  Representing as usual (see \eqref{eq:xihatangularexpansion} and \eqref{eq:0xi} above)
  \[
   \widehat{\xi}(\pp)\;=\;\frac{1}{\sqrt{4\pi}}f^{(\xi)}(|\pp|)\,,\qquad  \widehat{\eta}(\pp)\;=\;\frac{1}{\sqrt{4\pi}}f^{(\eta)}(|\pp|)
  \]
  in terms of the corresponding radial components, switching to the \emph{re-scaled} radial components $\theta^{(\xi)}$ and $\theta^{(\eta)}$ defined in \eqref{ftheta-1}, and representing the l.h.s.~of \eqref{eq:TMSeq0} by means of \eqref{eq:Tlxi-theta}, equation \eqref{eq:TMSeq0} takes the form
  \begin{equation}\label{radialTMS0}
    \theta^{(\xi)}(x)-\frac{4}{\pi\sqrt{3}}\int_{\mathbb{R}}\ud y\,\theta^{(\xi)}(y) \log \frac{\,2\cosh(x-y)+1\,}{\,2\cosh(x-y)-1\,}\;=\;\frac{1}{2\pi^2\sqrt{\lambda}}\frac{\theta^{(\eta)}(x)}{\cosh x}
    \end{equation}
  as an identity for $x\geqslant 0$. With the short-hands
  \[
   \theta\;\equiv\; \theta^{(\xi)}\,,\qquad \vartheta\;\equiv\;\frac{1}{2\pi^2\sqrt{\lambda}}\frac{\theta^{(\eta)}}{\cosh x}
  \]
  one writes
  \begin{equation}\label{radialTMS0radial}
    \theta(x)-\frac{4}{\pi\sqrt{3}}\int_{\mathbb{R}}\ud y\,\theta(y) \log \frac{\,2\cosh(x-y)+1\,}{\,2\cosh(x-y)-1\,}\;=\;\vartheta(x)
    \end{equation}
   in the unknown $\theta$. 
   Should one like to interpret \eqref{radialTMS0radial} as an identity on the whole real line, one has to assume for consistency that also $\vartheta$, as $\theta^{(\eta)}$, is prolonged by odd parity.  \eqref{radialTMS0radial} is referred to as the \emph{radial Ter-Martirosyan-Skornyakov equation}\index{Ter-Martirosyan-Skornyakov!radial equation} (for the $\ell=0$ sector).

   Applying  \eqref{eq:fourierconvolution}-\eqref{eq:gamma-distribution} one sees that \eqref{radialTMS0radial} is equivalently re-written in the Fourier-transformed version
  \begin{equation}\label{radialTMS0radial-FOURIER}
   \widehat{\gamma}(s)\,\widehat{\theta}(s)\;=\;\widehat{\vartheta}(s)\,,
  \end{equation}
  understood in general as a distribution equation in the distribution unknown $\widehat{\theta}$.

   In  \eqref{radialTMS0radial} the functional space for the unknown $\theta$ is determined by the space for the original unknown $\xi$ through formula \eqref{eq:mellinnorms}. The same holds for the functional space for $\vartheta$, recalling that $\vartheta=\frac{1}{2\pi^2\sqrt{\lambda}}\frac{\theta^{(\eta)}(x)}{\cosh x}$ for some datum $\eta$. For the present applications (Sect.~\ref{sec:symmTMSubdd}-\ref{sec:adjointBirman}) one needs $\eta\in H^{s}_{\ell=0}(\mathbb{R}^3)$ for some $s>-\frac{1}{2}$, in which case \eqref{eq:mellinnorms} implies that $\theta^{(\eta)}/(\cosh x)^{\frac{1}{2}-s}$ is an $L^2$-function and hence, by the H\"{o}lder inequality,
   \[
    \vartheta\;\sim\;\frac{1}{(\cosh x)^{\frac{1}{2}+s}}\,\frac{\theta^{(\eta)}}{(\cosh x)^{\frac{1}{2}-s}}\;\in\; L^p(\mathbb{R})\quad\forall p\in[1,2]\,.
   \]
   Therefore, as $|x|\to\infty$, the above $\vartheta$ has an $L^2$-behaviour dumped by a multiplicative exponential decay.

   When $\vartheta$ is smooth and has rapid decrease, one can translate the solution to \eqref{radialTMS0radial-FOURIER} back to the $x$-coordinate.

   \begin{lemma}\label{lem:0tms-generalsol}
    The general solution to \eqref{radialTMS0radial} when $\vartheta$ is smooth and with rapid decrease is 
    \begin{equation}\label{eq:0tms-generalsol}
      \theta(x)\;=\;c\,\sin s_0 x - \frac{1}{2s_0}\,\sin (s_0 |x|)*\Big(\frac{\widehat{\vartheta}}{\:\widehat{\gamma}_+}\Big)^{\!\vee}\,,\qquad c\in\mathbb{C}\,.
    \end{equation}
    The function \eqref{eq:0tms-generalsol} belongs to $L^\infty(\mathbb{R})$ and is asymptotically $\cos$-periodic as $|x|\to +\infty$ with period $2\pi/s_0$.
%
%
   \end{lemma}

   In the proof, as well as in the sequel, the distributional identities
   \begin{eqnarray}
    \big(\delta(s-s_0)-\delta(s+s_0)\big)^\vee(x)\;&=&\;\frac{2\ii}{\sqrt{2\pi}}\,\sin s_0 x\,, \label{eq:distri-deltaFouriersin}\\
    \Big(PV\frac{1}{s-s_0}-PV\frac{1}{s+s_0}\Big)^\vee(x)\;&=&\;
    -\sqrt{2\pi}\sin s_0|x| \label{eq:distriPVsignsin}
   \end{eqnarray}
   will be used, where $PV$ stands for the principal value distribution. \eqref{eq:distriPVsignsin} in particular follows from $\big(PV\frac{1}{s-s_0}\big)^\vee(x)=e^{\ii s_0 x}\big(PV\frac{1}{s}\big)^\vee(x) =\ii\sqrt{\frac{\pi}{2}}\,e^{\ii s_0 x}\,\mathrm{sign}\,x$.

   \begin{proof}[Proof of Lemma \ref{lem:0tms-generalsol}]
    The Fourier-transformed version \eqref{radialTMS0radial-FOURIER} of \eqref{radialTMS0radial} is equivalent to
   \[
    (s-s_0)(s+s_0)\,\widehat{\theta}(s)\;=\;\frac{\widehat{\vartheta}(s)}{\widehat{\gamma}_+(s)}\,,\qquad s\in\mathbb{R}\,,
   \]
   owing to \eqref{eq:gammaplus}. Dividing by $\widehat{\gamma}_+(s)$ has not altered the set of solutions because $\widehat{\gamma}_+(s)>0$ $\forall s\in\mathbb{R}$. Observe that by construction $\widehat{\theta}(-s)=-\widehat{\theta}(s)$, $\widehat{\vartheta}(-s)=-\widehat{\vartheta}(s)$, and  $\widehat{\gamma}_+(-s)=\widehat{\gamma}_+(s)$, thus in the identity above both sides change sign when $s\mapsto -s$, consistently. Moreover, since $\widehat{\gamma}_+$ is smooth, $\widehat{\gamma}_+(s)\sim s^{-2}$ as $|s|\to\infty$, and $\vartheta$ is smooth and with rapid decrease, then so too is $\widehat{\vartheta}/\widehat{\gamma}_+$. Therefore, such a distributional equation has general solution
  \[
   \widehat{\theta}(s)\;=\;c\big(\delta(s-s_0)-\delta(s+s_0)\big)+\frac{\widehat{\vartheta}(s)}{\widehat{\gamma}_+(s)}\,\frac{1}{2s_0}\Big(PV\frac{1}{s-s_0}-PV\frac{1}{s+s_0}\Big)\,,\quad c\in\mathbb{C}\,.
  \]
  The linear combination of $\delta(s-s_0)$ and $\delta(s+s_0)$ had to be anti-symmetric, owing to the odd parity of $\widehat{\theta}$. Taking the inverse Fourier transform by means of \eqref{eq:distri-deltaFouriersin}-\eqref{eq:distriPVsignsin}, one obtains
 \[
  \theta(x)\;=\;\frac{2\ii c}{\sqrt{2\pi}}\,\sin s_0 x - \frac{1}{2s_0}\,\sin (s_0 |x|)*\Big(\frac{\widehat{\vartheta}}{\:\widehat{\gamma}_+}\Big)^{\!\vee}\,.
 \]
 An obvious re-scaling of the arbitrary constant $c$ finally yields \eqref{eq:0tms-generalsol}. 

   The functions $\vartheta$, $\widehat{\vartheta}$, $\widehat{\vartheta}/\widehat{\gamma}_+$, and $(\widehat{\vartheta}/\widehat{\gamma}_+)^\vee$ are all functions of rapid decrease in the respective variable. Young's inequality then gives
   \begin{equation*}
    \Big\|\sin (s_0 |x|)*\Big(\frac{\widehat{\vartheta}}{\:\widehat{\gamma}_+}\Big)^\vee\Big\|_{L^\infty(\mathbb{R}_x)}\;\leqslant\;\Big\|\Big(\frac{\widehat{\vartheta}}{\:\widehat{\gamma}_+}\Big)^{\!\vee}\Big\|_{L^1(\mathbb{R}_x)}\;<\;+\infty.
   \end{equation*}
   The whole function \eqref{eq:0tms-generalsol} is therefore bounded.

   Next, set 
   \[
    \begin{split}
     h\;&:=\;(\widehat{\vartheta}/\widehat{\gamma}_+)^\vee \, ,\\
     A\;&:=\;\sin (s_0 |x|)*h\,.
    \end{split}
   \]
   Since $h$ is smooth and with rapid decrease, then $A$ is differentiable at any order, and a simple computation yields
   \[
     A''(x)\;=\;-s_0^2\, A(x)+2s_0 h(x)\,.
   \]
   Thus, using again the rapid decrease of $h$, asymptotically when $|x|\to +\infty$ the function $A$ satisfies $A''(x)+s_0^2\, A(x)=0$, whence its asymptotic periodicity
   \[
    A(x)\stackrel{|x|\to +\infty}{=} \Big(c_1^{(\vartheta)}\cos s_0 x + c_2^{(\vartheta)}\sin s_0 x\Big)\big(1 + o(1)\big)
   \]
   for some constants $c_1^{(\vartheta)},c_2^{(\vartheta)}\in\mathbb{C}$ that vanish when $\vartheta\equiv 0$.
   \end{proof}

  \subsection{Symmetric, unbounded below, TMS extension}\label{sec:symmTMSubdd}\index{Ter-Martirosyan-Skornyakov!symmetric extension}

  Consider the subspace
  \begin{equation}\label{eq:Dtilde0}
   \widetilde{\mathcal{D}}_0\;:=\;\left\{
   \xi\in H^{-\frac{1}{2}}_{\ell=0}(\mathbb{R}^3)\left|
   \begin{array}{c}
    \textrm{$\xi$ has re-scaled radial component} \\
     \theta=\sin (s_0 |x|)*\big(\widehat{\vartheta}/\widehat{\gamma}_+\big)^{\!\vee} \\
     \textrm{for }\;\vartheta\in C^\infty_{c,\mathrm{odd}}(\mathbb{R}_x)
   \end{array}
   \!\right.\right\}\,.
  \end{equation}
  In \eqref{eq:Dtilde0} the subscript `odd' indicates functions with odd parity. The constant $s_0\approx 1.0062$ is the unique positive root of $\widehat{\gamma}(s)=0$ as defined in \eqref{eq:gamma-distribution}, and $\widehat{\gamma}_+$ is defined in \eqref{eq:gammaplus}. The correspondence between $\xi$ and its re-scaled radial component $\theta$ is given by \eqref{eq:0xi}-\eqref{ftheta-1}.

  In the above definition of $\widetilde{\mathcal{D}}_0$ it is tacitly understood that the re-scaled radial components are all taken with the same parameter $\lambda>0$ in the definition \eqref{ftheta-1}. This does not mean that $\widetilde{\mathcal{D}}_0$ is a $\lambda$-dependent subspace, as one can easily convince oneself: the choice of $\lambda$ only fixes the convention for representing the element of $\widetilde{\mathcal{D}}_0$ in terms of the corresponding $\theta$.

  \begin{lemma}\label{lem:D0tildedomainproperties}~
  
  \begin{enumerate}[(i)]
  \item One has
  \begin{equation}\label{eq:thetavanishesatzero}
   \xi\,\in\,\widetilde{\mathcal{D}}_0\qquad\Rightarrow\qquad\theta^{(\xi)}(0)\;=\;0\,.
  \end{equation}
   \item One has
   \begin{equation}\label{eq:thetaxiD0}
   \xi\,\in\,\widetilde{\mathcal{D}}_0\qquad\Rightarrow\qquad\widehat{\gamma}(s)\,\widehat{\theta^{(\xi)}}(s)\;=\;-2 s_0\,\widehat{\vartheta}(s)\;\;\forall s\in\mathbb{R}\,.
  \end{equation}
   \item $\widetilde{\mathcal{D}}_0$ is dense in $H^{\frac{1}{2}-\varepsilon}_{\ell=0}(\mathbb{R}^3)$ for every $\varepsilon>0$.
   \item $\widetilde{\mathcal{D}}_0 \cap H^{\frac{1}{2}}_{\ell=0}(\mathbb{R}^3)=\{0\}$.
      \item For every $\lambda>0$ and $s\in\mathbb{R}$, $T_\lambda^{(0)}\widetilde{\mathcal{D}}_0\subset H^{s}_{\ell=0}(\mathbb{R}^3)$.
      \item For $\lambda>0$, $T_\lambda^{(0)}$ is injective on $\widetilde{\mathcal{D}}_0$, that is,
      \begin{equation}
       \big(\;\xi\in \widetilde{\mathcal{D}}_0\;\;\textrm{ and }\;\;T_\lambda^{(0)}\xi\equiv 0\;\big)\qquad\Rightarrow\qquad \xi\equiv 0\,.
      \end{equation}
   \item For every $\lambda>0$ one has 
   \begin{equation}
    \big\langle\xi, T_\lambda^{(0)}\eta\big\rangle_{L^2}\;=\;\big\langle T_\lambda^{(0)}\xi, \eta\big\rangle_{L^2}\qquad\forall \xi,\eta\in\widetilde{\mathcal{D}}_0\,.
   \end{equation}
  \end{enumerate} 
  \end{lemma}

  \begin{proof}
   Part (i) follows from the fact that in $\theta(0)=\int_{\mathbb{R}}\sin(s_0|y|)\big(\widehat{\vartheta}/\widehat{\gamma}_+\big)^{\!\vee}(y)$ the term $\big(\widehat{\vartheta}/\widehat{\gamma}_+\big)^{\!\vee}$ has odd parity, whereas $\sin(s_0|y|)$ is even.

   Part (ii) follows at once from the definition \eqref{eq:Dtilde0}, by means of Lemma \ref{lem:0tms-generalsol} and \eqref{radialTMS0radial-FOURIER}.
  
   (iii) Let $\xi\in\widetilde{\mathcal{D}}_0$ and $\varepsilon>0$. By means of \eqref{eq:mellinnorms} and Lemma \ref{lem:0tms-generalsol} one gets
   \[
    \big\|\xi\big\|_{H^{\frac{1}{2}-\varepsilon}(\mathbb{R}^3)}^2\;\approx\;\int_{\mathbb{R}}\,\frac{\,|\theta(x)|^2}{(\cosh x)^{2\varepsilon}}\,\ud x\;\leqslant\;\|\theta\|_{L^\infty}\!\int_{\mathbb{R}}\,\frac{\ud x}{(\cosh x)^{2\varepsilon}}\;<\;+\infty\,.
   \]
   In fact, \eqref{eq:mellinnorms} also implies that the density of the $\xi$'s of $\widetilde{\mathcal{D}}_0$ in $H^{\frac{1}{2}-\varepsilon}_{\ell=0}(\mathbb{R}^3)$ is equivalent to the density of the associated $\theta$'s in $L^2_{\mathrm{odd}}(\mathbb{R},(\cosh x)^{-2\varepsilon}\ud x)$. If in the latter Hilbert space a function $\theta_0$ was orthogonal to all such $\theta$'s, then
   \[
    \begin{split}
     0\;&=\;\int_{\mathbb{R}}\frac{\,\overline{\theta_0(x)}}{(\cosh x)^{2\varepsilon}}\,\,\theta(x)\,\ud x\;=\;\int_{\mathbb{R}}\frac{\,\overline{\theta_0(x)}}{(\cosh x)^{2\varepsilon}}\,\,\Big(\sin( s_0| x|)*\big(\widehat{\vartheta}/\widehat{\gamma}_+\big)^{\!\vee}\Big)(x)\,\ud x \\
     &=\;\int_{\mathbb{R}}\Big(\sin (s_0 |x|)*\frac{\,\overline{\theta_0(x)}}{(\cosh x)^{2\varepsilon}}\Big)(x)\,\big(\widehat{\vartheta}/\widehat{\gamma}_+\big)^{\!\vee}(x)\,\ud x\qquad\forall \vartheta\in C^\infty_{c,\mathrm{odd}}(\mathbb{R}_x)\,.
    \end{split}
   \]
   (The above change in the integration order is allowed, via Fubini-Tonelli, thanks to the rapid decrease of the functions $(\cosh x)^{-2\varepsilon}$ and $(\widehat{\vartheta}/\widehat{\gamma}_+)^\vee$.) This would imply 
   \[
    \int_{\mathbb{R}}\sin(s_0|x-y|) h_0(y)\,\ud y\;=\;0\qquad\textrm{for a.e.~} x\in\mathbb{R}\,,\qquad h_0\;:=\;\frac{\theta_0}{(\cosh x)^{2\varepsilon}}\,,
   \]
  whence $h_0\equiv 0$ and then $\theta_0\equiv 0$.
   
   (iv) Let $\xi\in\widetilde{\mathcal{D}}_0\setminus\{0\}$. Since $\theta$ is not identically zero and is asymptotically $\cos$-periodic (Lemma \ref{lem:0tms-generalsol}), then \eqref{eq:mellinnorms} implies
   \[
    \big\|\xi\big\|_{H^{\frac{1}{2}}(\mathbb{R}^3)}^2\;\approx\;\int_{\mathbb{R}}|\theta(x)|^2\,\ud x\;=\;+\infty\,.
   \]

   (v) Let $\xi\in\widetilde{\mathcal{D}}_0$ and $\lambda>0$. Owing to formula \eqref{eq:Tlambda0xi-with-theta} of Lemma \ref{lem:xithetaidentities},
    \begin{equation*}
     \begin{split}
      \big\|T_\lambda^{(0)}\xi\big\|_{H^{s}(\mathbb{R}^3)}^2\;&\approx\;\int_{\mathbb{R}}\ud x (\cosh x)^{1+2s}\bigg|\,\theta(x)-\frac{4}{\pi\sqrt{3}}\int_{\mathbb{R}}\ud y\,\theta(y) \log \frac{\,2\cosh(x-y)+1\,}{\,2\cosh(x-y)-1\,}\bigg|^2,
     \end{split}
    \end{equation*}
   and owing to Lemma \ref{lem:0tms-generalsol},
   \begin{equation*}
    \theta(x)-\frac{4}{\pi\sqrt{3}}\int_{\mathbb{R}}\ud y\,\theta(y) \log \frac{\,2\cosh(x-y)+1\,}{\,2\cosh(x-y)-1\,}\;=\;2s_0\,\vartheta(x)\,.
    \end{equation*}
    Therefore,
    \[
     \big\|T_\lambda^{(0)}\xi\big\|_{H^{s}(\mathbb{R}^3)}^2\;\lesssim\;\int_{\mathbb{R}}(\cosh x)^{1+2s}|\vartheta(x)|^2\,\ud x\;<\;+\infty\,,
    \]
   because $\vartheta$ is smooth and has compact support.

   (vi) Based on what argued for \eqref{eq:TMSeq0}-\eqref{radialTMS0radial}, and on Lemma \ref{lem:0tms-generalsol}, it is straightforward to see that if $\xi\in\widetilde{\mathcal{D}}_0$, and hence $\theta^{(\xi)}=\sin (s_0 |x|)*\big(\widehat{\vartheta}/\widehat{\gamma}_+\big)^{\!\vee}$ for some $\vartheta\in C^\infty_{c,\mathrm{odd}}(\mathbb{R}_x)$, then the re-scaled radial function $\theta^{(\eta)}$ of the charge $\eta:=T_\lambda^{(0)}\xi$ is
   \[
    \theta^{(\eta)}\;=\;-4\pi^2s_0\sqrt{\lambda}\,(\cosh x)\,\vartheta\,.
   \]
   Therefore, if $T_\lambda^{(0)}\xi\equiv 0$, then $\theta^{(\eta)}\equiv 0$, thus also $\vartheta\equiv 0$, implying that $\xi\equiv 0$.
   
   (vii) Owing to \eqref{eq:xiTxi-with-theta},
   \begin{equation*}
     \int_{\mathbb{R}^3} \overline{\,\widehat{\xi}(\pp)}\, \big(\widehat{T_\lambda^{(0)}\eta}\big)(\pp)\,\ud\pp\,=\,\frac{\,8\pi^2}{3\sqrt{3}}\int_{\mathbb{R}}\ud s\, \widehat{\gamma}(s)\,\overline{\widehat{\theta^{(\xi)}}(s)}\,\widehat{\theta^{(\eta)}}(s)\,=\,\int_{\mathbb{R}^3} \overline{\big(\widehat{T_\lambda^{(0)}\xi}\big)(\pp)}\,\widehat{\eta}(\pp)\,\ud\pp
    \end{equation*}
   as an identity between \emph{finite} quantities. Indeed,
   \[
   \begin{split}
    \int_{\mathbb{R}}\ud s\, \widehat{\gamma}(s)\,\overline{\widehat{\theta^{(\xi)}}(s)}\,\widehat{\theta^{(\eta)}}(s)\;&=\;-2s_0\int_{\mathbb{R}}\ud x\,\overline{\,\vartheta^{(\xi)}(x)}\,\theta^{(\eta)}(x) \\
    &=\;-2s_0\int_{\mathbb{R}}\ud x\,\overline{\,\vartheta^{(\xi)}(x)}\,\Big(\sin (s_0 |x|)*\big(\widehat{\vartheta^{(\eta)}}/\widehat{\gamma}_+\big)^{\!\vee}\Big)(x)
   \end{split}
   \]
  (having applied part (ii) in the first identity and \eqref{eq:Dtilde0} in the second), whence
  \[
   \begin{split}
    \Big| \int_{\mathbb{R}}\ud s\, \widehat{\gamma}(s)\,\overline{\widehat{\theta^{(\xi)}}(s)}\,\widehat{\theta^{(\eta)}}(s)\Big|\;&\leqslant\;\big\|\vartheta^{(\xi)}\big\|_{L^1}\Big\|\sin (s_0 |x|)*\big(\widehat{\vartheta^{(\eta)}}/\widehat{\gamma}_+\big)^{\!\vee}\Big\|_{L^\infty} \\
   &\leqslant\;\big\|\vartheta^{(\xi)}\big\|_{L^1}\Big\|\Big(\frac{\widehat{\vartheta^{(\eta)}}}{\:\widehat{\gamma}_+}\Big)^{\!\vee}\Big\|_{L^1}\;<\;+\infty
   \end{split}
  \]
 (again by Young's inequality).
%
  \end{proof}

  \begin{remark}\label{rem:whensymmetric2}
    Lemma \ref{lem:D0tildedomainproperties} supplements the picture previously emerged from Lemma \ref{lem:Tlambdaproperties}(i) and (v), and discussed in Remark \ref{rem:whensymmetric}, concerning the validity of the identity
     \begin{equation*}
 \int_{\mathbb{R}^3} \overline{\,\widehat{\xi}(\pp)}\, \big(\widehat{T_\lambda\eta}\big)(\pp)\,\ud\pp\;=\;\int_{\mathbb{R}^3} \overline{\,\widehat{T_\lambda\xi}(\pp)}\, \widehat{\eta}(\pp)\,\ud\pp\,.
 \end{equation*}
    Indeed, such a symmetry property was previously established for $\xi,\eta\in H^{\frac{1}{2}}(\mathbb{R}^3)$, whereas Lemma \ref{lem:D0tildedomainproperties}(v) now guarantees it also for $\xi,\eta\in\widetilde{\mathcal{D}}_0$, namely a non-$H^{\frac{1}{2}}$ domain (as shown by Lemma \ref{lem:D0tildedomainproperties}(iii)).     
  \end{remark}

  \begin{remark}
   Along the same line of the previous Remark, Lemma \ref{lem:D0tildedomainproperties} also supplements the analysis of the problem, considered in Lemma \ref{lem:Tlambdaproperties}(ii)-(iii) and Remark \ref{rem:Tl-failstomap}, of finding charges $\xi\in H^{-\frac{1}{2}}_{\ell=0}(\mathbb{R}^3)$ satisfying $T^{(0)}_\lambda\xi\in H^{\frac{1}{2}}(\mathbb{R}^3)$, which is in turn crucial to construct $W_\lambda^{-1}T^{(0)}_\lambda$. As observed already, high regularity of $\xi(\yy)$ (hence fast decay of $\widehat{\xi}(\pp)$) is of no avail in the sector $\ell=0$. Now Lemma \ref{lem:D0tildedomainproperties} implies that the appropriate feature for having $T_\lambda\xi\in H^{\frac{1}{2}}(\mathbb{R}^3)$ is a suitable \emph{oscillation} of $\widehat{\xi}(\pp)$ combined with some $|\pp|$-decay compatible with $H^{-\frac{1}{2}^-}$-regularity (which, loosely speaking, corresponds to some localisation of $\xi(\yy)$ close to $\yy=0$). This is seen by the fact that $\theta^{(\xi)}$ must satisfy \eqref{eq:thetaxiD0}, hence (Lemma \ref{lem:0tms-generalsol}) $\theta^{(\xi)}(x)$ is $\cos$-periodic in $x$, with consequent (non-periodic) oscillation in $\widehat{\xi}(\pp)$ via \eqref{eq:0xi} and \eqref{ftheta-2}.
  \end{remark}

  \begin{lemma}\label{lem:expectationunbounded}
   Let $\lambda>0$. Then
   \begin{equation}\label{eq:expectationunbounded}
    \inf_{\xi\in \widetilde{\mathcal{D}}_0}\frac{\;\langle\xi,T_\lambda^{(0)}\xi\rangle_{L^2}}{\;\;\|\xi\|^2_{H^{-\frac{1}{2}}}}\;=\;-\infty\,.
   \end{equation}   
  \end{lemma}

  \begin{proof}
   Let $\xi\in \widetilde{\mathcal{D}}_0$, and let $\theta$ be its re-scaled radial component \eqref{eq:0xi}-\eqref{ftheta-1}. 
   The numerator in \eqref{eq:expectationunbounded} is indeed finite and real (Lemma \ref{lem:D0tildedomainproperties}(vii), Remark \ref{rem:whensymmetric2}).
   For $\varepsilon>0$ let $\xi_\varepsilon$ be the element of $\widetilde{\mathcal{D}}_0$ whose re-scaled radial component $\theta_\varepsilon$ is defined by $\theta_\varepsilon(x):=\theta(\varepsilon x)$. Then $\widehat{\theta}_\varepsilon(s)=\varepsilon^{-1}\widehat{\theta}(s/\varepsilon)$.
   Therefore,
   \[
    \begin{split}
     \lim_{\varepsilon\downarrow 0}\,\big\langle\xi_\varepsilon,T_\lambda^{(0)}\xi_\varepsilon\big\rangle_{L^2}\;&=\;\frac{\,8\pi^2}{3\sqrt{3}}\,\lim_{\varepsilon\downarrow 0}\,\frac{1}{\,\varepsilon^2}\int_{\mathbb{R}}\ud s\, \widehat{\gamma}(s)\,|\widehat{\theta}(s/\varepsilon)|^2 \\
     &=\;\frac{\,8\pi^2}{3\sqrt{3}}\,\lim_{\varepsilon\downarrow 0}\,\frac{1}{\varepsilon}\int_{\mathbb{R}}\ud s\, \widehat{\gamma}(\varepsilon s)\,|\widehat{\theta}(s)|^2\;=\;-\infty\,,
    \end{split}
   \]
%
   having used \eqref{eq:xiTxi-with-theta} in the first identity, and the fact that $\int_{\mathbb{R}}\ud s|\widehat{\theta}(s)|^2=+\infty$ (which follows from Lemma \ref{lem:D0tildedomainproperties}(iv) and \eqref{eq:mellinnorms}) and $ \widehat{\gamma}(0)=1-\frac{4\pi}{3\sqrt{3}}<0$. On the other hand,
    \[
     \|\xi_\varepsilon\|^2_{H^{-\frac{1}{2}}}\;\approx\;\int_{\mathbb{R}}\frac{|\theta_\varepsilon(x)|^2}{(\cosh x)^2}\,\ud x\;=\;\int_{\mathbb{R}}\frac{|\theta(\varepsilon x)|^2}{(\cosh x)^2}\,\ud x\;\xrightarrow[]{\varepsilon\downarrow 0}\;0\,,
    \]
   owing to \eqref{eq:mellinnorms}, \eqref{eq:thetavanishesatzero}, and dominated convergence. As a consequence,
   \[
    \lim_{\varepsilon\downarrow 0}\frac{\;\langle\xi_\varepsilon,T_\lambda^{(0)}\xi_\varepsilon\rangle_{L^2}}{\;\;\|\xi_\varepsilon\|^2_{H^{-\frac{1}{2}}}}\;=\;-\infty\,,
   \] 
    which proves \eqref{eq:expectationunbounded}.  
  \end{proof}

  For $\lambda>0$, let us now consider the operator
    \begin{equation}\label{eq:Alambdatilde-0}
   \begin{split}
       \widetilde{\mathcal{A}_{\lambda}^{(0)}}\;&:=\;3W_\lambda^{-1}T_\lambda^{(0)} \, , \\
       \mathcal{D}\big( \widetilde{\mathcal{A}_{\lambda}^{(0)}}\big)\;&:=\; \widetilde{\mathcal{D}}_0 \, ,
   \end{split}
  \end{equation}
  with respect to the Hilbert space $H^{-\frac{1}{2}}_{W_\lambda,\ell=0}(\mathbb{R}^3)$.
 The definition \eqref{eq:Alambdatilde-0} is well-posed, owing to Lemma \ref{lem:D0tildedomainproperties}(v) and Lemma \ref{lem:Wlambdaproperties}(ii). The operator $\widetilde{\mathcal{A}_{\lambda}^{(0)}}$ is the counterpart, for the sector of angular momentum $\ell=0$, of the operator $\widetilde{\mathcal{A}_{\lambda}^{(\ell)}}$ defined in \eqref{eq:Alambdatildenot0}.

  \begin{lemma}\label{lem:symmetricAtilde0} With respect to the Hilbert space $H^{-\frac{1}{2}}_{W_\lambda,\ell=0}(\mathbb{R}^3)$, the operator $\widetilde{\mathcal{A}_{\lambda}^{(0)}}$ is densely defined, symmetric, and not semi-bounded.
  \end{lemma}

  \begin{proof}
   The density of $\widetilde{\mathcal{A}_{\lambda}^{(0)}}$ follows from Lemma \ref{lem:D0tildedomainproperties}(iii) and the canonical Hilbert space isomorphism  $H^{-\frac{1}{2}}_{W_\lambda,\ell=0}(\mathbb{R}^3)\cong H^{-\frac{1}{2}}(\mathbb{R}^3)$. 
      Symmetry follows from the finiteness and reality of
   \[
    \big\langle \xi,\widetilde{\mathcal{A}_{\lambda}^{(0)}}\xi\big\rangle_{H^{-\frac{1}{2}}_{W_\lambda}}\;=\;3\big\langle\xi,T_\lambda^{(0)}\xi\big\rangle_{L^2}\;=\;\frac{\,8\pi^2}{\sqrt{3}}\int_{\mathbb{R}}\ud s\, \widehat{\gamma}(s)\,\big|\widehat{\theta}^{(\xi)}(s)\big|^2\qquad\forall\xi\in\widetilde{\mathcal{D}}_0\,,
   \]
   having used \eqref{eq:Alambdatilde-0} and \eqref{eq:W-scalar-product} in the first identity, and \eqref{eq:xiTxi-with-theta} in the second. (The finiteness of the above quantities is argued precisely as done in the proof of Lemma \ref{lem:D0tildedomainproperties}(vii).)   
      The unboundedness of $\widetilde{\mathcal{A}_{\lambda}^{(0)}}$ from above is obvious, and from below it follows directly from Lemma \ref{lem:expectationunbounded}, using again the isomorphism  $H^{-\frac{1}{2}}_{W_\lambda,\ell=0}(\mathbb{R}^3)\cong H^{-\frac{1}{2}}(\mathbb{R}^3)$.   
  \end{proof}

  Analogously to the operator $\widetilde{\mathcal{A}_{\lambda}^{(\ell)}}$ from \eqref{eq:Alambdatildenot0} when $\ell\in\mathbb{N}$, $\widetilde{\mathcal{A}_{\lambda}^{(0)}}$ is a well-defined labelling (Birman) extension parameter \index{Birman extension parameter} for a Ter-Martirosyan-Skornyakov \emph{symmetric} extension \index{Ter-Martirosyan-Skornyakov!symmetric extension} of our original operator of interest $\mathring{H}$, with inverse scattering length $\alpha=0$, of course only as far as the $\ell=0$ sector is concerned. Indeed, based on Lemma \ref{lem:symmetricAtilde0}, one applies the considerations of Remark \ref{rem:remonsym}(ii). Moreover, as found for $\widetilde{\mathcal{A}_{\lambda}^{(\ell)}}$, it will be soon shown (Subsect.~\ref{sec:adjointBirman}) that  $\widetilde{\mathcal{A}_{\lambda}^{(0)}}$ is not self-adjoint in $H^{-\frac{1}{2}}_{W_\lambda,\ell=0}(\mathbb{R}^3)$, and therefore it does not identify a Ter-Martirosyan-Skornyakov \emph{self-adjoint} extension of $\mathring{H}$ (with $\ell=0$).

  \emph{Unlike} $\widetilde{\mathcal{A}_{\lambda}^{(\ell)}}$, however, $\widetilde{\mathcal{A}_{\lambda}^{(0)}}$ is not bounded from below, hence does not admit a Friedrichs extension: the identification of its self-adjoint extension(s), if any, requires a different approach than Proposition \ref{prop:Alambdaellnot0}, discussed in the following Subsection \ref{sec:multiselfadj-0}.

  Moreover, the Ter-Martirosyan-Skornyakov \emph{symmetric} extension of $\mathring{H}$ identified by $\widetilde{\mathcal{A}_{\lambda}^{(0)}}$ for the sector $\ell=0$ is itself unbounded from below (and above) on the bosonic space $\cH_\mathrm{b}$. This follows from the computation made in Remark \ref{rem:TMSsym}, namely
  \[
   \big\langle g , (H_{\mathcal{A}_{\lambda}^{(0)}} +\lambda\mathbbm{1})g\big\rangle_{\cH_\mathrm{b}}\;=\;\big\langle \xi,\widetilde{\mathcal{A}_{\lambda}^{(0)}}\xi\big\rangle_{H^{-\frac{1}{2}}_{W_\lambda}}
  \]
  for functions $g=u_\xi^\lambda$ (with an innocent abuse of notation in writing $H_{\mathcal{A}_{\lambda}^{(0)}}$, as one is only referring here to the Hamiltonian on functions with $\xi$-charge in the sector $\ell=0$): when $\xi\in\widetilde{\mathcal{D}}_0$ the expectation 
     \[
 \frac{\big\langle g , H_{\mathcal{A}_{\lambda}^{(0)}}g\big\rangle_{\cH_\mathrm{b}}}{\|g\|_{\cH_\mathrm{b}}^2}\;=\; \frac{\big\langle \xi,\widetilde{\mathcal{A}_{\lambda}^{(0)}}\xi\big\rangle_{H^{-\frac{1}{2}}_{W_\lambda}}}{\;\|\xi\|^2_{H^{-\frac{1}{2}}_{W_\lambda}}}-\lambda
 \]
  can be made, at fixed $\lambda$, arbitrarily negative, owing to \eqref{eq:expectationunbounded}.

  \begin{remark}\label{rem:merepurpose}
   For the mere purpose of realising $W_\lambda^{-1}T_\lambda^{(0)}$ as a densely defined and symmetric operator in $H^{-\frac{1}{2}}_{W_\lambda,\ell=0}(\mathbb{R}^3)$, one could have selected the \emph{larger} domain 
   \begin{equation}\label{eq:Dolarger}
    \widetilde{\mathcal{D}}_0^\prime\;:=\;\widetilde{\mathcal{D}}_0\dotplus\mathrm{span}\{\xi_{s_0}\}\,,
   \end{equation}
  where $\xi_{s_0}$ has re-scaled radial component
  \begin{equation}
   \theta_{s_0}(x)\;:=\;\sin s_0 x
  \end{equation}
  for some tacitly declared $\lambda>0$. All conclusions of Lemma \ref{lem:D0tildedomainproperties} remain straightforwardly valid for $\widetilde{\mathcal{D}}_0^\prime$, but for the injectivity of $T_\lambda^{(0)}$ on $\widetilde{\mathcal{D}}_0^\prime$, for 
  \begin{equation}
   T_\lambda^{(0)}\xi_{s_0}\;=\;0\,,
  \end{equation}
 as follows from (the proof of) Lemma \ref{lem:0tms-generalsol}. The conclusions of Lemma \ref{lem:expectationunbounded} apply to $\widetilde{\mathcal{D}}_0^\prime$ as well, and therefore one has a version of Lemma \ref{lem:symmetricAtilde0} also for the operator
       \begin{equation}\label{eq:Alambdatilde-0prime}
   \begin{split}
       \widetilde{\mathcal{B}_{\lambda}^{(0)}}\;&:=\;3W_\lambda^{-1}T_\lambda^{(0)} \, , \\
       \mathcal{D}\big( \widetilde{\mathcal{B}_{\lambda}^{(0)}}\big)\;&:=\; \widetilde{\mathcal{D}}_0^\prime\,.
   \end{split}
  \end{equation}
  \emph{The issue with such $\widetilde{\mathcal{B}_{\lambda}^{(0)}}$}, as compared to $\widetilde{\mathcal{A}_{\lambda}^{(0)}}$, is that it is symmetric on a domain that is \emph{too large} for the problem of finding self-adjoint operators of Ter-Martirosyan-Skornyakov type: the operator $\mathring{H}_{\widetilde{\mathcal{A}_{\lambda}^{(0)}}}$ (in the sector $\ell=0$) admits self-adjoint TMS extensions on $L^2_{\mathrm{b}}(\mathbb{R}^3\times\mathbb{R}^3\,\ud\yy_1,\ud\yy_2)$, the operator $\mathring{H}_{\widetilde{\mathcal{B}_{\lambda}^{(0)}}}$ does not. This discussion will be continued and completed in Remark \ref{rem:choicedomains}.  
  \end{remark}

  \subsection{Adjoint of the Birman parameter}\label{sec:adjointBirman}

  Here the operator $ \big(\widetilde{\mathcal{A}_{\lambda}^{(0)}}\big)^\star$ will be characterised.  Recall that `$\star$' indicates the adjoint with respect to $H^{-\frac{1}{2}}_{W_\lambda,\ell=0}(\mathbb{R}^3)$, whereas `$*$' is reserved for the adjoint in $L^2(\mathbb{R}^3)$.

  \begin{lemma}\label{lem:deficiency1-1}
   Let $\lambda>0$. The densely defined and symmetric operator $\widetilde{\mathcal{A}_{\lambda}^{(0)}}$ defined in \eqref{eq:Alambdatilde-0} on the Hilbert space $H^{-\frac{1}{2}}_{W_\lambda,\ell=0}(\mathbb{R}^3)$ has deficiency indices $(1,1)$.\index{deficiency indices} Thus, for every $\mu>0$ there exist two non-zero functions $\xi_{\ii\mu},\xi_{-\ii\mu}\in H^{-\frac{1}{2}}_{W_\lambda,\ell=0}(\mathbb{R}^3)$ such that
   \begin{equation}\label{eq:twodefsub}
  \begin{split}
   \ker\Big(\big(\widetilde{\mathcal{A}_{\lambda}^{(0)}}\big)^\star-\ii\mu\mathbbm{1}\Big)\;&=\;\mathrm{span}\{\xi_{\ii\mu}\} \\
   \ker\Big(\big(\widetilde{\mathcal{A}_{\lambda}^{(0)}}\big)^\star+\ii\mu\mathbbm{1}\Big)\;&=\;\mathrm{span}\{\xi_{-\ii\mu}\}\,.
  \end{split}
  \end{equation}
  Moreover, $\big(\widetilde{\mathcal{A}_{\lambda}^{(0)}}\big)^\star$ acts on such eigenvectors as
  \begin{equation}\label{eq:A0tildestareigenv}
   \big(\widetilde{\mathcal{A}_{\lambda}^{(0)}}\big)^\star\xi_{\pm \ii \mu}\;=\;3 W_\lambda^{-1}T_\lambda^{(0)}\,	\xi_{\pm \ii \mu}\;=\;\pm \ii\mu\,\xi_{\pm \ii \mu}\,.
  \end{equation}
  \end{lemma}

  \begin{corollary}\label{cor:A0tildeadj}
   Under the above assumptions,
   \begin{equation}\label{eq:A0tildestar}
   \begin{split}
    \mathcal{D}\big(\big(\widetilde{\mathcal{A}_{\lambda}^{(0)}})^\star\big)\;&=\;\widetilde{\mathcal{D}}_{0}' \dotplus\mathrm{span}\{\xi_{\ii\mu}\}\dotplus \mathrm{span}\{\xi_{-\ii\mu}\} \\
     \big(\widetilde{\mathcal{A}_{\lambda}^{(0)}}\big)^\star\;&=\;3 W_\lambda^{-1}T_\lambda^{(0)}\,,
   \end{split}
  \end{equation}
  where $\widetilde{\mathcal{D}}_{0}'$ is the domain of the operator closure of $ \widetilde{\mathcal{A}_{\lambda}^{(0)}}$ with respect to $H^{-\frac{1}{2}}_{W_\lambda,\ell=0}(\mathbb{R}^3)$, that is, $\widetilde{\mathcal{D}}_{0}'$ is the closure of $\widetilde{\mathcal{D}}_{0}$ in the graph norm of  $\widetilde{\mathcal{A}_{\lambda}^{(0)}}$.
  \end{corollary}

  In practice the explicit characterisation of $\widetilde{\mathcal{D}}_{0}'$ is not needed. This will be clear in due time from the usage of Lemma \ref{lem:adjointzero} in the proof of Theorem \ref{thm:spectralanalysis}.
   
    Convenient asymptotics for the elements of $\big(\widetilde{\mathcal{A}_{\lambda}^{(0)}}\big)^\star$ will be now determined as well.

  \begin{lemma}\label{lem:adjointasymtotics}
   Let $\lambda>0$ and $z\in\mathbb{C}\setminus\mathbb{R}$ (in practice $z=\ii\mu$ for $\mu\in\mathbb{R}\setminus\{0\}$, so as to cover both deficiency subspaces \eqref{eq:twodefsub}). Let $\xi\in\ker\big(\big(\widetilde{\mathcal{A}_{\lambda}^{(0)}}\big)^\star-z\mathbbm{1}\big)$.
   Then, 
   \begin{equation}\label{eq:adjointasymtotics}
   \begin{split}
    \widehat{\xi}(\pp)\;&=\;A_\xi\frac{\,\sin({s_0\log|\pp|})+W_{\lambda,\xi}(z)\,\cos({s_0\log|\pp|})\,}{\pp^2}(1+o(1)) \\
    &\qquad\textrm{as }|\pp|\to+\infty\,,\;|z|/\lambda\to 0
   \end{split}
   \end{equation}
    for two constants $A_\xi,W_{\lambda,\xi}(z)\in\mathbb{C}$ with
    \begin{equation}\label{eq:Wconstproperty}
     W_{\lambda,\xi}(\overline{z})\;=\;\overline{W_{\lambda,\xi}(z)}\,.
    \end{equation}
 Here $s_0\approx 1.0062$ is the unique positive root of $\widehat{\gamma}(s)=0$ as defined in \eqref{eq:gamma-distribution}.
   \end{lemma}

  \begin{proof}[Proof of Lemma \ref{lem:deficiency1-1}] Let $\mu>0$. Let $\xi_+\in H^{-\frac{1}{2}}_{W_\lambda,\ell=0}(\mathbb{R}^3)$ with
  \[
   \xi_+\;\in\;\ker\Big(\big(\widetilde{\mathcal{A}_{\lambda}^{(0)}}\big)^\star-\ii\mu\mathbbm{1}\Big)\;=\;\mathrm{ran}\big(\widetilde{\mathcal{A}_{\lambda}^{(0)}})+\ii\mu\mathbbm{1}\big)^{\perp_\lambda}\,,
  \]
  and let $\theta_+$ be the re-scaled radial component of $\xi_+$. Then
  \[\tag{$\star$}\label{eq:tagstar}
  \begin{split}
   0\;&=\;\big\langle\xi_+,\big(\widetilde{\mathcal{A}_{\lambda}^{(0)}})+\ii\mu\mathbbm{1}\big)\xi\big\rangle_{H^{-\frac{1}{2}}_{W_\lambda}} \\
   &=\;3\big\langle \xi_+,T_\lambda^{(0)}\xi\big\rangle_{H^{-\frac{1}{2}},H^{\frac{1}{2}}}+\ii\mu\big\langle \xi_+,W_\lambda^{(0)}\xi\big\rangle_{H^{-\frac{1}{2}},H^{\frac{1}{2}}}\qquad\forall\xi\in\widetilde{\mathcal{D}}_0\,.
  \end{split}
  \]
  Here one applied \eqref{eq:W-scalar-product} as usual, together with \eqref{eq:Alambdatilde-0} and Lemmas \ref{lem:Wlambdaproperties}(ii) and  \ref{lem:D0tildedomainproperties}(v).

   The two duality products appearing on the r.h.s.~above have been computed in (the proof of) Lemma \ref{lem:xithetaidentities} and in Lemma \ref{lem:xiWxi-zero}:
   \begin{equation*}
   \begin{split}
    \big\langle \xi_+,&T_\lambda^{(0)}\xi\big\rangle_{H^{-\frac{1}{2}},H^{\frac{1}{2}}}\;=\;\frac{\;8\pi^2}{\,3\sqrt{3}}\bigg(\int_{\mathbb{R}}\overline{\theta_+(x)}\,\theta(x)\,\ud x\\
   &\quad -\frac{4}{\pi\sqrt{3}}\iint_{\mathbb{R}\times\mathbb{R}}\overline{\theta_+(x)}\,\theta(y)\log \frac{\,2\cosh(x-y)+1\,}{\,2\cosh(x-y)-1\,}\,\ud x\,\ud y\bigg) \, ,\\
   \big\langle \xi_+,&W_\lambda^{(0)}\xi\big\rangle_{H^{-\frac{1}{2}},H^{\frac{1}{2}}}\;=\;\frac{4\pi^2}{\,\lambda\sqrt{3}\,}\int_{\mathbb{R}}\frac{\,\overline{\theta_+}(x)\,\theta(x)}{\,(\cosh x)^2}\,\ud x \\
   &\quad +\frac{32\pi}{\lambda}\iint_{\mathbb{R}\times\mathbb{R}}\frac{\overline{\theta_+(x)}\,\theta(y)}{\,(2\cosh(x+y)+1)\,(2\cosh(x-y)-1)\,}\,\ud x\,\ud y\,.
    \end{split}
  \end{equation*}
  Here $\theta$ denotes the re-scaled radial component of $\xi\in\widetilde{\mathcal{D}}_0$.

  The above quantities can be re-written after taking the Fourier transform, by means of \eqref{eq:xiTxi-with-theta} and \eqref{eq:xiWxi-zero-FOURIER}:
  \[
   \begin{split}
     \big\langle \xi_+,T_\lambda^{(0)}\xi\big\rangle_{H^{-\frac{1}{2}},H^{\frac{1}{2}}}\;&=\;\frac{\;8\pi^2}{\,3\sqrt{3}}\int_{\mathbb{R}}\widehat{\gamma}(s)\,\overline{\widehat{\theta}_+(s)}\,\widehat{\theta}(s)\,\ud s\,  \, ,\\
     \big\langle \xi_+,W_\lambda^{(0)}\xi\big\rangle_{H^{-\frac{1}{2}},H^{\frac{1}{2}}}\;&=\;\frac{2\pi^2}{\,\lambda\sqrt{3}\,}\int_{\mathbb{R}}\Big(\frac{s}{\,\sinh\frac{\pi}{2}s}*\overline{\widehat{\theta}_+}\Big)(s)\,\widehat{\theta}(s)\,\ud s \\
     &\!\!\!\!\!\!\!\!\!\!\!\!\!\!\!\!\!\!+\frac{16\pi}{3\lambda}\iint_{\mathbb{R}\times\mathbb{R}}\overline{\widehat{\theta}_+(t)}\,\widehat{\theta}(s)\,\frac{\sinh\frac{\pi}{6}(s+t)}{\sinh \frac{\pi}{2}(s+t)}\,\frac{\sinh\frac{\pi}{3}(s-t)}{\sinh \frac{\pi}{2}(s-t)}\,\ud s\,\ud t\,.
   \end{split}
  \]

  Adding them up into the eigenvector equation \eqref{eq:tagstar}, and using the density $\widetilde{\mathcal{D}}_0$ together with Fubini-Tonelli theorem, leads to the following equation in the unknown $\theta_+$:
  \begin{equation}\label{eq:eigenequation-thetaplus}
   \begin{split}
    & \bigg(\theta_+(x)-\frac{4}{\pi\sqrt{3}}\int_{\mathbb{R}}\theta_+(y)\,\log \frac{\,2\cosh(x-y)+1\,}{\,2\cosh(x-y)-1\,}\,\ud y\bigg) \\
    & =\;\frac{ \ii \mu}{ 2 \lambda }\bigg(\frac{\theta_+(x)}{\:(\cosh x)^2}+\frac{8\sqrt{3}}{\pi}\int_{\mathbb{R}}\frac{\theta_+(y)}{\,(2\cosh(x+y)+1)\,(2\cosh(x-y)-1)\,}\,\ud y\bigg)
   \end{split}
  \end{equation}
  (for a.e.~$x\in\mathbb{R}$), or, equivalently, 
  \begin{equation}\label{eq:eigenequation-thetaplusF}
   \begin{split}
      \widehat{\gamma}(s)\,\widehat{\theta}_+(s)\;&= \;\frac{\,\ii\mu}{4\lambda}\Big(\Big(\frac{s}{\,\sinh\frac{\pi}{2}s}*\widehat{\theta}_+\Big)(s) \\
      & +\frac{8}{\,\pi\sqrt{3}}\int_{\mathbb{R}}\frac{\sinh\frac{\pi}{6}(s+t)}{\sinh \frac{\pi}{2}(s+t)}\,\frac{\sinh\frac{\pi}{3}(s-t)}{\sinh \frac{\pi}{2}(s-t)}\,\widehat{\theta}_+(t)\,\ud t\Big)\,.
   \end{split}
  \end{equation}

  The integration order's exchange in the double integrals was possible thanks to the fast decay of the integral kernels. This also demonstrates, unfolding \eqref{eq:eigenequation-thetaplusF} backwards, that $\xi_+$ satisfies
  \[
   3T_\lambda^{(0)}\xi_+\;=\;\ii \mu W_\lambda^{(0)}\xi_+\,,
  \]
  therefore $T_\lambda^{(0)}\xi_+\in\mathrm{ran}\,W_\lambda^{(0)}=H^{\frac{1}{2}}_{\ell=0}(\mathbb{R}^3)$ (Lemma \ref{lem:Wlambdaproperties}(ii)) and 
  \[
   3W_\lambda^{-1}T_\lambda^{(0)}\xi_+\;=\;\ii \mu\xi_+\;=\;\big(\widetilde{\mathcal{A}_{\lambda}^{(0)}}\big)^\star\xi_+\,.
  \]
  Thus, \emph{$\big(\widetilde{\mathcal{A}_{\lambda}^{(0)}}\big)^\star$ acts on the eigenvector $\xi_+$ precisely as $3W_\lambda^{-1}T_\lambda^{(0)}$}.

  Now, on the one hand (Sect.~\ref{sec:I-symmetric-selfadj}) the dimension of the deficiency subspace considered at the beginning of this proof is independent of $\mu>0$, and therefore the dimension of the space of solutions to \eqref{eq:eigenequation-thetaplus} (equivalently, \eqref{eq:eigenequation-thetaplusF}) does not depend on $\mu>0$. In fact, owing to von Neumann's conjugation criterion (Theorem \ref{thm:II-vNconj}), such a dimension is the same even when one takes instead $\mu<0$, namely when one considers the other deficiency subspace. On the other hand, mirroring the reasoning of Lemma \ref{lem:0tms-generalsol}'s proof, \eqref{eq:eigenequation-thetaplusF} can be re-written as
  \[
    \widehat{\theta}_+(s)\;=\;\widehat{\theta}_0(s)+ \ii\,\frac{\mu}{\lambda}\,\mathcal{L}\widehat{\theta}_+(s)
   \]
   with
   \[
    \widehat{\theta}_0(s)\;:=\;c\big(\delta(s-s_0)-\delta(s+s_0)\big)\,,\qquad c\in\mathbb{C}
   \]
   and 
   \[
    \begin{split}
     \mathcal{L}\widehat{\theta}_+(s)\;:=\;&\frac{1}{\,8 s_0\widehat{\gamma}_+(s)}\Big(\big({\textstyle\frac{s}{\,\sinh\frac{\pi}{2}s}}*\widehat{\theta}_+\big)(s) \\
     &+{\textstyle\frac{8}{\,\pi\sqrt{3}}}\int_{\mathbb{R}}{\textstyle\frac{\sinh\frac{\pi}{6}(s+t)}{\sinh \frac{\pi}{2}(s+t)}\,\frac{\sinh\frac{\pi}{3}(s-t)}{\sinh \frac{\pi}{2}(s-t)}}\,\widehat{\theta}_+(t)\,\ud t\Big)\;\times \\
     &\;\;\;\times\Big(PV\frac{1}{s-s_0}-PV\frac{1}{s+s_0}\Big)
    \end{split}
   \]
   ($s_0\approx 1.0062$ being the unique positive root of $\widehat{\gamma}(s)=0$).
   
  From this one sees that the dimension of the space of solutions to \eqref{eq:eigenequation-thetaplusF} is precisely equal to one, namely it is dictated by the dimensionality of the solution space for the pivot equation $ \widehat{\gamma}\,\widehat{\theta}_+=0$. The previous formulas provide also an iterative expansion of the form
  \[
    \widehat{\theta}_+\;=\;\sum_{k=0}^{N-1}\Big(\frac{\ii\mu}{\lambda}\Big)^k\mathcal{L}^k\widehat{\theta}_0+\Big(\frac{\ii\mu}{\lambda}\Big)^N\mathcal{L}\widehat{\theta}_+\,,\qquad N\in\mathbb{N}
  \]
  in the regime $\mu/\lambda\ll 1$, which does not alter the dimension of the space of solutions.

  In conclusion,
  \[
   \dim \ker\Big(\big(\widetilde{\mathcal{A}_{\lambda}^{(0)}}\big)^\star-\ii\mu\mathbbm{1}\Big)\;=\;\dim \ker\Big(\big(\widetilde{\mathcal{A}_{\lambda}^{(0)}}\big)^\star+\ii\mu\mathbbm{1}\Big)\;=\;1\,,
  \]
  the analysis of the second deficiency subspace being clearly the same as above, upon exchanging $\mu$ with $-\mu$. This proves that $\widetilde{\mathcal{A}_{\lambda}^{(0)}}$ has deficiency indices $(1,1)$.  
  \end{proof}

  \begin{proof}[Proof of Corollary \ref{cor:A0tildeadj}]
   Formula \eqref{eq:A0tildestar} is an immediate consequence of Lemma \ref{lem:deficiency1-1} and \eqref{eq:Alambdatilde-0}, as an application of von Neumann's formula \eqref{eq:II-vNformula} (Proposition \ref{prop:II-vNformula}). For sure $\big(\widetilde{\mathcal{A}_{\lambda}^{(0)}})^\star$ does act on $\widetilde{\mathcal{D}}_{0}'$ as $3 W_\lambda^{-1}T_\lambda^{(0)}$ because this is the action of $\big(\widetilde{\mathcal{A}_{\lambda}^{(0)}})^\star$ both on  $\widetilde{\mathcal{D}}_{0}$ (owing to the definition \eqref{eq:Alambdatilde-0}) and on the deficiency subspaces (owing to \eqref{eq:A0tildestareigenv}).  
  \end{proof}

  \begin{proof}[Proof of Lemma \ref{lem:adjointasymtotics}]
   Let $\theta$ be the re-scaled radial function associated with $\xi$ through \eqref{eq:0xi}-\eqref{ftheta-1}. 
   
   Continuing the discussion from Lemma \ref{lem:deficiency1-1}'s proof (where the present $\theta$ was denoted by $\theta_+$), $\theta$ is the unique solution, up to complex multiplicative constant, to
   \[\tag{i}\label{eq:asymptag2}
    \widehat{\theta}(s)\;=\;\widehat{\theta}_0(s)+ \frac{z}{\lambda}\,\mathcal{L}\widehat{\theta}(s)\,.
   \]
   In the iteration
     \[
    \widehat{\theta}\;=\;\sum_{k=0}^{N-1}\Big(\frac{z}{\lambda}\Big)^k\mathcal{L}^k\widehat{\theta}_0+\Big(\frac{z}{\lambda}\Big)^N\mathcal{L}\widehat{\theta}\,,\qquad N\in\mathbb{N}\,,
  \]
  and in the considered regime $|z|/\lambda\ll 1$, the leading expression for the solution is
  \[\tag{ii}\label{eq:asymptag3new}
   \widehat{\theta}\;=\;\widehat{\theta}_0+ \frac{z}{\lambda}\,\mathcal{L}\widehat{\theta}_0\,,
  \]
  up to $O((z/\lambda)^2)$-corrections as $|z|/\lambda\to 0$.
  
%

  One then works out \eqref{eq:asymptag3new} choosing explicitly $c=\frac{\sqrt{2\pi}}{2\ii}$ for the convenience of having
  \[
   \widehat{\theta}_0(s)\;=\;\frac{\sqrt{2\pi}}{2\ii}(\delta(s-s_0)-\delta(s+s_0)) \, ,\qquad\textrm{ and hence }\qquad \theta_0(x)=\sin s_0 x
  \]
  (see \eqref{eq:distri-deltaFouriersin}). This will fix $\theta$, and hence $\xi$, up to a complex multiplicative constant. The computation for $\mathcal{L}\widehat{\theta}_0$ then gives
  \[
   \mathcal{L}\widehat{\theta}_0\;=\;-\widehat{\Lambda}(s)\Big(PV\frac{1}{s-s_0}-PV\frac{1}{s+s_0}\Big)
  \]
  with
  \[
   \begin{split}
    \widehat{\Lambda}(s)\;:=\;&-\frac{1}{\,8 s_0\widehat{\gamma}_+(s)}\Big(\big({\textstyle\frac{s}{\,\sinh\frac{\pi}{2}s}}*\widehat{\theta}_0\big)(s)+{\textstyle\frac{8}{\,\pi\sqrt{3}}}\int_{\mathbb{R}}{\textstyle\frac{\sinh\frac{\pi}{6}(s+t)}{\sinh \frac{\pi}{2}(s+t)}\,\frac{\sinh\frac{\pi}{3}(s-t)}{\sinh \frac{\pi}{2}(s-t)}}\,\widehat{\theta}_0(t)\,\ud t\Big) \\
    =\;&\frac{\ii\sqrt{\pi}}{\,8s_0\widehat{\gamma}_+(s)\sqrt{2}\,}\Big(\frac{s-s_0}{\,\sinh\frac{\pi}{2}(s-s_0)}-\frac{s+s_0}{\,\sinh\frac{\pi}{2}(s+s_0)} \\
    &-\frac{32}{\pi\sqrt{3}}\,\frac{\sinh\frac{\pi}{6}s_0\,\sinh\frac{\pi}{6}s}{\,\,(1+2\cosh\frac{\pi}{3}(s-s_0))\,(1+2\cosh\frac{\pi}{3}(s+s_0))}\,\Big).
   \end{split}
  \]
  It is easy to check that the function $\widehat{\Lambda}$ is smooth, rapidly decreasing, purely imaginary, and with odd parity (Figure \ref{fig:Lambdafunction}). Then $\Lambda$ is also smooth and rapidly decreasing, and is real-valued and with odd parity.  Thus, \eqref{eq:asymptag3new} takes the form
 \[\tag{iii}\label{eq:asymptag3}
   \widehat{\theta}\;=\; \widehat{\theta}_0-\frac{z}{\lambda}\,\widehat{\Lambda}(s)\,\Big(PV\frac{1}{s-s_0}-PV\frac{1}{s+s_0}\Big)\,.
 \]
  
  \begin{figure}[t!]
  \begin{center}
\includegraphics[width=8cm]{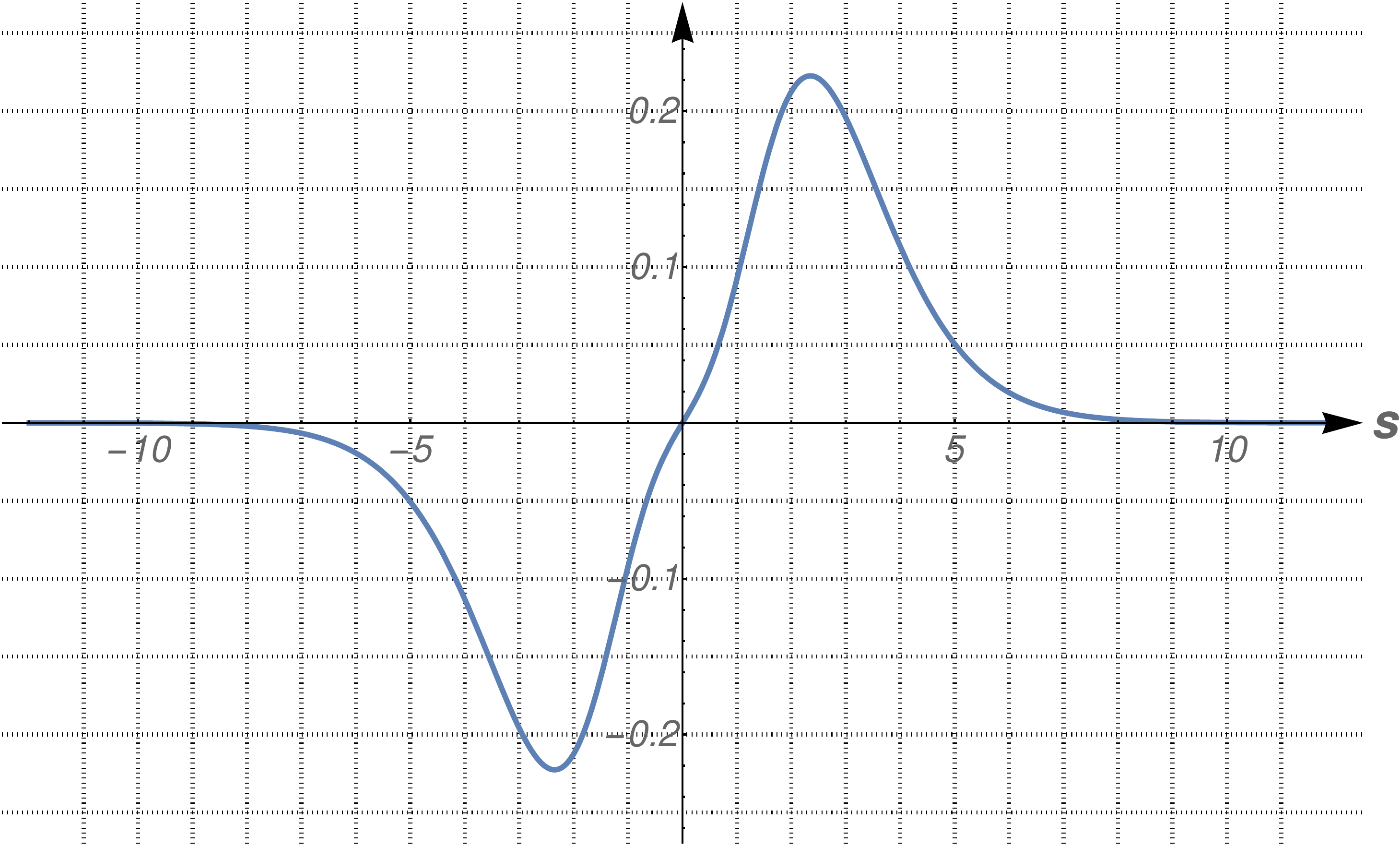}
\caption{Plot of the imaginary part of the function $\widehat{\Lambda}(s)$.}\label{fig:Lambdafunction}
\end{center}
\end{figure}

 The inverse Fourier transform of \eqref{eq:asymptag3} and \eqref{eq:distri-deltaFouriersin}-\eqref{eq:distriPVsignsin} then yield 
 \[\tag{iv}\label{eq:asymptag4}
  \theta(x)\;=\;\sin s_0 x+\frac{z}{\lambda}\,\big(\sin (s_0|\cdot|) *\Lambda\big)(x)\,.
 \]
 For such expression one can repeat the very same reasoning of Lemma \ref{lem:0tms-generalsol}'s proof and deduce that asymptotically as $|x|\to +\infty$ (and $|z|/\lambda\to 0$) 
 \[\tag{v}\label{eq:asymptag5}
  \theta(x)\;=\;\big(\sin s_0 x+\frac{z}{\lambda}\,a_\xi\cos( s_0 x+\sigma_\xi)\big)(1+o(1))
 \]
 for some constants $a_\xi\in\mathbb{R}$ and $\sigma_\xi\in[0,2\pi)$ depending on $\xi$. In particular, $a_\xi$ is surely real, because \eqref{eq:asymptag4} expresses a real-valued function.

 With the asymptotics \eqref{eq:asymptag5} for the re-scaled radial function $\theta$, one reconstructs the asymptotics for $\xi$ by means of \eqref{eq:0xi}-\eqref{ftheta-1}. Up to $o(1)$-corrections as $|x|\to +\infty$ (hence $|\pp|\to +\infty)$,
 \[
  \begin{split}
   \sin s_0 x\;&=\;\sin \Big(s_0\log\Big({\textstyle\sqrt{\frac{3\pp^2}{4\lambda}}+\sqrt{\frac{3\pp^2}{4\lambda}+1}}\Big)\Big) \\
   &\approx\;\sin \Big(s_0\log{|\pp|\textstyle\sqrt{\frac{3}{\lambda}}}\Big)\\
   &=\;\sin\big(s_0\log{\textstyle\sqrt{\frac{3}{\lambda}}}\big)\,\cos(s_0\log|\pp|)+\cos\big(s_0\log{\textstyle\sqrt{\frac{3}{\lambda}}}\big)\,\sin(s_0\log|\pp|)
  \end{split}
 \]
 and 
\[
  \begin{split}
   \cos (s_0 x+\sigma_\xi)\;&=\;\cos \Big(s_0\log\Big({\textstyle\sqrt{\frac{3\pp^2}{4\lambda}}+\sqrt{\frac{3\pp^2}{4\lambda}+1}}\Big)+\sigma_\xi\Big) \\
   &\approx\;\cos \Big(s_0\log{|\pp|\textstyle\sqrt{\frac{3}{\lambda}}}\Big) \\
   &=\;\cos\big(s_0\log{\textstyle\sqrt{\frac{3}{\lambda}}}\big)\,\cos(s_0\log|\pp|)-\sin\big(s_0\log{\textstyle\sqrt{\frac{3}{\lambda}}}\big)\,\sin(s_0\log|\pp|)\,.
  \end{split}
 \]
 Plugging the latter identities into \eqref{eq:asymptag5} yields
 \[
  \begin{split}
   \theta(x)\;&=\;\Big(\cos\big(s_0 \log{\textstyle\sqrt{\frac{3}{\lambda}}}\big)- \frac{z}{\lambda} \,a_\xi\,\sin\big(s_0 \log{\textstyle\sqrt{\frac{3}{\lambda}}}\big)\Big)\sin (s_0\log|\pp|) \\
   &\qquad +\Big(\sin\big(s_0 \log{\textstyle\sqrt{\frac{3}{\lambda}}}\big)+ \frac{z}{\lambda} \,a_\xi\,\cos\big(s_0 \log{\textstyle\sqrt{\frac{3}{\lambda}}}\big)\Big)\cos (s_0\log|\pp|)
  \end{split}
 \]
 up to $o(1)$-corrections as $|\pp|\to +\infty$ and $|z|/\lambda\to 0$. Inserting this into \eqref{eq:0xi}-\eqref{ftheta-1}, that is,
  \[
  \widehat{\xi}(\pp)\;=\;\frac{\,\theta\Big(\log\Big(\sqrt{\frac{3\pp^2}{4\lambda}}+\sqrt{\frac{3\pp^2}{4\lambda}+1}\Big)\Big)}{\,\sqrt{3\pi}\,|\pp|\sqrt{\frac{3}{4}\pp^2+\lambda}}\,,
 \]
 yields finally \eqref{eq:adjointasymtotics} with
    \begin{equation*}
    W_{\lambda,\xi}(z)\;:=\;\frac{\,\sin\big(s_0 \log\sqrt{\frac{3}{\lambda}}\big)+\frac{z}{\lambda} \,a_\xi\,\cos\big(s_0 \log\sqrt{\frac{3}{\lambda}}\big)\,}{\,\cos\big(s_0 \log\sqrt{\frac{3}{\lambda}}\big)- \frac{z}{\lambda} \,a_\xi\,\sin\big(s_0 \log\sqrt{\frac{3}{\lambda}}\big)\,}\,.
   \end{equation*}
 In \eqref{eq:adjointasymtotics} the overall multiplicative constant was re-instated.

 The above expression for $W_{\lambda,\xi}(z)$ shows that $\overline{W_{\lambda,\xi}(z)}=W_{\lambda,\xi}(\overline{z})$, thanks to the fact that $a_\xi\in\mathbb{R}$. Had one expressed the asymptotic periodicity \eqref{eq:asymptag5} above in terms of the sinus function, which amounts in practice to re-define the shift $\sigma_\xi$, one would have come up with an analogous expression for $W_{\lambda,\xi}(z)$ with the same symmetry property \eqref{eq:Wconstproperty}. The proof is thus completed. 
  \end{proof}

  Based on the discussion of this Subsection, it is convenient to introduce the following nomenclature for the charges belonging to the domain of $\big(\widetilde{\mathcal{A}_{\lambda}^{(0)}}\big)^\star$. Fixed $\mu>0$ and representing $\xi\in\mathcal{D}\Big(\widetilde{\mathcal{A}_{\lambda}^{(0)}}\Big)^\star$ through \eqref{eq:A0tildestar} as 
  \[
   \xi\;=\;\widetilde{\xi}+ c_+\xi_{\ii\mu}+c_-\xi_{-\ii\mu}
  \]
  for some $\widetilde{\xi}\in\widetilde{\mathcal{D}}_{0}'$ and $c_\pm\in\mathbb{C}$, one refers to
  \begin{equation}\label{eq:xiregxising}
   \xi_{\mathrm{reg}}\;:=\;\widetilde{\xi}\,,\qquad\xi_{\mathrm{sing}}\;:=\;c_+\xi_{\ii\mu}+c_-\xi_{-\ii\mu}
  \end{equation}
 as, respectively, the \emph{regular} and the \emph{singular} component of $\xi$. Regular and singular parts are unambiguously defined because the sum in \eqref{eq:A0tildestar} is direct. 

 To visualise the actual difference in behaviour, take for concreteness $\xi_{\mathrm{reg}}\in\widetilde{\mathcal{D}}_{0}$: then its re-scaled radial function $\theta_{\mathrm{reg}}$ satisfies
 \[
  \widehat{\theta}_{\mathrm{reg}}(s)\;=\;-\frac{\widehat{\vartheta}(s)}{\widehat{\gamma}_+(s)}\,\Big(PV\frac{1}{s-s_0}-PV\frac{1}{s+s_0}\Big)
 \]
 for some $\vartheta\in C^\infty_{c,\mathrm{odd}}(\mathbb{R}_x)$. On the contrary, $\xi_{\mathrm{sing}}$ has the behaviour of $\xi_{\pm\ii\mu}$ and in the course of the proof of Lemma \ref{lem:adjointasymtotics} it was shown that the associated re-scaled radial functions $\theta_{\pm\ii\mu}$ satisfy
 \[
   \widehat{\theta}_{\pm\ii\mu}(s)\;\approx\; \frac{\sqrt{2\pi}\,}{2\ii}\big(\delta(s-s_0)-\delta(s+s_0)\big)\mp\frac{\ii\mu}{\lambda}\,\widehat{\Lambda}(s)\Big(PV\frac{1}{s-s_0}-PV\frac{1}{s+s_0}\Big)
 \]
 up to an overall multiplicative constant and up to $O((z/\lambda)^2)$-corrections as $|z|/\lambda\to 0$.

    \subsection{Multiplicity of TMS self-adjoint realisations}\label{sec:multiselfadj-0}\index{Ter-Martirosyan-Skornyakov!self-adjoint extension}

    The operator $\widetilde{\mathcal{A}_{\lambda}^{(0)}}$ defined in \eqref{eq:Alambdatilde-0} admits a one-real-parameter family of self-adjoint extensions with respect to the Hilbert space $H^{-\frac{1}{2}}_{W_\lambda,\ell=0}(\mathbb{R}^3)$. They are characterised as follows.

    \begin{proposition}\label{prop:Aellzero-selfadj}
     Let $\lambda>0$. The self-adjoint extensions of $\widetilde{\mathcal{A}_{\lambda}^{(0)}}$ on $H^{-\frac{1}{2}}_{W_\lambda,\ell=0}(\mathbb{R}^3)$ form the family
     \begin{equation}
      \big\{\mathcal{A}_{\lambda,\beta}^{(0)}\,\big|\,\beta\in\mathbb{R}\big\}
     \end{equation}
    with
    \begin{equation}\label{eq:Azerolambda}
    \begin{split}
     \mathcal{D}\big(\mathcal{A}_{\lambda,\beta}^{(0)}\big)\;&:=\;\mathcal{D}_{0,\beta} \, , \\
     \mathcal{A}_{\lambda,\beta}^{(0)}\;&:=\;3 W_\lambda^{-1} T_\lambda^{(0)}\,,
   \end{split}
   \end{equation}
   where
   \begin{equation}\label{eq:domainD0beta}
    \mathcal{D}_{0,\beta}\;=\;\left\{
    \begin{array}{c}
     \xi\in\mathcal{D}\Big(\widetilde{\mathcal{A}_{\lambda}^{(0)}}\Big)^{\!\star} \;\;\textrm{with singular part satisfying} \\
     \widehat{\xi}_{\mathrm{sing}}(\pp)\,=\,c\,\displaystyle\frac{\,\cos({s_0\log|\pp|})+\beta \sin({s_0\log|\pp|})\,}{\pp^2}\,(1+o(1)) \\
     \textrm{as $|\pp|\to +\infty$\quad for some $c\in\mathbb{C}$}
    \end{array}
    \right\}.
   \end{equation}     
    \end{proposition}

     Each of the $\mathcal{A}_{\lambda,\beta}^{(0)}$ is a legitimate TMS parameter \index{Ter-Martirosyan-Skornyakov!extension parameter} in the sector of zero angular momentum, according to the discussion of Section \ref{sec:TMSextension-section}, precisely as the operator $\mathcal{A}_{\lambda}^{(\ell)}$ defined in \eqref{AFop-ellnot0} is a TMS parameter in the sector $\ell\in\mathbb{N}$.

    \begin{proof}[Proof of Proposition \ref{prop:Aellzero-selfadj}]
     Let $\mu>0$.
     By a standard application of von Neumann's extension theory (Theorem \ref{thm:vonN_thm_selfadj_exts}) to the operator $\widetilde{\mathcal{A}_{\lambda}^{(0)}}$ with deficiency subspaces \eqref{eq:twodefsub} and adjoint \eqref{eq:A0tildestareigenv}-\eqref{eq:A0tildestar}, the self-adjoint extensions of $\widetilde{\mathcal{A}_{\lambda}^{(0)}}$ form the family 
      \[
       \big\{\mathcal{A}_{\lambda,U_\nu}^{(0)}\,\big|\,\nu\in[0,2\pi)\big\}\,,
      \]
 where
   \[
    \begin{split}
     \mathcal{D}\big(\mathcal{A}_{\lambda,U_\nu}^{(0)}\big)\;&:=\;\left\{\xi\:=\:\widetilde{\xi}+c(\xi_{\ii\mu}+e^{\ii \nu}\,\xi_{-\ii\mu})\left|
     \begin{array}{c}
      \widetilde{\xi}\in\widetilde{\mathcal{D}}_0' \\
      c\in\mathbb{C}
     \end{array}\!\!
     \right.\right\} \\
     \mathcal{A}_{\lambda,U_\nu}^{(0)}\,\xi\;&:=\;3 W_\lambda^{-1} T_\lambda^{(0)}\xi\;=\;3 W_\lambda^{-1} T_\lambda^{(0)}\widetilde{\xi}+c\big(\ii\mu\xi_{\ii\mu}-\ii\mu e^{\ii\nu}\xi_{-\ii\mu}\big)\,.
    \end{split}
   \]
   Here one has tacitly and non-restrictively assumed that the functions $\xi_{\pm\ii\mu}$ are normalised in $H^{-\frac{1}{2}}_{W_\lambda,\ell=0}(\mathbb{R}^3)$. The notation $U_\nu$ is to remind that the  map $\xi_{\ii\mu}\mapsto e^{\ii\nu}\xi_{-\ii\mu}$ induces a unitary isomorphism $U_\nu$ between the two deficiency subspaces \eqref{eq:twodefsub}.

   One now characterises the $\xi$'s in $ \mathcal{D}\big(\mathcal{A}_{\lambda,U_\nu}^{(0)}\big)$ in terms of the large-$|\pp|$ asymptotics of the corresponding $\widehat{\xi}_{\mathrm{sing}}=c(\widehat{\xi}_{\ii\mu}+e^{\ii \nu}\,\widehat{\xi}_{-\ii\mu})$ (see definition \eqref{eq:xiregxising}). Owing to Lemma \eqref{lem:adjointasymtotics}, at the leading order as $\mu/\lambda\to 0$ and $|\pp|\to +\infty$, and up to an overall multiplicative constant, one has
   \[
   \begin{split}
    |\pp|^{-2}\,\widehat{\xi}_{\mathrm{sing}}(\pp)\;&=\;\sin({s_0\log|\pp|})+w_{\xi,\lambda,\mu}\cos({s_0\log|\pp|}) \\
    &\qquad +e^{\ii\nu}\big(\sin({s_0\log|\pp|})+\overline{w_{\xi,\lambda,\mu}}\,\cos({s_0\log|\pp|})\big)\,,
   \end{split}
   \]
   having set
   \[
    w_{\xi,\lambda,\mu}\;:=\;W_{\lambda,\xi}(\ii\mu)
   \]
 from formula \eqref{eq:adjointasymtotics} and having used the property
  \[
   W_{\lambda,\xi}(-\ii\mu)\;=\;\overline{W_{\lambda,\xi}(\ii\mu)}\;=\;\overline{w_{\xi,\lambda,\mu}}
  \]
 from \eqref{eq:Wconstproperty}. Thus, within such an approximation, and suitably re-defining the overall multiplicative constant,
  \[
   \begin{split}
    |\pp|^{-2}\,\widehat{\xi}_{\mathrm{sing}}(\pp)\;&=\;\cos({s_0\log|\pp|})+\frac{1+e^{\ii\nu}}{\,w_{\xi,\lambda,\mu}+e^{\ii\nu}\overline{w_{\xi,\lambda,\mu}}\,}\,\sin({s_0\log|\pp|})\,.
   \end{split}
  \]
 At fixed $\nu$, the above (asymptotic) condition selects all charges from $\mathcal{D}\Big(\Big(\widetilde{\mathcal{A}_{\lambda}^{(0)}}\Big)^\star\Big)$ that constitute the domain of the `$\nu$-th extension'.
 
 As
 \[
  \beta\;:=\;\frac{1+e^{\ii\nu}}{\,w_{\xi,\lambda,\mu}+e^{\ii\nu}\overline{w_{\xi,\lambda,\mu}}\,}\;\in\mathbb{R}\,,
 \]
 one can switch from $\nu$-parametrisation to the $\beta$-parametrisation, re-defining
 \[
  \mathcal{A}_{\lambda,\beta}^{(0)}\;:=\;\mathcal{A}_{\lambda,U_\nu}^{(0)}\,.
 \]
 This leads to the final thesis.    
    \end{proof}

   \begin{remark}
    It is worth underlying that the overall construction so far has involved \emph{two distinct extension schemes}: the Kre{\u\i}n-Vi\v{s}ik-Birman scheme for the self-adjoint extensions of the operator $\mathring{H}$, classified in Theorem \ref{thm:generalclassification}, and von Neumann's scheme for the self-adjoint extensions of the operator $\widetilde{\mathcal{A}_{\lambda}^{(0)}}$, classified in Proposition \ref{prop:Aellzero-selfadj}. (As $\widetilde{\mathcal{A}_{\lambda}^{(0)}}$ is not semi-bounded, the Kre{\u\i}n-Vi\v{s}ik-Birman scheme is not applicable.) In either case one speaks of singular component of a generic element of the adjoint as that component belonging to the deficiency subspaces, and in either case each extension corresponds to a suitable restriction of the domain of the adjoint. However, in the Kre{\u\i}n-Vi\v{s}ik-Birman scheme such a restriction selects a subspace of the adjoint's domain by means of a constraint between the singular and the regular component of its elements (as commented at the beginning of Section \ref{sec:two-body-short-scale-sing}), whereas in von Neumann's scheme the restriction of self-adjointness is a constraint within the singular components only (see \eqref{eq:xiregxising} and \eqref{eq:domainD0beta} above).    
   \end{remark}

\section{The canonical model and other well-posed variants}\label{sec:canonicalmodel}

  Merging the findings of Section \ref{sec:higherell} and \ref{sec:lzero} within the general scheme of Section \ref{sec:TMSextension-section} one finally obtains a class of models for the bosonic trimer 
  which are mathematically well-posed (i.e., self-adjoint) and physically meaningful (i.e., of Ter-Martirosyan-Skornyakov type), and which in a sense are canonical, as will be commented further on.

  \subsection{Canonical model at unitarity and at given three-body parameter}\label{sec:constructioncanonical}

  Let $\beta\in\mathbb{R}$ and $\lambda>0$.
  With respect to the decomposition \eqref{eq:bigdecompW}, namely
  \begin{equation}\label{eq:bigdecompW2}
 \begin{split}
 H^{-\frac{1}{2}}_{W_\lambda}(\mathbb{R}^3)\;\cong\;\bigoplus_{\ell=0}^\infty \,H^{-\frac{1}{2}}_{W_\lambda,\ell}(\mathbb{R}^3) \,,
 \end{split}
\end{equation}
  define
  \begin{equation}\label{eq:globalAlambda-1}
   \mathcal{A}_{\lambda,\beta}\;:=\; \mathcal{A}_{\lambda,\beta}^{(0)}\:\oplus\: \bigoplus_{\ell=1}^\infty\mathcal{A}_\lambda^{(\ell)}
  \end{equation}
 in the usual sense of direct sum of operators on an orthogonal direct sum of Hilbert spaces (Sect.~\ref{sec:I-preliminaries} and \ref{sec:I_invariant-reducing-ssp}). $\mathcal{A}_\lambda^{(\ell)}$, with $\ell\in\mathbb{N}$, is defined in \eqref{AFop-ellnot0}, taking here $\alpha=0$, and $\mathcal{A}_{\lambda,\beta}^{(0)}$ is defined in \eqref{eq:Azerolambda}-\eqref{eq:domainD0beta}. Observe that the condition $\lambda>\lambda_\alpha$ required in the definition \eqref{AFop-ellnot0} is automatically satisfied here, as $\alpha=0$.

  The self-adjointness of each summand in \eqref{eq:globalAlambda-1} with respect to the corresponding Hilbert space $H^{-\frac{1}{2}}_{W_\lambda,\ell}(\mathbb{R}^3)$ is proved, respectively, in Propositions \ref{prop:Alambdaellnot0} and \ref{prop:Aellzero-selfadj}. Therefore, altogether $\mathcal{A}_{\lambda,\beta}$ is self-adjoint on $H^{-\frac{1}{2}}_{W_\lambda}(\mathbb{R}^3)$.

  Upon setting
  \begin{equation}\label{eq:finalDbeta}
   \mathcal{D}_\beta\;:=\;\mathcal{D}_{0,\beta}\;\boxplus\;\op_{k=1}^\infty\mathcal{D}_\ell\,,
  \end{equation}
  the definition \eqref{eq:globalAlambda-1} is equivalent to
  \begin{equation}\label{eq:globalAlambda-2} 
   \begin{split}
    \mathcal{D}(\mathcal{A}_{\lambda,\beta})\;&:=\;\mathcal{D}_\beta \, ,\\
    \mathcal{A}_{\lambda,\beta}\;&:=\;3 W_\lambda^{-1}T_\lambda\,.
   \end{split}
  \end{equation}
  Recall from Section \ref{sec:I-preliminaries} that here the symbol `$\boxplus$' instead of `$\oplus$' is to indicate that the sum is orthogonal with respect to the Hilbert space orthogonal direct sum \eqref{eq:bigdecompW2}, but the summands are non necessarily closed subspaces of $H^{-\frac{1}{2}}_{W_\lambda}(\mathbb{R}^3)$. Actually \eqref{eq:finalDbeta} is nothing but the explicit expression for the domain of the direct sum operator \eqref{eq:globalAlambda-1} (with respect to the decomposition \eqref{eq:bigdecompW2}): the domain $\mathcal{D}_{0,\beta}$ of $\mathcal{A}_{\lambda,\beta}^{(0)}$ is defined in \eqref{eq:domainD0beta}, and the domain $\mathcal{D}_\ell$ of $\mathcal{A}_{\lambda}^{(\ell)}$ is defined in \eqref{eq:domainDell}. Observe that $\mathcal{D}_\beta$ is $\lambda$-independent, because so are its $\ell$-components. The second line of \eqref{eq:globalAlambda-2} is due to the fact that each of the summands in \eqref{eq:globalAlambda-1} is an operator acting on the corresponding $\ell$-sector as $W_\lambda^{(\ell)}T_\lambda^{(\ell)}$ (Propositions \ref{prop:Alambdaellnot0} and \ref{prop:Aellzero-selfadj}), and in turn $T_\lambda$ and $W_\lambda$ are reduced with respect to the decomposition \eqref{eq:bigdecompW2} with component, respectively, $T_\lambda^{(\ell)}$ and $W_\lambda^{(\ell)}$(as seen in \eqref{eq:decompTTell} and \eqref{eq:WlambdaWlambdaell}).

  It is instructive to re-cap what $\mathcal{D}_\beta$ altogether is:
  \begin{equation}\label{eq:Dbetaaltogether}
   \mathcal{D}_\beta\;=\;\left\{ 
   \begin{array}{c}
    \displaystyle\xi=\sum_{\ell=0}^\infty\xi^{(\ell)}\in\;\bigoplus_{\ell=0}^\infty \,H^{-\frac{1}{2}}_{W_\lambda,\ell}(\mathbb{R}^3)\,\cong\, H^{-\frac{1}{2}}_{W_\lambda}(\mathbb{R}^3) \\
    \textrm{such that} \\
    \xi^{(\ell)}\in H^{\frac{1}{2}}_\ell(\mathbb{R}^3)\;\textrm{ and }\; T_\lambda^{(\ell)}\xi^{(\ell)}\in H^{\frac{1}{2}}_\ell(\mathbb{R}^3)\;\textrm{ for }\;\ell\in\mathbb{N}\,, \\
    \xi^{(0)}\in\mathcal{D}\Big(\widetilde{\mathcal{A}_{\lambda}^{(0)}}\Big)^{\!\star} \;\;\textrm{with singular part satisfying} \\
    \widehat{\xi}_{\mathrm{sing}}(\pp)\,=\,c\,\displaystyle\frac{\,\cos({s_0\log|\pp|})+\beta \sin({s_0\log|\pp|})\,}{\pp^2}\,(1+o(1)) \\
     \textrm{as $|\pp|\to +\infty$\quad for some $c\in\mathbb{C}$}
   \end{array}
   \right\},
  \end{equation}
   where the subspace $\mathcal{D}\Big(\Big(\widetilde{\mathcal{A}_{\lambda}^{(0)}}\Big)^{\!\star\Big)}\subset H^{-\frac{1}{2}}_{W_\lambda,\ell=0}(\mathbb{R}^3)$ is defined in \eqref{eq:A0tildestar}. Moreover, following from the analogous properties of each $\ell$-component,
   \begin{equation}\label{eq:Dbetaproperties}
    \textrm{$\mathcal{D}_\beta$ is dense in $H^{-\frac{1}{2}}_{W_\lambda}(\mathbb{R}^3)$}\qquad\textrm{and}\qquad T_\lambda\mathcal{D}_\beta\;\subset\;H^{\frac{1}{2}}(\mathbb{R}^3)\,.
   \end{equation}

   Being self-adjoint on the deficiency subspace of $\mathring{H}+\lambda\mathbbm{1}$ (more precisely, on a unitarily equivalent version of it), $\mathcal{A}_{\lambda,\beta}$ identifies a self-adjoint extension $\mathring{H}_{\mathcal{A}_{\lambda,\beta}}$ of $\mathring{H}$ in the sense of the general classification of Theorem \ref{thm:generalclassification}.

   In turn, since $\mathcal{A}_{\lambda,\beta}$ acts as $3 W_\lambda^{-1}T_\lambda$, its domain $\mathcal{D}_\beta$ satisfies \eqref{eq:Dbetaproperties}, according to Theorem \ref{thm:globalTMSext} the operator $\mathring{H}_{\mathcal{A}_{\lambda,\beta}}$ is a Ter-Martirosyan-Skornyakov extension \index{Ter-Martirosyan-Skornyakov!self-adjoint extension} of $\mathring{H}$, namely a physical extension.

   Such extension can be defined as follows.

   \begin{theorem}\label{thm:H0beta}
    Let $\beta\in\mathbb{R}$ and $\lambda>0$. Define
    \begin{equation}\label{eq:H0betadomaction}
     \begin{split}
      \mathcal{D}(\mathscr{H}_{0,\beta})\;&:=\;\left\{g=\phi^\lambda+u_\xi^\lambda\left|\!
  \begin{array}{c}
   \phi^\lambda\in H^2_\mathrm{b}(\mathbb{R}^3\times\mathbb{R}^3)\,,\;\xi\in\mathcal{D}_\beta\,, \\
   \displaystyle\phi^\lambda(\yy_1,\mathbf{0})\,=\,(2\pi)^{-\frac{3}{2}} (T_\lambda\xi)(\yy_1)
  \end{array}
  \!\!\!\right.\right\} \, , \\
  (\mathscr{H}_{0,\beta}+\lambda\mathbbm{1})g\;&:=\;(-\Delta_{\yy_1}-\Delta_{\yy_2}-\nabla_{\yy_1}\cdot\nabla_{\yy_2}+\lambda\mathbbm{1})\phi^\lambda\,,
     \end{split}
    \end{equation}
    where the subspace $\mathcal{D}_\beta\subset H^{-\frac{1}{2}}(\mathbb{R}^3)$ is given by \eqref{eq:Dbetaaltogether}.
  \begin{enumerate}[(i)]
   \item The decomposition of $g$ in terms of $\phi^\lambda$ and $\xi$ is unique, at fixed $\lambda$. The subspace $ \mathcal{D}(\mathscr{H}_{0,\beta})$ is $\lambda$-independent.
   \item $\mathscr{H}_{0,\beta}$ is self-adjoint on $L^2_\mathrm{b}(\mathbb{R}^3\times\mathbb{R}^3,\ud\yy_1,\ud\yy_2)$ and extends $\mathring{H}$ given in \eqref{eq:domHring-initial}.
   \item For each $g\in  \mathcal{D}(\mathscr{H}_{0,\beta})$ one has
   \begin{equation}\label{eq:allBPTMS}
    \begin{split}
     \phi^\lambda(\yy_1,\mathbf{0})\;&=\;(2\pi)^{-\frac{3}{2}} (T_\lambda\xi)(\yy_1)\,, \\
     \int_{\mathbb{R}^3}\widehat{\phi^\lambda}(\pp_1,\pp_2)\,\ud\pp_2\;&=\;(\widehat{T_\lambda\xi})(\pp_1)\,, \\
     \int_{\!\substack{ \\ \\ \pp_2\in\mathbb{R}^3 \\ |\pp_2|<R}}\widehat{g}(\pp_1,\pp_2)\,\ud\pp_2\;&=\;4\pi R\,\widehat{\xi}(\pp_1)+o(1)\qquad\textrm{as }R\to +\infty\,.
    \end{split}
   \end{equation}
  All such conditions are equivalent, and each of them expresses the Bethe-Peierls alias Ter-Martirosyan-Skornyakov condition. \index{contact condition!Bethe-Peierls} \index{contact condition!Ter-Martirosyan-Skornyakov} In particular, the first version of \eqref{eq:allBPTMS} is an identity in $H^{\frac{1}{2}}(\mathbb{R}^3)$. 
  \item $\mathscr{H}_{0,\beta}$ is not semi-bounded.
  \end{enumerate}
  \end{theorem}

   \begin{proof}
    As argued already, the operator $\mathring{H}_{\mathcal{A}_{\lambda,\beta}}$ matches the conditions of Theorem \ref{thm:globalTMSext}(ii) for the considered $\lambda$, therefore it is a Ter-Martirosyan-Skornyakov self-adjoint extension of $\mathring{H}$ with inverse scattering length $\alpha=0$. 
    Renaming $\mathring{H}_{\mathcal{A}_{\lambda,\beta}}\equiv\mathscr{H}_{0,\beta}$, Theorem \ref{thm:globalTMSext} guarantees that such $\mathscr{H}_{0,\beta}$ is $\lambda$-independent (only the explicit decomposition of its domain's elements $g$ depends on $\lambda$), with
       \begin{equation*}
  \mathcal{D}(\mathscr{H}_{0,\beta})\;=\;
  \left\{g=\phi^\lambda+u_\xi^\lambda\left|\!
  \begin{array}{c}
   \phi^\lambda\in H^2_\mathrm{b}(\mathbb{R}^3\times\mathbb{R}^3)\,,\;\xi\in\mathcal{D}_\beta\,, \\
   \displaystyle\int_{\mathbb{R}^3}\widehat{\phi^\lambda}(\pp_1,\pp_2)\,\ud\pp_2\,=\,(\widehat{T_\lambda\xi})(\pp_1)
  \end{array}
  \!\!\!\right.\right\}.
  \end{equation*}
  The various BP/TMS conditions for $\mathscr{H}_{0,\beta}$ and their equivalence are then guaranteed by Lemma \ref{eq:oneTMSfunction}(iii), and the unboundedness from below (and above) of $\mathscr{H}_{0,\beta}$ follows from the fact that $\mathscr{H}_{0,\beta}$ extends, in the $\ell=0$ sector, a symmetric operator that is not semi-bounded (see Lemma \ref{lem:symmetricAtilde0} and the observations right after).  
   \end{proof}

   \begin{remark}
    Owing to the bosonic symmetry, if $g\in  \mathcal{D}(\mathscr{H}_{0,\beta})$, then \eqref{eq:allBPTMS} has equivalent versions in the other variables, e.g.,
    \begin{equation}
     \phi^\lambda(\yy,\mathbf{0})\;=\; \phi^\lambda(\mathbf{0},\yy)\;=\;\phi^\lambda(\yy,\yy)\;=\;(2\pi)^{-\frac{3}{2}} (T_\lambda\xi)(\yy)
    \end{equation}
    (see \eqref{eq:Hbosonic}).
   \end{remark}

   \begin{remark}
    In the sectors of definite angular momentum $\ell\in\mathbb{N}$ the Birman operator $\mathcal{A}_{\lambda,\beta}$ labelling the Hamiltonian $\mathring{H}_{\mathcal{A}_{\lambda,\beta}}\equiv\mathscr{H}_{0,\beta}$ is strictly positive, and correspondingly $\mathscr{H}_{0,\beta}$ is lower semi-bounded (Theorem \ref{thm:generalclassification}(ii)). In this case one can express the quadratic form of $\mathscr{H}_{0,\beta}$ according to Theorem \ref{thm:generalclassification}(iii). Explicitly, combining \eqref{eq:HFform}, \eqref{eq:decomposition_of_form_domains_Tversion}, and \eqref{AFform-ellnot0}, one finds that for all $g$'s of the form
    \begin{equation}\label{eq:g-for-the-form}
     g\;=\;\phi^\lambda+u_\xi^\lambda\,,\qquad \phi^\lambda\in H^1_{\mathrm{b}}(\mathbb{R}^3\times\mathbb{R}^3)\,,\quad\xi\in H^{\frac{1}{2}}_\ell(\mathbb{R}^3)
    \end{equation}
    for some $\lambda>0$ and some $\ell\in\mathbb{N}$ (thus, excluding $\ell=0$) the evaluation of the quadratic form of $\mathscr{H}_{0,\beta}$ gives
    \begin{equation}\label{eq:formhighsectors}
    \begin{split}
     \mathscr{H}_{0,\beta}[g]\;&=\;\frac{1}{2}\Big(\big\|(\nabla_{\yy_1}+\nabla_{\yy_2})\phi^\lambda\big\|^2_{L^2}+\big\|\nabla_{\yy_1}\phi^\lambda\big\|^2_{L^2}+\big\|\nabla_{\yy_2}\phi^\lambda\big\|^2_{L^2}\Big) \\
     &\qquad+\lambda\Big(\|\phi^\lambda\big\|^2_{L^2}-\big\|\phi^\lambda+u_\xi^\lambda\big\|^2_{L^2}\Big)+3\int_{\mathbb{R}^3} \overline{\,\widehat{\xi}(\pp)}\, \big(\widehat{T_\lambda\xi}\big)(\pp)\,\ud\pp
    \end{split}
    \end{equation}
  (the $L^2$-norms being norms in $L^2(\mathbb{R}^3\times\mathbb{R}^3)$). Of course, the above expression is the same for all $\beta$'s, since the parameter $\beta$ only characterises the properties of the Hamiltonian $\mathscr{H}_{0,\beta}$ in the sector $\ell=0$. On the $g$'s of \eqref{eq:g-for-the-form} one then has $\mathscr{H}_{0,\beta}[g]\geqslant 0$, and by self-adjointness the form \eqref{eq:formhighsectors} is closed. Through a quadratic form analysis, the form \eqref{eq:formhighsectors} was proposed and proved to be closed and semi-bounded in the recent work \cite{Basti-Teta-2015}. 
   \end{remark}

  The double index in $\mathscr{H}_{0,\beta}$ is to indicate that \emph{two parameters} have been selected in order to identify the operator within the general class of self-adjoint extensions of $\mathring{H}$, namely the parameter $\alpha=0$ in the Ter-Martirosyan-Skornyakov condition, \index{contact condition!Ter-Martirosyan-Skornyakov} and the parameter $\beta\in\mathbb{R}$ in the choice of the charge domain $\mathcal{D}_\beta$. 

  Explicitly, $\alpha=0$ and $\beta$ select the following prescriptions:
  \begin{eqnarray}
  &  &\!\!\!\!\!\!\!\!\!\!\!\!\!\!\!\!\!\!\!\!\!\!\!\!\!\!\!\!\!\!\int_{\!\substack{ \\ \\ \pp_2\in\mathbb{R}^3 \\ |\pp_2|<R}}\widehat{g}(\pp_1,\pp_2)\,\ud\pp_2\stackrel{R\to +\infty}{=} 4\pi R+\,\widehat{\xi}(\pp_1)+o(1)   \qquad\qquad\quad\;\; (\textrm{TMS}_{\alpha=0}) \label{eq:TMSalphazero}\\
  & & \!\!\!\!\!\!\!\!\!\!\!\!\!\!\!\!\!\!\!\!\!\!\!\!\!\!\!\!\!\!\widehat{\xi}^{(0)}_{\mathrm{sing}}(\pp)\stackrel{|\pp|\to +\infty}{=}c\,\displaystyle\frac{\,\cos({s_0\log|\pp|})+\beta \sin({s_0\log|\pp|})\,}{\pp^2}\,(1+o(1)) \quad (\textrm{III}_\beta) \label{eq:IIIbeta}\,.
  \end{eqnarray}
  As discussed in Subsect.~\ref{sec:symmTMSubdd}-\ref{sec:adjointBirman} and Proposition \ref{prop:Aellzero-selfadj}, the TMS condition alone, indicated here with ($\textrm{TMS}_{\alpha=0}$), is \emph{not} enough to qualify the self-adjointness of the model: an additional $\beta$-driven condition is needed, present only for charges in the $\ell=0$ sector.

  Besides, ($\textrm{III}_\beta$) in \eqref{eq:IIIbeta} is meant to express the following difference. \eqref{eq:TMSalphazero} is a \emph{two-body} condition, constraining the trimer's wave-function when \emph{two} of the identical bosons come on top of each other, which is explicitly seen from the first version of \eqref{eq:allBPTMS} or also from its consequence
   \begin{equation*}
 g_{\mathrm{av}}(\yy_1;|\yy_2|)\,\stackrel{|\yy_2|\to 0}{\sim}\frac{1}{|\yy_2|}\,\xi(\yy_1) + o(1)  
 \end{equation*}
  (see \eqref{eq:g-TMS-BP-generic} above). Instead, \eqref{eq:IIIbeta} is interpreted as a \emph{three-body} condition,\index{contact condition!three-body} regulating the behaviour of the trimer's wave-function in the vicinity of the \emph{triple} coincidence configuration. An amount of mathematical heuristics on such a three-body interpretation is presented in \cite[Remark 2.8]{CDFMT-2015} and \cite[Section 8]{MO-2017}. Notably, one can see from the spectral analysis that follows (Subsect.~\ref{sec:spectralThomas}) that such a $\beta$ has precisely the role of the three-body parameter introduced by the physicists (as was referred to in Section \ref{sec:BOSCHAPTintro}: see \cite{Gribov-1959,Danilov-1961,Naidon-Endo-Review_Efimov_Physics-2017}) and then emerged in the mathematical analysis of \cite{Albe-HK-Wu-1981}.

  From this perspective, each $\mathscr{H}_{0,\beta}$ is a \emph{canonical} Hamiltonian for the bosonic trimer with zero-range interaction: it is defined by a canonical choice, namely the Friedrichs extension of the TMS parameter, in all sectors $\ell\neq 0$, and by a $\beta$-extension of the TMS parameter in the sector $\ell=0$. In retrospect, also in the latter sector the construction was canonical, in that the choice of the initial domain of symmetry $\widetilde{\mathcal{D}}_0$ (formula \eqref{eq:Dtilde0}) is the natural one guaranteeing the well-posedness condition $T_\lambda^{(0)}\widetilde{\mathcal{D}}_0\subset H^{\frac{1}{2}}_{\ell=0}(\mathbb{R}^3)$ (Lemmas \ref{lem:0tms-generalsol} and \ref{lem:D0tildedomainproperties}(v)).


   \subsection{Spectral analysis and Thomas collapse}\label{sec:spectralThomas}\index{Thomas effect (collapse onto the centre)}

   For the Hamiltonian $\mathscr{H}_{0,\beta}$, $\beta\in\mathbb{R}$, consider the eigenvalue problem
   \begin{equation}
    \mathscr{H}_{0,\beta}\,g\;=\;E\,g\,,\qquad E<0\,.
   \end{equation}
   As $\mathscr{H}_{0,\beta}$ is a non-trivial self-adjoint extension of the positive symmetric operator $\mathring{H}$, one is indeed concerned with the \emph{negative bound states} of $\mathscr{H}_{0,\beta}$.

   \begin{theorem}\label{thm:spectralanalysis}
    Let $\beta\in\mathbb{R}$ and let $\mathscr{H}_{0,\beta}$ the operator introduced in Theorem \ref{thm:H0beta}. The negative eigenvalues of $\mathscr{H}_{0,\beta}$ relative to eigenfunctions with spherically symmetric singular charge constitute the sequence $(E_{\beta,n})_{n\in\mathbb{Z}}$ with
    \begin{equation}\label{eq:EVbeta}
     E_{\beta,n}\;=\;-3\,e^{-\frac{2}{\,s_0}\,\mathrm{arccot}\beta}\,e^{\frac{2\pi}{s_0}n}\,.
    \end{equation}
    The constant $s_0\approx 1.0062$ is the unique positive root of $\widehat{\gamma}(s)=0$ as defined in \eqref{eq:gamma-distribution}.
   Each such eigenvalue is simple and corresponds to an eigenfunction of the form $g_{\beta,n}=u_{\xi_{\beta,n}}^{(-E_{\beta,n})}$ with
    \begin{equation}\label{eq:EFbeta}
    \widehat{\xi}_{\beta,n}(\pp)\;=\;c_{\beta,n}\,\frac{\,\sin s_0\Big(\log\Big(\sqrt{\frac{3\pp^2}{\,4|E_{\beta,n}|\,}}+\sqrt{\frac{3\pp^2}{\,4|E_{\beta,n}|\,}+1}\Big)\Big)}{\,|\pp|\sqrt{\frac{3}{4}\pp^2+|E_{\beta,n}|}}
   \end{equation}
   with normalisation factor $c_{\beta,n}\in\mathbb{C}$.
   \end{theorem}

   Recall that the nomenclature `singular charge' is reserved for the function $\xi$ uniquely associated to $g$ in the general decomposition \eqref{eq:DHstardecomposed} (Lemmas \ref{lem:Hstaretc} and \ref{lem:chargexiofg}).

   \begin{corollary}\label{cor:spectralanalysis}~
    \begin{enumerate}[(i)]
     \item Each $\mathscr{H}_{0,\beta}$ admits an infinite sequence of negative bound states with energies $E_{\beta,n}$ accumulating to $-\infty$ as $n\to+\infty$, and accumulating to zero from below as $n\to -\infty$.
     \item Different realisations $\mathscr{H}_{0,\beta_1}$ and $\mathscr{H}_{0,\beta_2}$, namely $\beta_1\neq\beta_2$, have disjoint sequences of negative bound states, but with the same universal geometric law
     \[
      \frac{E_{\beta,n+1}}{E_{\beta,n}}\;=\;\exp \frac{2\pi}{s_0}\;\approx\;515\qquad\forall n\in\mathbb{Z}
     \]
       irrespective of $\beta$.
     \item Denoting by
     \[
      \sigma_{\mathrm{p}}^-(\mathscr{H}_{0,\beta})\;:=\;\{ E_{\beta,n}\,|\,n\in\mathbb{Z}\}
     \]
    the negative point spectrum of $\mathscr{H}_{0,\beta}$ in the sector $\ell=0$, one has
    \[
     \bigcup_{\beta\in\mathbb{R}}\sigma_{\mathrm{p}}^-(\mathscr{H}_{0,\beta})\;=\;\mathbb{R}^-\,.
    \]
    \end{enumerate}
   \end{corollary}

  Prior to proving Theorem \ref{thm:spectralanalysis} and its corollary, a few important comments on its content are in order.

  The presence of an infinite sequence of eigenvalues for the three-body Hamiltonian which accumulate to $-\infty$ is referred to as the \emph{Thomas effect}, or \emph{Thomas collapse},\index{Thomas effect (collapse onto the centre)} with reference to the phenomenon that, as mentioned in Section \ref{sec:BOSCHAPTintro}, was first discovered theoretically by Thomas in 1935 \cite{Thomas1935} through an analysis of the three-body problem in which the Bethe-Peierls contact condition was formally implemented in each two-body channel. The collapse, or `fall to the centre', refers to the circumstance that the corresponding three-body wave-function was argued to shrink around the triple coincidence point. This is precisely what can be seen from the eigenfunctions \eqref{eq:EFbeta} (Remark \ref{rem:eigenfunctions} below).

  The presence of an infinite sequence of negative eigenvalues for the three-body Hamiltonian which accumulate  to zero is referred to as the \emph{Efimov effect},\index{Efimov effect} with reference to the same phenomenon predicted theoretically in the early 1970's by Efimov \cite{Efimov-1971,Efimov-1973} for three-body quantum systems with two-body \emph{resonant} interaction of \emph{finite range} \cite{Yafaev-discspectr3body-1974, Klaus-Simon-1979-Efimov, Ovchinnikov-Sigal-1979-Efimov, Albe-HK-Wu-1981, Tamura-1990-Efimov, Tamura-JFA1991-Efimov, Tamura-1993-Efimov-asymptotics, Sobolev-1993-Efimov, Lakaev-1993-Efimov, Coutinho-Perez-Wreszinski-1995-ThomasEffect,Makarov-1996-Efimov, Albeverio-Lakaev-Djumanova-2009-Efimov, Basti-Teta-2017-Efimov}.

  Each Hamiltonian $\mathscr{H}_{0,\beta}$ thus displays both the Thomas and the Efimov effect.

  Moreover, the negative point spectra of the  $\mathscr{H}_{0,\beta}$'s fibre the whole negative half line and their disjoint union fills $\mathbb{R}^-$. This is an indirect signature of the fact that the $\mathscr{H}_{0,\beta}$'s are a one-parameter family of extensions of the same symmetric operator.

  The above properties of the negative point spectra of the $\mathscr{H}_{0,\beta}$'s, significantly formula \eqref{eq:EVbeta}, coincide with those emerging from the formal diagonalisation argument of physicists' `zero-range methods'\index{zero-range methods} \cite{Braaten-Hammer-2006,Naidon-Endo-Review_Efimov_Physics-2017} previously surveyed in Section \ref{sec:BOSCHAPTintro}, \emph{where $\beta$ is precisely the physically grounded `three-body parameter'} \index{contact condition!three-body} \cite[Section 4]{Naidon-Endo-Review_Efimov_Physics-2017}. On this basis, as anticipated in the discussion of \eqref{eq:TMSalphazero}-\eqref{eq:IIIbeta}, here too one shall refer to $\beta$ as the three-body parameter in the Hamiltonian. In Remark \ref{rem:eigenfunctions} below such a nomenclature will be substantiated with rigorous mathematical arguments.

  In view of the proof of Theorem \ref{thm:spectralanalysis}, it is convenient to single out this simple fact.

   \begin{lemma}\label{lem:adjointzero}
    Let $A$ be a densely defined and symmetric operator on a Hilbert space $\mathfrak{h}$ and assume that $A$ admits self-adjoint extensions, i.e., $\mathrm{dim}\ker(A^*-z\mathbbm{1})=\mathrm{dim}\ker(A^*-\overline{z}\mathbbm{1})>0$ for $z\in\mathbb{C}\setminus\mathbb{R}$. Let $A_{\mathrm{U}}$ be the 	self-adjoint extension of $A$ with the notation of von Neumann's extension scheme (Theorem \ref{thm:vonN_thm_selfadj_exts}), that is, $A_{\mathrm{U}}=A^*|_{\mathcal{D}(A_{\mathrm{U}})}$ with
    \[
     \mathcal{D}(A_{\mathrm{U}})\;=\;\mathcal{D}(\overline{A})\dotplus(\mathbbm{1}+U)\ker(A^*-z\mathbbm{1})
    \]
    for some unitary $U:\ker(A^*-z\mathbbm{1})\stackrel{\cong}{\to}\ker(A^*-\overline{z}\mathbbm{1})$. Decompose a generic $\xi\in \mathcal{D}(A_{\mathrm{U}})$ accordingly as $\xi=\xi_0+c(\xi_+ + U\xi_+)$ for some $\xi_0\in\mathcal{D}(\overline{A})$, $\xi_+\in\ker(A^*-z\mathbbm{1})$, $c\in\mathbb{C}$.
    Assume in addition that $A$ is injective and that for some non-zero $\xi\in \mathcal{D}(A_{\mathrm{U}})$ one has $A^*\xi=0$. Then $\xi_0=0$.
   \end{lemma}

   \begin{proof}
    By assumption $0=A^*\xi=\overline{A}\xi_0+c(z\xi_+ + \overline{z} U\xi_+)$. Moreover,
    \[
     \langle \overline{A}\xi_0,\xi_+\rangle_{\mathfrak{h}}\;=\;\langle \overline{A}\xi_0,U\xi_+\rangle_{\mathfrak{h}}\;=\;0\,,
    \]
    meaning that $\overline{A}\xi_0$ and $c(z\xi_+ + \overline{z} U\xi_+)$ are orthogonal in $\mathfrak{h}$. Therefore, both such vectors must vanish, and in particular $\overline{A}\xi_0=0$. By injectivity of $A$ (and hence of $\overline{A}$) the conclusion follows.   
   \end{proof}

   \begin{proof}[Proof of Theorem \ref{thm:spectralanalysis}]
    Decompose $g\in\mathcal{D}(\mathscr{H}_{0,\beta})$ according to \eqref{eq:H0betadomaction} with decomposition parameter
    \[
     \lambda\;:=\;-E\,,
    \]
    that is, $g=\phi^\lambda+u_\xi^\lambda$. Then \eqref{eq:H0betadomaction}, combined with $ \mathscr{H}_{0,\beta}g=-\lambda g$, implies $\phi^\lambda\equiv 0$ and $T_\lambda\xi\equiv 0$. The eigenfunctions have then necessarily the form $g=u_\xi^\lambda$ for $\xi\in\mathcal{D}_\beta$ such that $T_\lambda\xi= 0$.

    As $T_\lambda$ is reduced with respect to the decomposition \eqref{eq:bigdecompW2} with components $T_\lambda^{(\ell)}$ (as seen in \eqref{eq:decompTTell}), the latter equation is equivalent to the collection of equations $T_\lambda^{(\ell)}\xi^{(\ell)}= 0$, $\ell\in\mathbb{N}_0$.
    The focus here is on the eigenfunctions relative to charges $\xi$ belonging to the sector $\ell=0$, namely the physically relevant ones.

    To this aim, consider the problem
    \[
     T_\lambda^{(0)}\xi^{(0)}\;=\; 0\,,\qquad \xi^{(0)}\in\mathcal{D}_{0,\beta}\,,
    \]
    henceforth expressing the unknown $\xi^{(0)}$ simply as $\xi$.
    Such equation, owing to Lemma \ref{lem:0tms-generalsol}, is solved by those $\xi$'s in the subspace $\mathcal{D}_{0,\beta}$ such that the corresponding re-scaled radial function $\theta$'s, in the notation \eqref{eq:0xi}-\eqref{ftheta-1}, have the form
    \[
     \theta(x)\;=\;c\,\sin s_0 x\,,\qquad c\in\mathbb{C}.
    \]
   In this case \eqref{eq:0xi} and \eqref{ftheta-2} give
    \[
  \widehat{\xi}(\pp)\;=\;c\,\frac{\,\sin s_0\Big(\log\Big(\sqrt{\frac{3\pp^2}{4\lambda}}+\sqrt{\frac{3\pp^2}{4\lambda}+1}\Big)\Big)}{\,\sqrt{3\pi}\,|\pp|\sqrt{\frac{3}{4}\pp^2+\lambda}}\,.
 \]

   Now, in order for such $\xi$ to belong to $\mathcal{D}_{0,\beta}$, $\xi$ must only have singular component, that is, $\xi=\xi_{\mathrm{sing}}$ in the notation \eqref{eq:A0tildestar} and \eqref{eq:xiregxising}. This follows from Lemma \ref{lem:adjointzero} applied to the operator $\widetilde{\mathcal{A}_{\lambda}^{(0)}}$ defined in \eqref{eq:Alambdatilde-0} and to its extension $\mathcal{A}_{\lambda,\beta}^{(0)}$ defined in \eqref{eq:Azerolambda}. For the former, injectivity is proved in Lemma \ref{lem:D0tildedomainproperties}(vi) (using also the bijectivity property of $W_\lambda$, Lemma \ref{lem:Wlambdaproperties}(ii)). For the latter, $\Big(\widetilde{\mathcal{A}_{\lambda}^{(0)}}\Big)^\star\xi=\mathcal{A}_{\lambda,\beta}^{(0)}\xi=3W_\lambda^{-1}T_\lambda^{(0)}\xi=0$. Lemma \ref{lem:adjointzero} is then applicable, and yields $\xi-\xi_{\mathrm{sing}}=\xi_{\mathrm{reg}}=0$.

   It then remains to impose that the above solution $\xi$ satisfy the asymptotics \eqref{eq:IIIbeta} for the considered $\beta$.

   With simple computations analogous to those made in the proof of Lemma \ref{lem:adjointasymtotics} one finds
   \[
    \widehat{\xi}(\pp)\,=\,c'\Big(\cos\big(s_0\log{\textstyle\sqrt{\frac{3}{\lambda}}}\big)\,\frac{\,\sin (s_0\log|\pp|)}{\pp^2}+\sin\big(s_0\log{\textstyle\sqrt{\frac{3}{\lambda}}}\big)\,\frac{\,\cos (s_0\log|\pp|)}{\pp^2}\Big)(1+o(1))
   \]
   as $|\pp|\to +\infty$, for some $c'\in\mathbb{C}$. The comparison with \eqref{eq:IIIbeta} then implies
   \[
    \cos\big(s_0\log{\textstyle\sqrt{\frac{3}{\lambda}}}\big)\;=\;\beta\,\sin\big(s_0\log{\textstyle\sqrt{\frac{3}{\lambda}}}\big)\,.	
   \]
   The latter condition selects the admissible values for $\lambda$, and hence $E=-\lambda$: explicitly, only the values $E_{\beta,n}=-\lambda_n$ with
   \[
    \lambda_n\;=\;3\,e^{-\frac{2}{\,s_0}\,\mathrm{arccot}\beta}\,e^{\frac{2\pi}{s_0}n}\,,\qquad n\in\mathbb{Z}\,.
   \]
   This establishes \eqref{eq:EVbeta}, and moreover it is clear from the above discussion that the corresponding eigenfunctions are all of the form $u_{\xi_n}^{(-\lambda_n)}$ and that each eigenvalue $E_{\beta,n}$ is non-degenerate.    
   \end{proof}

     \begin{proof}[Proof of Corollary \ref{cor:spectralanalysis}]
    Parts (i) and (iii), as well as the geometric formula of part (ii), all follow at once from \eqref{eq:EVbeta} of Theorem \ref{thm:spectralanalysis}. The fact that
    \[
     \sigma_{\mathrm{p}}^-(\mathscr{H}_{0,\beta})\cap\sigma_{\mathrm{p}}^-(\mathscr{H}_{0,\beta'})\;=\;\emptyset\,,\qquad \beta\neq\beta'\,,
    \]
   can be seen as follows. If $E_{\beta,n}=E_{\beta',n'}$ for some $n,n'\in\mathbb{Z}$, then
   \[
   \frac{1}{\pi}\big(\mathrm{arccot}\beta-\mathrm{arccot}\beta'\big)\;=\;k
   \]
  for some $k=n-m\in\mathbb{Z}$, as follows straightforwardly from \eqref{eq:EVbeta}. For the properties of the $\mathrm{arccot}$-function, this is only possible when $k=0$, in which case $\beta=\beta'$.   
   \end{proof}

\begin{figure}[t!]
    \begin{center}
\includegraphics[width=8cm]{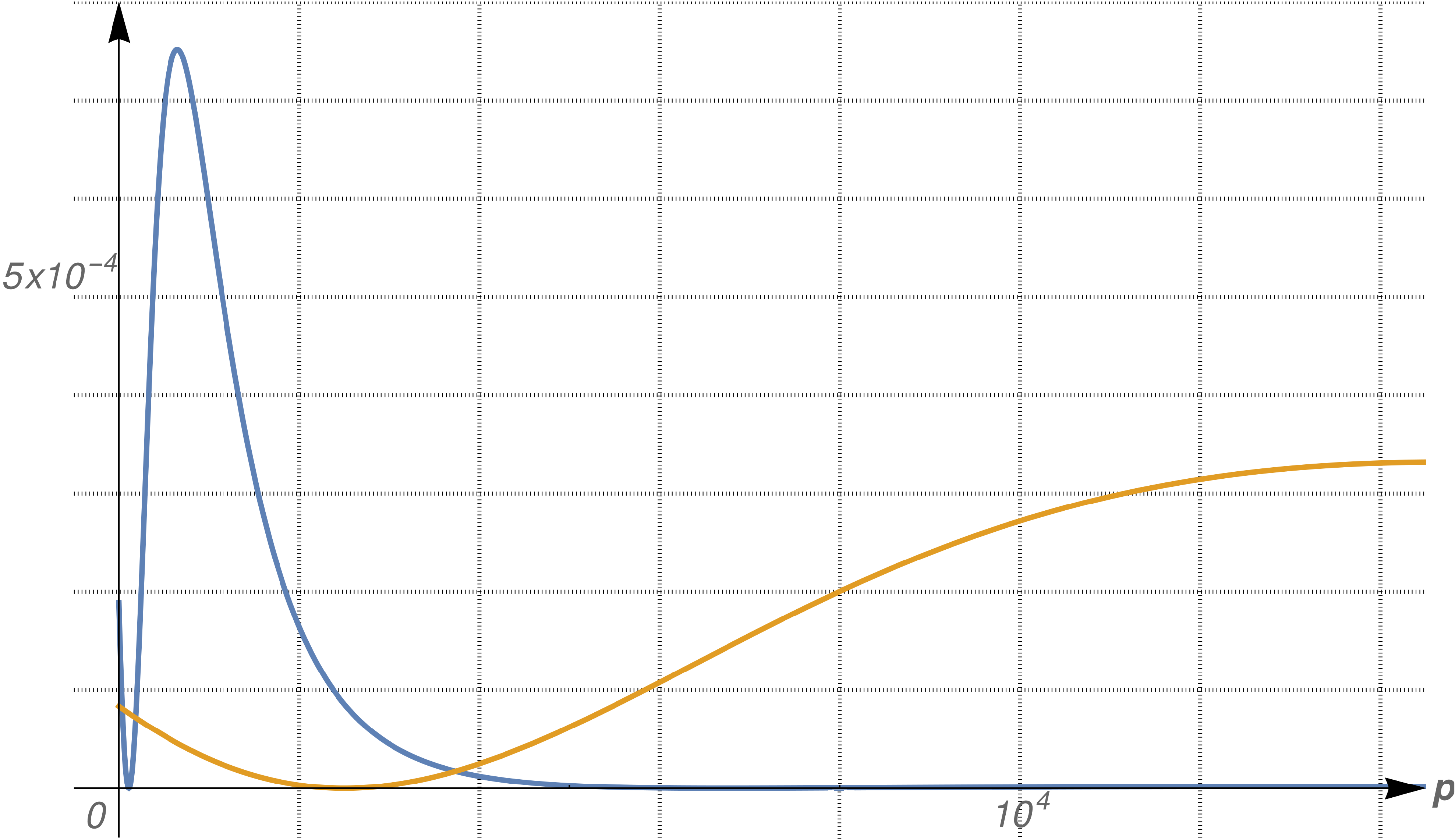}
\caption{Plot of the radial distribution profile $\varrho_{\beta,n}$ for the two charges of the eigenfunctions of $\mathscr{H}_{0,\beta}$, with $\beta=1$, relative to the quantum numbers $n=2$ (blue curve) and $n=3$ (orange curve: this has been multiplicatively magnified by a factor 10, for a clearer rendering). Correspondingly, $E_{\beta,3}<E_{\beta,2}<0$. The charge relative to the more negative eigenvalue is more de-localised in momentum, hence more localised around zero in space. Discussion in Remark \ref{rem:eigenfunctions}.}\label{fig:eigenfunctions}
\end{center}
 \end{figure}

   \begin{remark}\label{rem:eigenfunctions}
    At given $\beta\in\mathbb{R}$, the eigenfunctions $g_{\beta,n}=u_{\xi_{\beta,n}}^{(-E_{\beta,n})}$ have charges $\xi_{\beta,n}$ that tend more and more to be localised around $\yy=0$ as $E_{\beta,n}\to-\infty$, and on the contrary more and more de-localised in space as $E_{\beta,n}\uparrow 0$. In the former case $g_{\beta,n}(\yy_1,\yy_2)$ is generated by a `charge distribution'
    \begin{equation*}
  \xi_{\beta,n}(\yy_1)\delta(\yy_2)+\delta(\yy_1)\xi_{\beta,n}(\yy_2)+\delta(\yy_1-\yy_2)\xi_{\beta,n}(-\yy_2)
 \end{equation*}
   (up to a multiplicative constant, see \eqref{eq:livingonhyperplanes}) that tends to concentrate at the triple coincidence point $\yy_1=\yy_2=\mathbf{0}$ as $E_{\beta,n}\to-\infty$. This is precisely the fall-to-the-centre phenomenon associated with the Thomas effect.\index{Thomas effect (collapse onto the centre)} All this can be seen from the explicit expression of the eigenfunctions \eqref{eq:EFbeta}. To visualise it one may consider the radial distribution $\varrho_{\beta,n}$ of the charge $\xi_{\beta,n}$ in momentum coordinate, namely
   \[
    \varrho_{\beta,n}(p)\;=\;\frac{p^2\,|f_{\beta,n}(p)|^2}{\displaystyle\int_0^{+\infty}\!\ud p\,p^2\,|f_{\beta,n}(p)|^2\,}\,,
   \]
   where
   \[
   \begin{split}
    \widehat{\xi}_{\beta,n}(\pp)\;&=\;\frac{1}{\sqrt{4\pi}}\,f_{\beta,n}(|\pp|)\,,\\
    f_{\beta,n}(p)\;&=\;c_{\beta,n}\,\frac{\,2\,\sin s_0\Big(\log\Big(\sqrt{\frac{3\pp^2}{\,4|E_{\beta,n}|\,}}+\sqrt{\frac{3\pp^2}{\,4|E_{\beta,n}|\,}+1}\Big)\Big)}{\,\sqrt{3}\,|\pp|\sqrt{\frac{3}{4}\pp^2+|E_{\beta,n}|}}\,.
   \end{split}
   \]
  Figure \ref{fig:eigenfunctions} shows indeed that the more negative $E_{\beta,n}$ (namely, the larger $n>0$), the more flattened $\varrho_{\beta,n}(p)$, meaning the more localised in space $\xi_{\beta,n}(\yy)$.
   \end{remark}

\begin{remark}\label{rem:whereas}
 Whereas Theorem \ref{thm:spectralanalysis} and Corollary \ref{cor:spectralanalysis} focus on the negative point spectrum of the Hamiltonian $\mathscr{H}_{0,\beta}$, it is not difficult to determine that its essential spectrum is precisely
 \begin{equation}
  \sigma_{\mathrm{ess}}(\mathscr{H}_{0,\beta})\;=\;[0,+\infty)\,.
 \end{equation}
 This can be obtained by suitably adjusting to the present setting the reasoning developed in \cite[Sections 5 and 6]{BMO-2017}. That $[0,+\infty)\subset\sigma_{\mathrm{ess}}(\mathscr{H}_{0,\beta})$ can be seen by means of a Weyl sequence (Sect.~\ref{sec:I-parts-of-spectrum}) of the same type as the standard Weyl sequences in $H^2_{\mathrm{b}}(\mathbb{R}^3\times\mathbb{R}^3)$ used to show that $\sigma_{\mathrm{ess}}(\mathring{H}_{\mathrm{F}})=[0,+\infty)$, suitably modified so as to vanish at the coincidence hyperplanes (see \cite[Proposition 5.1]{BMO-2017}). For the opposite inclusion, one can reproduce a version of the Minlos-Yoshitomi decomposition 
 (exploited first in \cite{Yoshitomi_MathSlov2017} as a modification of \cite[Eq.~(2.18)-(2.20)]{Minlos-2011-preprint_May_2010} by means of a suitable localisation in momentum space) for the present $T_\lambda$, as done in \cite[Section 6]{BMO-2017} for the fermionic $T_\lambda$ (technically, in this respect fermionic and bosonic case are completely analogous), based on which by compactness arguments one can show that the spectral projection relative to each interval $[a,b]\subset(-\infty,0)$ is finite-dimensional and hence $(-\infty,0)$ does not contain essential spectrum (Sect.~\ref{sec:I-parts-of-spectrum}). 
\end{remark}

 \begin{remark}\label{rem:choicedomains}
 As a follow-up to Remark \ref{rem:merepurpose}, in retrospect one can comment over the choice \eqref{eq:Dtilde0} of the initial charge domain $\widetilde{\mathcal{D}}_0$ used to realise $\widetilde{\mathcal{A}_\lambda^{(0)}}=W_\lambda^{-1}T_\lambda^{(0)}$ as a densely defined and symmetric operator on $H^{-\frac{1}{2}}_{W_\lambda,\ell=0}(\mathbb{R}^3)$. 
 \begin{enumerate}[(i)]
  \item As observed already, also on the larger domain $\widetilde{\mathcal{D}}_0'$ from \eqref{eq:Dolarger} could one make $W_\lambda^{-1}T_\lambda^{(0)}$ symmetric on $H^{-\frac{1}{2}}_{W_\lambda,\ell=0}(\mathbb{R}^3)$. In either case, the resulting  $\widetilde{\mathcal{A}_\lambda^{(0)}}$ is an admissible Birman parameter for a symmetric Ter-Martirosyan  Skornyakov extension $\mathscr{H}_0\equiv\mathring{H}_{\widetilde{\mathcal{A}_\lambda^{(0)}}}$ of the initial operator $\mathring{H}$ (in the sector $\ell=0$). The difference is in the smaller ($\widetilde{\mathcal{D}}_0$) or larger ($\widetilde{\mathcal{D}}_0'$) domain of charges, with corresponding re-scaled radial functions of the form 
 \[
  \begin{split}
   \theta\;&=\;\sin (s_0 |x|)*\big(\widehat{\vartheta}/\widehat{\gamma}_+\big)^{\!\vee}\qquad\qquad\qquad\quad\!\!\textrm{for charges in $\widetilde{\mathcal{D}}_0$}, \\
   \theta\;&=\;c\,\sin s_0 x+\sin (s_0 |x|)*\big(\widehat{\vartheta}/\widehat{\gamma}_+\big)^{\!\vee}\qquad\;\;\:\textrm{for charges in $\widetilde{\mathcal{D}}_0'$},
  \end{split}
 \]
 with $\vartheta\in C^\infty_{c,\mathrm{odd}}(\mathbb{R}_x)$ and $c\in\mathbb{C}$.
 \item Choosing the \emph{larger} domain, one solves the eigenvalue problem
 \[
  \mathscr{H}_0 \,g\;=\;-\lambda g\,,\qquad \lambda>0
 \]
 (in the sector $\ell=0$) with the same reasoning as in the proof of Theorem \ref{thm:spectralanalysis}, and thus finds solutions $g=u_{\xi_\lambda}^{\lambda}$ with
 \[
 \begin{split}
  \theta(x)\;&=\;\sin s_0 x\,, \\
   \widehat{\xi}_\lambda(\pp)\;&=\;\frac{\,\sin s_0\Big(\log\Big(\sqrt{\frac{3\pp^2}{4\lambda}}+\sqrt{\frac{3\pp^2}{4\lambda}+1}\Big)\Big)}{\,\sqrt{3\pi}\,|\pp|\sqrt{\frac{3}{4}\pp^2+\lambda}}\,,
 \end{split}
 \]
 up to an overall multiplicative constant. All such solutions are now admissible, in that all the above $\xi_\lambda$'s belong to $\widetilde{\mathcal{D}}_0'$ irrespective of $\lambda>0$. This proves that $\mathscr{H}_0$ has a continuum of eigenvalues, which is \emph{incompatible with self-adjointness}.
 \item In fact, a laborious but instructive computation (originally alluded to in \cite[Section III.3]{Flamand-Cargese1965}) shows that imposing the orthogonality of any two such $u_{\xi_{\lambda}}^{\lambda}$ and $u_{\xi_{\lambda'}}^{\lambda'}$ in $L^2_\mathrm{b}(\mathbb{R}^3\times\mathbb{R}^3,\ud\yy_1\ud\yy_2)$ does partition $\mathbb{R}^+$ into the disjoint union
 \[
  \mathbb{R}^+\;=\;\bigcup_{\beta\in\mathbb{R}}\sigma_\beta\,,\qquad \sigma_\beta\;:=\;\Big\{\lambda_{\beta,n}=3\,e^{-\frac{2}{\,s_0}\,\mathrm{arccot}\beta}\,e^{\frac{2\pi}{s_0}n}\,\Big|\,n\in\mathbb{Z}\Big\}\,,
 \]
 where each sequence $(-\lambda_{\beta,n})_{n\in\mathbb{Z}}$ is an admissible sequence of simple eigenvalues, with orthogonal eigenfunctions by construction, for a self-adjoint operator on $L^2_\mathrm{b}(\mathbb{R}^3\times\mathbb{R}^3,\ud\yy_1\ud\yy_2)$ (precisely, for the operator $\mathscr{H}_{0,\beta}$ from Theorem \ref{thm:H0beta}).
 \item In connection to (ii), and in view of Theorems \ref{thm:H0beta} and \ref{thm:spectralanalysis}, one sees that the inclusion
 \[
  \widetilde{\mathcal{D}}_0\;\subset\; \mathcal{D}_{0,\beta} \;\subset\;\widetilde{\mathcal{D}}_0'
 \] 
 involves three distinct admissible choices for the singular charges domain (in the sector $\ell=0$) for symmetric TMS extensions of $\mathring{H}$, \index{Ter-Martirosyan-Skornyakov!symmetric extension} among which only $\mathcal{D}_{0,\beta}$ produces a self-adjoint extension. \index{Ter-Martirosyan-Skornyakov!self-adjoint extension}
 \end{enumerate}
 \end{remark}

 \subsection{Variants}\label{sec:variants}
 
 The canonical model(s) $\mathscr{H}_{0,\beta}$, $\beta\in\mathbb{R}$, have variants that do not alter the $\ell=0$ sector construction, where the essential physics takes place.

 As already observed at the beginning of Section \ref{sec:higherell}, there is an amount of arbitrariness in the definition of the trimer's Hamiltonian in sectors of higher angular momentum.

 The construction developed in Section \ref{sec:higherell} is canonical in that it provides the Friedrichs realisation of an operator of Ter-Martirosyan-Skornyakov type with inverse negative scattering length $\alpha$. (It has of course also a considerable degree of instructiveness, from the technical point of view.)

 Such a construction, combined with the analysis in the sector $\ell=0$ (Sect.~\ref{sec:lzero}) led to the self-adjoint Ter-Martirosyan-Skornyakov Hamiltonians of the form $\mathring{H}_{\mathcal{A}_{\lambda,\beta}}$ with Birman parameter $\mathcal{A}_{\lambda,\beta}$ given by \eqref{eq:globalAlambda-1}.

 Equally admissible (self-adjoint and TMS) alternatives are given by modified Birman parameters of the form 
  \begin{equation}
   \mathcal{A}_{\lambda,\beta}\;:=\; \mathcal{A}_{\lambda,\beta}^{(0)}\:\oplus\: \bigoplus_{\ell=1}^\infty\mathcal{A}_{\lambda,\alpha_\ell}^{(\ell)}
  \end{equation}
 where
 \begin{equation}\label{AFop-ellnot0_ell}
    \begin{split}
     \mathcal{D}\big(\mathcal{A}_{\lambda,\alpha_\ell}^{(\ell)}\big)\;&:=\;\mathcal{D}_\ell\;=\;\big\{\xi\in H_\ell^{\frac{1}{2}}(\mathbb{R}^3)\,\big|\,T_\lambda^{(\ell)}\xi\in H_\ell^{\frac{1}{2}}(\mathbb{R}^3)\big\} \\
     \mathcal{A}_{\lambda,\alpha_\ell}^{(\ell)}\;&:=\;3 W_\lambda^{-1}\big(T_\lambda^{(\ell)}+\alpha_\ell\mathbbm{1}\big)\,,
    \end{split}
   \end{equation}
 thus on the same charge domain $\mathcal{D}_\ell$ that guarantees self-adjointness (Proposition \ref{prop:Alambdaellnot0}), but with scattering lengths that depend on the angular sector.

 In fact, it would be physically acceptable also to ignore in the first place the interaction in sectors of non-zero angular momentum, thus focusing on the Hamiltonians of interest only as effective models in the sector $\ell=0$. This is obtained by taking the trivial (Friedrichs) extension of $\mathring{H}$ whenever $\ell\neq 0$: particles in a three-body state with charges that do not belong to the zero angular momentum sector just move with free dynamics. In this case the final Birman parameter's domain, instead of \eqref{eq:finalDbeta}, becomes
 \begin{equation}\label{eq:finalDbetaNOINT}
   \mathcal{D}_{0,\beta}\;\boxplus\;\bigoplus_{k=1}^\infty\{0\}\,.
  \end{equation}
 The TMS condition \index{contact condition!Ter-Martirosyan-Skornyakov} remains only in the sector $\ell=0$. The Hamiltonian is just the free kinetic operator on the other sectors.

\section{Ill-posed models}\label{sec:illposed}


In Section \ref{sec:BOSCHAPTintro} it was argued that for three-body quantum systems with contact interaction physical zero-range methods determine eigenfunctions and eigenvalues of a formal Hamiltonian that otherwise remains unspecified. It was also argued that mathematical approaches are aimed at constructing a self-adjoint Hamiltonian of Ter-Martirosyan-Skornyakov type: first one declares the operator or its quadratic form, then one performs the subsequent spectral analysis on it. Of course, on the physical side there is the advantage of an ultimate agreement check with the experiments.

As a matter of fact, one can track down, through the mathematical literature on the subject, certain recurrent sources of ill-posed models, failing to provide a three-body Hamiltonian that at the same time be self-adjoint and exhibit the Bethe-Peierls / Ter-Martirosyan-Skornyakov contact condition. \index{contact condition!Bethe-Peierls} \index{contact condition!Ter-Martirosyan-Skornyakov}

On the mathematical technical level, the model's well-posedness lies in the correct choice of the domain of self-adjointness, among those domains that in addition reproduce the desired short-scale physical asymptotics. A wrong choice of the (operator or form) domain fails to yield self-adjointness and produces incorrect spectral data. In this informal sense one speaks of incomplete or ill-posed models.

In some circumstances an explicit signature of some sort of incompleteness of the mathematical model is the quantitative discordance in the spectral analysis with numerical and experimental evidence from physics. This has been the case significantly for three-body systems with a pair of identical fermions: the recent works \cite{CDFMT-2015,MO-2016,MO-2017} mentioned already in Section \ref{sec:BOSCHAPTintro} were essentially aimed at clarifying this perspective, on which further comment are going to be made in the course of this Section.

In other occurrences the ill-posedness of the model is more subtle and less evident, and the case of the bosonic trimer is typical in this sense.

For clarity of presentation, such occurrences are grouped here into two categories, discussed, respectively, in Subsect.~\ref{sec:illbc} and \ref{sec:illdomain}.

\subsection{Ill-posed boundary condition}\label{sec:illbc}

The operator-theoretic programme aims at realising a Hamiltonian of zero-range interaction as a suitable self-adjoint extension of $\mathring{H}$, the free Hamiltonian initially restricted to wave-functions supported away from the coincidence configuration $\Gamma$ (see \eqref{eq:domHring-initial}), by selecting an extension domain where instead the wave-functions behave at $\Gamma$ with a precise, physically grounded boundary condition (of Bethe Peierls or of Ter-Martirosyan Skornyakov type).

As demonstrated in Section \ref{sec:TMSextension-section} (Theorem \ref{thm:globalTMSext}), such a two-fold requirement is possible if and only if one restricts $\mathring{H}^*$ to those functions $g\in\mathcal{D}(\mathring{H}^*)$ with singular charges $\xi$ from a distinguished subspace $\mathcal{D}\subset H^{-\frac{1}{2}}(\mathbb{R}^3)$:
\begin{itemize}
 \item[1.] $\mathcal{D}$ must be dense in $H^{-\frac{1}{2}}(\mathbb{R}^3)$ (for generic self-adjoint extensions of $\mathring{H}$ the charge domain need not be dense: Theorem \ref{thm:generalclassification});
 \item[2.] $\mathcal{D}$ must be mapped by $T_\lambda+\alpha\mathbbm{1}$ into $H^{\frac{1}{2}}(\mathbb{R}^3)$ for some (and hence for all) $\lambda>0$;
 \item[3.] $\mathcal{D}$ must be a domain of self-adjointness for $W_\lambda^{-1}(T_\lambda+\alpha\mathbbm{1})$ in the Hilbert space given by $H^{-\frac{1}{2}}(\mathbb{R}^3)$ equipped with the twisted (equivalent) scalar product $\langle\cdot,W_\lambda\cdot\rangle$ (see \eqref{eq:W-scalar-product}).
\end{itemize}

There are no other possibilities (Theorem \ref{thm:globalTMSext}).

Condition 2.~above makes $W_\lambda^{-1}(T_\lambda+\alpha\mathbbm{1})$ well-posed, because $W_\lambda$ is a \emph{bijection} of $H^{-\frac{1}{2}}(\mathbb{R}^3)$ \emph{onto} $H^{\frac{1}{2}}(\mathbb{R}^3)$ (Lemma \ref{lem:Wlambdaproperties}(ii)), and eventually leads to the desired boundary condition, namely 
\begin{equation}\label{eq:again3/2}
 \begin{split}
  & \textrm{for every $\xi\in\mathcal{D}$ there is $\phi^\lambda\in H^2_{\mathrm{b}}(\mathbb{R}^3\times\mathbb{R}^3)$ with} \\
  & \phi^\lambda(\yy,\mathbf{0})\;=\;(2\pi)^{-\frac{3}{2}} (T_\lambda+\alpha\mathbbm{1})\xi(\yy)\qquad\textrm{for a.e.~$\yy\in\mathbb{R}^3$}\,,
 \end{split}
\end{equation}
or any of the equivalent versions \eqref{eq:g-largep2-TMS0}-\eqref{eq:phi-largep2-star-yversion-TMS0}. From the perspective of \eqref{eq:again3/2} the requirement $(T_\lambda+\alpha\mathbbm{1})\mathcal{D}\subset H^{\frac{1}{2}}(\mathbb{R}^3)$ is needed because by standard trace arguments \eqref{eq:again3/2} is a $H^{\frac{1}{2}}$-identity and would not have sense if $(T_\lambda+\alpha\mathbbm{1})\xi$ had \emph{strictly less} than $H^{\frac{1}{2}}$-regularity.

In a number of past studies the choice of the charge domain $\mathcal{D}$ left instead the boundary condition \eqref{eq:again3/2} ambiguous.

The first semi-rigorous mathematical treatment of the bosonic trimer was given by Minlos and Faddeev in the work \cite{Minlos-Faddeev-1961-1}, and there the choice was (with our current notation) $\widetilde{\mathcal{D}}=\mathcal{F}^{-1}C^\infty_c(\mathbb{R}^3_{\pp})$. That is, a \emph{symmetric} extension of $\mathring{H}$ of Ter-Martirosyan-Skornyakov type \index{Ter-Martirosyan-Skornyakov!symmetric extension} was suggested as follows: the extension's domain consists of those functions whose singular charges $\xi$ are all those with smooth and compactly supported Fourier transform $\widehat{\xi}$. In fact, the first two seminal works \cite{Minlos-Faddeev-1961-1,Minlos-Faddeev-1961-2} by Minlos and Faddeev had rather the form of very brief announcements with only sketches of the main reasoning and proofs; yet the space of charges was clearly declared therein and moreover, shortly after, Flamand \cite{Flamand-Cargese1965} presented a detailed review of \cite{Minlos-Faddeev-1961-1} with the same explicit domain declaration.

Subsequent choices to be mentioned are $\widetilde{\mathcal{D}}=\mathcal{F}^{-1}C^\infty_c(\mathbb{R}^3_{\pp})$ in \cite{Minlos-1987,Minlos-Shermatov-1989,Minlos-2011-preprint_May_2010,Minlos-2014-I_RusMathSurv}, $\widetilde{\mathcal{D}}=H^1(\mathbb{R}^3)$ in \cite{Minlos-2012-preprint_30sett2011,Minlos-2014-II_preprint-2012,Moser-Seiringer-2017,Figari-Teta-2020}, and $\widetilde{\mathcal{D}}=H^{\frac{3}{2}-\varepsilon}(\mathbb{R}^3)$, $\varepsilon>0$, in \cite{Shermatov-2003}. (The above-mentioned works \cite{Minlos-Shermatov-1989,Shermatov-2003,Minlos-2011-preprint_May_2010,Minlos-2012-preprint_30sett2011,Minlos-2014-I_RusMathSurv,Minlos-2014-II_preprint-2012,Moser-Seiringer-2017} are actually 
focused on the \emph{fermionic} counterpart setting; yet, also in that case one has to face the very same technical problem of providing a well-posed definition of $W_\lambda^{-1}(T_\lambda+\alpha\mathbbm{1})$ and of making the boundary condition \eqref{eq:again3/2} unambiguous, up to non-essential changes of numerical coefficients in $T_\lambda$ and $W_\lambda$ from the bosonic to the fermionic analysis.)

Now, such proposals for $\widetilde{\mathcal{D}}$ are problematic. In the case $\widetilde{\mathcal{D}}=\mathcal{F}^{-1}C^\infty_c(\mathbb{R}^3_{\pp})$, hence $\widetilde{\mathcal{D}}\subset H^s(\mathbb{R}^3)$ $\forall s\in\mathbb{R}$, the sectors $\ell\in\mathbb{N}$ are unambiguously described through analogues of Lemma \ref{lem:Atildenot0} and Proposition \ref{prop:Alambdaellnot0} (where the present choice was $\widetilde{\mathcal{D}}=H^\frac{3}{2}_{\ell}(\mathbb{R}^3)$, namely the lowest Sobolev space that is entirely mapped with continuity by $T_\lambda^{(\ell)}$ into the desired $H^\frac{1}{2}_{\ell}(\mathbb{R}^3)$), and one realises $W_\lambda^{-1}(T_\lambda^{(\ell)}+\alpha\mathbbm{1})$ self-adjointly on the domain $\mathcal{D}_\ell=\big\{\xi\in H_\ell^{\frac{1}{2}}(\mathbb{R}^3)\,\big|\,T_\lambda^{(\ell)}\xi\in H_\ell^{\frac{1}{2}}(\mathbb{R}^3)\big\}$ (Proposition \ref{prop:Alambdaellnot0}). On the contrary, choosing $\widetilde{\mathcal{D}}=H^1_{\ell}(\mathbb{R}^3)$, $\ell\in\mathbb{N}$, poses the problem of whether $(T_\lambda^{(\ell)}+\alpha\mathbbm{1})\widetilde{\mathcal{D}}\subset H^\frac{1}{2}_{\ell}(\mathbb{R}^3)$, which is not true in general.

Moreover, even the most stringent choice $\widetilde{\mathcal{D}}=\mathcal{F}^{-1}C^\infty_c(\mathbb{R}^3_{\pp})$ does not guarantee the well-posedness of the sector $\ell=0$. It was already observed (Remark \ref{rem:Tl-failstomap}) that if $\xi\in\mathcal{F}^{-1}C^\infty_c(\mathbb{R}^3_{\pp})$, then $T_\lambda^{(0)}\xi$ belongs to $H^{\frac{1}{2}-\varepsilon}(\mathbb{R}^3)$ $\forall\varepsilon>0$, but not to $H^{\frac{1}{2}}(\mathbb{R}^3)$.

\subsection{Incomplete criterion of self-adjointness}\label{sec:illdomain}

The next source of ill-posedness may be tracked down in the problem of determining a domain $\mathcal{D}\supset\widetilde{\mathcal{D}}$ of self-adjointness for $W_\lambda^{-1}(T_\lambda+\alpha\mathbbm{1})$ with respect to the Hilbert space $H^{-\frac{1}{2}}_{W_\lambda}(\mathbb{R}^3)$, once a domain $\widetilde{\mathcal{D}}$ of symmetry is selected.

Because of the special form of the scalar product \eqref{eq:W-scalar-product} in $H^{-\frac{1}{2}}_{W_\lambda}(\mathbb{R}^3)$, it is straightforward to see (Lemma \ref{lem:symsym}) that, as long as $\widetilde{\mathcal{D}}$ is dense in $L^2(\mathbb{R}^3)$, the symmetry on $\widetilde{\mathcal{D}}$ of $W_\lambda^{-1}(T_\lambda+\alpha\mathbbm{1})$ with respect to $H^{-\frac{1}{2}}_{W_\lambda}(\mathbb{R}^3)$ is equivalent to the symmetry on $\widetilde{\mathcal{D}}$ of $T_\lambda$ with respect to $L^2(\mathbb{R}^3)$.

Based on such a suggestive property, an amount of previous investigations \cite{Minlos-Faddeev-1961-1,Minlos-Faddeev-1961-2,Flamand-Cargese1965,Minlos-1987,Minlos-Shermatov-1989,Menlikov-Minlos-1991,Menlikov-Minlos-1991-bis,Minlos-TS-1994,Shermatov-2003,Minlos-2011-preprint_May_2010,Minlos-2010-bis,Minlos-2012-preprint_30sett2011,Minlos-2014-I_RusMathSurv,Minlos-2014-II_preprint-2012,Figari-Teta-2020} adopted the claim that, for a dense subspace $\mathcal{D}$ of $L^2(\mathbb{R}^3)$, $W_\lambda^{-1}(T_\lambda+\alpha\mathbbm{1})$ on $\mathcal{D}$ is self-adjoint with respect to $H^{-\frac{1}{2}}_{W_\lambda}(\mathbb{R}^3)$ if and only if $T_\lambda$ on $\mathcal{D}$ is self-adjoint with respect to $L^2(\mathbb{R}^3)$.

In fact, this is not true (Lemma \ref{lem:exampleMinloswrong}) and the link between the two self-adjointness problems is more subtle (Lemma \ref{lem:two-selfadj-problems}).

That the emergent Hamiltonian obtained by realising the Birman parameter $W_\lambda^{-1}(T_\lambda+\alpha\mathbbm{1})$ self-adjointly on $L^2(\mathbb{R}^3)$ (instead of $H^{-\frac{1}{2}}_{W_\lambda}(\mathbb{R}^3)$) yields inconsistencies, has been known for a few years with reference to the \emph{fermionic} problem (a trimer consisting of two identical fermions of mass $m$ and a third particle of different type, and with inter-particle zero-range interaction). In that setting, a quantitative difference emerges between the mass thresholds of self-adjointness in the various $\ell$-sectors computed in \cite{Minlos-2011-preprint_May_2010,Minlos-2012-preprint_30sett2011,Minlos-2014-I_RusMathSurv,Minlos-2014-II_preprint-2012} by solving the self-adjointness problem in $L^2(\mathbb{R}^3)$, and certain spectral mass thresholds having the same conceptual meaning and obtained by formal theoretical computations and numerics within the physicists' zero-range methods \cite{Werner-Castin-2006-PRA,Kartavtsev-Malykh-2007,Castin-Tignone-2011}. The work \cite{CDFMT-2015} by Correggi, Dell'Antonio, Finco, Michelangeli, and Teta gave a first mathematical explanation of the situation, in the unitary regime $\alpha=0$, by means of a quadratic form construction of self-adjoint Hamiltonians of Ter-Martirosyan-Skornyakov type, showing that certain non-$L^2$-charges in $H^{-\frac{1}{2}}(\mathbb{R}^3)$ were needed for a correct domain of self-adjointness. Right after, Michelangeli and Ottolini \cite{MO-2016,MO-2017} addressed the same issue, recognising that indeed the correct self-adjointness problem for the Birman parameter is only with respect to the Hilbert space $H^{-\frac{1}{2}}_{W_\lambda}(\mathbb{R}^3)$.

For a three-body systems of \emph{three identical bosons} there is of course no mass parameter, hence no counterpart of the type of inconsistencies described above for the fermionic case.

Moreover, deceptively enough, the study of the self-adjoint extensions of $T_\lambda$ with respect to $L^2(\mathbb{R}^3)$, with initial domain, say, $H^1(\mathbb{R}^3)$, yields conclusions that are qualitatively very similar to the correct analysis of the self-adjoint realisations of $W_\lambda^{-1}T_\lambda^{(0)}$ with respect to $H^{-\frac{1}{2}}_{W_\lambda,\ell=0}(\mathbb{R}^3)$.

More precisely, in analogy to the present discussion of Lemma \ref{lem:deficiency1-1}, one can easily check that the $L^2$-computation of the deficiency spaces, namely of the solutions $\xi$ to $T_\lambda^{(0)}\xi=\ii\mu\xi$ in $L^2(\mathbb{R}^3)$ for $\mu>0$, yields 
\begin{equation*}
    \theta_+(x)-\frac{4}{\pi\sqrt{3}}\int_{\mathbb{R}}\theta_+(y)\,\log \frac{\,2\cosh(x-y)+1\,}{\,2\cosh(x-y)-1\,}\,\ud y \;=\;\frac{\ii \mu}{\,2\pi^2\sqrt{\lambda}}\,\frac{\theta_+(x)}{\,\cosh x}
  \end{equation*}
 (see \eqref{radialTMS0} for a comparison). 
 The above homogeneous equation replaces \eqref{eq:eigenequation-thetaplus}, and is equivalent to
 \[
  \widehat{\gamma}(s)\,\widehat{\theta}_+(s)\;=\;\frac{\ii \mu}{\,4\pi^2\sqrt{\lambda}}\,\Big(\frac{1}{\,\cosh\frac{\pi}{2}s}*\widehat{\theta}_+\Big)(s)\,,
 \]
which replaces \eqref{eq:eigenequation-thetaplusF}. By the same reasoning of the proof of Lemma \ref{lem:deficiency1-1}, the latter equation has a unique solution, up to multiplicative pre-factor, meaning that the deficiency indices are $(1,1)$. Then, mimicking the proof of Lemma \ref{lem:adjointasymtotics}, one finds a completely analogous large-momentum asymptotics for the singular elements of the adjoint, which leads to a structure of $L^2$-self-adjoint realisations of $T_\lambda^{(0)}$ that mirrors that of Proposition \ref{prop:Aellzero-selfadj}.

Nevertheless, each such domain of $L^2$-self-adjointness for $T_\lambda$ is not enough to guarantee that the corresponding three-body Hamiltonian is self-adjoint.

\section{Regularised models}\label{sec:regularisedmodels}

The Hamiltonian $\mathscr{H}_{0,\beta}$ constructed as canonical model in Theorem \ref{thm:H0beta} is regarded as \emph{unstable}, owing to its infinite sequence of bound state energy levels accumulating to $-\infty$ (Thomas collapse). \index{Thomas effect (collapse onto the centre)}

In retrospect, this feature is due to the combination of the \emph{zero-range character} of the modelled interaction and the \emph{bosonic symmetry} of the model. As a comparison, the analogous construction for a three-body system with zero-range interaction consisting of two identical fermions and a particle of different type produces a Hamiltonian that is lower semi-bounded in a suitable regime of masses \cite{CDFMT-2012}.

Thus, in order to have a stable model one hypothesis must be removed, among the vanishing of the interaction range and the bosonic symmetry. Such an observation was made by Thomas himself in his work on the tritium  \cite{Thomas1935}, which is remarkable if one considers that at the time of \cite{Thomas1935} neither the precise nature of the nuclear interaction nor the connection between spin and statistics had been understood yet.

This poses the problem of constructing \emph{regularised} models for the bosonic trimer, which do not display the spectral instability analysed in Theorem \ref{thm:spectralanalysis}, and yet describe an interaction of zero range that retains certain spectral features such as the continuous spectrum all above some threshold, or the occurrence of negative eigenvalues accumulating at the continuum threshold (Efimov effect), or the typical short-range profile of the wave-functions.

Abstractly speaking, one can perform a regularisation with an ad hoc energy cut-off on the canonical model $\mathscr{H}_{0,\beta}$, or also with a modified Hamiltonian in the form of a proper Schr\"{o}dinger operator with two-body potentials of small but finite (i.e., non-zero) effective range.

Somewhat intermediate between these two directions is a construction, formerly contemplated by Minlos and Faddeev with no further analysis, of a contact interaction Hamiltonian similar to $\mathscr{H}_{0,\beta}$, but with a regularisation that has the overall effect of switching the interaction off in the vicinity of the triple coincidence configuration. The three identical bosons are allowed by the statistics to occupy that region, in which now the regularisation make them asymptotically free. This removes the instability of the canonical model. (Subsect.~\ref{sec:MFregularisation}-\ref{sec:MinFadzero}).

Further types of regularisations have been proposed, which are conceptually analogous to the idea of Minlos and Faddeev in that they introduce a non-constant, effective scattering length that tends to be suppressed (meaning, no interaction, particles are free) when the three bosons get close to the point of triple coincidence. Whereas the Minlos-Faddeev regularisation implements such an idea in position coordinates, one can analogously work in momentum coordinates, making the effective scattering length vanish at large relative momenta. For comparison,such high energy cut-off will be discussed in Subsect.~\ref{sec:highenergycutoff}.

\subsection{Minlos-Faddeev regularisation}\label{sec:MFregularisation}

This is the ultra-violet regularisation originally proposed in \cite[Section 6]{Minlos-Faddeev-1961-1} (see also \cite[Section VI.2]{Flamand-Cargese1965} and \cite{Albe-HK-Wu-1981}), and on which a number of results with the quadratic form approach have been recently announced in \cite{Figari-Teta-2020}.


In practice, this is a modification of the canonical model (Theorem \ref{thm:H0beta}) along the following line: the ordinary Birman parameter \index{Birman extension parameter} $3W_\lambda^{-1}(T_\lambda+\alpha\mathbbm{1})$, that selects (via Theorems \ref{thm:generalclassification} and \ref{thm:globalTMSext}) self-adjoint extensions of Ter-Martirosyan-Skornyakov type of the minimal operator $\mathring{H}$ defined in \eqref{eq:domHring-initial}, is replaced by 
\begin{equation}\label{eq:newBirmanpar}
 3\,W_\lambda^{-1}(T_\lambda+\alpha\mathbbm{1}+K_\sigma)\,,\qquad \sigma\,>\,0\,,
\end{equation}
where, for generic $\sigma\in\mathbb{R}$,
\begin{equation}\label{eq:Ksigmaposition}
\begin{split}
  (K_\sigma\xi)(\yy)\;&:=\;\frac{\,\sigma_0+\sigma\,}{|\yy|}\,\xi(\yy)  \, ,\\
 \sigma_0\;&:=\;2\pi\sqrt{3}\,\Big(\frac{4\pi}{3\sqrt{3}}-1\Big)\,.
\end{split}
\end{equation}

The motivation is clear from the large momentum asymptotics \eqref{eq:g-largep2-star} valid for generic $g\in\mathcal{D}(\mathring{H})$, namely
 \begin{equation*}
  \int_{\!\substack{ \\ \\ \pp_2\in\mathbb{R}^3 \\ |\pp_2|<R}}\widehat{g}(\pp_1,\pp_2)\,\ud\pp_2\;=\;4\pi R\,\widehat{\xi}(\pp_1)+\Big({\textstyle\frac{1}{3}}(\widehat{W_\lambda\eta})(\pp_1)-(\widehat{T_\lambda\xi})(\pp_1)\Big)+o(1)
 \end{equation*}
 as $R\to +\infty$. Indeed, when a self-adjoint extension is selected, out of the family \eqref{eq:family}, labelled by the Birman parameter \eqref{eq:newBirmanpar} densely defined in $H^{-\frac{1}{2}}(\mathbb{R}^3)$, then in the asymptotics above one has $\eta=3W_\lambda^{-1}(T_\lambda+\alpha\mathbbm{1}+K_\sigma)\xi$ (as prescribed by formula \eqref{eq:domDHA} of Theorem \ref{thm:generalclassification}), whence
 \begin{equation*}
  \int_{\!\substack{ \\ \\ \pp_2\in\mathbb{R}^3 \\ |\pp_2|<R}}\widehat{g}(\pp_1,\pp_2)\,\ud\pp_2\;=\;4\pi R\,\widehat{\xi}(\pp_1)+\alpha\widehat{\xi}(\pp_1)+(\widehat{K_\sigma\xi})(\pp_1)+o(1)\,,
 \end{equation*}
 and also (see Corollary \ref{cor:largepasympt-star})
  \begin{equation*}
 (2\pi)^{\frac{3}{2}} c_g\,g_{\mathrm{av}}(\yy_1;|\yy_2|)\,\stackrel{|\yy_2|\to 0}{=}\,\frac{4\pi}{|\yy_2|}\xi(\yy_1)+\Big(\alpha+ \frac{\,\sigma_0+\sigma\,}{|\yy_1|}\Big)\xi(\yy_1)+ o(1)\,.
 \end{equation*}
 Thus, the new self-adjoint Hamiltonian has a domain of functions that display a modified short-scale asymptotics, as compared to the zero-range Bethe-Peierls condition: \index{contact condition!Bethe-Peierls} the modification consists of the inverse negative scattering length $\alpha$ being replaced by a position-dependent value
 \begin{equation}
  \alpha_{\mathrm{eff}}\;:=\;\alpha+(\sigma_0+\sigma)/|\yy|\,,
 \end{equation}
 where $|\yy|$ is the distance of the third particle from the point towards which the other two are getting closer and closer. 
 Therefore, $\alpha_{\mathrm{eff}}\to +\infty$ when \emph{all three particles} collapse to the same spatial position, meaning that the scattering length vanishes in such a limit. As vanishing scattering length means absence of interaction, the overall effect is a three-body regularisation that prevents the collapse of the system along an unbounded sequence of negative energy levels.

 By construction, $K_\sigma$ commutes with the rotations in $\mathbb{R}^3$ and therefore is reduced as
 \begin{equation}\label{eq:Ksigmarotations}
  K_\sigma\;=\;\bigoplus_{\ell\in\mathbb{N}}K_\sigma^{(\ell)}
 \end{equation}
 with respect to the orthogonal Hilbert space decomposition \eqref{eq:bigdecompW} of $H^{-\frac{1}{2}}_{W_\lambda}(\mathbb{R}^3)$. For all practical purposes (in view also of the discussion of Subsect.~\ref{sec:variants}) it suffices to implement the Minlos-Faddeev regularisation in the sector $\ell=0$, thus only inserting $K_\sigma^{(0)}$ in \eqref{eq:newBirmanpar}, as the canonical model is already stable in the sectors of higher angular momentum. That is the version of the regularisation studied here.

 With definition \eqref{eq:Ksigmaposition} and the above considerations in mind, the same conceptual path of Section \ref{sec:lzero} can be now re-done, adapting it to the new Birman parameter \eqref{eq:newBirmanpar}. \index{Birman extension parameter}

 To this aim, one needs updated expressions of the quantities of interest in terms of the re-scaled radial components associated to the charges.
 
 First of all, taking the Fourier transform in \eqref{eq:Ksigmaposition} yields
    \begin{equation}\label{eq:Ksigmamomentum}
     (\widehat{K_\sigma\xi})(\pp)\;=\;\frac{\sigma_0+\sigma}{2\pi^2}\,\int_{\mathbb{R}^3}\frac{\widehat{\xi}(\qq)}{\,|\pp-\qq|^2}\,\ud\qq\,.
    \end{equation}
 
\begin{figure}[t!]
 \begin{center}
\includegraphics[width=8cm]{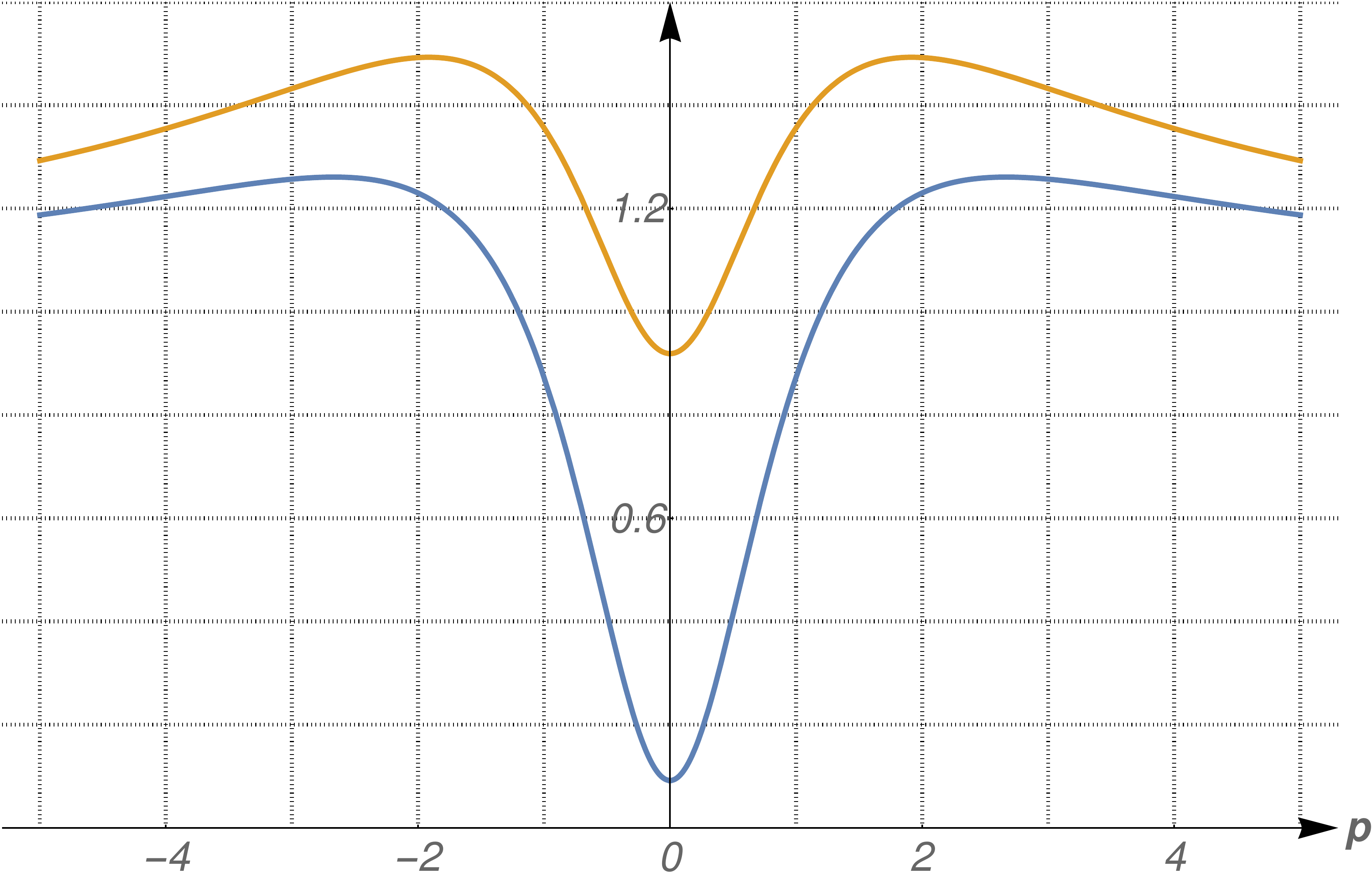}
\caption{Plot of the function $\widehat{\gamma}_\sigma(s)$ defined in \eqref{eq:gamma-s-distribution}, with parameter $\sigma=1$ (blue) and $\sigma=10$ (orange).}\label{fig:gammasigma}
\end{center}
\end{figure}

 One also introduces two auxiliary functions, namely
 \begin{equation}\label{eq:bigTheta}
  \begin{split}
    \Theta^{(\theta)}_\sigma(x)\;&:=\;\theta(x)-\frac{4}{\pi\sqrt{3}}\int_{\mathbb{R}}\ud y\,\theta(y) \log \frac{\,2\cosh(x-y)+1\,}{\,2\cosh(x-y)-1\,}+ \\
     &\qquad\qquad\qquad +\frac{\,\sigma_0+\sigma\,}{\pi^3\sqrt{3}}\int_{\mathbb{R}}\ud y\,\theta(y)\,\log\Big|\coth\frac{x-y}{2}\Big|
  \end{split}
 \end{equation}
  for given $\theta$, and
  \begin{equation}\label{eq:gamma-s-distribution}
   \widehat{\gamma}_\sigma(s)\;:=\;1+\frac{1}{\,\sqrt{3}\,s\,\cosh\frac{\pi}{2}s}\Big(\frac{\,\sigma_0+\sigma\,}{\pi^2}\,\sinh\frac{\pi}{2}s-8\sinh\frac{\pi}{6}s\Big)\,.
  \end{equation}
  Observe that $\widehat{\gamma}_{-\sigma_0}$ is precisely the function $\widehat{\gamma}$ defined in \eqref{eq:gamma-distribution}. It is easy to see that $\widehat{\gamma}_\sigma$ is a smooth even function of $\mathbb{R}$, that converges asymptotically to 1 as $|s|\to +\infty$, and that for $\sigma\in[0,2\pi\sqrt{3})$ has \emph{absolute} minimum at $s=0$ of magnitude
  \begin{equation}\label{eq:gammasigmazero}
   \widehat{\gamma}_\sigma(0)\;=\;\widehat{\gamma}(0)+\frac{\,\sigma_0+\sigma}{\,2\pi\sqrt{3}\,}\;=\;\frac{\,\sigma}{\,2\pi\sqrt{3}\,}
  \end{equation}
  (see Figure \ref{fig:gammasigma}). Indeed, $\sigma_0=-2\pi\sqrt{3}\,\widehat{\gamma}(0)$.

   Lemma \ref{lem:xithetaidentities} has the following analogue.

  \begin{lemma}\label{lem:regularisedlemma1}
   Let $\lambda>0$, $s,\sigma\in\mathbb{R}$, and $\xi$ be as in \eqref{eq:0xi}. One has the identities
         \begin{eqnarray}
         \big(\big(T_\lambda^{(0)}+K_\sigma^{(0)}\big)\xi\big)\,{\textrm{\large $\widehat{\,}$\normalsize}}\,(\pp)\;&=&\;\frac{1}{\sqrt{4\pi}\,|\pp|}\,\frac{4\pi^2}{\sqrt{3}\,}\,\Theta^{(\theta)}_\sigma(x)\,, \label{eq:Tlxi-theta-withK}\\
      \big\|\big(T_\lambda^{(0)}+K_\sigma^{(0)}\big)\xi\big\|_{H^{s}(\mathbb{R}^3)}^2\!\!&\approx&\!\!\int_{\mathbb{R}}\ud x\,(\cosh x)^{1+2s}\,\big|\Theta^{(\theta)}_\sigma(x)\big|^2\,, \label{eq:Tlambda0xi-with-theta-andK} \\
      \widehat{\Theta^{(\theta)}_\sigma}(s)\;&=&\;\widehat{\gamma}_\sigma(s)\,\widehat{\theta}(s)\,, \label{eq:ThetaTransformed} \\
       \int_{\mathbb{R}^3} \overline{\,\widehat{\xi}(\pp)}\, \big(\big(T_\lambda^{(0)}+K_\sigma^{(0)}\big)\xi\big)\,{\textrm{\large $\widehat{\,}$\normalsize}}\,(\pp)\,\ud\pp\;&=&\;\frac{\,8\pi^2}{3\sqrt{3}}\int_{\mathbb{R}}\ud s\, \widehat{\gamma}_\sigma(s)\,|\widehat{\theta}(s)|^2\,, \label{eq:xiTxi-with-theta-andK}
    \end{eqnarray}
    with $x$ and $\theta$ given by \eqref{ftheta-1}, $\Theta^{(\theta)}_{\sigma}$ given by \eqref{eq:bigTheta} with respect to the present $\theta$, and $\widehat{\gamma}_\sigma$ given by \eqref{eq:gamma-s-distribution}.
    In \eqref{eq:Tlxi-theta-withK} it is understood that $x\geqslant 0$, and \eqref{eq:Tlambda0xi-with-theta-andK} is meant as an equivalence of norms (with $\lambda$-dependent multiplicative constant).   
  \end{lemma}

  \begin{proof}
   Consider the $K_\sigma$-term. From \eqref{eq:Ksigmamomentum}, in complete analogy with \eqref{eq:fellsector},
   \[
    (\widehat{K_\sigma^{(\ell)}\xi^{(\ell)}})(\pp)\;=\;\frac{\,\sigma_0+\sigma\,}{\pi}\sum_{n=-\ell}^\ell Y_{\ell,n}(\Omega_{\pp})\int_{\mathbb{R}^+}\ud q\,q^2\,f_{\ell,n}^{(\xi)}(q)\int_{-1}^1\ud t\,\frac{P_\ell(t)}{\,\pp^2+q^2-2|\pp|qt\,}\,,
   \]
   having used the expansion \eqref{eq:xihatangularexpansion} for $\xi$, namely
   \[
  \widehat{\xi}(\pp)\;=\;\sum_{\ell=0}^\infty\sum_{n=-\ell}^\ell f_{\ell,n}^{(\xi)}(|\pp|) Y_{\ell,n}(\Omega_{\pp})\;=\;\sum_{\ell=0}^\infty\widehat{\xi^{(\ell)}}(\pp)\,.
\]
  Thus, for a spherically symmetric charge $\widehat{\xi}(\pp)=\frac{1}{\sqrt{4\pi}}f(|\pp|)$,
   \[
    (\widehat{K_\sigma^{(0)}\xi})(\pp)\;=\;\frac{1}{\sqrt{4\pi}}\,\frac{1}{p}\,\frac{\,\sigma_0+\sigma\,}{\pi}\int_{\mathbb{R}^+}\ud q\,qf(q)\,\log\frac{p+q}{\,|p-q|\,}\,,\qquad p=|\pp|\,.
   \]
  Let $\theta$ be the re-scaled radial function \eqref{ftheta-1} associated with $\xi$ (and $f$) above. With $p=\frac{2\sqrt{\lambda}}{\sqrt{3}}\,\sinh x$ and $q=\frac{2\sqrt{\lambda}}{\sqrt{3}}\,\sinh y$ one has
  \[
   \frac{p+q}{\,|p-q|\,}\;=\;\frac{\sinh\frac{x+y}{2}\,\cosh\frac{x-y}{2}}{\,|\sinh\frac{x-y}{2}\,\cosh\frac{x+y}{2}|\,}\;=\;\Big|\coth\frac{x-y}{2}\Big|\,\tanh\frac{x+y}{2}\,,
  \]
  which, together with \eqref{ftheta-2}, gives
  \[
   \begin{split}
    \int_{\mathbb{R}^+}\ud q\,qf(q)\,\log\frac{p+q}{\,|p-q|\,}\;&=\;\frac{4}{3}\int_{\mathbb{R}^+}\ud y\,\theta(y)\Big(\log\Big|\tanh\frac{x+y}{2}\Big|+\log\Big|\coth\frac{x-y}{2}\Big|\Big) \\
    &=\;\frac{4}{3}\int_{\mathbb{R}}\ud y\,\theta(y)\,\log\Big|\coth\frac{x-y}{2}\Big|\,,
   \end{split}
  \]
  the last identity following from the odd-parity extension of $\theta$ over $\mathbb{R}$. Therefore,
  \[
    (\widehat{K_\sigma^{(0)}\xi})(\pp)\;=\;\frac{1}{\sqrt{4\pi}}\,\frac{1}{|\pp|}\,\frac{\,4(\sigma_0+\sigma)\,}{3\pi}\int_{\mathbb{R}}\ud y\,\theta(y)\,\log\Big|\coth\frac{x-y}{2}\Big|\,.
   \]
   Combining this with \eqref{eq:Tlxi-theta} yields \eqref{eq:Tlxi-theta-withK}.

  Formula \eqref{eq:Tlambda0xi-with-theta-andK} is proved from \eqref{eq:Tlxi-theta-withK} with the very same reasoning used for the analogous formula \eqref{eq:Tlambda0xi-with-theta} in Lemma \ref{lem:xithetaidentities}.

  Concerning \eqref{eq:ThetaTransformed}, observe that the first two summands in the expression \eqref{eq:bigTheta} of $\Theta^{(\theta)}_\sigma$ have precisely Fourier transform $\widehat{\gamma}(s)\widehat{\theta}(s)$, as determined already in the proof of Lemma \ref{lem:xithetaidentities}, with $\widehat{\gamma}$ defined in \eqref{eq:gamma-distribution}. Focus first on the third summand, namely the one with the $\sigma$-dependent pre-factor. One has
  \[
   \begin{split}
    \Big(\log\Big|\coth\frac{x}{2}\Big|\,\Big){\!\!\textrm{\huge ${\,}^{\widehat{\,}}$\normalsize}}\,(s)\;&=\;\sqrt{\frac{2}{\pi}}\int_{\mathbb{R}^+}\ud x\,\cos sx\,\log\coth\frac{x}{2} \\
    &=\;\sqrt{\frac{2}{\pi}}\int_{\mathbb{R}^+}\ud x\,\cos sx\,\big(\log(1+e^{-x})-\log(1-e^{-x})\big)\,,
   \end{split}
  \]
 and  (see, e.g., \cite[I.1.5.(13)-(14)]{Erdelyi-Tables1})
 \[
  \begin{split}
   \int_{\mathbb{R}^+}\ud x\,\cos sx\,\log(1+e^{-x})\;&=\;\frac{1}{\,2s^2}-\frac{\pi}{\,2s}\,\frac{1}{\sinh \pi s}\,, \\
   \int_{\mathbb{R}^+}\ud x\,\cos sx\,\log(1-e^{-x})\;&=\;\frac{1}{\,2s^2}-\frac{\pi}{\,2s}\,\coth \pi s\,,
  \end{split}
 \]
 whence
 \[
  \Big(\log\Big|\coth\frac{x}{2}\Big|\,\Big){\!\!\textrm{\huge ${\,}^{\widehat{\,}}$\normalsize}}\,(s)\;=\;\sqrt{\frac{\pi}{2}}\,\frac{\,\tanh\frac{\pi}{2}s}{s}\,.
 \]
 Therefore, the Fourier transform of the last summand in the expression \eqref{eq:bigTheta} of $\Theta^{(\theta)}_\sigma$ is
 \[
 \frac{\,\sigma_0+\sigma\,}{\pi^3\sqrt{3}}\,\Big(\theta*\log\Big|\coth\frac{x}{2}\Big|\Big){\!\!\textrm{\huge ${\,}^{\widehat{\,}}$\normalsize}}\,(s)\;=\;\frac{\,\sigma_0+\sigma\,}{\,\pi^2\sqrt{3}\,}\,\frac{\,\tanh\frac{\pi}{2}s}{s}\:\widehat{\theta}(s)\,.
 \]
 Adding this term to $\widehat{\gamma}(s)\widehat{\theta}(s)$ yields \eqref{eq:gamma-s-distribution} and hence \eqref{eq:ThetaTransformed}.

 Last, concerning \eqref{eq:xiTxi-with-theta-andK}, in complete analogy with the proof of \eqref{eq:xiTxi-with-theta}, one finds
  \[
   \begin{split}
    \int_{\mathbb{R}^3} \overline{\,\widehat{\xi}(\pp)}\, \big(\big(T_\lambda^{(0)}+K_\sigma^{(0)}\big)\xi\big)\,{\textrm{\large $\widehat{\,}$\normalsize}}\,(\pp)\,\ud\pp\;&=\;\int_{\mathbb{R}^+}\ud p\,p^2\,\overline{f(p)}\,\frac{4\pi^2}{\,p\sqrt{3}\,}\,\Theta^{(\theta)}_\sigma(x(p)) \\
    &=\;\frac{8\pi^2}{3\sqrt{3}\,}\int_{\mathbb{R}}\ud x\,\overline{\theta(x)}\,\Theta^{(\theta)}_\sigma(x) \\
    &=\;\frac{\,8\pi^2}{3\sqrt{3}}\int_{\mathbb{R}}\ud s\, \widehat{\gamma}_\sigma(s)\,|\widehat{\theta}(s)|^2
   \end{split}
  \]
  having applied \eqref{eq:Tlxi-theta-withK} in the first identity, \eqref{ftheta-2}-\eqref{ftheta-3-eq:pxchangevar} and the odd parity in the second, and Parseval's identity in the third.
  \end{proof}

  \begin{corollary}\label{cor:equivalentH12norm}
   For $\lambda,\sigma>0$, the map
   \[
    \xi\;\longmapsto\;\bigg(\int_{\mathbb{R}^3} \overline{\,\widehat{\xi}(\pp)}\, \big(\big(T_\lambda^{(0)}+K_\sigma^{(0)}\big)\xi\big)\,{\textrm{\large $\widehat{\,}$\normalsize}}\,(\pp)\,\ud\pp\bigg)^{\!\frac{1}{2}}
   \]
   defines an equivalent norm in $H^{\frac{1}{2}}_{\ell=0}(\mathbb{R}^3)$.
  \end{corollary}

  \begin{proof}
   Because of \eqref{eq:xiTxi-with-theta-andK} and the fact that $\widehat{\gamma}_\sigma$ is uniformly bounded and strictly positive when $\sigma>0$, up to inessential pre-factors each $\xi$ is mapped to $\|\theta^{(\xi)}\|_{L^2(\mathbb{R})}$ and hence to $\|\xi\|_{H^{\frac{1}{2}}(\mathbb{R}^3)}$, owing to \eqref{eq:mellinnorms}.   
  \end{proof}

 \subsection{Regularisation in the sector $\ell=0$}\label{sec:MinFadzero}

 Next, one now makes the Birman parameter explicit for a self-adjoint extension of $\mathring{H}$ when the Minlos-Faddeev regularisation is implemented with respect to the canonical construction of extensions of Ter-Martirosyan-Skornyakov type. As argued in the previous Subsection, one only needs to replace the Birman parameter \index{Birman extension parameter} $\mathcal{A}_{\lambda,\beta}^{(0)}$ (defined in \eqref{eq:Azerolambda}-\eqref{eq:domainD0beta}) of the sector $\ell=0$ with a \emph{regularised} version, henceforth denoted by $\mathcal{R}_{\lambda,\sigma}^{(0)}$. 
 Then, with such $\mathcal{R}_{\lambda,\sigma}^{(0)}$ at hand, and with all other canonical Birman parameters $\mathcal{A}_\lambda^{(\ell)}$, $\ell\in\mathbb{N}$, one constructs the associated self-adjoint Hamiltonian by means of Theorem \ref{thm:generalclassification}.

 To begin with, for $\sigma>0$ define (in analogy to \eqref{eq:Dtilde0}) the subspace
  \begin{equation}\label{eq:newD0regularised}
   \widetilde{\mathsf{D}}_{0,\sigma}\;:=\;\left\{
   \xi\in H^{-\frac{1}{2}}_{\ell=0}(\mathbb{R}^3)\left|
   \begin{array}{c}
    \textrm{$\xi$ has re-scaled radial component} \\
     \theta=\big(\widehat{\Theta}/\widehat{\gamma}_\sigma)\big)^{\!\vee} \\
     \textrm{for }\;\Theta\in C^\infty_{c,\mathrm{odd}}(\mathbb{R}_x)
   \end{array}
   \!\right.\right\}\,.
  \end{equation}
  Here the subscript `odd' indicates functions with odd parity and $\widehat{\gamma}_\sigma$ is defined in \eqref{eq:gamma-s-distribution}.  The correspondence between $\xi$ and its re-scaled radial component $\theta$ is given by \eqref{eq:0xi}-\eqref{ftheta-1}. It is tacitly understood that the re-scaled radial components are all taken with the same parameter $\lambda>0$ in the definition \eqref{ftheta-1}: this does not mean $\widetilde{\mathsf{D}}_{0,\sigma}$ is a $\lambda$-dependent subspace, as one can easily convince oneself, the choice of $\lambda$ only fixes the convention for representing its elements in terms of the corresponding $\theta$.

  \begin{lemma}\label{lem:DD0properties} Let $\lambda,\sigma>0$.
  \begin{enumerate}[(i)]
   \item $\widetilde{\mathsf{D}}_{0,\sigma}$ is dense in $H^{\frac{1}{2}}_{\ell=0}(\mathbb{R}^3)$.
   \item $(T_\lambda^{(0)}+K_\sigma^{(0)})\widetilde{\mathsf{D}}_{0,\sigma}\subset H^{s}_{\ell=0}(\mathbb{R}^3)$ for every $s\in\mathbb{R}$.
  \end{enumerate}
  \end{lemma}

  \begin{proof} (i) For $\xi\in  \widetilde{\mathsf{D}}_{0,\sigma}$, the identity $\widehat{\gamma}_\sigma\widehat{\theta}=\widehat{\Theta}$ and \eqref{eq:mellinnorms} imply
  \[
  \begin{split}
   \|\xi\|_{H^{\frac{1}{2}}(\mathbb{R}^3)}\;\approx\;\|\theta\|_{L^2(\mathbb{R})}\;&=\;\|\widehat{\theta}\|_{L^2(\mathbb{R})}\;\leqslant\;\|\widehat{\gamma}_\sigma^{-1}\|_{L^\infty(\mathbb{R})}\|\widehat{\Theta}\|_{L^2(\mathbb{R})} \\
   &=\;(\widehat{\gamma}_\sigma(0))^{-1}\|\Theta\|_{L^2(\mathbb{R})}\;<\;+\infty\,,
  \end{split}
  \]
  owing to \eqref{eq:gammasigmazero} and to the fact that $\Theta$ is smooth and with compact support.
  Because of \eqref{eq:mellinnorms}, the density of the $\xi$'s of $\widetilde{\mathsf{D}}_{0,\sigma}$ in $H^{\frac{1}{2}}_{\ell=0}(\mathbb{R}^3)$ is equivalent to the density of the associated $\theta$'s in $L^2_{\mathrm{odd}}(\mathbb{R})$. If in the latter Hilbert space a function $\theta_0$ was orthogonal to all such $\theta$'s, then
  \[
   0\;=\;\int_{\mathbb{R}}\overline{\theta_0(x)}\,\theta(x)\,\ud x\;=\;\int_{\mathbb{R}}\overline{\widehat{\theta}_0(s)}\,\frac{\widehat{\Theta}(s)}{\widehat{\gamma}_\sigma(s)}\,\ud s\;=\;\int_{\mathbb{R}}\overline{\big(\widehat{\theta}_0/\widehat{\gamma}_\sigma\big)^{\!\vee}\!(x)}\,\Theta(x)\,\ud x
  \]
  for all $\Theta\in C^\infty_{c,\mathrm{odd}}(\mathbb{R})$: as $\widehat{\gamma}_\sigma$ is uniformly bounded and strictly positive, this implies $\theta_0\equiv 0$.
  
  (ii) On the one hand $\widehat{\gamma}_\sigma\widehat{\theta}=\widehat{\Theta}$ by the assumption that $\xi\in  \widetilde{\mathsf{D}}_{0,\sigma}$, on the other hand $\widehat{\gamma}_\sigma\widehat{\theta}=\widehat{\Theta^{(\theta)}_\sigma}$, owing to \eqref{eq:ThetaTransformed}, whence $\Theta^{(\theta)}_\sigma=\Theta\in C^\infty_{c,\mathrm{odd}}(\mathbb{R})$. Plugging this information into \eqref{eq:Tlambda0xi-with-theta-andK} yields the conclusion.   
  \end{proof}

  Next, for $\lambda,\sigma>0$ define
  \begin{equation}
   \begin{split}
    \mathcal{D}\big(\widetilde{\mathcal{R}}_{\lambda,\sigma}^{(0)}\big)\;&:=\;\widetilde{\mathsf{D}}_{0,\sigma} \, , \\
    \widetilde{\mathcal{R}}^{(0)}_{\lambda,\sigma}\;&:=\; 3W_\lambda^{-1}\big(T_\lambda^{(0)}+K_\sigma^{(0)}\big)\,.
   \end{split}
  \end{equation}
 Lemma \ref{lem:DD0properties}, together with Corollary \ref{cor:equivalentH12norm}, guarantees that this is a well-posed definition for a densely defined, symmetric, and coercive operator on $H^{-\frac{1}{2}}_{W_\lambda,\ell=0}(\mathbb{R}^3)$. As such, having a strictly positive lower bound, $\widetilde{\mathcal{R}}^{(0)}_{\lambda,\sigma}$ has the Friedrichs extension (Theorem \ref{thm:Friedrichs-ext}). That will be the actual Birman parameter in this context. \index{Birman extension parameter} In analogy with Proposition \ref{prop:Alambdaellnot0}, one proves the following.

 \begin{proposition}\label{prop:NEW-Birman-param-selfadj}
  Let $\lambda,\sigma>0$. Define
  \begin{equation}
   \mathsf{D}_{0,\sigma}\;:=\;\big\{\xi\in H^{\frac{1}{2}}_{\ell=0}(\mathbb{R}^3)\,\big|\,\big(T_\lambda^{(0)}+K_\sigma^{(0)}\big)\xi\in H^{\frac{1}{2}}_{\ell=0}(\mathbb{R}^3)\big\}\,.
  \end{equation}
  The operator
  \begin{equation}\label{eq:RlambdasigmaOPERATOR}
   \begin{split}
    \mathcal{D}\big(\mathcal{R}_{\lambda,\sigma}^{(0)}\big)\;&:=\;\mathsf{D}_{0,\sigma} \\
    \mathcal{R}_{\lambda,\sigma}^{(0)})\;&:=\;3W_\lambda^{-1}\big(T_\lambda^{(0)}+K_\sigma^{(0)}\big)
   \end{split}
  \end{equation}
 is the Friedrichs extension of $\widetilde{\mathcal{R}}^{(0)}_{\lambda,\sigma}$ with respect to $H^{-\frac{1}{2}}_{W_\lambda,\ell=0}(\mathbb{R}^3)$. Its sesquilinear form is
 \begin{equation}\label{eq:RlambdasigmaFORM}
   \begin{split}
    \mathcal{D}\big[\mathcal{R}_{\lambda,\sigma}^{(0)}\big]\;&=\;H^{\frac{1}{2}}_{\ell=0}(\mathbb{R}^3) \\
    \mathcal{R}_{\lambda,\sigma}^{(0)}[\eta,\xi]\;&=\;3\int_{\mathbb{R}^3} \overline{\,\widehat{\xi}(\pp)}\, \big(\big(T_\lambda^{(0)}+K_{\sigma}^{(0)}\big)\xi\big)\,{\textrm{\large $\widehat{\,}$\normalsize}}\,(\pp)\,.
   \end{split}
\end{equation}
 \end{proposition}

 \begin{proof}
  Denote temporarily by $\mathcal{R}_{\mathrm{F}}$ the Friedrichs extension of $\widetilde{\mathcal{R}}^{(0)}_{\lambda,\sigma}$ with respect to $H^{-\frac{1}{2}}_{W_\lambda,\ell=0}(\mathbb{R}^3)$, and set
  \[
   \|\xi\|_{\mathcal{R}}\;:=\; \Big(\big\langle\xi,\widetilde{\mathcal{R}}^{(0)}_{\lambda,\sigma}\xi\big\rangle_{H^{-\frac{1}{2}}_{W_\lambda}}\Big)^{\frac{1}{2}}\;=\;\big(3\big\langle \xi,\big(T_\lambda^{(0)}+K_\sigma^{(0)}\big)\xi\big\rangle_{L^2}\big)^{\frac{1}{2}}\,.
  \]
  Owing to Corollary \ref{cor:equivalentH12norm}, the latter induces an equivalent $H^{\frac{1}{2}}$-norm on $\widetilde{\mathsf{D}}_{0,\sigma} $. As prescribed by the Friedrichs construction (Theorem \ref{thm:Friedrichs-ext}(i) and (iii)), $\mathcal{R}_{\mathrm{F}}$ has form domain
  \[
   \mathcal{D}[\mathcal{R}_{\mathrm{F}}]\;=\;\overline{\mathcal{D}\big(\widetilde{\mathcal{R}}^{(0)}_{\lambda,\sigma}\big)\,}^{\|\,\|_{\mathcal{R}}}\;=\;\overline{\widetilde{\,\mathsf{D}}_{0,\sigma}\,}^{\|\,\|_{H^{\frac{1}{2}}}}\;=\;H^{\frac{1}{2}}_{\ell=0}(\mathbb{R}^3)
  \]
  (the last identity following from Lemma \ref{lem:DD0properties}(i)), and for $\xi,\eta\in H^{\frac{1}{2}}_{\ell=0}(\mathbb{R}^3)$
  \[
   \mathcal{R}_{\mathrm{F}}[\eta,\xi]\;=\;\lim_{n\to\infty}\big\langle\eta_n,\widetilde{\mathcal{R}}^{(0)}_{\lambda,\sigma}\xi_n\big\rangle_{H^{-\frac{1}{2}}_{W_\lambda}}\;=\;3\lim_{n\to\infty}\big\langle \eta_n,\big(T_\lambda^{(0)}+K_\sigma^{(0)}\big)\xi_n\big\rangle_{L^2}
  \]
  for any two sequences $(\xi_n)_n$ and $(\eta_n)_n$ in $\widetilde{\mathsf{D}}_{0,\sigma} $ such that $\xi_n\to \xi$ and $\eta_n\to\eta$ in $H^{\frac{1}{2}}_{\ell=0}(\mathbb{R}^3)$. Since the pairing $(\eta_n,\xi_n)\mapsto \big\langle \eta_n,\big(T_\lambda^{(0)}+K_\sigma^{(0)}\big)\xi_n\big\rangle_{L^2}$ is an $H^{\frac{1}{2}}$-pairing (Corollary \ref{cor:equivalentH12norm}), one finds
  \[
   \mathcal{R}_{\mathrm{F}}[\eta,\xi]\;=\;3\lim_{n\to\infty}\big\langle \eta_n,\big(T_\lambda^{(0)}+K_\sigma^{(0)}\big)\xi_n\big\rangle_{L^2}\;=\;3\,\big\langle \eta,\big(T_\lambda^{(0)}+K_\sigma^{(0)}\big)\xi\big\rangle_{L^2}\,.
  \]
  Formula \eqref{eq:RlambdasigmaFORM} is thus proved. The operator $\mathcal{R}_{\mathrm{F}}$ is derived from its quadratic form in the usual matter: a straightforward adaptation of the analogous argument used in the proof of Proposition \ref{prop:Alambdaellnot0} shows that $\mathcal{R}_{\mathrm{F}}$ is indeed the operator \eqref{eq:RlambdasigmaOPERATOR}. 
 \end{proof}

 With the new Birman parameter \eqref{eq:RlambdasigmaOPERATOR} for the sector $\ell=0$, the construction of the canonical model $\mathscr{H}_{0,\beta}$ is modified as follows (see the discussion of Subsect.~\ref{sec:constructioncanonical}).

 Instead of the self-adjoint extension $\mathring{H}_{\mathcal{A}_{\lambda,\beta}}$ obtained by means of Theorem \ref{thm:generalclassification} with Birman parameter
  \begin{equation*}
   \mathcal{A}_{\lambda,\beta}\;=\; \mathcal{A}_{\lambda,\beta}^{(0)}\:\oplus\: \bigoplus_{\ell=1}^\infty\mathcal{A}_\lambda^{(\ell)}
  \end{equation*}
 (see \eqref{eq:globalAlambda-1}), one considers another operator from the family \eqref{eq:family}, namely the self-adjoint extension $\mathring{H}_{\mathcal{R}_{\lambda,\sigma}}$ with modified Birman parameter \index{Birman extension parameter}
 \begin{equation}\label{eq:globalRlambda-1}
   \mathcal{R}_{\lambda,\sigma}\;:=\;\mathcal{R}_{\lambda,\sigma}^{(0)}\:\oplus\: \bigoplus_{\ell=1}^\infty\mathcal{A}_\lambda^{(\ell)}\,.
 \end{equation}
 By definition, the domain of $\mathcal{R}_{\lambda,\sigma}$ is
 \begin{equation}
  \mathcal{D}(\mathcal{R}_{\lambda,\sigma})\;=\;\left\{ 
   \begin{array}{c}
    \displaystyle\xi=\sum_{\ell=0}^\infty\xi^{(\ell)}\in\;\bigoplus_{\ell=0}^\infty \,H^{-\frac{1}{2}}_{W_\lambda,\ell}(\mathbb{R}^3)\,\cong\, H^{-\frac{1}{2}}_{W_\lambda}(\mathbb{R}^3) \\
    \textrm{such that} \\
    \xi^{(\ell)}\in H^{\frac{1}{2}}_\ell(\mathbb{R}^3)\;\textrm{ and }\; T_\lambda^{(\ell)}\xi^{(\ell)}\in H^{\frac{1}{2}}_\ell(\mathbb{R}^3)\;\textrm{ for }\;\ell\in\mathbb{N}\,, \\
    \xi^{(0)}\in H_{\ell=0}^{\frac{1}{2}}(\mathbb{R}^3)\;\textrm{ and }\;\big(T_\lambda^{(0)}+K_\sigma^{(0)}\big)\xi\in H^{\frac{1}{2}}_{\ell=0}(\mathbb{R}^3)
   \end{array}
   \right\},
 \end{equation}
 to be compared with the previous domain \eqref{eq:Dbetaaltogether}.

 For the sake of a more compact, unified notation, one writes
 \begin{equation}\label{eq:compactTTT}
  (\mathbf{T}_{\lambda,\sigma}\xi)(\yy)\;:=\;\big(\big(T_\lambda^{(0)}+K_\sigma^{(0)}\big)\xi^{(0)})(\yy)+\sum_{\ell=1}^\infty \big(T_\lambda^{(\ell)}\xi^{(\ell)}\big)(\yy)
 \end{equation}
 for $\xi=\sum_{\ell=0}^\infty\xi^{(\ell)}\in\bigoplus_{\ell=0}^\infty \,H^{-\frac{1}{2}}_{W_\lambda,\ell}(\mathbb{R}^3)\cong H^{-\frac{1}{2}}_{W_\lambda}(\mathbb{R}^3)$. Thus,
 \begin{equation}\label{eq:Tcompact1}
  \begin{split}
   \mathcal{D}(\mathcal{R}_{\lambda,\sigma})\;&=\;\big\{\xi\in H^{\frac{1}{2}}(\mathbb{R}^3)\,\big|\,\mathbf{T}_{\lambda,\sigma}\xi\in H^{\frac{1}{2}}(\mathbb{R}^3)\big\}\, , \\
   \mathcal{R}_{\lambda,\sigma}\;&=\;3\,W_\lambda^{-1} \mathbf{T}_{\lambda,\sigma} \, ,
  \end{split}
 \end{equation}
 and 
 \begin{equation}\label{eq:Tcompact2}
  \big(\mathbf{T}_{\lambda,\sigma}-T_\lambda\big)\xi\;=\; K_\sigma^{(0)}\xi^{(0)}\,.
 \end{equation}

%
%

 \begin{theorem}\label{thm:regularised-models}
  Let $\sigma>0$ and $\lambda>0$. Define
 \begin{equation}\label{eq:Hsigmadomaction}
     \begin{split}
      \mathcal{D}(\mathscr{H}_\sigma)\;&:=\;\left\{g=\phi^\lambda+u_\xi^\lambda\left|\!
  \begin{array}{c}
   \phi^\lambda\in H^2_\mathrm{b}(\mathbb{R}^3\times\mathbb{R}^3)\,, \\
   \xi\in H^{\frac{1}{2}}(\mathbb{R}^3)\,\textrm{ with }\,\mathbf{T}_{\lambda,\sigma}\xi\in H^{\frac{1}{2}}(\mathbb{R}^3)\,,\\
   \displaystyle\phi^\lambda(\yy,\mathbf{0})\,=\,(2\pi)^{-\frac{3}{2}} (\mathbf{T}_{\lambda,\sigma}\xi)(\yy)
  \end{array}
  \!\!\!\right.\right\} \, ,\\
  (\mathscr{H}_\sigma+\lambda\mathbbm{1})g\;&:=\;(-\Delta_{\yy_1}-\Delta_{\yy_2}-\nabla_{\yy_1}\cdot\nabla_{\yy_2}+\lambda\mathbbm{1})\phi^\lambda\,.
     \end{split}
    \end{equation}
    \begin{enumerate}[(i)]
   \item The decomposition of $g$ in terms of $\phi^\lambda$ and $\xi$ is unique, at fixed $\lambda$. The subspace $\mathcal{D}(\mathscr{H}_\sigma)$ is $\lambda$-independent.
   \item $\mathscr{H}_\sigma$ is self-adjoint on $L^2_\mathrm{b}(\mathbb{R}^3\times\mathbb{R}^3,\ud\yy_1,\ud\yy_2)$ and extends $\mathring{H}$ given in \eqref{eq:domHring-initial}.
   \item For each $g\in  \mathcal{D}(\mathscr{H}_\sigma)$ one has
   \begin{equation}\label{eq:allBPTMS-regularised}
    \begin{split}
     \phi^\lambda(\yy_1,\mathbf{0})\;&=\;(2\pi)^{-\frac{3}{2}} (\mathbf{T}_{\lambda,\sigma}\xi)(\yy_1)\,, \\
     \int_{\mathbb{R}^3}\widehat{\phi^\lambda}(\pp_1,\pp_2)\,\ud\pp_2\;&=\;(\widehat{\mathbf{T}_{\lambda,\sigma}\xi})(\pp_1)\,, \\
     \int_{\!\substack{ \\ \\ \pp_2\in\mathbb{R}^3 \\ |\pp_2|<R}}\widehat{g}(\pp_1,\pp_2)\,\ud\pp_2\;&=\;4\pi R\,\widehat{\xi}(\pp_1)+(\widehat{K_\sigma^{(0)}\xi^{(0)}})(\pp_1)+o(1)
    \end{split}
   \end{equation}
   (where $R\to +\infty$). All such conditions are equivalent. In particular, the first version of \eqref{eq:allBPTMS-regularised} is an identity in $H^{\frac{1}{2}}(\mathbb{R}^3)$. 
  \item $\mathscr{H}_\sigma$ is non-negative and with quadratic form
  \begin{equation}\label{eq:HsigmaQuadrForm}
   \begin{split}
    \mathcal{D}[\mathscr{H}_\sigma]\;&=\;\left\{g=\phi^\lambda+u_\xi^\lambda\left|\!
  \begin{array}{c}
   \phi^\lambda\in H^1_\mathrm{b}(\mathbb{R}^3\times\mathbb{R}^3)\,,\;\xi\in H^{\frac{1}{2}}(\mathbb{R}^3)
  \end{array}
  \!\!\!\right.\right\} \\
     \mathscr{H}_\sigma[g]\;&=\;\frac{1}{2}\Big(\big\|(\nabla_{\yy_1}+\nabla_{\yy_2})\phi^\lambda\big\|^2_{L^2}+\big\|\nabla_{\yy_1}\phi^\lambda\big\|^2_{L^2}+\big\|\nabla_{\yy_2}\phi^\lambda\big\|^2_{L^2}\Big) \\
     &\qquad\qquad+\lambda\Big(\|\phi^\lambda\big\|^2_{L^2}-\big\|\phi^\lambda+u_\xi^\lambda\big\|^2_{L^2}\Big) \\
     &\qquad\qquad +3\int_{\mathbb{R}^3} \overline{\,\widehat{\xi}(\pp)}\, \big(\widehat{\mathbf{T}_{\lambda,\sigma}\xi}\big)(\pp)\,\ud\pp\,,
   \end{split}
  \end{equation}
  the $L^2$-norms being norms in $L^2(\mathbb{R}^3\times\mathbb{R}^3)$.
  \end{enumerate}
 \end{theorem}

 \begin{proof}
  All claims follow from plugging \eqref{eq:Tcompact1}-\eqref{eq:Tcompact2} (hence, in particular, \eqref{eq:RlambdasigmaOPERATOR}-\eqref{eq:RlambdasigmaFORM}) into the general classification formulas of Theorem \ref{thm:generalclassification}, owing to the self-adjointness of the Birman parameter guaranteed by Proposition \ref{prop:NEW-Birman-param-selfadj}.  
 \end{proof}

 Each of the asymptotics \eqref{eq:allBPTMS-regularised} for $g\in \mathcal{D}(\mathscr{H}_\sigma)$ expresses an ultra-violet regularised Bethe-Peierls alias Ter-Martirosyan-Skornyakov condition, \index{contact condition!regularised} that in view of Corollary \ref{cor:largepasympt-star} can be thought of as
 \begin{equation}\label{eq:allBPTMS-regularised-2}
 (2\pi)^{\frac{3}{2}} g_{\mathrm{av}}(\yy_1;|\yy_2|)\,\stackrel{|\yy_2|\to 0}{=}\,\frac{4\pi}{|\yy_2|}\xi(\yy_1)+\frac{\,\sigma_0+\sigma\,}{|\yy_1|}\,\xi(\yy_1)+ o(1)\,.
 \end{equation}
 As the simultaneous limit $|\yy_1|\to 0$, $|\yy_2|\to 0$ in the expression above suggests, the regularisation effectively amounts to distorting the ordinary Bethe-Peierls short-scale asymptotics by means of a position-dependent scattering length
 \begin{equation}
  a_{\mathrm{eff}}(\yy)\;:=\;-\frac{4\pi|\yy|}{\,\sigma_0+\sigma\,}
 \end{equation}
 (see \eqref{eq:a-alpha} above) that vanishes when all three bosons come to occupy the same point. At this effective level, a (small) three-body correction prevents the triple collision.

 In fact, this Minlos-Faddeev regularisation is rather radical, because it completely eliminates the negative spectrum (see also Remark \ref{rem:noEV} below). Yet, $\mathscr{H}_\sigma$ is not merely the reduced component of the canonical Hamiltonian $\mathscr{H}_{0,\beta}$ onto the positive spectral subspace: the signature of the physical short-scale behaviour is retained in \eqref{eq:allBPTMS-regularised} and \eqref{eq:allBPTMS-regularised-2}, with the $|\yy_2|^{-1}$ leading singularity as $|\yy_2|\to 0$ at fixed $\yy_1$, only with a distorted sub-leading singularity driven by an effective scattering length that vanishes as $|\yy_1|\to 0$.

  \begin{remark}\label{rem:noEV}
  In contrast with the computation of the negative eigenvalues of the Hamiltonian $\mathscr{H}_{0,\beta}$ (proof of Theorem \ref{thm:spectralanalysis}), the analogous computation for $\mathscr{H}_\sigma$ would lead to the equation $\big(T_\lambda^{(0)}+K_\sigma^{(0)}\big)\xi=0$ for some $\xi$ in the sector $\ell=0$. Owing to \eqref{eq:Tlxi-theta-withK} and \eqref{eq:ThetaTransformed}, this is the same as $\widehat{\gamma}_\sigma\widehat{\theta}=0$, where $\theta$ is the re-scaled radial function associated with $\xi$. The difference is thus
  \[
   \begin{array}{lcl}
    \;\,\widehat{\gamma}\,\widehat{\theta}\;=\;0 & & \textrm{for the \emph{canonical} eigenvalue problem} \, , \\
    \widehat{\gamma}_\sigma\,\widehat{\theta}\;=\;0 & & \textrm{for the \emph{regularised} eigenvalue problem}\,.
   \end{array}
  \]
  Because of the presence of roots of $\widehat{\gamma}(s)=0$, the first equation turns out to have non-trivial admissible solutions. Instead, $\widehat{\gamma}_\sigma(s)\geqslant\widehat{\gamma}_\sigma(0)>0$ and for the second equation one necessarily has $\theta\equiv 0$ (absence of negative eigenvalues).  
 \end{remark}

 \begin{remark}
 The charge term in the quadratic form expression \eqref{eq:HsigmaQuadrForm} is explicitly given by
 \begin{equation}
  \begin{split}
   \int_{\mathbb{R}^3} \overline{\,\widehat{\xi}(\pp)}\, \big(\widehat{\mathbf{T}_{\lambda,\sigma}\xi}\big)(\pp)\,\ud\pp\;&=\;\sum_{\ell=1}^\infty\int_{\mathbb{R}^3} \overline{\,\widehat{\xi^{(\ell)}}(\pp)}\, \big(\widehat{T_{\lambda}^{(\ell)}\xi^{(\ell)}}\big)(\pp)\,\ud\pp \\
   &\qquad +\int_{\mathbb{R}^3} \overline{\,\widehat{\xi^{(0)}}(\pp)}\, \big(\big(T_\lambda^{(0)}+K_\sigma^{(0)}\big)\xi^{(0)}\big)\,{\textrm{\large $\widehat{\,}$\normalsize}}\,(\pp)\,\ud\pp
  \end{split}
 \end{equation}
 (as follows from \eqref{eq:RlambdasigmaFORM} and \eqref{eq:compactTTT}). As $\mathscr{H}_\sigma$ is self-adjoint and non-negative, its quadratic form \eqref{eq:HsigmaQuadrForm} is obviously closed and non-negative.
  An announcement that the quadratic form \eqref{eq:HsigmaQuadrForm} is closed and lower semi-bounded on $L^2_\mathrm{b}(\mathbb{R}^3\times\mathbb{R}^3)$, and thus induces a self-adjoint Hamiltonian for the regularised three-body interaction in the bosonic trimer, has been recently made in \cite{Figari-Teta-2020}. 
 \end{remark}

 \subsection{High energy cut-off}\label{sec:highenergycutoff}

 As mentioned already, an alternative, conceptually equivalent way of making the scattering length effectively vanish in the vicinity of the triple coincidence point is to realise this effect at large relative momenta.

 An example of this type of high energy cut-off has been recently proposed in \cite[Section 3]{Basti-Figari-Teta-Rendiconti2018} by means of quadratic form methods. In the following this possibility will be studied in the general operator-theoretic framework of the present analysis, working out more precisely a modification of \cite{Basti-Figari-Teta-Rendiconti2018} that allows for explicit computations.
 
  For a clearer readability and comparison with Subsect.~\ref{sec:MFregularisation}-\ref{sec:MinFadzero}, the same notation used therein will be kept here for the counterpart regularised quantities, of course re-defined them now in a different way.

 The original proposal of \cite{Basti-Figari-Teta-Rendiconti2018} goes along the following line: the ordinary Birman parameter $3W_\lambda^{-1}(T_\lambda+\alpha\mathbbm{1})$, that selects (via Theorems \ref{thm:generalclassification} and \ref{thm:globalTMSext}) self-adjoint extensions of Ter-Martirosyan-Skornyakov type of the minimal operator $\mathring{H}$ defined in \eqref{eq:domHring-initial}, is replaced by 
\begin{equation*}
\begin{split}
 & 3\,W_\lambda^{-1}(T_\lambda+\alpha\mathbbm{1}+K_{\sigma,\rho})\,,\qquad \sigma,\rho\,>\,0\,,  \\
 & (\widehat{K_{\sigma,\rho}\xi})(\pp)\;:=\;\sigma\,\mathbf{1}_{\{|\pp|\geqslant\rho\}}\,\pp^2\,\widehat{\xi}(\pp)\,.
\end{split}
\end{equation*}
 This gives rise, via \eqref{eq:g-largep2-star}, to the modified large momentum asymptotics
 \begin{equation*}
 \begin{split}
  \int_{\!\substack{ \\ \\ \pp_2\in\mathbb{R}^3 \\ |\pp_2|<R}}\widehat{g}(\pp_1,\pp_2)&\,\ud\pp_2\stackrel{R\to+\infty}{=}
  4\pi R\,\widehat{\xi}(\pp_1)+\alpha_{\mathrm{eff}}(\pp)\widehat{\xi}(\pp_1)+o(1) \\
  & \alpha_{\mathrm{eff}}(\pp)\;:=\;\alpha+\sigma\,\mathbf{1}_{\{|\pp|\geqslant\rho\}}\,\pp^2\,,
 \end{split}
 \end{equation*}
 for the elements $g$ in the domain of the corresponding self-adjoint extension, 
 again with the interpretation of an effective parameter $\alpha_{\mathrm{eff}}(\pp)\to +\infty$ as $|\pp|\to +\infty$.

 Reasoning in terms of quadratic forms, it is simple to check (as done in \cite[Section 3]{Basti-Figari-Teta-Rendiconti2018}) that $\sigma$ and $\rho$ can be adjusted on $\alpha$ and $\lambda$ so that the map
 \[
  \xi\;\longmapsto\;\bigg(\int_{\mathbb{R}^3}\overline{\widehat{\xi}(\pp)}\,\big((T_\lambda+\alpha\mathbbm{1}+K_{\sigma,\rho})\xi\big)\,{\textrm{\large $\widehat{\,}$\normalsize}}\,(\pp)\bigg)^{\!\frac{1}{2}}
 \]
 is an equivalent $H^1$-norm: in fact, the effect of the additional $K_{\sigma,\rho}$ is to rise the multiplicative part of $T_\lambda$ with an $H^1$-term (added to the original $H^{\frac{1}{2}}$-term), which controls the integral part of $T_\lambda$. This way, a quadratic form on $L^2_{\mathrm{b}}(\mathbb{R}^3_{\yy_1}\times\mathbb{R}^3_{\yy_2})$ of the same type of \eqref{eq:decomposition_of_form_domains_Tversion} can be constructed, namely with regular functions from $H^1_{\mathrm{b}}(\mathbb{R}^3_{\yy_1}\times\mathbb{R}^3_{\yy_2})$ and charges from $H^1(\mathbb{R}^3)$, in which the charge term (the quadratic form of the Birman parameter) is precisely
 \[
  \int_{\mathbb{R}^3}\overline{\widehat{\xi}(\pp)}\,\big((T_\lambda+\alpha\mathbbm{1}+K_{\sigma,\rho})\xi\big)\,{\textrm{\large $\widehat{\,}$\normalsize}}\,(\pp)\,.
 \]
 Standard arguments then show that the form is closed and non-negative, hence it is the energy form of a self-adjoint Hamiltonian for the bosonic trimer with zero-range interactions.

 The Friedrichs construction for the Birman parameter $3\,W_\lambda^{-1}(T_\lambda+\alpha\mathbbm{1}+K_{\sigma,\rho})$ associated to the above form would lead to a somewhat implicit expression for its operator domain.
 Furthermore, for the purpose of having a closed and semi-bounded charge form it suffices to add to $T_\lambda$ an additional  $H^{\frac{1}{2}}$-term, instead of the $H^1$-term proposed in \cite{Basti-Figari-Teta-Rendiconti2018}, which be large enough so as to shift the form up above zero. That is the version of the large momentum (high energy) cut-off presented here, as the computations within the operator-theoretic scheme are explicit.

 For $\lambda,\sigma>0$ set 
 \begin{equation}\label{eq:heregularisationchoice}
 \begin{split}
  (\widehat{K_{\lambda,\sigma}\,\xi})(\pp)\;&:=\;2\pi^2(\sigma_0+\sigma)\sqrt{\frac{3}{4}\pp^2+\lambda}\,\widehat{\xi}(\pp)  \, ,\\
  \sigma_0\;&:=\;-\widehat{\gamma}(0)\;=\;\frac{4\pi}{3\sqrt{3}}-1\,,
  \end{split}
 \end{equation}
 as well as (with $x,s\in\mathbb{R}$)
 \begin{equation}\label{eq:newregularisedquantities}
  \begin{split}
   \Theta^{(\theta)}_{\lambda,\sigma}(x)\;&:=\;(1+\sigma_0+\sigma)\theta(x)-\frac{4}{\pi\sqrt{3}}\int_{\mathbb{R}}\ud y\,\theta(y) \log \frac{\,2\cosh(x-y)+1\,}{\,2\cosh(x-y)-1\,} \, , \\
   \widehat{\gamma}_\sigma(s)\;&:=\;\widehat{\gamma}(s)+\sigma_0+\sigma\,,
  \end{split}
 \end{equation}
 for given $\theta$,
 where $ \widehat{\gamma}$ is defined in \eqref{eq:gamma-distribution}. By construction $\widehat{\gamma}_\sigma$ is the shifted version of the $\widehat{\gamma}$-curve from Fig.~\ref{fig:gammagammaplus}, thus a smooth even function on $\mathbb{R}$ that is uniformly bounded and strictly positive, with absolute minimum $\widehat{\gamma}_\sigma(0)=\sigma$.

 Analogously to \eqref{eq:Ksigmarotations}, $K_{\lambda,\sigma}=\bigoplus_{\ell\in\mathbb{N}}K_{\lambda,\sigma}^{(\ell)}$ with respect to the usual decomposition in sectors of definite angular momentum, and arguing as in Subsect.~\ref{sec:variants} and \ref{sec:MFregularisation} it is sufficient here to implement the regularisation in the meaningful sector $\ell=0$.

 A straightforward modification of the reasoning for Lemmas \ref{lem:xithetaidentities} and \ref{lem:regularisedlemma1} and for Corollary \ref{cor:equivalentH12norm} yields the following.

  \begin{lemma}\label{lem:regularisedlemma2}
   Let $\lambda,\sigma>0$, $s\in\mathbb{R}$, and $\xi$ be as in \eqref{eq:0xi}. One has the identities
         \begin{eqnarray}
         \big(\big(T_\lambda^{(0)}+K_{\lambda,\sigma}^{(0)}\big)\xi\big)\,{\textrm{\large $\widehat{\,}$\normalsize}}\,(\pp)\;&=&\;\frac{1}{\sqrt{4\pi}\,|\pp|}\,\frac{4\pi^2}{\sqrt{3}\,}\,\Theta^{(\theta)}_{\lambda,\sigma}(x)\,, \label{eq:Tlxi-theta-withK2}\\
      \big\|\big(T_\lambda^{(0)}+K_{\lambda,\sigma}^{(0)}\big)\xi\big\|_{H^{s}(\mathbb{R}^3)}^2\!\!&\approx&\!\!\int_{\mathbb{R}}\ud x\,(\cosh x)^{1+2s}\,\big|\Theta^{(\theta)}_{\lambda,\sigma}(x)\big|^2\,, \label{eq:Tlambda0xi-with-theta-andK2} \\
      \widehat{\Theta^{(\theta)}_{\lambda,\sigma}}(s)\;&=&\;\widehat{\gamma}_\sigma(s)\,\widehat{\theta}(s)\,, \label{eq:ThetaTransformed2} \\
       \int_{\mathbb{R}^3} \overline{\,\widehat{\xi}(\pp)}\, \big(\big(T_\lambda^{(0)}+K_{\lambda,\sigma}^{(0)}\big)\xi\big)\,{\textrm{\large $\widehat{\,}$\normalsize}}\,(\pp)\,\ud\pp\;&=&\;\frac{\,8\pi^2}{3\sqrt{3}}\int_{\mathbb{R}}\ud s\, \widehat{\gamma}_\sigma(s)\,|\widehat{\theta}(s)|^2\,, \label{eq:xiTxi-with-theta-andK2}
    \end{eqnarray}
    with $x$ and $\theta$ given by \eqref{ftheta-1}, and $\Theta^{(\theta)}_{\lambda,\sigma}$ and $\widehat{\gamma}_\sigma$ given by \eqref{eq:newregularisedquantities}, where the definition of $\Theta^{(\theta)}_{\lambda,\sigma}$ is now taken with respect to the present $\theta$.
    In \eqref{eq:Tlxi-theta-withK2} it is understood that $x\geqslant 0$, and \eqref{eq:Tlambda0xi-with-theta-andK2} is meant as an equivalence of norms (with $\lambda$-dependent multiplicative constant).
    Moreover, the map
   \begin{equation}
    \xi\;\longmapsto\;\bigg(\int_{\mathbb{R}^3} \overline{\,\widehat{\xi}(\pp)}\, \big(\big(T_\lambda^{(0)}+K_{\lambda,\sigma}^{(0)}\big)\xi\big)\,{\textrm{\large $\widehat{\,}$\normalsize}}\,(\pp)\,\ud\pp\bigg)^{\!\frac{1}{2}}
   \end{equation}
   defines an equivalent norm in $H^{\frac{1}{2}}_{\ell=0}(\mathbb{R}^3)$.
  \end{lemma}

  One can now define, for $\sigma>0$,
  \begin{equation}\label{eq:newD0regularised2}
   \begin{split}
    \widetilde{\mathsf{D}}_{0,\sigma}\;&:=\;\left\{
   \xi\in H^{-\frac{1}{2}}_{\ell=0}(\mathbb{R}^3)\left|
   \begin{array}{c}
    \textrm{$\xi$ has re-scaled radial component} \\
     \theta=\big(\widehat{\Theta}/\widehat{\gamma}_\sigma)\big)^{\!\vee} \\
     \textrm{for }\;\Theta\in C^\infty_{c,\mathrm{odd}}(\mathbb{R}_x)
   \end{array}
   \!\right.\right\}\,, \\
   \mathsf{D}_{0,\sigma}\;&:=\;\big\{\xi\in H^{\frac{1}{2}}_{\ell=0}(\mathbb{R}^3)\,\big|\,\big(T_\lambda^{(0)}+K_{\lambda,\sigma}^{(0)}\big)\xi\in H^{\frac{1}{2}}_{\ell=0}(\mathbb{R}^3)\big\}\,,
   \end{split}
  \end{equation}
  with the same remarks made for \eqref{eq:newD0regularised}, and prove the following, based on Lemma \ref{lem:regularisedlemma2}, and easily mimicking the reasoning that let to Lemma \ref{lem:DD0properties} and Proposition \ref{prop:NEW-Birman-param-selfadj}.

  \begin{proposition}\label{prop:againandagain} Let $\sigma>0$.  
  \begin{enumerate}[(i)]
    \item $\widetilde{\mathsf{D}}_{0,\sigma}$ is dense in $H^{\frac{1}{2}}_{\ell=0}(\mathbb{R}^3)$.
   \item $(T_\lambda^{(0)}+K_{\lambda,\sigma}^{(0)})\widetilde{\mathsf{D}}_{0,\sigma}\subset H^{s}_{\ell=0}(\mathbb{R}^3)$ for every $s\in\mathbb{R}$ and $\lambda>0$.
    \item $\widetilde{\mathsf{D}}_{0,\sigma}\subset \mathsf{D}_{0,\sigma}$.
    \item For every $\lambda>0$ the operator
    \begin{equation}
   \begin{split}
    \mathcal{D}\big(\widetilde{\mathcal{R}}_{\lambda,\sigma}^{(0)}\big)\;&:=\;\widetilde{\mathsf{D}}_{0,\sigma} \, , \\
    \widetilde{\mathcal{R}}^{(0)}_{\lambda,\sigma}\;&:=\; 3W_\lambda^{-1}\big(T_\lambda^{(0)}+K_{\lambda,\sigma}^{(0)}\big)\,.
   \end{split}
  \end{equation}
  is densely defined, symmetric, and coercive on $H^{-\frac{1}{2}}_{W_\lambda,\ell=0}(\mathbb{R}^3)$.
  \item For every $\lambda>0$ the operator
    \begin{equation}\label{eq:RlambdasigmaOPERATOR2}
   \begin{split}
    \mathcal{D}\big(\mathcal{R}_{\lambda,\sigma}^{(0)}\big)\;&:=\;\mathsf{D}_{0,\sigma} \, ,\\
    \mathcal{R}_{\lambda,\sigma}^{(0)})\;&:=\;3W_\lambda^{-1}\big(T_\lambda^{(0)}+K_{\lambda,\sigma}^{(0)}\big)
   \end{split}
  \end{equation}
 is the Friedrichs extension of $\widetilde{\mathcal{R}}^{(0)}_{\lambda,\sigma}$ with respect to $H^{-\frac{1}{2}}_{W_\lambda,\ell=0}(\mathbb{R}^3)$. Its sesquilinear form is
 \begin{equation}\label{eq:RlambdasigmaFORM2}
   \begin{split}
    \mathcal{D}\big[\mathcal{R}_{\lambda,\sigma}^{(0)}\big]\;&=\;H^{\frac{1}{2}}_{\ell=0}(\mathbb{R}^3) \, , \\
    \mathcal{R}_{\lambda,\sigma}^{(0)}[\eta,\xi]\;&=\;3\int_{\mathbb{R}^3} \overline{\,\widehat{\xi}(\pp)}\, \big(\big(T_\lambda^{(0)}+K_{\lambda,\sigma}^{(0)}\big)\xi\big)\,{\textrm{\large $\widehat{\,}$\normalsize}}\,(\pp)\,.
   \end{split}
\end{equation}
    \end{enumerate}
  \end{proposition}

  Proposition \ref{prop:againandagain} finally shows that the operator
  \begin{equation}\label{eq:globalRlambda-2}
   \mathcal{R}_{\lambda,\sigma}\;:=\;\mathcal{R}_{\lambda,\sigma}^{(0)}\:\oplus\: \bigoplus_{\ell=1}^\infty\mathcal{A}_\lambda^{(\ell)}
 \end{equation}
 is an admissible Birman parameter \index{Birman extension parameter} labelling a self-adjoint extension $\mathring{H}_{\mathcal{R}_{\lambda,\sigma}}$ of $\mathring{H}$ according to the general classification and construction of Theorem \ref{thm:generalclassification}. With a more compact notation one writes
  \begin{equation}\label{eq:Tcompact1-2version}
  \begin{split}
   \mathcal{D}(\mathcal{R}_{\lambda,\sigma})\;&=\;\big\{\xi\in H^{\frac{1}{2}}(\mathbb{R}^3)\,\big|\,\mathbf{T}_{\lambda,\sigma}\xi\in H^{\frac{1}{2}}(\mathbb{R}^3)\big\} \,, \\
   \mathcal{R}_{\lambda,\sigma}\;&=\;3\,W_\lambda^{-1} \mathbf{T}_{\lambda,\sigma}
  \end{split}
 \end{equation}
 and 
 \begin{equation}\label{eq:Tcompact2-2version}
  \big(\mathbf{T}_{\lambda,\sigma}-T_\lambda\big)\xi\;=\; K_{\lambda,\sigma}^{(0)}\xi^{(0)}\,,
 \end{equation}
 where
%
%
  \begin{equation}\label{eq:compactTTT-2version}
  (\mathbf{T}_{\lambda,\sigma}\xi)(\yy)\;:=\;\big(\big(T_\lambda^{(0)}+K_{\lambda,\sigma}^{(0)}\big)\xi^{(0)})(\yy)+\sum_{\ell=1}^\infty \big(T_\lambda^{(\ell)}\xi^{(\ell)}\big)(\yy)
 \end{equation}
 for $\xi=\sum_{\ell=0}^\infty\xi^{(\ell)}\in\bigoplus_{\ell=0}^\infty \,H^{-\frac{1}{2}}_{W_\lambda,\ell}(\mathbb{R}^3)\cong H^{-\frac{1}{2}}_{W_\lambda}(\mathbb{R}^3)$.

 All this leads to a new version of Theorem \ref{thm:regularised-models}, \emph{with exactly the same statement}, of course now referred to the present $\mathbf{T}_{\lambda,\sigma}$ defined in \eqref{eq:compactTTT-2version} (and not to its counterpart \eqref{eq:compactTTT} considered in the previous Subsection).

 Thus, \emph{two} distinct classes have been identified of regularised self-adjoint Hamiltonians of zero-range interaction for the bosonic trimer:
 \begin{itemize}
  \item the operator $\mathscr{H}_\sigma^{\mathrm{MF}}$, $\sigma>0$, namely the Hamiltonian with the \emph{Minlos-Faddeev regularisation}, obtained as $\mathscr{H}_\sigma$ from Theorem \ref{thm:regularised-models} with the modified $\mathbf{T}_{\lambda,\sigma}$ fixed in \eqref{eq:compactTTT};
  \item the operator $\mathscr{H}_\sigma^{\mathrm{he}}$, $\sigma>0$, namely the Hamiltonian with \emph{high energy regularisation}, obtained as $\mathscr{H}_\sigma$ from Theorem \ref{thm:regularised-models} with the modified $\mathbf{T}_{\lambda,\sigma}$ fixed in \eqref{eq:compactTTT-2version}.  
 \end{itemize}

 Both types of Hamiltonians are non-negative, with only essential spectrum given by $[0,+\infty)$, and both retain a physically meaningful short-scale structure. 
 With the Minlos-Faddeev regularisation, each $g\in\mathcal{D}(\mathscr{H}_\sigma^{\mathrm{MF}})$ and the corresponding regular part $\phi^\lambda$ of $g$ (for fixed $\lambda>0$) satisfy
 \begin{equation}\label{eq:allBPTMS-regularised-again1}
    \begin{split}
      \int_{\!\substack{ \\ \\ \pp_2\in\mathbb{R}^3 \\ |\pp_2|<R}}\widehat{g}(\pp_1,\pp_2)\,\ud\pp_2\;&=\;4\pi R\,\widehat{\xi}(\pp_1)+\frac{\sigma_0+\sigma}{2\pi^2}\,\int_{\mathbb{R}^3}\frac{\widehat{\xi^{(0)}}(\qq)}{\,|\pp_1-\qq|^2}\,\ud\qq+o(1)\,, \\
      \phi^\lambda(\yy_1,\mathbf{0})\;&=\;(2\pi)^{-\frac{3}{2}}\Big( (T_\lambda\xi)(\yy_1)+\frac{\,\sigma_0+\sigma\,}{|\yy_1|}\,\xi^{(0)}(\yy_1)\Big)\,,
    \end{split}
   \end{equation}
 where $\xi$ is the singular charge of $g$, $\xi^{(0)}$ is the spherically symmetric component of $\xi$, and $R\to +\infty$. With the high energy regularisation, each $g\in\mathcal{D}(\mathscr{H}_\sigma^{\mathrm{he}})$ and the corresponding regular part $\phi^\lambda$ of $g$ satisfy
 \begin{equation}\label{eq:allBPTMS-regularised-again2}
    \begin{split}
      \int_{\!\substack{ \\ \\ \pp_2\in\mathbb{R}^3 \\ |\pp_2|<R}}\widehat{g}(\pp_1,\pp_2)\,\ud\pp_2\;&=\;4\pi R\,\widehat{\xi}(\pp_1)+2\pi^2(\sigma_0+\sigma)\sqrt{\frac{3}{4}\pp_1^2+\lambda}\,\widehat{\xi^{(0)}}(\pp_1)+o(1)\,,\!\!\!\!\!\!\!\!\!\!\! \\
      \int_{\mathbb{R}^3}\widehat{\phi^\lambda}(\pp_1,\pp_2)\,\ud\pp_2\;&=\;(\widehat{T_{\lambda,\sigma}\xi})(\pp_1)+2\pi^2(\sigma_0+\sigma)\sqrt{\frac{3}{4}\pp_1^2+\lambda}\,\widehat{\xi^{(0)}}(\pp_1)\,.
    \end{split}
   \end{equation}



%
%
%

\begin{partbacktext}
\part{Appendices}

\end{partbacktext}
\appendix
%
%
%


\chapter{Physical requirements prescribing self-adjointness of quantum observables}
\label{appendix_SA_in_physics} 


 This Appendix is modelled on the recent discussion \cite{CM-selfadj-2021} concerning the emergence of self-adjointness for quantum observables as a consequence of precise requirements of physical nature made in the set-up of quantum theory. Here the aim isn't to review those mathematical properties, with crucial physical interpretation, implied by self-adjointness: this is ordinarily done in the specialised literature \cite{Amrein-HilberSpMethods-2009,Birman-Solomjak-1980book,DeOliveira-SpectralTheory-Book2009,Davies_LinearOpSpectra-book2007,Jiri-Exner-Havlicek-2008,Kato-perturbation,rs1,rs2,rs3,rs4,schmu_unbdd_sa,Weidmann-LinearOperatosHilbert} already mentioned in Chapter \ref{chaper-preliminaries}. The perspective is exactly the converse: to demonstrate the emergence of self-adjointness (which thus become \emph{necessary} in quantum mechanics, and not just \emph{sufficient}) as an implication of certain mathematical conditions that encode crucial physical aspects of the theory. All this provides an implicit underlying motivation for the applications contained in the second part of the present monograph.

\section{Levels of mathematical formalisation of quantum mechanics}

 Quantum mechanics is formalised at various levels of increasing mathematical abstraction, by inference from experimental observations and by analogy with other physical theories, and depending also on the class of systems and phenomena that one or another formalism is meant to cover.

 The mathematical foundations of quantum mechanics are discussed in depth throughout an extensive and established literature spanning all such levels of formalisation, including in Dirac \cite{Dirac-PrinciplesQM}, von Neumann \cite{vonNeumann-MathFoundQM}, Ludwig \cite{Ludwig-1954-1983}, Mackey \cite{Mackey-QM-1963-2004}, D'Espagnat \cite{DEspagnat-1971}, Bratteli and Robinson \cite{Bratteli-Robinson-1,Bratteli-Robinson-2}, Strocchi \cite{Strocchi-MathQM}, Dell'Antonio \cite{DellAnt_QM1-2015,DellAnt_QM1-2016}, Barrett \cite{Barrett-2019}, Moretti \cite{Moretti_QM-2019}, D\"{u}rr and Lazarovici \cite{Duerr-Lazarovici-2020}, among others.

 For the purposes of this monograph and of the present Appendix, a simplification can be made by merely distinguishing between the mathematics of quantum mechanics of systems with \emph{finite} and with \emph{infinite} degrees of freedom, for concreteness quantum systems with finite or infinite numbers of particles.

 With both cases an amount of standard familiarity is assumed here for the reader, and one refers to the above-mentioned references.

 For short, the former is the case where the \emph{states}\index{state (quantum)} of the systems are described by normalised vectors $\psi$ of a (finite- or infinite-dimensional) Hilbert space $\cH$ (\emph{pure} states), or more generally normalised density matrix\index{density matrix} operators $\varrho$ acting on $\cH$ (\emph{mixed} states), \emph{observables}\index{observables (quantum)} are self-adjoint operators on $\cH$, and the \emph{dynamics} is implemented as a strongly continuous one-parameter unitary group\index{strongly continuous unitary group} $(e^{-\ii t H})_{t\in\mathbb{R}}$ acting on $\cH$, whose self-adjoint generator is the Hamiltonian $H$ of the system, so that a pure state $\psi$ (respectively, a mixed state $\varrho$) evolves at time $t$ into the state $e^{-\ii t H}\psi$ (respectively, into $e^{-\ii t H}\varrho e^{\ii t H}$).

 In contrast, the latter is the case where observables\index{observables (quantum)} are declared first, as self-adjoint elements of a (commutative or non-commutative) $C^*$-algebra $\mathcal{A}$, then states are normalised, positive, linear functionals $\omega$ on $\mathcal{A}$, and the time evolution $t\mapsto\alpha_t(A)$ of an observable $A$ is given by a continuous one-parameter group $(\alpha_t)_{t\in\mathbb{R}}$ of $*$-automorphisms of $\mathcal{A}$. The Hilbert space picture is recovered from the $C^*$-algebraic framework through the following Gelfand-Na\u{\i}mark-Segal\index{Gelfand-Na\u{\i}mark-Segal representation}\index{theorem!Gelfand-Na\u{\i}mark-Segal} representation: when the system is in a state $\omega$, a Hilbert space $\cH_\omega$, a normalised vector $\Omega_\omega\in\cH_\omega$, and a $*$-morphism $\pi_\omega:\mathcal{A}\to\mathcal{B}(\cH_\omega)$ can be uniquely identified (up to unitary equivalence) such that the vectors $\pi_\omega(A)\Omega_\omega$ span a dense of $\cH_\omega$ when $A$ runs over $\mathcal{A}$, and $\omega(A)=\langle\Omega_\omega,\pi_\omega(A)\Omega_\omega\rangle_{\cH_\omega}$ $\forall A\in\mathcal{A}$.

 The two descriptions are in many respects not interchangeable. For one thing, the $C^*$-algebraic description retains a higher conceptual position, because of the more natural centrality attributed to observables, namely the actual physical quantities that can be measured, because of the natural quasi-local structure of observables on systems of infinite degrees of freedom, and because of the emergence of an underlying Hilbert space in a suitable representation.

 Furthermore, the Hilbert space description is unsuited to formalise a system of infinite many particles or spins, as it would require an infinite tensor product of Hilbert spaces. As long as the number $N$ of such particles or spins is finite, each one described for concreteness by the same one-body Hilbert space $\cH$, the Hilbert space of the $N$-body system is the $N$-fold tensor product $\cH^{\otimes N}$ (or possibly some subspace of it selected by the exchange symmetry of the system), and the $C^*$-algebra naturally associated to it is $\mathcal{A}=\mathcal{B}(\cH^{\otimes N})$. When $N=\infty$, a natural notion of infinite tensor product Hilbert space $\cH^{\otimes \infty}$ \index{tensor product (of Hilbert spaces)!infinite} is completely understood as well, starting from the seminal work \cite{vonNeumann-1937-1939} on the subject, written by von Neumann in 1937 (and published in 1939); however, the space $\cH^{\otimes \infty}$ is hard to control in view of its physical applications, in particular it is not separable and contains a physically unnatural vastness of orthogonal vectors, and moreover the class $\mathcal{B}(\cH^{\otimes \infty})$ is much larger than the algebra of bounded operators generated (algebraically or by limiting process) by bounded operators on each $\cH$.

 The above-mentioned two mathematical frameworks have an important technical difference, which is most relevant for the present discussion: the Hilbert space set-up may accommodate \emph{unbounded} self-adjoint operators acting on $\cH$, whereas the $C^*$-algebraic set-up does not, as to each element $a\in\mathcal{A}$ one associates a finite norm $\|a\|_{\mathcal{A}}$. (To be precise, also in the $C^*$-algebraic framework one sees unbounded self-adjoint operators indirectly emerge, \emph{affiliated} to the von Neumann algebra obtained in a concrete Gelfand-Na\u{\i}mark-Segal representation. Also, in such a representation a possibly unbounded Hamiltonian $H_\omega$ may emerge as the self-adjoint generator of the unitary group $(e^{-\ii t H_\omega})_{t\in\mathbb{R}}$ obtained by representing on $\cH_\omega$ a $*$-automorphism group acting on $\mathcal{A}$. Yet, the primary objects, the self-adjoint elements of $\mathcal{A}$, are obviously all bounded.)

 This has a relevance in that everywhere defined and bounded self-adjoint operators on Hilbert space are simply characterised by \emph{hermiticity} (i.e., symmetry), with no need of further specifications involving domain issues (Sect.~\ref{sec:I-symmetric-selfadj}). Symmetry is a basic feature of quantum observables dictated by the sole necessity of real expectations (ultimately, real numbers measured in experiments). Thus, in order to discuss what features of the physical theory prescribe mathematically the \emph{self-adjointness}, and not the mere \emph{symmetry}, of quantum observables, one must set the analysis within the infinite-dimensional Hilbert space formalism, for example the space $\cH=L^2(\Lambda)$ for a (spinless) quantum particle confined in a non-trivial spatial region $\Lambda\subset\mathbb{R}^d$, $d\in\mathbb{N}$. This will be the explicit or tacitly underlying setting of the rest of this Appendix.

 Incidentally, this is also the 
 precise stage at which quantum observables\index{observables (quantum)} enter the traditional physical introductions on the mathematical framework of quantum mechanics -- such as, among others, those by Dirac \cite{Dirac-PrinciplesQM}, Landau \cite{Landau-Lifshitz-3}, Cohen-Tannoudji et al.~\cite{Cohen-Tannoudji-1977-2020}, Sakurai \cite{sakurai_napolitano_2017}, Weinberg \cite{weinberg_2015}.

 The way observables\index{observables (quantum)} are introduced in this context is rather operational in nature, through \emph{first quantisation}\index{first quantisation} from classical mechanics. First quantisation, rigorously speaking, is a nebulous concept (at a more fundamental level position and momentum operators emerge as generators in the Schr\"{o}dinger representation\index{Schr\"{o}dinger representation} of the Weyl $C^*$-algebra\index{Weyl $C^*$-algebra} \cite[Chapter 3]{Strocchi-MathQM}): quite pragmatically, it provides a physically grounded operational recipe to construct quantum observables from their classical counterparts (up to non-commutative ordering) in the form of linear functional operators acting in the Hilbert space of states for the considered quantum system \cite[Section I.2]{vonNeumann-MathFoundQM}, \cite[\S 15 and \S 17]{Landau-Lifshitz-3}, \cite[Sections 1.3, 1.4, 3.3]{weinberg_2015}, \cite[Section III.5]{Cohen-Tannoudji-1977-2020}, \cite[Section 1.6]{sakurai_napolitano_2017}.

 For the considerations of this Appendix one should then have in mind typical playgrounds such as $\cH=L^2(\mathbb{R}^d)$ or $\cH=L^2(\Lambda)$ for some region $\Lambda\subset\mathbb{R}^d$: then first quantisation associates the quantum position observable with the multiplication by the spatial coordinate $x$ and the quantum momentum observable with the differential operator $-\ii\hslash\nabla_x$. This allows one to pass from a classical Hamiltonian function $H(p,q)$ on the phase space, to the quantum Hamiltonian counterpart $H(-\ii\hslash\nabla_x,x)$ on $L^2(\mathbb{R}^d)$ (up to ordering, due to the non-commutativity of position and momentum, although the above correspondence is unambiguous for classical Hamiltonians of the form $H(p,q)=\frac{1}{2m}p^2+V(q)$). This yields typical actions of quantum observables such as
\begin{equation}\label{eq:op-list}
 \begin{array}{ccl}
  \textrm{multiplication by $x$} & \qquad & \textrm{(position operator)}, \\
  \displaystyle-\ii\nabla_x  & & \textrm{(momentum operator)}, \\
   \displaystyle-\Delta_x & & \textrm{(kinetic energy operator)}, \\
  \displaystyle\big(-\ii\nabla_x-A(x)\big)^2 & & \textrm{(magnetic kinetic energy operator)}, \\
  \displaystyle-\Delta_x+V(x)  & & \textrm{(Schr\"{o}dinger operator)}, \\
  \sqrt{\displaystyle-\Delta_x+1}+V(x) & & \textrm{(semi-relativistic Schr\"{o}dinger operator)},
 \end{array}
\end{equation}
and so forth, where inessential physical constants have been re-scaled out.

 \section{Physical requirement for symmetric observables. Connection with unboundedness.}

 The requirement that quantum observables be associated to symmetric operators on Hilbert space is clearly a consequence of (in fact, equivalent to) the physical requisite that each observables must have \emph{real expectations} on every state. As the expectation of the observable $A$ on the state $\psi$ is the number $\langle \psi, A\psi\rangle$, requiring $\langle \psi, A\psi\rangle\in\mathbb{R}$ $\forall\psi\in\mathcal{D}(A)$ is tantamount as the symmetry of $A$ (Sect.~\ref{sec:I-symmetric-selfadj}).

 Noticeably, this does \emph{not} impose $A$ to be densely defined. At this stage the density of $\mathcal{D}(A)$ in the underlying Hilbert space $\cH$ is rather motivated by including in the theory a `sufficiently large' set of admissible physical states, such as all the various examples of natural $L^2$-dense subspaces where the operators whose action is of type \eqref{eq:op-list} are well defined (typically, regular enough functions supported away from the singular loci of the potentials, and vanishing sufficiently fast at infinity). Of course, the density of $\mathcal{D}(A)$ of a generic operator $A$ on $\cH$ is also essential in order to give unambiguous meaning to the adjoint $A^*$  (Sect.~\ref{sec:I-adjoint}).

 The requirement of real expectations does \emph{not} prescribe quantum observables to be necessarily bounded either -- and symmetry and self-adjointness are only equivalent for bounded operators (Sect.~\ref{sec:I-symmetric-selfadj}). In fact, unboundedness unavoidably manifests for many concrete operator actions, irrespectively of the considered domain of symmetry, as for the operators whose action is of type \eqref{eq:op-list} on $L^2(\mathbb{R}^d)$. Unboundedness of certain physical observables is also imposed by the fulfillment of an uncertainty relation, as was noticed first by Wintner \cite{Wintner-1947-pq-qp} in 1947, and re-proved soon after by Wielandt \cite{Wielandt-1949-pq-qp} in 1949.

 \begin{lemma}[Winter-Wielandt]\index{theorem!Winter-Wielandt}
  Two everywhere defined symmetric operators $P$ and $Q$ on a Hilbert space $\cH$ cannot satisfy the identity $QP-PQ=\ii\mathbbm{1}$ (or, in fact, the identity $QP-PQ= z\mathbbm{1}$ for $z\in\mathbb{C}\setminus\{0\}$).
 \end{lemma}

 \begin{proof}
  Here is the classical algebraic argument by Wielandt \cite{Wielandt-1949-pq-qp}, whereas Wintner \cite{Wintner-1947-pq-qp} followed another simple reasoning  based on the spectra of $PQ$ and $QP$. 
  If $QP-PQ=\ii\mathbbm{1}$, then $Q^2P-PQ^2=Q(PQ+\ii\mathbbm{1})-PQ^2=2\ii Q$, and inductively $Q^nP-PQ^n=\ii n Q^{n-1}$, whence
 \[
  n\|Q^{n-1}\|_{\mathrm{op}}\;\leqslant\; 2 \,\|Q^{n-1}\|_{\mathrm{op}}\|Q\|_{\mathrm{op}}\|P\|_{\mathrm{op}}\qquad \forall n\in\mathbb{N}\,.
 \]
If $\|Q^{n-1}\|_{\mathrm{op}}=0$ for some $n$, then solving the above hierarchy backwards would yield $Q=\mathbb{O}$, which is incompatible with $QP-PQ=\ii\mathbbm{1}$. Then necessarily it is always $\|Q^{n-1}\|_{\mathrm{op}}>0$, whence $\|Q\|_{\mathrm{op}}\|P\|_{\mathrm{op}}\geqslant\frac{1}{2}n$. As $n\in\mathbb{N}$ is arbitrary, at least one among $Q$ and $P$ cannot be bounded, thus contradicting the assumption.
 \end{proof}

 The above emergence of unboundedness can be made quantitative, as observed by Popa \cite{Popa-1982-XD-DX} in 1981.

 \begin{lemma}[Popa]\label{Popas-example}\index{theorem!Popa}
   Assume that on a given Hilbert space $\cH$ two everywhere defined and bounded operators $Q$ and $P$ satisfy $\big\|[Q,P]-\ii\mathbbm{1}\big\|_{\mathrm{op}}\leqslant\varepsilon$  for some $\varepsilon>0$. Then
 \[
  \|Q\|_{\mathrm{op}}\|P\|_{\mathrm{op}}\;\geqslant\;\frac{1}{2}\,\log\frac{1}{\varepsilon}\,.
 \]
 \end{lemma}

 \begin{proof}
   Simplify for $\ii$ in ``$QP-PQ=\ii\mathbbm{1}$'' and re-interpret it as ``$[X,D]=\mathbbm{1}$''. Thus, by assumption, $\big\|[X,D]-\mathbbm{1}\big\|_{\mathrm{op}}\leqslant\varepsilon$.
  By multiplying $D$ by a constant and dividing $X$ by the same constant, one may normalise $\|X\|_{\mathrm{op}}=\frac{1}{2}$. The estimate to prove is $\|D\|_{\mathrm{op}}\geqslant\log\frac{1}{\varepsilon}$.
 Set $E:=[X,D]-\mathbbm{1}$. Then $\|E\|_{\mathrm{op}}\leqslant\varepsilon$ and, by standard induction,
 \[
  [X,D^n]\;=\;nD^{n-1}+D^{n-1}E+D^{n-2}ED+\cdots+ED^{n-1}\qquad\forall n\in\mathbb{N}\,.
 \]
 The triangular inequality then yields
 \[
  \begin{split}
   n\|D^{n-1}\|_{\mathrm{op}}\;&\leqslant\;\|[X,D^n]\|_{\mathrm{op}}+ n\varepsilon\|D\|^{n-1} \\
  &\leqslant\; \|D^n\|_{\mathrm{op}}+ n\varepsilon\|D\|^{n-1}\qquad\quad\forall n\in\mathbb{N}\,.
  \end{split}
 \]
 Dividing both sides by $n!$ and summing over $n$ gives
 \[
  \sum_{n=1}^{\infty}\frac{\;\|D^{n-1}\|_{\mathrm{op}}}{(n-1)!}\;\leqslant\;\sum_{n=1}^{\infty}\frac{\;\|D^{n}\|_{\mathrm{op}}}{n!}+\varepsilon\sum_{n=1}^{\infty}\frac{\;\|D^{n-1}\|_{\mathrm{op}}}{(n-1)!}\,,
 \]
 i.e.,
  \[
  \sum_{n=0}^{\infty}\frac{\;\|D^n\|_{\mathrm{op}}}{n!}\;\leqslant\;\sum_{n=1}^{\infty}\frac{\;\|D^{n}\|_{\mathrm{op}}}{n!}+\varepsilon \sum_{n=0}^{\infty}\frac{\;\|D^n\|_{\mathrm{op}}}{n!}\,.
 \]
 This implies $1\leqslant\varepsilon\,e^{\|D\|_{\mathrm{op}}}$, whence the conclusion $\|D\|_{\mathrm{op}}\geqslant\log\frac{1}{\varepsilon}$. 
 \end{proof}

\begin{remark}
 Popa's bound quantifies the tendency of $Q$ or $P$ to have larger and larger norms in terms of the norm displacement between their commutator and a multiple of the identity. The bound $\|Q\|_{\mathrm{op}}\|P\|_{\mathrm{op}}\geqslant\frac{1}{2}\log\frac{1}{\varepsilon}$ is expected to be optimal: examples of everywhere defined and bounded operators $Q$ and $P$ with $\big\|[Q,P]-\ii\mathbbm{1}\big\|_{\mathrm{op}}\leqslant\varepsilon$ and $\|Q\|_{\mathrm{op}}\|P\|_{\mathrm{op}}=O(\varepsilon^{-2})$ were constructed in \cite{Popa-1982-XD-DX}, and recent examples where $\big\|[Q,P]-\ii\mathbbm{1}\big\|_{\mathrm{op}}\leqslant\varepsilon$ and $\|Q\|_{\mathrm{op}}\|P\|_{\mathrm{op}}=O(\log^5\frac{1}{\varepsilon})$ were constructed in \cite{Tao-2019-commutators-identity}. 
 \end{remark}

  It is also worth stressing that it is the symmetry of observables that makes for them incompatible to be simultaneously unbounded and everywhere defined.

  \begin{theorem}[Hellinger-Toeplitz]\label{thm:hellingertoeplitz}\index{theorem!Hellinger-Toeplitz}
Let $A$ be an everywhere defined linear operator on a Hilbert space $\cH$ with $\langle \psi,A\phi\rangle=\langle A\psi,\phi\rangle$ for all $\psi$ and $\phi$ in $\cH$. Then $A$ is bounded. 
\end{theorem}

The Hellinger-Toeplitz theorem is a standard consequence of the closed graph theorem, which is in turn a consequence of the completeness of $\cH$ via the Baire category theorem (see, e.g., \cite[Section III.5]{rs1}). Completeness is indeed a feature of Hilbert spaces (among other topological structures) that is apparently innocent when its mathematical definition is laid down, but whose actual relevance in the conceptual structure of quantum mechanics is more indirect (and yet crucial). 

 \begin{example}\label{ex:unbdd-everywhere}
 Let $\cH$ be an infinite-dimensional separable Hilbert space, and let $\mathcal{B}$ be an \emph{algebraic basis} of $\cH$, with all elements non-restrictively normalised to 1. Observe that the existence of $\mathcal{B}$ requires the axiom of choice and that $\mathcal{B}$ is necessarily uncountable. Moreover, let $\mathcal{B}'=\{\psi_n|n\in\mathbb{N}\}$ be a countable subset of $\mathcal{B}$ and let $\phi_\circ\in\cH$ with $\|\phi_\circ\|=1$. Define the linear operator $A$ by linear extension of
 \[
 A\psi\;:=\;
 \begin{cases}
  n\phi_\circ & \textrm{if $\psi=\psi_n$ for some $n$} \\
  \; 0 & \textrm{if }\,\psi\in \mathcal{B}\setminus\mathcal{B}'\,.
 \end{cases}
 \]
 By construction $A$ is everywhere defined. It is also unbounded, for $\|A\psi_n\|=n$. On the other hand, $\langle\psi_n,A\psi_n\rangle=n\langle\psi_n,\phi_\circ\rangle$, whereas $\langle A\psi_n,\psi_n\rangle=n\langle\phi_\circ,\psi_n\rangle$. Now it is easy to choose $\phi_\circ$ and $\psi_n$ so as to conclude that $A$ is not symmetric. 
\end{example}

  \section{Self-adjointness of the quantum Hamiltonian inferred from the Schr\"{o}dinger evolution}\label{sec:A1-selfadjH}

The most relevant type of quantum observable for which to discuss the emergence and the role of self-adjointness is surely the Hamiltonian of a given quantum system.

In the typical physical introduction to quantum mechanics the Hamiltonian emerges as a distinguished operator in connection with the time evolution of the system: by \emph{physical reasonings} \cite[Section 27]{Dirac-PrinciplesQM}, \cite[Section 2.1]{sakurai_napolitano_2017}, \cite[Section 3.6]{weinberg_2015}, one argues that a quantum system evolves in time along a trajectory $\psi(t)$ of states of the considered Hilbert space $\cH$ of the form $\psi(t)=U(t)\psi(0)$, such that
\begin{itemize}
 \item the norm (hence the total probability) is preserved along time, in the future and in the past, whatever the initial state;
 \item the evolution from an initial state $\psi(0)$ at time $t=0$ to a final state $\psi(T)$ at time $T$ is the same as two subsequent evolutions, the first one of duration $t$ starting from $\psi(0)$, and the second one of duration $T-t$ starting from $\psi(t)$;
 \item the trajectory of $\psi(t)$ is continuous in time.
\end{itemize}
This is formalised as:
\begin{itemize}
 \item $\|U(t)\psi\|=\|\psi\|$ $\forall\psi\in\cH$ and $\forall t\in\mathbb{R}$, and $\ran\,U(t)=\cH$ $\forall t\in\mathbb{R}$
 (\emph{unitarity});
 \item $U(0)=\mathbbm{1}$ and $U(t_1)U(t_2)=U(t_1+t_2)$ $\forall t_1,t_2\in\mathbb{R}$ (\emph{group composition});
 \item $\|U(t)\psi-\psi\|\xrightarrow{t\to 0}0$ $\forall\psi\in\cH$ (\emph{strong continuity}).
\end{itemize}
 For short, $\{U(t)\,|\,t\in\mathbb{R}\}$ is a strongly continuous one-parameter unitary group\index{strongly continuous unitary group} on $\cH$. Moreover, to such $U(t)$'s one associates an operator $H$ obtained from the $O(t)$-term of a \emph{formal analytic expansion} of $U(t)$ as $t\to 0$ and finally argues that $\psi(t)$ is determined by the celebrated time-dependent Schr\"{o}dinger equation\index{Schr\"{o}dinger equation}
\begin{equation}\label{eq:schreq}
 \ii\frac{\ud}{\ud t}\psi(t)\;=\;H\psi(t)
\end{equation}
that governs the evolution of the quantum system. Besides, first-quantisation-like arguments prescribe the explicit form of $H$ from its classical counterpart in those cases when the system is described by a wave-function $\psi(t;x)$ of space-time coordinates only: for concreteness, when $\cH=L^2(\mathbb{R}^d)$ 
the Hamiltonian of a one-particle quantum system has the form $H=-\Delta_x+V(x)$, a Schr\"{o}dinger operator with real-valued potential $V$. The formal symmetry of such $H$ follows as usual from integration by parts tested on a suitable class of functions that be regular and fast decaying to a degree that depends on the potential $V$.

 The required properties on the quantum evolution necessarily imply (in fact, are equivalent to the fact) that the Hamiltonian $H$ governing the Schr\"{o}dinger equation \eqref{eq:schreq} is \emph{self-adjoint}, and not merely \emph{symmetric}.

\begin{theorem}\label{thm:selfadj-SchrEq}
 Let $\cH$ be a complex Hilbert space and let $H$ be a \emph{symmetric} operator acting on $\cH$ with domain $\mathcal{D}(H)\subset\cH$. The following conditions are equivalent.
 \begin{enumerate}[(i)]
  \item There exists a strongly continuous one-parameter unitary group  $\{U(t)\,|\,t\in\mathbb{R}\}$ acting on $\cH$ such that
 \begin{itemize}
  \item[$\bullet$] for every $t\in\mathbb{R}$ one has $U(t)\mathcal{D}(H)\subset\mathcal{D}(H)$,
  \item[$\bullet$] for every $\psi_0\in\mathcal{D}(H)$ the collection of vectors $\psi(t):=U(t)\psi_0$ defined for every $t\in\mathbb{R}$ constitute a solution to the initial value problem 
  \begin{equation}\label{eq:SchrIVP}
   \begin{cases}
    \ii\displaystyle\frac{\ud}{\ud t}\psi(t)\;=\;H\psi(t)\,, \\
    \psi(0)\;=\;\psi_0
   \end{cases}
  \end{equation}
 \end{itemize}
  \item $\mathcal{D}(H)$ is dense in $\cH$ and for every $\psi_0\in\mathcal{D}(H)$ there exists a unique solution $\psi(\cdot)\in C^1(\mathbb{R},\cH)$, with $\psi(t)\in\mathcal{D}(H)$ $\forall t\in\mathbb{R}$, to the problem \eqref{eq:SchrIVP}.
  \item The operator $H$ is self-adjoint.
 \end{enumerate}
  When either condition above is satisfied,
    \begin{equation}\label{eq:actdomH}
   \begin{split}
    \mathcal{D}(H)\;&=\;\left\{\psi\in\cH\,\left|\,\exists\,\frac{\ud}{\ud t}\Big|_{t=0}U(t)\psi:=\lim_{t\to 0}\frac{U(t)-\mathbbm{1}}{t}\psi\in\cH \right.\right\}, \\
    H\psi\;&=\;\ii\frac{\ud}{\ud t}\Big|_{t=0}U(t)\psi\,,
   \end{split}
  \end{equation}
  and 
  \begin{equation}\label{eq:HU=UH=dU}
   HU(t)\psi_0\;=\;U(t)H\psi_0\;=\;\ii\frac{\ud}{\ud t}U(t)\psi_0\qquad\forall \psi_0\in\mathcal{D}(H)\,,\;\forall t\in\mathbb{R}\,.
  \end{equation}
\end{theorem}

 Theorem \ref{thm:selfadj-SchrEq}, whose proof is postponed for the moment, is formulated so as to show the emergence of the mathematical requirement of self-adjointness from certain physical requisites of the quantum evolution, and in a form that avoids a somewhat frequent confusion with the underlying theorem of Stone (established by Stone \cite{Stone-1930-III,Stone-1932-thm} in 1930, and generalised by von Neumann \cite{vonNeumann-1932-Stone} in 1932 from strongly continuous to weakly measurable unitary groups), which only expresses the opposite implication.

 \begin{theorem}[Stone]\label{thm:Stone}\index{theorem!Stone}
  Let $\{U(t)\,|\,t\in\mathbb{R}\}$ be a strongly continuous one-parameter unitary group on a Hilbert space $\cH$. Then there exists a unique self-adjoint operator $A$ on $\cH$ such that $U(t)=e^{\ii t A}$ for all $t\in\mathbb{R}$.  
 \end{theorem}

  \begin{proof}
  
  \emph{First step:} the subspace $\mathcal{D}\subset\cH$ of all finite linear combinations of vectors $\varphi_f\in\cH$ of the form $\varphi_f=\int_{-\infty}^{+\infty}f(t)U(t)\varphi\,\ud t$ for some $\varphi\in\cH$ and some $f\in C^\infty_0(\mathbb{R})$ (a Riemann integral, since $U(t)$ is strongly continuous) is dense in $\cH$. This is seen by approximating a generic $\varphi\in\cH$ by $\varphi_{j_n}$ along a standard sequence of positive mollifiers, i.e., $j_n(t):=n j(nt)$ $\forall n\in\mathbb{N}$, where $j\in C^\infty_0(\mathbb{R})$, $j\geqslant 0$, $\mathrm{supp}(j)\subset[-1,1]$, and $\int_{-1}^1j(t)\ud t=1$. Indeed, 
  \[
   \begin{split}
    \|\varphi_{j_n}-\varphi\|\;&=\;\left\|\int_{\mathbb{R}}j_n(t)\big(U(t)\varphi-\varphi\big)\,\ud t\right\| \\
    &\leqslant\;\left(\int_{\mathbb{R}}j_n(t)\,\ud t\right)\sup_{t\in[-\frac{1}{n},\frac{1}{n}]}\|U(t)\varphi-\varphi\|\;\xrightarrow{\;n\to\infty\;}\;0\,.
   \end{split}
  \]

  \emph{Second step:} the operator $A\varphi_f:=-\mathrm{i}\,\varphi_{-f'}$, $\mathcal{D}(A):=\mathcal{D}$ is (densely defined and) symmetric. Indeed, 
  \[
   \begin{split}
    \frac{U(s)-\mathbbm{1}}{s}\varphi_f\;&=\;\int_{\mathbb{R}}f(t)\,\frac{U(s+t)-U(t)}{s}\,\varphi\,\ud t\;=\;\int_{\mathbb{R}}\frac{f(\tau-s)-f(\tau)}{s}\,U(\tau)\varphi\,\ud\tau \\
    &\;\xrightarrow{\;s\to 0\;}\;-\int_{\mathbb{R}}f'(\tau)\,U(\tau)\varphi\,\ud\tau\;=\;\varphi_{-f'}
   \end{split}
  \]
  ($\frac{f(\tau-s)-f(\tau)}{s}\xrightarrow{s\to 0} -f'(\tau)$ uniformly, as $f\in C^\infty_0(\mathbb{R})$), whence, for every $\varphi_f,\varphi_g\in\mathcal{D}$,
  \[
   \begin{split}
    \langle A\varphi_f,\varphi_g\rangle\;&=\;\lim_{s\to 0}\,\left\langle -\ii\frac{U(s)-\mathbbm{1}}{s} \varphi_f,\varphi_g\right\rangle\;=\;\lim_{s\to 0}\,\left\langle \varphi_f,\ii\frac{U(-s)-\mathbbm{1}}{s} \varphi_g\right\rangle \\
    &=\;\langle \varphi_f,-\ii\varphi_{-g'}\rangle\;=\;\langle \varphi_f,A\varphi_g\rangle\,.
   \end{split}
  \]

  \emph{Third step:} both $A$ and $U(t)$, $\forall t\in\mathbb{R}$, leave $\mathcal{D}$ invariant and commute on $\mathcal{D}$. Indeed, $U(t)\mathcal{D}\subset\mathcal{D}$ because $U(t)\varphi_f=\int_{\mathbb{R}}f(s)U(t+s)\varphi\,\ud s=\varphi_g$ with $g(s):=f(s-t)$, $A\mathcal{D}\subset\mathcal{D}$ because $A\varphi_f=-\mathrm{i}\,\varphi_{-f'}$ (as $f$ belongs to $C^\infty_0(\mathbb{R})$, so does $f'$), and moreover
  \[
   \begin{split}
    U(t)A\varphi_f\;&=\;-\ii\, U(t)\varphi_{-f'}\;=\;-\ii\int_{\mathbb{R}}f'(s)U(t+s)\,\varphi\,\ud s \;=\;-\ii\int_{\mathbb{R}}f'(\tau-t)U(\tau)\,\varphi\,\ud \tau \\
    &=\;A\int_{\mathbb{R}}f(\tau-t)U(\tau)\,\varphi\,\ud \tau\;=\;A\int_{\mathbb{R}}f(s)U(t+s)\,\varphi\,\ud s\;=\;AU(t)\varphi_f\,.
   \end{split}
  \]

  \emph{Fourth step:} $\ker(A^*\pm\ii\mathbbm{1})=\{0\}$ and therefore (Sect.~\ref{sec:I-symmetric-selfadj}) $A$ is essentially self-adjoint. To see this, let $u\in\mathcal{D}(A^*)$ satisfy $A^*u=\ii u$ (the argument for the eigenvalue $-\ii$ is analogous). Then, for fixed arbitrary $\varphi\in\mathcal{D}(A)$, the function $h(t):=\langle U(t)\varphi,u\rangle$ satisfies
  \[
    h'(t)\;=\;\left\langle\frac{\ud}{\ud t} U(t)\varphi,u\right\rangle\;=\;\langle\ii AU(t)\varphi,u\rangle\;=\;\langle U(t)\varphi,u\rangle\;=\;h(t)
  \]
  (the second identity following from $U'(t)\varphi=\ii AU(t)\varphi$, as was found in the second step), whence $h(t)=h(0)e^t$. Since $|h(t)|\leqslant\|U(t)\|_{\mathrm{op}}\|\varphi\|\|u\|=\mathrm{const.}$ $\forall t\in\mathbb{R}$, necessarily $0=h(0)=\langle\varphi,u\rangle$. Then, $\varphi$ being arbitrary from a dense, $u=0$.

  \emph{Final step:} $U(t)=e^{\mathrm{i}t\overline{A}}$ $\forall t\in\mathbb{R}$. To this aim, define $V(t):=e^{\mathrm{i}t\overline{A}}$ with the functional calculus of the self-adjoint operator $\overline{A}$ (Sect.~\ref{sec:I_funct_calc}), and set $w(t):=U(t)\varphi-V(t)\varphi$ for $\varphi\in\mathcal{D}$. For any $t\in\mathbb{R}$ one has $U(t)\varphi\in\mathcal{D}=\mathcal{D}(A)\subset\mathcal{D}(\overline{A})$ (third step) and $\frac{\ud}{\ud t}U(t)\varphi=\ii AU(t)\varphi=\ii \overline{A}U(t)\varphi$ (second step). Analogously, as is going to be repeated at the beginning of the proof of   
  Theorem \ref{thm:selfadj-SchrEq}, standard properties of the functional calculus and ordinary dominated convergence immediately imply $V(t)\varphi\in\mathcal{D}(\overline{A})$ and $\frac{\ud}{\ud t}V(t)\varphi=\ii\overline{A} V(t)\varphi$. All this shows that the function $w$ is differentiable and 
  \[
   w'(t)\;=\;\ii \overline{A}U(t)\varphi-\ii \overline{A}U(t)\varphi\;=\;\ii \overline{A} w(t)\qquad\forall t\in\mathbb{R}\,.
  \]
  Therefore,
  \[
  \begin{split}
    \frac{\ud}{\ud t}\|w(t)\|^2\;&=\;\langle w'(t),w(t)\rangle+\langle w(t),w'(t)\rangle\;\\
    =\;\langle \ii \overline{A} w(t),w(t)\rangle+\langle w(t),\ii \overline{A} w(t)\rangle\;=\;0\,,
  \end{split}
  \]
  meaning that $w(t)=w(0)=0$ $\forall t\in\mathbb{R}$. Thus, $U(t)\varphi=V(t)\varphi=e^{\mathrm{i}t\overline{A}}$ $\forall\varphi\in\mathcal{D}$, whence the conclusion, on account of the density of $\mathcal{D}$.
  \end{proof}

 \begin{proof}[Proof of Theorem \ref{thm:selfadj-SchrEq}]~
 
  \underline{(iii) $\Rightarrow$ (i)}. Define $U(t):=e^{-\ii t H}$ by means of the functional calculus of self-adjoint operators (Sect.~\ref{sec:I_funct_calc}). Standard properties of the functional calculus and ordinary dominated convergence immediately imply that $\{U(t)\,|\,t\in\mathbb{R}\}$ is a strongly continuous one-parameter unitary group, with also
  \[
   U(t)\mathcal{D}(H)\,\subset\,\mathcal{D}(H)\qquad\textrm{and}\qquad HU(t)\psi_0\,=\,U(t)H\psi_0\quad \forall t\in\mathbb{R}\,,\; \forall \psi_0\in\mathcal{D}(H)
  \]
(see, e.g., \cite[Theorem VIII.7]{rs1} and \cite[Proposition 6.1]{schmu_unbdd_sa}). The auxiliary operator
  \begin{equation*}\tag{*}\label{eq:defA}
   \begin{split}
    \mathcal{D}(A)\;&:=\;\left\{\psi\in\cH\,\left|\,\exists\,\frac{\ud}{\ud t}\Big|_{t=0}U(t)\psi:=\lim_{t\to 0}\frac{U(t)-\mathbbm{1}}{t}\psi\in\cH \right.\right\}, \\
    A\psi\;&:=\;\ii\frac{\ud}{\ud t}\Big|_{t=0}U(t)\psi
   \end{split}
  \end{equation*}
 is symmetric, since 
\[
 \langle\psi,A\psi\rangle\;=\;\lim_{t\to 0}\left\langle\psi,\ii\frac{U(t)-\mathbbm{1}}{t}\psi\right\rangle\;=\;\lim_{t\to 0}\left\langle-\ii\frac{U(-t)-\mathbbm{1}}{t}\psi,\psi\right\rangle\;=\;\langle A\psi,\psi\rangle
\]
for every $\psi\in\mathcal{D}(A)$. For each $\psi\in\mathcal{D}(H)$,
\[
 \Big\|\,\ii\frac{U(t)-\mathbbm{1}}{t}\psi-H\psi\Big\|^2\;=\;\int_{\mathbb{R}}\Big|\,\ii\frac{e^{-\ii t\lambda}-1}{t}-\lambda\Big|^2\ud\langle\psi,E^{(H)}(\lambda)\psi\rangle\;\xrightarrow[]{\:t\to 0\:}\;0\,,
\]
where $\ud E^{(H)}$ is the spectral measure of $H$ (Sect.~\ref{sec:I_spectral_theorem}-\ref{sec:I_funct_calc}), owing to the properties of the functional calculus, dominated convergence, and mean value theorem. Thus, $\psi\in\mathcal{D}(A)$ and
\[
 H\psi\;=\;\ii\frac{\ud}{\ud t}\Big|_{t=0}U(t)\psi\;=\;A\psi\,.
\]
This means that the symmetric operator $A$ is an extension of the self-adjoint operator $H$: by maximality of symmetry of any self-adjoint operator (Sect.~\ref{sec:I-symmetric-selfadj}), necessarily $A=H$. \eqref{eq:actdomH} is therefore established. Moreover, for $t\in\mathbb{R}$ and $\psi_0\in\mathcal{D}(H)$,
\[
	\begin{split}
		 \ii\frac{\ud}{\ud t}U(t)\psi_0\;&=\;\ii\lim_{\tau\to 0}\frac{U(t+\tau)-U(t)}{\tau}\psi_0\;\\
		 &=\;\ii \,U(t)\lim_{\tau\to 0}\frac{U(\tau)-\mathbbm{1}}{\tau}\psi_0\;=\;U(t)H\psi_0\,.
	\end{split}
\]
This, together with the already proved identity $HU(t)\psi_0=U(t)H\psi_0$, establishes \eqref{eq:HU=UH=dU}. In turn, \eqref{eq:HU=UH=dU} yields finally \eqref{eq:SchrIVP}.

 \underline{(iii) $\Rightarrow$ (ii)}. The density of $\mathcal{D}(H)$ is part of the assumption of self-adjointness of $H$.  Moreover, \eqref{eq:SchrIVP} clearly implies that $\psi(\cdot)\in C^1(\mathbb{R},\cH)$ and $\psi(t)\in\mathcal{D}(H)$ for every $t\in\mathbb{R}$). Concerning the uniqueness, should there exist two such solutions $\psi_1(\cdot)$ and $\psi_2(\cdot)$, then $\phi(t):=\psi_1(t)-\psi_2(t)$ would satisfy $\phi(0)=0$, as well as
\[\tag{**}\label{eq:argumentuniqueness}
\begin{split}
  \frac{\ud}{\ud t}\|\phi(t)\|^2\;&=\;\Big\langle \frac{\ud}{\ud t}\phi(t),\phi(t)\Big\rangle+\Big\langle\phi(t),\frac{\ud}{\ud t}\phi(t)\Big\rangle \\
  &=\;\langle-\ii H\phi(t),\phi(t)\rangle+\langle\phi(t),-\ii H\phi(t)\rangle\;=\;0\,.
\end{split}
\]
Thus, $\|\phi(t)\|=\|\phi(0)\|=0$, meaning $\psi_1(t)=\psi_2(t)$ for every $t\in\mathbb{R}$.

  \underline{(ii) $\Rightarrow$ (i)}. The existence and uniqueness assumptions allow to unambiguously define the maps $U(t)$ satisfying $\psi(t)= U(t)\psi_0$ for any $t\in\mathbb{R}$ and $\psi_0\in\mathcal{D}(H)$. All such maps are linear, since the equation is linear. By construction, $U(0)=\mathbbm{1}|_{\mathcal{D}(H)}$ and $U(t)\mathcal{D}(H)\subset\mathcal{D}(H)$ $\forall t\in\mathbb{R}$. For fixed $\psi_0\in\mathcal{D}(H)$ and arbitrary $t,s\in\mathbb{R}$ set 
  \[
   \psi_t\;:=\; U(t)\psi_0\,,\qquad \psi_{t+s}\;:=\; U(t+s)\psi_0\,,\qquad \widetilde{\psi}_{t,s}\;:=\;U(s)\psi_t\;=\;U(s)U(t)\psi_0\,.
  \]
  For any fixed $t$, both $s\mapsto \psi_{t+s}\in\mathcal{D}(H)$ and $s\mapsto \widetilde{\psi}_{t,s}\in\mathcal{D}(H)$ solve in $C^1(\mathbb{R},\cH)$ the initial value problem 
  \begin{equation*}
   \begin{cases}
    \ii\displaystyle\frac{\ud}{\ud t}\xi(t)\;=\;H\xi(t)\,, \\
    \xi(0)\;=\;\psi_t\,\in\,\mathcal{D}(H)\,,
   \end{cases}
  \end{equation*}
  whence (by the uniqueness assumption) $ \psi_{t+s}=\widetilde{\psi}_{t,s}$ $\forall t,s\in\mathbb{R}$, i.e., $U(s)U(t)=U(s+t)$ $\forall t,s\in\mathbb{R}$. Now, based on the sole \emph{symmetry} of $H$ and the fact that by construction $\ii\frac{\ud}{\ud t}U(t)\psi_0=HU(t)\psi_0$, the same argument \eqref{eq:argumentuniqueness} can be repeated, showing that $\|U(t)\psi_0\|=\|\psi_0\|$ $\forall t\in\mathbb{R}$, $\forall \psi_0\in\mathcal{D}(H)$. Therefore, as $\mathcal{D}(H)$ is dense in $\cH$, each $U(t)$ lifts to a unitary operator on $\cH$, denoted with the same symbol, and $\{U(t)\,|\,t\in\mathbb{R}\}$ is a one-parameter unitary group on $\cH$. Strong continuity follows from the assumption $\psi(\cdot)\in C^1(\mathbb{R},\cH)$.
  
 \underline{(i) $\Rightarrow$ (iii)}. Owing to Stone's theorem (Theorem \ref{thm:Stone}), $U(t)=e^{-\ii t A}$ $\forall t\in\mathbb{R}$ for some uniquely determined self-adjoint $A$. By the very same arguments developed for the implication (iii) $\Rightarrow$ (i) (replacing now $H$ with $A$), the domain and the action of $A$ are given by \eqref{eq:defA} (which is not a definition now); moreover,
\[
 \ii\frac{\ud}{\ud t}U(t)\psi_0\;=\;A U(t)\psi_0\qquad\forall \psi_0\in\mathcal{D}(A)\,,\;\forall t\in\mathbb{R}\,.
\]
This and \eqref{eq:SchrIVP} then imply that each $\psi_0$ from $\mathcal{D}(A)$ also belongs to $\mathcal{D}(H)$, with $A\psi_0=H\psi_0$. Therefore, the symmetric operator $H$ extends the self-adjoint operator $A$, which by maximality implies $H=A$. $H$ is thus necessarily self-adjoint.
 \end{proof}

  \emph{Strong continuity} in Theorem \ref{thm:selfadj-SchrEq}, namely the strong continuity of a one-parameter unitary group $\{U(t)\,|\,t\in\mathbb{R}\}$, is equivalent to \emph{weak continuity}: indeed, by unitarity,
  \begin{equation}
   \| U(t)\varphi-\varphi\|^2\;=\;2\big( \|\varphi\|^2-\mathfrak{Re}\langle\varphi ,U(t)\varphi\rangle\big)\qquad \forall t\in\mathbb{R}\,,\;\forall \varphi\in\cH\,.
  \end{equation}
  When $\cH$ is \emph{separable}, the strong continuity of the unitary group is also equivalent to its \emph{weak measurability}: this remarkable fact (see, e.g., \cite[Theorem VIII.9]{rs1}) was proved, as mentioned, by von Neumann \cite{vonNeumann-1932-Stone}.

  \section{Self-adjointness of the quantum Hamiltonian inferred from the series expansion of the evolution propagator}\markboth{Physical requirements prescribing self-adjointness of quantum observables}{Self-adjointness inferred from the propagator's series expansion}\label{sec:A1-selfadj-expitHexp}

  A related type of physical prescription that implies the self-adjointness, and not the mere symmetry, of the quantum Hamiltonian $H$ (and again is in fact equivalent to the self-adjointness) is the possibility of expressing the Schr\"{o}dinger evolution of an initial state $\psi_0$, which in the notation of Theorem \ref{thm:selfadj-SchrEq} is the trajectory described by $\psi(t)=e^{-\ii t H}\psi_0$, as an actual power series
  \begin{equation}\label{eq:A1-exp-expansion}
   e^{-\ii t H}\psi_0\;=\;\sum_{n=0}^\infty\frac{(-\ii)^n t^n}{n!} H^n\psi_0
  \end{equation}
  for a sufficiently large selection of initial states.

  Clearly, \eqref{eq:A1-exp-expansion} is particularly ``tempting'' in various reasonings at the physical heuristic level, including for explicit numerical approximations, and yet it is false for generic $\psi_0\in\cH$ and a generic operator $H$ acting on $\cH$, even when for its interpretation of quantum observable $H$ is assumed to be densely defined and symmetric. In fact, even before inquiring the convergence of the series on the r.h.s.~of \eqref{eq:A1-exp-expansion}, $H$ may be such that \emph{no} non-zero $\psi_0$ belongs simultaneously to the the domain of all powers of $H$, thus making the sum itself meaningless. For example, the symmetric operator $Q$ of multiplication by $x$ on $L^2(\mathbb{R})$ with dense domain (of essential self-adjointness) given by the \emph{really simple functions}\index{really simple functions} on $\mathbb{R}$ (i.e., the step-wise constant functions given by finite linear combinations of characteristic functions of intervals of finite length \cite[Sections 1.17-1.18]{Lieb-Loss-Analysis}), is such that $\mathcal{D}(Q)\cap\mathcal{D}(Q^2)=\{0\}$, because obviously the multiplication by $x$ destroys the step-function structures: for such $Q$ the expressions $Q^n\psi_0$ with $n\geqslant 2$ only make sense on $\psi_0=0$. Of course, \eqref{eq:A1-exp-expansion} is correct for any $\psi_0\in\cH$ and $t\in\mathbb{R}$ when $H$ is everywhere defined and bounded.

  To circumvent such limitations, it is natural to restrict \eqref{eq:A1-exp-expansion} to a meaningful selection of $\psi_0$'s. This justifies the definition of \emph{analytic vectors}:\index{analytic vectors} given an operator $A$ on Hilbert space $\cH$, an element $\psi\in\cH$ is called an analytic vector for $A$ when  
\begin{equation}\label{eq:A1-analytic}
\begin{split}
 &\psi\in\mathcal{D}(A^n)\quad\forall n\in\mathbb{N}\qquad\textrm{and}\qquad  \\
 &\sum_{n=0}^\infty\frac{\|A^n\psi\|}{n!}t^n\;<\;+\infty\quad\textrm{for some }t>0\,.
\end{split}
\end{equation}
The requisite \eqref{eq:A1-analytic} is clearly designed to be a sufficient condition for the convergence of the series \eqref{eq:A1-exp-expansion}. In view of \eqref{eq:A1-exp-expansion}, requiring that the a priori symmetric and densely defined operator $H$ admits a dense of analytic vectors in $\cH$ turns out to be equivalent to the self-adjointness of $H$, under the additional technical condition that $H$ be closed. This is the content of a deep theorem established by Nelson \cite{Nelson-analytic-1959} in 1959 (the proof of which can be found, in \cite[Theorem X.39]{rs2} or \cite[Theorem 7.16]{schmu_unbdd_sa}).

\begin{theorem}[Nelson's analytic vector theorem]\label{thm:nelson}\index{theorem!Nelson (analytic vector)}
 Let $A$ be a \emph{symmetric} and \emph{closed} operator on a Hilbert space $\cH$. Then these two conditions are equivalent:
 \begin{enumerate}[(i)]
  \item $\mathcal{D}(A)$ contains a dense set of analytic vectors for $A$;
  \item $A$ is self-adjoint.
 \end{enumerate}
\end{theorem}

 The closedness condition in Nelson's theorem is crucial: for example, the operator $Q$ on $L^2(\mathbb{R})$ acting as multiplication by $x$ on the domain $C^\infty_c(\mathbb{R})$ is clearly densely defined and symmetric, and moreover its whole domain consists of analytic vectors because
 \[
 	\begin{split}
  \sum_{n=0}^\infty\frac{\;\|x^n f\|_{L^2(\mathbb{R})}}{n!}\,\leqslant&\,\|f\|_{L^\infty(\mathbb{R})}\,|\mathrm{supp}\,f|^{\frac{1}{2}}\sum_{n=0}^\infty\frac{1}{n!}\Big(\sup_{x\in\mathrm{supp}\,f}|x|\Big)^n\,<\,+\infty \\& \forall f\in C^\infty_c(\mathbb{R})\,,
  \end{split}
 \]
 yet $Q$ is not closed, and indeed $Q$ is only essentially self-adjoint, not self-adjoint: its operator closure $\overline{Q}$ is the multiplication by $x$ on the domain $\{ f\in L^2(\mathbb{R})|xf\in L^2(\mathbb{R})\}$.

  \section{Non-uniqueness of the Schr\"{o}dinger dynamics when self-adjointness is not declared}\label{sec:non-uniquedynamics}

  As a complement to Sections \ref{sec:A1-selfadjH}-\ref{sec:A1-selfadj-expitHexp}, it is worth highlighting that in the lack of an explicit declaration of a self-adjoint Hamiltonian $H$ (in practice, with reference to typical first-quantisation operators of the form \eqref{eq:op-list}, in the lack of an explicit declaration of a self-adjoint domain for $H$ on Hilbert space), the ubiquitous physical claim that the Schr\"{o}dinger equation governed by $H$ ``encodes'' all the information to determine \emph{the} forward-in-time trajectory $\{\psi(t)\,|\,t\geqslant 0\}$ starting from a given initial state of the system turns out to be profoundly misleading, since the Schr\"{o}dinger equation then typically displays the quite unphysical phenomenon of non-unique solutions.

  This can be argued at various levels. To begin with, like in concrete settings where the Schr\"{o}dinger equation is an actual time-dependent partial differential equation on $\mathbb{R}^d$ (this is precisely the case with the differential operators from \eqref{eq:op-list}), such an equation \emph{alone} cannot characterise the quantum dynamics because of the typical presence of multiple non-Hilbertian solutions.

  \begin{example}
    With respect to the Hilbert space $\cH=L^2(\mathbb{R})$, the Schr\"{o}dinger equation
\begin{equation*}
 \ii\frac{\partial}{\partial t}\psi(t,x)\,=\,-\frac{\partial^2}{\partial x^2}\psi(t,x)+V(x)\psi(t,x)
\end{equation*}
 with analytic potential $V:\mathbb{R}\to\mathbb{R}$ can be re-interpreted as $P(-\ii\frac{\partial}{\partial t},-\ii\frac{\partial}{\partial x},x)\psi=0$ in terms of the second order differential operator $P(\xi_1,\xi_2,x):=\xi_1+\xi_2^2+V(x)$. One then sees that in the standard sense of linear partial differential operator theory the initial value problem for $P(-\ii\frac{\partial}{\partial t},-\ii\frac{\partial}{\partial x},x)\psi=0$ with initial condition $\psi(0,x)=\psi_0(x)$ for some ``nice'' $\psi_0\in\mathcal{S}(\mathbb{R})\subset L^2(\mathbb{R})$ is a \emph{characteristic initial value problem} in the sense that the initial plane $t=0$ is \emph{characteristic} for the equation (see, e.g., \cite[Chapter III]{Hormander-LinearPDOp-1976}). As such, the initial value problem has an \emph{infinite number} of solutions. A mechanism for an infinity of non-zero solutions with $V\equiv 0$ and with zero initial value at $t=0$, is discussed in \cite[Section 8.9]{Hormander-LinearPDOp-1976}: however, \emph{none} of such (non-zero) solutions belongs to $L^2(\mathbb{R})$ for any $t>0$. 
In other words: not imposing the solutions to be square-integrable allows for an unphysical multiplicity of solutions for the Schr\"{o}dinger dynamics.
  \end{example}

 As a further level, one argues that even when only Hilbertian solutions are to be admitted, the lack of declaration of the self-adjointness domain of the differential operator $H$ in the Schr\"{o}dinger equation, which is the domain where the trajectory of the Schr\"{o}dinger dynamics is searched for, produces once again multiple distinct solutions stemming from the same initial state.

 \begin{example}
  With respect to the Hilbert space $\cH=L^2(0,1)$, the self-adjoint operators 
  \begin{align*}
H_D&=-\frac{\ud^2}{\ud x^2}           &  \mathcal{D}(H_D) &=\big\{ \psi\in H^2(0,1)\,\big|\, \psi(0)=0=\psi(1) \big\}\,,     \\
H_P&=-\frac{\ud^2}{\ud x^2}           &  \mathcal{D}(H_P) &=\left\{\psi\in H^2(0,1)\,\left|\,
\begin{array}{c}
 \psi(0)=\psi(1) \\
 \psi'(0)=\psi'(1)
\end{array}\! \right.\right\}
\end{align*}
 provide two distinct self-adjoint realisations, respectively with Dirichlet and with periodic boundary conditions, of the same differential operator $-\frac{\ud^2}{\ud x^2}$, the `free kinetic energy quantum Hamiltonian'. They have clearly different domains, as well as different spectra:
 \begin{itemize}
 \item the eigenvalues of $H_D$ are the numbers $E^{(D)}_n=\pi^2 n^2$, $n\in\mathbb{N}$, each of which is non-degenerate and with normalised eigenfunction $\psi^{(D)}_n(x)=\sqrt{2}\sin\pi n x$;
 \item the eigenvalues of $H_P$ are the numbers $E^{(P)}_n=4\pi^2 n^2$, $n\in\mathbb{N}_0$, each of which is double-degenerate apart from the non-degenerate ground state $n=0$, and with normalised eigenfunctions $\psi^{(P)}_n(x)=\sqrt{2}\sin 2\pi n x$, $\psi^{(P)}_{-n}(x)=\sqrt{2}\cos 2\pi n x$\,.
\end{itemize}
 In either case determining the ``free Schr\"{o}dinger dynamics'' originating from the initial state $\psi_0\in C^\infty_0(0,1)\subset\mathcal{D}(H_D)\cap\mathcal{D}(H_P)$ amounts to solving the initial value differential problem
 \begin{equation}\label{eq:SchrIVP-01}
  \begin{cases}
  \ii\frac{\partial}{\partial t}\psi(t,x)\,=\,-\frac{\partial^2}{\partial x^2}\psi(t,x)\,, \\
  \quad \,\psi(0,x)\,=\,\psi_0(x)\qquad\qquad t\in\mathbb{R}\,,\; x\in(0,1)
 \end{cases}
\end{equation}
imposing $\psi(t,\cdot)\in L^2(\mathbb{R})$ at any time $t$. Yet, \eqref{eq:SchrIVP-01} alone is not enough to characterise $\psi(t,x)$ at later times, because of the non-uniqueness of the $L^2$-solutions: representing the same $\psi_0$ in the eigenstate basis for $H_D$ and $H_P$, that is, respectively,
\begin{align*}
 \psi_0\,&=\,\sum_{n\in\mathbb{N}} \alpha_n\psi^{(D)}_n\,, & \alpha_n&:=\langle\psi^{(D)}_n,\psi_0\rangle_{L^2}\,, \\
  \psi_0\,&=\,\sum_{n\in\mathbb{N}_0} \big(\beta_n\psi^{(P)}_n+\beta_{-n}\psi^{(P)}_{-n}\big)\,,  &\beta_{\pm n}&:=\langle\psi^{(D)}_{\pm n},\psi_0\rangle_{L^2}\,,
\end{align*}
one finds two distinct solutions to \eqref{eq:SchrIVP-01}, the Dirichlet one, $\psi_D(t,\cdot)\in\mathcal{D}(H_D)$, and the periodic one, $\psi_P(t,\cdot)\in\mathcal{D}(H_P)$, explicitly given by 
\[
 \begin{split}
  \psi_D(t,x)\,&=\,\sqrt{2}\sum_{n\in\mathbb{N}} e^{-\ii t \pi^2 n^2}\alpha_n\sin\pi n x\,, \\
  \psi_P(t,x)\,&=\,\sqrt{2}\sum_{n\in\mathbb{N}_0} e^{-4\ii t \pi^2 n^2} \big(\beta_n\sin 2\pi n x+\beta_{-n}\cos 2\pi n x\big)
 \end{split}
\]
 (each of which is \emph{unique} in the respective domain, according to Theorem \ref{thm:selfadj-SchrEq}). Only the declaration of the precise domain of self-adjointness for the differential operator $H=-\frac{\ud^2}{\ud x^2}$ resolve such an unphysical ambiguity, in which case the unique solution finally becomes $e^{-\ii t H}\psi_0$.
 \end{example}

\section{Self-adjointness of quantum observables with an orthonormal basis of eigenvectors}\label{sec:OBN-estates}~

 One further typical physical requirement that is central in the mathematical structure of quantum mechanics and eventually implies the self-adjointness, and not the mere symmetry, of certain quantum observables, is the following -- one can refer directly to the subtle reasoning presented by Dirac in \cite[Section 10]{Dirac-PrinciplesQM}, or more recently, by Weinberg \cite[Section 3.3]{weinberg_2015}.

When discussing the measurement mechanism\index{measurement mechanism (quantum)} of a quantum observable $A$, one comes to infer that the measurement performed with the system in a particular state makes it jump onto an eigenstate of $A$, the result of the measurement being the corresponding eigenvalue; moreover, the original state must be dependent on such eigenstates, in the sense of being expressed by a linear combination of them. Merging this with the operational assumption that the measurement can be performed on \emph{any} state, one should conclude, as written for instance by Dirac in \cite[Section 10]{Dirac-PrinciplesQM}, that the eigenstates of $A$ must form a complete set in the Hilbert space, and this provides an additional constraint on those operators that represent a quantum observable.

This is a paradigmatic line of reasoning in physical introductions to quantum mechanics. In more precise terms, when the underlying Hilbert space $\cH$ is infinite-dimensional and the observable $A$ is (symmetric and) possibly unbounded, the physical request is formulated as follows. If on \emph{any} state $\psi$ of $\mathcal{D}(A)$ it is possible to perform a measurement of $A$ in the sense of a Stern-Gerlach-like experiment,\index{Stern-Gerlach experiment} namely a filter with output given by one of the eigenstates of $A$, and hence if the generic initial $\psi$ must admit an expansion in eigenstates, then owing to the density of the domain $\mathcal{D}(A)$ the eigenstates of $A$ must constitute an orthonormal basis of $\cH$.

Clearly, this applies to quantum observables for which the measurement is feasible in the `filter' sense of a Stern-Gerlach\index{Stern-Gerlach experiment} apparatus, like the usual harmonic oscillator, or the free kinetic operators $H_D$ and $H_P$ with Dirichlet or periodic boundary conditions considered in Subsection \ref{sec:non-uniquedynamics} (instead, observables like the free kinetic energy $-\Delta_x$ on $L^2(\mathbb{R}^d)$ self-adjointly defined on the domain $H^2(\mathbb{R}^d)$, or the momentum $-\ii\nabla_x$ self-adjointly defined on $H^1(\mathbb{R}^d)$, are not covered: they have no eigenstates at all).

For such observables, the above physical reasoning produces a requirement that does characterise them as self-adjoint, and not merely symmetric, operators.

\begin{theorem}\label{thm:onbEV-sa}
 Let $A$ be a \emph{closed symmetric} operator acting on the Hilbert space $\cH$. Assume that $\mathcal{D}(A)$ contains an orthonormal basis of $\cH$ whose elements are eigenvectors of $A$. Then $A$ is self-adjoint.
\end{theorem}

\begin{proof}

 Let $(\psi_n)_{n\in\mathbb{N}}$ be the considered orthonormal basis of eigenvectors of $A$ and let $(\lambda_n)_{n\in\mathbb{N}}$ be the collection of the corresponding eigenvalues, all counted with multiplicity. As $A$ is symmetric, the $\lambda_n$'s are all real. Besides, since $(\psi_n)_{n\in\mathbb{N}}$ is an orthonormal basis for $\cH$ and is contained in $\mathcal{D}(A)$, then $\mathcal{D}(A)$ is dense in $\cH$.

 The subspace $\mathrm{ran}(A+\ii\mathbbm{1})$ is dense in $\cH$, since it contains all the vectors
 \[
  \psi_n\;=\;(\lambda_n+\ii)^{-1}(A+\ii\mathbbm{1})\psi_n\;\in\;\mathrm{ran}(A+\ii\mathbbm{1})\,.
 \]
 In addition, $\mathrm{ran}(A+\ii\mathbbm{1})$ is also \emph{closed} in $\cH$. To check the latter claim, let $(\eta_m)_{m\in\mathbb{N}}$ be a sequence in $\mathrm{ran}(A+\ii\mathbbm{1})$ that converges to some $\eta\in\cH$: arguing as follows, one sees that $\eta\in \mathrm{ran}(A+\ii\mathbbm{1})$. Indeed, writing $\eta_m=(A+\ii\mathbbm{1})\xi_m$ for some $\xi_m\in\mathcal{D}(A)$, the symmetry of $A$ implies
 \[
  \|\eta_m-\eta_{m'}\|^2\;=\;\|(A+\ii\mathbbm{1})(\xi_m-\xi_{m'})\|^2\;=\;\|A(\xi_m-\xi_{m'})\|^2+\|\xi_m-\xi_{m'}\|^2\,.
 \]
 Thus,
 \[
  \|\xi_m-\xi_{m'}\|\;\leqslant\;\|\eta_m-\eta_{m'}\|\,,
 \]
 which shows that $(\xi_m)_{m\in\mathbb{N}}$ is a Cauchy sequence in $\cH$ and hence, by completeness of $\cH$, converges to some $\xi\in\cH$. Thus, $\xi_m\to\xi$ and $(A+\ii\mathbbm{1})\xi_m\to\eta$ as $m\to\infty$: as $A$ is a closed operator, and so too is therefore $A+\ii\mathbbm{1}$ (Sect.~\ref{sec:I-bdd-closable-closed}), then necessarily $\xi\in\mathcal{D}(A)$ and $\eta=(A+\ii\mathbbm{1})\xi\in\mathrm{ran}(A+\ii\mathbbm{1})$.

 The range of $(A+\ii\mathbbm{1})$ being simultaneously a dense and closed subspace of $\cH$, one concludes that $\mathrm{ran}(A+\ii\mathbbm{1})=\cH$. Analogously, $\mathrm{ran}(A-\ii\mathbbm{1})=\cH$. This allows one to apply a basic criterion of self-adjointness (Sect.~\ref{sec:I-symmetric-selfadj}) and conclude that $A$ is self-adjoint.
%
%
%
%
%
%
\end{proof}

\section{Self-adjointness of quantum Hamiltonians inferred from their physical stability}\label{sec:closedsemibdd-sa}~

 The \emph{stability}\index{stability (of a quantum Hamiltonian)} of a considered quantum system is another fundamental physical requirement that translates mathematically into the prescription that certain quantum Hamiltonians be self-adjoint, and not merely symmetric.

 The stability requirement is usually made at one of two possible levels of the physical modelling, in order to select quantum systems that cannot attain arbitrary negative energies.

 The first is the operator level, when one prescribes the candidate quantum Hamiltonian, a densely defined symmetric operator $H$ on a Hilbert space $\cH$, to have on normalised states expectations that stay \emph{uniformly above} a given threshold; that is, the system cannot attain arbitrarily negative energies. Mathematically, this reads $\langle \psi,H\psi\rangle\geqslant\mathfrak{m}\|\psi\|^2$ $\forall\psi\in\mathcal{D}(H)$, for some $\mathfrak{m}\in\mathbb{R}$, and corresponds therefore to the lower semi-boundedness of $H$ (Sect.~\ref{sec:I-symmetric-selfadj}). When such physical requirement is made, a canonical construction solely based on the density of the domain, the symmetry, and the lower semi-boundedness of $H$, associates to $H$ a \emph{self-adjoint} operator $H_{\mathrm{F}}$ that extends $H$, has spectrum starting at the same greatest lower bound of $H$ and lies all above it, and, among all other possible extensions of $H$, if any, is the only one whose quadratic form is precisely the closure of the form $\psi\mapsto\langle \psi,H\psi\rangle$, all other extensions having larger form domain and lower expectations. This is the outcome of the Friedrichs construction described in Theorem \ref{thm:Friedrichs-ext} and $H_{\mathrm{F}}$ is the Friedrichs extensions of $H$. Thus, even if the model is initially set up by means of a symmetric-only densely defined candidate Hamiltonian, the stability requirement uniquely realises the latter as a self-adjoint (and not merely symmetric) operator, the Hamiltonian  of the system, through a minimal addition of further admissible states, and with the same energy expectations and the same finite bottom for the energy. The self-adjointness is inescapable.

 In other circumstances the modelling is more naturally framed in terms of \emph{expectations} of a (candidate) Hamiltonian that is only implicitly or formally identified. A pedagogical example is the quantum model for the Hydrogen atom formalised in terms of its energy expectations
 \[
  \psi\;\mapsto\;\int_{\mathbb{R}^3}|\nabla\psi(x)|^2\,\ud x-\int_{\mathbb{R}^3}\frac{\;|\psi(x)|^2}{\;|x|}\,.
 \]
 In practice physical heuristics yield a \emph{symmetric} energy form $\psi\mapsto\mathcal{E}[\psi]$, initially defined on a minimal, physically reasonable dense domain of the considered Hilbert space, to which one wants to meaningfully associate a Hamiltonian with those very energy expectations. Here too, basic physical requirements turn out to uniquely identify the Hamiltonian as a self-adjoint, and not merely symmetric operator.

 The case of interest is when on the (densely defined, symmetric) form $\mathcal{E}[\cdot]$ the following two prescriptions of physical relevance are made. First, with the same motivation as above, the form $\mathcal{E}[\cdot]$ is required to be stable, i.e., lower semi-bounded. Thus, $\mathcal{E}[\psi]\geqslant\mathfrak{m}\|\psi\|^2$ $\forall\psi\in\mathcal{D}[\mathcal{E}]$ for some $\mathfrak{m}\in\mathbb{R}$. In addition, a suitable type of continuity is also required on $\psi\mapsto\mathcal{E}[\psi]$, in one of these two formulations:
 \begin{enumerate}[(a)]
  \item one requires
  \begin{equation*}
  \mathcal{E}\Big[\lim_{n\to\infty}\psi_n\Big]\,\leqslant\,\liminf_{n\to\infty}\mathcal{E}[\psi_n]
 \end{equation*}
  along any convergent sequence $(\psi_n)_{n\in\mathbb{N}}$ in $\cH$ (having tacitly extended $\mathcal{E}[\cdot]$ on the whole $\cH$ by setting $\mathcal{E}[\psi]=+\infty$ for all those vectors that do not belong to $\mathcal{D}[\mathcal{E}]$, in order for the above inequality to be consistent);
  \item one requires that when $\mathcal{E}[\cdot]$ is evaluated along a sequence $(\psi_n)_{n\in\mathbb{N}}$ of very close states in $\mathcal{D}[\mathcal{E}]$ up to a final state $\psi\in\cH$ with $\|\psi_n-\psi\|\to 0$ as $n\to\infty$, and the considered sequence of states also satisfies $\mathcal{E}[\psi_n-\psi_m]\to 0$ as $n,m\to\infty$, \emph{then} on the limit state $\psi$ too it is possible to evaluate the expectation $\mathcal{E}[\psi]$, i.e., $\psi\in\mathcal{D}[\mathcal{E}]$, and indeed $\mathcal{E}[\psi_n-\psi]\to 0$ as $n\to\infty$.
 \end{enumerate}
 Physically, both (a) and (b) are rather self-explanatory, if one thinks, for instance, of a given measurement apparatus providing the evaluations $\mathcal{E}[\psi]$ on the states of a system that is brought to occupy very close configurations along the considered sequence $(\psi_n)_{n\in\mathbb{N}}$, say, by suitably tiny shifts in space. 
 Mathematically (Sect.~\ref{sec:I_forms}), for a lower semi-bounded quadratic form $\mathcal{E}[\cdot]$ conditions (a) and (b) above are actually equivalent (in particular, (a) means that $\mathcal{E}[\cdot]$ is lower semi-continuous), and both amount to saying that $\mathcal{E}[\cdot]$ is a \emph{closed form}.

 The self-adjointness of the Hamiltonian is inescapable in this framework as a consequence of an amount of basic properties reviewed in Section \ref{sec:I_forms} and cast for convenience into the following statement.

 \begin{theorem}\label{thm:closed-sa}
  Let $\cH$ be a complex Hilbert space and let $\mathcal{E}[\cdot]$ be a quadratic form on $\cH$ such that the domain $\mathcal{D}[\mathcal{E}]$ is a dense subspace of $\cH$ and that the form is lower semi-bounded and closed. Then the operator $H$ associated to $\mathcal{E}$ through the definition
  \begin{equation}\label{eq:fromFormToOp}
    \begin{split}
  \mathcal{D}(H)\;&:=\;\left\{\psi\in\mathcal{D}[\mathcal{E}]\,\left|
  \begin{array}{c}
   \exists\,\xi_\psi\in\cH\,\textrm{ such that} \\
   \mathcal{E}[\varphi,\psi]=\langle\varphi,\xi_\psi\rangle\;\forall\varphi\in\mathcal{D}[\mathcal{E}]
  \end{array}
   \right.\right\}\,, \\
  H\psi\;&:=\;\xi_\psi
 \end{split}
  \end{equation}
 is self-adjoint, and $\mathcal{E}[\psi]=\langle\psi,H\psi\rangle$ for every $\psi\in\mathcal{D}(H)$. 
 \end{theorem}

 Thus, requiring that the expectations of an observable behave as a densely defined, lower semi-bounded, closed quadratic form, forces the underlying linear operator \eqref{eq:fromFormToOp} associated with the form to be self-adjoint, and not merely symmetric.
%
%
%


\chapter{References to pedagogical examples}
\label{appendix_pedagogical_examples} 


 Whereas in the present monograph the applications, presented in Chapters \ref{chapter-Hydrogenoid}-\ref{chapter-bosonic_trimer}, of the classical self-adjoint extension schemes of Chapter \ref{chaper-extension-schemes} are all part of active ongoing research, it maybe beneficial for the non-specialist reader to also be re-directed to standard references in the literature where basic, yet instructive pedagogical examples are discussed in depth. This appendix has such a purpose, and it also includes references to settings that are not so straightforward, yet by now well established and elucidative. Besides, the deep value of such pedagogical examples is not at all to be underestimated, in view of the insight and inspiration they provide for more complicated and realistic models.

\section{Quantum particle on real line}

Self-adjoint realisations in the von Neumann scheme:
\begin{itemize}
  \item \cite[Example 188.2]{Smirnov-HigherMaths-5-orig1959} (concerning the particle quantum momentum);
  \item \cite[Examples 4.2.5 and 4.7.3]{Jiri-Exner-Havlicek-2008};
   \item \cite[Section 3.2.3]{Amrein-HilberSpMethods-2009} (particle momentum);
 \item \cite[Section 6.1.1]{GTV-2012} (particle momentum), \cite[Section 6.2.1]{GTV-2012} (concerning the free particle Hamiltonian).
\end{itemize}

\section{Quantum particle on half-line}

Self-adjoint realisations in the von Neumann scheme:
\begin{itemize}
 \item \cite[Section X.1, Example 2]{rs2};
 \item \cite[\S 13, Example]{faris-1975};
 \item \cite[Section 8.2, Example 1]{Weidmann-LinearOperatosHilbert} (concerning the particle quantum momentum);
  \item \cite[Examples 4.2.5, 4.7.3, 4.8.5, and 4.9.6]{Jiri-Exner-Havlicek-2008};
     \item \cite[Section 3.2.3]{Amrein-HilberSpMethods-2009} (particle momentum);
 \item \cite[Section 6.1.2]{GTV-2012} (particle momentum), \cite[Section 6.2.2]{GTV-2012} (free particle Hamiltonian).
\end{itemize}

\noindent Self-adjoint realisations in the Kre{\u\i}n-Vi\v{s}ik-Birman scheme:
\begin{itemize}
 \item \cite[Exercise 13.6]{schmu_unbdd_sa}, \cite[Example 14.9]{schmu_unbdd_sa} (with boundary triplet techniques that are reminiscent of the KVB approach), \cite[Example 14.15]{schmu_unbdd_sa};
 \item \cite[Section 7.1]{GMO-KVB2017}.
\end{itemize}

\section{Quantum particle on interval}

 Self-adjoint realisations in the von Neumann scheme:
\begin{itemize}
  \item \cite[Example 188.1]{Smirnov-HigherMaths-5-orig1959} (concerning the particle quantum momentum);
 \item \cite[Section X.1, Example 1]{rs2} (particle momentum);
  \item \cite[Section 8.2, Example 2 and statement (8.7)]{Weidmann-LinearOperatosHilbert} (particle momentum);
  \item \cite[Example 4.12]{Weidmann-book1987} (particle momentum), \cite[Example 4.13]{Weidmann-book1987} (free particle Hamiltonian); 
  \item \cite[Examples 4.2.5, 4.7.3, 4.7.11, and 4.9.1]{Jiri-Exner-Havlicek-2008};
     \item \cite[Section 3.2.3]{Amrein-HilberSpMethods-2009} (particle momentum);
      \item \cite[Examples 13.1 and 13.4]{schmu_unbdd_sa};
 \item \cite[Section 6.1.3]{GTV-2012} (particle's momentum), \cite[Section 6.2.3]{GTV-2012} (free particle Hamiltonian);
 \item \cite[Section 2.6, Example]{Teschl-MMQM-2014} (particle momentum).
\end{itemize}

\noindent Self-adjoint realisations in the Kre{\u\i}n-Vi\v{s}ik-Birman scheme:
\begin{itemize}
  \item \cite[Examples 2.2, 2.3, 2.4, 3.3, and 5.1]{Alonso-Simon-1980};
 \item \cite[Example 14.10]{schmu_unbdd_sa} (with boundary triplet techniques that are reminiscent of the KVB approach) \cite[Example 14.14]{schmu_unbdd_sa}, \cite[Exercises 14.10-14.13]{schmu_unbdd_sa};
 \item \cite[Section 10.1]{Ashbaugh-Gesztesy-Mitrea-Teschl-2009};
 \item \cite[Section 7.2]{GMO-KVB2017};
 \item \cite[Sections 3.1 and 3.4]{M-2020-Friedrichs}.
\end{itemize}

\section{Quantum particle with point interaction of $\delta$-type in one dimension}

Self-adjoint realisations in the von Neumann scheme:
\begin{itemize}
  \item \cite[Section X.2 Example 3]{rs2};
  \item \cite[Chapters I.3, II.2, III.2]{albeverio-solvable};
  \item \cite[Section 1.2.1]{albverio-kurasov-2000-sing_pert_diff_ops}
  \item \cite[Section 7.3.3]{GTV-2012}.
\end{itemize}

\section{Quantum particle with point interaction of $\delta'$-type in one dimension}

Self-adjoint realisations in the von Neumann scheme:
\begin{itemize}
  \item \cite[Chapters I.4, II.3, III.3]{albeverio-solvable}.
\end{itemize}

\section{Quantum particle with point interaction in two dimensions}

Self-adjoint realisations in the von Neumann scheme:
\begin{itemize}
 \item \cite[Chapters I.5, II.4, III.4]{albeverio-solvable}.
\end{itemize}

\section{Quantum particle with point interaction in three dimensions}

Self-adjoint realisations in the von Neumann scheme:
\begin{itemize}
 \item \cite[Chapters I.1, II.1, III.1]{albeverio-solvable};
 \item \cite[Section 1.5.1]{albverio-kurasov-2000-sing_pert_diff_ops}.
\end{itemize}

\noindent Self-adjoint realisations in the Kre{\u\i}n-Vi\v{s}ik-Birman scheme:
\begin{itemize}
 \item \cite[Section 3]{MO-2016};
 \item \cite[Section 3.2]{M-2020-Friedrichs}.
\end{itemize}

\section{Quantum particle with point interaction on domain}

Self-adjoint realisations in the von Neumann scheme:
\begin{itemize}
 \item \cite[Section 2]{Exner-Mantile-2007}.
\end{itemize}

\noindent Self-adjoint realisations in the Kre{\u\i}n-Vi\v{s}ik-Birman scheme:
\begin{itemize}
 \item \cite[Example 14.16]{schmu_unbdd_sa};
 \item \cite[Section III]{LM-FaberKrahn-2020}.
\end{itemize}

\section{Friedrichs lower bound}\label{sec:appendixFriedrichsTOP}

Examples of densely defined, lower semi-bounded, non-essentially self-adjoint operators admitting an infinite multiplicity of self-adjoint extensions with the same Friedrichs lower bound:
\begin{itemize}
 \item \cite[Sections 3.1 and 3.2]{M-2020-Friedrichs} (quantum particle on an interval);
 \item \cite[Sections 3.1 and 3.2]{M-2020-Friedrichs} (Hydrogenoid-type Hamiltonians).
\end{itemize}

\noindent Examples of densely defined, lower semi-bounded, non-essentially self-adjoint operators whose top extensions only consist of the Friedrichs extension:
\begin{itemize}
 \item \cite[Sections 3.4]{M-2020-Friedrichs} (quantum particle on an interval).
\end{itemize}

\backmatter


\def\cprime{$'$}

\printindex


\end{document}